\documentclass[12pt,twoside,a4paper]{book}
\pdfoutput=1
\usepackage{graphicx}
\usepackage{txfonts}
\usepackage{fancyhdr}
\usepackage[figuresright]{rotating}
\usepackage{lscape}
\usepackage{longtable}
\usepackage{rotating}
\usepackage{multirow}
\usepackage{calc}
\usepackage{float}                      % force floats where we want them
\usepackage{dcolumn}                    % align columns on decimal point
\usepackage{upgreek}
\newcolumntype{d}[1]{D{.}{.}{#1}}       % create decimal column type d
\usepackage{natbib}                     % for author-year citations
\usepackage{pdfpages}
\bibpunct{(}{)}{;}{a}{}{,}              % to follow the A&A style
\renewcommand{\bibname}{References}

\usepackage{hyperref}

\newcommand\getsto{\mathrel{\mathchoice {\vcenter{\offinterlineskip
\halign{\hfil
$\reset@font\displaystyle##$\hfil\cr\gets\cr\to\cr}}}
{\vcenter{\offinterlineskip\halign{\hfil$\reset@font\textstyle##$\hfil\cr\gets
\cr\to\cr}}}
{\vcenter{\offinterlineskip\halign{\hfil$\reset@font\scriptstyle##$\hfil\cr\gets
\cr\to\cr}}}
{\vcenter{\offinterlineskip\halign{\hfil$\reset@font\scriptscriptstyle##$\hfil\cr
\gets\cr\to\cr}}}}}

\newcommand\cor{\mathrel{\mathchoice {\hbox{$\widehat=$}}{\hbox{$\widehat=$}}
{\hbox{$\reset@font\scriptstyle\hat=$}}
{\hbox{$\reset@font\scriptscriptstyle\hat=$}}}}

\newcommand\lid{\mathrel{\mathchoice {\vcenter{\offinterlineskip\halign{\hfil
$\reset@font\displaystyle##$\hfil\cr<\cr\noalign{\vskip1.2pt}=\cr}}}
{\vcenter{\offinterlineskip\halign{\hfil$\reset@font\textstyle##$\hfil\cr<\cr
\noalign{\vskip1.2pt}=\cr}}}
{\vcenter{\offinterlineskip\halign{\hfil$\reset@font\scriptstyle##$\hfil\cr<\cr
\noalign{\vskip1pt}=\cr}}}
{\vcenter{\offinterlineskip\halign{\hfil$\reset@font\scriptscriptstyle##$\hfil\cr
<\cr
\noalign{\vskip0.9pt}=\cr}}}}}

\newcommand\gid{\mathrel{\mathchoice {\vcenter{\offinterlineskip\halign{\hfil
$\reset@font\displaystyle##$\hfil\cr>\cr\noalign{\vskip1.2pt}=\cr}}}
{\vcenter{\offinterlineskip\halign{\hfil$\reset@font\textstyle##$\hfil\cr>\cr
\noalign{\vskip1.2pt}=\cr}}}
{\vcenter{\offinterlineskip\halign{\hfil$\reset@font\scriptstyle##$\hfil\cr>\cr
\noalign{\vskip1pt}=\cr}}}
{\vcenter{\offinterlineskip\halign{\hfil$\reset@font\scriptscriptstyle##$\hfil\cr
>\cr
\noalign{\vskip0.9pt}=\cr}}}}}

\newcommand\sol{\mathrel{\mathchoice {\vcenter{\offinterlineskip\halign{\hfil
$\reset@font\displaystyle##$\hfil\cr\sim\cr<\cr}}}
{\vcenter{\offinterlineskip\halign{\hfil$\reset@font\textstyle##$\hfil\cr\sim\cr
<\cr}}}
{\vcenter{\offinterlineskip\halign{\hfil$\reset@font\scriptstyle##$\hfil\cr\sim\cr
<\cr}}}
{\vcenter{\offinterlineskip\halign{\hfil$\reset@font\scriptscriptstyle##$\hfil\cr
\sim\cr<\cr}}}}}

\newcommand\sog{\mathrel{\mathchoice {\vcenter{\offinterlineskip\halign{\hfil
$\reset@font\displaystyle##$\hfil\cr\sim\cr>\cr}}}
{\vcenter{\offinterlineskip\halign{\hfil$\reset@font\textstyle##$\hfil\cr\sim\cr
>\cr}}}
{\vcenter{\offinterlineskip\halign{\hfil$\reset@font\scriptstyle##$\hfil\cr
\sim\cr>\cr}}}
{\vcenter{\offinterlineskip\halign{\hfil$\reset@font\scriptscriptstyle##$\hfil\cr
\sim\cr>\cr}}}}}

\newcommand\lse{\mathrel{\mathchoice {\vcenter{\offinterlineskip\halign{\hfil
$\reset@font\displaystyle##$\hfil\cr<\cr\simeq\cr}}}
{\vcenter{\offinterlineskip\halign{\hfil$\reset@font\textstyle##$\hfil\cr
<\cr\simeq\cr}}}
{\vcenter{\offinterlineskip\halign{\hfil$\reset@font\scriptstyle##$\hfil\cr
<\cr\simeq\cr}}}
{\vcenter{\offinterlineskip\halign{\hfil$\reset@font\scriptscriptstyle##$\hfil\cr
<\cr\simeq\cr}}}}}

\newcommand\gse{\mathrel{\mathchoice {\vcenter{\offinterlineskip\halign{\hfil
$\reset@font\displaystyle##$\hfil\cr>\cr\simeq\cr}}}
{\vcenter{\offinterlineskip\halign{\hfil$\reset@font\textstyle##$\hfil\cr
>\cr\simeq\cr}}}
{\vcenter{\offinterlineskip\halign{\hfil$\reset@font\scriptstyle##$\hfil\cr
>\cr\simeq\cr}}}
{\vcenter{\offinterlineskip\halign{\hfil$\reset@font\scriptscriptstyle##$\hfil\cr
>\cr\simeq\cr}}}}}

\newcommand\grole{\mathrel{\mathchoice {\vcenter{\offinterlineskip\halign{\hfil
$\reset@font\displaystyle##$\hfil\cr>\cr\noalign{\vskip-1.5pt}<\cr}}}
{\vcenter{\offinterlineskip\halign{\hfil$\reset@font\textstyle##$\hfil\cr
>\cr\noalign{\vskip-1.5pt}<\cr}}}
{\vcenter{\offinterlineskip\halign{\hfil$\reset@font\scriptstyle##$\hfil\cr
>\cr\noalign{\vskip-1pt}<\cr}}}
{\vcenter{\offinterlineskip\halign{\hfil$\reset@font\scriptscriptstyle##$\hfil\cr
>\cr\noalign{\vskip-0.5pt}<\cr}}}}}

\newcommand\leogr{\mathrel{\mathchoice {\vcenter{\offinterlineskip\halign{\hfil
$\reset@font\displaystyle##$\hfil\cr<\cr\noalign{\vskip-1.5pt}>\cr}}}
{\vcenter{\offinterlineskip\halign{\hfil$\reset@font\textstyle##$\hfil\cr
<\cr\noalign{\vskip-1.5pt}>\cr}}}
{\vcenter{\offinterlineskip\halign{\hfil$\reset@font\scriptstyle##$\hfil\cr
<\cr\noalign{\vskip-1pt}>\cr}}}
{\vcenter{\offinterlineskip\halign{\hfil$\reset@font\scriptscriptstyle##$\hfil\cr
<\cr\noalign{\vskip-0.5pt}>\cr}}}}}

\newcommand\loa{\mathrel{\mathchoice {\vcenter{\offinterlineskip\halign{\hfil
$\reset@font\displaystyle##$\hfil\cr<\cr\approx\cr}}}
{\vcenter{\offinterlineskip\halign{\hfil$\reset@font\textstyle##$\hfil\cr
<\cr\approx\cr}}}
{\vcenter{\offinterlineskip\halign{\hfil$\reset@font\scriptstyle##$\hfil\cr
<\cr\approx\cr}}}
{\vcenter{\offinterlineskip\halign{\hfil$\reset@font\scriptscriptstyle##$\hfil\cr
<\cr\approx\cr}}}}}

\newcommand\goa{\mathrel{\mathchoice {\vcenter{\offinterlineskip\halign{\hfil
$\reset@font\displaystyle##$\hfil\cr>\cr\approx\cr}}}
{\vcenter{\offinterlineskip\halign{\hfil$\reset@font\textstyle##$\hfil\cr
>\cr\approx\cr}}}
{\vcenter{\offinterlineskip\halign{\hfil$\reset@font\scriptstyle##$\hfil\cr
>\cr\approx\cr}}}
{\vcenter{\offinterlineskip\halign{\hfil$\reset@font\scriptscriptstyle##$\hfil\cr
>\cr\approx\cr}}}}}

\newcommand\diameter{{\ifmmode\mathchoice
{\ooalign{\hfil\hbox{$\reset@font\displaystyle/$}\hfil\crcr
{\hbox{$\reset@font\displaystyle\mathchar"20D$}}}}
{\ooalign{\hfil\hbox{$\reset@font\textstyle/$}\hfil\crcr
{\hbox{$\reset@font\textstyle\mathchar"20D$}}}}
{\ooalign{\hfil\hbox{$\reset@font\scriptstyle/$}\hfil\crcr
{\hbox{$\reset@font\scriptstyle\mathchar"20D$}}}}
{\ooalign{\hfil\hbox{$\reset@font\scriptscriptstyle/$}\hfil\crcr
{\hbox{$\reset@font\scriptscriptstyle\mathchar"20D$}}}}
\else{\ooalign{\hfil/\hfil\crcr\mathhexbox20D}}%
\fi}}

\newcommand\sq{\ifmmode\squareforqed\else{\unskip\nobreak\hfil
\penalty50\hskip1em\null\nobreak\hfil\squareforqed
\parfillskip=0pt\finalhyphendemerits=0\endgraf}\fi}
\newcommand\squareforqed{\hbox{\rlap{$\sqcap$}$\sqcup$}}

\newcommand{\romn}[1] {{\mathrm #1}}

\newcommand\fp{\hbox{$.\!\!^{\reset@font\reset@font\scriptscriptstyle\romn p}$}}

\fancypagestyle{plain}{%
\fancyhf{}
\fancyfoot[LE,RO]{\thepage} 

}

\renewcommand{\baselinestretch}{1.5}		% this sets to one and a half line spacing

\makeatletter
\def\@endpart{\vfil\newpage}
\makeatother

\let\mypdfximage\pdfximage
\protected\def\pdfximage{\immediate\mypdfximage}

\newenvironment{tab}[1]{
  \begin{table}[#1]
  \renewcommand{\baselinestretch}{1.1}
  \centering\small}{
  \end{table}}

\newenvironment{fig}{
  \begin{figure}
  \renewcommand{\baselinestretch}{1.1}
  \centering\small}{
  \end{figure}}

\newenvironment{tabland}{
  \begin{sidewaystable}
  \renewcommand{\baselinestretch}{1.1}
  \small}{
  \end{sidewaystable}}

\newenvironment{abs}[2]{
  \newpage
  \thispagestyle{plain}
  \markboth{}{}
  \addcontentsline{toc}{chapter}{#1}
  \begin{center}
 {\huge\bf #2}
  \end{center}
  \vspace{10mm}}{
  \clearpage}

\newenvironment{refs}[1]{
  \newpage
  \thispagestyle{plain}
  \markboth{}{}
  \addcontentsline{toc}{chapter}{#1}
  \begin{center}
 {\huge\bf #1}
  \end{center}
  \vspace{10mm}
  \renewcommand{\baselinestretch}{1.1}}{
  }

\newcommand{\degree}{\ensuremath{^\circ}} 
\newcommand{\lowres}{_lowres}
\begin{document}

%	Title page

\vspace*{10mm}

\begin{center}

% thesis title goes here!

{\huge\bf
Simulations and Observations of\\
the Microwave Universe
}

\vspace*{35mm}

A thesis submitted to The University of Manchester\\
 for the degree of Doctor of Philosophy \\
in the Faculty of Engineering and Physical Sciences 

\vspace*{35mm}

\bf\large

2009

\vspace*{35mm}

Michael Peel

Jodrell Bank Centre for Astrophysics

School of Physics and Astronomy 

\end{center}
\thispagestyle{empty}

\newpage

%_______________________________________________________________________

\tableofcontents

Word count: $\sim$45,000

\cleardoublepage

%_______________________________________________________________________

\addcontentsline{toc}{chapter}{List of Figures}
\listoffigures
\newpage

\addcontentsline{toc}{chapter}{List of Tables}
\listoftables
\cleardoublepage

%_______________________________________________________________________

%	Abstract

\begin{abs}{Abstract}{Abstract}

\noindent Simulations and observations of the microwave sky are of great importance for understanding the Universe that we reside in. Specifically, knowledge of the Cosmic Microwave Background (CMB) and its foregrounds -- including the Sunyaev-Zel'dovich (SZ) effect from clusters of galaxies and radio point sources -- tell us about the Universe on its very largest scales, and also what the Universe is made of.

We describe the creation of software to carry out large numbers of virtual sky simulations. The simulations include the CMB, SZ effect and point sources, and are designed to examine the effects of point sources and the SZ effect on present and recent observations of the CMB. Utilizing sets of 1,000 simulations, we find that the power spectrum resulting from the SZ effect is expected to have a larger standard deviation by a factor of 3 than would be expected from purely Gaussian realizations. It also has a distribution that is significantly skewed towards increased values for the power spectrum, especially when small map sizes are used. The effects of the clustering of galaxy clusters, residual point sources and uncertainties in the gas physics are also investigated, as are the implications for the excess power measured in the CMB power spectrum by the CBI and BIMA experiments. We also investigate the possibility of using the One Centimetre Receiver Array (OCRA) receivers to observe the CMB and measure this high-multipole excess.

An automated data reduction package has been created for the OCRA receivers, which has been used in end-to-end simulations for OCRA-p observations of point sources. We find that these simulations are able to realistically simulate the noise present in real observations, and that the introduction of $1/f$ noise into the simulations significantly reduces the predicted ability of the instruments to observe weak sources by measuring the sources for long periods of time.

The OCRA-p receiver has been used to observe point sources in the Very Small Array fields so that they can be subtracted from observations of the CMB power spectrum. We find that these point sources are split between steep and flat spectrum sources. We have also observed 550 CRATES flat spectrum radio sources, which will be useful for comparison to {\it Planck} satellite observations.

Finally, the assembly and commissioning of the OCRA-F receiver is outlined. This receiver is now installed on the Toru\'n 32-m telescope, and is currently being calibrated prior to starting observations in the next few months.

This thesis was submitted by Michael Peel to The University of Manchester for the degree of Doctor of Philosophy in the Faculty of Engineering and Physical Sciences on the 18$^\mathrm{th}$ December 2009.

\end{abs}

%_______________________________________________________________________

%	Declaration

\begin{abs}{Declaration}{Declaration}
\begin{center}

I declare that no portion of the work referred to in the thesis has been submitted 
in support of an application for another degree or qualification of this or any other 
university or other institute of learning.

\end{center}
\end{abs}

%_______________________________________________________________________

%	Copyright Statement

\begin{abs}{Copyright Statement}{Copyright Statement}

\renewcommand{\labelenumi}{(\roman{enumi})}

\begin{enumerate}
\item
The author of this thesis (including any appendices and/or schedules to this 
thesis) owns any copyright in it (the ``Copyright'') and they have given The 
University of Manchester the right to use such Copyright for any 
administrative, promotional, educational and/or teaching purposes. 

\item
Copies of this thesis, either in full or in extracts, may be made only in 
accordance with the regulations of the John Rylands University Library of 
Manchester.  Details of these regulations may be obtained from the Librarian.  
This page must form part of any such copies made.   

\item
The ownership of any patents, designs, trade marks and any and all other 
intellectual property rights except for the Copyright (the ``Intellectual Property 
Rights'') and any reproductions of copyright works, for example graphs and 
tables (``Reproductions''), which may be described in this thesis, may not be 
owned by the author and may be owned by third parties.  Such Intellectual 
Property Rights and Reproductions cannot and must not be made available for 
use without the prior written permission of the owner(s) of the relevant 
Intellectual Property Rights and/or Reproductions.

\item
Further information on the conditions under which disclosure, publication and 
exploitation of this thesis, the Copyright and any Intellectual Property Rights 
and/or Reproductions described in it may take place is available from the Head 
of the School of Physics and Astronomy (or the Vice-President) 
\end{enumerate}

\end{abs}
%_______________________________________________________________________

%	Tell your examiner a little bit about yourself!

\begin{abs}{The Author}{The Author}

\noindent
The Author graduated from the University of Manchester in 2006 with an M.Phys in Physics and Astronomy. Since then he has been carrying out research at the Jodrell Bank Centre for Astrophysics. The results of this research are described in this thesis.
\end{abs}

%_______________________________________________________________________

%	Acknowledgements

\begin{abs}{Acknowledgments}{Acknowledgments}

\noindent
No-one stands alone, and that is especially true here. First, I would like to thank my supervisors, Professor Ian Browne and Dr Richard Battye, for all of their support, advice and guidance over the last 3 years.

I would also like to thank the members of the OCRA collaboration at the Toru\'n Radio Astronomy Observatory, the University of Bristol and the University of Manchester. This collaboration includes Professor Mark Birkinshaw, Professor Ian Browne (again), Professor Richard Davis, Roman Feiler, Dr Marcin Gawro\'nski, Professor Andrzej Kus, Dr Katy Lancaster, Dr Stuart Lowe, Bogna Pazderska, Eugeniusz Pazderski and Professor Peter Wilkinson.

The assembly of OCRA-F depended vitally on the technical staff at Jodrell Bank Observatory, including Colin Baines, Jason Marshall, Don Lawson, Eddie Blackhurst and John Edgley, and those who designed the instrument, including Dr Dan Kettle, Neil Roddis and John Kitching. My thanks also go to Professor Phil Diamond for supporting the final construction work on OCRA-F, thus enabling it to finally reach the telescope.

My thanks also to Dr Scott Kay, Dr Clive Dickinson and Dr Bob Watson, with whom I have collaborated on some of the work described in this thesis, and also Dr Anthony Holloway and Dr Bob Dickson for their support with computer equipment and for tirelessly setting up and dismantling the AccessGrid video conference equipment every few weeks.

Of course, I would like to thank all of the staff, postdocs and students at Jodrell Bank for making this a great place to work (funding threats notwithstanding...) over the last 3 years. There are too many names to mention, so, to those that I have lived with, gone to the pub with, gone climbing with, talked over lunch or tea with, and all the rest: thank you.
% For those of you who are even reading the source code of my thesis (why??): thank you too. :-)

% To name some names:
% To those I started this with: Dandan Xu, Dr Christobal Espinoza, Steven Pediani, Mike Preece, Mark Purver, Lyshia Quinn, Jenny Williams, Val\'erio Ribeiro, Satoru Sakai
% To those that came before: Dr Stuart Lowe, Dr Tony Rushton, Dr Magda Todorovic, Dr Rob Beswick, Dr R\'ois\'in N\'i Chuim\'in, Dr David Tideswell, Danny Wong-McSweeney
% To those that came after: Jen Gupta, Sarah Bryan, Chris Tibbs, Liz Guzman-Ramirez, Nadya Kunawicz, Kerry Hebden, Marta Alves
% To those that were already here: Dr Tim O'Brien, Linda Bennett, 
% And to those that came from, or went to, other paths: Dr Cormac Purcell, Dr Tess Jaffe, Dr Nick Rattenbury, Dr Janine Van Eymeren, Dr Paul Woods, Dr Edward Boyce, Dr Eric Lagadec, Dr Jimmy Green, Dr Alan Duffy
% And to those that I have missed from these lists
% Thank you all.

Finally, my thanks go to my parents, my sister and my relatives for supporting me through the last three years.

The author acknowledges the support of an STFC studentship. The programs written as part of this work include functions and libraries provided by Healpix, GNU Scientific Library, Libnova and S.R. Lowe. This research has made use of the NASA/IPAC Extragalactic Database (NED) which is operated by the Jet Propulsion Laboratory, California Institute of Technology, under contract with the National Aeronautics and Space Administration. This thesis was typeset with \LaTeX.

\end{abs}

%_______________________________________________________________________

%	Put your supporting publications list here

\begin{refs}{Supporting Publications}

\begin{trivlist}

\item
{\bf Statistics of the Sunyaev-Zel'dovich Effect power spectrum} \\
Michael W. Peel, Richard A. Battye and Scott T. Kay, 2009, MNRAS, 397, 2189 (Chapters 2-3)

\item
{\bf 30~GHz observations of sources in the VSA fields} \\
M.\,P.~Gawro\'nski, M.W.~Peel, K.~Lancaster et al., 2009, MNRAS (sub.), arXiv:0909.1189 (Chapter 5)

\end{trivlist}

\end{refs}

\clearpage

%_______________________________________________________________________

%	Abbreviations

\begin{refs}{List of Abbreviations}

The following abbreviations are used in this document:

\begin{itemize}
\item ACBAR -- Arcminute Cosmology Bolometer Array Receiver
\item ACT -- Atacama Cosmology Telescope
\item AMI -- Arcminute Microkelvin Imager
\item APRICOT -- All Purpose Radio Imaging Cameras On Telescopes
\item BEM -- Back End Module
\item BIMA -- Berkeley-Illinois-Maryland Association (interferometer)
\item BMR -- Browne's Millihertz Rotator
\item CBI -- Cosmic Background Imager
\item CMB -- Cosmic Microwave Background
\item CRATES -- Combined Radio All-sky Targeted Eight GHz Survey
\item DAQ -- Data AcQuisition system
\item Dec -- Declination
\item FARADAY - Focal-plane Arrays for Radio Astronomy, Design Access and Yield
\item FEM -- Front End Module
\item FWHM -- Full Width at Half Maximum
\item ICM -- IntraCluster Medium
\item JBO -- Jodrell Bank Observatory
\item LFI -- Low Frequency Instrument (of the {\it Planck} satellite)
\item LNA -- Low Noise Amplifier
\item MMIC - Monolithic Microwave Integrated Circuit
\item OCRA -- One Centimetre Receiver Array\\
OCRA-p -- OCRA-prototype\\
OCRA-F -- OCRA-FARADAY
\item RA -- Right Ascension
\item RF -- Radio Frequency
\item RT -- Ryle Telescope
\item SPT -- South Pole Telescope
\item SZ -- Sunyaev-Zel'dovich (as in the SZ effect)
\item SZA -- Sunyaev-Zel'dovich Array
\item TCfA -- Toru\'n Centre for Astronomy
\item VNA -- Vector Network Analyser
\item VSA -- Very Small Array
\item WMAP -- Wilkinson Microwave Anisotropy Probe
\end{itemize}

\end{refs}

\newpage
\vspace{15mm}
\begin{center}
{\it At Jodrell Bank, someone decided it was time for a nice relaxing cup of tea.}\\
 -- Douglas Adams
\end{center}
\cleardoublepage

\chapter{Introduction}\label{intro}

Since prehistory, humanity has looked up at the stars and used their eyes to record the patterns of stars and the movement of the planets. Over four hundred years ago, the first telescopes were turned to the skies, giving us a closer look at the known astronomical objects and the ability to search for fainter, undiscovered objects. A vital component of all of these observations for many thousands of years was the human eye -- however that can only see light at visible wavelengths. Photography replaced the human eye for astronomical observations in the 19th century, but again was focused on the small range of frequencies surrounding the optical regime.

When Karl Jansky and Reber first used radio frequency technology for astronomy in 1930-40, humanity gained a new view of the Universe. This new window enables us to look back to the very earliest stages of our Universe, and see the photons that make up the Cosmic Microwave Background (CMB). By observing this carefully (measuring its temperature to an accuracy of one part in a million) we can deduce the fundamental properties of our observable universe, telling us how much mass our Universe contains and of what type, and also what scales it is distributed on.

Matter on the very largest scales -- that of clusters of galaxies -- casts shadows on the CMB via the Sunyaev-Zel'dovich effect (hereafter the SZ effect). Very careful measurements of these shadows provide information about the behaviour of matter on these very large scales. By surveying them, we can weigh the Universe, and study its growth over time. We can also learn about the individual galaxy clusters at the same time, learning about the distribution of gas within the clusters. The formation of structures within our universe is summarized in Section \ref{sec:intro_structureformation}. Clusters of galaxies, their emission at optical, X-ray and radio frequencies, and the cosmology that can be done by observing them, are then reviewed in Section \ref{sec:intro_clusters}. Sources of confusion for observations of galaxy clusters at radio frequencies are then summarized in Section \ref{sec:intro_confusion}.

Whilst optical observations now use charged-coupled devices (CCDs) that have many millions of pixels, radio astronomy traditionally uses single pixels, or techniques such as interferometry, to painstakingly build up maps of the sky over a long period of time. This is especially problematic at the shortest wavelengths -- those around a centimetre or millimetre in size -- where the primary beams of large telescopes (and hence the map pixels for observations from a single telescope) cover very small areas of the sky. The advent of radio cameras -- arrays of radio frequency pixels -- promises to make large scale surveys possible much more quickly than with single pixels. The One Centimetre Receiver Array (OCRA) currently consists of a pair of prototype instruments aimed towards the development of a $\sim$100 pixel radio camera at wavelengths of 30~GHz (1~cm). These instruments are summarized in Section \ref{sec:ocra}. Finally, an overview of this Thesis is given in Section \ref{sec:intro_structure}.

\section{Structure formation} \label{sec:intro_structureformation}
\begin{fig}
\centering
% http://www.mpa-garching.mpg.de/galform/virgo/millennium/
\includegraphics[scale=0.19]{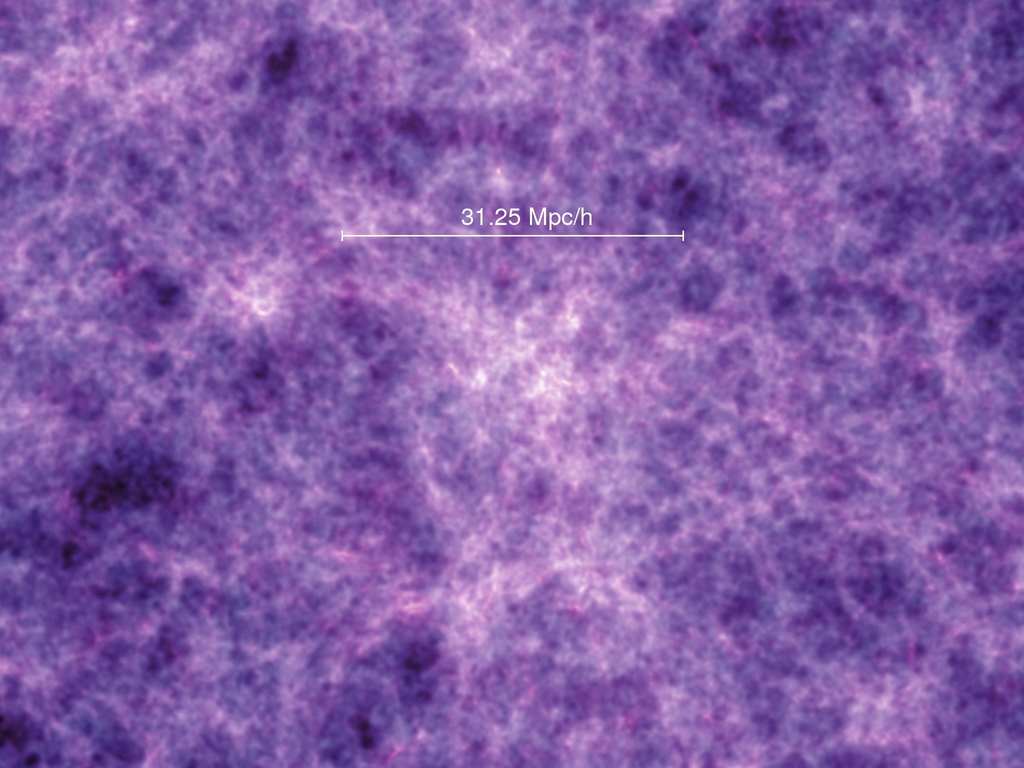}
\includegraphics[scale=0.19]{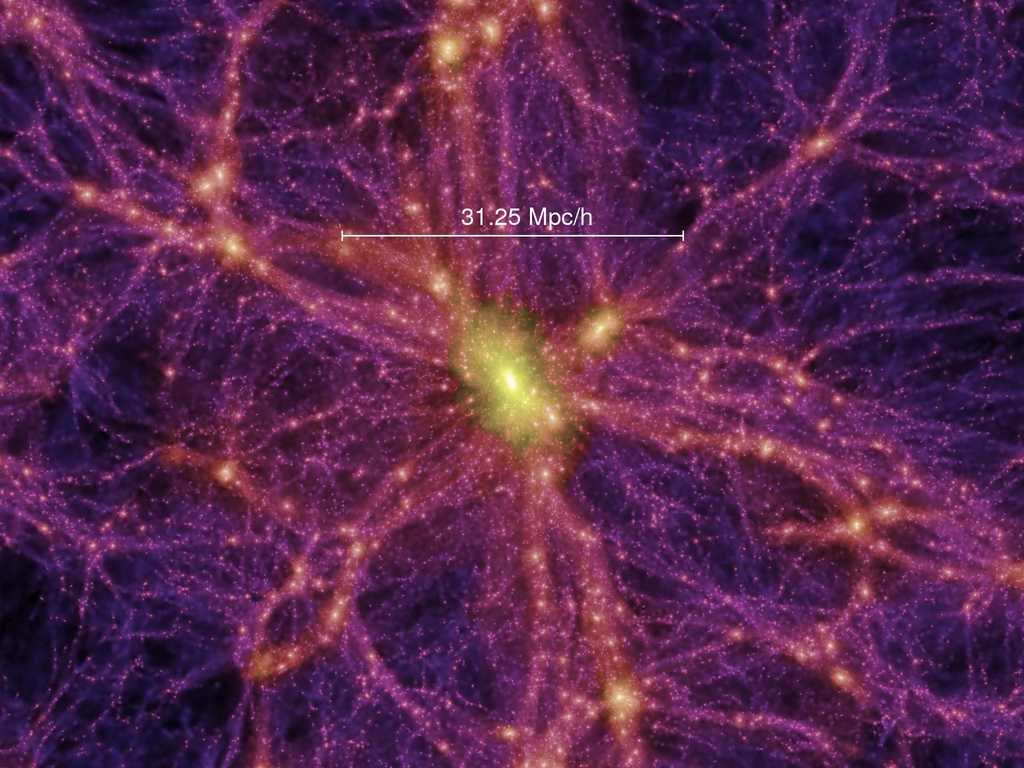}
\caption[Slices through the Millennium Simulation at $z=18.3$ and 0]{Slices through the ``Millennium Simulation'', showing the distribution of dark matter and its evolution to form a cluster of galaxies. Left: redshift 18.3; small structures are present but very little large scale structure has formed. Right: redshift 0; a massive galaxy cluster sits at the centre of a junction of the cosmic web. Credit: \citet{2005Springel}.}
\label{fig:millenniumsim}
\end{fig}

The origin of the structure within our Universe lies within the first $10^{-36}$~s of the Big Bang. 
It is thought that inflation in the very early universe amplified quantum fluctuations, which formed the seeds of the anisotropic structure that we see around us. These fluctuations gave rise to the temperature anisotropies present in the CMB, and gravitational waves from this era may be detectable in the structure of the polarization of the CMB radiation.

Gravitational instabilities then caused the growth of structures within our Universe, with matter evolving linearly and establishing a ``cosmic web'' with filamentary structures separated by voids. The gravitational instabilities lead to the formation of small galaxies first, with their mergers giving rise to larger galaxies as well as groups and clusters of galaxies. This evolution is driven by dark matter, with normal matter (predominantly in the form of gas) only becoming important on small scales within galaxies. The process of structure formation has been well studied using N-body simulations, for example the ``Millennium Simulation'' \citep{2005Springel} as depicted in Figure \ref{fig:millenniumsim}. For a review of inflation and structure formation, see e.g. \citet{2000Liddle}.

\section{Clusters of galaxies}\label{sec:intro_clusters}

The largest structures to form, and hence the most recent, are clusters of galaxies. A galaxy cluster is defined as a set of over 50 galaxies in close ($\lesssim2 \mathrm{Mpc}$) proximity to each other; sets of galaxies smaller than this are termed ``galaxy groups''. Pioneering work on the classification of galaxy clusters was done by \citet{1958Abell}, leading to the definition of a ``richness'' factor that depends on the number of constituent galaxies that can be allotted to each cluster, spanning from Class 0 (30-49 galaxies) to Class 5 (300+ galaxies).

Galaxy clusters emit radiation across the entire electromagnetic spectrum. Optical and infrared observations can locate individual galaxies; however, only about 10 per cent of the baryonic mass of a cluster is contained within the galaxies. The rest forms the IntraCluster Medium (ICM; also known as the intergalactic medium or the intracluster plasma). The majority of this is hot gas concentrated in the center of the cluster, where it is heated predominantly by energy from the infall of matter into the gravitational well of the galaxy cluster (although other effects also provide sources of energy). Due to its temperature, it emits thermal radiation peaking at X-ray wavelengths. It can also be observed via its effect on photons from the CMB -- the SZ effect. The SZ effect has cosmological importance as it provides a tracer for locating clusters of galaxies, the distribution of which can provide constraints on fundamental cosmological parameters.

The baryonic mass of a cluster only makes up around a seventh of its total mass \citep[e.g.][]{2005Ostriker}. The remainder consists of dark matter, probably in the form of a cluster halo and a series of sub-haloes, which are most likely associated with individual galaxies. The presence of this dark matter can be inferred both from the application of the virial theorem to the motions of galaxies within the cluster as well as gravitational lensing. Its distribution can potentially be related to the distribution of galaxies (and hence radio sources) within the cluster.

\subsection{Optical and infrared observations}
\begin{fig}
\centering
\includegraphics[scale=0.2]{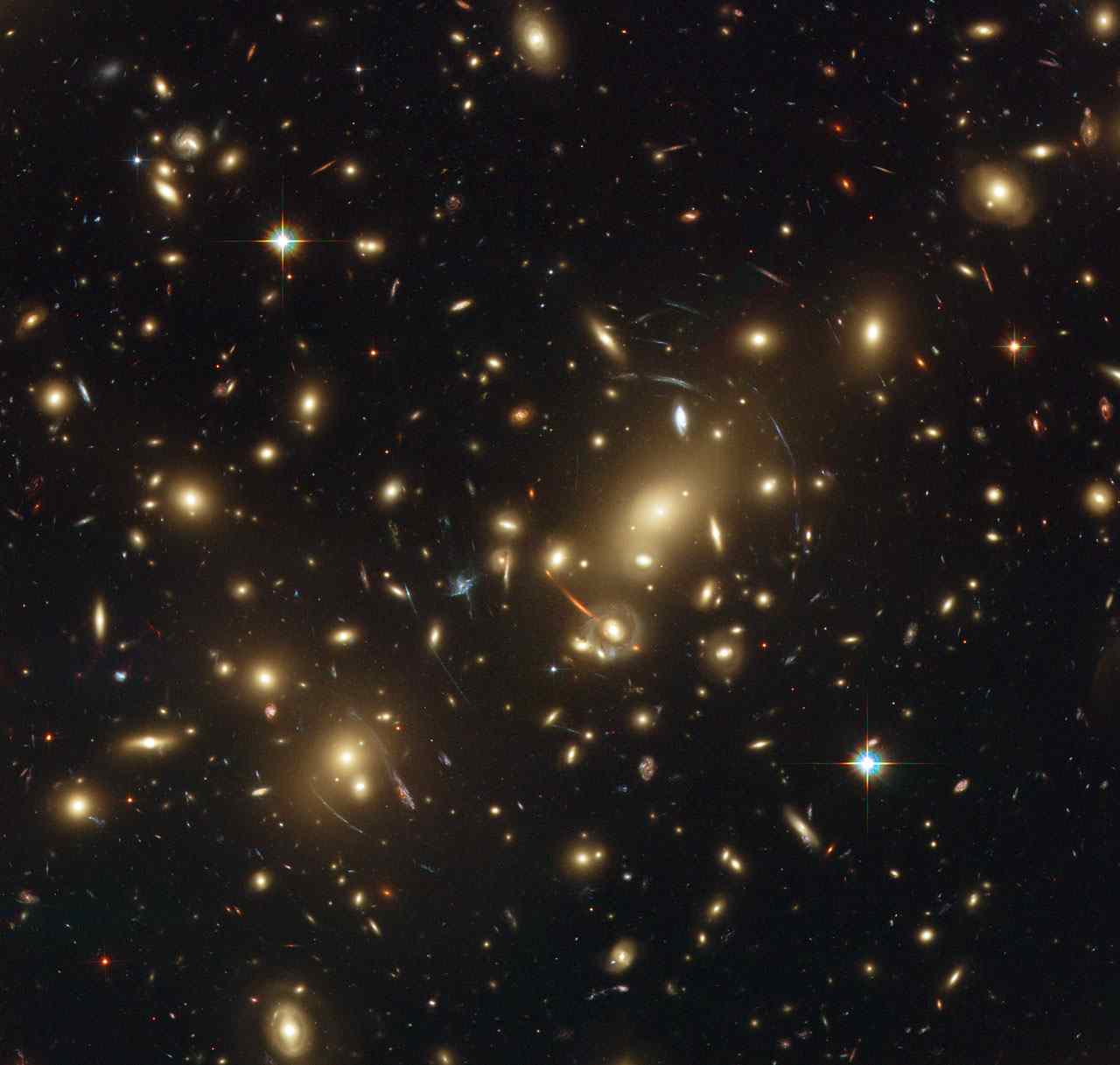}
\caption[The Galaxy Cluster Abell 2218 at optical wavelengths]{The Galaxy Cluster Abell 2218 at optical wavelengths, as imaged by the Hubble Space Telescope. A bright, central galaxy is visible, surrounded by arcs caused by weak lensing of a background galaxy. Credit: NASA, ESA, and Johan Richard (Caltech, USA). License: Public Domain.}
\label{fig:optical}
\end{fig}

Clusters of galaxies have been identified using optical wavelengths since the late $18^{\mathrm{th}}$ century \citep[see][for a historical review]{2000Biviano}. The classical catalogue for the subject is that published by \citet{1958Abell} \citep[also][]{1989Abell}, which focuses on clusters with more than 50 component galaxies that are fairly close to our Galaxy, i.e. with redshifts up to $z \approx 0.2$.

Optical observations show that there is frequently a Brightest Cluster Galaxy (BCG) or a first-ranked galaxy at the centre of galaxy clusters. These massive, luminous galaxies only appear in clusters -- never in the field -- and are mostly evolved and elliptical. Around 20 per cent are surrounded by a low surface brightness envelope; these are also called cD galaxies \citep{2001Oegerle,2006Seigar}.

Cluster galaxies can also be split up into several groups depending on their positions on a colour-magnitude diagram; these are the red and blue sequences. The red galaxies in the cluster are old and passively evolving, and form a tight sequence in colour-magnitude diagrams \citep{2004Lucia}. They can be used to locate clusters (e.g. the Red-Sequence Cluster Survey, \citealp{2000Gladders}).

Spectroscopy forms an important part of the optical observations of galaxies in clusters, as it allows the redshifts of the clusters to be measured to high precision ($\Delta z \sim 0.01$), as well as providing measurements of the velocity dispersion amongst the galaxies. Large-scale surveys aimed at obtaining redshifts for large numbers of galaxies are currently in progress, for example the Sloan Digital Sky Survey \citep[SDSS; ][]{2000SDSS} which started in 2000 and is still ongoing, and the WiggleZ Dark Energy Survey \citep{2009Drinkwater}.

Galaxy clusters are complicated, diffuse objects in the optical regime (see Figure \ref{fig:optical}). It often proves tricky to differentiate between cluster galaxies and foreground galaxies \citep[hence the selection methods used by][]{1958Abell}. In general, spectroscopic observations of the galaxies are needed to identify their redshifts, so that their line-of-sight distances are known. N-body techniques such as friends-of-friends can also be applied to select the cluster galaxies \citep{2006Clewley}.

\subsection{X-ray emission}
\begin{fig}
\centering
\includegraphics[scale=0.4]{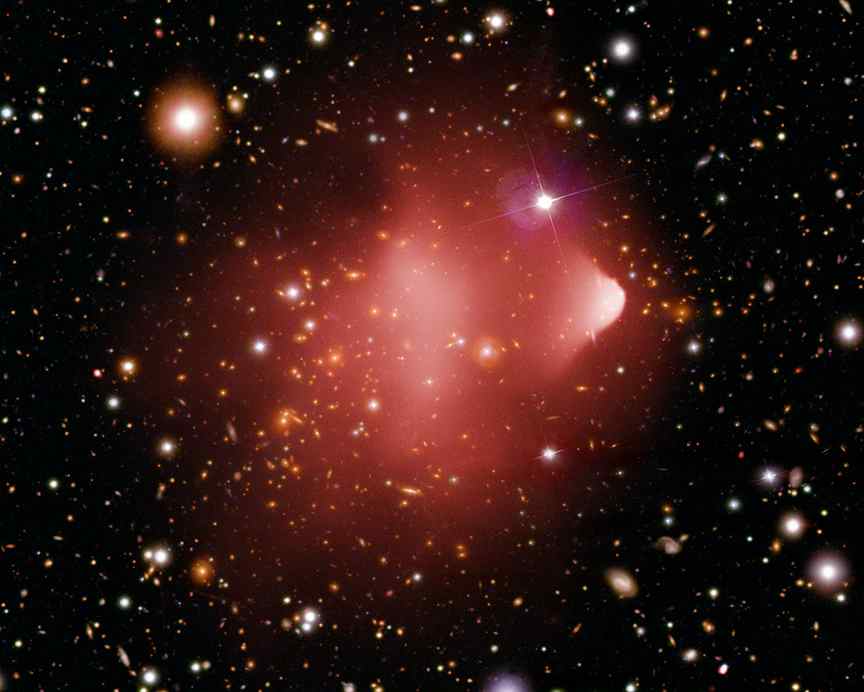}
\caption[The Bullet Cluster at optical and X-ray wavelengths]{The Bullet Cluster (1E 0657-56) in X-ray and optical. Credit: X-ray: NASA/CXC/CfA/M.Markevitch et al.; Optical: NASA/STScI; Magellan/U.Arizona/D.Clowe et al.; License: reuse allowed for non-commercial educational and public information purposes.}
\label{fig:xray}
\end{fig}

Due to the temperature of the gas in the ICM, thermal emission is given off as X-rays mostly via thermal bremsstrahlung, but also through line radiation. The spectral surface brightness at energy $E$ along a particular line of sight $l$ is given by \citep{1999Birkinshaw}
\begin{equation}
b(\theta, E) = \frac{1}{4 \pi (1+z)^3}\int n_{\mathrm{e}}(\theta)^2 \Lambda (E (\theta),T_{\mathrm{e}} (\theta)) dl,
\end{equation}
where $z$ is the redshift of the cluster, $n_{\mathrm{e}}(\theta)$ is the electron number density at a distance $\theta$ from the center of the cluster and $\Lambda (E, T_{\mathrm{e}})$ is the spectral emissivity at energy $E(\theta)$ of a gas at temperature $T_{\mathrm{e}}(\theta)$; $\Lambda (E, T_{\mathrm{e}})$ is generally a complicated function. A related parameter is the cooling function, which gives the total energy emitted from a gas at temperature $T_{\mathrm{e}}$.

The intensity of X-rays emitted depends on the square of the electron density, hence X-ray observations predominantly see the emission from cores of the galaxy clusters rather than the outskirts of the cluster. The gas in the cluster center can be subject to complicating effects including heating and cooling \citep[see e.g.][]{2005Motl}, and the X-ray emission is also sensitive to clumps and shocks in the ICM (see Figure \ref{fig:xray}).

\subsection{The Sunyaev-Zel'dovich effect} \label{szeffect}
The Sunyaev-Zel'dovich (SZ) effect is due to the inverse Compton scattering of photons from the CMB when they pass through the hot electron gas in the ICM. The photons gain energy from the electrons and are hence shifted up in frequency, meaning that the SZ effect is manifested as a decrement in the CMB at low frequencies and an increment at high frequencies, with a null at 217GHz. It was predicted by \citet{1970Sunyaev, 1972Sunyaev}, based on earlier work by \citet{1966Weymann} and \citet{1969Zeldovich}. It can be parameterized by the Compton $y(\theta)$ parameter, which represents the integrated pressure along the line of sight $l$ at a distance $\theta$ from the center of the cluster, by \citep{1999Birkinshaw}
\begin{equation}
y(\theta) = \frac{\sigma_{\mathrm{T}} k_{\mathrm{B}}}{m_{\mathrm{e}} c^2} \int n_{\mathrm{e}} (\theta) T_{\mathrm{e}}(\theta) \phantom{1} dl,
\end{equation}
where $\sigma_{\mathrm{T}}$ is the Thomson scattering cross-section, $k_{\mathrm{B}}$ is Boltzmann's constant, $m_{\mathrm{e}}$ is the electron mass, $c$ is the speed of light, $n_{\mathrm{e}}(\theta)$ is again the number density of electrons and $T_{\mathrm{e}}(\theta)$ is the temperature of the electrons. Note that this equation does not include $z$ -- the effect is redshift independent. Relativistic corrections to this equation can be found in e.g. \citet{1998Itoh}.

The SZ effect resulting from the hot gas in a cluster or group of galaxies produces a detectable flux density $S$ at frequency $\nu$ by
\begin{equation}
S_{\mathrm{\nu}} = \frac{2 \nu^2 k_{\mathrm{B}} T_{\mathrm{CMB}}}{c^{2}} f(x) \int y(\theta) d\Omega,
\end{equation}
where the integral is over the observed section of sky and $f(x) = x^2 e^x g(x) / (e^x-1)^2$ represents the frequency dependence of the effect, in which $g(x) = (x / \tanh(x / 2)) - 4$ and the dimensionless frequency $x = h \nu / k_{\mathrm{B}}T_{\mathrm{CMB}}$. $T_{\mathrm{CMB}}$ is the present-day temperature of the CMB.

The SZ effect has a major advantage with locating clusters compared to optical and X-ray methods in that the magnitude of the effect does not depend on redshift, and as such it should be possible to observe clusters of galaxies out to high redshifts relatively easily. Additionally, the SZ effect only depends on $n_\mathrm{e}$, not $n_\mathrm{e}^2$ as in X-ray, so the outer part of the cluster atmosphere can also be observed, and less complications will arise from substructure and cooling/heating effects within the cluster \citep{2005Motl}.

Since the first observations of the SZ effect in the 1970s (\citealp{1978Birkinshaw}), a number of experiments have measured the effect in galaxy clusters (e.g. \citealp{1993Jones,2001Grego,2008Wu,2009Plagge,2009Hincks}, Lancaster et al. in prep.), although typically with sample sizes in the low double-digits. Observations with the {\it Planck} satellite, and ground-based experiments like the South Pole Telescope \citep{2004Ruhl}, AMI \citep{2008Zwart} and OCRA \citep{2000Browne}, promise to increase the number of SZ detected clusters dramatically in the near future.

The SZ effect produced by clusters can be used for cosmology primarily via number counts. The expected number of detections depends on the total amount of matter within our Universe ($\Omega_\mathrm{m}$), as well as the normalization of the matter power spectrum ($\sigma_8$) and the equation of state for dark energy ($w$). See Section \ref{sec:distribution_clusters} for details. Comparison of X-ray emission with the SZ effect signal can also provide measures of the cluster distance, and hence constraints on the Hubble constant \citep[e.g.][]{1991Birkinshaw,2006Cunha}. The frequency dependence of the SZ effect can also be used to measure $T_\mathrm{CMB}$ at different redshifts \citep[e.g.][]{2009Luzzi}.

Only the thermal SZ effect is considered here. A kinetic SZ effect also exists; see e.g. Section 5 of the review by \citet{1999Birkinshaw}. This effect can become significant when clusters have high velocities relative to the CMB.

\section{Sources of contamination and confusion} \label{sec:intro_confusion}

The main source of contamination for SZ observations of galaxy clusters at microwave frequencies is point sources -- radio-loud galaxies that lie in the direction of the galaxy cluster. These can either be physically associated with the cluster or they can be randomly distributed in front of, or behind, the cluster. At lower frequencies where the SZ effect causes a decrement in the CMB, these point sources can ``fill in'' the decrement and hence reduce the signal. Point sources that are present (in projection) to one side of the centre of a galaxy cluster can also present experimental difficulties, depending on the configuration of the instrument making the observations.

CMB anisotropies themselves can also be confused with SZ detections, again depending on the configuration and properties of the instrument. Whilst CMB anisotropies are dominant over the SZ effect only at large scales, if the beam of the instrument is not well matched to the scale of the galaxy cluster being observed (typically $\sim1$ arcmin) then the observations can include the effects of CMB anisotropies. This is also true of switched beam systems where the beams are separated by more than $\sim 1$ arcmin.

Galactic emission presents a large foreground for galaxy clusters. This is especially the case along the plane of our Galaxy. However Galactic emission also extends to high and low latitudes, and can present issues if the emission has a complex spatial distribution on arcminute scales. Fortunately, at these latitudes the emission is thought to be smooth enough that this is unlikely to be a major problem.

Coming down to earth, our own atmosphere presents a major observational challenge for ground-based observations, particularly at sea level. At microwave frequencies, water vapour in our atmosphere attenuates and emits radiation. As this water vapour is not evenly distributed across the sky (as evidenced by clouds) as it drifts through the telescope beam, this creates time-variable noise which, if not carefully subtracted, causes contamination in the observations.

\section{The One Centimetre Receiver Array} \label{sec:ocra}
The One Centimetre Receiver Array (OCRA) programme \citep{2000Browne} is focused on developing multi-pixel arrays of continuum receivers at microwave frequencies. Two receivers have thus far been constructed -- OCRA-p and OCRA-F -- both of which operate at a wavelength around 1~cm (30~GHz). OCRA-p is a 2-beam prototype that has been observing on the Toru\'n 32~m telescope in Poland since 2005, and OCRA-F is an 8-beam receiver array that is just starting observations from the same location.

The first receiver in the OCRA program was the OCRA prototype (OCRA-p). This is a two-beam pseudo-correlation receiver based upon the Planck LFI receiver chain, and is similar in design to the WMAP 23~GHz receiver. The two beams are combined together using a hybrid, then combinations of the signals passed through two Low Noise Amplifiers (LNAs) and a pair of phase switches. The signals are then separated by another hybrid, further amplified and square-law detected to measure the RF power. The detected signals are subtracted from each other to get the difference in signal between the two beams. This reduces the effect of $1/f$ noise from the LNAs and the atmosphere. A full technical description of the implementation, commissioning process and initial observations with OCRA-p are detailed in \citet{2006Lowe} (see also \citealp{2007Lowe}).

OCRA-p has been used to survey radio point sources, including the CJF sample \citep{2007Lowe}, the Very Small Array fields \citep[][also see Section \ref{section:vsasources}]{2009Gawronski}, the SENSE sample \citep{2009Gupta} and the CRATES sample (see Section \ref{section:crates}). It has also been used to observe the SZ effect from clusters of galaxies \citep{2006Lancaster}, novae \citep[RS Ophuchi;][]{2009Eyres} and planetary nebulae \citep{2009Pazderska}.

The second generation of receiver is OCRA FARADAY (OCRA-F). This receiver currently has 8 beams, with the capacity for expansion to 16 beams; these are arranged in pairs. The receiver builds upon OCRA-p, following the same receiver chain pattern but using Monolithic Microwave Integrated Circuits (MMICs) in place of traditional components (see \citealp{2007Kettle}). The assembly, testing and commissioning of OCRA-F are described in this thesis (Chapter \ref{ocraf}). OCRA-F will be used to do small scale blind surveys for point sources and for studying the SZ effect. It will also, within certain limitations, be able to create maps of extended emission.

The ultimate goal of the OCRA program is to construct a 100 beam array, which can then be used for large scale blind surveys of point sources and the SZ effect \citep{2000Browne}. The receiver technology required for such an instrument will be studied in the next few years via the EC Framework 7 APRICOT (All Purpose Radio Imaging Cameras On Telescopes) project within RadioNet, with the aim of combining spectroscopic and continuum measurements into one receiver package.

\section{Thesis structure} \label{sec:intro_structure}

This chapter has provided an overview of structure formation, galaxy clusters in the optical, X-ray and microwave regimes, and the main sources of confusion for the SZ effect. It has also summarized the OCRA series of receivers. 

The next chapter describes the creation of simulated maps of the CMB, SZ effect and point sources (``Virtual Skies''). These simulations are then utilized in Chapter 3, which examines the statistics of the power spectrum of the SZ effect and concludes with the possibility of measuring the CMB power spectrum at high multipoles using OCRA-F.

Data reduction software that has been written specifically to analyse measurements from the OCRA receiver is described in Chapter 4. The chapter also describes the assembly of this data reduction package with pre-existing software to simulate a telescope equipped with an OCRA receiver. Virtual Sky maps can then used as the input for this software, such that end-to-end simulations of OCRA observations can be performed. The performance of the OCRA receivers for observing point sources is then evaluated using these end-to-end simulations.

Chapter 5 then uses the data reduction package to analyse observations by OCRA-p as part of two surveys -- a survey of 121 radio sources within the five fields of the Super Extended Very Small Array, and a sample of 550 flat-spectrum radio sources selected from the CRATES sample.

Chapter 6 describes the assembly of the OCRA-F receiver as well as the testing and commissioning of the receiver both in the lab and on the Toru\'n 32-m telescope, culminating in a description of its first light observations. Finally, Chapter 7 concludes the thesis, and looks to the future work that can be done based on that described herewithin.

\chapter{Creating a Virtual Sky}\label{virtualsky}

Several components need to be generated to create a virtual map of the sky at microwave frequencies. The first of these is the CMB, which forms the background to everything else. Then, the distribution of galaxy clusters needs to be generated, determining their positions on the sky as well as their redshifts and masses. Once this is known, a cluster model needs to be used to create a map of the SZ effect. Finally, point sources need to be added, requiring the determination of their positions on the sky and their flux densities.

The effects of other foregrounds, including the galaxy, are not included as the observations focus on areas of the sky where galactic foregrounds are expected to be low. Additionally, these foregrounds are expected to be fairly smooth on the scales under consideration here.

These simulations of the microwave sky can be used to study the power spectra of the different components, particularly the SZ effect, and for carrying out simulations of observations (amongst other things). In this chapter, the methods of creating the maps are described; their use and results are given in Chapter \ref{vs_results}. Similar work creating maps of the SZ effect has been carried out by e.g. \citet{2006Schafer}, \citet{2007Geisbusch}, \citet{2007Holder} and \citet{2007Sehgal,2009Sehgal}.

Maps are created both of small sections of the sky (``flat sky" maps) and of the whole sky (using HEALPix; \citealp{2005Gorski}). The process of creating these maps is essentially the same; this chapter primarily focuses the former, with differences in the creation of the latter described where appropriate.

Throughout this chapter, unless otherwise stated, we use a canonical frequency of 30~GHz, to ease comparisons with potential observations with the OCRA receiver, as well as the observations of the high-multipole excess by CBI and BIMA (see Chapter \ref{sec:cbi_excess}). We use a $\Lambda \mathrm{CDM}$ cosmology throughout this thesis, with the ratio of the matter density compared to the critical density $\Omega_\mathrm{m} = 0.3$, the same ratio for dark energy $\Omega_\mathrm{\Lambda} = 0.7$ and the ratio of baryons to critical density $\Omega_\mathrm{b} = 0.05$. The Hubble constant $H_0 = 100 \phantom{.} h_{100} \phantom{.} \mathrm{km s^{-1} Mpc^{-1}}$, where $h _{100}= 0.7$. We use three values of $\sigma_8$: 0.75, 0.825 and 0.9, with 0.825 as the default. These span the best-fitting values from {\it WMAP} after 1, 3 and 5 years of observations \citep{2003Spergel,2007Spergel,2009Komatsu}.

The process of generating maps of the CMB is described in Section \ref{sec:cmb}. Section \ref{sec:distribution_clusters} describes the construction of cluster catalogues, and Section \ref{clustermodel} describes the cluster model; these two are then combined to create maps of the SZ effect, examples and power spectra for which are given in Section \ref{sec:example_realisations}. Section \ref{sec:theoretical_ps} describes the calculation of the theoretical power spectra for the SZ effect, and compares this with that from the realisations. Point sources are then added to the model in Section \ref{sec:pointsources}, before the section is summarized in Section \ref{sec:virtualsky_summary}.

\section{Cosmic Microwave Background} \label{sec:cmb}
Maps of the small-scale CMB anisotropies can be computed using the standard assumption that the CMB is a Gaussian random field and hence is completely described by its angular power spectrum, $C_l$. We generate the expected power spectrum of the gravitationally lensed CMB from {\sc CAMB}\footnote{November 2008 version; available from \url{http://www.camb.info}} \citep{2000Lewis}. It is known that the gravitational lensing component is non-Gaussian \citep[see for example][]{2000Zaldarriaga}, however as this component is sub-dominant we assume that it is Gaussianly distributed.

For flat sky maps, we create the CMB realizations in Fourier space. The power at each Fourier mode $l$ is calculated by $F_{m_{\mathrm{x}},m_{\mathrm{y}}} = a + ib$, where $a$ and $b$ are independent  random numbers, chosen from a Gaussian distribution with zero mean and a standard deviation of $\sqrt{C_{l} / (2 \delta_l^2)}$, where $\delta_l = 2 \pi / \theta_\mathrm{map}$ is the resolution of the map in Fourier space. We use the {\it Fastest Fourier Transform in the West}\footnote{Version 3.1.2; available from \url{http://www.fftw.org}} ({\sc FFTW3}; \citealp{2005Frigo}) package to carry out the required 2-dimensional complex-to-real Fourier transforms between the power spectra and the maps.

The relationship between spherical harmonics and Fourier modes is simply $l = |k|$, where $k = 2 \pi m / \theta_{\mathrm{map}}$ is the $\mathrm{m}^\mathrm{th}$ Fourier mode of the map. $\theta_{\mathrm{map}}$ is the width of the map in radians, and $m = \sqrt{m_\mathrm{x}^2 + m_\mathrm{y}^2}$, in which $m_\mathrm{x}$ and $m_\mathrm{y}$ are the integer Fourier modes in the x- and y-directions of the map.

We use the 2-dimensional real to complex discrete Fourier transform, such that the map $M_\mathrm{x,y}$ only has a real component, while Fourier space $F_\mathrm{m_x,m_y}$ has both real and complex components. The Fourier transform from Fourier to real space is \citep[][p.40]{2006Frigo}
\begin{equation}
M_{x,y} = \frac{\Delta_l^2}{2 \pi} \sum_{j=0}^{N_x-1} \sum_{p=0}^{N_y-1} F_{m_x,m_y} e^{\frac{2 \pi j m_x i}{N_x}} e^{\frac{2 \pi p m_y i}{N_y}},
\end{equation}
whilst the inverse is
\begin{equation}
F_{m_x, m_y} = \frac{2 \pi}{\Delta_l^2 N_{\mathrm{pix}}} \sum_{j=0}^{N_x-1} \sum_{p=0}^{N_y-1} M_{x, y} e^{\frac{-2 \pi j m_x i}{N_x}} e^{\frac{-2 \pi p m_y i}{N_y}},
\end{equation}
where $\Delta_l = 2 \pi / \theta_\mathrm{map}$, $N_\mathrm{pix}$ is the total number of pixels in the map, i.e. $N_x \times N_y$, and $i = \sqrt{-1}$.

The power spectrum $C_l$ of the map is calculated by
\begin{equation}
\frac{l (l+1) C_{l}}{2 \pi} = \frac{\Delta_l^2}{2 \pi} \sum k^{2} \left( \Re^2 (F_{m_x,m_y}) + \Im^2 (F_{m_x,m_y}) \right),
\end{equation}
where $\Re(F_{m_x,m_y})$ and $\Im (F_{m_x,m_y})$ refer to the real and imaginary components of $F_{m_x,m_y}$. The sum is over all modes $(m_{\mathrm{x}},m_{\mathrm{y}})$ of the map with the same multipole $l$.

\begin{fig}
\centering
  \includegraphics[scale=0.55]{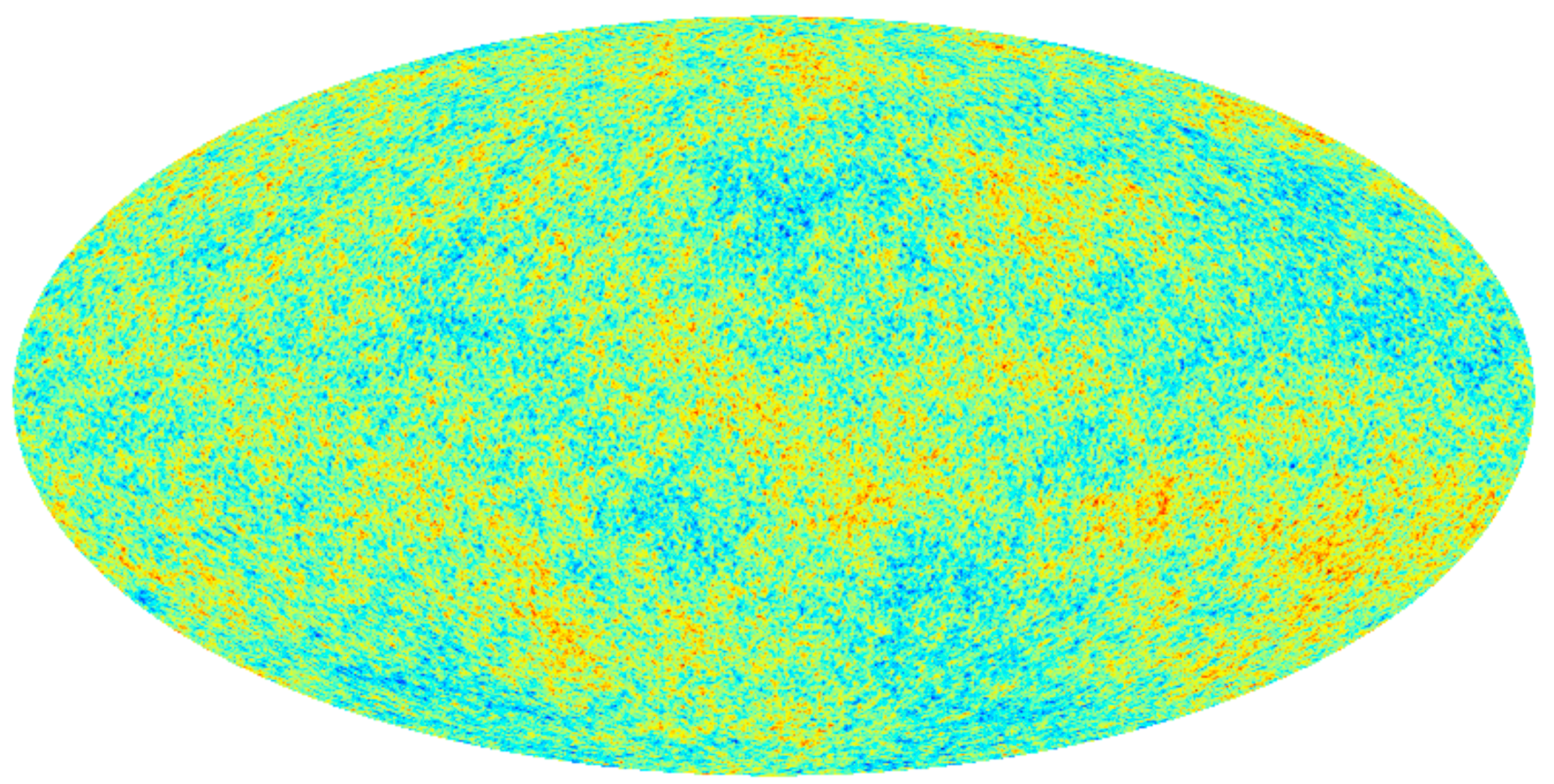}
	\caption[An example full sky realization of the CMB]{An example full sky realization of the CMB, with ``hot'' areas in red and ``cold'' in blue, using a Mollweide projection of a HEALPix map \citep{2005Gorski}. The colour is linear, between $\pm 520 \upmu \mathrm{K}$.}
	\label{fig:cmb}
\end{fig}

For full sky maps, assuming that the CMB has a Gaussian distribution (i.e. the power spectrum contains all of the information about the CMB), the $C_{l}$ values can be converted to a random set of complex spherical harmonic coefficients ($a_{lm}$) via
\begin{equation} \label{eq:cmbstart}
\Re (a_{lm}) = \sqrt{- 2 \ln (\alpha) C_{l}} \cos (2 \pi \beta),
\end{equation}
\begin{equation}
\Im (a_{lm}) = 0,
\end{equation}
for $m=0$, and
\begin{equation}
\Re (a_{lm}) = \sqrt{- \ln (\alpha) C_{l}} \cos (2 \pi \beta),
\end{equation}
\begin{equation} \label{eq:cmbend}
\Im (a_{lm}) = \sqrt{- \ln (\alpha) C_{l}} \sin (2 \pi \beta),
\end{equation}
for $1 \leq m \leq l_\mathrm{max}$, with the maximum value of $l$ being established by the resolution of the map. $\alpha$ and $\beta$ are uniform random variables, the value of which is computed separately for each value of $m$ and $l$. The coefficients for negative values of $m$ can be found by $a_{l,-m} = a_{l,m}$*, where * denotes complex conjugation. These $a_{l,m}$ values can then be used to create a HEALPix \citep{2005Gorski} map via a Fourier transform. See Figure \ref{fig:cmb} for a map of an example realisation.

When calculating the binned power spectrum, we assume that $l(l+1)C_l / (2 \pi)$ is flat over a region of size $\Delta l$ such that the power within a multipole bin centred on $\bar{l}_i$ can be calculated by
\begin{equation} \label{eq:binning}
B_i = \frac{1}{\Delta l} \sum_{l \in \mathrm{bin}} \frac{l(l+1) {C_l}}{2 \pi},
\end{equation}
where the summation is over $\bar{l}_i - \frac{1}{2} \Delta l \leq l < \bar{l}_i + \frac{1}{2} \Delta l$. Values of $C_l$ that have not been sampled due to the finite grid are assumed to have the value of the closest $C_l$ that has been sampled. This is obviously imperfect where the $C_l$ values have a large gradient or are ill- or irregularly-sampled, but is likely to be sufficient for the purposes at hand.

The expected mean in each multipole bin can be calculated from the input power spectrum, and the expected variance is calculated by
\begin{equation} \label{eq:cv}
\delta B_{i} = \frac{1}{\Delta l}\sqrt{\sum_{l \in \mathrm{bin}} \left(\frac{l(l+1)C_l}{2 \pi} \right)^2 \frac{2 }{(2 l + 1) f_\mathrm{sky}}},
\end{equation}
where $f_\mathrm{sky}$ is the fraction of the sky, $\Delta l$ is the width of the multipole bin (here, we use a fiducial bin of $\Delta l = 1000$) and the sum is over all multipoles within the bin. The quantity $\delta B_i \sim \bar{l}_i^{-1/2} f_\mathrm{sky}^{-1/2}$ is sometimes called the cosmic variance since it represents the limit on how well one can possibly measure $B_i$ as imposed by the Gaussian statistics. The aim of this section is to understand this issue in the case of the non-Gaussian anisotropies created by the SZ effect.

\begin{fig}
\centering
\includegraphics[scale=0.8]{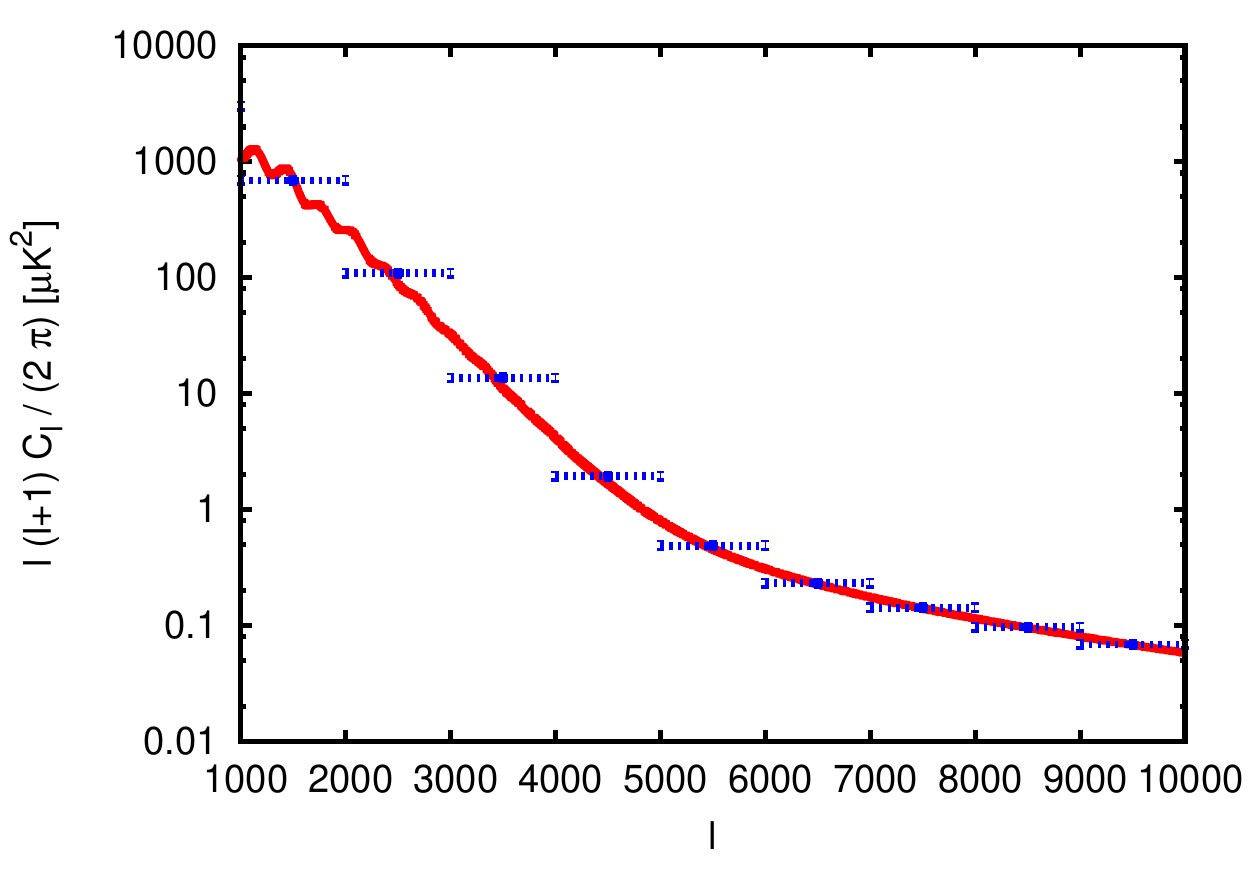}
\caption[Theoretical and binned power spectra of the CMB]{The theoretical (red line) and binned (blue points) power spectrum of the lensed CMB between $l=$1000 and 10000, averaged over 1000 maps and binned with $\Delta l = 1000$. The binned power spectrum agrees well with the input power spectrum.}
\label{fig:cmb_powerspectrum}
\end{fig}

\begin{fig}
\centering
\includegraphics[scale=0.8]{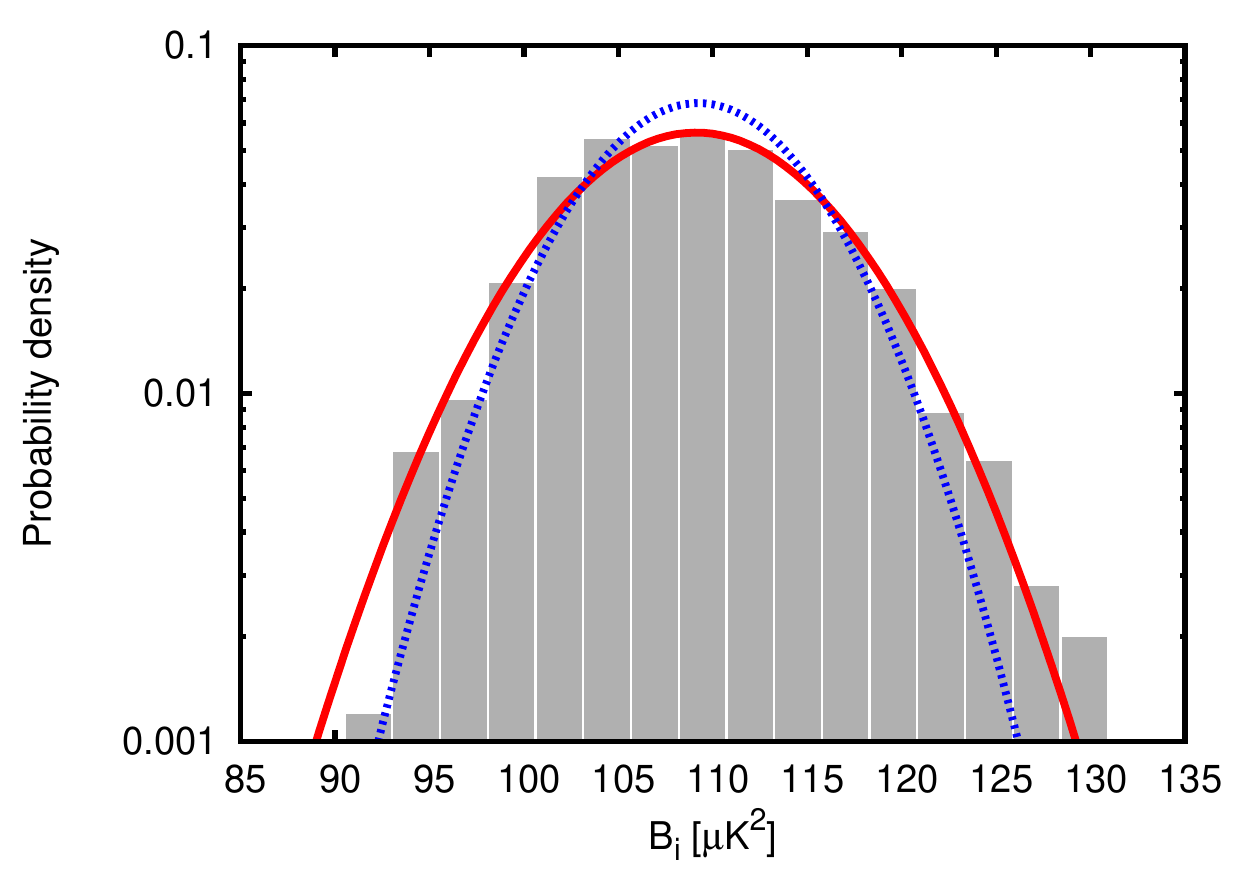}
\caption[The statistics of the CMB realizations in the multipole bin 2000-3000]{The statistics of the CMB realizations in the multipole bin 2000-3000. The solid red line is the best fit Gaussian distribution, whilst the dotted blue line is the distribution for this bin based on the analytic formula for the variance (Equation \ref{eq:cv}). The distribution is well fit by a Gaussian distribution with a slightly higher standard deviation than the analytical fomula.}
\label{fig:cmb_statistics}
\end{fig}

Figure \ref{fig:cmb_powerspectrum} shows the CMB power spectrum using bins of 1000 multipoles, and Figure \ref{fig:cmb_statistics} shows the statistics within the multipole bin 2000-3000. The statistics are consistent with a Gaussian distribution with a slightly larger standard deviation than expected from cosmic variance due to inefficiencies in the sampling of the spectrum at different multipoles. The distribution is also slightly skewed, as the finite sampling leads to a $\chi^2$ distribution rather than a Gaussian one.

\section{Distribution of galaxy clusters} \label{sec:distribution_clusters}
The positions of galaxy clusters can be specified in three dimensions -- $\theta$ and $\phi$ on the sky, and the redshift $z$. To these, a fourth parameter needs to be added -- the mass of the cluster -- to fully specify the distribution. These four parameters can then used to generate and position a cluster model on a map of the sky.

There are several ways of creating simulations of the large scale structure of the universe. Most common are N-body simulations; these are generally used for looking at the distributions of clusters \citep[e.g.][]{1998Retzlaff,2002Evrard} and their evolution over time. As a result, they can provide expected mass functions and number counts, e.g. \citet{2001Jenkins}. Additionally, they can be used to investigate the distribution of dark matter within haloes.

Additional detail can be simulated by adding hydrodynamical simulations to N-body simulations. These consist of solving the hydrodynamical equations with various physical effects, such as radiative cooling, adiabatic compression and shock heating, and are therefore useful for simulating the distributed material such as the intracluster medium \citep[see e.g.][]{2002Muanwong}.

Virtual cluster catalogues can be obtained from these simulations, which can then be used directly to create maps by applying models of clusters at each cluster position; for a recent example of this, see \citet{2006Schafer}. However, there are a limited number of these catalogues available due to the amount of computing power required for each simulation; for example, the gigaparticle $(10^9)$ simulation run by \citet{2002Evrard} took around 7 months to run on a 512-CPU computer. An alternative approach is to use theoretical equations calibrated by the simulations to create similar maps comparatively easily and much quicker; this approach is described in Section \ref{sec:massfunctions}. Maps can also be created using N-body simulation output (Section \ref{sec:nbody}), as well as using {\sc Pinocchio} simulations (Section \ref{sec:pinocchio}).

\subsection{Analytical formulae} \label{sec:massfunctions}
The dependence of the number of clusters with $z$ is given by
\begin{equation}
\label{eq:numclusters}
\frac{dN}{dz} = \Delta \Omega \frac{dV}{dz d \Omega}(z) \int^{\infty}_{M_{\mathrm{lim}}(z)} \frac{dN}{dM} (M,z) dM,
\end{equation}
where $\Delta \Omega$ is the solid angle of the sky being looked at, $dN/dM (M,z)$ is the comoving number density of clusters with mass $M$ and redshift $z$, and also called the mass function. $M_{\mathrm{lim}} (z)$ is the minimum mass that is being considered (this is usually the minimum detectable mass, which generally depends on redshift). This can easily be adapted to give the number of clusters in a given mass range.

\begin{equation}
\frac{dV}{dz d\Omega} = \frac{r(z)^{2}}{H(z)} = \frac{ \left( \int_{0}^{z} H^{-1}(z') dz' \right)^{2} }{H(z)}
= \frac{\left( \int_{0}^{z} E^{-1}(z') dz' \right)^{2} }{E(z) H_{0}^{3} }
\end{equation}
is the comoving volume in a flat universe, in which $r(z)$ is the coordinate distance. $H(z)$ is the Hubble parameter, which is given by the Friedmann equation
\begin{equation} \label{eq:hubble}
	E^{2}(z) = \frac{H^2(z)}{H_0^2} = \Omega_{\mathrm{m}} (1 + z)^3 + \Omega_{\mathrm{\Lambda}}
\end{equation}

The standard Press-Schechter mass function is \citep{1974Press}
\begin{equation}
\frac{dN}{dM}(z,M) = - \sqrt{\frac{2}{\pi}} \frac{\rho_{\mathrm{m}}(t_{0})}{M} \frac{\delta_{\mathrm{c}}}{D(z) \sigma (M)^{2}} \frac{d\sigma (M)}{dM} \exp \left( - \frac{\delta_{\mathrm{c}}^{2}}{2 D(z)^{2 }\sigma (M)^{2}}\right),
\end{equation}
in which $\delta_{\mathrm{c}} = 1.686$, $\rho_{\mathrm{m}}(t_{0})$ is the present value of the matter density and $\sigma (M)$ is the current over-density with mass $M$; this can be calculated from the density power spectrum. $D(z)$ is the growth factor, which is normalized by $D(0) = 1$; the dependence on redshift can be found by solving the perturbation equation for matter fluctuations \citep[see e.g.][]{2003Battye}, $D(z) = \delta_{m}(z) / \delta_{m}(0)$, where $\delta_{m}(z)$ is found by numerically solving
\begin{equation}
\delta_{m}'' + \frac{3}{2a} [1 - \omega(a) (1 - \Omega_{m}(a))] \delta_{m}' - \frac{3}{2a^{2}} \Omega_{m}(a) \delta_{m} = 0.
\end{equation}

The over-density as a function of mass is calculated using a spherical top hat window function and the linear power spectrum (based upon the transfer function $T(k)$),
\begin{equation} \label{eq:sigmar}
\sigma^2 (R) = \int_{k_{\mathrm{min}}}^{k_{\mathrm{max}}} W^{2} (kR) 4 \pi k^{2} P(k) \phantom{.} dk.
\end{equation}
The transfer function is calculated using CAMB \citep{2000Lewis}. The conversion between mass $M$ and radius $R$ is via $M = 4 \pi \Omega_{m} \rho_{crit} (t_{0}) R^{3} / 3$. Equation \ref{eq:sigmar} is numerically differentiated to get $d \sigma / d M$. The window function used is
\begin{equation}
W(x) = 3 \left( \frac{\sin(x)}{x^{3}} - \frac{\cos(x)}{x^{2}} \right),
\end{equation}
and the power spectrum is calculated by
\begin{equation}
P(k) = A k^{n_{s}} T^{2}(k)
\end{equation}
where $A$ is a normalization constant that is calculated using $\sigma_{8}$, the density contrast at the scale of $8 h_{100}^{-1} \mathrm{Mpc}$.

The Press-Schechter mass function was extended by \citet[][henceforth ST99]{1999Sheth} to take into consideration large-scale biasing of the mass function; the ``extended Press-Schechter'' formula for the mass function is
\begin{equation}
\frac{dN}{dM} = A \left( 1 + \frac{1}{\nu'^{q}} \right) \sqrt{\frac{\nu'}{2 \pi}} e^{-\frac{\nu'}{2}} \frac{\Omega_{\mathrm{m}} \rho_{\mathrm{crit}} (t_{0})}{M_{\mathrm{vir}}^{2}} 2 \frac{d \ln \sigma^{-1}(M)}{d \ln M},
\end{equation}
where $\nu' = 0.707 \nu$, in which $\nu = \left(\delta_{\mathrm{sc}} / \sigma (M) \right) ^{2}$; $\delta_{\mathrm{sc}} = 1.686 / D(z)$, and $A = 0.322$ and $q = 0.3$.
% NB: 2 is from the differentiation of v.

The total number of clusters on a map of a given size $\Delta \Omega$ on the sky can be found by integrating Equation \ref{eq:numclusters} over both mass and redshift; to make the map more realistic, this number can then be Poisson distributed. A 2D probability distribution as a function of mass and redshift is calculated using Equation \ref{eq:numclusters}, and a catalogue of individual cluster masses and redshifts is then obtained by using a random number generator to sample from this probability distribution.

\subsection{N-body simulations} \label{sec:nbody}
The simplest method of positioning clusters on the sky is to Poisson distribute them, i.e. by generating the positions from a uniform distribution. However, this will not create the clustering of the clusters that is expected from theory and N-body simulations.

There are several sources of this clustering. First, early galaxy clusters are formed at peaks in the density fluctuations set up by quantum fluctuations in the early universe; this will impart a specific distribution upon them. This is most important with the distribution of high-mass clusters, which will have subsequently evolved approximately linearly with time via gravity. Second, clusters will gravitationally attract each other in a non-linear way. This is more important with smaller clusters or groups than large ones, as they will tend to cluster around the larger clusters. The third, and most complicated, source of clustering is due to mergers and close encounters between clusters. As these effects are included within N-body and hydrodynamical simulations (down to a scale size that depends on the resolution of the simulations and the physical effects that they incorporate), the cluster catalogues from them can be used to create maps containing realistic clustering of the galaxy clusters.

N-body simulations can also be used to calibrate the analytical mass functions. \citet{2002Evrard} provide a fit for the mass function at $z = 0$, which can be generalized in redshift to give
\begin{equation}
\frac{dN}{dM} = \left( \frac{M_{200}}{M_{vir}} \right) A\frac{\Omega_\mathrm{m} \rho_{\mathrm{crit}} (t_{0})}{M_{200}^{2}} \frac{d \ln \sigma^{-1} (M_{vir})}{d \ln M} \mathrm{exp} [- |\ln [\sigma^{-1} (M_{vir})] - \ln[D(z) ] + B|^{\epsilon} ],
\end{equation}
where $M_{200}$ is the cluster mass within an overdensity of 200 times the critical density (see Equation \ref{eq:r200}; calculated at z=0), $A$, $B$ and $\epsilon$ are parameters that can be found by fitting to the simulation. They provide values for these parameters for both a $\Lambda \mathrm{CDM}$ cosmology and a nonrelativistic cold dark matter-dominated cosmology ($\Omega_{\mathrm{m}} = 1, \Omega_{\mathrm{\Lambda}} = 0$; denoted $\tau \mathrm{CDM}$). To account for $\Omega_{\mathrm{M}}$ varying with redshift in $\Lambda \mathrm{CDM}$, where it approaches 1 at high redshifts, they interpolate between the two sets of parameters by
\begin{equation}
A (\Omega_{\mathrm{m}} (z)) = (1 - x) A(\tau \mathrm{CDM}) + x A(\Lambda \mathrm{CDM}),
\end{equation}
with $x = (1 - \Omega_{\mathrm{m}}(z)) / 0.7$, using similar equations for $B$ and $\epsilon$.

\citet{2002Evrard} also provide a parameterized equation for the overdensity as a function of mass,
\begin{equation}
\ln \sigma ^{-1} (M_{200}) = - \ln \sigma_{15} + a \ln M_{200} + b (\ln M_{200})^{2},
\end{equation}
$\sigma (M_{200} ) = \exp \left( - \ln \sigma ^{-1} (M_{200}) \right)$, in which $\ln \sigma_{15} = \ln \sigma_{8} + \mathrm{const.}$, which is 0.578 for $\sigma_{8} = 0.9$, and for $\Lambda \mathrm{CDM}$ the parameters $a = 0.281$ and $b = 0.0123$. $M_{200}$ is the mass of the cluster within an overdensity of 200 in units of $10^{15} h^{-1} M_{\odot}$ The overdensity as a function of mass can be differentiated to give
\begin{equation}
\frac{d \ln \sigma^{-1} (M_{200})}{d \ln (M)} = b + 2c \ln M_{200}.
\end{equation}

\subsection{{\sc Pinocchio} simulations} \label{sec:pinocchio}
The PINpointing Orbit-Crossing Collapsed HIerarchical Objects ({\sc Pinocchio}) algorithm\footnote{Version 2.1.2 beta, available from \\ \url{http://adlibitum.oats.inaf.it/monaco/Homepage/Pinocchio/}} \citep{2002Monacoa,2002Monaco,2002Taffoni} can also be used to generate mock cluster catalogues. {\sc Pinocchio} uses Lagrangian perturbation theory to predict the collapse of matter into haloes. Galaxy cluster positions and masses generated by {\sc Pinocchio} will be associated with the large-scale cosmological structure, and hence will have a non-zero two-point correlation function (they will be clustered), whereas distributions created from analytical mass functions with random cluster positions will be Poisson distributed with zero two-point correlations. {\sc Pinocchio} does not simulate the structure within haloes, thus offering a significantly faster method of creating cluster catalogues compared with N-body techniques. In contrast to N-body simulations, requiring many thousands of CPU-hours to run, a reasonable resolution {\sc Pinocchio} run takes less than a day on a single CPU.

We have run 100 {\sc Pinocchio} simulations for three values of $\sigma_8$ -- 0.7, 0.825 and 0.9 -- each simulating comoving cubes with sides of 500~Mpc with a cell size of 1.25~Mpc (thus 400~cells to a side). Halo catalogues were created at 50~redshifts between $z=5$ and $z=0$, the centres of which are spaced 160~Mpc apart in comoving distance. The initial power spectrum used to generate the distribution of matter in the simulation is from \citet{1992Efstathiou}.

A set of 1000 lightcone catalogues for each value of $\sigma_8$ were generated from the simulation output. For each of the 50 redshifts, a halo catalogue is randomly selected from the 100~simulations. This halo catalogue is then translated by random amounts, independently in all three dimensions using a periodic box, as well as being rotated by a randomly chosen integer multiple of $90\degree$ independently along each of the three axes. A 160~Mpc deep slice of the simulation is then taken, and clusters within a $3\degree \times 3\degree$ block centred on the middle of the slice are included in the lightcone. The limit of $3 \degree$ to a side is imposed by the size of the simulation at the highest redshift; the box size subtends an angle of $3 \degree$ at $z=5$.

\begin{fig}
\centering
\includegraphics[scale=0.68]{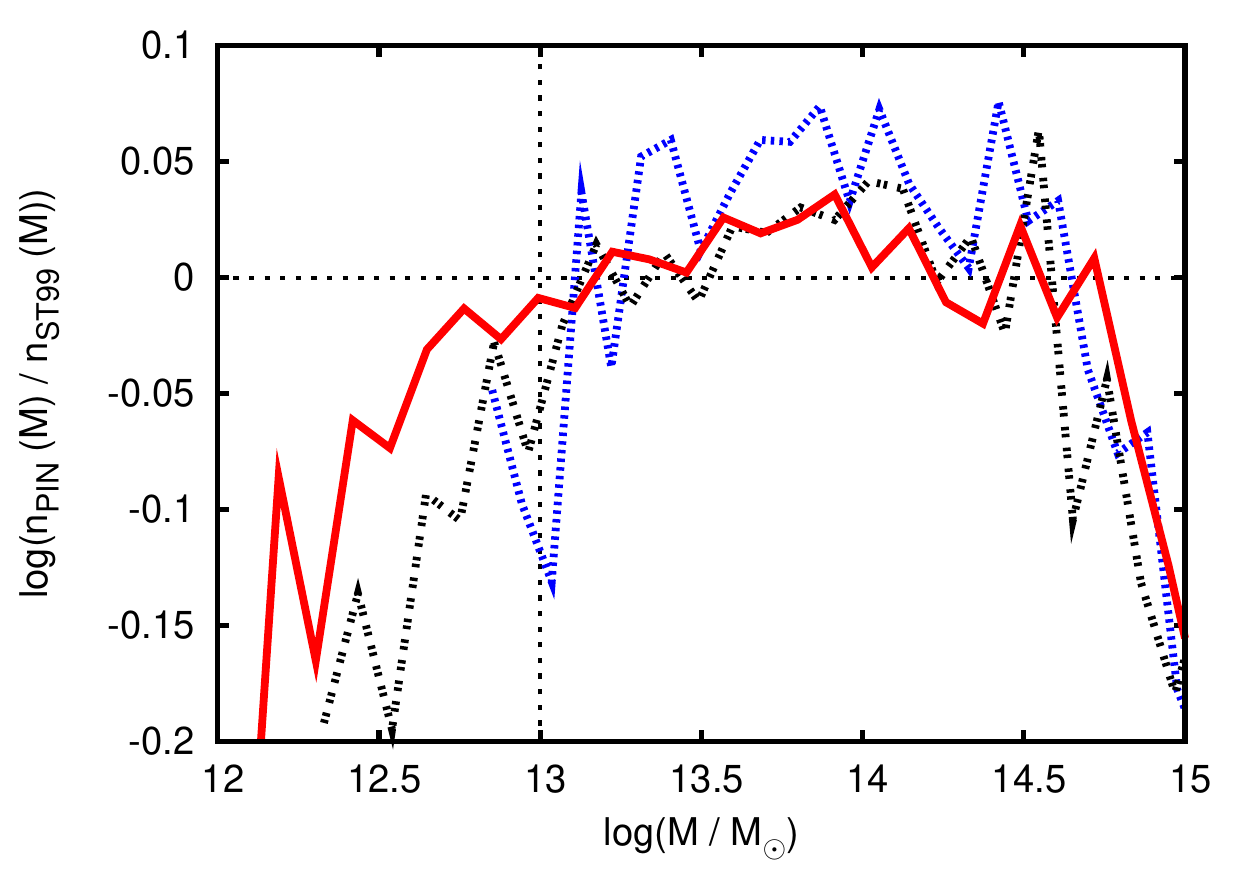}
\caption[The ratio of the {\sc Pinocchio} mass function to ST99 for different resolutions of the same {\sc Pinocchio} simulation]{The ratio of the {\sc Pinocchio} mass function to ST99 for different resolutions of the same {\sc Pinocchio} simulation. The resolutions are 2.5~Mpc (blue short dashed line), 1.66~Mpc (black dotted line) and 1.25~Mpc (red solid line; our adopted resolution). The latter two are consistent with each other down to $M_\mathrm{vir} \sim 10^{13} M_\odot$. At $M_\mathrm{vir}=10^{13} M_\odot$ the mass function from the highest resolution simulation is in good agreement with that from ST99 for a resolution of 1.25~Mpc. At the highest masses the scatter is caused by small number statistics, but a systematic trend downwards is still present.}
\label{fig:pinocchio_convergence}
\end{fig}

\begin{fig}
\centering
\includegraphics[scale=0.68]{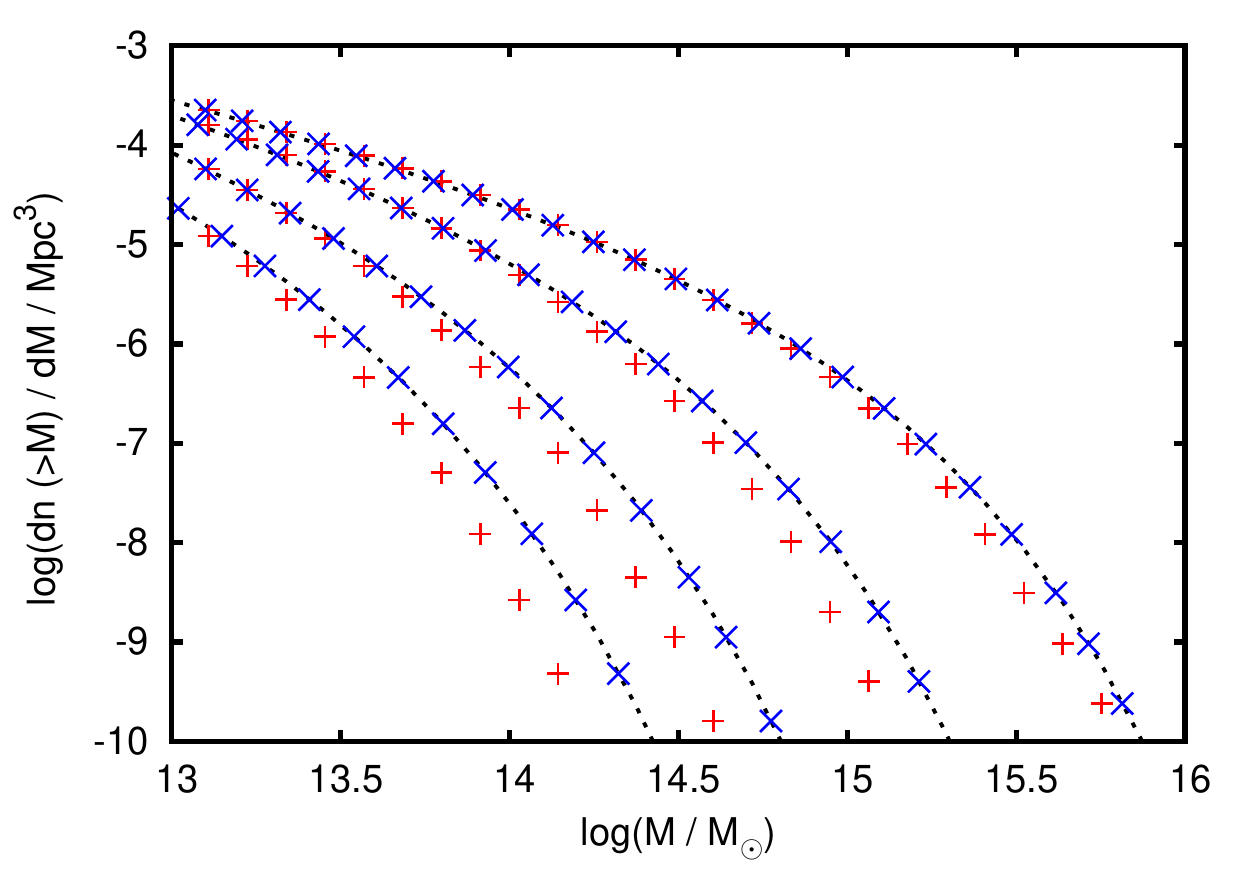}
\caption[The comoving differential number count of clusters greater than a given mass]{The comoving differential number count of clusters greater than a given mass analytically from ST99 (dashed lines), from the mean mass function of all {\sc Pinocchio} catalogues with $\sigma_8=0.9$ (red +) and from the corrected {\sc Pinocchio} catalogues (blue crosses). The number counts for four redshifts are given; from the top down these are z=0, 1, 2 and 3. The correction brings the two mass functions into agreement.}
\label{fig:pinocchio_discrepancy}
\end{fig}

The choice of simulation resolution means that the simulations have a high degree of completeness down to $M_\mathrm{vir} = 10^{13} M_\odot$; that is, the comoving number density of clusters $n(M) = dn/dM$ agrees between the {\sc Pinocchio} simulations and that expected from ST99, as shown in Figure \ref{fig:pinocchio_convergence}. However, {\sc Pinocchio} systematically underestimates the number of the largest mass clusters when compared to the analytical prediction of ST99; this becomes more pronounced at higher redshifts (see Figure \ref{fig:pinocchio_discrepancy}). This is unexpected since {\sc Pinocchio} is designed to reproduce ST99, and causes a significant discrepancy in the amplitude of the power spectrum of the SZ effect compared with maps created from random cluster catalogues generated from the ST99 number count predictions.

We assume that the ST99 mass function is the more accurate of the two, given the extensive comparisons that have been made between this mass function and numerical simulations (see e.g. \citealp{2006Warren}), although \citet{2006Warren} show that this may itself be systematically low at the highest masses compared to N-body simulations. We calculate a mass correction table for each cosmology such that the number of clusters greater than a given mass at each redshift (calculated from the average of all of the simulations with the same cosmology) matches the predicted value from ST99. The appropriate correction is then applied to all clusters in the lightcone prior to the creation of the maps.

\begin{fig}
\centering
\includegraphics[scale=0.5]{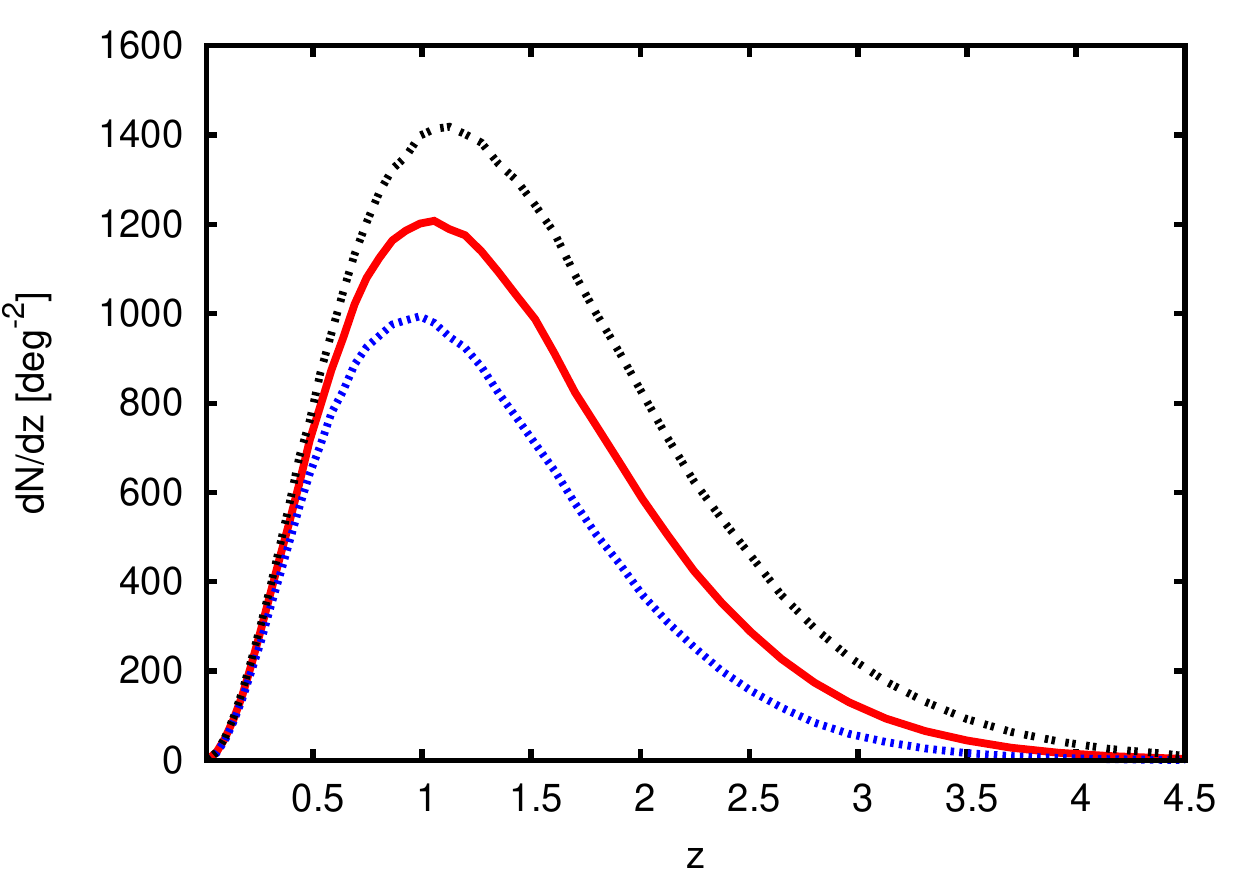}
\includegraphics[scale=0.5]{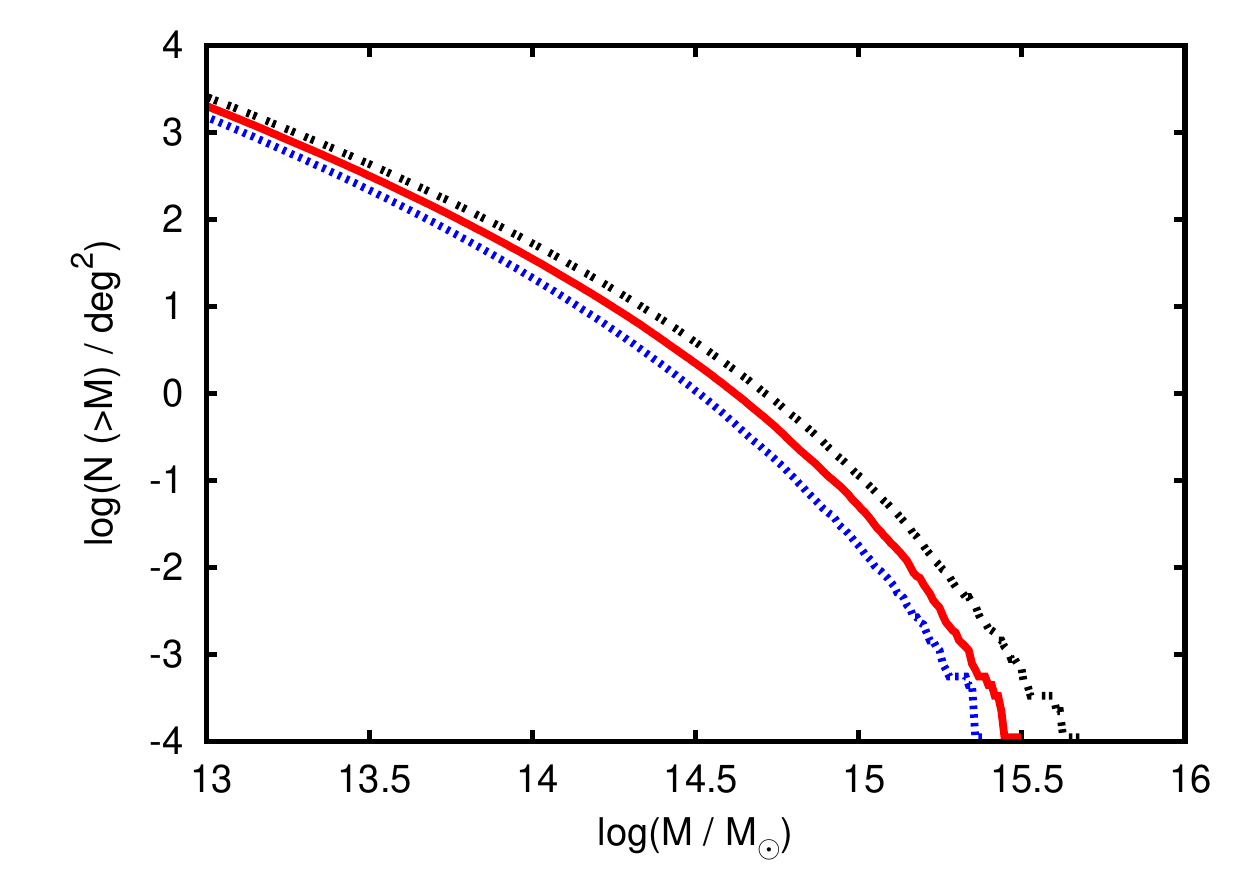}
\caption[The distribution of all clusters as a function of redshift (left) and mass (right) within the modified lightcone catalogues above $10^{13} M_\odot$]{The distribution of all clusters as a function of redshift (left) and mass (right) within the modified lightcone catalogues above $10^{13} M_\odot$, for each of the three values of $\sigma_8$ (from the top, 0.9 with black dots, 0.825 with the red solid line and 0.7 with blue large dots). Values from ST99 would overlay the points plotted; they are essentially the same (by construction).}
\label{fig:pinocchio_redshift_mass}
\end{fig}

The effect of this correction is shown in Figure \ref{fig:pinocchio_discrepancy}, which gives the differential number count of clusters greater than a given mass before and after the correction has been applied. The correction brings the {\sc Pinocchio} spectrum into agreement with that generated from a set of 1000 cluster catalogues with mass and redshift values drawn from the ST99 mass function and randomly determined positions. The average mass and redshift distributions of the corrected catalogues for all three cosmologies are given in Figure \ref{fig:pinocchio_redshift_mass}.

%%%

\section{Galaxy cluster model} \label{clustermodel}
Although clusters are complicated in the optical regime, for the purpose of modeling the SZ effect they can be represented very simply by assuming that the clusters are relaxed, virialised objects. Although clusters undergo major mergers, these are relatively infrequent, occurring in around 10 to 20 percent of the population depending on the redshift \citep{2007Kay}.

The cluster model we utilize is spherically symmetric and consists of two components: the dark matter and the gas. The former is used to calculate the surface mass density of the cluster, used for the inclusion of point sources, and the latter gives the thermal SZ effect. A kinetic SZ effect also exists, which is created by the relative motion of the clusters to the CMB, but this is likely to be negligible compared to the thermal SZ effect at the frequencies discussed in this paper and hence is not considered here.

\subsection{Gravitational collapse}
The virial radius, $r_{\mathrm{vir}}$, is used to define the size of a cluster. This can be calculated by using the spherical infall model \citep{1972Gunn}, which gives
\begin{equation} \label{eq:rvir}
r_{\mathrm{vir}} = \left( \frac{3 M_{\mathrm{vir}}}{4 \pi \Delta_{\mathrm{c}}(z) \rho_{\mathrm{crit}}(z)} \right) ^{1/3},
\end{equation}
where $M_{\mathrm{vir}}$ is the virial mass of the cluster, $\rho_{\mathrm{crit}}(z) = 3H^2(z) / \left(8 \pi G \right)$ is the critical density for the Universe to be flat and $H(z)$ is the Hubble parameter, given by the Friedmann equation, Equation \ref{eq:hubble}, for a flat universe (which we will assume throughout). The quantity $\Delta_{\mathrm{c}}$ is the mean matter density within the virial radius of the cluster in units of the critical density,
\begin{equation}
\Delta_{\mathrm{c}}(z) = \frac{\rho_{\mathrm{cluster}} (z)}{\rho_{\mathrm{crit}} (z)}.
\end{equation}
The value for $\Delta_{\mathrm{c}}$ depends on the cosmological parameters; in an Einstein-de Sitter universe it is exactly $18 \pi^2 \approx 178$. For our adopted cosmology, we use the fit given by \citet{1998Bryan},
\begin{equation}
\Delta_\mathrm{c} = 18 \pi^2 + 82 (\Omega_m(z) - 1) - 39 (\Omega_m(z) - 1)^2,
\end{equation}
where $\Omega_m(z) = \Omega_m (1 + z)^3 / E^2(z)$ is the matter density at redshift $z$.

The radius at which the cluster has an overdensity of 200 can also be calculated, via
\begin{equation} \label{eq:r200}
r_{\mathrm{200}}(z) = \left( \frac{3 M_{\mathrm{200}}}{4 \pi 200 \rho_{\mathrm{crit}}(z)} \right) ^{1/3},
\end{equation}
and similarly for other overdensity values.

Finally, the cluster radius needs to be converted to an angular size so that the cluster can be projected on to a virtual sky map. This is done by dividing the virial radius by the angular diameter distance, $\theta_{\mathrm{vir}} = r_{\mathrm{vir}} / d_{\mathrm{A}}(z)$ in which
\begin{equation}
d_{\mathrm{A}}(z) = \frac{c \int_{0}^{z} H^{-1}(z^\prime) dz^\prime}{(1 + z)}.
\end{equation}

\subsection{Dark matter}
We calculate the surface mass density of each galaxy cluster for the purpose of distributing point sources on to the map (see Section \ref{sec:pointsources}). For the dark matter density profile, we use the Navarro, Frenk and White (NFW) profile,
\begin{equation}
\rho_\mathrm{DM} \left( \frac{r}{r_\mathrm{s}} \right) = \frac{\rho_{\mathrm{crit}} \Delta_{\mathrm{c}}}{ \frac{r}{r_\mathrm{s}} \left( 1+ \frac{r}{r_\mathrm{s}} \right)^2}.
\end{equation}
This was found by fitting profiles to dark matter haloes from N-body/hydrodynamical simulations run by \citet{1995Navarro, 1996Navarro,1997Navarro}, and is now the standard profile for modeling dark matter haloes. The scale radius for the NFW profile is $r_{\mathrm{s}} = r_{\mathrm{vir}} / c_\mathrm{DM}$, where $c_\mathrm{DM}$ is the concentration parameter. Values for the concentration parameter depend weakly on the cluster mass with a certain amount of scatter \citep[see for example][]{2007Neto,2008Duffy}. However, as we are only using the surface mass density maps indirectly these effects are unimportant here, so we assume a constant value, $c_\mathrm{DM}=5$, which is appropriate for clusters.

The surface mass density at distance $\varphi_s = \theta / \theta_{\mathrm{s}}$ from the centre of the cluster, where $\theta_{\mathrm{s}} = r_{\mathrm{s}} / d_{\mathrm{A}}$ and $\theta$ is the angular distance from any given position in the sky to the centre of the cluster, is given by
\begin{equation}
\Sigma (\varphi_s) = \Sigma_{0} \zeta_{\mathrm{DM}} (\varphi_s).
\end{equation}
In this, the central surface mass density of a cluster can be calculated using (see for example \citealp{2001Lokas})
\begin{equation}
\Sigma_{0} = \frac{M_{\mathrm{vir}}}{2 \pi r_{s}^{2} (\ln(1 + c) - (c / (1 + c)))},
\end{equation}
and the NFW profile can be projected on to a 2D sky to give \citep{1996Bartelmann}
\begin{equation}
\zeta_{\mathrm{DM}} = \frac{2}{{\varphi_s^2 - 1}}\left\{ {\begin{array}{*{20}c}
 {1 - \frac{2}{{\sqrt {\varphi_s^2 - 1} }}\arctan \left( {\sqrt {\frac{{\varphi_s - 1}}{{\varphi_s + 1}}} } \right)} & {\varphi_s > 1}, \\
 {1 - \frac{2}{{\sqrt {1 - \varphi_s^2 } }}\arctan \left( {\sqrt {\frac{{1 - \varphi_s}}{{1 + \varphi_s}}} } \right)} & {\varphi_s < 1}, \\
 1 & {\varphi_s = 1}. \\
\end{array}} \right.
\end{equation}

\subsection{SZ effect}
Our gas model consists of two parts: a profile, and a normalization for the level of the SZ effect, which is dependent on the mass and redshift of the cluster. The method is analogous to that used for creating maps of the dark matter distribution.

We use the isothermal $\beta$-model \citep{1976Cavaliere} for the cluster profile,
\begin{equation} \label{eq:betamodel}
\xi_\mathrm{gas} \left(\frac{r}{r_\mathrm{c}} \right) = \left( 1+ \frac{r^2}{r_\mathrm{c}^2} \right)^{-\frac{3\beta}{2}},
\end{equation}
in which $\beta$ is a dimensionless parameter that measures the outer slope of the profile, $r_{\mathrm{c}} = r_{\mathrm{vir}} / c_\mathrm{gas}$ is the core radius that describes the turn-over point between the core and the power-law slope. The concentration of the SZ effect around the cluster centre is controlled by $c_\mathrm{gas}$. Fiducial values of $c_\mathrm{gas}=10$ and $\beta = 2/3$ are used, following \citet{2003Battye}. We truncate the profile at the virial radius, $r_{\mathrm{vir}}$, which prevents potential divergence in the gas mass \citep[see for example][]{1999Birkinshaw}. This may however underestimate the total SZ effect from an individual cluster, where infalling gas is shocked outside of the virial radius and prevented from falling inwards (see for example \citealp{1990Evrard,2005Kocsis}).

The change in the temperature of the CMB from the SZ effect can be calculated using
\begin{equation} \label{eq:conv_y_t}
\Delta T(\varphi_c) = y(\varphi_c) g(x) T_{\mathrm{CMB}}
\end{equation}
where $g(x) = (x / \tanh(x / 2)) - 4$, the dimensionless frequency $x = h_\mathrm{P} \nu / k_{\mathrm{B}}T_{\mathrm{CMB}}$, $k_{\mathrm{B}}$ is Boltzmann's constant, $h_\mathrm{P}$ is Planck's constant and $\nu$ is the frequency of interest. The present-day temperature of the CMB as measured by the FIRAS instrument on {\it COBE} is $T_{\mathrm{CMB}} = 2.728$K \citep{1996Fixsen}. The SZ effect along the line of sight from a single cluster depends on the integrated SZ effect from the cluster, $Y(M,z)$, and the cluster profile $\zeta(\varphi_c)$ as
\begin{equation} \label{eq:y_theta}
y(\varphi_c) = \frac{Y \zeta(\varphi_c) }{2 \pi \int_0^{\varphi_\mathrm{vir}} \zeta(\varphi_c^\prime ) \varphi_c^\prime d\varphi_c^\prime }
\end{equation}
where $\varphi_c = \theta / \theta_\mathrm{c}$ is the distance from the centre of the cluster in units of $\theta_{\mathrm{c}} = r_{\mathrm{c}} / d_{\mathrm{A}}$.

The integrated Y parameter is a measure of the total power of the SZ effect and is used as the normalization of our cluster profile. It is directly related to the total thermal energy of the gas. We assume a power-law relation between $Y$ and the cluster mass,
\begin{equation}
Y = \frac{Y_{*} h^{-1}}{d_A^2(z)} \left( \frac{M_\mathrm{vir}}{10^{14} h^{-1} M_\odot} \right)^\gamma \left( \frac{\Delta_c(z)}{\Delta_c(0)} E^2(z) \right)^{1/3} \left(1 + z \right)^\alpha,
\label{eq:clustermodel}
\end{equation}
where $Y_{*}, \gamma$ and $\alpha$ fix the normalisation, slope and redshift evolution respectively. For our fiducial values we adopt $Y_{*} = 2 \times 10^{-6}$ Mpc$^{2}$; the isothermal value of $\gamma = 5/3$ and $\alpha = 0$ such that the clusters are approximately self-similar. These choices are motivated by the results from recent cosmological simulations, which demonstrate that the relationship between $Y$ and $M$ is a power-law with small intrinsic scatter, being close to that predicted by the self-similar model at all redshifts, and relatively insensitive to the effects of cooling and heating of the intracluster medium (see, for example, \citealp{2004Silva}; \citealp{2005Motl}; \citealp{2006Nagai}).

Using the recent {\it Millennium Gas} simulations, Kay et al. (in prep.) obtain best-fitting values of $Y_*=2.3\times 10^{-6}$ Mpc$^2$ and $\gamma=1.64$ for a non-radiative simulation at $z=0$, and $Y_*=1.9\times 10^{-6}$ Mpc$^2$ and $\gamma=1.76$ for a simulation where the gas was preheated and allowed to cool radiatively. Both fits were applied to clusters with $M_{\rm vir}>10^{14}h^{-1}{\rm M}_{\odot}$, the objects that dominate the power spectrum over the range of multipoles of interest in this paper. These values agree well with the fiducial values above.

Finally, the $\beta$ profile can be projected from three dimensions to a 2D sky to become \citep{2003Battye}
\begin{equation}
\zeta_\mathrm{SZ}(\varphi_c) = \left( 1 + \varphi_c^2 \right)^{\frac{1}{2} - \frac{3\beta}{2}} \frac{J \left[\left(\frac{c^2 -\varphi_c^{2}}{1 + \varphi_c^{2}} \right)^{1/2}, \beta \right]}{J [c, \beta]},
\end{equation}
where
\begin{equation}
J[a, b] = \int_0^a \left( 1 + x^2 \right)^{-\frac{3 b}{2}} dx.
\end{equation}

\section{Example realizations and power spectra} \label{sec:example_realisations}

\begin{fig}
\centering
\includegraphics[scale=0.2,viewport=135 145 750 750,clip]{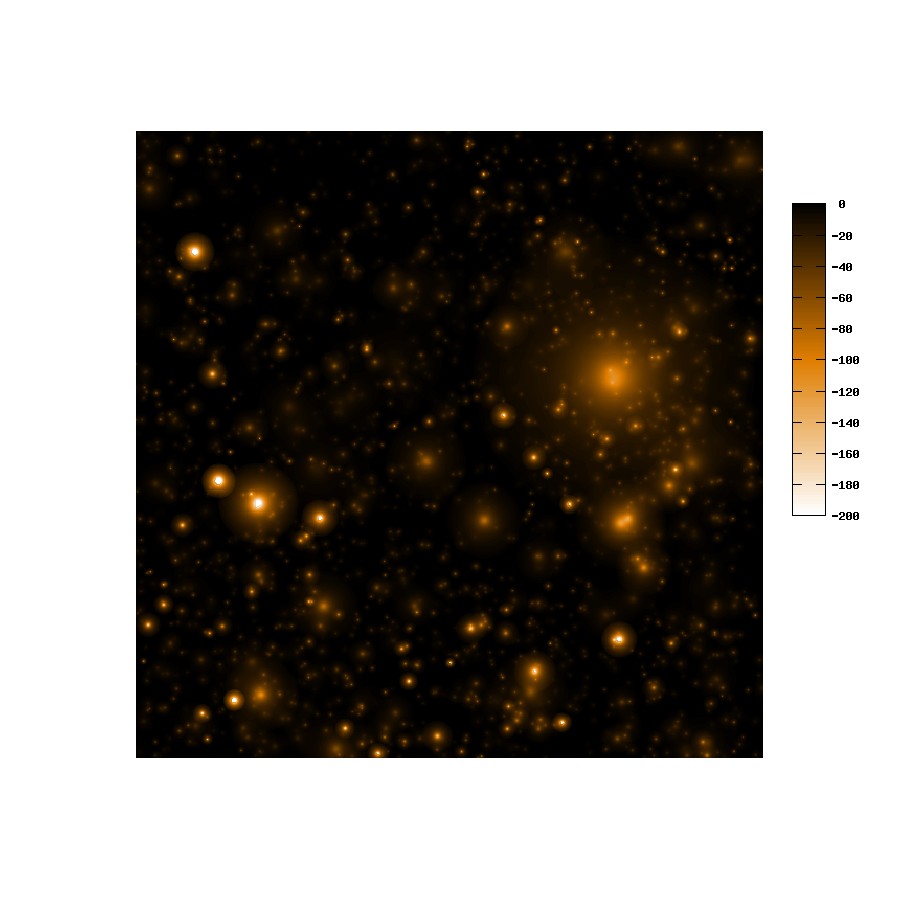}
\includegraphics[scale=0.2,viewport=135 145 750 750,clip]{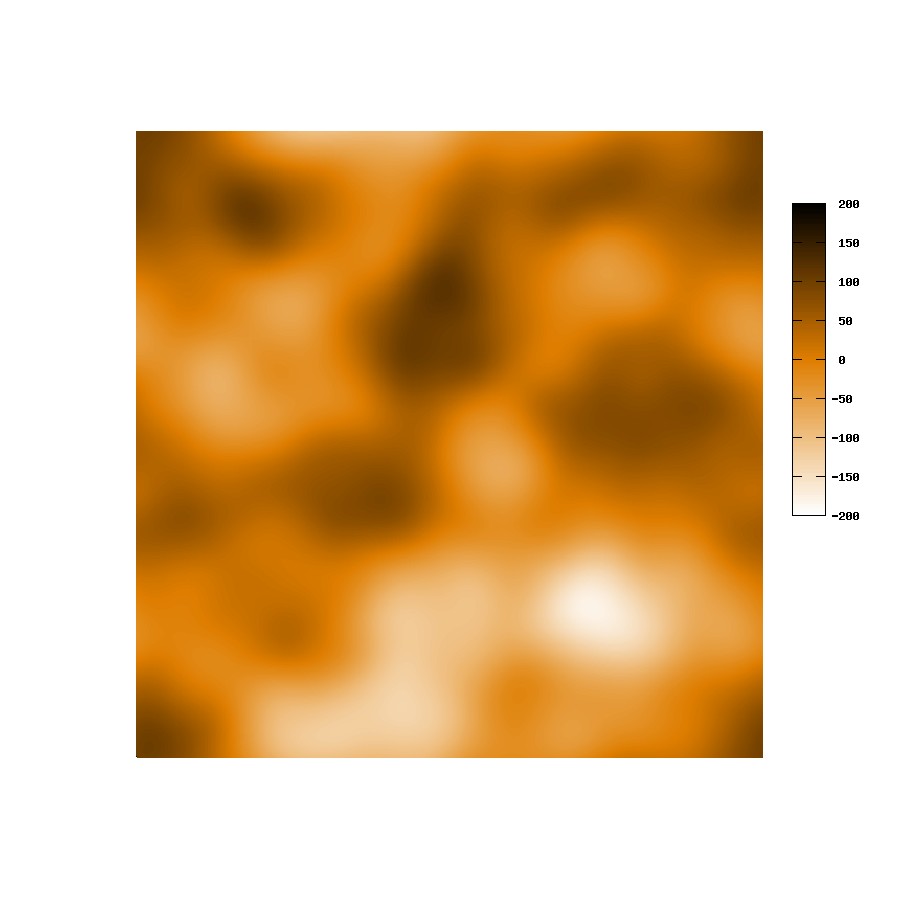}
\includegraphics[scale=0.2,viewport=135 145 750 750,clip]{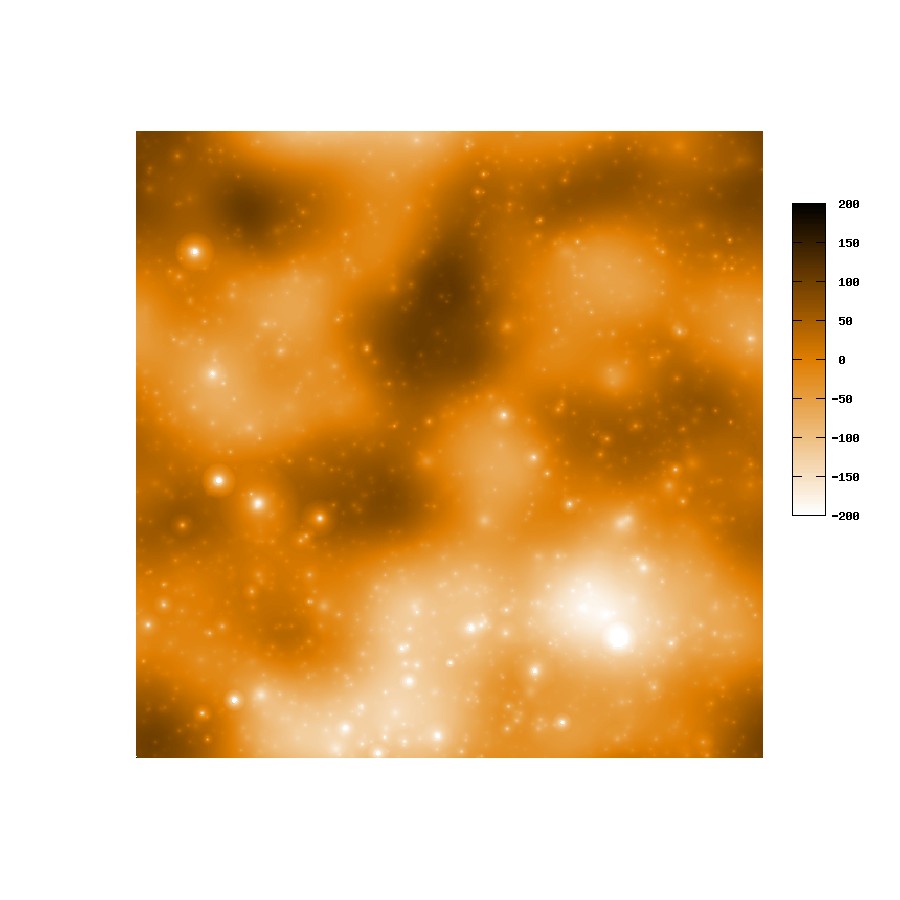}
\caption[A realization from the $\sigma_8 = 0.825$ cosmology using a clustered {\sc Pinocchio} lightcone with mass corrections]{A realization from the $\sigma_8 = 0.825$ cosmology using a clustered {\sc Pinocchio} lightcone with mass corrections. The maps are $1\degree \times 1\degree$, with a resolution of 6 arcseconds. From left to right the maps are of the temperature decrement from the SZ effect, the CMB and the two combined. White represents colder areas, with black representing hotter areas. The colours range from 0 to -200 $\upmu$K for the first image and 200 to -200 $\upmu$K for the other two.}
\label{fig:map_set}
\end{fig}

Figure \ref{fig:map_set} shows $1\degree \times 1\degree$ maps of the SZ effect and the CMB separately and combined for a single realization from the $\sigma_8 = 0.825$ cosmology, using the {\sc Pinocchio} simulations. In the SZ map, an $M_\mathrm{vir} = 5.4 \times 10^{14} M_\odot$ galaxy cluster at a redshift of $z=0.13$ lies in the top right, with a virial radius of 14 arcmin. The SZ map contains 1933 galaxy groups and clusters in total, with a maximum decrement in a single pixel of $450 \upmu$K and an average of $8.5 \upmu$K per pixel across the map.

For the $3\degree \times 3\degree$ realizations with $\sigma_8 = 0.75$ and a resolution of 18 arcseconds, there are $13~470\pm300$ galaxy groups and clusters per realization, where the standard deviation is from the scatter between the realizations. These objects cause an average decrement per pixel of $4.3\pm0.5 \upmu$K in each map (again with a standard deviation from the scatter). For $\sigma_8 = 0.825$ this increases to $18~040\pm350$ galaxy groups and clusters and an average of $6.5\pm1.3 \upmu$K, and for $\sigma_8 = 0.9$ this further increases to 23~000$\pm$400 galaxy groups and clusters and an average of {$9.3\pm 1.3 \upmu$K. The increase in the number of objects, and hence the total amount of power from the SZ effect, is due to the increase in the clustering of matter as $\sigma_8$ increases.

\begin{fig}
\centering
\includegraphics[scale=0.5]{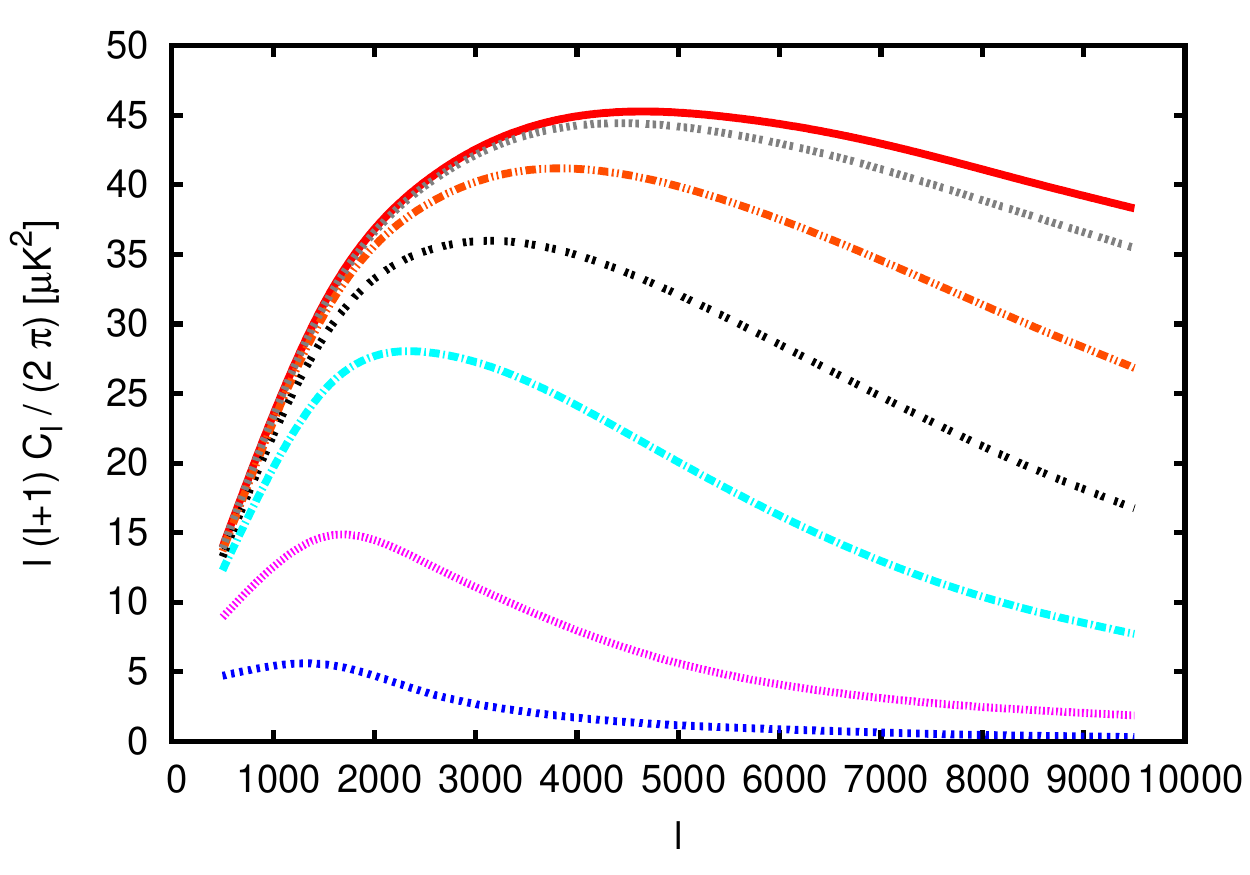}
\includegraphics[scale=0.5]{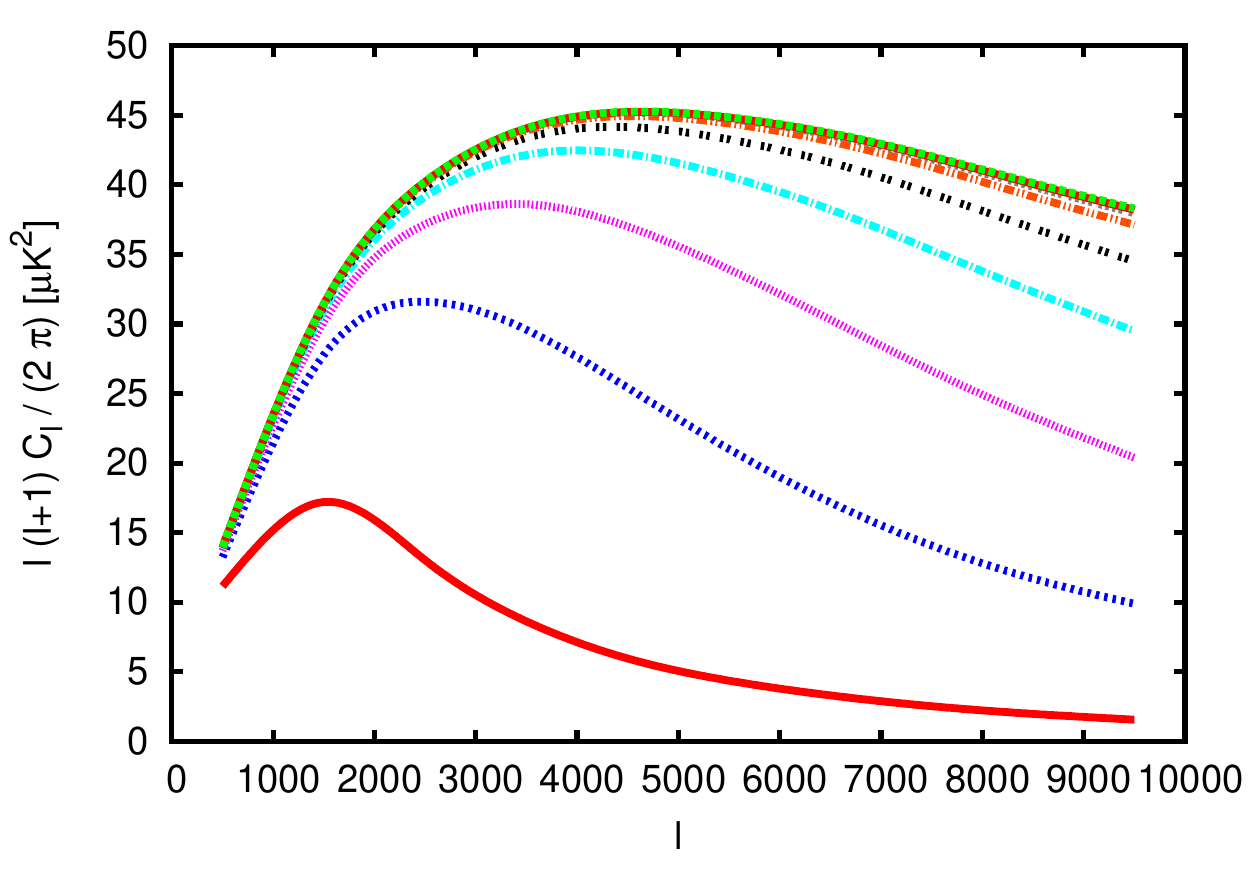}
\caption[The mean power spectra from the SZ effect in the $\sigma_8=0.825$ cosmology imposing various minimum mass and maximum redshift limits]{Left panel: The mean power spectra from the SZ effect in the $\sigma_8=0.825$ cosmology imposing various minimum mass limits: from the bottom up, $1 \times 10^{15}$, $5 \times 10^{14}$, $2 \times 10^{14}$, $1 \times 10^{14}$, $5 \times 10^{13}$, $2 \times 10^{13}$ and $1 \times 10^{13} M_\odot$. Right panel: The same, but imposing a maximum redshift cutoff of (from the bottom up) $z=$ 0.5, 1.0, 1.5, 2.0, 2.5, 3.0, 3.5, 4.0 and 4.5. These results show that the spectrum is approximately converged for $M>10^{13} M_\odot$ and $z<3.5$.}
\label{fig:convergence_testing}
\end{fig}

To confirm that the power spectra calculated from these realizations are converged such that no significant amount of power comes from clusters with lower masses or higher redshifts, the mean spectra from realizations with various minimum mass and maximum redshift cutoffs imposed are shown in Figure \ref{fig:convergence_testing} for $\sigma_8 = 0.825$. The spectra are converged to within a few $\upmu$K$^2$ over the multipole range of interest when clusters down to $10^{13} M_\odot$ and out to a redshift $z=3.5$ are included (the effects of reducing the {\it maximum} mass cutoff will be discussed in Section \ref{sec:upper_mass}). We use $10^{13} M_\odot$ and $z=4.5$ as the fiducial limits for the realizations shown in the rest of this paper. Although the average SZ effect across the map is converged by the same redshift, it is not converged at a minimum mass of $10^{13} M_\odot$; lower mass objects than considered here will significantly contribute towards the average SZ effect.

As an additional check, we compare the ratio of the mean power spectra between realizations with different values of $\sigma_8$. \citet{2002Komatsu} found that $C_l$ scales as $\sigma_8^\alpha$, where $\alpha \approx 7$. Fitting for $\alpha$ for our realizations and averaging over the multipole bins, we find that between the $\sigma_8 = 0.75$ and $0.825$ realizations, $\alpha \approx 7.1$. Between $\sigma_8 = 0.90$ and $0.825$ the value is $\alpha \approx 6.9$. Thus there is fairly good agreement, although the relationship appears not to be a perfect power law over this range of $\sigma_8$.

\section{Theoretical power spectra} \label{sec:theoretical_ps}
Rather than calculating the mean power spectrum from maps of the SZ effect, it can be evaluated directly from the number density and cluster profile using the halo formalism \citep[see for example][]{2001Cooray,2002Komatsu} under the assumption that the clusters are Poisson distributed. The mean binned spectra can be calculated by
\begin{equation}
B_i = \frac{1}{N}\sum_{l \in \mathrm{bin}} \frac{l(l+1)}{2 \pi} \int_{\mathrm{z_{min}}}^{\mathrm{z_{max}}} \int_\mathrm{M_{min}}^\mathrm{M_{max}} \frac{dV}{dz} \frac{dn}{dM} y_{l}^2 dM dz,
\end{equation}
where $i$ is the bin number and the sum is over the multipoles $l$ in the bin.
\begin{equation}
\frac{dV(z)}{dz} = \frac{c^3 \left( \int_0^z E(z^\prime) dz^\prime \right)^2}{E(z) H_0^3}
\end{equation}
is the comoving volume of the Universe at a given redshift, where $E(z)$ is as in Equation \ref{eq:hubble} and $dn / dM$ is the comoving cluster number density. The parameter $y_{l}$ is the value of the Fourier transform of the cluster model at the multipole $l$, given by
\begin{equation}
y_{l}(M,z) = 2 \pi g(x) \int_0^{\varphi_\mathrm{vir}} \varphi_c y(\varphi_c) J_0 \left( (l+0.5) \varphi_c \right) d \varphi_c,
\end{equation}
where $g(x)$ determines the frequency dependence of the SZ effect, as per Equation \ref{eq:conv_y_t}, $y(\varphi_c)$ is from Equation \ref{eq:y_theta} and $J_0$ is zeroth-order cylindrical Bessel function.

Using the covariance matrix, the standard deviation can also be calculated in a similar way. Here, the angular trispectrum $T_{ll^\prime}$ is used, defined by \citep{2001Cooray,2002Komatsu}
\begin{equation}
T_{ll^\prime} = \int_{\mathrm{z_{min}}}^{\mathrm{z_{max}}} \int_\mathrm{M_{min}}^\mathrm{M_{max}} \frac{dV}{dz} \phantom{.} \frac{dn}{dM} \phantom{.} y_{l}^2 \phantom{.} y_{l^\prime}^2 \phantom{.} dM dz.
\end{equation}
This is then combined with the expected Gaussian cosmic variance from the mean spectrum to calculate the covariance matrix $M_{ll^\prime}$ via
\begin{equation}
M_{ll^\prime} = \frac{1}{f_\mathrm{sky}} \left( \frac{2 C_l C_{l\prime}}{(2l+1) \Delta l}\delta_{ll^\prime} + \frac{T_{ll^\prime}}{4 \pi} \right),
\end{equation}
where $\Delta l$ is the width of the multipole bin and $f_\mathrm{sky}$ is the fraction of the sky being considered. The standard deviation can then be calculated from the diagonal of the covariance matrix, that is, $\sigma_l = l(l+1) \sqrt{M_{ll}} / 2 \pi$. This effectively includes both Gaussian and Poissonian terms. Note that \citet{2007Zhang} add an additional component to the calculation of the variance to model the clustering of galaxy clusters but find that the Poissonian term dominates over the clustering term; for this reason we do not consider it here.

\begin{fig}
\centering
\includegraphics[scale=0.68]{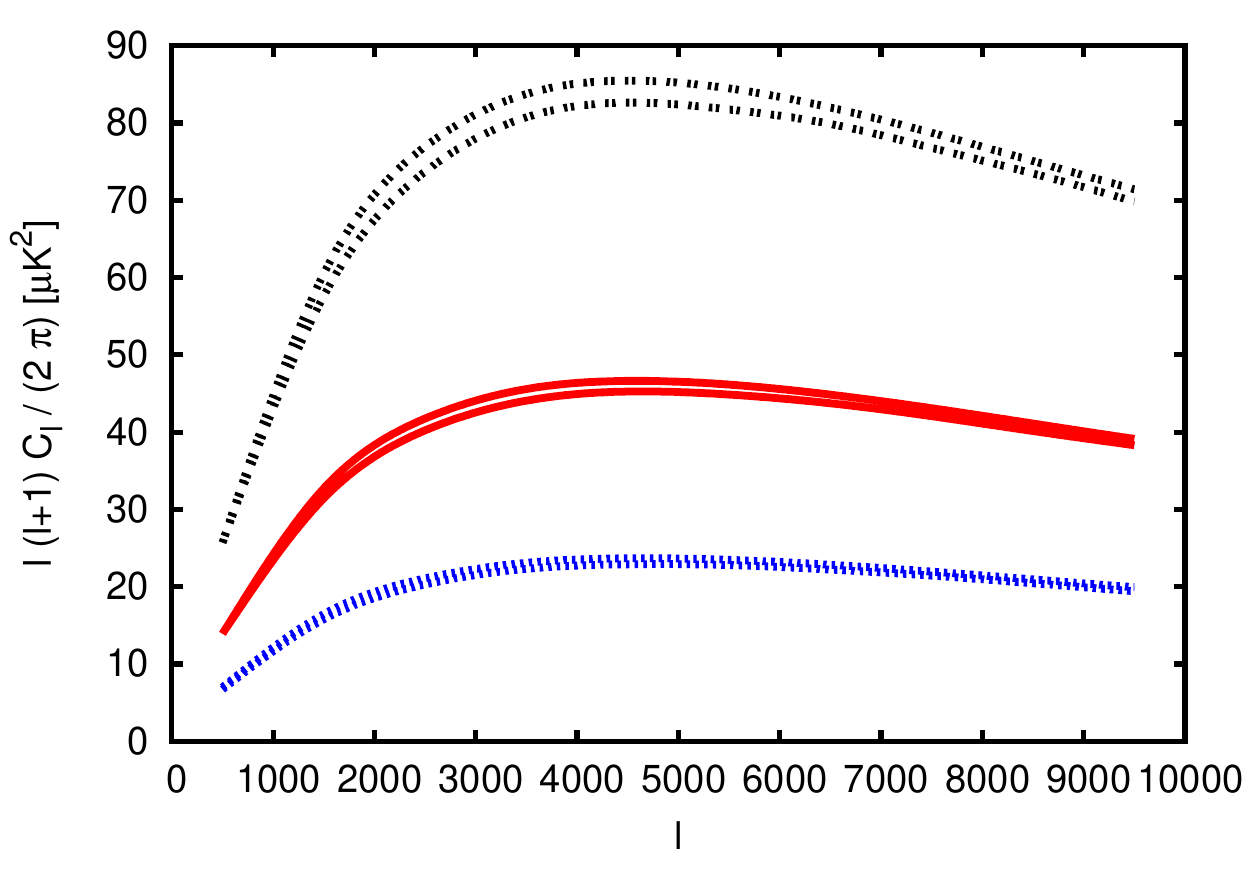}
\caption[The power spectra expected from the SZ effect as calculated using the analytical formula and the modified {\sc Pinocchio} realizations for each cosmology]{The power spectra expected from the SZ effect as calculated using the analytical formula and the modified {\sc Pinocchio} realizations for each cosmology. The blue dashed line shows $\sigma_8 = 0.75$, the red solid line $0.825$ and the black dotted line $0.9$. The higher values of the pairs are from the theoretical spectrum, the lower from the realizations; there is consistency between the shapes of the curves and the analytical formula is $2-4$ per cent higher than the realizations.}
\label{fig:comp_theory_mean}
\end{fig}

Figure \ref{fig:comp_theory_mean} compares the values computed using this method with those from the realizations for the three different cosmologies. The analytical method predicts $\sim2-4$ per cent more power than the realizations, which is likely due to inefficiencies in the creation of the maps, but it appears that there is broad consistency here.

\section{Point sources at microwave frequencies} \label{sec:pointsources}
Foreground emission from extragalactic point sources will contaminate the map. Synchrotron is the dominant process at frequencies below around 90~GHz, with emission from dust becoming important at higher frequencies. Although these sources are physically extended, this is generally much smaller than the beam size of the telescopes used to measure the CMB and SZ effect, so that they are well approximated as point sources.

We calculate the total number of point sources with a flux density between $S_{\mathrm{min}}$ and $S_{\mathrm{max}}$ within an area $\Delta \Omega$ on the sky by
\begin{equation}
N_{\mathrm{tot}} = \Delta \Omega \int_{S_{\mathrm{min}}}^{S_{\mathrm{max}}} \frac{dN}{dS_{\nu}} dS_{\nu},
\end{equation}
where $dN/dS_\nu$ is the differential source counts as a function of flux density. To calculate the flux densities of individual sources, we randomly sample from a probability distribution defined by a normalized $dN/dS_\nu$ between the flux density limits.

The flux densities can be converted to the thermodynamic temperatures at a given frequency $\nu$ using
\begin{equation}
T_\nu =S_{\nu} / \left( \theta_{\mathrm{pixel}}^{2} \frac{dB}{dT} \right),
\end{equation}
where $\theta_{\mathrm{pixel}}^{2}$ is the area of the pixel containing the point source and 
\begin{equation}
\frac{dB}{dT} = \frac{2k}{c^{2}} \left( \frac{kT_{\mathrm{CMB}}}{h} \right) ^{2} \frac{x^{4} e^{x}}{(e^{x} -1)^{2}},
\end{equation}
is the differential of the Planck function with respect to temperature, where $x$ is the dimensionless frequency, given by $x = h_\mathrm{P} \nu / k_{\mathrm{B}} T_{\mathrm{CMB}}$.

The theoretical power spectrum expected from the point sources can be calculated by \citep{2003White}
\begin{equation} \label{eq:ps_spectrum}
C_l = \left(\frac{dB}{dT} \right)^{-2} \int_{S_{\mathrm{min}}}^{S_{\mathrm{max}}} S_{\nu}^{2} (dN/dS_{\nu}) dS_{\nu},
\end{equation}
where $S_{\nu}$ is the flux of the point source at frequency $\nu$.

\subsection{Low frequency point sources} \label{sec:lowfreqps}
The majority of surveys for radio sources have been done at low frequencies; there are many less at high frequencies (i.e. the GHz range) as sensitive telescopes have a much smaller field of view at these frequencies, meaning that the surveys take much longer.

The most common way of parameterising the $dN/dS$ of radio sources is by using a single power law of the form
\begin{equation} \label{eq:dndsradio}
\frac{dN}{dS_{\nu}} = \frac{N_{0}}{S_{0}} \left( \frac{S_{\nu}}{S_{0}} \right) ^{-\alpha}.
\end{equation}
\citet{2005Cleary} provide a fit to observations by the VSA, at 33GHz where $S_{0} = 70 \mathrm{mJy}$, $\alpha = 2.34^{+0.25}_{-0.26}$ and $N_{0} / S_{0} = 10.6^{+2.3}_{-2.2} \phantom{.} \mathrm{mJy^{-1} sr^{-1}}$.

\begin{fig}
\centering
\includegraphics[scale=0.68]{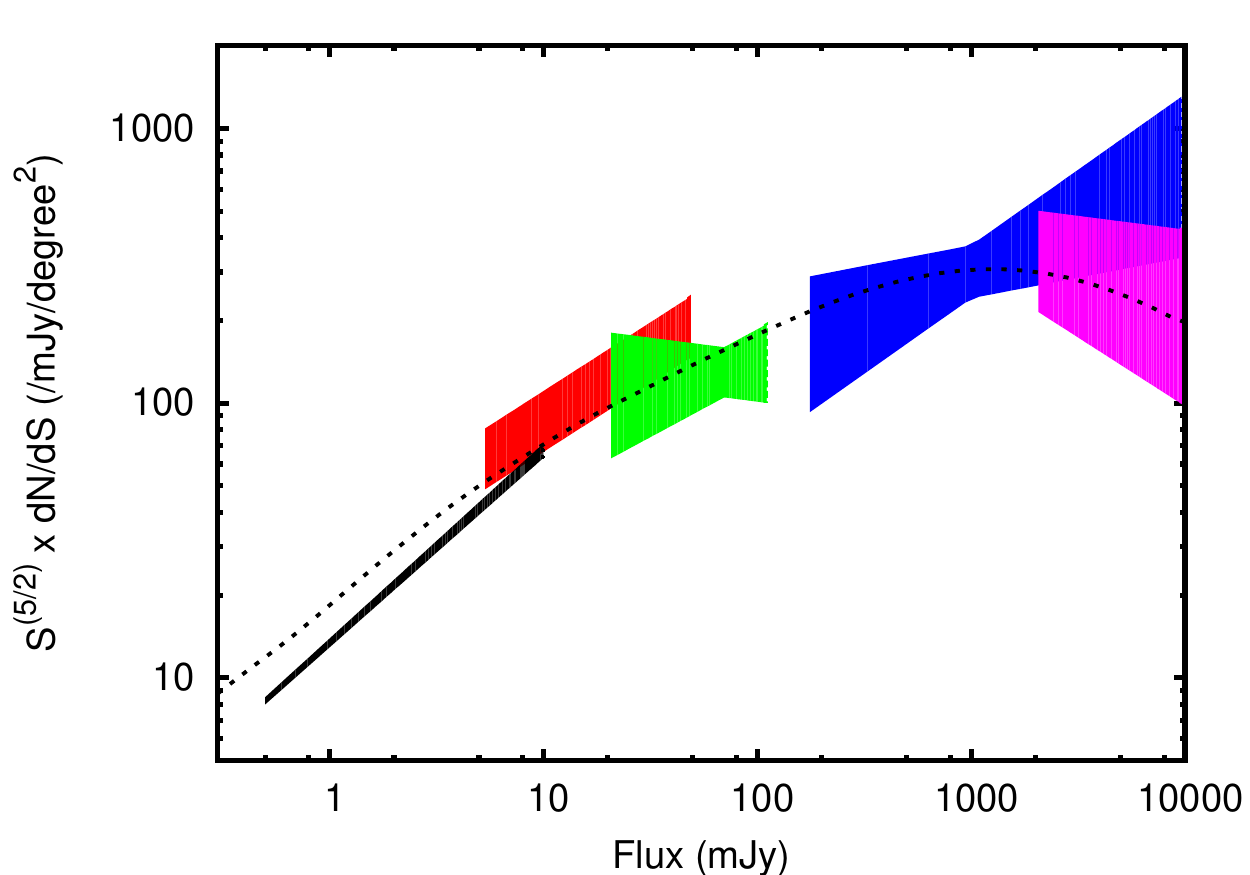}
\caption[Comparison of the 30~GHz differential source counts]{Comparison of the 30~GHz differential source counts from \citet{1998Toffolatti} rescaled by a factor of 0.7 (black dotted line) with the $1 \sigma$ ranges from measurements by (right-to-left) {\it WMAP} Ka band \citep[][pink]{2003Bennett}, DASI \citep[][blue]{2002Kovac}, VSA \citep[][green]{2005Cleary}, CBI \citep[][red]{2003Mason} and GBT \citep[][black]{2009Mason}}
\label{fig:pointsource_comparison}
\end{fig}

Here, we use the 30~GHz model for $dN/dS_\nu$ from \citet[][henceforth T98]{1998Toffolatti}. This is an extrapolation of the observed number counts at 1.4~GHz, which have been measured to much lower flux densities than the number counts at the frequencies of interest here. We have normalized the model by a factor of 0.7 \citep[following][]{2005Cleary} to bring it into closer agreement with measurements of the number count at 30~GHz (see Figure \ref{fig:pointsource_comparison}). Although this still results in an over-prediction of the number of sources around 1~mJy compared with recent observations using the Green Bank Telescope \citep{2009Mason}, we note that \cite{2009Muchovej} find a higher number of sources than T98 using the Sunyaev-Zel'dovich Array at the same flux density level.

We use $S_{\nu} / S_{\mathrm{30~GHz}} =$ $ (\nu / 30~\mathrm{GHz})^{\alpha}$ to scale the flux densities to the desired frequency $\nu$. The spectral index $\alpha$ is determined for each source using a Gaussian distribution about a mean $\overline \alpha = -0.3$, with $\sigma_{\alpha} = 0.36$, for each source, which is the distribution for 15 to 30~GHz measured for the 9C sample (Cleary, private communication). We include point sources with flux densities between $10^{-4.5}$ and $10^{5}$~mJy at 30~GHz; sources above this flux density range are very rare. Sources with weaker flux densities are not well characterized and should have a negligible contribution to the power spectra.

\subsection{High frequency point sources}
Although negligible at the fiducial frequency of 30~GHz, dusty galaxies become important at higher frequencies. For these, we use the fit to SCUBA observations provided by \citet{2003Borys}. This is in the form of a double power law,
\begin{equation}
\frac{dN}{dS_{\nu}} = \frac{N_{0}}{S_{0}} \left[ \left( \frac{S}{S_{0}} \right) ^{\alpha} + \left( \frac{S}{S_{0}} \right) ^{\beta} \right]^{-1},
\end{equation}
where the parameters at 350~GHz are $S_{0} = 1.8 \mathrm{mJy}$, $\alpha = 1.0$, $\beta = 3.3$ and $N_{0} = 1.5 \times 10^{4} \phantom{.} \mathrm{deg}^{-2}$. This is extrapolated to other frequencies using $S_{\nu} = \nu^{\gamma}$, where $\gamma = 2.5$. We use the flux range $10^{1}$ to $10^{5}$~mJy at 350~GHz.

\subsection{Spatial distribution}
\begin{fig}
\centering
  \includegraphics[scale=0.19,viewport=135 145 740 740,clip]{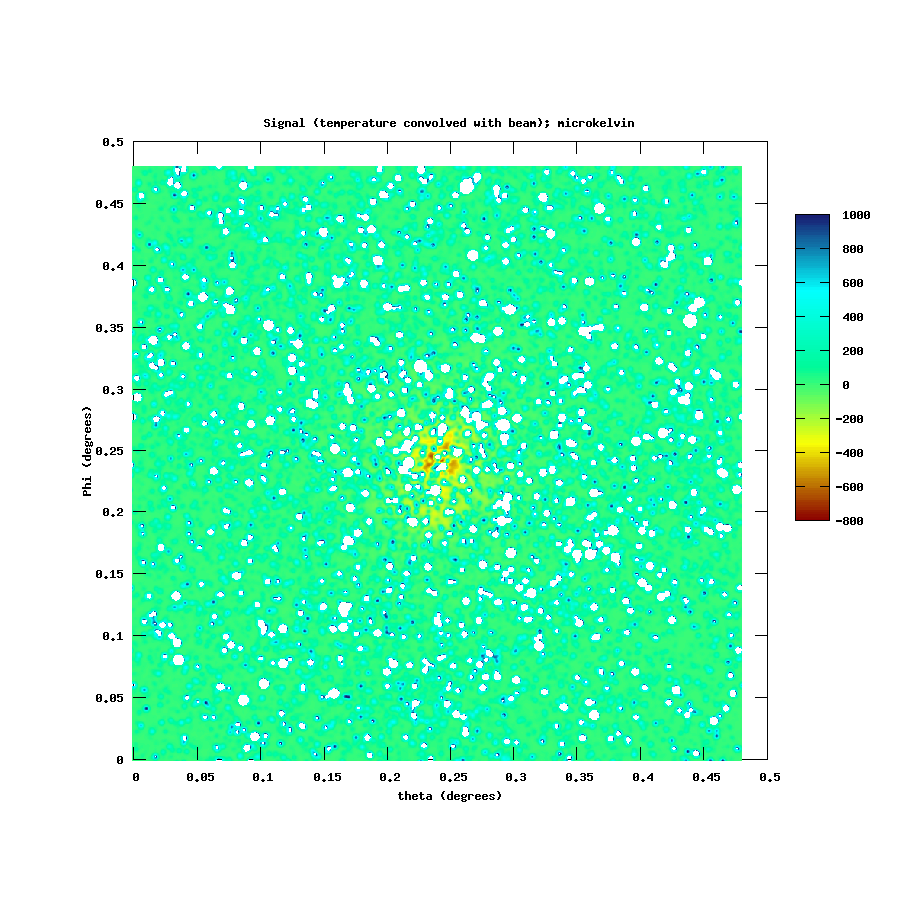}
  \includegraphics[scale=0.19,viewport=135 145 740 740,clip]{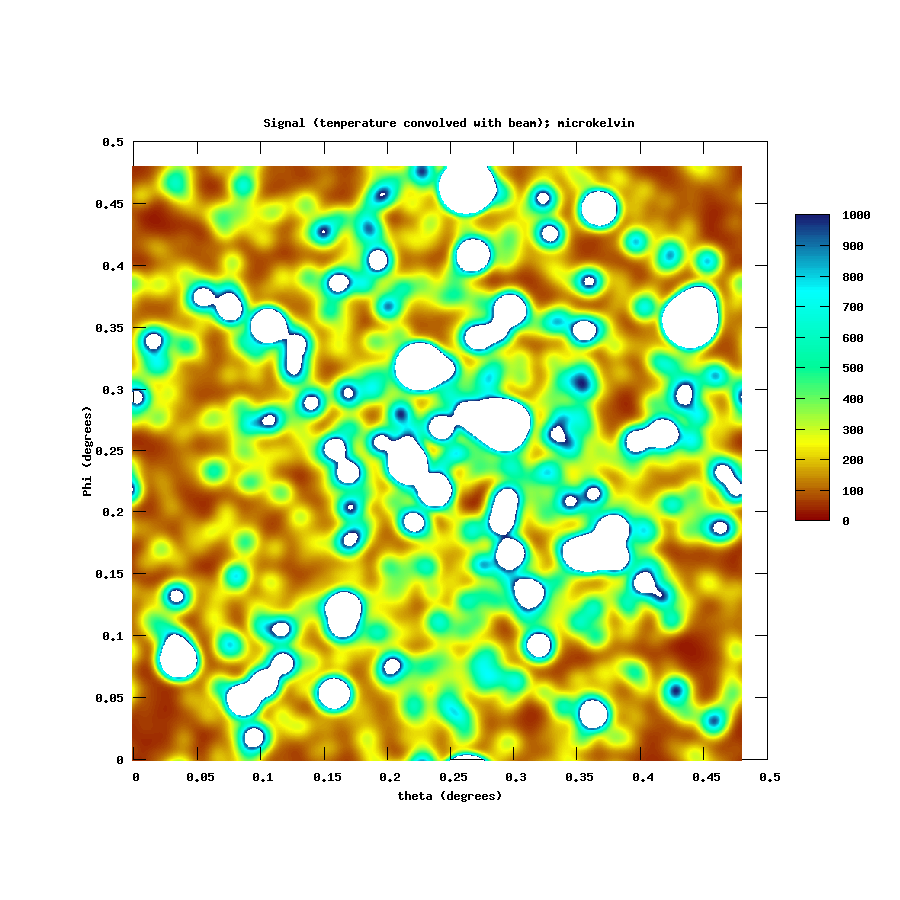}
  \includegraphics[scale=0.19,viewport=135 145 740 740,clip]{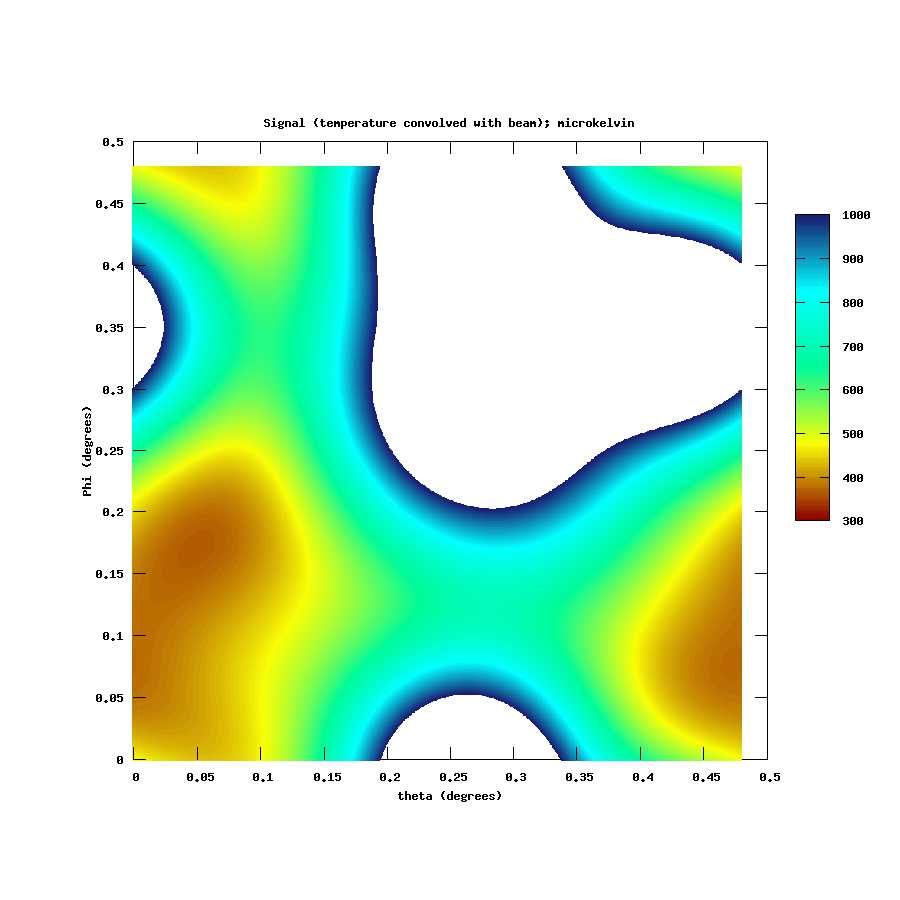}
  \\
  \includegraphics[scale=0.19,viewport=135 145 740 740,clip]{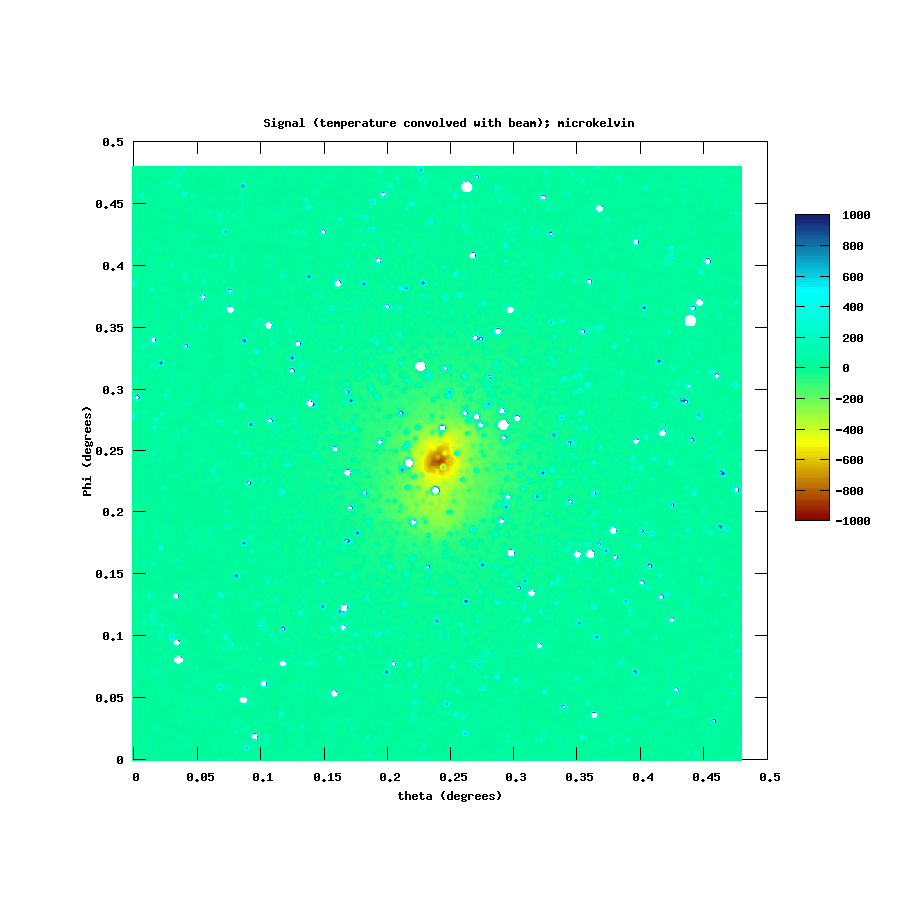}
  \includegraphics[scale=0.19,viewport=135 145 740 740,clip]{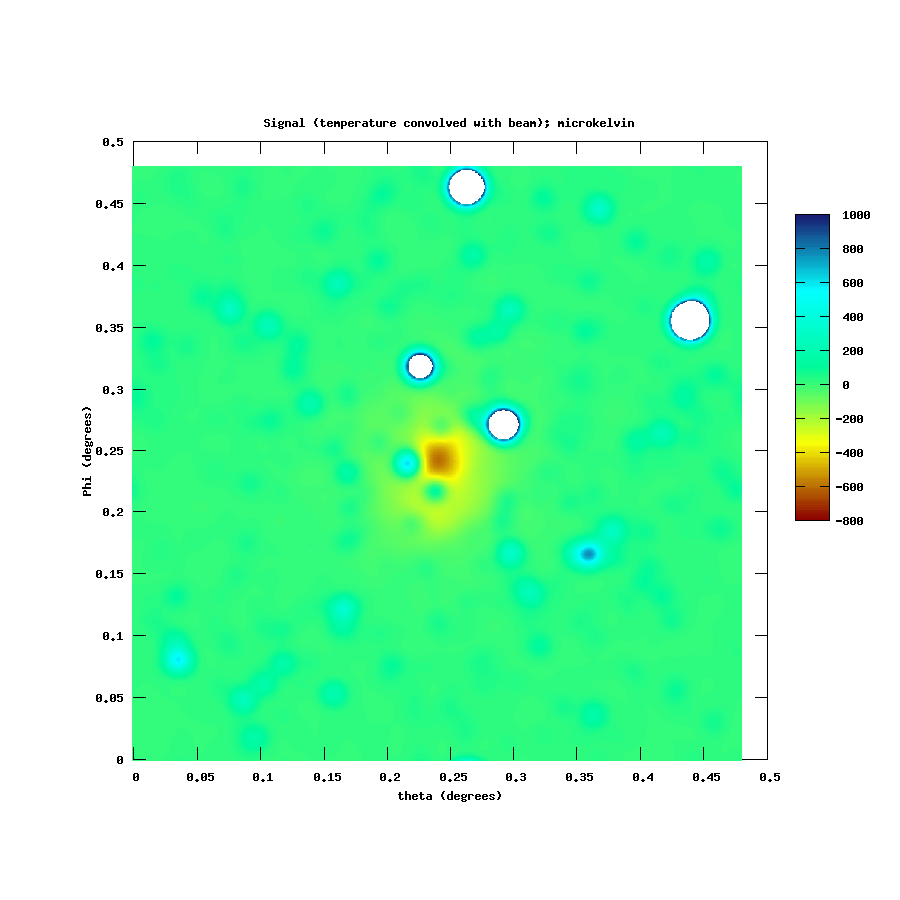}
  \includegraphics[scale=0.19,viewport=135 145 740 740,clip]{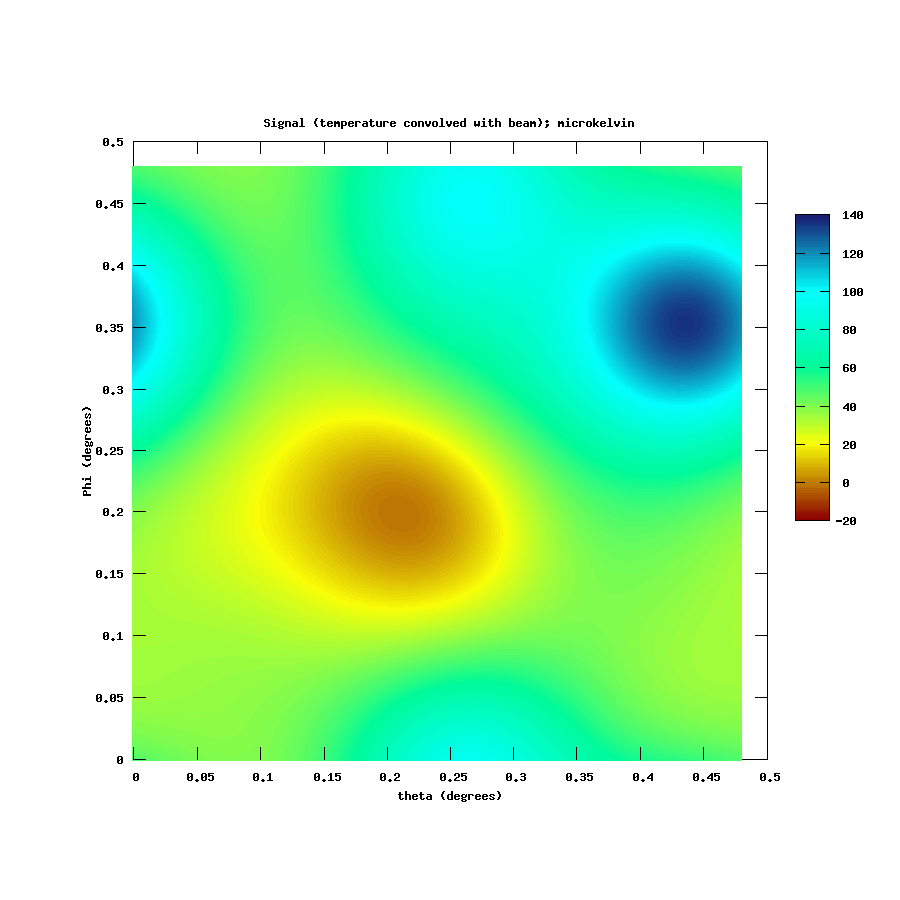}
  \\
  \includegraphics[scale=0.19,viewport=135 145 740 740,clip]{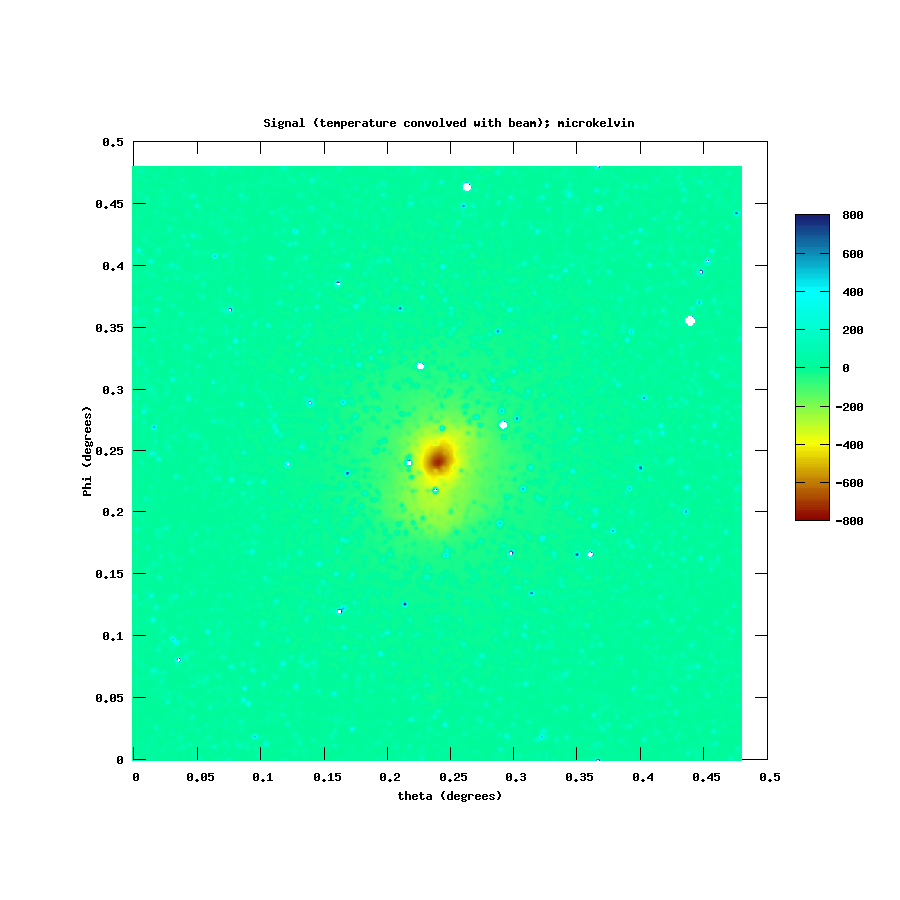}
  \includegraphics[scale=0.19,viewport=135 145 740 740,clip]{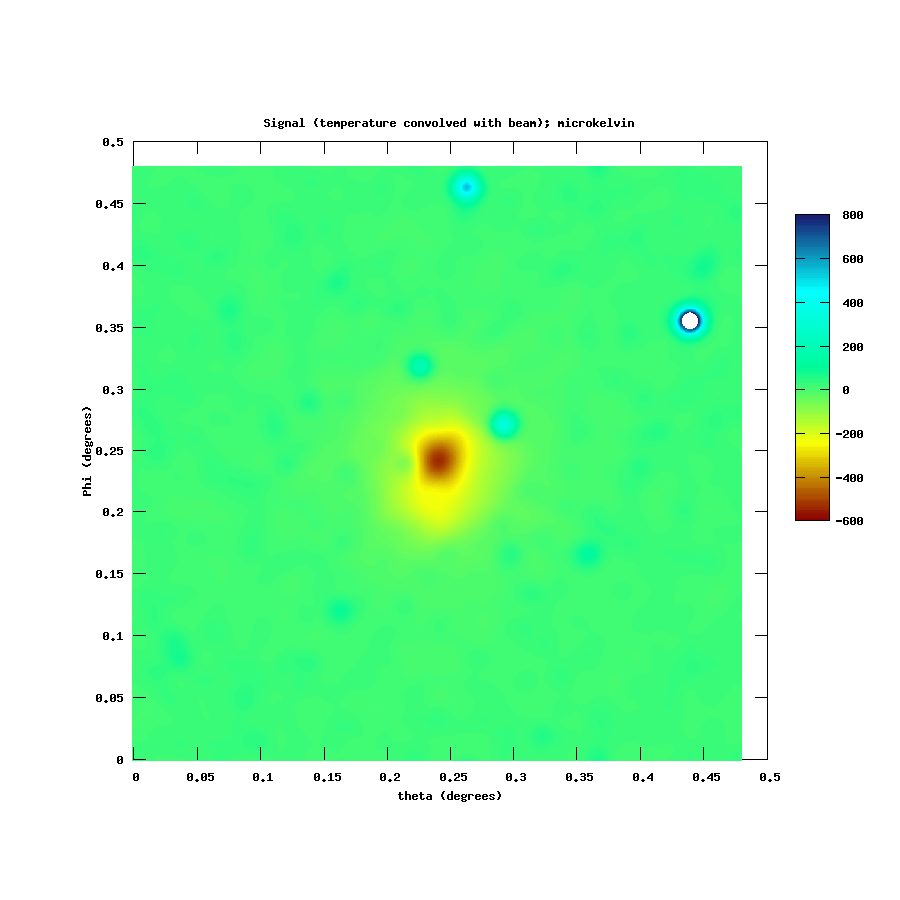}
  \includegraphics[scale=0.19,viewport=135 145 740 740,clip]{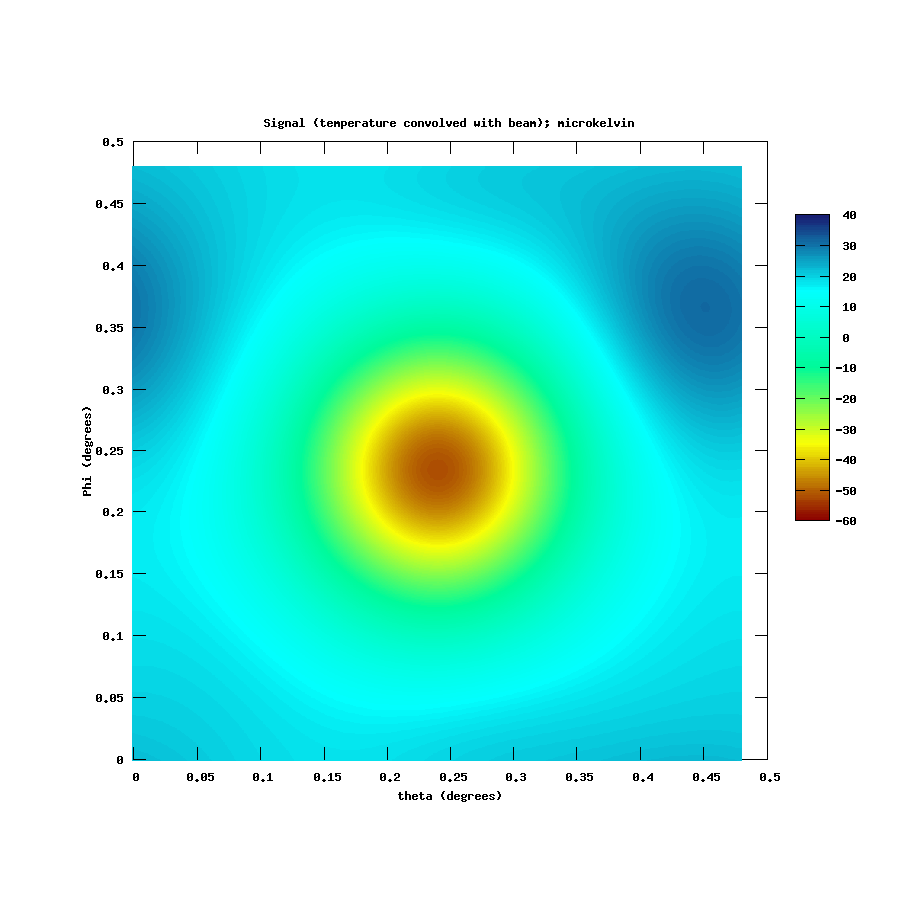}
  \\
  \includegraphics[scale=0.19,viewport=135 145 740 740,clip]{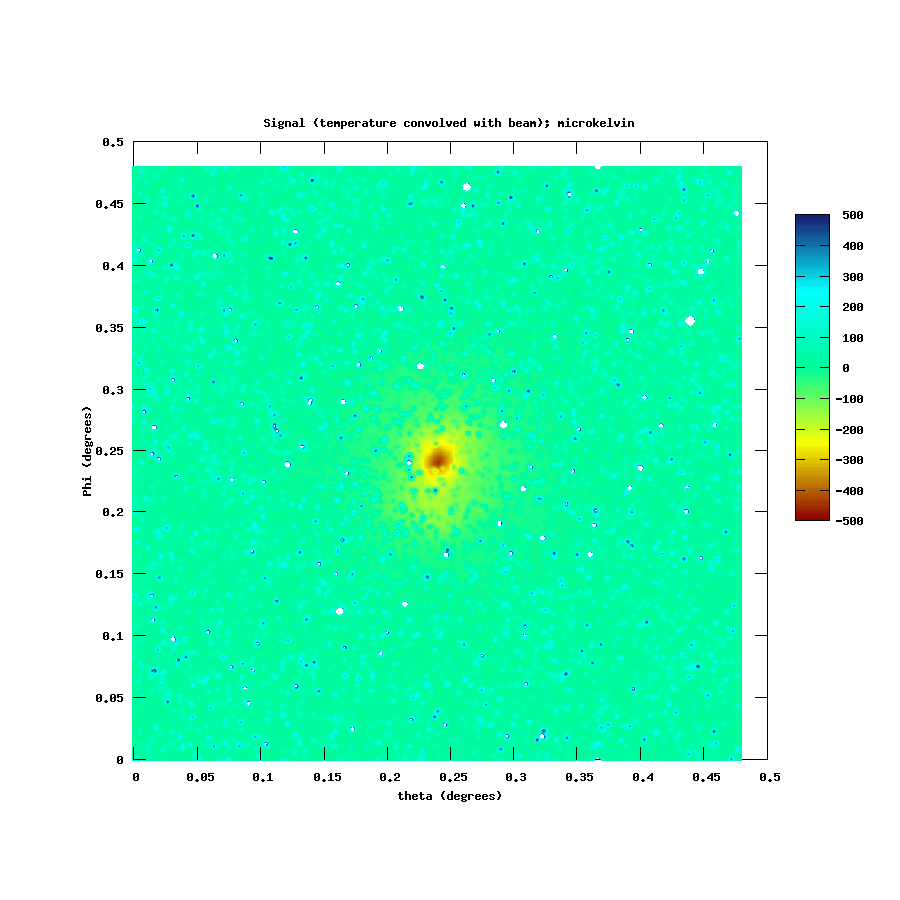}
  \includegraphics[scale=0.19,viewport=135 145 740 740,clip]{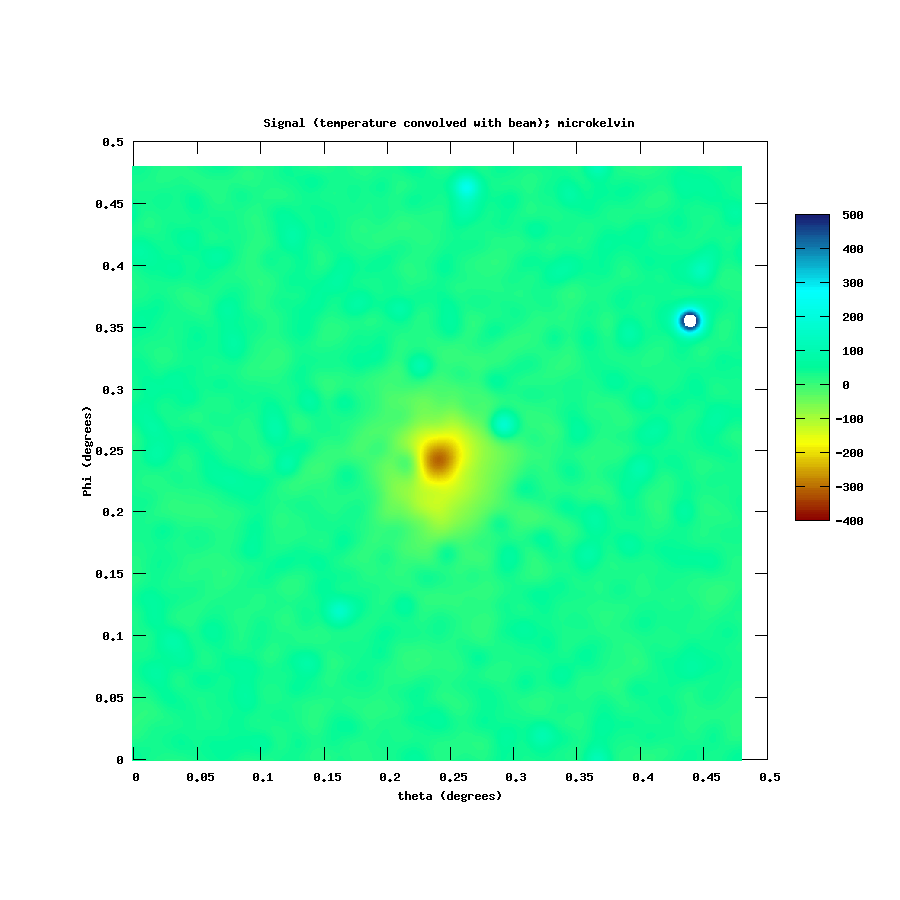}
  \includegraphics[scale=0.19,viewport=135 145 740 740,clip]{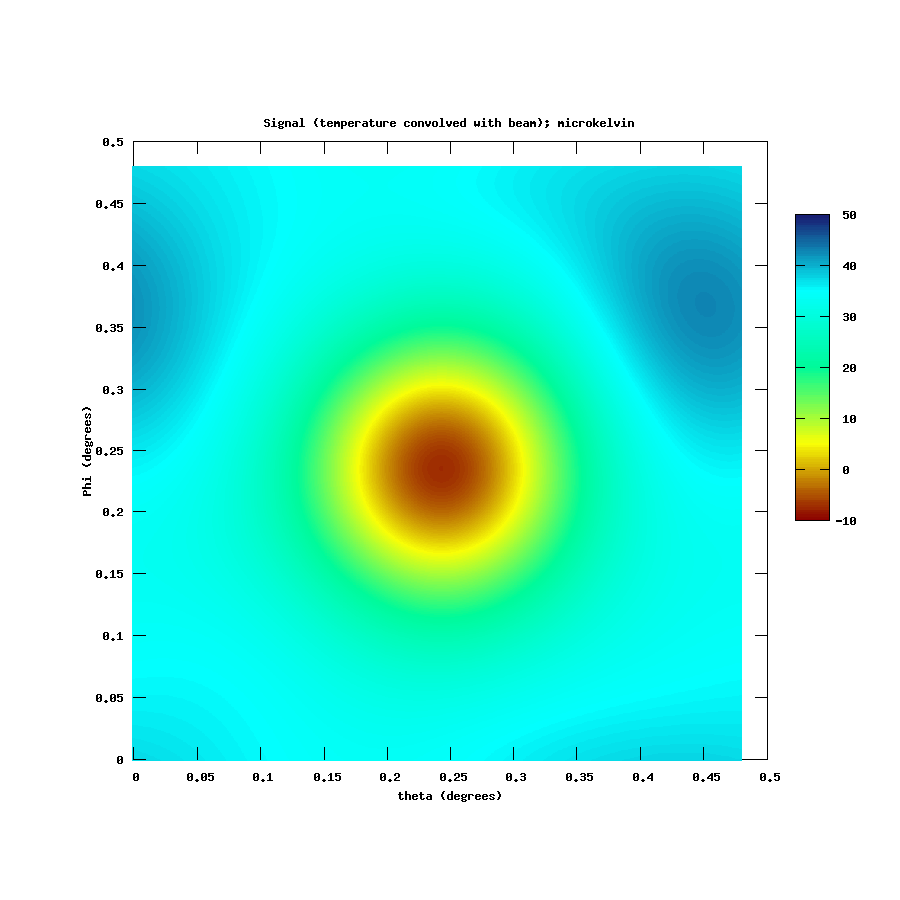}
  \\
  \includegraphics[scale=0.19,viewport=135 145 740 740,clip]{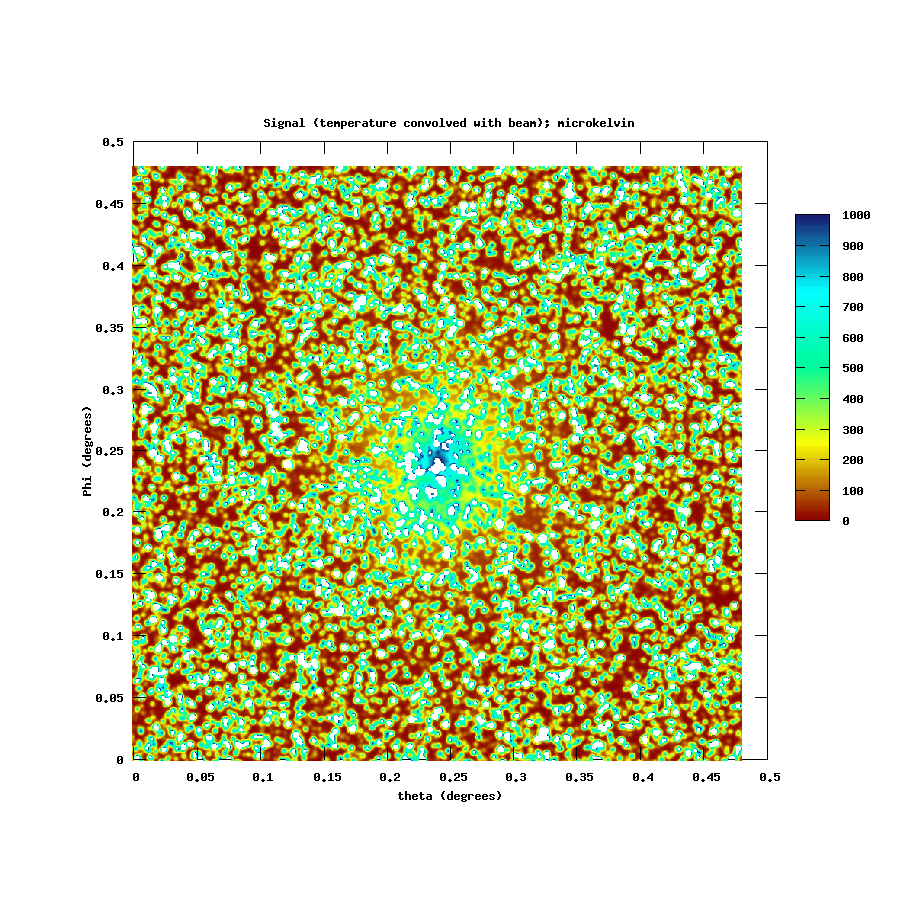}
  \includegraphics[scale=0.19,viewport=135 145 740 740,clip]{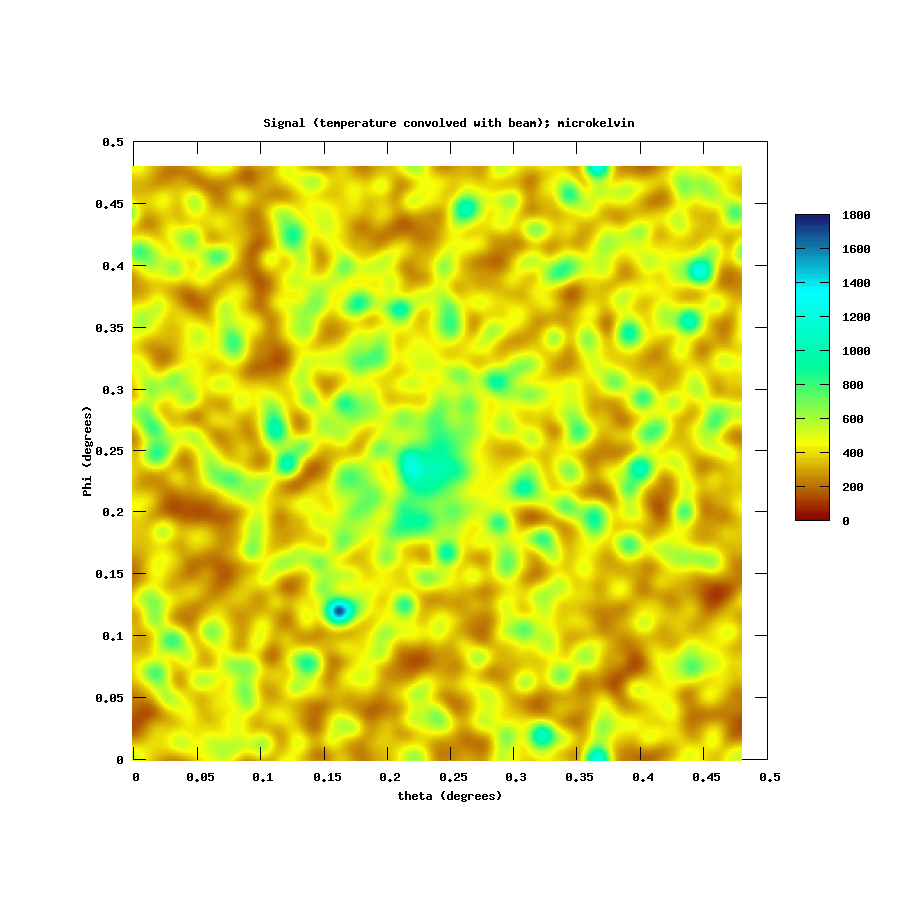}
  \includegraphics[scale=0.19,viewport=135 145 740 740,clip]{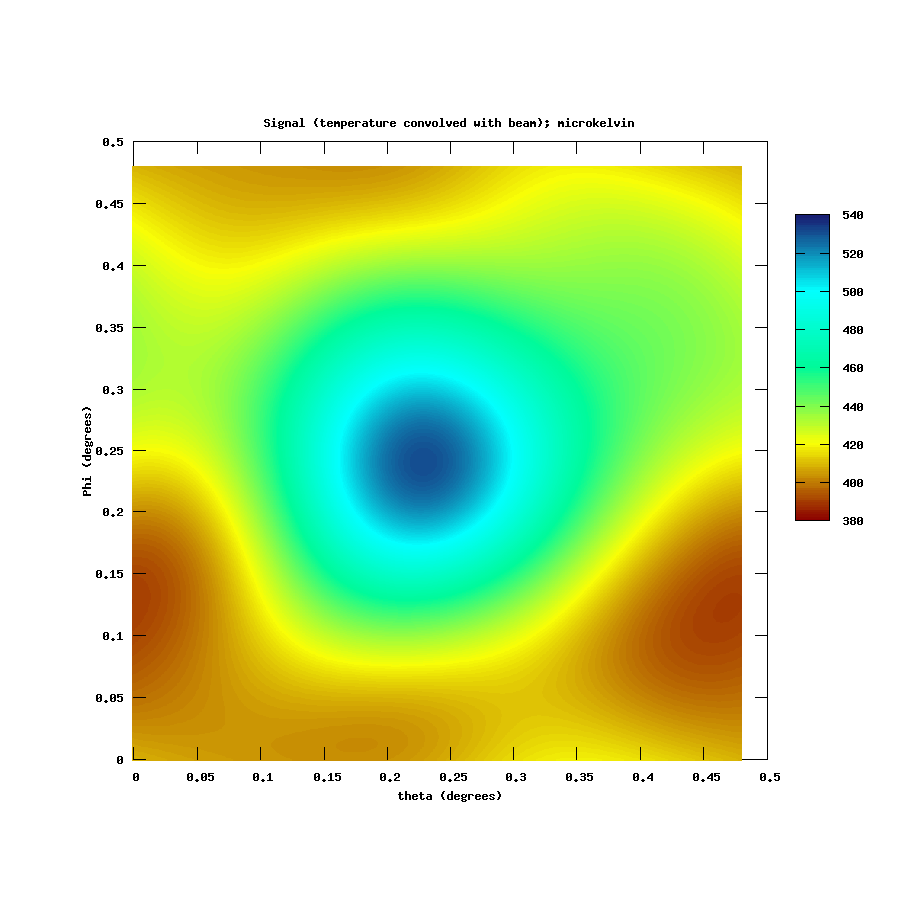}
	\caption[Point source contamination of a $10^{15} h^{-1} M_{\odot}$ cluster]{Point source contamination of a $10^{15} h^{-1} M_{\odot}$ cluster from \citet{2004Kay}, positioned at $z = 0.2$. The maps are $0.48\degree \times 0.48\degree$. Across: beam size (12 arcsec, 1 arcmin, 10 arcmin). Down: frequency (10, 30, 90, 150, 350~GHz). The colour scales vary between maps; red is colder whilst blue is hotter.}
	\label{fig:pointsource}
\end{fig}

In addition to the total number of point sources and their flux densities, we need positions. The simplest method is to distribute them randomly over the map, so that the probability of finding a source at a particular position, $P(x,y)$, is constant across the map. However, galaxies emitting synchrotron radiation are thought to reside preferentially within galaxy clusters. Observations at 30~GHz by \citet{2007Coble} suggest that the number counts of these sources increase in the direction of known massive clusters; \citet{2007Lin} come to a similar conclusion at 1.4~GHz. To account for this, we correlate the positions of these sources with the galaxy clusters by using a power of the surface mass density $\Sigma (x,y)$ of the clusters. In particular we use the probability distribution
\begin{equation}
P(x,y) = \frac{\Sigma (x,y)^{1/3}}{\int \Sigma(x^\prime,y^\prime)^{1/3} \phantom{.} dx^\prime \phantom{.} dy^\prime}.
\end{equation}
The specific choice of $\Sigma (x,y)^{1/3}$ brings the number counts towards clusters into reasonable agreement with those estimated by \citet{2007Coble}.

As the radio point sources are thought to reside in galaxies within clusters, these are distributed according to the surface density, and their number is enriched by a factor of 3 when a single cluster is being mapped. Infrared sources, however, are at high redshift, and any clustering present in their distribution will be unrelated to the galaxy cluster. Thus, they are not enriched and are randomly distributed.

Figure \ref{fig:pointsource} shows the effect of point sources on the SZ effect from a simulated large, nearby cluster over a range of frequencies between 10 and 350~GHz, and at resolutions of 12 arcsec, 1 and 10 arcmin. The radio point sources are enriched towards the centre of the cluster, depending on the surface matter density. Infrared point sources are also included, but are randomly distributed. With a beam of 10 arcmin, the cluster can be detected at 90~GHz and above, however it is heavily point source confused. With a one arcmin beam, the cluster can be seen at 30~GHz and above, but point source contamination is obvious. At 12 arcsec resolution, individual features within the cluster can start to be resolved, however individual point sources are obvious at all frequencies. The large number of infrared point sources is obvious at 350~GHz, requiring high resolution to distinguish between point sources and extended emission from the SZ effect.

\subsection{Polarization}

The polarization of point sources is of great interest for instruments that aim to observe the B-mode polarization within the CMB. \citet{2009Jackson} have observed a set of WMAP point sources using the VLA at a range of frequencies, and have measured the percentage of their flux density that is polarized. Battye et al. (in prep.) have analysed the statistical properties of the sources using the results of these observations. As part of this, the model described above was used to predict the expected polarized point source power spectra for a variety of both total intensity and polarized flux density cuts. The component of this work that the Author was involved in is described here.

Equation \ref{eq:ps_spectrum} can be modified to provide the polarized point source power spectrum up to a flux density cut in total intensity, yielding
\begin{equation} \label{eq:pps_spectrum}
C_l^P = \left(\frac{dB}{dT} \right)^{-2} \left<\Pi^2\right> \int_{S_{\mathrm{min}}}^{S_{\mathrm{max}}} S_{\nu}^{2} (dN/dS_{\nu}) dS_{\nu},
\end{equation}
where
\begin{equation}
\left< \Pi^2 \right> = \int_0^1 d\Pi \Pi^2 P(\Pi).
\end{equation}
We model the observed histogram of polarization percentages as a log-normal distribution,
\begin{equation}
P(\Pi) = \frac{A}{\Pi} \exp \left( -\frac{ \left[ \ln (\Pi / \Pi_0) \right]^2}{2 \sigma^2} \right).
\end{equation}
The values for the median $\Pi_0$ and the standard deviation $\sigma$ can be found from the observations by two different ways. Only considering the point sources where polarization has been detected yields values of $100 \Pi_0 = 2.5$ and $\sigma = 0.67$, however this distribution is biased towards point sources with high polarization percentages. Taking the undetected point sources into account yields $100 \Pi_0 = 2.25$ and $\sigma = 0.74$; This gives a mean of $100\langle\Pi\rangle=2.95$ and an RMS of $100\langle\Pi^2\rangle^{1/2}=3.9$, which is comparable with the observations. We adopt these values for the remainder of this section. The function for both parameter sets and the binned observation results are plotted in Figure \ref{fig:pointsource_polprob}; there is good agreement between the two.

\begin{fig}
\centering
\includegraphics[scale=0.4]{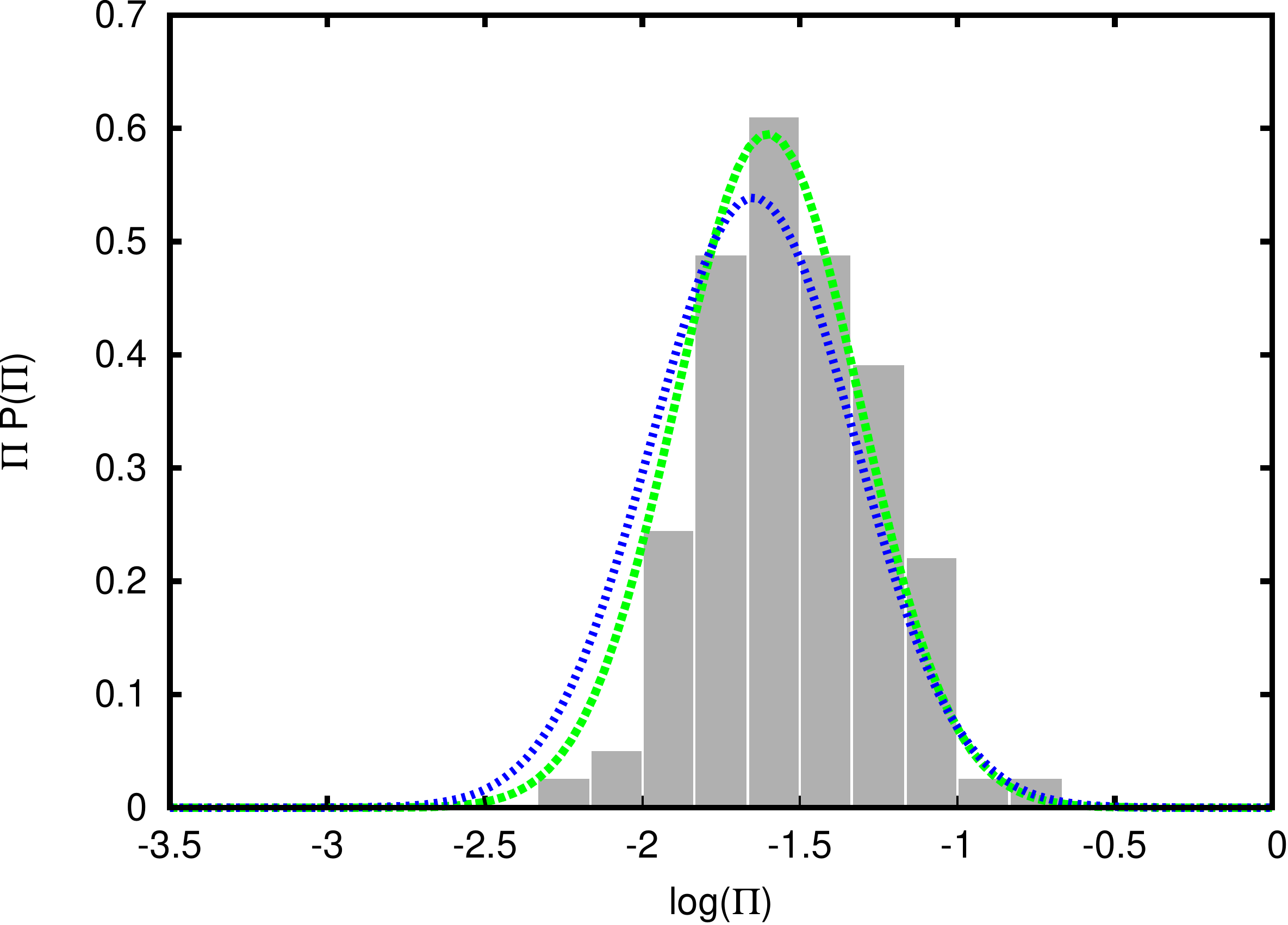}
\caption[Probability distribution of the point source polarization percentage]{The model of the point source polarization percentage probability distribution for the parameters calculated using the complete (blue line) and detected (green line) point source samples, compared with the histogram of fractional polarization of WMAP sources at 22~GHz from Battye et al. (in prep.)}
\label{fig:pointsource_polprob}
\end{fig}

\begin{fig}
\centering
\includegraphics[scale=0.4]{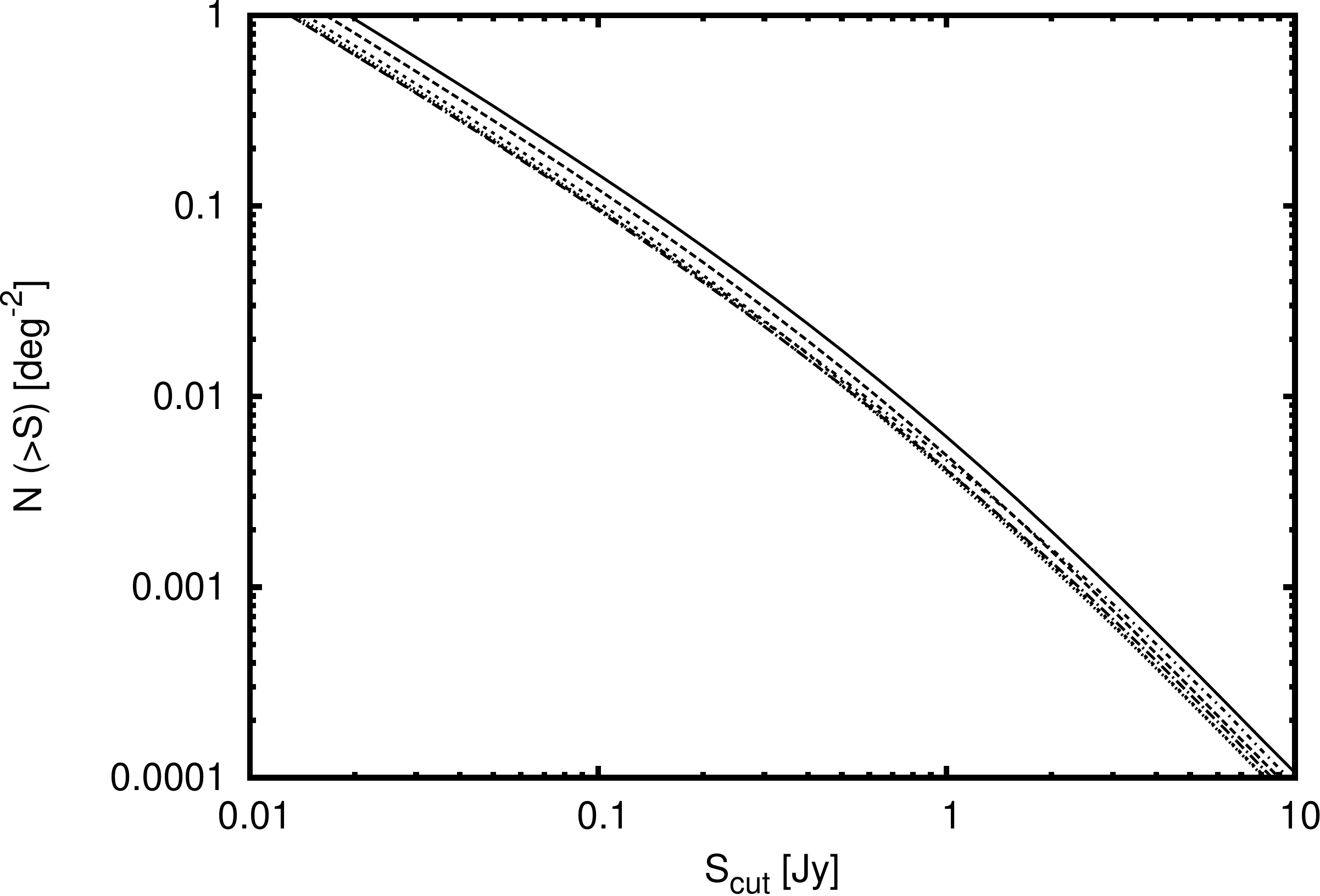}
\caption[The number of radio point sources greater than a given flux density]{The number of radio point sources per square degree greater than a given flux density at frequencies of (top-to-bottom) 30, 44, 70, 100, 150 and 220~GHz. There is no dramatic change in the number of sources over this range of frequencies.}
\label{fig:pointsource_dnds}
\end{fig}

As per Section \ref{sec:lowfreqps}, the model of \citet{1998Toffolatti} at 30~GHz, rescaled by a factor of 0.7, is used for the number of point sources with given flux densities. The number counts are extrapolated to other frequencies using \citep{1984Condon}
\begin{equation}
n(S, \nu) = \left(\frac{\nu}{30~\mathrm{GHz}} \right)^q n(S, 30~\mathrm{GHz}),
\end{equation}
where
\begin{equation}
q=\bar\alpha(\chi-1)+{1\over 2}\sigma_{\alpha}^2(\chi-1)^2\log_{10}\left({\nu\over 30~{\rm GHz}}\right)
\end{equation}
Here, $\bar \alpha=-0.44$ is the mean spectral index, with a scatter of $\sigma_\alpha=0.47$, as measured between 22 and 43~GHz by Battye et al. (in prep.), and $\chi(S) = -d(\log n)/d(\log S)$ is the power law slope of the source count. The number of point sources above a given flux density per square degree is shown in Figure \ref{fig:pointsource_dnds} for a range of frequencies between 30 and 220~GHz. Due to the presence of flat spectrum sources, there is not a significant change in the number of point sources with a given flux density over the range of frequencies considered here. However, we caution that the observations were made at frequencies up to 43~GHz, so extrapolation to 220~GHz will not provide the most accurate results. We have also not taken into account the steepening of the spectral index distribution observed both in this sample and others \citep[e.g.][]{2006Ricci,2007Massardi}, which will have led to an overestimate in these simulations. Hence the results should be considered to be an upper limit at the highest frequencies.

\begin{fig}
\centering
 \includegraphics[scale=0.35]{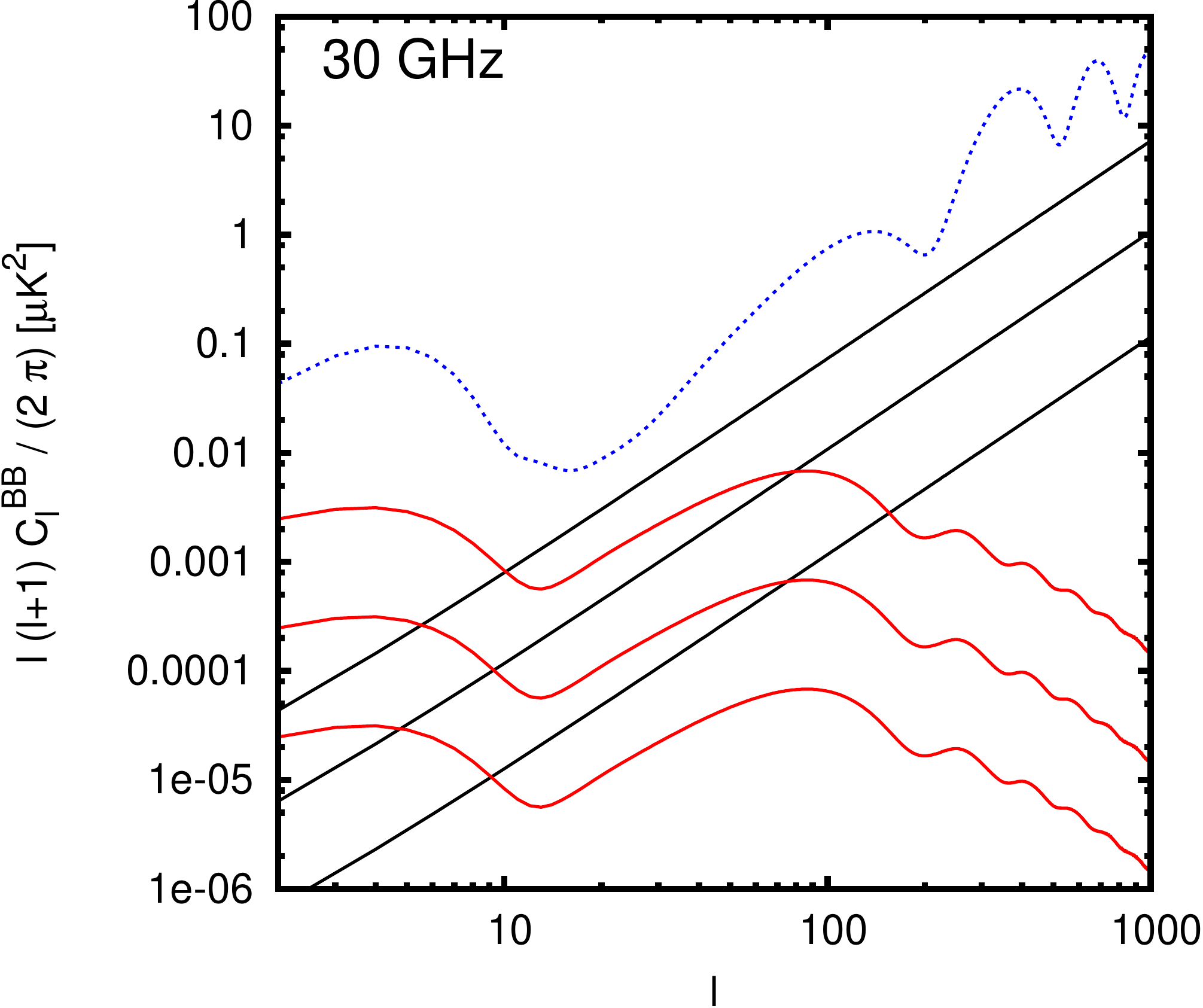}
 \includegraphics[scale=0.35]{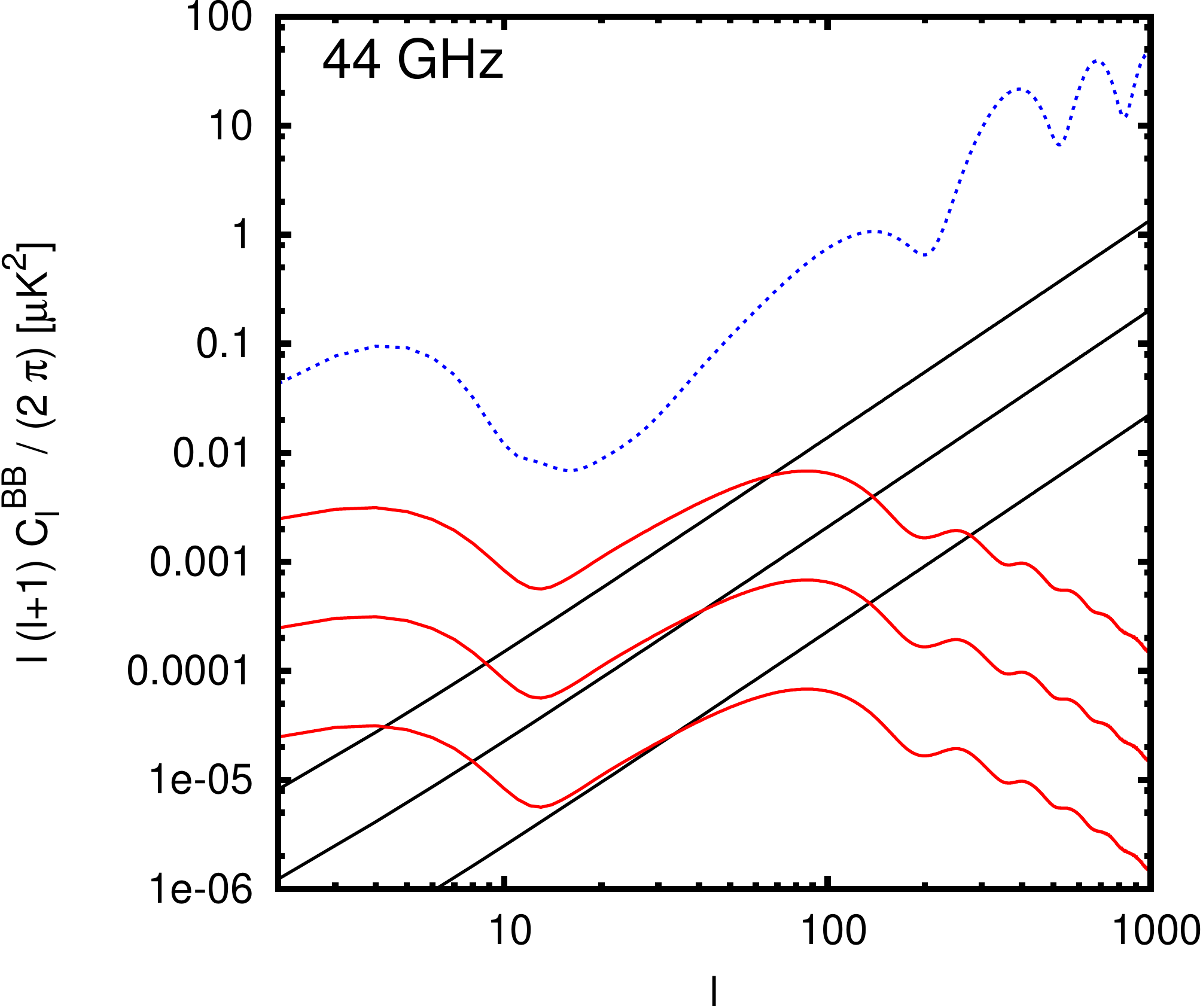}\\
  \includegraphics[scale=0.35]{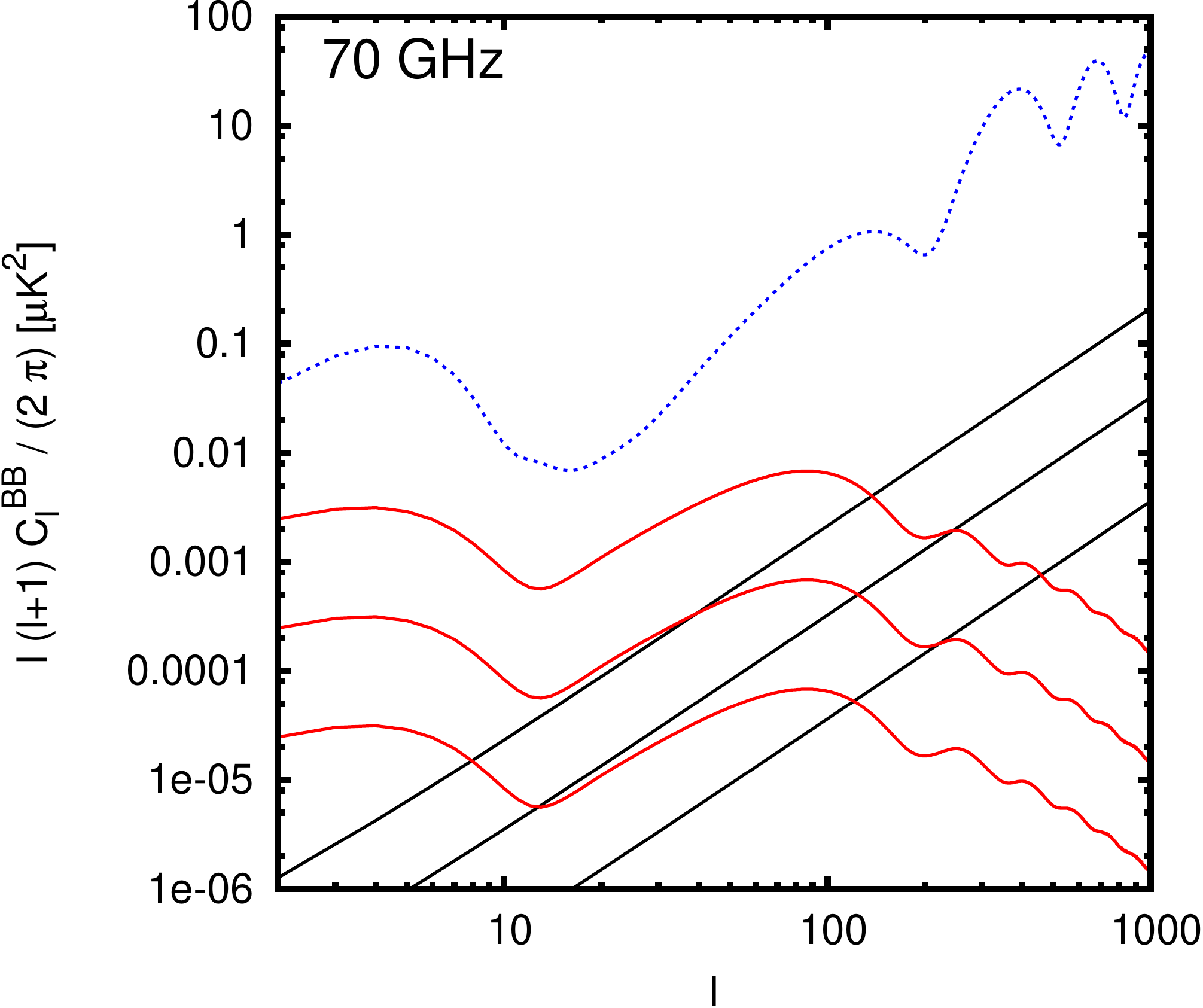}
 \includegraphics[scale=0.35]{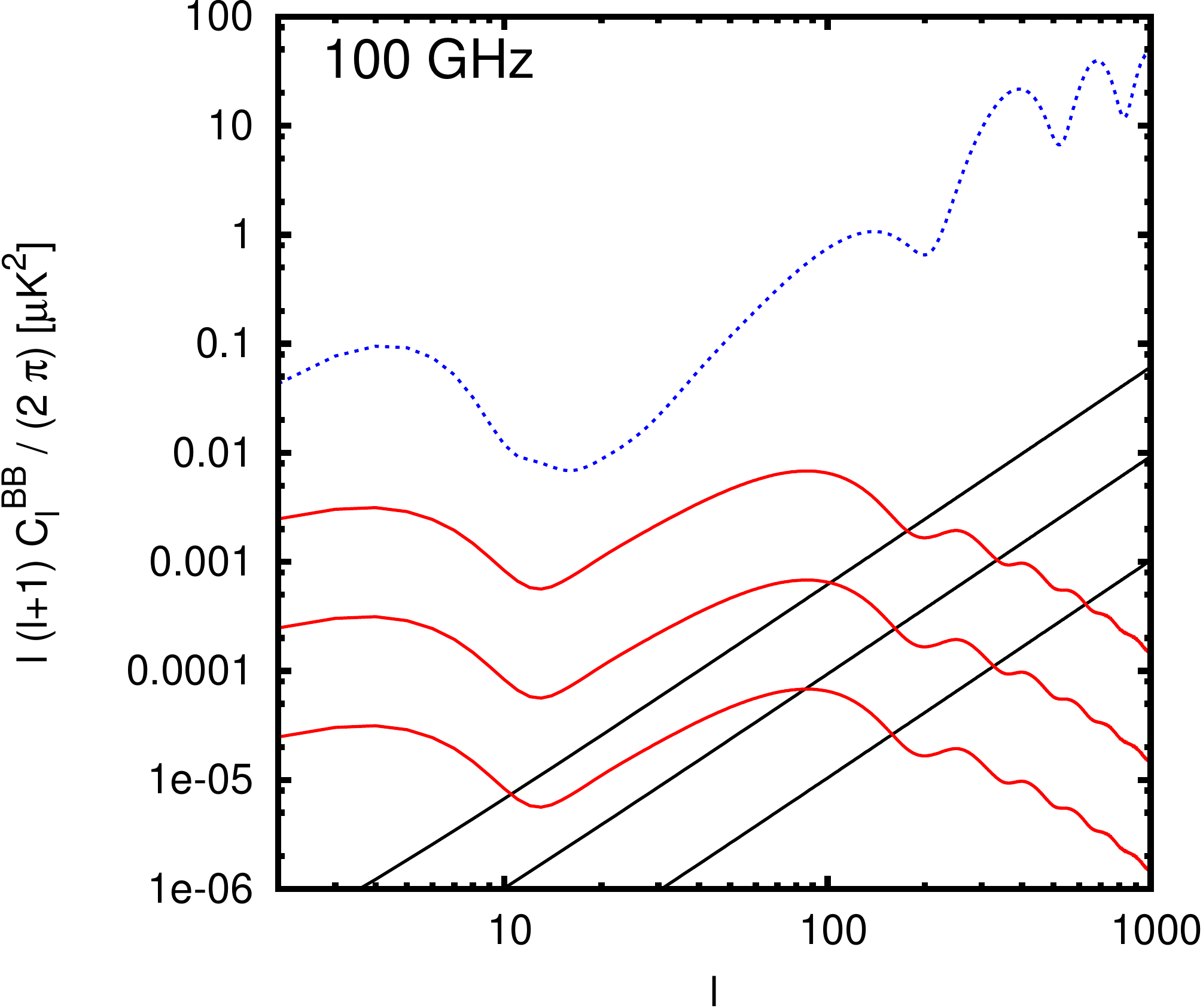}\\
 \includegraphics[scale=0.35]{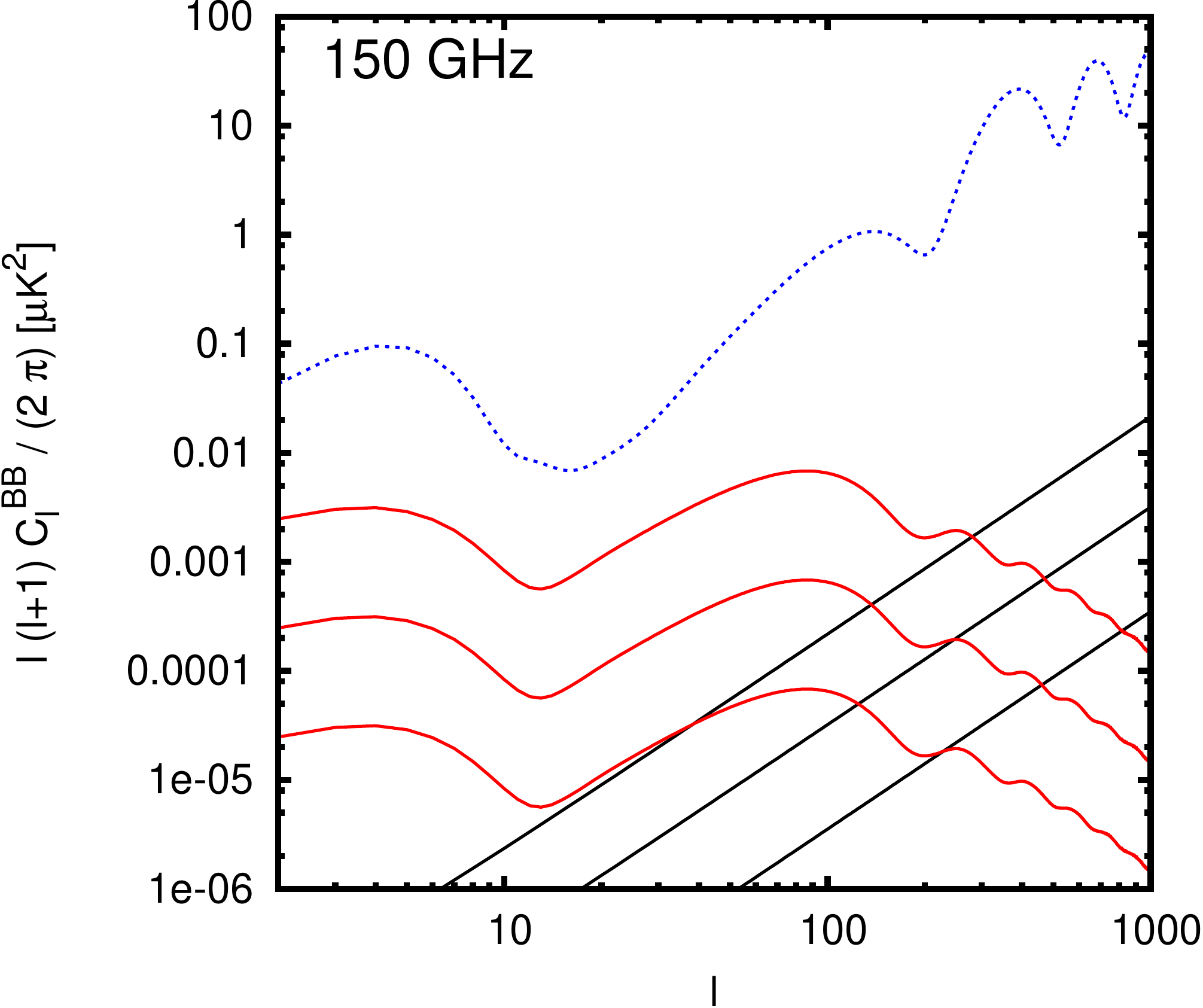}
 \includegraphics[scale=0.35]{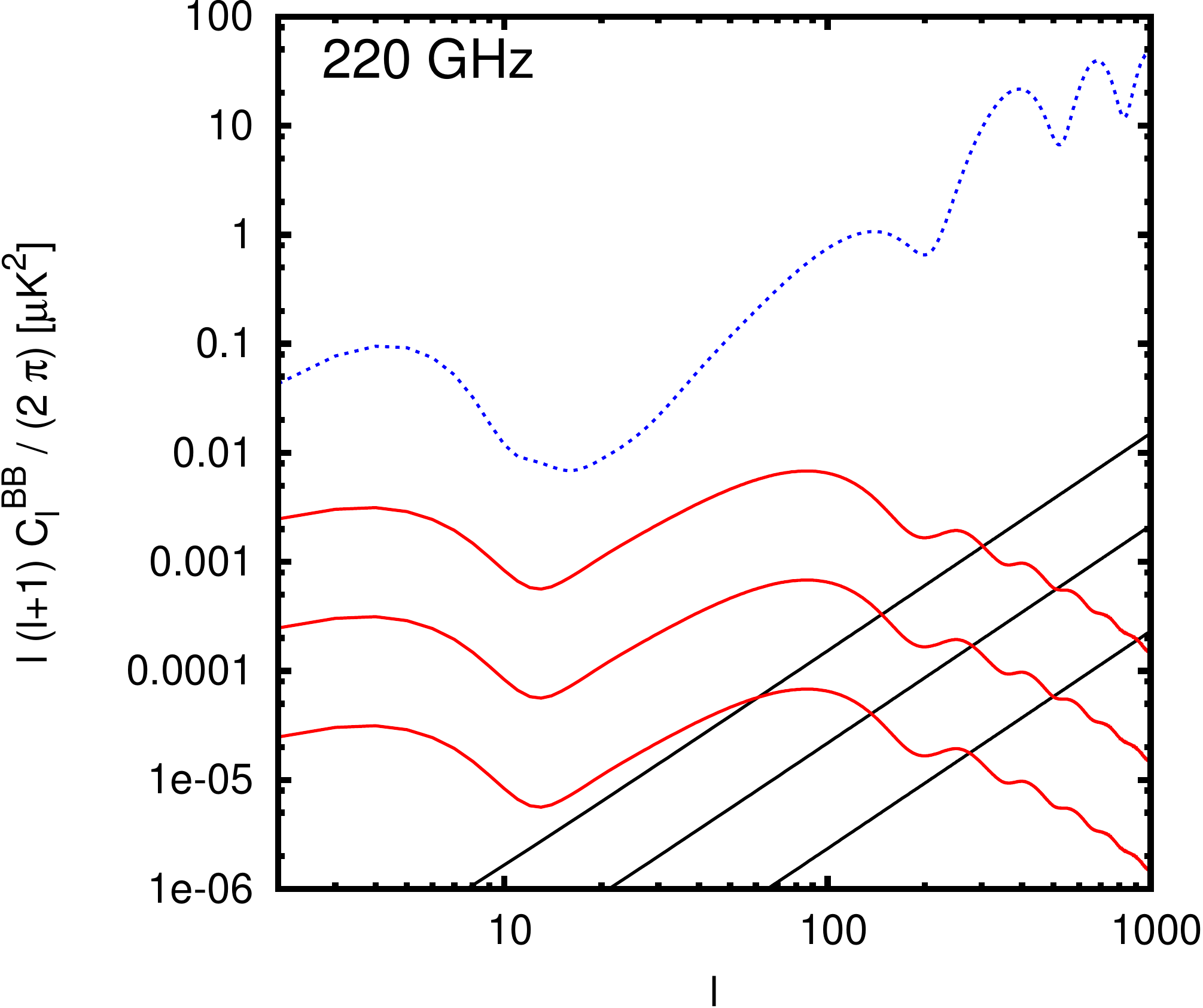}

	\caption[Polarized point source spectra for a range of frequencies and flux density cuts]{Polarized point source spectra for a range of frequencies and flux density cuts. The frequencies are 30~GHz (top left), 44~GHz (top right), 70~GHz (middle left), 100~GHz (middle right), 150~GHz (bottom left) and 220~GHz (bottom right). The point source flux density cuts were 1~Jy, 100~mJy and 10~mJy (black lines). Also shown are the expected B-mode power spectra for tensor-to-scalar ratios $r=0.1, 0.01$ and 0.001 (red lines), as well as the E-mode power spectrum (blue dashed line).}
	\label{fig:ps_spectra}
\end{fig}

Infrared sources are expected to have a polarization percentage of around 1 per cent, based on observations of Arp 220 that find an upper limit of 1.5 per cent \citep{2007Seiffert}. At the frequencies of interest here, the polarized emission of infrared point sources is subdominant to that from the radio point sources, and hence it does not contribute significantly to the power spectra. As a result, we do not include infrared sources in these calculations.

Assuming that the polarized power is evenly distributed between E and B modes, $C_l^{BB} = C_l^P/2$, we calculate the power spectra expected from point sources according to Equation \ref{eq:pps_spectrum}. These are plotted in Figure \ref{fig:ps_spectra} for 30, 44, 70, 100, 150 and 220~GHz for point source flux density cuts of 1~Jy, 100~mJy and 10~mJy. We also show the expected B-mode polarization spectra for tensor-to-scalar ratios $r=0.1, 0.01$ and $0.001$, and for comparison we also plot the E-mode power spectrum. For observations at frequencies lower than 100~GHz, point source subtraction will be necessary to reach $r=0.1$. Point source subtraction will be necessary to measure $r=0.01$ at all frequencies considered here.

\begin{fig}
\centering
\includegraphics[scale=0.4]{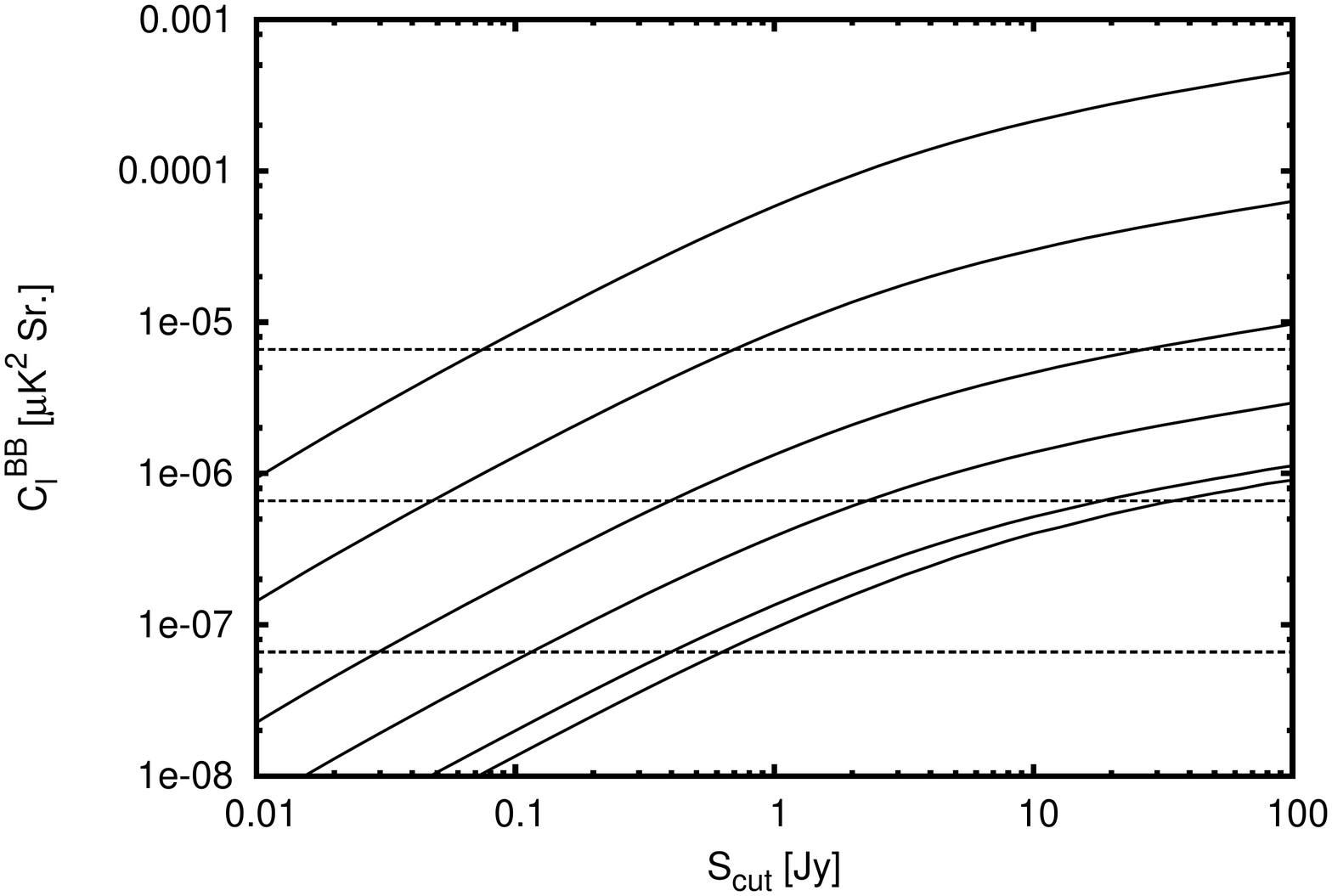}
\caption[The projected point source noise spectrum, $C_{\ell}^{BB}$, as function of $S_{\rm cut}$ for a range of frequencies]{The projected point source noise spectrum, $C_{\ell}^{BB}$, as function of $S_{\rm cut}$ for a range of frequencies (30~GHz - solid line, 44~GHz dotted line, 70~GHz dotted line, 100~GHz fine dotted line, 150~GHz long-dash dotted line, 220~GHz short-dash dotted line. The 3 horizontal lines correspond to the power spectrum amplitudes at the maximum ($\ell=80$) of the B-mode spectrum for $r=0.1$, $0.01$ and $0.001$ from top to bottom. }
\label{fig:pointsource_pol_scut}
\end{fig}

\begin{fig}
\centering
\includegraphics[scale=0.4]{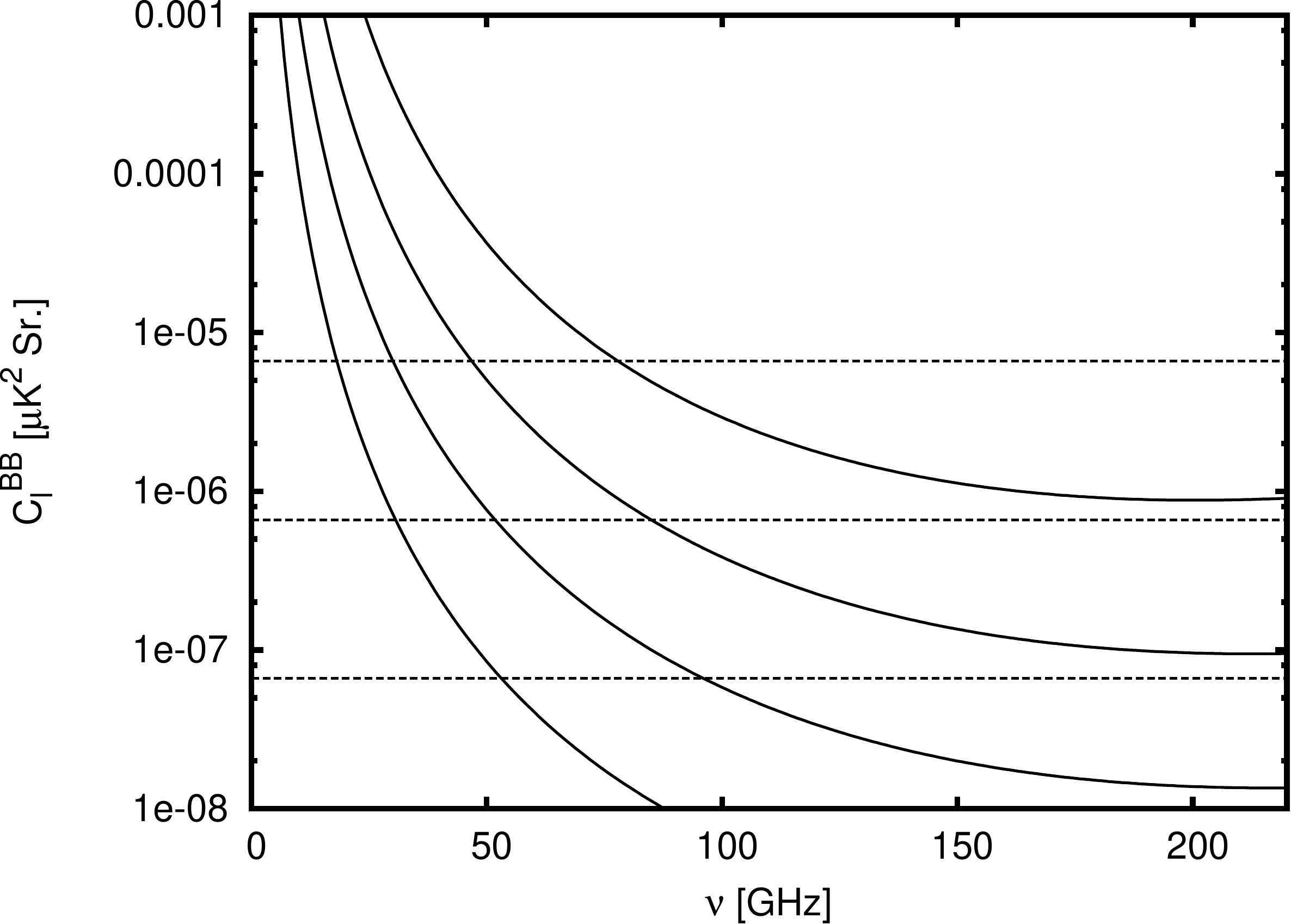}
\caption[The value of the point source noise spectrum for $S_\mathrm{cut}=100$, 1, 0.1 and 0.01~Jy as a function of frequency]{The value of the point source noise spectrum for $S_\mathrm{cut}=100$, 1, 0.1 and 0.01~Jy as a function of frequency.}
\label{fig:pointsource_pol_freq}
\end{fig}

Figure \ref{fig:pointsource_pol_scut} shows the level of the power spectrum due to the point sources as a function of $S_\mathrm{cut}$ for a range of frequencies. It also shows the level of the maximum for the B-mode power spectrum (i.e. the power at $l=80$) for r=0.1, 0.01 and 0.001, showing the required level of source subtraction depending on the required detection limit of $r$ and the frequency of the observations. Figure \ref{fig:pointsource_pol_freq} is similar, but shows lines of constant $S_\mathrm{cut}$ as a function of frequency. Note that at high frequencies, the point source contribution to the power spectrum flattens.

These results show that subtraction of point sources will be vital for upcoming low frequency B-mode experiments that are sensitive to values of $r$ of less than 0.1, and less than 0.01 for all frequencies.

\section{Summary} \label{sec:virtualsky_summary}

This chapter has described the creation of Virtual Skies - simulated microwave frequency maps consisting of the Cosmic Microwave Background, galaxy clusters via the SZ effect, and foreground point sources. It has described the process of creating simulated cluster catalogues -- by analytical formulae, N-body simulations and using {\sc Pinocchio} simulations -- and the application of a galaxy cluster model to turn these catalogues into maps of the SZ effect. Example realisations, and power spectra of the SZ effect, were covered, and were also compared with theoretical power spectra.

The process of adding point sources -- both low- and high-frequency -- to the maps was then covered, including distributing the point sources according to the surface matter density of the galaxy clusters. The polarization of point sources was also discussed, looking at the effect of these point sources on observations of B-mode power spectra in the polarization of the CMB. These results show that point source subtraction will be necessary in order to measure the B-mode power spectra, particularly if the tensor-to-scalar ratio is small ($<0.1$).

With the model established, the next chapter will look at the scientific uses for these maps, focusing on the statistics of the SZ effect.
\chapter{Statistics of the SZ power spectrum} \label{sec:cbi_excess} \label{vs_results}
The discovery of the CMB by \citet{1965Penzias} helped provide the foundations of modern cosmology. The discovery of the primordial anisotropies on large scales by {\it COBE} \citep{1992Smoot}, followed by the precision measurements by the Wilkinson Microwave Anisotropy Probe \citep[{\it WMAP}; see for example][]{2009Hinshaw} and other experiments, pinned down the values of the principle cosmological parameters. With the forthcoming measurements by the {\it Planck} satellite, the primordial anisotropies at large scales will be measured as accurately as possible \citep{2005Planck}, with cosmic variance preventing any more precise measurements of the cosmological parameters from these anisotropies. Interest is now turning towards measurements of the primordial polarization and the total power on smaller scales. 

At these smaller scales, ``secondary" anisotropies start to dominate over the primordial signal. A significant component of this secondary anisotropy will be caused by the SZ effect, which is expected to be dominated by groups and clusters of galaxies due to the high temperatures of the intracluster medium. At microwave frequencies ($\sim 30$~GHz) the SZ effect will contaminate the power spectrum due to the primary CMB anisotropies at spherical multipoles of $l\sim1000$ and above, making it difficult to extract cosmological information from the primary anisotropies at these scales; by $l\sim2000$ it is expected that it will dominate the power spectrum.

Several experiments have measured the power spectrum in this region of multipoles, and some of these show hints of excess power over what would be expected from the primary anisotropies alone. The Cosmic Background Imager \citep[CBI;][]{2003Mason,2004Readhead,2009Sievers}, which observed multipoles up to $l=4000$ at 30~GHz, was the first to measure a possible excess, and this has been confirmed by subsequent analysis. This was followed by the Berkeley-Illinois-Maryland Association (BIMA) interferometer \citep{2006Dawson} observing around $l=5000$ at 28.5~GHz, and at higher frequencies by the Arcminute Cosmology Bolometer Array Receiver (ACBAR), which has weakly detected a possible excess at around $l=2500$ at 150~GHz \citep{2008Reichardt}. Curiously, observations at similar multipoles by the Sunyaev-Zel'dovich Array at 30~GHz \citep[SZA;][]{2009Sharp} and QUaD at 100 and 150~GHz \citep{2009Friedman} do not show any excess power.

The amount of signal from the SZ effect depends on the amplitude of the power spectrum of the density fluctuations in the Universe. This power spectrum is normalized by $\sigma_8$, which is the variance of the fluctuations on scales of $8 h^{-1}$~Mpc. To explain the excess measured by the CBI, \citet{2009Sievers} find that $\sigma_8$ must be between 0.9 and 1.0. However, observations of the primordial CMB anisotropies by {\it WMAP} in combination with other instruments and methods yield $\sigma_8 = 0.812 \pm 0.026$ \citep{2009Komatsu}.

An important issue to consider when estimating the range of possible values of $\sigma_8$ that agree with the data is the statistical properties of the SZ effect, which will be non-Gaussian. Here we investigate those properties to see whether the excess could be the result of measuring part of the sky where the power from the SZ effect is higher than average, which would bias the measurement of $\sigma_8$ to higher than the average value. To do this, we use $3\degree \times 3\degree$ sky maps with a resolution of 18~arcseconds (600 pixels to a side), which allow us to accurately probe the multipole range $l = 1000-10~000$.

Using the virtual sky maps described in Section 2, the binned power spectra from the SZ effect are calculated using bins of $\Delta l = 1000$. We analyse the histograms of these power spectra and investigate the normalized skew as well as the mean and standard deviation. Higher orders than the skewness, such as the kurtosis, tend to be significantly affected by outlying data points and hence are not considered here.

Similar work investigating the statistics of SZ effect realizations has been carried out by \citet{2002Zhang} and \citet{2009Shaw}.

\section{Fiducial results}
\begin{fig}
\centering
  \includegraphics[scale=0.34,viewport=50 0 330 245,clip]{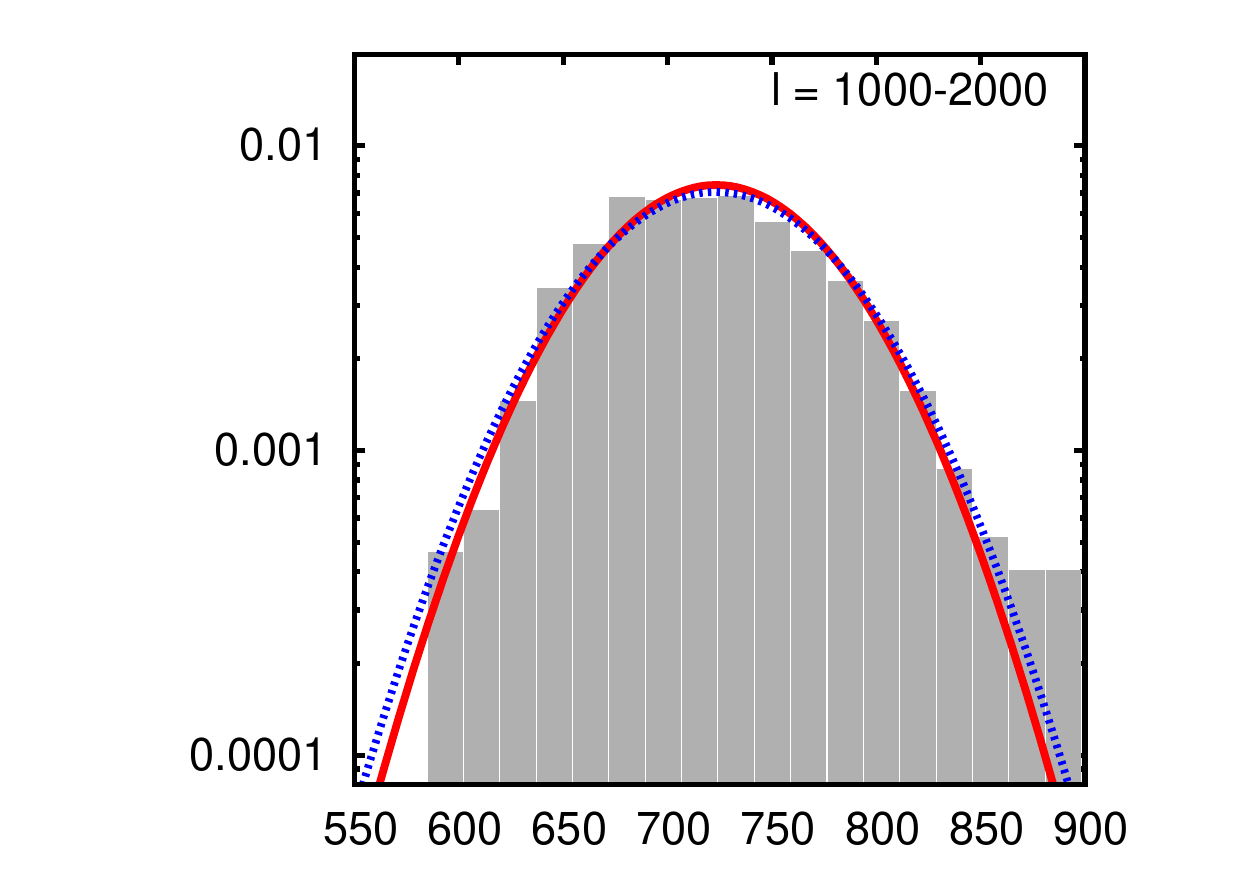}
  \includegraphics[scale=0.34,viewport=50 0 330 245,clip]{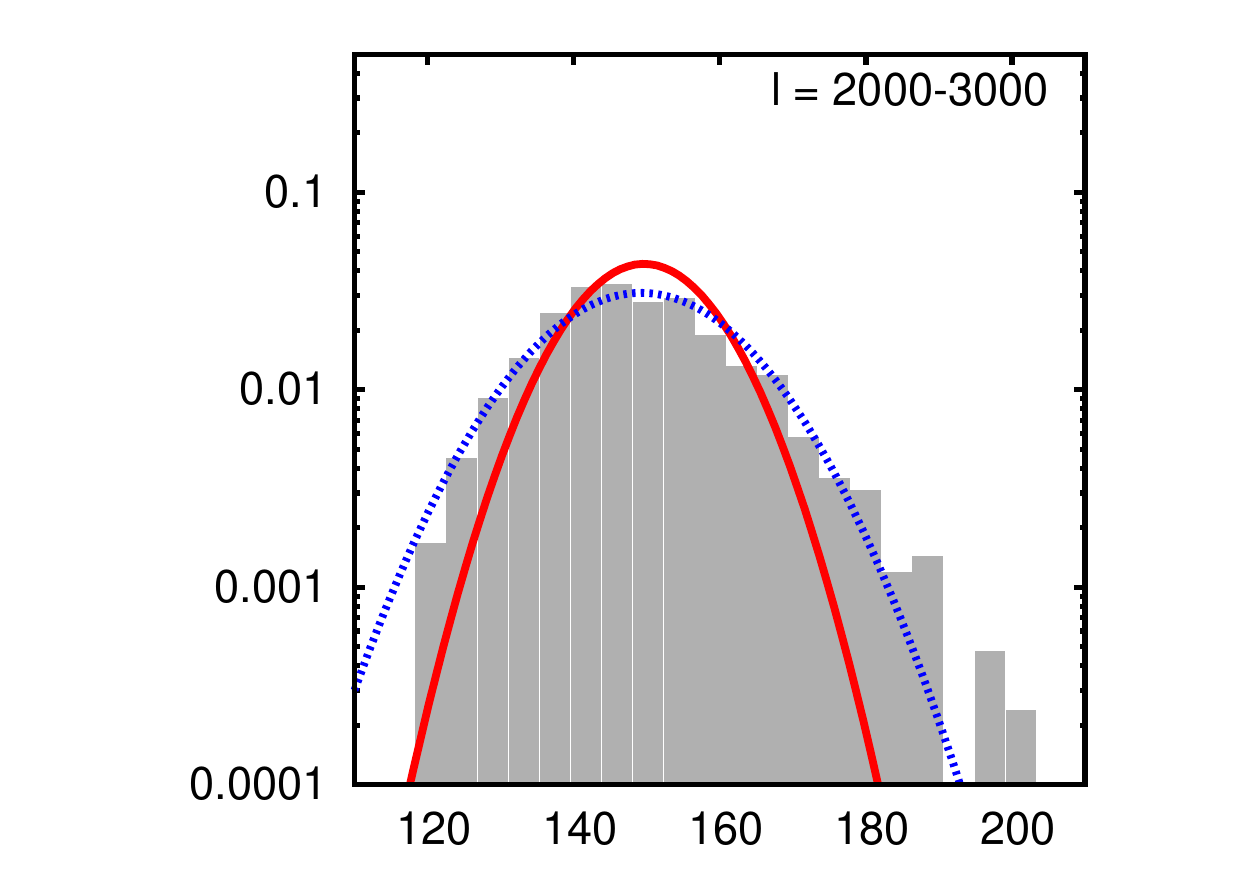}
  \includegraphics[scale=0.34,viewport=50 0 330 245,clip]{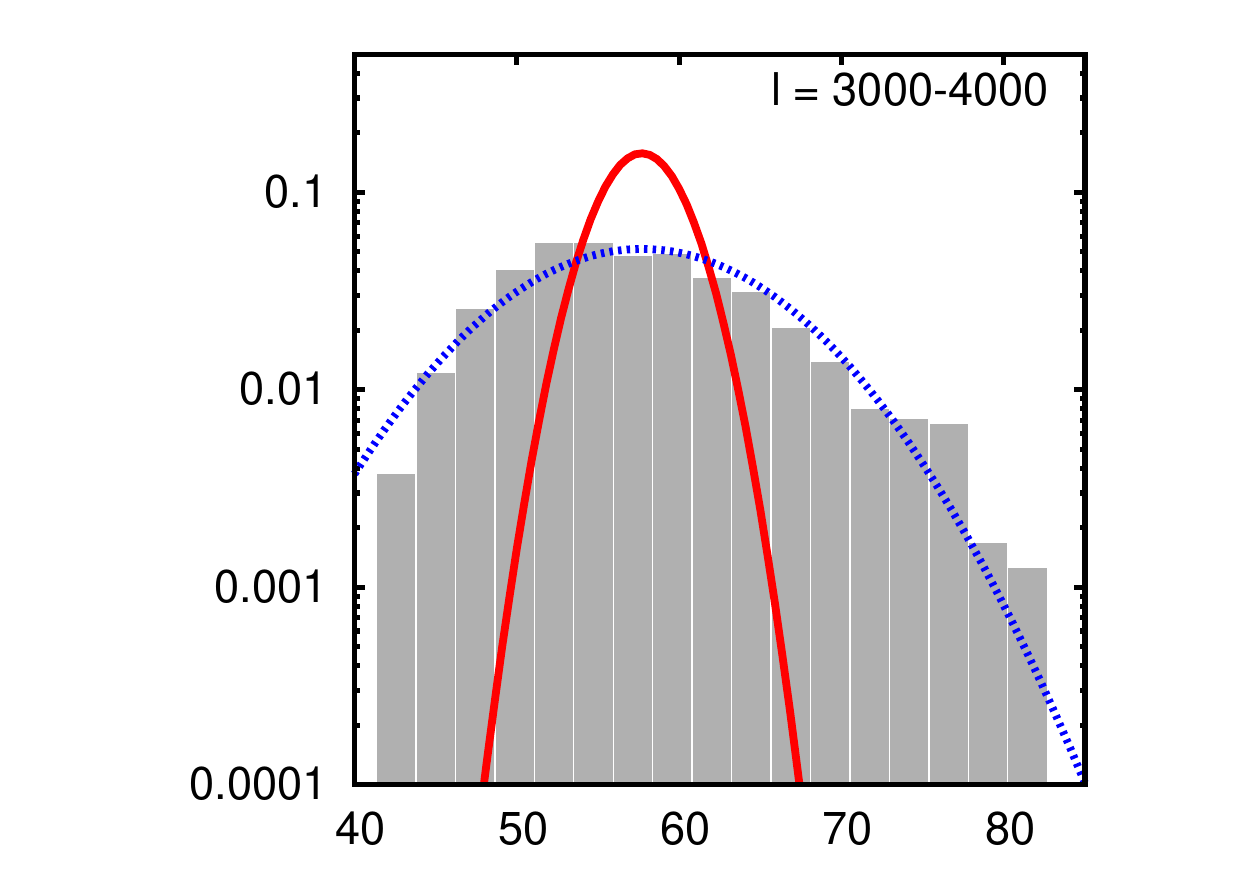}
  \includegraphics[scale=0.34,viewport=50 0 330 245,clip]{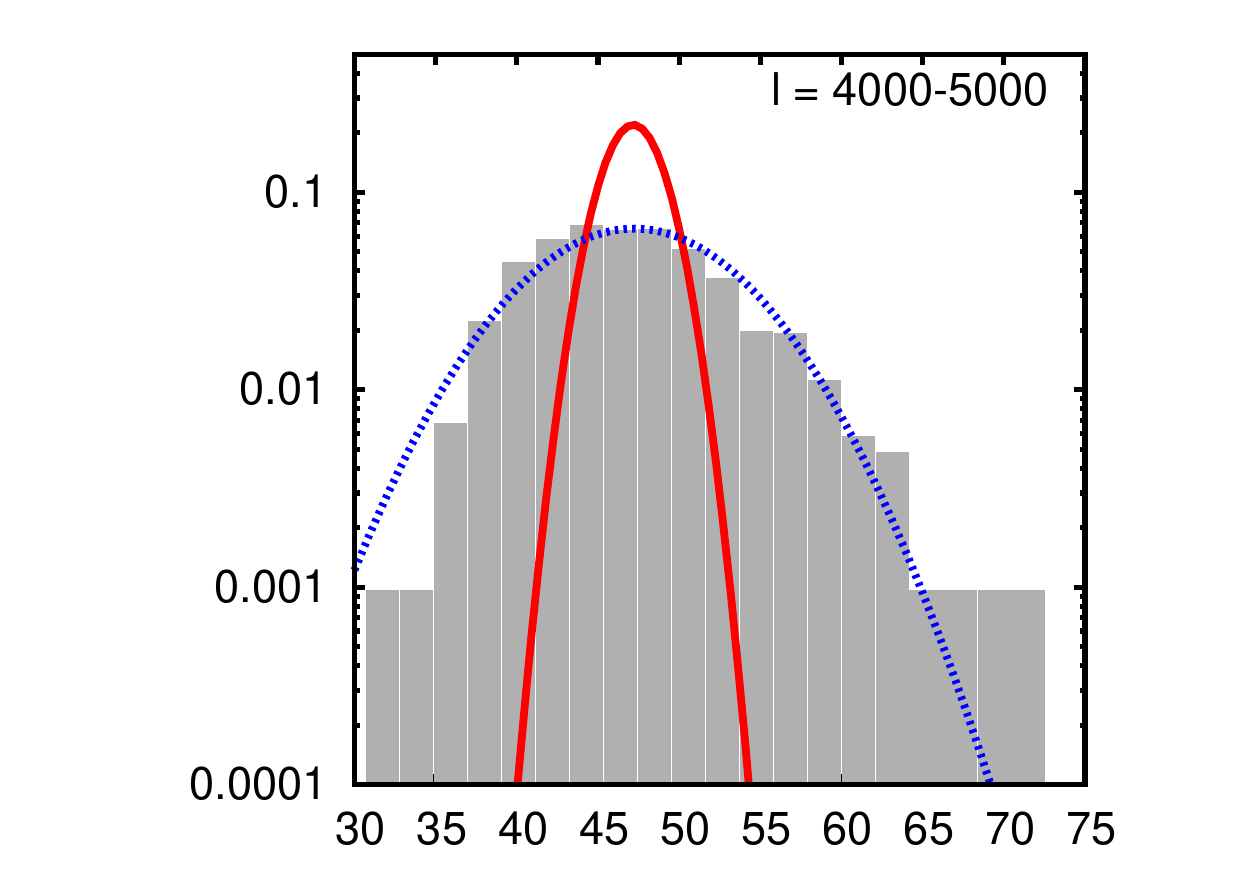}
  \\
  \includegraphics[scale=0.34,viewport=50 0 330 245,clip]{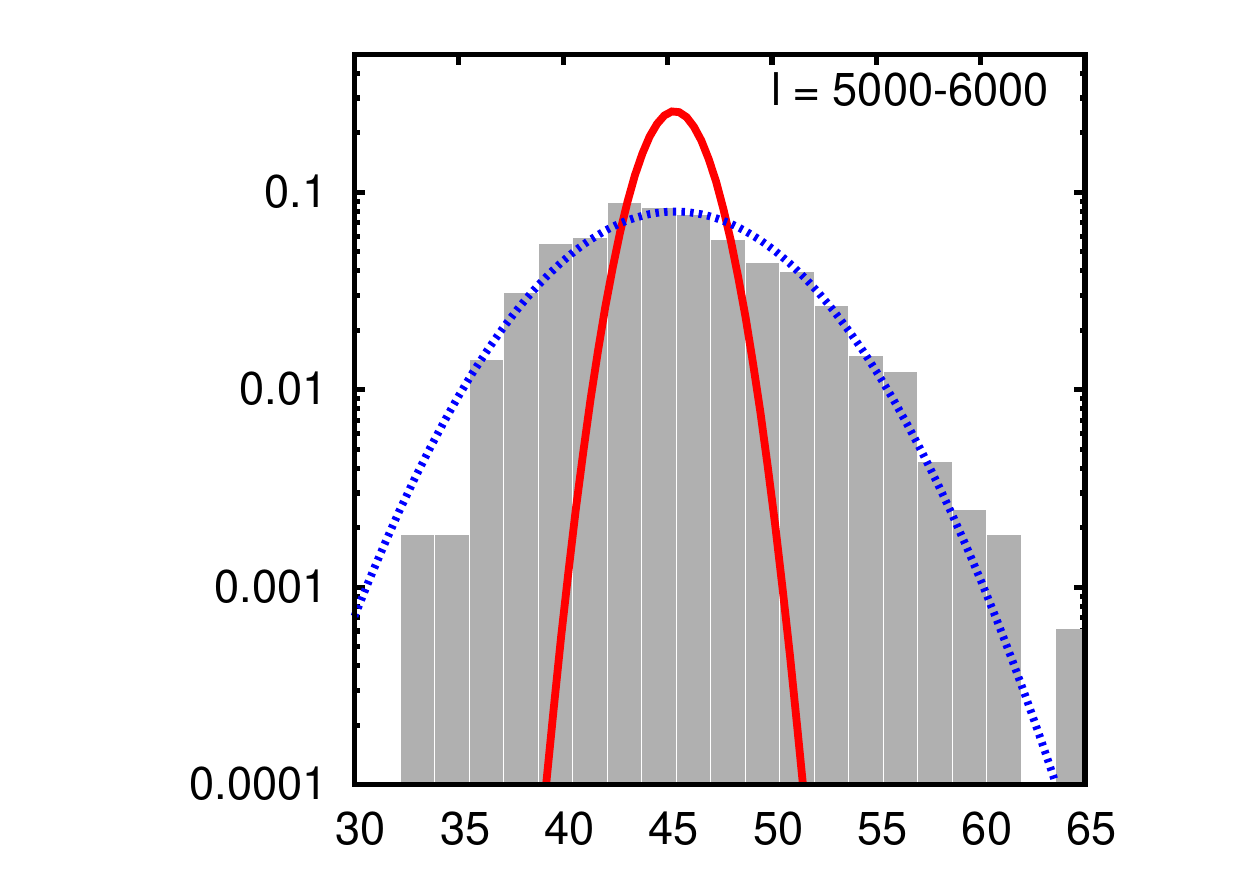}
  \includegraphics[scale=0.34,viewport=50 0 330 245,clip]{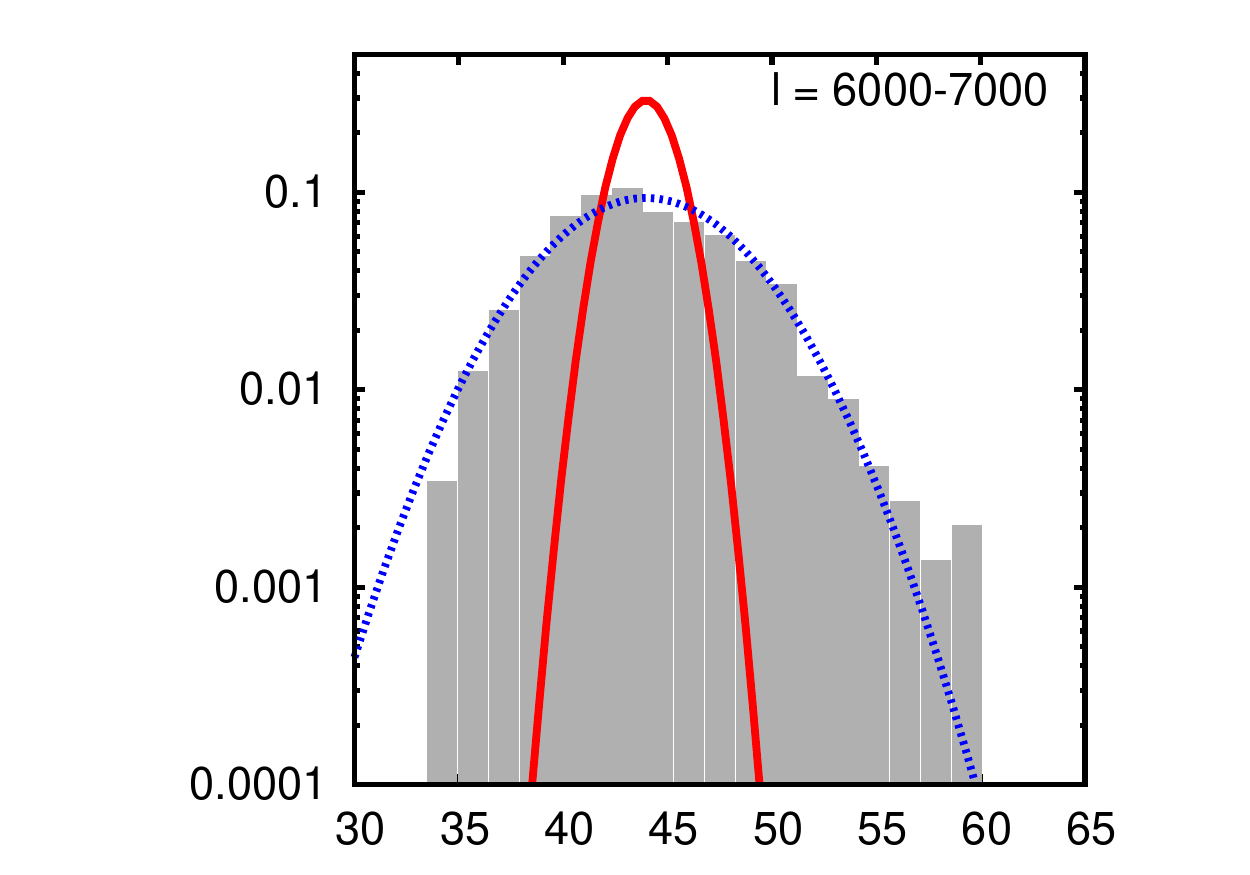}
  \includegraphics[scale=0.34,viewport=50 0 330 245,clip]{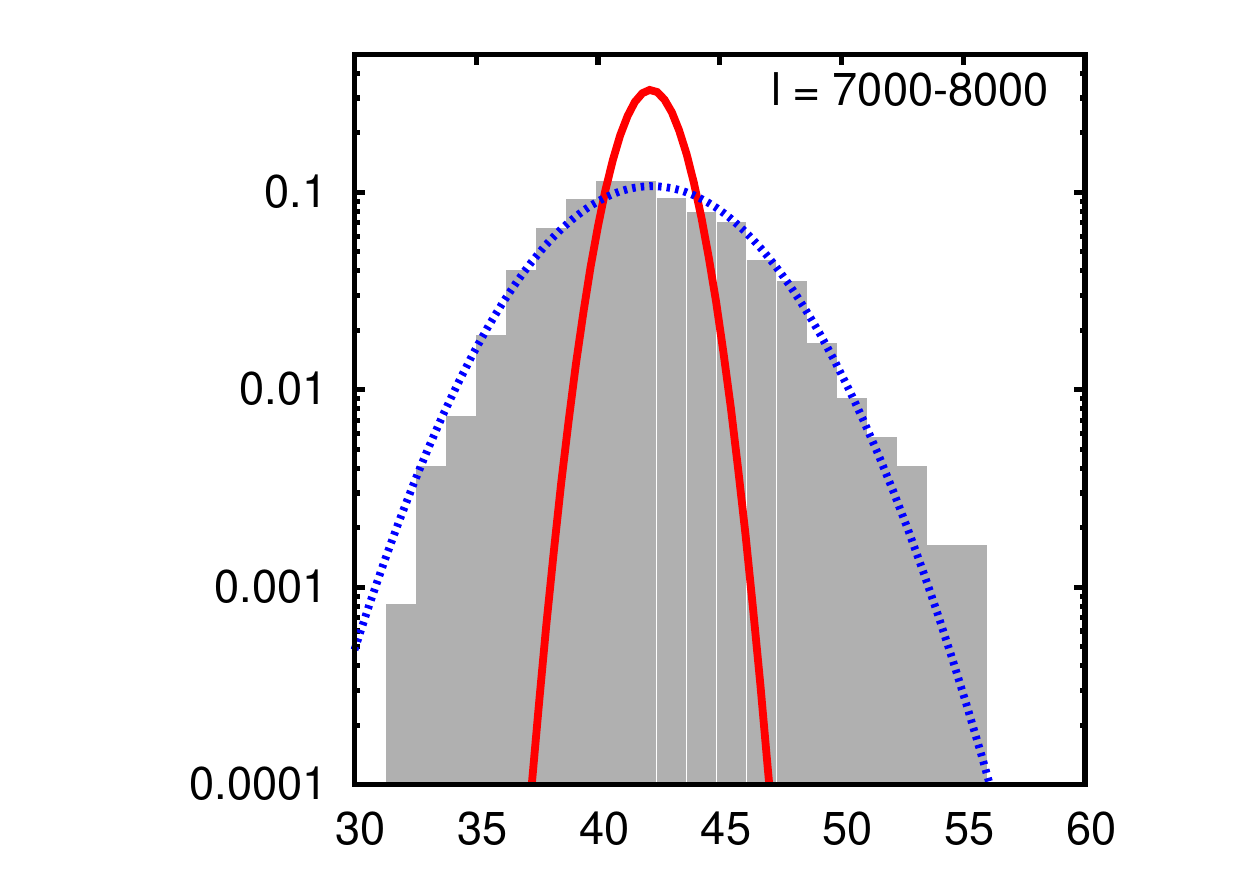}
  \includegraphics[scale=0.34,viewport=50 0 330 245,clip]{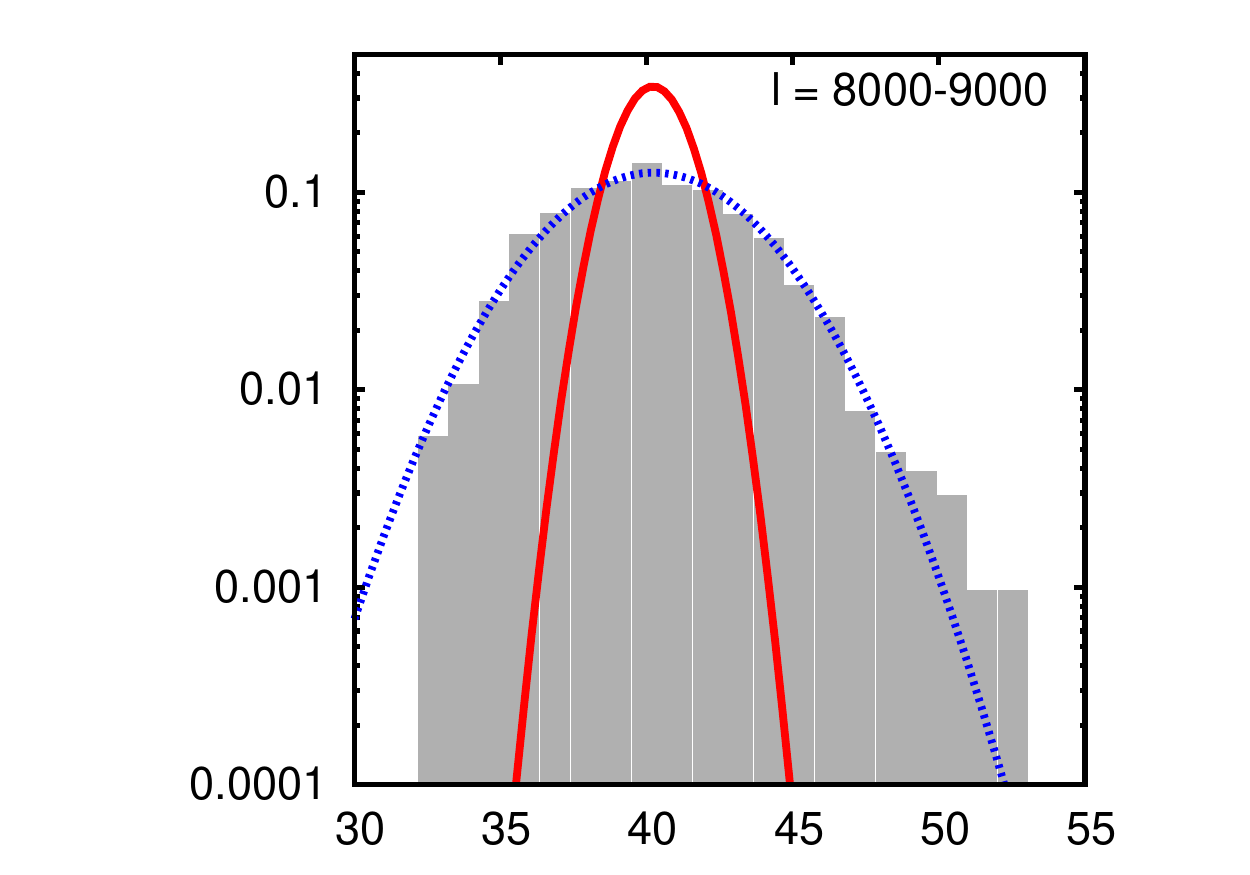}
	\caption[The statistics of the combined CMB and SZ effect power spectra for the $\sigma_8 = 0.825$ cosmology for a thousand $3\degree \times 3\degree$ maps]{The statistics of the combined CMB and SZ effect power spectra at 30~GHz for the $\sigma_8 = 0.825$ cosmology for a thousand $3\degree \times 3\degree$ maps. In each histogram, the x-axis is the binned power $B_i$ in the maps in $\upmu$K$^2$, and the y-axis is the probability density of that power in a map. The histograms on the top row show bins of 1000 multipoles between $l=$1000 and 4000; the second row 5000-9000. A Gaussian curve generated from the mean and standard deviation of the distribution is shown by the dotted blue line. The red solid line shows the equivalent for a set of 1000 realizations generated using the mean power spectrum from the combination of the CMB and SZ effect but assuming that all the power is Gaussianly distributed. This agrees at the lowest multipoles, but quickly becomes too narrow, indicative of the increased variance due to the SZ effect.}
	\label{fig:cmb_sz_plots_3deg}
\end{fig}
The resulting histograms for the combined $3\degree \times 3\degree$ CMB and SZ effect maps with $\sigma_8 = 0.825$ between $l=1000$ and $9000$ are shown in Figure \ref{fig:cmb_sz_plots_3deg}. To quantify the distributions, we compute the mean $\bar{B}_i$ for each bin $i$, as well as the standard deviation $\delta B_i$ and normalized skewness $s_i$. The latter is defined by
\begin{equation}
\mathrm{s_i} = \frac{1}{N}\sum_n \left(\frac{B_i^n - \bar{B}_i}{\sigma} \right)^3,
\end{equation}
where $B_i^n$ is the value of the binned $l(l+1) C_l / (2 \pi)$ from the nth realization, $\bar{B}_i$ is the mean of those values, N is the number of realizations and the sum is over all realizations. The computed values for all three values of $\sigma_8$, for both the power spectrum of the SZ effect on its own and that from the combined CMB and SZ effect, are given in Table \ref{tab:sz_statistics}.

\begin{tabland}{htb}
\begin{tabular}{c|cccc|cccc|cccc}
& \multicolumn{4}{c|}{$\sigma_8 = 0.75$} & \multicolumn{4}{c|}{$\sigma_8 = 0.825$} & \multicolumn{4}{c}{$\sigma_8 = 0.9$}\\
\hline
Multipoles & $\bar{B}_i$ & $\delta B_i$ & $s_i$ & $\delta B_i / \delta B_{i,\mathrm{G}}$ & $\bar{B}_i$ & $\delta B_i$ & $s_i$ & $\delta B_i / \delta B_{i,\mathrm{G}}$ & $\bar{B}_i$ & $\delta B_i$ & $s_i$ & $\delta B_i / \delta B_{i,\mathrm{G}}$\\
\hline % SZ only
1000-2000 & 16 & 7.4 & 2.6 & 7.3 & 31 & 12 & 1.8 & 6.0 & 58 & 23 & 2.4 & 6.3\\
2000-3000 & 20 & 5.5 & 1.4 & 5.1 & 40 & 9.3 & 0.94 & 4.4 & 74 & 17 & 1.0 & 4.3\\
3000-4000 & 22 & 4.3 & 0.86 & 4.4 & 44 & 7.5 & 0.72 & 4.0 & 81 & 14 & 0.72 & 3.8\\
4000-5000 & 23 & 3.4 & 0.66 & 3.9 & 45 & 6.0 & 0.65 & 3.5 & 83 & 11 & 0.56 & 3.5\\
5000-6000 & 23 & 2.8 & 0.39 & 3.5 & 45 & 5.0 & 0.49 & 3.3 & 82 & 9.1 & 0.52 & 3.3\\
6000-7000 & 22 & 2.3 & 0.45 & 3.3 & 44 & 4.3 & 0.50 & 3.0 & 80 & 7.7 & 0.41 & 3.0\\
7000-8000 & 21 & 2.0 & 0.45 & 3.1 & 42 & 3.7 & 0.40 & 3.0 & 77 & 6.6 & 0.46 & 2.8\\
8000-9000 & 20 & 1.7 & 0.41 & 3.0 & 40 & 3.2 & 0.40 & 2.6 & 73 & 5.7 & 0.36 & 2.7\\
9000-10~000 & 19 & 1.5 & 0.38 & 2.9 & 38 & 2.8 & 0.41 & 2.7 & 70 & 4.9 & 0.33 & 2.6\\
\hline % CMB + SZ
1000-2000 & 710 & 54 & 0.37 & 1.0 & 720 & 56 & 0.40 & 1.0 & 750 & 62 & 0.26 & 1.1\\
2000-3000 & 130 & 9.3 & 0.24 & 1.1 & 150 & 13 & 0.59 & 1.4 & 180 & 19 & 0.60 & 1.9\\
3000-4000 & 36 & 4.5 & 0.77 & 2.7 & 58 & 7.7 & 0.70 & 3.1 & 94 & 14 & 0.72 & 3.3\\
4000-5000 & 25 & 3.4 & 0.64 & 3.6 & 47 & 6.1 & 0.68 & 3.4 & 85 & 11 & 0.55 & 3.4\\
5000-6000 & 23 & 2.8 & 0.39 & 4.1 & 45 & 5.0 & 0.50 & 3.2 & 82 & 9.1 & 0.51 & 3.2\\
6000-7000 & 22 & 2.4 & 0.45 & 3.3 & 44 & 4.3 & 0.49 & 3.1 & 80 & 7.7 & 0.40 & 3.1\\
7000-8000 & 22 & 2.0 & 0.44 & 3.3 & 42 & 3.7 & 0.40 & 3.1 & 77 & 6.6 & 0.46 & 2.9\\
8000-9000 & 21 & 1.7 & 0.40 & 2.9 & 40 & 3.2 & 0.39 & 2.7 & 74 & 5.7 & 0.36 & 2.6\\
9000-10~000 & 19 & 1.5 & 0.37 & 2.9 & 38 & 2.8 & 0.41 & 2.8 & 70 & 4.9 & 0.33 & 2.7\\
\end{tabular}
\caption[Statistics of the SZ effect for $3\degree \times 3\degree$ maps]{The statistics of the SZ effect (top section) and combined CMB and SZ effect (bottom section), generated using the $3\degree \times 3\degree$ clustered galaxy cluster catalogues from the modified {\sc Pinocchio} simulations. The mean ($\bar{B}_i$) and the standard deviation ($\delta B_i$) within bin $i$ are in units of $\upmu$K$^2$; the skew ($s$) is dimensionless. The ratio of the standard deviation to that from the Gaussianly distributed simulations with the same power spectrum is given by $\delta B_i / \delta B_{i,\mathrm{G}}$. There is a considerable increase in the standard deviation from the SZ effect compared to Gaussian distributions, and the distribution is also skewed.}
\label{tab:sz_statistics}
\end{tabland}

\begin{fig}
\centering
\includegraphics[scale=0.68]{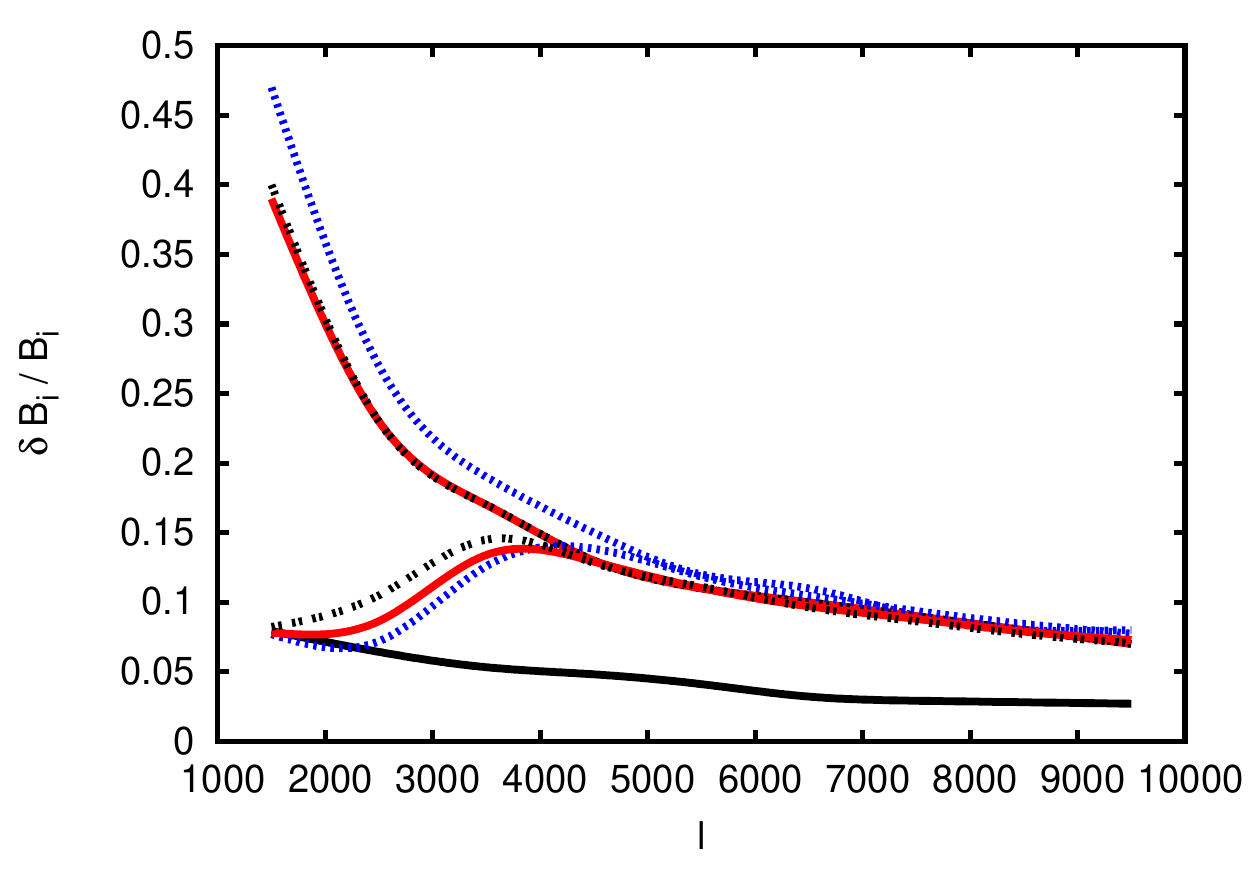}
\caption[The ratio of the standard deviation to the mean for the CMB only, the SZ effect only and the combined CMB and SZ effect for the three different values of $\sigma_8$]{The ratio of the standard deviation to the mean for the CMB only (solid black line), the SZ effect only (top three lines) and the combined CMB and SZ effect (central lines) for the three different values of $\sigma_8$ (0.75: blue dotted line, 0.825: red solid line, 0.9: black double-dotted line) from the $3\degree \times 3\degree$ maps. The values for the CMB scale as $l^{-1/2}$, as per equation \ref{eq:cv}, whereas those from the SZ effect scale as $l^{-1}$ but with a higher amplitude.}
\label{fig:cmb_sz_stats}
\end{fig}

The SZ effect significantly increases the standard deviation of the realization statistics for all but the lowest multipoles. This is shown graphically in Figure \ref{fig:cmb_sz_stats}, where the ratio of the standard deviation to the mean is given for the CMB and SZ effects separately, and then combined. The SZ effect has a much higher ratio than for the CMB on its own. When combined with the CMB, the realizations have a low standard deviation at the lowest multipoles, where the CMB is dominant, but quickly return to higher values at the higher multipoles where the CMB has decreased in power due to Silk damping and the SZ effect is closer to its peak power. The point at which the transition occurs is slightly different for the three values of $\sigma_8$; this is due to the different power levels from the SZ effect, with the $\sigma_8 = 0.9$ cosmology having more power and hence transitioning at lower multipoles.

Figure \ref{fig:cmb_sz_plots_3deg_150ghz} shows the equivalent histograms to Figure \ref{fig:cmb_sz_plots_3deg} but at 150~GHz. The SZ effect at this frequency is weaker by a factor of 2 than at 30~GHz, reducing the power spectrum by a factor of 4. As a result, the CMB dominates the power spectrum to higher multipoles and the increase in standard deviation due to the SZ effect does not become important until $l \sim 4000$. The highest multipoles considered here are not greatly affected by this change in frequency as these remain dominated by the SZ effect.

\begin{fig}
\centering
  \includegraphics[scale=0.34,viewport=50 0 330 245,clip]{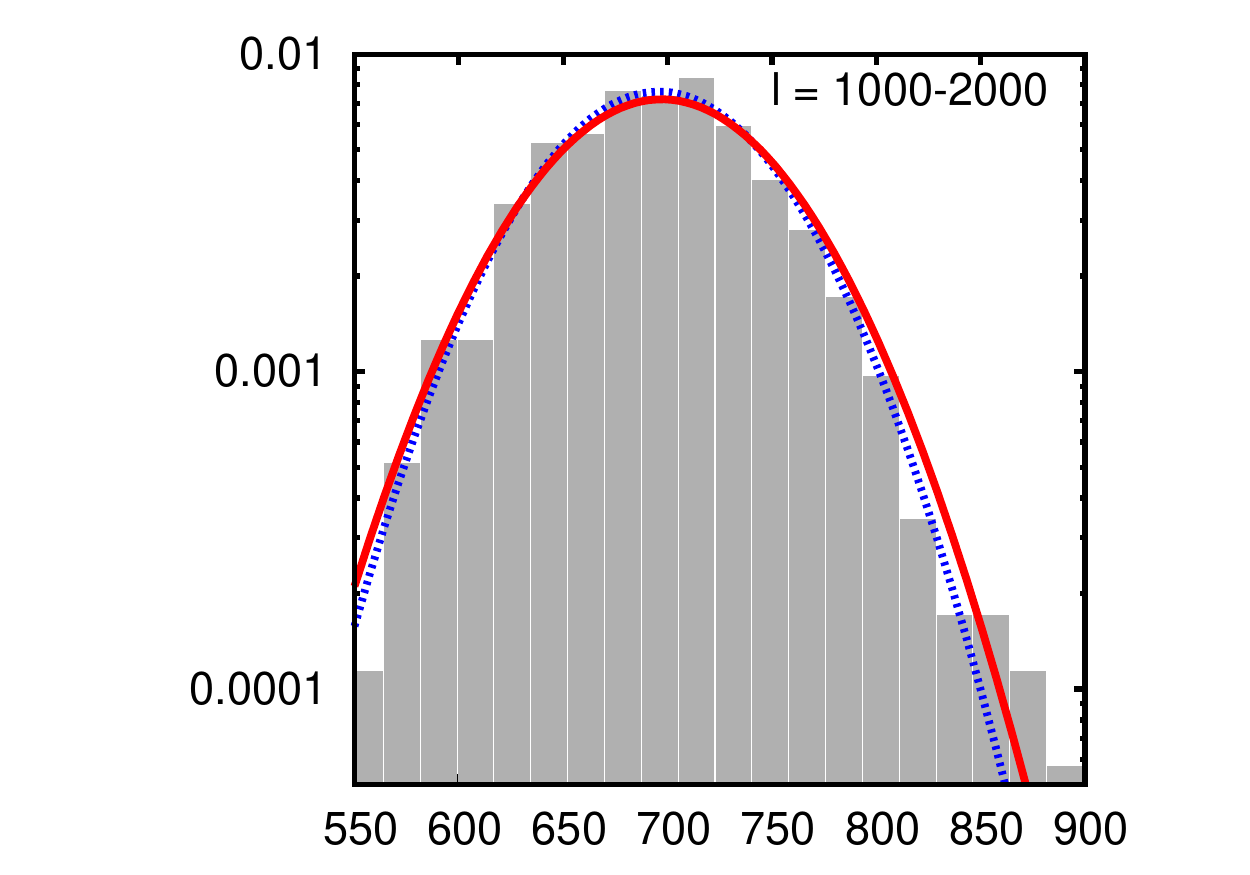}
  \includegraphics[scale=0.34,viewport=50 0 330 245,clip]{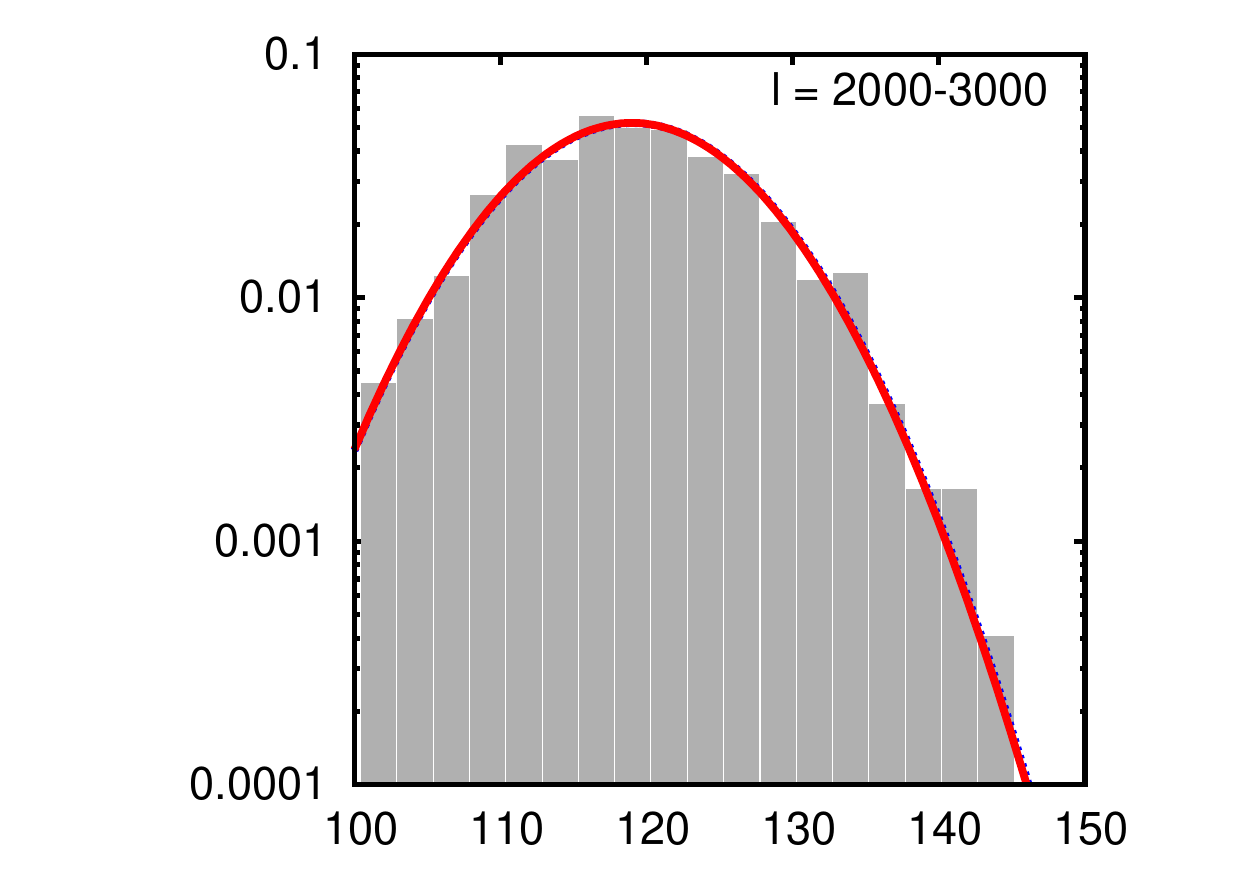}
  \includegraphics[scale=0.34,viewport=50 0 330 245,clip]{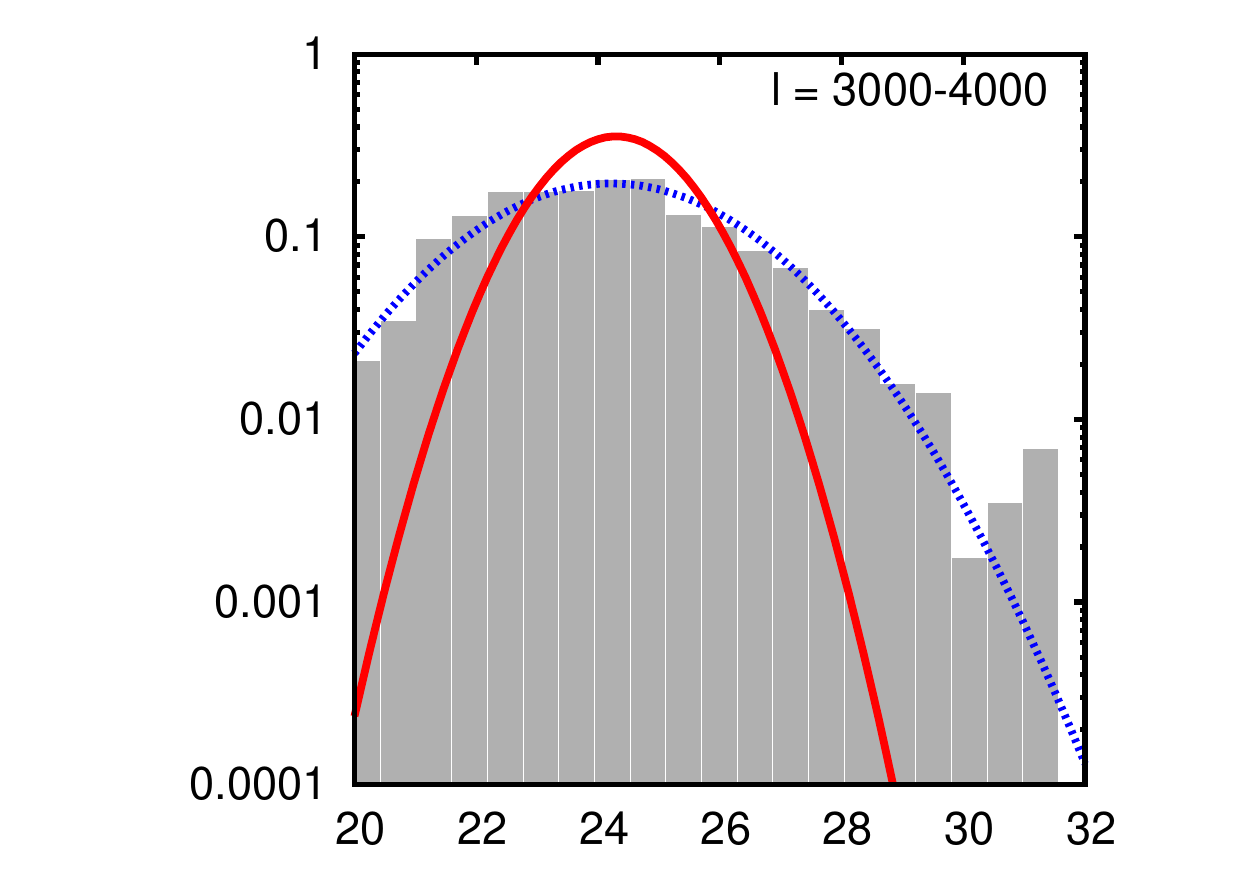}
  \includegraphics[scale=0.34,viewport=50 0 330 245,clip]{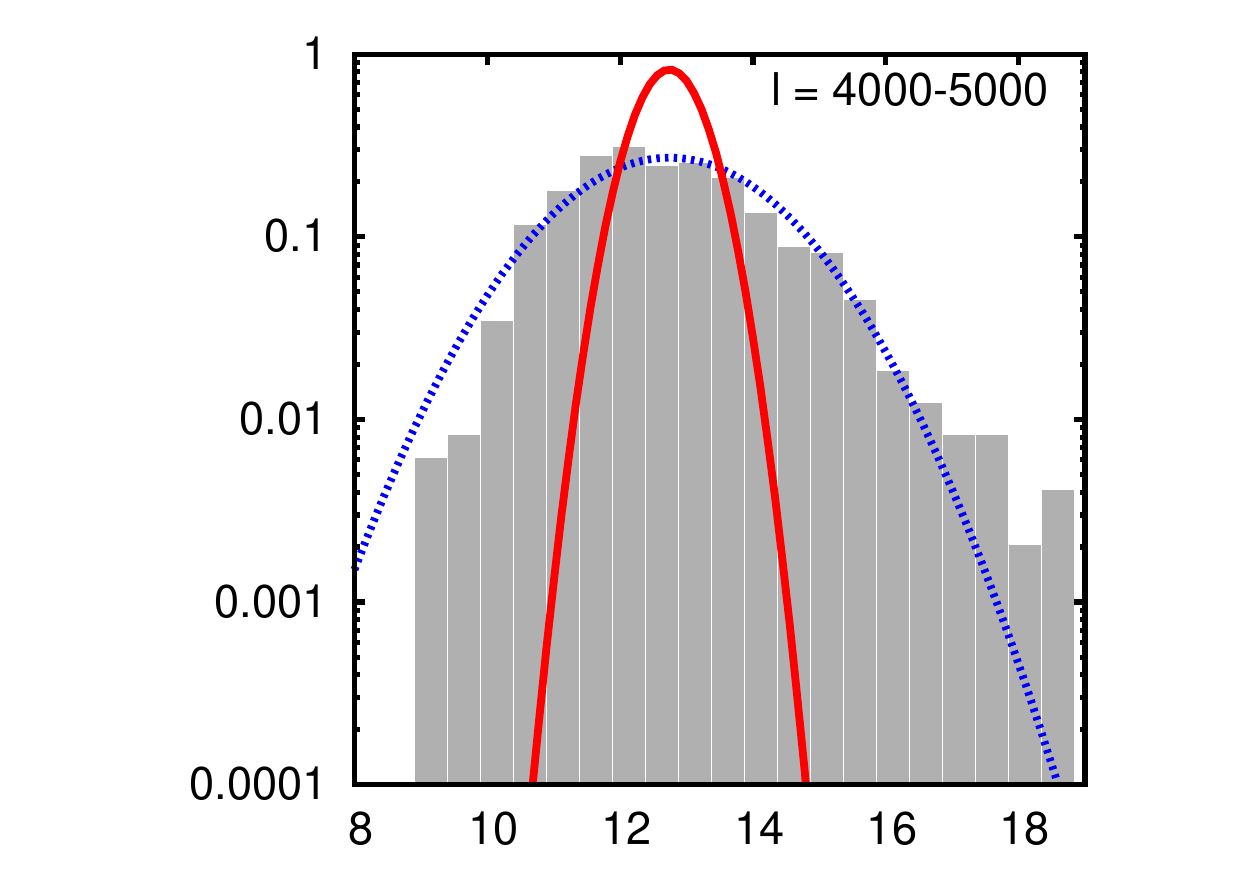}
  \\
  \includegraphics[scale=0.34,viewport=50 0 330 245,clip]{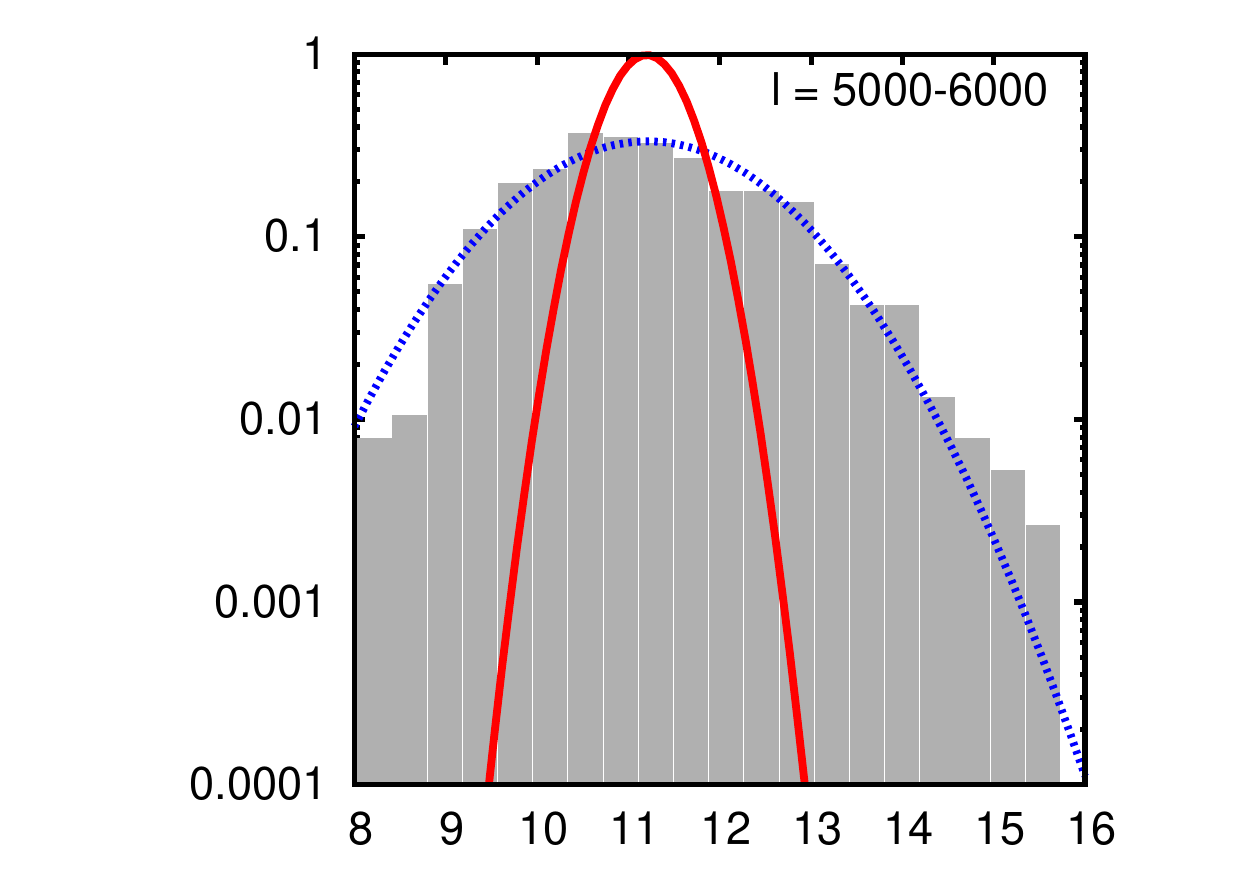}
  \includegraphics[scale=0.34,viewport=50 0 330 245,clip]{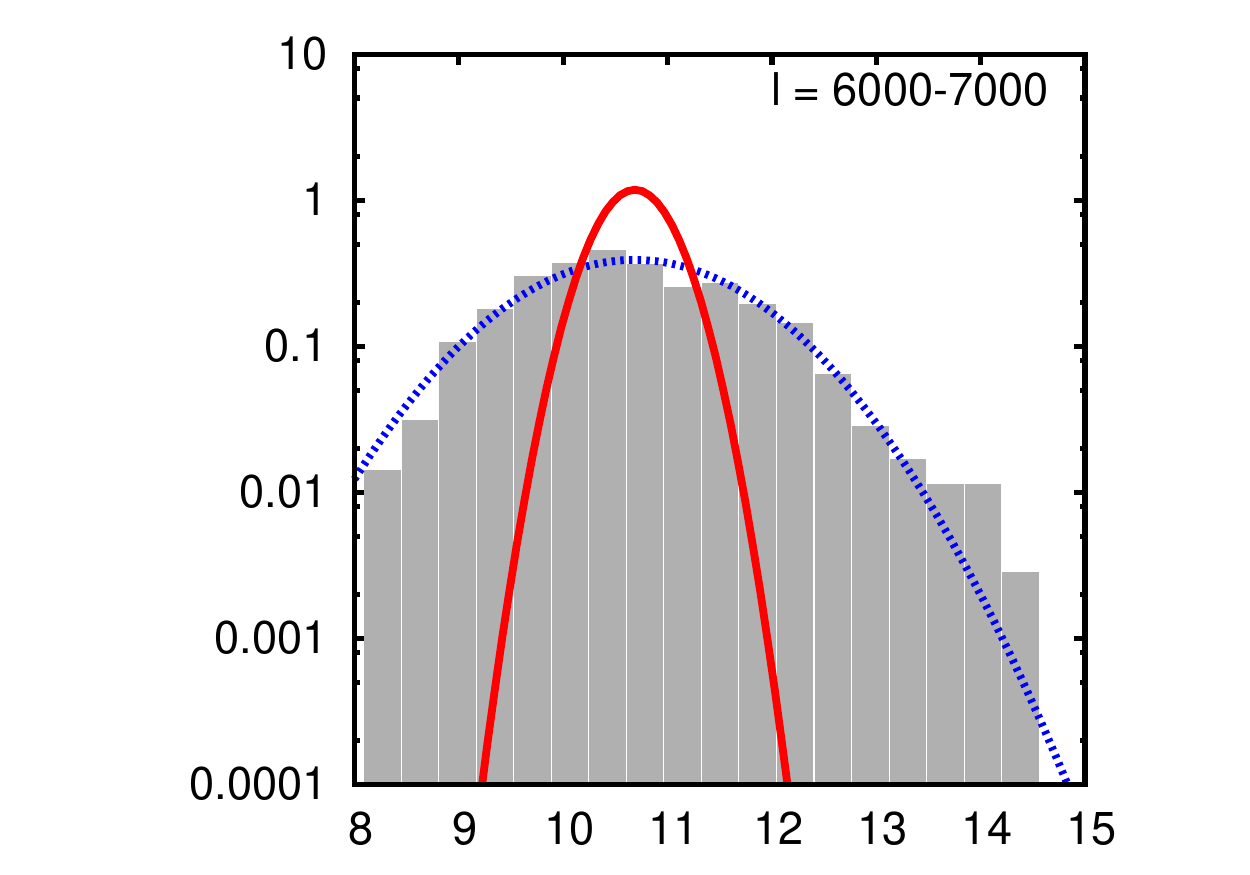}
  \includegraphics[scale=0.34,viewport=50 0 330 245,clip]{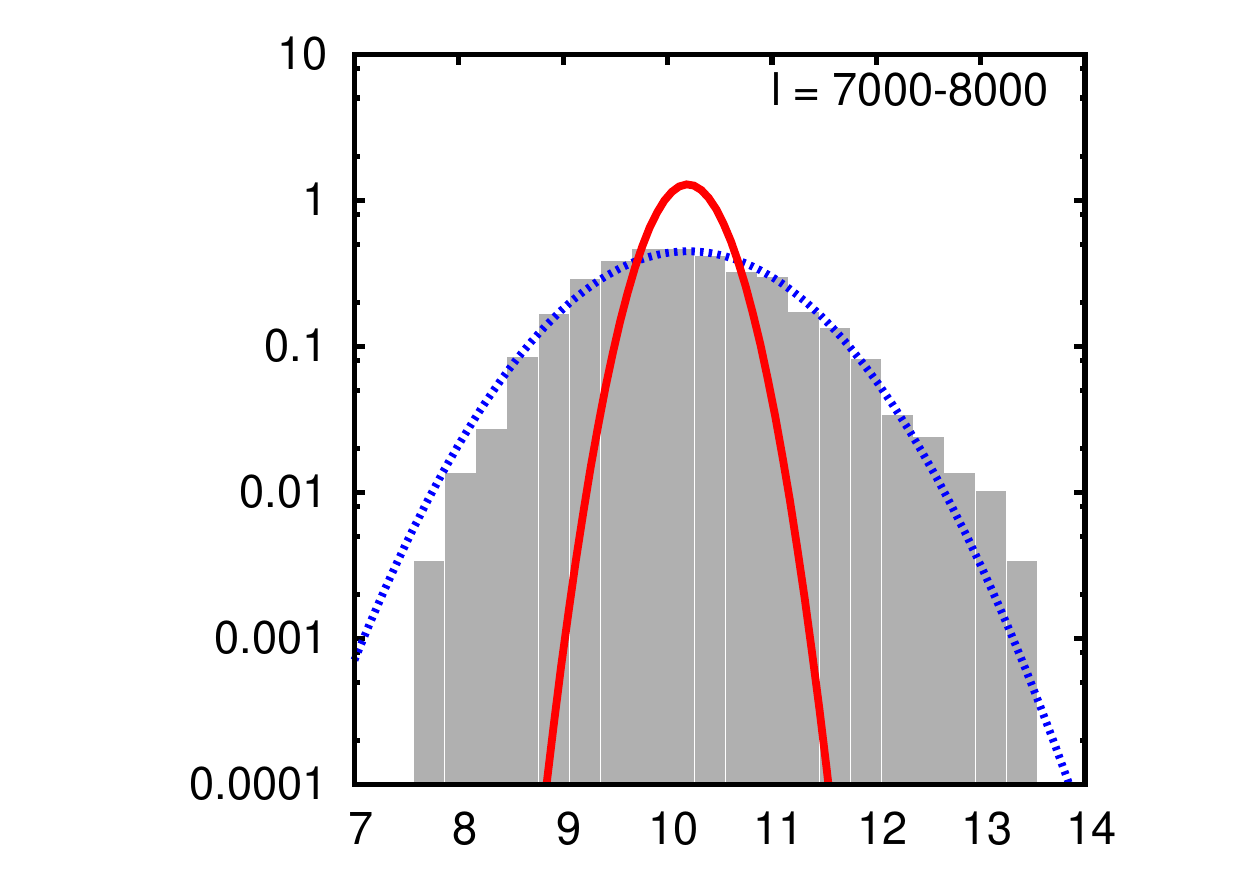}
  \includegraphics[scale=0.34,viewport=50 0 330 245,clip]{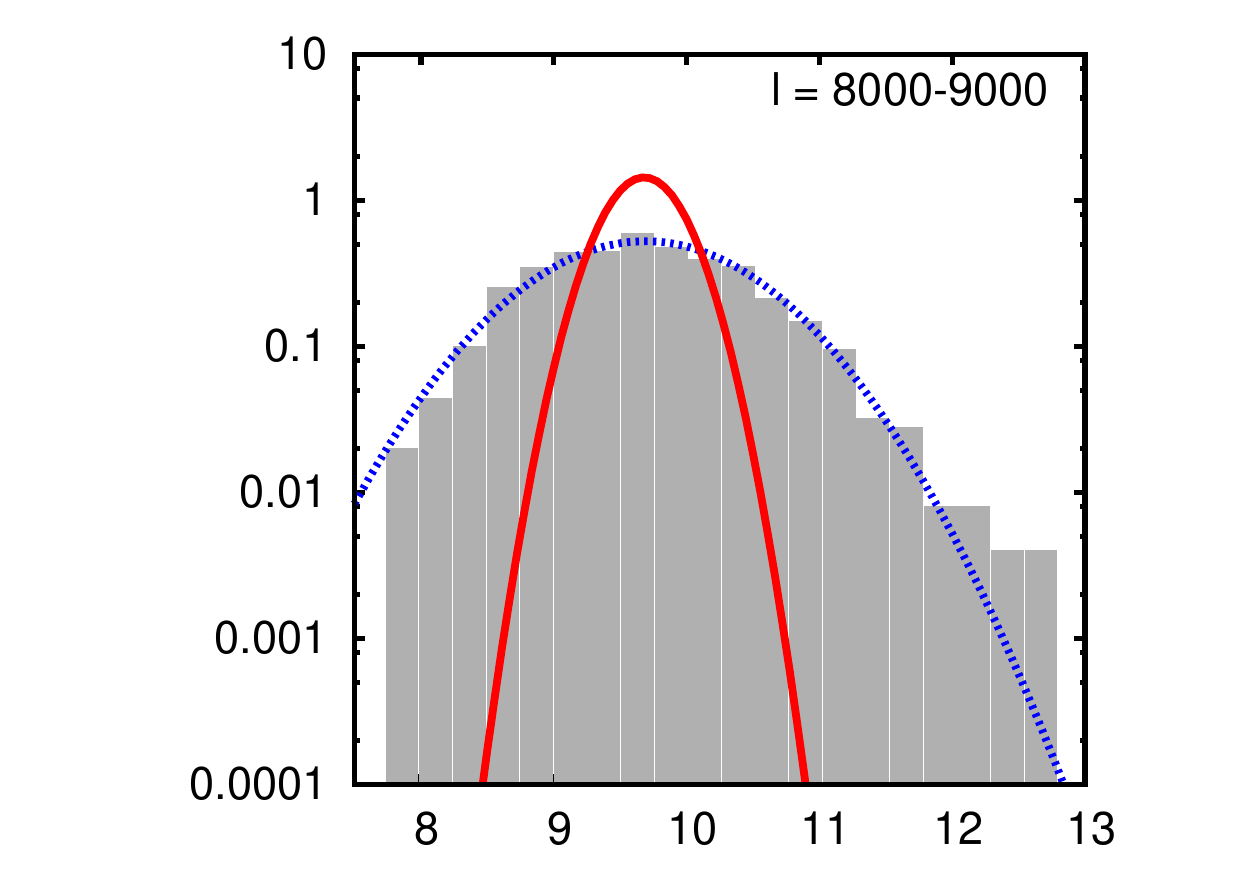}
	\caption[As Figure \ref{fig:cmb_sz_plots_3deg}, but for 150~GHz]{As Figure \ref{fig:cmb_sz_plots_3deg}, but for 150~GHz. The CMB dominates to higher multipoles than at 30~GHz., such that the broadening of the distributions due to the SZ effect is less important at intermediate multipoles.}
	\label{fig:cmb_sz_plots_3deg_150ghz}
\end{fig}

To quantify more precisely this increase in standard deviation, we create realizations with the same mean power spectrum but Gaussianly distributing all of the power in the map, as if all of the power is due to the CMB. The Gaussian fit to these distributions is shown by the red solid lines in Figures \ref{fig:cmb_sz_plots_3deg} and \ref{fig:cmb_sz_plots_3deg_150ghz}. At the lowest multipoles, where the CMB is dominant, the statistics from this method are in good agreement with those from the realizations. At higher multipoles, as expected, the standard deviation is much lower for Gaussian statistics than for the SZ effect. The ratios of the standard deviation from the two methods are given in Table \ref{tab:sz_statistics}; the standard deviation is much smaller, typically by a factor of 3 for $l>3000$, when compared with the non-Gaussian realizations.

In Section \ref{sec:example_realisations}, the mean power spectra were found to scale approximately as $C_l \propto \sigma_8^\alpha$, where $\alpha \sim 7$, as per \citet{2002Komatsu}. Assuming that the standard deviation varies with $\sigma_8$ in the same fashion, we find that $\alpha_\mathrm{SD} \approx 6.0$ between $\sigma_8 = 0.75$ and 0.825, and $\approx 6.8$ between 0.9 and 0.825. Thus, there is a strong dependence on $\sigma_8$ for the standard deviation, although not quite as strong as for the mean power spectrum. The curve for $\sigma_8 = 0.75$ is slightly higher than for the other two values of $\sigma_8$ in Figure \ref{fig:cmb_sz_stats}; this is an illustration of the non-universality of $\alpha_\mathrm{SD}$.

In Section \ref{sec:theoretical_ps}, the mean theoretical spectrum calculated by an analytical formula was compared to the mean from the realizations and was found to agree well. The standard deviation can also be computed; see the Appendix for the method. The standard deviation calculated solely from the Gaussian components differs from that calculated from the realizations by a factor between 3 and 7 depending on the multipole; when the angular trispectrum component is added the two agree to within 20 per cent when compared with the results from the {\sc Pinocchio} realizations presented in Figure \ref{fig:cmb_sz_stats}. The origin of the remaining discrepancy will be discussed in Section \ref{sec:clustering}.

We find that the SZ effect results in a positive skew in the statistics, resulting from an overabundance of high-power realizations when compared with the expectations from Gaussian statistics due to the presence of rare, massive clusters and associated smaller clusters. This is shown in Figure \ref{fig:skewness}; the lowest multipoles for the SZ effect on its own are especially skewed. With the addition of the CMB, the low multipole bins become Gaussian as the CMB dominates, but some skewness remains at the higher multipoles where the SZ effect is dominant. There is a certain amount of noise in the measurements of the skew here, however; an increased number of realizations would measure this more accurately. \citet{2007Zhang} have also found that the SZ effect yields a positively skewed distribution using an analytical approach for calculating the probability distribution for a single multipole.
\begin{fig}
\centering
\includegraphics[scale=0.68]{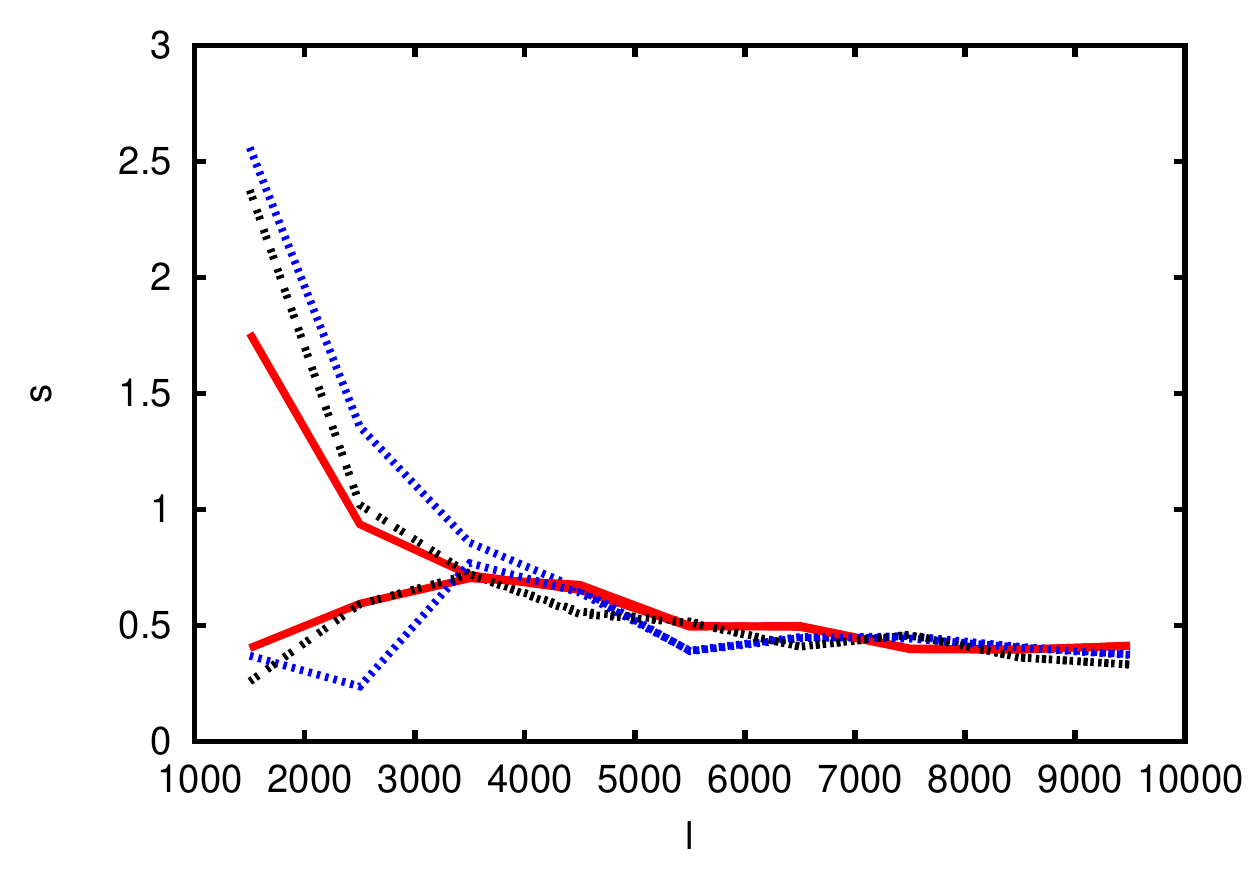}
\caption[The skewness $s$ for the $3\degree \times 3\degree$ maps for the three values of $\sigma_8$]{The skewness $s$ for the $3\degree \times 3\degree$ maps for the three values of $\sigma_8$: $0.75$ is shown by the blue dotted line, $0.825$ by the red solid line and $0.9$ by the black double-dotted line. The top set of lines are for the SZ only; the bottom set are when combined with the CMB. There is little difference between the two above $l\sim4000$. At the lowest multipoles, the SZ effect significantly skews the distribution, and this skewness remains important at intermediate to high multipoles.}
\label{fig:skewness}
\end{fig}

\section{Correlation matrix}
For purely Gaussian maps, for example the CMB, the power on different scales is uncorrelated such that the off-diagonal terms of the covariance matrix are zero. However, this may not be the case for the SZ effect. The covariance matrix can be calculated by
\begin{equation}
C_{ij} = \frac{1}{N} \sum_{n} (B_i^n - \bar{B}_i) (B_j^n - \bar{B}_j).
\end{equation}
This is then normalized to give the correlation matrix by
\begin{equation}
\hat{C}_{ij} = \frac{C_{ij}}{\sqrt{C_{ii} C_{jj}}}.
\end{equation}

The correlation matrix is depicted graphically for the CMB and SZ effect individually and combined in Figure \ref{fig:sz_covariance}. The matrix is diagonal for the CMB, however the off-diagonal terms are significant for the SZ effect, in agreement with the predictions of \citet{2001Cooray}. This is due to the power from individual clusters spanning many thousands of multipoles, correlating the multipole bins considered here. This effect is reduced for widely separated multipole bins as individual clusters are not dominant at all multipoles; for example, as shown by Figure \ref{fig:convergence_testing} the largest clusters only provide the bulk of the SZ effect at the lowest multipoles, and smaller (mostly uncorrelated) clusters provide the power at the highest multipoles.

\begin{fig}	
\centering
\includegraphics[scale=0.55,viewport=80 30 300 215,clip]{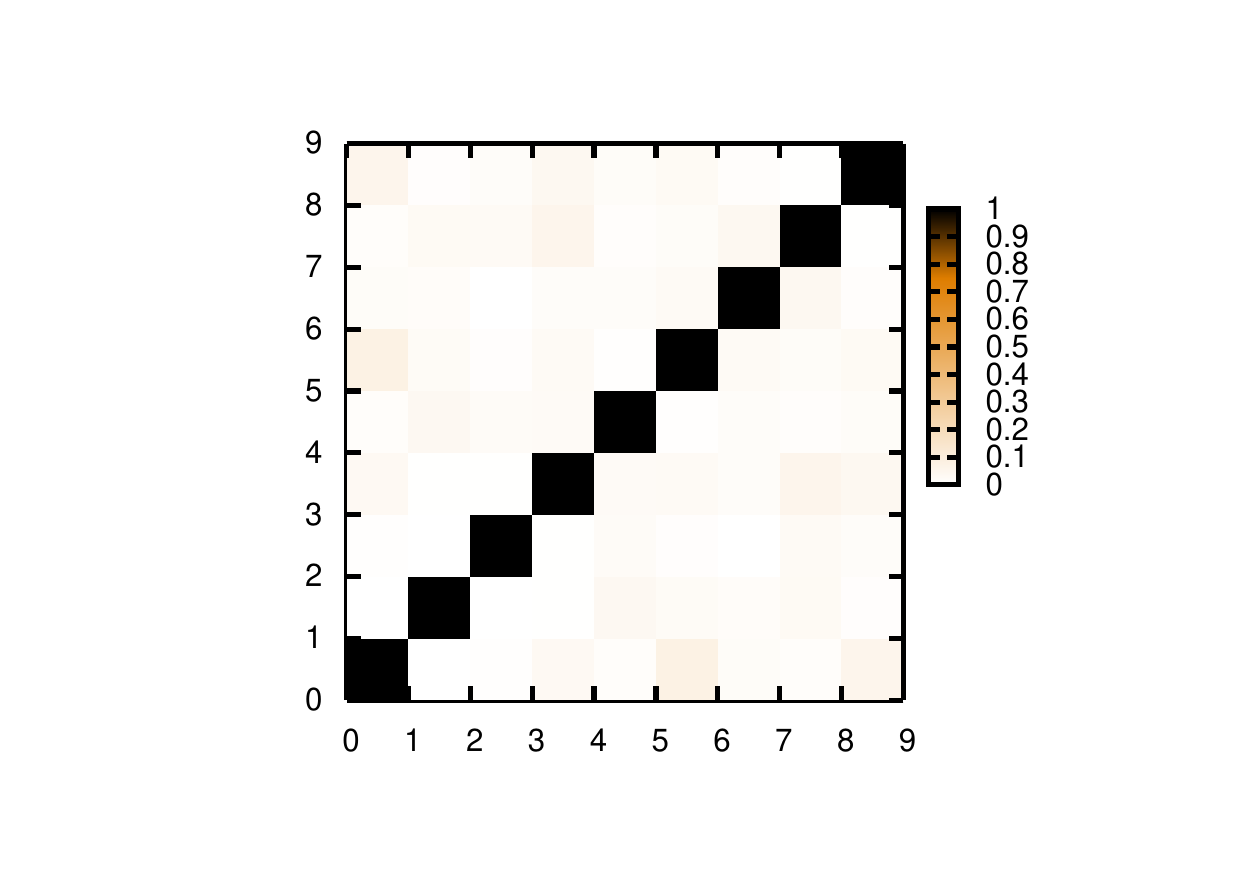}
\includegraphics[scale=0.55,viewport=80 30 300 215,clip]{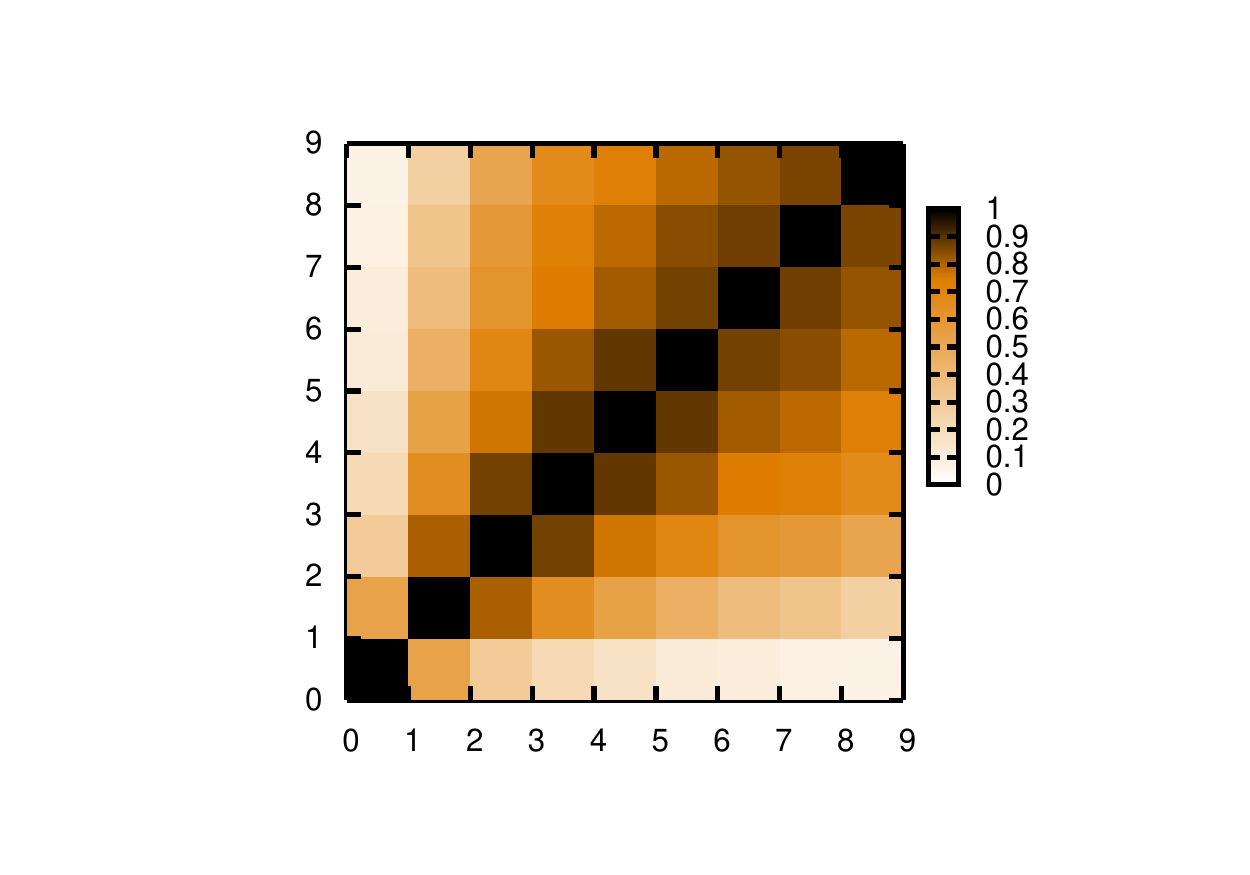}
\includegraphics[scale=0.55,viewport=80 30 300 215,clip]{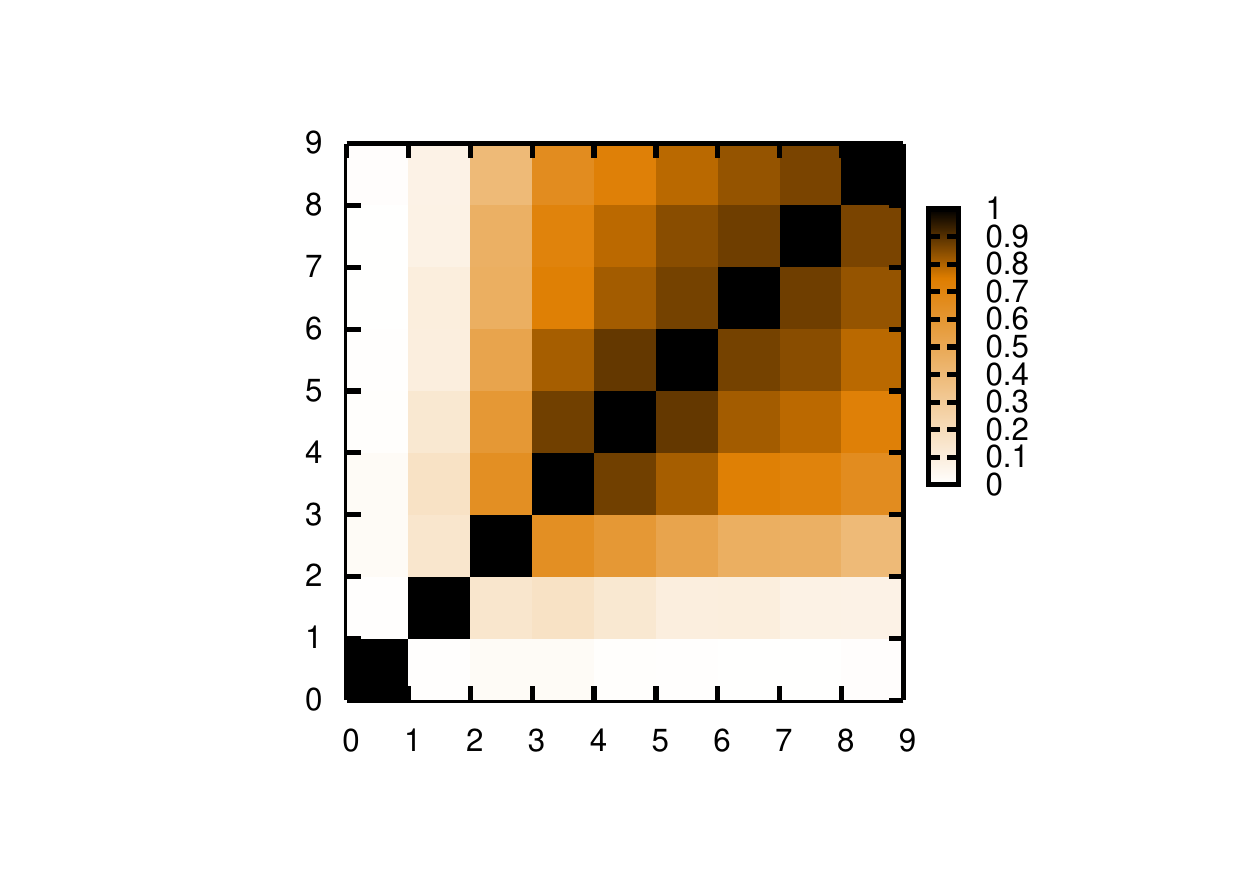}
\caption[The correlation matrix for 1000 realizations of the CMB, SZ effect and the combined CMB and SZ effect]{The correlation matrix for 1000 realizations of the CMB (left), SZ effect (middle) and the combined CMB and SZ effect (right), using $3\degree \times 3\degree$ maps with bins of $\Delta l = 1000$. The bin numbers are given in the $x$ and $y$ axis (for example, 0 is the multipole range $0-1000$). The CMB has negligible off-diagonal terms, however these terms are no longer negligible when the SZ effect is accounted for.}
\label{fig:sz_covariance}
\end{fig}

\section{Dependency on map size}
\begin{fig}
\centering
  \includegraphics[scale=0.34,viewport=50 0 330 245,clip]{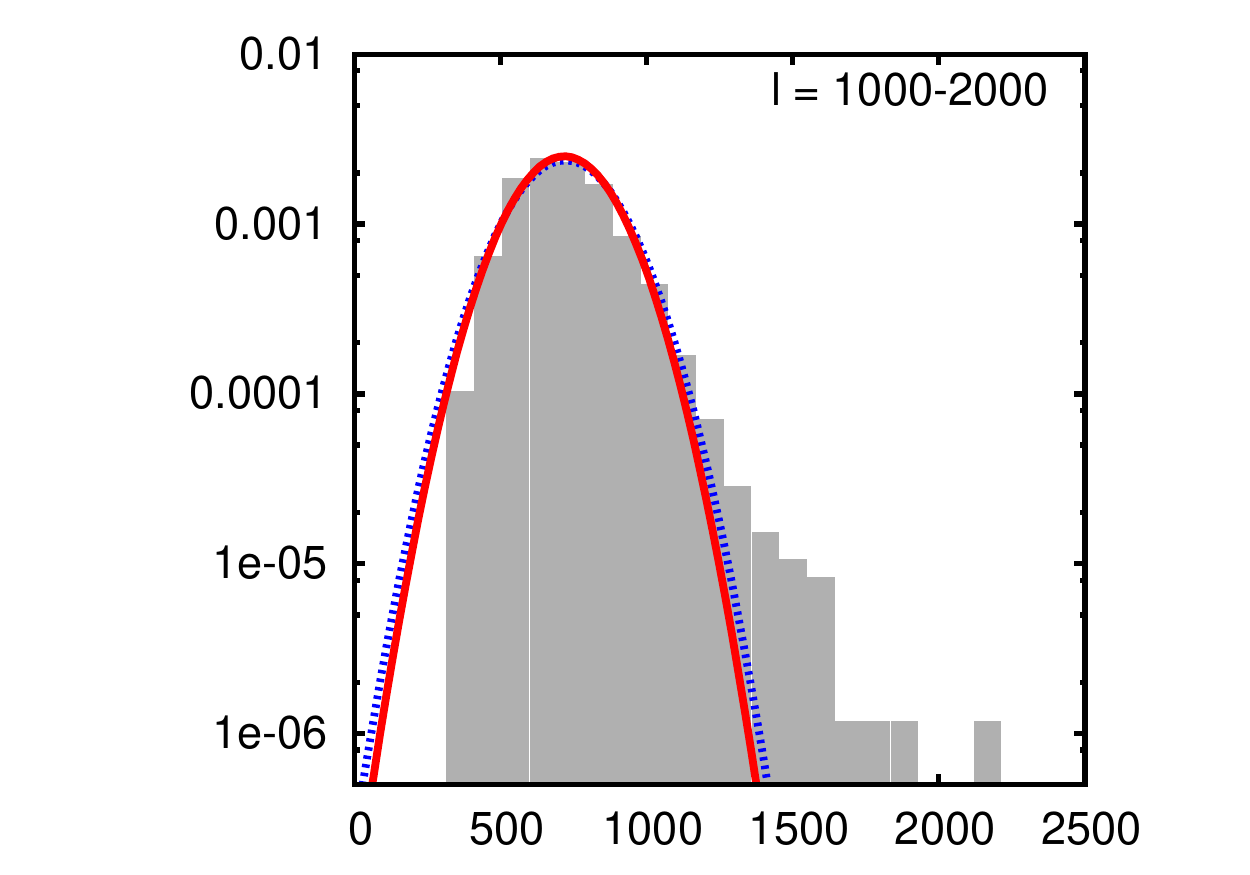}
  \includegraphics[scale=0.34,viewport=50 0 330 245,clip]{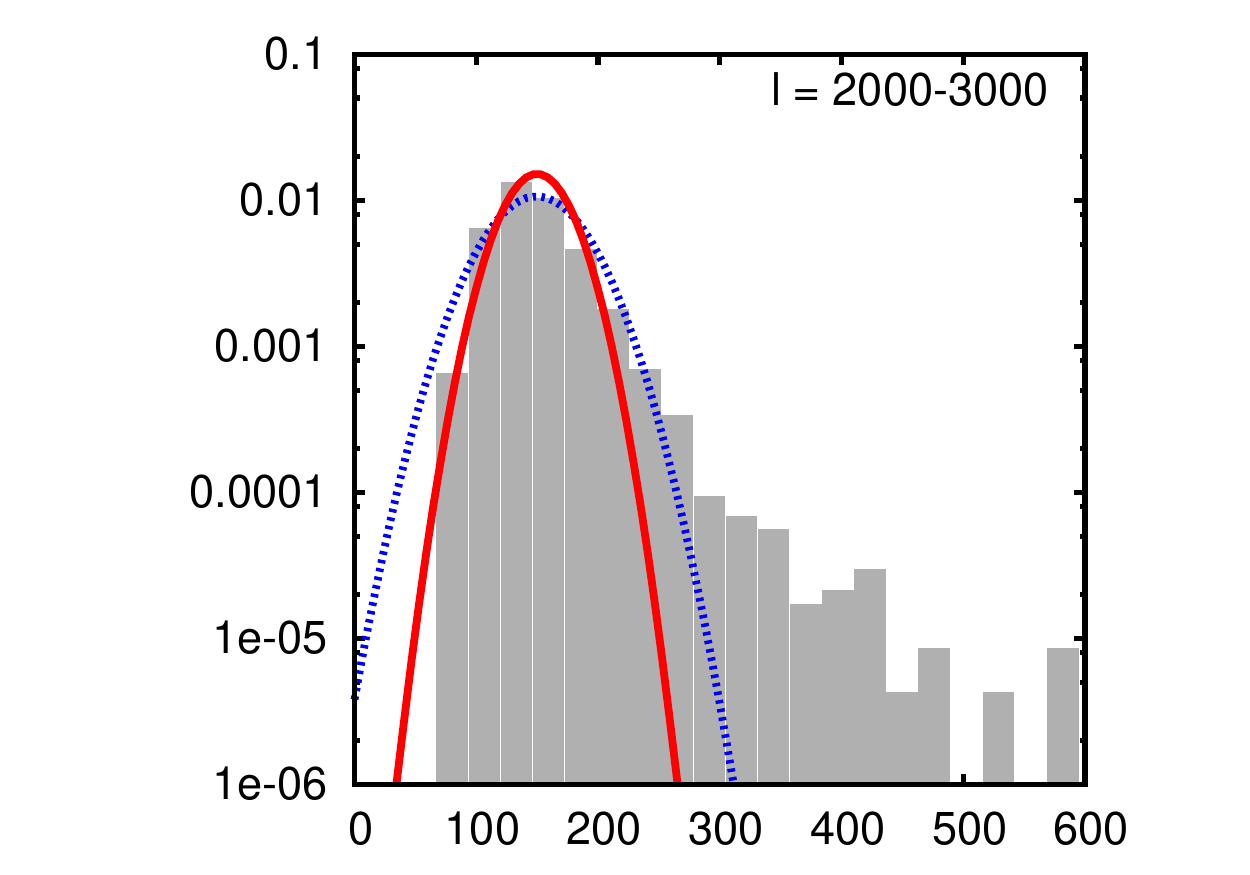}
  \includegraphics[scale=0.34,viewport=50 0 330 245,clip]{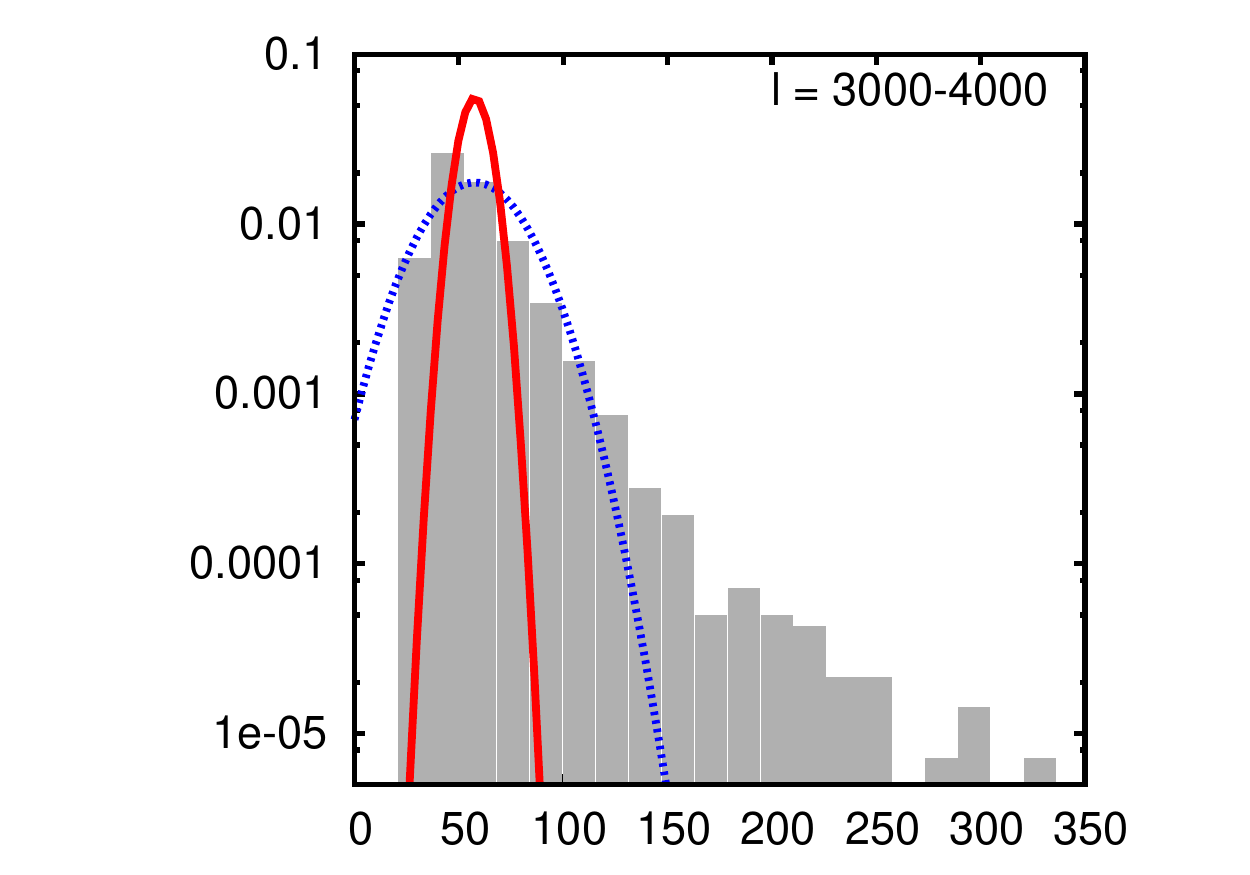}
  \includegraphics[scale=0.34,viewport=50 0 330 245,clip]{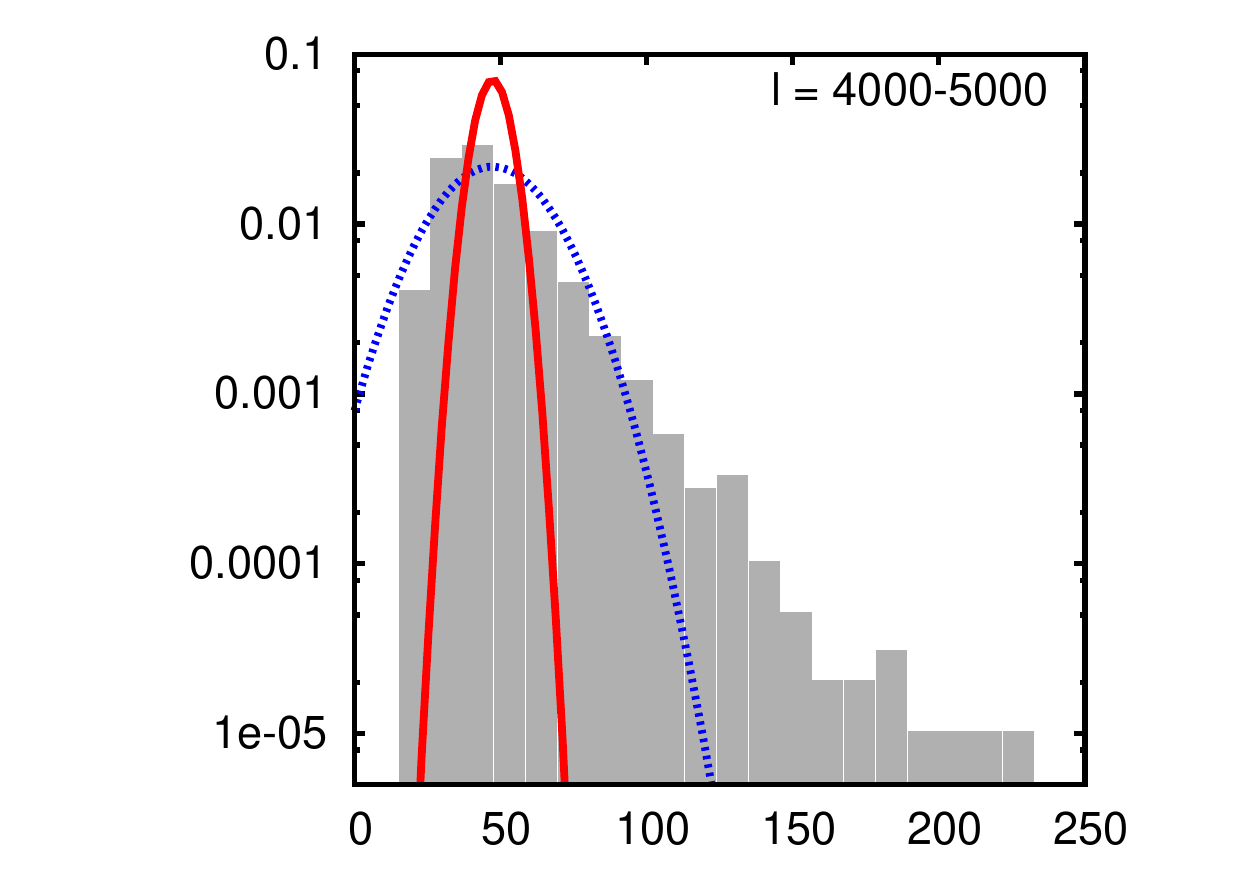}
  \\
  \includegraphics[scale=0.34,viewport=50 0 330 245,clip]{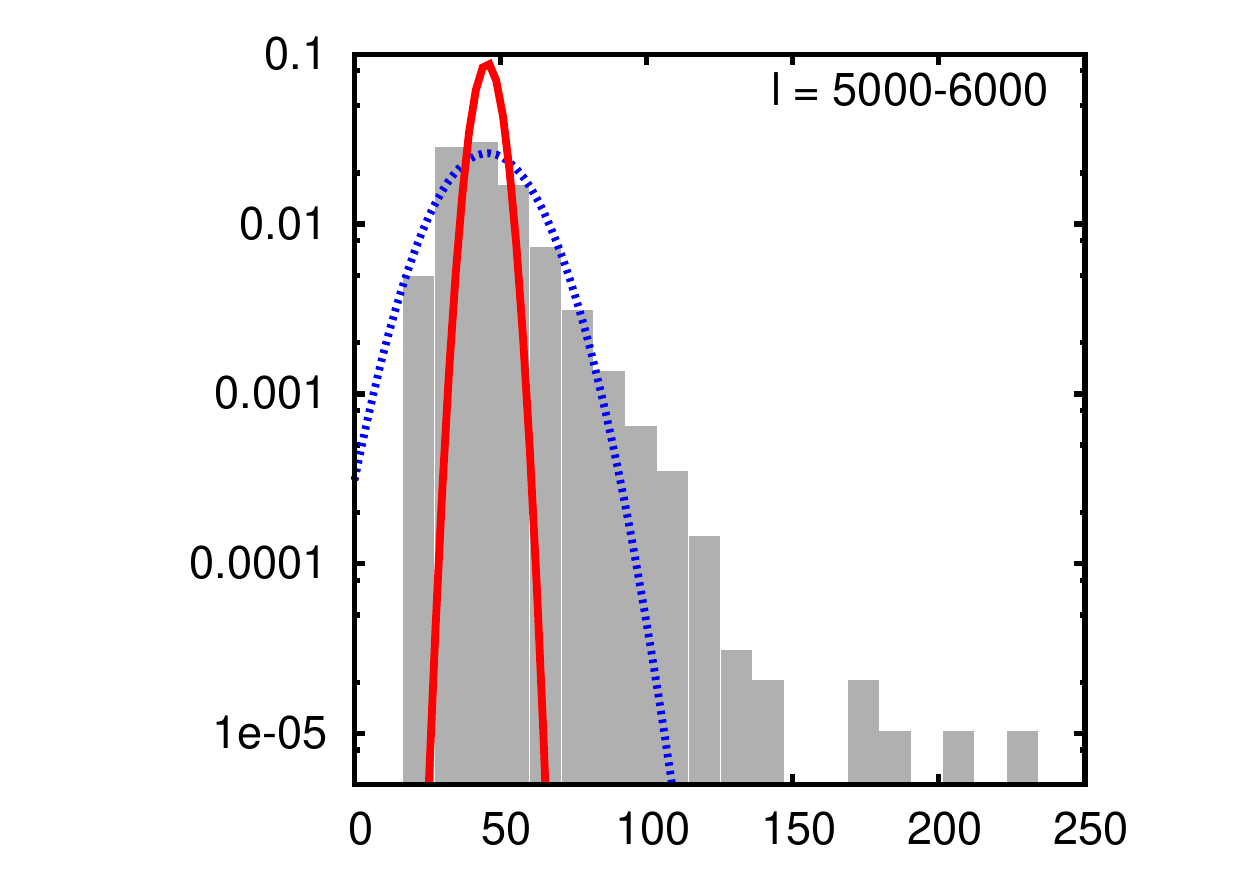}
  \includegraphics[scale=0.34,viewport=50 0 330 245,clip]{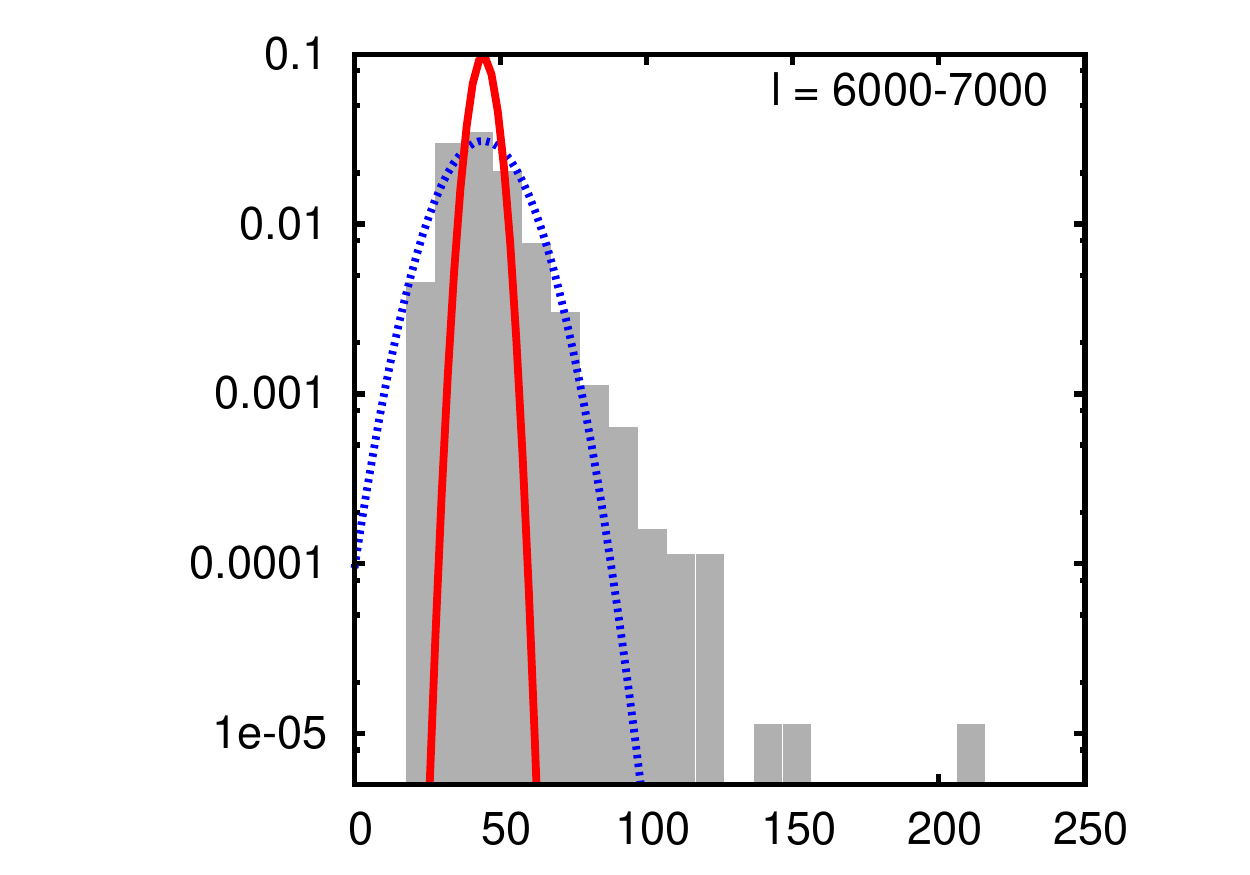}
  \includegraphics[scale=0.34,viewport=50 0 330 245,clip]{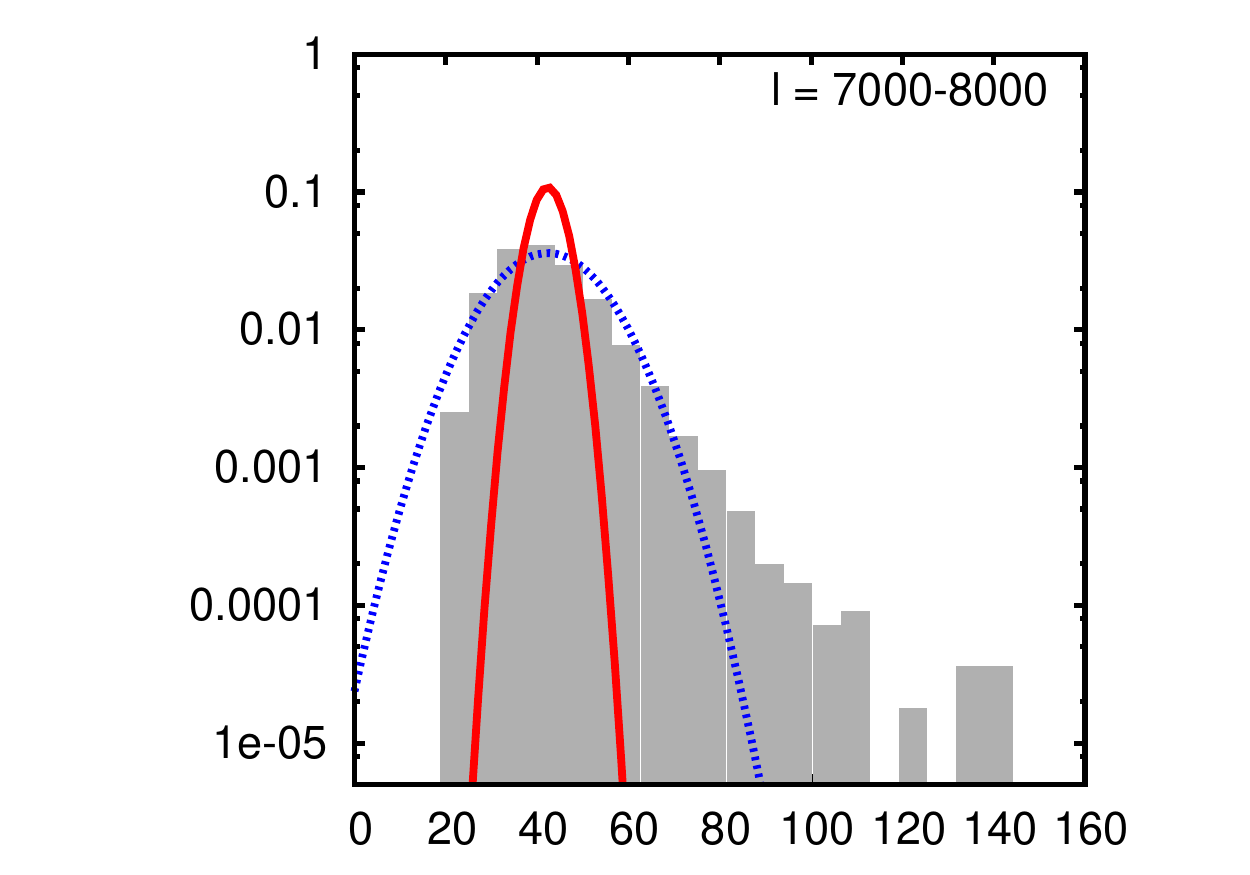}
  \includegraphics[scale=0.34,viewport=50 0 330 245,clip]{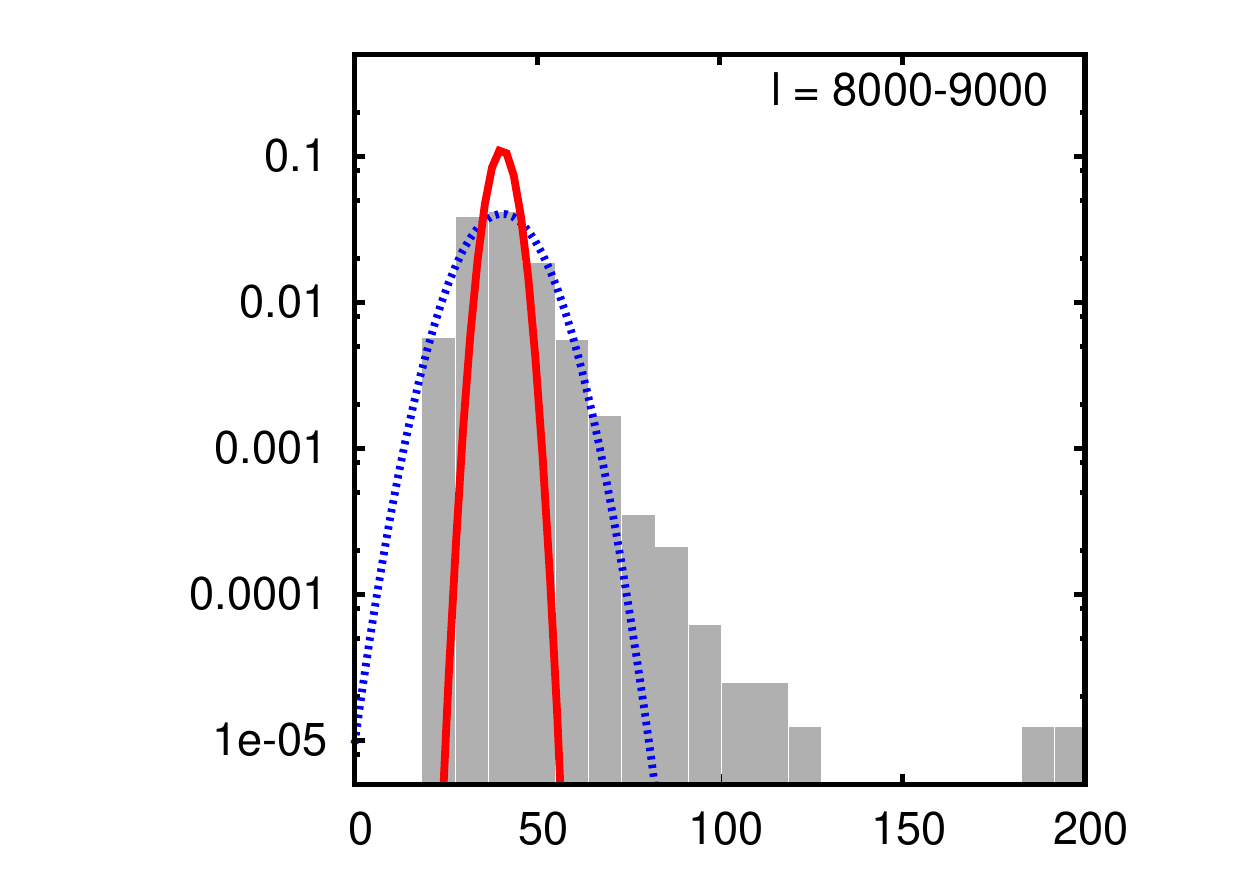}
	\caption[As Figure \ref{fig:cmb_sz_plots_3deg}, but for 9000 $1\degree \times 1\degree$ maps]{As Figure \ref{fig:cmb_sz_plots_3deg}, but for 9000 $1\degree \times 1\degree$ maps. The first few multipole bins are dominated by the CMB, hence have a distribution that is close to Gaussian; the later bins become increasingly more skewed to higher values. Extremal values are more likely than in the $3\degree \times 3\degree$ maps.}
	\label{fig:cmb_sz_plots_1deg}
\end{fig}

The statistics of the SZ effect power spectrum will depend on the size of the map being considered. We investigate this dependency by considering maps of $1\degree \times 1\degree$ and $2\degree \times 2\degree$, in addition to the $3\degree \times 3\degree$ maps that have been investigated thus far. These smaller maps are created by selecting and mapping only the clusters from the catalogues that lie within these areas. For the $1\degree \times 1\degree$ maps, this provides 9000 independent realizations, whereas for the $2\degree \times 2\degree$ maps we continue using 1000 realizations. The mean power spectra for these three different map sizes remain the same.

The histograms for the $1\degree \times 1\degree$ realizations containing both the CMB and the SZ effect are shown in Figure \ref{fig:cmb_sz_plots_1deg}. As before, for the lowest multipoles the CMB dominates, so the histogram for the lowest bin is approximately Gaussian. For the higher multipoles, however, the histograms are significantly skewed towards positive values, much more so than for the $3\degree \times 3\degree$ maps. This qualitatively agrees with \citet{2007Zhang}, who found that larger map sizes decrease the skewness of the probability distribution due to averaging over a larger number of clusters.

For the CMB, the standard deviation from the realizations scales linearly with $1/f_\mathrm{sky}^{1/2}$. The ratio of the standard deviation to the mean for the three maps sizes is shown in Figure \ref{fig:mapsize}; this scales as $1/f_\mathrm{sky}^{1/2}$ for both the SZ effect on its own and the combined CMB and SZ effect. This is as predicted by the analytical formula. The transition between the CMB- and SZ-dominated regimes is clear for all three map sizes. The skew for the SZ effect also scales close to $1/f_\mathrm{sky}^{1/2}$. If the dependence is parameterized as $1/f_\mathrm{sky}^{\alpha/2}$ and $\alpha$ is calculated independently for each multipole bin, then $\alpha \approx 1.2$ at the lowest multipoles, changing to $\approx 0.9$ at the highest multipoles. This scaling indicates that the skewness can be reduced to $<0.1$ for maps larger than $400$ square degrees.

\begin{fig}
\centering
\includegraphics[scale=0.55]{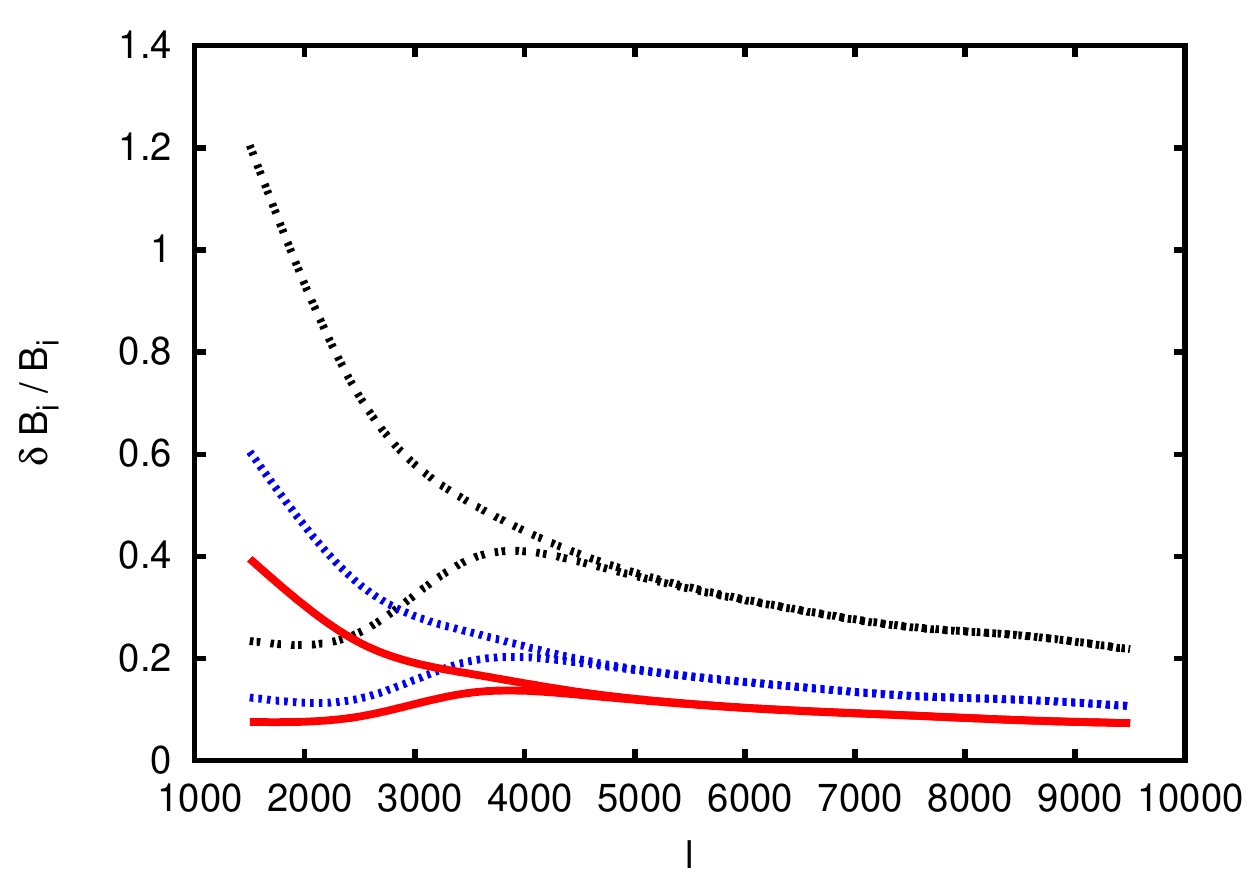}
\includegraphics[scale=0.55]{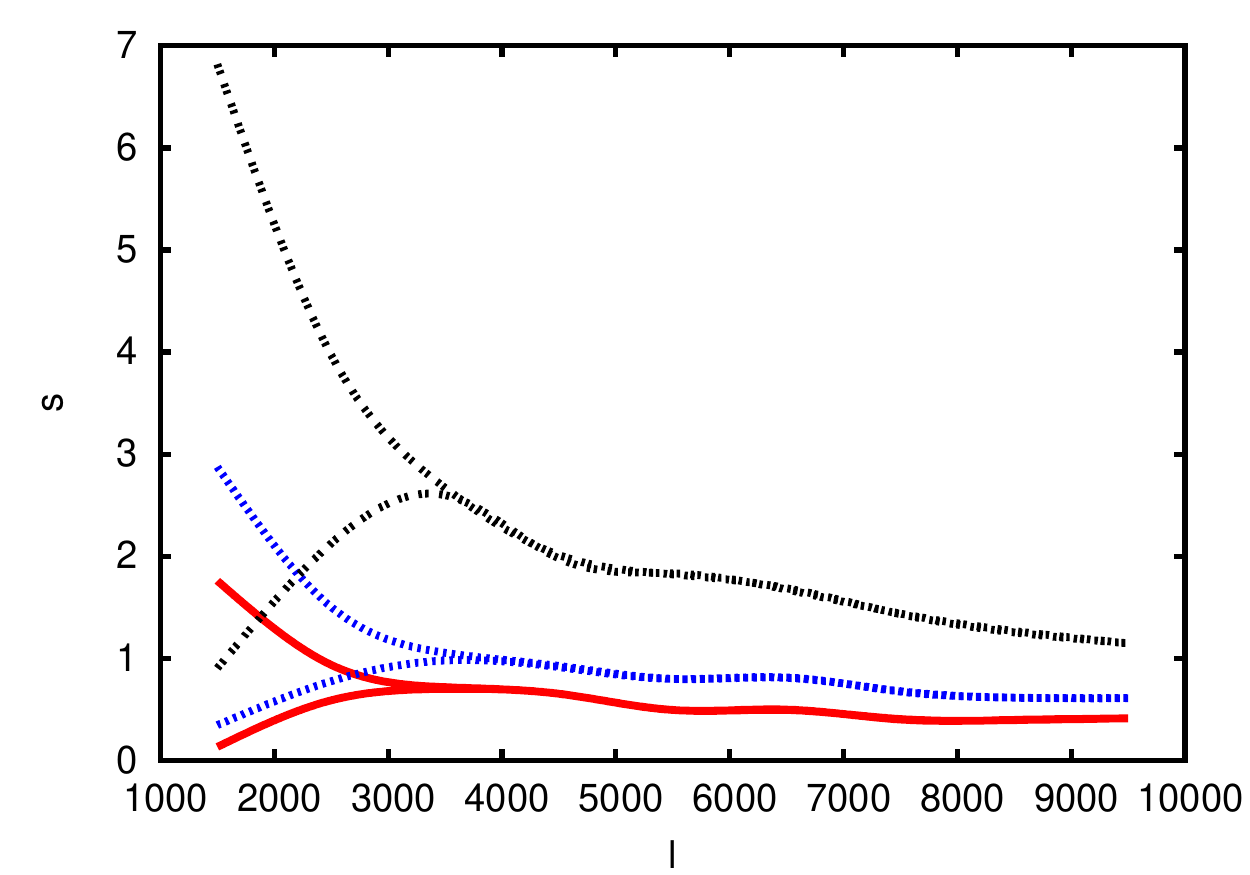}
\caption[$\delta B_i / \bar{B}_i$ and skewness for map sizes of $3\degree \times 3\degree$, $2\degree \times 2\degree$ and $1\degree \times 1\degree$]{Left: $\delta B_i / \bar{B}_i$ from the SZ effect on its own (top lines) and for the CMB and SZ effect combined (bottom lines) with $\sigma_8 = 0.825$ for three different map sizes: $3\degree \times 3\degree$  (red solid line), $2\degree \times 2\degree$ (blue dotted line) and $1\degree \times 1\degree$ (black double-dotted line). The values scale as $1/f_\mathrm{sky}^{1/2}$. Right: As the left plot, but plotting the skewness for the three map sizes; this also scales roughly as $1/f_\mathrm{sky}^{1/2}$.}
\label{fig:mapsize}
\end{fig}

As we are only considering the clusters within the map size, we will not be including any effects from clusters that are positioned outside the map but extend into the map. This was true for the larger maps considered earlier, however this may become more important here as large clusters can extend up to a quarter of a degree in the maps. As the means from the three map sizes differ by less than a percent, however, and the standard deviations agree with predictions, this appears to be negligible.

\section{Effects of clustering} \label{sec:clustering}
To check whether the clustering of the galaxy clusters can be responsible for the non-Gaussianity in the statistics, we randomize the positions of the clusters on the sky. This effectively modifies the angular correlation function to that expected of a Poissonian distribution, although this still includes the effects of clustering due to fluctuations that are larger than the map size. For the $3\degree \times 3\degree$ maps, there is no significant change in any of the values of the mean, variance or normalized skew between the statistics of the randomized and the clustered maps. However, when the values of  $\delta B_i / \bar{B}_i$ are compared to those from the analytical calculations (as described in the Appendix), a $10-15$ per cent excess is present at high multipoles.

Simply randomising the positions on the sky effectively destroys the clustering between galaxy clusters on scales smaller than the fiducial map size of $3\degree \times 3\degree$. However, this will not remove the effects of clustering on the mean number density of clusters within the maps. To illustrate the effect of this, we create sets of $1\degree \times 1\degree$ maps where we randomise the clusters by two methods: first, selecting the clusters within a $1\degree \times 1\degree$ area then randomising their positions (analogous to the randomization performed on the $3\degree \times 3\degree$ maps), and second, randomising the whole map and then selecting $1\degree \times 1\degree$ patches.

\begin{fig}
\centering
\includegraphics[scale=0.55]{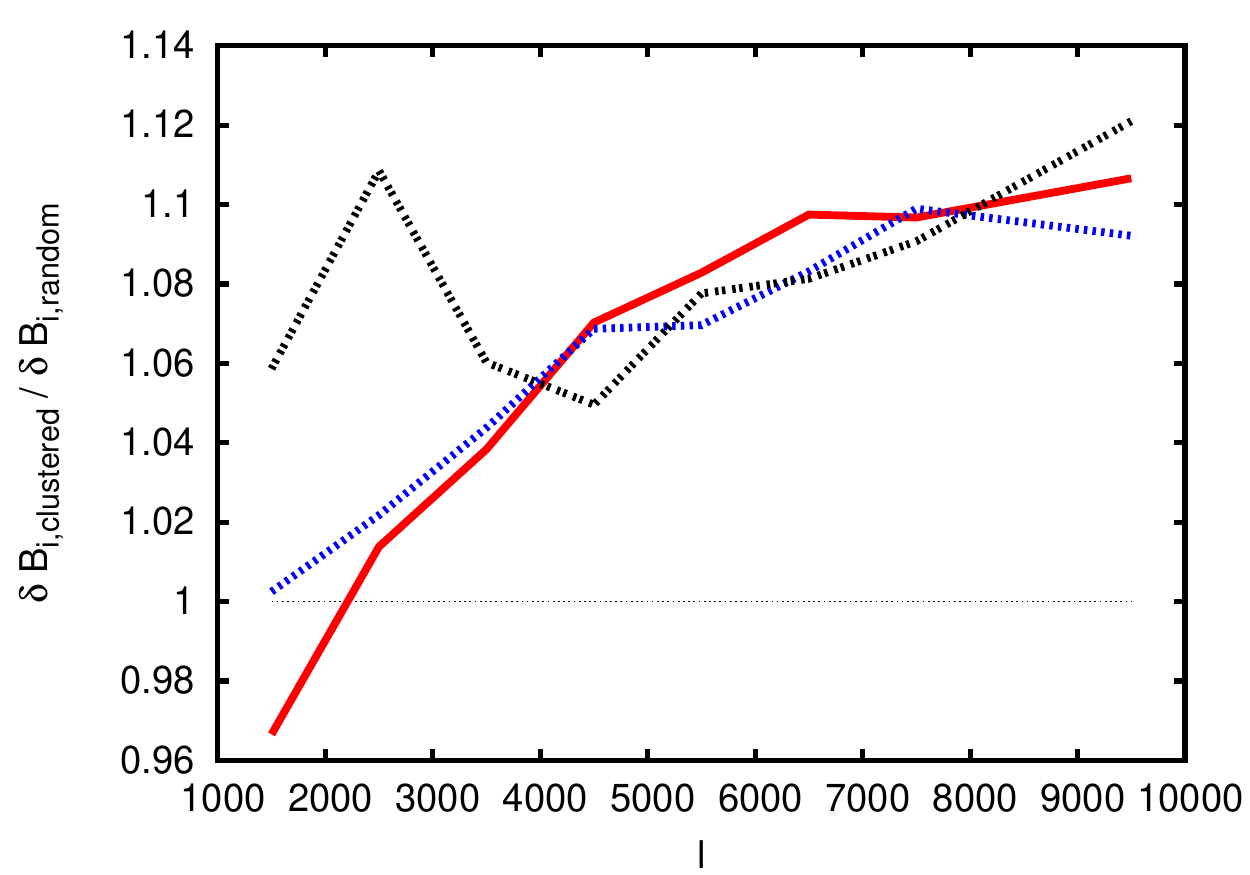}
\includegraphics[scale=0.55]{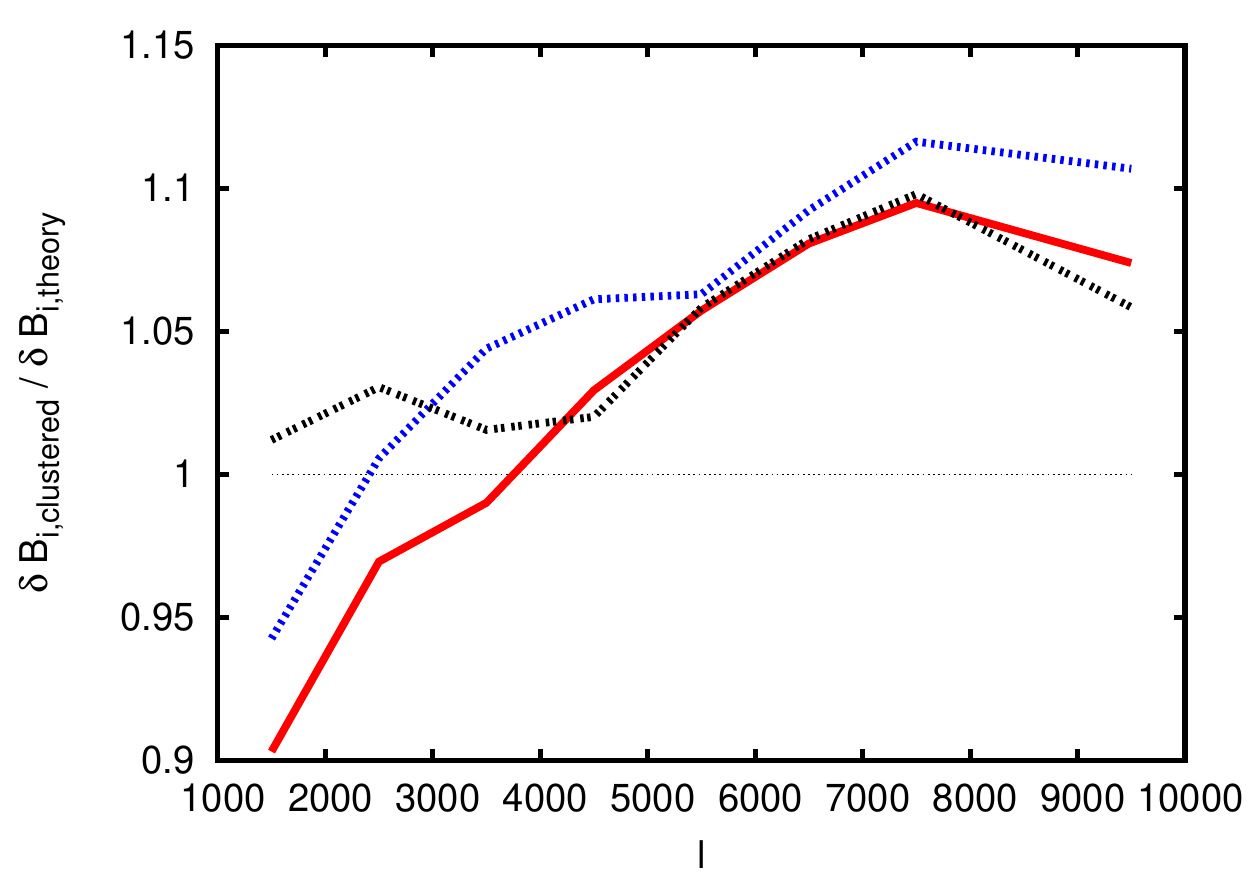}
\caption[The ratio of $\delta B_i / \bar{B}_i$ from clustered and randomised $1\degree \times 1\degree$ maps, as well as the analytical expectation]{Left: The ratio of $\delta B_i / \bar{B}_i$ from clustered and randomised $1\degree \times 1\degree$ maps for the three different values of $\sigma_8$ (the red solid line is $\sigma_8 = 0.825$, the blue dotted line $0.9$ and the black double-dotted line $0.75$). Clustering increases the standard deviation by $\sim 10$ per cent. Right: The same as the left figure, but comparing the clustered realizations to the analytical expectation. The increase of $\sim 10$ per cent remains.}
\label{fig:compare_randomisation_methods}
\end{fig}

Selecting the clusters that lie within the $1\degree \times 1\degree$ area and then randomizing their positions provides largely the same statistical properties as the clustered realizations, as was the case for the $3\degree \times 3\degree$ realizations. However, if the clusters are randomized prior to selection, then the ratio of $\delta B_i / \bar{B}_i$ is close to agreement with the analytical estimate, differing by a few percent. The effect of randomising and then selecting the clusters is shown in Figure \ref{fig:compare_randomisation_methods}, which gives the ratio of $\delta B_i$ between the clustered and random realizations ($\bar{B}_i$ is the same for the two methods) and between the clustered realizations and the analytical values. Clustering increases $\delta B_i / \bar{B}_i$ by $\sim 10$ per cent at the highest multipoles, where the standard deviation is increased as the galaxy clusters group together in certain realizations and are absent in others, broadening the standard deviation. The lowest multipoles are not well sampled and are highly skewed, such that their standard deviation alone does not adequately represent the distribution.

\section{Upper mass limit} \label{sec:upper_mass}
\begin{fig}
\centering
\includegraphics[scale=0.68]{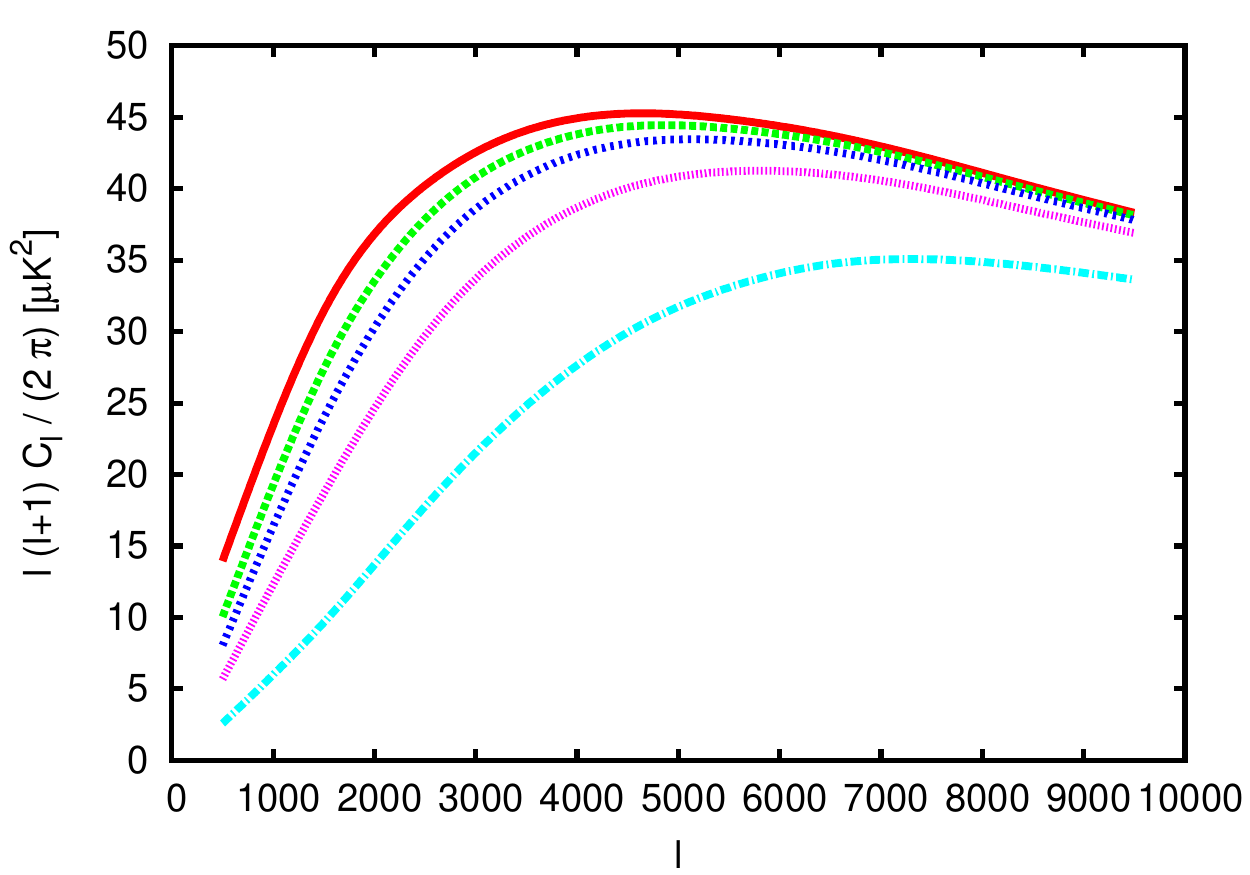}
\includegraphics[scale=0.68]{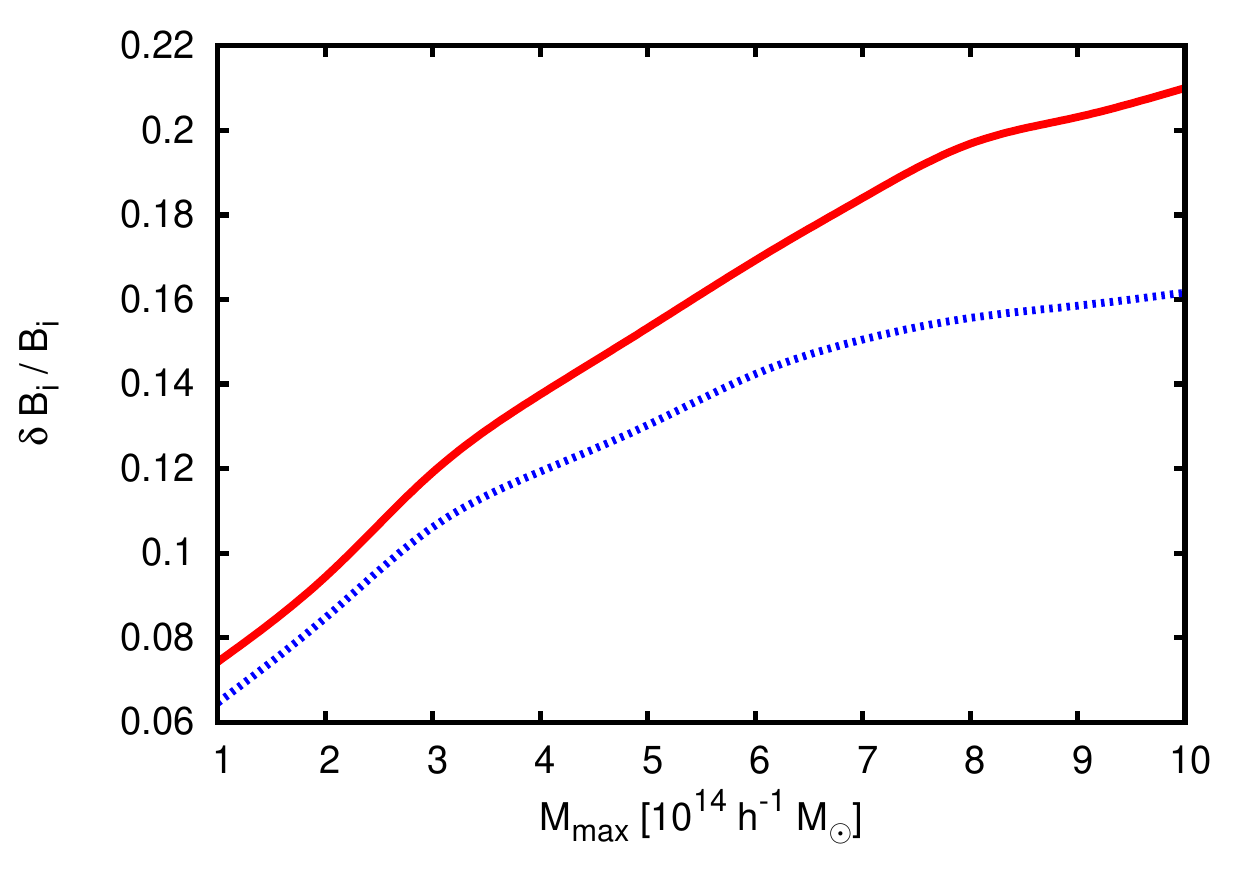}
\includegraphics[scale=0.68]{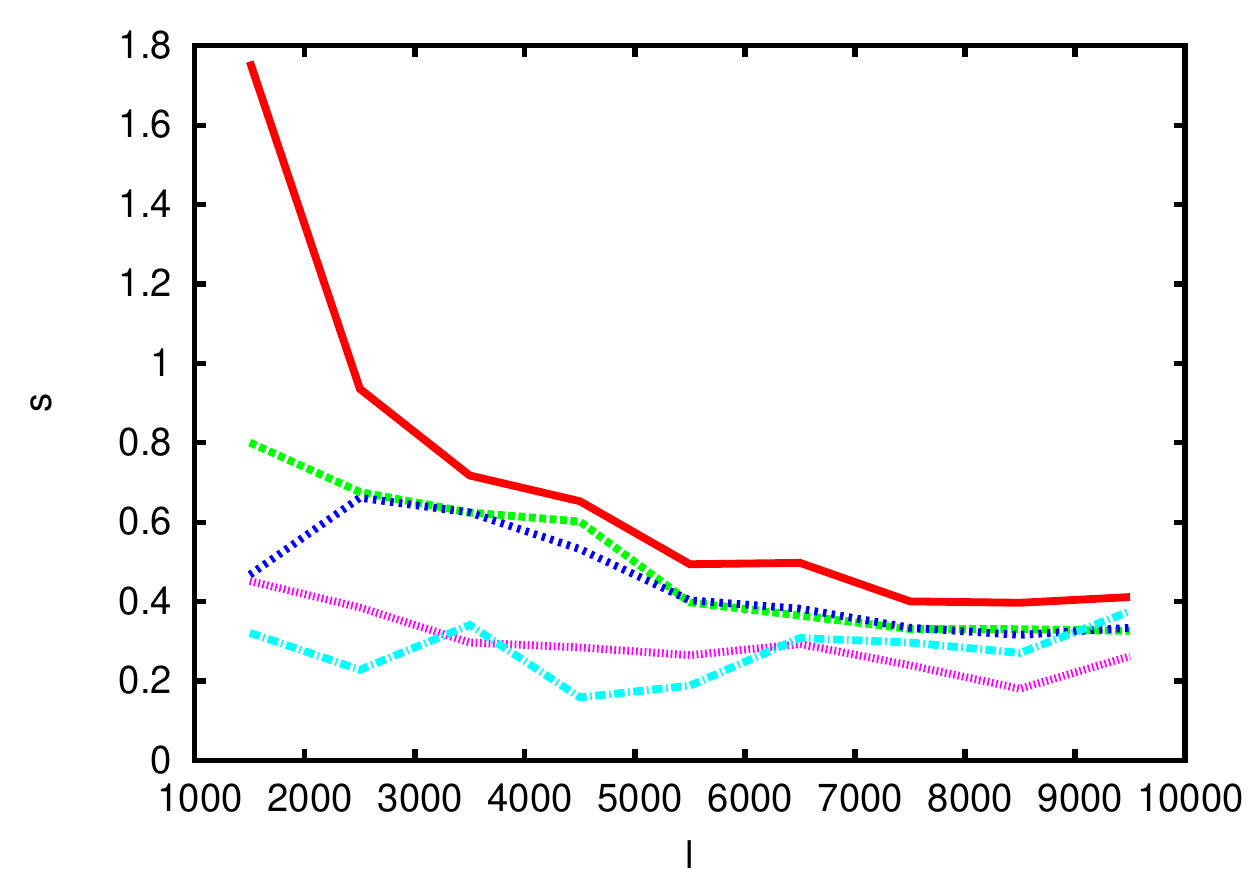}
\caption[The mean power spectra, $\delta B_i / \bar{B}_i$  and skewness for a range of maximum mass cuts]{Top: The mean power spectra from the SZ effect in the $\sigma_8=0.825$ cosmology with all clusters (red solid line) and after removing all clusters greater than $8 \times$ (green large-dashed), $6 \times$ (blue small-dashed), $4 \times$ (pink dotted) and $2 \times 10^{14} h^{-1} M_\odot$ (turquoise dot-dashed), from the top down. Middle: $\delta B_i / \bar{B}_i$ from the SZ effect as a function of the maximum mass in the realizations in two bins: $l=2000-3000$ (red solid line) and $l=3000-4000$ (blue dotted line). Bottom: the skewness $s$ of the distributions for the same mass cuts as the mean power spectra. All three of these quantities are significantly affected by the largest mass clusters in the realizations.}
\label{fig:sz_masscuts}
\end{fig}

The power spectrum of the SZ effect measured in different parts of the sky is very sensitive to the mass of the largest clusters within that field. Fields used for CMB observations might be preselected to avoid such objects, hence biasing our measurement of the power spectrum. In order to test the dependence on this, we apply a series of maximum mass cuts to the {\sc Pinocchio} catalogues; the mean spectra after these cuts are shown in Figure \ref{fig:sz_masscuts}. Note that the effect of minimum mass cuts was discussed in Section \ref{sec:example_realisations}.

The mean power spectrum is largely converged at maximum masses $\sim 6-8 \times 10^{14} h^{-1} M_\odot$, with the exact values depending on the multipole -- higher mass clusters contribute most to the lowest multipoles. The figure also shows the ratio of the standard deviation to the mean in the multipole bins $l=2000-3000$ and $3000-4000$ as a function of the maximum mass; for both multipole bins, the ratio increases as the maximum mass is increased. The skew also increases with the maximum mass of the cluster, approximately doubling between $M_\mathrm{max} = 10^{14} h^{-1} M_\odot$ and $10^{15} h^{-1} M_\odot$. Thus, it is the largest mass clusters that provide the largest amounts of non-Gaussianity within the power spectrum.

\section{Parameter dependence} \label{sec:parameter_dependence}
The five free parameters used in the cluster model (see equations \ref{eq:betamodel} and \ref{eq:clustermodel}) are not yet constrained to high accuracy, with different simulations of galaxy clusters finding different values. Observations to measure the SZ effect are still in their early stages, and no large scale SZ surveys of clusters yielding many clusters have been carried out yet. To see the effect of the uncertainty in these parameters on our results, we vary the parameters over a wide range of possible values.

\subsection{Parameters of the Y-M relation}
The overall amplitude of the SZ effect is determined by $Y_*$, which is fiducially set to $2 \times 10^{-6}$ Mpc$^2$. As discussed earlier, Kay et al. (in prep.) find that this can be $1.9$ or $2.3 \times 10^{-6}$ Mpc$^2$ for non-radiative or preheated gas. Changing this parameter simply scales the mean and standard deviation of the spectrum from the SZ effect, with the mean and standard deviation being multiplied by $\left( Y_{*,\mathrm{new}} / Y_{*,\mathrm{fiducial}} \right)^2$. This is because the increase does not depend on the structure or distribution of the clusters. There is no effect on normalized quantities such as $\delta B_i / \bar{B}_i$ and $s$ from the SZ effect only. Due to the increase in the power from the SZ effect, though, the cross-over point between the CMB and SZ effect will be shifted to lower multipoles as $Y_*$ is increased, which will increase $\delta B_i / \bar{B}_i$ and $s$ at those lower multipoles.

The evolution of the SZ effect as a function of mass is governed by $\gamma$. If the gas is assumed to only be gravitationally heated, then $\gamma = 5/3$. This slope could be steeper, reflecting a reduced amount of gas mass in smaller clusters, or it could be flatter due to an increase in temperature in smaller clusters from extra, non-gravitational energy. Via X-ray observations of the inner region of galaxy clusters, \citet{2007Arnaud} find $\gamma = 1.82^{+0.09}_{-0.08}$ for the $Y_X$-M relationship\footnote{$Y_X = M_\mathrm{gas} T_X$ is an X-ray proxy for the integrated SZ effect $Y$ that has recently been shown to correlate well with $M$ with little scatter \citep{2006Kravtsov}.}, in agreement with the preheating result from Kay et al. (in prep.). As this is measured for the inner region, whilst the SZ effect extends much further out in the cluster atmosphere, this is likely to be an upper limit. As such, we use $\gamma = 1.5$ and $1.8$ as the expected range for this parameter. The effect of these values on the power spectrum and $\delta B_i / \bar{B}_i$ is shown in the top row of Figure \ref{fig:param_dependence}.

\begin{fig}
\centering
\includegraphics[scale=0.45]{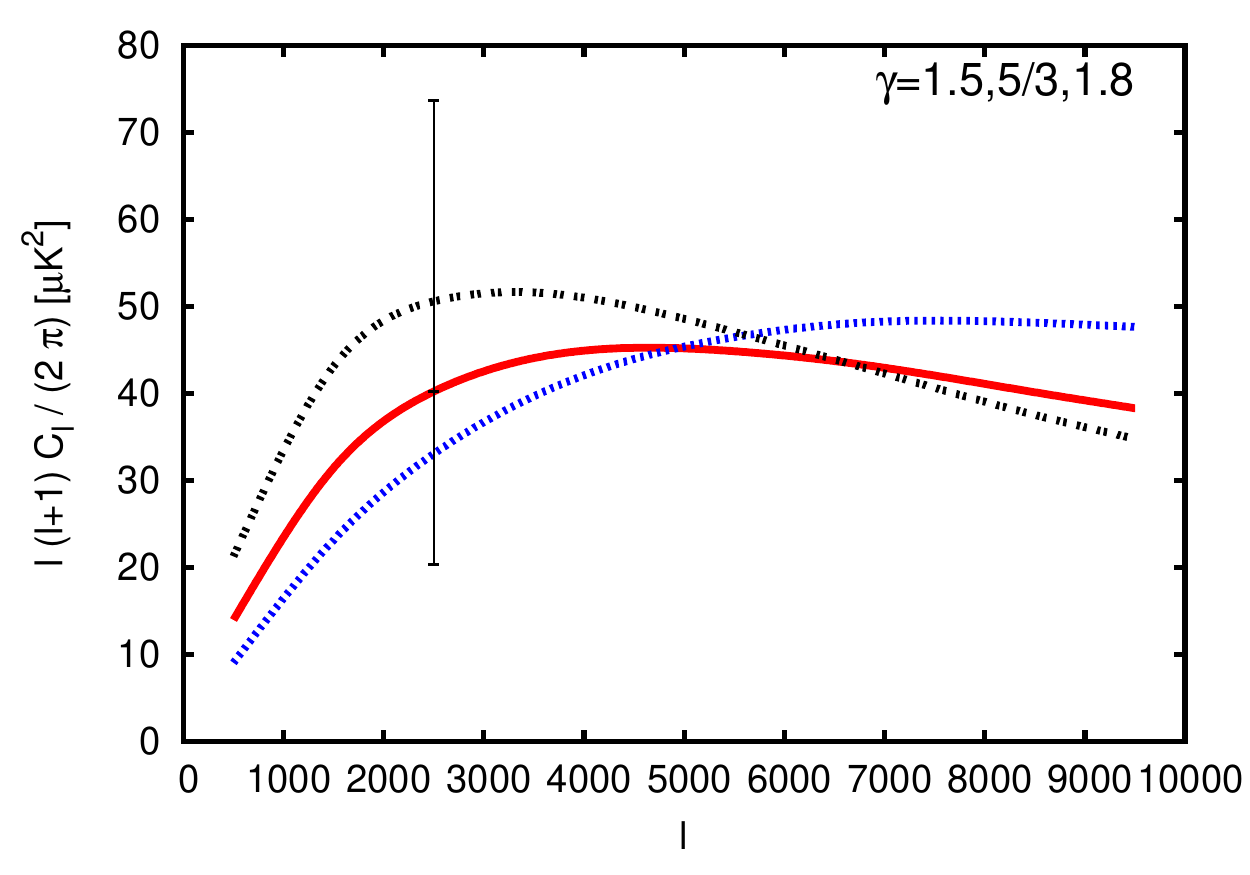}
\includegraphics[scale=0.45]{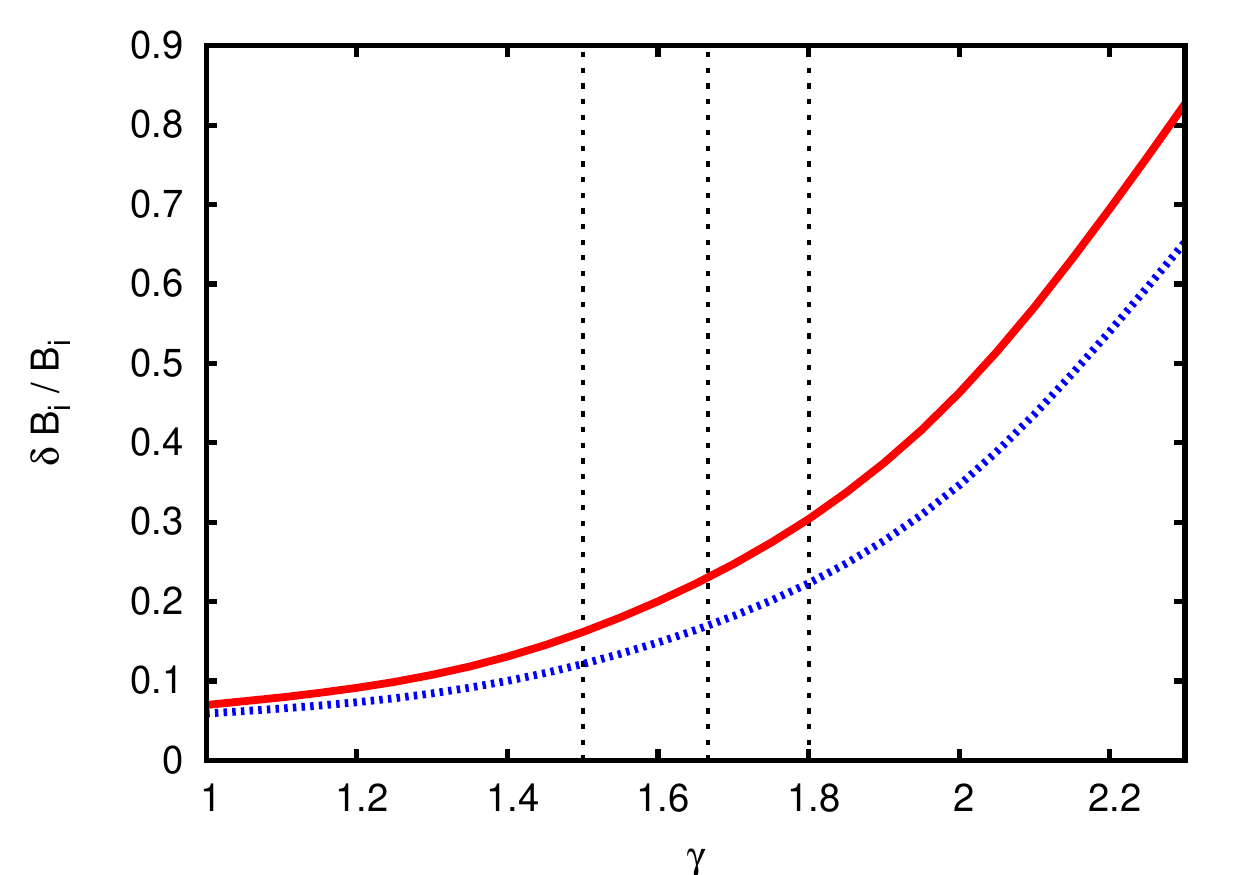}\\
\includegraphics[scale=0.45]{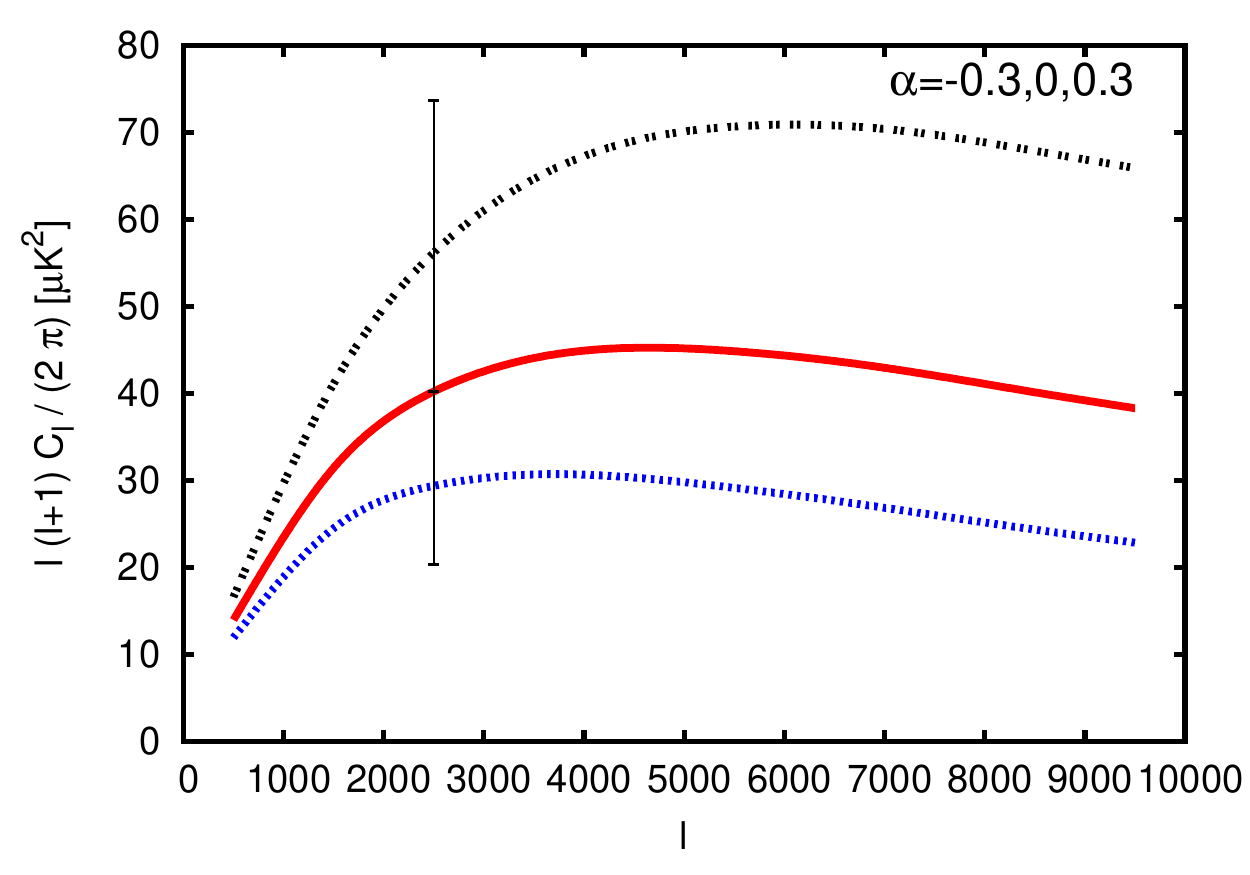}
\includegraphics[scale=0.45]{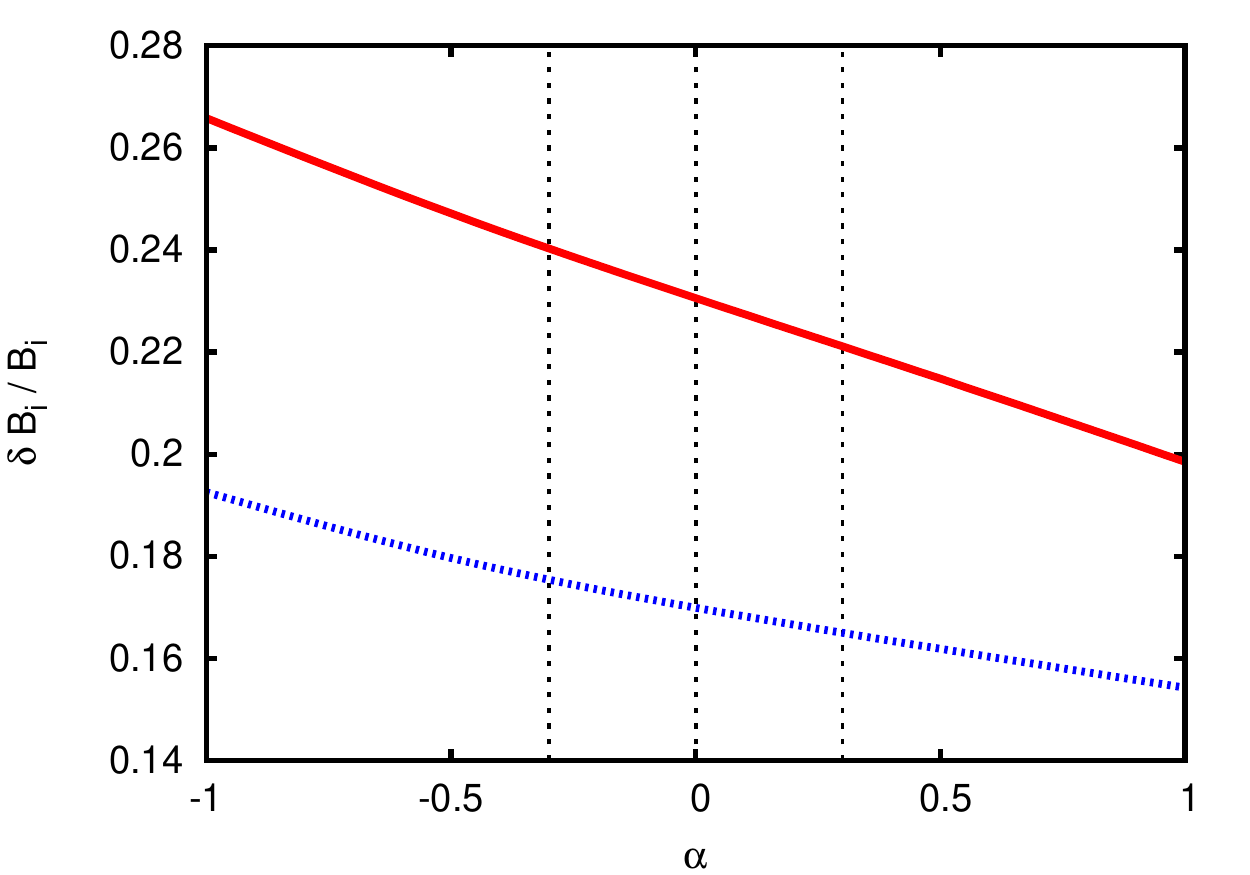}\\
\includegraphics[scale=0.45]{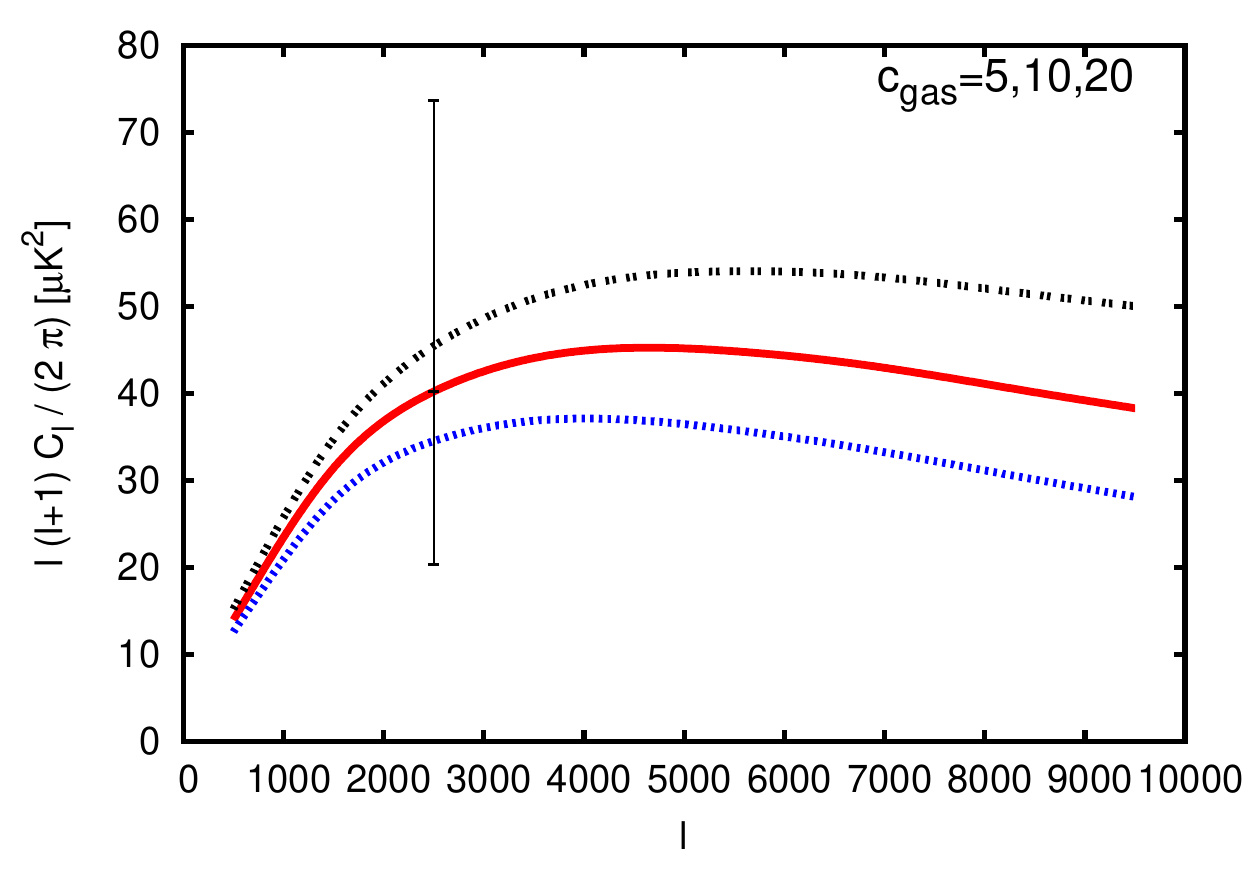}
\includegraphics[scale=0.45]{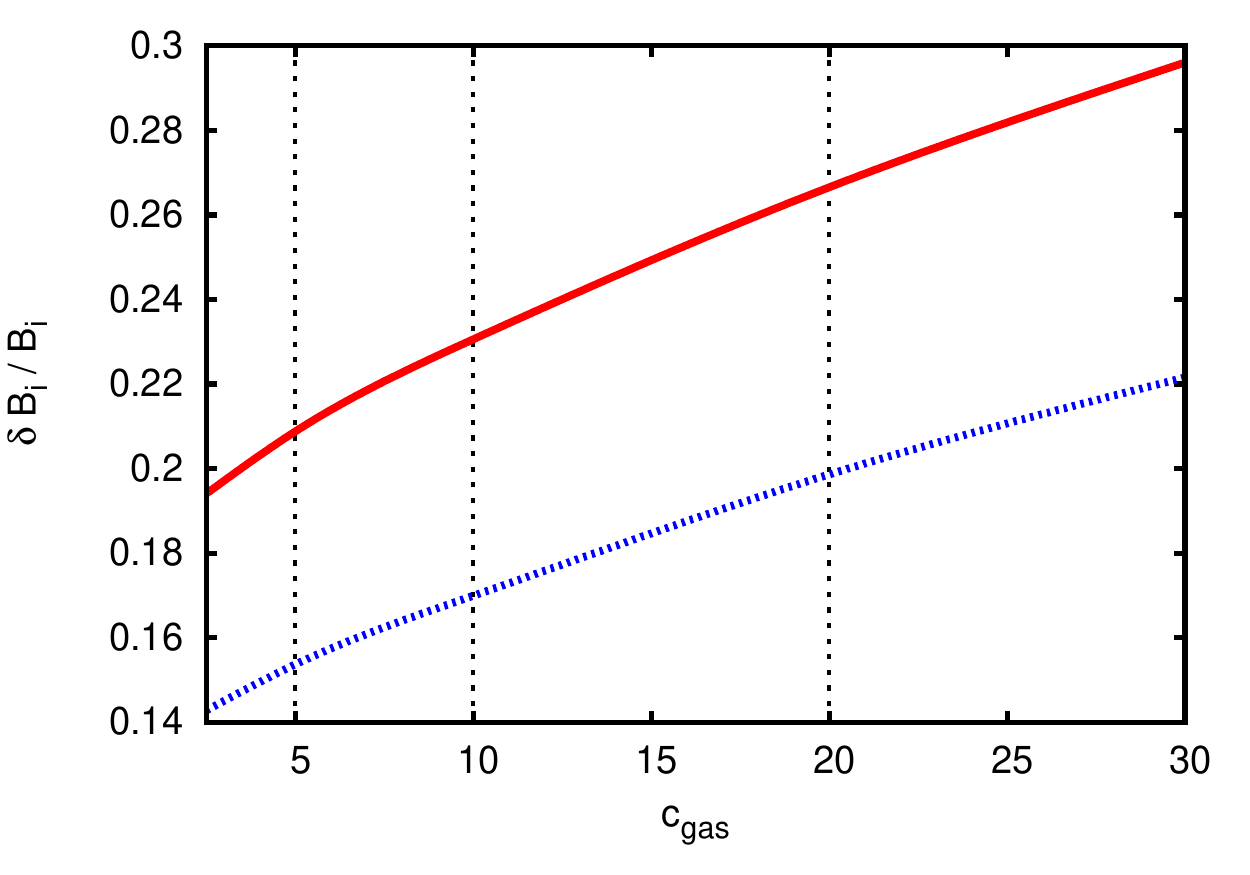}\\
\includegraphics[scale=0.45]{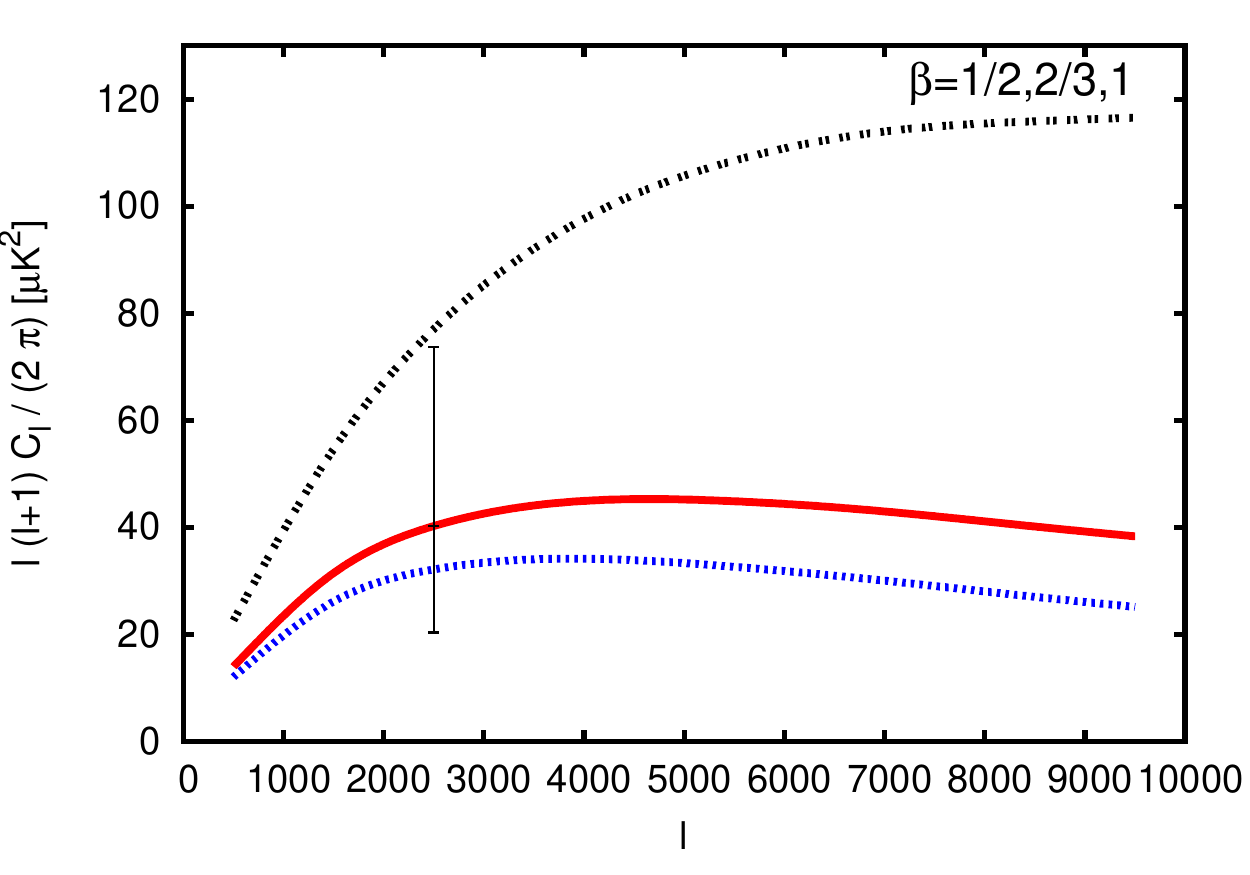}
\includegraphics[scale=0.45]{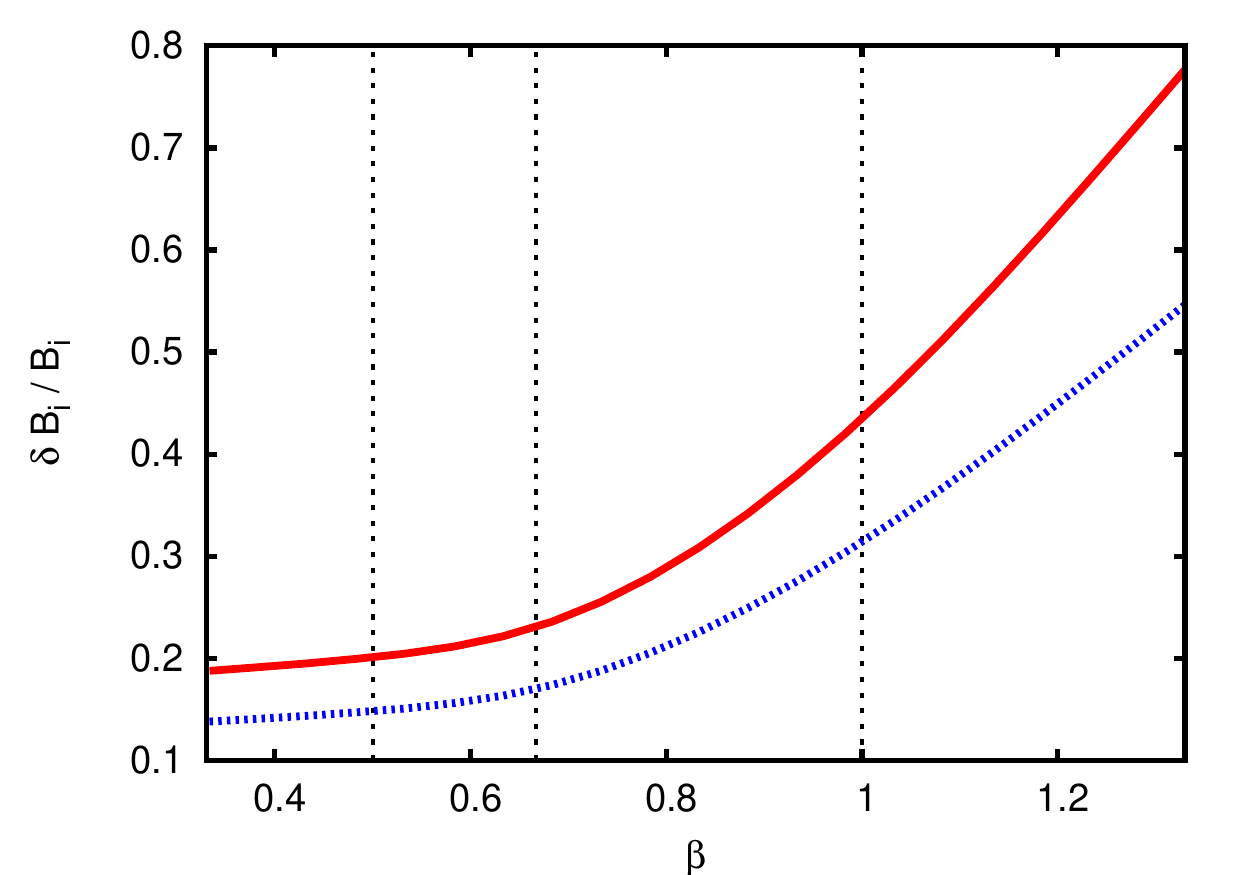}
\caption[The effect on the power spectrum from changing the models of the parameters]{The effect on the power spectrum from changing the models of the parameters. From top to bottom the parameters are $\gamma$ (the mass dependence), $\alpha$ (the redshift dependence), $c_\mathrm{gas}$ (the gas concentration) and $\beta$ (the outer gas profile). The left hand plots show the mean power spectrum as a function of multipole $l$. The fiducial value is shown by the red solid line, with the blue dotted line showing the lower value and the black double-dotted line showing the higher value (see text for the parameter values). The data point is the range of values from the three values of $\sigma_8$ in the 2000-3000 multipole bin. The right hand plots show $\delta B_i / \bar{B}_i$ vs. the parameter values. The solid red line is for the multipole bin $l=2000-3000$, and the blue dotted line is for $l=3000-4000$. The dashed vertical lines show the three values used in the left-hand plots. The central vertical line is the fiducial value; the $\delta B_i / \bar{B}_i$ for this value are between $0.23-0.27$ and $0.17-0.19$ for the $l=2000-3000$ and $3000-4000$ bins respectively for the three values of $\sigma_8$.}
\label{fig:param_dependence}
\end{fig}

As $\gamma$ is increased, the peak in the SZ power spectrum shifts to lower multipoles, and $\delta B_i / \bar{B}_i$ increases. This is because the power from clusters with a mass over $10^{14} h^{-1} M_\odot$ will be increased as $\gamma$ increases, whilst that from smaller mass clusters will decrease. The larger clusters are dominant at lower multipoles due to their greater angular size, so the peak of the SZ effect will naturally shift to lower multipoles. The effect is similar to simply increasing the mass of all the clusters. Section \ref{sec:upper_mass} showed that $\delta B_i / \bar{B}_i$ increases as larger clusters are included in the maps; the same effect is present here, with high values of $\gamma$ providing the largest increase in the effective mass, hence the largest values of $\delta B_i / \bar{B}_i$.

The SZ effect is expected to be self-similar, that is, it will depend on the redshift of the cluster via the evolution of the mean matter density within the virial radius of the cluster as well as the Friedmann equation. The parameter $\alpha$ models evolution beyond that from self-similarity, and is expected to be close to zero. Using numerical simulations, \citet{2004Silva} find no evidence of evolution from self-similarity.

To illustrate the effect of changing this parameter, we choose a modest range of values: $\alpha = -0.3$ and $0.3$; the results are presented in the second row of Figure \ref{fig:param_dependence}. Higher values of $\alpha$ will increase the SZ effect from clusters at the highest redshift. As the clusters at higher redshifts have a smaller angular size, this affects the power in the higher multipoles more than that in the lower ones. There are more clusters at higher redshifts, so when these provide increased power compared with the low redshift clusters the maps will become more similar to each other - it will become more difficult for a single cluster to stand out - hence the standard deviation of the maps will decrease.

\subsection{Parameters of the gas profile}
The final two parameters govern the profile of the SZ effect. The concentration of the gas within the cluster is determined by $c_\mathrm{gas}$, and  the slope of the cluster profile is governed by $\beta$ (see equation \ref{eq:betamodel}). For $\beta$ we use a fiducial value of $2/3$, which is in agreement with observations of X-ray emission from nearby clusters by \citet{2006Vikhlinin} that find $\beta \sim 0.6-0.9$ at $r_{500}$ (approximately half of the virial radius). Additionally, \citet{2008Croston} find a range of $\beta \sim 0.37-0.81$ for a range of radii between 0.3 and 0.8 $R_{500}$. However, steeper values than these are expected in the outer regions where the gas pressure profile is expected to trace the NFW profile; this is demonstrated by \citet{2007Hallman}, who use simulations of clusters to compare X-ray and SZ profiles, finding that $\beta=0.88$ for X-ray but $1.13$ for the SZ effect. As a result, we investigate the effects of $\beta = 1/2$ and $1$. For $c_\mathrm{gas}$, we use a fiducial value of $10$ and also try $5$ and $20$. Duffy et al. (in prep.) indicates that $c_\mathrm{gas}$ is similar to $c_\mathrm{DM}$ and depends on the cluster mass by around 10 per cent above $10^{14} M_\odot$, such that the values will lie around $5-10$.

The effects of $c_\mathrm{gas}$ and $\beta$ are shown in the bottom two rows of Figure \ref{fig:param_dependence}. As these parameters only change the profile of the SZ effect, the average pixel value within the maps remains constant. The SZ effect is more concentrated within the centre of the cluster for higher values of $\beta$ and $c_\mathrm{gas}$, with the effect of increasing the mean power spectra across all multipoles, but particularly at the higher multipoles, in a similar way to the effect of point sources on the power spectrum (see Section \ref{sec:powerspectrum_ps}). As the parameters are increased, the ratio of the standard deviation to the mean also increases, as the power from the central part of the largest (and rarest) clusters becomes more important at these multipoles.

The assumption that all clusters are the same is unlikely to be true; in reality, there will be a certain amount of scatter within the cluster parameters, which may increase the standard deviation further when looking at smaller areas of the sky. Clusters are also likely to deviate from spherical symmetry. The extent of this scatter and its impact on the SZ effect is not yet known.

\subsection{Comparison to the effects of $\sigma_8$}
The changes in the gas physics have a comparable effect on the mean power spectrum as the range of $\sigma_8$ considered in this paper, especially for the outer gas profile $\beta$ and the evolution of the $Y-M$ relationship with redshift ($\alpha$). At the same time, the gas physics can substantially increase the standard deviation of the power spectrum to a much larger degree than is possible with $\sigma_8$, providing a much greater scatter in the power from different maps. These effects are obviously different to those from $\sigma_8$ when the whole of the power spectrum can be considered, but will be difficult to separate when only a few bins are measured.

\section{Power spectrum statistics with point sources} \label{sec:powerspectrum_ps}
\begin{tab}{htpb}
\begin{tabular}{c|ccc|ccc|ccc}
& \multicolumn{3}{c}{CMB+SZ}& \multicolumn{3}{c}{CMB+SZ+PS} & \multicolumn{3}{c}{CMB+SZ+CPS}\\
Multipole bin & $\bar{B}_i$ & $\delta B_i$ & $s$  & $\bar{B}_i$ & $\delta B_i$ & $s$  & $\bar{B}_i$ & $\delta B_i$ & $s$ \\
\hline % 3x3 degree
1000-2000 & 720 & 56 & 0.40 & 730 & 56 & 0.20 & 720 & 57 & 0.25\\
2000-3000 & 150 & 13 & 0.59 & 160 & 13 & 0.61 & 150 & 13 & 0.65\\
3000-4000 & 58 & 7.7 & 0.70 & 85 & 8.1 & 0.57 & 75 & 7.7 & 0.73\\
4000-5000 & 47 & 6.1 & 0.68 & 91 & 6.9 & 0.40 & 82 & 6.3 & 0.56\\
5000-6000 & 45 & 5.0 & 0.50 & 110 & 6.2 & 0.29 & 100 & 6.0 & 0.30\\
6000-7000 & 44 & 4.3 & 0.49 & 140 & 6.1 & 0.17 & 130 & 6.0 & 0.23\\
7000-8000 & 42 & 3.7 & 0.40 & 160 & 6.6 & 0.06 & 160 & 6.3 & 0.23\\
8000-9000 & 40 & 3.2 & 0.39 & 200 & 7.4 & 0.14 & 190 & 7.0 & 0.09\\
9000-10~000 & 38 & 2.8 & 0.41 & 230 & 8.0 & 0.0 & 230 & 8.2 & -0.03\\
\hline % 1x1 degree
1000-2000 & 720 & 170 & 0.90 & 730 & 160 & 0.72 & 720 & 160 & 0.71\\
2000-3000 & 150 & 38 & 2.1 & 160 & 40 & 2.0 & 150 & 37 & 1.9\\
3000-4000 & 58 & 23 & 2.6 & 84 & 24 & 2.2 & 74 & 22 & 2.4\\
4000-5000 & 47 & 18 & 2.0 & 91 & 20 & 1.5 & 80 & 19 & 1.7\\
5000-6000 & 45 & 15 & 1.8 & 110 & 19 & 1.2 & 100 & 17 & 1.2\\
6000-7000 & 44 & 13 & 1.7 & 140 & 19 & 0.84 & 120 & 18 & 0.82\\
7000-8000 & 42 & 11 & 1.4 & 170 & 20 & 0.49 & 150 & 19 & 0.60\\
8000-9000 & 40 & 9.9 & 2.0 & 200 & 22 & 0.30 & 190 & 21 & 0.43\\
9000-10~000 & 38 & 8.3 & 1.1 & 230 & 24 & 0.34 & 220 & 23 & 0.32\\
\hline % 150GHz, 3x3 degrees
1000-2000 & 700 & 52 & 0.21 & 700 & 55 & 0.37 & 700 & 55 & 0.11\\
2000-3000 & 120 & 7.7 & 0.26 & 120 & 7.7 & 0.16 & 120 & 7.8 & 0.18\\
3000-4000 & 24 & 2.0 & 0.51 & 33 & 2.3 & 0.34 & 33 & 2.3 & 0.47\\
4000-5000 & 13 & 1.5 & 0.66 & 28 & 1.9 & 0.43 & 27 & 1.8 & 0.39\\
5000-6000 & 11 & 1.2 & 0.47 & 33 & 1.8 & 0.29 & 33 & 1.8 & 0.18\\
6000-7000 & 11 & 1.0 & 0.49 & 41 & 1.8 & 0.21 & 41 & 2.0 & 0.02\\
7000-8000 & 10 & 0.89 & 0.39 & 51 & 2.2 & -0.10 & 51 & 2.3 & 0.03\\
8000-9000 & 10 & 0.76 & 0.39 & 62 & 2.5 & 0.12 & 62 & 2.6 & 0.13\\
9000-10~000 & 9 & 0.67 & 0.42 & 75 & 3.0 & 0.16 & 75 & 3.0 & 0.14\\
\end{tabular}
\caption[Statistics of the combined CMB, SZ and point source maps]{The statistics of the $\sigma_8 = 0.825$ realizations containing CMB, SZ and point sources. The mean $\bar{B}_i$ and the standard deviation $\delta B_i$ within the bin $i$ are in $\upmu$K$^2$; the skew $s$ is dimensionless. The top set are for $3\degree \times 3\degree$ maps at 30~GHz using 0.2~mJy point sources; the middle for $1\degree \times 1\degree$ maps also at 30~GHz and 0.2~mJy point sources, and the bottom set for $3\degree \times 3\degree$ maps at 150~GHz with 2~mJy point sources. The three sets of values given are for no point sources (left), randomly positioned point sources (middle) and clustered point sources (right). Adding point sources increases the mean and standard deviation at high multipoles, as expected. Clustered point sources fill in some of the decrement from the SZ effect, reducing the mean and standard deviation.}
\label{tab:ps_stats_0.2mjy}
\end{tab}

\begin{fig}	
\centering
\includegraphics[scale=0.80,viewport=80 30 300 215,clip]{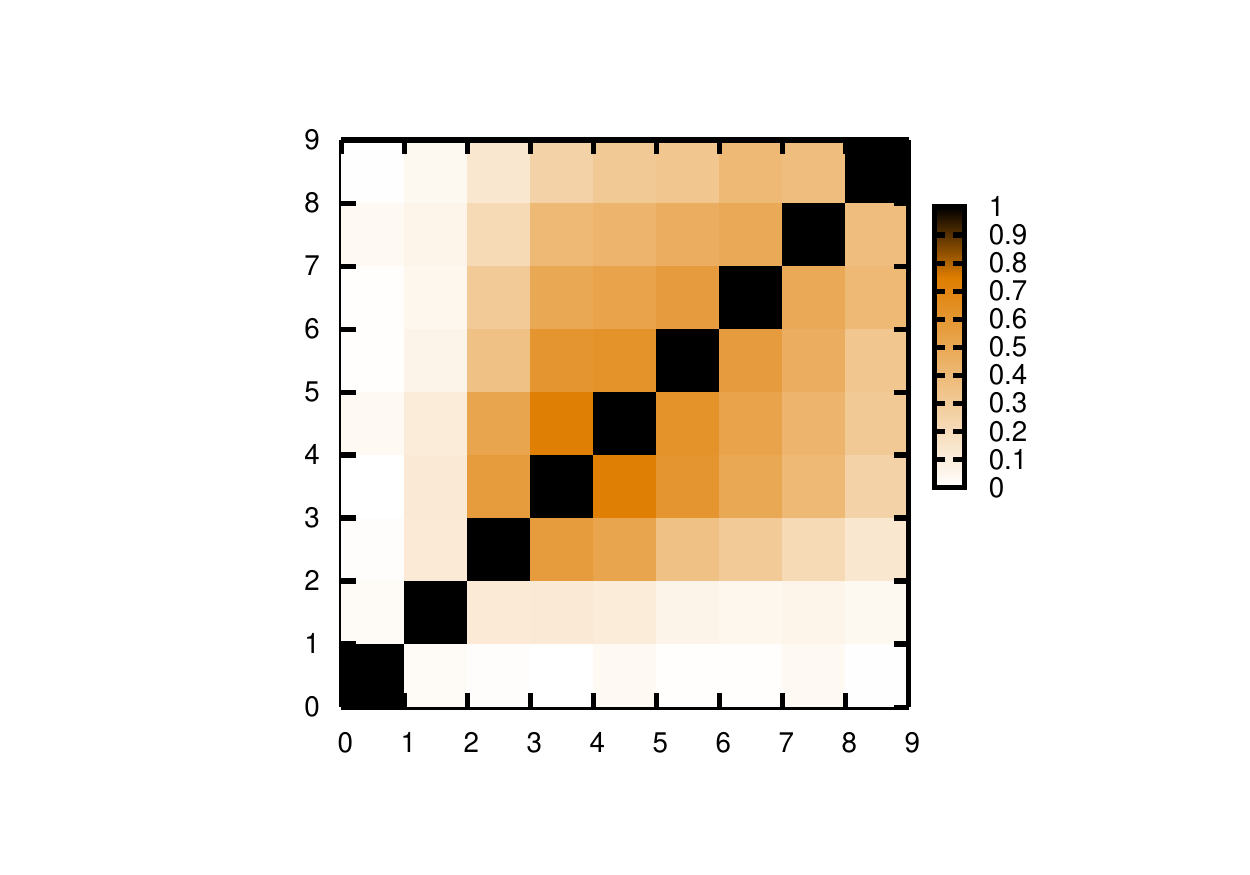}
\caption[The correlation matrix for 1000 realizations of the combined CMB, SZ effect and 0.2~mJy point sources at 30~GHz, using $3\degree \times 3\degree$ maps]{The correlation matrix for 1000 realizations of the combined CMB, SZ effect and 0.2~mJy point sources at 30~GHz, using $3\degree \times 3\degree$ maps. The addition of point sources decreases the off-diagonal terms (compare with Figure \ref{fig:sz_covariance}).}
\label{fig:ps_covariance}
\end{fig}

Table \ref{tab:ps_stats_0.2mjy} gives the power spectrum statistics for maps containing point sources below 0.2~mJy in addition to the CMB and SZ effect. This value for the flux density was chosen to represent the residual point sources after subtraction from CMB maps due to the uncertainty in the point source model. The addition of point sources naturally increases the mean and standard deviation of the spectra at higher multipoles. The point sources also decrease the normalised skewness and the off-diagonal terms of the correlation matrix at 30~GHz (see Figure \ref{fig:ps_covariance}) as they can reduce the amount of power from the SZ effect by filling in the decrements \citep{2002Holder}. Clustering the point sources increases the chance of this happening, resulting in a slightly lower mean values for the power spectrum and a slightly reduced standard deviation. The effect at 150~GHz is similar to that at 30~GHz, except that the effect of clustering the low frequency point sources is reduced as the unclustered infrared sources are becoming more important.

We do not attempt to account for the effects of imperfectly removing strong point sources from the realizations, which may result in a distribution of effective point sources with a flux density below the cut-off flux density. This could be an important issue, especially if the point sources lie in galaxy clusters, in which case the residual point source flux density could fill in the decrement from the SZ effect.

\section{Implications for the high multipole excess}

The original, and highest significance, measurement of a possible excess at large multipoles comes from the Cosmic Background Imager. An excess was first announced by \citet{2003Mason} and was refined by the addition of more observations by \citet{2003Pearson} and \citet{2004Readhead}. The final results were announced by \citet{2009Sievers} using the full 5 years of observations with the CBI. In total, the CBI has observed five $5\degree \times 5\degree$ fields and one $5\degree \times 6\degree$ field using mosaicked, shallow observations. Additionally, one $5\degree \times 0.75\degree$ and three $0.75\degree \times 0.75\degree$ ``deep'' fields were observed, which have reduced noise as they have been observed for longer. Following from \citet{2005Bond}, \citet{2009Sievers} find that if this excess is due to the SZ effect, then $\sigma_8$ must be between 0.9 and 1.0.

We do not attempt to carry out a direct comparison between our results and the CBI measurements as we omit a number of factors that will be important in this comparison. These include the difference in map sizes and the combination method - we investigate the statistics of $3\degree \times 3\degree$ and $0.75\degree \times 0.75\degree$ maps separately, whereas the CBI measurement results from the combination of a number of different areas of the sky and different map sizes. We do not take into account the effects of the $u,v$ coverage, beam shapes, nor the window function for the observations. We also exclude any effects of noise, point source subtraction and any other foregrounds such as galactic emission or atmospheric effects. Finally, we only briefly cover the effects of gas physics and limits on the maximum cluster masses on the statistics. As such, our results are only indicative.

\begin{fig}
\centering
\includegraphics[scale=0.68]{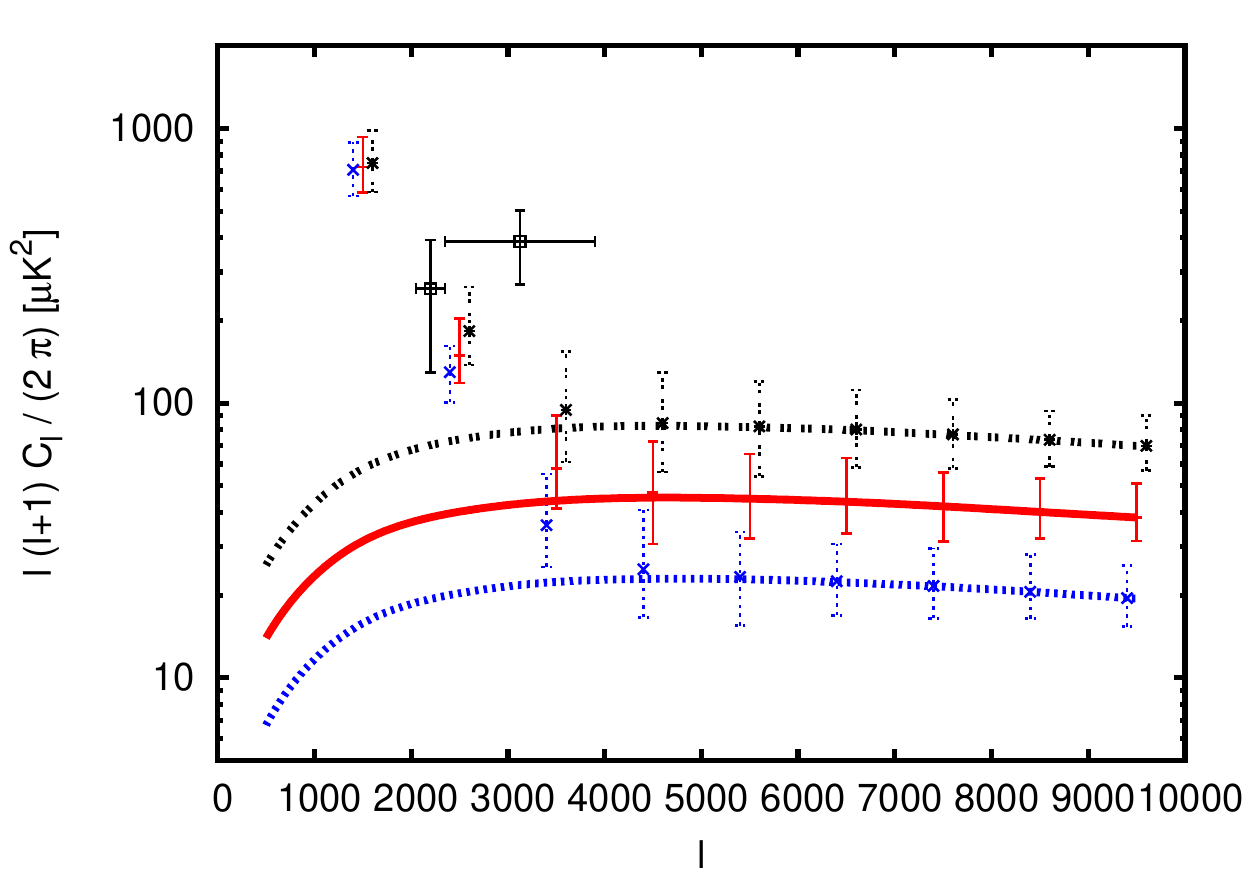}
\caption[The power spectra from the CMB+SZ effect for the three cosmologies, with the error bars showing the minimum and maximum values from all 1000 realizations with a map size of $3\degree \times 3\degree$]{The power spectra from the CMB+SZ effect for the three cosmologies, with the error bars showing the minimum and maximum values from all 1000 realizations with a map size of $3\degree \times 3\degree$. The black double-dotted points (offset by 100 multipoles for clarity) are for $\sigma_8 = 0.9$, the red solid points for $\sigma_8 = 0.825$ and the blue dotted points (also offset by 100 multipoles) are for $\sigma_8 = 0.75$. The two black data points are the latest values for the highest multipole bins from CBI \citep{2009Sievers}. The three lines are (from top to bottom) the  mean power spectra from the SZ effect in with $\sigma_8=0.9$ (black double-dotted line), $0.825$ (red solid line) and $0.7$ (blue dotted line).}
\label{fig:cmb_sz_powerspectra}
\end{fig}

We concentrate on two of the bins measured by the CBI, namely $l=2050-2350$ and $l=2350-3900$; these are the bins where the SZ effect will be most significant. For these bins, the CBI has measured $261 \pm 132$~$\upmu$K$^2$ and $387 \pm 117$~$\upmu$K$^2$ respectively. They find similar values, but with larger error bars, when only the mosaicked fields are considered (and the same for only considering the deep fields), and find low significance excesses in each of the four deep fields separately. The CBI data points are shown in Figure \ref{fig:cmb_sz_powerspectra}, along with the power spectra from the SZ effect on its own for each three values of $\sigma_8$, as well as the mean values from the combined CMB and SZ realizations for $3\degree \times 3\degree$ maps with the spectrum binned with bin widths of 1000; the minimum and maximum values from the realizations are shown by the error bars. The highest CBI data point is clearly in excess over what would be expected from the mean and scatter of these realizations.

\begin{fig}
\centering
\includegraphics[scale=0.68]{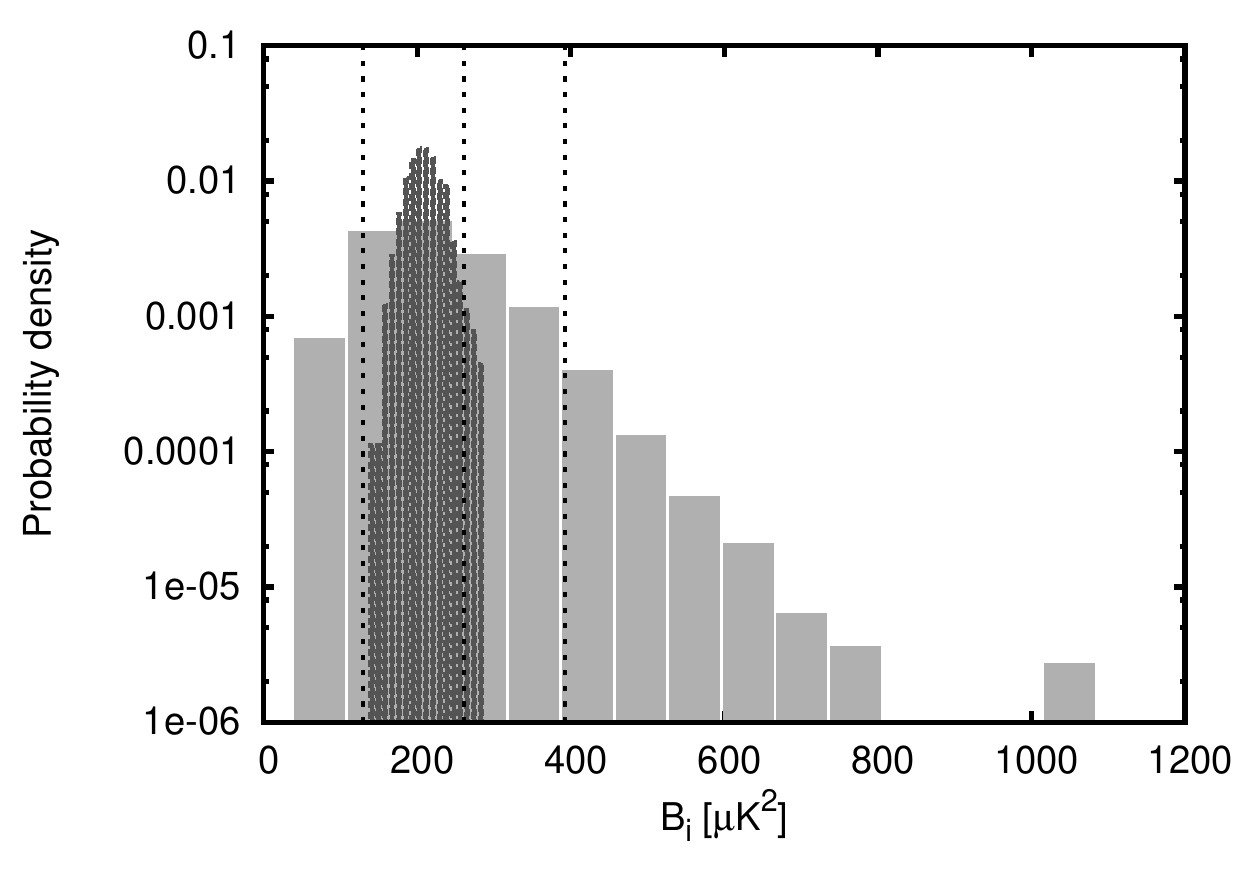}
\caption[The statistics of the combined CMB and SZ effect realizations with $\sigma_8 = 0.825$ for the multipole bin 2050-2350]{The statistics of the combined CMB and SZ effect realizations with $\sigma_8 = 0.825$ for the multipole bin 2050-2350. The large grey histogram is for the $0.75\degree \times 0.75\degree$ realizations (bin width of 75~$\upmu$K$^2$); the small, darker histogram is for the $3\degree \times 3\degree$ realizations (bin width of 9~$\upmu$K$^2$). The CBI value for this bin is $261 \pm 132$~$\upmu$K$^2$ (mean and $\pm 1 \sigma$ represented by the three dashed vertical lines); 26.5 per cent of the $0.75\degree\times0.75\degree$ realizations have this central power or higher, and 5 per cent have $376.5$~$\upmu$K$^2$ or higher.}
\label{fig:cbi_histogram_lowerbin}
\end{fig}

\begin{fig}
\centering
\includegraphics[scale=0.68]{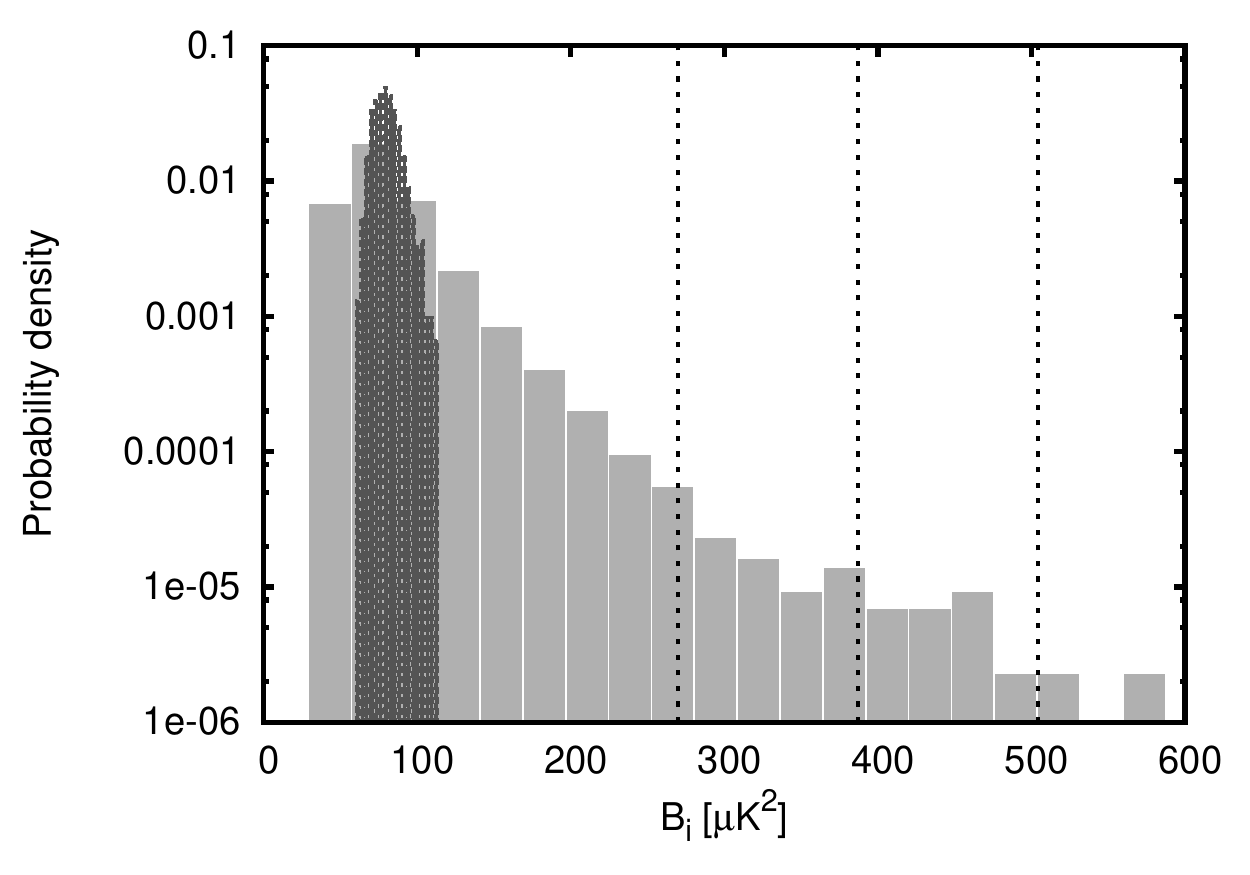}
\caption[As Figure \ref{fig:cbi_histogram_lowerbin}, but for the multipole bin 2350-3900 with bin widths of 28 and 3~$\upmu$K$^2$ for the $1\degree \times 1\degree$ and $3\degree \times 3\degree$ realizations respectively]{As Figure \ref{fig:cbi_histogram_lowerbin}, but for the multipole bin 2350-3900 with bin widths of 28 and 3~$\upmu$K$^2$ for the $1\degree \times 1\degree$ and $3\degree \times 3\degree$ realizations respectively. The CBI value for this bin is $387 \pm 117$~$\upmu$K$^2$ (mean and $\pm 1 \sigma$ represented by the three dashed vertical lines); 0.36 per cent of the $0.75\degree \times 0.75\degree$ maps have this central power or higher, with 5 per cent having $137.7$~$\upmu$K$^2$ or higher.}
\label{fig:cbi_histogram_upperbin}
\end{fig}

\begin{tab}{h}
\begin{tabular}{lcc|cccc|cccc}
& & & \multicolumn{4}{c|}{$3\degree \times 3 \degree$} & \multicolumn{4}{|c}{$0.75\degree \times 0.75\degree$}\\
\hline
Components & $\sigma_8$ & $S_\mathrm{max}$ & $B_i$ & $\delta B_i$ & $s_i$ & \% & $B_i$ & $\delta B_i$ & $s_i$ & \%\\
\hline
CMB only & -- & -- & 170 & 17 & 0.19 & 0 & 180 & 65 & 0.77 & 12.1\\
CMB, SZ & 0.75 & -- & 190 & 20 & 0.27 & 0 & 200 & 74 & 1.1 & 18.7\\
CMB, SZ & 0.825 & -- & 210 & 23 & 0.34 & 1.4 & 220 & 86 & 1.5 & 26.5\\
CMB, SZ & 0.9 & -- & 240 & 30 & 0.40 & 24.6  & 250 & 110 & 3.0 & 38.1\\
CMB, ST99 SZ & 0.825 & -- & 210 & 23 & 0.46 & 2.7 & 220 & 88 & 1.5 & 26\\
CMB, SZ, PS & 0.825 & 0.2 & 220 & 24 & 0.41 & 5.6 & 230 & 89 & 1.3 & 30.6\\
CMB, SZ, PS & 0.825 & 1.0 & 270 & 28 & 0.24 & 60.8 &  280 & 110 & 1.1 & 51.0\\
CMB, SZ, CPS & 0.825 & 0.2 & 210 & 22 & 0.29 & 2.7 & 220 & 86 & 1.4 & 27.2\\
CMB, SZ, CPS & 0.825 & 1.0 & 260 & 28 & 0.39 & 43.5 & 270 & 100 & 1.2 & 47.2\\
CMB, SZ, $Y_*=2.3$ & 0.825 & -- & 220 & 25 & 0.5 & 7.6 & 200 & 120 & 0.74 & 27.2\\
CMB, SZ, $\beta=1$ & 0.825 & -- & 230 & 42 & 2.1 & 14.7 & 250 & 170 & 10 & 34.9\\
\hline

CMB only & -- & -- & 37 & 2.0 & 0.17 & 0 & 37 & 6.6 & 0.41 & 0\\
CMB, SZ & 0.75 & -- & 59 & 5.2 & 0.71 & 0 & 59 & 20 & 4.7 & 0.07\\
CMB, SZ & 0.825 & -- & 80 & 8.6 & 0.70 & 0 & 80 & 33 & 3.5 & 0.36\\
CMB, SZ & 0.9 & -- & 120 & 15 & 0.73 & 0 & 120 & 59 & 5.3 & 2.4\\
CMB, ST99 SZ & 0.825 & -- & 80 & 8.5 & 1.3 & 0 & 80 & 35 & 5.4 & 0.14\\
CMB, SZ, PS & 0.825 & 0.2 & 100 & 8.8 & 0.66 & 0 & 100 & 34 & 3.1 & 0.5\\
CMB, SZ, PS & 0.825 & 1.0 & 200 & 12 & 0.24 & 0 & 200 & 49 & 1.5 & 10.3\\
CMB, SZ, CPS & 0.825 & 0.2 & 93 & 8.2 & 0.83 & 0 & 91 & 32 & 3.4 & 0.36\\
CMB, SZ, CPS & 0.825 & 1.0 & 190 & 12 & 0.26 & 0 & 190 & 46 &1.5 & 6.0\\
CMB, SZ, $Y_*=2.3$ & 0.825 & -- & 94 & 11 & 0.66 & 0 & 94 & 43 & 3.6 & 0.18\\
CMB, SZ, $\beta=1$ & 0.825 & -- & 120 & 31 & 2.0 & 0 & 120 & 120 & 9.1 & 2.4\\
\end{tabular}
\caption[Simulated statistics of the CBI excess bins]{Values for the mean and standard deviation at 30~GHz for the $l=2050-2350$ (top) and $l=2350-3900$ (bottom) CBI bins for $3\degree \times 3\degree$ (1000 realizations) and $0.75\degree \times 0.75\degree$ (16000 realizations) for various cosmologies and components, as well as the minimum and maximum values found and the normalized skew. The mean $\bar{B}_i$ and the standard deviation $\delta B_i$ within the bin $i$ are in $\upmu$K$^2$; the skew $s$ is dimensionless and and the values of $S_\mathrm{max}$ given are in mJy. The components are the Cosmic Microwave Background (CMB), the SZ effect (SZ), point sources (PS) and clustered point sources (CPS). The CBI data points are $261 \pm 132$ and $387 \pm 117$ $\upmu$K$^2$ for the lower and higher bin respectively; the final column gives the percentage of realizations at or above the central values of these measurements.}
\label{tab:excess_bin_values}
\end{tab}

To investigate the excess more systematically, we sample the power spectra for the same bins as used by the CBI. The histograms of the power spectrum from the CMB and SZ effect within the multipole bin $l=2050-2350$ for $\sigma_8 = 0.825$ and map sizes of $0.75\degree \times 0.75\degree$ (representative of the deep fields) and $3\degree \times 3\degree$ are shown in Figure \ref{fig:cbi_histogram_lowerbin}. The measured value for the CBI power spectrum is compatible with the simulations for this multipole range; 26.5 per cent of the realizations have the same or higher value as the central value from the CBI measurement. The same histogram for the $l=2350-3900$ bin is shown in Figure \ref{fig:cbi_histogram_upperbin}. In this histogram, the distribution is significantly skewed towards higher values, however the probability of getting the CBI excess from the realizations is low, with only 0.36 per cent of the realizations having the same or more power than the central CBI value. This would be reduced further if high-mass clusters are removed from the maps.

Table \ref{tab:excess_bin_values} gives the values for the various components and maps sizes for the two CBI high-multipole bins. Although the mosaicked fields extend to $5\degree$ across, we use $3\degree \times 3\degree$ fields due to the limits of the simulation box size at the highest redshifts. For the lower multipole bin, we find that the central value can be easily obtained using the CMB and SZ effect with $\sigma_8 = 0.9$, and realizations remain with the same amount of power for $\sigma_8 = 0.825$. Due to the large error on the measurement, however, it is in agreement with a large range of values of $\sigma_8$. For the higher multipole bin, there are no realizations with the central power for any of our three values of $\sigma_8$ using $3\degree \times 3\degree$ realizations. There are $0.75\degree \times 0.75\degree$ realizations that match the central value, even down to $\sigma_8 = 0.75$, but these are very unlikely, making up less than a percentage of the realizations for the lower values of $\sigma_8$.

Adding point sources to the realizations at the level of 0.2 or 1 mJy significantly increases the mean of the realizations in both bins, greatly increasing the number of realizations with the central power or higher as measured by the CBI in both bins. Clustering the point sources, however, slightly decreases this probability.

Due to the uncertainty in the parameters controlling the power and profile of the SZ effect (see Section \ref{sec:parameter_dependence}), the mean power spectra given here are uncertain to around 50 per cent. If this effect is in the positive direction, then this would significantly increase the number of realizations which agree with the CBI excess measurements. We change two of the parameter values, setting $Y_*=2.3 \times 10^{-6}$ Mpc$^{2}$ and $\beta = 1$ in two different sets of realizations. We find that the change in $Y_*$ has little effect, however the change in $\beta$ has a dramatic effect on the standard deviation and the skew of the distributions, as well as on the mean.

As shown earlier in Figure \ref{fig:sz_masscuts}, a large part of the skew and increased standard deviation compared to the mean comes from the largest clusters in the maps. Although there is no formal mass limit for the CBI fields, they were selected to avoid known large clusters, making it unlikely that any local, high mass clusters are present in the fields. This will have the effect of reducing the number of realizations matching the power measured by the CBI in accordance with the results of Section \ref{sec:upper_mass}.

In addition to the CBI, the Berkeley-Illinois-Maryland Association (BIMA) interferometer operating at 28.5~GHz has also measured an excess. \citet{2006Dawson} surveyed eighteen 6.6 arcminute fields with a total area of $\sim 0.2$ square degrees, and found $220^{+140}_{-120}$~$\upmu$K$^2$ with an average multipole of $l=5237$ and a FWHM of 2870. We model this as a constant bin between $l=3800$ and $6670$. For a thousand realizations with $\sigma_8 = 0.825$ and the same map size as BIMA's $6.6$ arcminute field of view, we find a mean within this bin of 41 $\upmu$K$^2$, a standard deviation of 82 $\upmu$K$^2$ and a normalized skew of 5.5. Using just the mean and standard deviation, this makes the central value of the measurement a $\sim$ 2.2 $\sigma$ excess. We find that 3.3 per cent of the realizations have greater than 220 $\upmu$K$^2$.

The Sunyaev-Zel'dovich Array (SZA) has also measured the power spectrum at higher multipoles, putting an upper limit of $149$~$\upmu$K$^2$ (95 per cent confidence) on the power spectrum from secondary anisotropies between $l=2000-6000$ \citep{2009Sharp} using forty-three 12 arcminute fields. As 68 per cent of their multipole coverage is between $l=2929-5883$, we use a constant bin with a width of $l=2900-5900$. We find a mean of $50$~$\upmu$K$^2$, with a standard deviation of $64$ and a skew of 5.8. 5.6 per cent of the realizations have a power greater than $149$~$\upmu$K$^2$. This is compatible with the measurement from SZA.

At higher frequencies, ACBAR has also measured an excess of $34 \pm 20$~$\upmu$K$^2$ at 150GHz in the range $l = 1950-3000$; this 1.7 $\sigma$ excess is compatible with the CBI and BIMA measurements if the signal has the spectral distribution of the SZ effect \citep{2008Reichardt}. These measurements cover nearly 700 square degrees over 10 fields; we are limited by the map size of the {\sc Pinocchio} simulations, which are 9 square degrees, so cannot make a prediction for these fields. For comparison, however, we find that the SZ effect alone contributes $9.6$~$\upmu$K$^2$ with a standard deviation of $2.2$~$\upmu$K$^2$ and a skew of 0.94 within the bin $l=1950-3000$ for $3\degree \times 3\degree$ maps. We note that QUaD \citep{2009Friedman} also measure the 150~GHz power spectrum at high multipoles; they find values that are compatible with $\Lambda$CDM on its own with no SZ effect required.

\section{CMB power spectrum observations with OCRA} \label{sec:cmb_ocra}
The OCRA receivers are sensitive to multipoles $\ell \sim 4000$, and hence could potentially be used to investigate the high multipole excess. In order to investigate this, simulations were run using maps of the CMB and SZ effect created as described in Chapter \ref{virtualsky}. These were convolved with the 2-beam window function of OCRA using \citep{1995White}:
\begin{equation}
W_\ell = B_\ell^2 (\sigma) 2 \left( 1-P_\ell (\cos(\alpha_0)) \right) \label{eq:windowfunction},
\end{equation}
in which $\sigma$ is the beam width, calculated via $\sigma = \theta_\mathrm{FWHM}/\sqrt{8 \ln 2}$ where the $\theta_\mathrm{FWHM}=1.2$ arcmin for OCRA, $P_l$ are the zeroth-order Legendre polynomials, and $\alpha_0$ is the distance between the beams; here 3.0 arcmin. The window function from the beam is calculated by
\begin{equation}
B_\ell(\sigma) = \exp \left(-\frac{1}{2} \ell (\ell +1) \sigma^2 \right).
\end{equation}
The window functions for a range of different values of $\sigma$ and $\alpha_0$ are shown in Figure \ref{fig:2beam_windowfunctions}. For OCRA, the peak in the multipole sensitivity is around 4000. The CBI measurement of the high multipole excess is primarily around $l\approx2500$; if this is due to either of the SZ effect or point sources, then the effect should be comparable or stronger at the multipole range of OCRA.

\begin{fig}
\centering
\includegraphics[scale=0.68]{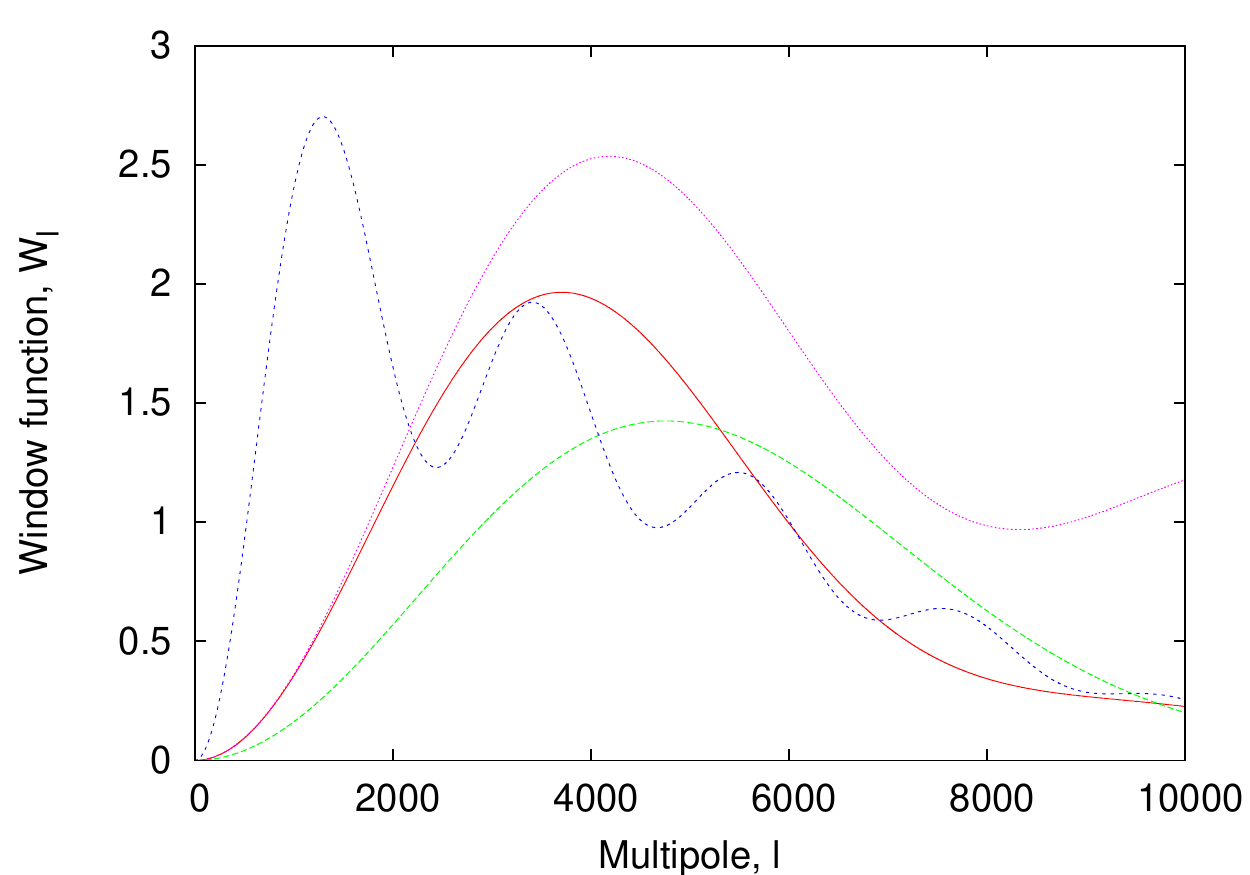}
\caption[The window functions of 2-beam receivers]{The window functions of 2-beam receivers, calculated using Equation \ref{eq:windowfunction}. The red solid line is for an OCRA system with $\sigma = 1.2$ arcmin FWHM beams separated by $\alpha_0 = 3$ arcmin. The green dashed line is for $\sigma = 1.2$, $\alpha_0 = 2$; the blue short-dashed line $\sigma = 1.2$, $\alpha_0 = 10$ and the magenta dotted line $\sigma = 0.6$, $\alpha = 3.0$.}
\label{fig:2beam_windowfunctions}
\end{fig}

Noise is then added to the map using Gaussian random numbers with a mean of zero and a variance defined by the noise per pixel, calculated by
\begin{equation}
\sigma_\mathrm{pixel} = \sigma_\mathrm{beam} \sqrt{ \frac{ \Omega_\mathrm{beam}}{\Omega_\mathrm{pixel}} } 
\end{equation}
where $\Omega_\mathrm{beam}$ is the area of the beam, calculated by $\Omega_\mathrm{beam} = \pi \phantom{.} \theta_\mathrm{FWHM}^2 / \left( 4 \ln2 \right)$. The noise is then subtracted from the power spectrum using $<C_l^\mathrm{Noise}> = \sigma_\mathrm{pixel}^2 \Omega_\mathrm{pixel}$. This removes the average noise power, but the random fluctuations in the noise levels remain. The window function is then removed by dividing by $W_\ell$, and the power spectrum is binned as before.

The noise can be analytically calculated in a similar way to cosmic variance (see equation \ref{eq:cv}) from the mean power spectrum of the CMB and the weighted power spectrum of the Gaussian noise via
\begin{equation} \label{eq:cv_noise}
\delta B_{i} = \frac{1}{\Delta l}\sqrt{\sum_{l \in \mathrm{bin}} \left(\frac{l(l+1)}{2 \pi} \right)^2 \frac{2 }{(2 l + 1) f_\mathrm{sky}} \left( C_l + \frac{\sigma_\mathrm{beam}^2\Omega_\mathrm{beam}}{W_l} \right)^2},
\end{equation}
Note that this does not include the increase in cosmic variance due to the SZ effect, or any noise that is not Gaussian. This equation is based upon that provided by \citet{1995Knox}. Note that this analytic expression does not take into consideration any correlations in the noise between samples.

\begin{fig}
\centering
\includegraphics[scale=0.68]{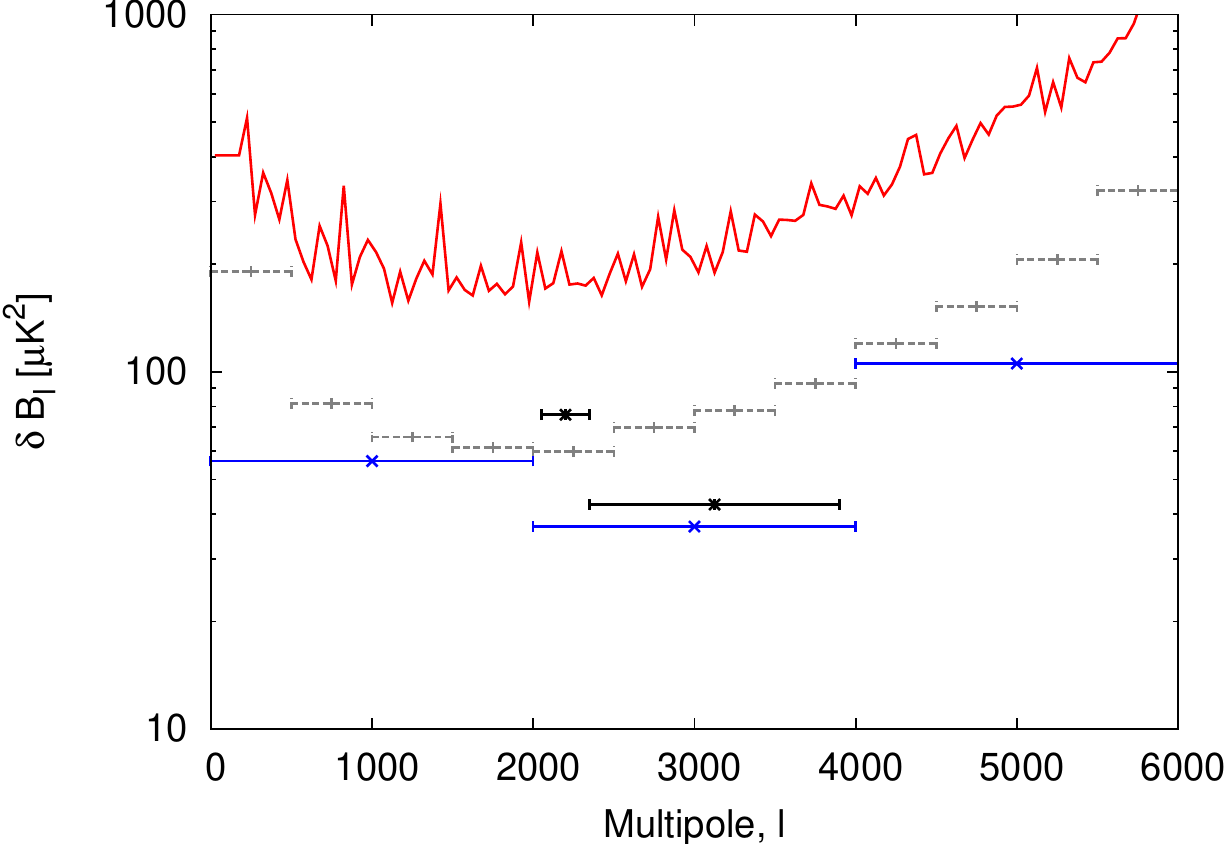}
\caption[The error on the power spectrum from a $3\degree \times 3\degree$ OCRA map consisting only of noise]{The binned error on the power spectrum for a $3\degree \times 3\degree$ OCRA map consisting only of noise (100~$\upmu$K). The red line shows the error on the power spectrum using multipole bins of 50; the grey points use a multipole bin of 500, and the blue points use a multipole bin of 2000. The two black points match the two highest multipole bins used by the CBI.}
\label{fig:noisy_powerspectrum}
\end{fig}

\begin{fig}
\centering
\includegraphics[scale=0.6]{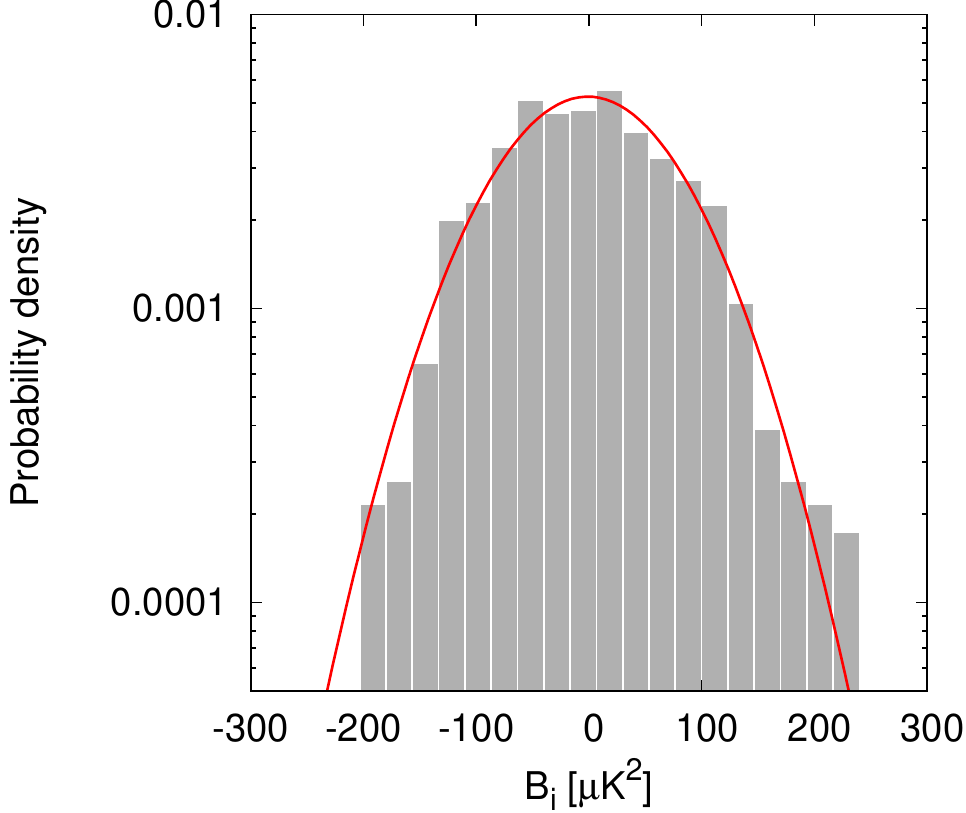}
\includegraphics[scale=0.6]{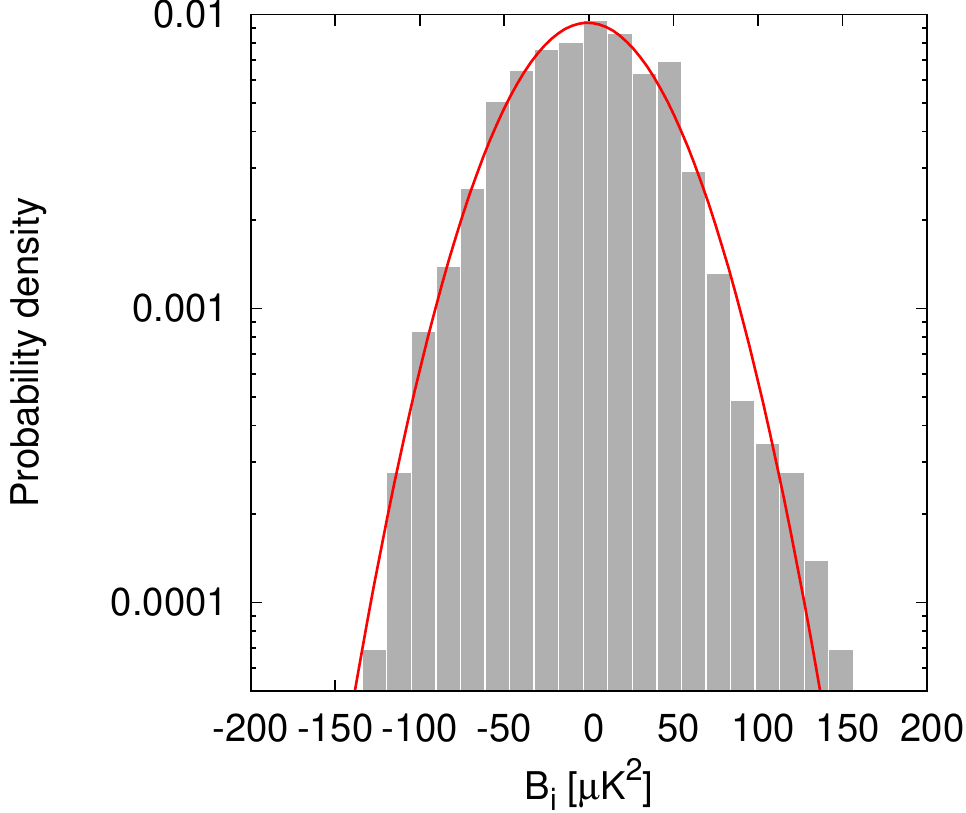}
\caption[The histograms for the CBI points shown in Figure \ref{fig:noisy_powerspectrum}]{The histograms for the CBI points shown in Figure \ref{fig:noisy_powerspectrum}. The left-hand figure is the histogram for the multipole range $2050-2350$; the right-hand figure for $2350-3900$.}
\label{fig:noisy_cbihistograms}
\end{fig}

As a ``null'' test, the mean power spectra and the scatter on that mean is measured for 1000 realisations where the maps just consist of noise, i.e. there is no signal from the CMB, galaxy clusters or point sources present. The fiducial results for a $3\degree \times 3\degree$ field observed to $100$~$\upmu \mathrm{K}$ per beam ($1 \sigma$ standard deviation) are shown in Figure \ref{fig:noisy_powerspectrum}, using multipole bin sizes of 50, 500 and 2000, as well as for the two CBI multipole bins. The noise levels increase significantly at the highest multipoles ($>5000$), as well as the lowest multipoles ($<1000$).

Based on noise alone, a single $3\degree \times 3\degree$ map observed to 100~$\upmu \mathrm{K}$ per beam would have an uncertainty of 76~$\upmu \mathrm{K}^2$ for the multipole bin 2050-2350, and 43~$\upmu \mathrm{K}^2$ for the multipole bin 2350-3900, based on the standard deviation of these realisations. These errors provide a reasonable degree of improvement over those obtained by the CBI observations (132~$\upmu\mathrm{K}^2$ and 117~$\upmu \mathrm{K}^2$ respectively). The histograms for these two bins are shown in Figure \ref{fig:noisy_cbihistograms}, along with a Gaussian fit. As with the CMB (see Section \ref{sec:cmb}), the histogram is slightly skewed to higher values due to the finite sampling.

\begin{fig}
\centering
\includegraphics[scale=0.68]{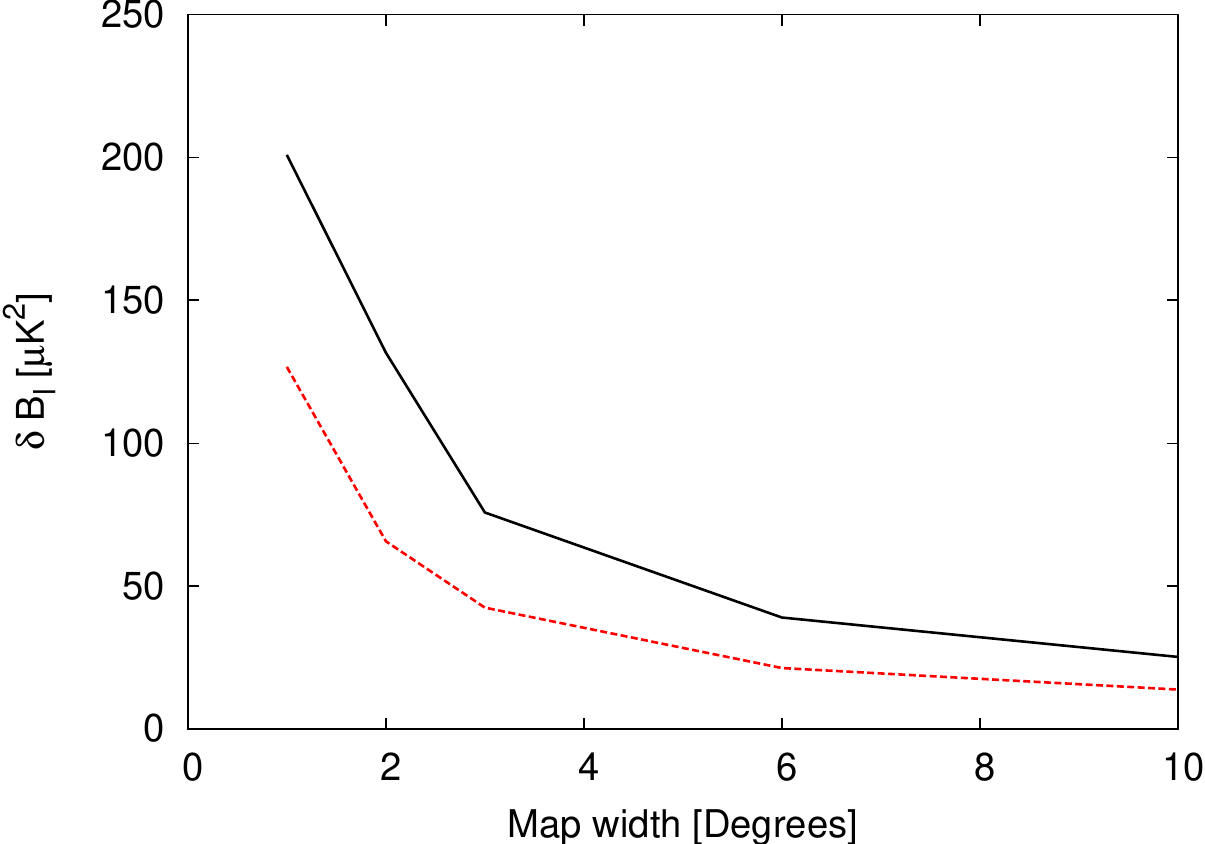}
\caption[Power spectrum noise levels versus map size for the CBI multipole bins]{The power spectrum noise level as a function of map width, for a noise of $100~\upmu \mathrm{K}^2$ per beam. The black line is for the multipole range $2050-2350$; the red line for $2350-3900$.}
\label{fig:noise_fn_mapsize}
\end{fig}

The dependence of the predicted uncertainty in the two highest multipole bins observed by the CBI is shown in Figure \ref{fig:noise_fn_mapsize} for fields between 1$\degree$ and 10$\degree$ on a side, all with 100 $\upmu$K noise per beam. The noise scales approximately as $f_\mathrm{sky}^{-1/2}$. Additionally, the same simulations are carried out for $1\degree \times 1\degree$ fields using noise levels between $50$ and $1000$ $\upmu K$ per beam; the uncertainty scales as $\sigma^2$ as expected.

\begin{fig}
\centering
\includegraphics[scale=0.68]{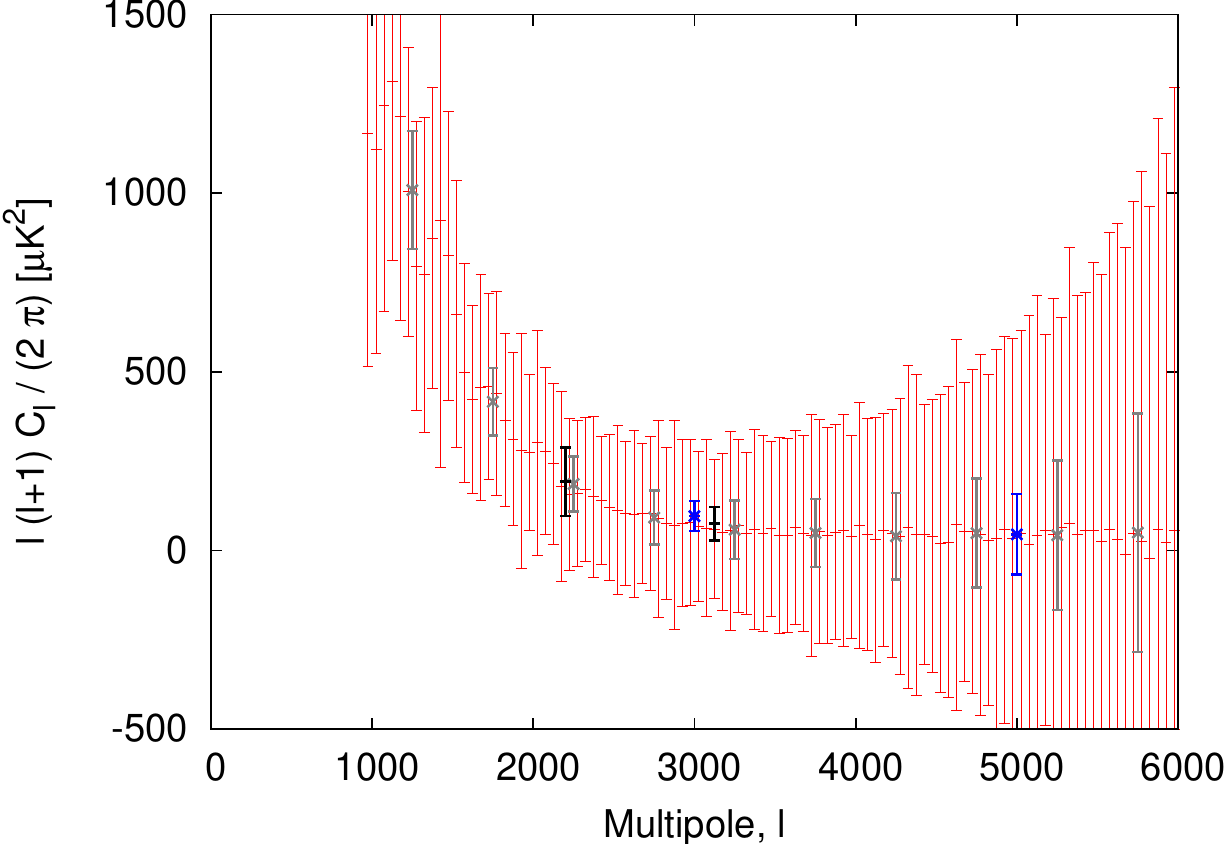}
\caption[The binned power spectrum of a $3\degree \times 3\degree$ OCRA map consisting of noise and signal]{The binned power spectrum of a $3\degree \times 3\degree$ OCRA map consisting of noise and signal. The red points are the power spectrum with multipole bins of 50; the green points use a multipole bin of 500, and the grey points use a multipole bin of 2000. The two black points match the two highest multipole bins used by the CBI.}
\label{fig:noise_signal_powerspectrum}
\end{fig}

\begin{fig}
\centering
\includegraphics[scale=0.6]{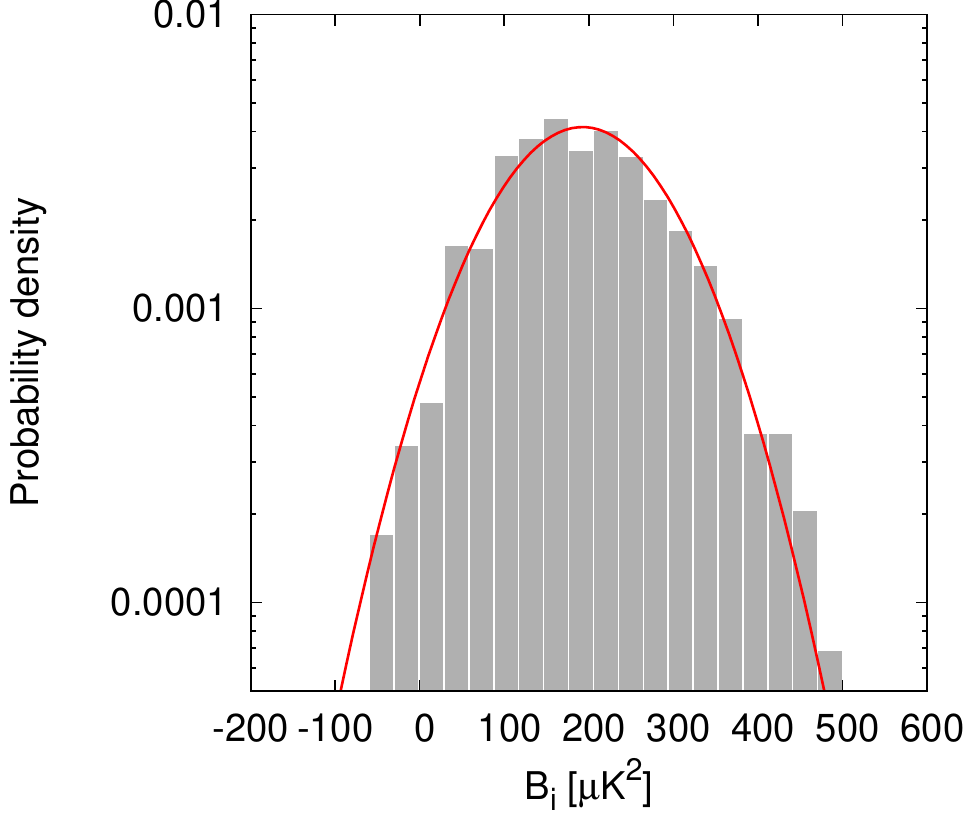}
\includegraphics[scale=0.6]{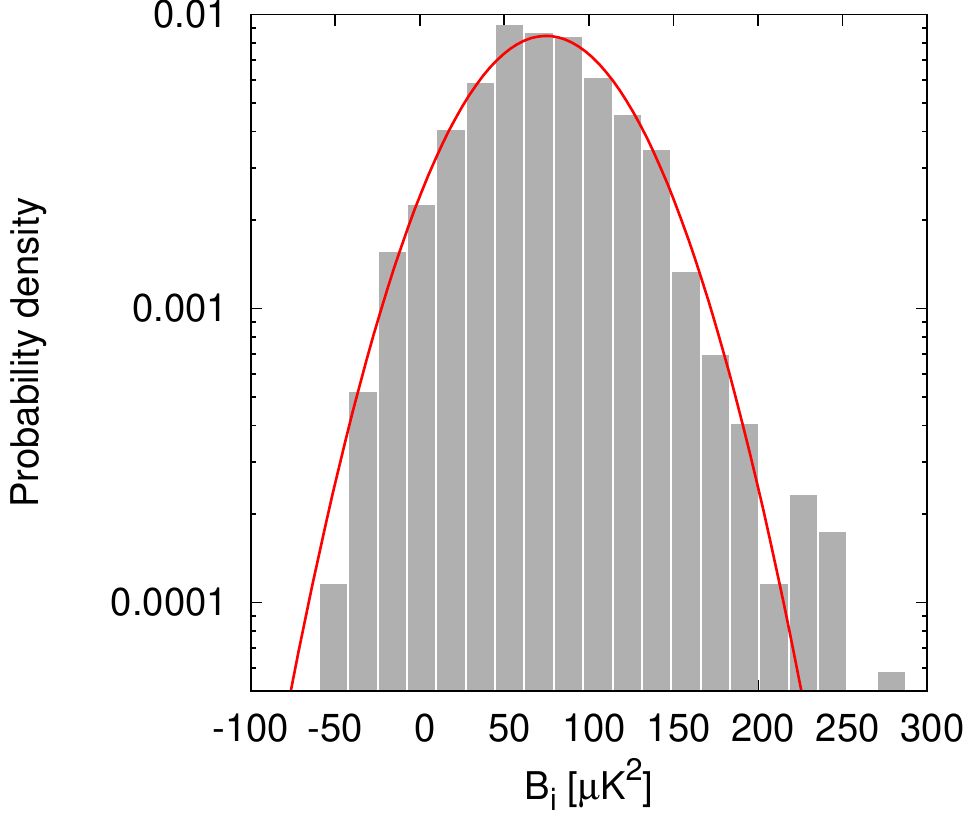}
\caption[The histograms for the CBI points shown in Figure \ref{fig:noisy_cbihistograms}]{The histograms for the CBI points shown in Figure \ref{fig:noisy_cbihistograms}. The left-hand figure is the histogram for the multipole range $2050-2350$; the right-hand figure for $2350-3900$.}
\label{fig:noise_signal_cbihistograms}
\end{fig}

When the astrophysical signal from the CMB and SZ effect (with $\sigma_8 = 0.825$) are added in, the error increases to $96~\upmu \mathrm{K}^2$ for the multipole bin 2050-2350, and $57~\upmu \mathrm{K}^2$ for the multipole bin 2350-3900, where the increase is due to cosmic variance. When the signal is entirely from the CMB, but with the same mean power spectrum, then these numbers decrease to $94$ and $46~\upmu \mathrm{K}^2$, reflecting the previously described increase in cosmic variance due to the SZ effect. These numbers remain competitive with the CBI observations, however lower noise measurements or larger map sizes would be preferable to obtain an improved measurement. The values for the two CBI bins, as well as the multipole bin $2000-4000$, for various map sizes and noise levels are given in Table \ref{tab:ocra_excess_bin_values}. As described above, this signal does not match the level that the CBI has measured, but this should not matter for these indicative simulations. We caution that the error on the mean will depend on the standard deviation, which is why the mean power spectra differ between different map sizes and noise levels.

\begin{tab}{h}
\begin{tabular}{c|cc|cc|cc|cc|c}
& \multicolumn{2}{c}{Configuration} & \multicolumn{2}{c}{Noise only} & \multicolumn{2}{c}{CMB+SZ} & \multicolumn{2}{c}{CMB only}&Analytical\\
\hline
Multipoles & Map size & Noise level & $B_i$ & $\delta B_i$ & $B_i$ & $\delta B_i$ & $B_i$ & $\delta B_i$ & $\delta B_i$\\
\hline
2050 & $1 \degree \times 1 \degree$ & 100 $\upmu$K & 0.0 & 200 & 200 & 255 & 200 & 255 & 250\\
-- & $1 \degree \times 1 \degree$ & 200 $\upmu$K & 0.6 & 840 & 190 & 850 & 160 & 860 & 860\\
2350 & $3 \degree \times 3 \degree$ & 100 $\upmu$K & -0.9 & 76 & 190 & 96 & 200 & 94 & 84\\
& $3 \degree \times 3 \degree$ & 200 $\upmu$K & -7.8 & 290 & 200 & 320 & 200 & 320 & 290\\
\hline
2350& $1 \degree \times 1 \degree$ & 100 $\upmu$K & 0.8 & 130 & 81 & 140 & 70 & 130 & 120\\
-- & $1 \degree \times 1 \degree$ & 200 $\upmu$K & -17 & 490 & 63 & 510 & 56 & 490 & 450\\
3900 & $3 \degree \times 3 \degree$ & 100 $\upmu$K & -0.8 & 43 & 75 & 57 & 73 & 46 & 39\\
& $3 \degree \times 3 \degree$ & 200 $\upmu$K & -4.0 & 175 & 67 & 190 & 73 & 180 & 150\\
\hline
2000 & $1 \degree \times 1 \degree$ & 100 $\upmu$K & -0.5 & 110 & 100 & 120 & 96 & 120 & 100\\
-- & $1 \degree \times 1 \degree$ & 200 $\upmu$K & -20 & 440 & 82 & 460 & 71 & 450 & 390\\
4000 & $3 \degree \times 3 \degree$ & 100 $\upmu$K & -0.6 & 37 & 96 & 42 & 96 & 41 & 34\\
& $3 \degree \times 3 \degree$ & 200 $\upmu$K & -4.5 & 150 & 92 & 160 & 97 & 160 & 130\\
\hline
4000 & $1 \degree \times 1 \degree$ & 100 $\upmu$K & -14 & 330 & 40 & 310 & 30 & 300 & 270\\
-- & $1 \degree \times 1 \degree$ & 200 $\upmu$K & -35 & 1200 & -2 & 1200 & 19 & 1200 & 1100\\
6000 & $3 \degree \times 3 \degree$ & 100 $\upmu$K & 0.3 & 110 & 45 & 110 & 42 & 110 & 90\\
& $3 \degree \times 3 \degree$ & 200 $\upmu$K & 4.8 & 440 & 48 & 440 & 57 & 420 & 360\\
\hline
6000 & $1 \degree \times 1 \degree$ & 100 $\upmu$K & 1.3 & 1400 & 71 & 1400 & 41 & 1400 & 1200\\
-- & $1 \degree \times 1 \degree$ & 200 $\upmu$K & 300 & 5700 & -57 & 5800 & -14 & 5800 & 4900\\
8000 & $3 \degree \times 3 \degree$ & 100 $\upmu$K & 2.3 & 480 & 51 & 490 & 27 & 480 & 410\\
& $3 \degree \times 3 \degree$ & 200 $\upmu$K & -81 & 2100 & 61 & 2000 & 160 & 2000 & 1600\\
\end{tabular}
\caption[Estimated error bars for potential OCRA measurements of the CBI excess]{Values for the mean and standard deviation at 30~GHz for the $l=2050-2350$ and $l=2350-3900$ CBI bins, and for multipole bins of $2000-4000$, $4000-6000$ and $6000-8000$ for different map sizes and noise levels, for both noise only (left), in the presence of CMB and SZ (middle) and in the presence of CMB only but with the same mean power spectrum as the combined CMB and SZ (right). The expected analytical values from equation \ref{eq:cv_noise} are given in the final column. The mean $\bar{B}_i$ and the standard deviation $\delta B_i$ within the bin $i$ are in $\upmu$K$^2$  The CBI data points are $261 \pm 132$ and $387 \pm 117$~$\upmu$K$^2$ for the lower and higher bin respectively. For a $3\degree \times 3 \degree$ field observed to 100~$\upmu$K OCRA can measure the CBI multipole bins to a factor of 2 lower in noise, and also provides power spectrum information at higher multipoles.}
\label{tab:ocra_excess_bin_values}
\end{tab}

It is worth noting that the code is capable of carrying out these simulations for a wide range of different beam sizes and separations; this has not been studied here as these values are fixed for the OCRA instrument. Additionally, a range of different values for the input signal could be used, e.g. depending on the different values of $\sigma_8$, however this has not been investigated as this should not significantly affect the noise levels that the experiment can obtain. Point sources are currently not 	considered; an interesting extension to this work would be to add these into the simulations. Finally, these simulations have only considered Gaussian noise; $1/f$ noise will be important on the large scales (longer timescales), with the exact amount depending on the scan strategy used. Due to this dependency on the mapping strategies, the end-to-end simulations described in the next chapter are most suited to considering this.

An important consideration in assessing these simulations is the required observation time. OCRA-p has a theoretical noise level of around $10~\mathrm{mJy~s}^{1/2}$ (see Section \ref{sec:ocrap_capabilities}), which converts into $\sim 1.3~\mathrm{mK~s}^{1/2}$. In order to obtain a measurement in a beam to $100~\upmu$K, approximately 49~s of integration time are required. With a $1.2$~arcminute beam, one square degree is $\sim 2500$ beams. Assuming that OCRA-F has similar performance, and taking into account the four pairs of receivers within OCRA-F (which speeds up the survey by a factor of four), the total amount of telescope time required would be $\sim1\times10^5$ seconds, or 30 hours of integration time, per square degree.

This obviously depends linearly on the number of beams that the receiver has; completing OCRA-F (16 beams) would halve this time requirement. The time estimate excludes overheads -- including calibration and $T_\mathrm{sys}$ measurements -- however using the receiver in survey mode, rather than carrying out pointed observations, should significantly reduce the observation overhead compared with OCRA-p observations thus far. $1/f$ gain fluctuations and atmospheric fluctuations will also present observational challenges. Finally, it should also be noted that this observation time requires good weather, which restricts the number of days that are available for these observations. However, from these simulations it appears that the time requirements for OCRA-F to contribute towards observations of the CBI excess are not prohibitive.

\section{Conclusions}

We have simulated the microwave sky, including the CMB, SZ effect and point sources, and have used {\sc Pinocchio} to generate large numbers of cluster catalogues with realistic distributions for three different values of $\sigma_8$. Using these maps, we have investigated the statistics of the power spectrum between multipoles of 1000 and 10~000. We find that the inclusion of the SZ effect increases the standard deviation of the power spectrum by a factor of 3 over that expected from cosmic variance, in agreement with the predictions from an analytical calculation based on the halo formalism. The mean and standard deviation vary as $1/f_\mathrm{sky}^{1/2}$ as expected, and scale approximately as $\sigma_8^7$ over the range of values sampled here. We also find that the distributions are non-Gaussian, and are skewed by large mass clusters, with the degree of this skewness increasing as the map size is decreased.  Additionally, we find that correlations between galaxy clusters play a small role in the statistics of the power spectrum at the level of $\sim$10 per cent.

Several instruments have measured an excess at high multipoles, which may be due to the SZ effect with a large value of $\sigma_8$. We cannot explain the central values of these measurements with the range of $\sigma_8$ investigated here, however the increased standard deviation and the presence of skewness in the distribution means that these measurements could be explained by a lower value of $\sigma_8$ than has been suggested so far. There is also a large uncertainty in the parameters describing the cluster gas physics, which can have a large effect on the mean of the distributions, comparable to that from the different values of $\sigma_8$, and can also significantly effect the standard deviation and the skew of the distributions.

The next generation of CMB instruments are currently being commissioned, and are expected to provide more data at multipoles comparable to those probed by the CBI. The {\it Planck} satellite will measure the power spectrum from the whole sky out to multipoles of 2500 within the next few years, and instruments such as the South Pole Telescope \citep[SPT;][]{2004Ruhl}, the Arcminute Microkelvin Imager \citep[AMI;][]{2008Zwart} and the Atacama Cosmology Telescope \citep[ACT;][]{2004Fowler} will observe large numbers of galaxy clusters using the SZ effect. These measurements will provide much more information on the SZ effect and may provide a resolution to the discrepancy in $\sigma_8$ from the measurements to date. Our results should also be relevant to these observations. Additionally, future OCRA observations may be able to carry out observations of the CBI excess to either confirm or refute the existence of this high-multipole excess

\chapter{End-to-end simulations of OCRA}\label{ocra_sims}

In order to get the best results from real observations, the observational methods and the contributions to the instrument noise need to be well understood. \citet{2006Lowe} constructed a computer simulation (UMBRELLA, an Upper Millimeter Band Receiver Emulator for Large Linked Arrays) of the OCRA-p instrument (see \S \ref{sec:ocra}), involving a model of the atmosphere, the telescope and the receiver system. UMBRELLA was first used to perform mock drift scan observations of point sources. With the further work described in this chapter, it can now be used to carry out mock observations of SZ clusters from Virtual Sky simulations to complement the extensive observations of SZ clusters that have been carried out using OCRA-p (\citealp{2006Lancaster}, Lancaster et al. in prep.). It has also been extended to simulate the more complex OCRA-F instrument, and hence it could be used to find the best observing methods for trial blind surveys. This would also enable the quantification of the detection efficiency of clusters with different masses and redshifts using OCRA-F.

In order to analyse the simulated data (and, of course, the real OCRA data) an automated data reduction software package is required. The creation of this software is described in the next section. Simulated noise power spectra have been created with UMBRELLA to investigate the effects of Gaussian noise, $1/f$ gain fluctuations and atmospheric fluctuations on the double-differenced noise power spectrum (Section \ref{sec:sim_noisepowerspectra}). The data reduction software has then been linked with UMBRELLA,  and the combination has been used to simulate and analyse point source observations, and to investigate the data reduction methods that can be applied to these observations (Sections \ref{sec:sim_ps} and \ref{sec:dd_comparison}). The chapter is summarized in Section \ref{sec:sim_futurework}, which also discusses the future work that can be done based upon the tools developed here. The overall plan of the end-to-end simulations is shown in Figure \ref{fig:endtoend_overallplan}.

\begin{fig}
\centering
\includegraphics[scale=0.7]{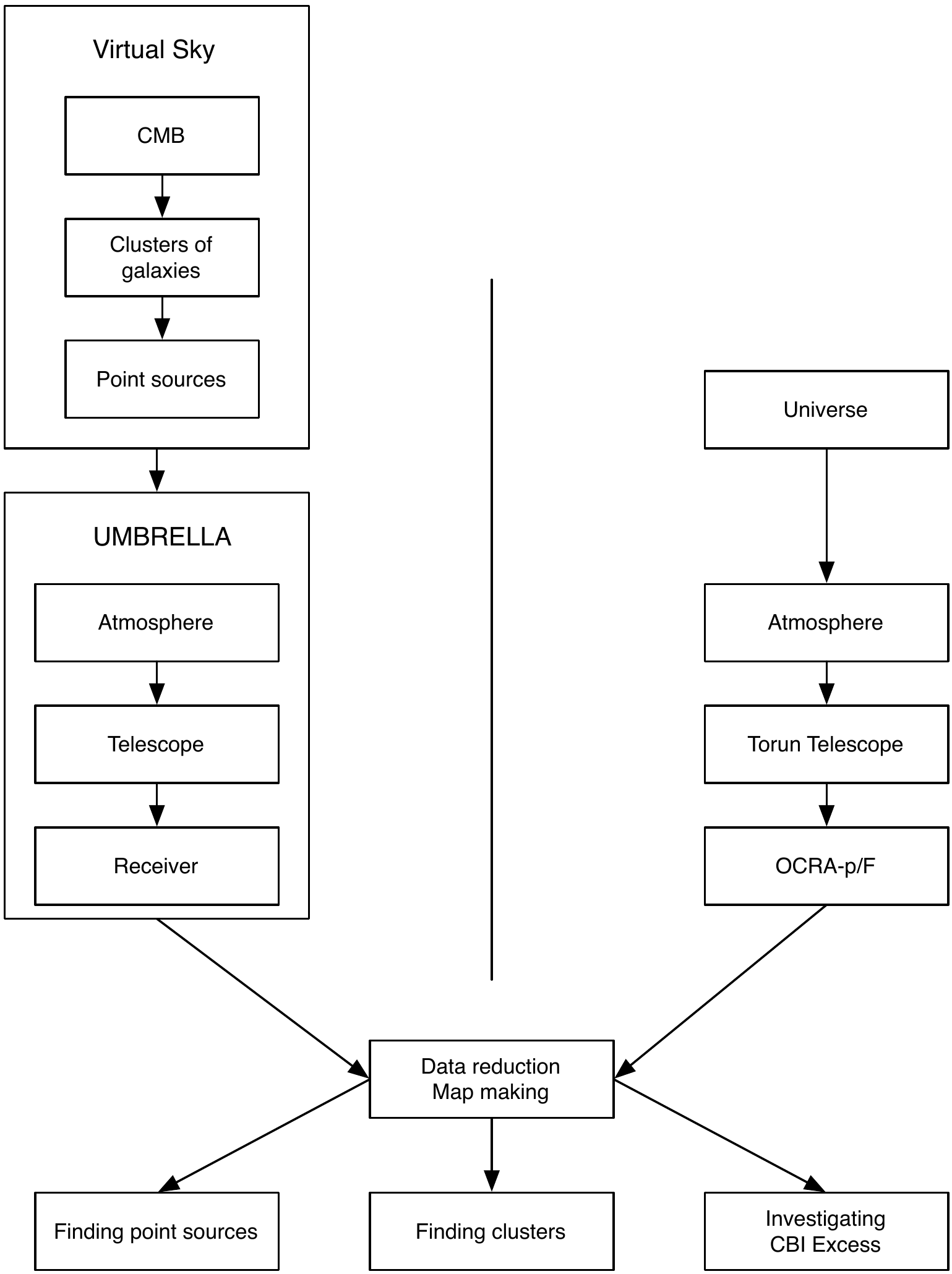}
\caption[Overall plan of end-to-end simulations for OCRA]{An overall plan for end-to-end simulations with OCRA, compared with the real thing. Left: simulations, consisting of Virtual Sky and UMBRELLA. Right: reality; the Universe, atmosphere, telescope and receiver. Both come together at the data reduction stage, which will also include map making. The output from both can then be used to search for point sources and galaxy clusters, and also potentially to observe the CBI excess.}
\label{fig:endtoend_overallplan}
\end{fig}

\section{Data reduction} \label{sec:dred}

There are currently two observational methods utilized by OCRA-p. ``Cross-scan'' measurements are used for strong sources, and ``on-off'' measurements for weak sources. In both of these methods, a calibration diode is observed. This calibration diode is located in the receiver cabin of the telescope, and transmits into OCRA-p by means of a wire protruding into one of the receiver horns. The methods of reducing data from each of these observing techniques to give a measured flux density for the source is covered in the next two sections. The calibration method is then described in Section \ref{sec:dred_cal}. The calculation of the final flux densities -- including the identification of bad data -- is described in Section \ref{sec:dred_result}. The data reduction process is used both for the end-to-end simulations, described later on in this chapter, and also for the observations taken with OCRA-p (see Section \ref{chapter:ocrap_obs}). As such, the data reduction process is described in general terms rather than solely focusing on the elements that are relevant to the simulations.

\subsection{Cross-scan measurements}
Strong ($>50$~mJy) sources can be observed by scanning the two beams of OCRA-p across the source. ``Cross-scan'' measurements are carried out by first scanning across the source position in elevation, such that the source passes through only one of the beams, which are separated in azimuth. These are immediately analysed and a pointing correction in elevation is calculated and applied for that position on the sky. The telescope then scans across the source in azimuth with both beams. Following on from this, a short ($\sim 10$~s) measurement of a blank section of sky is made, followed by a short ($\sim 10$~s) measurement of a calibration diode. An example of this type of measurement is show in Figure \ref{fig:example_qscan}. All measurements of calibrator sources are performed using this method.

The data are automatically binned by the Data AcQuisition system (DAQ) into 1 second samples, and a scatter from the measurements within that second is also calculated. The measurements are recorded along with the time since the start of the observation and the difference in azimuth and elevation from the assumed source position at the time of the measurement.

\begin{fig}
\begin{center}
   \includegraphics[scale=1.0]{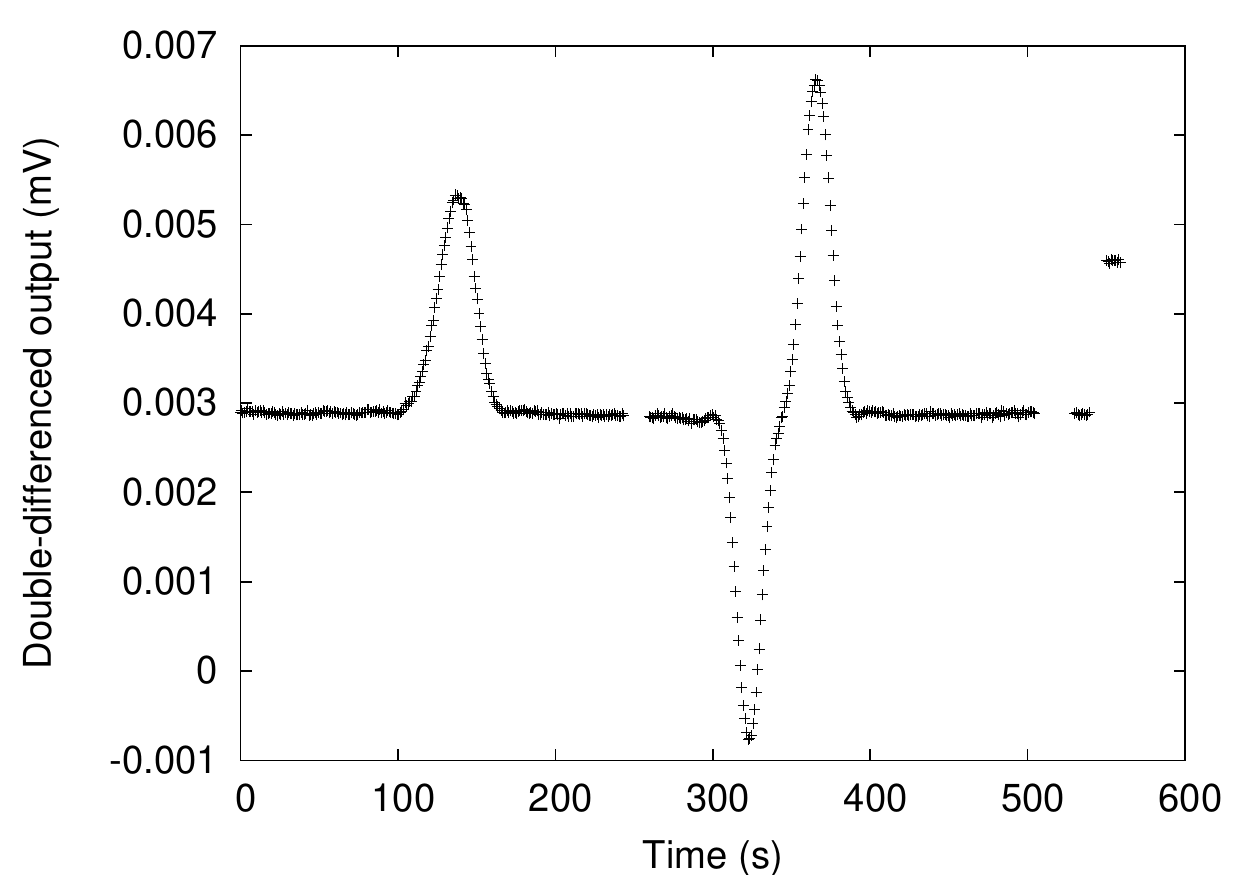}
\caption[An example cross-scan of NGC~7027]{An example cross-scan of NGC~7027; first in elevation, then azimuth, followed by a background and calibration diode measurement}
\label{fig:example_qscan}
\end{center}
\end{fig}

The pointing correction from the elevation scan is fitted automatically at the telescope, with the peak position used for the azimuthal scan. Thus, strong sources act as their own pointing calibrator and can be observed regardless of any initial offsets in the telescope pointing. The peak value from the azimuthal scan is then used as a measurement of the flux density of the source. In reality, the signal will be noisy, so a fit is applied to the data to determine the source flux density. We use a nine parameter non-linear least squares fit, involving a quadratic background and two gaussians:
\begin{equation}
f(t) = B_0 + B_1 t + B_2 t^2 + A_1 \exp \left(\frac{- (t - t_1)^2}{2 \sigma_1^2} \right) + A_2 \exp \left(\frac{- (t - t_2)^2}{2 \sigma_2^2} \right)
\label{eq:2gaussian}
\end{equation}
Here, $B_i$ represents the quadratic background terms, $A_1$ and $A_2$ the amplitude of the Gaussian peaks, $t_1$ and $t_2$ the times at which the peaks occur and $\sigma_1$ and $\sigma_2$ represent the widths of the two peaks. In the case of a perfect measurement, $A_1 = A_2$ and $\sigma_1 = \sigma_2$. The average of $A_1$ and $A_2$ is used to calculate the measured flux density of the source, with a measurement error calculated from the root mean square (rms) of the fitting errors on these two parameters.

As a cross-check, a fit is also applied to the elevation scans in a similar way to the azimuthal scans, except using a 5-parameter 1 gaussian fit:
\begin{equation}
f(t) = B_0 + B_1 t + A_1 \exp \left(\frac{- (t - t_1)^2}{2 \sigma_1^2} \right)
\label{eq:1gaussian}
\end{equation}
The amplitude of the elevation scan compared with that from the azimuth scan can then be used to check whether the scan was successful (the amplitude from the azimuth scan should always be greater or equal to the amplitude from the elevation scan).

\subsection{On-off measurements} \label{sec:onoff_methods}
\begin{fig}
\begin{center}
   \includegraphics[scale=1.0]{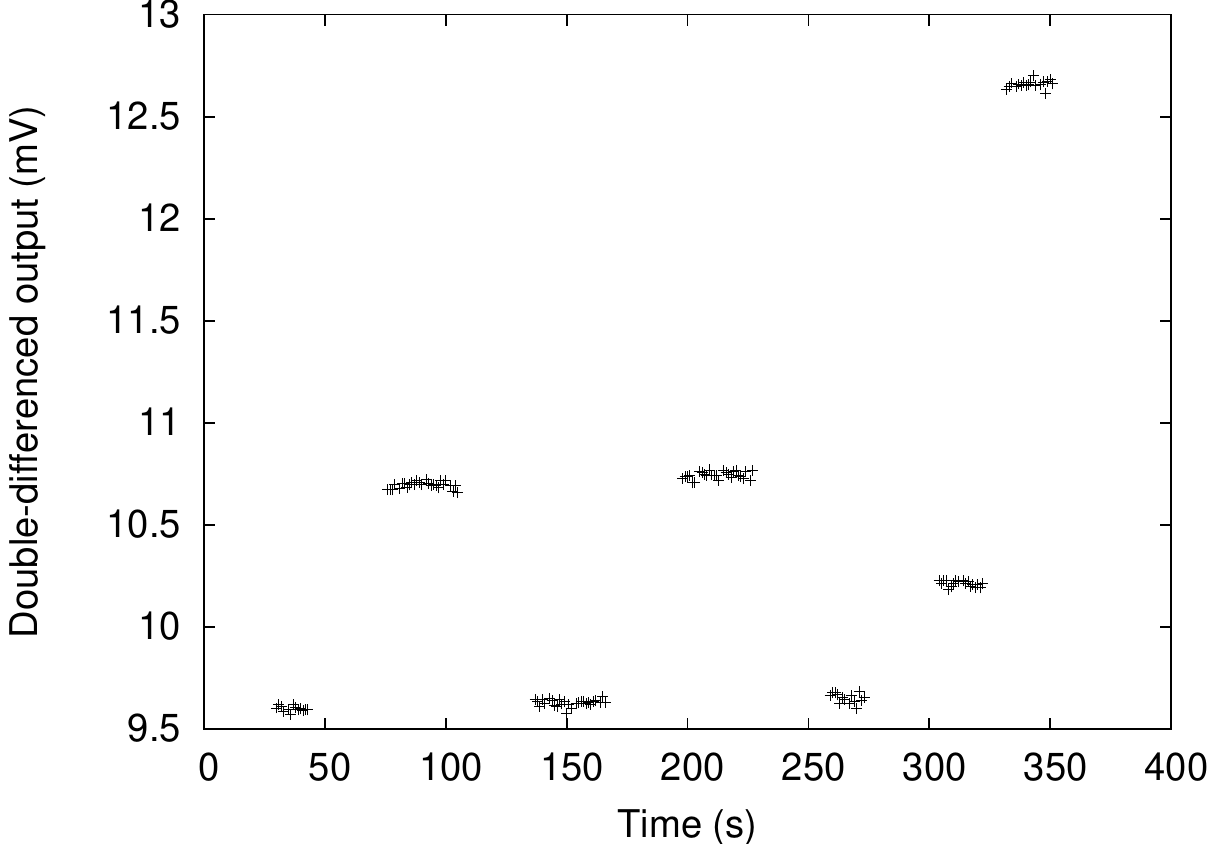}
   \includegraphics[scale=1.0]{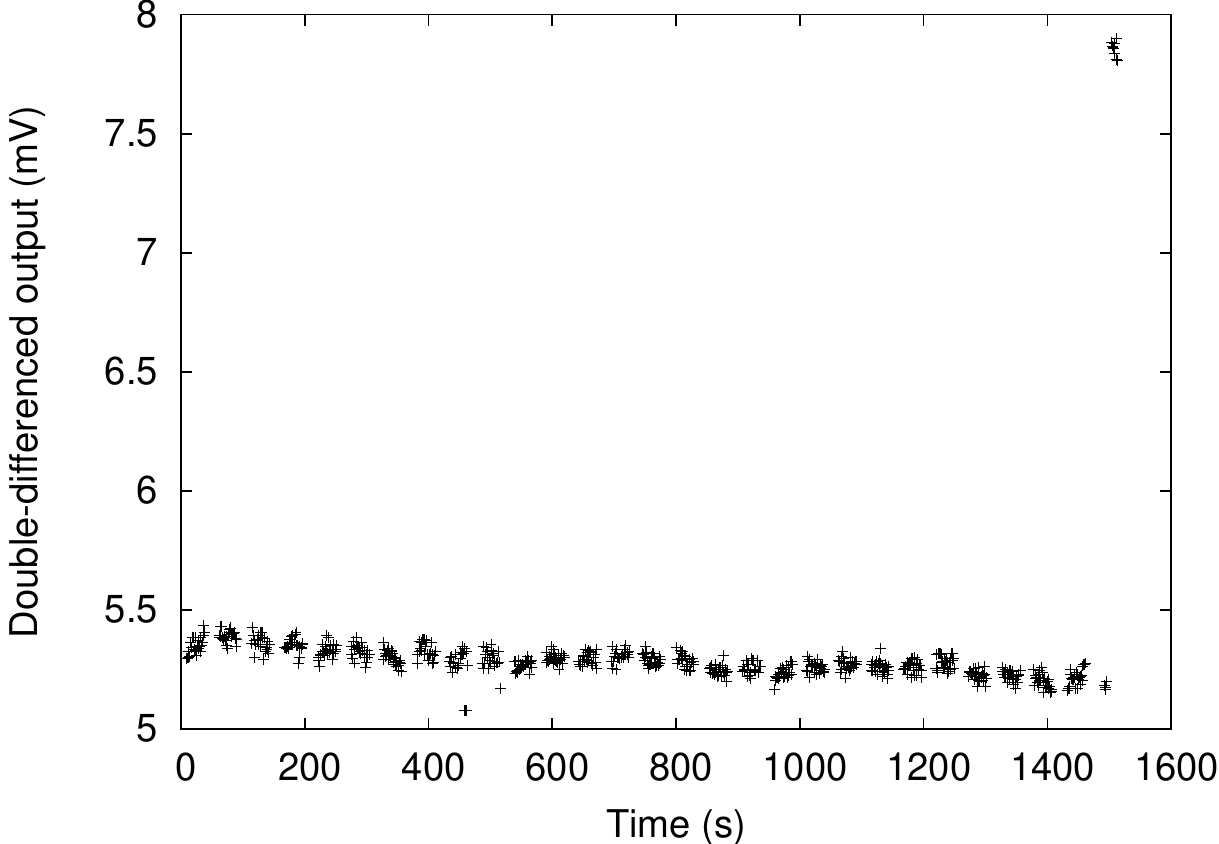}
\caption[Example on-off measurements of a strong and weak source]{Example on-off measurements, with a final noise diode calibration. Top: the strong source 1656+477 ($S_\mathrm{30~GHz} \approx 700$mJy), observed as part of the CRATES source sample. The observation pattern is (off)-(on)-(off)-(on)-(off)-(background)-(cal). Bottom: the SZ cluster 0016+16, observed by \citet{2006Lancaster}.}
\label{fig:example_onoff}
\end{center}
\end{fig}

For weaker ($<50$~mJy) sources, cross-scan measurements are not sensitive enough, so ``on-off'' measurements are performed. These consist of a series of measurements where one of the beams of the receiver is directed to the source position, with the other on blank sky, thus measuring the power difference between the source and the background value; after 30 seconds or so the beams are swapped over. This is repeated several times during a measurement cycle, or many times in the case of a measurement of a very weak source such as an SZ cluster. Subsequently the background level is measured, followed by the calibration diode.

As with the cross-scan observations, the data are automatically binned into 1 second samples. The azimuth and zenith angle of the telescope, as well as the time of the measurement, are also recorded. A marker is recorded for each sample, saying whether the measurement is of the background (1), the calibration diode (3), with the ``positive" beam on source (4) or with the ``negative" beam on source (6). Data are not currently recorded whilst the telescope is moving between positions.

The pattern of the on-off scans -- i.e. which order the different types of measurements are made in, and how many measurements are made within a single session -- has evolved over time and also depend on the sources being observed. The reduction methods have been kept general to cope with all formats.

For each measurement state $i$, the average (denoted $S_i$) and the scatter ($\sigma_i$) of the data are calculated. The average and scatter is also calculated for both the first ($S_{i,1}$ and $\sigma_{i,1}$) and second ($S_{i,2}$ and $\sigma_{i,2}$) halves of the measurement state, to enable the removal of any long-period drifts caused by atmospheric or gain fluctuations that are present in the measurement. To remove any spikes in the data, data points that are more than $3 \sigma$ away from the mean are rejected with the mean and scatter subsequently recalculated.

The source flux density is determined using a series of different methods, which are described below. The efficiency of these double-difference methods are compared in Section \ref{sec:dd_comparison} using model inputs, and are also contrasted using real data in Section \ref{section:vsasources}.

\subsubsection{Double differencing}
The standard method of ``double-differencing" the measurement to calculate the source flux density is to use
\begin{equation}
S_\mathrm{DD} = \frac{1}{N} \sum 0.5 \left( \pm S_{i-1} \mp S_i \right), \label{eq:dd}
\end{equation}
with an error calculated by 
\begin{equation}
\sigma_\mathrm{DD} = \frac{1}{\sqrt{N}} \sum \sqrt{ \left(0.5 \sigma_{i-1} \right)^2 + \left(0.5 \sigma_{i} \right)^2}.
\end{equation}
Here, the sum is over all measurements made with either the positive or negative beam on the source. If the pair of measurement states are positive then negative, this corresponds to $+S_{i-1} - S_{i}$; otherwise it corresponds to $-S_{i-1} + S_{i}$. $N$ is the total number of state pairs that are summed over.

\subsubsection{Symmetric double differencing}
The double-differencing method works well where there is a flat background, but becomes unreliable once linear drifts are present in the data. To cope with these drifts, symmetric differencing can be performed, subtracting the average of the two negative states either side of the positive state, 
\begin{equation}
S = \frac{1}{N} \sum 0.5 \left( S_i - 0.5 (S_{i-1} + S_{i+1}) \right) \label{eq:sdd}
\end{equation}
where the sum is over all states that have the positive beam on source, with measurements using the negative beam on either side. N is the number of states where this is the case. The error is calculated by:
\begin{equation}
\sigma = \frac{1}{\sqrt{N}} \sum \sqrt{ (0.5 \sigma_i)^2 + (0.25 \sigma_{i-1})^2 + (0.25 \sigma_{i+1})^2}
\end{equation}
Where there exist two adjacent measurements (e.g. neg-pos-neg-pos-neg), then this method will use the central measurement state twice; to avoid this, the negative states can be subdivided into two measurements, such that
\begin{equation}
S_\mathrm{SDD} = \frac{1}{N} \sum 0.5 \left( S_i - 0.5 (S_{i-1,2} + S_{i+1,1}) \right)
\end{equation}
where the sum is as before. The error is calculated by:
\begin{equation} \label{eq:doubledifference_error}
\sigma_\mathrm{SDD} = \frac{1}{\sqrt{N}} \sum \sqrt{ (0.5 \sigma_i)^2 + (0.25 \sigma_{i-1,2})^2 + (0.25 \sigma_{i+1,1})^2}.
\end{equation}
Essentially this is the same as the first symmetric double-difference method, but using the second half of the first negative measurement, and the first half of the second, either side of the positive measurement. In the case of a long series of measurements of a weak source, measurements $3 \sigma$ away from the mean are rejected, with the mean and scatter recalculated. This removes any large jumps in the data stream. This method was initially tested on SZ observations with OCRA-p by K. Lancaster and M. Birkinshaw (private communication) and found to perform better than the standard double-differencing method; it was subsequently adopted as the fiducial double-difference reduction method for OCRA-p data.

\subsubsection{Subtraction of a quadratic background}
The symmetric double-differencing technique is adequate for linear drifts, but will in turn break down if quadratic or higher order drifts are observed. These can be removed by subtracting a quadratic background $B_0 + B_1 t + B_2 t^2$ from the data, before calculating the double differenced flux density. The values that are used to find the quadratic background are the measured background (``bg'') values directly, and also the differences between positive and negative states. e.g. where an observation is bg1-pos1-neg1-pos2-neg2-bg2, the values used are (bg1, 0.5 (pos1+neg1), 0.5 (neg1+pos2), 0.5 (pos2 + neg2), bg2). The times of these values are the mid-points of the data used. All values that lie after the calibrator measurement are ignored.

The background is subtracted from the data, and the averages and scatter of the data within each measurement state are recomputed. The source flux density is then calculated from
\begin{equation}
S_\mathrm{QDD} = \frac{1}{N} \sum \pm S_i \label{eq:qdd},
\end{equation}
where $S_i$ is added if the measurement state is positive, or subtracted if the measurement state is negative. The error is calculated by
\begin{equation}
\sigma_\mathrm{QDD} = \frac{1}{\sqrt{N}} \sqrt{\sum \sigma_i^2}
\end{equation}

An alternative to this is to use the weighted sum:
\begin{equation}
S_\mathrm{QDDW} = \frac{\sum \pm S_i \sigma_i^{-2}}{\sum \sigma_i^{-2}} \label{eq:qddw}
\end{equation}
\begin{equation}
\sigma_\mathrm{QDDW} = \sigma_\mathrm{DD4}
\end{equation}

For longer measurements, higher order terms can be subtracted. Using the same technique as for a quadratic background, a polynomial background is subtracted, and the symmetric double-difference technique is used to measure the flux density. This is referred to here as $S_\mathrm{PSDD}$.

\subsubsection{Least Squares Fitting}
Finally, rather than dealing with each on-off state separately, the whole data file can be simultaneously fitted for using a least squares fitter. We use
\begin{equation}
f(t) = B_0 + B_1 t + B_2 t^2( ± S_\mathrm{LSDD}) (+ S_\mathrm{cal}) \label{eq:lsdd}
\end{equation}
where $B_i$ represents the quadratic background, and where the addition/subtraction of $S_\mathrm{dd}$ and $S_\mathrm{cal}$ depends on the measurement state. Assuming that these are the only terms present, then this should return the a good fit to the measurement; where additional terms are present then this will return a bad fit.

\subsection{Calibration} \label{sec:dred_cal}
\subsubsection{Gain-elevation correction}
\begin{fig}
\begin{center}
   \includegraphics[scale=1.0]{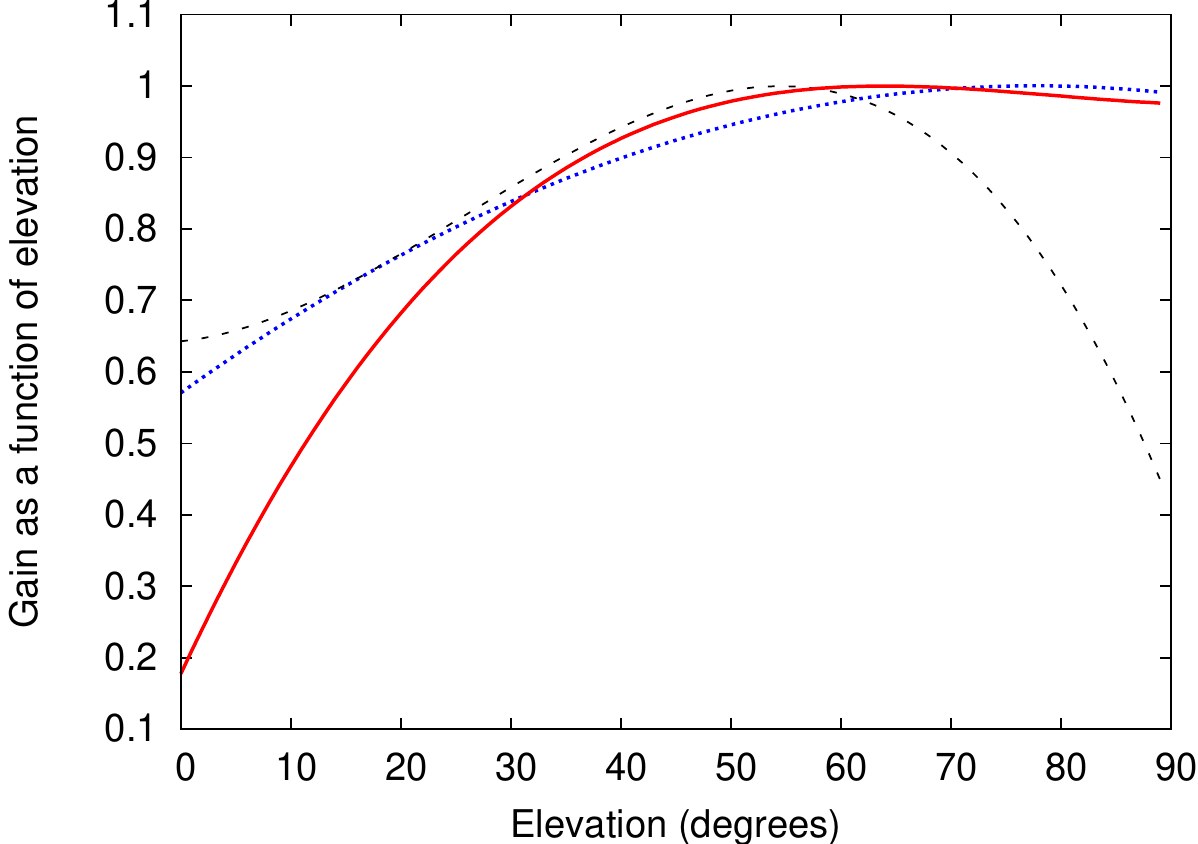}
\caption[Gain-elevation curves for OCRA-p]{Gain-elevation curves for OCRA-p. The black dashed line is the curve until summer 2006 (equation \ref{eq:gainelevation_orig}), the blue dotted line until August 2009 (equation \ref{eq:gainelevation}) and the red solid line from August 2009 (equation \ref{eq:gainelevation_new}). The latest curve shows the effect of relocating OCRA-p off-axis to accommodate OCRA-F.}
\label{fig:gainelevation}
\end{center}
\end{fig}

The gain of a radio telescope depends on the accuracy of the telescope surface, the angle at which the bowl of the telescope is inclined (the elevation of the telescope) and also the position of the receiver in the focal plane. All of these can be affected by gravitational deformations of the structure. A ``gain-elevation correction'' can be applied to correct for these effects. This correction has been measured at a series of elevation angles using observations of NGC~7027; a correction function can then be used to correct measurements of sources made at any elevation.

The gain-elevation correction values have changed several times during the time over which OCRA-p has been installed on the Toru\'n telescope. The initial correction function was (B. Pazderska, private communication):
\begin{equation} \label{eq:gainelevation_orig}
S_\mathrm{corrected} = S_\mathrm{measured} \times (2.7 + 0.25 h -0.0049 h^2 + 2.5\times 10^{-5} h^3 ) / 6.58.
\end{equation}
The surface accuracy was improved during the summer of 2006; following this the gain-elevation correction changed to (A. Kus, private communication):
\begin{equation} \label{eq:gainelevation}
S_\mathrm{corrected} = S_\mathrm{measured} / \left( 0.57 + 0.011\times h -7.12\times10^{-5} \times h^2 \right),
\end{equation}
where $h$ is the elevation of the telescope in degrees. In order to prepare for the installation of OCRA-F, OCRA-p was relocated to one side of the focal plane in mid August 2009. The gain-elevation function following this is (M. Gawronski, private communication):
\begin{equation} \label{eq:gainelevation_new}
S_\mathrm{corrected} = S_\mathrm{measured} \times (0.18 + 0.033 h - 0.00044 h^2 + 1.8\times 10^{-6}h^3).
\end{equation}
These three gain-elevation corrections are plotted in Figure \ref{fig:gainelevation}.

Additionally, following from the move of OCRA-p off-axis the ``negative'' beam of the instrument systematically under-measured the flux density of the source during cross-scan measurements. A correction function for this is used for all cross-scan measurements since then:
\begin{equation}
V_\mathrm{negative,corrected} = V_\mathrm{negative} \times \left(1.2 -0.0006 h -2.64 \times 10^{-6} h^ 2 -1.35\times 10^{-7} h^3 \right).
\end{equation}

\subsubsection{Atmospheric absorption correction}
A correction is also applied for the absorption of the signal in the atmosphere. This is achieved via the measurements of the system temperature at zenith and at 60 degrees from zenith. As the latter measurement contains twice the air mass of the first, the difference of these two measurements provides the temperature contribution of the partially transparent atmosphere, $T_\mathrm{atmosphere}$. The optical depth $\tau$ of the atmosphere can then be calculated from this by $\tau = T_\mathrm{atmosphere} / T_\mathrm{phys}$, where $T_\mathrm{phys}$ is the physical temperature of the atmosphere (typically 260-270~K). A multiplicative correction factor for the absorption of the signal from the astronomical source can then be calculated by
\begin{equation} \label{eq:atmosphere_correction}
S_\mathrm{corrected} = \left( 1 + \frac{\tau}{\cos (\theta)} \right) S_\mathrm{measured}
\end{equation}
where $\theta$ is the zenith angle of the measurement. The correction is typically $\sim 3-10$ per cent of the flux density of the source.

\subsubsection{Flux density calibration}
The reduction of data from the two observational methods starts with a value in volts, representative of the detected RF power coming out of the receiver. The relationship between volts and the actual source flux density depends on the properties and environment of the receiver and telescope, and can be measured by observing a bright astrophysical source with a known flux density. The standard calibrator source used for OCRA is NGC~7027, a planetary nebula with a flux density of $5.47 \pm 0.04$~Jy at 30~GHz in 2008 \citep[][also see Section \ref{section:vsaobservations}]{2008Hafez}. Although NGC~7027 is slowly decaying in flux density over time, a constant value is assumed for each data reduction. The ratio of the known flux density to volts for this source can be used to convert measured voltages into flux densities for other sources.

Secondary calibrators are then used to reduce the time between the source observations and the primary calibration; not only would it be prohibitively expensive in telescope time to re-observe NGC 7027 every hour, but there are times when this source is not at a suitable elevation to be observed. One of a network of secondary calibrator sources is hence observed; these sources provide both telescope pointing calibration measurements and can also be used in flux density calibration.

To improve the calibration further, a calibrator diode is observed during each measurement. The calibration diode is assumed to remain stable between calibrator measurements, hence any variations in the measured voltage is assumed to originate from gain fluctuations within the receiver.  The effective flux density of the calibration diode is determined by measuring both it and a background for 10 seconds each, then working out the average of both and subtracting one from the other. The diode was not thought to be sufficiently stable during the CJF observations \citep{2007Lowe}, so the secondary calibrator sources were used for calibration. However, for subsequent observations it was thought to be more stable, so it was used as the default secondary flux density calibrator, although we do have to recalibrate it regularly. As such, the secondary calibrator sources are now only used for pointing corrections.

The measured voltages of NGC~7027 are smoothed over the period of a day by averaging over measurements of NGC~7027 taken 12 hours each side of the source measurement that is being calibrated. If there are no measurements of NGC~7027 within the smoothing period either side of the measurement, then the nearest measurement is used. The ratio of the NGC~7027 flux density to the gain-elevation and atmosphere-corrected voltage is then used to calibrate the calibration diode. The individual source measurements are then calibrated using the same ratio but calculated from either the calibration diode or the secondary calibrator sources. Finally, the individual measurements are corrected for gain-elevation and atmosphere.

\subsection{Calculation of source flux densities} \label{sec:dred_result}

\subsubsection{Flagging bad data}
Flagging of ``bad'' data -- cases where the measurement has known problems, for example contamination by the weather or by incorrect pointing -- can either be done manually or automatically. Although it is not too prohibitive in time to check through all of the OCRA-p measurements (circa 10~000), automatic flagging applies a common standard across all of the measurements, rather than relying on human judgement. However, manual flagging remains useful for segments of the data with specific issues (for example, if the calibration diode was highly variable).

For all measurement types, data are flagged when they were taken at an elevation lower than $30 \degree$. Below that elevation, the telescope performance is not well characterized at 30~GHz, and a large atmospheric correction is required.

Bad cross-scans can be identified by comparing the amplitudes and widths of the two gaussian peaks; these should ideally agree with each other, but are frequently not the same e.g. where the source has only passed through one beam due to poorly determined pointing offsets. Measurements where there are large differences in these can be automatically flagged; this is currently applied where the ratios differ by greater than 20 per cent. The amplitude of the azimuthal cross-scans should always be equal to, or higher than, the elevation scan; if it is not, then that indicates that the source was ``missed''.

It is more tricky to find erroneous on-off measurements. At present measurements are automatically flagged if the error on the measurement is above a threshold -- typically 7~mJy -- with additional manual flagging carried out as necessary.

\subsubsection{Source position checks}
In order to check that the correct source position was observed (and that the correct source name was associated with the measurement), the right ascension (RA) and declination (Dec) of the measurement are calculated from the recorded azimuth and elevation of the telescope during the observation. As this will yield the RA and Dec at the observation epoch, these are then precessed to J2000 coordinates using Libnova\footnote{Available from \url{http://libnova.sourceforge.net/}}.

After precession has been applied, there remain some differences between the actual source position and the position calculated from the file. There are three reasons for these differences. The first reason is that the wrong source was observed due to an observer error; these are either resolved by comparison of the list of measurements with the observation log, or removed. The second reason is that the azimuth, elevation and time used to calculate the position are taken from the start of the observations, which for cross-scans will be offset from the actual source position. However due to the large separation of sources strong enough to carry out cross-scan observations, this discrepancy is rarely important in the identification of incorrectly named sources. The third reason is that the recorded azimuth and elevation will not take into account the pointing corrections applied within the control system, which could contain residual errors.

\subsubsection{Final flux density values}
The final flux densities are calculated using the weighted mean and associated error using all of the measurements for a given source, with the weighing factor depending on the error on each measurement calculated as described above. This has an advantage over the simple average of the measurements in that any measurements with large estimated errors (e.g. those affected by atmospheric emission at a higher level than usual) are downweighted, with preference given to measurements with small errors.

\section{Simulated OCRA-p observations with UMBRELLA} \label{sec:sim_ocrap}
The UMBRELLA software package was modified to mimic the observational methods used by OCRA-p -- namely, cross-scan and on-off measurements -- and output the results in the same data formats as are used for real measurements (clearly marked to avoid confusion between the two). This means that the data reduction software can used in exactly the same way for both the simulated and real data. UMBRELLA was also modified to automatically write a batch script to automate the data reduction process. A simple shell script can then be used to run the two programs and analyse the results. In addition, UMBRELLA is now capable of simulating observations of Virtual Sky maps, although this option is not used within this section.

UMBRELLA has been configured to simulate OCRA-p mounted on a 32-m telescope, following the configuration used by \citet{2006Lowe}. The hybrids are modeled using an imperfect power splitting as measured in \citet{2006Lowe}, of 0.497:0.503. The difference between the two phase switch states is set to 170$\degree$, again matching that measured for OCRA-p. We use a nominal gain for all of the amplifiers of 27dB; the precise value for this gain is unimportant so long as the noise levels from the amplifiers are defined appropriately.

``Noise'' can be added into the simulated timestream in three different ways: Gaussian thermal noise from the amplifiers; $1/f$ gain fluctuations within the amplifiers; and atmospheric $1/f$-like absorption and emission fluctuations. Gaussian noise can be generated using a random number generator, and is relatively quick to calculate compared with the other two methods. $1/f$ gain fluctuations depend on the history of the time series (the method used to create a timestream of these fluctuations is described in Section 4.3.2 of \citealp{2006Lowe}), hence this takes longer to simulate. The most time consuming noise source to generate is the atmosphere; UMBRELLA uses a fractal atmospheric model as described in Section 5.2 of \citet{2006Lowe}, which is convolved in real space with the near-field beam of the telescope.

We use the settings from \citet{2006Lowe} for the Gaussian noise and $1/f$ gain fluctuations, which provide a knee frequency of $\sim 20$~Hz from a single output. For the atmospheric model, we use an atmospheric layer at a height of 500~m moving at 10~m/s, with an optical depth $\tau = 0.03$, corresponding to the best observing conditions experienced at the sea-level Toru\'n. The temperature of the atmosphere is set to 260~K, with variations on the scale of the aperture size initially chosen to be 0.5~K and a turbulent spectrum of 2/3 (see Section 5.1.2 of \citealp{2006Lowe}; \citealp{1941Kolmogorov,1941Kolmogorova}).

The relative levels of these three noise sources are important, as each will add different noise/fluctuation characteristics to the measurement. However, it is difficult to disentangle them. We use reasonable estimates of the noise levels in comparison to measurements taken with OCRA-p, but we caution that at this stage these should be taken as indicative. Further improvements to this depend on the ability to distinguish receiver $1/f$ gain fluctuations from atmospheric fluctuations. This should become easier using multiple receiver chains as in OCRA-F; the common mode in the signal from the different chains will be diagnostic of the atmosphere.

\subsection{Simulated noise power spectra} \label{sec:sim_noisepowerspectra}
\begin{fig}
\begin{center}
   \includegraphics[scale=0.5]{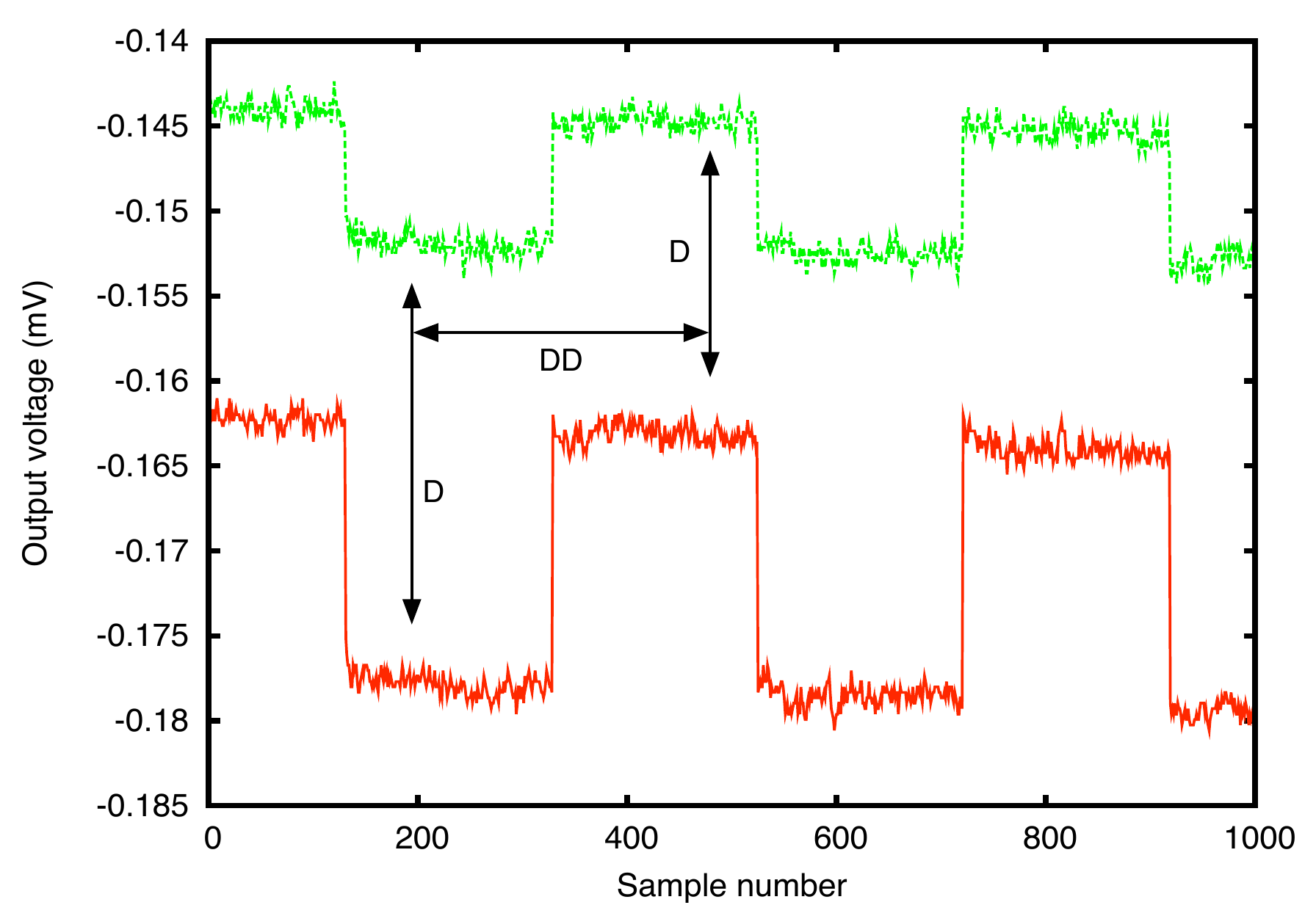}
\caption[An example of the double-differencing carried out on the switched output from OCRA]{An example of the double-differencing carried out on the switched output from OCRA. The ``difference'' D is taken between the two output signals, then the ``double-difference'' DD is taken between adjacent switch states.}
\label{fig:dd_example}
\end{center}
\end{fig}

The power spectra of the three different simulated components from Gaussian noise only and both Gaussian noise and $1/f$ gain fluctuations are shown in Figure \ref{fig:noise_powerspectrum_1}. The power spectrum of the noise from one of the two outputs is shown (the other output is analogous), as well as the ``differenced'' power spectrum, which is calculated by subtracting one output from the other. This greatly reduces the $1/f$ gain fluctuations from the front end amplifiers. The ``double-differenced'' power spectrum is also shown, which is calculated by subtracting one phase switch state from the other. This reduces the back end amplifier $1/f$ gain fluctuations. This differencing is shown schematically in Figure \ref{fig:dd_example}. Figure \ref{fig:noise_powerspectrum_2} shows the same power spectra for the combined Gaussian noise, $1/f$ gain fluctuations and atmospheric ``$1/f$'' fluctuations, and also data taken with OCRA-p on the Toru\'n telescope. The simulated timestreams are 5000 seconds in duration; the OCRA-p data is 600 seconds long.

\begin{fig}
\begin{center}
   \includegraphics[scale=0.5]{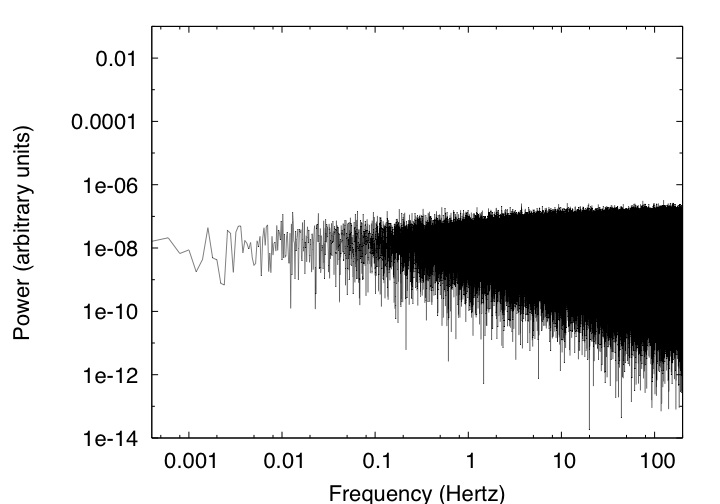}
   \includegraphics[scale=0.5]{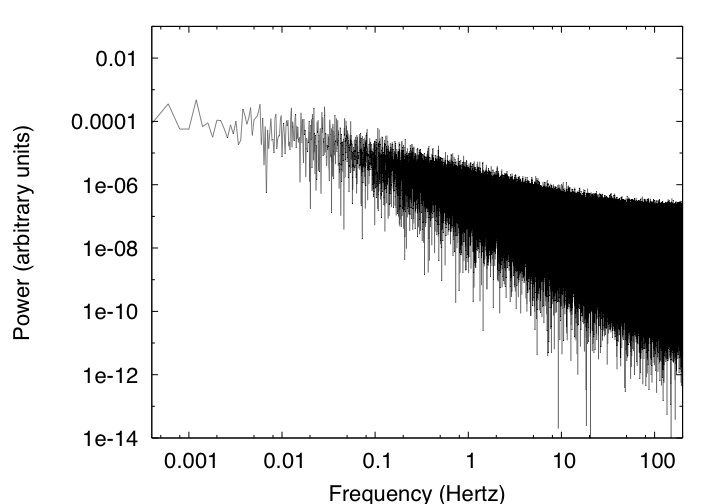}\\
   \includegraphics[scale=0.5]{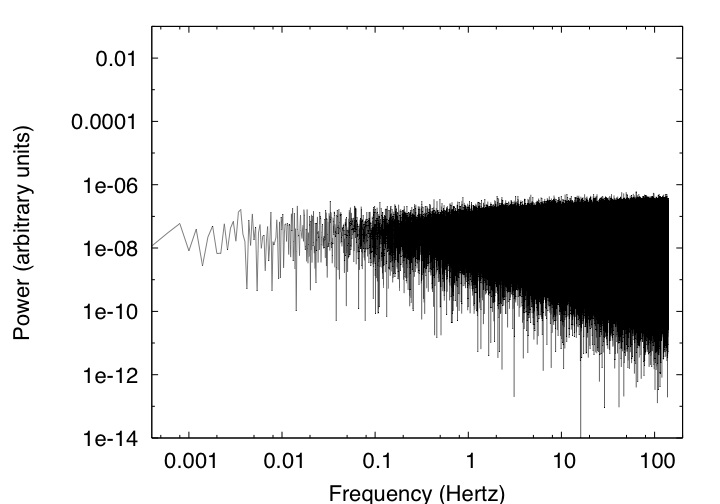}
   \includegraphics[scale=0.5]{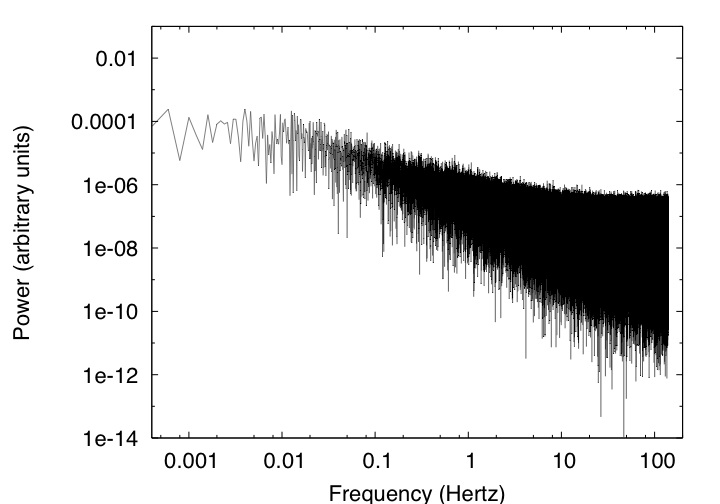}\\
   \includegraphics[scale=0.5]{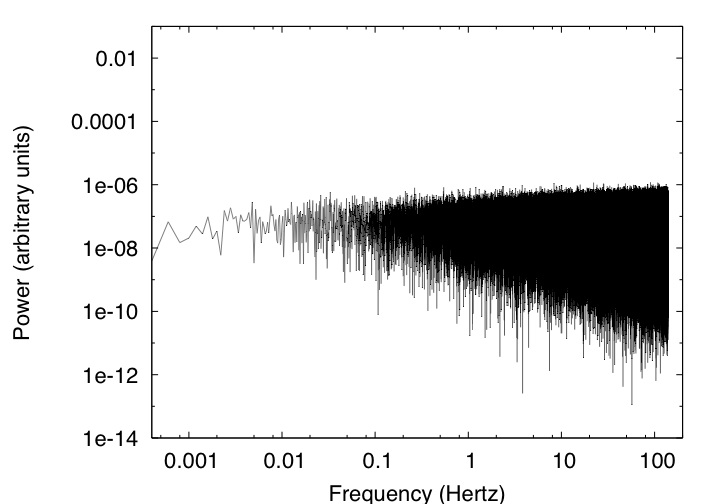}
   \includegraphics[scale=0.5]{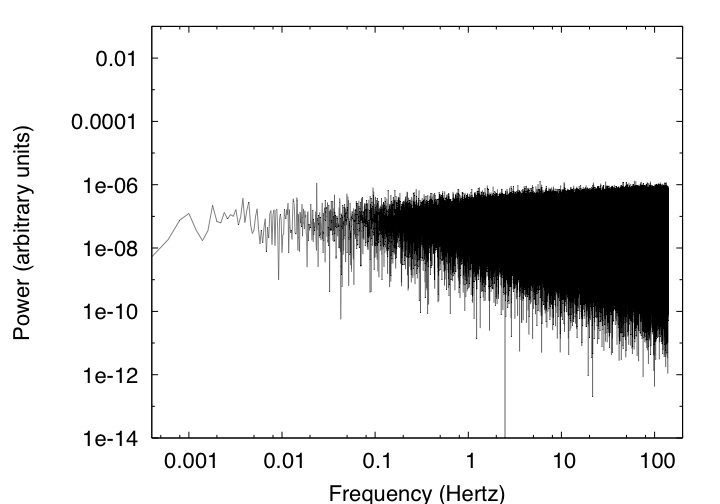}\\
\caption[Simulated receiver noise power spectra for Gaussian and $1/f$ noise]{Simulated receiver noise power spectra for Gaussian (left) and combined Gaussian and $1/f$ noise (right). Top row: raw output from one of the channels. Middle row: differenced output. Bottom row: double differenced output.}
\label{fig:noise_powerspectrum_1}
\end{center}
\end{fig}

\begin{fig}
\begin{center}
   \includegraphics[scale=0.5]{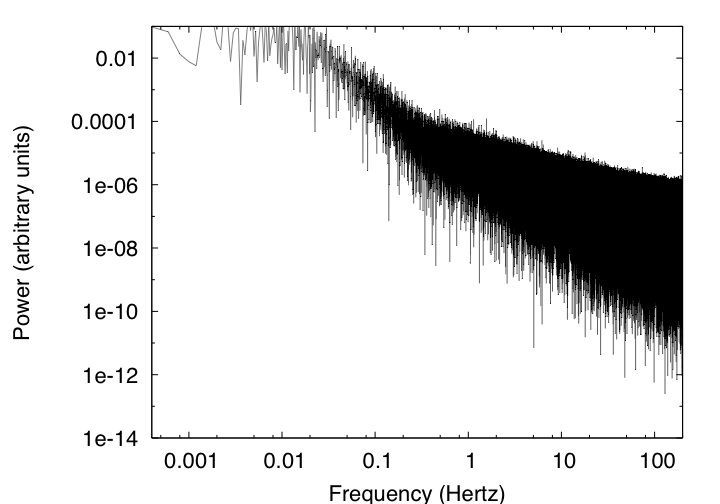}
   \includegraphics[scale=0.5]{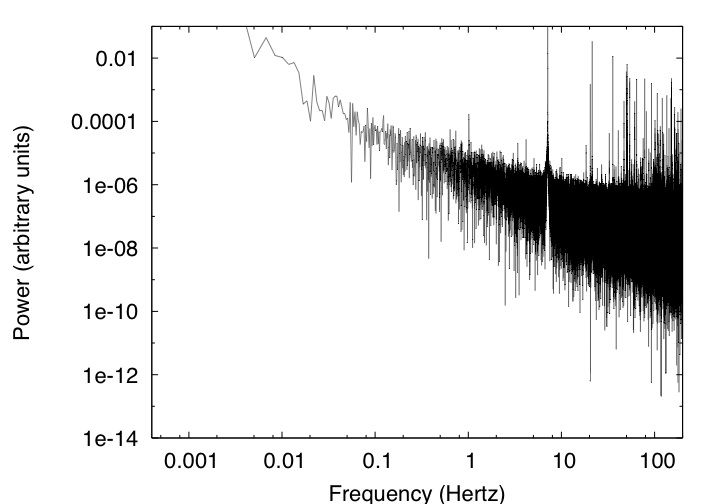}\\
   \includegraphics[scale=0.5]{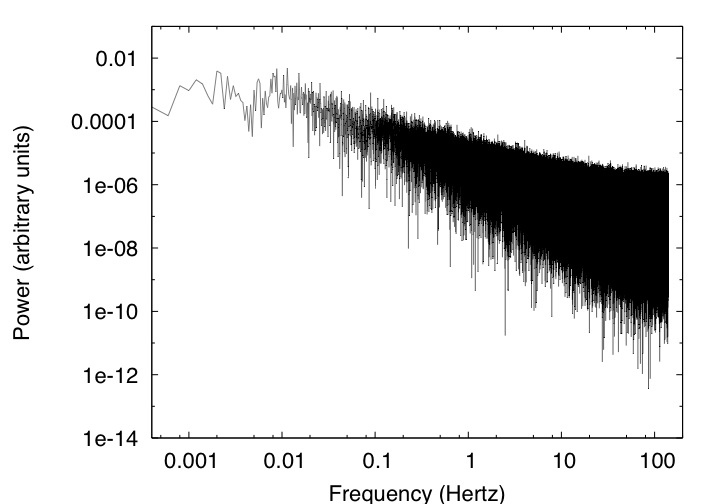}
   \includegraphics[scale=0.5]{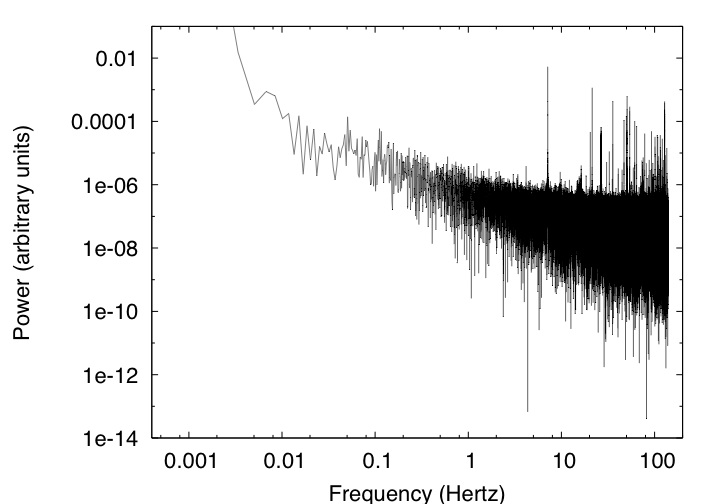}\\
   \includegraphics[scale=0.5]{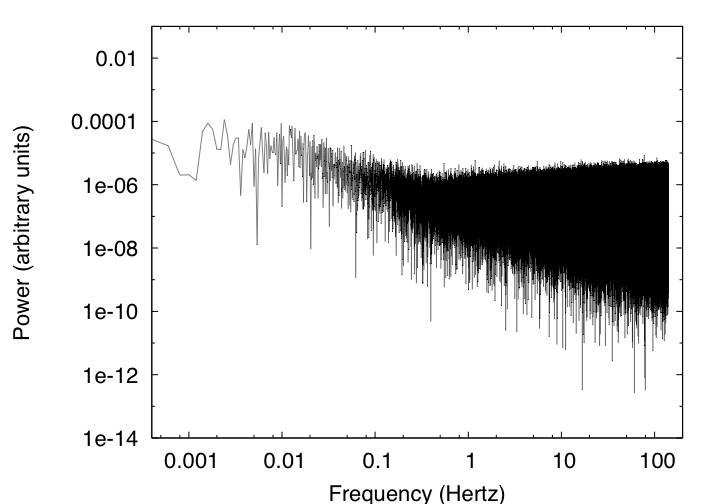}
   \includegraphics[scale=0.5]{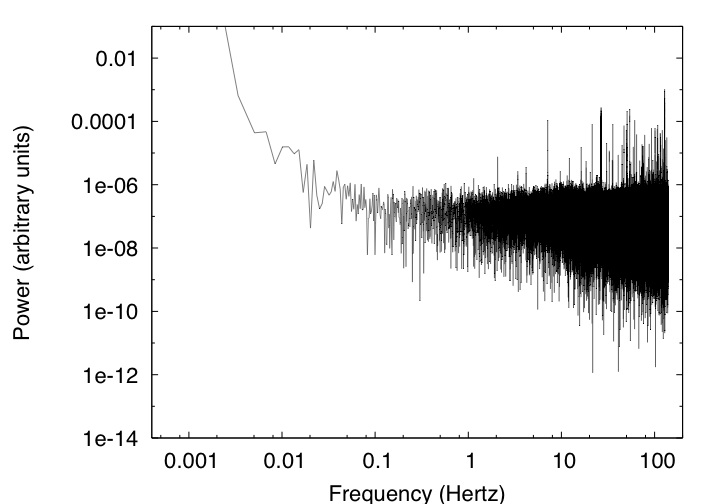}\\
\caption[Simulated receiver noise power spectra for combined Gaussian, $1/f$ and atmospheric noise, compared with measurements from OCRA-p]{Simulated receiver noise power spectra for combined Gaussian, $1/f$ and atmospheric noise (left), and measurements from OCRA-p (right). Top row: raw output from one of the channels. Middle row: differenced output. Bottom row: double differenced output.}
\label{fig:noise_powerspectrum_2}
\end{center}
\end{fig}

\begin{fig}
\begin{center}
   \includegraphics[scale=0.5]{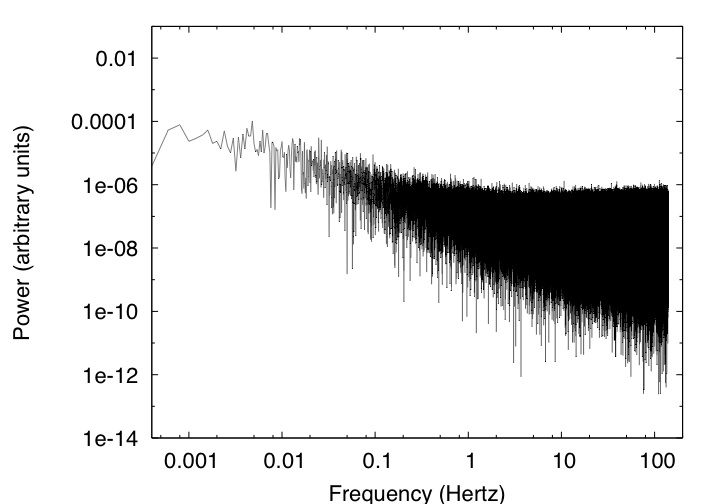}
\caption[Simulated double-differenced receiver noise power spectra adding $1/f$ noise into the double-differenced output]{Simulated double-differenced receiver noise power spectra after adding additional $1/f$ noise into the double-differenced output. Note that the knee frequency here is higher than the value seen from OCRA-p (cf. Figure \ref{fig:noise_powerspectrum_2}).}
\label{fig:noise_powerspectrum_daqnoise}
\end{center}
\end{fig}

We note that the simulations currently subtract the $1/f$ gain fluctuations from the amplifiers too perfectly by double-differencing, which leads to an overestimate of the capabilities of the receiver. This cancellation does not depend on the balance of the gain of the two amplifiers, nor the phases of the two phase switch states. Further exploration of the source of the $1/f$ gain fluctuations in the double-differenced signal from the amplifiers is needed; it is likely that the cause is not presently included in the simulations. Possibilities include the effects of imbalanced complex gain across the bandwidth of the instrument, or alternatively the isolation between the different modules within the receiver chain. For the purpose of these initial simulations, we therefore add $1/f$ noise directly to the double-differenced output, and do not simulate $1/f$ gain fluctuations within the amplifiers purely for computational efficiency at this stage of the development of the simulation package. The resulting double-differenced power spectrum is shown in Figure \ref{fig:noise_powerspectrum_daqnoise}; we use this to represent the combined residuals from the receiver and atmosphere where the index of the $1/f^\alpha$ power is $\alpha=1$. The $1/f$ knee frequency in these simulations is higher than observed with OCRA-p, and hence the results represent conservative estimates of the performance of the receiver.

The noise power spectra from the atmosphere simulations show a transition from Gaussian-dominated to atmosphere-dominated fluctuations in the noise power spectrum at $\sim 0.3$~Hz. The wind speed moves the position of the transition in frequency; lower wind speeds reduce the frequency (as the atmosphere layer is changing more slowly), and faster wind speeds increase the frequency. This agrees well with the transition being determined by the crossing time of the atmosphere across the beam of the telescope, where the fluctuations in the double-differenced signal are caused by the atmosphere in the parts of the beam that do not overlap. The optical depth of the atmosphere, $\tau$, simply scales the entire power spectrum up or down. The level of the temperature fluctuations softens the transition between Gaussian and atmosphere-dominated noise regimes. The height of the atmosphere layer increases the power within the from the atmosphere-dominated section as less of the atmospheric signal is common between the two beams. Finally, the power spectrum of the atmospheric fluctuations changes the ratio of the power on small scales compared to large scales, but does not appreciably change the knee frequency. As these parameters need additional refinement and comparison with measurements to reflect more accurately the real atmosphere, we do not use the atmospheric layer simulations further in this chapter.

\subsection{Simulated cross-scan measurements} \label{sec:sim_ps}
We have carried out 100 simulated cross-scan measurements of point sources for a range of flux densities (1500, 500, 250, 150, 50 and 20~mJy). We use the receiver imperfections as described above, and add Gaussian noise only and separately also add $1/f$ noise to the double-difference signal. In order to calibrate the measurements, we also perform a cross-scan measurement of a fake NGC~7027 with a flux density of 5470~mJy. All of the cross-scan measurements are 100~s in duration in elevation and 200~s in duration in azimuth, which is comparable to actual observations. The results from the simulations for the various input flux densities are given in Table \ref{tab:qscan_comparison}.

\begin{tab}{tb}
\begin{tabular}{r|rrr|rrr}
& \multicolumn{3}{c}{{\bf Gaussian noise}} & \multicolumn{3}{c}{{\bf $1/f$ noise}}\\
\hline
{\bf Input flux density} & {\bf Mean $\pm$ SD} & {\bf Min} & {\bf Max} & {\bf Mean $\pm$ SD} & {\bf Min} & {\bf Max}\\
\hline
1500 & 1500 $\pm$ 9 & 1479 & 1522 & 1478 $\pm$ 71 & 1341 & 1664\\
500 &  501 $\pm$ 4 & 491 & 509 & 508 $\pm$ 34 & 431 & 616\\
250 & 250 $\pm$ 3 & 243 & 257 & 272 $\pm$ 28 & 211 & 358\\
150 & 150$\pm$ 3 & 144 & 158 & 150 $\pm$ 22 & 94 & 211\\
50 & 51 $\pm$ 3 & 45 & 57 & 126 $\pm$ 156 & 59 & 1335\\
\end{tabular}
\caption[Comparison of input to output point source flux densities from simulated cross-scan observations]{Comparison of input to output point source flux densities from simulated cross-scan observations. Flux densities are given in milliJansky.}
\label{tab:qscan_comparison}
\end{tab}

When only Gaussian noise is present, successful observations of point sources can be made down to a flux density of $\sim 50$~mJy. Below this, the 1~s scatter becomes equivalent in size to the signal from the source, and the code used to fit the measurement is unable to find the peaks caused by the source. This causes erroneous fits to the data, resulting in a large fraction of the measurements being automatically flagged. This corresponds well to experience from observations with OCRA-p, where cross-scan measurements are only carried out on sources stronger than 50~mJy.

The introduction of $1/f$ noise increases the scatter in the measurements, and also increases the number of outliers at all flux densities. The 150~mJy source can be detected without problems; although it has some outlying measurements $\sim~50$~mJy away, these can be detected as bad measurements easily by eye, and hence manually flagged. The source with a flux density of 50~mJy cannot be identified within the timestream by the fitting program, resulting in erroneous fits to the data. 

\begin{fig}
\begin{center}
   \includegraphics[scale=0.5]{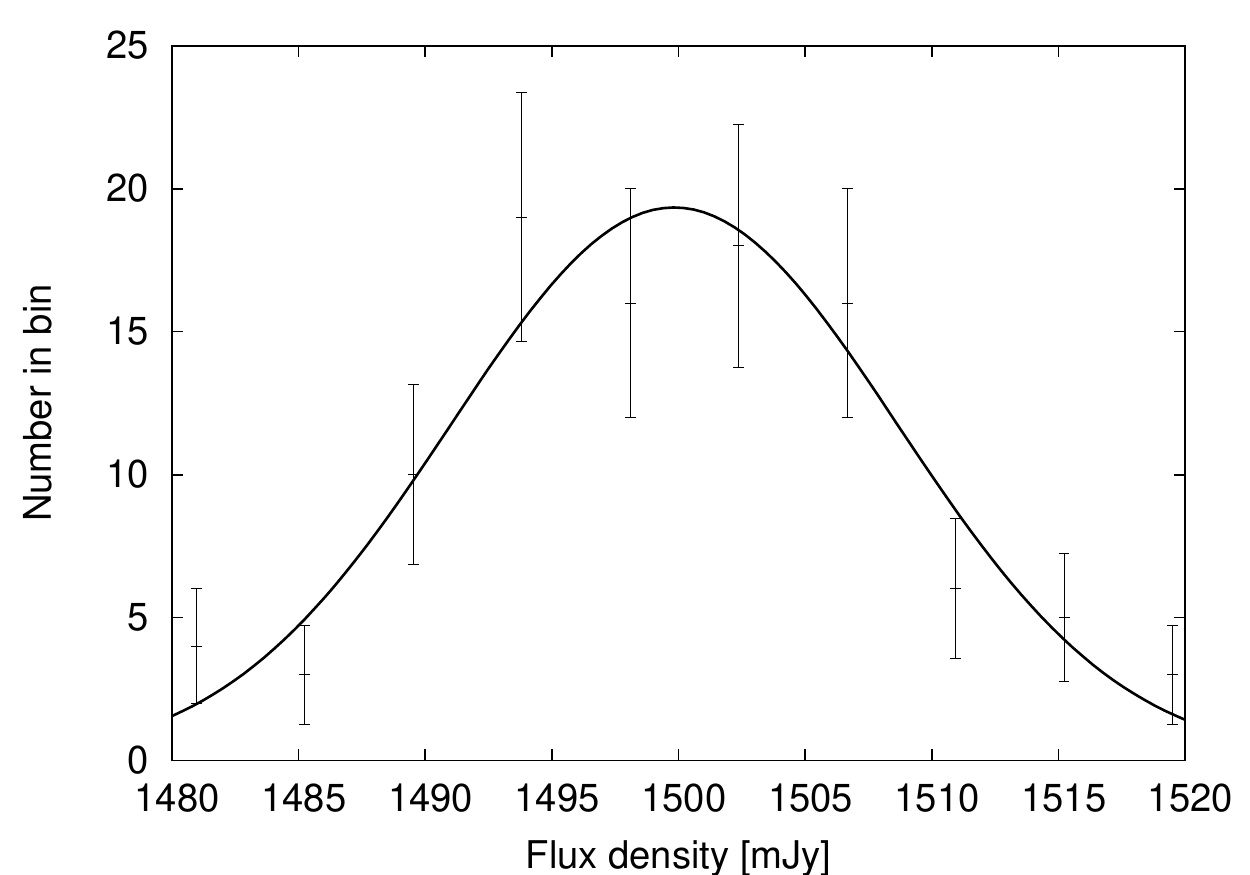}
   \includegraphics[scale=0.5]{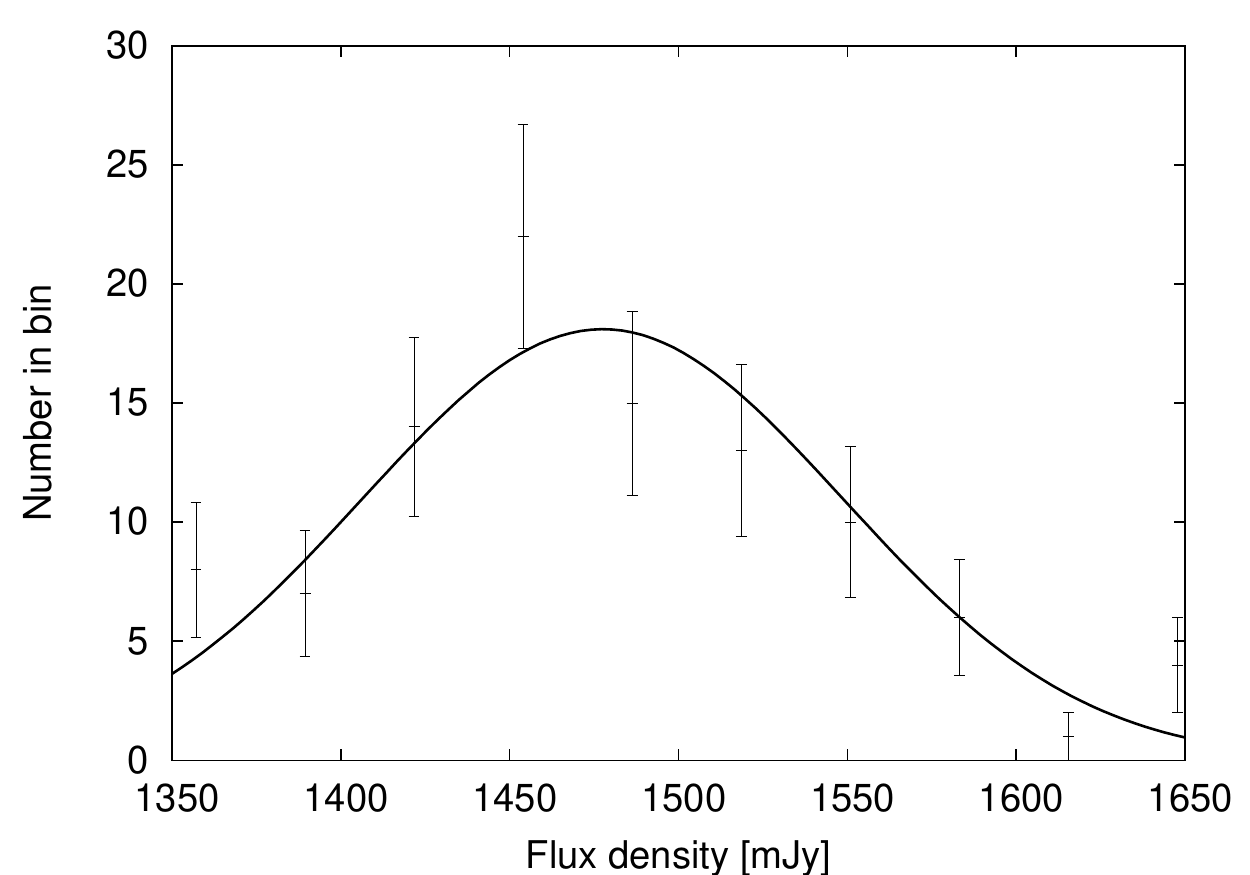}\\
   \includegraphics[scale=0.5]{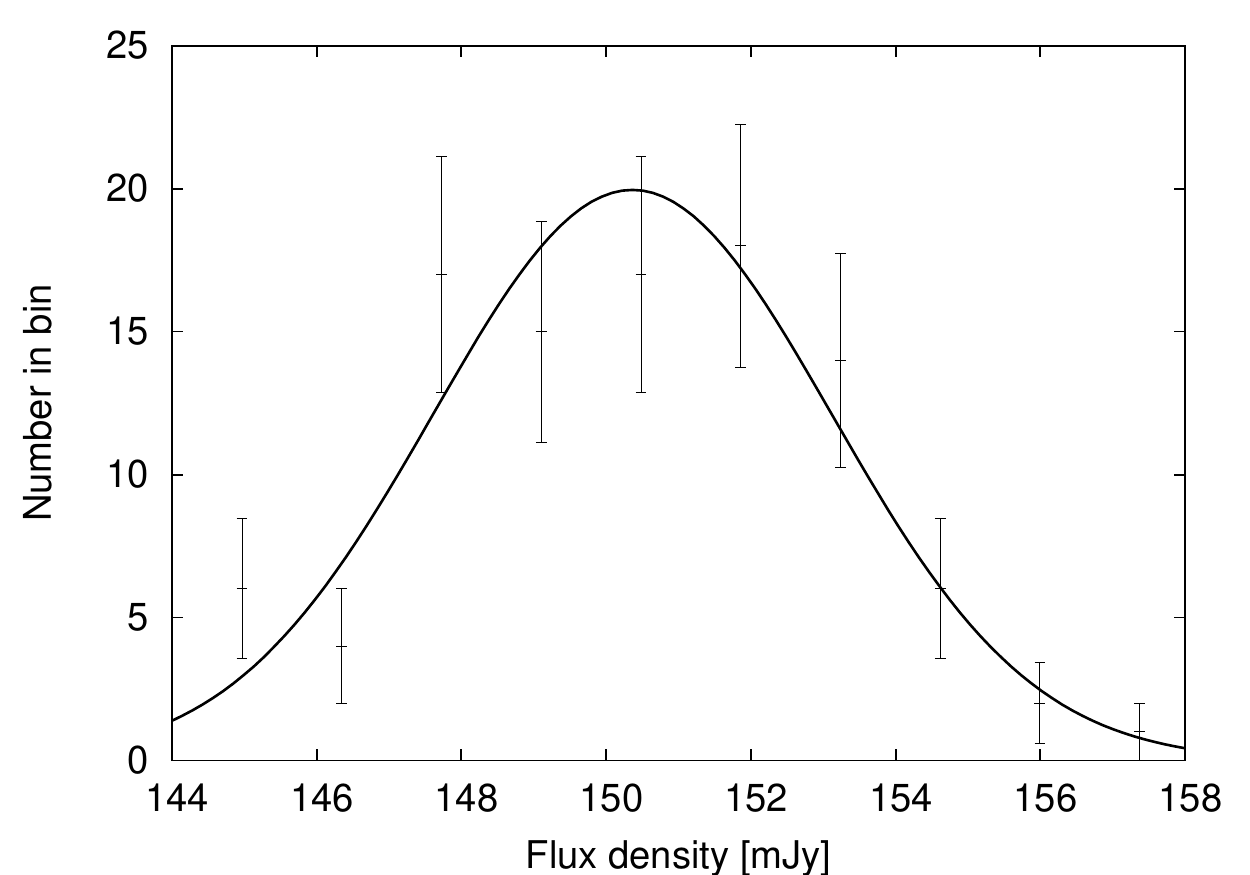}
   \includegraphics[scale=0.5]{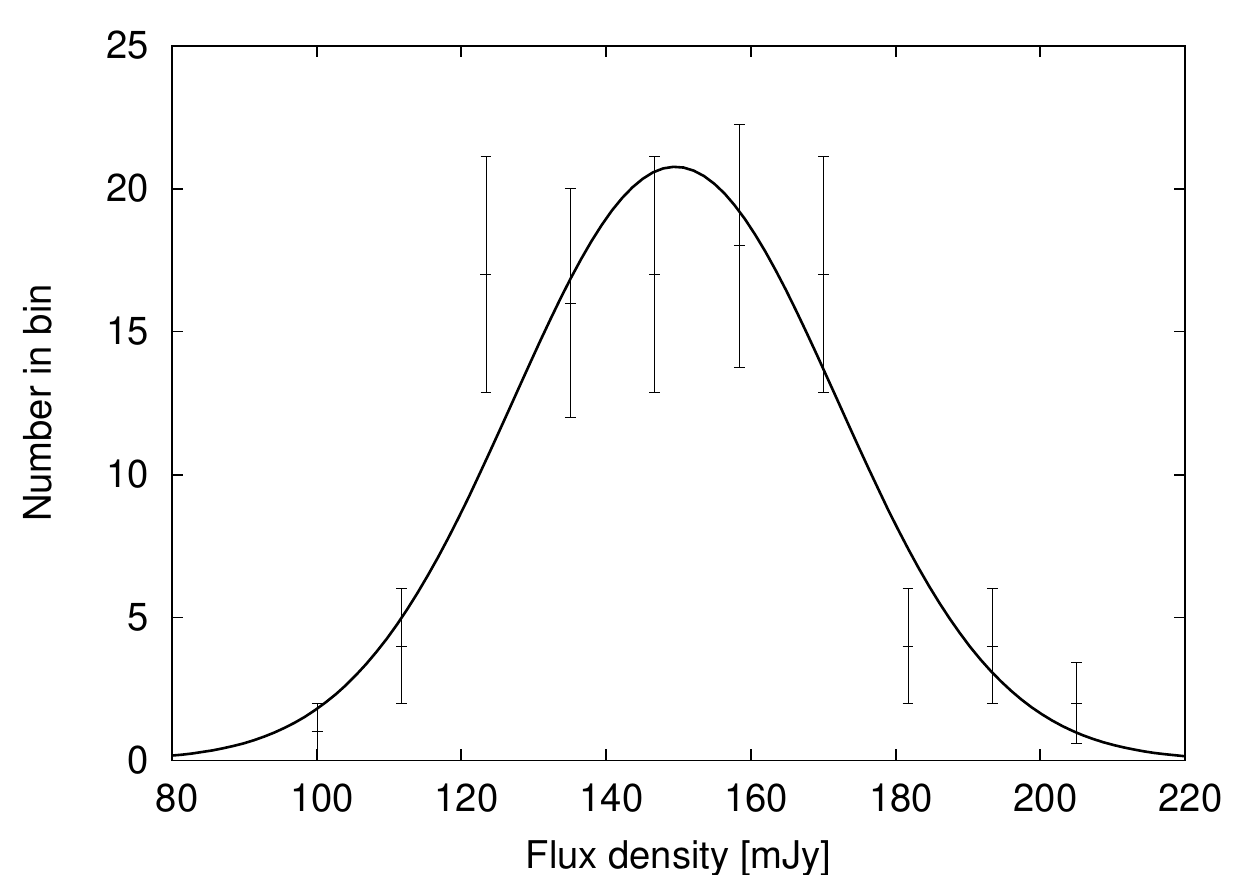}
\caption[Histogram of simulated cross-scan observations of 1500 and 150~mJy point sources in the presence of various types of noise]{Histogram of simulated cross-scan observations of 1500 (top) and 150~mJy (bottom) point source in the presence of only Gaussian noise (left) and Gaussian plus $1/f$ noise (right). In the case of Gaussian noise, the mean measurements are 1499$\pm$9~mJy and 150$\pm$3~mJy, where the error is from the scatter on the measurements. For the combined Gaussian and $1/f$ noise the measurements are 1477$\pm$71~mJy and 150$\pm$22~mJy. The histograms for the strong source are well described by Gaussian fits, but not as well for the weaker source (solid lines).}
\label{fig:qscan_fake0}
\end{center}
\end{fig}

Figure \ref{fig:qscan_fake0} shows histograms of the measured flux densities of 1500 and 150~mJy point sources.. At high flux densities, the statistics returned by these simulations are consistent with Gaussian distributions, however they deviate at lower flux densities where there is a platau in the statistics around the central value.

\begin{fig}
\begin{center}
   \includegraphics[scale=0.5]{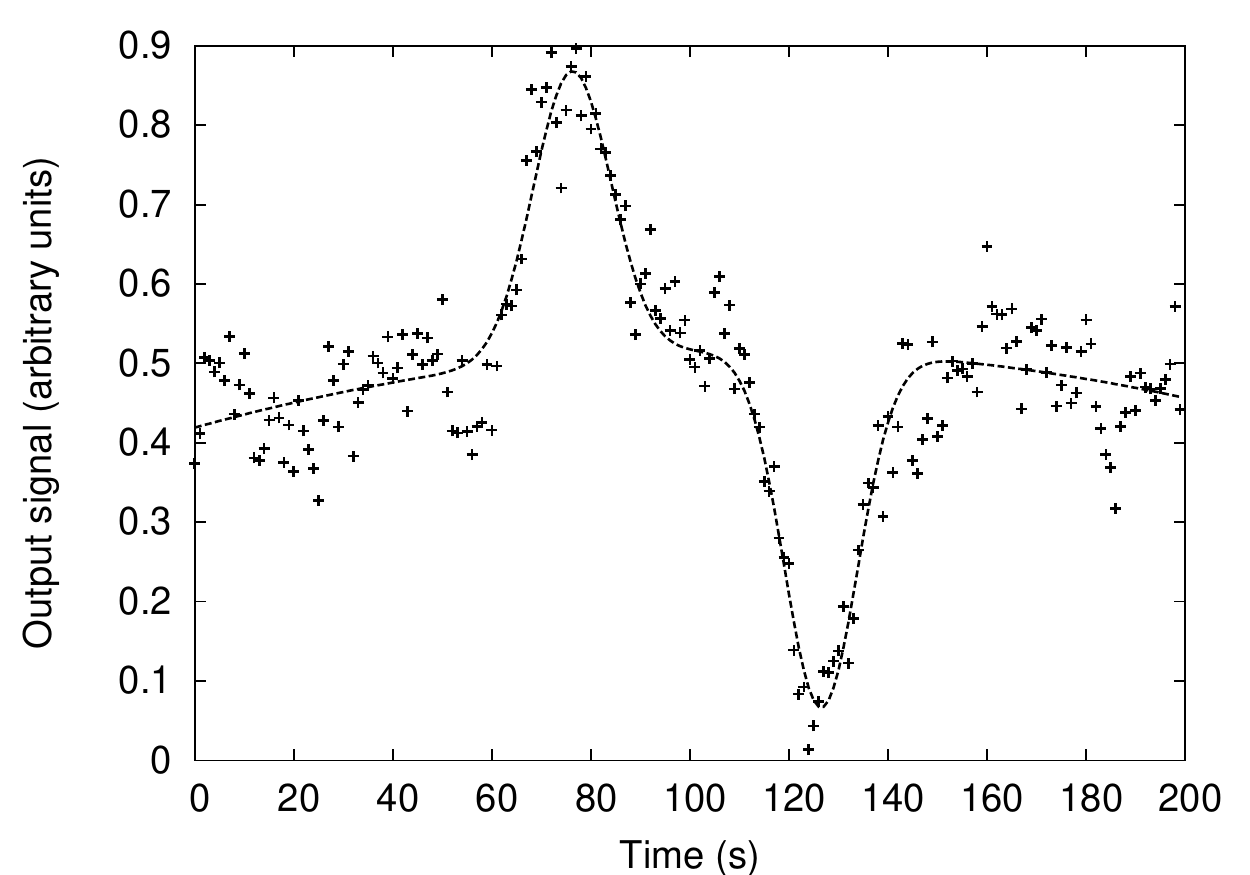}
   \includegraphics[scale=0.5]{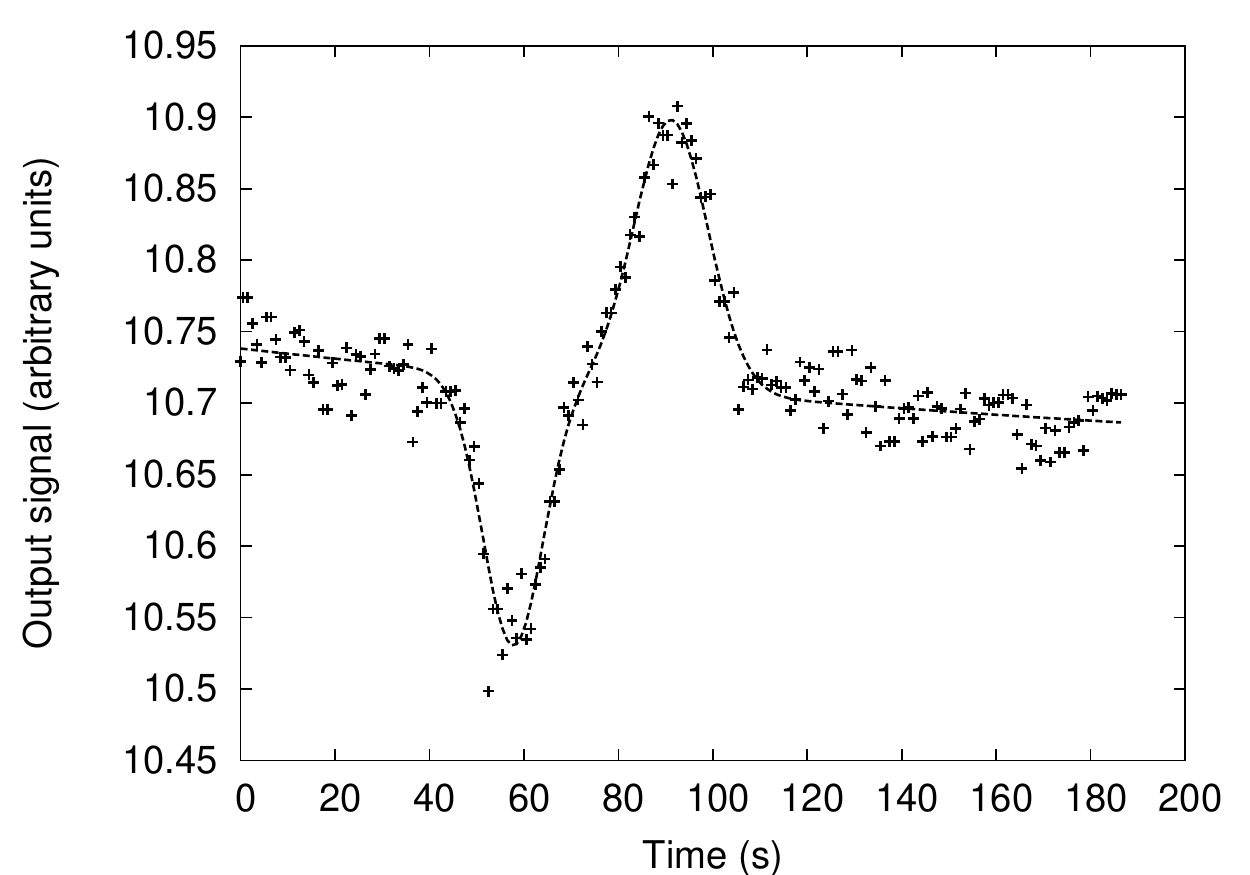}
\caption[Simulated and real cross-scans of a $\sim$250~mJy point source]{Simulated (left) and real (right) cross-scans of a $\sim$250~mJy point source. The real cross-scan is of the CRATES source 1810+5649 (290$\pm$30~mJy).}
\label{fig:qscan_example}
\end{center}
\end{fig}

The results of the simulations correspond well to measurements performed with OCRA in reasonable observational conditions. Figure \ref{fig:qscan_example} provides an example of a simulated cross-scan measurement of a 250~mJy point source compared with an OCRA-p observation of a $\sim290$~mJy point source; there is reasonable agreement between the two, although with higher noise within the simulations than in the real data. The Gaussian noise level on the double-differenced signal would be expected to be twice the point source sensitivity of the instrument due to factors of $\sqrt{2}$ for each of the two differences, hence $\sim 12$~mJy~s$^{1/2}$ for the OCRA-p receiver (see Section \ref{sec:ocrap_capabilities} for more details). Here the noise is approximately a factor of 2 higher than this due to the $1/f$ gain fluctuations still present within the signal.

We conclude that based upon the simulations, cross-scan measurements are suitable for sources stronger than 50-100~mJy, which agrees well with the experience from observations with OCRA-p. The limiting factor here is the analysis software used; smoothing the data over longer time periods might enable the detection of weaker sources, although the process of correcting the telescope pointing using the elevation scan would be less accurate. An alternative measurement technique for sources weaker than this is used: on-off measurements.

\subsection{Simulated on-off measurements} \label{sec:dd_comparison}
As with cross-scan measurements, we have performed 100 simulated observations of a range of point sources. As on-off measurements are more sensitive, we simulate 10, 5 and 1~mJy point source observations in addition to the source flux densities previously used. Matching the current observational strategy, we use 30~s measurements of each state, in the order neg-pos-neg-pos-neg-bg-cal. We use symmetric double-differencing as the default analysis method. The results from the simulations for the various input point source flux densities are given in Table \ref{tab:onoff_comparison}.

\begin{tab}{tb}
\begin{tabular}{r|rrr|rrr}
& \multicolumn{3}{c}{{\bf Gaussian noise}} & \multicolumn{3}{c}{{\bf $1/f$ noise}}\\
\hline
{\bf Input flux density} & {\bf Mean $\pm$ SD} & {\bf Min} & {\bf Max} & {\bf Mean $\pm$ SD} & {\bf Min} & {\bf Max}\\
\hline
1500 & 1463.8 $\pm$ 4.6 & 1452.6 & 1474.2 & 1396.5 $\pm$ 52.8 & 1274.2 & 1527.8\\
500 & 487.7 $\pm$ 2.2 & 482.4 & 493.7 & 486.8 $\pm$ 21.7 & 436.1 & 539.9\\
250 & 248.0 $\pm$ 1.4 & 244.4 & 251.6 & 235.1 $\pm$ 11.5 & 213.4 & 272.1\\
150 & 149.8 $\pm$ 1.2 & 146.3 & 152.8 & 152.3 $\pm$ 9.4 & 129.0 & 175.2\\
50 & 49.2 $\pm$ 1.0  & 47.0 & 51.3 & 52.9 $\pm$ 9.0 & 24.9 & 74.5\\
20 & 19.9 $\pm$ 1.0 & 17.4 & 23.1 & 19.7 $\pm$ 8.6 & -2.3 & 41.6\\
10 & 9.7 $\pm$ 1.0 & 7.3 & 12.8 & 11.4 $\pm$ 9.1 & -11.6 & 36.0\\
5 & 5.1 $\pm$ 1.0 & 2.9 & 7.7 & 3.7 $\pm$ 8.1 & -22.2 & 20.4\\
1 & 1.0 $\pm$ 1.0 & -1.5 & 3.2 & 1.1 $\pm$ 7.7 & -14.9 & 24.7\\
\end{tabular}
\caption[Comparison of input to output point source flux densities from simulated on-off observations]{Comparison of input to output point source flux densities from simulated on-off observations. Flux densities are given in milliJansky.}
\label{tab:onoff_comparison}
\end{tab}

We find that the flux density measurements from the simulated on-off measurements are systematically underestimated by 2.5 per cent as a result of the calibration process (a 1500~mJy source is measured to be 1463~mJy). This is due to the difference between the Bessel function beam used within these simulations and the fitted Gaussian curve to the calibration cross-scan measurement of NGC~7027. In reality, the beam will neither be a perfect Gaussian or Bessel function, and typical noise levels make the precise beam shape very difficult to measure. For the weak sources observed using on-off measurements, this difference should also be negligible.

\begin{fig}
\begin{center}
   \includegraphics[scale=0.5]{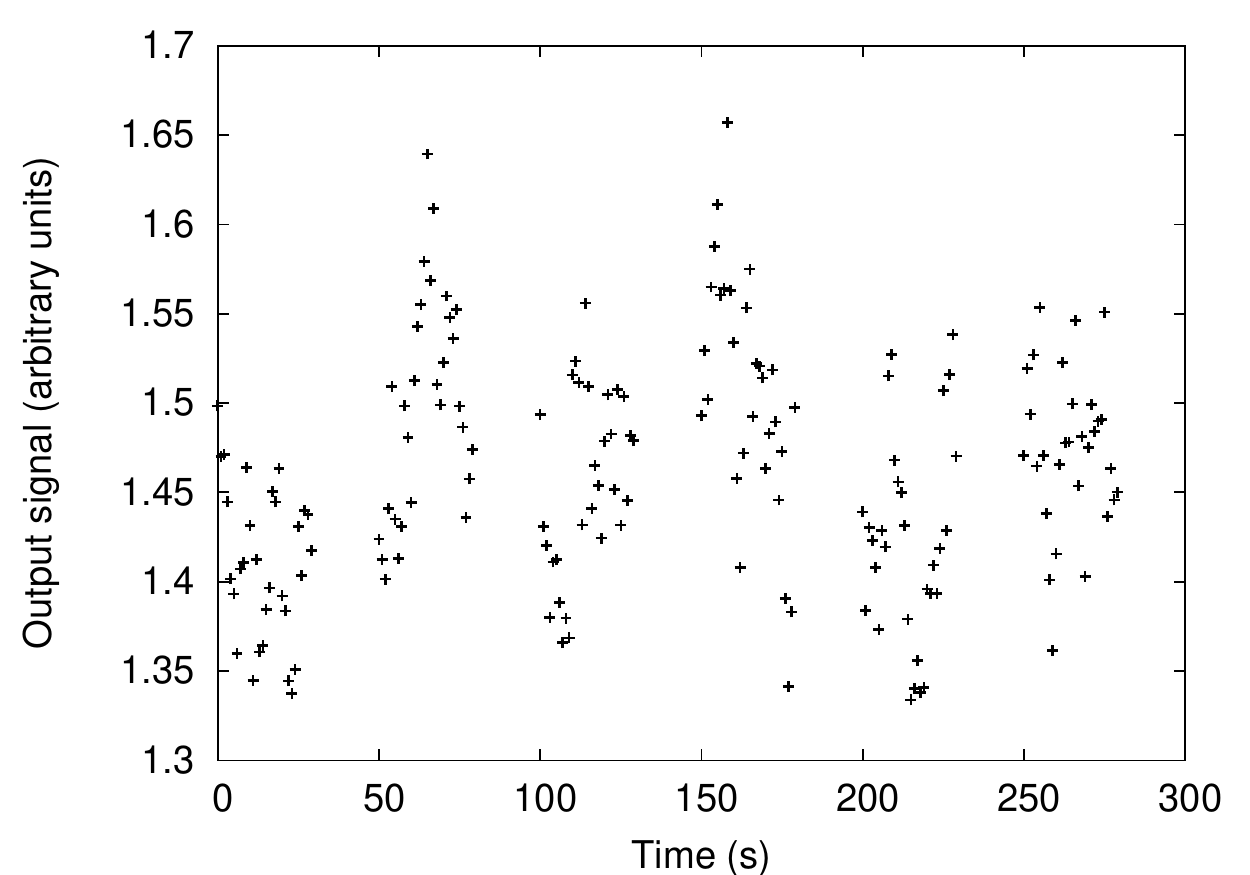}
   \includegraphics[scale=0.5]{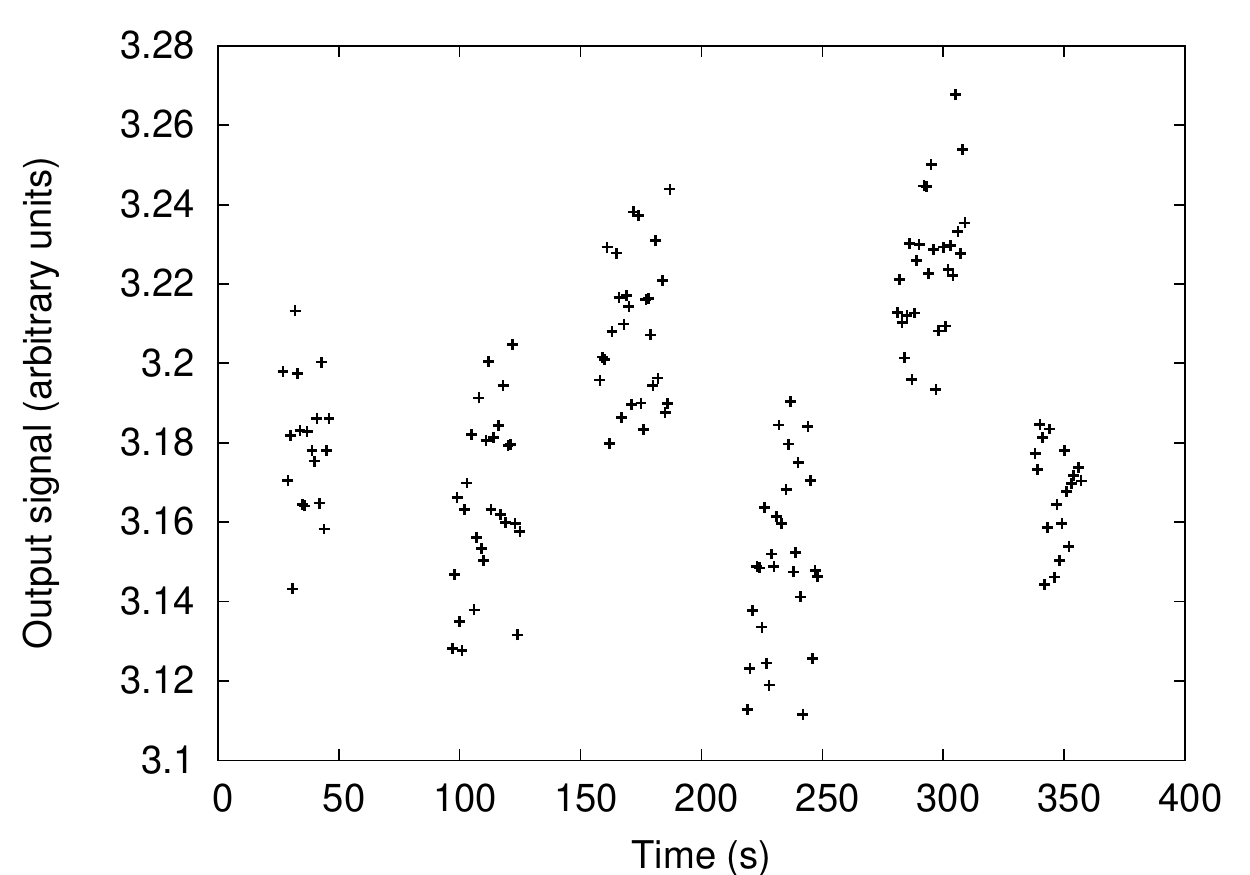}
\caption[Simulated and real on-off measurements of a $\sim$20~mJy point source]{Simulated (left) and real (right) on-off measurements of a $\sim$20~mJy point source. The real on-off measurement is of the VSA source J1233+5339 (21.4$\pm$2.4~mJy).}
\label{fig:onoff_example}
\end{center}
\end{fig}

The typical error on a simulated on-off measurement with Gaussian noise added is $\sim1.5$~mJy. Sources can be detected at any flux density value, either on an individual measurement for stronger sources or statistically for weaker sources, as expected for pure Gaussian noise. When $1/f$ noise is added, the error on a measurement increases to $\sim 3-5$~mJy, and the uncertainty in the mean measurement is significantly increased. However, we caution that the estimate used here for the level of the $1/f$ noise is overly pessimistic, as shown by the comparison of the simulated data to real data in Figure \ref{fig:onoff_example}.

When the observed point source is $\sim 5$~mJy, the simulated $1/f$ noise starts to cause severe problems, with measurements of $\pm$20~mJy present. From 100 measurements of a 5~mJy point source, the mean was 3.7~mJy from the symmetric double-differencing method and $\sim$3~mJy for the other methods. After 100 observations, the measurement of a 1~mJy point source is 1.17$\pm$0.77~mJy from the simulations with $1/f$ noise within them, compared with 1.04$\pm$0.13~mJy from the Gaussian measurements.

\begin{fig}
\begin{center}
   \includegraphics[scale=0.6]{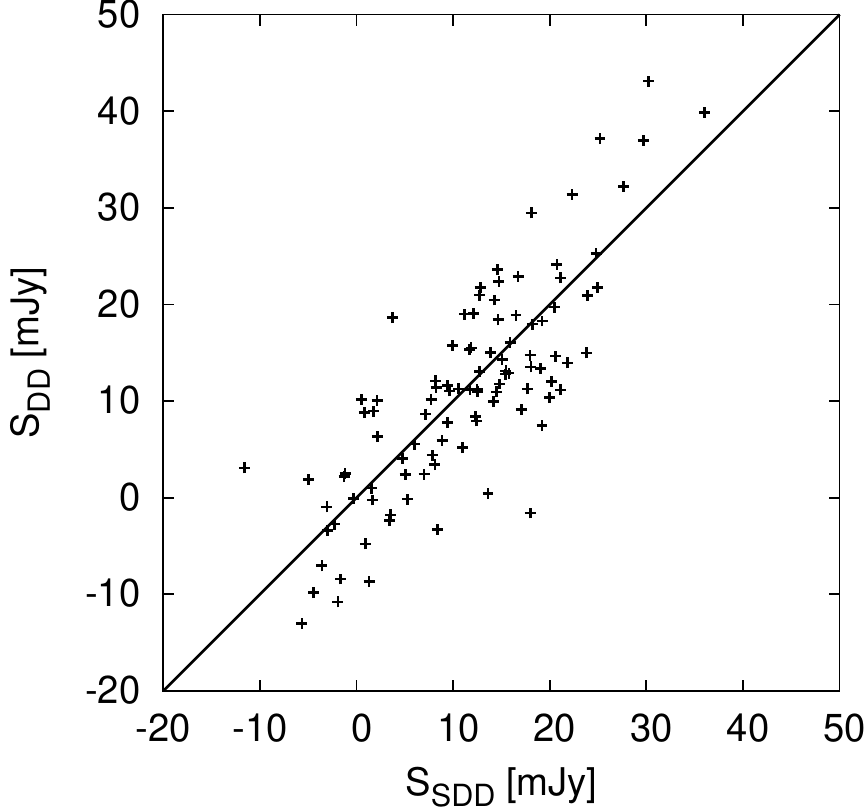}
   \includegraphics[scale=0.6]{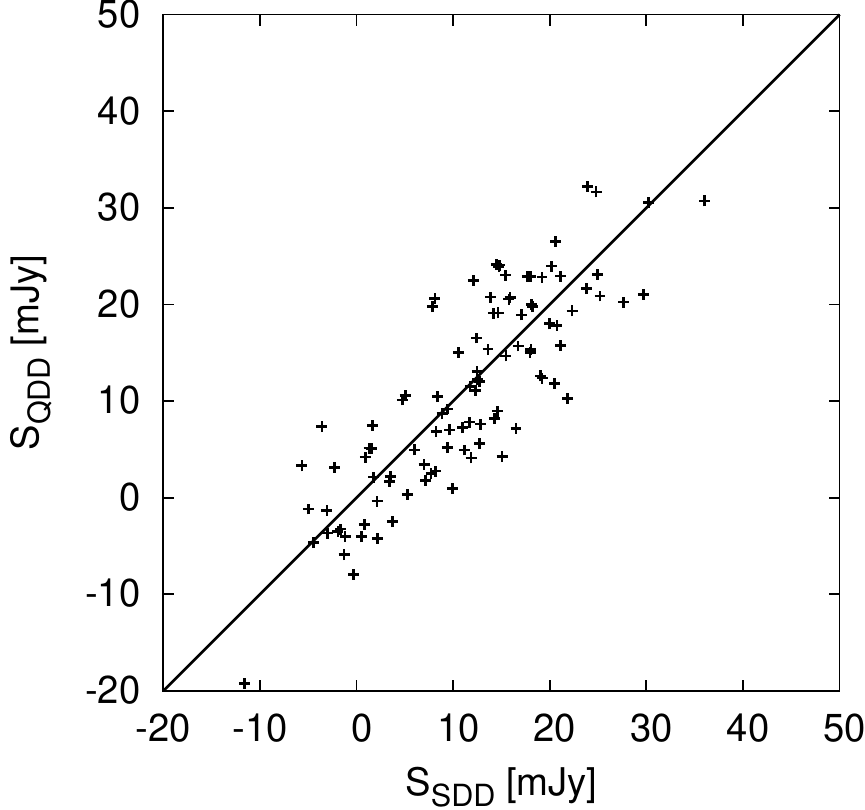}\\
   \includegraphics[scale=0.6]{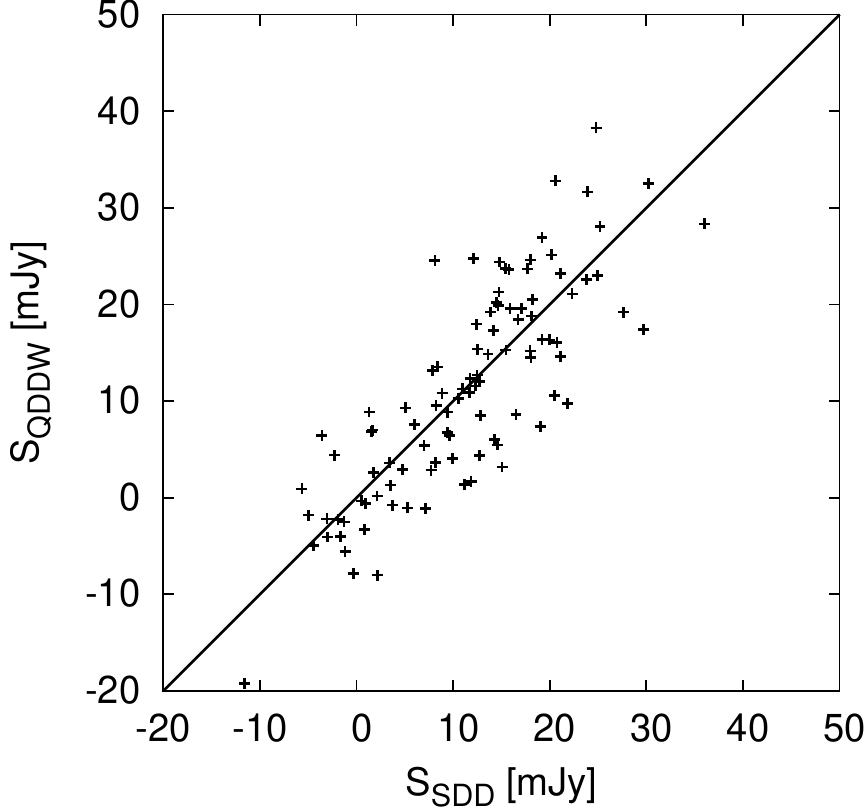}
   \includegraphics[scale=0.6]{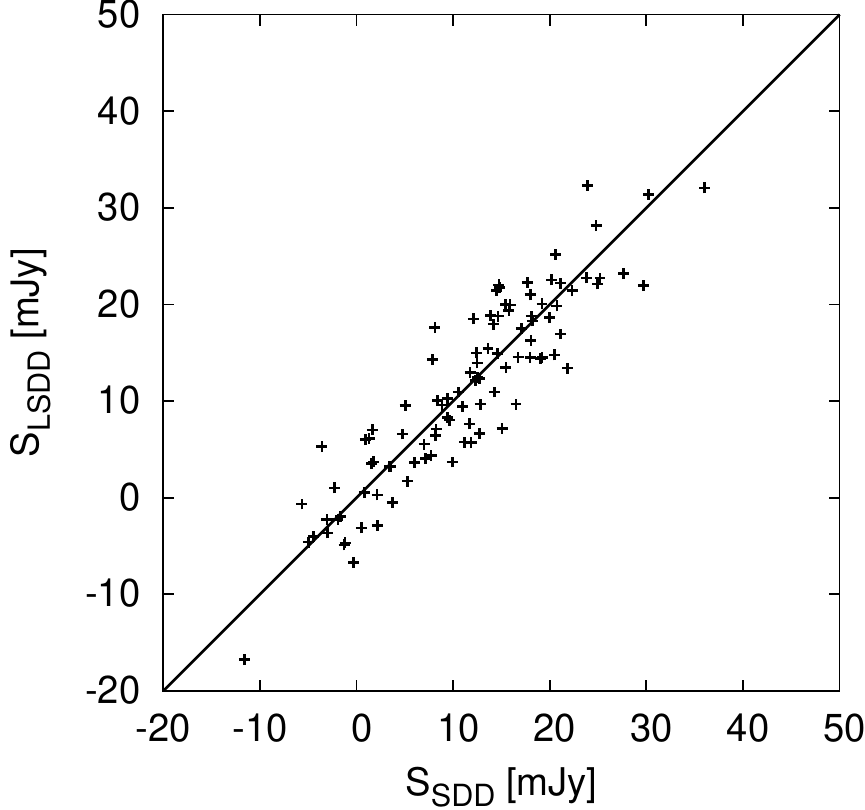}
\caption[Scatter plots of 100 observations of a 10~mJy point source in the presence of Gaussian and $1/f$ noise with different double-differencing methods]{Scatter plots of 100 observations of a 10~mJy point source with different double-differencing methods. Top left: symmetric double-differencing (Equation \ref{eq:sdd}) compared to standard double-differencing (Equation \ref{eq:dd}). Top right: standard versus double-differencing after the removal of a quadratic background (Equation \ref{eq:qdd}). Bottom left: the same, but using the weighted sum (Equation \ref{eq:qddw}). Bottom right: using a least squares fitter (Equation \ref{eq:lsdd}).}
\label{fig:onoff_ddtypes}
\end{center}
\end{fig}

Figure \ref{fig:onoff_ddtypes} shows scatter plots for 100 observations of a 10~mJy point source in the presence of Gaussian and $1/f$ noise, comparing the symmetric double-difference method with the standard double-difference, and also with the double-difference after the removal of a quadratic background with and without weighting the measurements. The improvement of the symmetric double-differencing over the standard version can be seen by the tilt in this scatter plot. The removal of a quadratic background (weighted or unweighted), nor using a least squares fitter, do not make a significant improvement over symmetric double-differencing for these measurements. However, they may offer improvements for analysing longer period observations.

\begin{fig}
\begin{center}
   \includegraphics[scale=0.5]{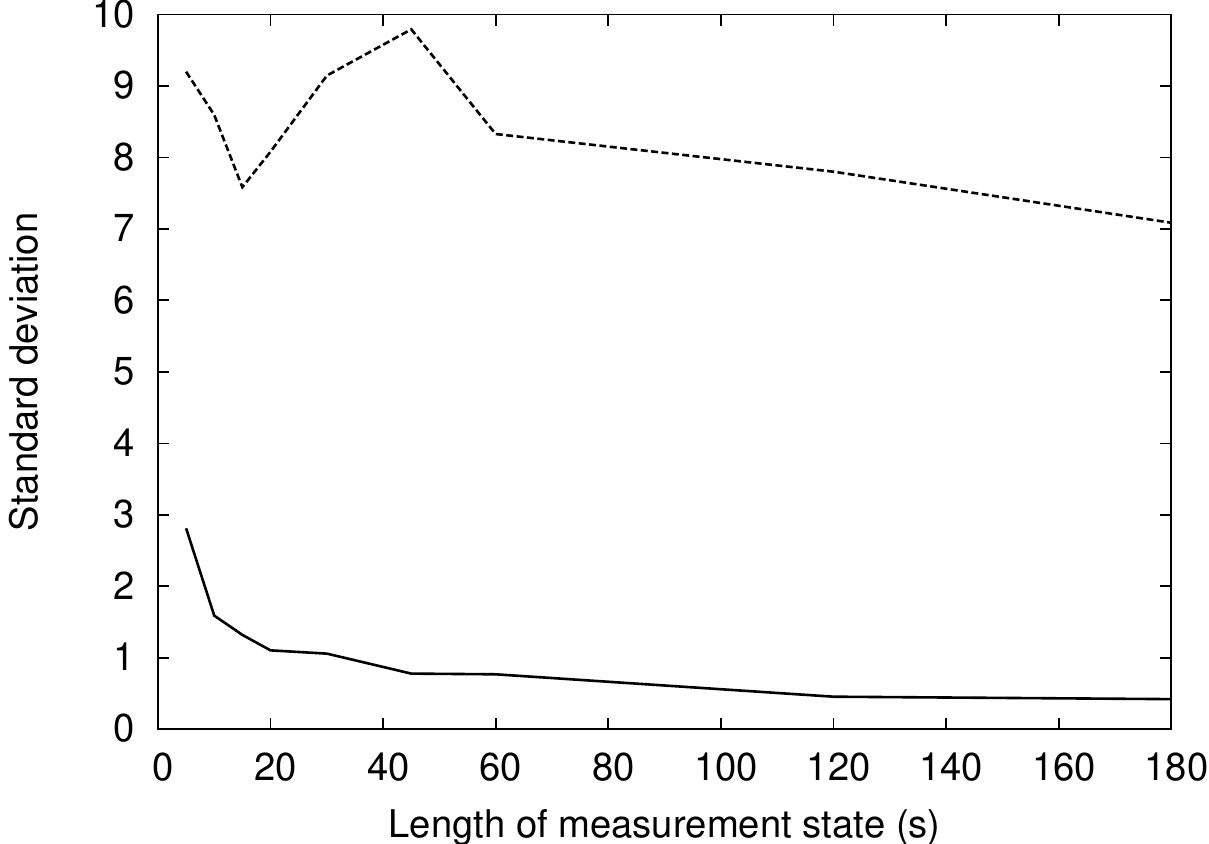}
   \includegraphics[scale=0.5]{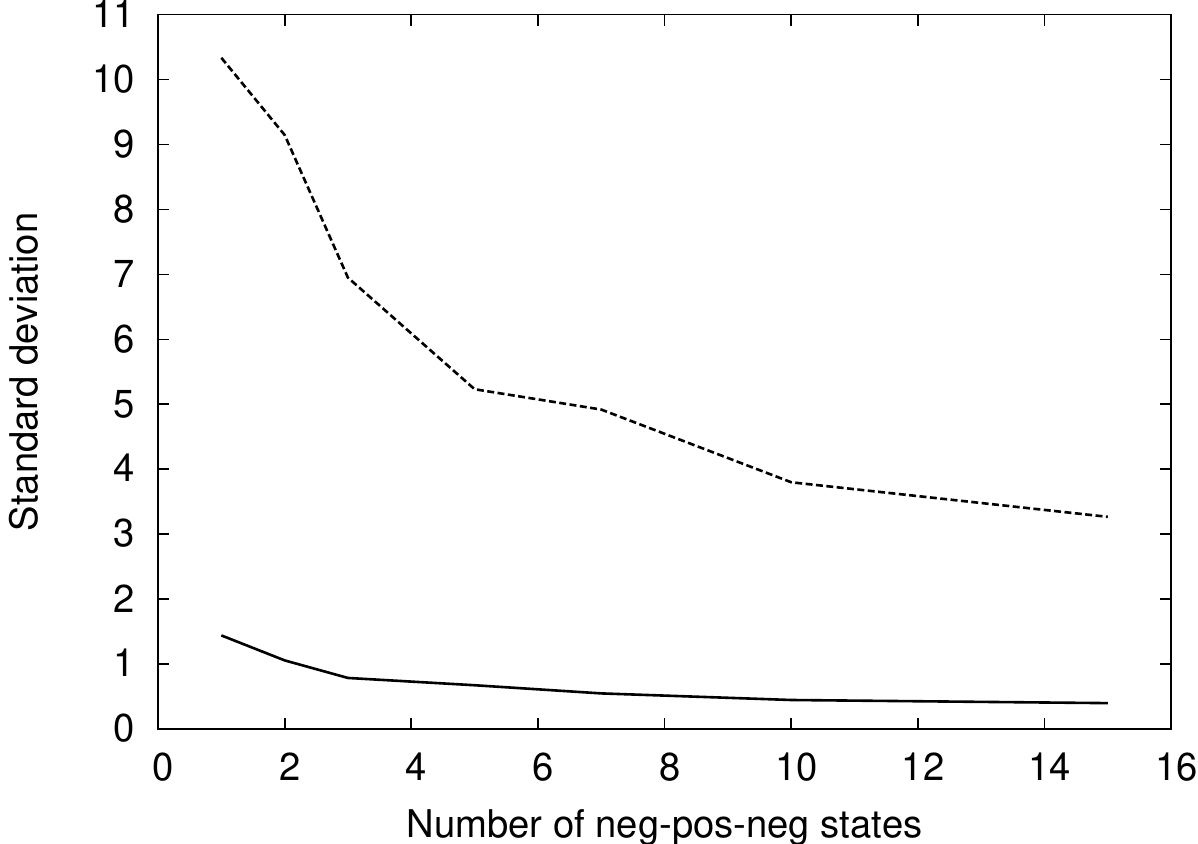}
\caption[Comparison of the noise on a measurement versus the length of time and number of states]{Comparison of the noise on a measurement versus the length of time and number of states.}
\label{fig:onoff_length_num}
\end{center}
\end{fig}

Figure \ref{fig:onoff_length_num} shows the standard deviation from the scatter for 100 observations of a 10~mJy point source when the length of each measurement state is changed (between 5 and 180 seconds) and when the number of measurement states is increased (in multiples of pos-neg states between 1 and 15). In both cases, where Gaussian noise only is present, the error on the measurement decreases as expected. However, when $1/f$ noise is present the benefits of increasing the length of the measurement states are negligible. The optimal length of the measurement state will depend critically upon the characteristics of the $1/f$ noise. The noise on the measurement can, however, be decreased by increasing the number of measurement states.

We note, however, that pointing inaccuracies of the telescope are not present within these simulations, which restrict the number of on-off states and the duration of the observations that can be made using OCRA-p on the telescope. As a result, these are indicative results requiring further comparison with observations.

\section{Summary and future work} \label{sec:sim_futurework}
The work described in this chapter has focused on setting up the tools to enable highly realistic end-to-end simulations of OCRA-p and OCRA-F. Data reduction software has been created and linked with the expanded UMBRELLA software, and point source observations using cross-scan and on-off measurements have been simulated.

The UMBRELLA software was designed from the start with support for multiple receiver pairs in mind, however the operational code was written to only simulate one receiver pair--OCRA-p. The software has been generalized by the author to work with multiple receiver pairs, requiring the creation of a new data storage class within the C++ code, and a modification of the input commands so that multiple receiver chains could be configured. The software has also been modified to read in Virtual Sky maps.

Now that this simulation software is in place, further work can be carried out by using this software (and also OCRA-p) to test different observational techniques and data reduction methods. This includes simulating observations of weaker sources, including the SZ effect, by long integrations. In the case of SZ clusters, it will be important to simulate the effects of correcting for the extended nature of these sources, as well as the removal of contaminating point sources. 

The process of creating maps with the OCRA-F instrument can now be simulated, and trials of suitable observation methods for blind surveys for point sources and the SZ effect can also be carried out using this software combined with mapping algorithms. Additionally, simulations of measurements of the CMB power spectrum with OCRA-F can also now be carried out, information from which can feed into actual observations with the instrument. When confronting the observational challenges posed by these very weak fields it is vital to understand all the imperfections of the instrument.

The software is also now capable of simulating even larger focal plane arrays, however this will consume a large amount of CPU time. In order for this to be achievable within a reasonable timespan the software needs to be optimized and parallelized so that it can run on multiple processors simultaneously. The simulations can also be further refined by comparison to more instrumental and observational measurements. These include measurements of the atmosphere using data from water vapour radiometers, further noise power spectra measurements with the OCRA instruments, and potentially also power spectra measurements from the WMAP and Planck spacecraft, as these receivers use very similar technology to that in the OCRA-p and OCRA-F receivers and hence will exhibit characteristic instrumental $1/f$ gain fluctuations without containing atmospheric emission.

One of the work packages within the FP7 Radionet APRICOT Join Research Activity will focus on simulations of large linked arrays over the course of the next 2 years. The software described here will feed into this project and be developed further as a result, including potentially being extended to simulate total power, polarization and spectral line observations.

\chapter{Point source observations with OCRA-p}\label{chapter:ocrap_obs}

The OCRA-p receiver has been used to observe a number of point sources from different catalogues since its installation on the 32-m Toru\'n telescope in 2003. The receiver is best suited for observing known sources, rather than surveying for new ones, as it only has two small (1.2 arcmin) beams to observe with. It is capable of observing sources down to the milliJansky level, depending on the weather conditions that it is observing in, as shown by the simulations in Section \ref{sec:sim_ps}. The actual observational capabilities of the receiver on the telescope are described in Section \ref{sec:ocrap_capabilities}.

OCRA-p has been used to observe the radio sources within the VSA super-extended array fields, which were selected via a blind survey with the Ryle Telescope (RT) at 15~GHz. It is important to know the flux densities of these sources so that they can be subtracted out from the maps of the CMB produced by the VSA. The results of the OCRA-p observations were published in \citet{2009Gawronski}. The Author was involve with a number of the elements of data reduction and generation of results within this paper; these are described in more detail in Section \ref{section:vsasources}.

Additionally, a subsample of 550 of the CRATES sample of flat spectrum radio sources ($>65$~mJy at 4.85GHz) were also observed. These sources will have significant flux densities at higher frequencies (compared with steep spectrum sources, which will be much weaker at high frequencies). Knowledge of these sources is important for CMB observations by the Planck satellite, which is most sensitive in the area of the sky where this subsample has been selected. The selection of the source subsample, the observations and the results are described in Section \ref{section:crates}.

The observations of the VSA and CRATES sources were all reduced using the software written by the author and described in Section \ref{sec:dred}; as such, the details of the data reduction methodology are not covered in this chapter.

\section{Capabilities of the OCRA-p receiver} \label{sec:ocrap_capabilities}
The performance of a radio telescope receiver depends on the ``system temperature'' $T_\mathrm{sys}$ (a measure of the Gaussian noise of the receiver-telescope system), the atmospheric temperature, and the bandwidth of the receiver. The standard formula to calculate the sensitivity for the receiver layout used within OCRA-p is
\begin{equation}
\Delta S_\mathrm{min} = \frac{2k}{A_\mathrm{eff}} \Delta T = \frac{2k}{A_\mathrm{eff}} \frac{\sqrt{2} T_\mathrm{sys}}{\sqrt{B \tau}}.
\end{equation}
Here, the factor of $2k / A_\mathrm{eff}$ converts between the ``temperature'' sensitivity and the flux density sensitivity, where $k$ is Boltzmann's Constant and $A_\mathrm{eff}$ is the effective collecting area of the telescope, which is the combination of the surface area and efficiency. The Toru\'n 32~m telescope is $\sim$45 per cent efficient at 30~GHz, yielding a conversion factor of 0.13 Janskys per Kelvin.

\begin{fig}
\centering
\includegraphics[scale=0.5]{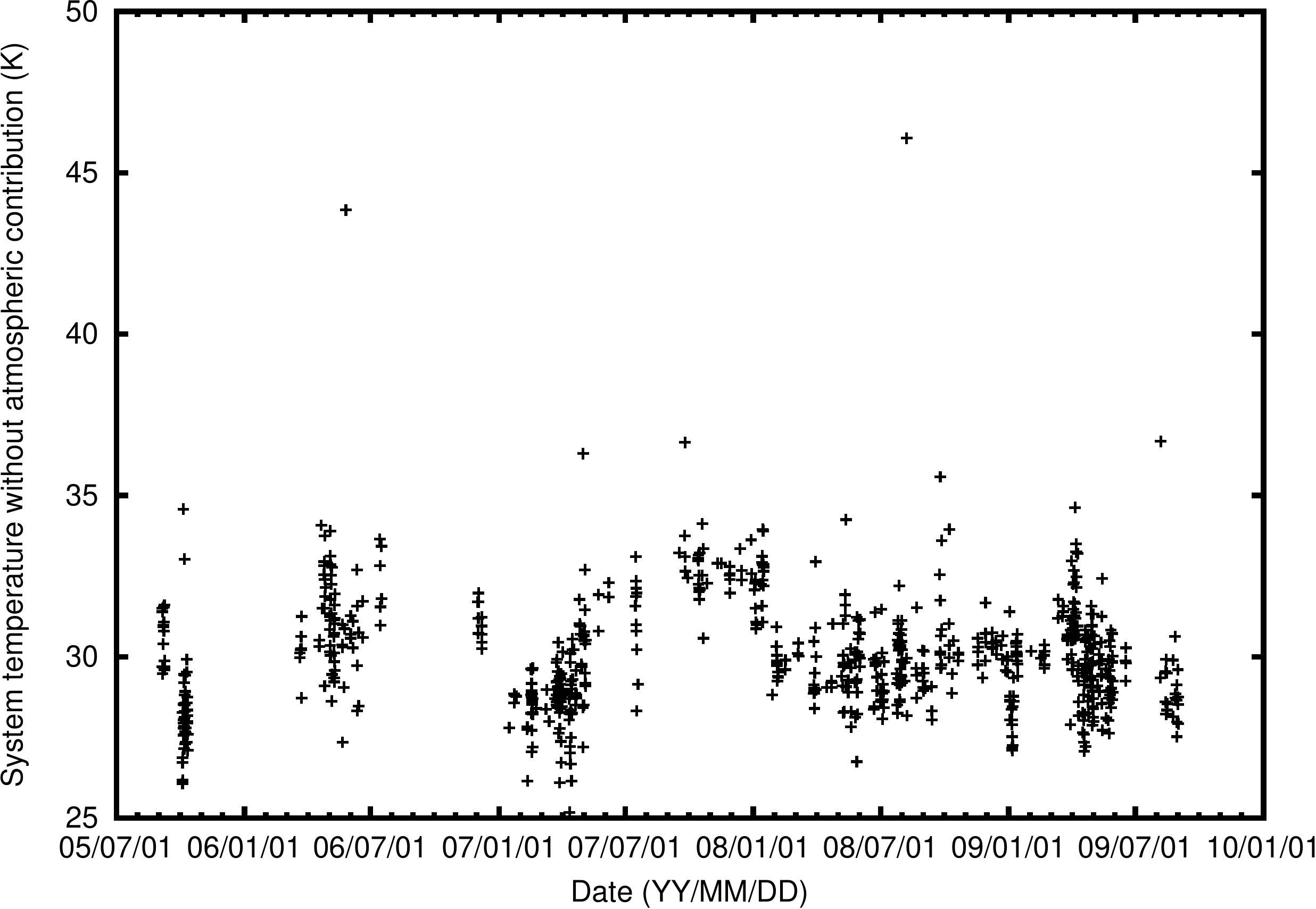}\\
\includegraphics[scale=0.5]{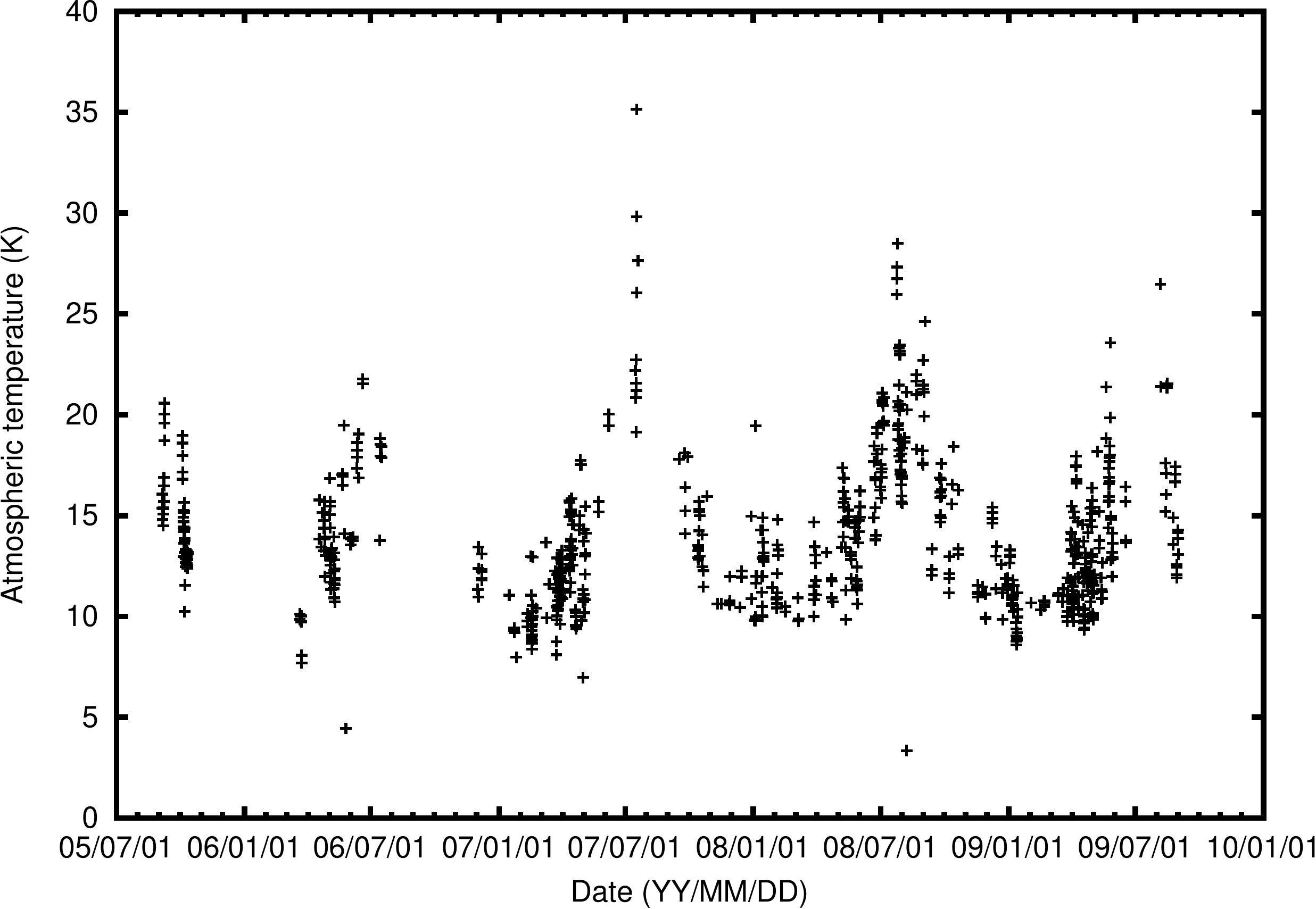}
\caption[The OCRA-p system temperature and atmospheric emission temperatures measured between 2005 and 2009]{Top: The measured system temperature, excluding the atmospheric contribution, of OCRA-p on the telescope between 2005 and 2009. Bottom: The measured atmospheric emission temperature at 30~GHz over the same time period. These measurements were taken during astronomical observation periods; as such they have been taken during the best weather at the telescope.}
\label{fig:system_temperature}
\end{fig}

The complete system temperature of OCRA-p, including atmospheric emission, can be measured by comparing the signal received when looking at a room temperature absorber with that received when looking at the sky. The difference of this measured at the zenith and at 30 degrees elevation measures the atmospheric emission component, and hence the system temperature can be calculated. This measurement of the system temperature is made routinely with OCRA-p so that the effects of atmospheric absorption can be corrected for in flux density measurements.

The system temperature of OCRA-p over time is shown in Figure \ref{fig:system_temperature}; this is fairly stable around 30~K over the time period 2005-2009. Figure \ref{fig:system_temperature} also shows the atmospheric emission temperature at 30~GHz as measured by OCRA-p over the same time period; this clearly peaks around August each year, with a minimum around April. The typical values for the atmospheric emission are between 10 and 20~K.

Using a typical value for the combined receiver and atmospheric temperature of 40~K, and a bandwidth of 6~GHz, this then yields a sensitivity of 0.7~mK s$^{1/2}$, or 6~mJy s$^{1/2}$. This means that a noise level of 1~mJy should be achievable in $\sim 30$ seconds. Note however that this only takes the Gaussian noise into account, and does not include the $1/f$ gain fluctuations in the amplifiers nor the $1/f$-like atmospheric noise. These increase the integration times required to reach a given flux density, such that it no longer scales as s$^{-0.5}$ (see Chapter \ref{ocra_sims}). In practice, the error on an individual on-off measurement with OCRA-p where each state lasts for $\sim30$~s is around 3-5~mJy (see later in this Chapter).

\section{Sources in the Very Small Array fields} \label{section:vsasources}

The Very Small Array (VSA) was a 33~GHz 14-element interferometer located at the Teide Observatory in Tenerife \citep{2003Watson}. It was used to measure the anisotropies in the Cosmic Microwave Background at a range of angular scales on the sky. Three instrumental setups were used: Compact, Extended and Super-Extended, each of which probed increasingly smaller angular scales.

In order to obtain high accuracy, uncontaminated CMB power spectrum, foreground point sources need to be subtracted out from the data. The VSA had a dedicated source subtracter, a separate interferometer with two 3.7-m dishes separated by 9.2~m \citep[see][]{2005Cleary}, which was used during observations with all three setups. However, point sources contaminate smaller angular scales much more than larger scales, such that source subtraction to lower flux densities becomes crucial in the process of measuring the small-scale CMB anisotropies (see Figure \ref{fig:30ghz_cbi_excess}). With the super-extended array configuration, it did not prove possible to use the source subtractor flux densities; instead, measurements by OCRA-p were used. OCRA-p, whilst not as well located as the source subtractor for observations at these frequencies, has the benefits of a much larger collecting area and a wider bandwidth, such that it can measure the 30~GHz flux densites of the contaminating sources more accurately.

\begin{fig}
\centering
\includegraphics[scale=1.0]{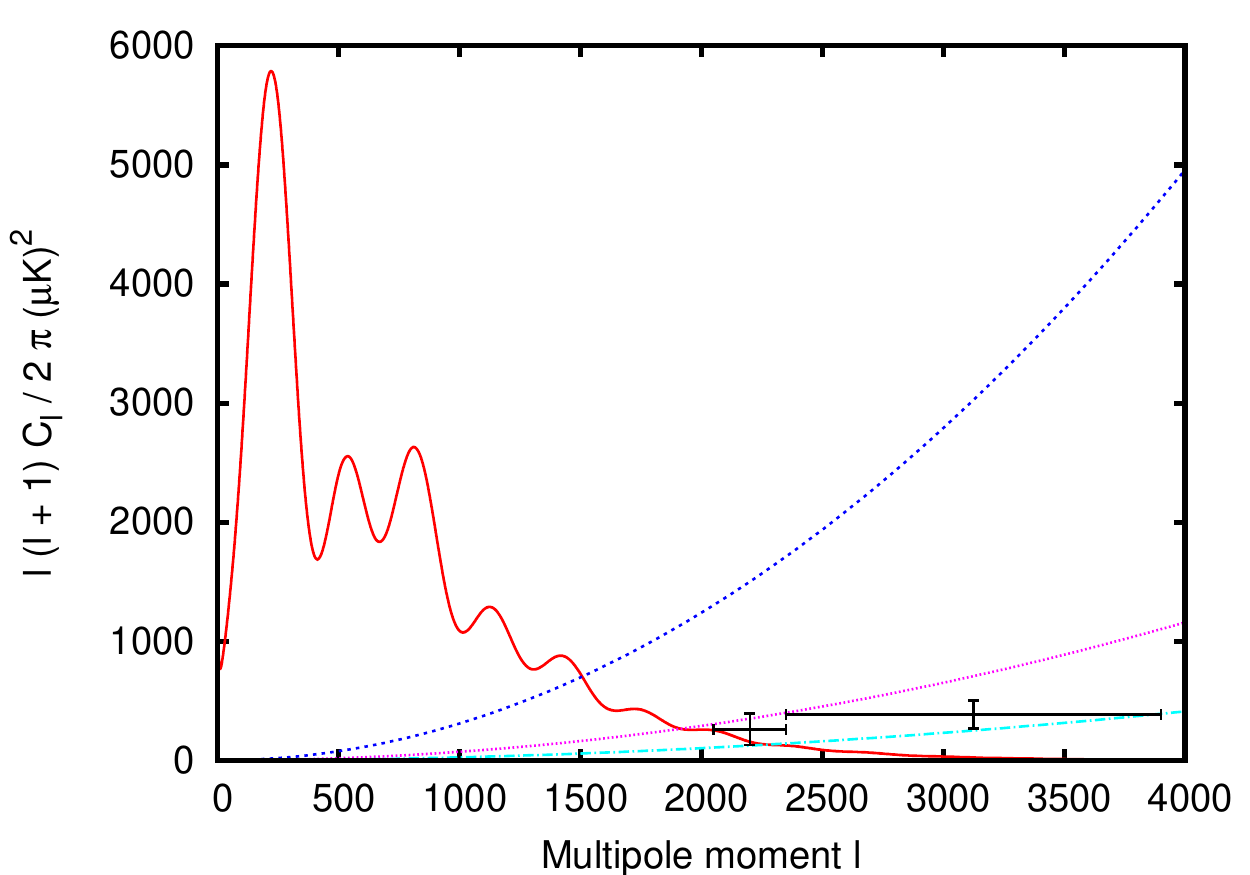}
\caption[The power spectrum at 30~GHz of the CMB and point sources]{The power spectrum at 30~GHz of the CMB (red; generated by {\sc CAMB}, \citealp{2000Lewis}) and the power spectrum of residual point sources at the level of 20~mJy (blue), 5~mJy (purple) and 2~mJy (turquoise) based on the model of \citet{1998Toffolatti} renormalized by 0.7. The CBI excess (\citealp{2009Sievers}; discussed in Section \ref{sec:cbi_excess}) is also shown (black).}
\label{fig:30ghz_cbi_excess}
\end{fig}

The five Super-Extended array fields were first surveyed by the Ryle Telescope (RT) at 15~GHz and the sources were identified and followed up by deeper, pointed observations with the RT. OCRA-p was then used to observe the detected sources at 30~GHz. Data analysis was carried out separately by the Author, M. Gawro\'nski and K. Lancaster, each using independent data reduction code. Results were compared, with anomalies between the results identified and resolved. The results were published in \citet{2009Gawronski}. This section describes the data reduction carried out by the Author, the comparison of the results to the other two reductions and also with previous measurements of these sources at comparable frequencies. It also discusses the variability and spectra of the sources, the 1.4-30~GHz spectral index distribution at different flux densities and the source surface density.

\subsection{Observations and calibration} \label{section:vsaobservations}
The five fields observed by the VSA in its super-extended configuration were first surveyed by the RT in Cambridge at 15~GHz. These observations (detailed in \citealp{2009Waldram,2009Gawronski}) consisted of a series of raster scans, followed by pointed observations on each detected source, yielding a list of point sources complete to 7~mJy, and with an error on each measurement of 5 per cent due to calibration. Due to the pointed observations, the thermal noise of the measurements is negligible compared with the calibration uncertainty.

\begin{tab}{tb}
\begin{tabular}{c|c|l}
\bf Epoch \rm & \bf Date \rm & \bf Format\\
\hline
1 & 10-11 February 2007* & + -- + -- cal +\\
 & 16-19 February 2007 &  -- + -- + cal \\
 \hline
2 & 23-28 March 2007 & bg -- + -- bg + cal\\
 & & bg -- + -- bg cal\\
 & 26-27 April 2007 & bg -- + -- + bg cal\\
 & 3-5 May 2007* & bg -- + -- + bg cal\\
 & 8 June 2007 & bg -- + -- + bg cal\\
 & 16-18 July 2007 & bg -- + -- + bg cal\\
  & & -- + -- + -- bg cal\\
 \hline
3 & 16, 24-25, 29 September 2007* & -- + -- + -- bg cal\\
 & 13-16 October 2007 &  -- + -- + cal\\
  & & -- + -- + -- bg cal\\
 \hline
4 & 13, 15, 29 December 2007 & -- + -- + -- bg cal \\
 & 2, 4, 5, 11* January 2008 & -- + -- + -- bg cal\\
 \hline
5 & 24 April 2008 & -- + -- + -- bg cal\\
 & 6, 9, 28 May 2008 & -- + -- + -- bg cal\\
 \hline
6 & 12-13 January 2009 & -- + -- + -- bg cal\\
\end{tabular}
\caption[Dates of measurements of the VSA sources]{Dates of measurements of the VSA sources, and the method of the on-off scans that was used. Due to bad weather or issues with the receiver, data from observation dates marked with an asterisk (*) were flagged; see Section \ref{section:vsaflagging} for details}
\label{tab:vsa_measurements}
\end{tab}

The 121 detected sources were then observed at 30~GHz with OCRA-p between 10~February 2007 and 28~May 2008, with some follow-up measurements taken on 12--13~January 2009 to complete the sample and to check for variability. A list of the observing sessions is given in Table \ref{tab:vsa_measurements}. The sources were observed using the on-off method (described in Section \ref{sec:onoff_methods}); the exact pattern of these observations evolved over the observations, as described in Table \ref{tab:vsa_measurements}. All measurements were reduced using the symmetric double difference method regardless of the observational pattern. This was due to the increased ability of this method to remove the effects of linear drifts in the signal compared to the standard double-differencing method, even though this does not necessarily use all of the recorded data.

The planetary nebula NGC~7027 was used for primary flux density calibration, and was observed at least once per day during the VSA source observations. This is the brightest planetary nebula in the radio sky. At 33~GHz, \citet{2008Hafez} report a flux density of $5.39 \pm 0.04$~Jy at an epoch of 2003.0 with a secular decrease of $-0.17 \pm 0.03$~per~cent. Extrapolating this to 2008.0 yields a flux density of $5.34 \pm 0.04$~Jy, which can then be scaled to 30~GHz using the quoted spectral index of -0.119 to give $5.47 \pm 0.04$~Jy. NGC~7027 has also been observed by \citet{2008Zijlstra}; from these measurements a value at 30~GHz of $5.37 \pm 0.28$~Jy at the epoch of 2008.0 can be derived, which is consistent with \citet{2008Hafez}.

A noise diode is used for secondary flux density calibration of the observations. The ratio of outputs obtained with cross-scans of NGC~7027 and the noise diode are averaged for a 24 hour period; this diode calibration is then applied to the individual measurements.

\begin{tab}{tb}
\begin{tabular}{c|c|c|l}
{\bf Name} & {\bf Position (J2000)} & {\bf Flux density (mJy)} & {\bf Notes} \\
\hline
NGC 7027 & 21h 07m 51.1s & $5470 \pm 40$ & Primary calibrator\\
 & +42d 10m 27.8s & \citep{2008Hafez} & \\
\hline
J0015+3216 & 00h 15m 06.2s & $277 \pm 24$ & Position calibrator, field 1L\\
B0012+3159 & +32d 16m 12.7s & & \\
\hline
J0245+2405 & 02h 45m 16.8s & $249 \pm 22$& Position calibrator, field 5E\\
B0242+238 & +24d 05m 35.2s & & \\
\hline
J0958+3224 & 09h 58m 20.9s & $635 \pm 54$ & Position calibrator, field 2G\\
B0955+3238 & +32d 24m 02.2s & & \\
\hline
J1219+4829 & 12h 19m 06.4s & $643 \pm 54$ & Position calibrator, field 7E\\
B1216+4847 & +48d 29m 56.4s & & \\
\hline
J1521+4336 & 15h 21m 49.6s & $415 \pm 36$ & Position calibrator, field 3L\\
B1520+4347 & +43d 36m 39.6s & & \\
\end{tabular}
\caption[Calibration sources used during the VSA source observations] {The calibrator sources observed during the VSA source measurements, along with their coordinates (as observed), flux densities (either referenced or the mean from the measurements, excluding calibration errors) and notes on their use within these observations.}
\label{calibrators}
\end{tab}

Additionally, observations were made of a pointing calibrator per field. These sources are the same as those used for flux density calibration by the RT/AMI collaboration in their measurements of the sources at 15GHz. The calibrators (listed in Table \ref{calibrators}) were observed every 30-45 minutes (or every 5-6 on-off measurements) during the observations, and were used solely to calibrate the pointing of the telescope; flux density calibration was carried out using the noise diode.

The change of the telescope gain with elevation was corrected for using a fit to observations of NGC~7027 made over a wide range of elevations (A. Kus, private communication), using Equation \ref{eq:gainelevation}. Atmospheric absorption is corrected for using measurements of the system temperature at the zenith and 60 degrees from zenith made at regular intervals during the observations. A simple ``flat earth" model in which the atmosphere is assumed to be a continuous slab with constant thickness and optical depth was used; see Equation \ref{eq:atmosphere_correction}. See Section \ref{sec:dred_cal} for more details about the calibration process.

\subsection{Data quality} \label{section:vsaquality}

\subsubsection{Flagging of bad data} \label{section:vsaflagging}

\begin{fig}
\begin{center}
   \includegraphics[scale=1.0]{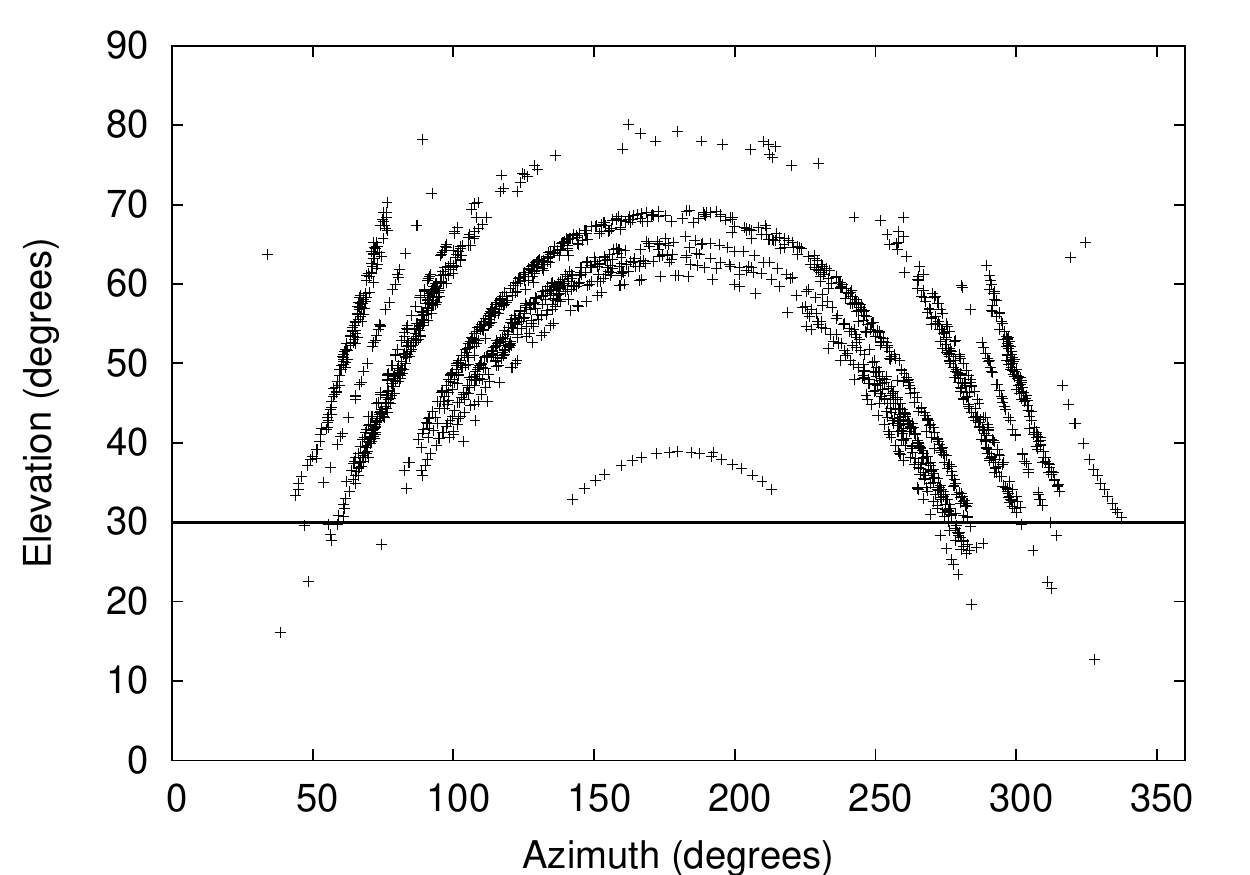}
\caption[Telescope azimuth and elevations for the VSA source measurements]{The azimuth and elevation of the telescope for all of the VSA source measurements. Measurements taken below the red line (30 degree elevation) were automatically flagged. The tracks of the 5 source fields, and some of the calibration sources, can clearly be seen.}
\label{fig:azel}
\end{center}
\end{fig}

In order to produce accurate flux density measurements, bad data must be removed where possible. This was done both automatically and manually, depending on the criteria described below. In total, 1350 measurements were made of the VSA sources, of which 336 measurements (25 per cent) were flagged, leaving 1014 good measurements (on average, 8.3 measurements per source). This flagging is due to atmospheric effects, calibration and telescope pointing issues.

Calibrator measurements where the ratio of the amplitudes and widths of the cross-scans were less than 0.8 or greater than 1 / 0.8 were automatically flagged, as was any measurement taken with an elevation angle of less than 30 degrees (see Figure \ref{fig:azel}). The calibration scans were also checked through manually, and any which visually appeared to be bad were manually flagged. Cross-scan measurements where the azimuth scan had a significantly lower intensity than the elevation scan were also flagged. Any measurements with an obviously erroneous calibration diode measurement (e.g. the calibration diode measurement was the same intensity as the background) were removed.

Individual on-off measurements were removed automatically where the error on the measurement was greater than 7~mJy, and manually where they had been flagged by the observer or the measurement had an anomalously low flux density for that source, something that would indicate problems with the telescope pointing during the measurement.

All measurements from 10-11~February 2007 were removed due to bad weather, and measurements taken between 16 and 24~September 2007 as well as those taken on 11~January 2008 were flagged due to technical issues with the receiver. Finally, data from 3-5~May 2007 were removed as NGC~7027 was not observed during that session, and there was a large jump in the output voltages from OCRA-p.

\subsubsection{Long time period variations in receiver}
\begin{fig}
\begin{center}
   \includegraphics[scale=1.0]{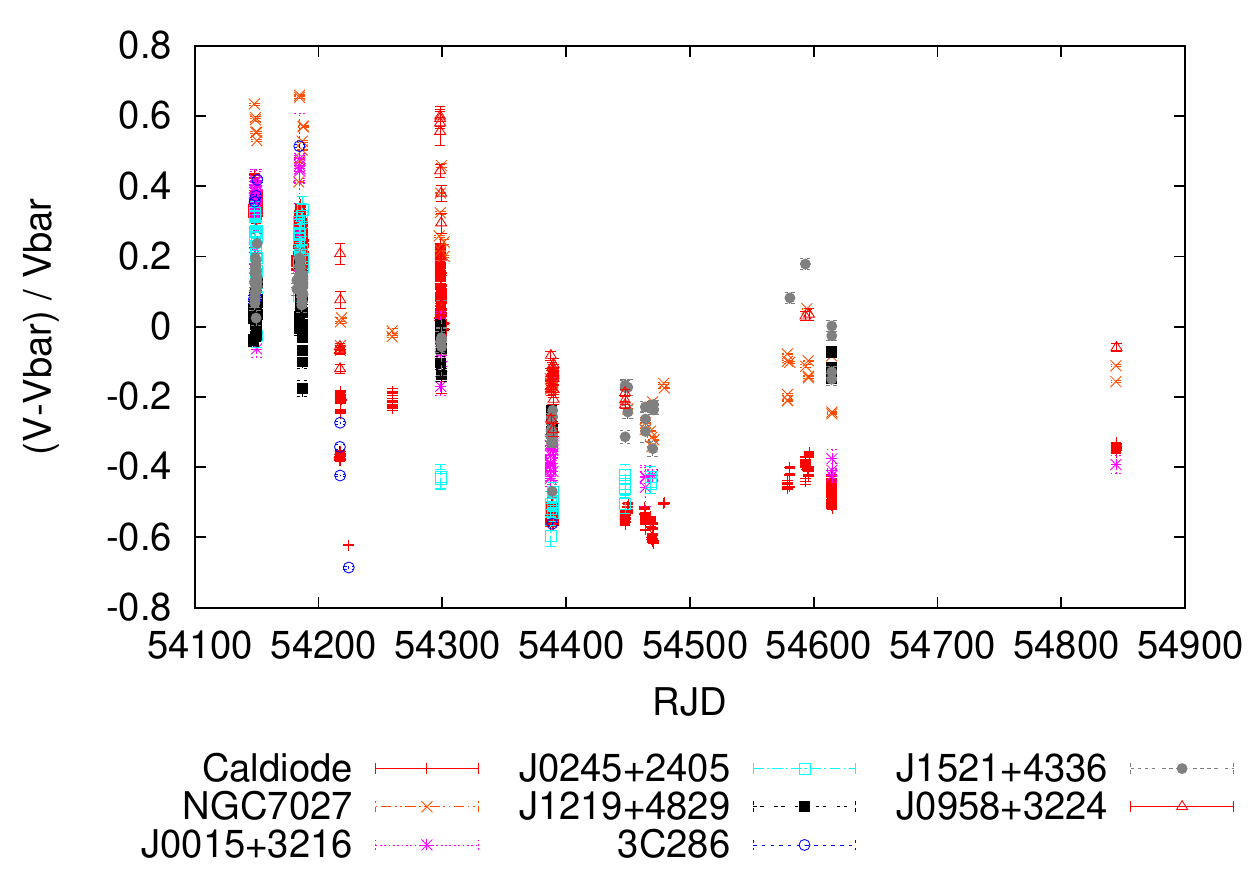}
   \includegraphics[scale=1.0]{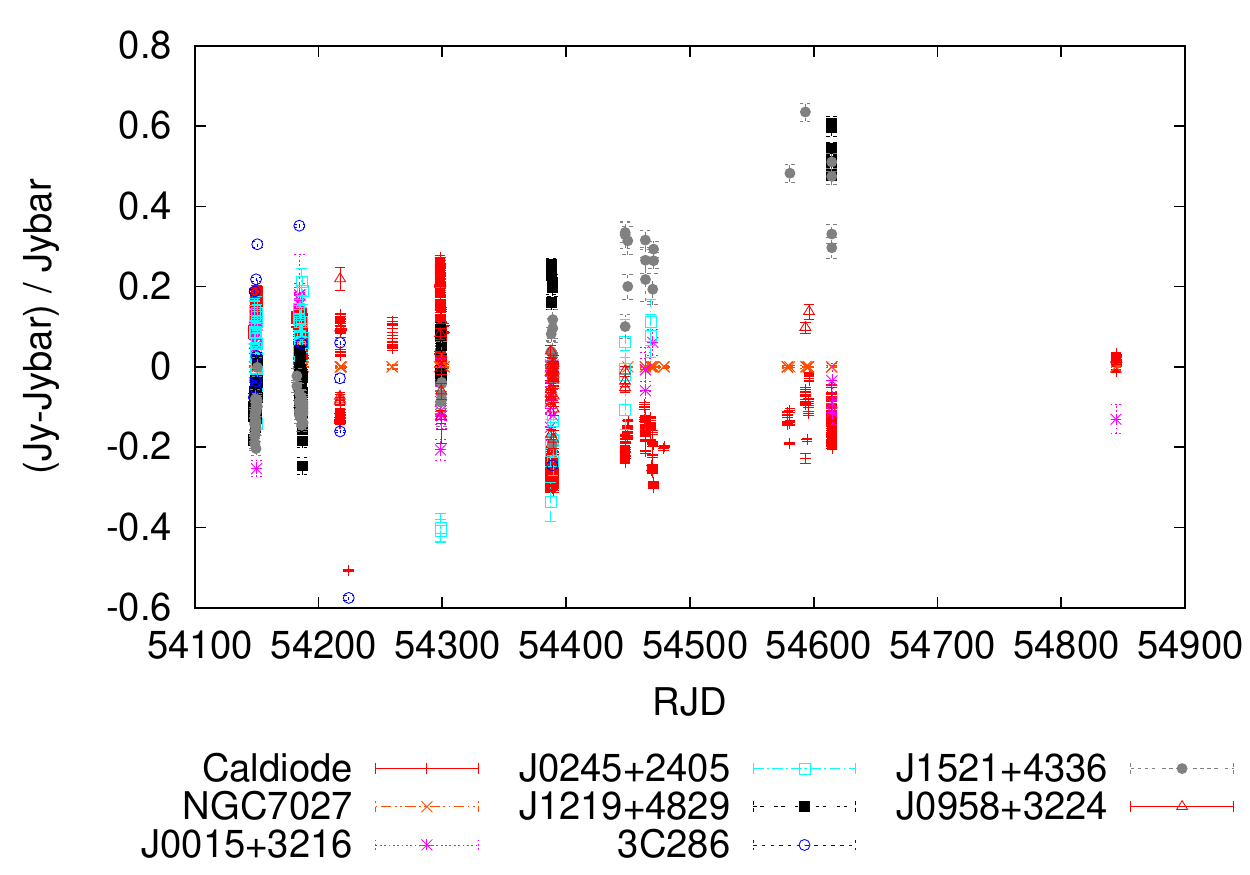}
\caption[Calibrator voltage and flux density over time for the VSA sources]{The top plot shows the difference between individual measurements minus the mean measurement of that source, divided by the mean measurement of that source, for all of the primary and secondary calibrators, as well as the calibrator diode and 3C286, in volts. The bottom plot shows the same, but after calibration using NGC~7027 smoothed over a day.}
\label{fig:vsa_cal_voltage_flux}
\end{center}
\end{fig}

Due to changes in the relative gains of the two receiver arms over time (including the correction applied to bring them to the same level prior to the double-difference stage), the raw double-difference voltages from the measurements can vary substantially over time, and over the course of the VSA source measurements they decreased by a factor of 2. This should be effectively removed by calibration. Figure \ref{fig:vsa_cal_voltage_flux} shows the fractional difference in the voltage and calibrated flux densities for the measurements of the secondary calibrator sources compared with the mean value for those sources. The systematic decrease in the voltage is evident in the top figure, but post-calibration it has effectively been removed.

The flux densities of two sources, however, increase over time post-calibration. J1521+4336 increases from $\sim350$~mJy to $\sim600$~mJy, and J1219+4829 from $\sim600$~mJy to $\sim1$~Jy. As these were secondary calibrators, and weren't used for flux density calibration, this does not present any issues for the other measurements. Additionally, 3C286, which was also observed during the VSA source observations, displays apparent variability, something that is expected as it is $\sim 12$ per cent polarized at 1.3~cm \citep{2003Perley} and OCRA-p is sensitive to a single linear polarization.

\subsubsection{Duplicate data points and delays}
\begin{fig}
\begin{center}
   \includegraphics[scale=1.0]{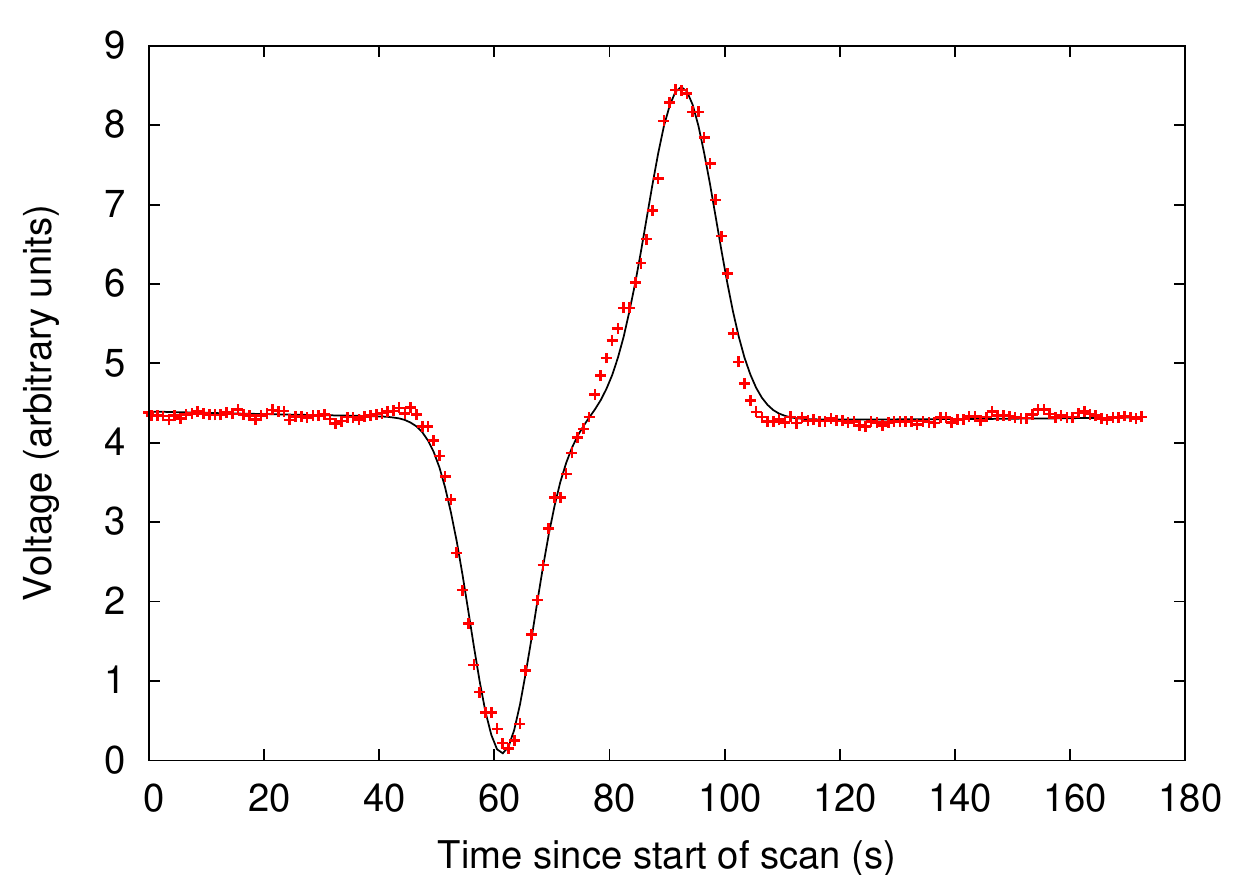}
\caption{An azimuth scan of NGC 7027, showing duplicated and displaced data points}
\label{fig:ngc7027.627_azimuth_scan}
\end{center}
\end{fig}

Some of the records for measurements of the VSA sources contain pairs of adjacent data points that have identical values. Following these, a series of points are displaced by 1 timestep, before a data point is missed out and the displacement ends. An example of this can be seen in Figure \ref{fig:ngc7027.627_azimuth_scan}. The cause of this is lags within the network system at Toru\'n; the measurement files are transmitted over the computer network to the control system, where they are tagged with the time and saved. When a measurement does not reach the control system in time, the previous value is recorded in its place. Due to the irregularity of this effect, and the varying time before the displacement ends, this effect cannot easily be corrected for. However, the effect of this on the fits to the measurements are negligible. The problem will be resolved for the system used for OCRA-F.

\subsection{Flux density values} \label{sec:vsa_sources_fluxes}
The source names and J2000.0 positions at 15~GHz for each of the VSA super-extended array fields are listed in Table \ref{tab:sourcelist_1l}--\ref{tab:sourcelist_7e}, along with their flux densities at 15~GHz from the RT observations and at 30~GHz from the OCRA-p observations from the Author's reduction of the data. We find flux densities for the sources at 1.4 and 4.8~GHz from the NVSS \citep{1998Condon}\footnote{Accessed via \url{http://www.cv.nrao.edu/nvss/}} and GB6 \citep{1996Gregory}\footnote{Accessed via \url{ http://heasarc.gsfc.nasa.gov/W3Browse/all/gb6.html}} catalogues per the description in \citet{2009Gawronski}. We calculate the two-point 1.4 to 30~GHz spectral index for the sources, defined such that $S_\nu \propto \nu^\alpha$. Sources marked with a `v' in the note column are thought to be variable on the basis of the 30~GHz observations (see Section \ref{sec:var}), and those marked with an `e' are extended in FIRST maps (\citealp{1995Becker}; see Section 6.4 of \citealp{2009Gawronski}). Notes on individual sources marked with a * are given in the Appendix of \citet{2009Gawronski}.

\begin{tabland}\small
\caption[Sources in the 1L VSA field]{Sources in the 1L VSA field, centered on J00 16 09.1, +28 16 40. See Section \ref{sec:vsa_sources_fluxes} for details.}
\label{tab:sourcelist_1l}
\begin{center}
\begin{tabular}{l|c|c|r|r|r|r|l|l}
Name & RA (J2000) & Dec (J2000) & $S_{1.4}$ (mJy) & $S_{4.8}$ (mJy) & $S_{15}$ (mJy) & $S_{30}$ (mJy) & $\alpha_{1.4}^{30}$ & Note\\
\hline
J0011+2830 & 00 11 46.71 & 28 30 23.1 & 128.8 $\pm$ 3.9 & 36 $\pm$ 5 & 8.8 & 4.6 $\pm$ 1.0 & -1.09$^{+0.08}_{-0.09}$\\
J0011+2811 & 00 11 50.57 & 28 11 08.4 & 10.0 $\pm$ 0.5 & -- & 3.6 & -3.7 $\pm$ 1.6 & $<-0.7$ & \\ % d was here
J0011+2757 & 00 11 51.72  & 27 57 18.3 & $<$1.5 & -- & 6.0 & 3.8 $\pm$ 0.9 & -- \\
J0012+2821 & 00 12 01.34 & 28 21 47.1 & 8.4 $\pm$ 0.5 & -- & 6.2 & -0.2 $\pm$ 0.7 & $<-0.4$ \\
J0012+2829 & 00 12 44.44 & 28 29 09.5 & 36.8 $\pm$ 1.2 & -- & 4.4 & 4.5 $\pm$ 1.4 & -0.68$^{+0.10}_{-0.13}$\\
J0013+2819 & 00 13 28.04 & 28 19 49.5 & 48.6 $\pm$ 1.5 & 23 $\pm$ 4 & 8.4 & 3.7 $\pm$ 1.6 & $<-0.5$\\
J0013+2834 & 00 13 32.68 & 28 34 48.8 & 39.8 $\pm$ 1.3 & 32 $\pm$ 5 & 23.8 & 14.7 $\pm$ 2.1 & -0.32$^{+0.05}_{-0.06}$ & v\\
J0014+2845 & 00 14 14.80 & 28 45 50.9 & 49.4 $\pm$ 2.1 & -- & 4.6 & 0.9 $\pm$ 1.6 & $<-0.7$ \\
J0014+2821 & 00 14 16.75 & 28 21 09.3 & 263.0 $\pm$ 7.9 & 56 $\pm$ 6 & 12.1 & 1.1 $\pm$ 1.8 & $<-1.2$ \\
J0014+2848 & 00 14 24.78 & 28 48 44.3 & 19.7 $\pm$ 0.7 & -- & 2.6 & 0.6 $\pm$ 3.3 & $<-0.2$ \\
(J0014+2852a) & 00 14 27.74 & 28 52 26.8 & 19.1 $\pm$ 1.0 & -- & 5.5 & 3.1 $\pm$ 1.1 & $<-0.3$ & *\\	% Confirm spectral index
(J0014+2852b) & 00 14 33.14 & 28 52 05.1 & 6.8 $\pm$ 0.5 & -- & 5.4 & 5.4 $\pm$ 1.0 & -0.09$^{+0.08}_{-0.10}$ & *\\	 % Confirm spectral index
J0014+2815 & 00 14 33.84 & 28 15 05.3 & 96.0 $\pm$ 2.9 & 135 $\pm$ 12 & 45.1 & 27.8 $\pm$ 2.7 & -0.40$^{+0.04}_{-0.04}$\\
J0015+2906 & 00 15 13.97 & 29 06 13.0 & 10.6 $\pm$ 0.5 & 20 $\pm$ 4 & 5.6 & 6.3 $\pm$ 1.3 & -0.17$^{+0.08}_{-0.09}$\\
J0015+2831 & 00 15 15.45 & 28 31 53.4 & 3.7 $\pm$ 1.0 &  -- & 3.0 & 4.5 $\pm$ 2.2 & $<1.5$ \\
J0015+2819 & 00 15 36.64 & 28 19 22.9 & 7.1 $\pm$ 0.5 & -- & 2.5 & 1.6 $\pm$ 1.5 & $<0.0$ \\
J0015+2843 & 00 15 42.48 & 28 43 39.8 & 35.7 $\pm$ 1.1 & -- & 7.5 & 6.0 $\pm$ 1.9 & -0.58$^{+0.10}_{-0.13}$\\
J0015+2718 & 00 15 58.02 & 27 18 53.8 & 2.1 $\pm$ 1.5 & -- & 3.1 & -2.5 $\pm$ 1.4 & $<0.4$\\
J0016+2804 & 00 16 02.36 & 28 04 27.4 & 50.2 $\pm$ 1.6 & 20 $\pm$ 4 & 2.7 & 2.1 $\pm$ 2.4 & $<-0.5$ \\
J0017+2733 & 00 17 04.58 & 27 33 42.3 & 11.5 $\pm$ 0.5 & -- & 6.2 & 2.9 $\pm$ 1.8 & $<-0.1$ \\
J0017+2750 & 00 17 23.77 & 27 50 27.5 & 302.0 $\pm$ 10.7 & 88 $\pm$ 8 & 20.9 & 11.3 $\pm$ 1.9 & -1.07$^{+0.06}_{-0.07}$\\
J0017+2835 & 00 17 32.50 & 28 35 12.4 & 82.8 $\pm$ 2.5 & 35 $\pm$ 5 & 9.3 & 5.7 $\pm$ 2.2 & $<-0.6$\\
J0017+2736 & 00 17 59.94 & 27 36 16.1 & 72.8 $\pm$ 2.2 & -- & 4.7 & 3.8 $\pm$ 1.4 & -1.0$^{+0.1}_{-0.2}$\\
J0018+2749 & 00 18 09.49 & 27 49 24.5 & 6.2 $\pm$ 0.4 & -- & 10.4 & 6.4 $\pm$ 2.0 & 0.0$^{+0.1}_{-0.1}$\\
J0018+2843 & 00 18 11.99 & 28 43 56.0 & 17.9 $\pm$ 0.7 & -- & 11.1 & 5.0 $\pm$ 1.2 & -0.42$^{+0.09}_{-0.11}$ & v\\
J0018+2846 & 00 18 12.78 & 28 46 37.6 & 153.7 $\pm$ 5.3 & 54 $\pm$ 6 & 15.0 & 9.5 $\pm$ 2.0 & -0.91$^{+0.07}_{-0.09}$\\
J0018+2731 & 00 18 32.72 & 27 31 02.9 & 150.8 $\pm$ 4.5 & 57 $\pm$ 6 & 21.2 & 11.3 $\pm$ 1.9 & -0.85$^{+0.06}_{-0.07}$\\
J0018+2801 & 00 18 42.15 & 28 01 07.9 & 7.5 $\pm$ 1.0 & -- & 3.7 & 1.5 $\pm$ 2.5 & $<0.2$ \\
J0018+2744 & 00 18 58.17 & 27 44 45.5 & 144.3 $\pm$ 4.3 & 42 $\pm$ 5 & 10.0 & 3.4 $\pm$ 1.5 & $<-0.9$ \\
J0019+2817 & 00 19 08.97 & 28 17 54.6 & 24.7 $\pm$ 0.8 & -- & 32.6 & 28.0 $\pm$ 4.0 & 0.04$^{+0.05}_{-0.06}$ & v\\

\end{tabular}
\end{center}
\end{tabland}
\begin{tabland}
\caption[Sources in the 2G VSA field]{Sources in the 2G VSA field, centered on J09 40 52.7, +31 46 21. See Section \ref{sec:vsa_sources_fluxes} for details.}
\label{tab:sourcelist_2g}
\begin{center}\small
\begin{tabular}{l|c|c|r|r|r|r|l|l}
Name & RA (J2000) & Dec (J2000) & $S_{1.4}$ (mJy) & $S_{4.8}$ (mJy) & $S_{15}$ (mJy) & $S_{30}$ (mJy) & $\alpha_{1.4}^{30}$ & Note\\
\hline
J0936+3203 & 09 36 37.43 & 32 03 31.6 & 191.2 $\pm$ 5.7 & 73 $\pm$ 7 & 22.2 & 4.8 $\pm$ 1.9 & -1.2$^{+0.1}_{-0.2}$ & v\\
J0936+3118 & 09 36 53.39 & 31 18 25.9 & 21.6 $\pm$ 0.8 & -- & 3.1 & -0.0 $\pm$ 2.5 & $<-0.3$ & e\\
J0936+3129 & 09 36 58.25 & 31 29 29.9 & 75.9 $\pm$ 8.3 &  31 $\pm$ 4 & 5.9 & 2.1 $\pm$ 3.3 & $<-0.5$ & e*\\
J0937+3206 & 09 37 06.42 & 32 06 55.4 & 116.0 $\pm$ 3.5 & 92 $\pm$ 9 & 49.3 & 55.6 $\pm$ 5.8 & -0.24$^{+0.04}_{-0.05}$ & e\\
J0937+3201 & 09 37 22.63 & 32 01 03.7 & $<$1.5 &  -- & 5.3 & -1.3 $\pm$ 2.5 & -- \\
J0937+3143 & 09 37 58.41 & 31 43 41.2 & 14.6 $\pm$ 0.6 & -- & 6.5 & -2.4 $\pm$ 3.0 & $<-0.2$ \\
J0938+3118 & 09 38 17.79 & 31 18 52.8 & 24.9 $\pm$ 0.9 & 22 $\pm$ 4 & 13.9 & 5.8 $\pm$ 2.2 & $<-0.2$ & e \\
J0939+3134 & 09 39 48.18 & 31 34 01.9 & 34.2 $\pm$ 1.1 & -- & 3.9 & 6.0 $\pm$ 2.6 & $<-0.3$ & e\\
J0939+3154 & 09 39 50.76 & 31 54 13.7 & 96.5 $\pm$ 2.9 & 30 $\pm$ 4 & 9.8 & 4.8 $\pm$ 2.2 & $<-0.7$ \\
J0939+3122 & 09 39 53.86 & 31 22 43.9 & 92.4 $\pm$ 3.2 & 22 $\pm$ 4 & 8.5 & -0.2 $\pm$ 2.3 & $<-0.8$ & e\\
J0940+3240 & 09 40 38.45 & 32 40 04.8 & 30.5 $\pm$ 1.0 & -- & 3.4 & -2.7 $\pm$ 1.9 & $<-0.7$ \\
J0940+3201 & 09 40 41.85 & 32 01 31.2 & 118.5 $\pm$ 3.6 & 34 $\pm$ 5 & 12.3 & 4.7 $\pm$ 1.6 & -1.1$^{+0.1}_{-0.1}$\\
J0941+3126 & 09 41 03.32 & 31 26 20.1 & 111.1 $\pm$ 3.7 & 67 $\pm$ 7 & 20.0 & 6.3 $\pm$ 3.1 & $<-0.6$ & e*\\
J0941+3221 & 09 41 05.86 & 32 21 45.3 & 381.0 $\pm$ 11.4 & 93 $\pm$ 9 & 20.2 & 11.8 $\pm$ 2.7 & -1.13$^{+0.08}_{-0.09}$\\
J0941+3057 & 09 41 16.03 & 30 57 29.2 & 61.2 $\pm$ 2.3 & 22 $\pm$ 4 & 4.4 & 0.6 $\pm$ 3.0 & $<-0.6$ & e\\
J0941+3146 & 09 41 46.95 & 31 46 47.9 & 3.3 $\pm$ 0.4 & -- & 3.1 & 1.1 $\pm$ 1.7 & $<0.3$ \\
J0941+3154 & 09 41 47.06 & 31 54 53.2 & 288.5 $\pm$ 9.4 & 99 $\pm$ 9 & 22.7 & 11.7 $\pm$ 1.8 & -1.05$^{+0.06}_{-0.06}$ & e\\
J0941+3226 & 09 41 47.47 & 32 26 46.1 & 76.1 $\pm$ 1.2 &  36 $\pm$ 5 & 5.4 & 2.8 $\pm$ 2.7 & $<-0.6$ & e*\\
J0942+3239 & 09 42 00.71 & 32 39 04.5 & 16.4 $\pm$ 0.6 & -- & 17.9 & 6.9 $\pm$ 3.0 & $<0.0$ \\
J0942+3206 & 09 42 08.09 & 32 06 40.7 & 189.3 $\pm$ 6.7 & 62 $\pm$ 7 & 21.4 & 11.3 $\pm$ 2.6 & -0.92$^{+0.08}_{-0.10}$ & e\\
J0942+3150 & 09 42 54.23 & 31 50 50.6 & 105.2 $\pm$ 3.9 & 37 $\pm$ 5 & 11.8 & 4.3 $\pm$ 3.0 & $<-0.6$ & e\\
J0943+3159 & 09 43 08.16 & 31 59 38.0 & 30.9 $\pm$ 1.0 & -- & 12.5 & 2.5 $\pm$ 2.6 & $<-0.3$ & ve \\
J0943+3132 & 09 43 44.12 & 31 32 07.1 & 33.3 $\pm$ 1.4 & -- & 3.4 & -1.1 $\pm$ 2.5 & $<-0.5$ & e\\
J0944+3115 & 09 44 11.60 & 31 15 21.0 & 56.2 $\pm$ 2.1 & 31 $\pm$ 4 & 24.3 & 32.5 $\pm$ 3.1 & -0.18$\pm$0.04 & e\\
\end{tabular}
\end{center}
\end{tabland}
\begin{tabland}
\caption[Sources in the 3L VSA field]{Sources in the 3L VSA field, centered on J15 38 35.3, +41 40 17. See Section \ref{sec:vsa_sources_fluxes} for details.}
\label{tab:sourcelist_3l}
\begin{center}\small
\begin{tabular}{l|c|c|r|r|r|r|l|l}
Name & RA (J2000) & Dec (J2000) & $S_{1.4}$ (mJy) & $S_{4.8}$ (mJy) & $S_{15}$ (mJy) & $S_{30}$ (mJy) & $\alpha_{1.4}^{30}$ & Note\\
\hline
J1535+4103 & 15 35 00.62 & 41 03 13.6 & 79.1 $\pm$ 2.4 & 19 $\pm$ 4 & 5.9 & -1.3 $\pm$ 2.2 & $<-0.8$ & e\\
(J1535+4143a) & 15 35 02.46 & 41 43 04.8 & 51.8 $\pm$ 2.2 & 67 $\pm$ 7 & 3.5 & 7.0 $\pm$ 1.1 & -0.65$^{+0.06}_{-0.07}$ & e*\\	% Check spectral index
(J1535+4143b) & 15 35 05.68 & 41 43 28.2 & 100.8 $\pm$ 3.7 & as above & 2.5 & 6.7 $\pm$ 2.3 & -0.90$^{+0.10}_{-0.13}$ & e*\\	% Check spectral index
(J1535+4142) & 15 35 14.72 & 41 42 54.0 & 85.2 $\pm$ 2.6 & as above & 4.0 & 4.6 $\pm$ 2.9 & $<-0.6$ & e*\\	% Check spectral index
J1536+4125 & 15 36 18.07 & 41 25 38.5 & 78.5 $\pm$ 2.4 & 36 $\pm$ 5 & 8.9 & 4.1 $\pm$ 2.4 & $<-0.6$\\
J1536+4219 & 15 36 49.85 & 42 19 57.5 & 60.5 $\pm$ 2.2 & -- & 6.3 & 2.4 $\pm$ 1.5 & $<-0.7$ & e\\
J1537+4104 & 15 37 11.91 & 41 04 13.0 & 4.8 $\pm$ 1.1 &  -- & 3.3 & 1.0 $\pm$ 1.4 & $<0.4$ & e\\
J1537+4212 & 15 37 14.86 & 42 12 53.6 & 16.4 $\pm$ 0.9 & -- & 9.8 & 9.7 $\pm$ 2.3 & -0.17$^{+0.09}_{-0.11}$\\
J1538+4114 & 15 38 13.93 & 41 14 06.9 & 12.4 $\pm$ 0.5 & -- & 7.5 & 2.6 $\pm$ 1.9 & $<-0.1$ & e\\
J1538+4158 & 15 38 15.57 & 41 58 28.4 & 29.9 $\pm$ 1.0 & -- & 5.8 & 1.0 $\pm$ 2.3 & $<-0.4$ \\
J1538+4105 & 15 38 18.59 & 41 05 48.4 & 60.7 $\pm$ 1.9 & 33 $\pm$ 5 & 25.4 & 28.3 $\pm$ 2.6 & -0.25$\pm 0.04$ & e\\
J1538+4044 & 15 38 32.10 & 40 44 55.3 & 31.3 $\pm$ 1.0 & 20 $\pm$ 4 & 3.2 & 4.0 $\pm$ 1.3 & $<-0.4$ & e\\
J1538+4215 & 15 38 47.13 & 42 15 27.1 & 230.5 $\pm$ 8.0 & 89 $\pm$ 8 & 25.6 & 15.1 $\pm$ 2.2 & -0.89$\pm 0.06$ & e\\
J1538+4225 & 15 38 55.81 & 42 25 27.0 & 57.4 $\pm$ 2.1 & 44 $\pm$ 5 & 31.9 & 27.7 $\pm$ 3.4 & -0.24$\pm 0.05$ & ve\\
J1539+4120 & 15 39 10.52 & 41 20 13.3 & 4.3 $\pm$ 0.4 & -- & 4.0 & 6.5 $\pm$ 2.1 & 0.1$^{+0.1}_{-0.2}$ & e\\
J1539+4217 & 15 39 25.63 & 42 17 28.3 & 49.0 $\pm$ 1.5 & 29 $\pm$ 4 & 31.7 & 32.2 $\pm$ 3.0 & -0.14$\pm 0.04$\\
J1539+4123 & 15 39 36.78 & 41 23 34.8 & 13.8 $\pm$ 0.6 & 26 $\pm$ 4 & 31.8 & 35.4 $\pm$ 3.4 & 0.31$^{+0.04}_{-0.05}$\\
J1539+4148 & 15 39 38.97 & 41 48 51.8 & 3.5 $\pm$ 0.4 & -- & 3.2 & 1.2 $\pm$ 1.4 & $<0.3$ & e\\
J1539+4143 & 15 39 51.38 & 41 43 26.4 & 23.2 $\pm$ 0.8 & -- & 17.3 & 14.9 $\pm$ 2.7 & -0.14$^{+0.07}_{-0.08}$\\
J1540+4221 & 15 40 42.21 & 42 21 37.3 & 62.0 $\pm$ 2.5 & 24 $\pm$ 4 & 2.5 & 2.1 $\pm$ 1.1 & $<-0.7$ & e\\
J1540+4138 & 15 40 43.09 & 41 38 17.4 & 19.4 $\pm$ 0.7 & 20 $\pm$ 4 & 45.6 & 25.4 $\pm$ 3.1 & 0.09$\pm 0.05$ & v\\
J1541+4114 & 15 41 01.24 & 41 14 27.5 & 48.3 $\pm$ 0.8 &  53 $\pm$ 6 & 37.8 & 16.3 $\pm$ 2.3 & -0.35$^{+0.05}_{-0.06}$ & ve*\\
J1541+4049 & 15 41 07.17 & 40 49 17.2 & 159.7 $\pm$ 5.3 & 37 $\pm$ 5 & 10.0 & 4.6 $\pm$ 1.1 & -1.16$^{+0.08}_{-0.10}$ & e\\
J1541+4122 & 15 41 47.41 & 41 22 24.2 & 7.1 $\pm$ 0.5 & -- & 3.5 & 1.2 $\pm$ 1.5 & $<0.0$ & e\\
J1542+4147 & 15 42 40.82 & 41 47 48.3 & 3.2 $\pm$ 0.4 & -- & 5.2 & 4.9 $\pm$ 1.4 & 0.1$\pm 0.1$\\
J1543+4152 & 15 43 14.37 & 41 52 24.0 & 4.2 $\pm$ 0.5 & -- & 5.6 & 2.1 $\pm$ 1.1 & $<0.2$\\
J1543+4139 & 15 43 33.09 & 41 39 32.7 & 37.9 $\pm$ 1.2 & -- & 5.6 & 0.4 $\pm$ 2.2 & $<-0.5$ \\	% Different
\end{tabular}
\end{center}
\end{tabland}
\begin{tabland}
\caption[Sources in the 5E VSA field]{Sources in the 5E VSA field, centered on J03 02 57.0, +26 11 44. See Section \ref{sec:vsa_sources_fluxes} for details.}
\label{tab:sourcelist_5e}
\begin{center}\small
\begin{tabular}{l|c|c|r|r|r|r|l|l}
Name & RA (J2000) & Dec (J2000) & $S_{1.4}$ (mJy) & $S_{4.8}$ (mJy) & $S_{15}$ (mJy) & $S_{30}$ (mJy) & $\alpha_{1.4}^{30}$ & Note\\
\hline
J0259+2627 & 02 59 55.32 & 26 27 29.2 & 13.8 $\pm$ 0.8 & -- & 9.4 & 5.4 $\pm$ 1.5 & -0.30$^{+0.10}_{-0.13}$\\
J0300+2654 & 03 00 15.58 & 26 54 52.2 & 5.3 $\pm$ 0.4 & -- & 5.4 & 2.6 $\pm$ 1.2 & $<0.1$ \\
J0301+2547 & 03 01 05.43 & 25 47 16.1 & 46.3 $\pm$ 1.8 & 23 $\pm$ 4 & 14.2 & 15.0 $\pm$ 2.2 & -0.37$\pm 0.06$\\
(J0301+2541) & 03 01 37.19 & 25 41 56.2 & 300.6 $\pm$ 9.8 & 92 $\pm$ 9 & 18.0 & 10.5 $\pm$ 2.5 & -1.11$^{+0.08}_{-0.10}$ & v*\\	% Check spectral index
J0301+2521 & 03 01 38.18 & 25 21 49.0 & 134.8 $\pm$ 4.1 & 44 $\pm$ 5 & 14.5 & 9.1 $\pm$ 2.1 & -0.88$^{+0.08}_{-0.09}$\\
(J0301+2542) & 03 01 39.80 & 25 42 30.1 & 23.5 $\pm$ 0.8 & 92 $\pm$ 9 & 7.2 & 4.7 $\pm$ 1.5 & -0.5 $\pm$ 0.1 & *\\	% Check spectral index
J0302+2549 & 03 02 19.94 & 25 49 49.0 & 6.7 $\pm$ 0.4 & -- & 7.3 & 6.8 $\pm$ 1.7 & 0.01$^{+0.09}_{-0.11}$\\
J0302+2607 & 03 02 42.78 & 26 07 53.8 & 77.7 $\pm$ 2.4 & 34 $\pm$ 5 & 11.7 & 7.0 $\pm$ 2.2 & $<-0.5$\\
J0302+2625 & 03 02 35.72 & 26 25 53.2 & 26.6 $\pm$ 0.9 & -- & 3.5 & 2.9 $\pm$ 1.6 & $<-0.4$\\
J0303+2641 & 03 03 31.21 & 26 41 20.9 & 20.9 $\pm$ 1.0 & -- & 5.0 & 3.1 $\pm$ 2.0 & $<-0.2$ \\
J0303+2645 & 03 03 34.79 & 26 45 37.4 & 365.0 $\pm$ 11.0 & 102 $\pm$ 10 & 34.1 & 15.4 $\pm$ 2.1 & -1.03$^{+0.05}_{-0.06}$\\
J0303+2700 & 03 03 58.95 & 27 00 51.0 & 3.0 $\pm$ 0.5 & -- & 6.0 & 3.3 $\pm$ 3.1 & $<0.7$ \\
J0303+2531 & 03 03 59.56 & 25 31 34.2 & 153.9 $\pm$ 4.6 & 122 $\pm$ 11 & 55.1 & 46.8 $\pm$ 5.4 & -0.39$^{+0.05}_{-0.05}$ & v\\
J0304+2659 & 03 04 23.60 & 26 59 11.0 & 6.5 $\pm$ 0.7 &  -- & 4.2 & 1.6 $\pm$ 1.2 & $<0.0$ & *\\
J0304+2640 & 03 04 34.02 & 26 40 03.3 & 52.1 $\pm$ 1.9 & 23 $\pm$ 4 & 6.1 & 5.0 $\pm$ 1.2 & -0.77$^{+0.08}_{-0.10}$\\
J0304+2623 & 03 04 40.82 & 26 23 10.6 & 53.1 $\pm$ 2.0 & -- & 4.6 & 4.7 $\pm$ 1.5 & -0.8$\pm 0.1$\\
J0304+2551 & 03 04 58.58 & 25 51 49.5 & 480.8 $\pm$ 14.4 & 116 $\pm$ 11 & 28.9 & 13.5 $\pm$ 1.6 & -1.16$^{+0.05}_{-0.05}$\\
J0305+2532 & 03 05 14.96 & 25 32 16.6 & 20.3 $\pm$ 0.7 & -- & 5.5 & 5.4 $\pm$ 1.5 & -0.43$^{+0.09}_{-0.11}$\\
J0306+2601 & 03 06 25.43 & 26 01 21.5 & 18.7 $\pm$ 0.7 & -- & 7.9 & 13.2 $\pm$ 2.1 & -0.11$^{+0.06}_{-0.07}$\\
J0306+2548 & 03 06 27.66 & 25 48 22.8 & 89.2 $\pm$ 2.7 & 57 $\pm$ 6 & 18.1 & 10.6 $\pm$ 1.8 & -0.70$^{+0.06}_{-0.07}$ & v\\
J0306+2627 & 03 06 32.73 & 26 27 26.3 & 95.2 $\pm$ 2.9 & 25 $\pm$ 4 & 9.1 & 1.9 $\pm$ 1.9 & $<-0.8$\\
J0307+2625 & 03 07 00.33 & 26 25 48.8 & 24.4 $\pm$ 0.8 & 20 $\pm$ 4 & 20.8 & 17.6 $\pm$ 4.3 & -0.11$^{+0.08}_{-0.10}$\\
\end{tabular}
\end{center}
\end{tabland}
\begin{tabland}
\caption[Sources in the 7E VSA field]{Sources in the 7E VSA field, centered on J12 32 21.8, +52 43 27. See Section \ref{sec:vsa_sources_fluxes} for details.}
\label{tab:sourcelist_7e}
\begin{center}\small
\begin{tabular}{l|c|c|r|r|r|r|l|l}
Name & RA (J2000) & Dec (J2000) & $S_{1.4}$ (mJy) & $S_{4.8}$ (mJy) & $S_{15}$ (mJy) & $S_{30}$ (mJy) & $\alpha_{1.4}^{30}$ & Note\\
\hline
J1227+5240 & 12 27 17.93 & 52 40 02.9 & 26.8 $\pm$ 0.9 & -- & 2.6 & 2.5 $\pm$ 1.6 & $<-0.4$ & e\\
J1227+5313 & 12 27 52.75 & 53 13 45.7 & 12.8 $\pm$ 0.5 & -- & 5.5 & 5.0 $\pm$ 1.5 & -0.30$^{+0.10}_{-0.13}$\\
J1228+5308 & 12 28 19.26 & 53 08 24.6 & 6.5 $\pm$ 0.4 & -- & 12.0 & 2.6 $\pm$ 1.3 & $<0.1$ \\
J1230+5230 & 12 30 08.85 & 52 30 51.6 & 74.8 $\pm$ 2.3 & 24 $\pm$ 4 & 7.5 & -1.2 $\pm$ 1.3 & $<-1.0$ & e\\
J1232+5202 & 12 32 18.49 & 52 02 19.2 & 93.0 $\pm$ 2.8 & 21 $\pm$ 4 & 9.2 & 4.1 $\pm$ 2.3 & $<-0.7$ & e\\
J1233+5339 & 12 33 11.08 & 53 39 56.9 & 62.6 $\pm$ 1.9 & 26 $\pm$ 5 & 24.9 & 21.4 $\pm$ 2.4 & -0.35$\pm 0.05$\\
J1233+5337 & 12 33 41.38 & 53 37 27.7 & 80.8 $\pm$ 2.5 & -- & 6.7 & 3.7 $\pm$ 0.9 & -1.00$^{+0.08}_{-0.10}$ &\\
J1234+5322 & 12 34 35.89 & 53 22 35.7 & 44.0 $\pm$ 1.7 & 24 $\pm$ 4 & 5.2 & 4.1 $\pm$ 1.1 & -0.77$^{+0.09}_{-0.11}$ & e\\
J1235+5317 & 12 35 00.71 & 53 17 58.9 & 5.7 $\pm$ 1.0  &  -- & 4.0 & 4.3 $\pm$ 1.0 & -0.1$\pm 0.1$ & *\\
J1235+5311 & 12 35 22.86 & 53 11 29.0 & $<$1.5  &  -- & 6.2 & 5.7 $\pm$ 2.2 & -- & *\\
J1235+5210 & 12 35 29.92 & 52 10 01.4 & 166.0 $\pm$ 5.0 & 41 $\pm$ 5 & 10.5 & 3.5 $\pm$ 1.3 & $<-1.0$\\
J1235+5228 & 12 35 30.53 & 52 28 27.3 & 87.5 $\pm$ 2.7 & 80 $\pm$ 8 & 47.0 & 27.8 $\pm$ 2.6 & -0.37$\pm 0.04$\\
J1235+5315 & 12 35 27.86 & 53 15 01.7 & 49.2 $\pm$ 2.1 & -- & 6.6 & 2.6 $\pm$ 1.3 & $<-0.6$ & e\\
J1237+5325 & 12 37 02.19 & 53 25 25.1 & 71.5 $\pm$ 2.6 & -- & 9.0 & 5.2 $\pm$ 2.8 & $<-0.5$ & e\\
J1237+5254 & 12 37 04.16 & 52 54 22.4 & 373.5 $\pm$ 14.4 & 126 $\pm$ 11 & 35.4 & 20.3 $\pm$ 2.2 & -0.95$\pm 0.05$ & e\\
J1237+5303 & 12 37 25.13 & 53 03 14.9 & 51.9 $\pm$ 1.9 & 24 $\pm$ 4 & 8.1 & 4.0 $\pm$ 1.3 & $<-0.6$& e\\
(J1238+5249a) & 12 38 32.28 & 52 49 15.8 & 93.1 $\pm$ 2.8 & -- & 8.3 & 3.6 $\pm$ 1.1 & -1.05$^{+0.09}_{-0.12}$ & *\\	% Check spectral index
(J1238+5249b) & 12 38 49.78 & 52 49 34.3 & 324.9 $\pm$ 9.8 & 103 $\pm$ 10 & 34.2 & 17.1 $\pm$ 2.0 & -0.95 $\pm$ 0.05 & e*\\	% Check spectral index
\end{tabular}
\end{center}
\end{tabland}

A significant number of measurements had to be discarded due to adverse effects of poor weather. Each source was observed multiple times, with individual measurements with large fit errors removed, giving an average of 8.3 observations per source. The final flux density for the source was calculated using the $1/\sigma^2$ weighted mean and corresponding standard error on the mean ($\sigma_\mathrm{w}$), with the errors on each measurement calculated as per Equation \ref{eq:doubledifference_error}.

The combined uncertainty from the combination of the calibration, atmospheric and gain-elevation corrections and atmospheric effects is $\sim 8$ per cent. The systematic error due to the error in the flux calibrator, NGC~7027, is negligible at 1 per cent. As such, the final errors on the 30~GHz flux densities are calculated by $\sigma = \sqrt{\sigma_\mathrm{w}^2 + (0.08 S)^2}$.

\subsubsection{Comparison to other reductions}
\begin{fig}
\begin{center}
   \includegraphics[scale=1.0]{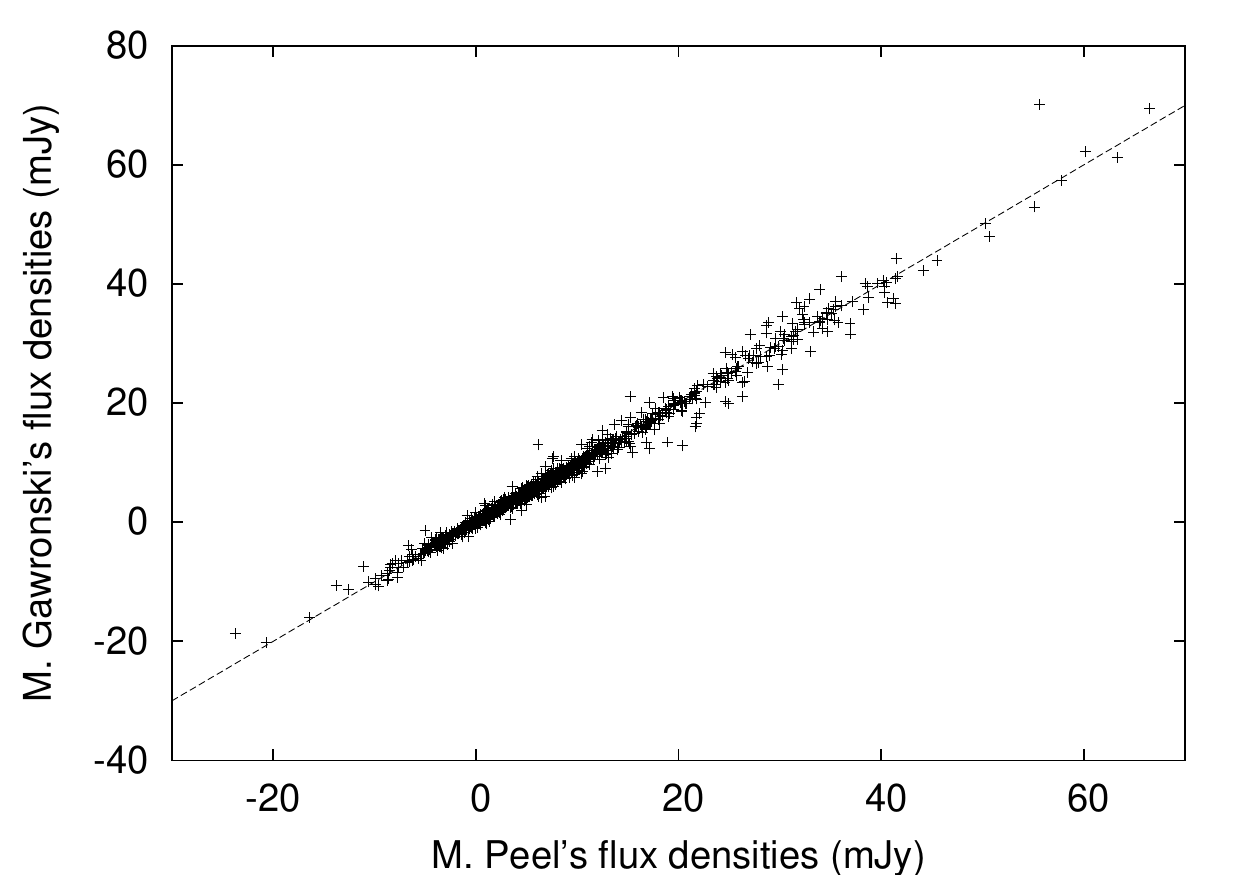}
   \includegraphics[scale=1.0]{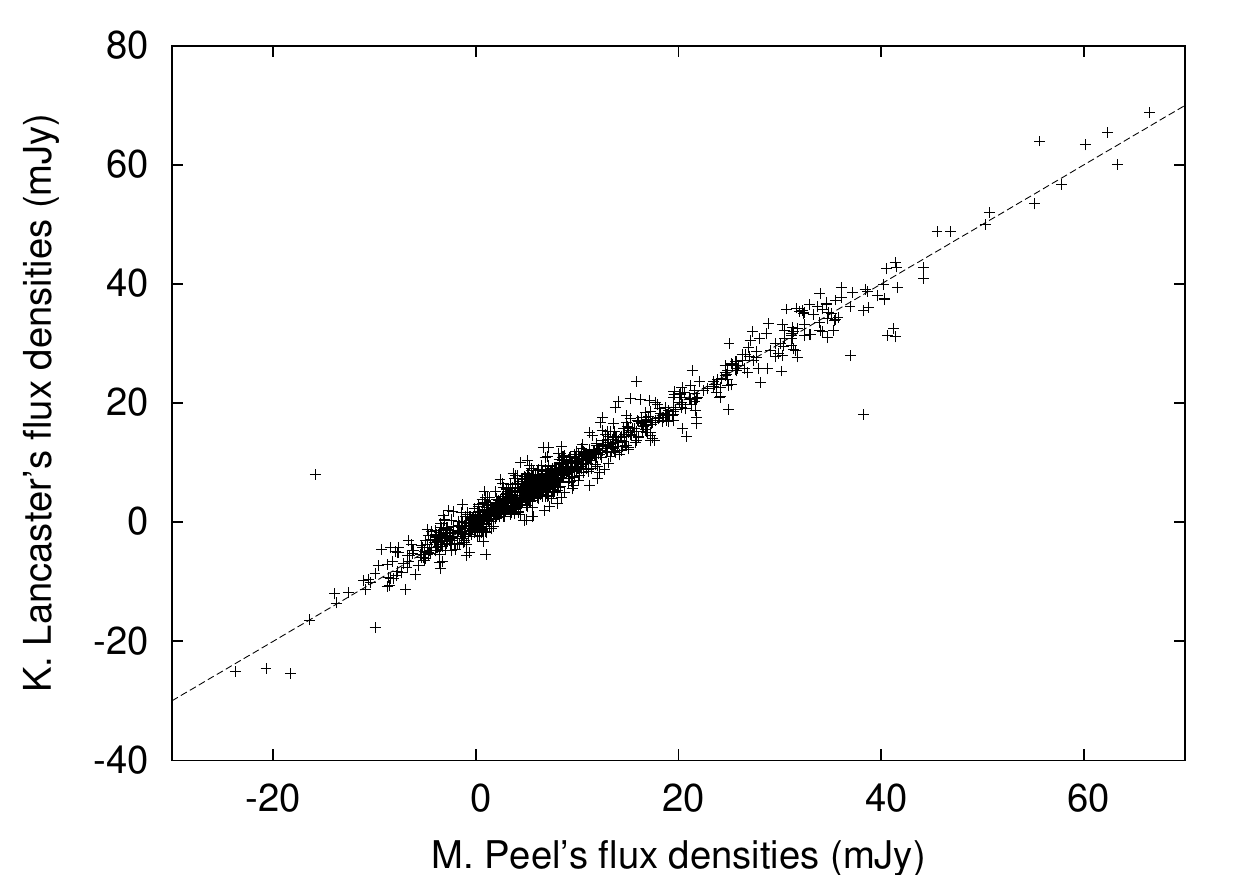}
\caption[Comparison of the different data reductions of the VSA sources]{Comparison of the flux densities for individual measurements. The Author's values are along the X-axis in both, compared M. Gawro\'nski's in the top panel and K. Lancaster's in the bottom. The line in both is $x=y$; the three data reductions fit well to this line, with scatter.}
\label{fig:vsa_compare_marcin_katy}
\end{center}
\end{fig}

To ensure the maximum reliability of the results for the VSA sources, the observations were reduced separately at the three institutions involved in the OCRA collaboration. Figure \ref{fig:vsa_compare_marcin_katy} compares the Author's values for the flux densities of the sources with those by M. Gawro\'nski (Toru\'n), and K. Lancaster (Bristol). The results agree well, with no systematic differences between the analyses. There is some scatter due to the slightly different implementations of the data reduction processes, particularly between the Author's and K. Lancaster's reductions. However this scatter is not significant compared with the error on the measurements.

\subsection{Comparison with previous measurements and source variability} \label{sec:var}
\begin{tabland}\small
\begin{center}
\caption[Comparison of the flux densities from RT, OCRA-p and VLA for the strong VSA sources]{Comparison of the 15 and 30~GHz flux densities from the RT and OCRA-p (in mJy) from common sources measured with the VLA at 1.4, 4.8, 15.2, 22 and 43~GHz by \citet{2004Bolton,2007Waldram} and those measured by the source subtractor \citep{2005Cleary,2008Cleary}.}
\label{tab:compare_to_cleary}
\begin{tabular}{cccccccccc}
& \multicolumn{5}{c}{VLA} & RT & OCRA & VSA\\
\hline
Name & $S_{1.4}$ & $S_{4.8}$ & $S_{15.2}$ & $S_{22}$ & $S_{43}$ & $S_{15}$ & $S_{30}$ & $S_{33}$ & Notes\\
\hline
J0013+2834 & 32.7 & 33.1 & 34.6 & 36.6 & 30.5 & 23.8 $\pm$ 1.2 & 14.7 $\pm$ 2.1 & 36.8 $\pm$ 1.8 & Variable\\
J0014+2815 & 80.4 & 60.1 & 45.5 & 37.7 & 23.8 & 45.1 $\pm$ 2.3 & 27.8 $\pm$ 2.7 & 34.8 $\pm$ 2.6 &\\
J0019+2817 & 25.9 & 23.8 & 17.4 & 27.0 & 34.9 & 32.6 $\pm$ 1.6 & 28.0 $\pm$ 4.0 & 23.9 $\pm$ 1.7 &  Variable (Fig \ref{fig:0016})\\
J0303+2531 & -- & -- & -- & -- & -- & 55.1 $\pm$ 2.8 & 46.8 $\pm$ 5.4 & 30.0 $\pm$ 2.1 &\\
J0937+3206 & 108.9 & 53.9 & 58.4 & 58.8 & 41.0 & 49.3 $\pm$ 2.5 & 55.6 $\pm$ 5.8 & --\\
J0944+3115 & -- & -- & -- & -- & -- & 24.3 $\pm$ 1.2 & 32.5 $\pm$ 3.1 & 25.3 $\pm$ 2.6 &\\
J1235+5228 & -- & -- & -- & -- & -- & 47.0 $\pm$ 2.4 & 27.8 $\pm$ 2.6 & 27.9 $\pm$ 1.8 &\\
J1539+4123 & -- & -- & -- & -- & -- & 31.8 $\pm$ 1.6 & 35.4 $\pm$ 3.4 & 33.7 $\pm$ 2.4 &\\
J1539+4217 & 53.3 & 40.0 & 34.2 & 37.0 & 26.3 & 31.7 $\pm$ 1.6 & 32.2 $\pm$ 3.0 & 34.0 $\pm$ 1.6 &\\
J1538+4225 & 42.0 & 40.5 & 41.6 & 42.2 & 29.0 & 31.9 $\pm$ 1.6 & 27.7 $\pm$ 3.4 & 41.0 $\pm$ 1.5 & Variable\\
J1540+4138 & 16.0 & 30.9 & 34.2 & 23.0 & 9.1 & 45.6 $\pm$ 2.3 & 25.4 $\pm$ 3.1 & -- &\\
J1541+4114 & 65.0 & 38.0 & 30.5 & 27.5 & 19.2 & 37.8 $\pm$ 1.9 & 16.3 $\pm$ 2.3 & 32.5 $\pm$ 2.3 & Variable\\
\end{tabular}
\end{center}
\end{tabland}

The stronger sources ($> 10$mJy at 15~GHz) within the 33 extended VSA fields were observed during the initial VSA campaign at 33~GHz using the VSA source subtractor, with those greater than 20~mJy at 33~GHz used for direct subtraction from the CMB data \citep{2004Dickinson}. A subsample of these was published by \citet{2005Cleary,2008Cleary}. The flux densities measured by the source subtractor for the sources in \citet{2005Cleary, 2008Cleary} that also lie within the super-extended array fields, as well as measurements of those sources made at a range of frequencies by \citet{2004Bolton} and \citet{2007Waldram} using the Very Large Array, are listed in Table \ref{tab:compare_to_cleary} along with the measurements by the RT and OCRA. A number of the sources show evidence of variability.

\begin{fig}
\centering
\includegraphics[scale=0.5]{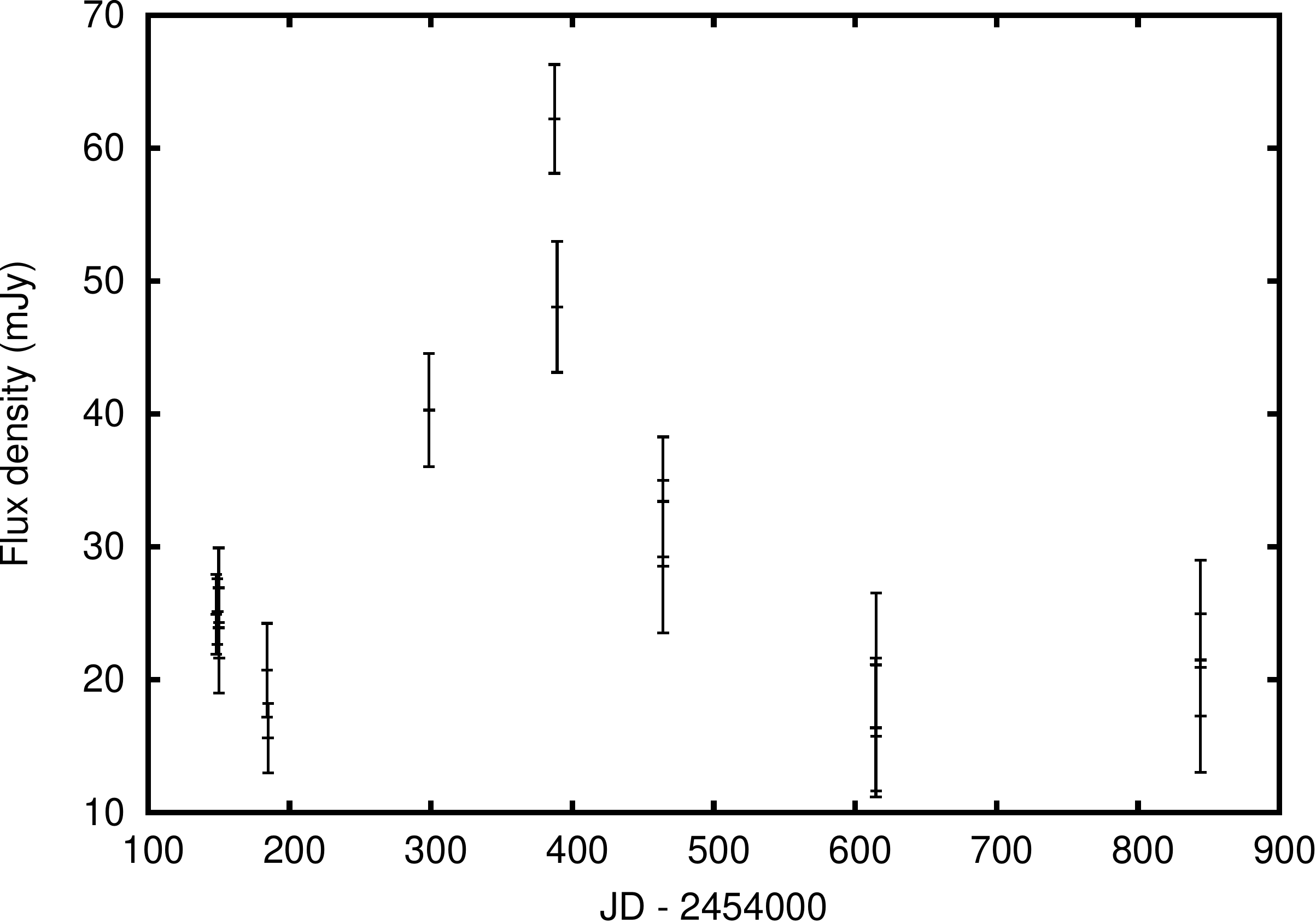}
\caption[The flux density measured for J0019+2817 as a function of time]{The flux density measured for J0019+2817 as a function of time. The source is clearly variable.}
\label{fig:0016}
\end{fig}

We note that OCRA-p is only sensitive to a single linear polarization, which could result in discrepancies between measurements from the VSA, VLA and OCRA for polarized sources; however these are likely to be small since the WMAP sources selected at 22~GHz (see \citealp{2009Jackson}) exhibit average linear polarizations of $\sim$3 per cent. Assuming that the polarization percentages of the VSA sources are similar to the WMAP sources, then the effects of such low polarization would not be noticed. The sources observed with the higher resolution interferometers (RT and VLA) may also be slightly resolved. Nevertheless we identify at least four sources (J0013+2834; J1538+4225; J1541+4114; J0019+2817) as likely to be intrinsically variable. One of these sources -- J0019+2817 -- shows clear variability from OCRA data alone (see Figure \ref{fig:0016}), although the mean flux density from the measurements agrees well with the source subtractor measurement.

\begin{fig}
\centering
\includegraphics[scale=0.5]{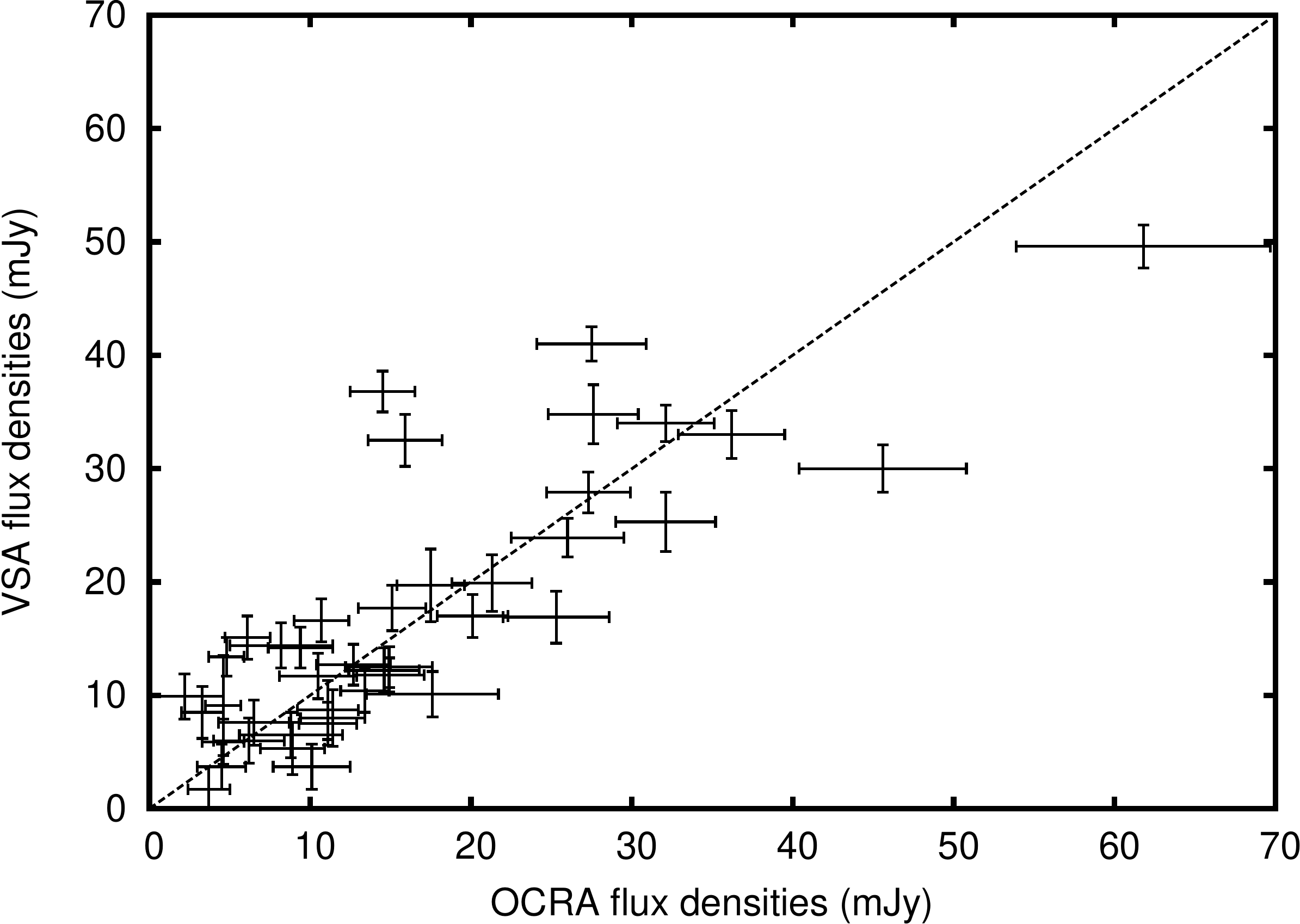}
\caption[Comparison of the OCRA-p flux densities against those from the VSA source subtractor]{Comparison of the OCRA-p flux densities against those from the VSA source subtractor.}
\label{fig:flux_comparison}
\end{fig}

A comparison of flux densities for 42 sources observed both with the VSA source subtractor and OCRA-p is shown in Figure \ref{fig:flux_comparison}. The flux density scales are consistent, however 10 out of the 42 sources show variations greater than $2 \sigma$ (combined error), some of which show variations greater than a factor of 2. These are denoted with a `v' in Tables~\ref{tab:sourcelist_1l}-\ref{tab:sourcelist_7e}.

\subsection{Source spectra} \label{sec:spectra}
\begin{fig}
\centering
\includegraphics[scale=0.5]{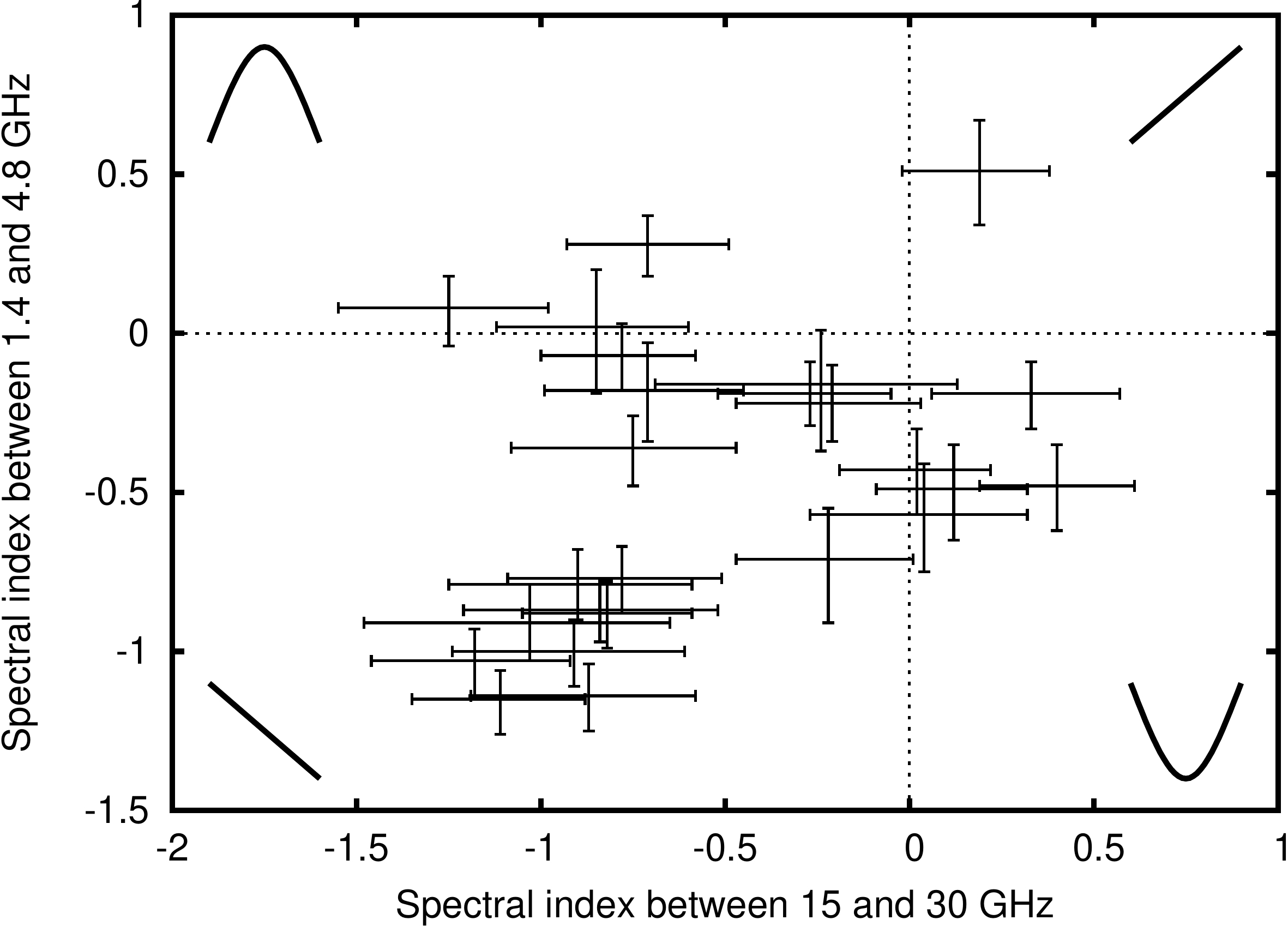}
\caption[$\alpha_{1.4}^{4.8}$ vs. $\alpha_{15}^{30}$ for all sources in the VSA source sample with a 30~GHz flux density greater than 10~mJy, and known flux densities at all four frequencies]{$\alpha_{1.4}^{4.8}$ vs. $\alpha_{15}^{30}$ for all sources in this sample with a 30~GHz flux density greater than 10~mJy, and known flux densities at all four frequencies. The diagrams in the corners schematically illustrate the spectral behaviour of the sources in each quadrant.}
\label{fig:2colour}
\end{fig}

Figure \ref{fig:2colour} shows the `2-colour' diagram for sources detected at all four frequencies (1.4, 4.8, 15, 30~GHz). We find few sources with peaked spectra (top left quadrant of Figure \ref{fig:2colour}), but a larger number of sources with spectra rising towards high frequencies (right hand quadrants of Figure \ref{fig:2colour}).

Figure \ref{fig:1p4to30} displays the distribution of spectral indices $S \propto \nu^\alpha$ between 1.4 and 30~GHz for 60 sources above the completeness limit of 7~mJy at 15~GHz (5 out of a total of 65 sources are excluded due to their complex morphology -- 0258+2530a/b, 0938+3140 and 1236+5305/6). There is clear evidence of bimodality, with a split around a spectral index of -0.5, dividing the sources into groups of sources with `flat' and `steep' spectra.

The distribution is markedly different from the 1.4 to 30~GHz spectral index distribution presented by \citet{2009Mason}. However, their source sample was selected at 1.4~GHz rather than 15~GHz as in this paper. The distribution in \citet{2009Mason} is not bimodal, but has a single peak around $-1.0$, implying that their sample is dominated by the steep spectrum sources. Our sample contains a larger fraction of flat and rising spectrum sources. This is not surprising given the difference of a factor ten in the sample selection frequency.

\subsection{The 1.4-30~GHz spectral index distribution vs flux density }
We are seeking to improve our knowledge of the source population at 30~GHz down to mJy flux densities and beyond (by extrapolation). In order to extrapolate one needs to know if there is a dependence of the spectral properties of the population on flux density at 30~GHz. We have therefore used the WMAP 5-year point source catalogue \citep{2009Wright} to select a complete strong source sample consisting of all sources with a 22~GHz flux above 1~Jy. We have then cross-identified these with the NVSS catalogue and calculated the spectral index distribution between 1.4 and 33~GHz. We chose 22~GHz as the selection frequency to be as close as possible to the 15~GHz selection frequency used in the present work.

\begin{fig}
\centering
\includegraphics[scale=0.5]{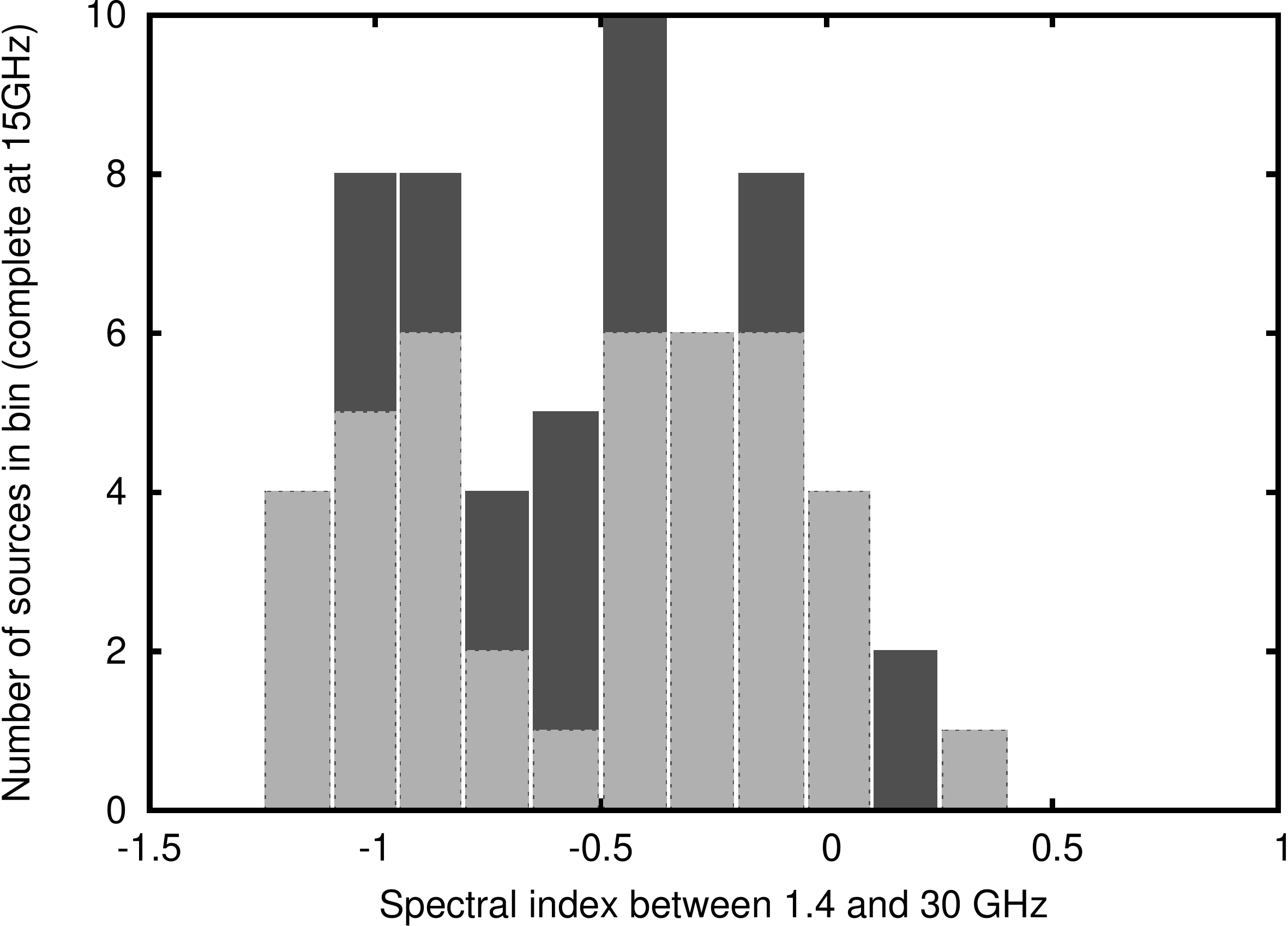}
\caption[The distribution of spectral indices between 1.4 and 30~GHz ($\alpha_{1.4}^{30}$) for 60 sources in the VSA source sample with a 15~GHz flux density greater than 7~mJy]{The distribution of spectral indices between 1.4 and 30~GHz ($\alpha_{1.4}^{30}$) for 60 sources in this sample with a 15~GHz flux density greater than 7~mJy. $3 \sigma$ detections at 30~GHz are shown in the light grey histogram; the dark grey histogram contains the $3 \sigma$ upper limits on the spectral indices.}
\label{fig:1p4to30}
\end{fig}

\begin{fig}
\centering
\includegraphics[scale=0.5]{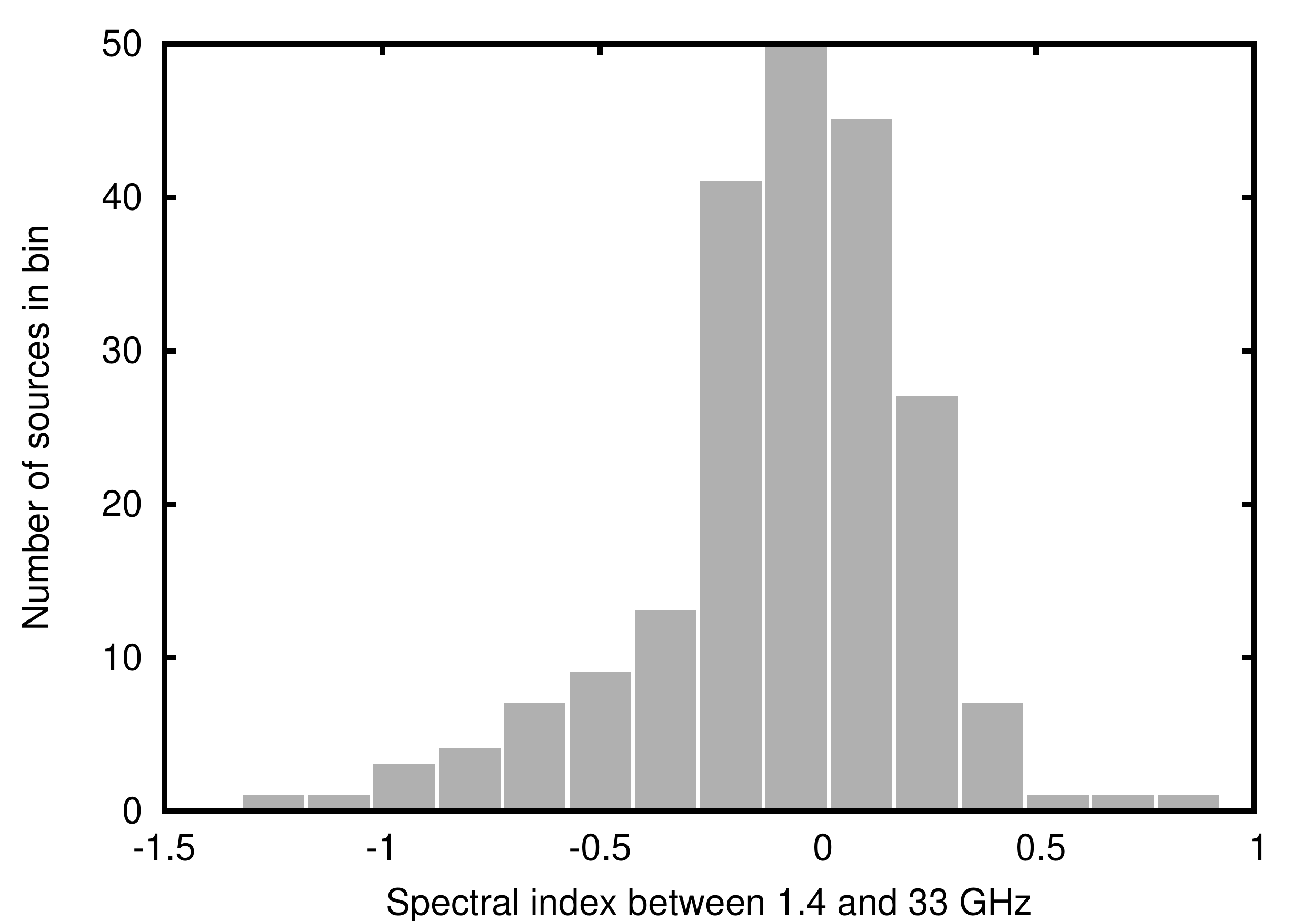}
\caption[The spectral index distribution of 22~GHz-selected WMAP sources between 1.4 and 33~GHz]{The spectral index distribution of 22~GHz-selected WMAP sources between 1.4 and 33~GHz. The majority of sources have flat spectra ($\alpha > -0.5$).}
\label{fig:wmap_indices}
\end{fig}

In this analysis, we exclude sources that are known to be extended on scales greater than $\sim30$~arcmin (WMAP J0322-3711/Fornax A; WMAP J1633+8226/NGC~6251) or confused in WMAP (WMAP J0223+4303/3C66 A and B). Sources are matched to single NVSS sources, except for WMAP J0108+1319/3C33 where four close NVSS sources are combined. In total, 211 sources are used in the analysis. Figure \ref{fig:wmap_indices} shows the spectral index distribution, which has a single peak centered on zero. All but 16 sources are classified as flat spectrum, i.e. have $\alpha > -0.5$. As expected, this is also very different to the distribution found by \citet{2009Mason}, which is dominated by steep spectrum sources.

There is no evidence for a large population of steep spectrum sources in the WMAP spectral index distribution. This contrasts with the VSA source distribution shown in Figure \ref{fig:1p4to30} where there is evidence for both flat and steep spectrum populations of comparable size. This difference could be ascribed to the selection frequencies (22~GHz cf. 15~GHz), however their closeness suggests that this should not be a major issue. Another consideration is that although we are selecting using the 22~GHz flux densities, in reality the WMAP point source catalogue is obtained using all of the bands \citep{2009Wright}. This should also not be an issue as the sources with 22~GHz flux density greater than 1~Jy are detected by WMAP at high significance. A final possibility, which we think is most likely, is that the dissimilar distributions are due to the difference in the flux densities of the sources in the samples. The population of steep spectrum sources increases as the flux densities decrease.

Such an effect is also seen in \citet{2009Waldram}, where they find a significant change in spectral index distribution with flux density amongst their 15~GHz selected sources. We note, however, that there is some overlap in sources between their samples and the one studied here.

\begin{fig}
\centering
\includegraphics[scale=0.29]{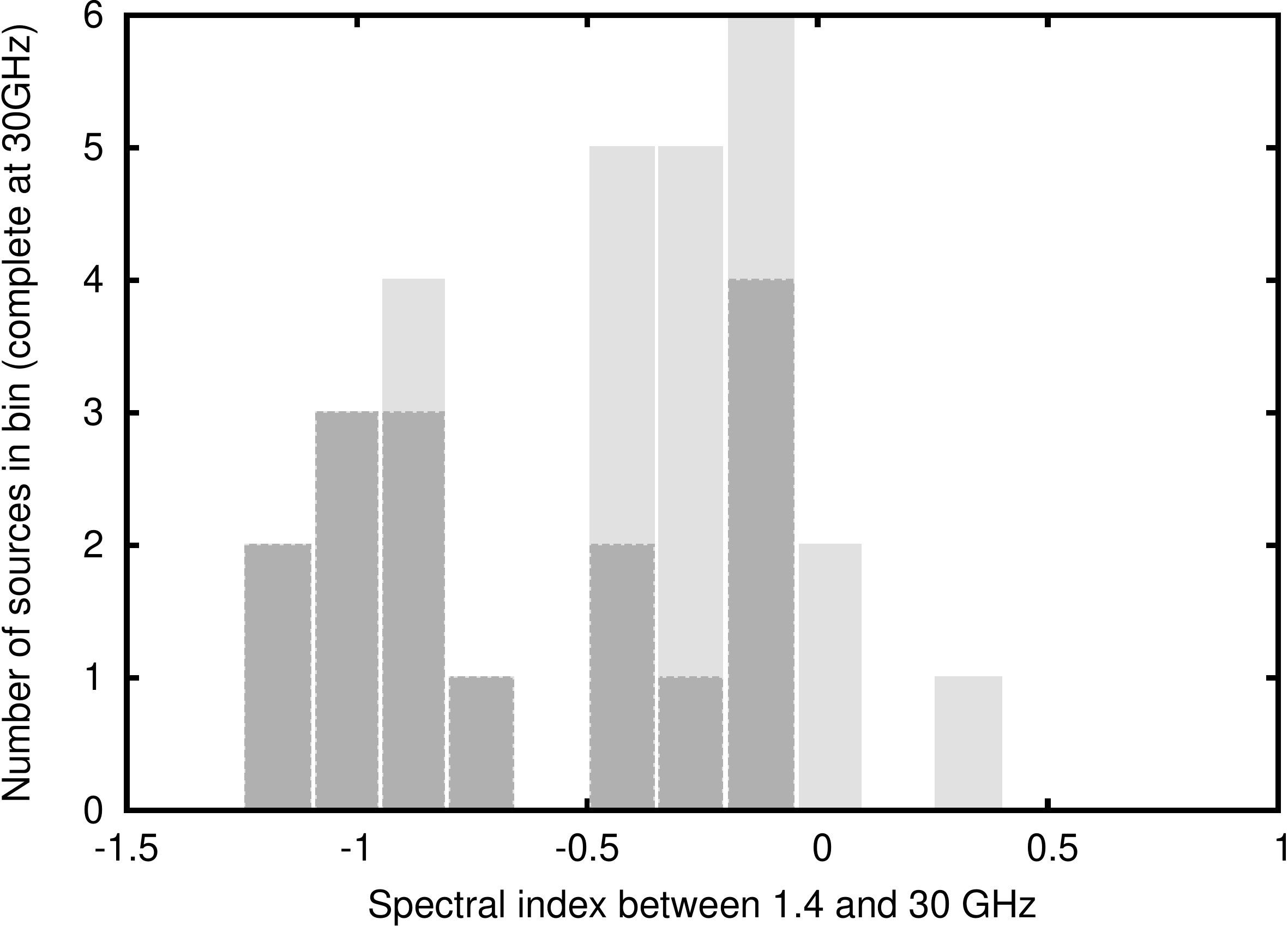}
\includegraphics[scale=0.29]{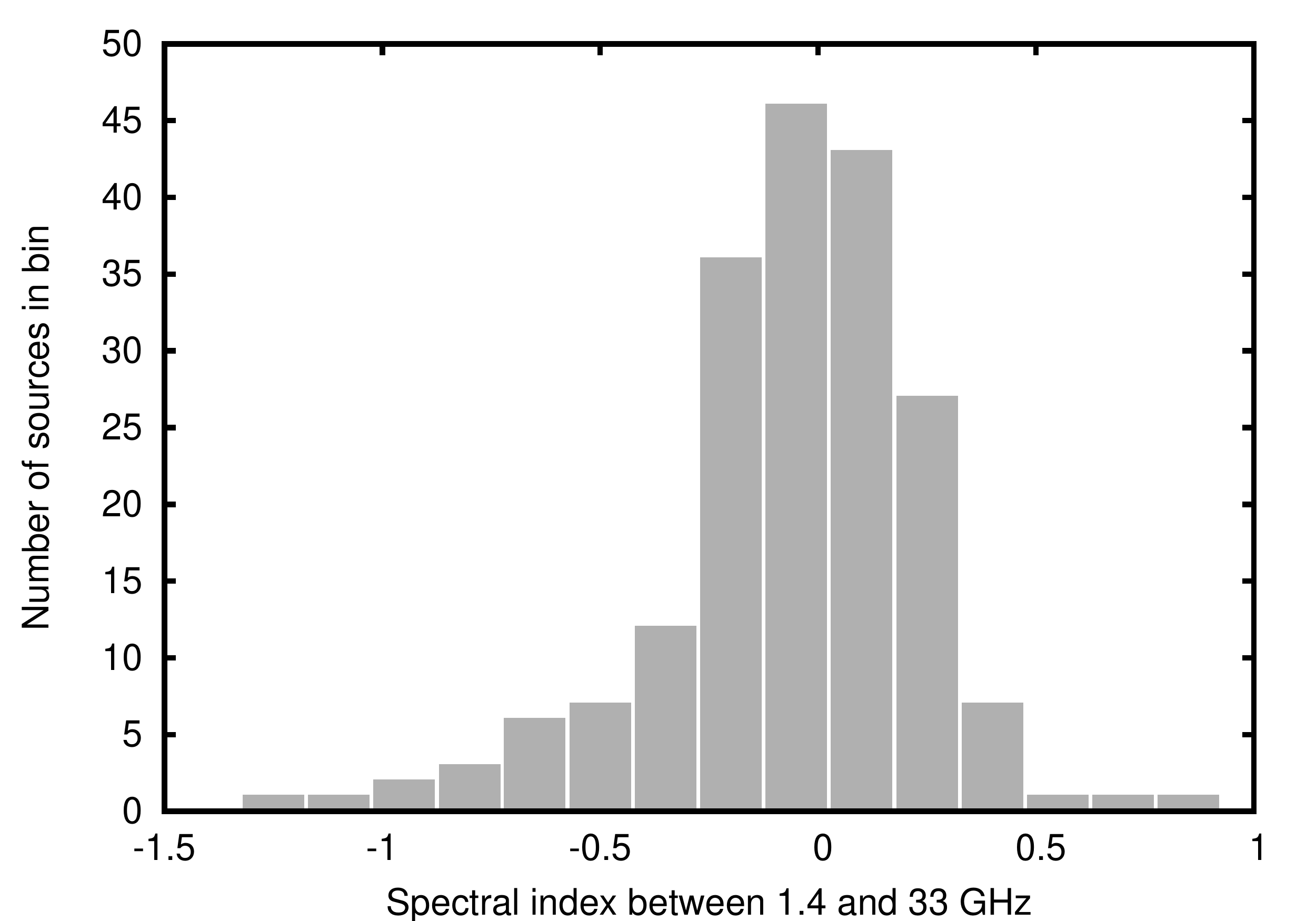}
\caption[The distribution of spectral indices between 1.4 and 30~GHz ($\alpha_{1.4}^{30}$) for 29 sources in this sample with a 30~GHz flux density greater than 10~mJy, compared to the distribution between 1.4 and 33~GHz of WMAP sources selected to be stronger than 1 Jansky at 33~GHz] {Left: The distribution of spectral indices between 1.4 and 30~GHz ($\alpha_{1.4}^{30}$) for 29 sources in this sample with a 30~GHz flux density greater than 10~mJy. The darker boxes represent sources with flux density between 10 and 20~mJy at 30~GHz; the lighter boxes represent the sources with greater than 20~mJy at 30~GHz. The distribution of the stronger sources closely resembles that of the WMAP sample. Right: The spectral index distribution between 1.4 and 33~GHz of WMAP sources selected to be stronger than 1 Jansky at 33~GHz. The distribution is essentially the same as for the 22~GHz selected sample, but with slightly fewer sources with spectral indices between 0 and -1.}
\label{fig:1p4to30_30}
\end{fig}

The bimodality of the spectral index distribution remains the same when the sources in the present sample are selected at 30~GHz. The left hand part of Figure \ref{fig:1p4to30_30} shows the spectral index distribution of the VSA sources selected to be complete at 30~GHz ($S_{30} > 10$~mJy), and the right hand part shows the spectral indices of WMAP sources selected to be greater than 1~Jy at 33~GHz. Splitting the VSA source sample complete at 30~GHz into high ($>$20~mJy) and low ($<$20~mJy) samples shows that the higher flux density sources have a spectral index distribution closer to the WMAP distribution; the lower flux density sources become increasingly more steep spectrum. A Kolmogorov-Smirnov comparison of the VSA ($S_{30} > 10$~mJy) and WMAP distributions shows that the samples are different at the 95 per cent confidence level. The semi-empirical model described by \citet{2005deZotti} predicts a cross-over between the dominance of flat and steep spectrum sources at about 30~mJy (their Figure 14). Our observations display a sharper cross-over than the models appear to suggest, and at the slightly lower flux density of 20~mJy, but are otherwise consistent with the predictions by \citet{2005deZotti}.

\subsection{Comparison of estimates of source surface densities at 30~GHz}
\begin{fig}
\centering
\includegraphics[scale=0.5]{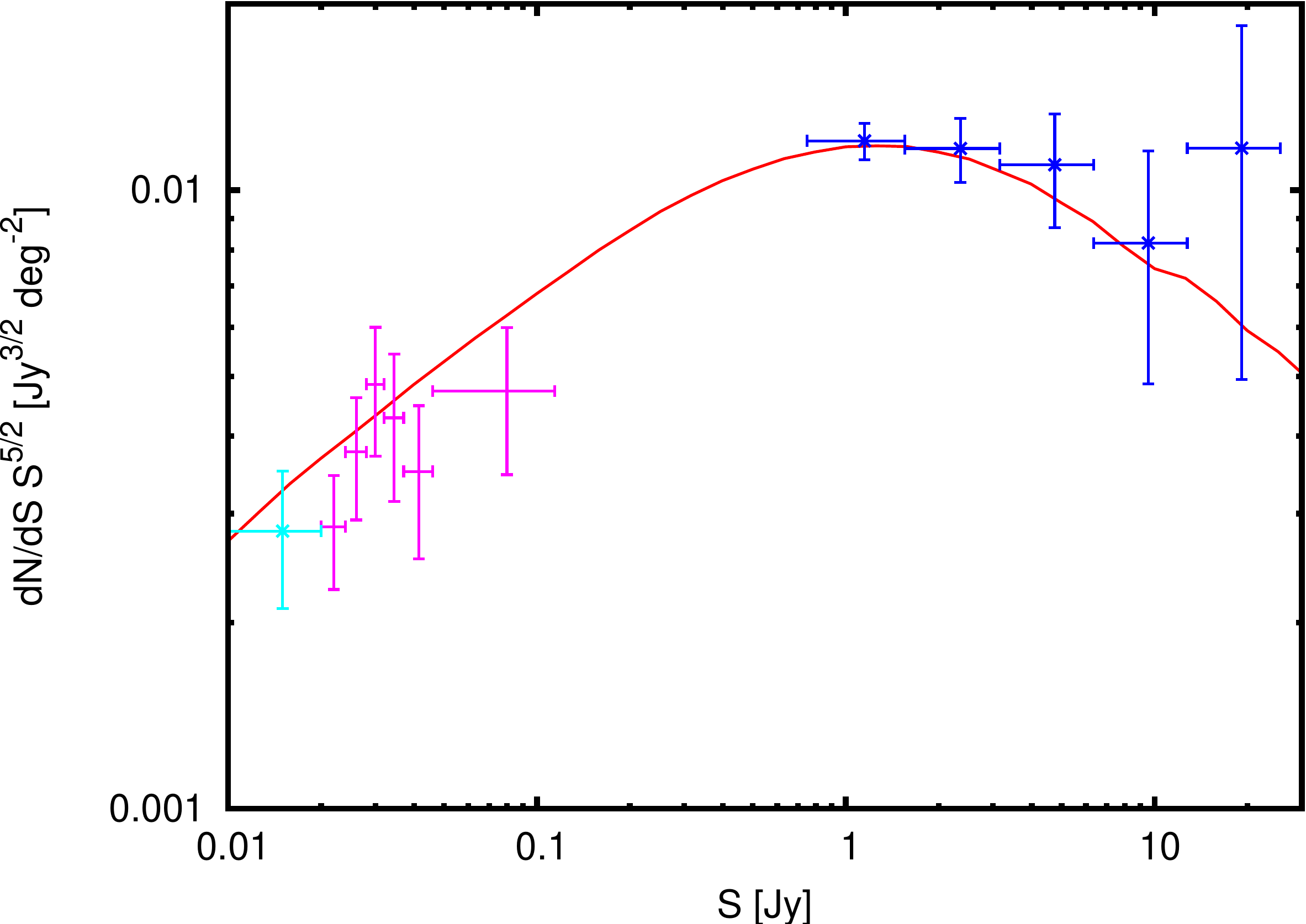}
\caption[Differential source counts at 30~GHz from WMAP, VSA and OCRA, compared with T98] {Differential source counts at 30~GHz. The solid line is the model of \citet{1998Toffolatti} at 30~GHz, normalized by 0.7. The blue data points are from the WMAP 5 year survey \citep{2009Wright}, the pink data points from the Very Small Array \citep{2005Cleary} and the turquoise point from the OCRA observations of the sources in the VSA fields, between 10 and 20~mJy. The OCRA point agrees with the model within the 1$\sigma$ error bars.}
\label{fig:vsa_sourcecounts}
\end{fig}

There are 31 sources in the five VSA fields detected above 10~mJy at 30~GHz. These effectively comprise a complete sample as the sources were selected from a deeper survey at the nearby frequency of 15~GHz. The survey area is 15.69~deg$^{2}$. The surface density of sources is thus $2.0 \pm 0.4$ sources per square degree (using the Poissonian error), which can be compared with the extrapolated value of $2.65 \pm 0.1$ in \citet{2009Mason}. \citet{2007Coble} also give an estimate of the surface density of 30~GHz sources in the field, which gives $2.2^{+2.5}_{-1.2}$ sources per square degree; they also point out that the density of sources in clusters of galaxies is significantly higher. Figure \ref{fig:vsa_sourcecounts} shows the differential source counts at 30~GHz from WMAP and VSA observations, as well as the differential source count for the sources in this sample between 10 and 20~mJy. The model of \citet{1998Toffolatti} is also shown, normalized by 0.7 (see Section \ref{sec:lowfreqps}). The OCRA data point is consistent with the model within the 1$\sigma$ error bars.

\subsection{Conclusions}
In order to aid the subtraction of individual sources from the VSA fields observed at $\sim$30~GHz, and to obtain a statistical estimate of the surface density of sources at 30~GHz, we have observed a sample of 121 sources using the OCRA-p receiver on the Toru\'n 32~m telescope; the sample was selected at 15~GHz with the RT. At 30~GHz, we detected 57 sources above a limiting flux density of $\sim 5$~mJy. This is the deepest follow-up of any complete sample of sources detected at 15~GHz by the RT.

At a flux density of 10~mJy, which is our estimated completeness limit, we derive a surface density of sources at 30~GHz of $2.0~\pm~0.4$ per square degree. This is consistent with the value obtained by \citet{2009Mason}, who observed a much larger sample of sources down to mJy levels but selected at a much lower frequency (1.4~GHz). The potential danger of using low frequency selected samples is that there may exist a significant population of sources with steeply rising spectra towards high frequencies that are not present in the low frequency surveys. As the two surface density estimates are consistent, this indicates that such a population is not obviously present at the 10~mJy level.

We have compared our flux density measurements with those from the VSA source subtractor and VLA measurements. These comparisons give confidence in our flux scale but reveal that a significant fraction of sources are variable on a timescale of a few years, some at the level of a factor of 2. This shows the importance of taking contemporaneous measurements of discrete sources in conjunction with measurements of the CMB.

We have also investigated the dependence of the spectral index distribution on flux density by comparing our measured spectral index distribution with that for much stronger sources (above 1~Jy) selected from the WMAP 22~GHz catalogue. We conclude that the proportion of steep spectrum sources increases with decreasing flux density. This is qualitatively consistent with models of source populations, for example \citet{2005deZotti}.

\section{CRATES source sample} \label{section:crates}

The Combined Radio All-sky Targeted Eight GHz Survey \citep[CRATES;][]{2007Healey} is an all sky sample of strong flat spectrum sources at 8.4~GHz. In the northern hemisphere the sample was selected from the 4.85~GHz GB6 catalogue \citep{1996Gregory} compared with 1.4~GHz fluxes from the NRAO VLA Sky Survey \citep[NVSS;][]{1998Condon}; it is essentially a subset of the Cosmic Lens All-Sky Survey \citep[CLASS;][]{2003Myersa,2003Browne}. CRATES combines other surveys for the southern hemisphere, as well as at declinations above $75 \degree$ where GB6 is incomplete. CRATES is currently the most complete large-area high frequency weak point source sample existing prior to observations by Planck; it has flux densities lower by over an order of magnitude than the WMAP source sample \citep{2009Wright}. Efforts to improve its completeness are ongoing \citep{2009Healey}.

The sample was originally selected to study blazars -- radio-loud Active Galactic Nuclei where we are looking down the jet axis. This angle of observation means that the sources are doppler boosted such that their received flux densities are increased.

The knowledge of the high frequency flux densities of these sources has several uses. Knowledge of the spectra of the sources is of use for investigations of the physics behind the sources. Higher frequency follow-ups are a possibility, and in the long term surveys with SCUBA2 will likely also detect some of these sources at high frequencies (300~GHz+), thus giving excellent spectral energy distributions.

The sources contained in CRATES are also the most likely ones to contaminate {\it Planck} measurements of the CMB. {\it Planck} will be most sensitive at the north ecliptic pole, hence this is the most logical place to observe a sub-sample of the CRATES sources. Potentially, the source flux density list could be used for subtraction from the 30~GHz LFI maps, and for cross-comparison with the source lists generated by Planck HFI. It would also be useful to know whether the sources will contaminate HFI measurements at 100, 143 and 217~GHz. This would require high-frequency follow-up observations.

OCRA has previously been used to observe the Caltech-Jodrell Bank flat-spectrum (CJF) sample, a set of 293 strong ($>$350~mJy at 4.85~GHz) flat spectrum radio sources; the observation of these by OCRA-p is described in \citet{2007Lowe} and \citet{2006Lowe}. A survey of a sample of CRATES sources essentially extends this work to lower flux densities. OCRA-F will follow up this work by doing a wide area blind survey to reasonable flux densities ($\sim$10~mJy) in this part of the sky.

\subsection{Subsample selection}
\begin{fig}
\centering
\includegraphics[scale=1.0]{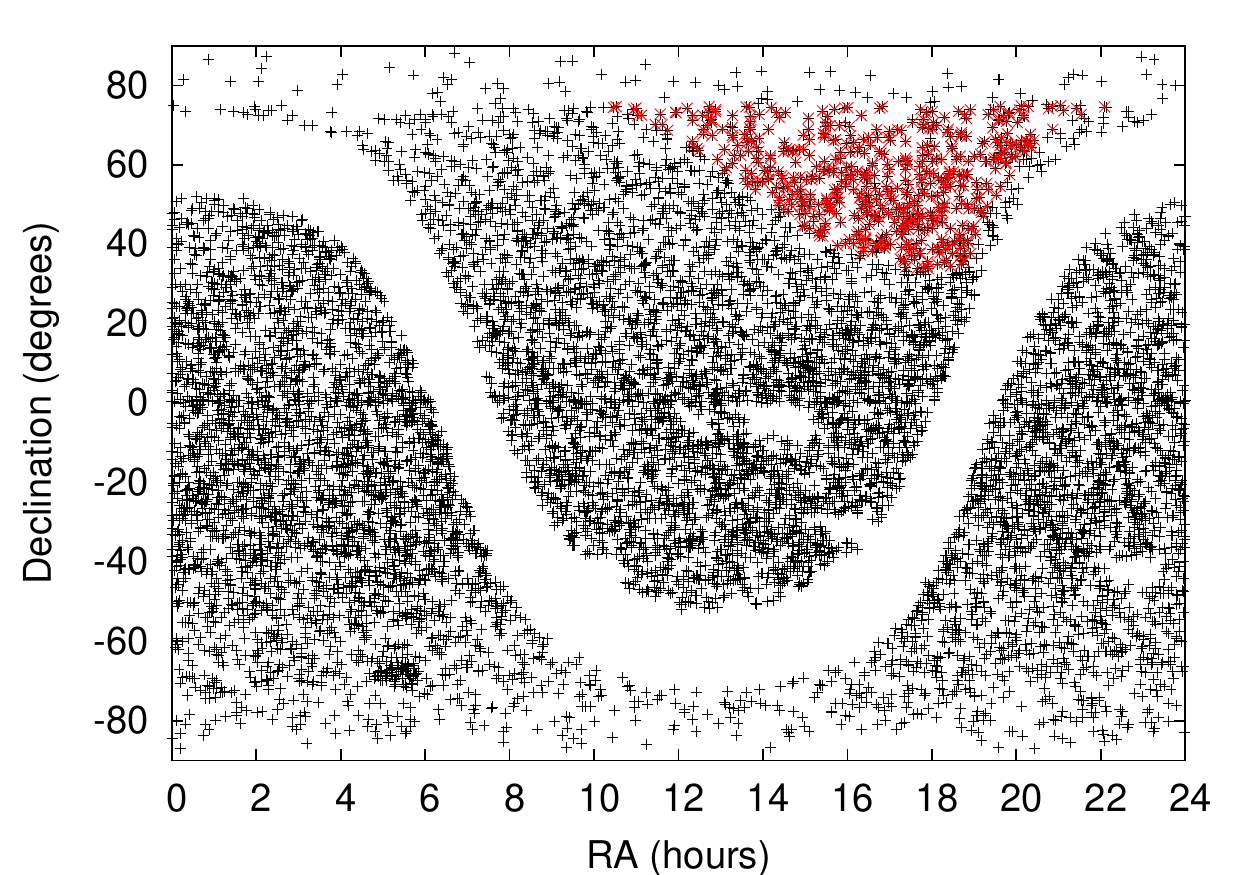}
\caption[The CRATES sources and the selected sample] {The distribution of CRATES sources in RA and Dec. The complete CRATES sample is shown in black. The sample observed by OCRA is shown in red. The cluster of sources in the bottom-left is the Large Magellanic Cloud, and the gap around Dec 0$\degree$ are where there are gaps in the 4.8~GHz observation coverage \citep{2007Healey}. The reduced number of sources above 75$\degree$ is caused by the upper limit in the GB6 observations.}
\label{fig:crates_sources}
\end{fig}

The CRATES sample consists of extragalactic sources ($|b| > 10 \degree$) that have spectral indices between 1.4 and 4.8~GHz of $\alpha > -0.5$, and have 4.8~GHz fluxes greater than 65 mJy. These limits yield 11~000 sources in the complete sample. The subsample observed with OCRA have a series of additional selection criteria. We require that the sources:
\begin{itemize}
\item are closer than 33 degrees from the North Ecliptic Pole (located at RA 18h, dec 66.5 degrees),
\item have a declination of less than 75 degrees (the GB6 survey limit),
\item are no closer than 15 degrees to the galactic plane ($|b| > 15$), and
\item have a measured 8.4~GHz flux density
\end{itemize}

This selection yields 693 sources. However, the CRATES catalog includes small-scale components within single sources, which are unobservable with the $\sim1$ arcmin OCRA beam. For sources with multiple components closer together than an arcmin we add their 8.4~GHz flux densities together. The position of the brightest component is used as the position of the source. The final source list contains 550 sources - a number that is feasible to observe with OCRA-p.

The RA range of the source sample is $10^\mathrm{h}~27^\mathrm{m}$ to $22^\mathrm{h}~6^\mathrm{m}$; the declination range is from $33\degree~51^\mathrm{m}$ to $75\degree$. The 550 CRATES sources within this sample are shown in red in Figure \ref{fig:crates_sources}, with all of the CRATES sources shown in black.

\subsection{Observations}
\begin{fig}
\begin{center}
   \includegraphics[scale=1.0]{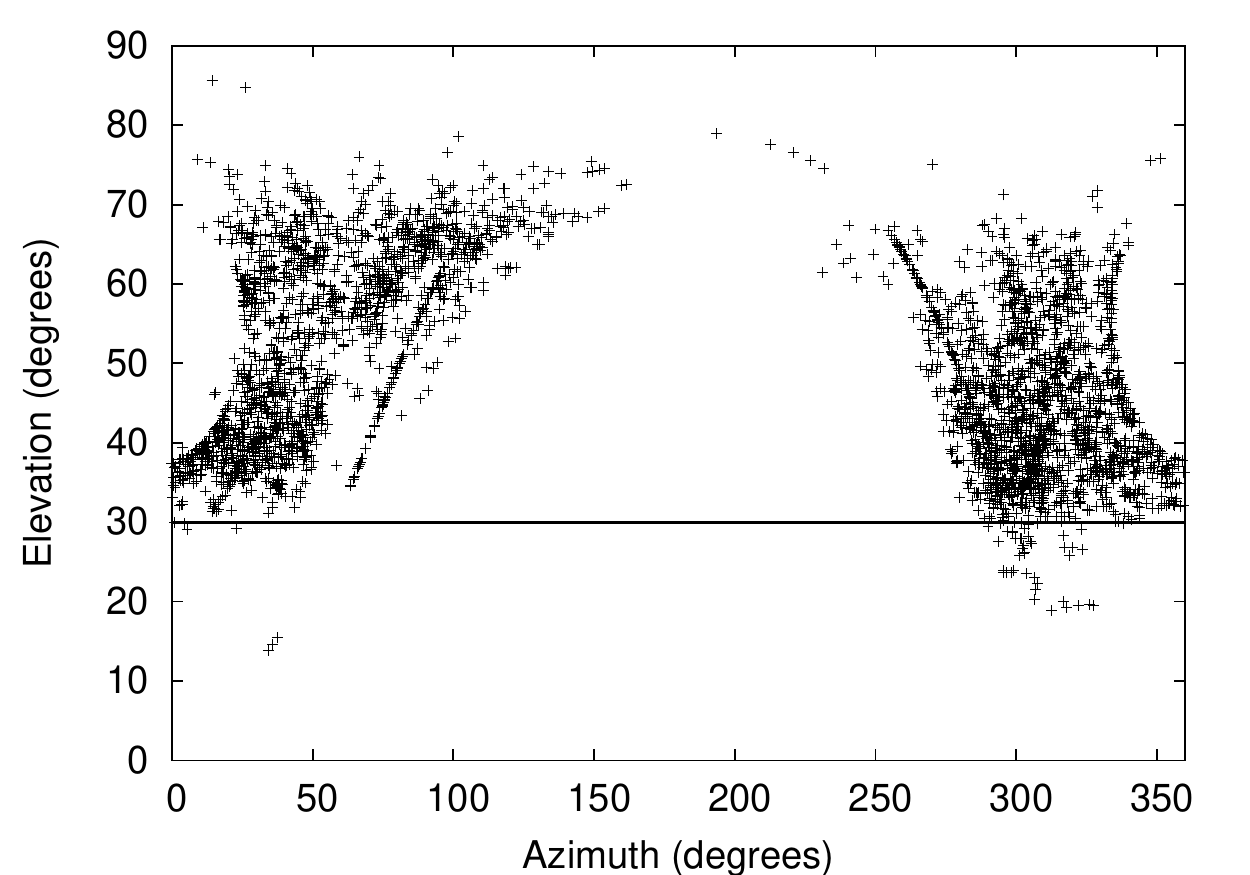}
\caption[Telescope azimuth and elevations for the CRATES source measurements]{The azimuth and elevation of the telescope during all of the CRATES source measurements.}
\label{fig:azel_crates}
\end{center}
\end{fig}

The observations were carried out using a mixture of cross-scans and on-off measurements, depending on the expected strength of the source based on an extrapolation from the 4.8 and 8.4~GHz flux densities. Those sources that were predicted to be stronger than 100~mJy were observed from the start using cross-scan measurements; those that were expected to be weaker than that were observed using on-off measurements. If a source was not detected using the cross-scan method, then it was re-observed using on-off measurements. As cross-scan measurements provide simultaneous pointing corrections for the telescope, these provide more robust flux density measurements; however on-off measurements are required for sources with a low flux density. In the case of these sources, the pointing corrections from nearby strong sources observed with cross-scan measurements can be used. This combination of the two observational techniques provides the most reliable flux density measurements from the OCRA-p instrument.

Observations commenced in November 2008. The majority of the observations were complete by October 2009, with only a few sources remaining for further observations. This section describes the results as of mid-December 2009. A total of 3597 measurements were made by this date -- an average of 6.5 measurements per source, with each source having at least 3 measurements. Objects with interesting spectra or discrepant flux density measurements are currently being reobserved. The observation positions in azimuth and elevation are shown in Figure \ref{fig:azel_crates}.

\subsection{Data quality}
Observations from several periods -- namely, 30 March, 2-3 April, 18 April and 29-30 April 2009 -- were flagged in their entirety due to calibration problems. Some shorter periods were also flagged, where the measurements following from a pointing calibration were systematically above or below the mean flux density for those sources.

Cross-scan measurements where the amplitudes of the peaks disagree by greater than 20 per cent, or where the widths of the peaks disagree by greater than 40 per cent, are automatically flagged. Measurements that are obviously affected by poor weather have then been manually flagged.

Although for the VSA sources all on-off measurements with an error on the measurement over 7~mJy  were automatically flagged, this does not work as well for the CRATES sources due to their stronger flux densities. As such, we automatically flag any measurements with an error on the measurement over 7~mJy where that error is greater than 15 per cent of the flux density of the measurement. Additionally, obviously erroneously low measurements (likely caused by the telescope not being positioned on the source) have been manually flagged.

In total, 1034 measurements have been flagged, out of 3597 in total. This leaves 2563 measurements, or an average of 4.7 per source.

\subsubsection{On-off vs. cross-scan measurements}
\begin{fig}
\centering
\includegraphics[scale=0.5]{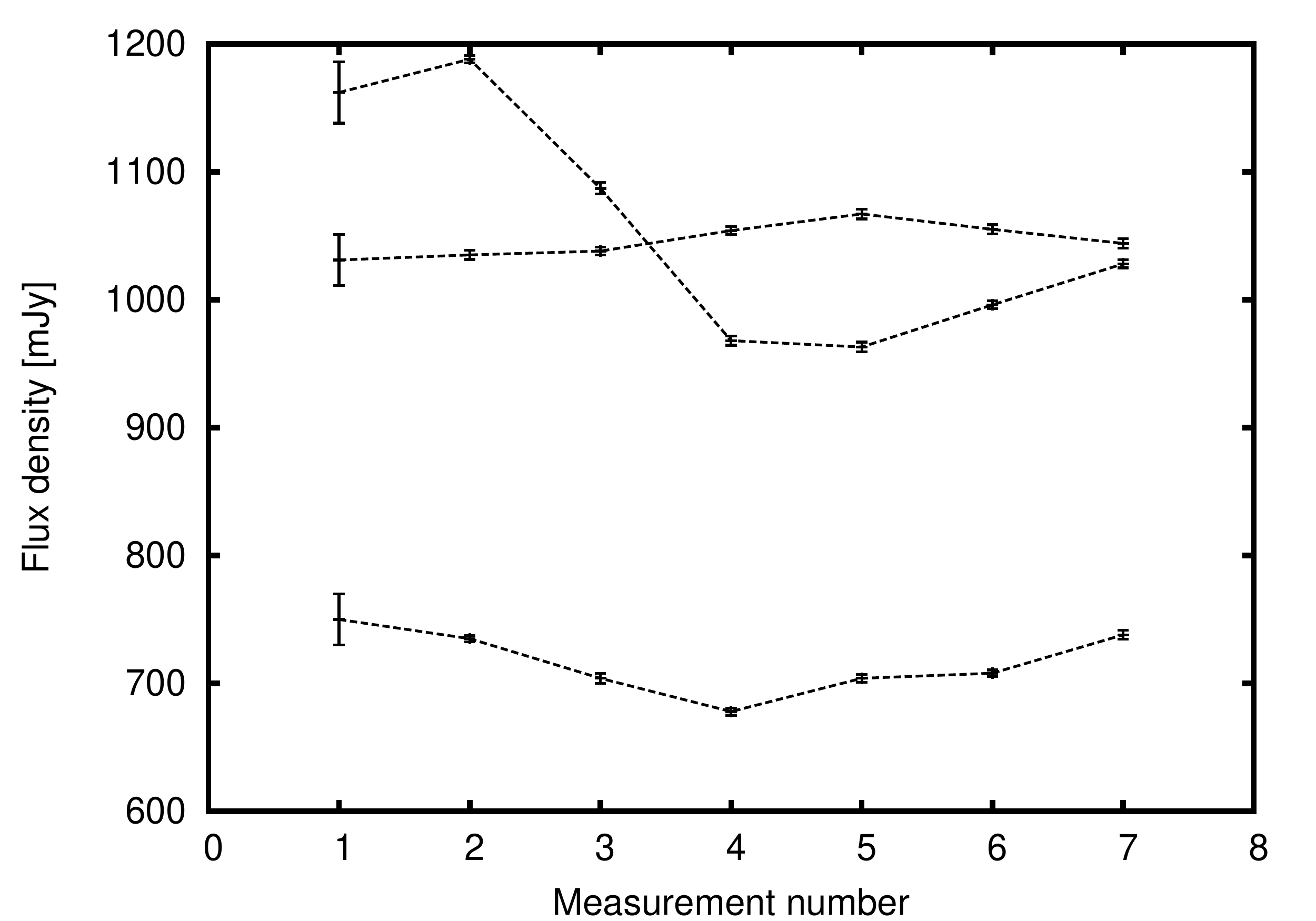}
\caption[Test cross-scan and on-off measurements of three strong CRATES sources]{Test cross-scan and on-off measurements of three strong CRATES sources. Measurement 1 is a cross-scan measurement; the remainder are on-off measurements. The error bars displayed are from the scatter of the 1s data points.}
\label{fig:crates_strong}
\end{fig}
In order to check the systematics due to the repeated pointing of the Toru\'n radio telescope during on-off measurements without recalibration of the telescope pointing corrections, three of the brightest CRATES sources -- 1435+638, 1656+477 and 1732+389 -- were observed first with a cross-scan measurement then with six on-off measurements in succession on 17 November 2008. The flux densities for these measurements are shown in Figure \ref{fig:crates_strong}. Although there do not appear to be systematic differences between the cross-scan measurements and the on-off measurements, for one source, 1732+389, the flux densities measured by on-offs fall by up to 20 per cent compared with the cross-scan measurement. The telescope pointing uncertainty is thought to be up to around 15 arcsec; a change of 20 per cent implies an offset of 20~arcsec. Although worrying, this particular measurement is likely to be an anomaly, caused by changes in wind, or a problem with the pointing correction table for the telescope at this particular position in the sky. Further investigation of this is needed, although similar instances within sets of on-off measurements can be identified by systematic offsets from the mean in the case of sources with high flux densities.

\begin{fig}
\centering
\includegraphics[scale=0.5]{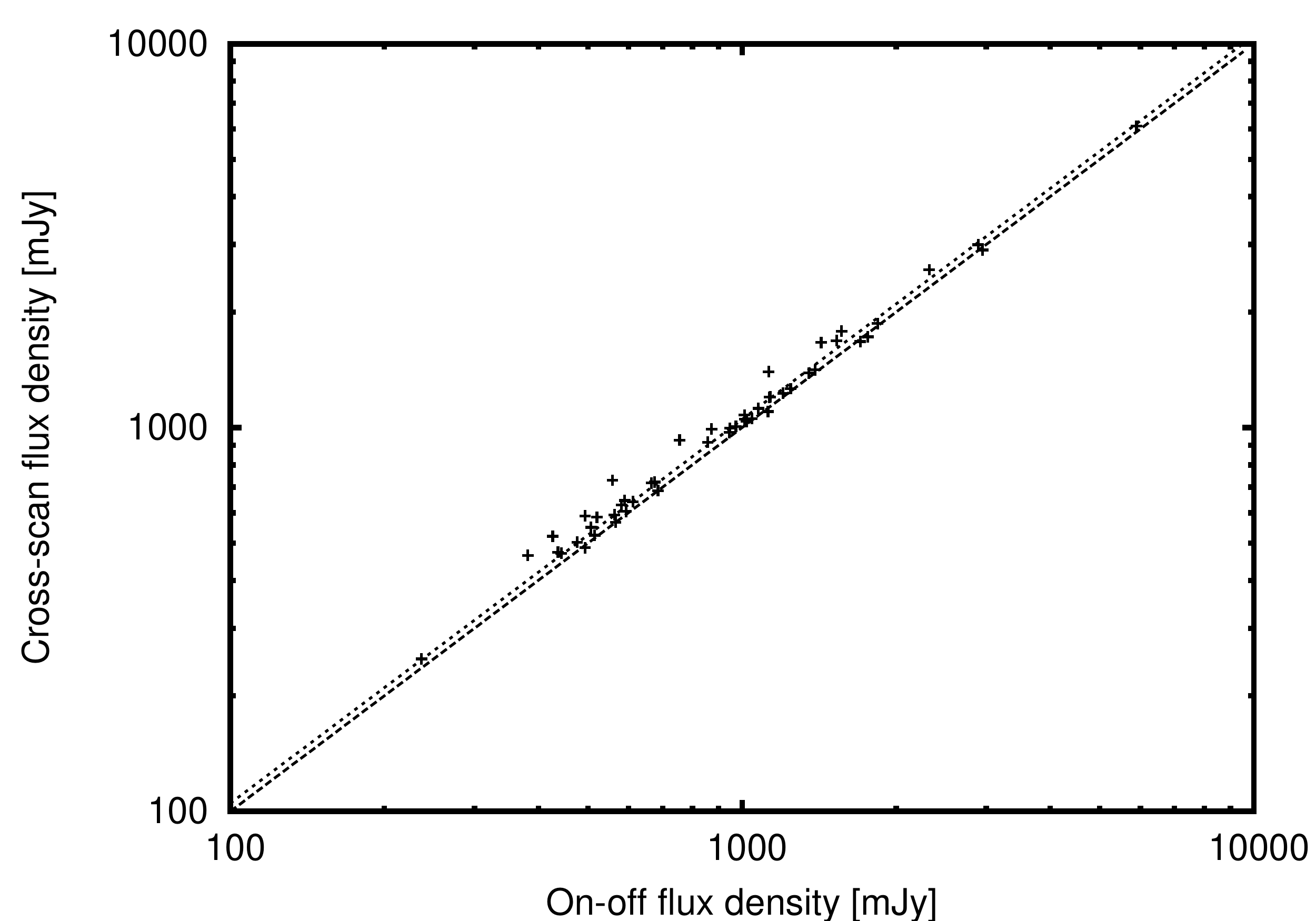}
\caption[Comparison of flux densities from pairs of cross-scan and on-off measurements on a number of CRATES sources]{Comparison of flux densities from pairs of cross-scan and on-off measurements on a number of CRATES sources. In each case, a cross-scan measurement was made of a source, immediately followed by an on-off measurement. The on-off measurements are systematically lower than the cross-scan measurements by $\sim5$ per cent. The two lines shown are $y=x$ (dashed line) and $y=1.05x$ (dotted line).}
\label{fig:crates_qscan_onoff}
\end{fig}
A discrepancy between flux density measurements of sources using the on-off measurement technique compared to cross-scan measurements was noticed during the observations of planetary nebulae for \citet{2009Pazderska}. To investigate this, a series of paired cross-scan and on-off measurements were made of a number of bright (200~mJy-6~Jy) CRATES sources between 18 November and 30 December 2008. These measurements are shown in Figure \ref{fig:crates_qscan_onoff}, after the removal of flagged measurements.

The on-off measurements are systematically lower than the cross-scan measurements; the scatter is always above the $y=x$ line, and the average of the offset is 5-6 per cent. The cross-scan measurements determine the pointing corrections for the telescopes, such that these measurements should always measure the flux density when the beam is on the source. The on-off measurements are ``blind'' -- that is, they rely the previous pointing corrections to measure the source, rather than remeasuring them. If those pointing corrections are not accurate, or drift over time, then this will result in the beam being positioned slightly to one side of the source, thus underestimating the flux density of the source. The drop of 5 per cent in the flux densities implies an offset of 10 arcsec of the telescope (based on the 72 arcsec OCRA-p beam), which is comparable to the known pointing uncertainty of the telescope.

This level of offset is within the calibration uncertainties for the measurements. Additionally, it will be below the Gaussian noise level for the weak VSA sources. As CRATES sources that are stronger than 100~mJy will typically be observed with cross-scan measurements rather than on-off measurements, this effect should also be fairly negligible here. However, we increase the flux density for on-off measurements for the CRATES sample by 5 per cent to compensate for the systematic offset when compared to cross-scan measurements, and also add 5 per cent of the measured flux density to the measurement error to account for the pointing uncertainty.

\subsection{Flux densities} \label{sec:crates_source_fluxes}

The flux densities of the CRATES source subsample as measured with OCRA-p are given in Table \ref{tab:crates_fluxes}. As with the VSA sources, the final error on the flux density for each source is calculated by $\sigma = \sqrt{\sigma_\mathrm{meas}^2 + (0.08 S)^2}$ where the 8 per cent of the flux density takes into account the uncertainty due to calibration, atmospheric and gain-elevation corrections and atmospheric effects. The 1.4, 4.8 and 8.4~GHz flux densities from the CRATES source catalogue are also listed. The spectra of the sources  between 26~MHz and 150~GHz are given in Figure \ref{fig:sourcespectra} within Appendix \ref{sec:crates_spectra}. These include data from a wide variety of sources, the references for which are given in Tables \ref{tab:crates_spectra_sources1} and \ref{tab:crates_spectra_sources2}, also within the Appendix.

From a visual comparison to NVSS \citep{1998Condon} and FIRST \citep{1995Becker} data, the majority of the CRATES sources are unresolved. A number show extension, multiple components or have other sources very close by; these are marked with an ``e'' in Table \ref{tab:crates_fluxes}. Sources where multiple CRATES components have been merged together are denoted ``NC'' where ``N'' is the number of components.

\subsection{Comparison to other measurements}
\begin{fig}
\centering
\includegraphics[scale=0.5]{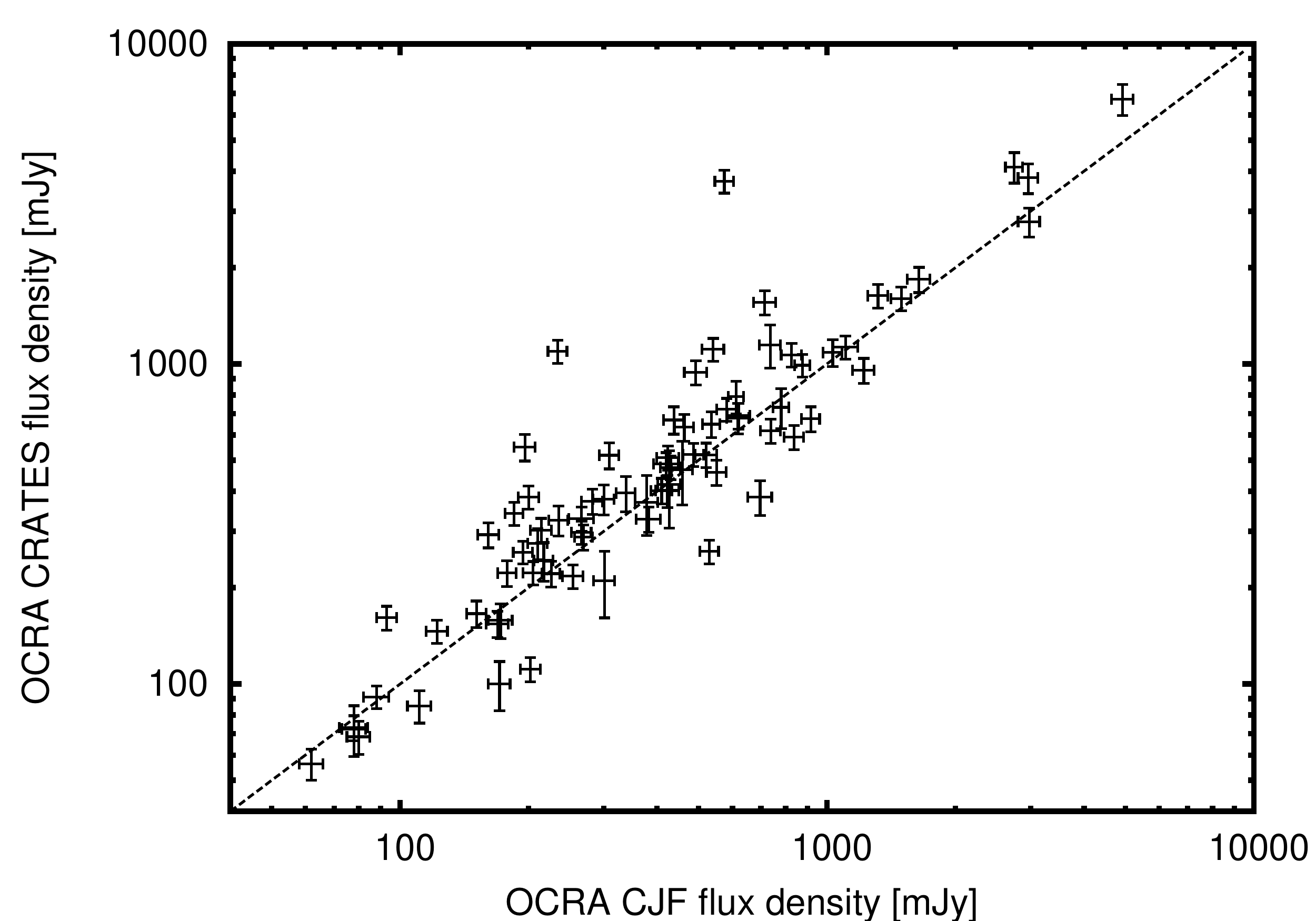}
\caption[Comparison of common sources between CJF and CRATES]{Comparison of common sources between CJF and CRATES. The measurements broadly agree, with scatter due to variability. Two sources stand out as having varied significantly in flux density.}
\label{fig:comparison_cj_crates}
\end{fig}

There are 77 sources in common between the CRATES subsample and the OCRA-p measurements of CJF sources made by \citet{2007Lowe}; these are plotted in Figure \ref{fig:comparison_cj_crates}. There is broad agreement between the two, although there is a large degree of scatter. Two of the sources have very discrepant flux densities. 2006+6424 was 234$\pm$12~mJy in \citet{2007Lowe}; it is now 1095$\pm$91~mJy. 1849+670 was 575$\pm$29~mJy, and is now 3722$\pm$304~mJy. Additionally, 1642+6856 has also increased significantly, from 2.75$\pm$0.13~Jy to 4.1$\pm$0.45~Jy with a large amount of scatter in the measured data points implying ongoing variability.

\begin{tab}{htb}
\begin{tabular}{c|c|c|c|l}
{\bf Name} & {\bf OCRA [Jy]} & {\bf WMAP K [Jy]} & {\bf WMAP Ka [Jy]} & {\bf Notes}\\
\hline
1343+6602 & 0.33$\pm$0.03 & 0.7$\pm$0.07 & 0.4$\pm$0.1& Paired with next, a\\
1344+6606 & 0.38$\pm$0.04 & as above & & \\
1419+5423 & 0.73$\pm$0.10 & 0.8$\pm$0.06 & 0.8$\pm$0.09& av\\
1436+6336 & 1.11$\pm$0.09 & 0.5$\pm$0.06 & -- & a\\
1549+5038 & 0.62$\pm$0.05 & 0.9$\pm$0.06& 0.8$\pm$0.09 & \\
1604+5714 & 0.57$\pm$0.08 & 0.7$\pm$0.04& 0.7$\pm$0.07 & a\\
1635+3808 & 2.78$\pm$0.29 & 3.9$\pm$0.05 & 4.3$\pm$0.08 & v\\
1637+4717 & 1.09$\pm$0.11 & 0.9$\pm$0.05& 1.0$\pm$0.08 & \\
1638+5720 & 1.63$\pm$0.14 & 1.3$\pm$0.04 & 1.3$\pm$0.07 &\\
1642+3948 & 6.72$\pm$0.74 & 6.5$\pm$0.05 & 6.0$\pm$0.08 & v\\
1657+5705 & 0.52$\pm$0.04 & 0.5$\pm$0.06 & 0.6$\pm$0.09 &\\
1700+6830 & 0.40$\pm$0.05 & 0.2$\pm$0.06 & 0.5$\pm$0.08\\
1716+6836 & 0.65$\pm$0.06 & 0.6$\pm$0.04 & 0.6$\pm$0.06 & \\
1727+4530 & 1.56$\pm$0.14 & 0.9$\pm$0.04 & 1.0$\pm$0.08 & v\\
1734+3857 & 1.13$\pm$0.09 & 1.2$\pm$0.05 & 1.3$\pm$0.08 & \\
1739+4737 & 0.69$\pm$0.06 & 0.8$\pm$0.05 & 0.8$\pm$0.06 & \\
1740+5211 & 1.07$\pm$0.09 & 1.2$\pm$0.04 & 1.2$\pm$0.07 & \\
1748+7005 & 0.48$\pm$0.04 & 0.6$\pm$0.03 & 0.7$\pm$0.06 & \\
1753+4409 & 0.46$\pm$0.04 & 0.7$\pm$0.06 & 0.6$\pm$0.1 & \\
1801+4404 & 1.15$\pm$0.18 & 1.2$\pm$0.04 & 1.4$\pm$0.07 & v\\
1806+6949 & 1.60$\pm$0.14 & 1.4$\pm$0.03 & 1.4$\pm$0.06 & v\\
1824+5651 & 1.84$\pm$0.17 & 1.5$\pm$0.04 & 1.3$\pm$0.07 & \\
1842+6809 & 0.89$\pm$0.08 & 1.1$\pm$0.03 & 1.2$\pm$0.05 & a\\
1849+6705 & 3.72$\pm$0.30 & 1.2$\pm$0.04 & 1.4$\pm$0.06 & av\\
1927+6117 & 0.59$\pm$0.05 & 1.0$\pm$0.04 & 1.0$\pm$0.07 & \\
1927+7358 & 3.81$\pm$0.41 & 3.5$\pm$0.04 & 3.2$\pm$0.06 & v\\
2009+7229 & 0.67$\pm$0.06 & 0.7$\pm$0.06 & 0.6$\pm$0.08 & \\
\end{tabular}
\caption[Comparison of WMAP and OCRA-p measurements of common CRATES sources]{Comparison of WMAP and OCRA-p measurements of common CRATES sources.}
\label{tab:crates_wmap}
\end{tab}

The subsample also has 27 sources in common with the 5 year WMAP point source catalogue \citep{2009Wright}. These are listed in Table \ref{tab:crates_wmap}, which gives the OCRA measurement and the WMAP K (22~GHz) and Ka-band (33~GHz) flux densities. The sources are also marked as ``a'' where there are multiple identifications for the WMAP sources and ``v'' where there is evidence for variability  in the WMAP observations, according to \citet{2009Wright}. There is a good correspondence between the WMAP variable sources and those with a high scatter within the OCRA measurements of a source, based on the flux density errors calculated using the scatter of the OCRA measurements. The majority of the discrepant flux density measurements are of those sources marked by WMAP as variable, with a few exceptions such as 1927+6117, which does not have a high enough flux density for WMAP to be able to detect its variability at high significance.

Two of the sources -- 1343+6602 and 1344+6606 -- are identified with the same WMAP source. The discrepancy between the combined and WMAP flux densities could be caused by the extension of the combined sources in the WMAP beam. Additionally, 1436+6336 does not have a 33~GHz flux density from WMAP, although it is detected at over 1~Jy by OCRA; this could potentially be due to source confusion.

\subsection{Spectral index distributions}
\begin{fig}
\centering
\includegraphics[scale=0.25]{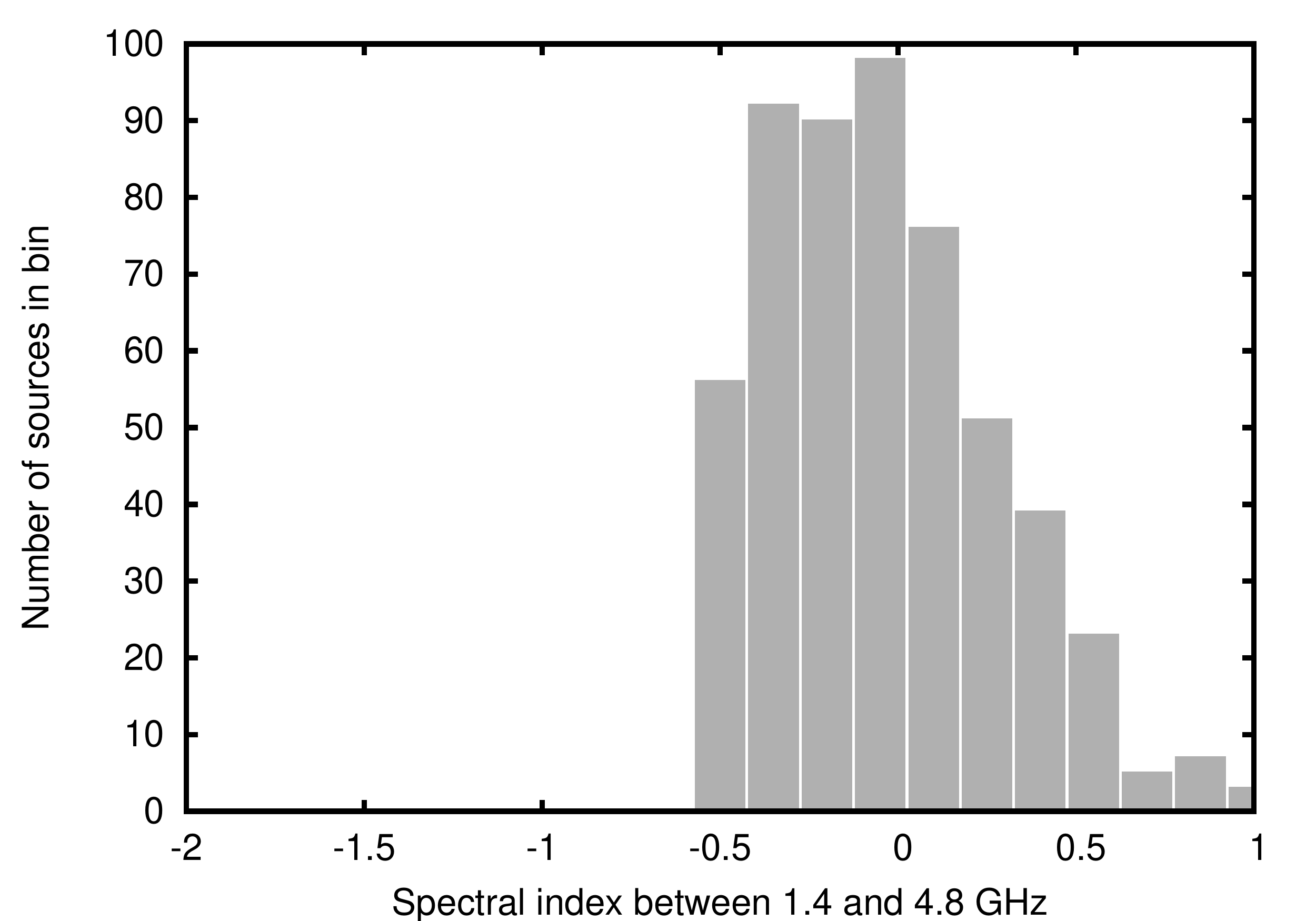}
\includegraphics[scale=0.25]{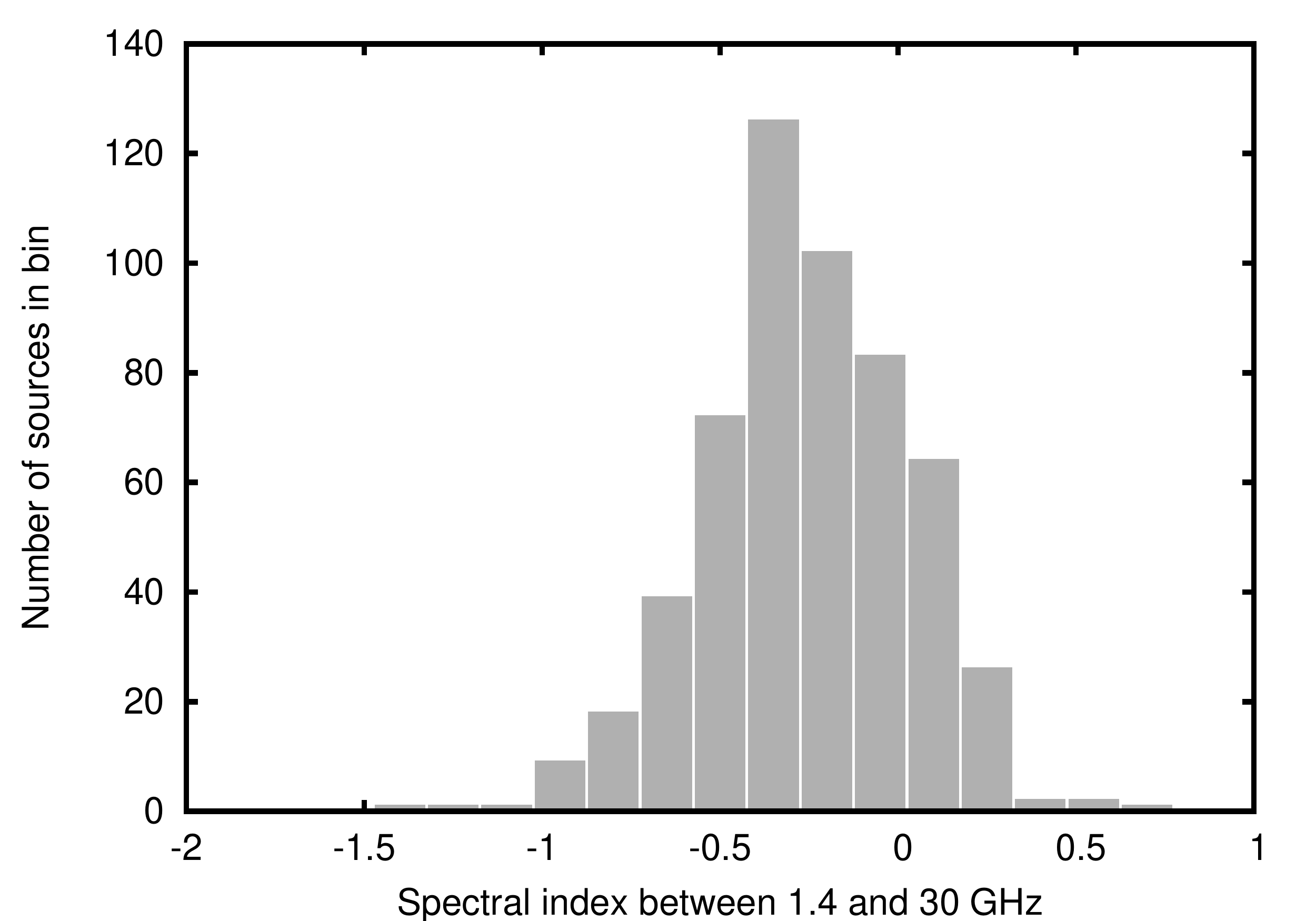}\\
\includegraphics[scale=0.25]{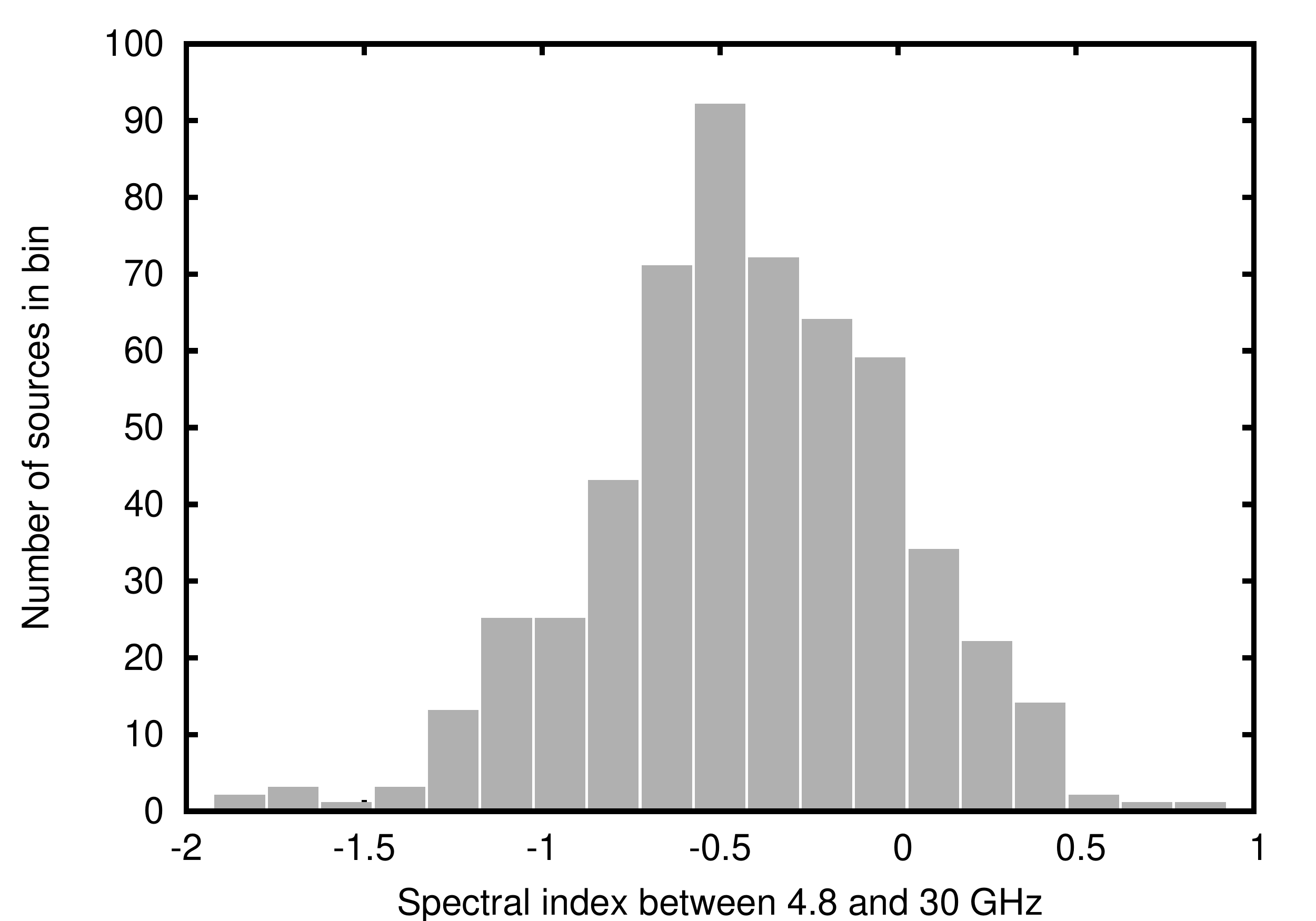}
\includegraphics[scale=0.25]{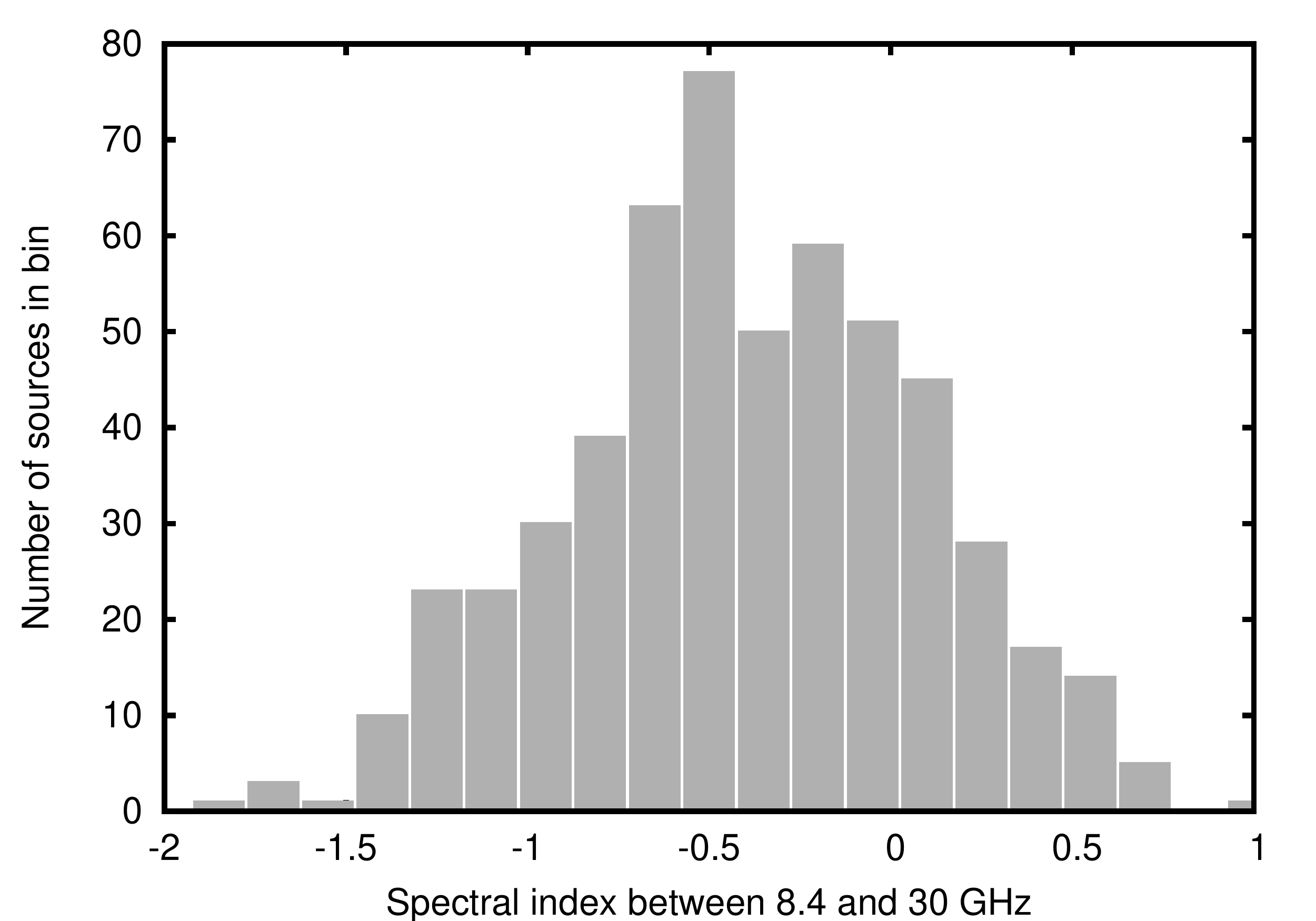}\\
\caption[Spectral index distributions for the CRATES source sample] {The spectral index distributions for the sources in the CRATES sample observed with OCRA-p. Top-left: 1.4-4.8~GHz; top-right: 1.4-30~GHz; bottom left: 4.8-30~GHz. Bottom right: 8.4-30~GHz.}
\label{fig:crates_spectralindex_histogram}
\end{fig}

\begin{fig}
\centering
\includegraphics[scale=0.5]{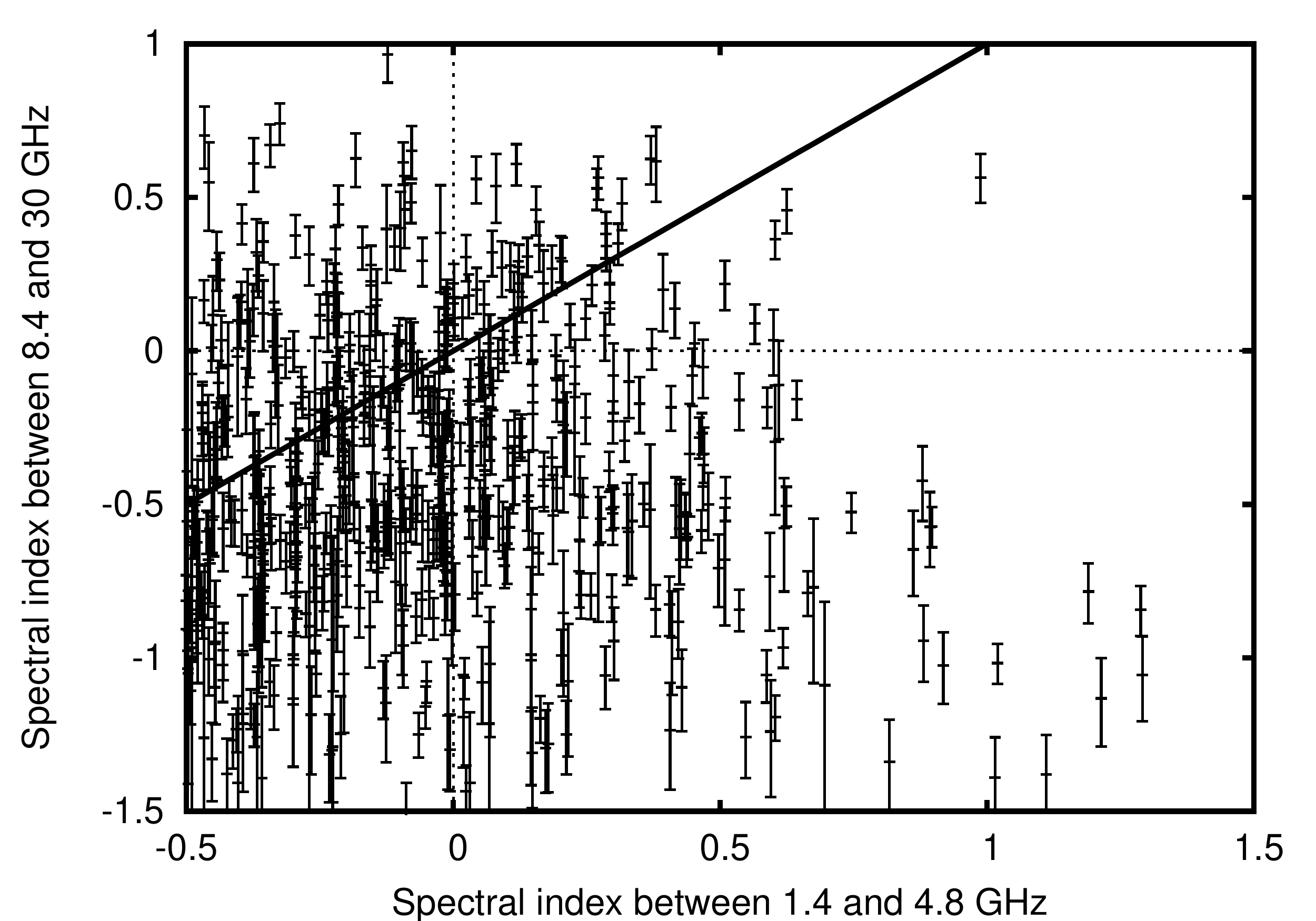}\\
\caption[2 colour diagram for the CRATES source sample] {A 2 colour diagram for the CRATES source sample, showing the spectral index from 1.4 to 4.8~GHz compared with the spectral index from 8.4 to 30~GHz. The solid line indicates $x=y$; 72 per cent of sources have steepened and lie below this line.}
\label{fig:crates_histogram_2colour}
\end{fig}

The spectral index distributions for the CRATES sources between 1.4 and 4.8; 1.4 and 30; 4.8 and 30 and 8.4 and 30~GHz are shown in Figure \ref{fig:crates_spectralindex_histogram}. The effects of the selection criterion of $\alpha_{1.4}^{4.8}>-0.5$ is obvious in the first of these. However, when the 30~GHz flux density is considered the selection cut becomes much less defined. The mean and spread for $\alpha_{1.4}^{4.8}$ is $-0.01\pm$0.37; this changes to $-0.26\pm$0.29 for $\alpha_{1.4}^{30}$, $-0.43\pm$0.43 for $\alpha_{4.8}^{30}$ and $-0.42\pm$0.53 for $\alpha_{8.4}^{30}$. This reflects the steepening or turning over of the source spectra at high frequencies, as seen by e.g. \citet{2006Ricci,2007Massardi}, although we note that there is a sample selection present within the CRATES subsample as sources with $\alpha_{1.4}^{4.8}<-0.5$ are excluded; any of those sources that flatten or upturn will be missing from this sample.

The 2-colour diagram depicting $\alpha_{1.4}^{4.8}$ v. $\alpha_{8.4}^{30}$ is shown in Figure \ref{fig:crates_histogram_2colour}. 72 per cent of the sources have steepened at $\alpha_{8.4}^{30}$ compared with $\alpha_{1.4}^{4.8}$. 31 (5.6 per cent) are Gigahertz-Peaked sources that peak between 4.8 and 8.4~GHz (defined by $\alpha_{1.4}^{4.8} > 0.5$ and $\alpha_{8.4}^{30} < -0.5$); 59 (10.7 per cent) are flat or rising ($\alpha_{1.4}^{4.8} > 0$ and $\alpha_{8.4}^{30} > 0$) and 64 (11.6 per cent) are inverted ($\alpha_{1.4}^{4.8} < 0$ and $\alpha_{8.4}^{30} > 0$).

We caution that, due to the high resolution of the 8.4~GHz interferometric observations compared with the other measurements, sources that are extended or have multiple components will likely have underestimated flux densities at 8.4~GHz. This will have increased the number of sources with rising flux density between 8.4 and 30~GHz. Ideally, the sources would have been observed with the same resolution at all of the frequencies using single dishes for this spectral index comparison to be completely representative of the behaviour of the complete source sample.

\subsection{Conclusions}

Using OCRA-p, we have surveyed 550 flat spectrum radio sources around the North Ecliptic Pole from the CRATES sample, which were selected based on their 1.4-4.8~GHz spectral index. These follow from the observation of the CJF sources by \citet{2006Lowe,2007Lowe}, and extend the work to lower flux densities.

We find reasonable agreement between the measurements presented here and those by \citet{2007Lowe} and also WMAP \citep{2009Wright} where sources are in common. A number of sources display variability between these three sets of observations, and also within the present measurements. This will present difficulties when subtracting these point sources from maps of the CMB.

As expected, we find that the spectral index distribution for these sources broadens at higher frequencies as the source spectral indices steepen. We find that there are a reasonable number (5.6 per cent) of Gigahertz-Peaked Sources, and also a number of inverted sources ($\sim 10$ per cent). We conclude that extrapolation from low frequency flux densities to higher frequencies assuming power law spectra is unreliable. This emphasizes the need for high-frequency blind surveys to low flux densities. Such surveys are currently being carried out by the ATCA at 20~GHz in the southern hemisphere (see e.g. \citealp{2007Sadler}) and AMI at 15~GHz in the northern hemisphere \citep{2004Waldram,2009Waldram}, and will be carried out by the OCRA-F instrument and its successors at 30~GHz in the near future.

The flux densities of the sources within the CRATES subsample described here will be useful for comparison to point source measurements by the {\it Planck} satellite, which will be most sensitive in the area of sky surveyed here. Due to the flat spectrum nature of these sources, they are the most likely sources to appear in all of the different observational bands of {\it Planck}. An Early Release Point Source Catalogue from {\it Planck} is expected in December 2010 \citep{2009Bouchet}, at which point such a comparison can be carried out.

\chapter{OCRA-FARADAY}\label{ocraf}

The OCRA FARADAY (OCRA-F) instrument is an 8-beam 30~GHz receiver, and is the second phase of the OCRA programme. It was designed and constructed as part of the EC-funded ``Focal-plane Arrays for Radio Astronomy, Design Access and Yield'' (FARADAY) project within the RadioNet consortium. The design of the instrument is based on OCRA-p, and marks the transition of the programme both from a single receiver module with a pair of horns to multiple receiver modules, and also from traditional components to integrated modules. The increase in the number of receivers greatly improves the survey ability of OCRA, both due to the increase in the number of receivers (and hence effective integration time), or alternatively an increased instantaneous field of view.

\begin{fig}
\begin{center}
	\includegraphics[scale=0.08]{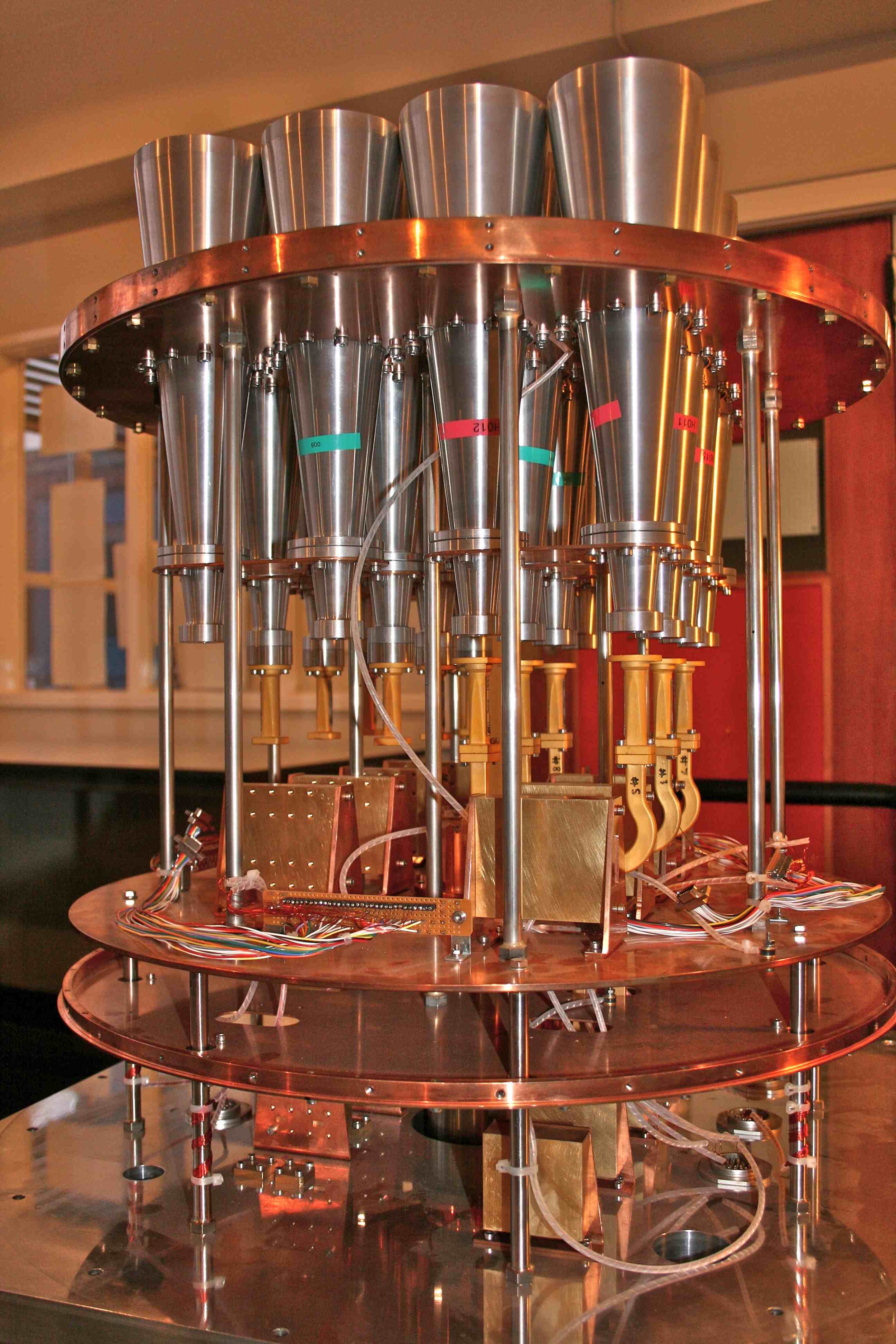}
\caption[OCRA-F in its test configuration]{OCRA-F in its test configuration, prior to disassembly. The three copper plates are (from top to bottom) the horn plate, the 20~K plate and the 50~K plate, with the base of the cryostat at the bottom of the image. The FEMs are located at the base of the horns, and the BEMs are located beneath the 50~K plate; dummy bodies were installed in the cryostat at the time that this picture was taken.}
\label{fig:ocraf_test_config}
\end{center}
\end{fig}

There are several interesting astrophysical questions that can be answered by using OCRA-F to carry out surveys of the microwave sky. The location of point sources down to milliJansky flux density levels would enable improved characterization of their statistical properties, making them easier to statistically remove from CMB observations. This information is also useful for SZ observations as it can be used to determine how many clusters will not be detected as a result of their presence, hence providing a correction factor to number counts of galaxy clusters for cosmology. The enhancement of point source numbers towards clusters is also of interest; recent difficulties have been encountered in searches for the SZ effect in nearby ($z<0.5$) clusters, which have yet to be definitively explained. This could potentially be due to this overdensity of sources towards clusters.

Another aim of OCRA-F is to conduct blind surveys for SZ clusters. Thus far only a few instruments have attempted this and only a small number of new clusters have been discovered this way to date \citep[see][]{2008Staniszewski,2009Hincks}. Potentially OCRA-F can be used to measure the power spectrum of the CMB and SZ effect on small angular scales (see Section \ref{sec:cmb_ocra}). OCRA-F can also carry out Galactic observations, studying clouds of gas and dust within the plane of our Galaxy.

OCRA-F will ultimately consist of 16 horns, arranged in 8 pairs. The first phase of the project is an 8-beam, 4-pair system, which was primarily funded by the EU. Whilst OCRA-p used traditional, Microwave Integrated Circuit (MIC)-based components, OCRA-F uses Monolithic Microwave Integrated Circuits (MMICs). As well as allowing more components into a smaller area, these should also allow for larger receiver arrays to be constructed more easily due to the reduced manpower needed to assemble the devices. It is also thought that MMIC technology is more repeatable than MIC technology, such that multiple devices should have similar values of gain and noise, although the degree to which this is true remains to be quantified.

At the start of the Author's involvement in the OCRA-F project, the receiver cryostat had been constructed in a test configuration, as shown by Figure \ref{fig:ocraf_test_config}, with 5 Front End Modules (FEMs) constructed and around 10 Back End Modules (BEMs). Tests had been carried out on the individual modules, with some testing done using a combination of a FEM and a BEM, however no complete receiver chain tests had been carried out. The Author assisted in the dismantling and reassembly of the cryostat into its final configuration, led the testing of the complete receiver chains and assisted in the installation and initial commissioning of the receiver on the Toru\'n 32-m telescope.

The design of the receiver is summarized in Section \ref{sec:design}. Section \ref{sec:ocraf_construction} then describes the individual components of OCRA-F, and the testing of them that has been carried out by the Author. This includes the three active components -- the FEMs, BEMs and detectors -- as well as the passive waveguide and filter components, and brief sections on the cryogenics and electronics of the receiver. The tests of partial and complete radiometer chain are then described in Section \ref{sec:radiometer_tests}. Problems were encountered with the foam vacuum support window; these are described in Section \ref{sec:foam}. The receiver was installed on the telescope in December 2009, and the initial steps of commissioning the receiver are described in Section \ref{sec:commissioning}. Section \ref{sec:future} gives a discussion of future possible improvements that could be made to OCRA-F, followed by a summary of the chapter in Section \ref{sec:ocraf_summary}.

\section{Design overview} \label{sec:design}

\begin{fig}
\begin{center}
	\includegraphics[scale=1.0]{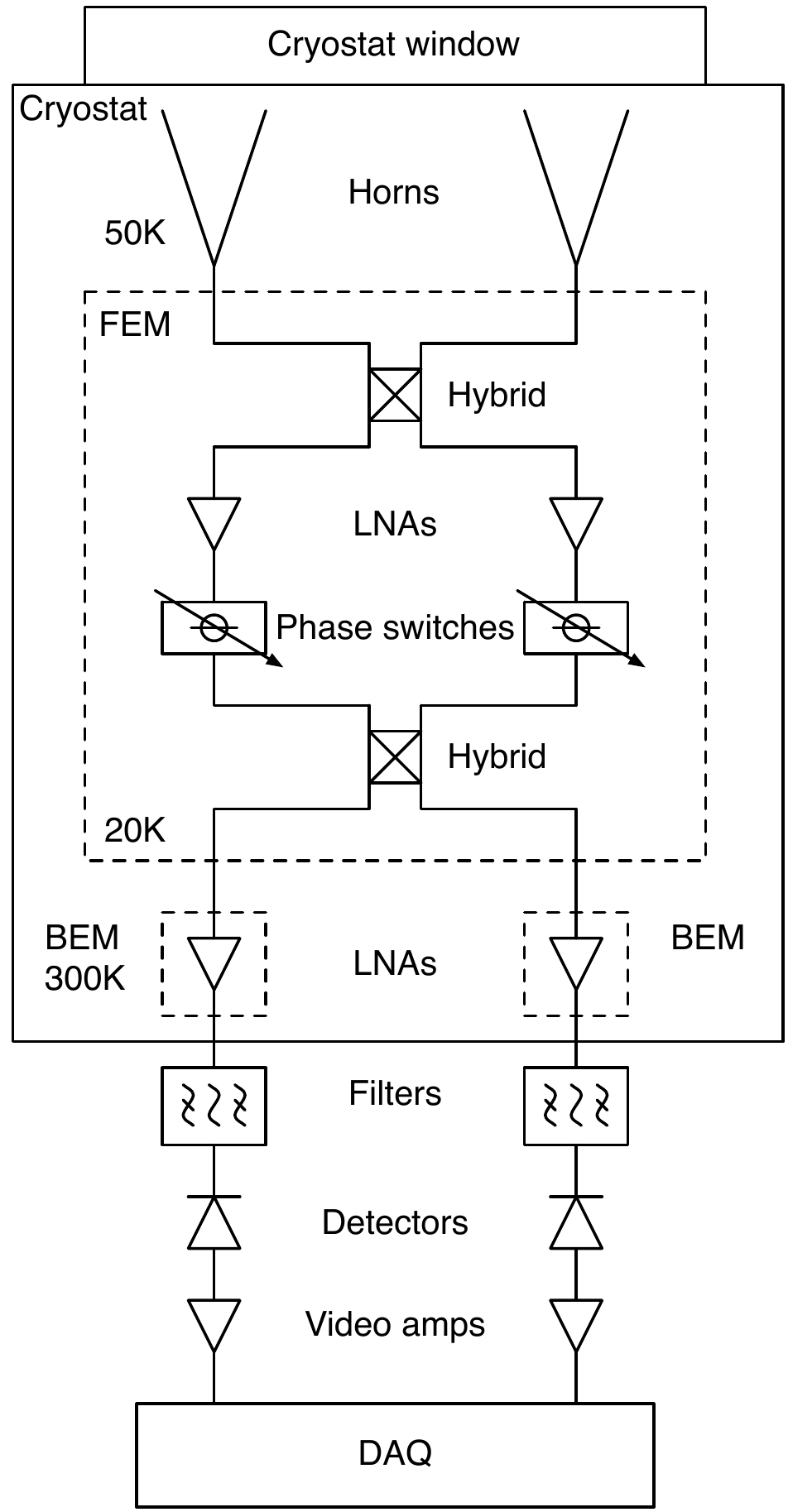}
\caption[Block diagram of an OCRA-F receiver chain]{A block diagram of an OCRA-F receiver chain (not to scale). The design is essentially the same as for OCRA-p (compare with Figure 2.2 of \citealp{2006Lowe}).}
\label{fig:ocraf_block_diagram}
\end{center}
\end{fig}

The design of a receiver chain in OCRA-F is essentially the same as that in OCRA-p (see \citealp{2006Lowe} for details). This design in turn is based on the receiver chains in the {\it Planck} Low Frequency Instrument \citep[LFI;][]{2000Mandolesi}, and is similar to the {\it WMAP} 23~GHz receivers \citep{2003Jarosik}. A block diagram of a chain is shown in Figure \ref{fig:ocraf_block_diagram}.

Unlike {\it Planck}, which differences between a cold load and the sky, OCRA has two horns such that the effects of fluctuations in atmospheric emission and attenuation (for convenience, this is hereafter referred to as atmospheric $1/f$ noise) that are common to the two beams can be cancelled out. The signals from the two horns are combined using a $90\degree$ hybrid, with the combination of the signals passed through a pair of LNAs and phase switches; the signals are then separated again using a second hybrid. Following this, further amplification is applied to the signals, which are then filtered to define the bandwidth of the system prior to the signal being detected. Video amplifiers then boost the detected signal such that it can be recorded using a Data AcQuisition system (DAQ).

It is important that the phase paths between the two hybrids are well matched; offsets in this phase will result in cross-mixing of the signals such that the system has very bad isolation, which in turn will greatly reduce the efficiency of the receiver. With OCRA-p, only the initial hybrid and LNAs were within the cryostat, with the phase switches and the second hybrid outside of the cryostat, such that the phase of the signal that exits the cryostat is important. With OCRA-F, however, the phase sensitive section of the receiver is within a single module, the FEM, all of which is within the cryostat and cooled to 20~K. This has both benefits and drawbacks. The increased modularization eases the assembly process of the receiver chain, however the balancing of the phase must be done during the construction of the FEM, and once set it is difficult to change as phase shifters cannot be inserted, so the only way of altering the phase is to alter the biases of the amplifiers.

It is worth noting that the BEMs, whilst situated within the cryostat, are at room temperature. This means that they have much higher noise than the FEMs, however this negligibly affects the overall system temperature of the receiver as it comes after the first stages of amplification.

\begin{tab}{tb}
\begin{tabular}{c|c}
{\bf Property} & {\bf Value}\\
\hline
Number of beams & 8; later 16\\
Resolution & 72 arcsec\\
Frequency range & 26-36GHz\\
Bandwidth & 10~GHz\\
System temperature & 50K (all contributions)\\
Nominal noise (per pair) & 7~mJy s$^{0.5}$\\
\end{tabular}
\caption[Specifications for OCRA-F on the Toru\'n telescope]{Specifications for OCRA-F on the Toru\'n 32-m telescope.}
\label{tab:ocraf_specs}
\end{tab}
The nominal specifications for the system are given in Table \ref{tab:ocraf_specs}. Note that the system temperature given is for all contributions, including the atmosphere: the receiver temperature should be around 30~K. This receiver temperature is higher than that of OCRA-p (15~K \citealp{2006Lowe}), however the bandwidth is larger to compensate, resulting in similar  noise performance of a receiver pair in OCRA-F to that of OCRA-p. The receiver temperature is comparable to that of the 31~GHz receiver on the Green Bank Telescope \citep[20-40~K across the band; see][]{2009Mason}. The best performance that has been achieved with 30~GHz receivers are those in Planck LFI, which have a system temperature of 10.7~K \citep{2010Mandolesi,2009Davis}; however these have been designed for operation in space, where atmospheric effects are not present.

\section{Construction and testing} \label{sec:ocraf_construction}
\begin{fig}
\begin{center}
	\includegraphics[scale=0.10]{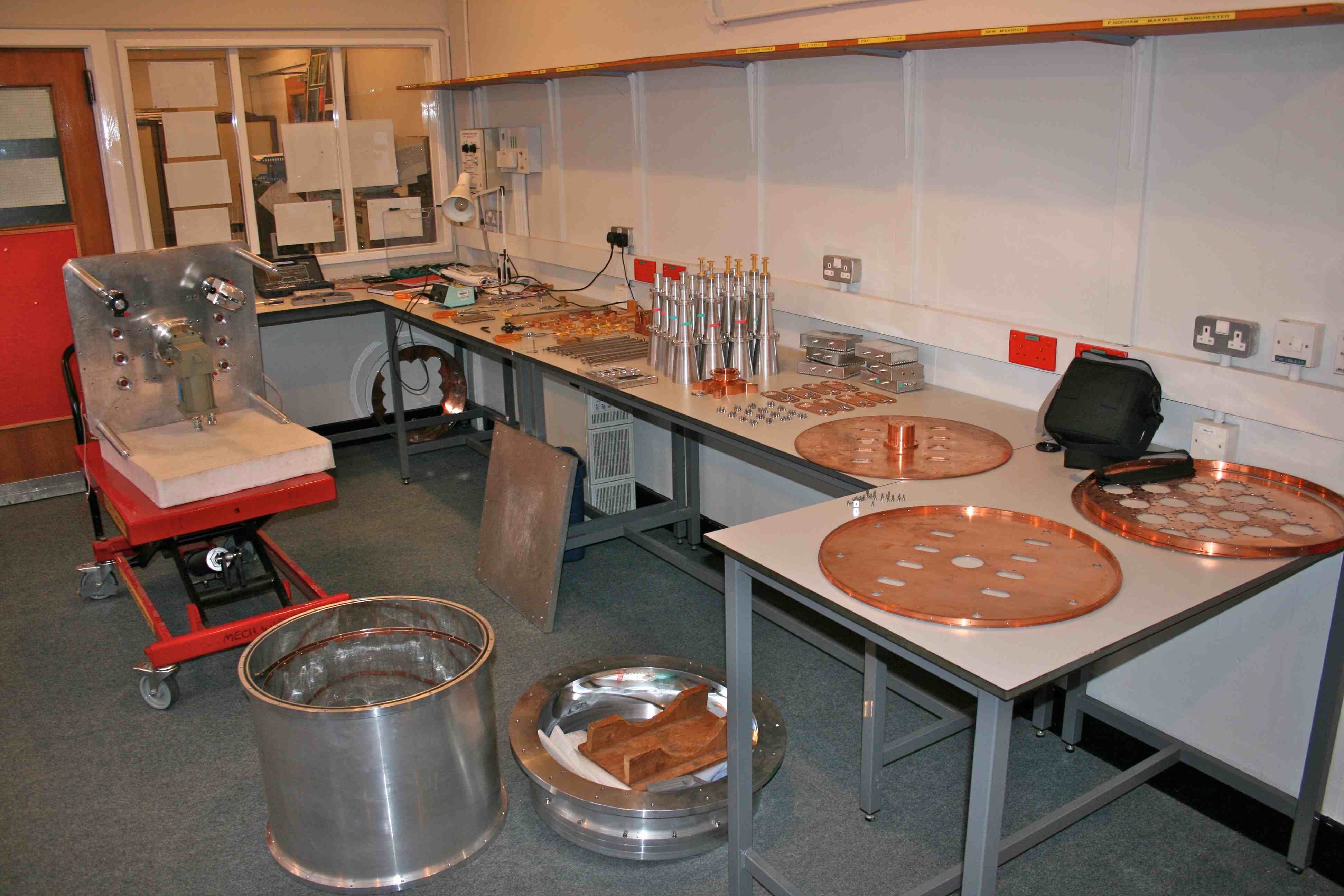}
\caption[OCRA-F prior to reassembly]{The components of OCRA-F on 15 January 2008, following the dismantling of it from its test configuration and prior to its reconstruction into its final configuration.}
\label{fig:ocraf_dismantled}
\end{center}
\end{fig}
The OCRA-F cryostat, which was initially assembled in a test configuration, was completely dismantled at the end of January 2008 (Figure \ref{fig:ocraf_dismantled}), at which point reconstruction from the bottom up was started. Such a complete reconstruction was required as the copper plates comprising the two cold stage plates and the horn plate needed to be plated with nickel. This increases the reflection (and hence decreases the absorption) of photons such that the radiation transfer of heat is decreased, hence improving the cryogenic performance of the cryostat. The build quality of the receiver was also unknown -- for example, the secureness of the screws and the state of the vacuum seals -- due to the long period of testing that the cryostat underwent. The reconstruction also allowed a series of small modifications to be made to the cryostat to improve its cryogenic properties.

\subsection{Active components}
\subsubsection{Front End Modules} \label{sec:ocraf_fems}

\begin{fig}
\begin{center}
	\includegraphics[scale=0.10]{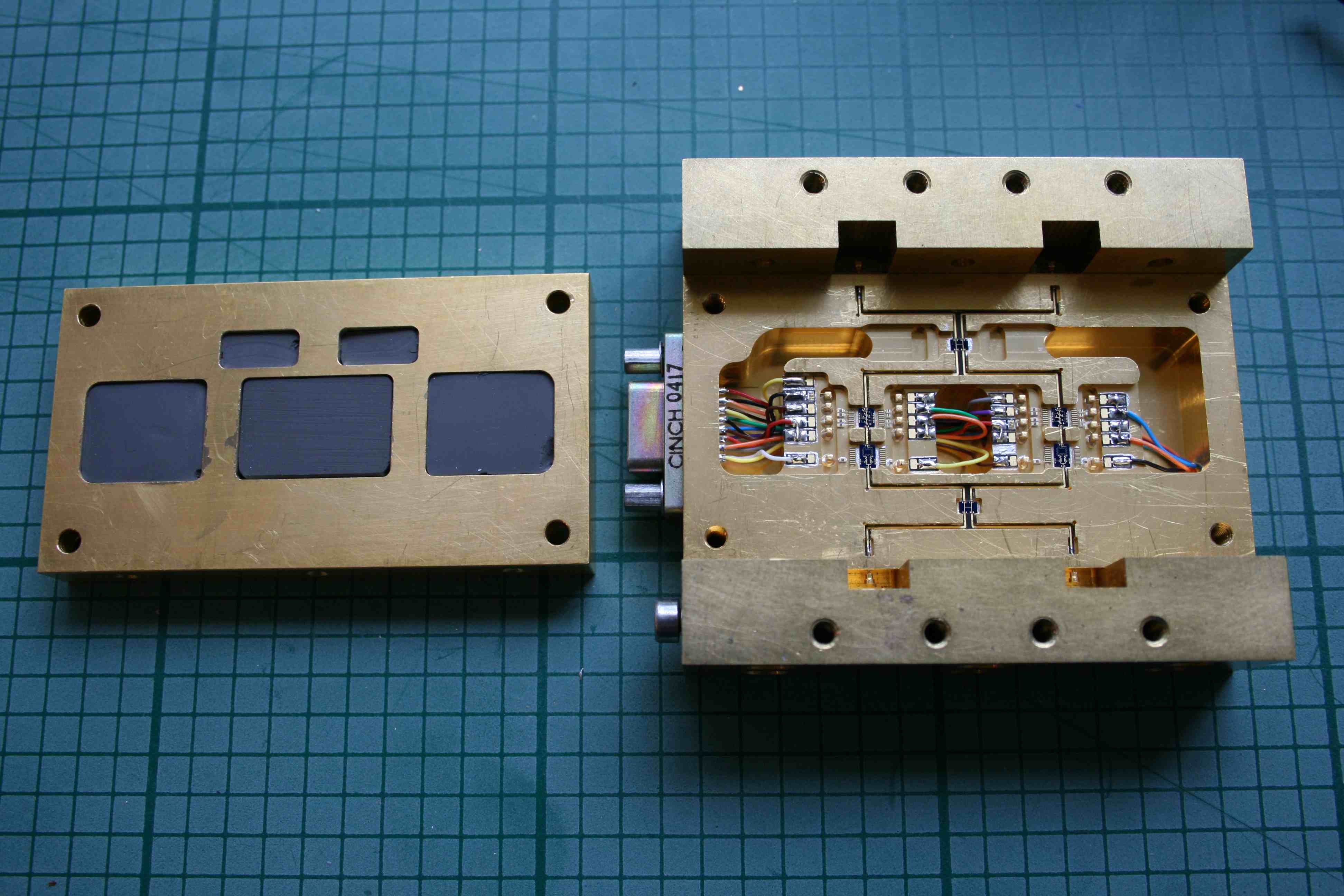}
\caption[An OCRA-F Front End Module]{An OCRA-F Front End Module (FEM~1). The lid is on the left hand side, and the main body on the right. The input is from the top and output at the bottom. The MMICs are the small black components; from top to bottom these are a hybrid, amplifiers, phase switches and a second hybrid.}
\label{fig:ocraf_fem}
\end{center}
\end{fig}

For astronomical observations, the most important parts of the receiver are the low noise amplifiers. Of these, the most critical are the first amplifiers, located in the Front End Modules (FEMs). The performance of the first stage of these amplifiers defines the majority of the noise of the receiver, with the gain of this stage down-weighting the noise in the signal added by the later stages.

As described in Section \ref{sec:design}, the FEMs consist of a pair of amplifiers and phase switches, encapsulated by two hybrids. FEM~1 is shown in Figure \ref{fig:ocraf_fem}. The amplifiers, phase switches and hybrids are all MMIC modules, designed and manufactured within the FARADAY framework; see \citet{2005Kettle,2005Kettlea,2007Kettlea,2007Kettle} for details.

\begin{tab}{tb}
\begin{tabular}{c|c|c|c|c|c|c|c|c}
\bf Property \rm & \multicolumn{2}{c|}{{\bf FEM~1}} & \multicolumn{2}{c|}{{\bf FEM~2}} & \multicolumn{2}{c|}{{\bf FEM~4}} & \multicolumn{2}{c}{{\bf FEM~5}} \\
\hline
Output & 1 & 2 & 1 & 2 & 1 & 2 & 1 & 2\\
\hline
Gain (dB) & 31 & 31 & 33 & 33 & 32 & 32 & 34 & 35\\
Noise (K) & 24 & 26 & 35 & 33 & 26 & 27 & 27 & 23\\
Isolation (per cent) swA & 2.6 & 1.8 & 5.8 & 4.5 & 3.1 & 3.1 & 3.0 & 4.1\\
Isolation (per cent) swB & 3.0 & 4.3 & 7.0 & 6.0 & 4.9 & 5.4 & 5.4 & 4.7\\
\end{tabular}
\caption[Properties of the OCRA-F FEMs]{The properties of the OCRA-F FEMs as of June 2007, measured by D. Lawson. The isolation measurements are for the two phase switch states of one arm; the other phase switch was in state A throughout. The measurements are the average across the band (25-36~GHz).}
\label{tab:ocraf_fem_properties}
\end{tab}

The FEMs were individually tested by D. Lawson prior to the Author's involvement in the project. A summary of the properties of the FEMs as measured by him are given in Table \ref{tab:ocraf_fem_properties}. Whilst 5 FEMs were constructed, FEM~3 could not be properly balanced -- that is, the gains and phases in the two arms were different, making it difficult to cancel out the $1/f$ noise of the amplifiers. This FEM has hence not been used in OCRA-F.

During testing of the radiometer chains (see Section \ref{sec:radiometer_tests}), one of the LNAs in FEM~1 stopped drawing current. After some investigation, it was discovered that the MMIC had broken due to the glue used to attach it to the body of the FEM. This was the first FEM to be constructed, and the adhesive used was most likely a rigid one that had previously caused problems like this. Later FEMs used a more flexible adhesive, so they should not be prone to this problem. For FEM~1, the LNA was replaced, with the gains becoming 31.8 dB in both LNAs, with noise temperatures of 24K.

\subsubsection{Back End Modules} \label{sec:bems}
\begin{fig}
\begin{center}
	\includegraphics[scale=0.05]{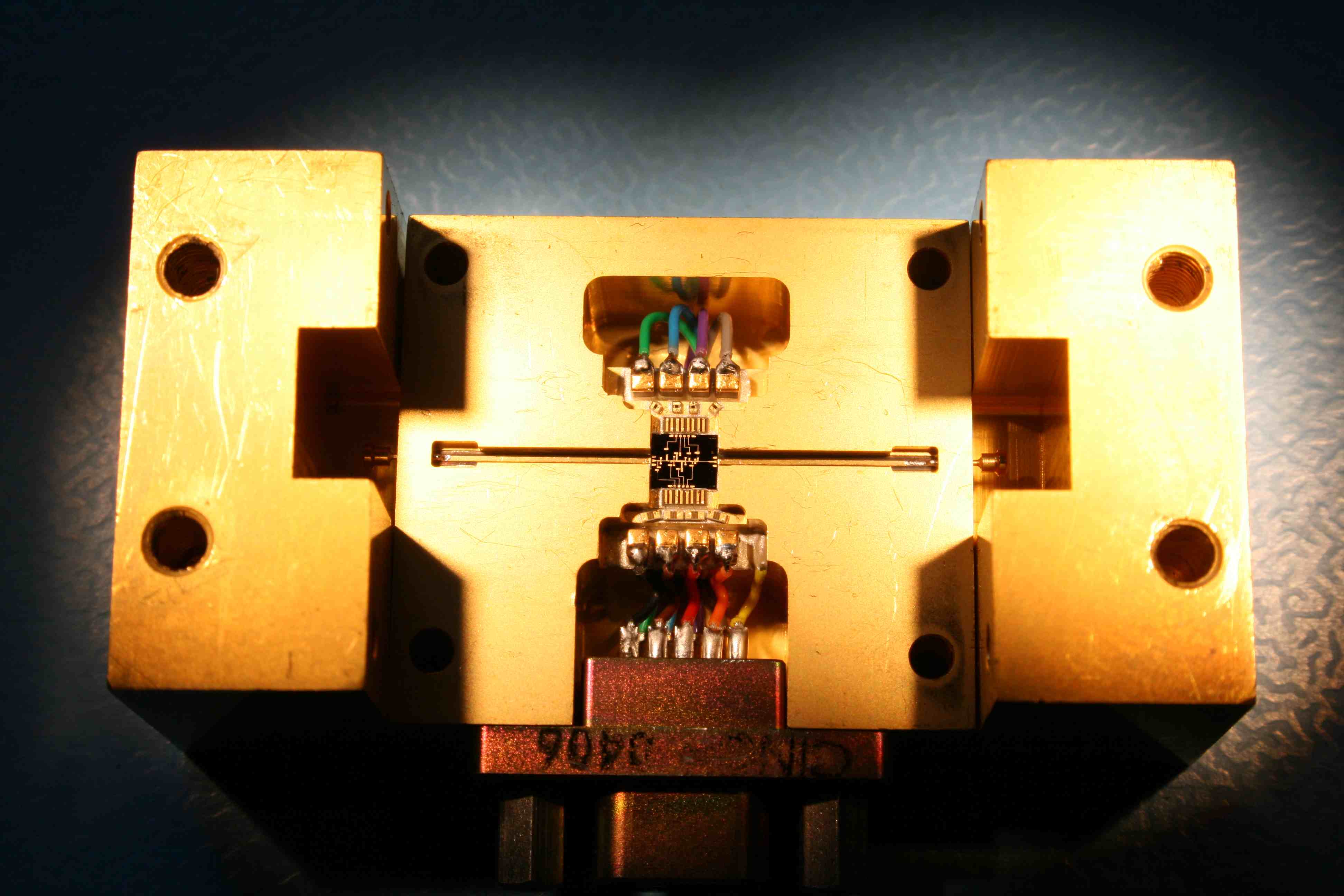}
\caption[An OCRA-F Back End Module]{The insides of one of the OCRA-F BEMs. The input is on the left and the output on the right. The amplifier MMIC can be seen in the centre of the body.}
\label{fig:ocraf_bem}
\end{center}
\end{fig}

\begin{fig}
\begin{center}
	\includegraphics[scale=0.55]{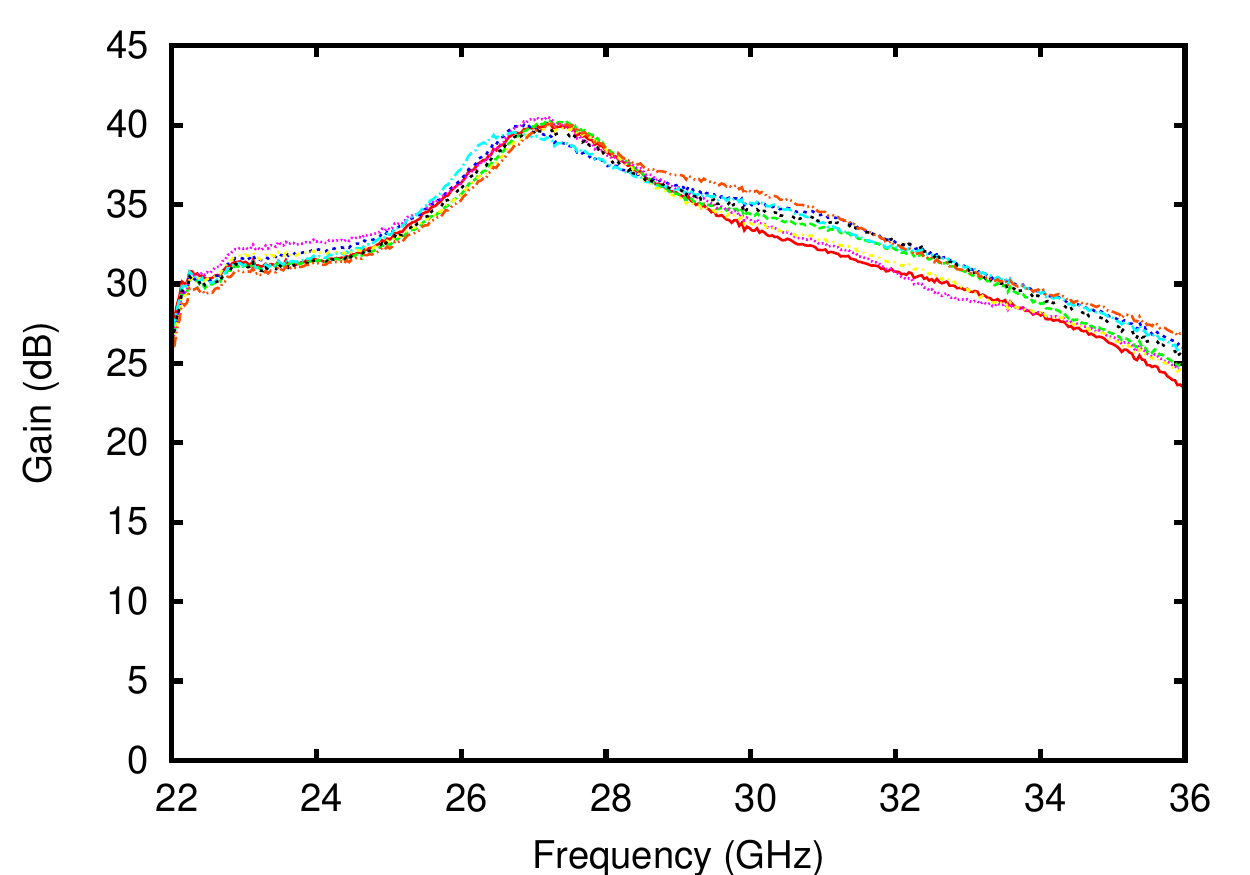}
	\includegraphics[scale=0.55]{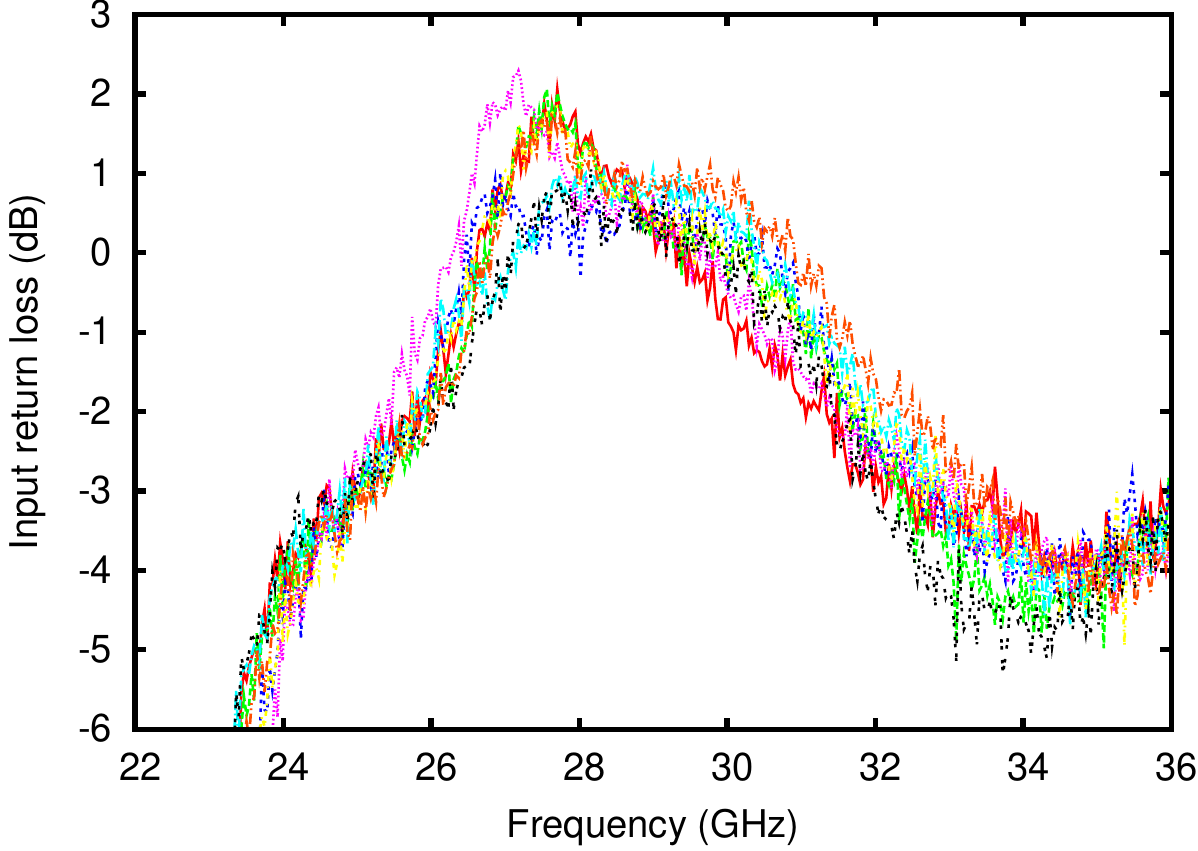}
\caption[Gain of the Back End Modules]{The gains (left) and input return losses (right) of the 8 BEMs used in OCRA-F over the range of frequencies that OCRA-F will observe at. The measurements were taken using standard bias settings to power the amplifiers.}
\label{fig:bem_gain}
\end{center}
\end{fig}

The Back End Modules (BEMs) are much simpler devices than the FEMs, consisting of a single MMIC amplifier (see Figure \ref{fig:ocraf_bem}) that is operated at room temperature. This makes them much quicker to construct and test.

The gain of the BEMs as a function of frequency is plotted in Figure \ref{fig:bem_gain}, using standard bias settings (drain voltages of 1.3~V, gate voltages chosen such that the current is 8~mA per stage). This peaks at $\sim$27~GHz, and at high frequencies it drops off by around 10dB, which reduces the effective bandwidth of the amplifiers. This is surprising, as the MMICs were designed to have a large range of frequencies ($\sim$10~GHz) over which the gain was approximately constant. This is not an issue for OCRA-F, however, as the combined FEM-BEM gain is considerably flatter (see later).

The input return loss (also plotted in Figure \ref{fig:bem_gain}) should ideally be around -30dB, however for the amplifiers used in the BEMs this is much closer to zero, and is positive at around 27~GHz in some of the BEMs. This is a concern, as it means that the BEMs will reflect a large amount of the power put into them by the FEMs, and can lead to the amplifiers interacting. The effect of this in practice will be discussed in Section \ref{sec:fem_bem_combination}.

\begin{fig}
\begin{center}
	\includegraphics[scale=0.55]{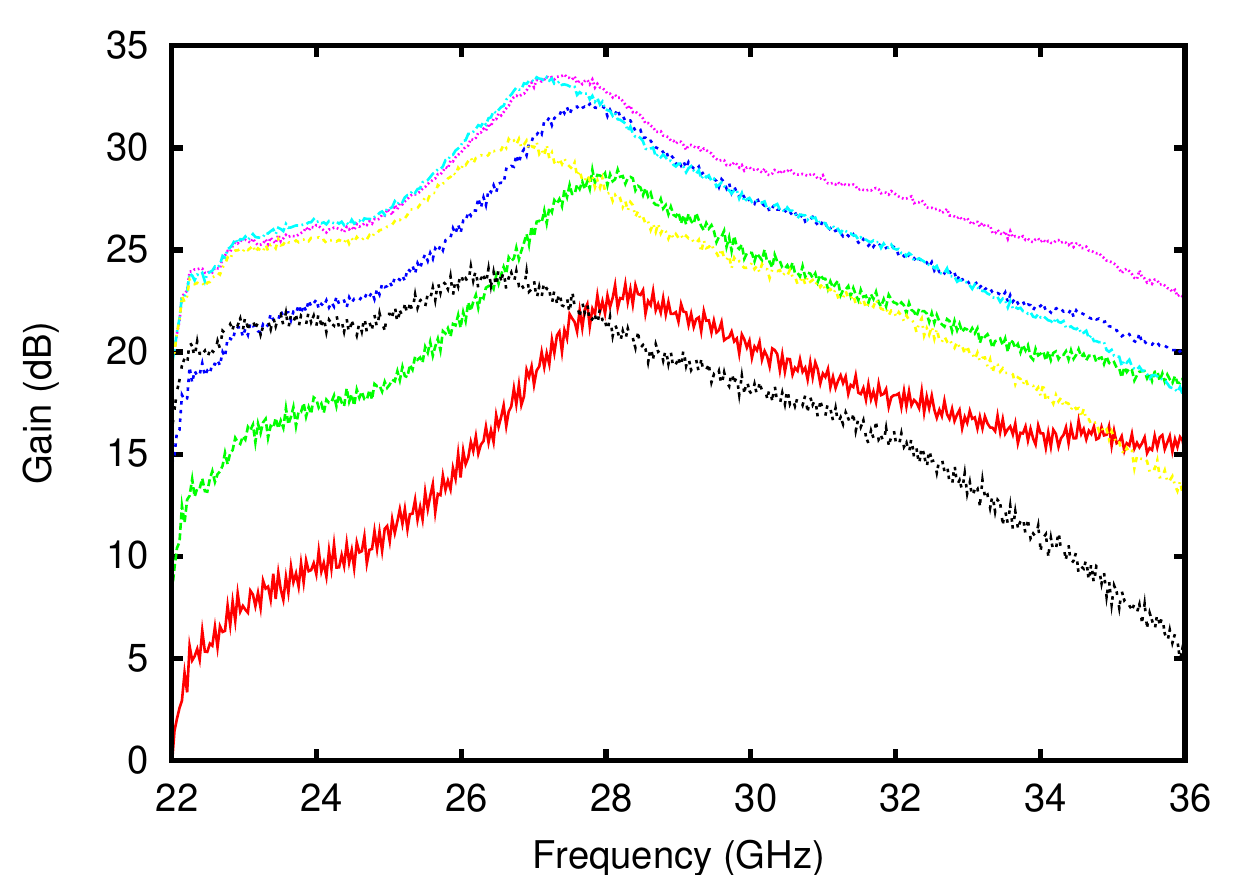}
	\includegraphics[scale=0.55]{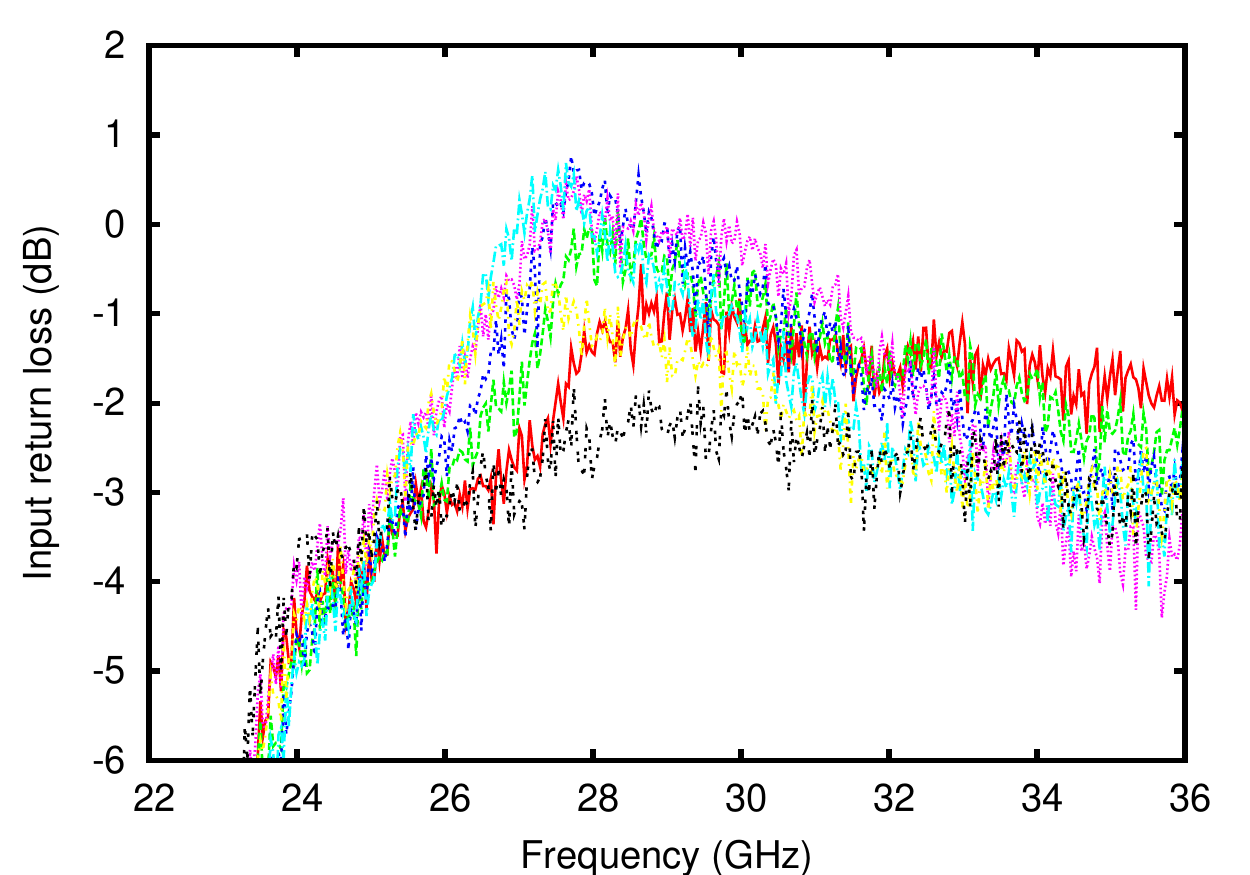}
\caption[Gain of BEM6 as a function of bias settings]{The gains (left) and input return losses (right) of BEM 6, using a constant drain voltage of 1.3~V and stepping the gate voltage by 0.05~V from -0.3-0~V. The peak in the gain shifts down in frequency as the gate voltage approaches zero.}
\label{fig:bem6_gain}
\end{center}
\end{fig}

Figure \ref{fig:bem6_gain} shows the gain and input return loss of BEM~6 as a function of gate voltage and hence current (keeping the drain voltage constant at 1.3~V).  The peak in the gain curve shifts down in frequency as the gate voltage is increased towards zero. The input return loss is positive when the gain is highest, and becomes more negative as the gain is decreased. As such, a bias setting that does not maximize the gain is preferable. We will return to this issue in Section \ref{sec:ocraf_tsys}.

\subsubsection{Detection} \label{sec:ocraf_detectors}
\begin{fig}
\begin{center}
	\includegraphics[scale=0.9]{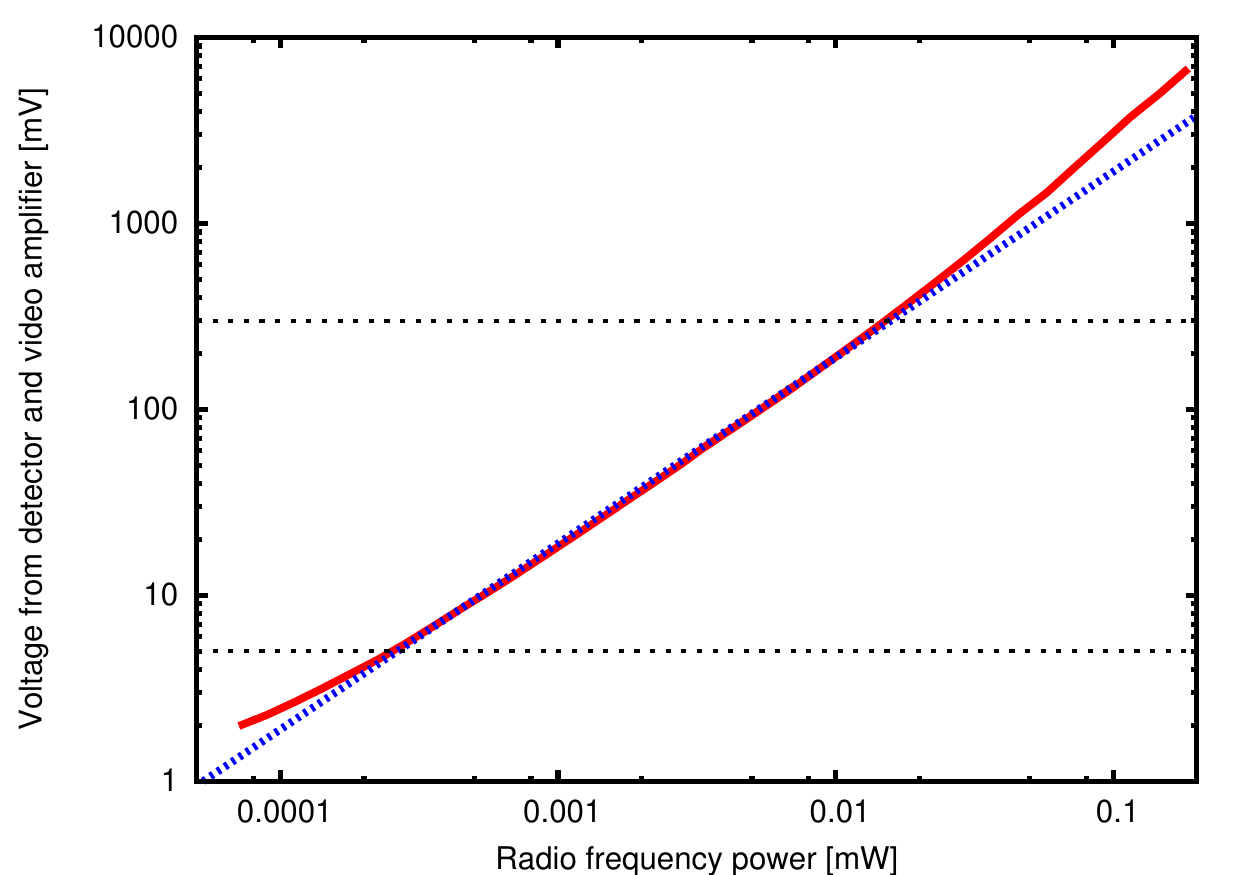}
\caption[Detector response as a function of input RF power]{Comparison of the detector and video amplifier output vs. RF power (measured using a power meter) for various input powers (red solid line) compared with a linear function (blue dotted line). The detector and video amplifier combination is approximately linear between 5 and 300~mV output, with non-linearity becoming significant below and above these values.}
\label{fig:detector_vs_power}
\end{center}
\end{fig}

Following these two sets of amplification, the signals are detected using commercial detectors, manufactured by Farran Technologies, with model number WDP-28. These detectors have also been used in OCRA-p. Final amplification prior to data acquisition is done by video amplifiers. These were constructed by E. Pazderski in late 2008, with temporary video amplifiers used for initial tests.

The detectors were tested by the Author using an RF signal generator and measuring the RF power using both an Anritsu ML2437A power meter equipped with a MA2474A power sensor, and also a detector and video amplifier combination, for a variety of input powers between approximately $-3$ and $-40$~dBm. The results of these test are shown in Figure \ref{fig:detector_vs_power}, which shows that the detectors are roughly linear between 5 and 300~mV ($-35$~dBm to $-14$~dBm), and become non-linear at higher and lower powers. The power meter should be linear to within 2.5 per cent, and sensitive between -70 and +20 dBm, according to its specification. Testing using the two different video amplifiers showed that whilst the temporary video amplifier had a lower gain, the properties of the video amplifiers are otherwise very similar. Hence it appears to be the detectors that are non-linear. We will return to this issue in Section \ref{sec:ocraf_tsys}.

\subsection{Waveguides}
\begin{fig}
\begin{center}
	\includegraphics[scale=0.05]{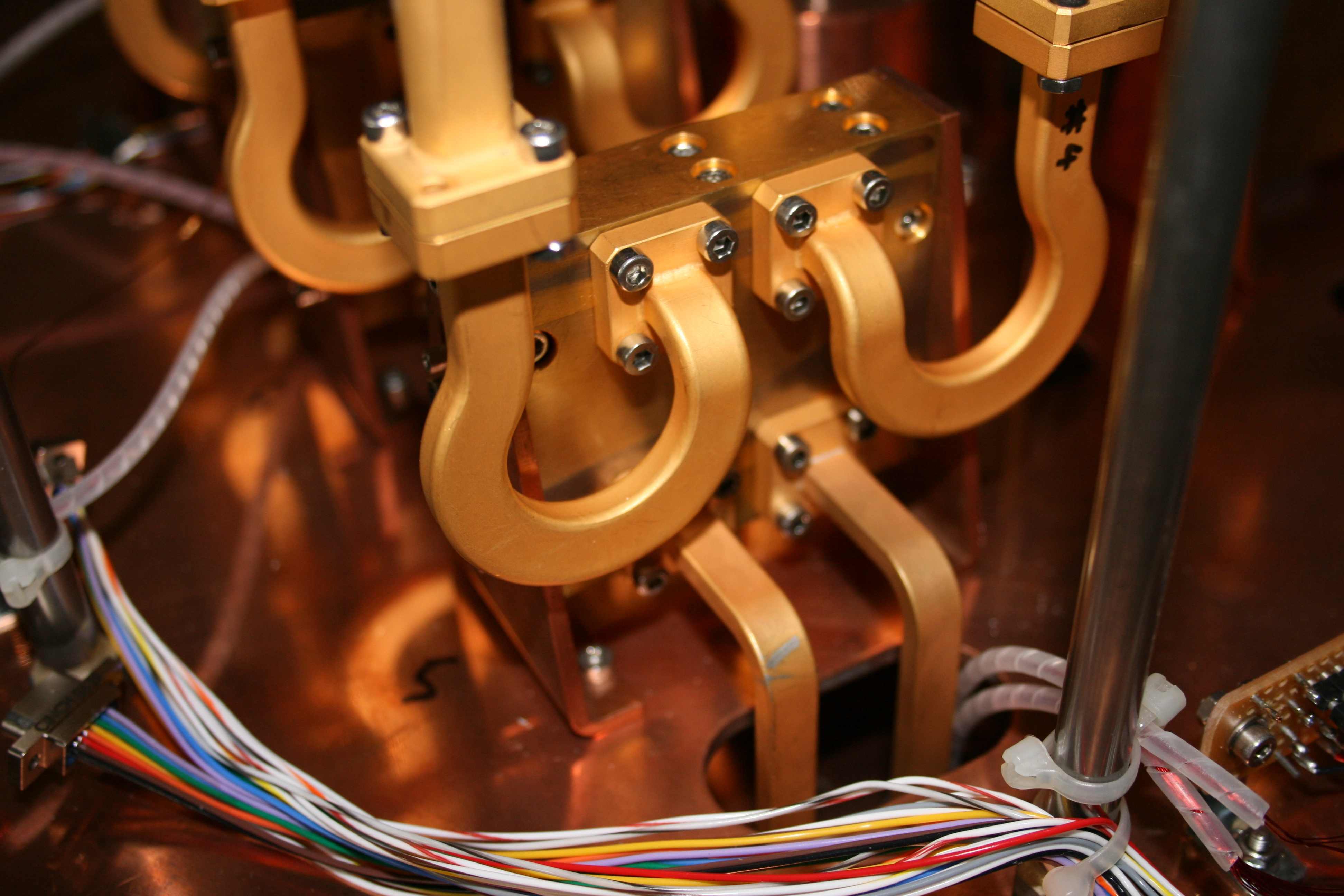}
	\includegraphics[scale=0.05]{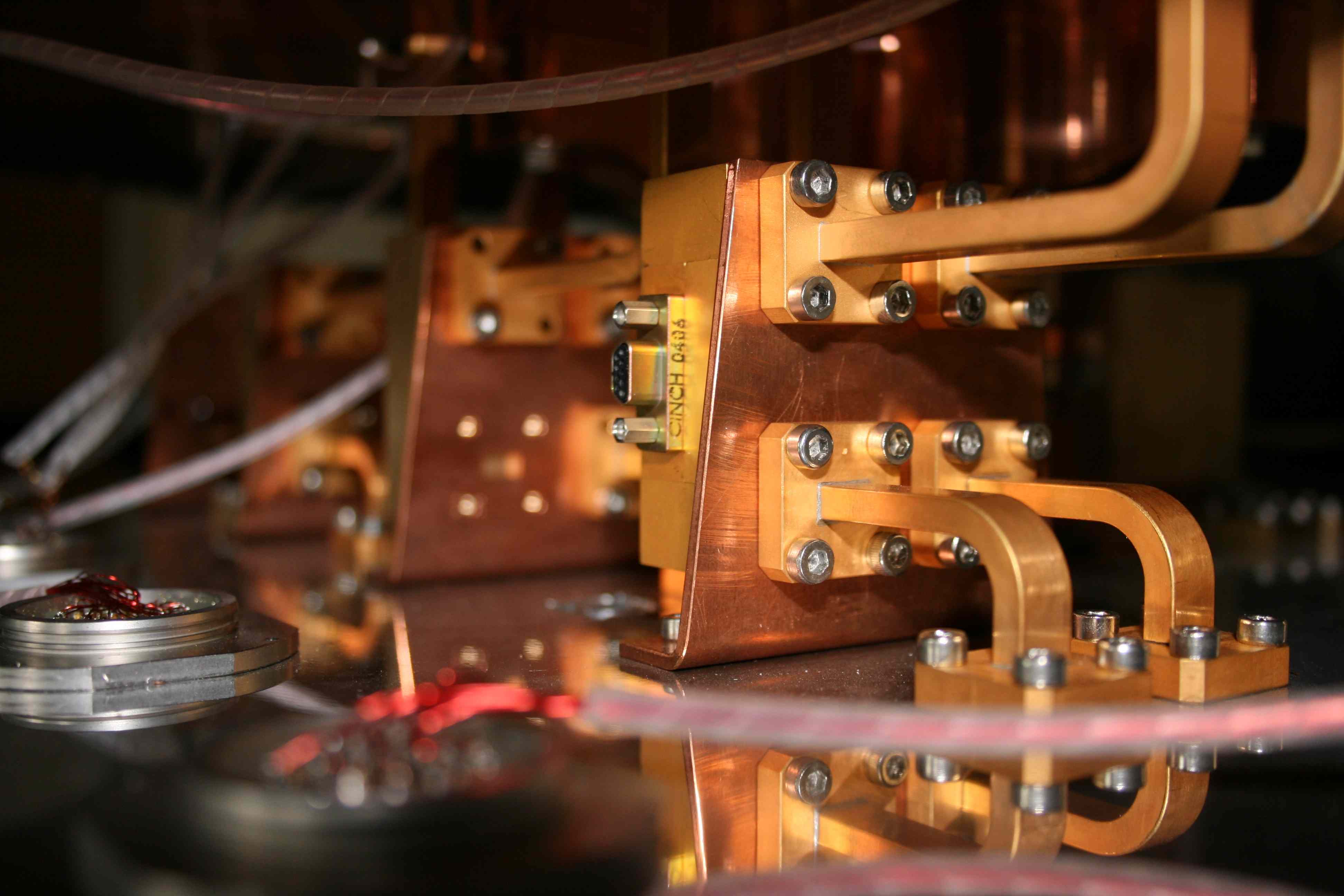}
\caption[The waveguides located within the OCRA-F cryostat]{The waveguides located within the OCRA-F cryostat. Left: the Z-shape waveguide connecting to the FEM, and the C-shape waveguide leading to the BEM. Right: the C-shape waveguide arriving at the BEM, and the L-shape waveguide leaving it.}
\label{fig:waveguides}
\end{center}
\end{fig}
Various sets of waveguides are used to guide the signal through the receiver. These start off with the horn, which has a thermal insulator attached prior to a circular-to-rectangular convertor. A Z-shape waveguide brings the signal into the FEMs, then a C-shape waveguide transfers the signal from the FEM to the BEMs. An L-shape waveguide then takes the output of the BEMs to the base of the cryostat where the filters are connected. These are shown in place within the cryostat in Figure \ref{fig:waveguides}.

Whilst the horns were not tested by the Author, the remaining components were. The tests were carried out by connecting the waveguides to an Anritsu 37397A Vector Network Analyser, which passed a known signal through them and measured the transmission and reflection of that signal, in both directions. This same device was used by D. Lawson and E. Blackhurst to measure the gain of the FEMs and the BEMs, and was also used by the Author to test the individual receiver chains (see Section \ref{sec:fem_bem_combination}). Prior to use, it was calibrated using a standard calibration kit to ensure that when the waveguide ports were connected together, the transmission between the ports was flat and at 0dB (within an uncertainty of $\sim$0.1dB), and that the return losses were around -30dB.

\subsubsection{Horn to waveguide thermal insulators and circular-to-rectangular convertors}
The thermal insulators are essentially stainless steel discs with a circular hole in the centre that acts as the waveguide. Several materials were considered for the insulators prior to the involvement of the Author, most notably gold-plated Nylon, however these were not deemed to be efficient waveguides. To optimise the insulation of the spacers, the surface area that connects them to the horns and the later waveguides was minimized by the removal of contact material not required to transmit the RF signal or maintain the spacer positions.

The RF performance of the insulators were tested using two rectangular-to-circular convertors to connect to the rectangular waveguide connectors of the VNA. Initial tests of the insulators showed a large number of spikes in their transmission (see Figure \ref{fig:spacergain14apr}). This is thought to be due to reflections from internal surfaces due to imperfect alignment of the components. To reduce this, the insulators were cleaned, and their siting was optimised. This reduced the spikes significantly, but did not completely eliminate them. However, as OCRA-F is a broadband receiver rather than a spectral receiver, these remaining spikes were deemed to be unimportant.

\begin{fig}
\begin{center}
   \includegraphics[scale=0.55]{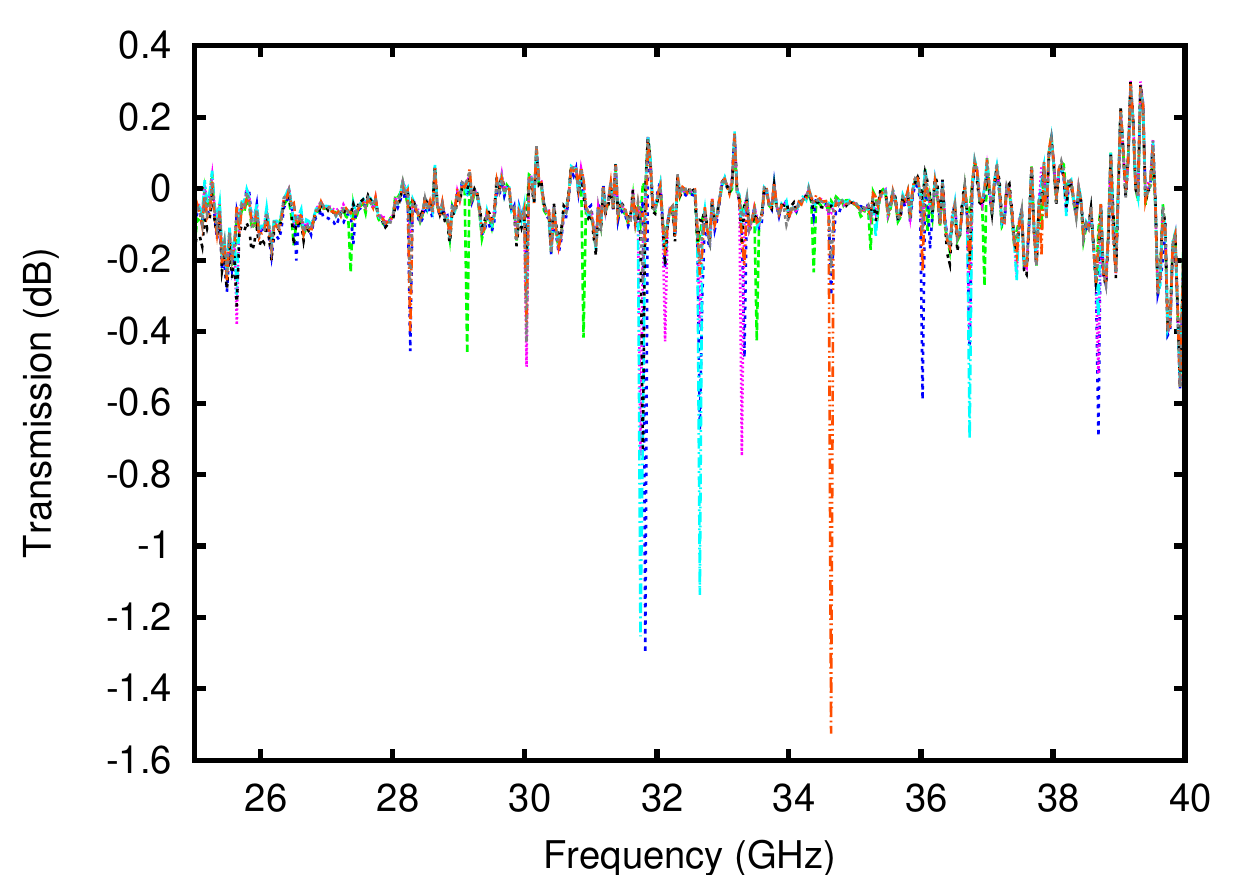}
   \includegraphics[scale=0.55]{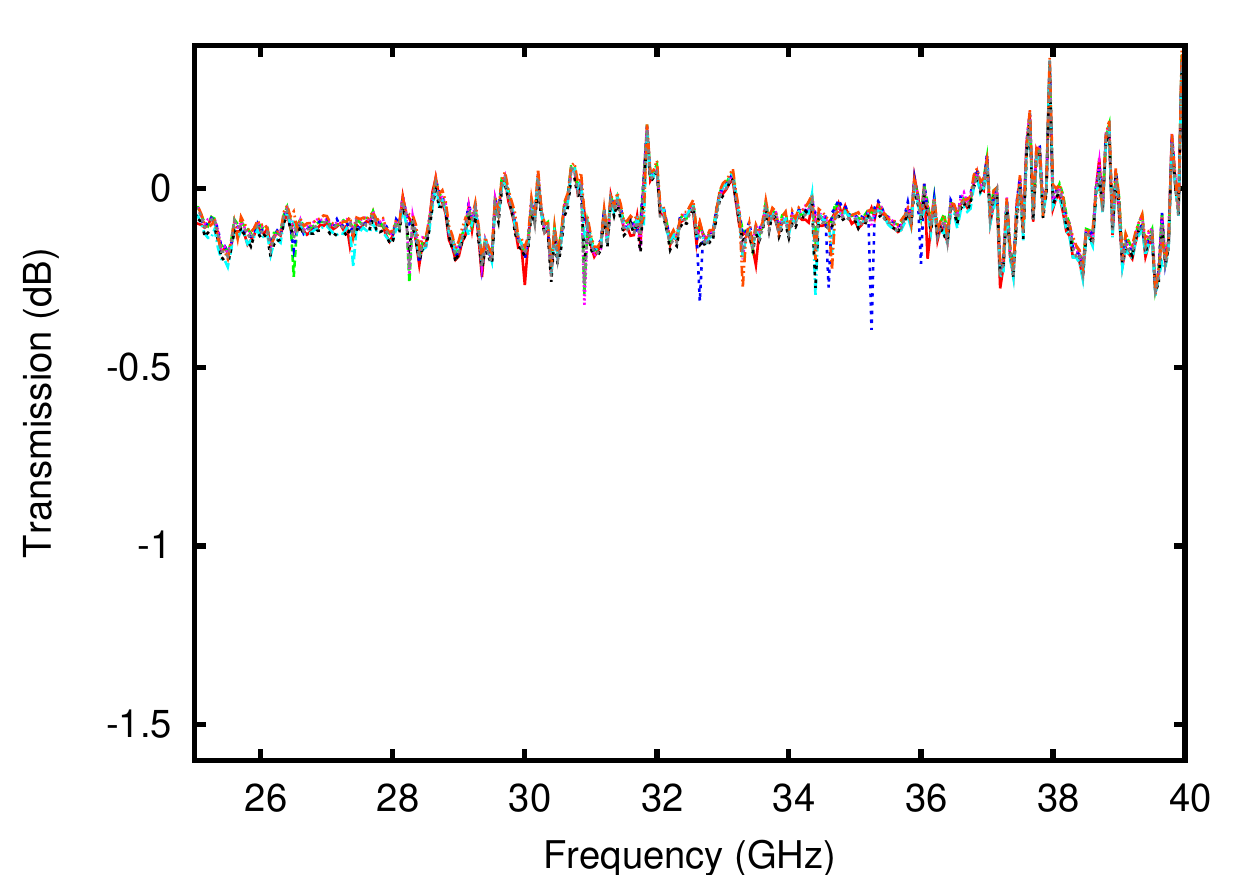}
\caption[Transmission properties of the thermal insulators between the horns and the circular-to-rectangular waveguides]{Transmission properties of the thermal insulators between the horns and the circular-to-rectangular waveguides prior to optimising (left) and after (right).}
\label{fig:spacergain14apr}
\end{center}
\end{fig}

At the same time as the thermal insulators were tested, the circular-to-rectangular convertors were also tested in pairs. Spikes also appeared in the gain as a function of frequency measured during these tests. These spikes are due to the same positioning issue as was present in the insulators.

\subsubsection{Interconnection waveguides}

The other waveguides within the cryostat were also tested to check their transmission performance. The C-shape waveguides between the FEM and the BEM have a loss of $\sim 0.3$dB (compared to $<$0.1dB for the other waveguides); this is unimportant, however, as they come after the initial amplification. It may even be beneficial -- it means that there is some loss (albeit not much) in the signal reflected by the BEMs. It is interesting to note that C-shape 7 was tested as more lossy, and subsequently broke.

The waveguides were made commercially using Nickel-Colbalt (NiCo), and were purchased in two batches. The first set of waveguides appear to have insufficient NiCo deposited on them during the electrolysis process, resulting in them being rather fragile. Two of the L-shape and one of the C-shape waveguides were broken either during transport or during installation; a Z-shape waveguide subsequently cracked during receiver testing. The second batch of waveguides have had more NiCo deposited on them, such that they look and feel more substantial -- to date, none of these have failed.

\subsubsection{Filters}
\begin{fig}
\begin{center}
   \includegraphics[scale=0.55]{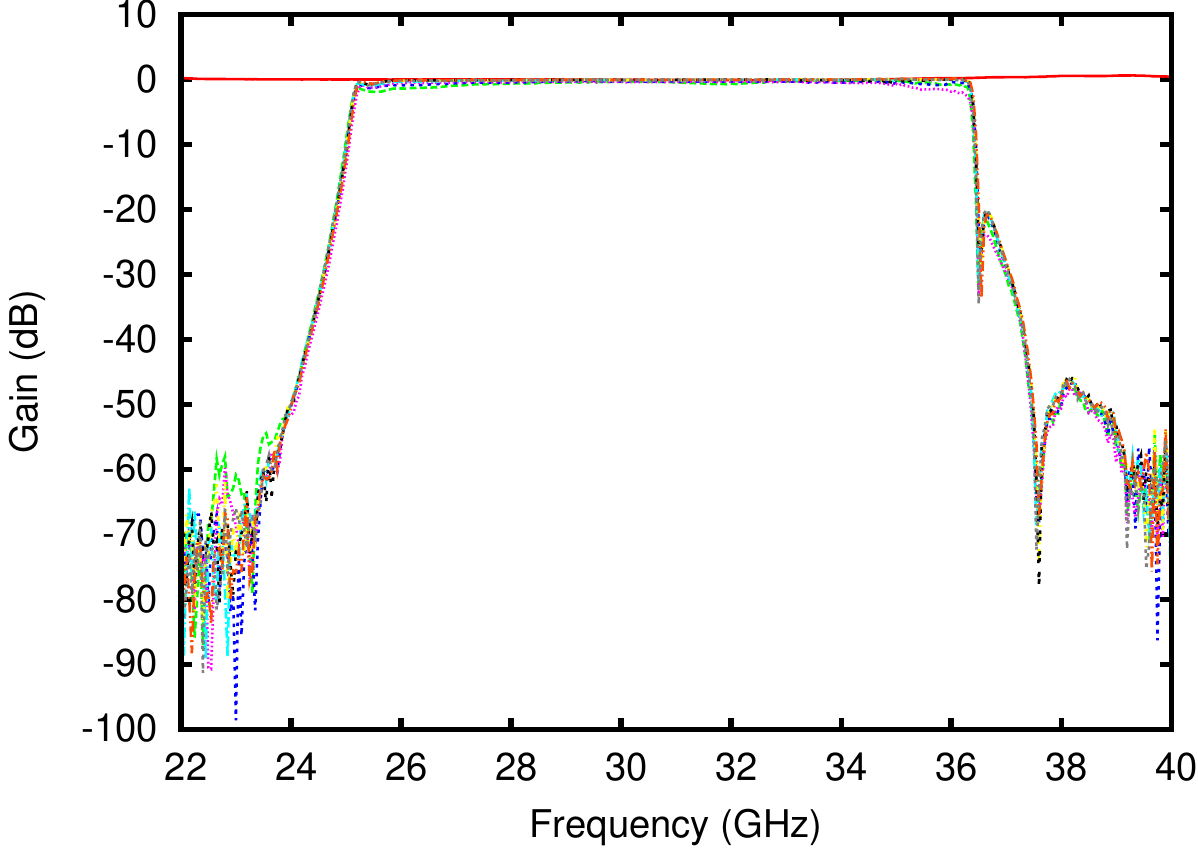}
   \includegraphics[scale=0.55]{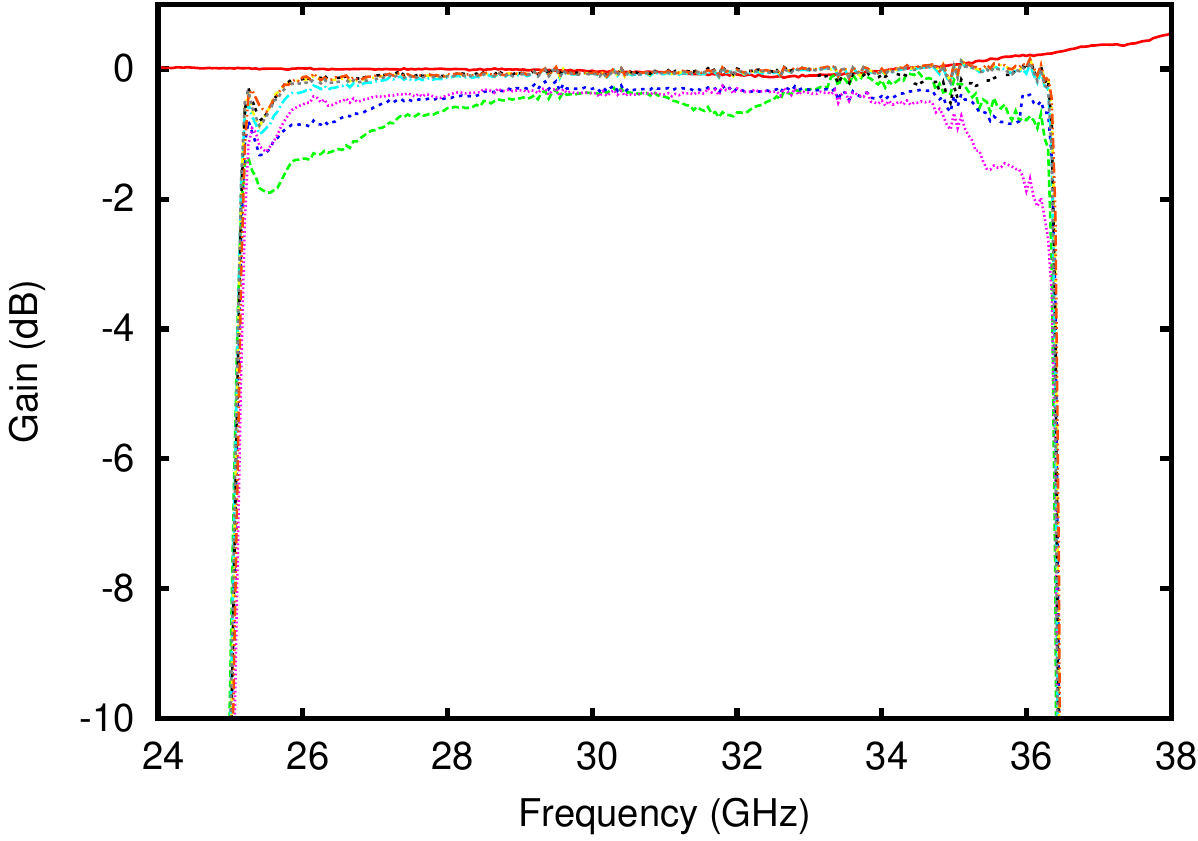}
\caption[Transmission properties of the OCRA-F filters]{Transmission properties of the OCRA-F filters as measured on 14 April 2008. The left-hand plot shows the complete range of transmission values between 22 and 40~GHz (the waveguide cut-offs can be seen at either end); the right-hand plot is zoomed in on the bandpass region between 24 and 38~GHz and shows 0 to -10dB.}
\label{fig:filters}
\end{center}
\end{fig}

In order to remove out-of-band noise, filters are used to define the frequency range of interest. The filters used within OCRA-F are manufactured by A1 Microwave, and are model number PB1332WB. They have a specified frequency range of 23 to 36~GHz. Figure \ref{fig:filters} shows the measured frequency range of the filters; the filters have very sharp cut-offs below 25 and above 36.5~GHz. Although different from the specified values, this range is suitable for use in OCRA-F as it increases the frequency separation from the 22~GHz atmospheric water vapour line, the most important source of atmospheric $1/f$ noise.

\subsection{Cryogenics}
When cold, the outer radiation shield and horn plate of the OCRA-F cryostat are cooled to $\sim$50~K, and the innermost plate upon which the FEMs are mounted is cooled to $\sim$20~K. As OCRA-F has a large cryostat (nearly 60cm across), these temperatures are not trivial to obtain. The cryogenic pump used within OCRA-F, CTI-Cryogenics model 1020 CP, is one of the most powerful available.

During the reassembly of the cryostat, various tweaks were made to the cryostat to improve its cryogenic performance. The Author was involved in the creation of a number of thermal super-insulation ``blankets'', which consist of multiple alternating layers of aluminium-coated mylar foil and loose nylon weave. These were positioned above and below the 20K plate, as well as around the outside of the radiation shield and on the 50K stages where they face loads of 300K. This thermal insulation prevents the colder components from ``seeing'' the hotter ones -- mainly the outer cryostat, which is at room temperature -- thus making it easier to cool the receiver, and to keep it cold.

To monitor the performance of the cryostat, four temperature diodes were positioned within it. The first of these is on the 50~K plate, the second on the cold head, the third is on the 20~K plate and the fourth on the horn plate (see Figure \ref{fig:ocraf_test_config}). Once cold, the cryostat obtains $\sim17$~K on the cold stage, giving a physical FEM temperature of $\sim20$~K. The horns and the 50~K plate are cooled to $\sim 45$~K. During initial testing, the cryostat was able to maintain these temperatures over periods of months; once mounted upon the telescope the temperatures of the receiver will be monitored routinely to quantify the temperature stability.

\subsection{Electrical connections}
During the reassembly of the cryostat, all of the wiring within it was replaced with manganine wire, which conducts less heat than copper wire and hence is preferable within a cryostat. However, manganine wire is higher resistance than standard, copper wire, and gives a resistance of 72 ohms between the connection at the power supply and the FEM connectors, hence causing a voltage drop between the two. This necessitates an increase in the drain voltages used to bias the amplifiers compared to the voltages used in laboratory tests (and hence those for which the balance of the module was determined) to compensate for this voltage drop and to bring the currents to the same level as used in the laboratory.

During the construction and initial testing of OCRA-F, the join between the manganine and connector wires for the FEMs was identified as a weak point: several wires broke at this point. In future, this join should probably be done by a breadboard, or the manganine wire should be connected straight to the connectors.

\section{Radiometer chain performance} \label{sec:radiometer_tests}
\subsection{FEM-BEM combination} \label{sec:fem_bem_combination}
The combination of the FEMs and BEMs were tested within the cryostat at room temperature prior to the installation of the horns on the receiver. The principle tests were carried out using the Vector Network Analyser to investigate the gain of the complete receiver. The tests were carried out using nominal biases for the FEMs at room temperature as previously used for tests of the modules, with drains of 1.5V and gates set such that the currents drawn by each pair of stages was 10mA, and also nominal biases for the BEMs (as per Section \ref{sec:bems} above). In addition to measuring the gain properties of the receiver chains, this also provided a test of the system assembly; several issues with the bias wiring were highlighted and fixed during these tests.

\begin{fig}
\begin{center}
   \includegraphics[scale=0.55]{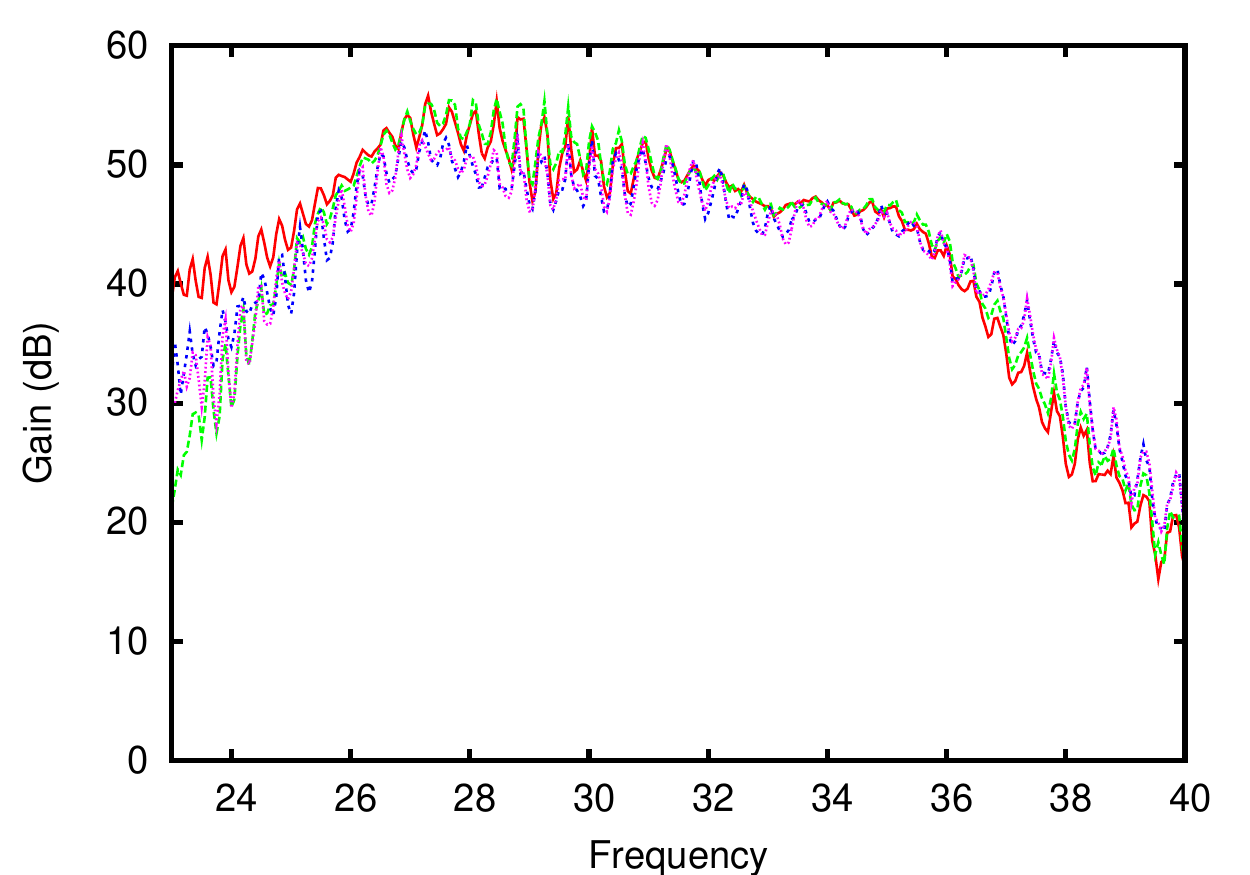}
   \includegraphics[scale=0.55]{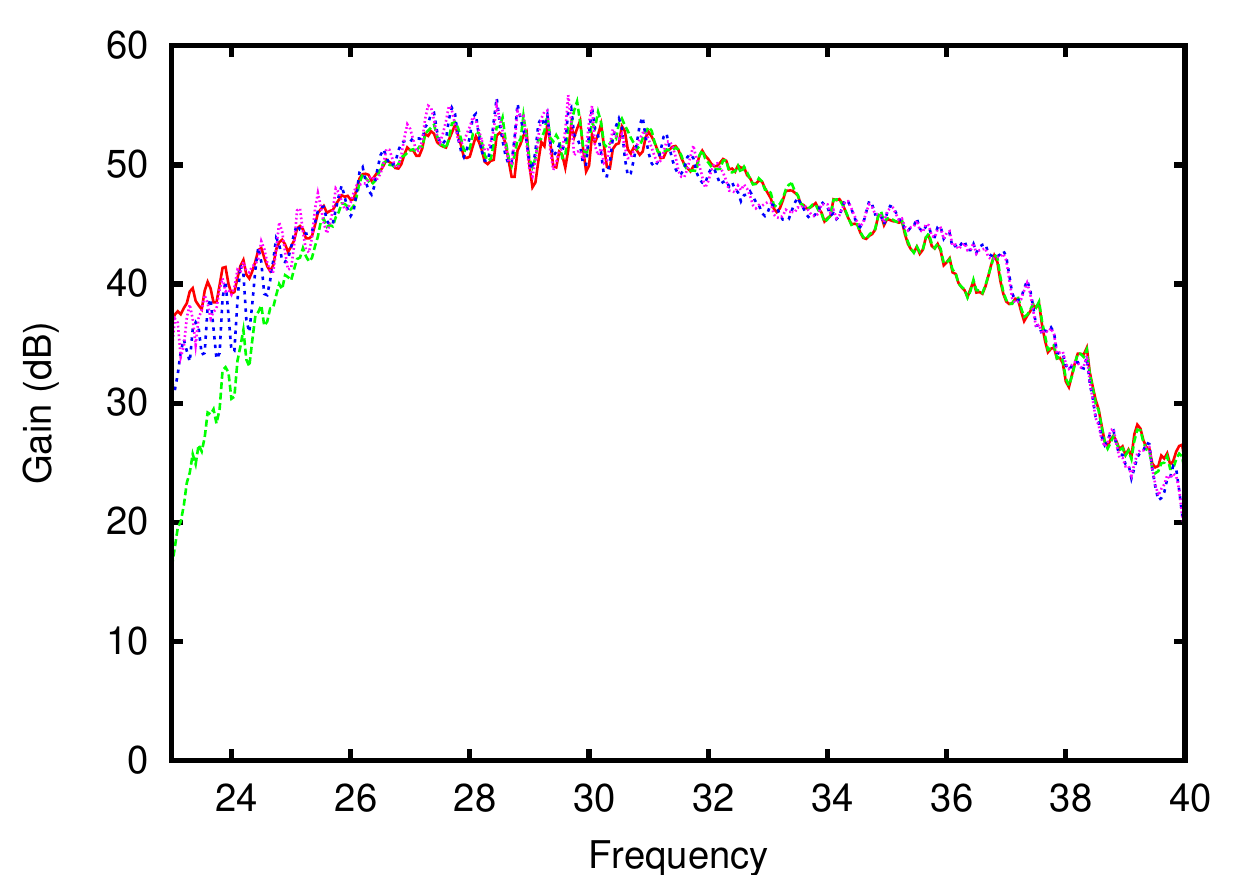}\\
   \includegraphics[scale=0.55]{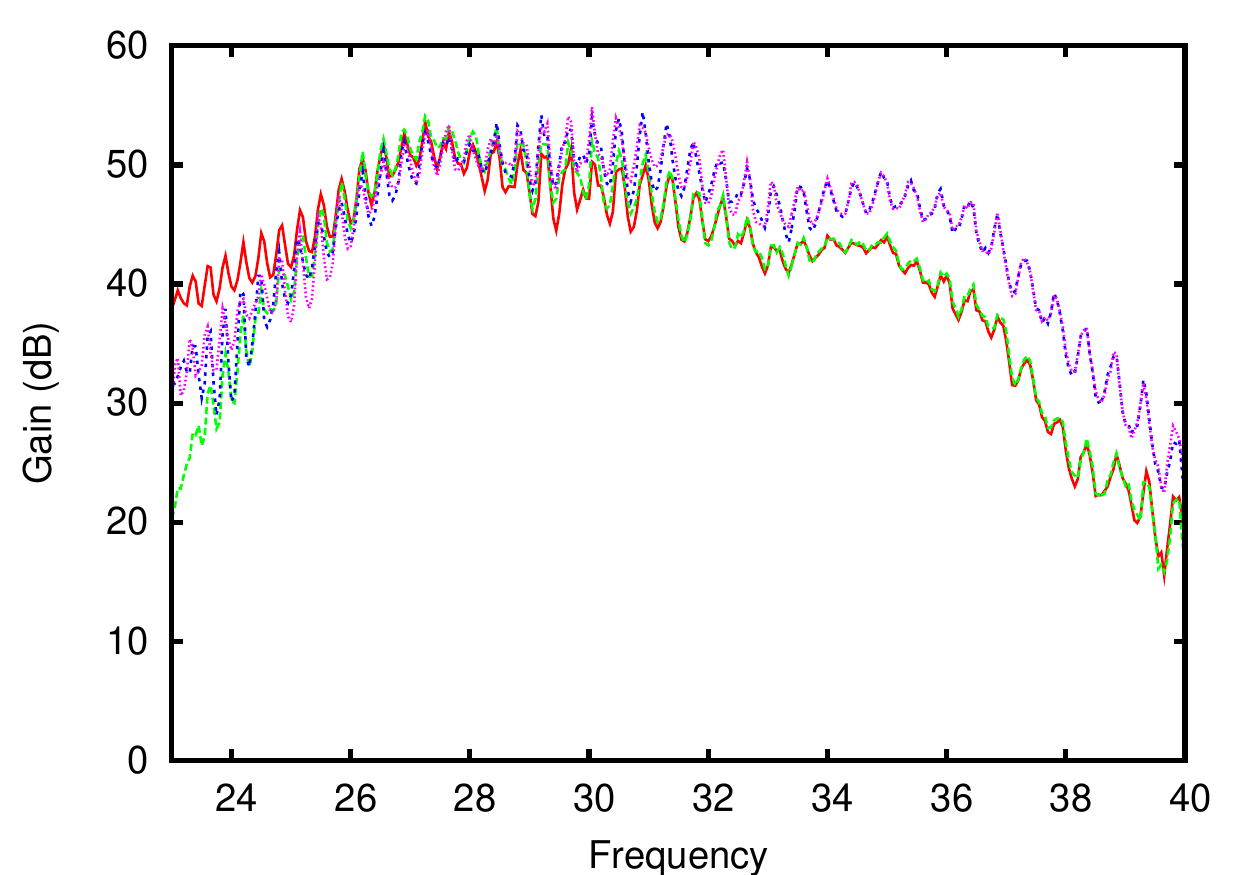}
   \includegraphics[scale=0.55]{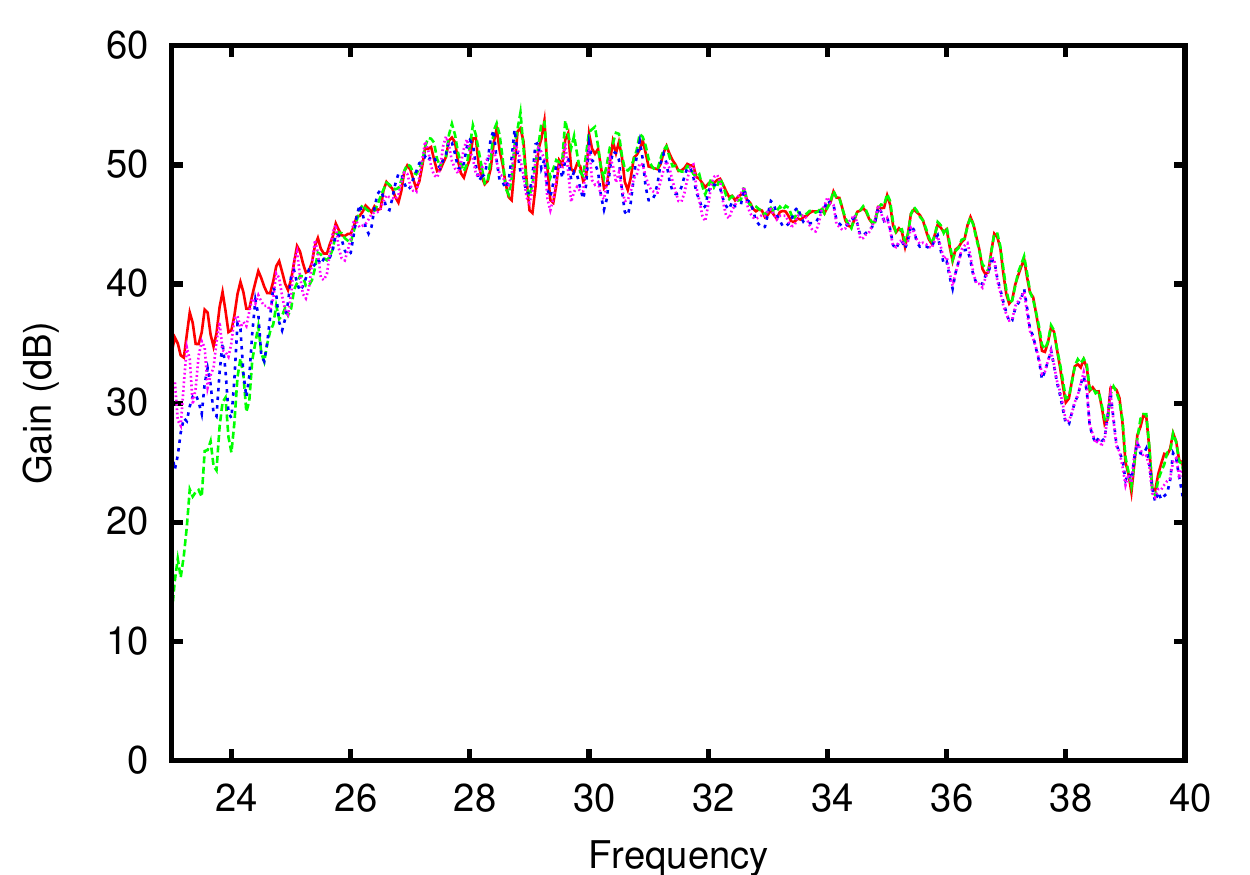}
\caption[Combined gain as a function of frequency from the OCRA-F FEM-BEM combinations]{Combined gain from the OCRA-F FEM-BEM combinations for Chains 1 (top left), 2 (top right), 3 (bottom left) and 4 (bottom right). The four lines are the 4 different phase switch settings; all signals were put into the same FEM input, and measured at the appropriate BEM output.}
\label{fig:fem_bem_gain}
\end{center}
\end{fig}
Figure \ref{fig:fem_bem_gain} shows the gains of the four chains, where the signals have been put into the first input to each of the FEMs and measured for all four phase switch settings at the appropriate BEM outputs. The gain as a function of frequency is broadly the same between all four chains, although with the chosen BEM bias settings the third receiver chain has a lower gain at high frequencies for one of the outputs.

All four chains have ripples in the gain curves; this is caused by the interaction between the FEM and the BEMs due to the bad input return loss of the BEMs (see Section \ref{sec:bems}). The effect is similar to that seen in the back end of OCRA-p during testing (see \citealp{2006Lowe}), and which was resolved via the insertion of an isolator before the back end amplifier in OCRA-p. In the case of OCRA-F, however, this is impractical as both the FEM and the BEM are located within the cryostat, such that there is no space to insert an isolator without modification to the waveguides. \footnote{It would be possible to install a small piece of resistive foam within the FEM-BEM waveguides to attenuate the RF signal, which would effectively do the same as an isolator. Tests using FEM~3 and a BEM at room temperature with a variable attenuator between the FEM and the BEM showed that as the attentuation is increased the gain ripples decrease, with around 6~dB attenuation required to remove the ripples fully.} However, the gain ripples are not critical as OCRA-F is a broadband continuum receiver.

\begin{tab}{tb}
\begin{tabular}{c|c|c|c|c|c|c|c|c}
\bf Property \rm & \multicolumn{2}{c|}{{\bf Chain~1}} & \multicolumn{2}{c|}{{\bf Chain~2}} & \multicolumn{2}{c|}{{\bf Chain~3}} & \multicolumn{2}{c}{{\bf Chain~4}} \\
\hline
Input & 1 & 2 & 1 & 2 & 1 & 2 & 1 & 2\\
\hline
Gain (dB) & 48.0 & 47.2 & 48.6 & 48.7 & 47.2 & 47.4 & 48.4 & 48.0\\
Bandwidth (GHz) & 7.9 & 7.9 & 7.4 & 7.4 & 7.3 & 7.2 & 8.0 & 8.0\\
Isolation (per cent) & 8.7 & 9.8 & 7.8 & 8.1 & 8.0 & 8.2 & 10.0 & 9.2\\
\end{tabular}
\caption[Room temperature properties of the OCRA-F receiver chains]{Room temperature properties of the OCRA-F receiver chains from measurements with the VNA. The bandwidth is close to the maximum value, cf. Table \ref{tab:ocraf_specs}}
\label{tab:fem_bem_properties}
\end{tab}

Table \ref{tab:fem_bem_properties} gives the room temperature values of the average gain, bandwidth and isolation of the OCRA-F FEM-BEM combination measured using the VNA, with the filters in place, and at nominal biases. The average gain was calculated using the mean of the recorded values between $\nu_\mathrm{min} = 25$ and $\nu_\mathrm{max} = 36$~GHz. The bandwidth $B$ was calculated via
\begin{equation}
B = \frac{\left( \sum_{\nu_\mathrm{min}}^{\nu_\mathrm{max}} S_{21,i}^2 \Delta \nu \right)^2}{\sum_{\nu_\mathrm{min}}^{\nu_\mathrm{max}} S_{21,i}^4 \Delta \nu},
\end{equation}
where $S_{21,i}$ is the measured gain between the input (1) and output (2) at each frequency step (squared such that it is a measurement of power), and $\Delta \nu$ is the size of the frequency step. This also shows that the gain ripples do not significantly affect the bandwidth. The isolation is calculated by measuring the output from both BEMs using the same input to the FEM and the same phase switch settings, then working out the percentage of the signal that is coming out of the ``wrong'' BEM compared to that from the ``right'' BEM, again between 25 and 36~GHz. Whilst this is larger than was measured by the FEMs alone, it should be remembered that this is using nominal bias settings for the amplifiers, rather than optimised ones. Improved bias settings to balance the gain and phase from the LNAs should improve the isolation.

Following from these tests, the horns were installed into the cryostat and the cryogenic performance of OCRA-F was tested. Tests of the complete receiver chains at cryogenic temperatures were subsequently carried out.

\subsection{System temperature measurements} \label{sec:ocraf_tsys}
It is important to measure the system temperature of the receiver as this determines the Gaussian noise level of the instrument. For OCRA-F, these measurements also revealed two important issues within the receiver -- that the detectors were non-linear (discovered by inconsistent results from different measurement techniques) and that the foam window in front of the receiver was ``noisy'' (producing a higher that expected system temperature). The measurements and discoveries are detailed below.

In order to carry out measurements of the system temperature, OCRA-F was relocated in December 2008 to the ``Don Dome'', a test facility away from buildings that consists of a metal reflective plate angled at $45 \degree$ at which the receiver is pointed, such the receiver sees the sky; further angled metal plates minimize the ground radiation that is seen by the receiver, and the complex is protected from the weather using radio transparent plastic sheeting (in practice, a poly-tunnel). This setup additionally allows the easy insertion of either room temperature or cooled sheets of absorber between the sky and the receiver.

System temperature tests were performed by measuring the output power from the receiver when looking at a 300~K (room temperature) sheet of microwave absorber, and also a sheet of absorber cooled to 77~K using liquid nitrogen. The system temperature $T_\mathrm{sys}$ can then be calculated using
\begin{equation}
T_\mathrm{sys} = \frac{T_\mathrm{hot} - Y T_\mathrm{cold}}{Y - 1}
\end{equation}
where $Y = V_\mathrm{hot} / V_\mathrm{cold}$ in the case of a measured voltage $V$ or $Y = 10^{P_\mathrm{hot} / 10}/10^{P_\mathrm{cold} / 10}$ in the case of a measured power in dBm.

The output power was measured in two ways: using the detectors and video amplifiers, or using a power meter. Initial tests with the detectors resulted in system temperature measurements of 20--30~K, with one output giving 50~K, which seemed reasonable. However, tests with the power meter yielded values closer to 60-70~K. After a lengthy investigation, it emerged that the detectors were being driven in their non-linear regime (see Section \ref{sec:ocraf_detectors}). Once the BEM amplification was decreased (by reducing the gate and drain voltages, and hence the current drawn) the two methods came into agreement and the same system temperatures of 60-70~K were measured using the detectors as was measured by the power meters. In practice, this non-linearity is only a problem for system temperature measurements due to the large range of input loads being seen (room temperature vs. a cold load). Astronomical sources do not display such a large range, so effects of this non-linearity on these measurements should be negligible.

The system temperature of OCRA-F should have been dominated by the noise from the initial amplifiers in the FEMs, which had been measured to be 25--30~K (see Section \ref{sec:ocraf_fems}, \citealp{2007Kettle}). It was thought that the increase in system temperature to 60-70~K may have been due to water vapour within the cryostat, however purging the cryostat with dry air had negligible effect. It soon emerged that the problem lay in the foam window, which was adding $\sim$20--30~K extra noise; this is discussed in the following section.

\subsection{Foam window} \label{sec:foam}
\begin{fig}
\begin{center}
	\includegraphics[scale=0.08]{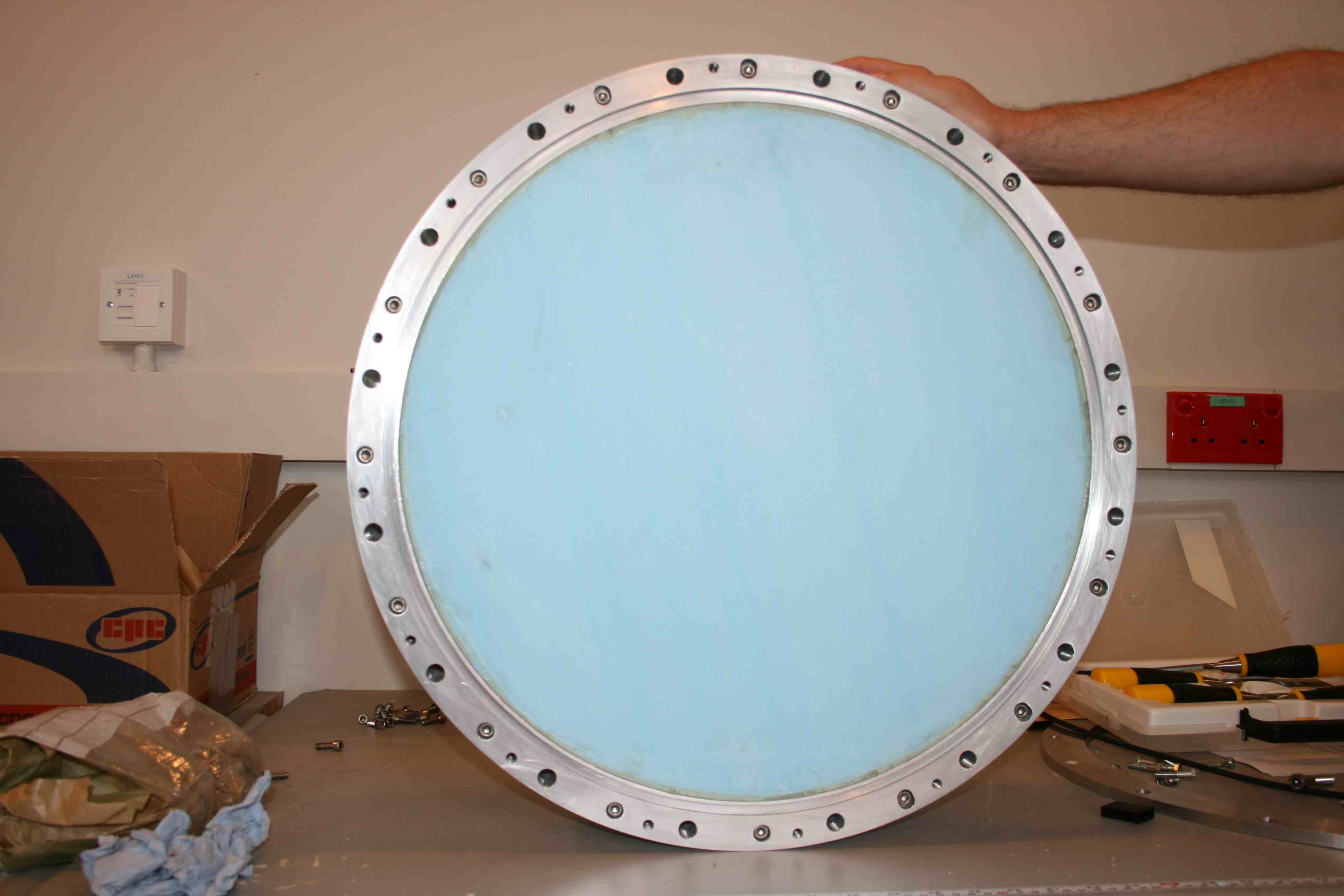}
\caption[The original foam window for OCRA-F]{The original foam window for OCRA-F, enclosed within its mounting ring. Note the hand on the top-right for scale.}
\label{fig:ocraf_foam}
\end{center}
\end{fig}

In order for a receiver in a vacuum chamber to see outside of the chamber, it has to look through a window. For most receivers this would consist of a thin mylar window, however as OCRA-F has such a large vacuum window (480mm; the largest constructed at JBO to date), a foam support is required for the mylar sheet.

This foam window, depicted in Figure \ref{fig:ocraf_foam}, was originally constructed using high density styrofoam manufactured by Dow Corning, and is retailed as Floormate 500-A. Initial tests found that this was able to support the pressure of the vacuum (around 1.7 tonnes) when suitably thick. Measurements with a VNA carried out by Neil Roddis on a piece of foam that had been baked and inserted into a waveguide suggested that its insertion loss was sufficiently small that it was no longer measurable. Based on this later evidence, the foam was used for the OCRA-F window. However, the tests with the complete OCRA-F receiver revealed that the system temperature was higher than expected by a factor of 2 (see Section \ref{sec:ocraf_tsys}). Subsequently, the noise properties of the foam were tested using a spare receiver system from the VSA, which provides a more conveniently sized test system with a lower system temperature than that of OCRA-F and its window. Tests of a smaller piece of a lower density version of the foam (Floormate 200-X, around 2/3 the thickness of the foam window) showed that the foam added $\sim$25K to the system temperature. This was not reduced by vacuum-baking the foam for several weeks, something that is as expected for a closed cell foam. As a result, it became apparent that different and lower loss foam was required for the window support.

A number of other high frequency experiments use foam either to support a vacuum window, or to support elements of the telescope. The most widely used type of foam is Zotefoam PPA-30, which is used by AMiBA, ACBAR, QUaD and SPT \citep[][respectively]{2003Chen,2003Runyan,2008Hinderks,2004Ruhl}, however it was not possible to obtain a suitable quantity of this foam. CBI and C-Bass use Plasterzote LDF45, which has less than 1 per cent loss at 30~GHz (Prof. M. Jones, private communication), unfortunately this low density foam does not have the mechanical strength to be used as a window.

Samples of alternative foams from the same manufacturer were obtained and tested using the VSA receiver system. Tests of 1~cm thick samples of Zotek F30, F40HT and F75HT, which were selected as they were made using inert nitrogen, showed an increase of $\sim$3K in the system temperature, thus these foams are not suitable for use for the window. Tests of Plastazote HD30, HD80 and LD45 were more promising; a 1~cm thick sample of these negligibly changed the receiver temperature. Finally, Plastazote PK80 was also tested; this gave an increased system temperature of $1.5 \pm 0.5$K for a 4~cm-thick piece (0.03dB loss).

Due to the availability of thick pieces of PK80 (formed by heat bonding multiple layers of foam together), this foam was used to make an OCRA-F window. Initial mechanical tests with this window however showed that the foam distorted easily, and the window collapsed inwards after a few hours of testing. As the full aperture of OCRA-F is not required at this point, because only 8 of the 16 horns are installed, a metal support plate was constructed to reduce the load on the foam. This support plate proved effective in reducing the distortion of the foam, and the new window with the lower loss foam support should be installed on OCRA-F in early 2010.

It is worth noting that other window materials exist aside from foam. ACT, for example, uses 4~mm thick Ultra High Molecular Weight Polyethylene \citep{2008Swetz} and the high frequency (90~GHz and above) instruments of ALMA use crystal quartz windows with plastic antireflective coatings \citep{2001Koller}. These were not investigated for use with OCRA-F for cost reasons, but may be of use for future OCRA instruments.

\section{Commissioning} \label{sec:commissioning}
OCRA-F was shipped to TCfA in July 2009. For the commissioning period a window made of the original (lossy) foam was used whilst a new window with mechanical support was being fabricated from the less lossy foam. Once at TCfA, it was integrated with the OCRA-F DAQ and a field rotator, and initial tests of the complete system were made. OCRA-F was then installed onto the 32~m telescope on 7 December 2009, and the process of commissioning the receiver was started.

The initial tests of the radiometer chains described thus far used a test DAQ, which could only record two inputs at once. A new DAQ capable of recording all 16 inputs from a complete OCRA-F was purchased and prepared in TCfA; this was connected to OCRA-F in August 2009. The DAQ consists of a dual-core computer system with a National Instruments DAQ card installed. This card interfaces with the output from the video amplifiers using a pair of National Instruments SCB-68 shielded breakout boxes. In addition to recording the output of the 8 OCRA-F channels, the DAQ also records the phase switch state and the time that the samples are taken at, as well as the telescope position.

Initial tests with the combined system showed that a large 50~Hz signal was present; switching the fluorescent lighting in the room off significantly reduced this. This should not be an issue for the receiver on the telescope, however care needs to be taken to ground the receiver appropriately to minimize 50~Hz signals from the mains electricity supply.

\begin{fig}
\begin{center}
	\includegraphics[scale=0.5]{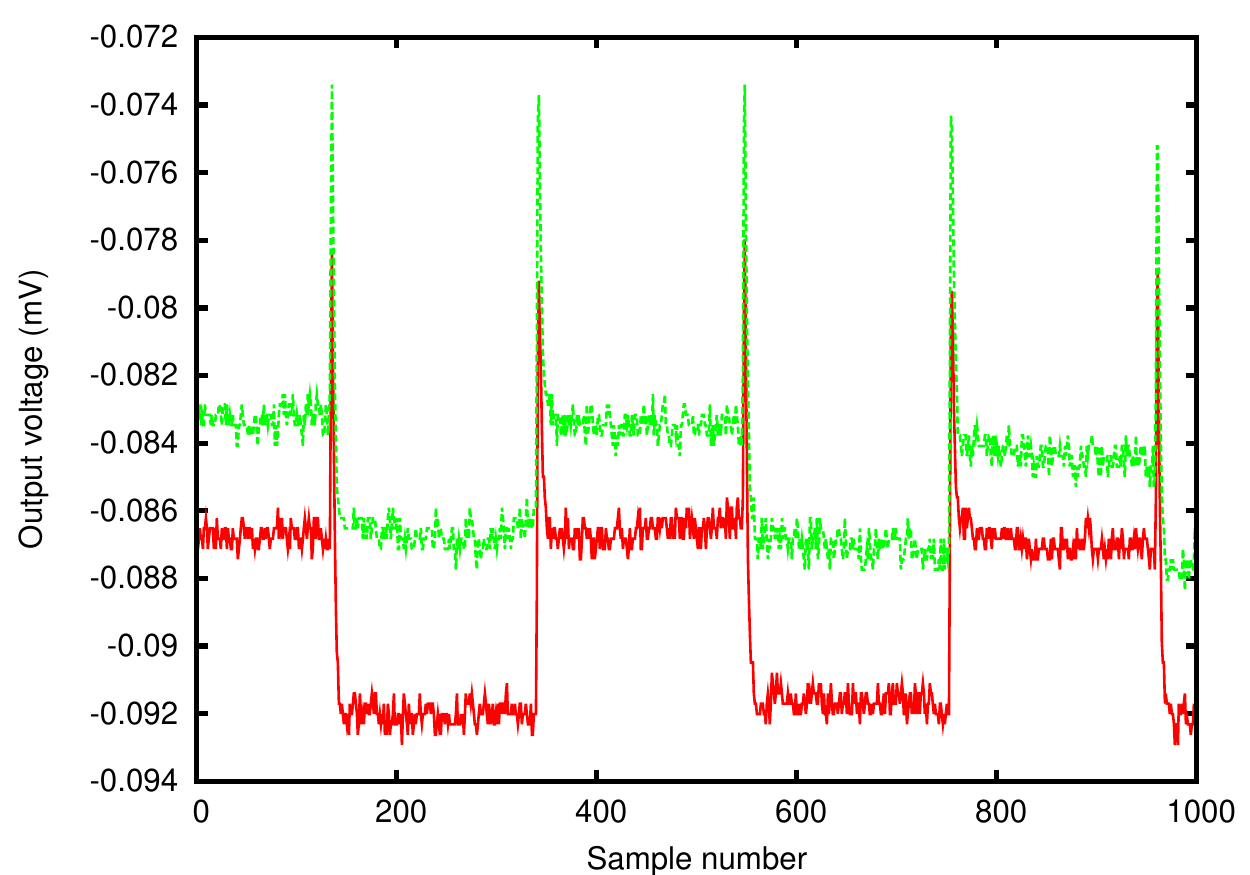}
	\includegraphics[scale=0.5]{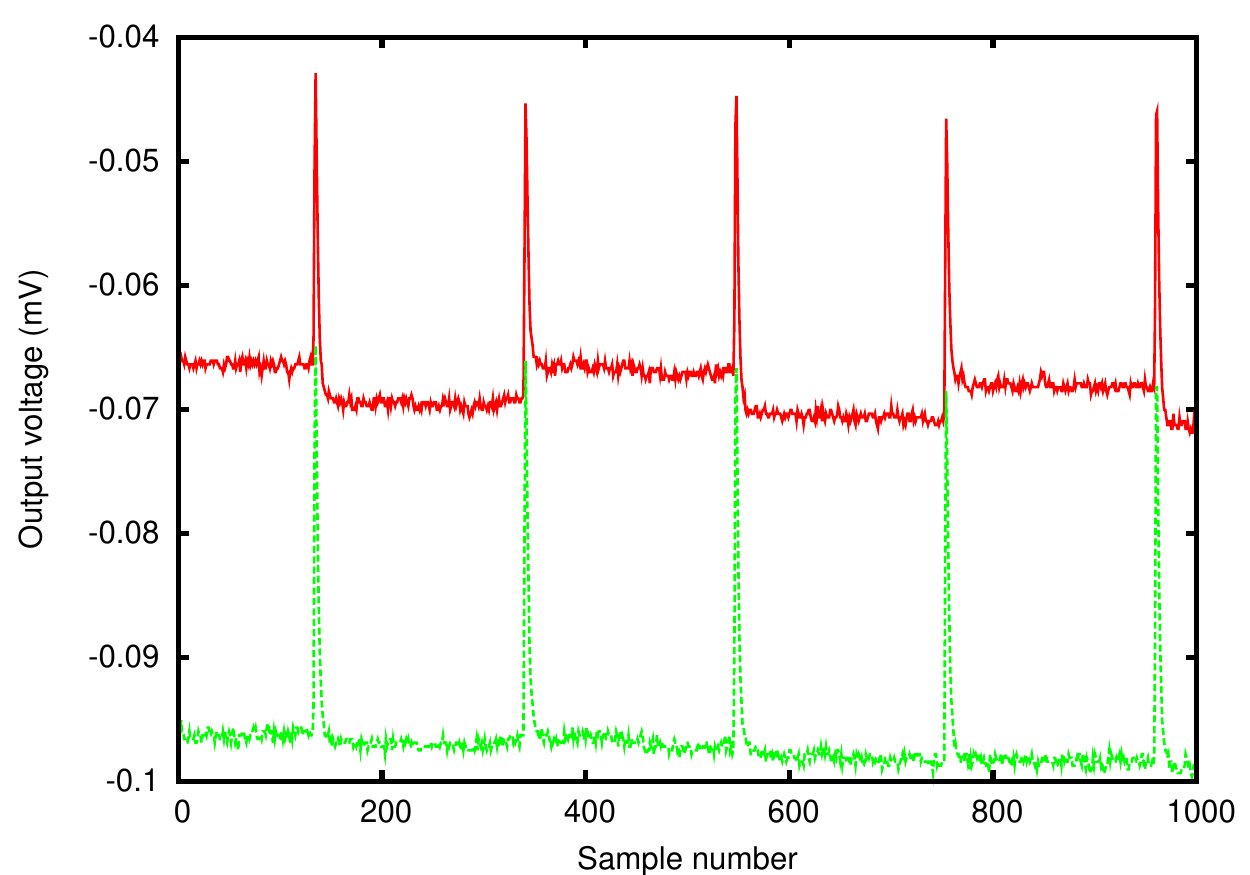}
	\includegraphics[scale=0.5]{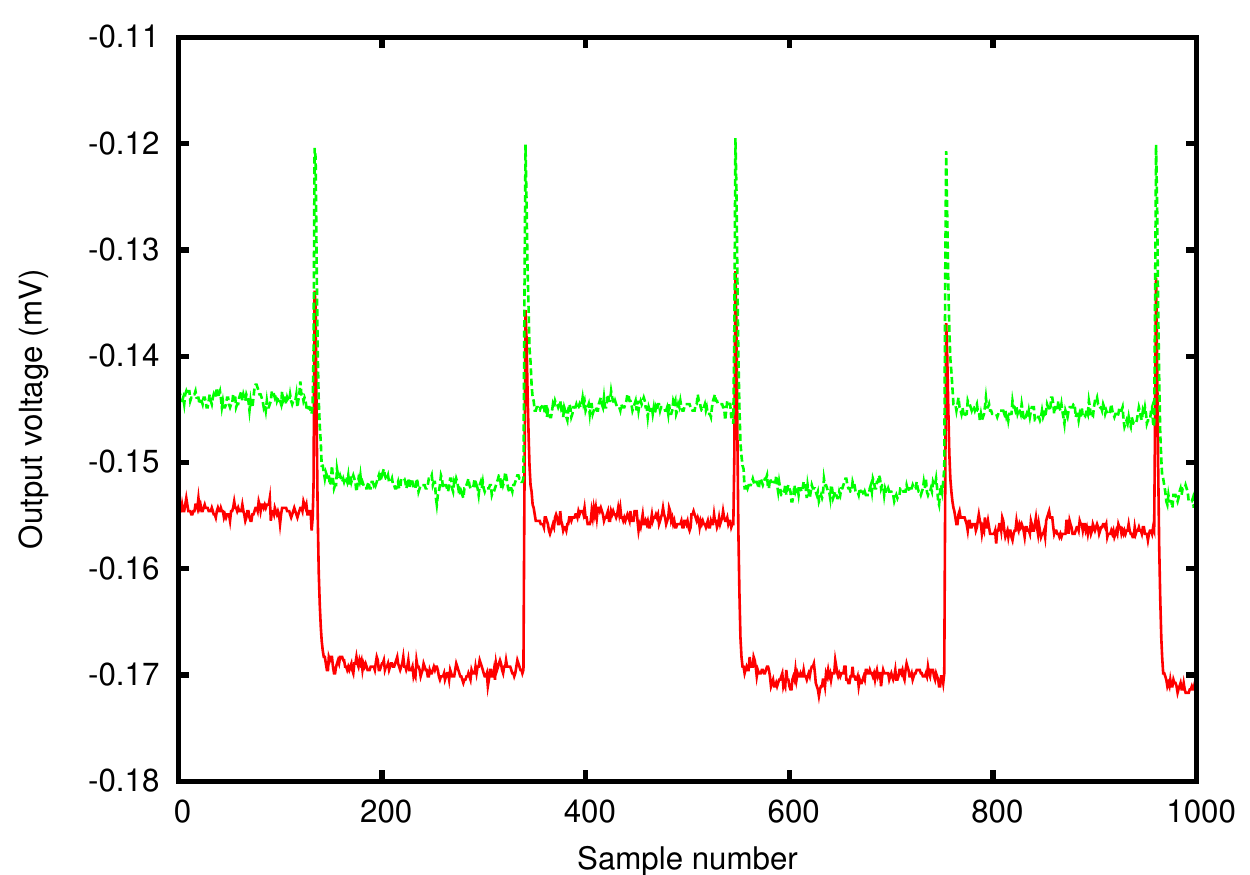}
	\includegraphics[scale=0.5]{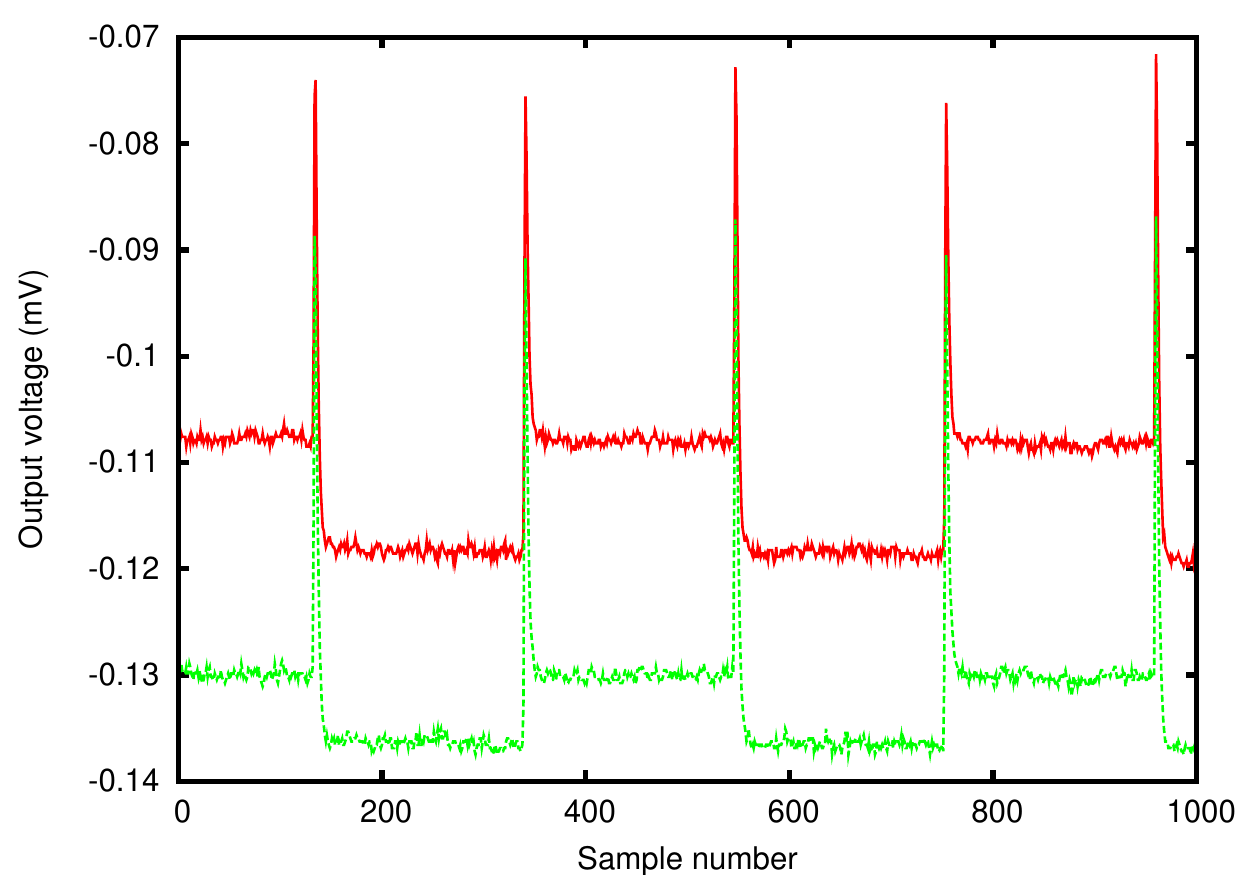}
\caption[Samples of the switched output from OCRA-F]{Samples of the switched output from the four receiver chains of OCRA-F. The output voltages depend on the gain during the measurements; these were taken before any fine-tuning of the amplifiers for performance. The phase switch is switching at 277 Hz; there are $\sim$200 samples per switch state.}
\label{fig:ocraf_output_switched}
\end{center}
\end{fig}

Figure \ref{fig:ocraf_output_switched} shows a sample of the switched output signal from all 8 outputs, taken whilst looking at a liquid nitrogen (77K) load. The first 7-9 of the $\sim 200$ samples in each switch state are systematically low in a characteristic `spike', caused by the finite time it takes for the phase switch to change from one state to the other, during which time it passes through an off state. These spikes will need to be removed (``blanked'') from the sampled output, reducing the integration time of the instrument by about 5 per cent (and hence increasing the noise level per second by the square root of that amount). Note that these measurements were taken in the laboratory prior to fine tuning of the amplifiers for performance, however the switching spikes depend solely on the phase switches and are independent of the amplifier settings.

\begin{fig}
\begin{center}
	\includegraphics[scale=0.5]{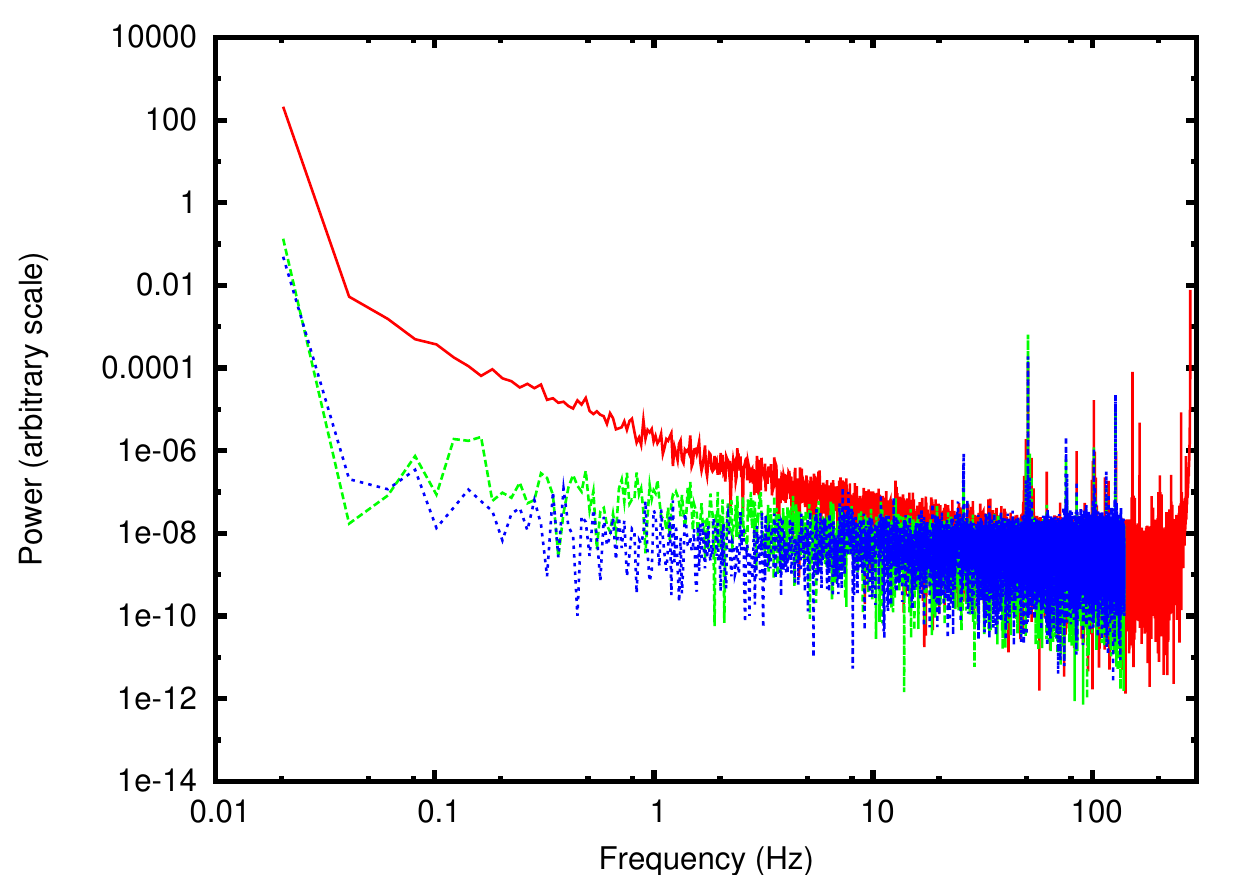}
	\includegraphics[scale=0.5]{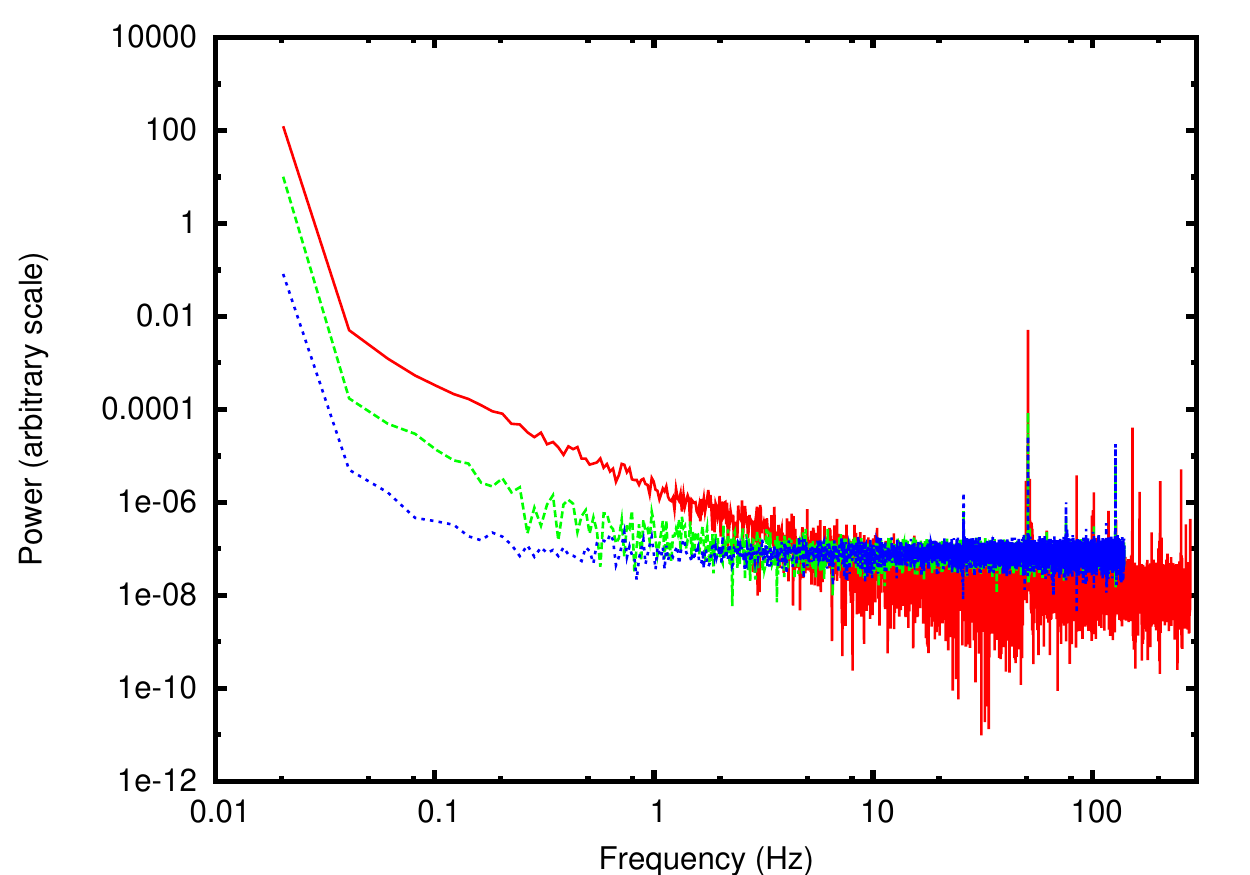}
	\includegraphics[scale=0.5]{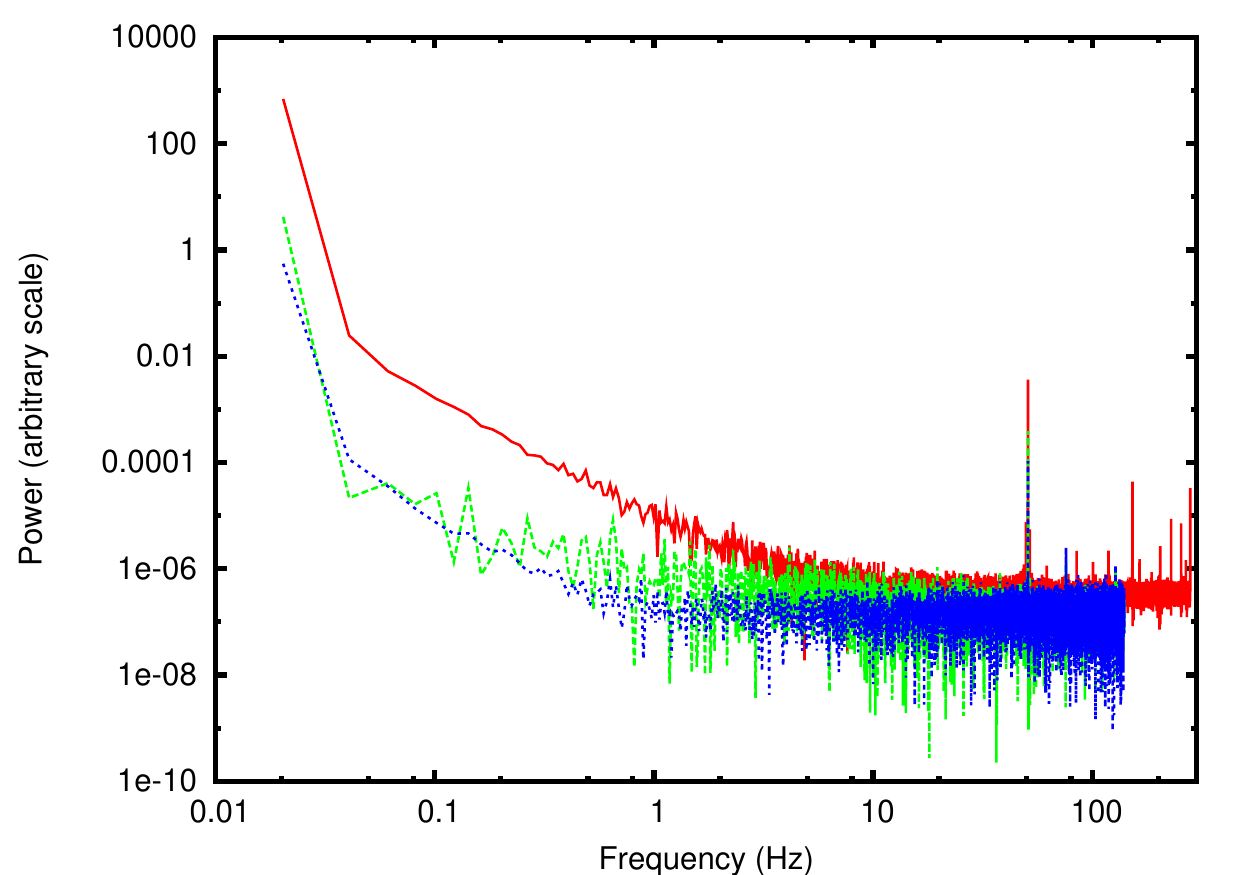}
	\includegraphics[scale=0.5]{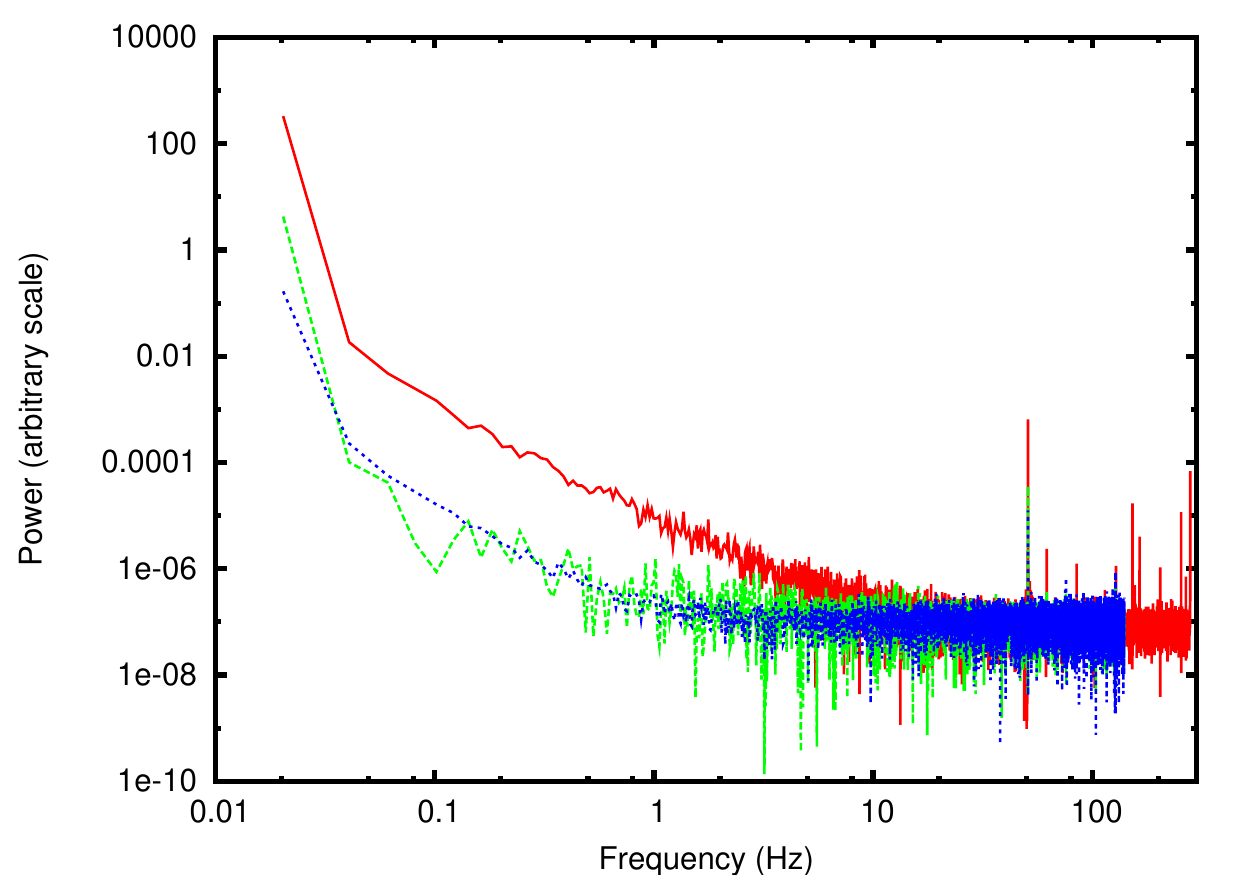}
\caption[The noise spectrum of the OCRA-F chains as measured in the lab]{The noise spectrum of the OCRA-F chains as measured in the lab. Red: raw output; green: differenced output; blue: double-differenced output.}
\label{fig:ocraf_output_noisespectrum}
\end{center}
\end{fig}

Two minute long timestreams from all 8 of the outputs of OCRA-F were taken in the laboratory at TCfA, with the receiver looking at a liquid nitrogen-soaked sheet of absorber. The Fourier transforms of these are shown in Figure \ref{fig:ocraf_output_noisespectrum}.  A peak at 50~Hz is still present at a fairly high level due to the lighting in the building, and possibly also the mains power supply. The knee frequency in the double-differenced output is currently between 0.1 and 1~Hz when looking at a 77~K load. Measurements of the double-differenced knee frequency by \citet{2007Kettle} looking at a $\sim 30$~K waveguide load showed this to be $\sim 8$mHz. This discrepancy is probably due to  the receivers not being fully tuned, the different load temperatures being measured and potentially also to the changing temperature of the liquid nitrogen load over time. The results are currently similar to those from OCRA-p installed on the telescope; see Figure \ref{fig:noise_powerspectrum_2}.

\begin{tab}{tb}
\begin{tabular}{c|c|c|c|c}
{\bf Output} & {\bf Chain 1} & {\bf Chain 2} & {\bf Chain 3} & {\bf Chain 4}\\
\hline
1 & 57~K & 62~K & 71~K & 71~K\\
2 & 46~K & 64~K & 81~K & 74~K\\
\end{tabular}
\caption[Measured system temperatures for OCRA-F prior to installation on the telescope]{Measured system temperatures for OCRA-F prior to installation on the telescope and fine tuning. The results are broadly consistent with $\sim$25-30~K for the amplifiers and the same from the window. Compare with the specifications in Table \ref{tab:ocraf_specs}.}
\label{tab:ocraf_tsys_lab}
\end{tab}

The comparison of the output voltage from the detectors when looking at a room temperature ($\sim$290~K) and liquid nitrogen-soaked ($\sim$77~K) absorber provided the system temperatures for the receiver chains simultaneously for the first time, albeit with the caveat noted above regarding the linearity of the detectors. These are given in Table \ref{tab:ocraf_tsys_lab}; the results are broadly consistent with $\sim$25-30~K for the amplifiers and the same from the window. Assuming that a further 15~K will be contributed from the atmosphere (as per Section \ref{sec:ocrap_capabilities}), then this means that OCRA-F will have at least a sensitivity of around 10~mJy~s$^{1/2}$ per receiver chain, using a total system temperature of 95~K. This will then be improved once the lower loss foam window is installed on the receiver.

Although the performance of OCRA-F was not as good as desired, due to the lack of fine tuning and the noisy window, the decision was made to install OCRA-F on the telescope and fine-tune afterwards. This approach has several advantages. First, it allows direct comparison of OCRA-F with OCRA-p, looking at the same atmosphere. Second, the atmosphere provides a colder and more stable cold load than liquid nitrogen-soaked absorber. This means that it can be used for longer and more easily repeatable measurements of the knee frequency can be made. Third, it means that measurements of strong astronomical sources can be performed with OCRA-F, which can provide direct measurements of the sensitivity of the receiver in flux density, as well as testing the shape of the beam of OCRA-F. OCRA-F was installed on the Toru\'n telescope on 7 December 2009, and cooled on the telescope the following day. Figure \ref{fig:ocraf_install} shows the receiver being lifted onto the telescope, and it installed in a field rotator.

\begin{fig}
\begin{center}
	\includegraphics[scale=0.07,angle=270]{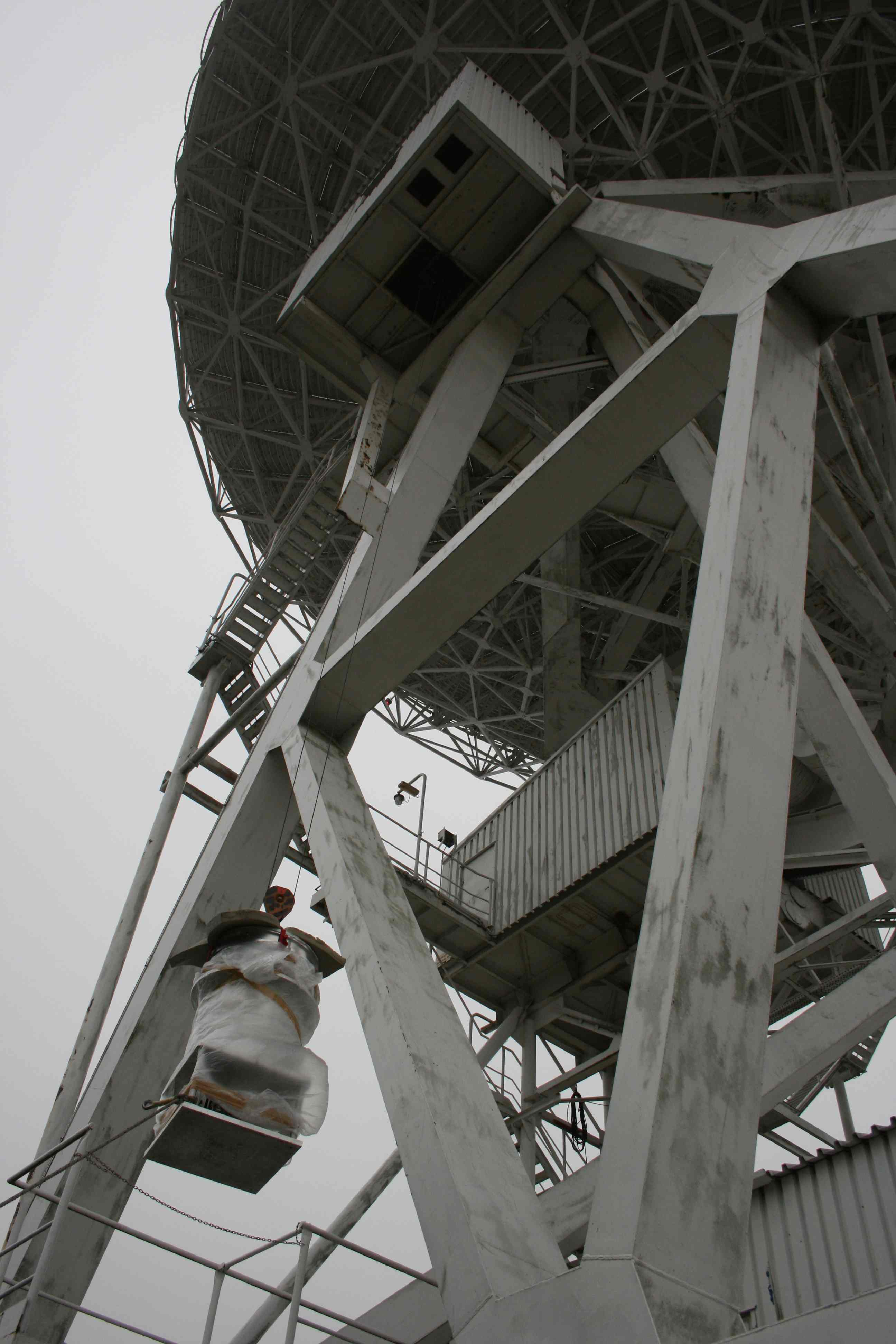}
	\includegraphics[scale=0.07,angle=270]{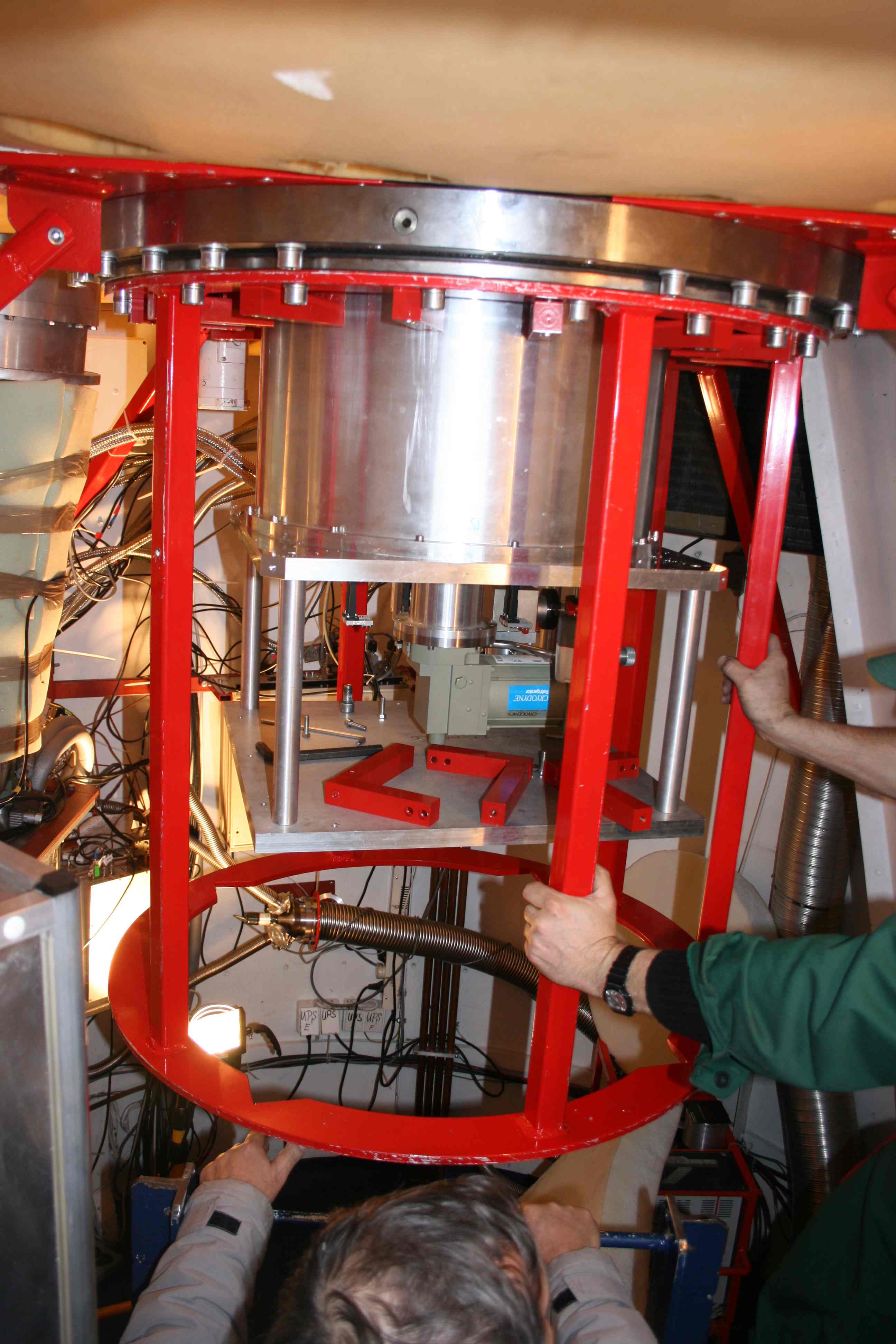}
\caption[OCRA-F being installed on the Toru\'n telescope]{OCRA-F being installed on the Toru\'n telescope. Left: being lifted up onto the telescope. Right: OCRA-F mounted in the field rotator.}
\label{fig:ocraf_install}
\end{center}
\end{fig}

Due to the rotation of the sky, the angle of the pair of beams on the sky changes over time, which can complicate both long integrations of sources (particularly SZ clusters) and the creation of maps of the sky. A field rotator, which rotates the receiver thus changing the angle of the beam pair with respect to the sky, can be used to remove these effects. Such a rotator was constructed at TCfA and consists of a rack-and-pinion ring attached to the ceiling of the secondary focus of the telescope that can be rotated using a small motor. The receiver is then connected to this using a load-bearing cage surrounding OCRA-F, with the power supply and data acquisition system installed at the base of the cage. The angle of the receiver will be measured by an encoder, whose output will then be recorded by the DAQ. At present the field rotator is at a fixed angle for the initial commissioning tests.

\begin{fig}
\begin{center}
	\includegraphics[scale=0.25]{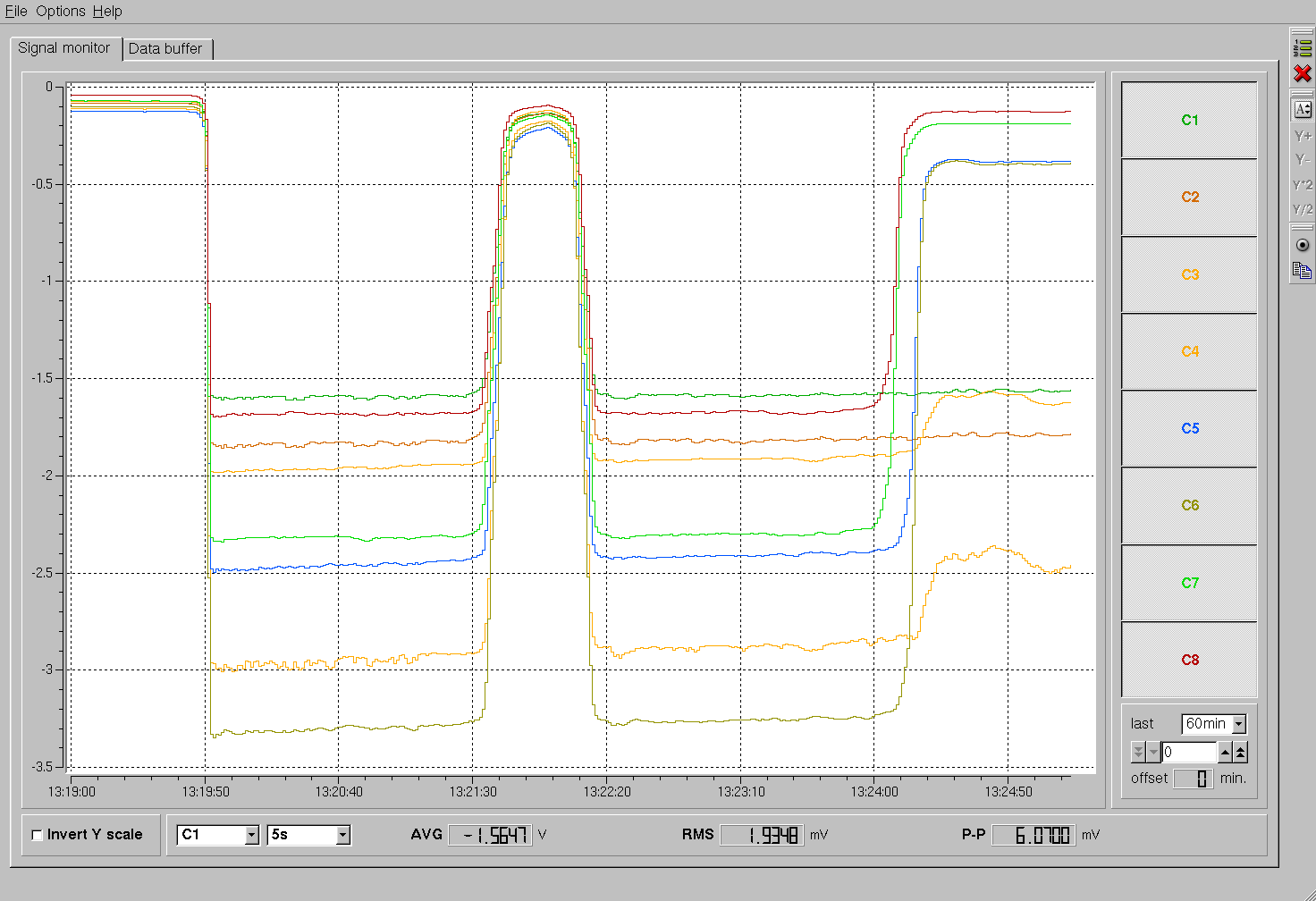}
\caption[First astronomical light observations with OCRA-F, looking at the Sun]{First astronomical light observations with OCRA-F, looking at the Sun. The vertical scale is reversed: more negative means more received signal.The receiver started off-source, moving onto the Sun at 13:19:50 then off at 21:30, become coming back on at 22:20. At 24:10 the receiver was positioned half on, half off, the Sun such that four of the eight outputs still see a large signal whilst the other four do not.}
\label{fig:sunlight}
\end{center}
\end{fig}

After installation, the telescope was pointed towards the Sun to test that the receiver was operational and to start the process of determining the beam offsets. Figure \ref{fig:sunlight} shows the output recorded by the DAQ during these observations, which are the first astronomical light observations with OCRA-F.

A large amount of work remains to be done with OCRA-F now that it is installed on the telescope. An ongoing series of noise tests and bias optimization needs to be carried out to get the most out of the receiver; in addition the new window constructed using the less noisy foam needs to be installed on the cryostat. A calibration diode system is currently being constructed by E. Pazderski, to be installed at the primary focus of the telescope; until this is installed OCRA-F will have to use astronomical sources for secondary calibration. Finally, there is a lot of integration work remaining to do, linking the instrument completely to the telescope control system. This work will be carried out over the next few months; OCRA-F should be fully operational by around April 2010.

\section{Potential improvements and extensions} \label{sec:future}
OCRA-F has provided a transition between using a single pair, and using several pairs of receivers. However, there is still a substantial way to go before construction of a 100-beam receiver becomes feasible. From the experience of constructing and testing OCRA-F, there are a number of improvements that could potentially be made to the receiver design. These are described in Section \ref{sec:ocraf_design}. The next step for the OCRA-F receiver is to increase the number of pixels to its design capacity of 16 beams; the steps towards doing this are discussed in Section \ref{sec:ocraf_16}. The receiver chain structure used within OCRA-F and OCRA-p could also be modified to carry out total power and/or spectral observations; these possibilities are discussed in Sections \ref{sec:ocraf_totalpower} and \ref{sec:ocraf_spectral}.

\subsection{Potential design improvements} \label{sec:ocraf_design}
It became apparent during the testing and commissioning of OCRA-F that there are a number of elements that could potentially be improved from the design of OCRA-F that would either simplify the physical layout of the receiver or provide improved performance. These are summarized below.

There are indications that the design of the FEMs could be improved to reduce the noise levels of the receiver: tests of the amplifiers for the 30~GHz receivers for the QUIJOTE experiment \citep{2008Rubino-Martin}, which currently uses the same MMIC LNAs as OCRA-F, showed that the MMICs can give temperatures of $\sim 15$~K; i.e. 10~K lower than those in OCRA-F. One suggestion is to reduce the length of the input lines in front of the amplifiers (E. Blackhurst, private communication).

It may also be possible to improve the performance of the amplifiers by using a hybrid MIC-MMIC design with a MIC transistor as the first stage within the amplifier, followed by a MMIC to provide the remaining amplification. The noise of an amplifier depends crucially on the first transistor within the amplifier; as MIC technology currently provides lower noise transistors than MMIC technology, a hybrid design would both enable improved noise performance whilst keeping the increased modularity, and hence ease of construction, provided by the MMIC chips.

Modularization of the FEM could assist with the testing of the module, and also reduce the number of MMICs used during the matching process of the amplifier pair. Within OCRA-F, all of the components are glued to the body of the FEM, whereas putting them on carriers -- or using separate modules for each component -- would enable the testing of individual components at any point in the construction and testing process. Modularization would also allow for improved matching between the components, and easier replacement in the case of failures. Some modularization took place to a limited extent with a later FEM constructed for the QUIJOTE instrument, albeit only with the hybrids rather than the vital amplifiers.

Interactions between the front and back end modules of OCRA-F could be mitigated by inserting an isolator between the two modules. Due to the layout and waveguide design used within OCRA-F, isolators cannot easily be retroactively fitted; they need to be considered during the design stages of such instruments. Additionally, optimal positioning of the modules and waveguide flanges can simplify the design within the cryostat. The amplifier modules in OCRA-F are orientated vertically, requiring expensive, curved interconnecting waveguides to transfer the RF signals. If the modules were instead positioned horizontally, with the output on the opposite face from the input, simple straight connecting waveguides could be used.

\subsection{Increased number of pixels} \label{sec:ocraf_16}
Although OCRA-F was constructed to have 16 pixels, only 8 of these are currently populated. The logical extension of OCRA-F would be to populate the other 8 pixels. This will require the construction of four more FEMs and BEMs, as well as the addition of the appropriate infrastructure within the cryostat (electrical wiring, etc.). The cryostat would ideally also be modified at the same time to have a more robust internal support structure. Additional filters, detectors and power supply modules would also be required, as would further investigation of the foam window to remove the necessity for the window support plate. The cost of such an extension would be minimal compared with developing a complete independent receiver system.

The ultimate aim for the OCRA program is to design and construct a 100 beam instrument \citep{2000Browne}. This will involve a considerable amount of work in terms of designing a cryostat (or a set of closely packed cryostats) that occupies so much of the focal plane array of a telescope. Experience from the relocation of OCRA-p to the side of the focal plane of the Toru\'n telescope to make space for OCRA-F has shown that it will be necessary to curve the array of horns to follow the focal surface for optimum performance, rather than having them in a flat array (this will of course be different from horns looking directly at the sky rather than being mounted on a telescope). An investigation into the design requirements and configuration of such ``radio cameras'' is currently being investigated by the EC Framework 7 APRICOT (All Purpose Radio Imaging Cameras On Telescopes) project within the RadioNet consortium.

\subsection{Total power radiometer chains} \label{sec:ocraf_totalpower}
The OCRA receivers currently difference one beam from the other to measure the differential power from the sky. This mitigates effects from atmospheric fluctuations, however it greatly increases the difficulty of carrying out measurements of extended emission on scales larger than the separation between the beams. With an array of receivers, the atmospheric effects could potentially be removed in software by looking for common atmospheric modes between adjacent beams instead. In order to continue removing the effects of $1/f$ gain fluctuations from the amplifiers, the difference between a cold load and a beam could be taken, as is done by the LFI receivers on the {\it Planck} satellite.

The degree to which this approach would be effective in removing atmospheric fluctuations can be simulated by extending the UMBRELLA software package to include cold loads in place of receiver horns. This would also enable an investigation of the analysis steps required to optimally utilize the atmospheric information provided by an array of total power receivers.

\subsection{Spectral bands and lines} \label{sec:ocraf_spectral}
The back end of the OCRA receivers could be easily modified with the introduction of filter banks. The output signal from the BEMs could be split into 4 to 5 frequency channels with the aid of a power splitter and a set of filters, which could then be measured simultaneously. This would provide an instantaneous measurement of the spectral indices of radio sources. It will also provide additional information on the atmospheric fluctuations in front of the receiver, as these fluctuations will be strongest at the lower end of the frequency range due to the water vapour line at 22~GHz. The frequency bins could be recombined in software to allow for more accurate measurements of weaker sources, so long as the frequency bins do not severely overlap and any correlations between the frequency bins are taken into account.

The current OCRA-F DAQ was designed to support 16 channels, of which 10 are currently in use (8 channels are used for the receiver outputs, one for the phase switch signal and one for the time). As a result, one of the OCRA-F chains could be converted to give $2\times4$ binned outputs, and the only modifications to the DAQ that would be required would be in software. The additional hardware required would be a pair of RF power splitters, two sets of filters, additional detectors and video amplifiers, all of which would be placed outside the cryostat.

The above would provide discrete frequency bins each with wide ($\sim 2$~GHz) bandwidth; a further improvement would be to use a local oscillator to reduce the frequency of the signals, which could then be sampled using a high-speed DAQ. The signal could then be Fourier transformed to provide spectral line measurements. Observations of spectral lines could then be carried out during periods when the weather does not permit continuum flux density measurements. An array of such receivers could potentially be used to create large-scale maps of spectral line emission for follow-up with high-resolution interferometers such as the Very Large Array (VLA) or the Atacama Large Millimeter/submillimeter Array (ALMA).

\section{Summary} \label{sec:ocraf_summary}
This chapter has described the testing of the components within the OCRA-F receiver and their assembly into the complete cryostat. Following this, the testing of the receiver in the field at JBO and at TCfA was described, as was the installation of OCRA-F on the Toru\'n 32-m telescope and the first steps in the process of commissioning the receiver. Potential improvements and extensions of the OCRA receivers, including adding total power, spectral bin and spectral line capabilities, were also discussed.

A number of issues were discovered and mitigated during the testing of the receiver chains. The gain of OCRA-F contains ripples caused by interactions between the FEM and the BEM; this can be mitigated by changing bias settings of the amplifiers, and it does not have a significant effect on the receiver performance due to the continuum nature of the receiver. The detectors become non-linear in the presence of a room temperature load and the maximum gain from the amplifiers; this is mitigated by reducing the gain of the amplifiers, and should not be a problem for measuring astronomical sources due to the low apparent temperature loads from these sources. Finally, the lossy vacuum window material used within OCRA-F increases its system temperature; a window made of a different, less-lossy type of foam has been constructed and tested, and will be installed on OCRA-F in the near future.

The OCRA-F receiver will be optimized over the next few months via testing on the telescope, and should be fully operational and carrying out astronomical surveys by around April 2010.

\chapter{Conclusions and future work}\label{conclusions}

\section{Virtual Sky simulations}
We have simulated the microwave sky, creating maps consisting of the Cosmic Microwave Background, galaxy clusters via the SZ effect and foreground point sources. Simulated cluster catalogues can be created by analytical formulae, N-body simulations and using {\sc Pinocchio} simulations. A galaxy cluster model can then be applied to turn these catalogues into maps of the SZ effect.  Point sources -- both low- and high-frequency -- can be added to the maps, including distributing the point sources according to the surface matter density of the galaxy clusters.

Making use of the tools set in place for these simulations, we examine the expected polarized power spectrum for point sources based upon polarized, high frequency VLA observations by \citet{2009Jackson} of a sample of WMAP-detected point sources. We find that point source subtraction will be vital for upcoming low frequency B-mode experiments that are sensitive to values of the tensor-to-scalar ratio $r$ of less than 0.1, and less than 0.01 for all frequencies.

Using virtual sky maps created from large numbers of cluster catalogues generated by {\sc Pinocchio} for three different values of $\sigma_8$, we have investigated the statistics of the power spectrum between multipoles of 1000 and 10000. We find that the inclusion of the SZ effect increases the standard deviation of the power spectrum by a factor of 3 over that expected from cosmic variance, in agreement with the predictions from an analytical calculation based on the halo formalism. The mean and standard deviation vary as $1/f_\mathrm{sky}^{1/2}$ as expected, and scale approximately as $\sigma_8^7$ over the range of values sampled here. We also find that the distributions are non-Gaussian, and are skewed by large mass clusters, with the degree of this skewness increasing as the map size is decreased.  Additionally, we find that correlations between galaxy clusters play a small role in the statistics of the power spectrum at the level of $\sim$10 per cent.

Several instruments have measured an excess at high multipoles, which may be due to the SZ effect with a large value of $\sigma_8$. We cannot explain the central values of these measurements with the range of $\sigma_8$ investigated here, however the increased standard deviation and the presence of skewness in the distribution means that these measurements could be explained by a lower value of $\sigma_8$ than has been suggested so far. There is also a large uncertainty in the parameters describing the cluster gas physics, which can have a large effect on the mean of the distributions, comparable to that from the different values of $\sigma_8$, and can also significantly effect the standard deviation and the skew of the distributions.

The next generation of CMB instruments are currently being commissioned, and are expected to provide more data at multipoles comparable to those probed by the CBI. The {\it Planck} spacecraft will measure the power spectrum from the whole sky out to multipoles of 2500 within the next few years, and instruments such as the South Pole Telescope \citep[SPT;][]{2004Ruhl}, the Arcminute Microkelvin Imager \citep[AMI;][]{2008Zwart} and the Atacama Cosmology Telescope \citep[ACT;][]{2004Fowler} will observe large numbers of galaxy clusters using the SZ effect. These measurements will provide much more information on the SZ effect and may provide a resolution to the discrepancy in $\sigma_8$ from the measurements to date. Our results should also be relevant to these observations.

Further virtual sky simulations can be carried out to investigate the behaviour of the SZ effect on lower multipoles using full sky maps.  Both these simulations and those on smaller map sizes can be used to estimate the increase in cosmological parameter error bars due to contamination of observations of the CMB by the SZ effect. Additionally, the maps can be combined with simulations of specific experiments to forecast their capabilities for observing the SZ effect, or when treating the SZ effect as a foreground to the CMB, and they can be used to investigate optimal approaches to detecting SZ clusters.

\section{OCRA simulations and observations}
We have extended the UMBRELLA simulations of the OCRA-p receiver mounted on a telescope observing through atmosphere, created by \citet{2006Lowe}, to incorporate automated data reduction (which can also be used for real observations) and to be able to observe Virtual Sky maps. Using this software to simulate the system from end to end, we have investigated the statistics of point source observations with OCRA-p, examining the capabilities of the receiver for observing using cross-scan and on-off measurements. We find that these simulations are able to realistically simulate the noise present in real observations, and that the introduction of $1/f$ noise into the simulations significantly reduces the predicted ability of the instruments to observe weak sources by integrating for long periods of time.

In order to aid the subtraction of individual sources from the VSA fields observed at $\sim$30~GHz, and to obtain a statistical estimate of the surface density of sources at 30~GHz, we have observed a sample of 121 sources using the OCRA-p receiver on the Toru\'n 32-m telescope; the sample was selected at 15~GHz with the RT. At 30~GHz, we detected 57 sources above a limiting flux density of $\sim 5$~mJy. This is the deepest follow-up of any complete sample of sources detected at 15~GHz by the RT.

At a flux density of 10~mJy, which is our estimated completeness limit, we derive a surface density of sources at 30~GHz of $2.0~\pm~0.4$ per square degree. This is consistent with the value obtained by \citet{2009Mason}, who observed a much larger sample of sources down to mJy levels but selected at a much lower frequency (1.4~GHz). The potential danger of using low frequency selected samples is that there may exist a significant population of sources with steeply rising spectra towards high frequencies that are not present in the low frequency surveys. As the two surface density estimates are consistent, this indicates that such a population is not obviously present at the 10~mJy level.

We have compared our flux density measurements with those from the VSA source subtractor and VLA measurements. These comparisons give confidence in our flux density scale but reveal that a significant fraction of sources are variable on a timescale of a few years, some at the level of a factor of 2. This shows the importance of taking contemporaneous measurements of discrete sources in conjunction with measurements of the CMB.

We have also investigated the dependence of the spectral index distribution on flux density by comparing our measured spectral index distribution with that for much stronger sources (above 1~Jy) selected from the WMAP 22~GHz catalogue. We conclude that the proportion of steep spectrum sources increases with decreasing flux density. This is qualitatively consistent with models of source populations, for example \citet{2005deZotti}.

Again using OCRA-p, we have also surveyed 550 flat spectrum radio sources around the North Ecliptic Pole from the CRATES sample, which were selected based on their 1.4-4.8~GHz spectral index. These follow from the observation of the CJF sources by \citet{2006Lowe,2007Lowe}, and extend the work to lower flux densities.

We find reasonable agreement between the measurements presented here and those by \citet{2007Lowe} and also WMAP \citep{2009Wright} where sources are in common between the samples. A number of sources display variability between these three sets of observations, and also within the present measurements. This will present difficulties when subtracting these point sources from maps of the CMB.

As expected, we find that the spectral index distribution for these sources broadens at higher frequencies as the source spectral indices steepen. We find that there are a reasonable number (5.6 per cent) of Gigahertz-Peaked Sources, and also a number of inverted sources ($\sim 10$ per cent). We conclude that extrapolation from low frequency flux densities to higher frequencies assuming power law spectra is unreliable. This emphasizes the need for high-frequency blind surveys to low flux densities. Such surveys are currently being carried out by the ATCA at 20~GHz in the southern hemisphere (see e.g. \citealp{2007Sadler}) and AMI at 15~GHz in the northern hemisphere \citep{2004Waldram,2009Waldram}, and will be carried out by the OCRA-F instrument and its successors at 30~GHz in the near future.

The flux densities of the sources within the CRATES subsample described here will be useful for comparison to point source measurements by the {\it Planck} satellite, which will be most sensitive in the area of sky surveyed here. Due to the flat spectrum nature of these sources, they are the most likely sources to appear in all of the different observational bands of {\it Planck}. An Early Release Point Source Catalogue from {\it Planck} is expected in December 2010 \citep{2009Bouchet}, at which point such a comparison can be carried out.

\section{Radio cameras}
The OCRA-F receiver has been tested at the individual component level, assembled into a complete cryostat and tested in the field at JBO and at TCfA. The receiver was installed on the Toru\'n 32-m telescope, and the first steps in the process of commissioning the receiver have been taken. The OCRA-F receiver will be optimized over the next few months via testing on the telescope, and should be fully operational and carrying out astronomical surveys by around April 2010.

A number of issues were discovered and mitigated during the testing of the receiver chains. The gain of OCRA-F contains ripples caused by interactions between the FEM and the BEM; this can be mitigated by changing bias settings of the amplifiers, and it does not have a significant effect on the receiver performance due to the continuum nature of the receiver. The detectors become non-linear in the presence of a room temperature load and the maximum gain from the amplifiers; this is mitigated by reducing the gain of the amplifiers, and should not be a problem for measuring astronomical sources due to the low apparent temperature loads from these sources. Finally, the lossy vacuum window material used within OCRA-F increases its system temperature; a window made of a different, less-lossy type of foam has been constructed and tested, and will be installed on OCRA-F in the near future.

Looking to the future, the next step towards a fully-fledged radio camera from the OCRA programme of receivers is the completion of OCRA-F with the addition of the remaining four receiver modules to bring it up to its design capacity of 16 beams. This will enable the development and real-life testing of observation methodologies for surveying the sky at these frequencies using a ground-based instrument.

Over the next few years, the APRICOT programme will be exploring the required technology for larger, multi-purpose arrays of radio receivers, looking also at polarization and spectral line studies. Information from this can then be used to start the development of the complete, 100-beam OCRA receiver.

With the installation of OCRA-F on the Toru\`n telescope, blind surveys for point sources can be started in the near future. Following from the CRATES observations, a blind survey of this section of the sky using OCRA-F will be started, with a target of finding all point sources at 30~GHz down to a flux density of $\sim$ 10~mJy over a large area of the sky.

Interweaved with these observations, OCRA-F will be used to observe a deep field to high sensitivity with the aim of detecting SZ clusters in a small blind survey. Depending on the performance of the receiver, this may also lead to observations of the CBI excess using the same field. It will be vital to carry out further simulations of the OCRA instruments, based on the combination of Virtual Sky, UMBRELLA and the automated data reduction software, to investigate the optimal approaches to carrying out these surveys.

As the number of microwave frequency receivers mounted on a telescope is increased, the amount of sky that surveys with that telescope can cover will dramatically increase, as will the sensitivity of those surveys. Ultimately, blind surveys of SZ clusters will be able to provide measurements of the cosmological parameters $\Omega_\mathrm{m}$, $\sigma_8$ and $w$ that are complementary to those from observations of the CMB and supernovae. This will ultimately increase our knowledge of dark matter and dark energy, as well as the evolution of our Universe.

\appendix
\chapter{Flux densities and spectra for sources within the CRATES subsample} \label{sec:crates_spectra}
\begin{center}
  \renewcommand{\baselinestretch}{1.1}
  \centering\small
\begin{longtable}[t]{l|c|c|r|r|r|r|l}

  \caption[Flux densities for the CRATES sources]{Flux densities (in mJy) for the CRATES sources. See Section \ref{sec:crates_source_fluxes} for details.}
\label{tab:crates_fluxes}\\

Name & RA (J2000) & Dec (J2000) & $S_{1.4}$ & $S_{4.8}$ & $S_{8.4}$ & $S_{30}$ & Note\\
\hline
\endfirsthead

Name & RA (J2000) & Dec (J2000) & $S_{1.4}$ & $S_{4.8}$ & $S_{8.4}$ & $S_{30}$ & Note\\
\hline
\endhead

\multicolumn{8}{r}{{Continued...}}
\endfoot

\endlastfoot

1027+7428 & 10 27 24.1 & 74 28 26.1 & 190.1 & 211.0 & 143.5 & 72.3$\pm$9.7& \\
1031+7441 & 10 31 22.0 & 74 41 58.3 & 213.2 & 250.0 & 103.9 & 129.9$\pm$13.2& \\
1056+7401 & 10 56 18.7 & 74 01 00.2 & 75.7 & 67.0 & 58.1 & 26.1$\pm$5.9& \\
1059+7421 & 10 59 49.2 & 74 21 01.8 & 83.5 & 70.0 & 51.0 & 63.0$\pm$8.5& \\
1101+7225 & 11 01 48.8 & 72 25 37.1 & 1281.2 & 858.0 & 366.7 & 942.1$\pm$81.4& \\
1107+7232 & 11 07 41.7 & 72 32 36.0 & 378.1 & 271.0 & 183.8 & 141.0$\pm$14.0& \\
1127+7011 & 11 27 03.5 & 70 11 57.2 & 77.4 & 71.0 & 114.5 & 38.0$\pm$3.9& \\
1134+7249 & 11 34 11.4 & 72 49 20.0 & 222.3 & 229.0 & 214.4 & 34.5$\pm$3.9& \\
\end{longtable}
\end{center}

%
% Table: Sources of other flux densities for the CRATES sources
%
\begin{tabland}
\begin{tabular}[t]{c|c|c|c|l}
\bf{Name} & \bf{Frequency} & \bf{Resolution} & \bf{\# Meas.} & \bf{Reference(s)}\\
\hline
8C & 38~MHz & $\sim 4.5$ arcmin & 86 & \citet{1995Hales}\\
6C & 151~MHz & $\sim 4.2$ arcmin & (22,65,37,18) & \citet{1988Hales,1990Hales,1991Hales,1993Hales}\\
WENSS & 325~MHz & $\sim 1$ arcmin & 298 & \citet{1997Rengelink}\\
Texas & 365~MHz & $\sim 10$ arcsec & 172 & \citet{1996Douglas}\\
B2 & 408MHz & & 16 & \citet{1973Colla}\\
Bologna 408 & 408~MHz & $2.6x4.8$ arcmin & 64 & \citet{1985Ficarra}\\
NVSS & 1.4~GHz & 0.75 arcmin & 550 & \citet{1998Condon,2007Healey}\\ % Used CRATES values for 1.4GHz
FIRST & 1.4~GHz &  & 58 & \citet{1995Becker}\\
GB1.4 & 1.4~GHz & $\sim 11.5$ arcmin & 266 & \citet{1992White}\\
VCS3,4,5 & 2.3, 8.6~GHz & & (13,11,68) & \citet{2005Petrov,2006Petrov,2007Kovalev}\\
GB6 & 4.8~GHz & $\sim 3.5$ arcmin & 550 & \citet{1996Gregory,2007Healey}\\ % Used CRATES values for 4.8GHz
VIPS & 5.0~GHz & VLBA & 165 & \citet{2007Helmboldt}\\
MG (II,III,IV) & 5.0~GHz &  & (28,28,116) & \citet{1990Langston,1990Griffith,1991Griffith}\\
CRATES/CLASS & 8.4~GHz & 0.2-0.3 arcsec & 550 & \citet{2007Healey,2003Myers}\\ % Used CRATES values for 8.4GHz
- & 8.4~GHz & 0.2 arcsec & 184 & \citet{1992Patnaik}\\
WMAP3 & 23-94~GHz & & 57 & \citet{2003Bennett}\\
\end{tabular}
\caption[Wide-area surveys from which other flux densities for the CRATES source samples have been found]{Wide-area surveys from which other flux densities have been found for the CRATES sources, as used within the source spectra plots. An additional reference is given to \citet{2007Healey} where the flux densities have been taken from CRATES.}
\label{tab:crates_spectra_sources1}
\end{tabland}

\newpage
\begin{center}
  \renewcommand{\baselinestretch}{1.1}
  \centering\small
\begin{longtable}[t]{l|c|l}

  \caption[Other sources of flux densities for the CRATES source sample]{Other sources of flux densities that have been found for the CRATES sources, as used within the source spectra plots.}
\label{tab:crates_spectra_sources2}\\

\bf{Frequencies} & \bf{\# Meas.} & \bf{Reference(s)}\\
\hline
\endfirsthead

\bf{Frequencies} & \bf{\# Meas.} & \bf{Reference(s)}\\
\hline
\endhead

\multicolumn{3}{r}{{Continued...}}
\endfoot

\endlastfoot

26.3~MHz & 2 & \citet{1975Viner}\\
26.3~MHz--14.9~GHz & 9 & \citet{1980Laing}\\
28.6~MHz--10.7~GHz & 74 & \citet{2009Kuehr}\\
38~MHz--5~GHz & 12 & \citet{1969Kellerman}\\
60~MHz & 1 & \citet{1968Aslanian}\\
80--160~MHz & 2 & \citet{1995Slee}\\
111~MHz  & 1 & \citet{2005TyulBashev}\\
151~Mhz--8.48~GHz & 8 & \citet{1998Stanghellini}\\
\end{longtable}
\end{center}

\clearpage
\begin{figure}
\centering
\includegraphics[scale=0.2]{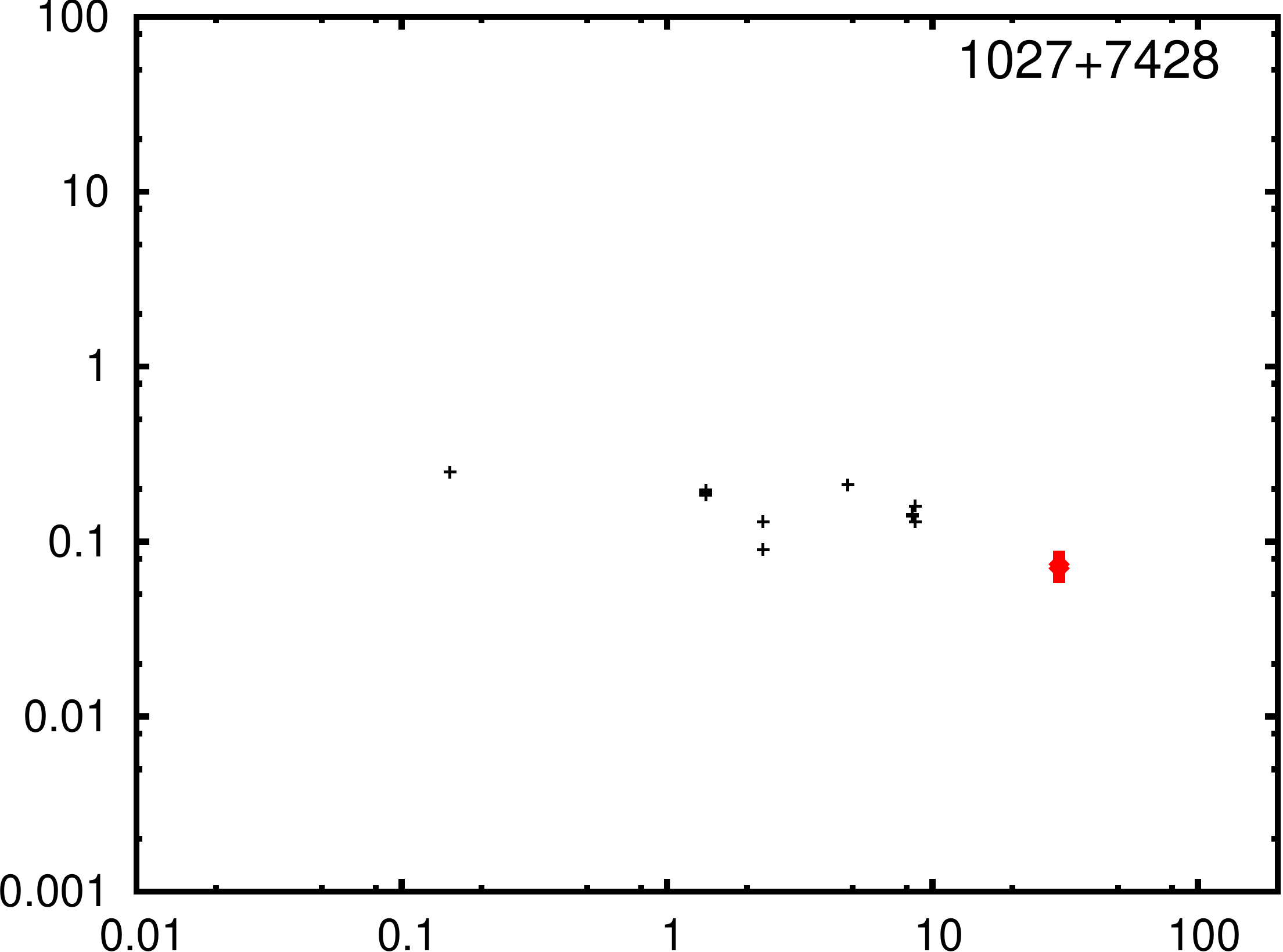}
\includegraphics[scale=0.2]{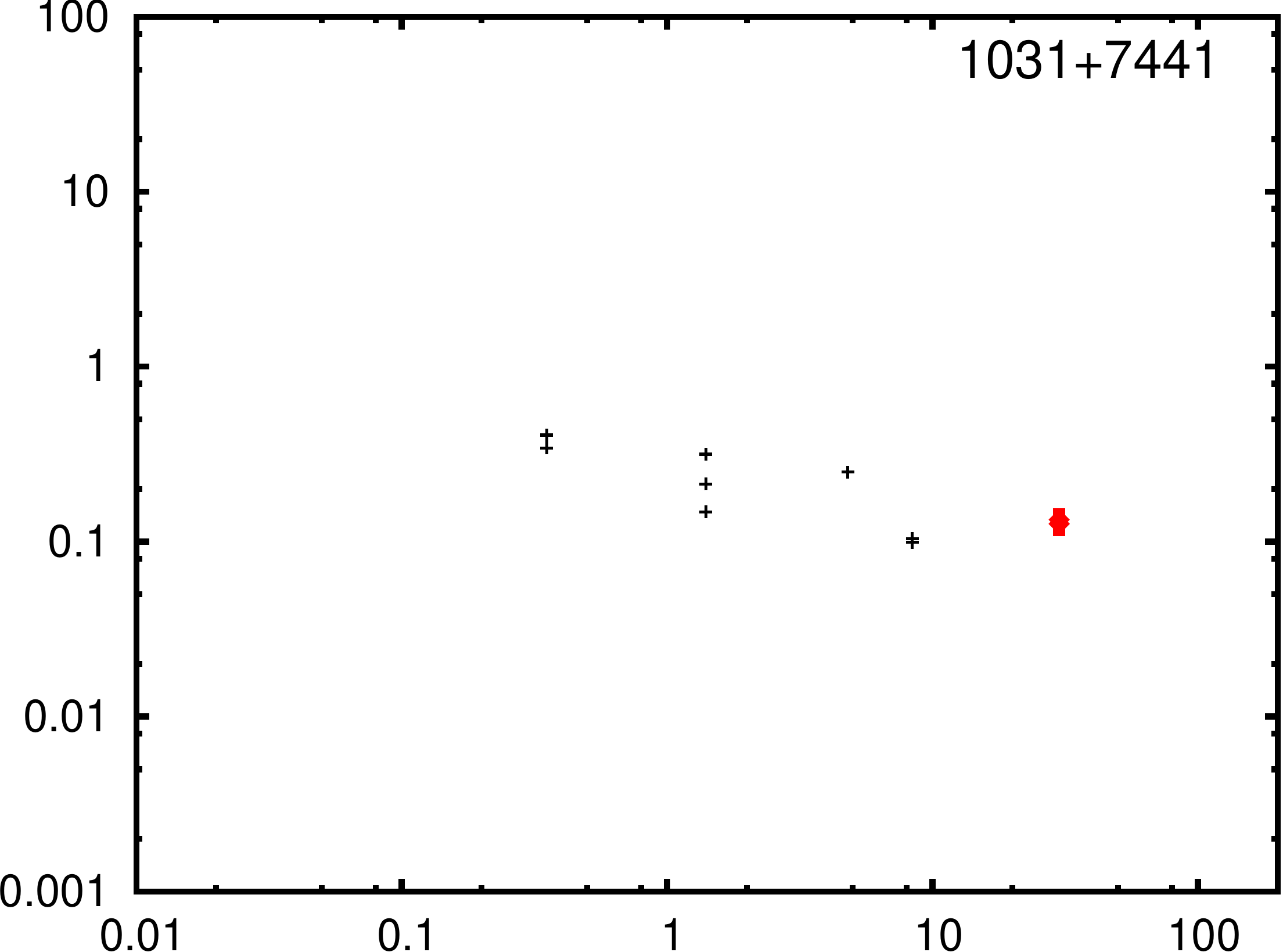}
\includegraphics[scale=0.2]{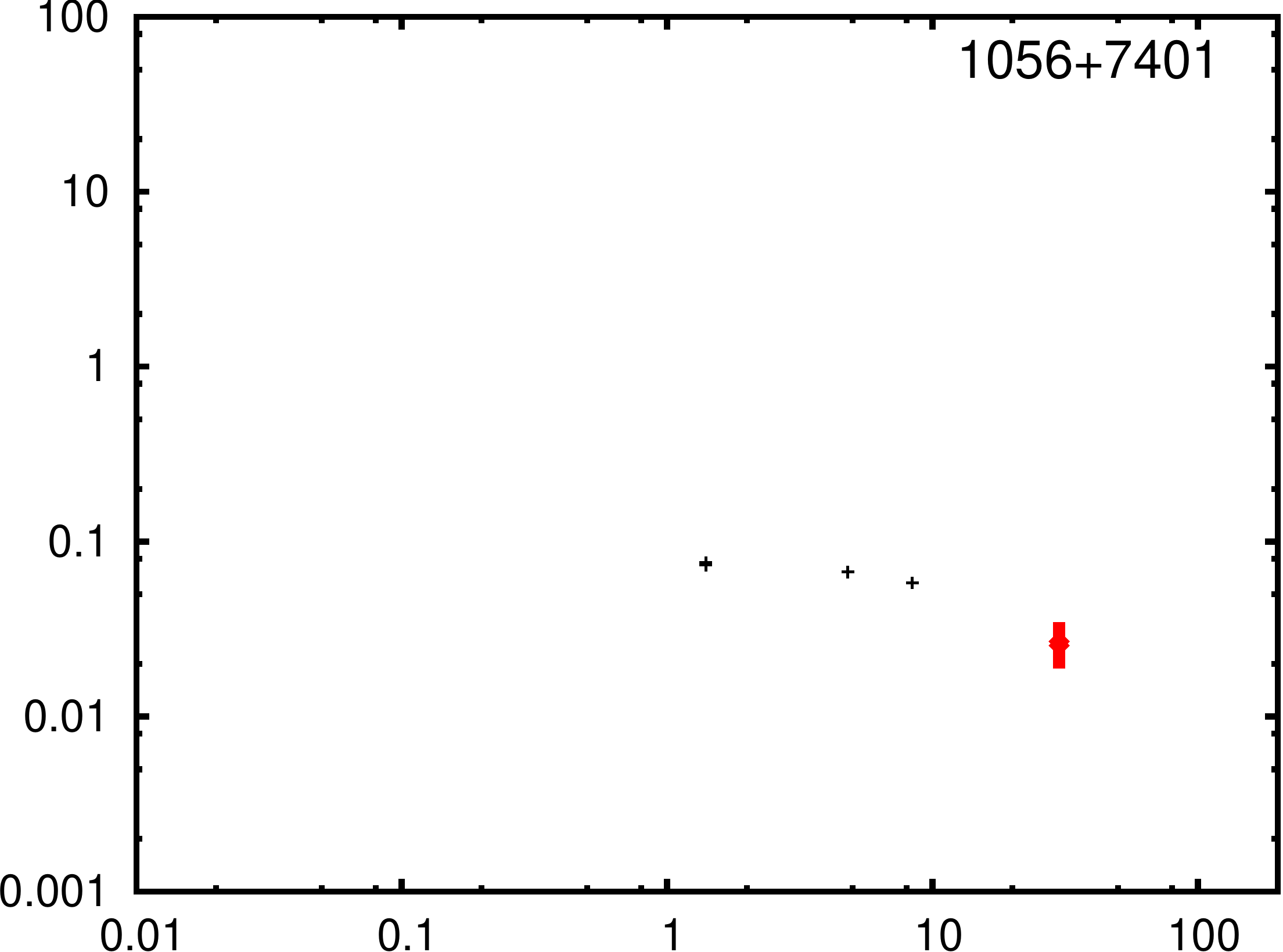}
\includegraphics[scale=0.2]{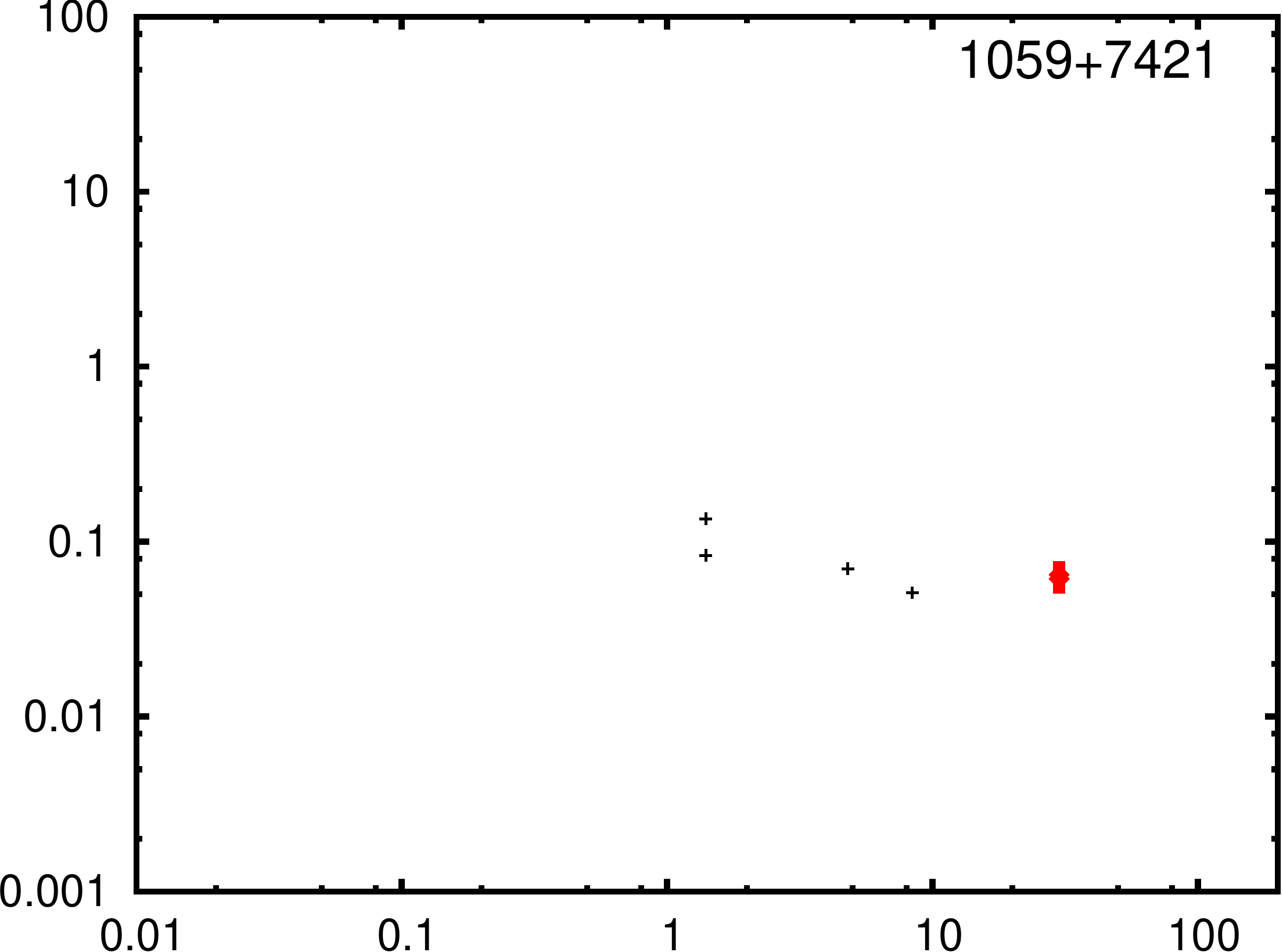}
\includegraphics[scale=0.2]{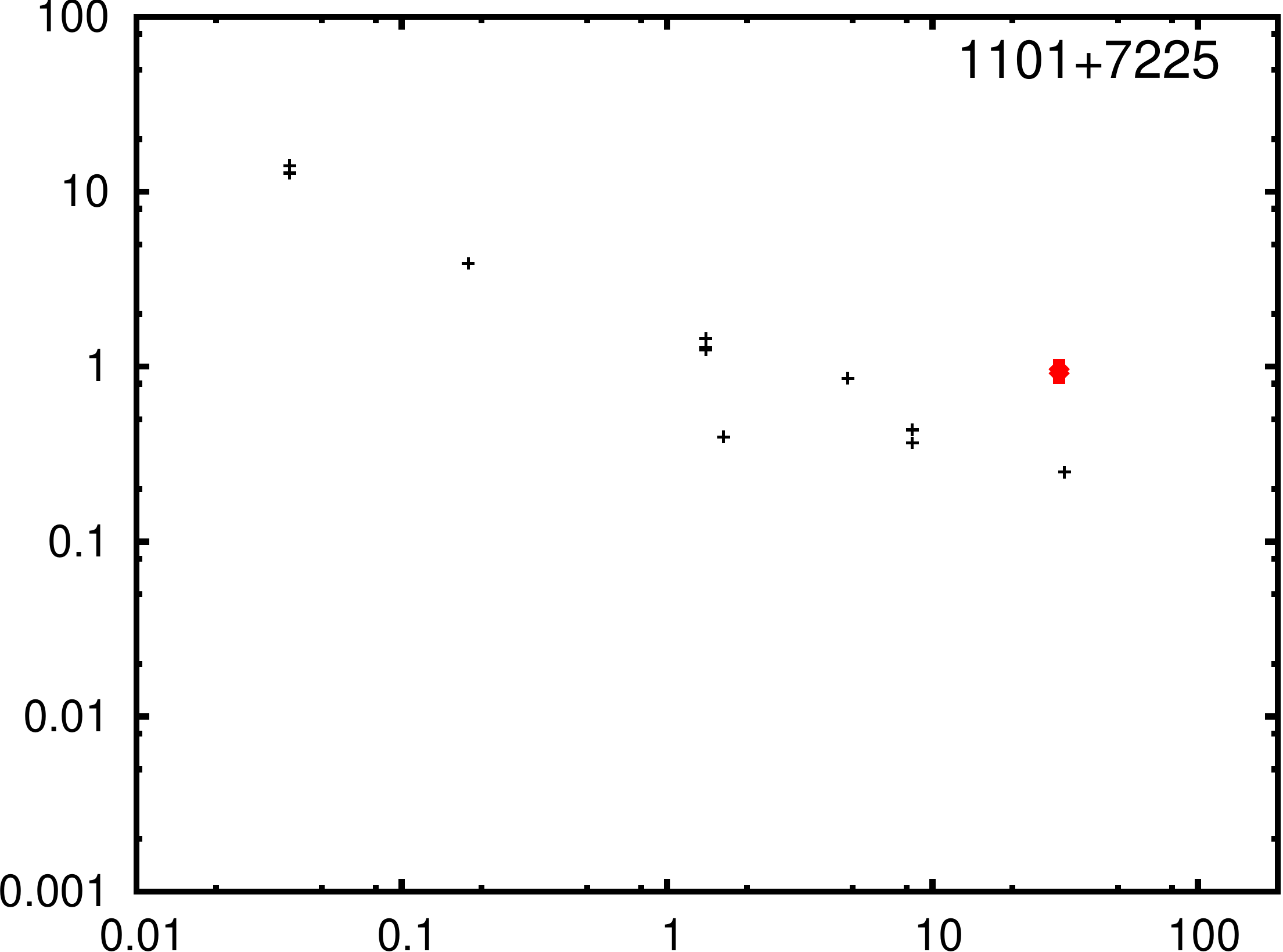}
\includegraphics[scale=0.2]{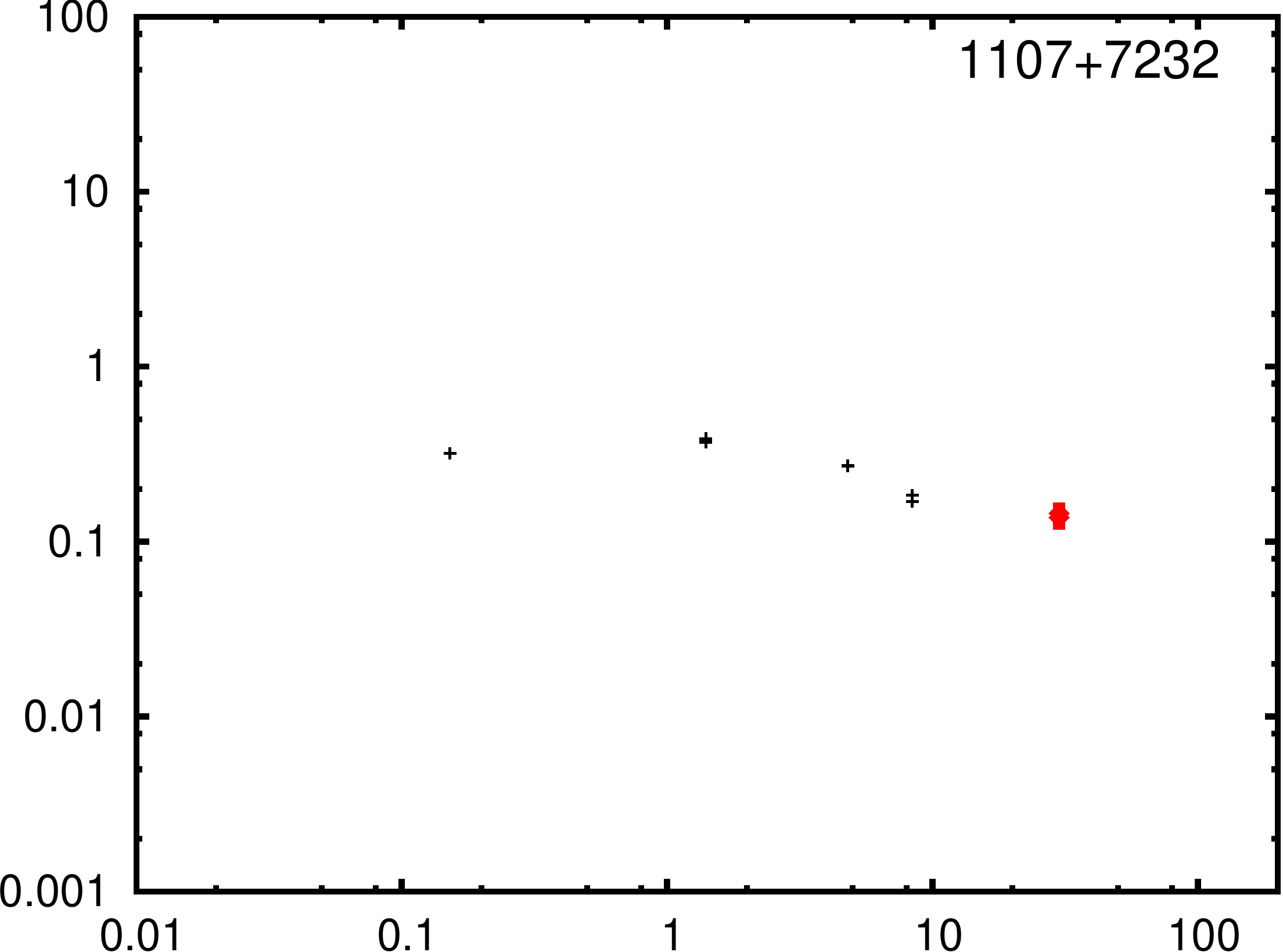}
\includegraphics[scale=0.2]{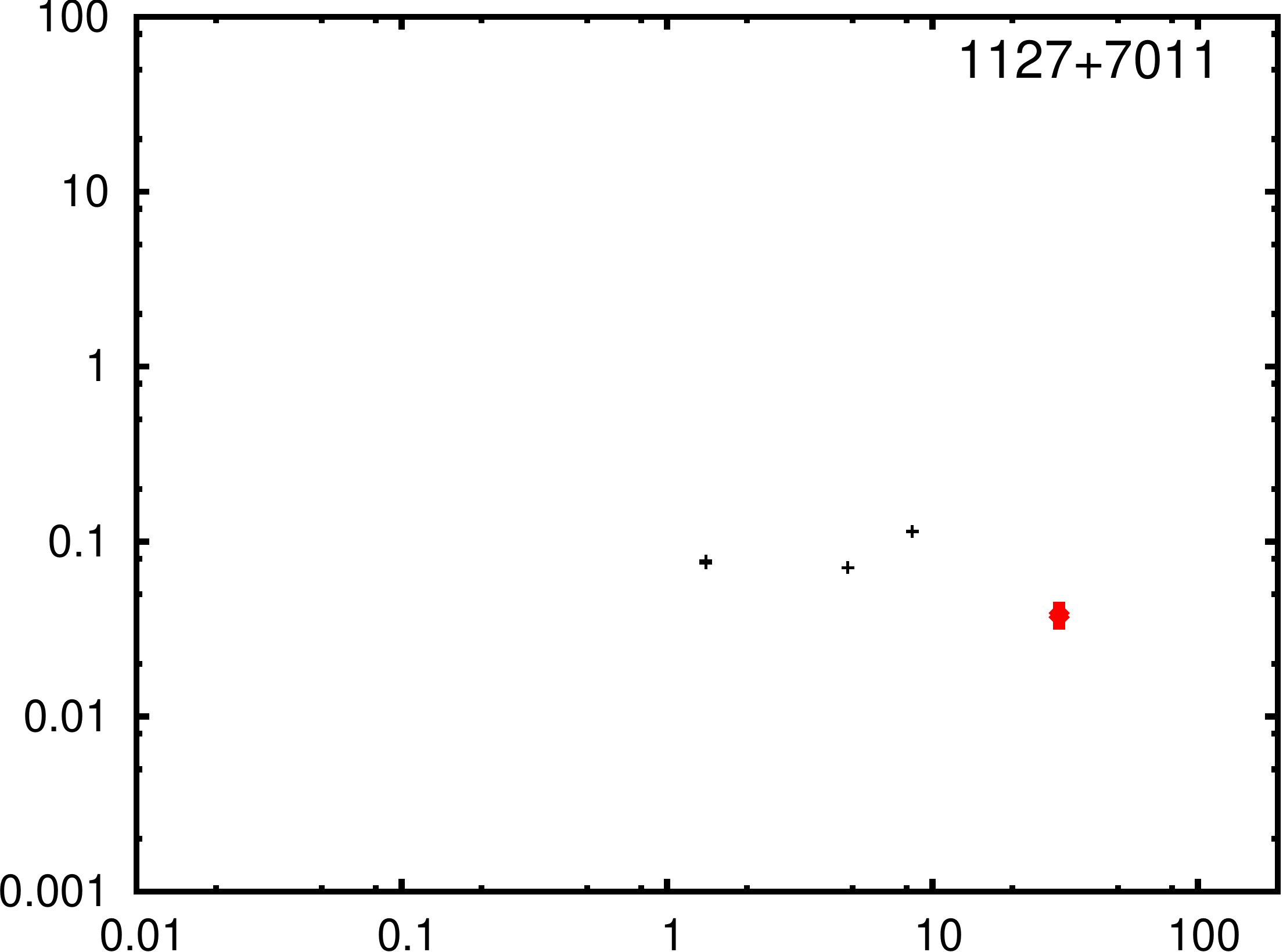}
\includegraphics[scale=0.2]{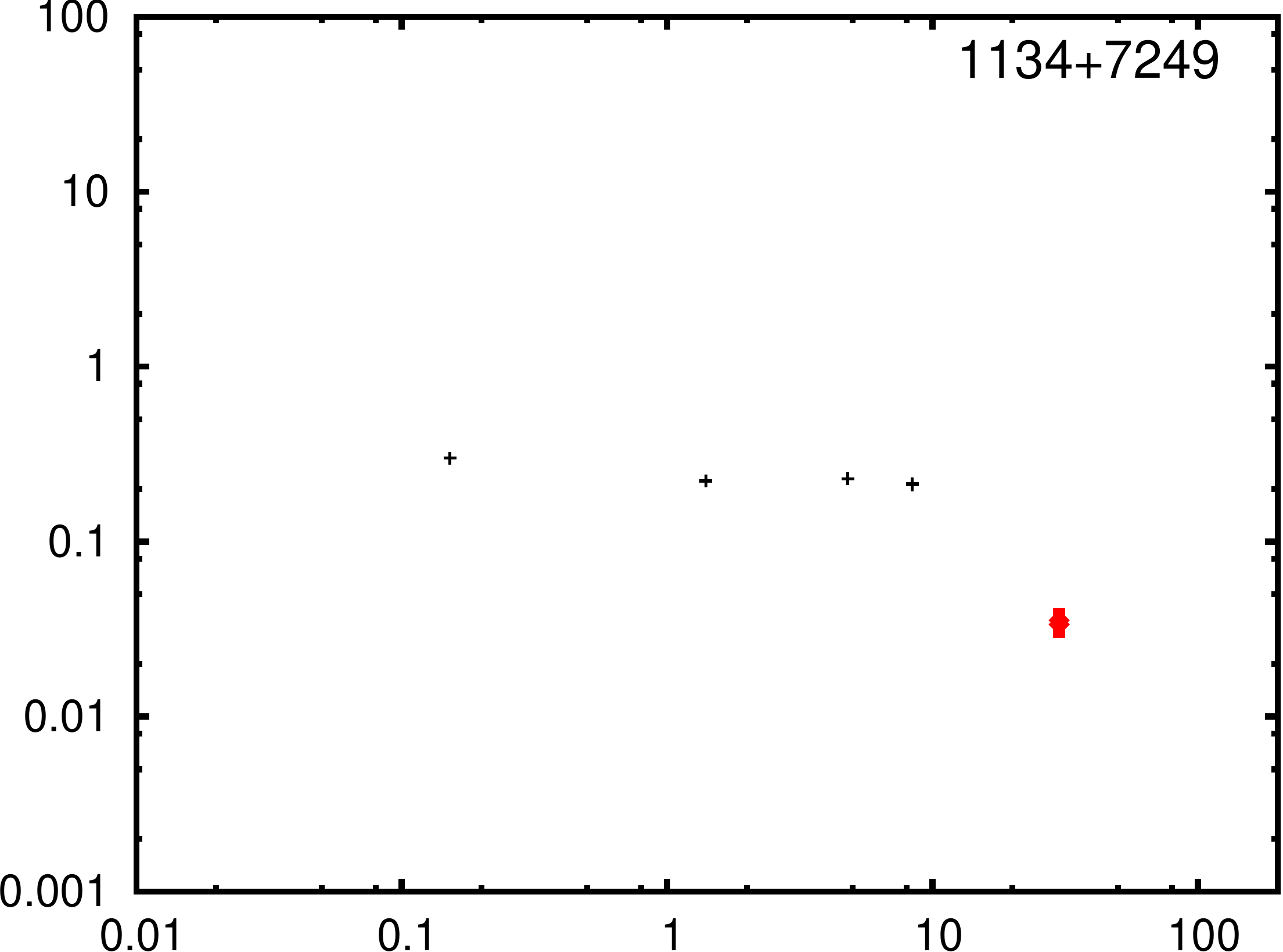}
\includegraphics[scale=0.2]{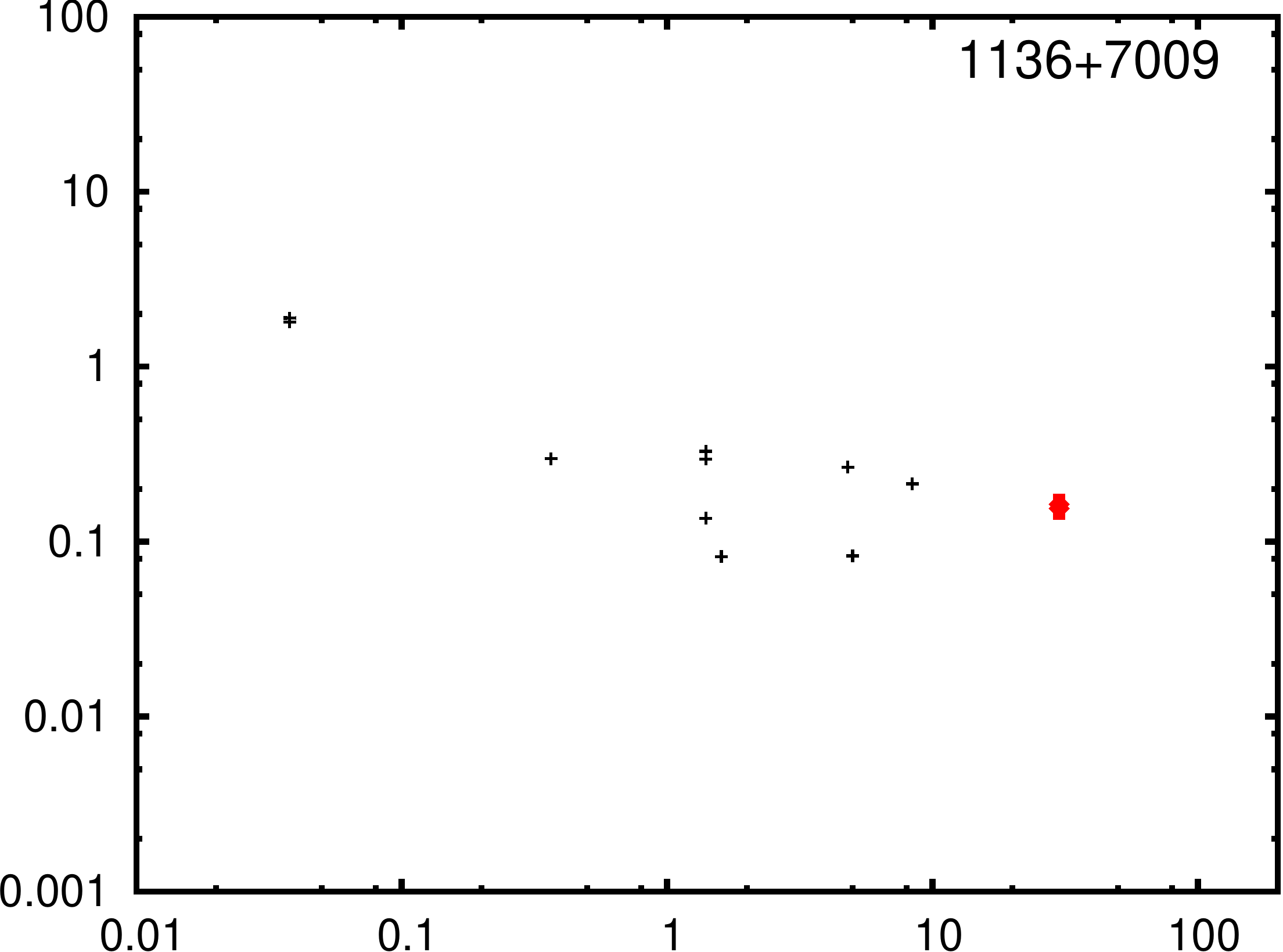}
\includegraphics[scale=0.2]{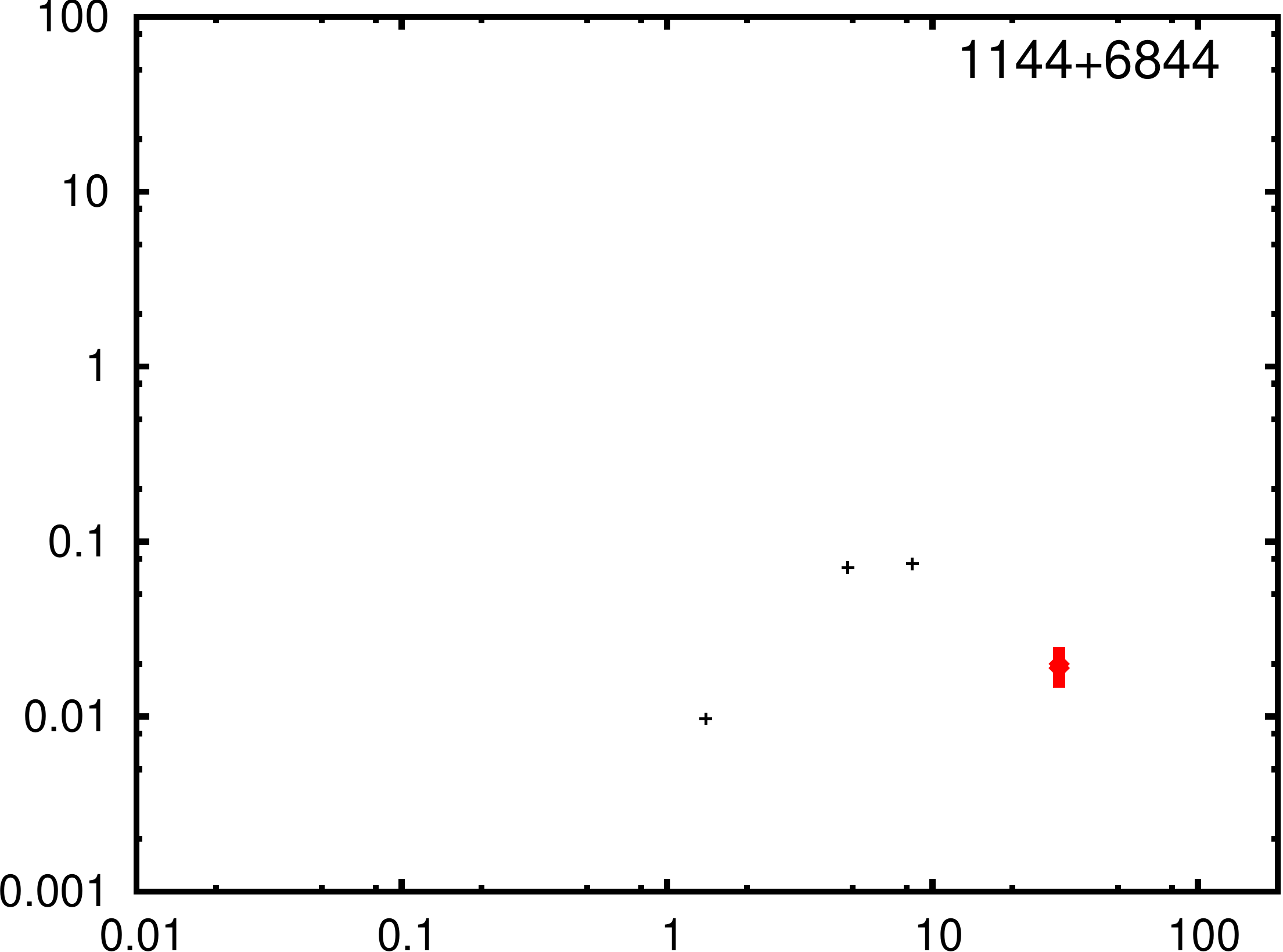}
\includegraphics[scale=0.2]{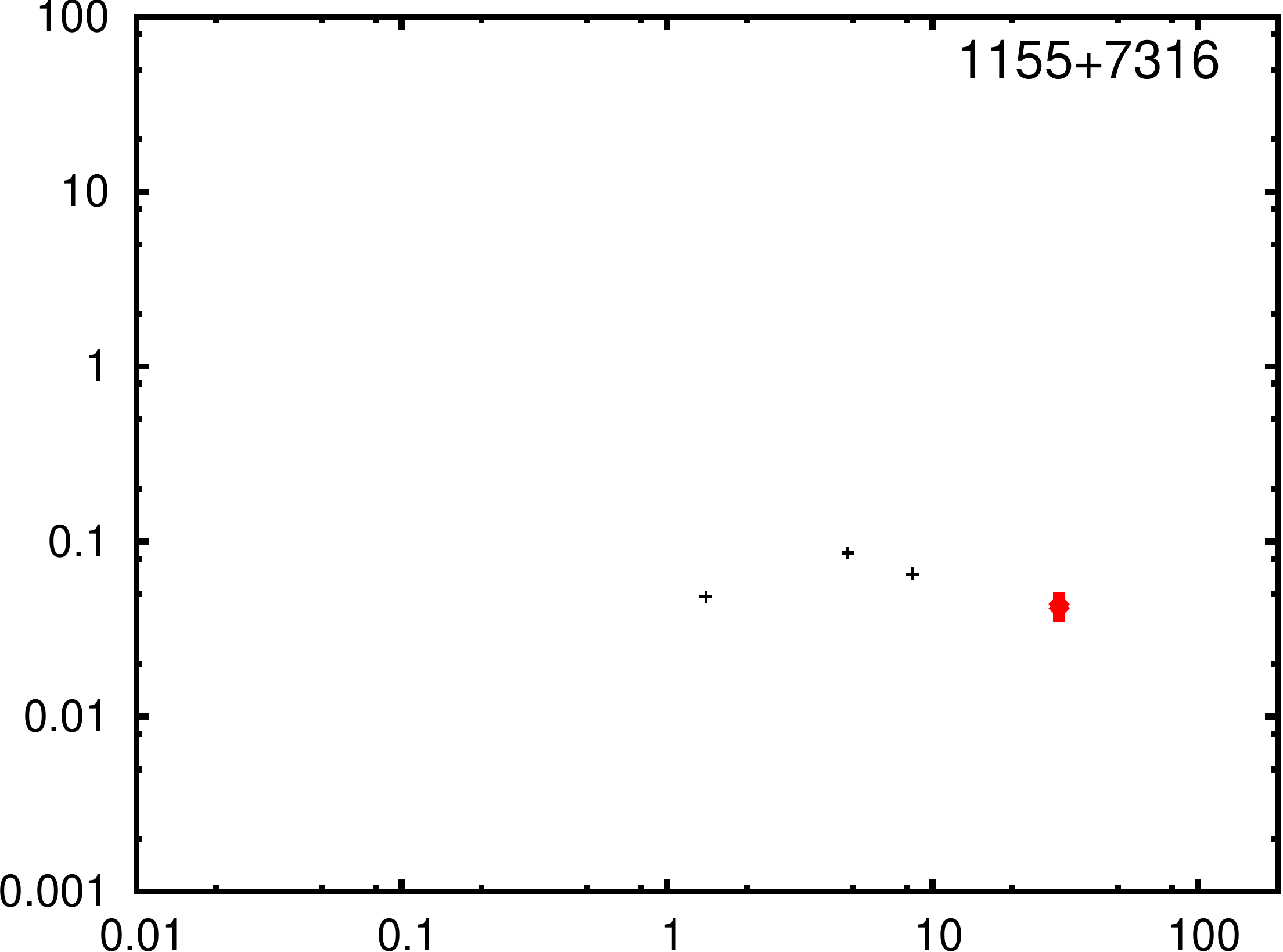}
\includegraphics[scale=0.2]{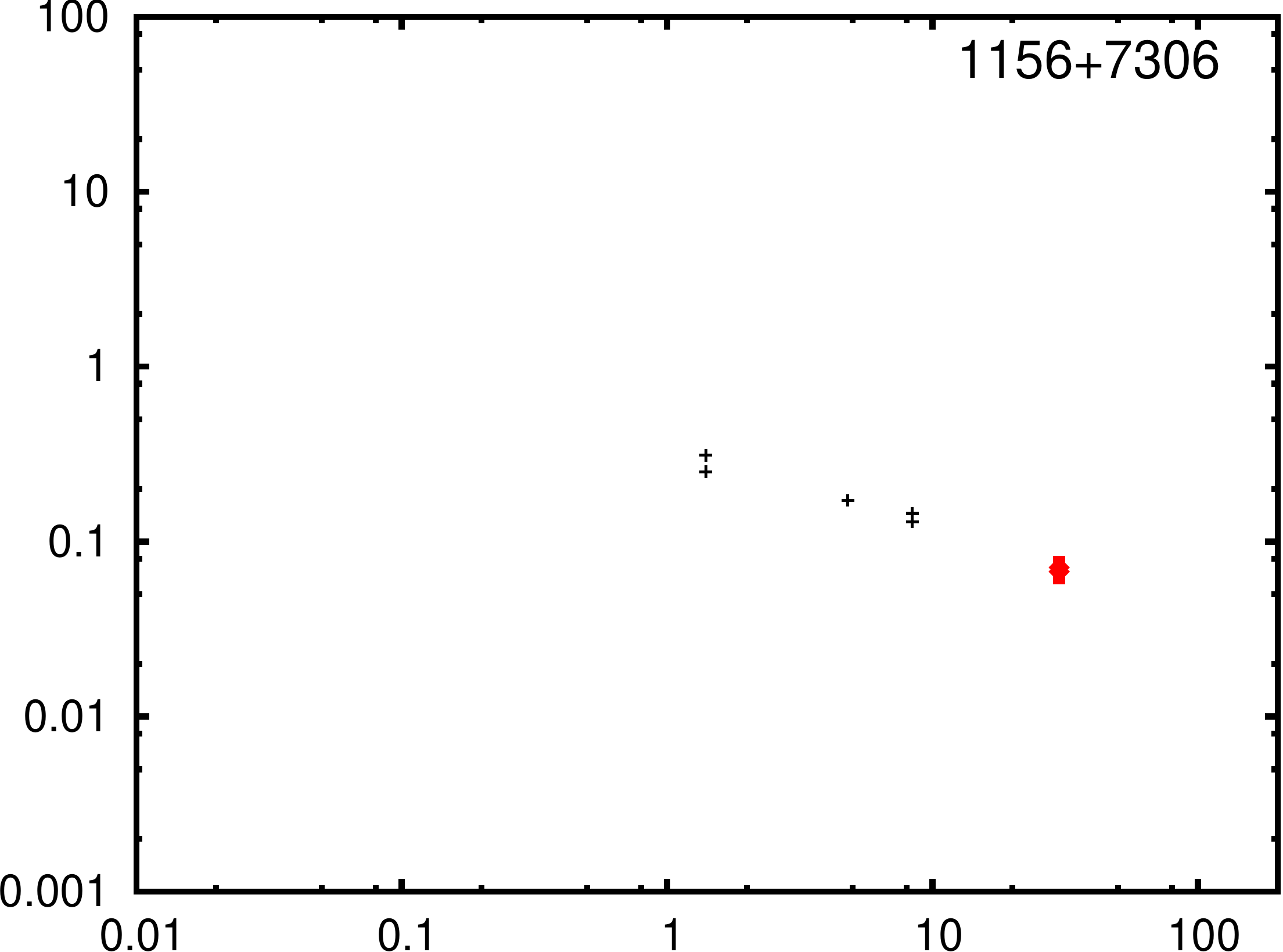}
\includegraphics[scale=0.2]{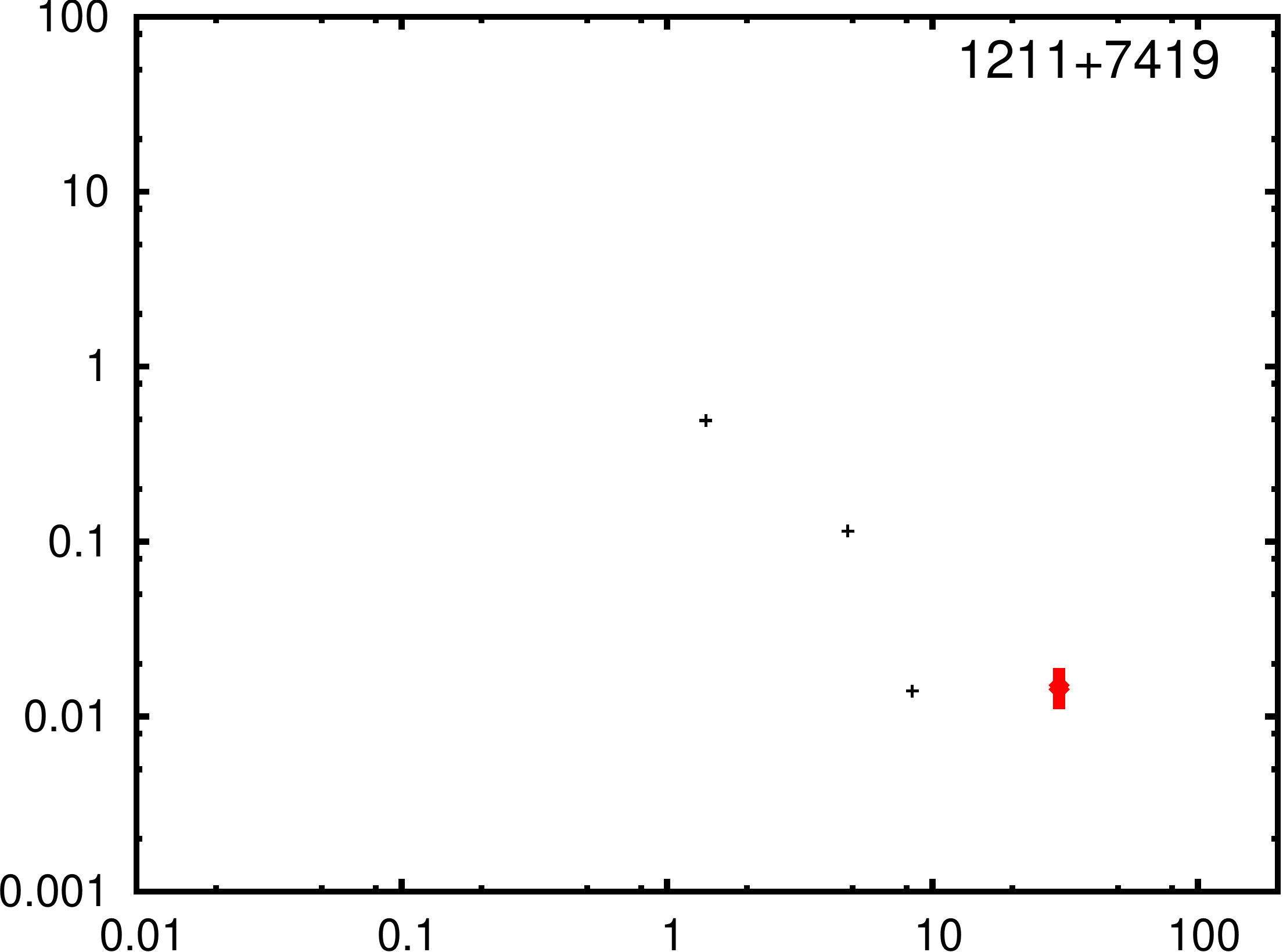}
\includegraphics[scale=0.2]{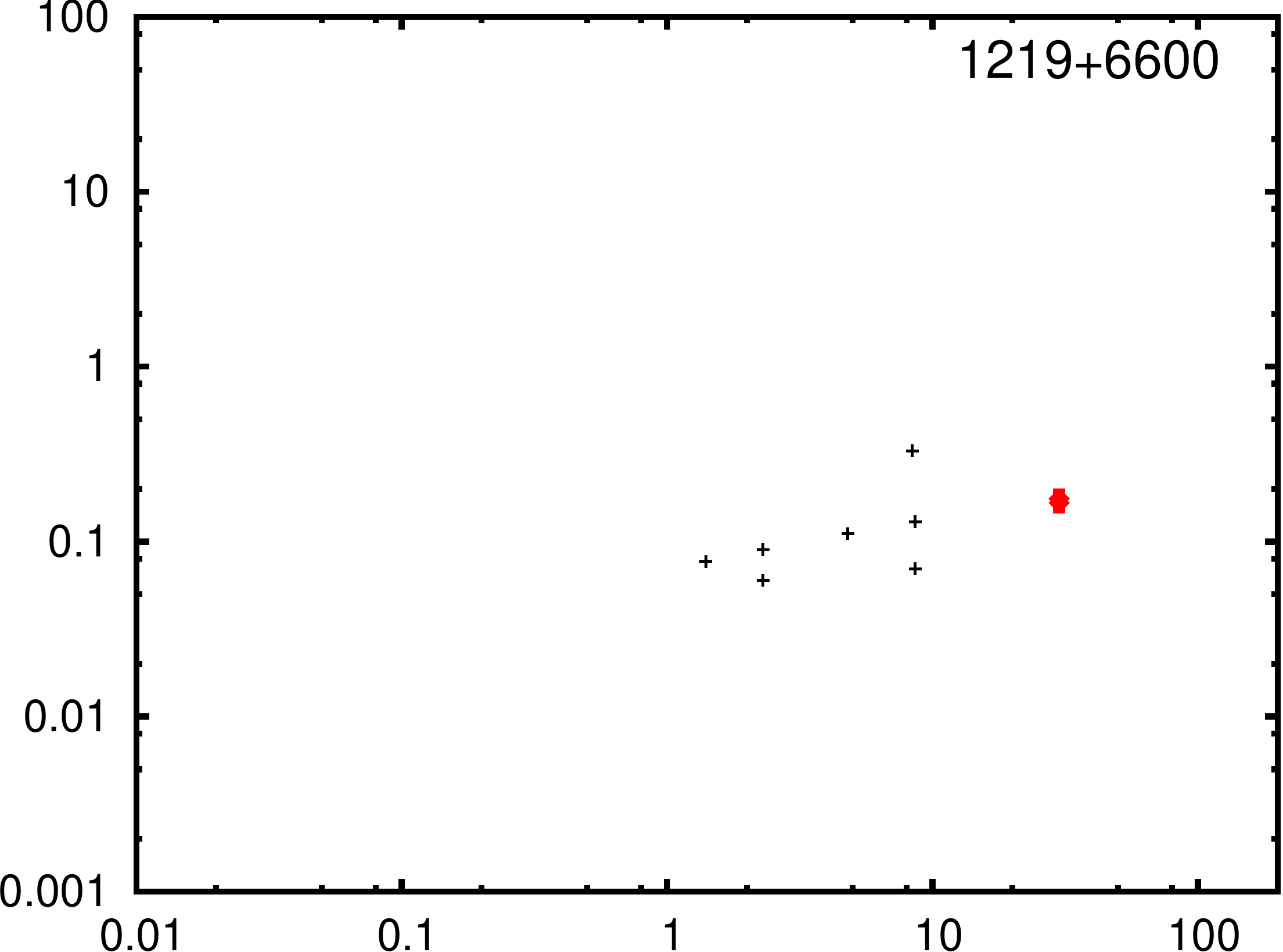}
\includegraphics[scale=0.2]{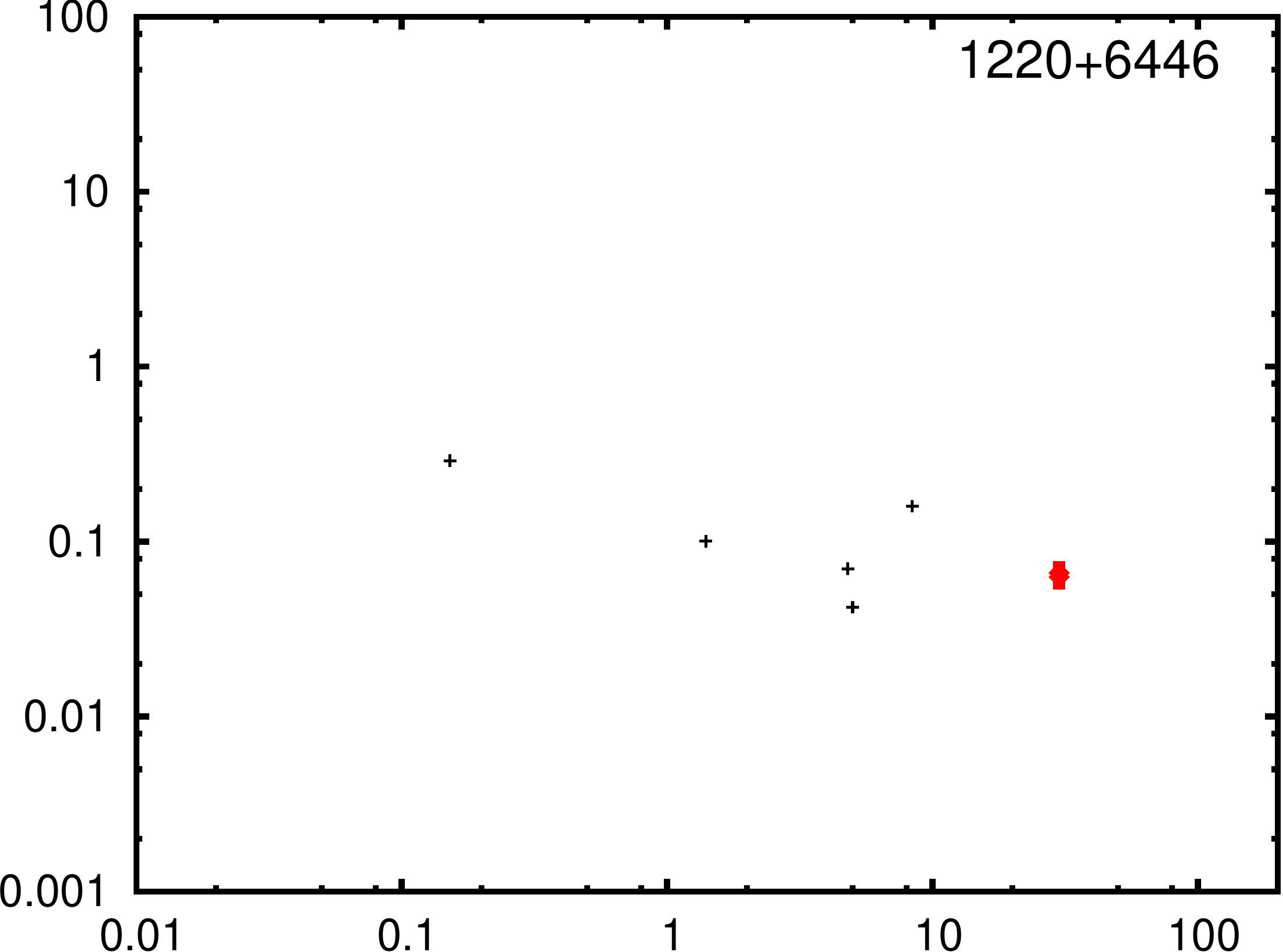}
\includegraphics[scale=0.2]{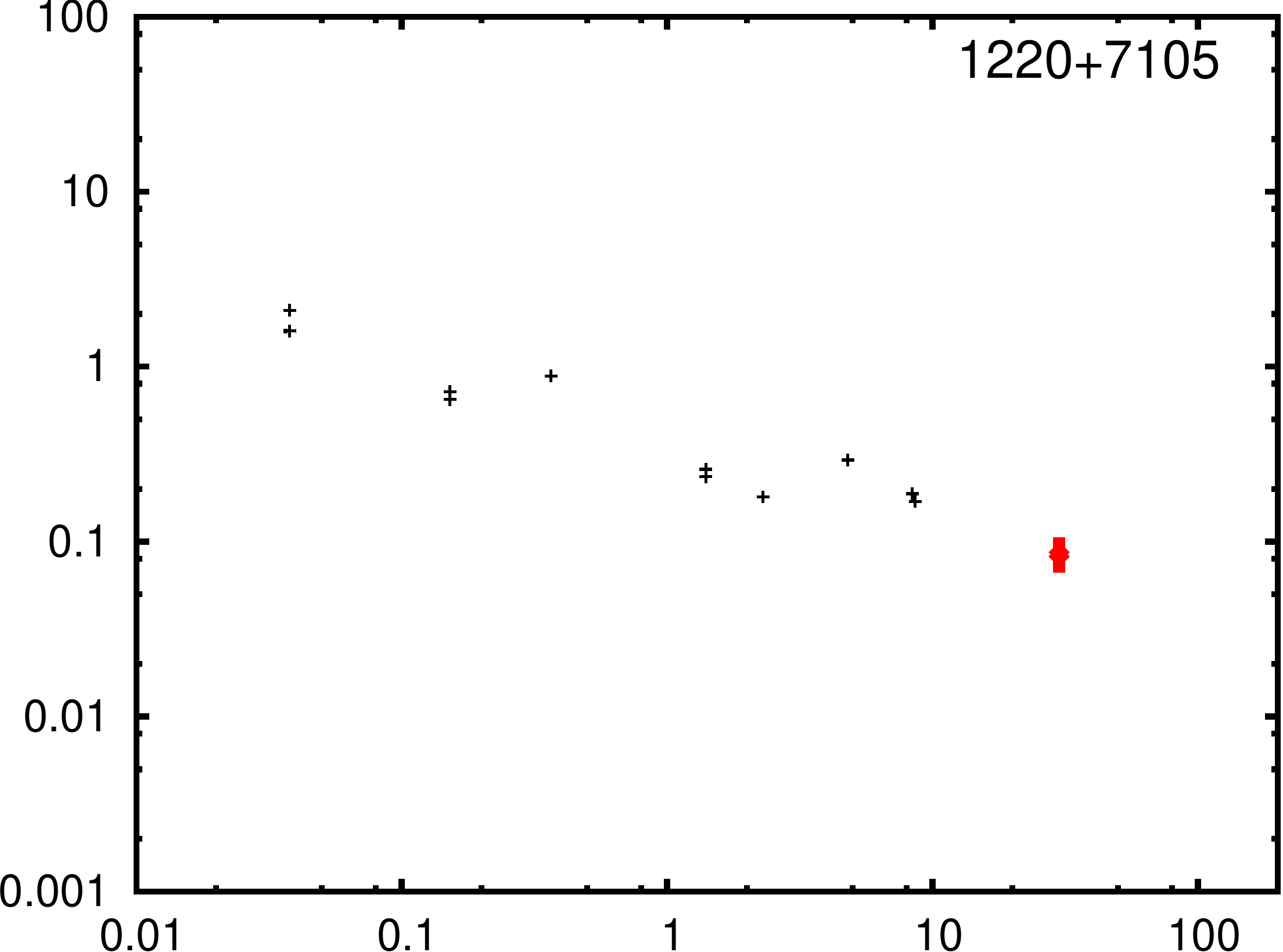}
\includegraphics[scale=0.2]{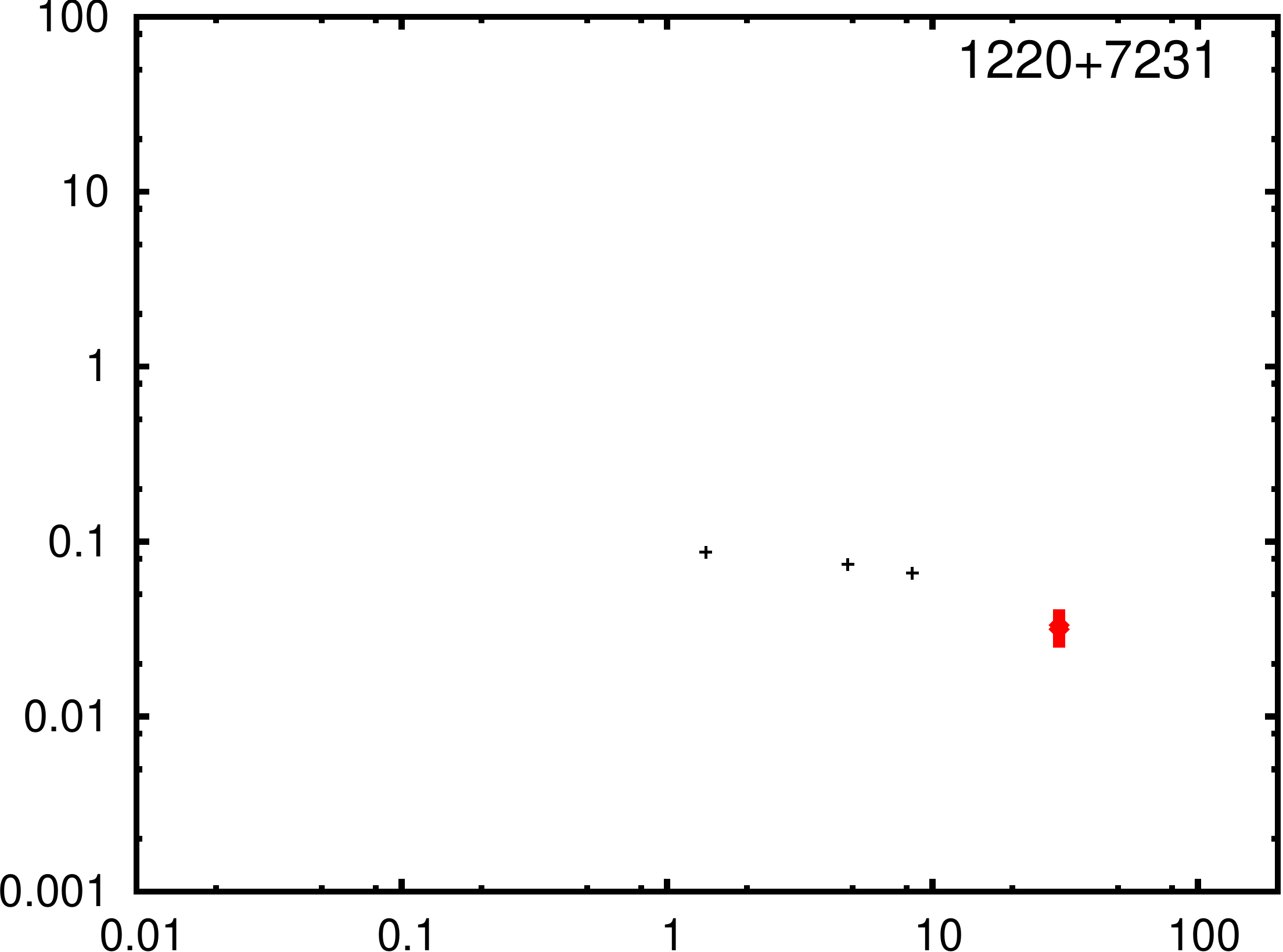}
\includegraphics[scale=0.2]{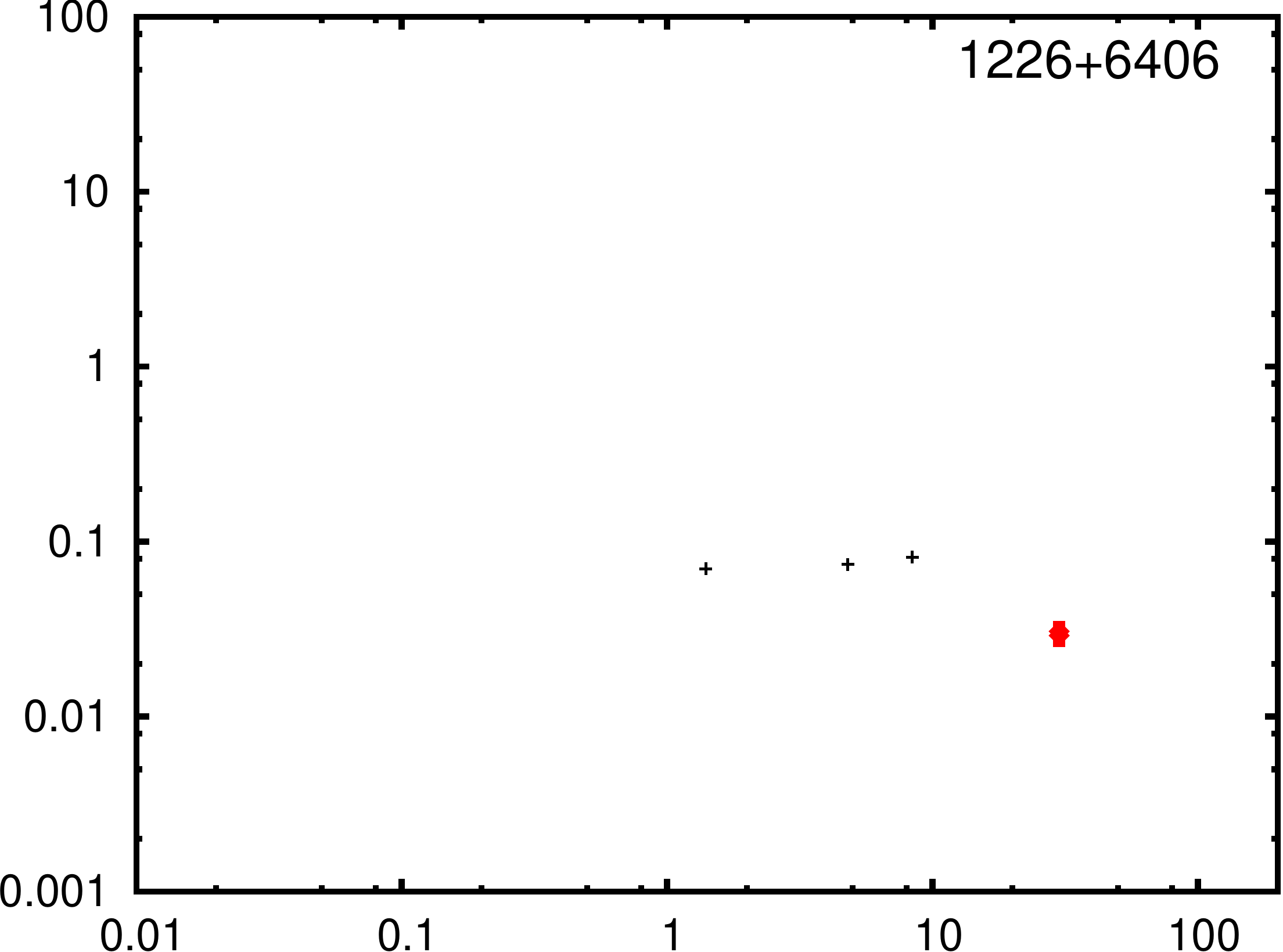}
\caption[Source spectra of the CRATES sources]{Source spectra for the CRATES sources. Flux density in Jansky (y-axis) vs. frequency in GHz (x-axis).}
\label{fig:sourcespectra}
\end{figure}
\clearpage\begin{figure}
\centering
\includegraphics[scale=0.2]{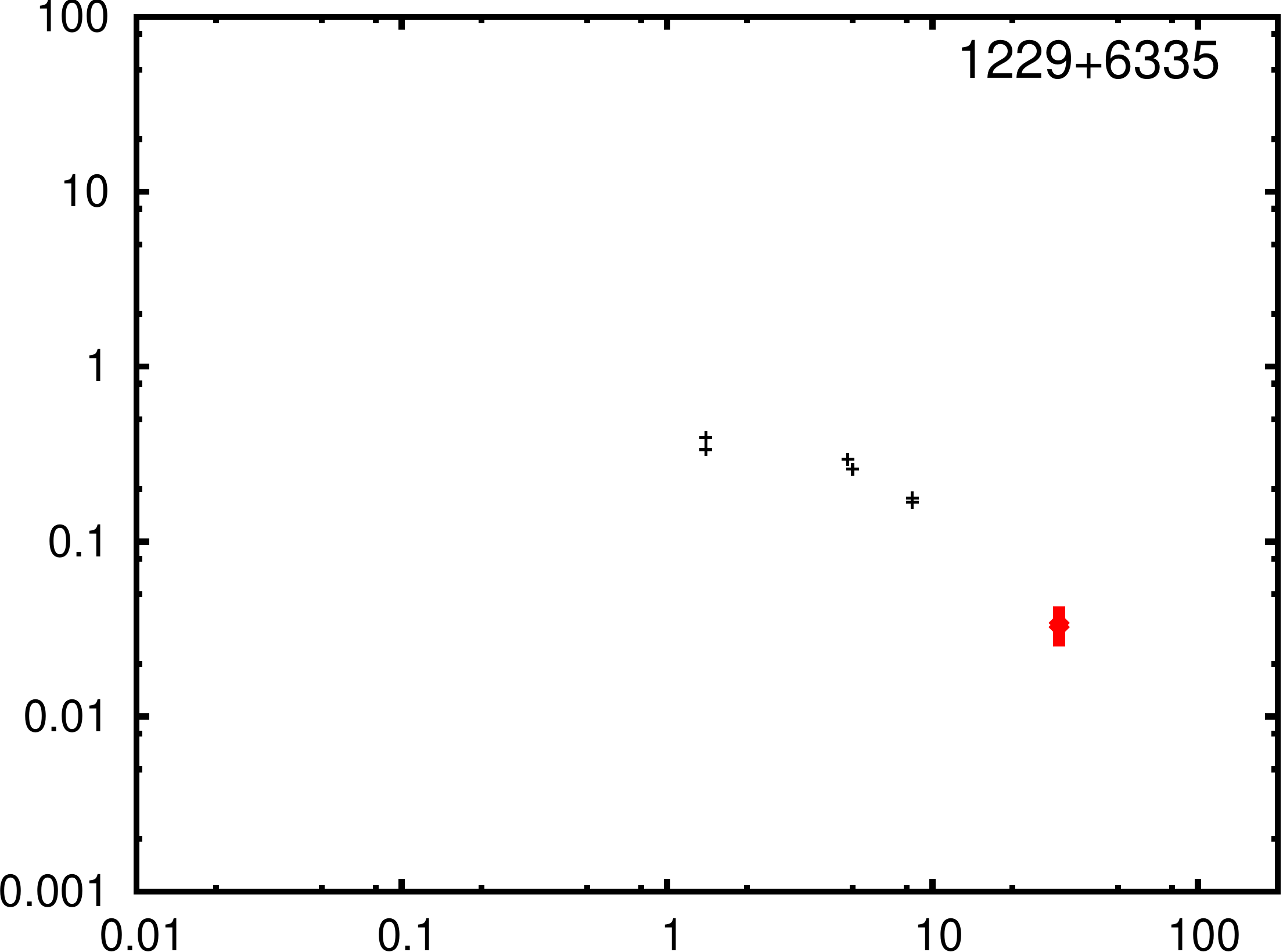}
\includegraphics[scale=0.2]{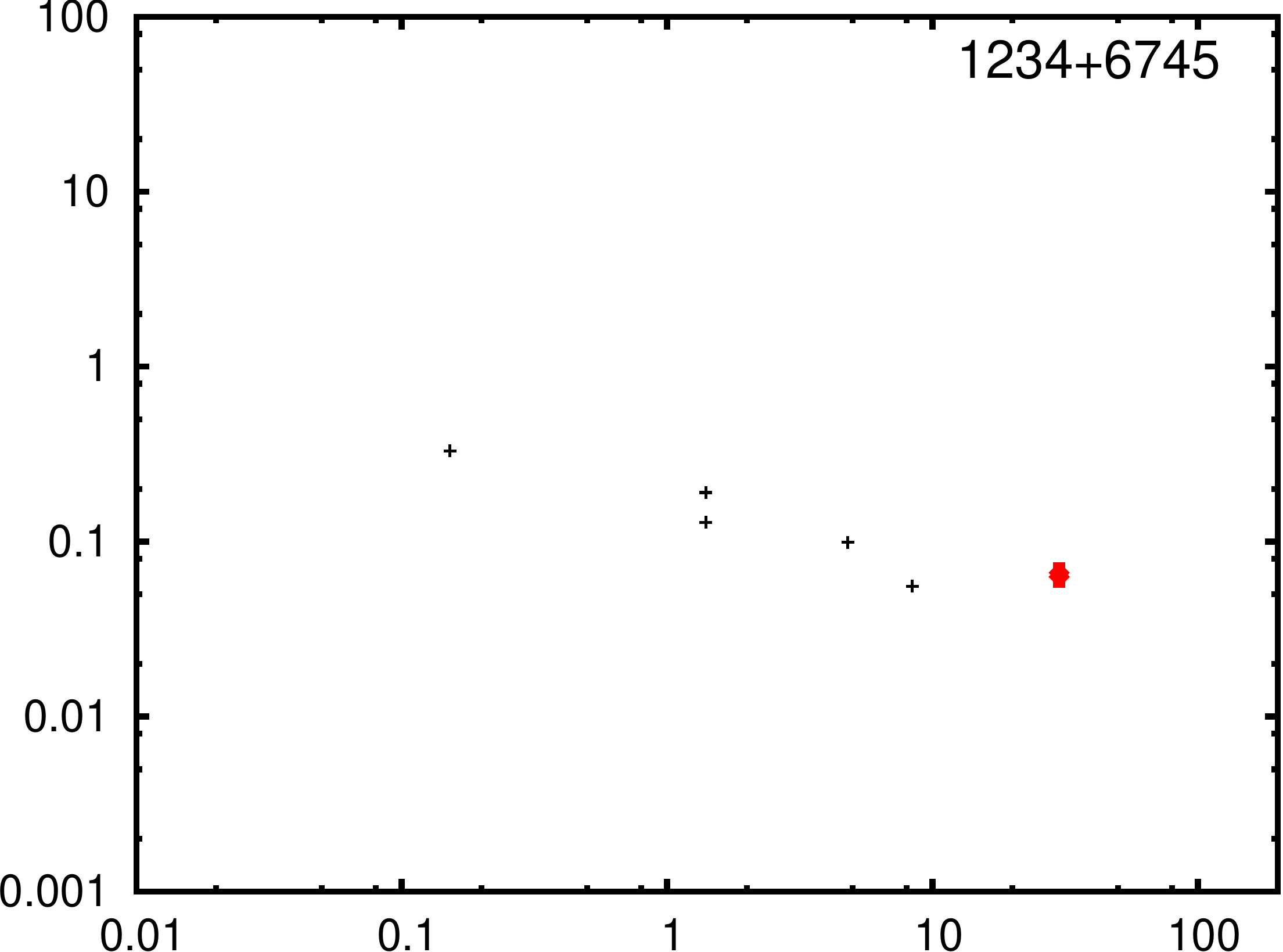}
\includegraphics[scale=0.2]{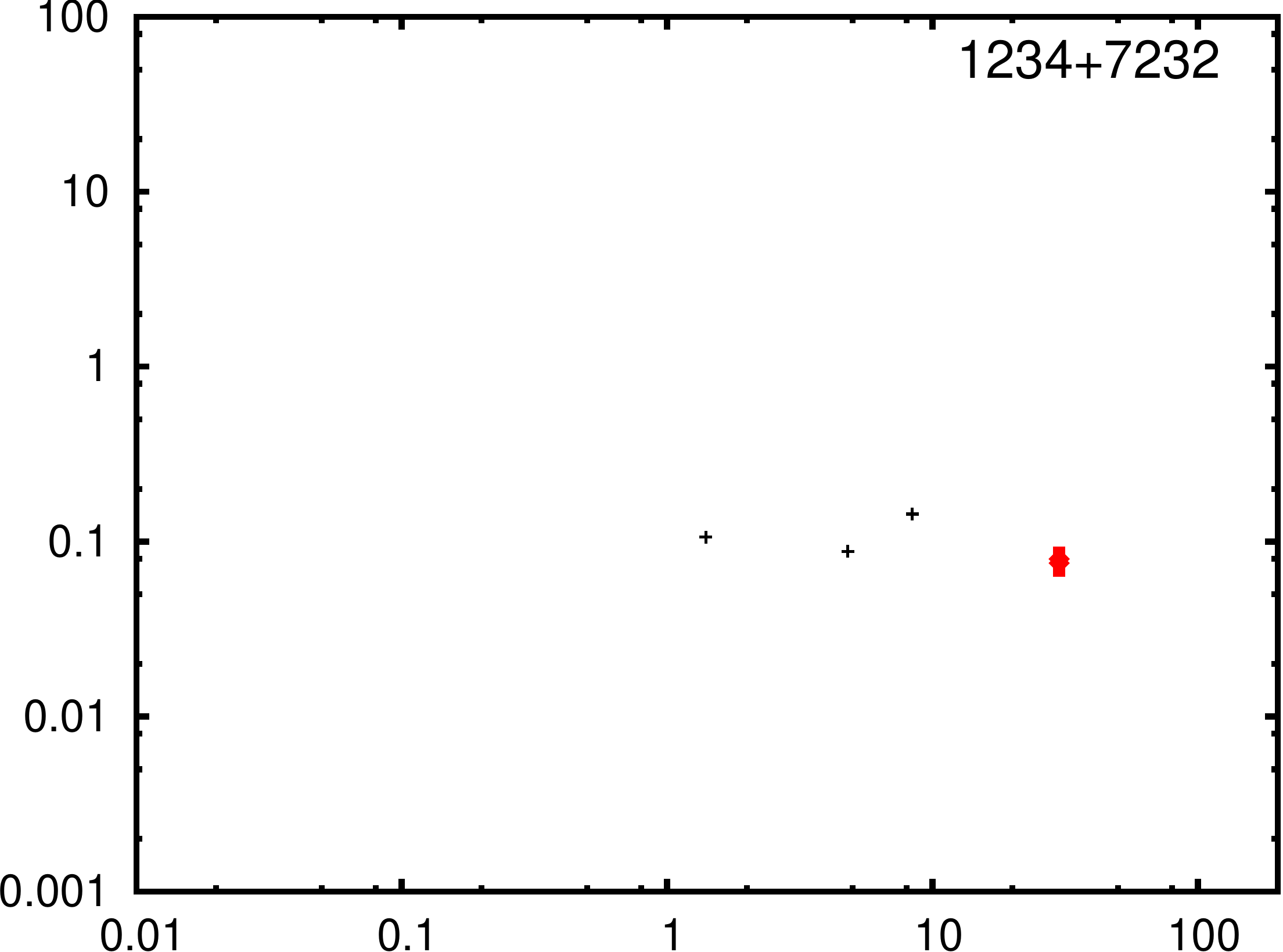}
\includegraphics[scale=0.2]{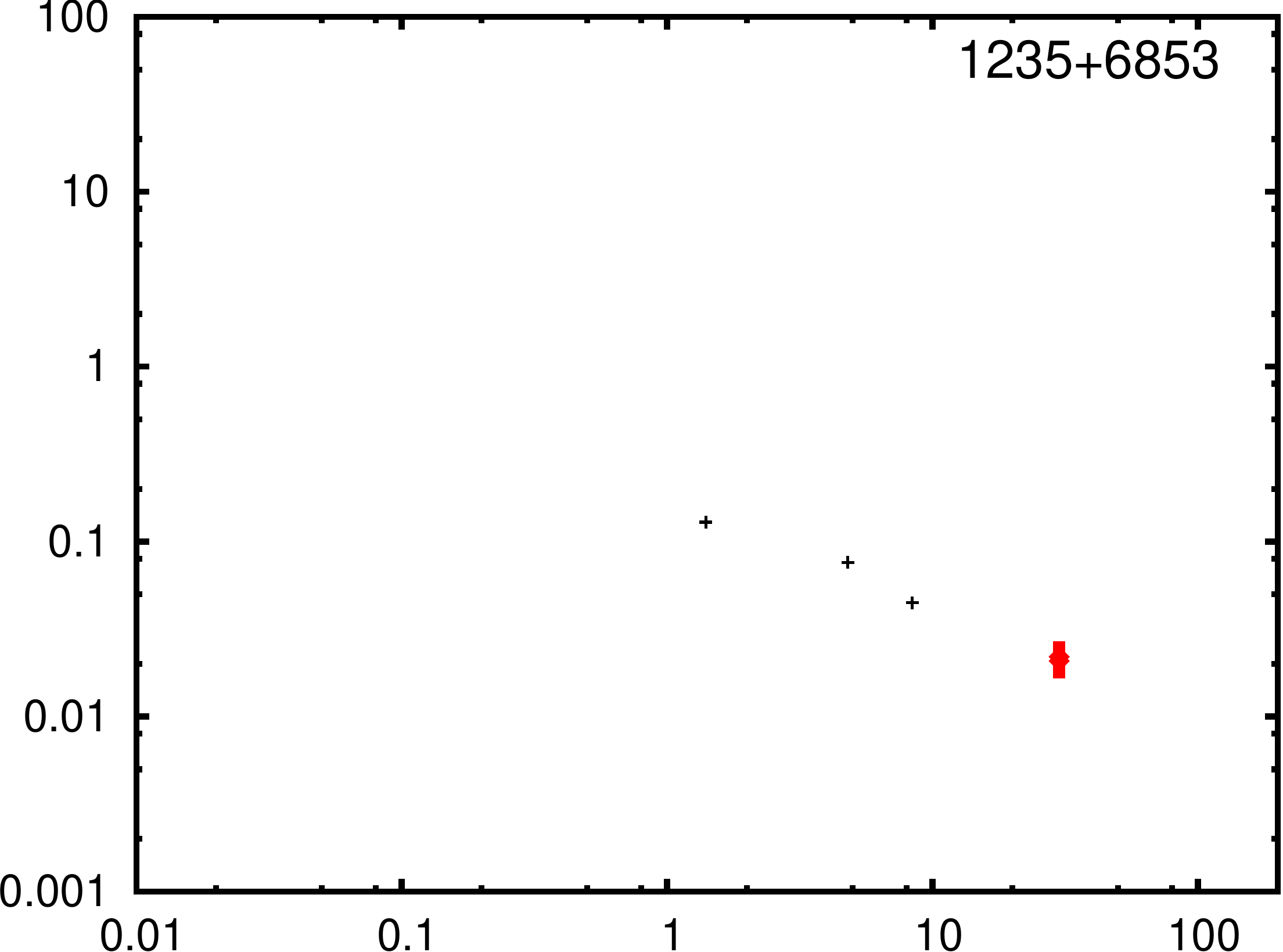}
\includegraphics[scale=0.2]{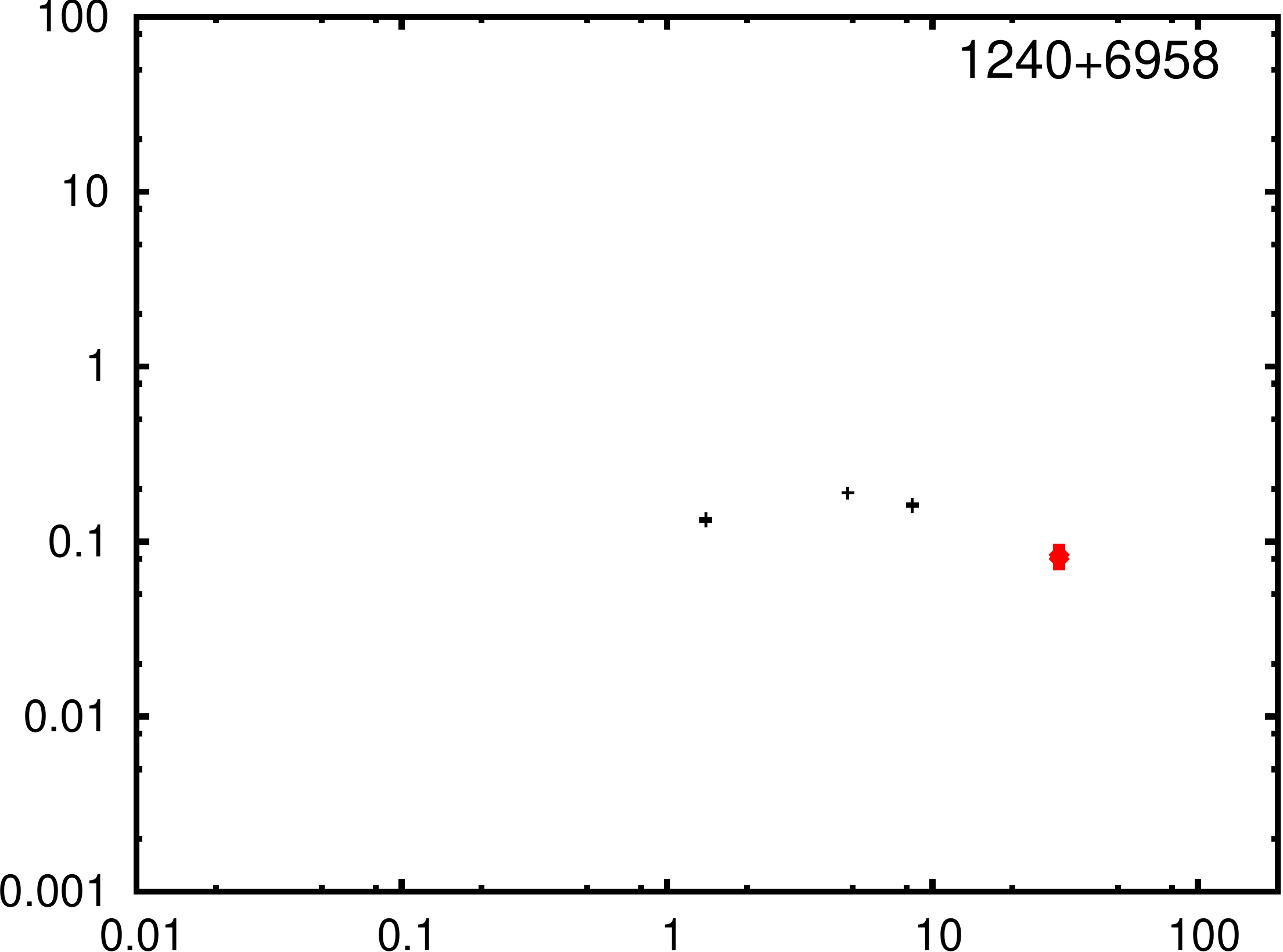}
\includegraphics[scale=0.2]{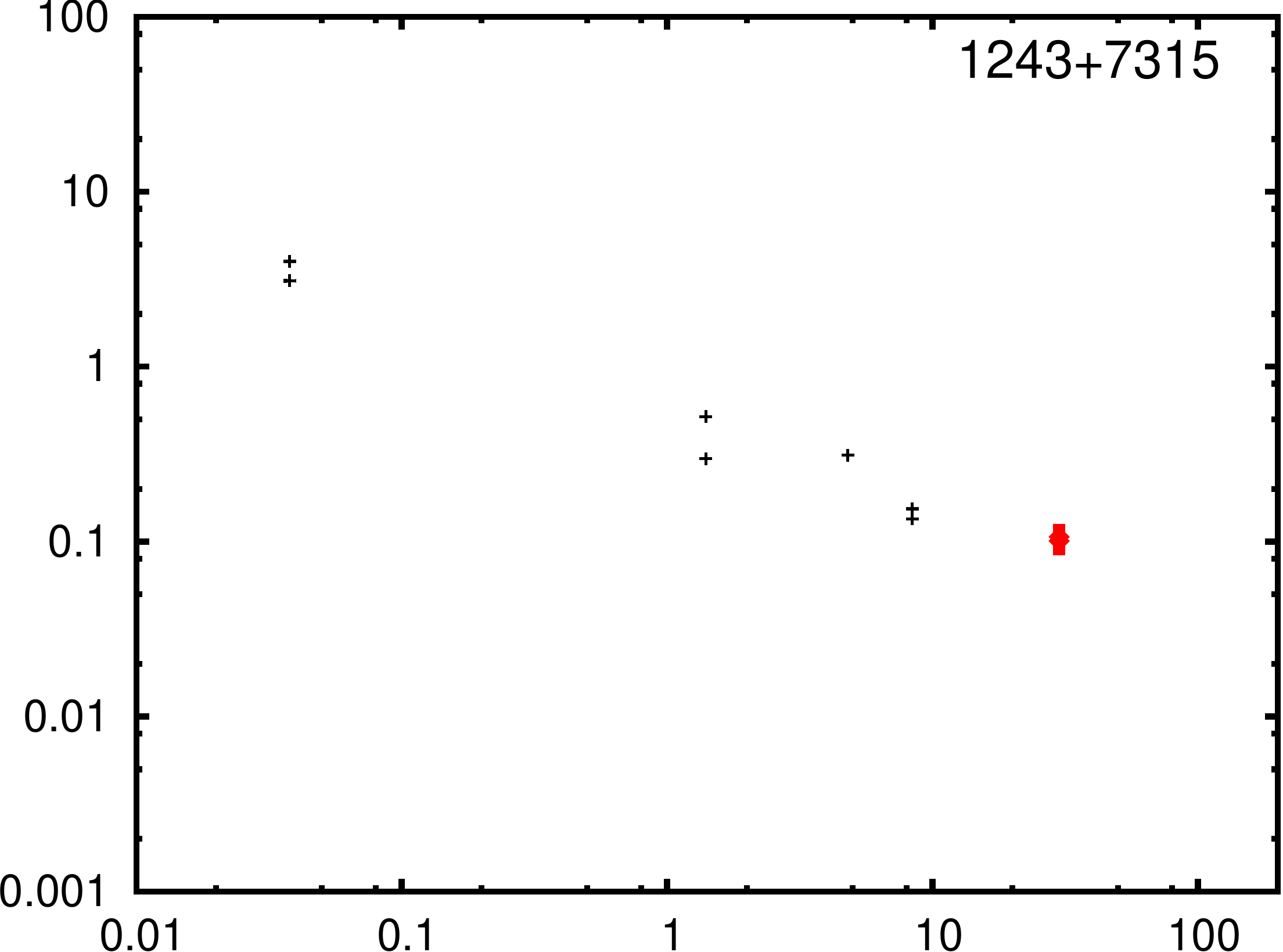}
\includegraphics[scale=0.2]{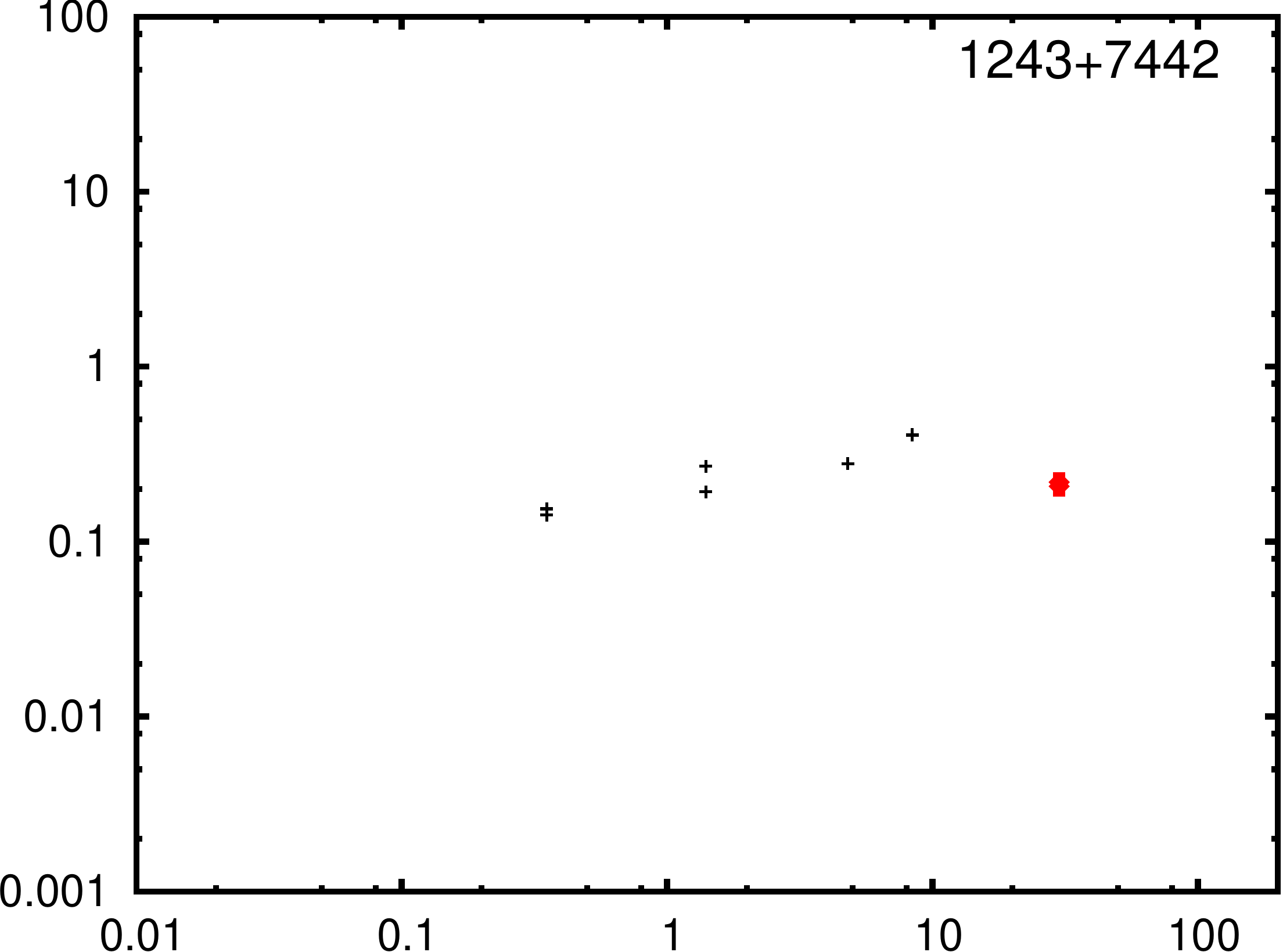}
\includegraphics[scale=0.2]{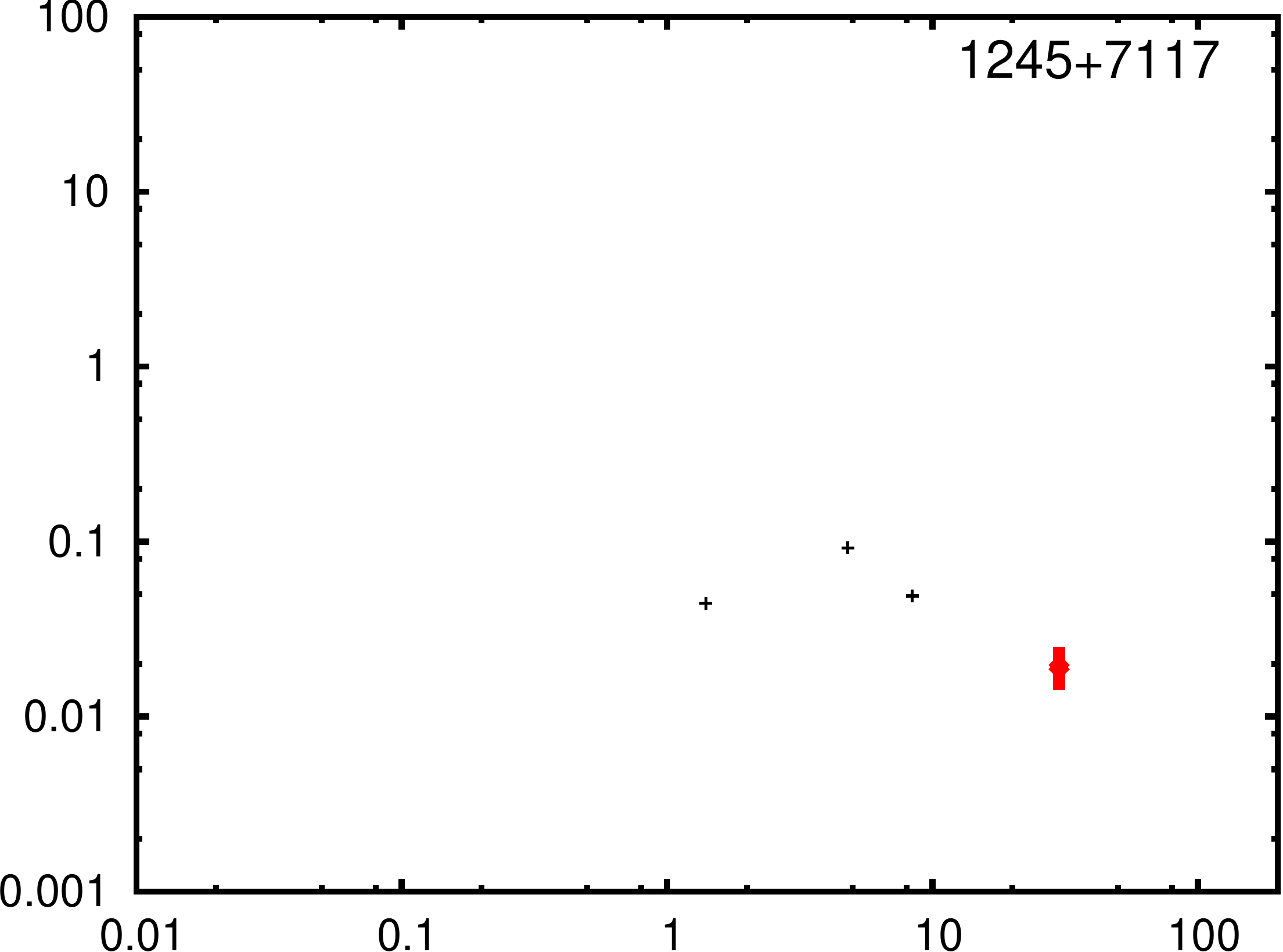}
\includegraphics[scale=0.2]{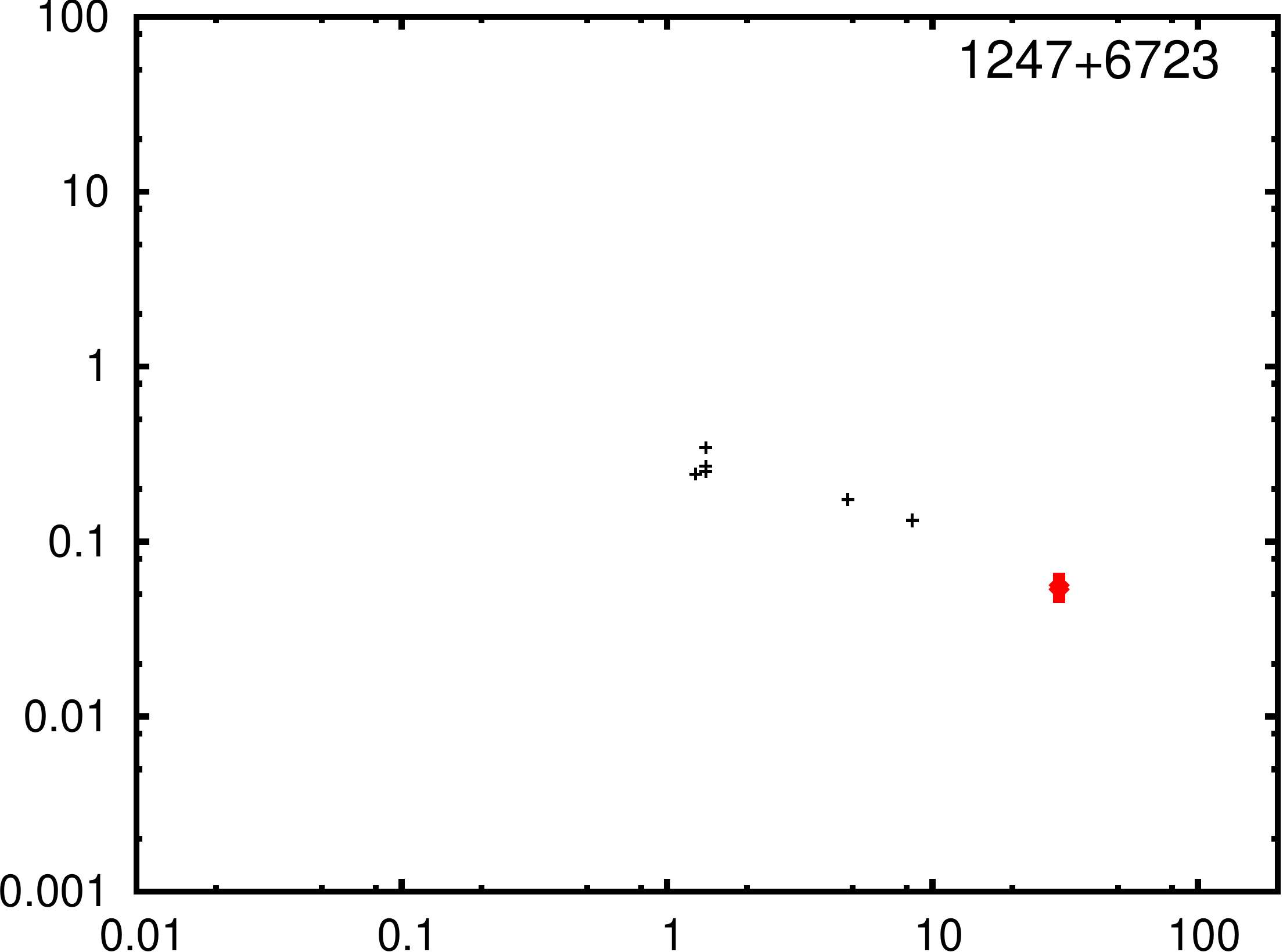}
\includegraphics[scale=0.2]{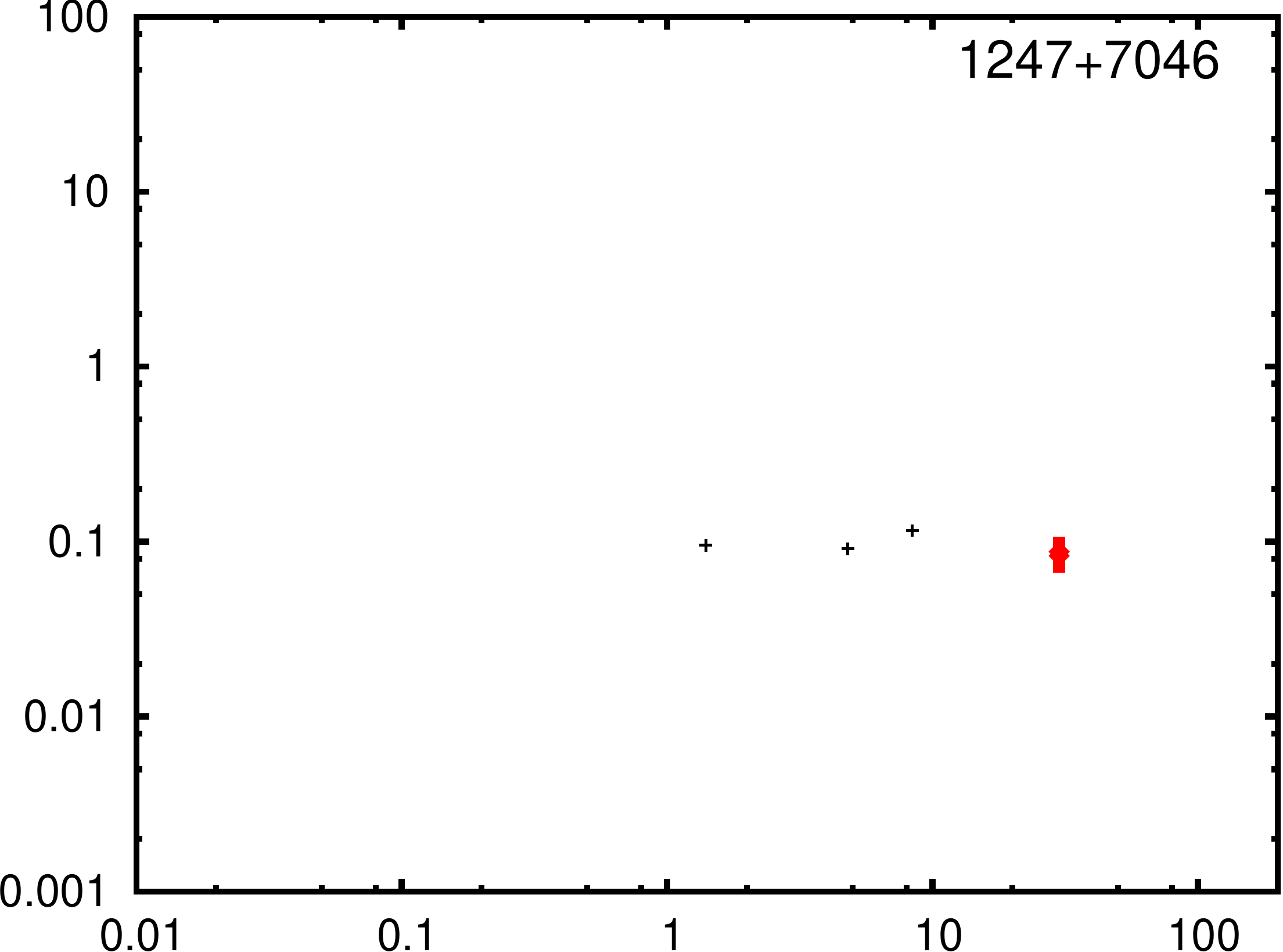}
\includegraphics[scale=0.2]{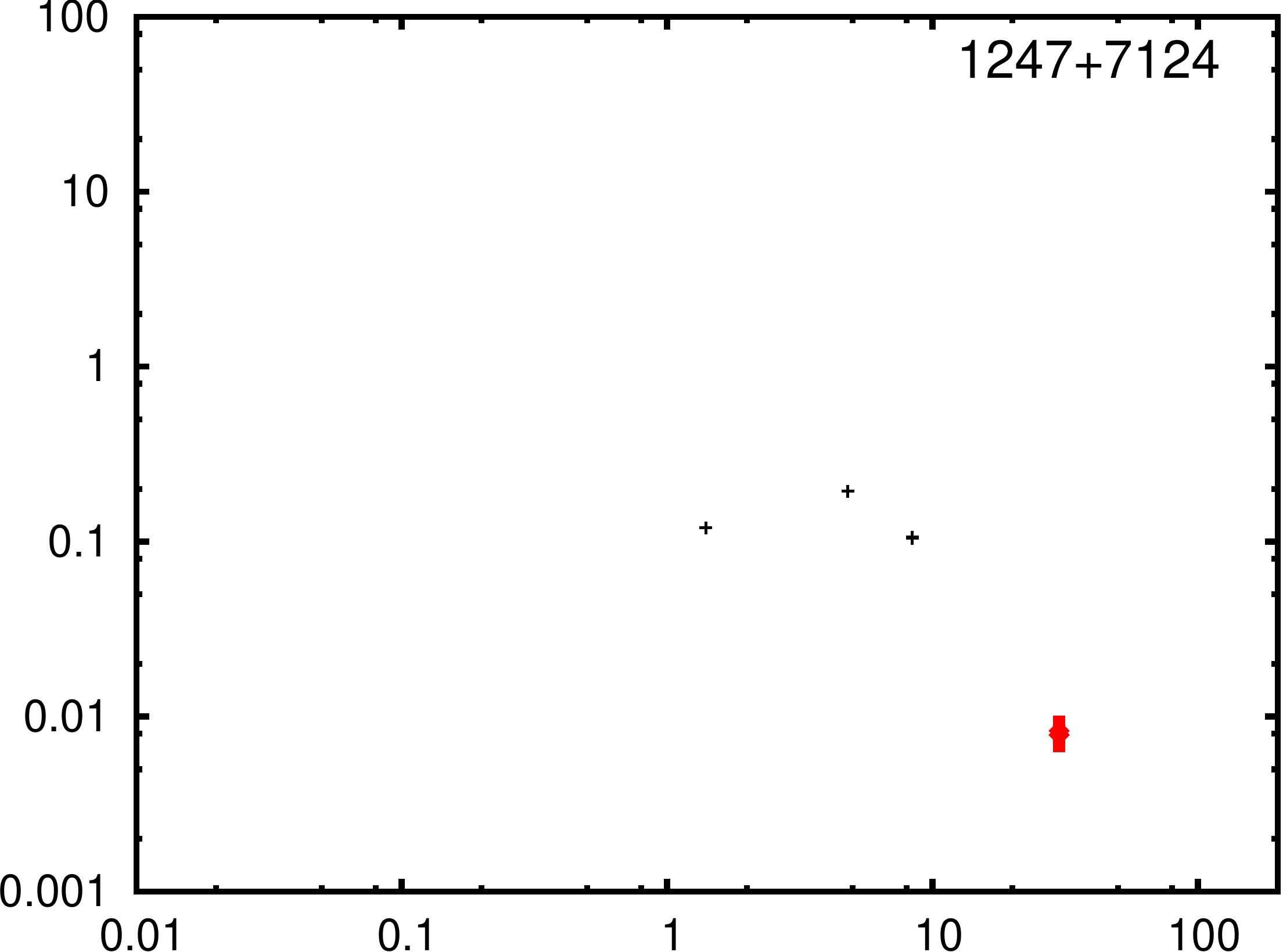}
\includegraphics[scale=0.2]{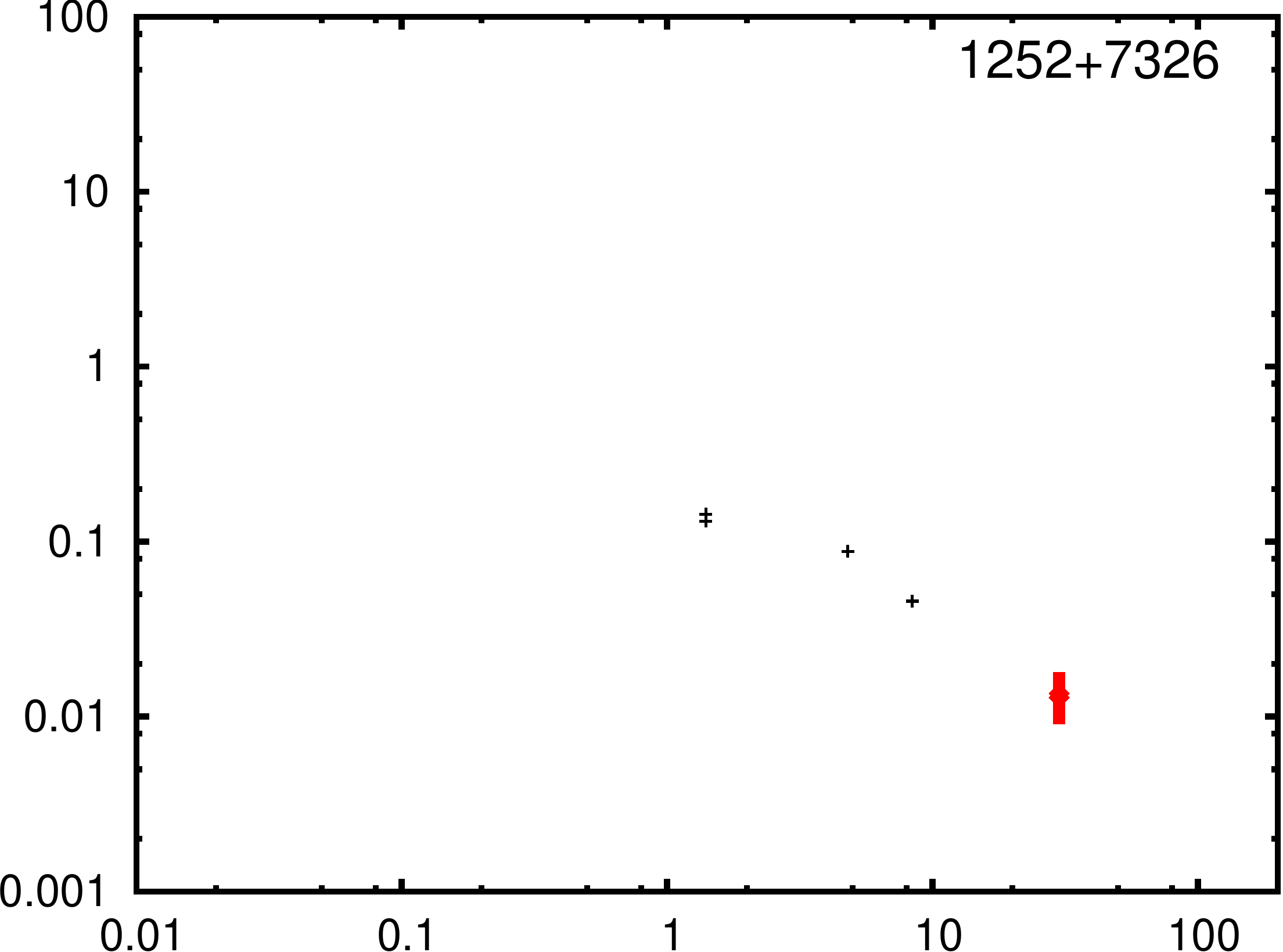}
\includegraphics[scale=0.2]{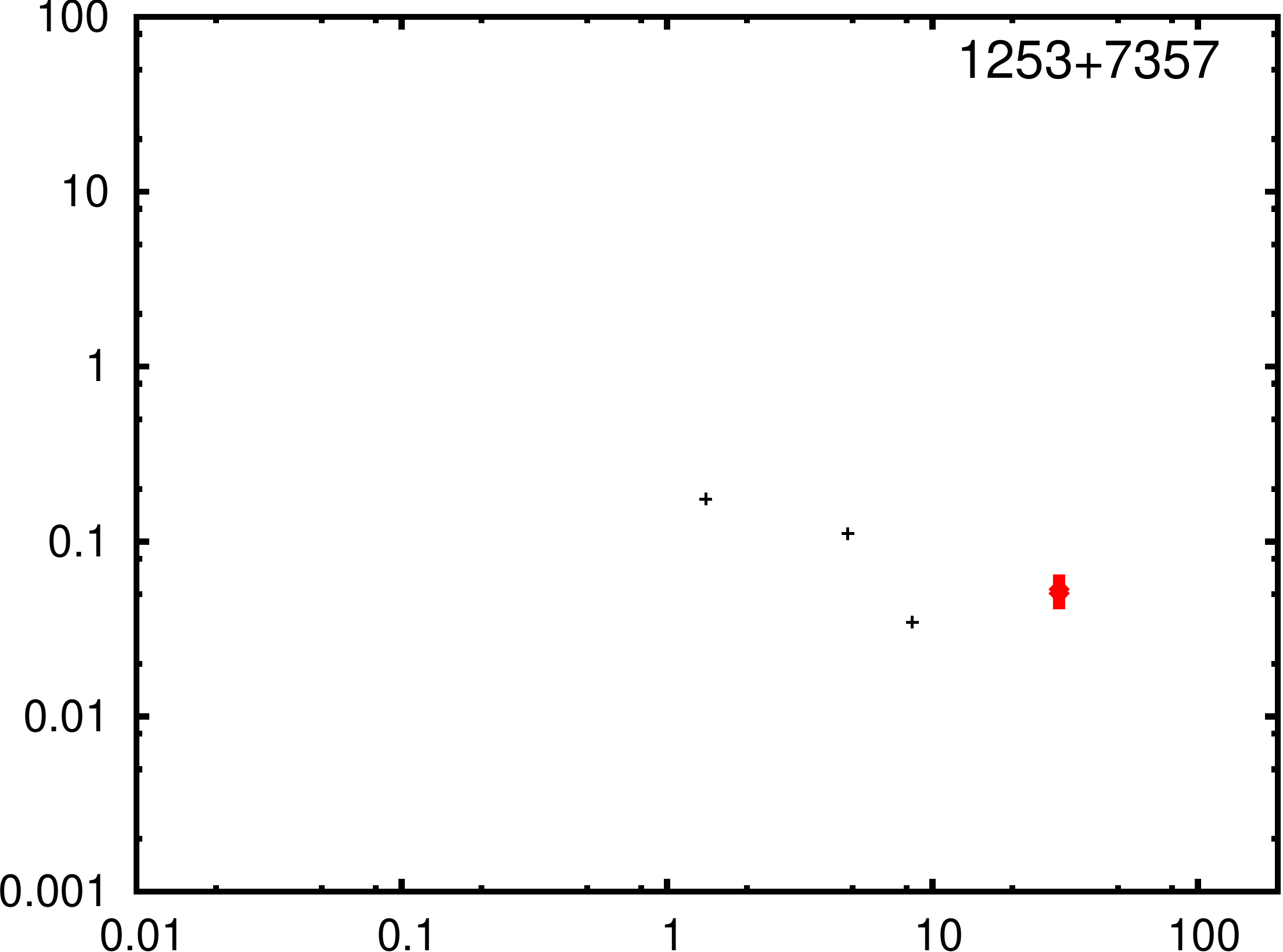}
\includegraphics[scale=0.2]{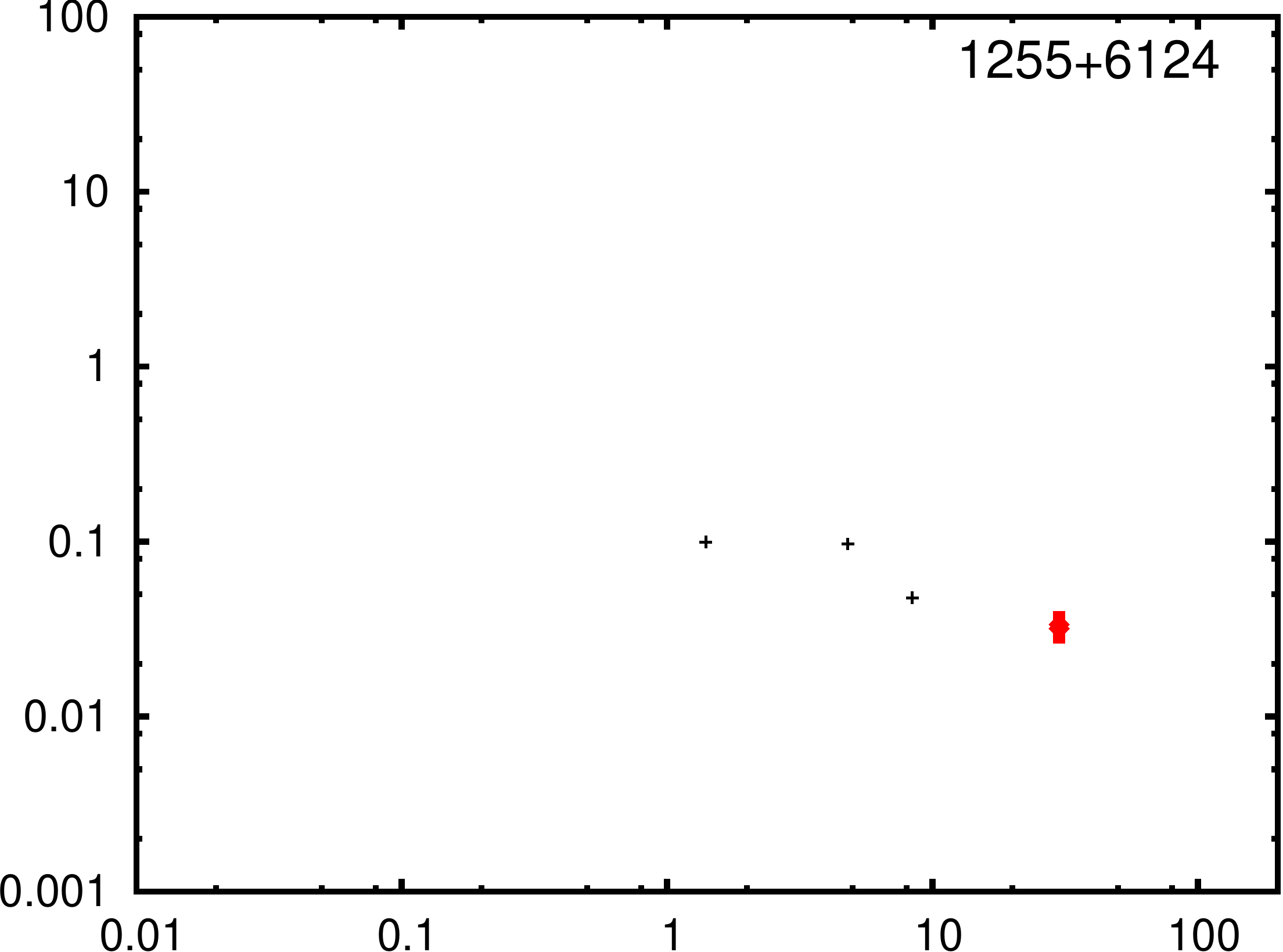}
\includegraphics[scale=0.2]{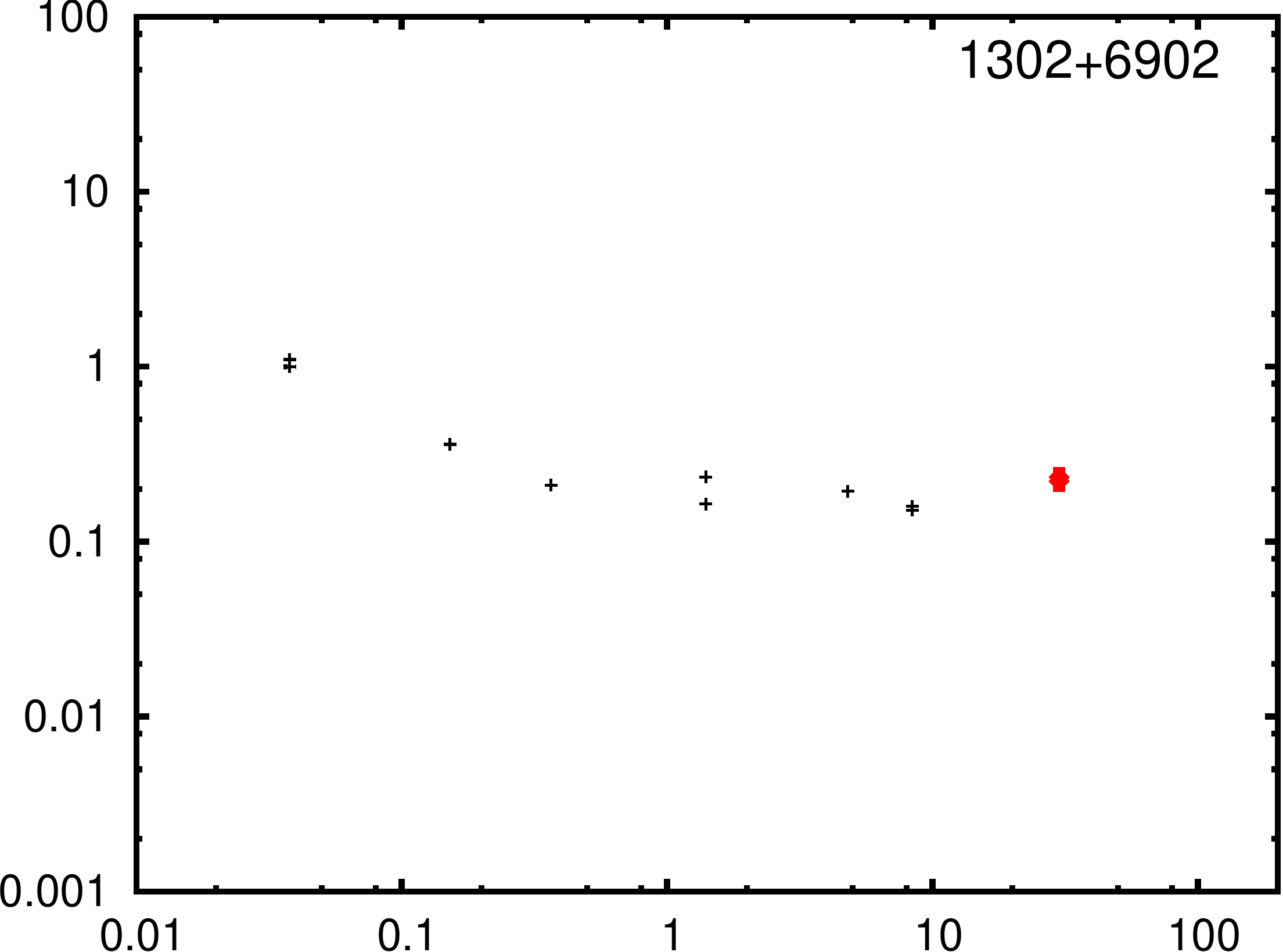}
\includegraphics[scale=0.2]{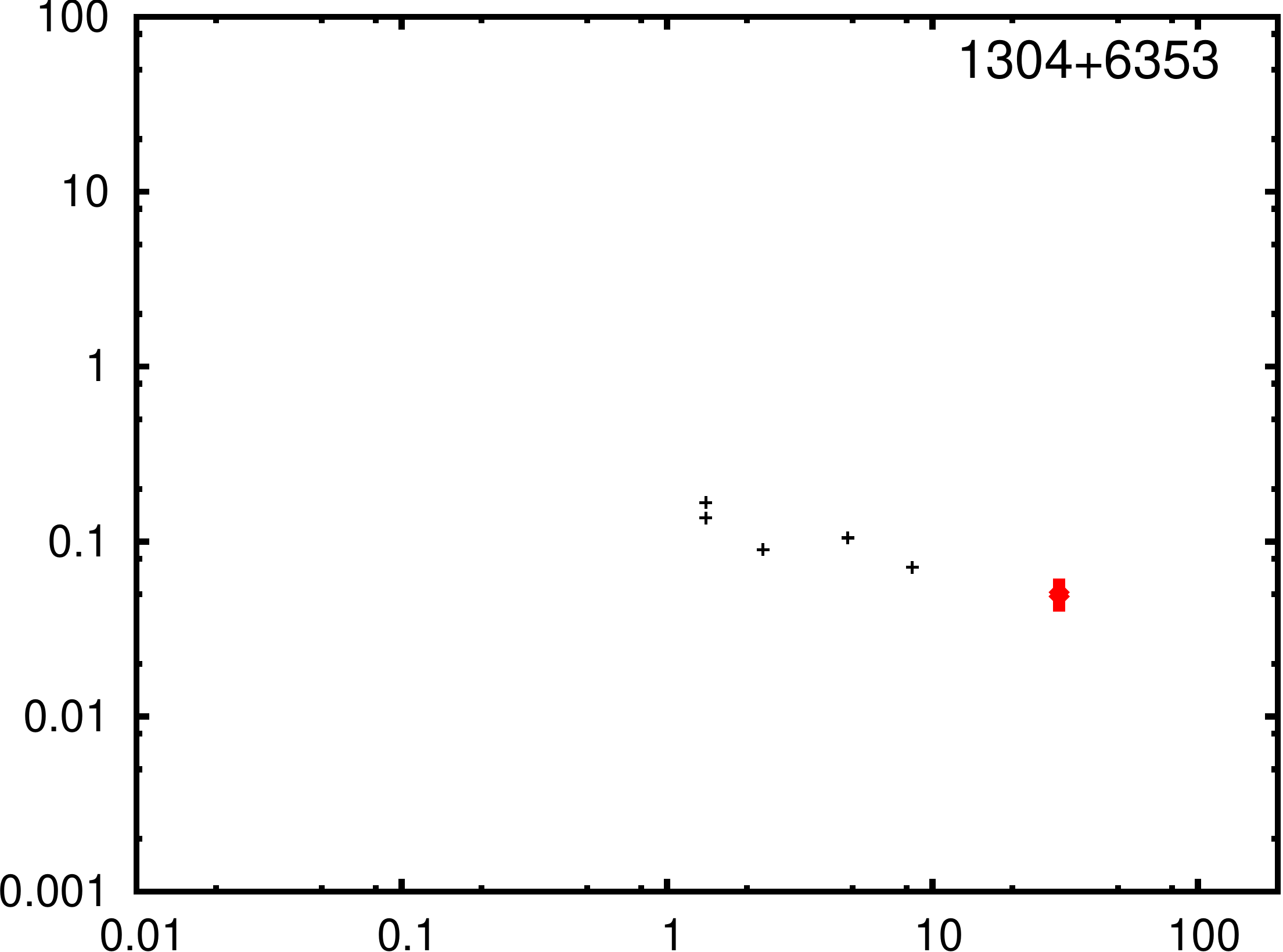}
\includegraphics[scale=0.2]{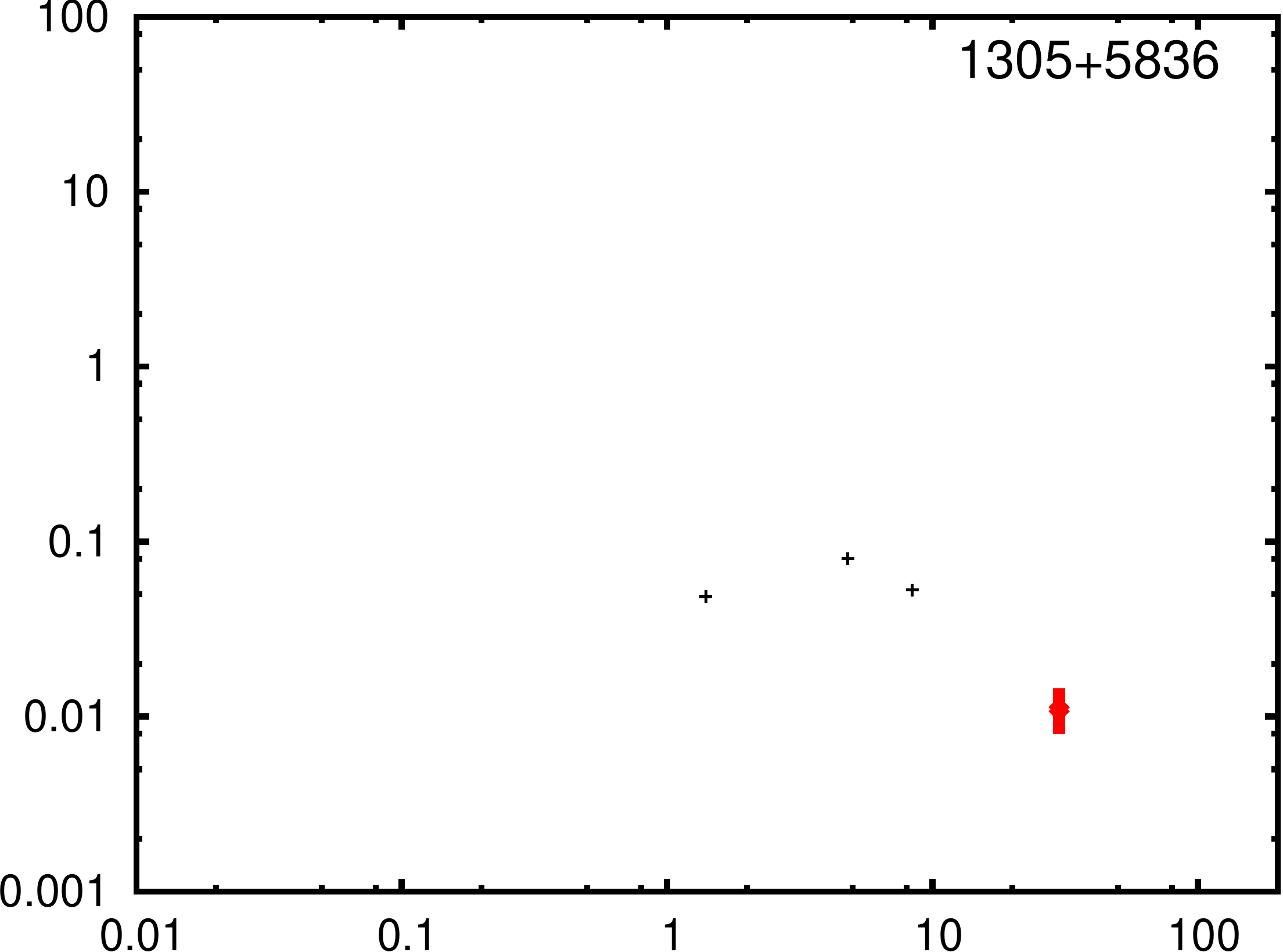}
\includegraphics[scale=0.2]{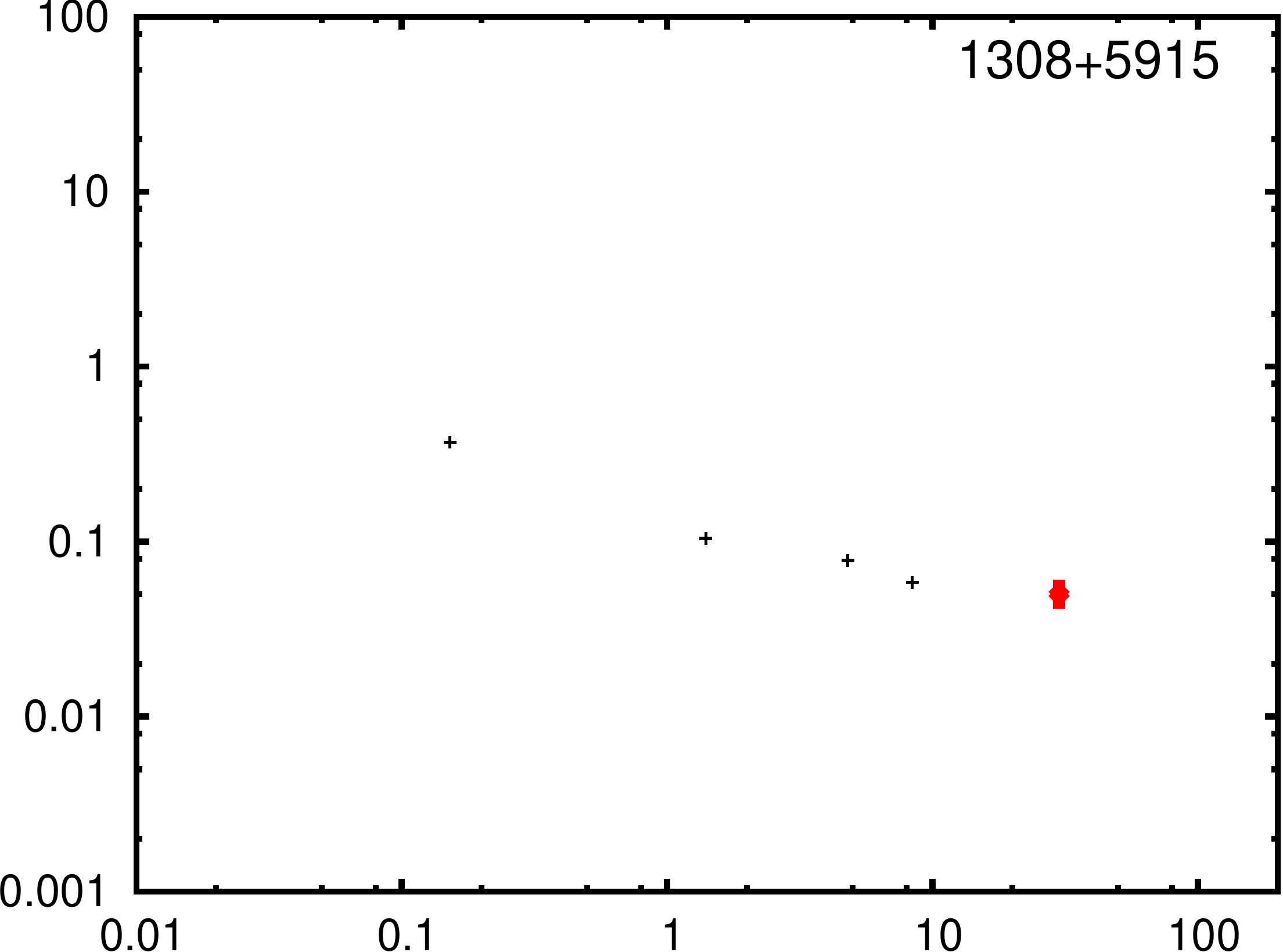}
\end{figure}
\clearpage\begin{figure}
\centering
\includegraphics[scale=0.2]{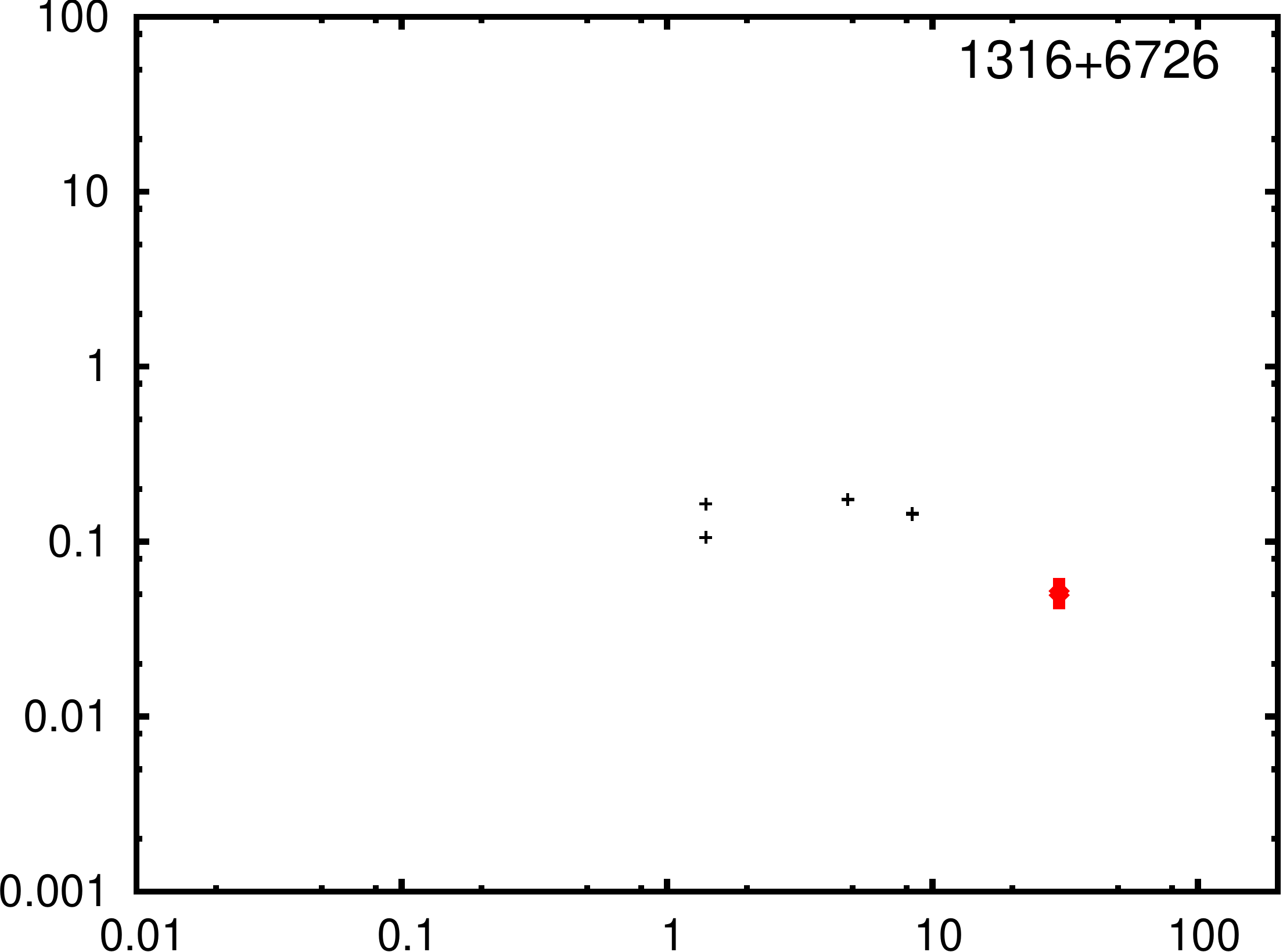}
\includegraphics[scale=0.2]{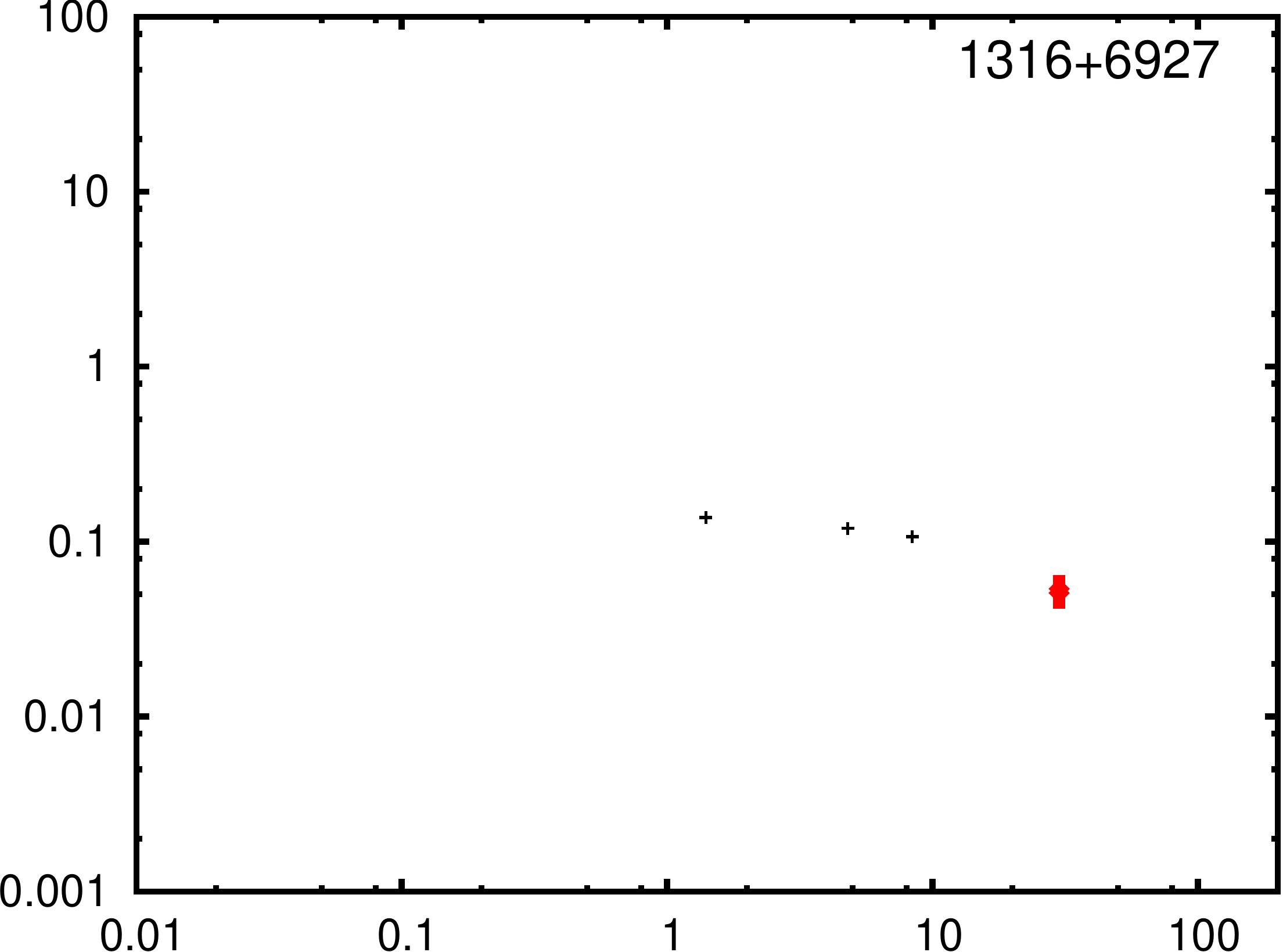}
\includegraphics[scale=0.2]{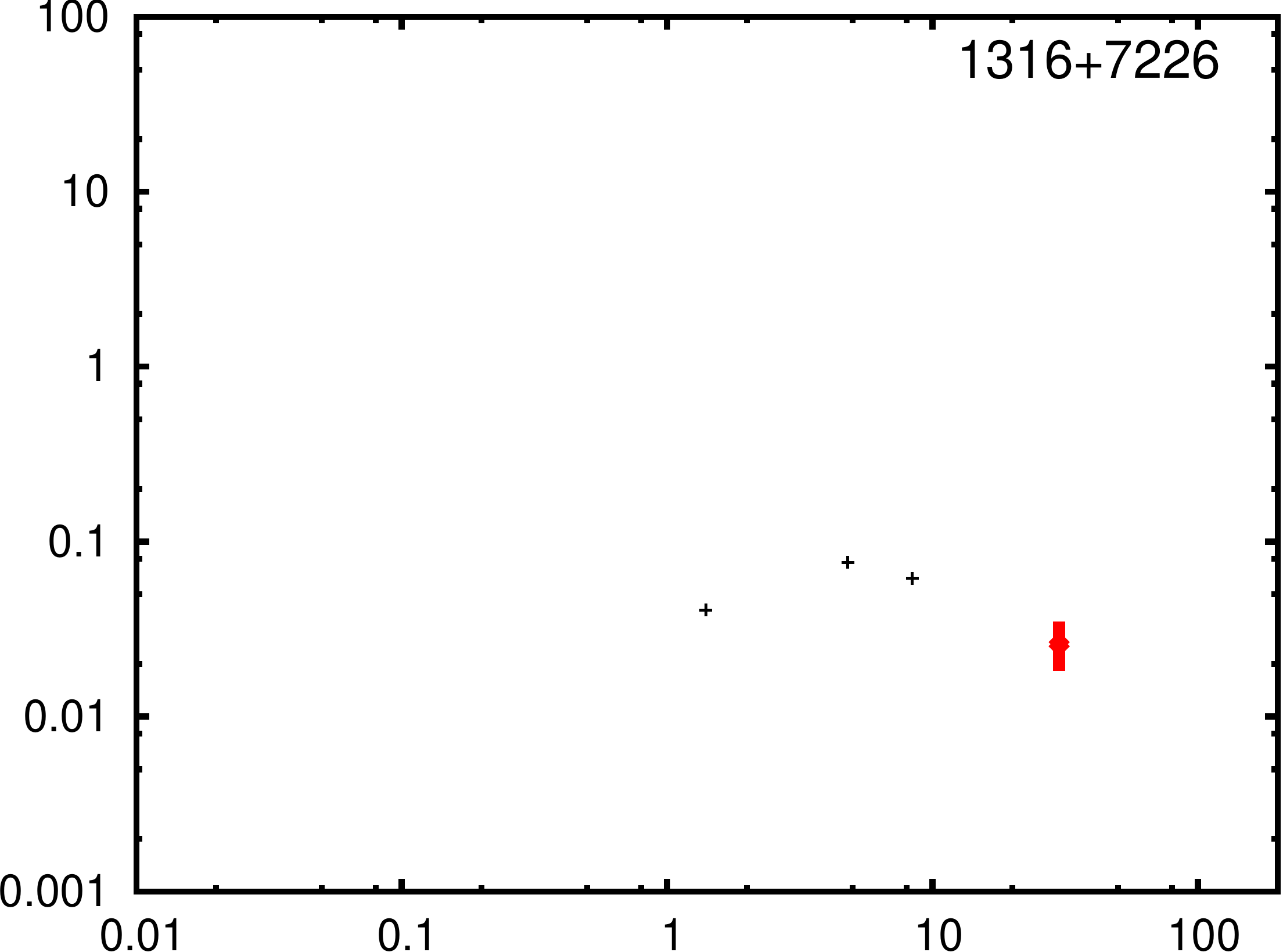}
\includegraphics[scale=0.2]{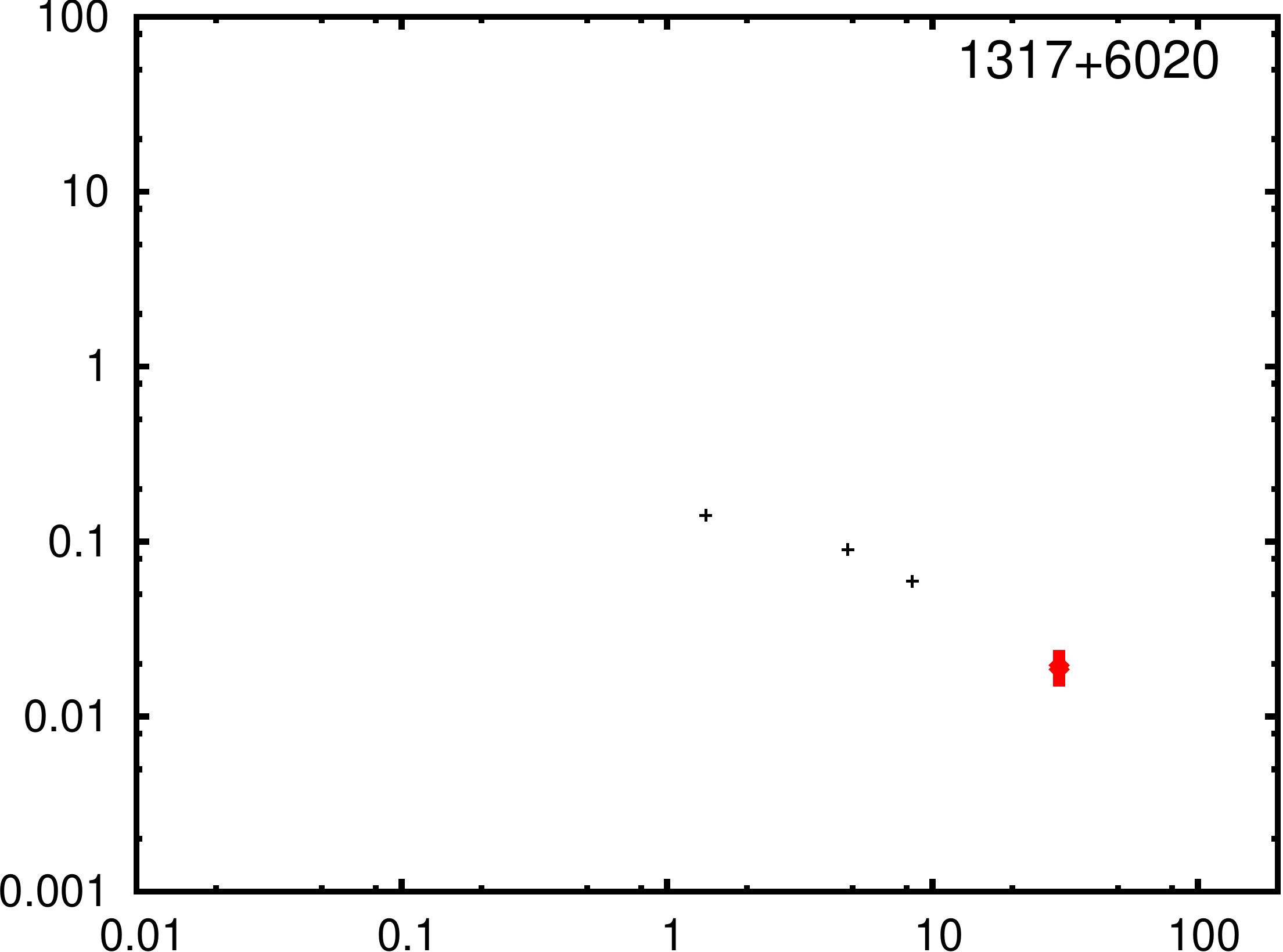}
\includegraphics[scale=0.2]{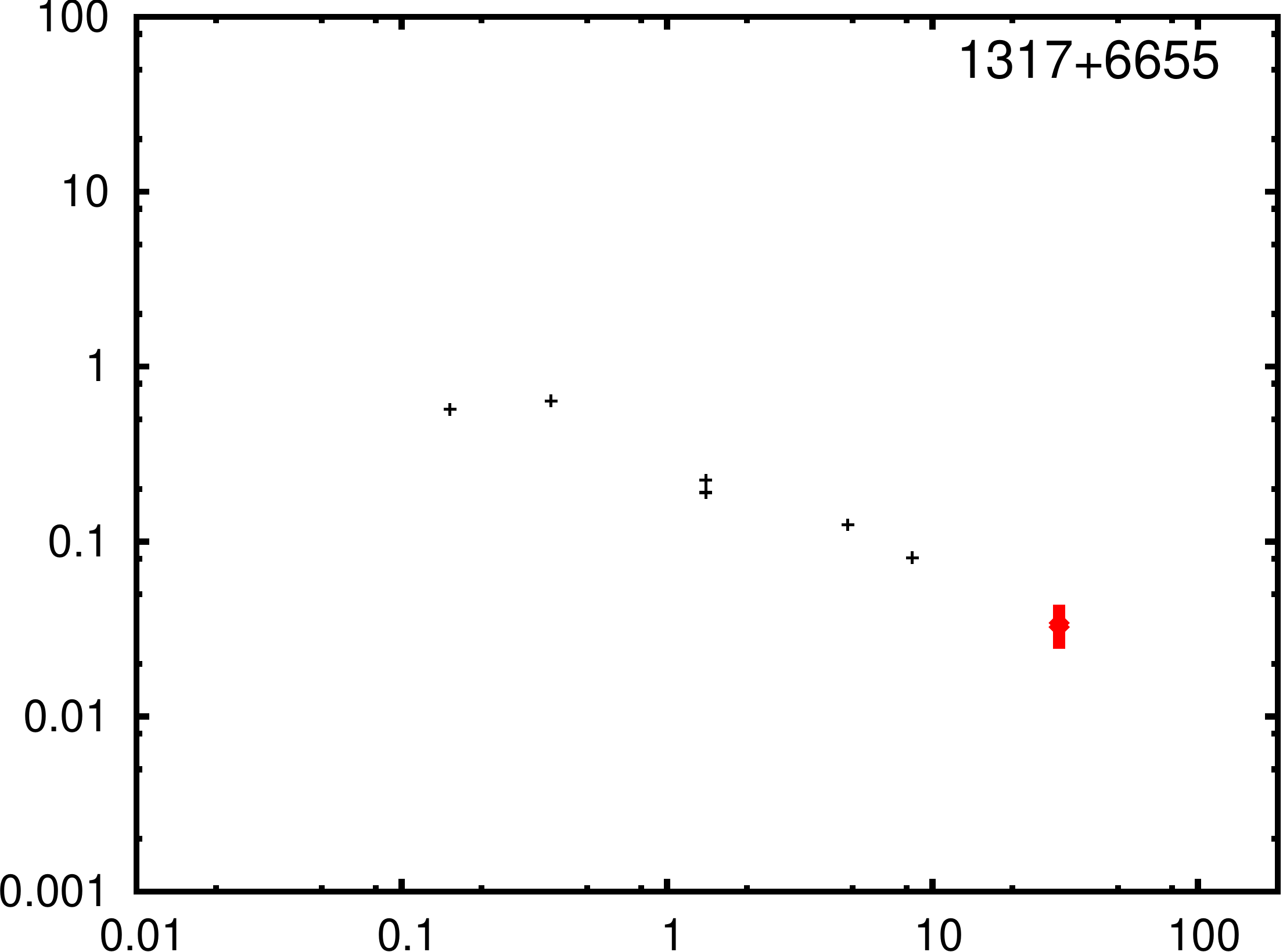}
\includegraphics[scale=0.2]{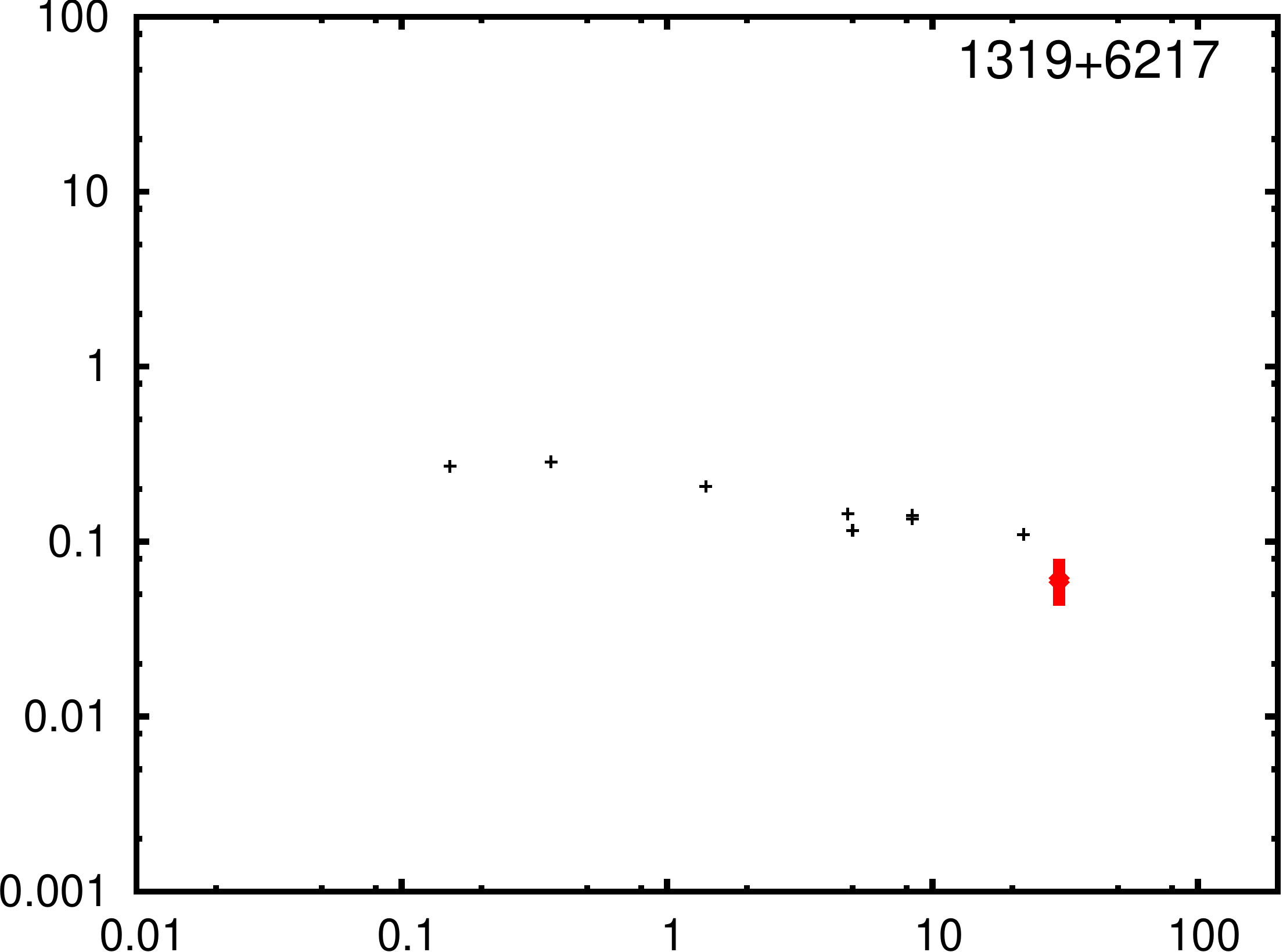}
\includegraphics[scale=0.2]{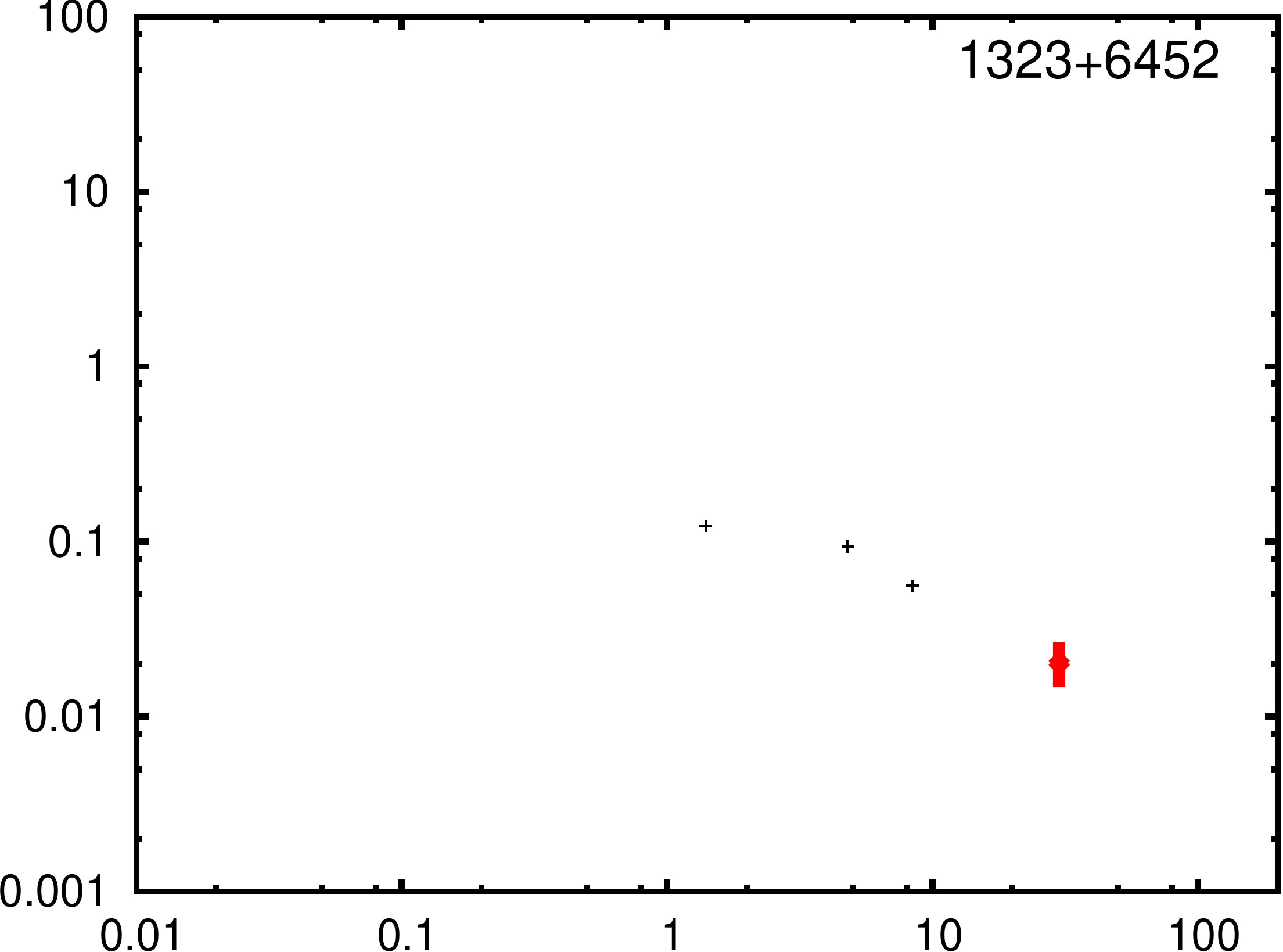}
\includegraphics[scale=0.2]{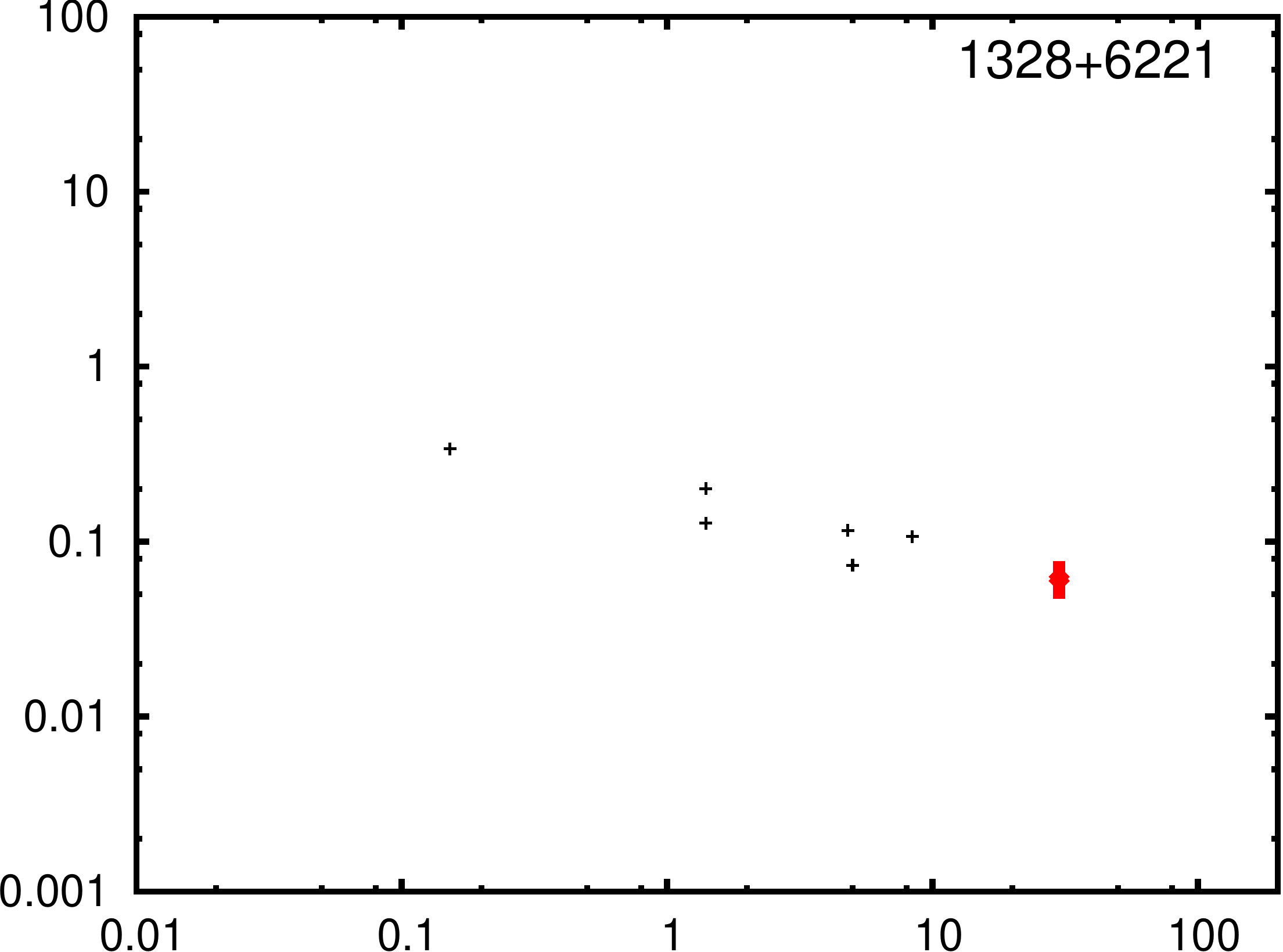}
\includegraphics[scale=0.2]{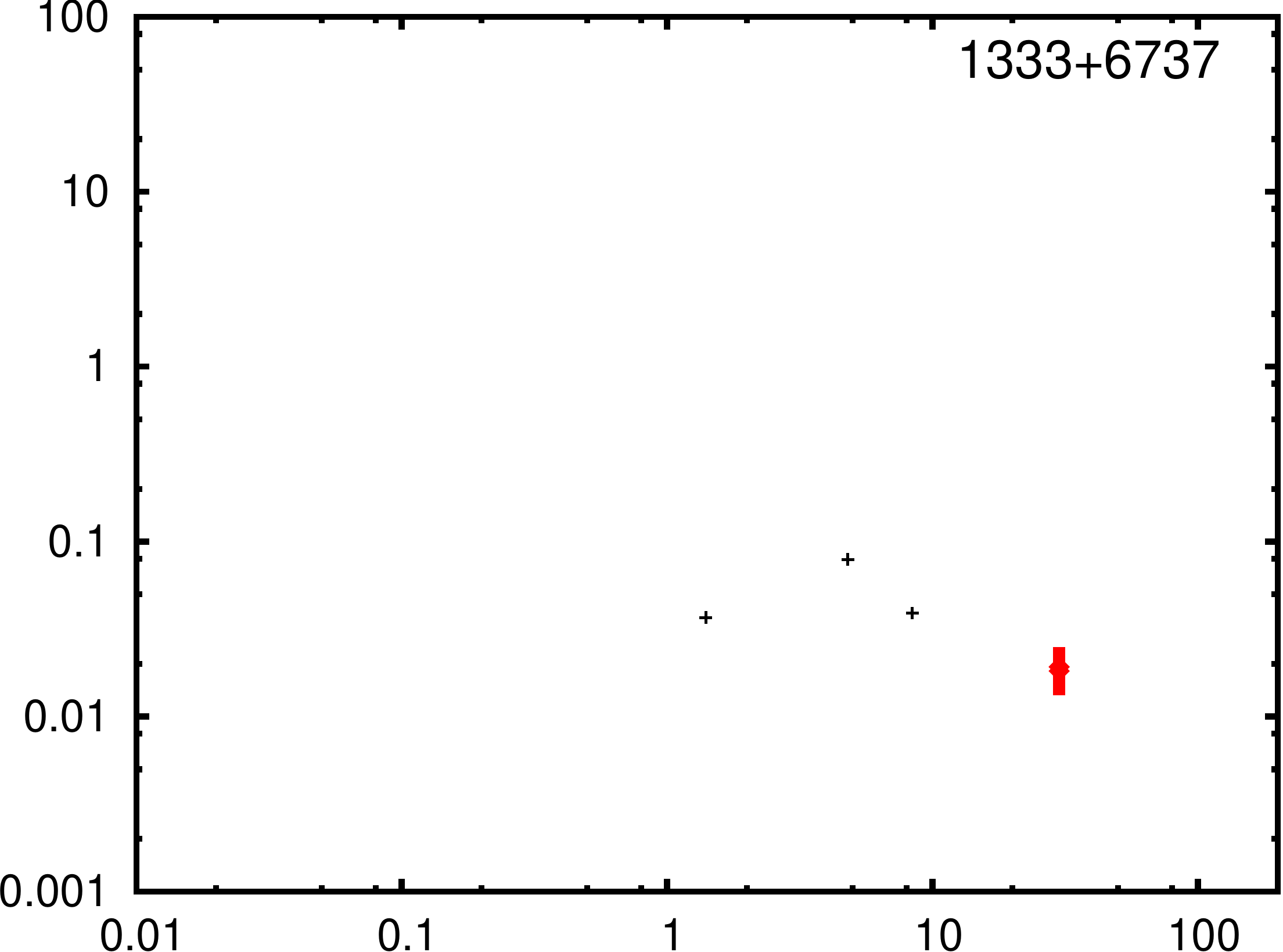}
\includegraphics[scale=0.2]{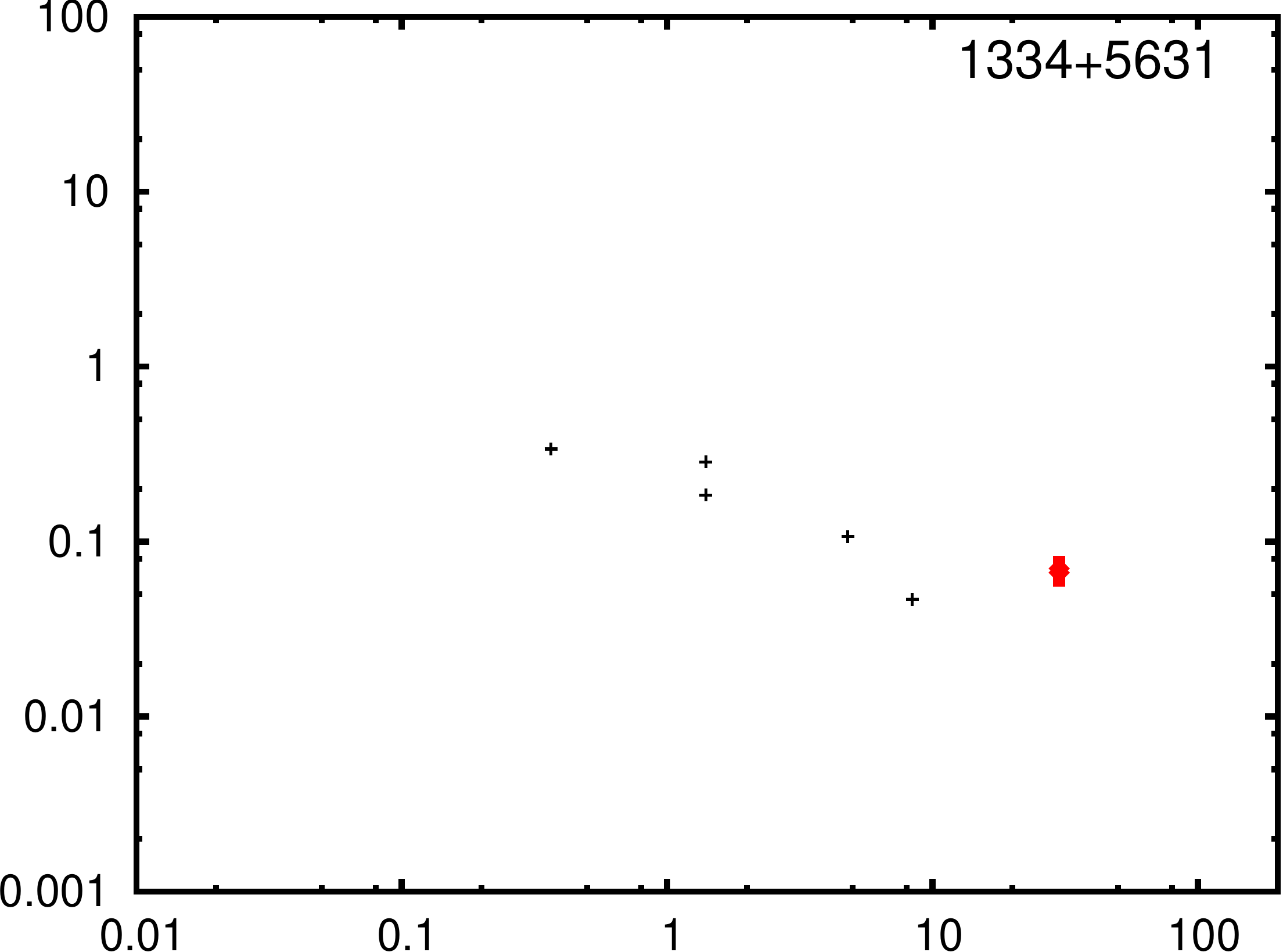}
\includegraphics[scale=0.2]{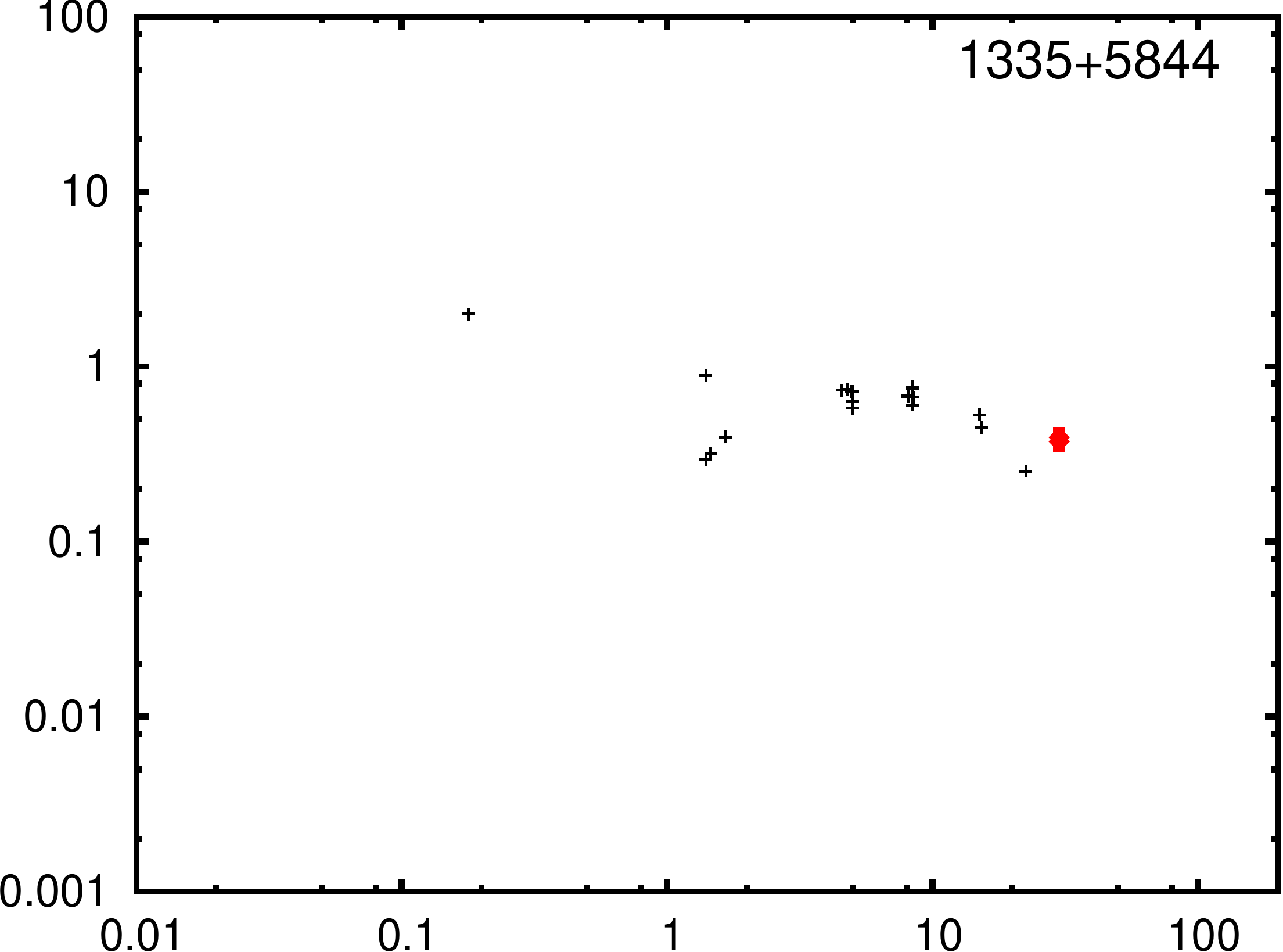}
\includegraphics[scale=0.2]{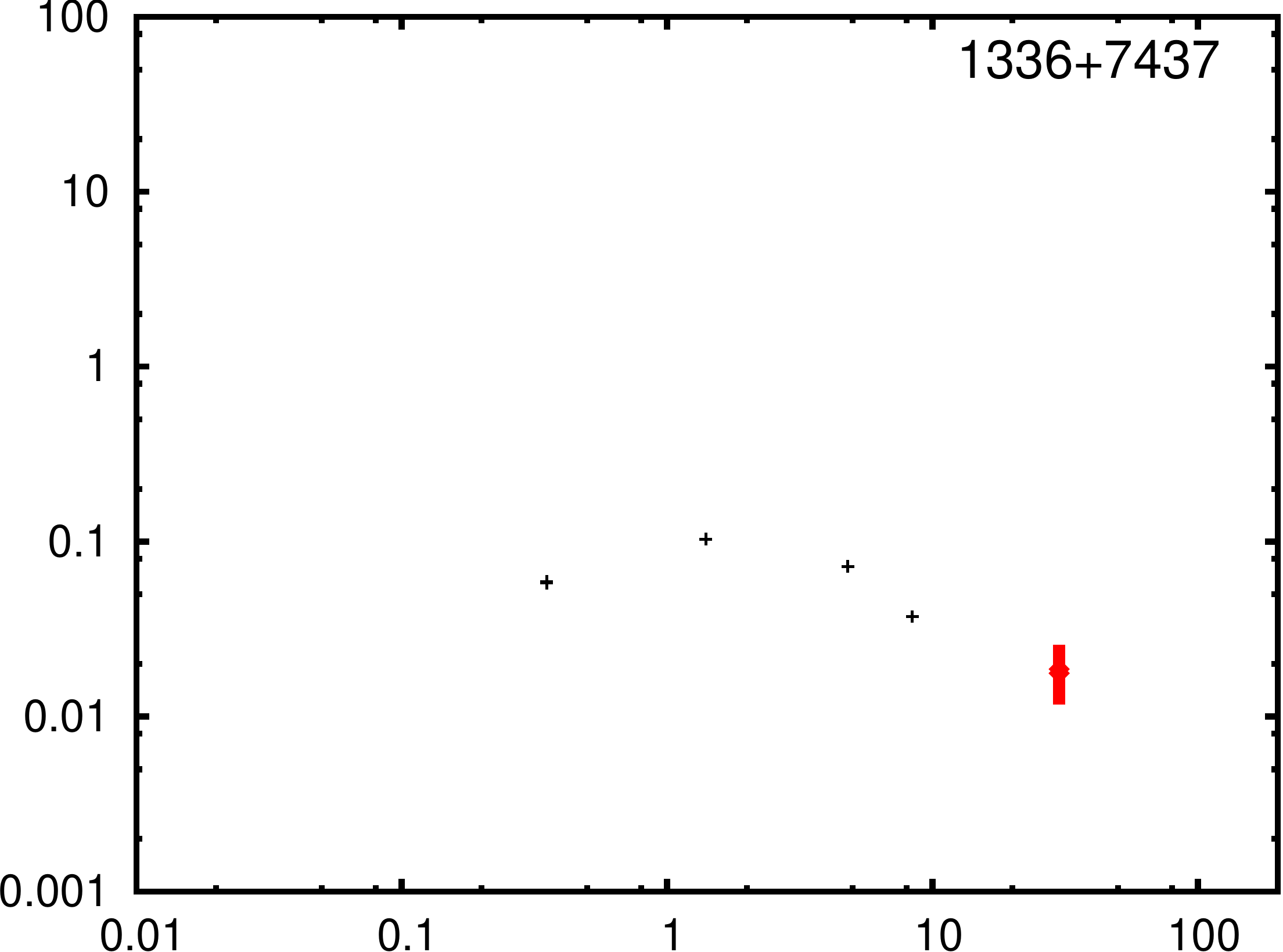}
\includegraphics[scale=0.2]{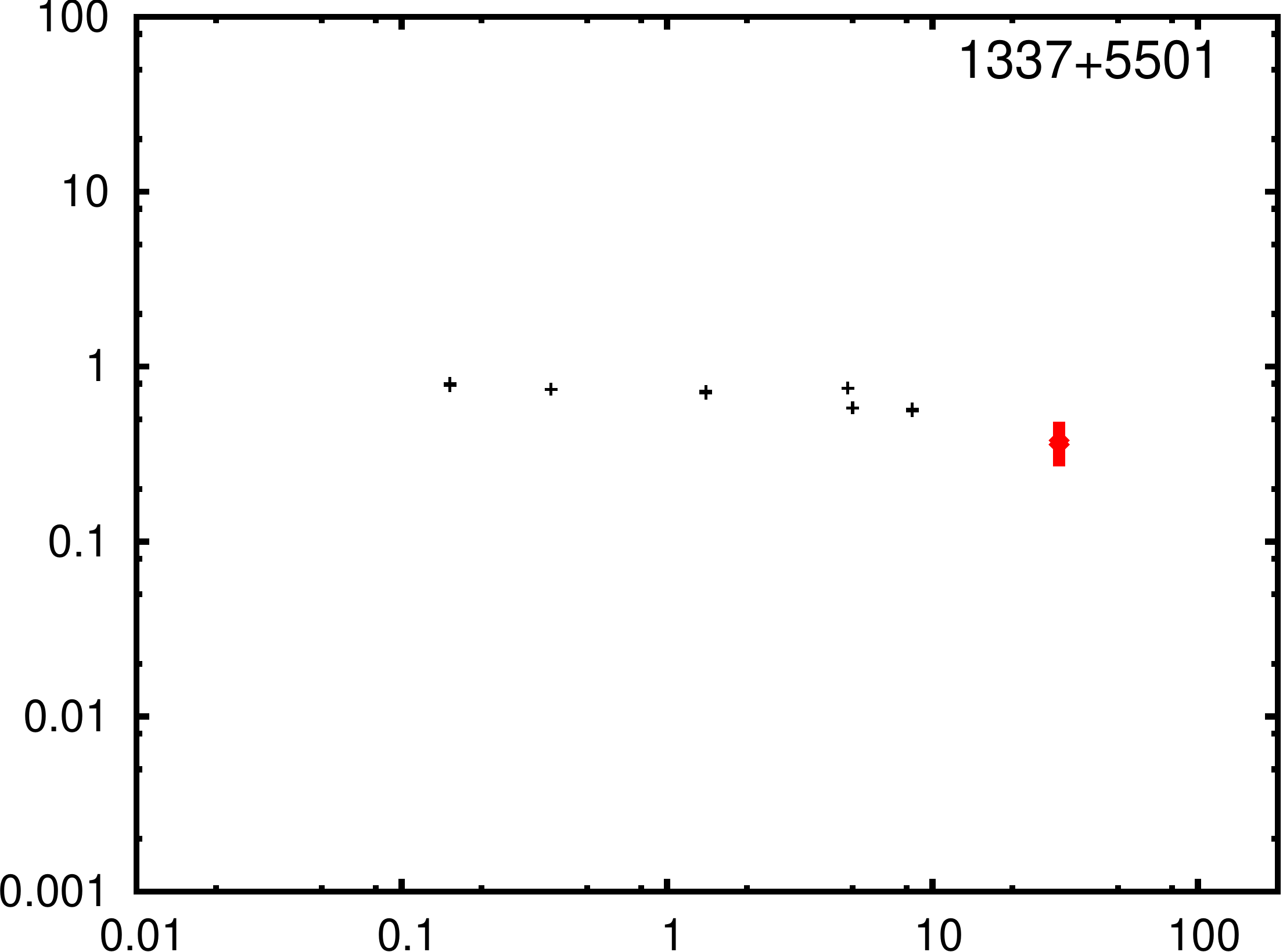}
\includegraphics[scale=0.2]{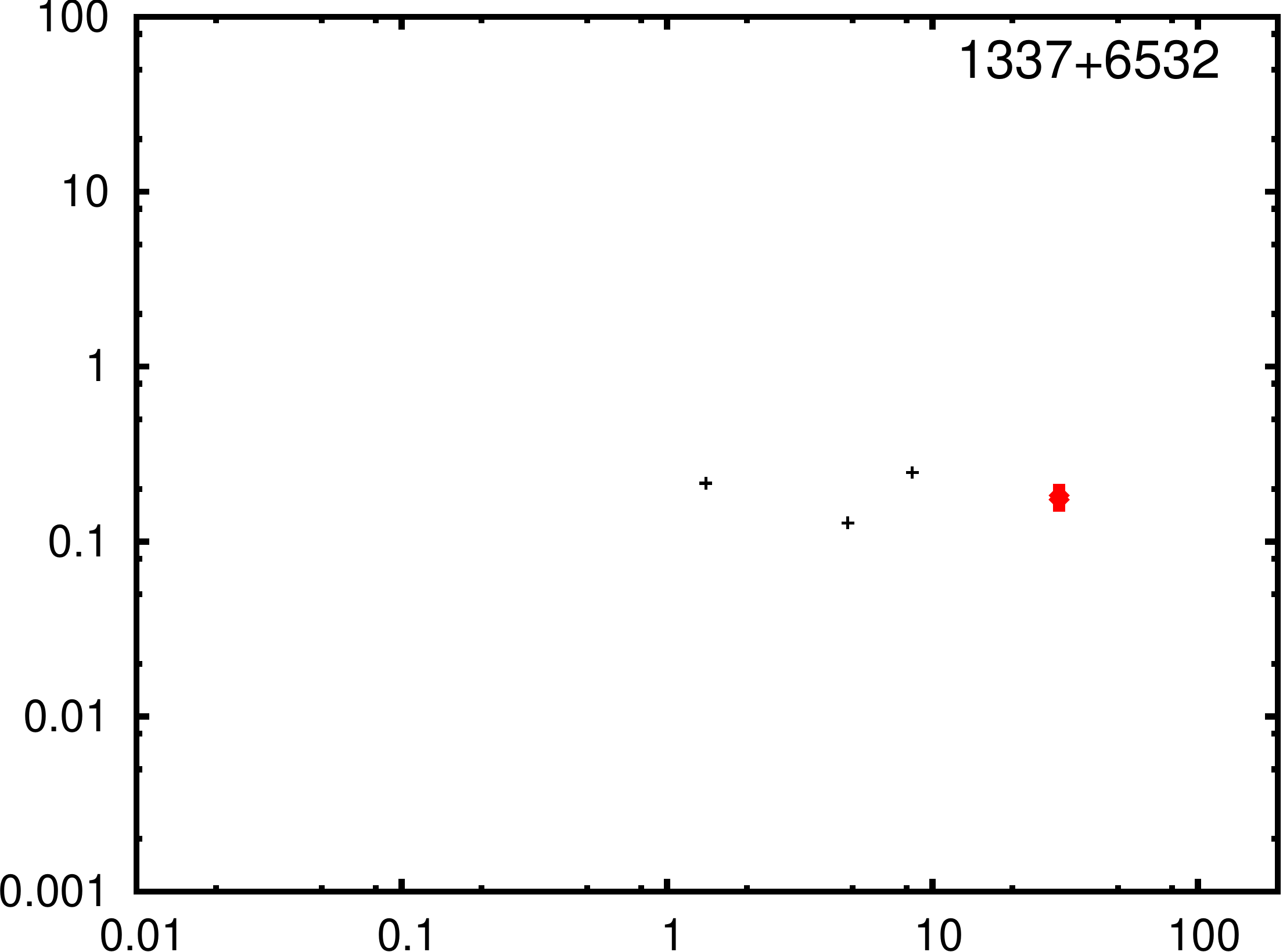}
\includegraphics[scale=0.2]{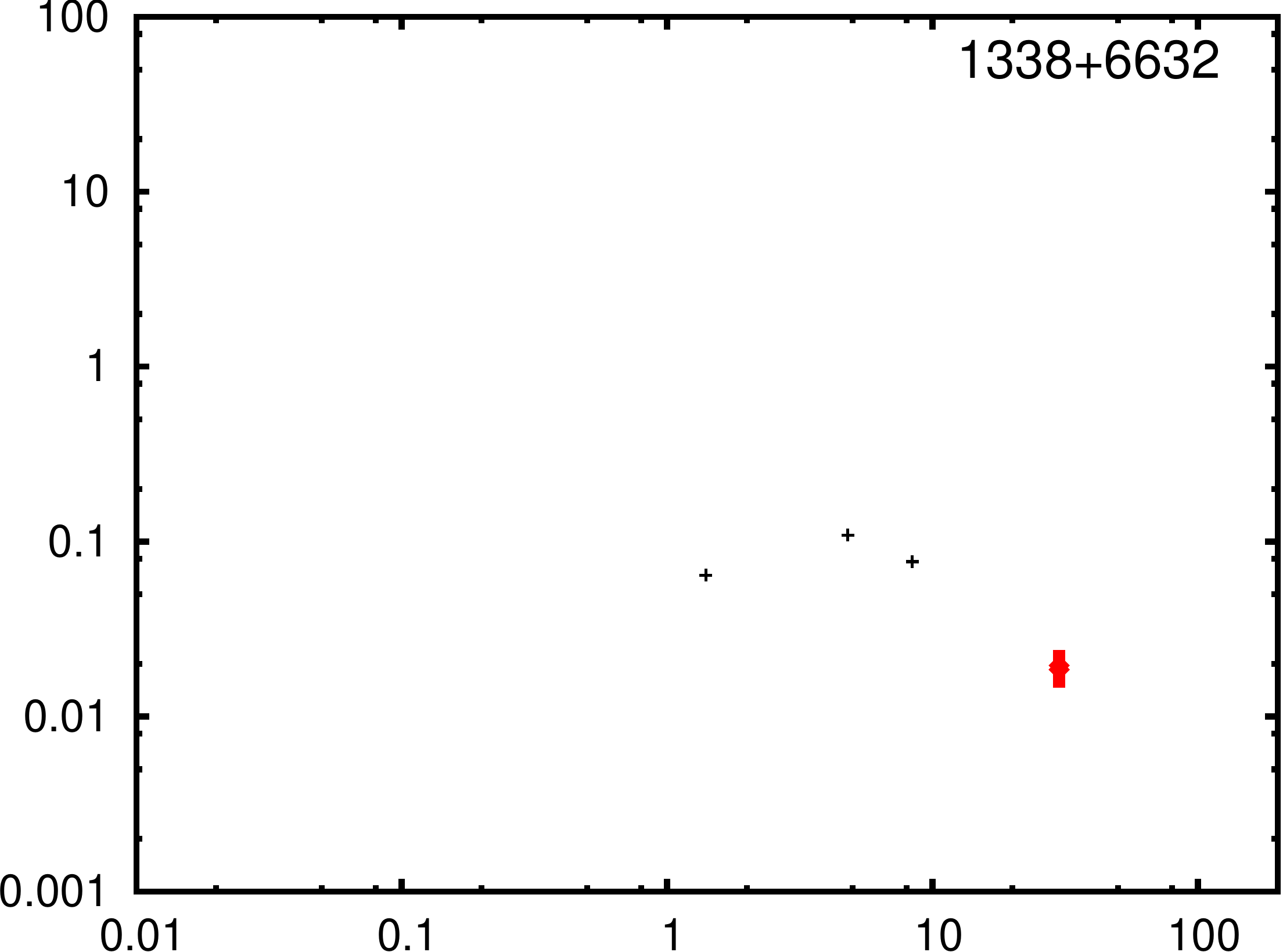}
\includegraphics[scale=0.2]{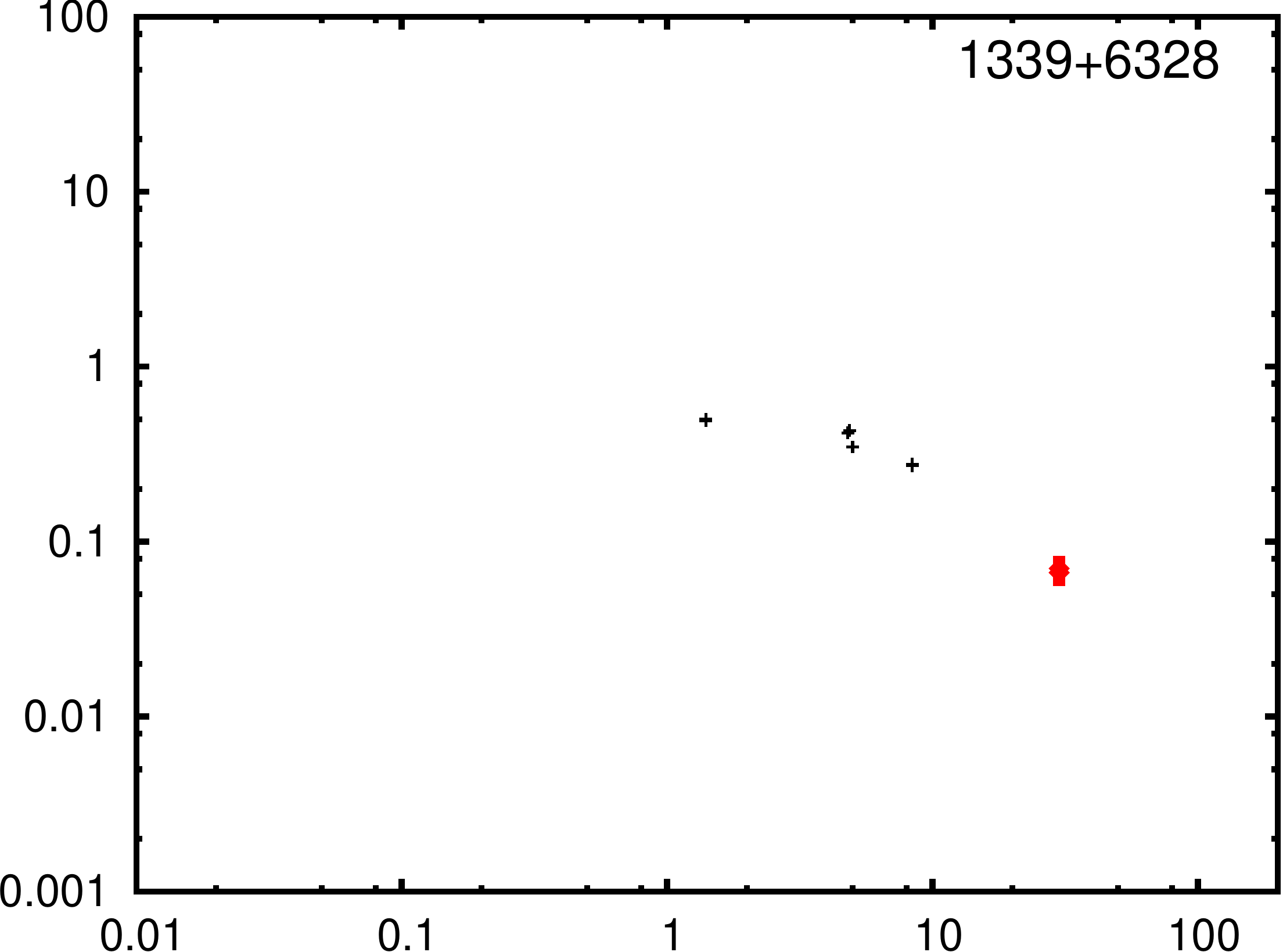}
\includegraphics[scale=0.2]{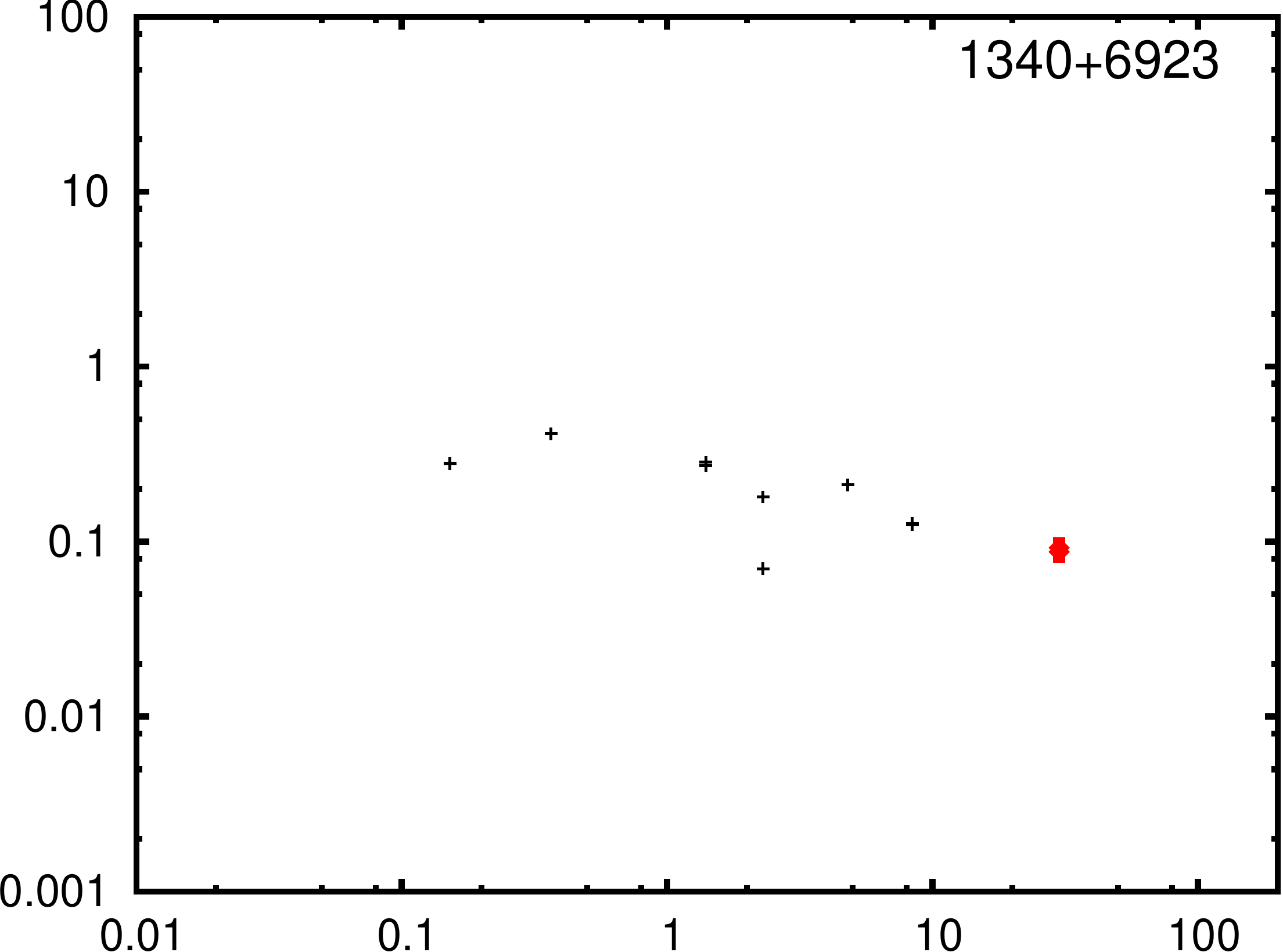}
\includegraphics[scale=0.2]{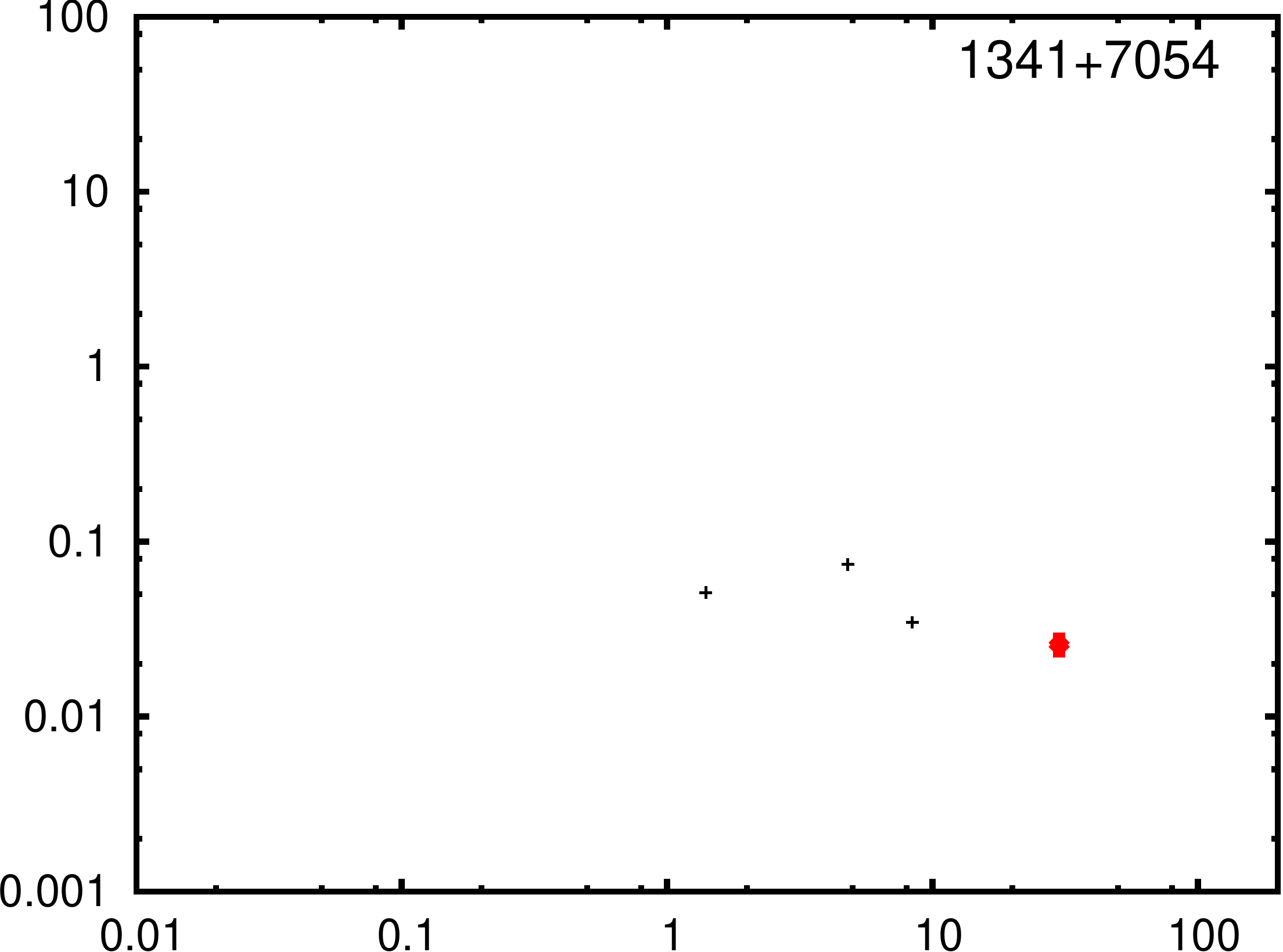}
\end{figure}
\clearpage\begin{figure}
\centering
\includegraphics[scale=0.2]{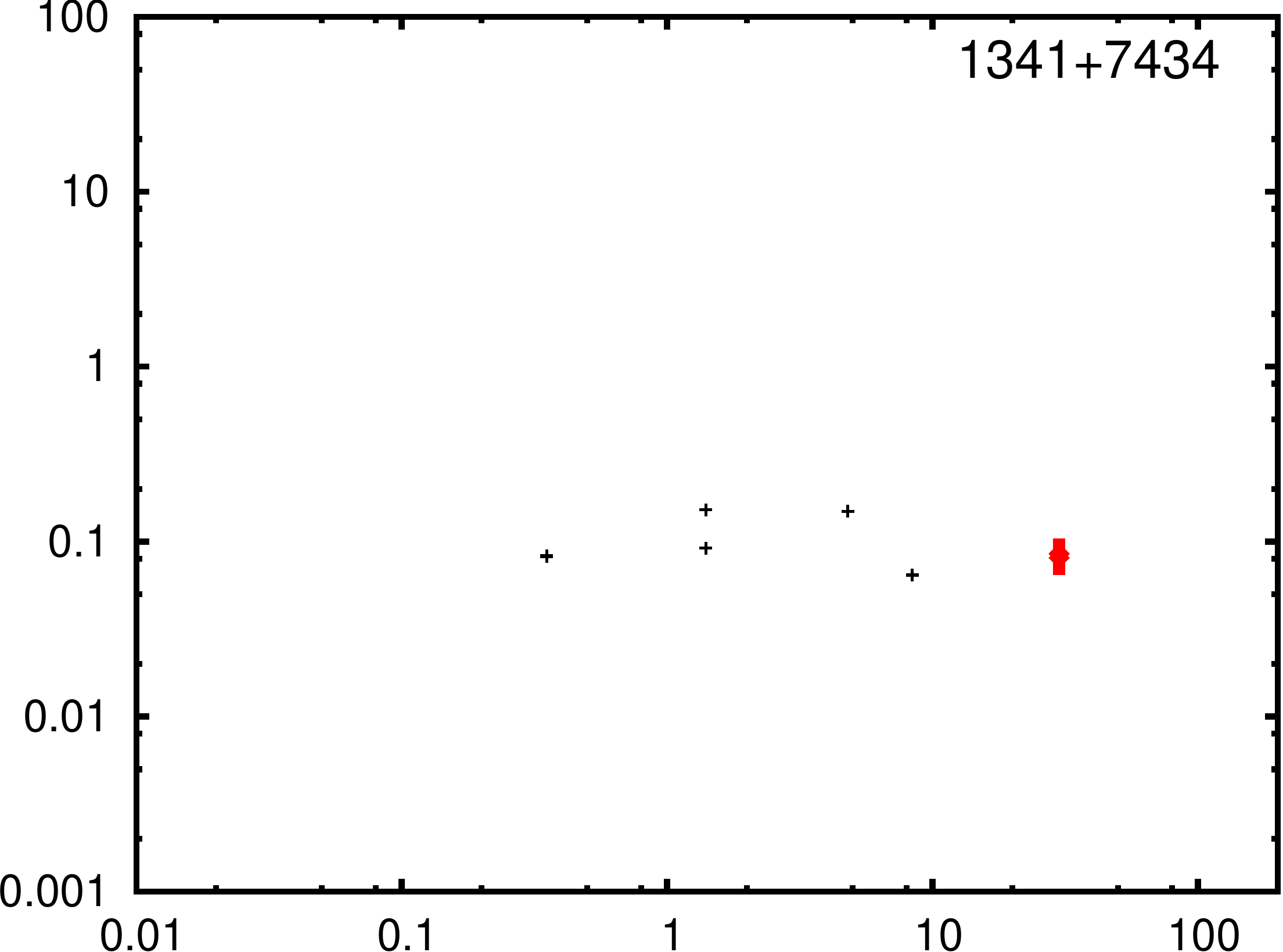}
\includegraphics[scale=0.2]{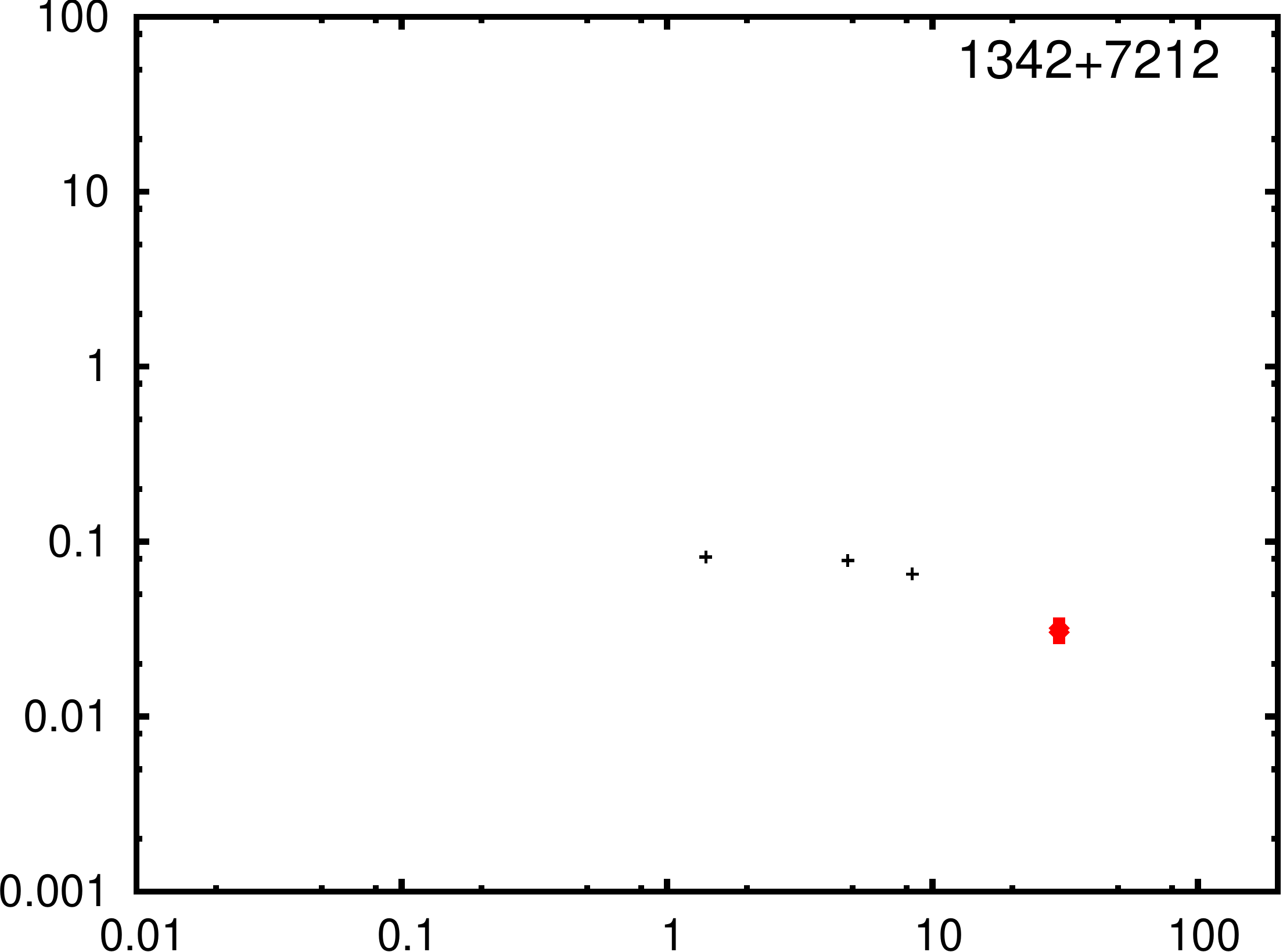}
\includegraphics[scale=0.2]{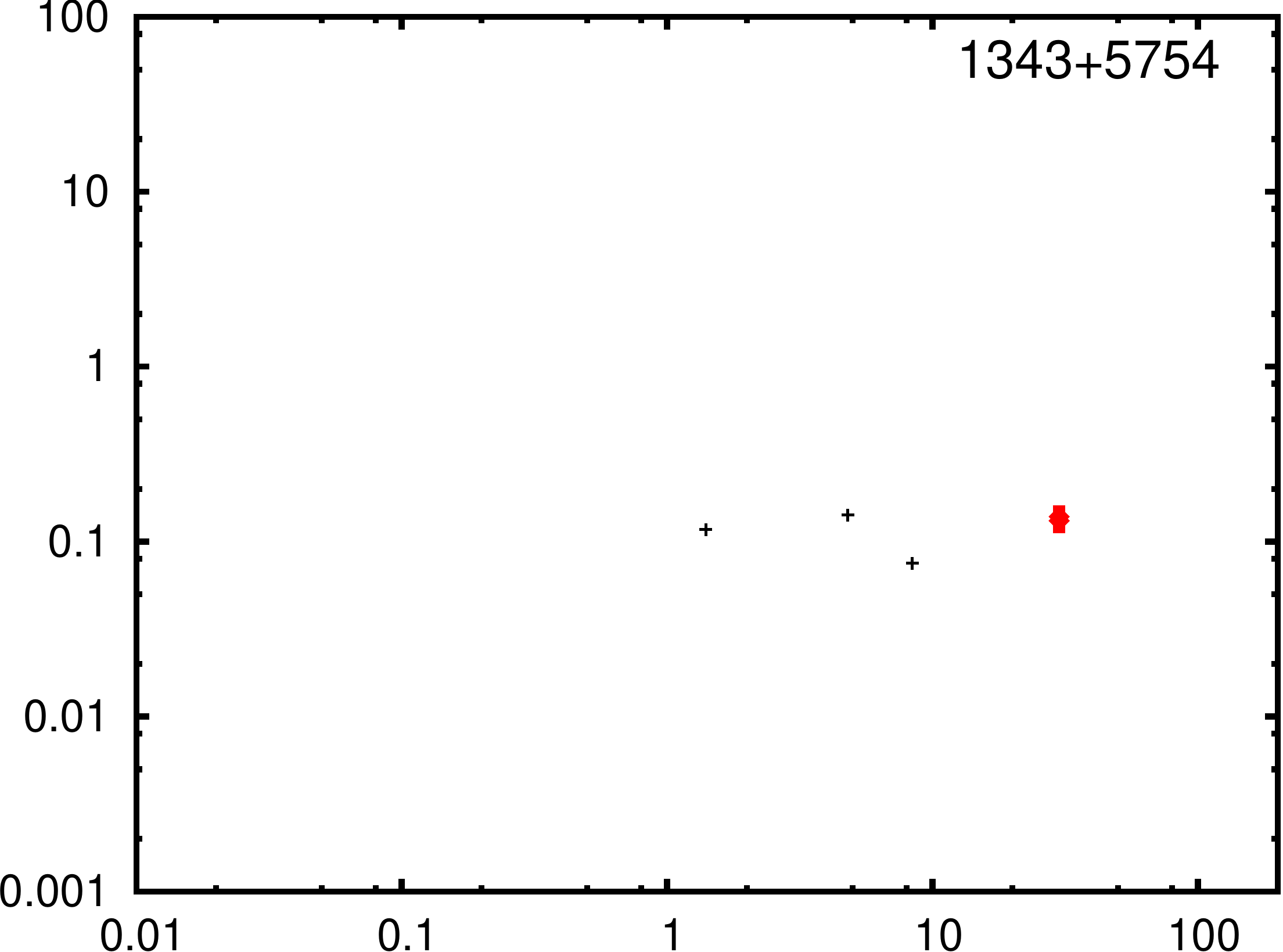}
\includegraphics[scale=0.2]{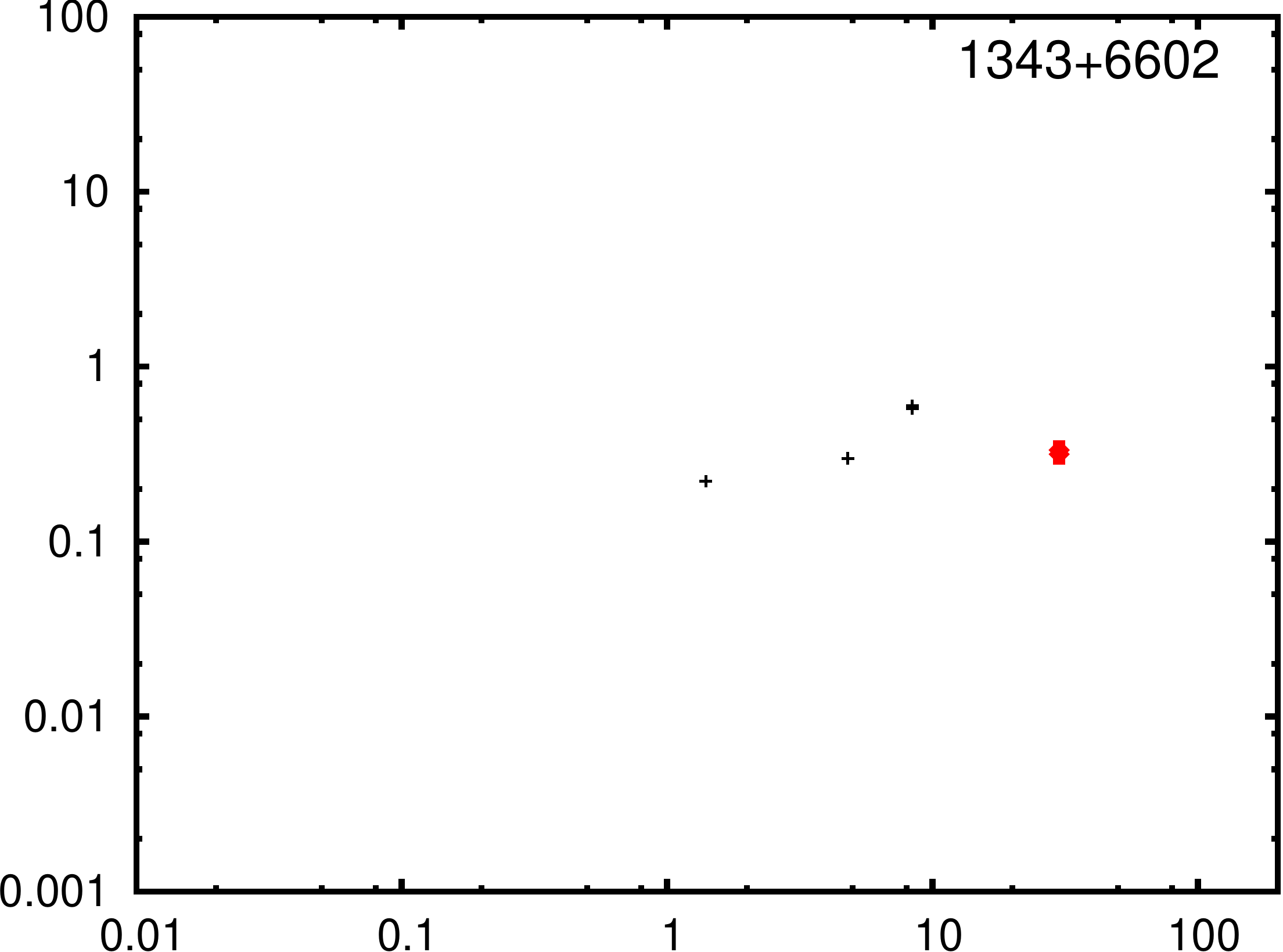}
\includegraphics[scale=0.2]{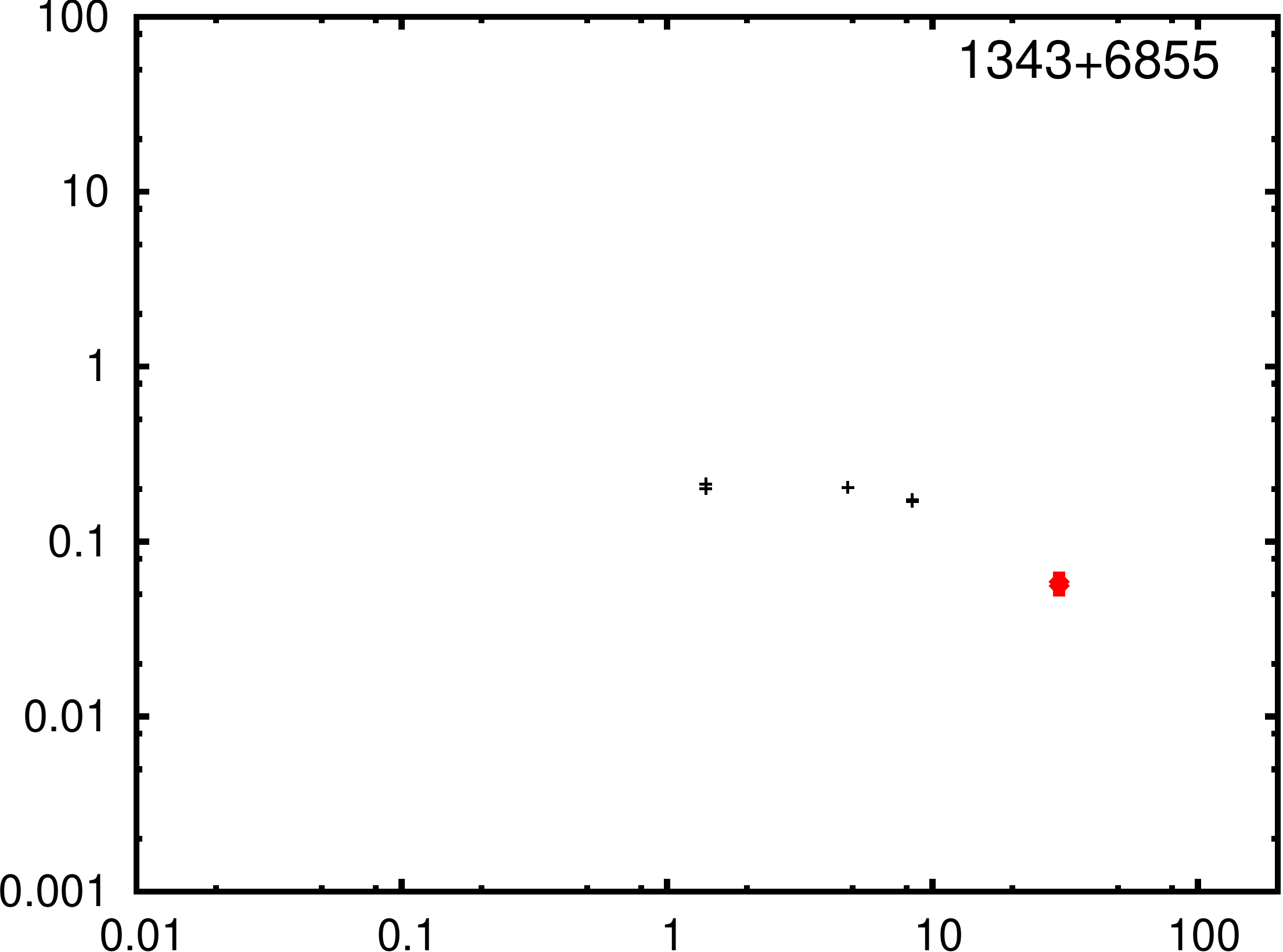}
\includegraphics[scale=0.2]{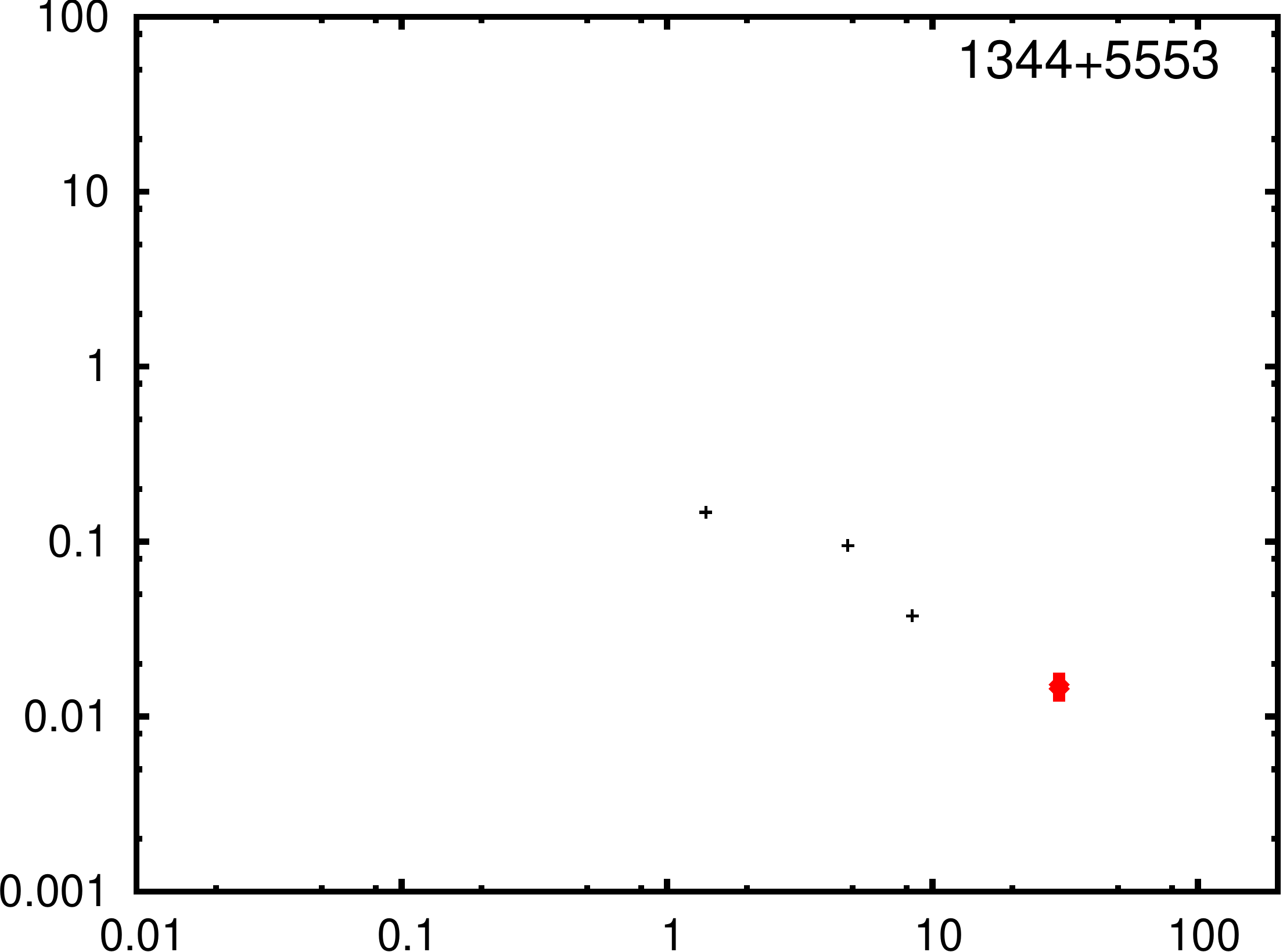}
\includegraphics[scale=0.2]{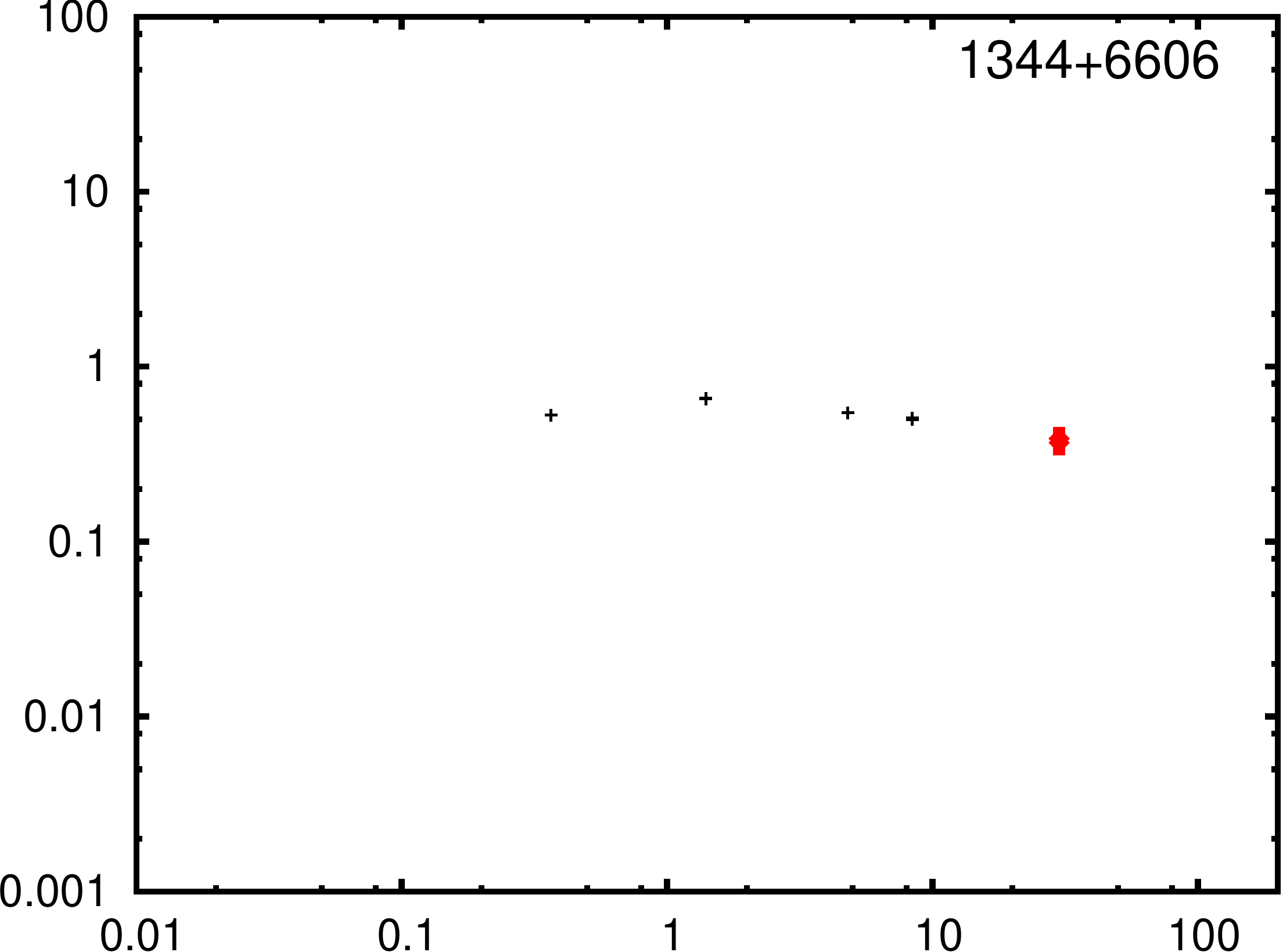}
\includegraphics[scale=0.2]{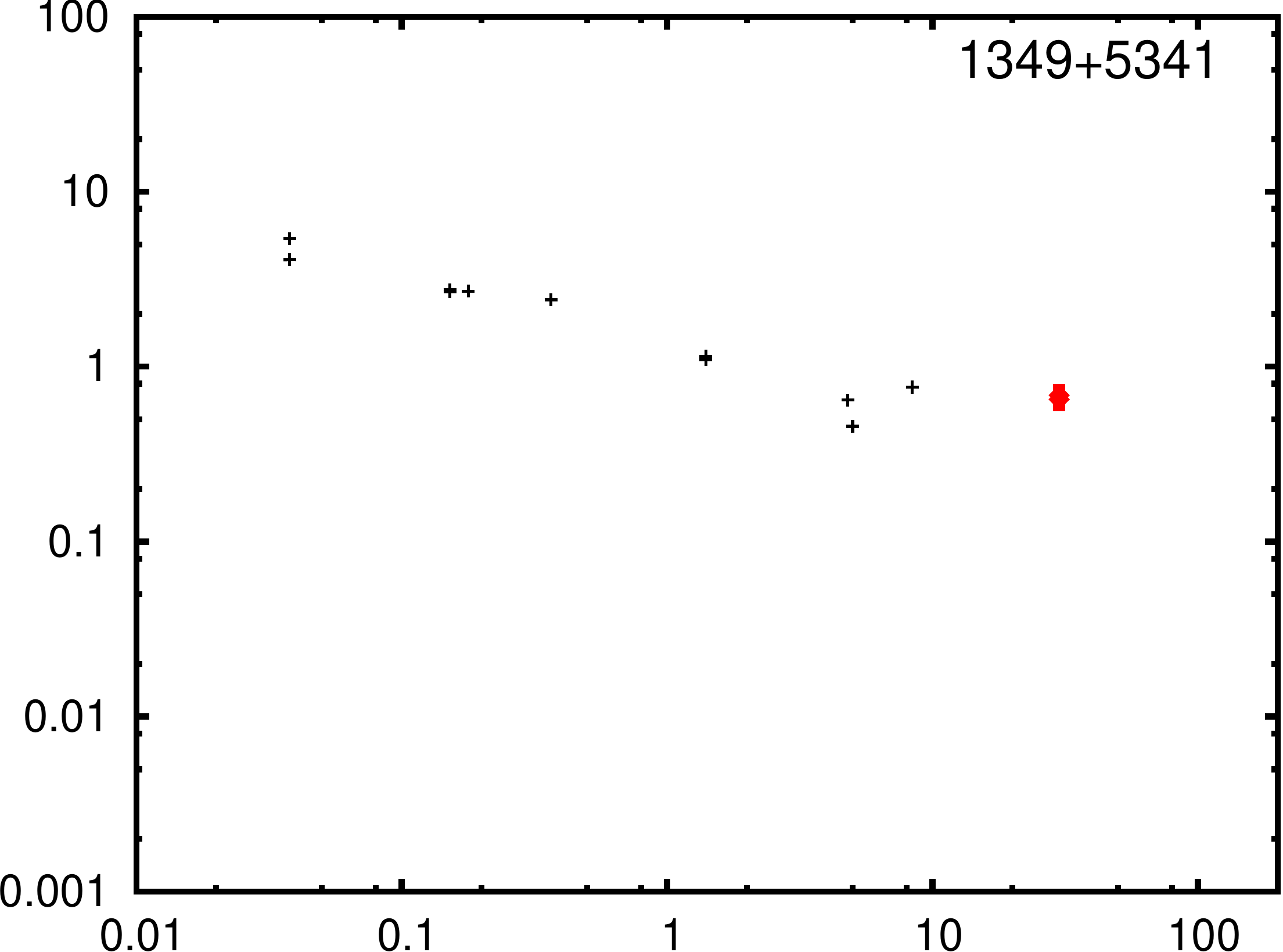}
\includegraphics[scale=0.2]{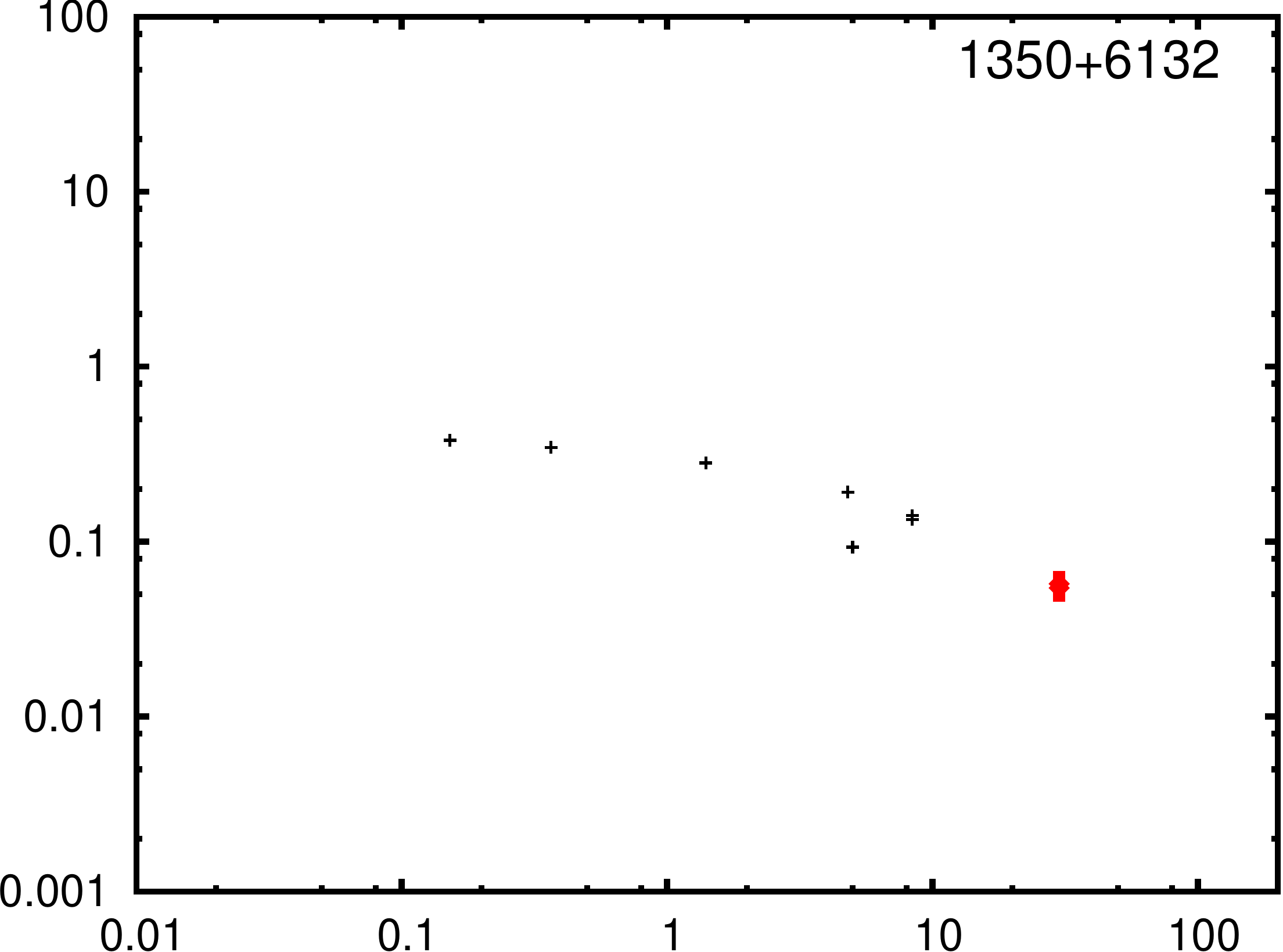}
\includegraphics[scale=0.2]{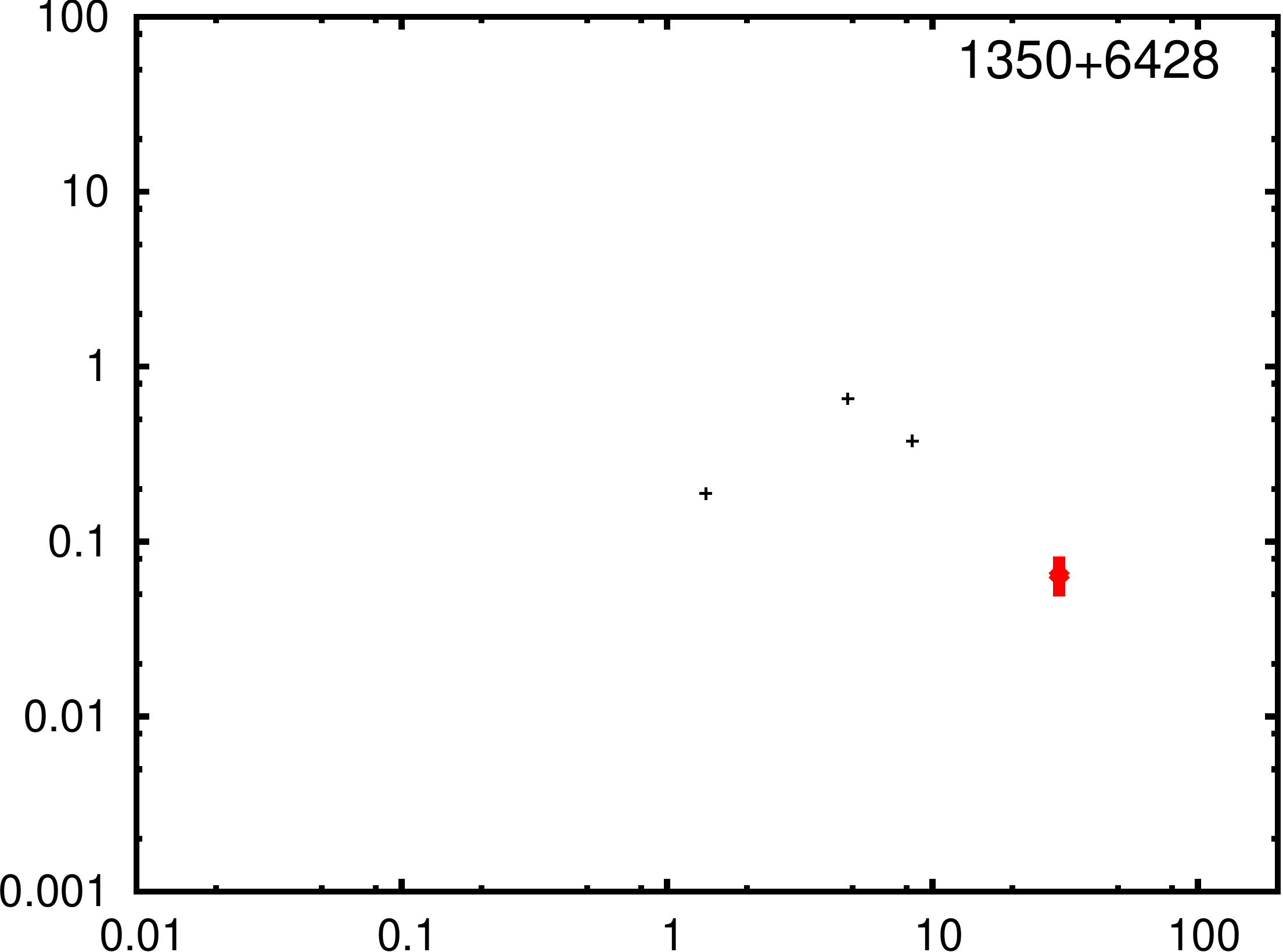}
\includegraphics[scale=0.2]{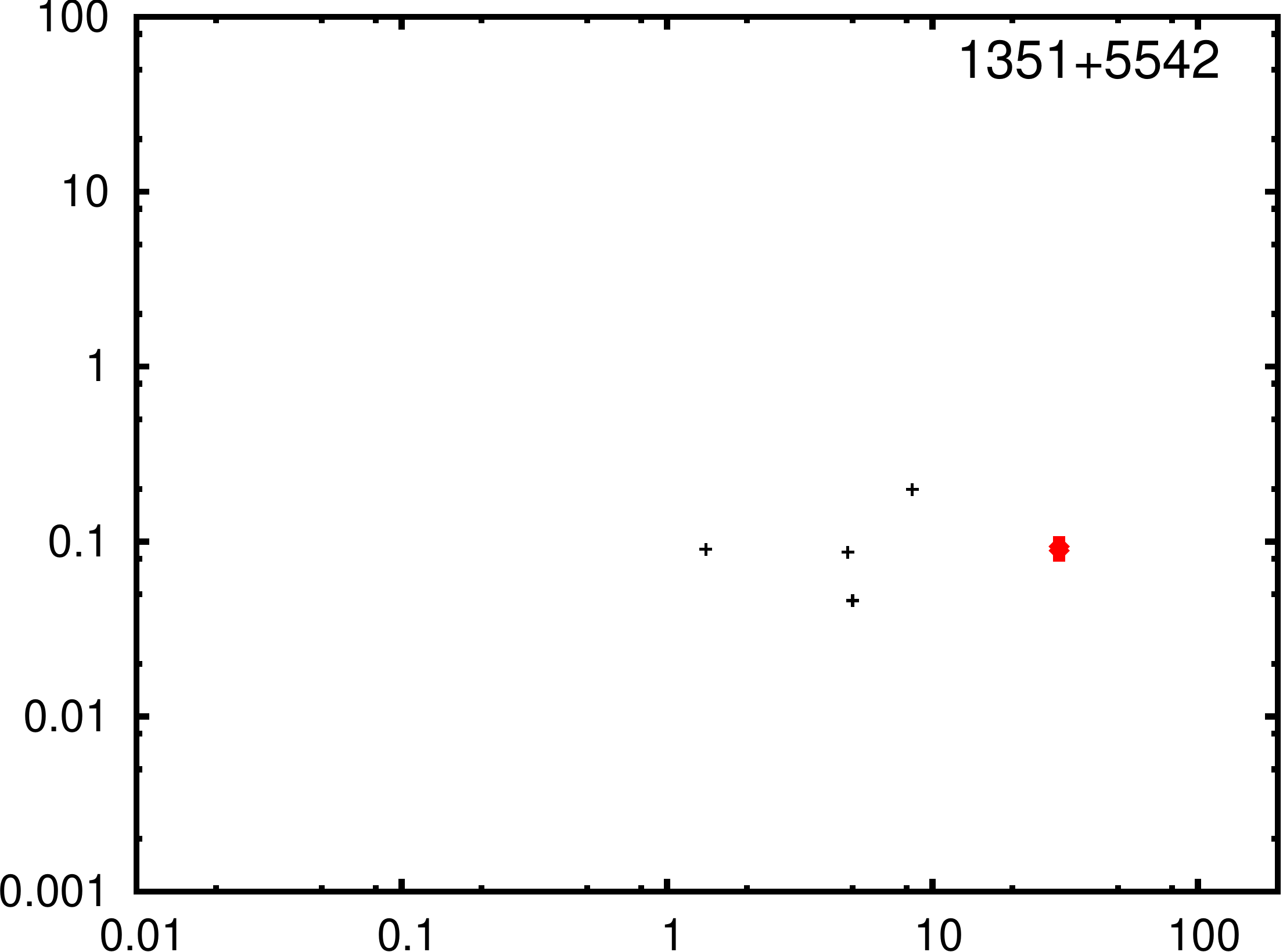}
\includegraphics[scale=0.2]{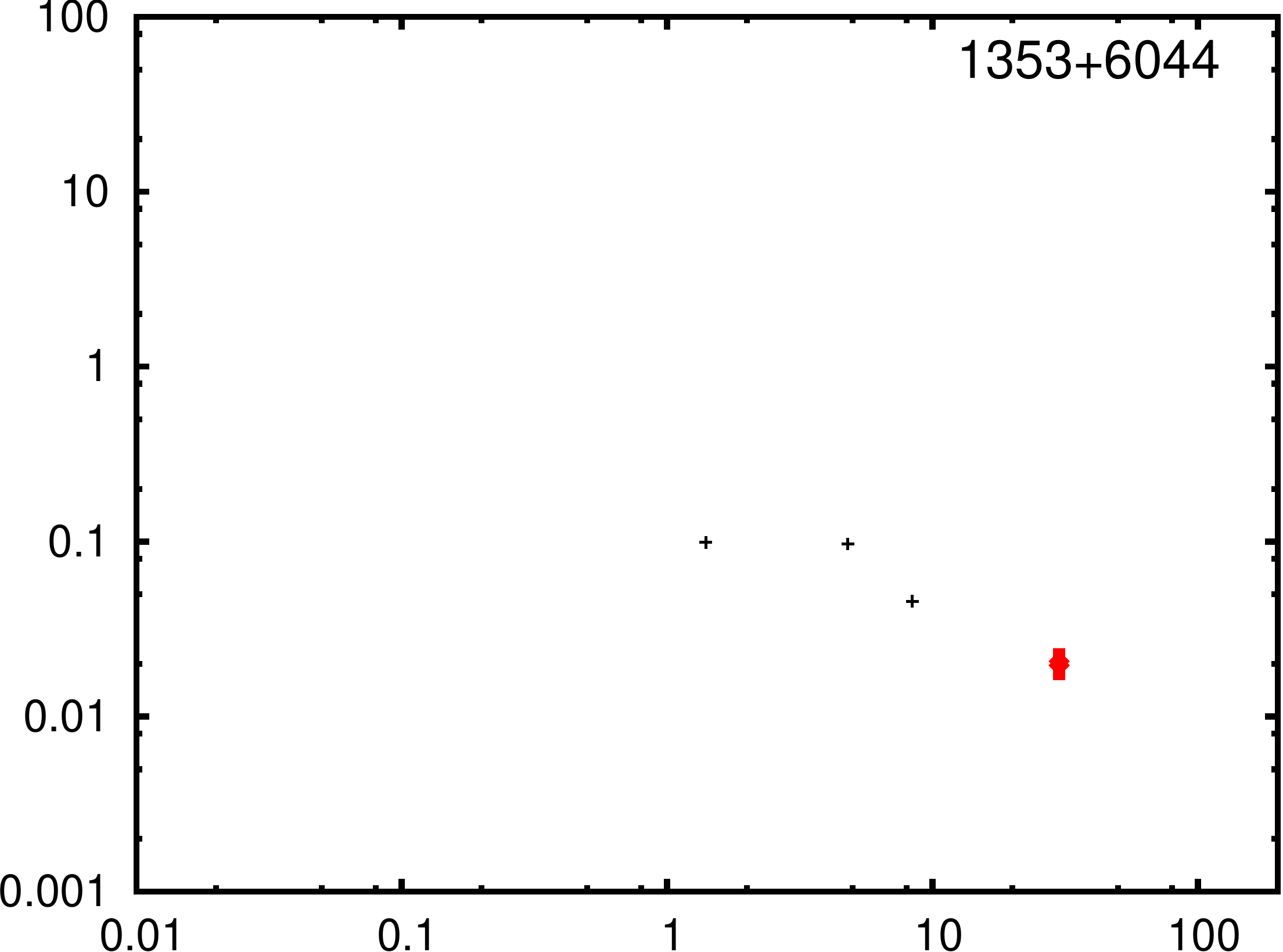}
\includegraphics[scale=0.2]{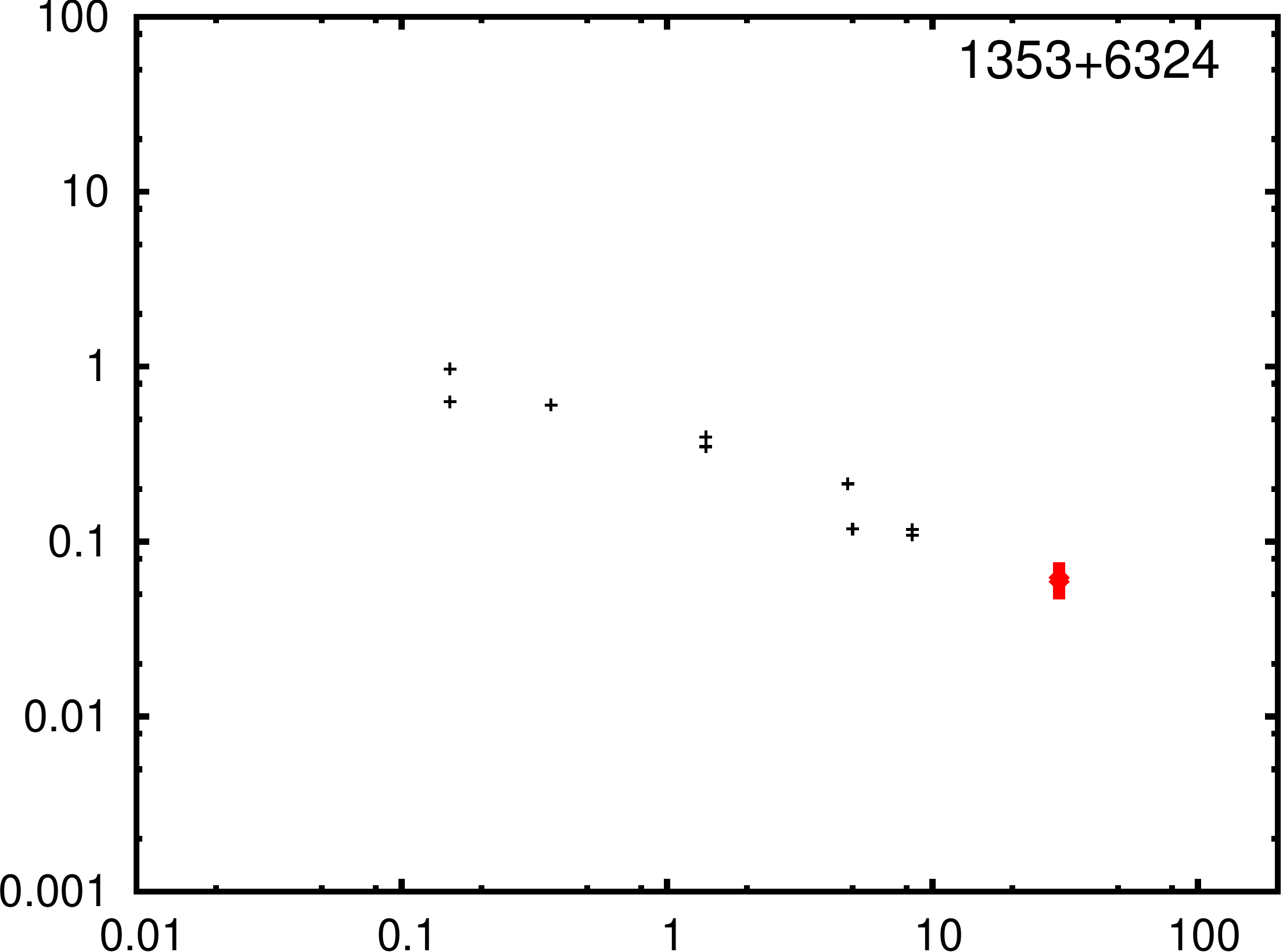}
\includegraphics[scale=0.2]{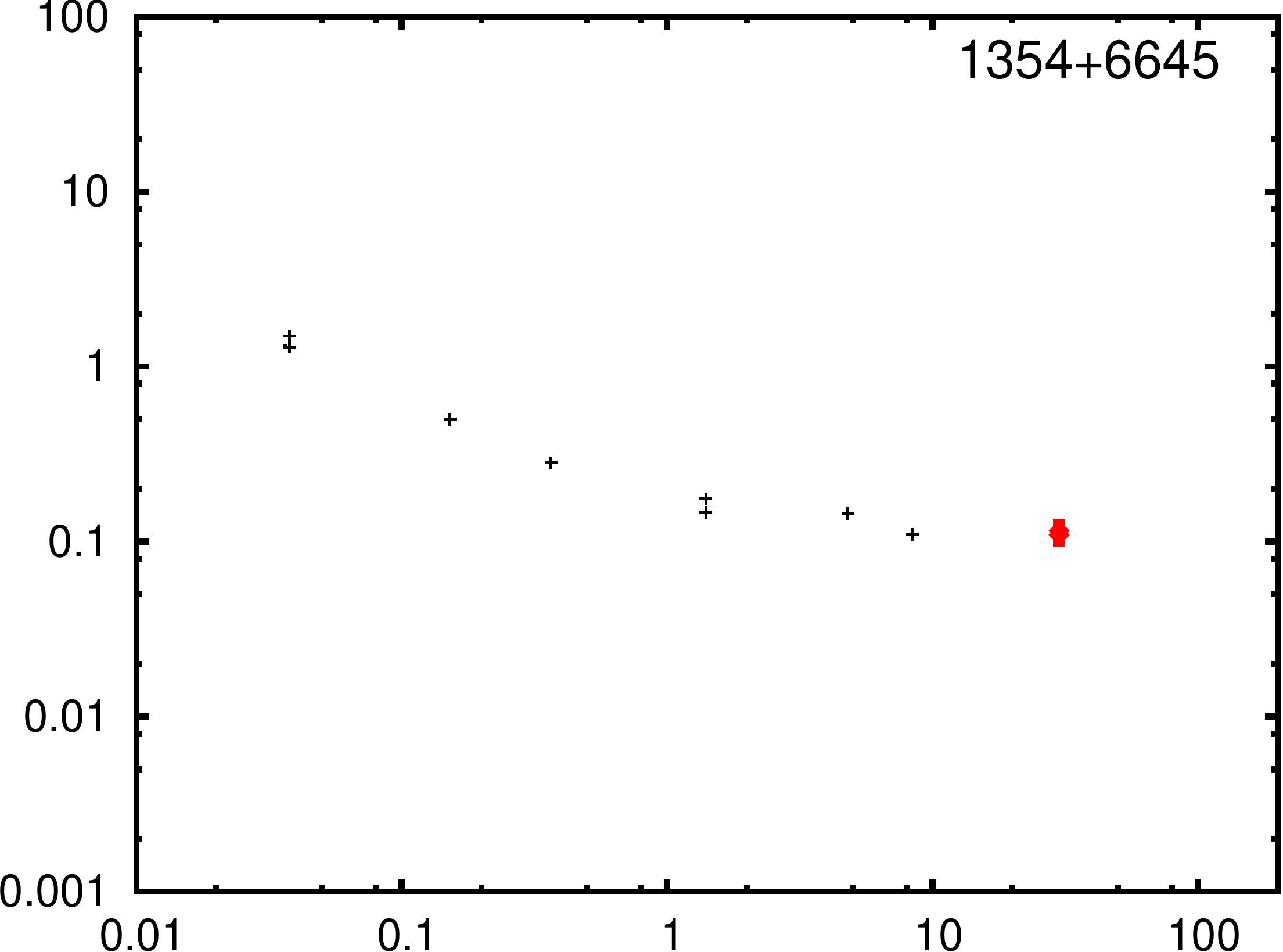}
\includegraphics[scale=0.2]{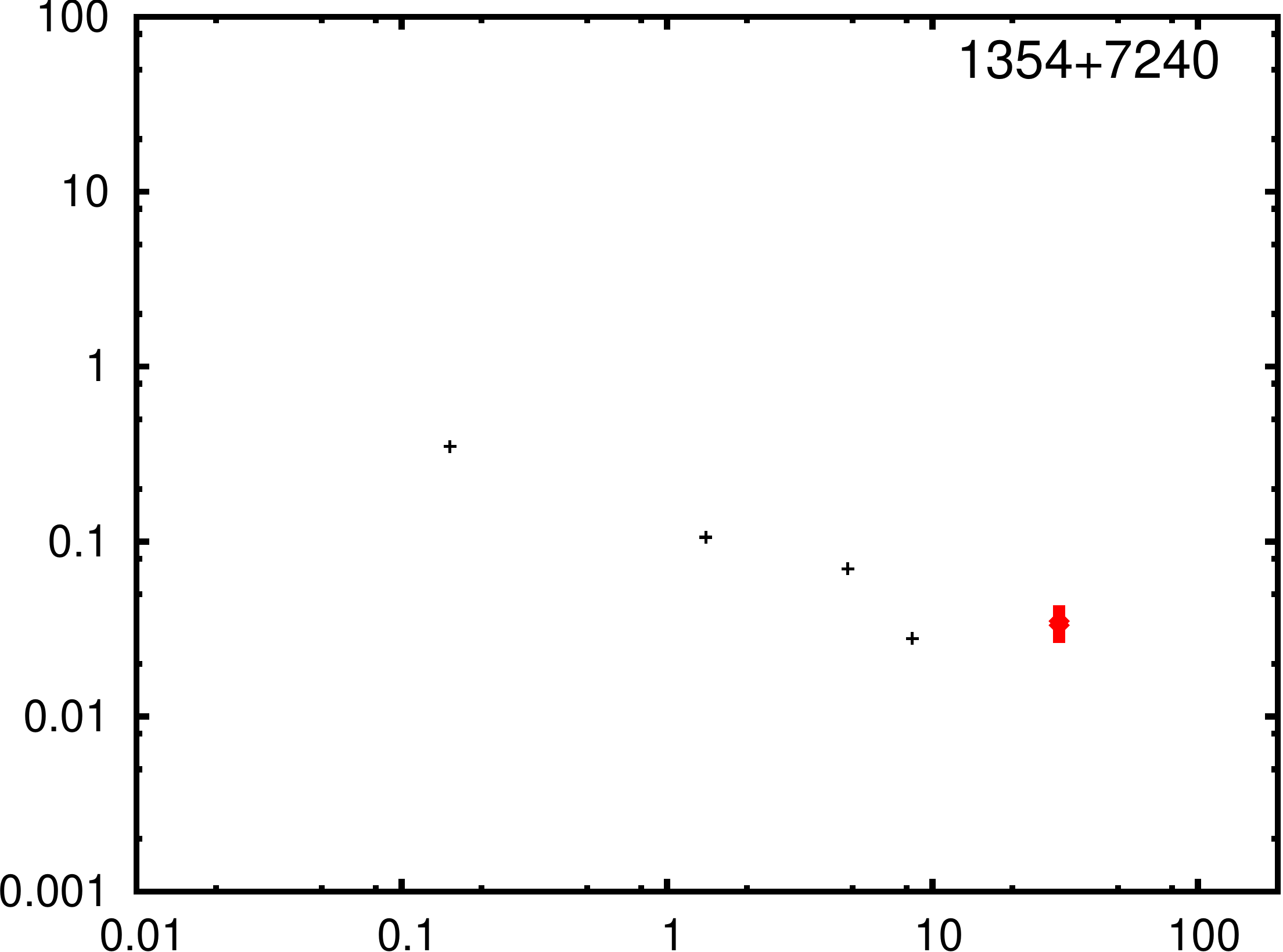}
\includegraphics[scale=0.2]{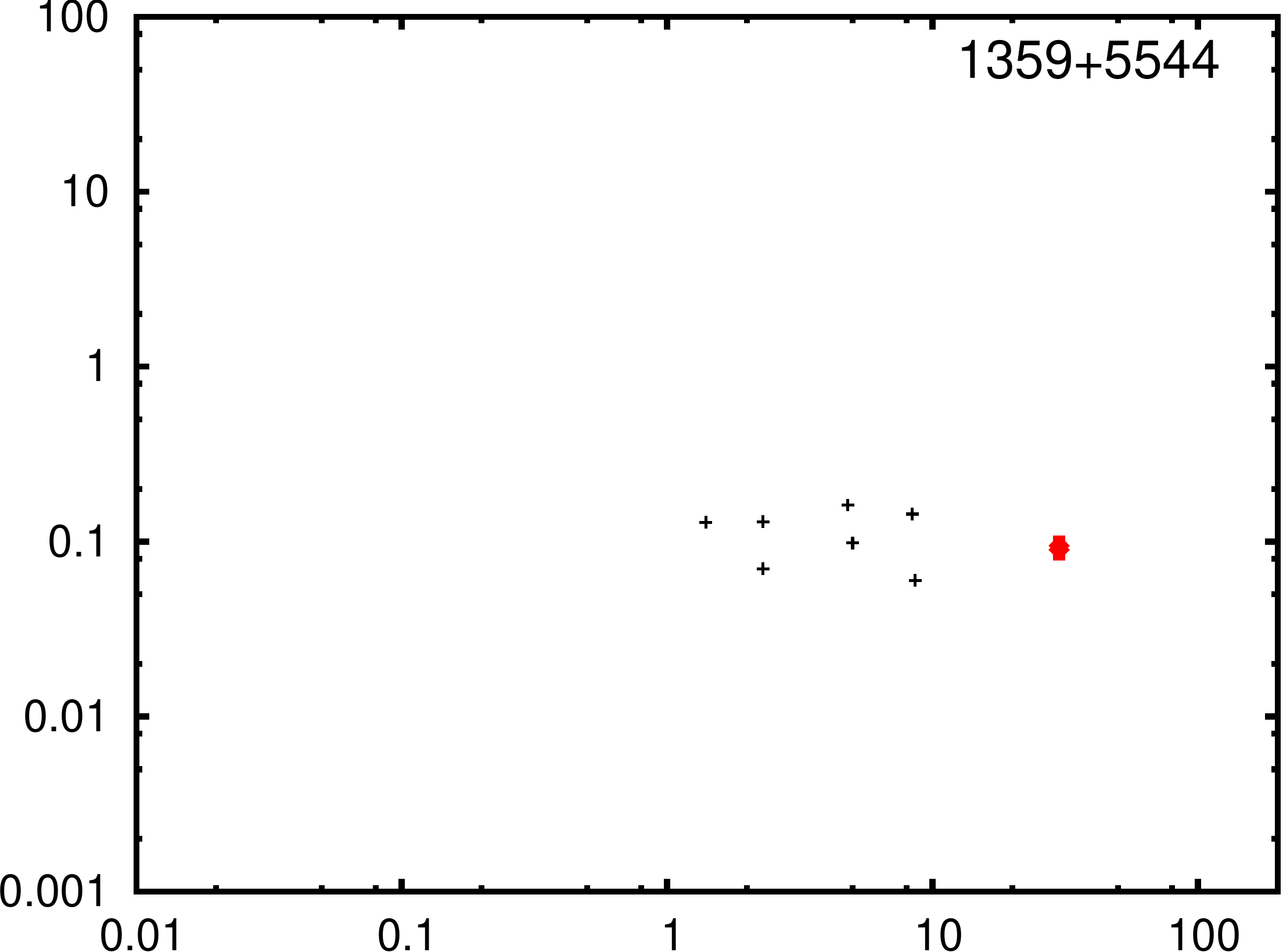}
\includegraphics[scale=0.2]{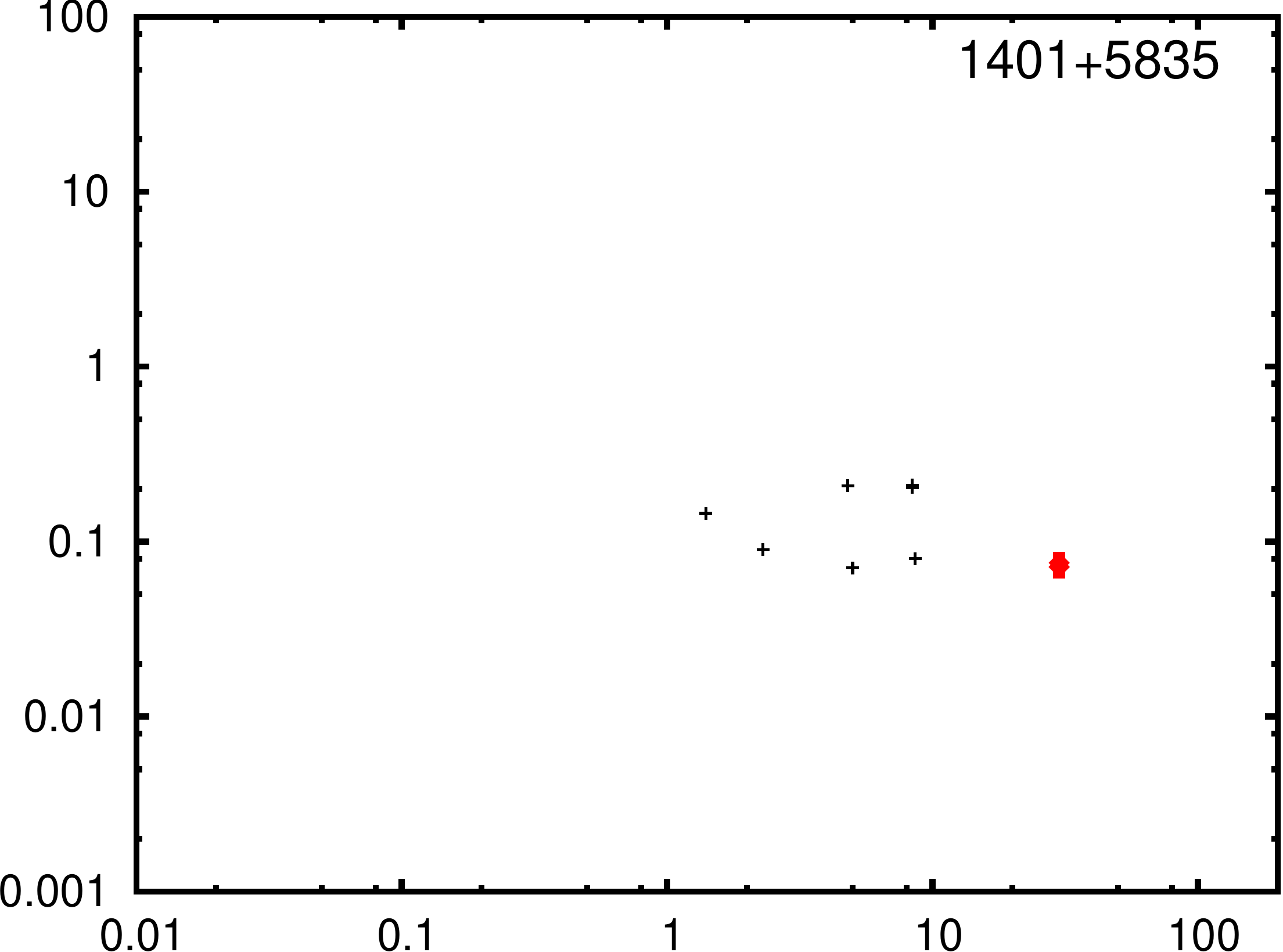}
\includegraphics[scale=0.2]{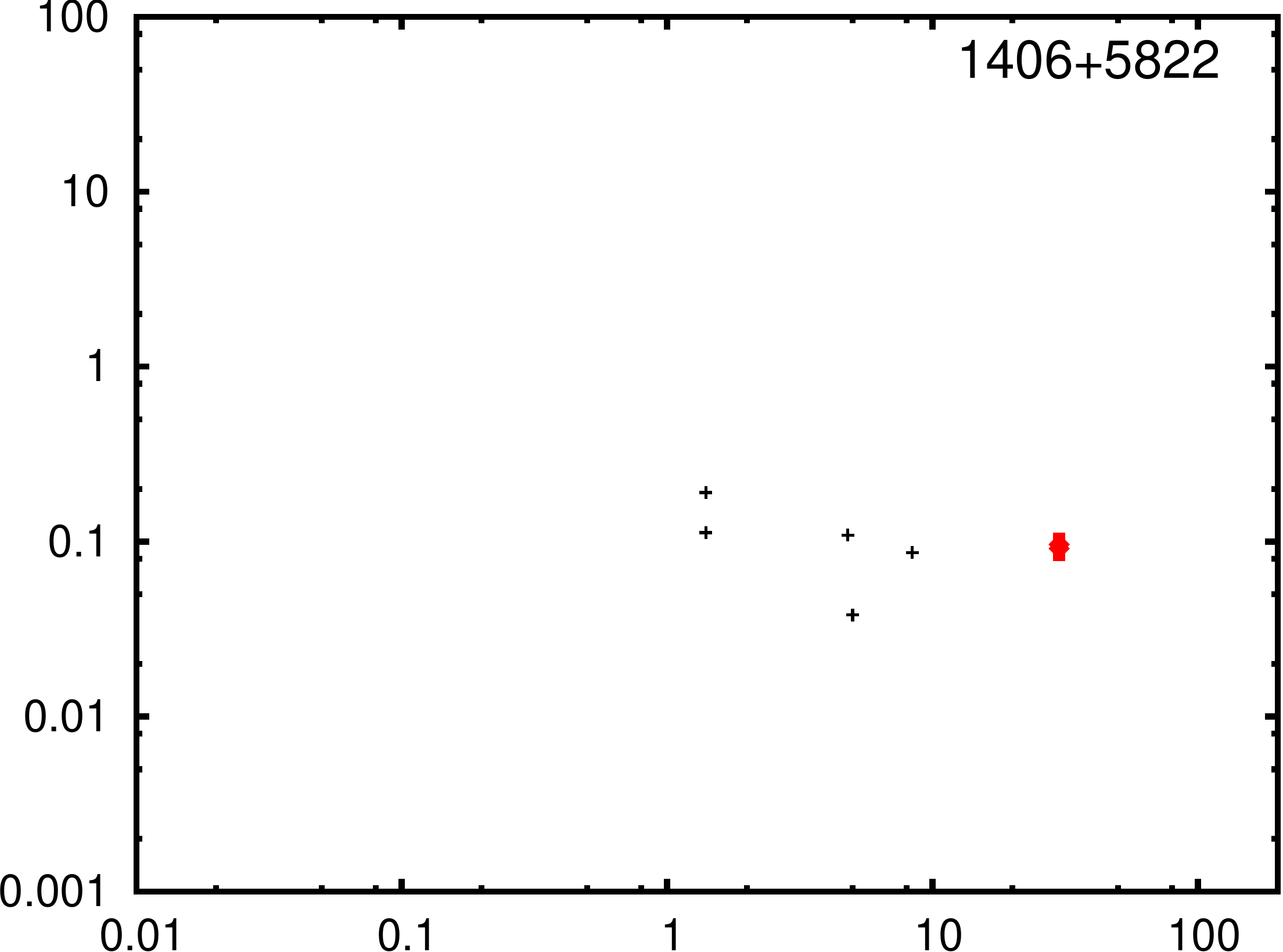}
\end{figure}
\clearpage\begin{figure}
\centering
\includegraphics[scale=0.2]{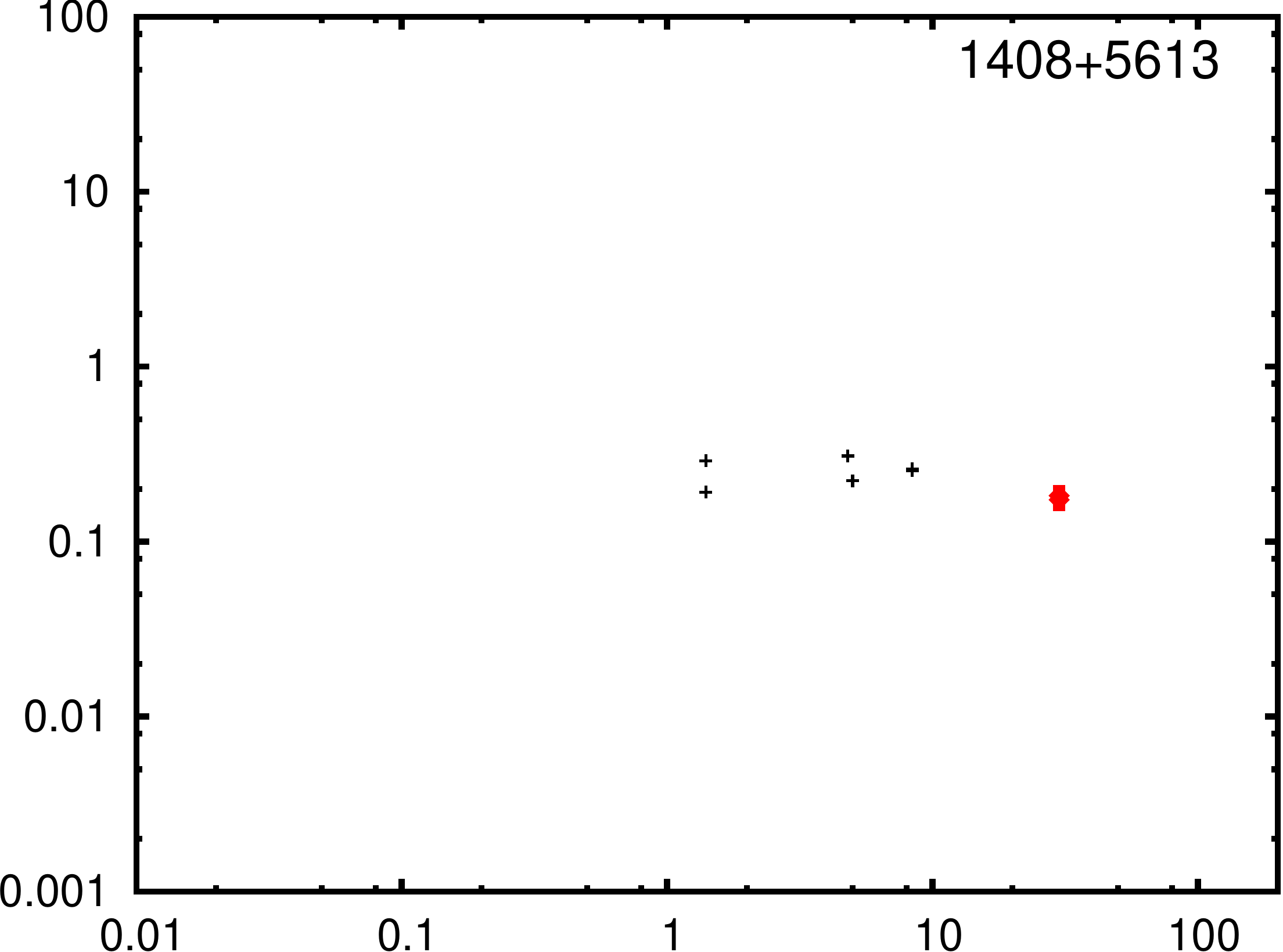}
\includegraphics[scale=0.2]{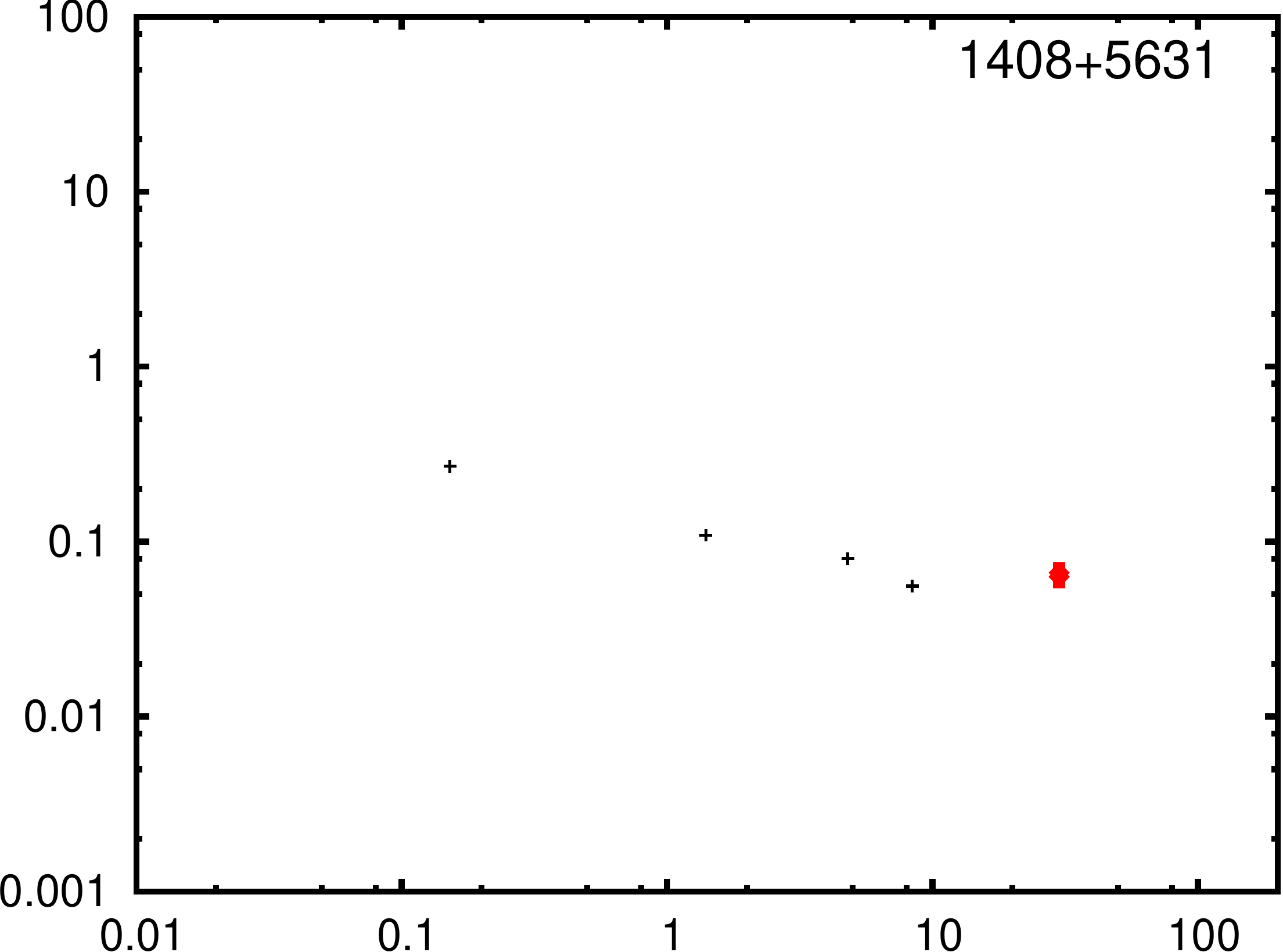}
\includegraphics[scale=0.2]{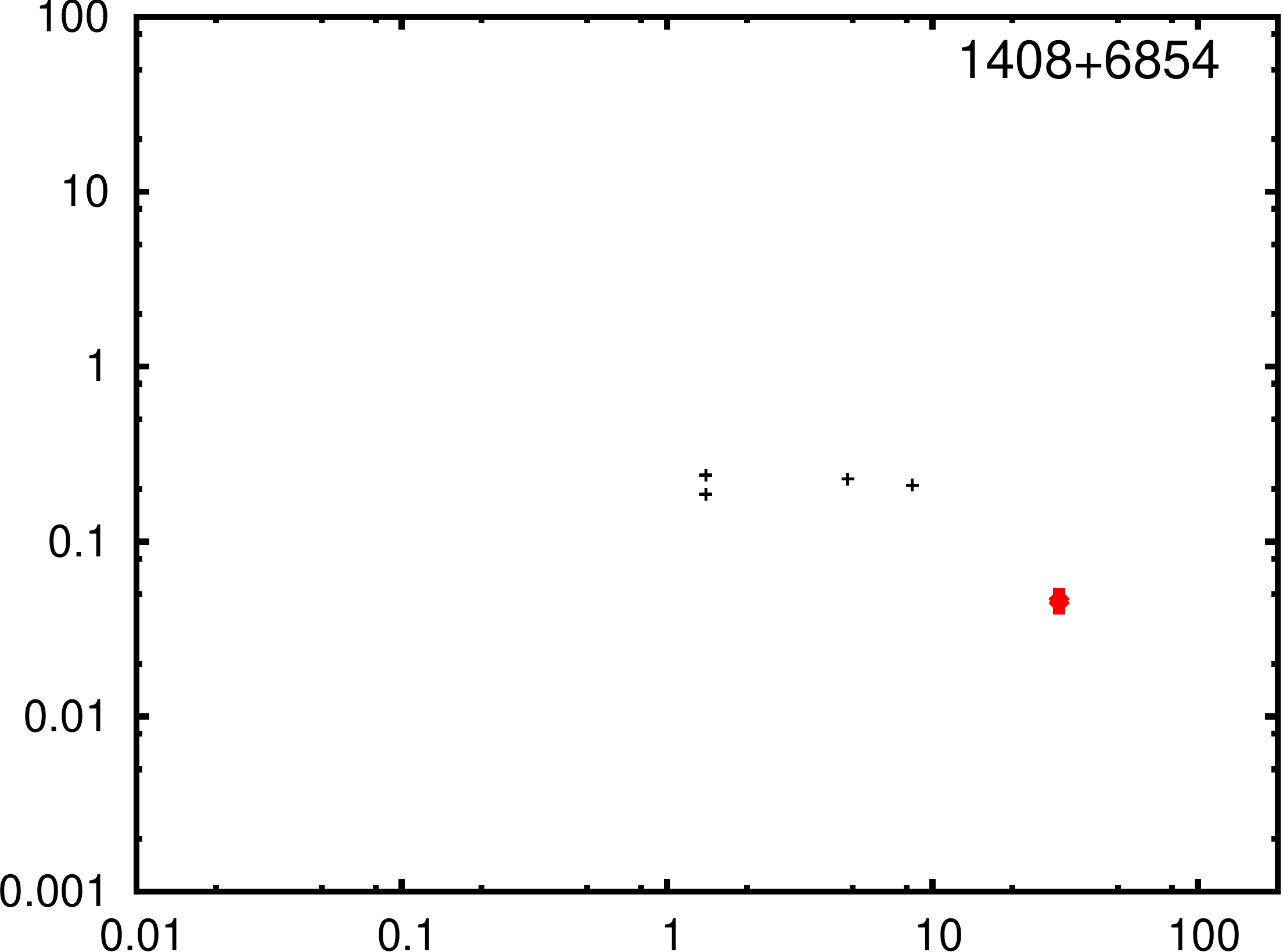}
\includegraphics[scale=0.2]{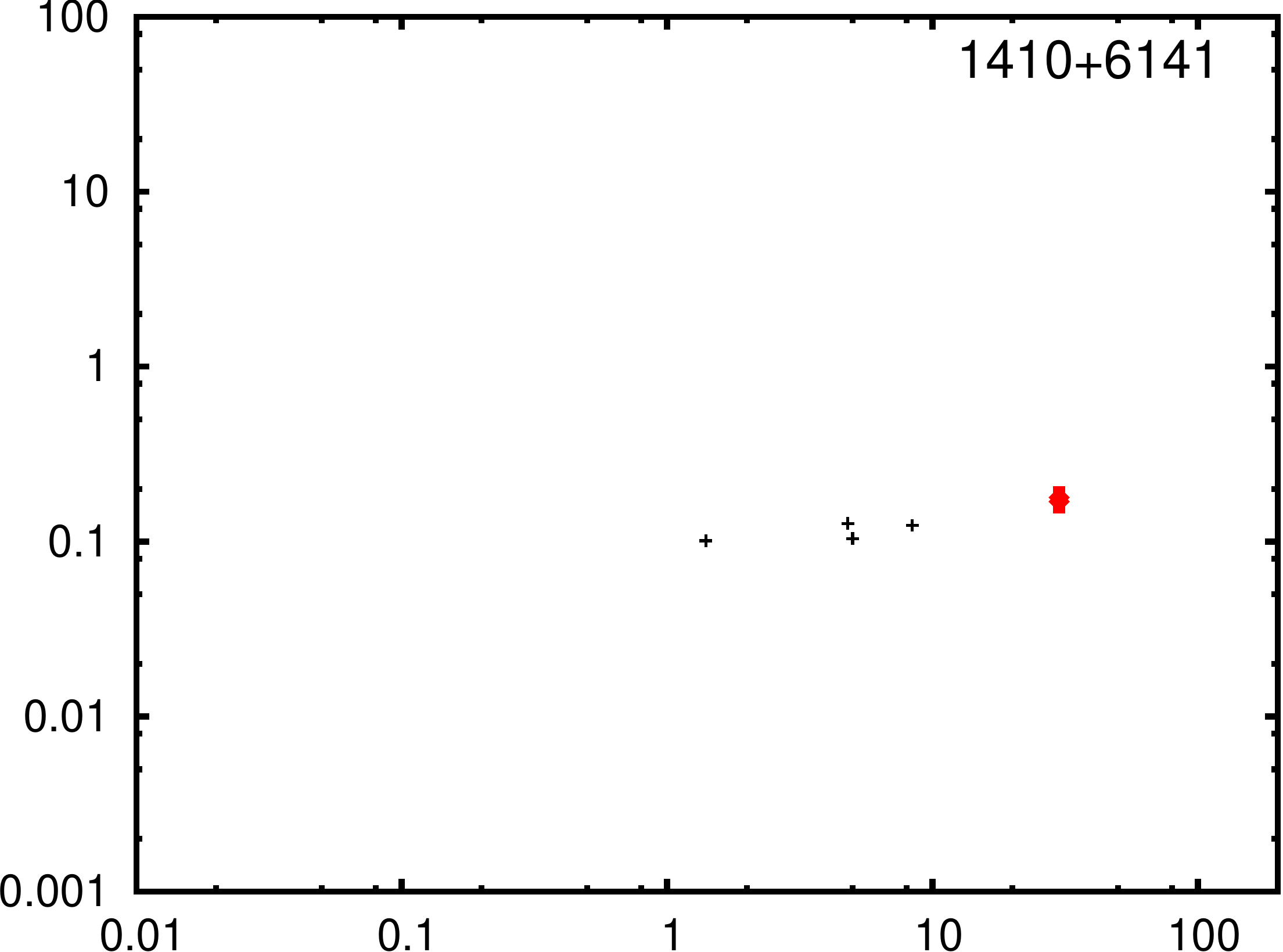}
\includegraphics[scale=0.2]{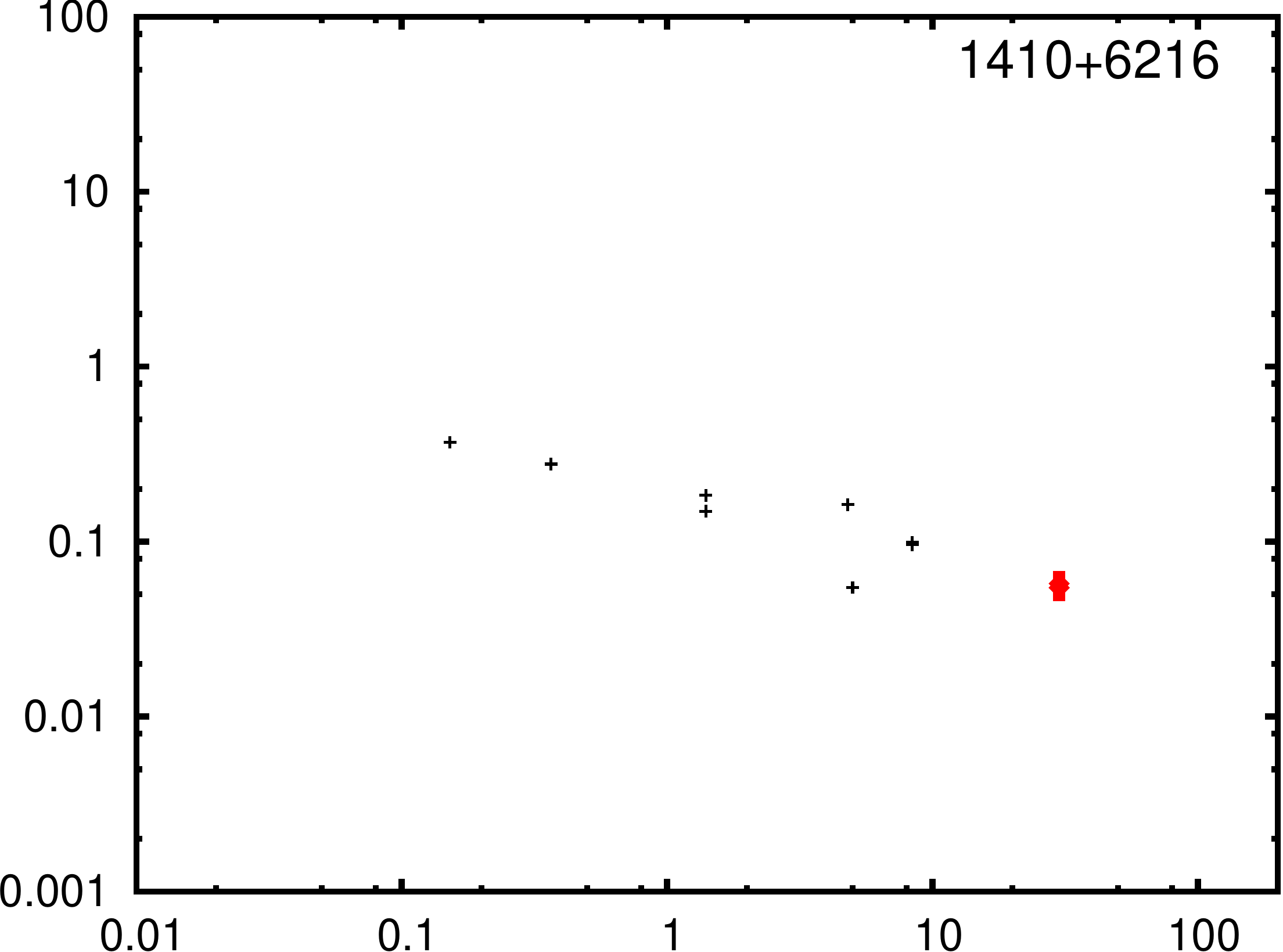}
\includegraphics[scale=0.2]{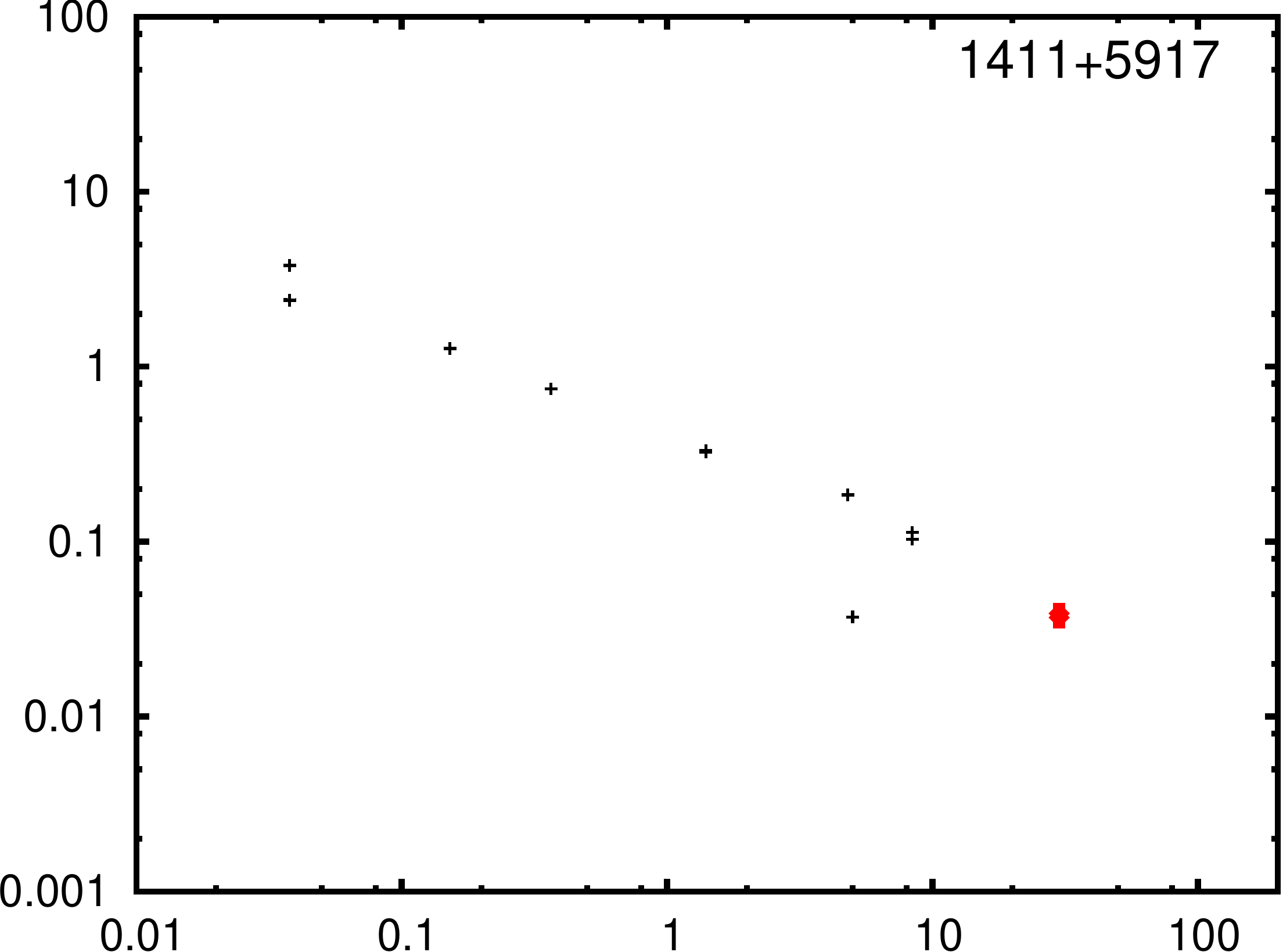}
\includegraphics[scale=0.2]{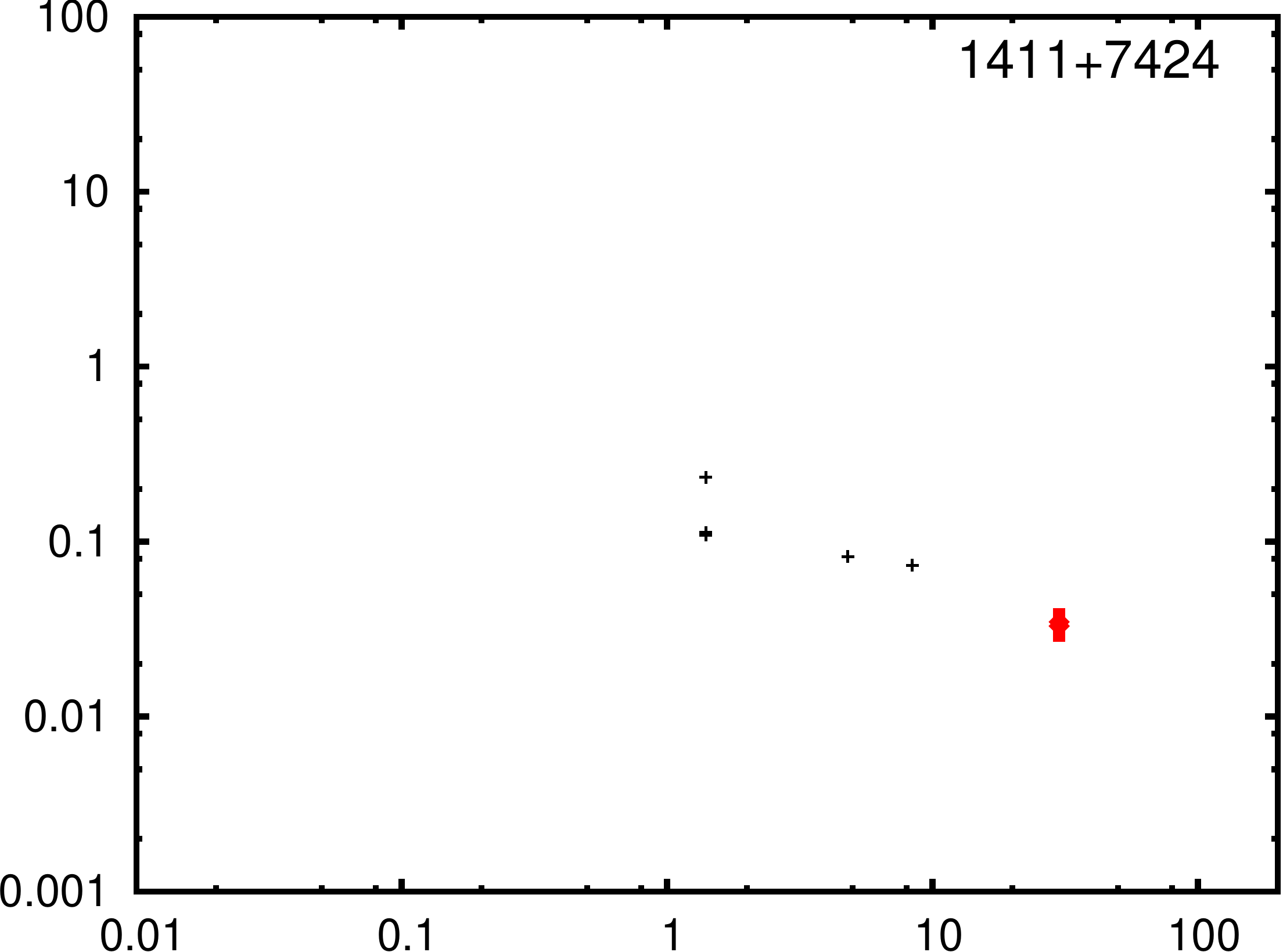}
\includegraphics[scale=0.2]{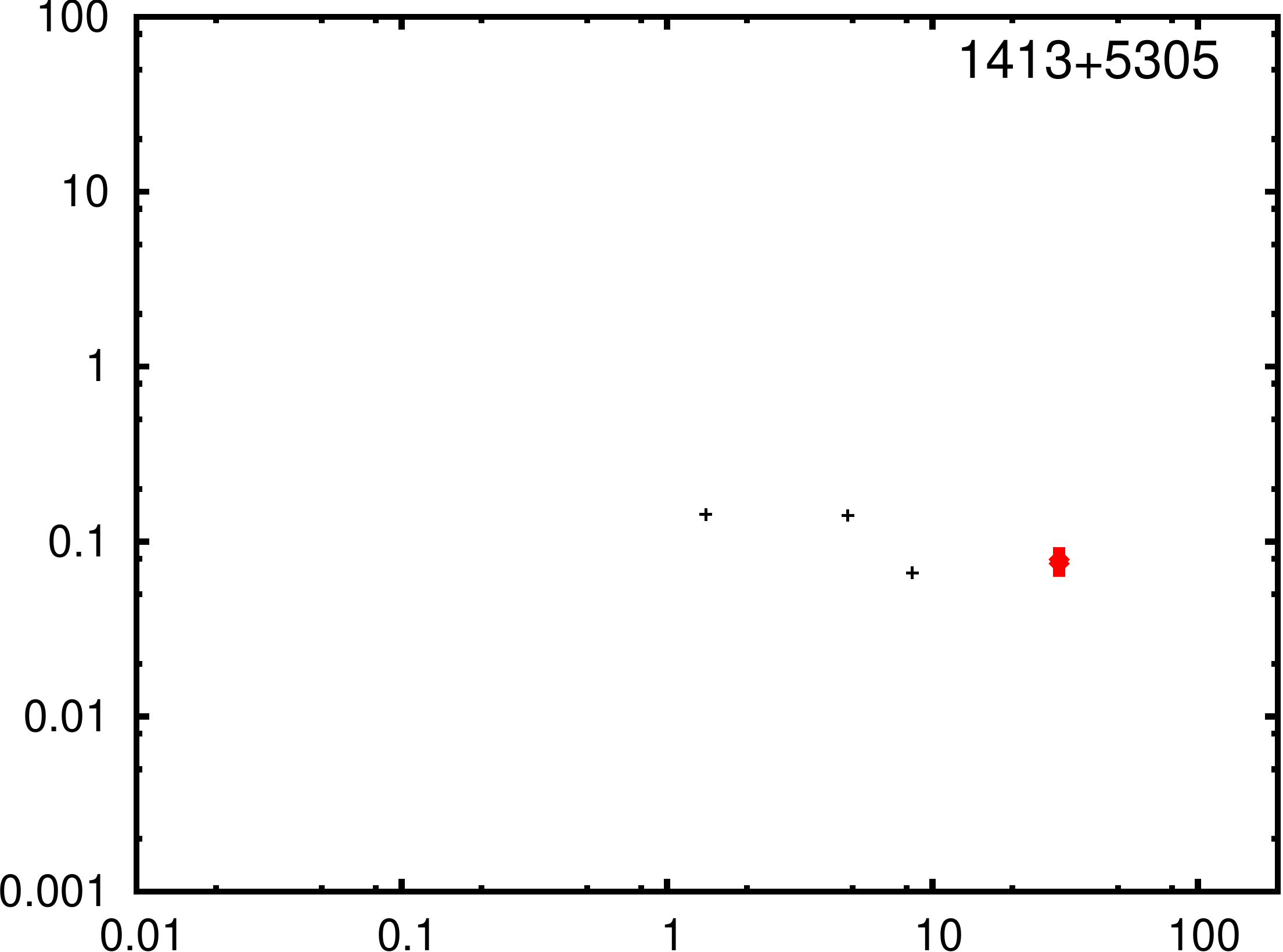}
\includegraphics[scale=0.2]{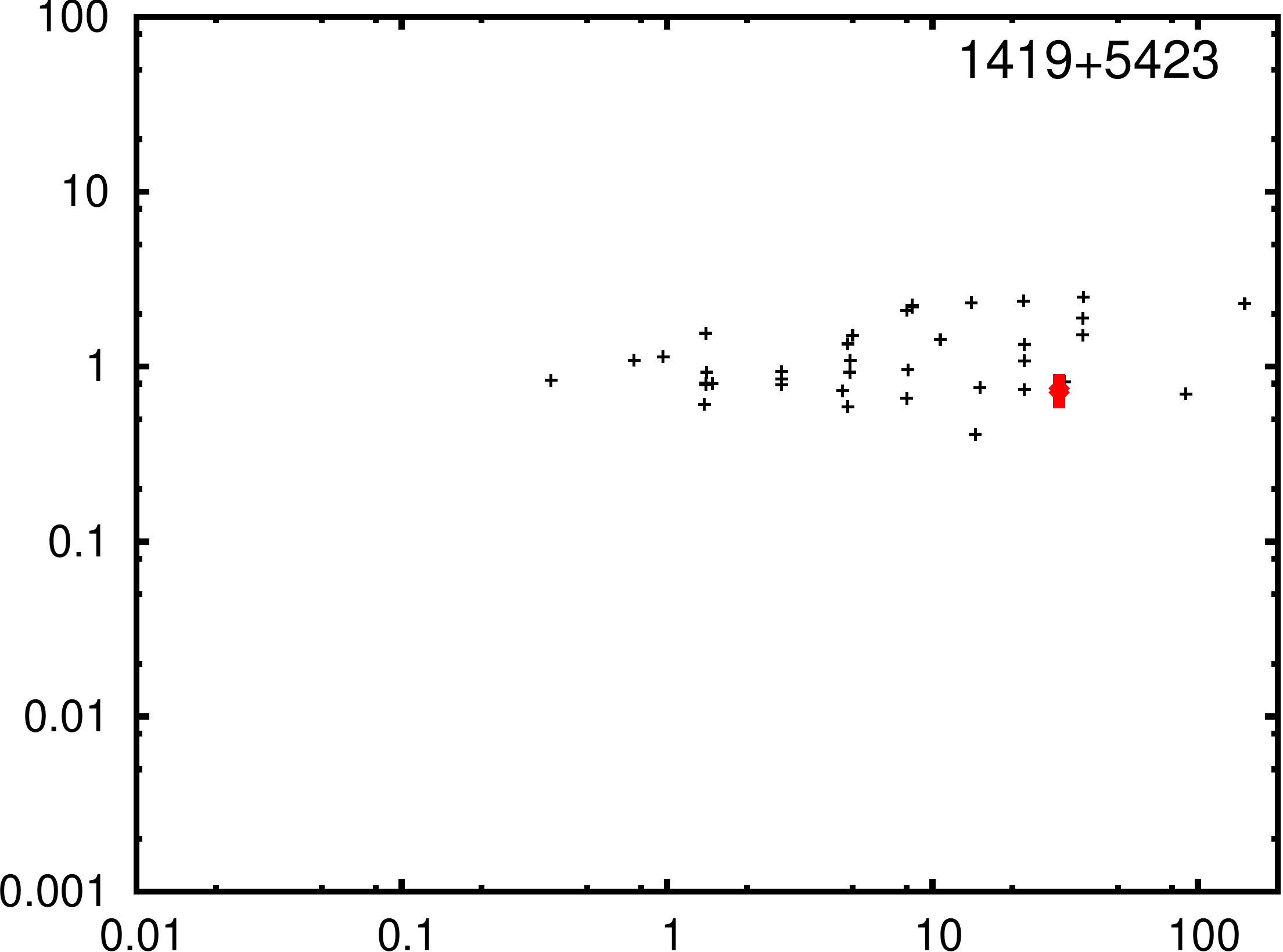}
\includegraphics[scale=0.2]{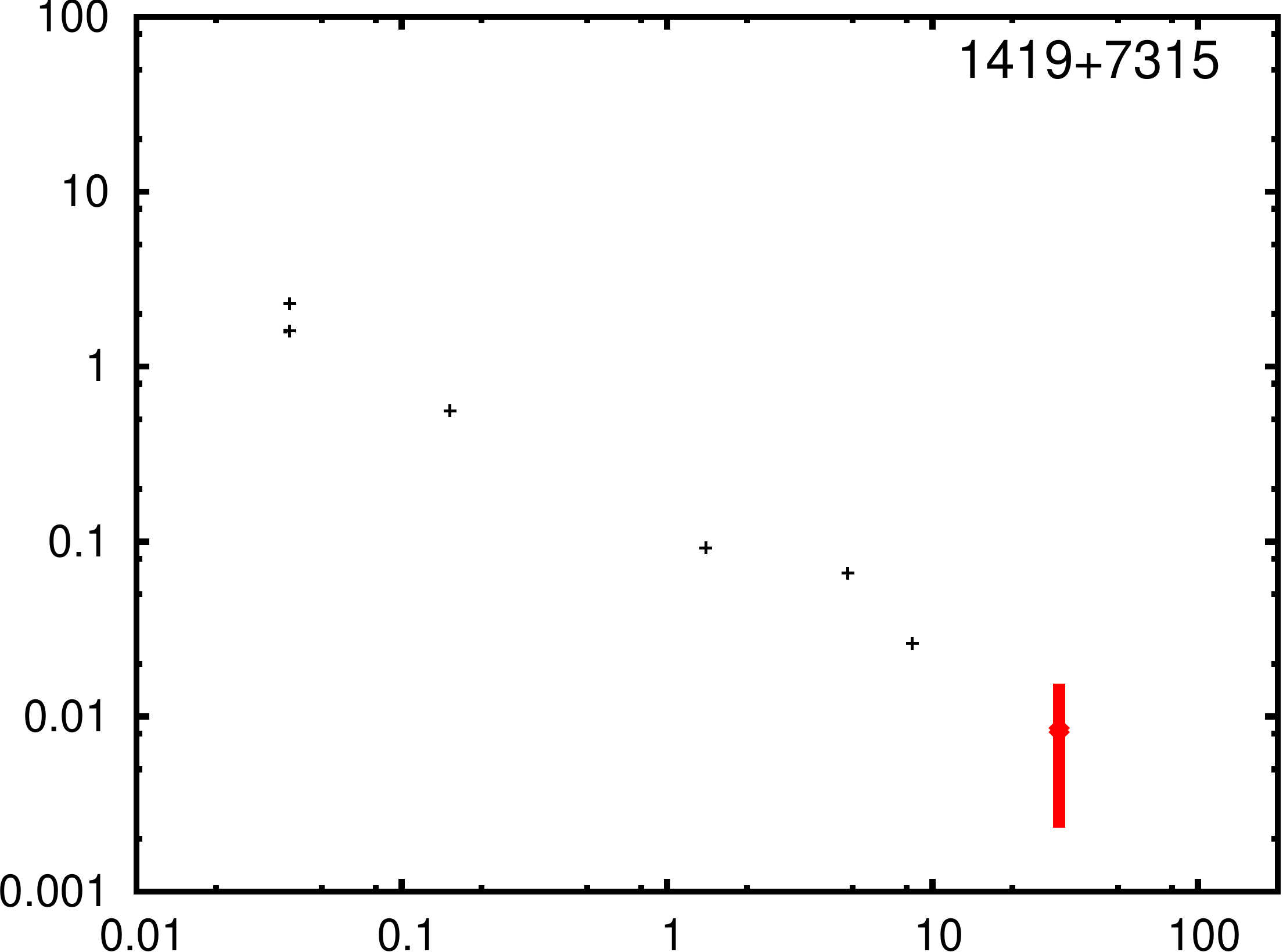}
\includegraphics[scale=0.2]{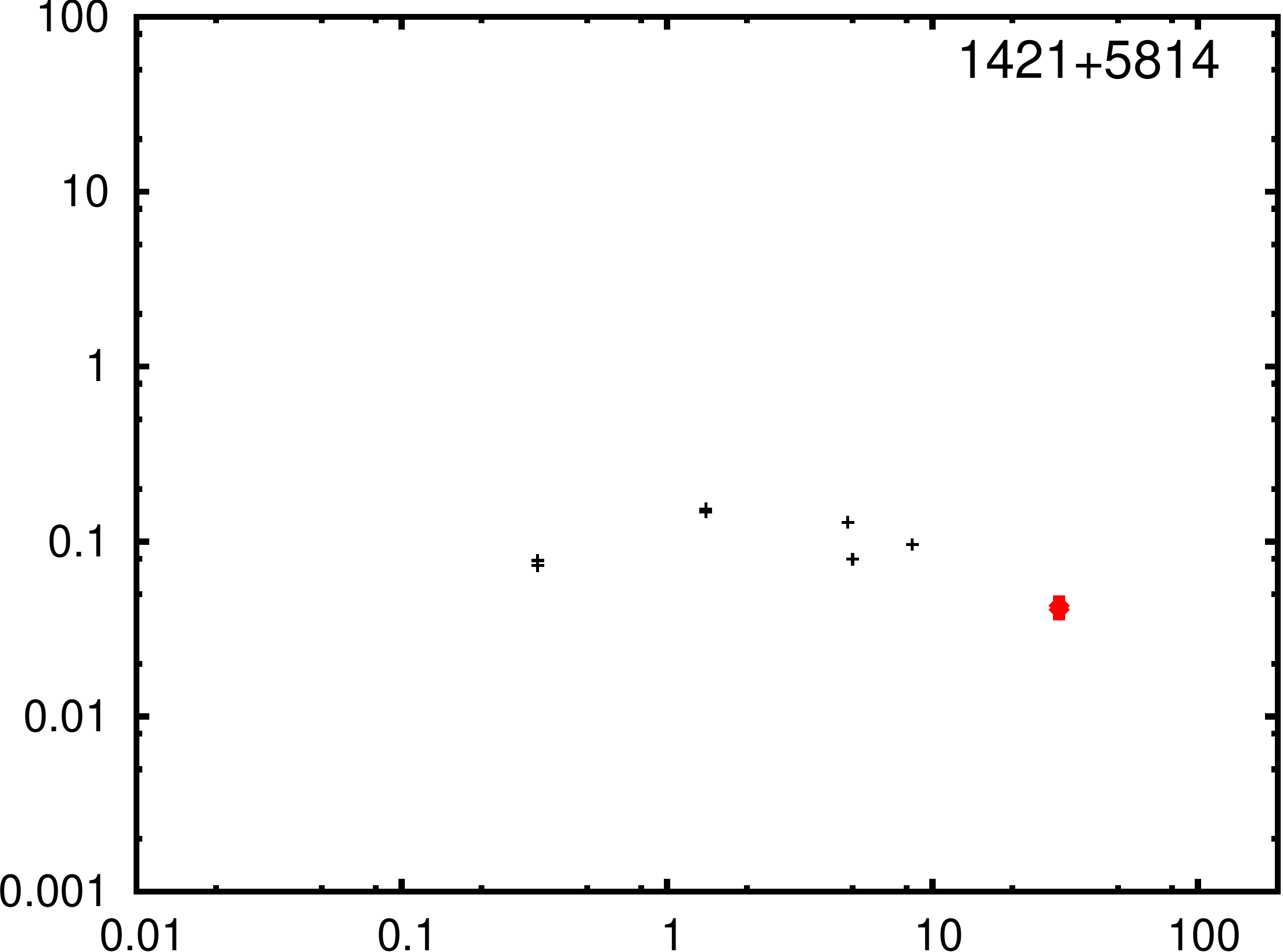}
\includegraphics[scale=0.2]{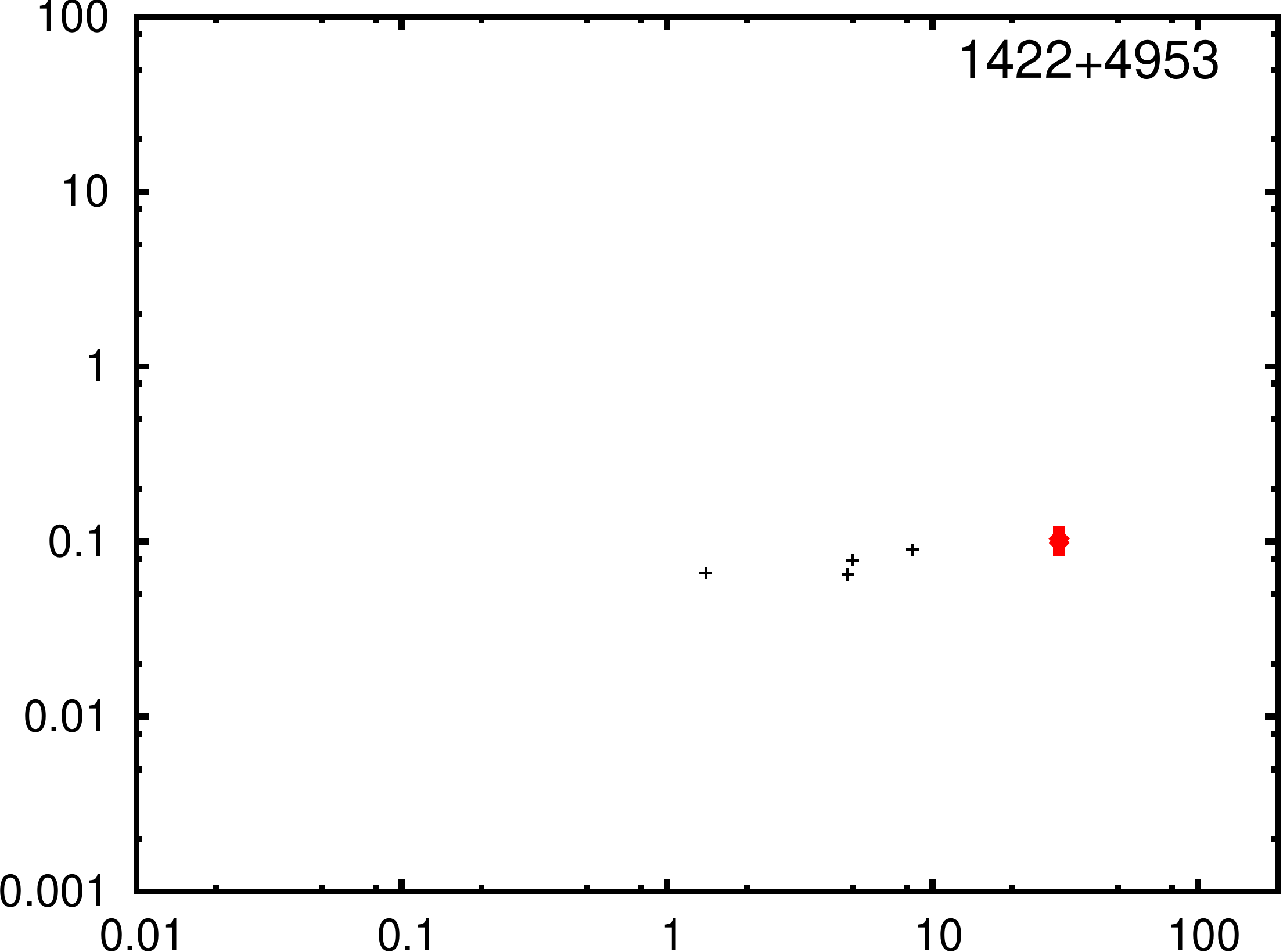}
\includegraphics[scale=0.2]{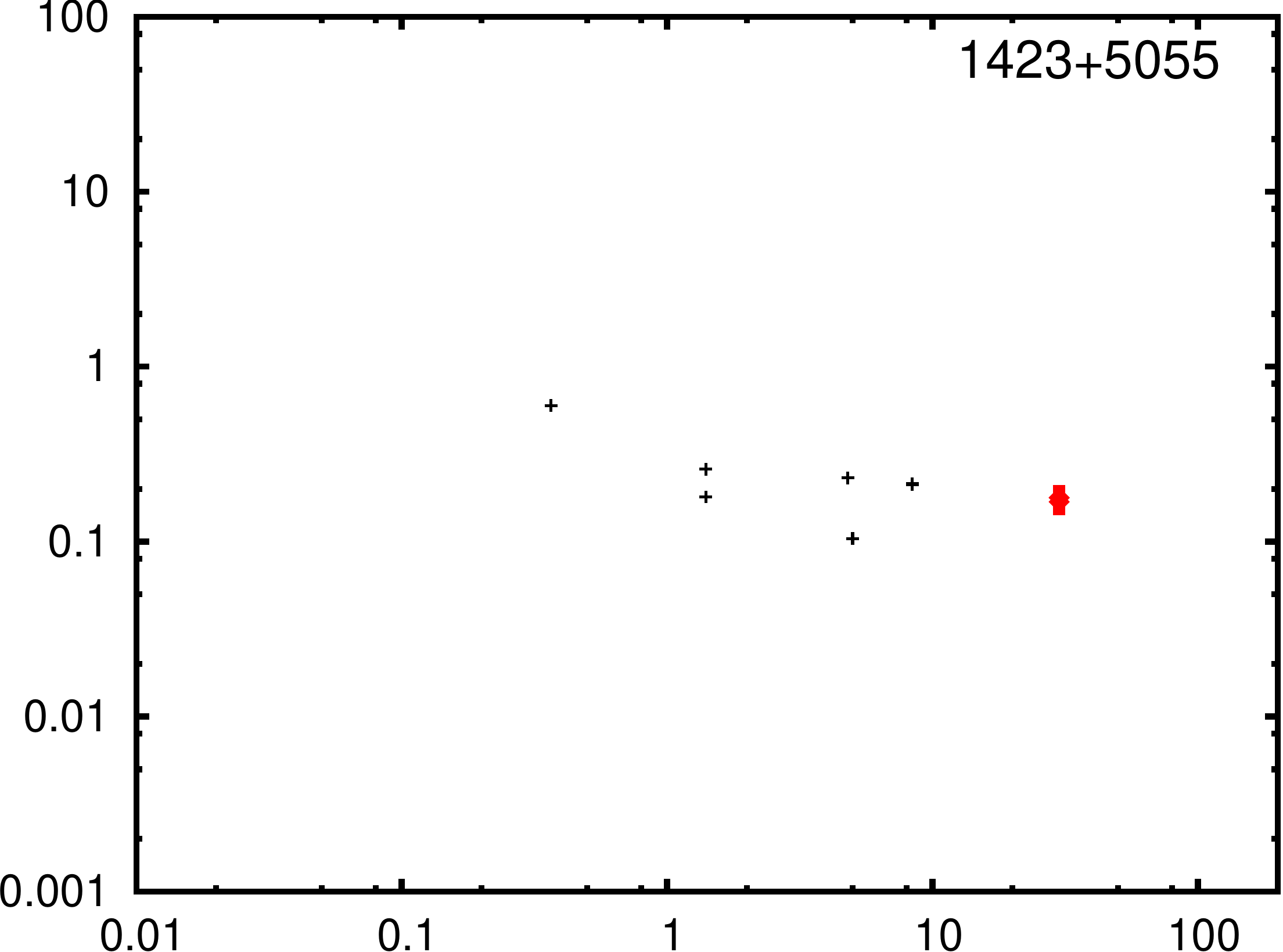}
\includegraphics[scale=0.2]{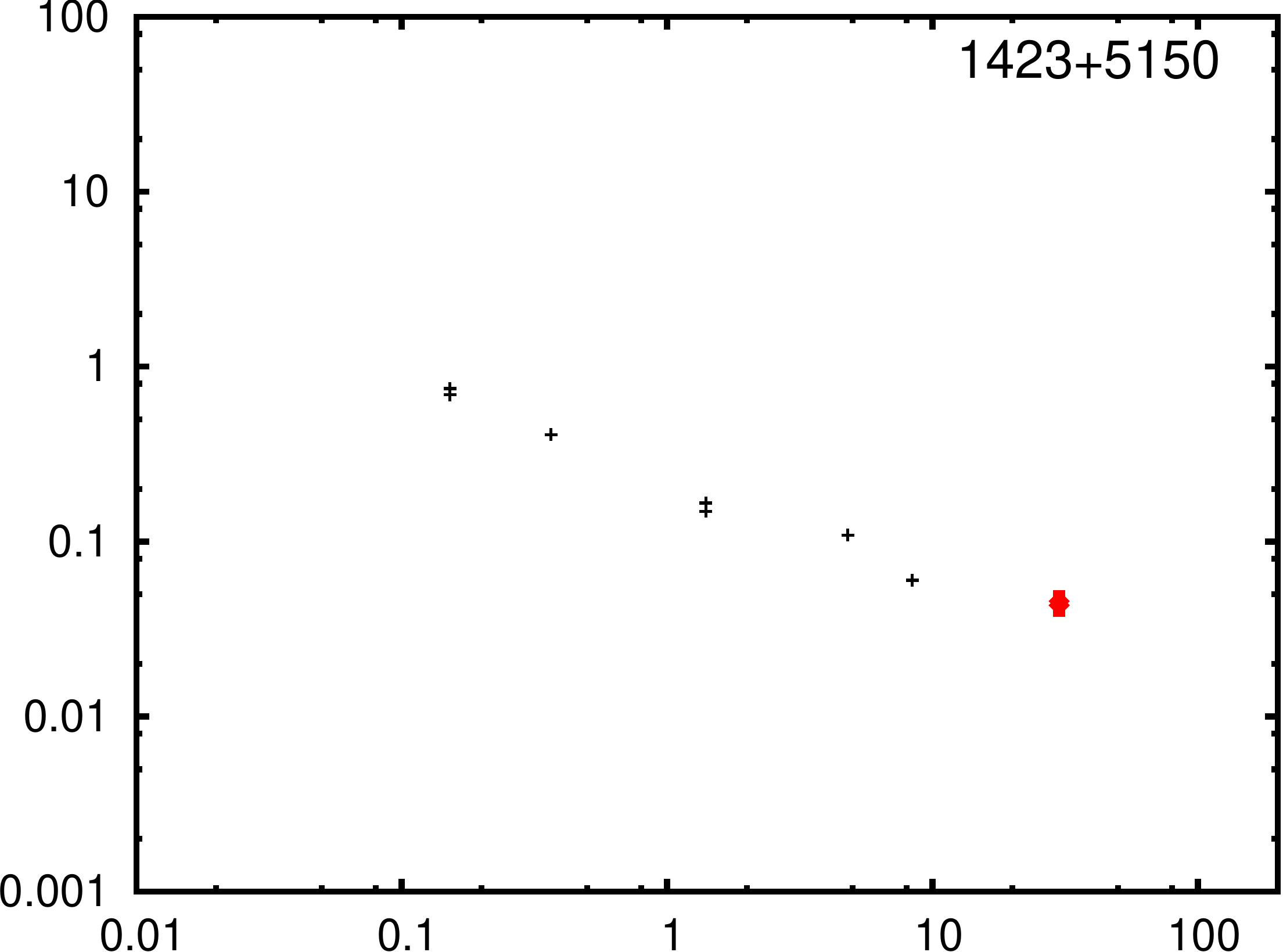}
\includegraphics[scale=0.2]{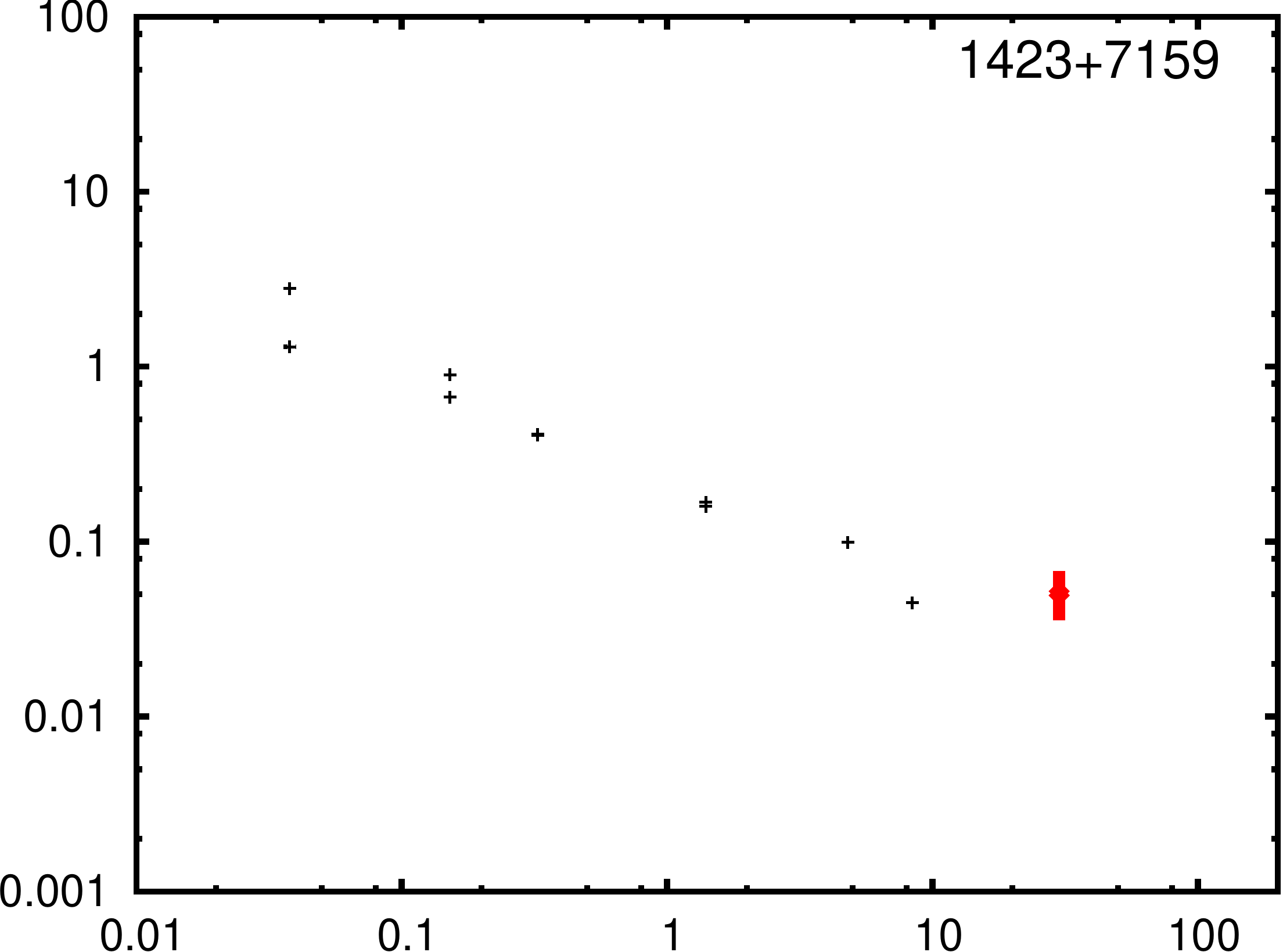}
\includegraphics[scale=0.2]{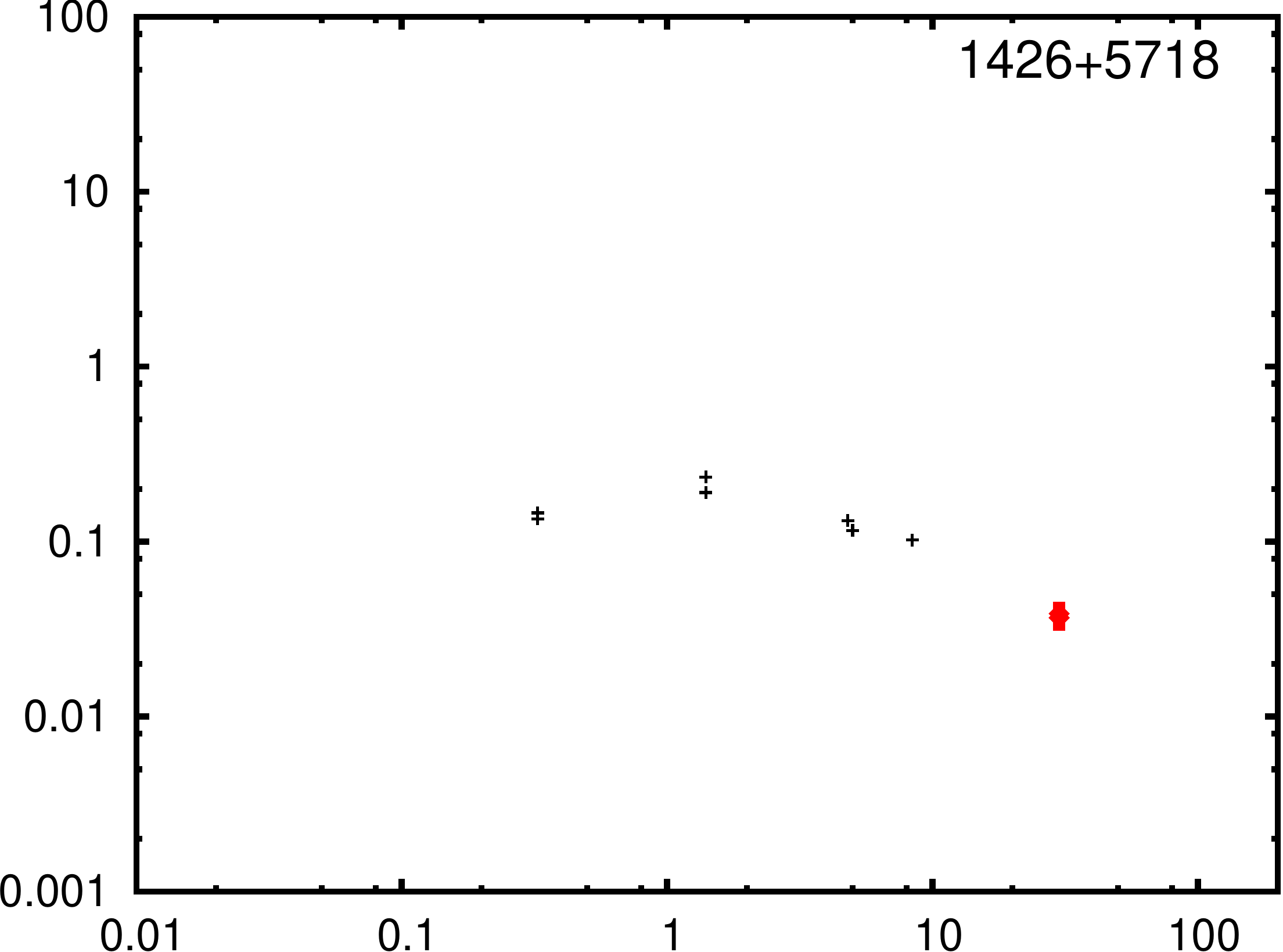}
\includegraphics[scale=0.2]{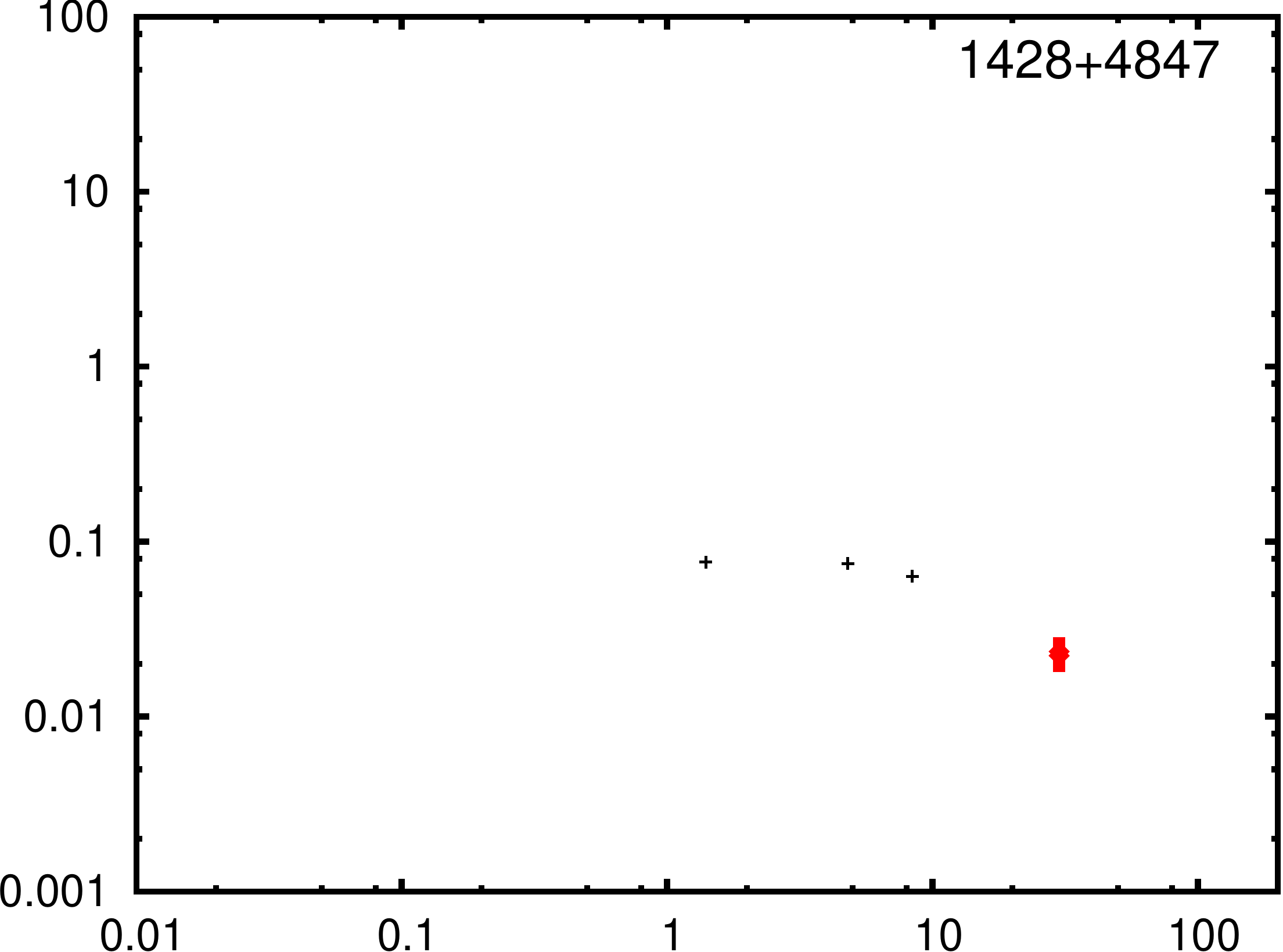}
\includegraphics[scale=0.2]{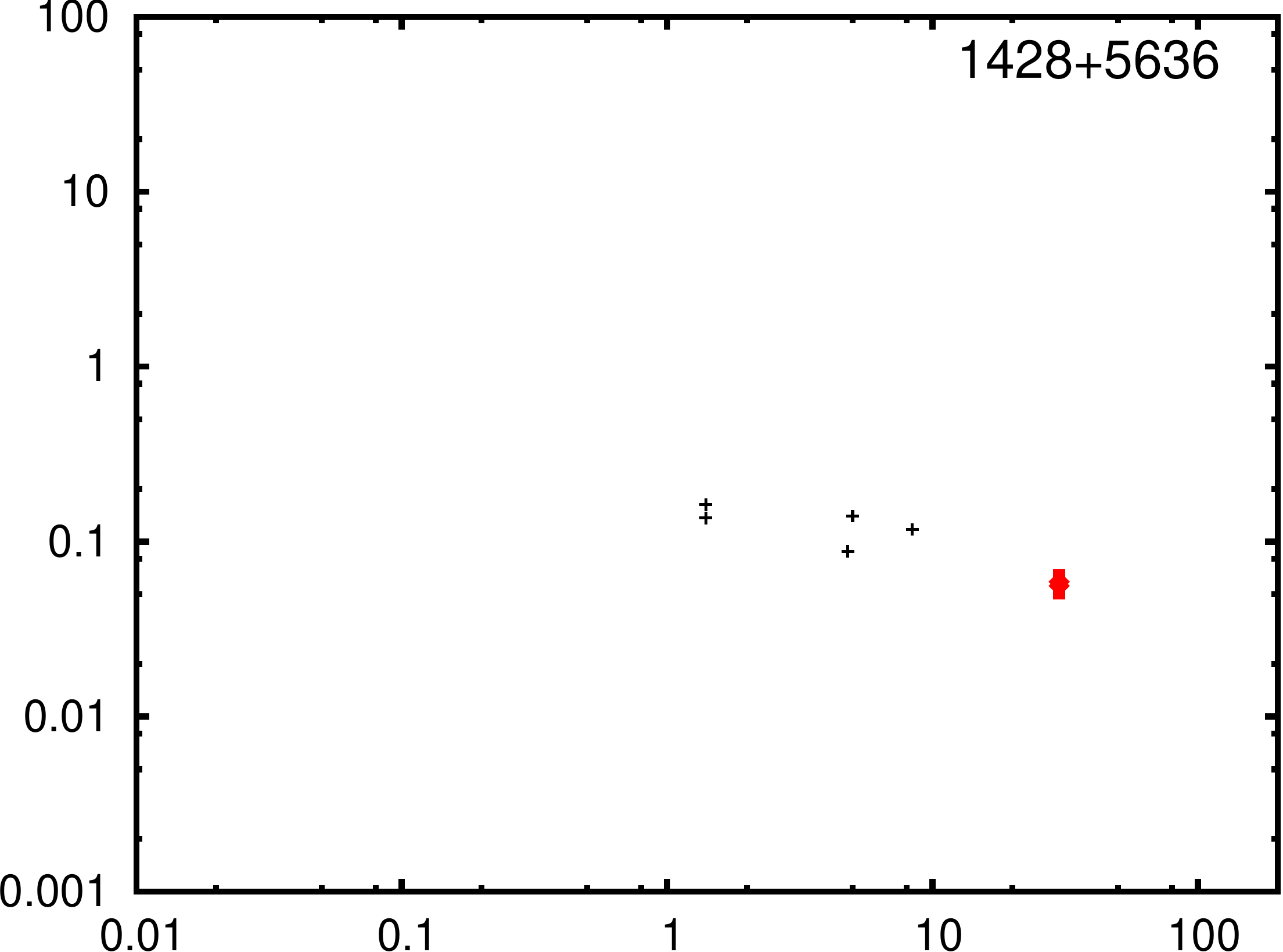}
\end{figure}
\clearpage\begin{figure}
\centering
\includegraphics[scale=0.2]{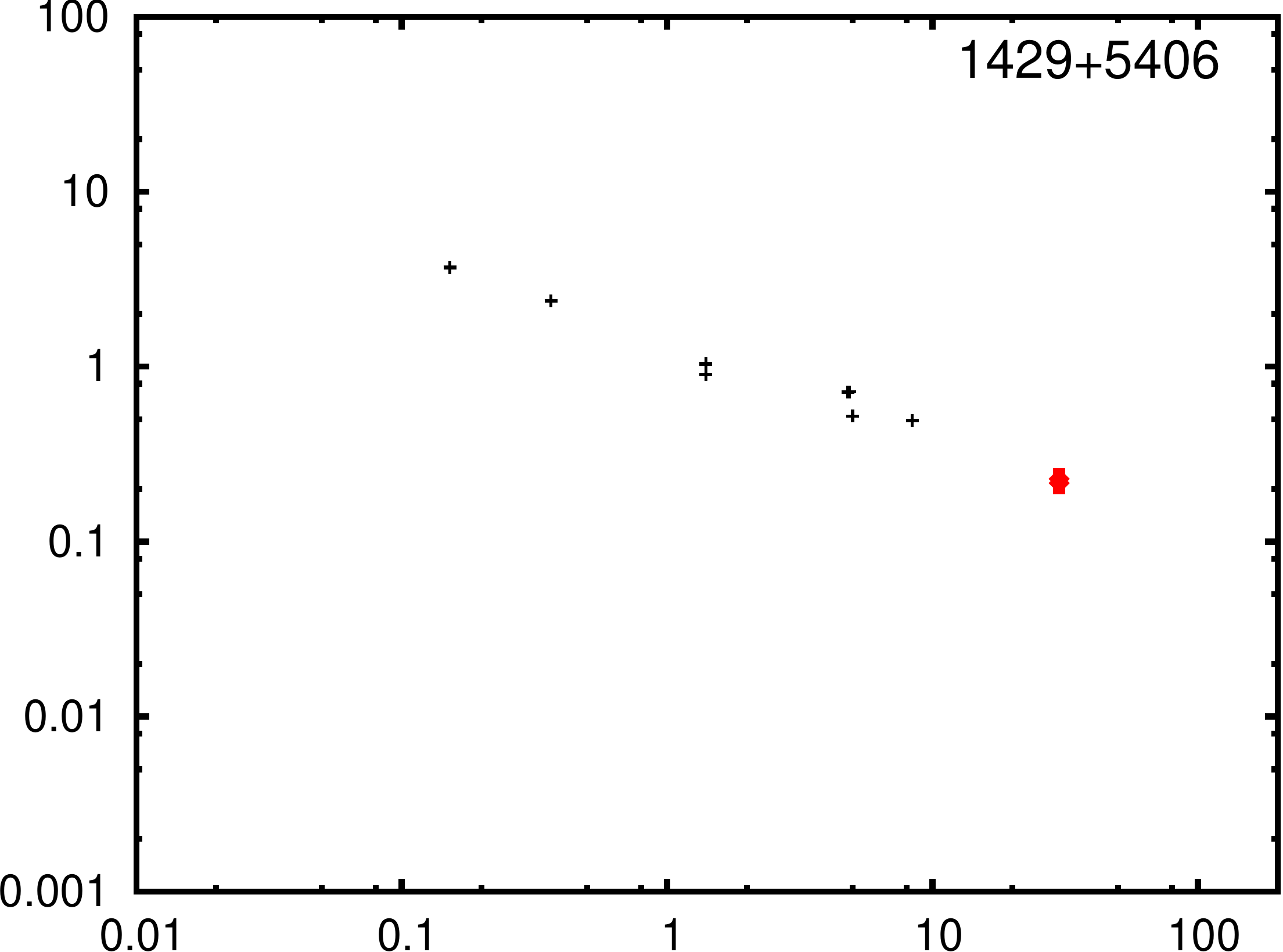}
\includegraphics[scale=0.2]{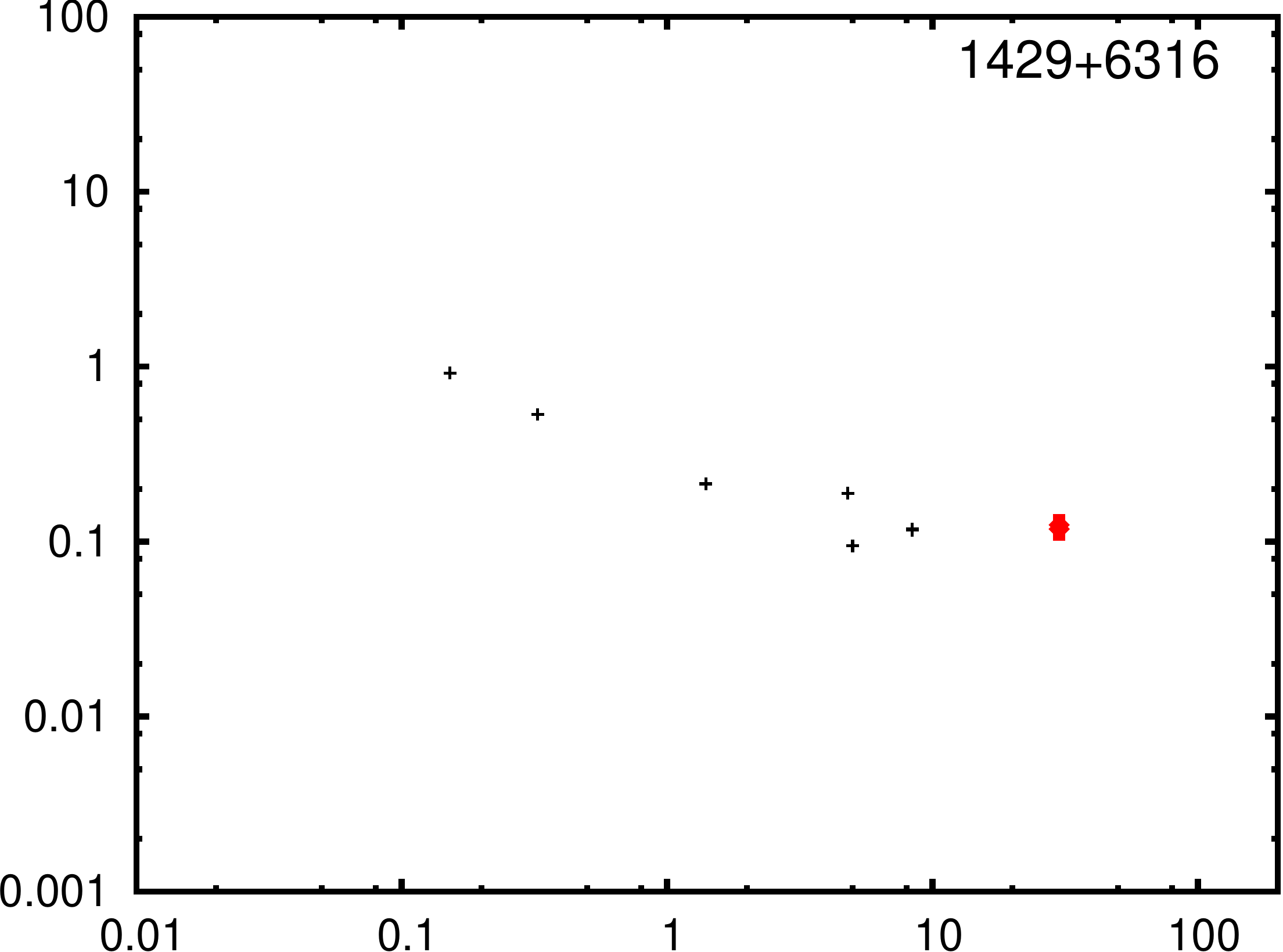}
\includegraphics[scale=0.2]{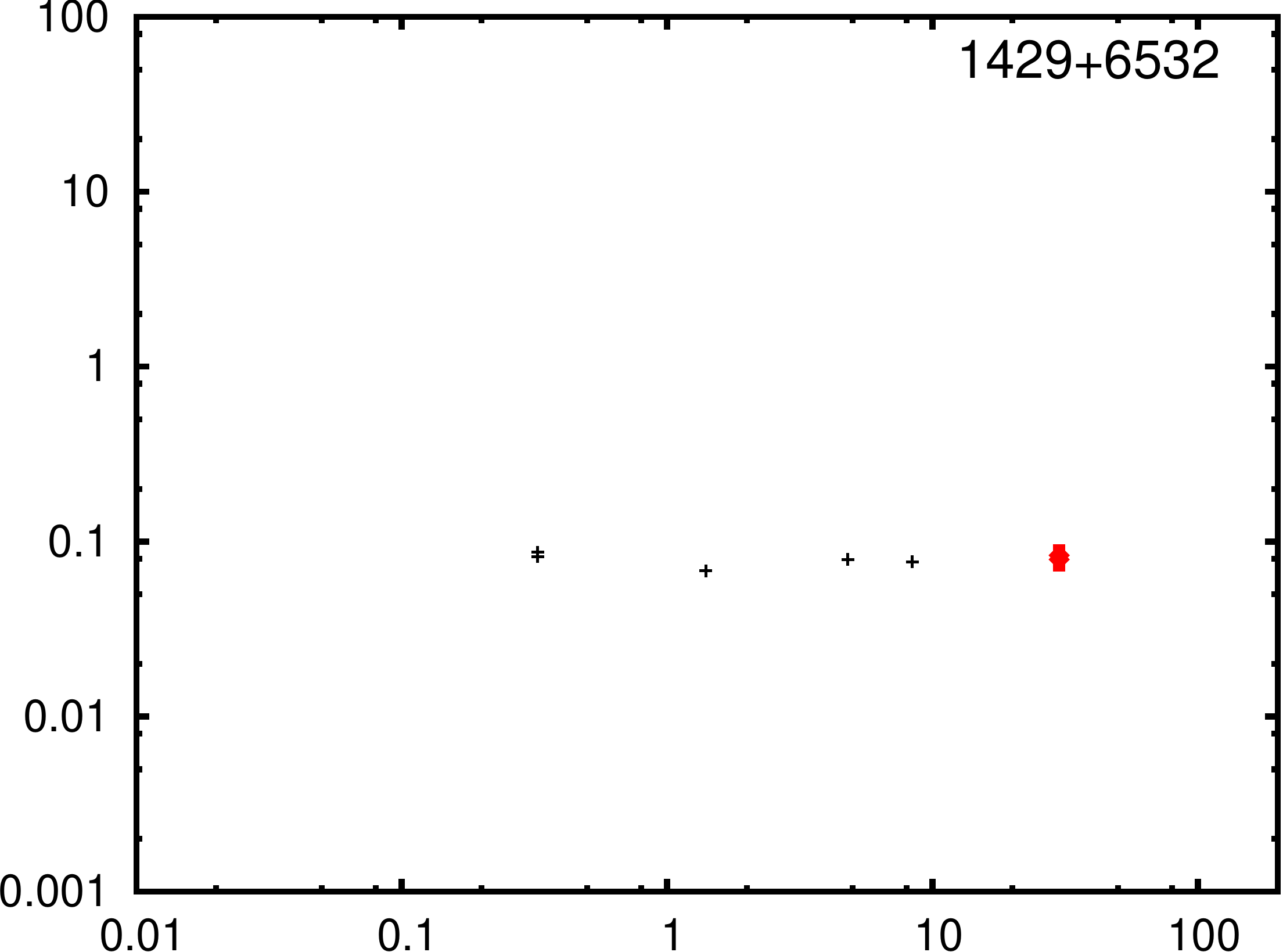}
\includegraphics[scale=0.2]{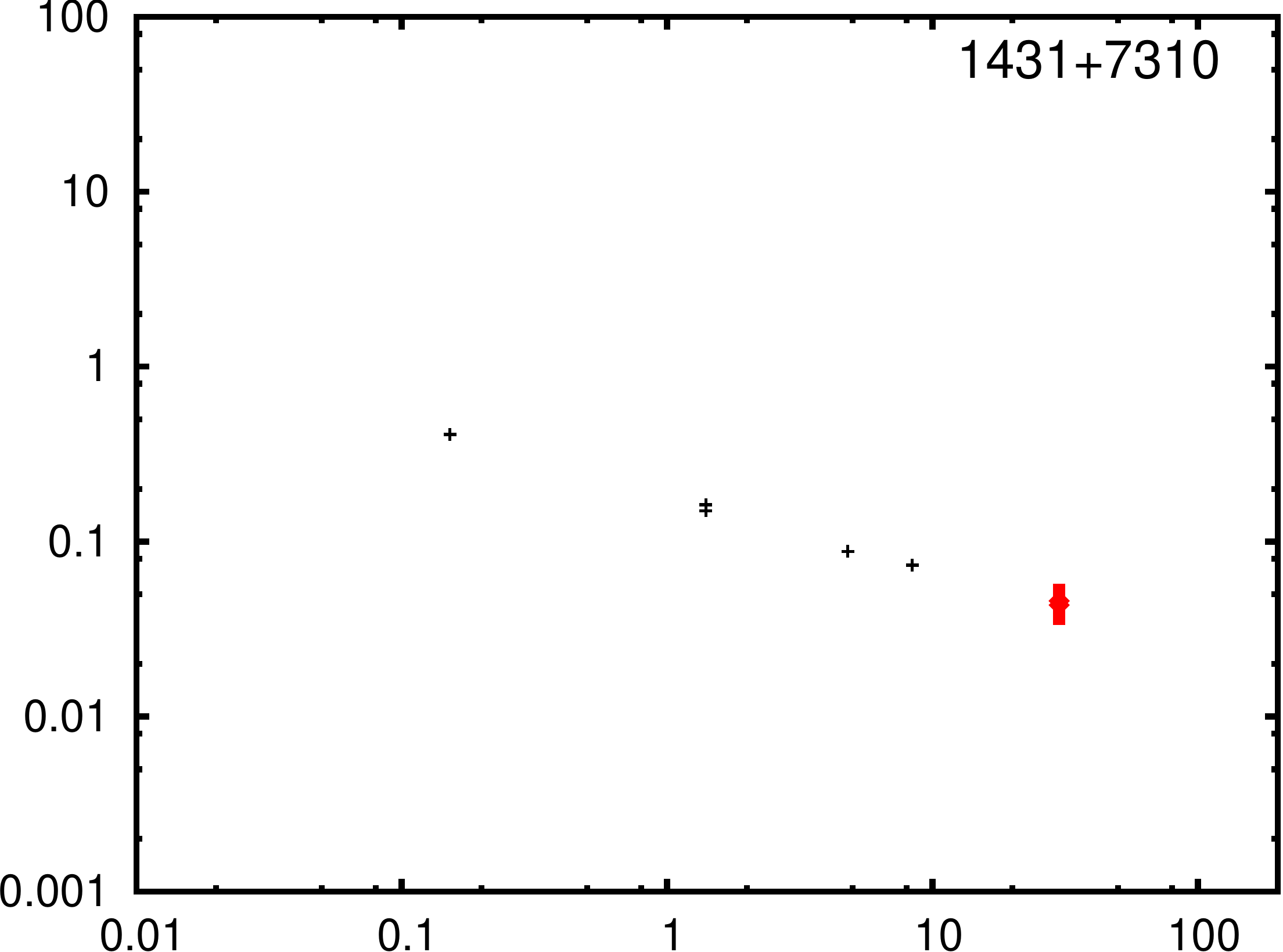}
\includegraphics[scale=0.2]{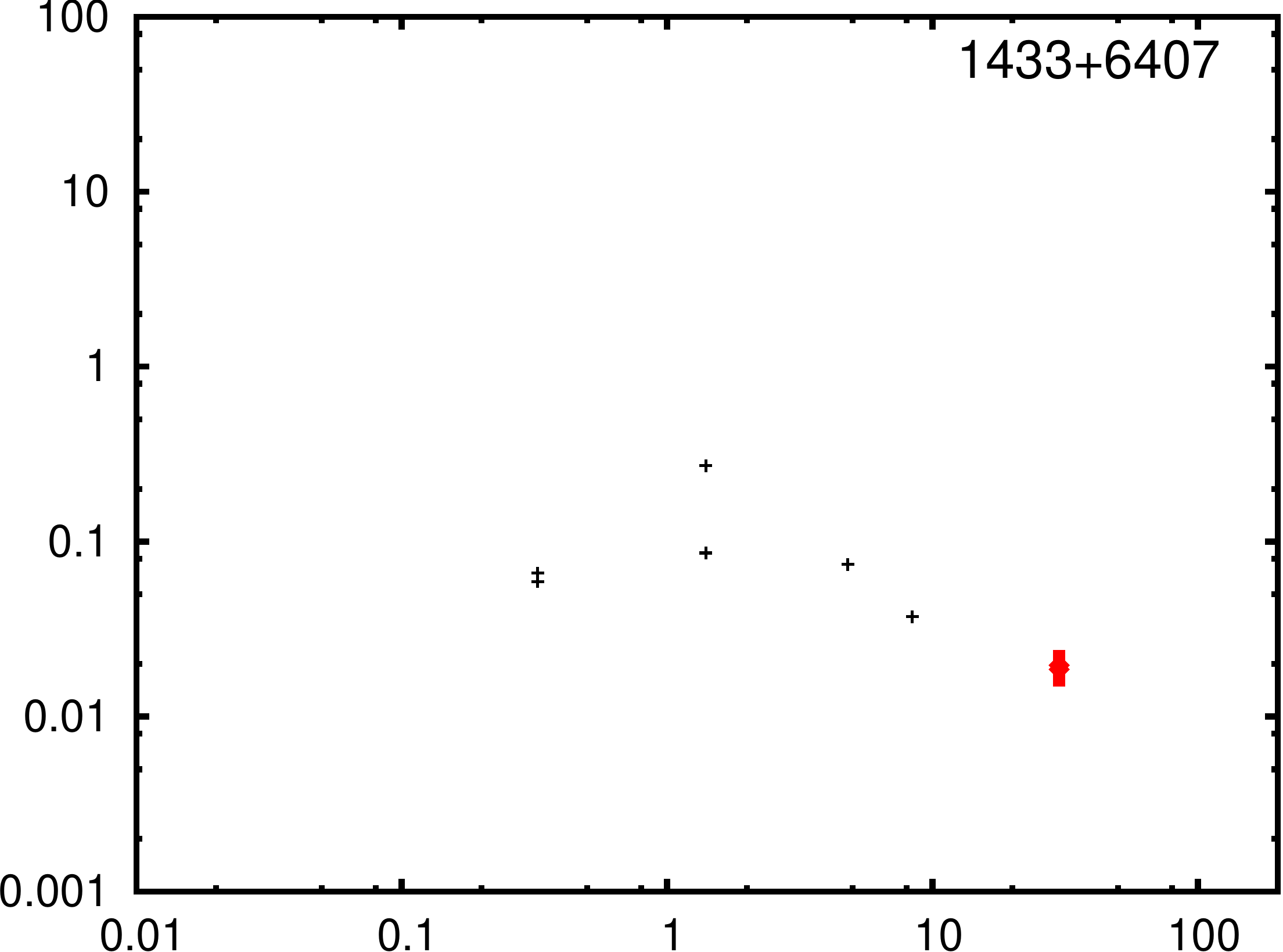}
\includegraphics[scale=0.2]{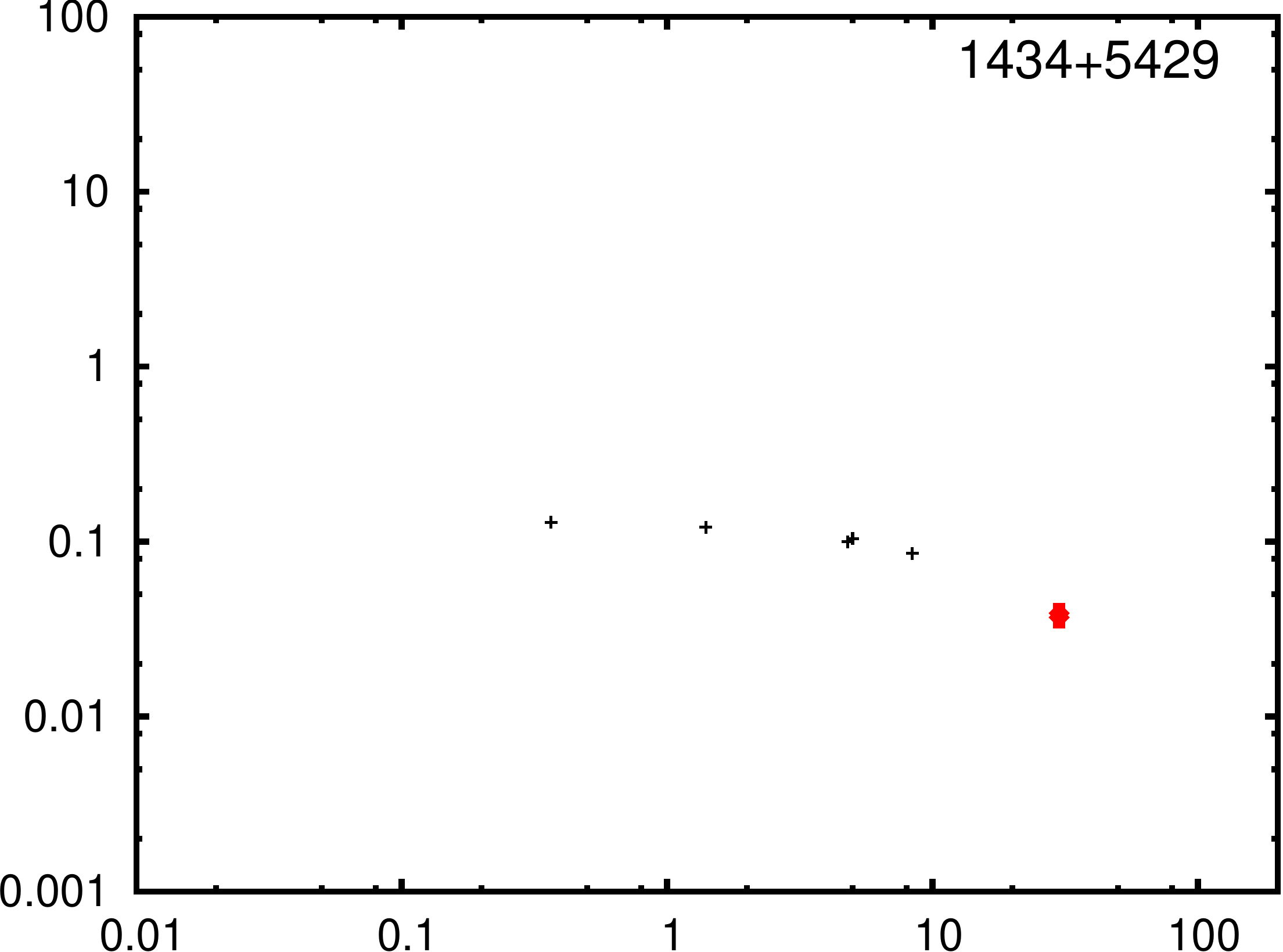}
\includegraphics[scale=0.2]{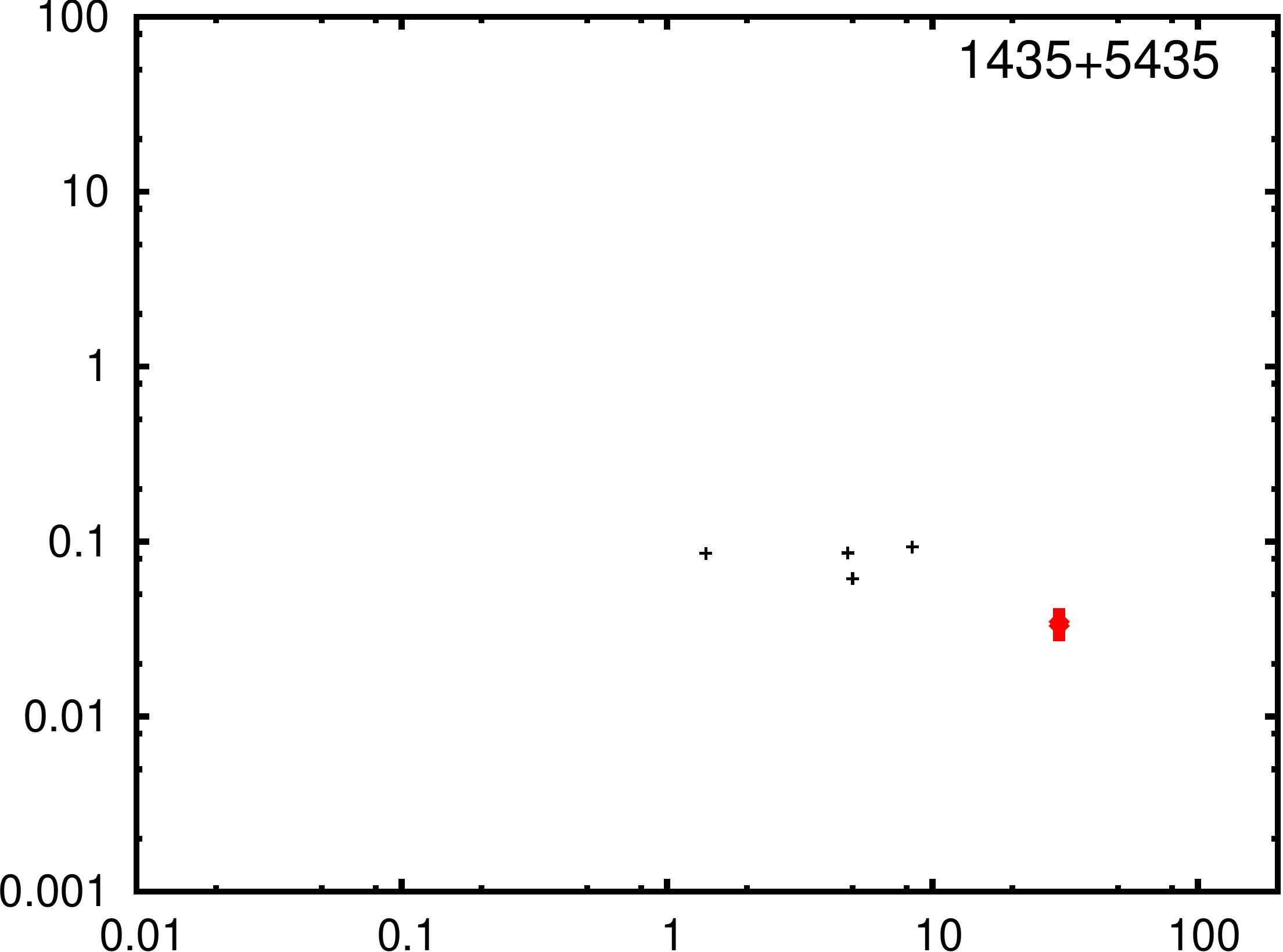}
\includegraphics[scale=0.2]{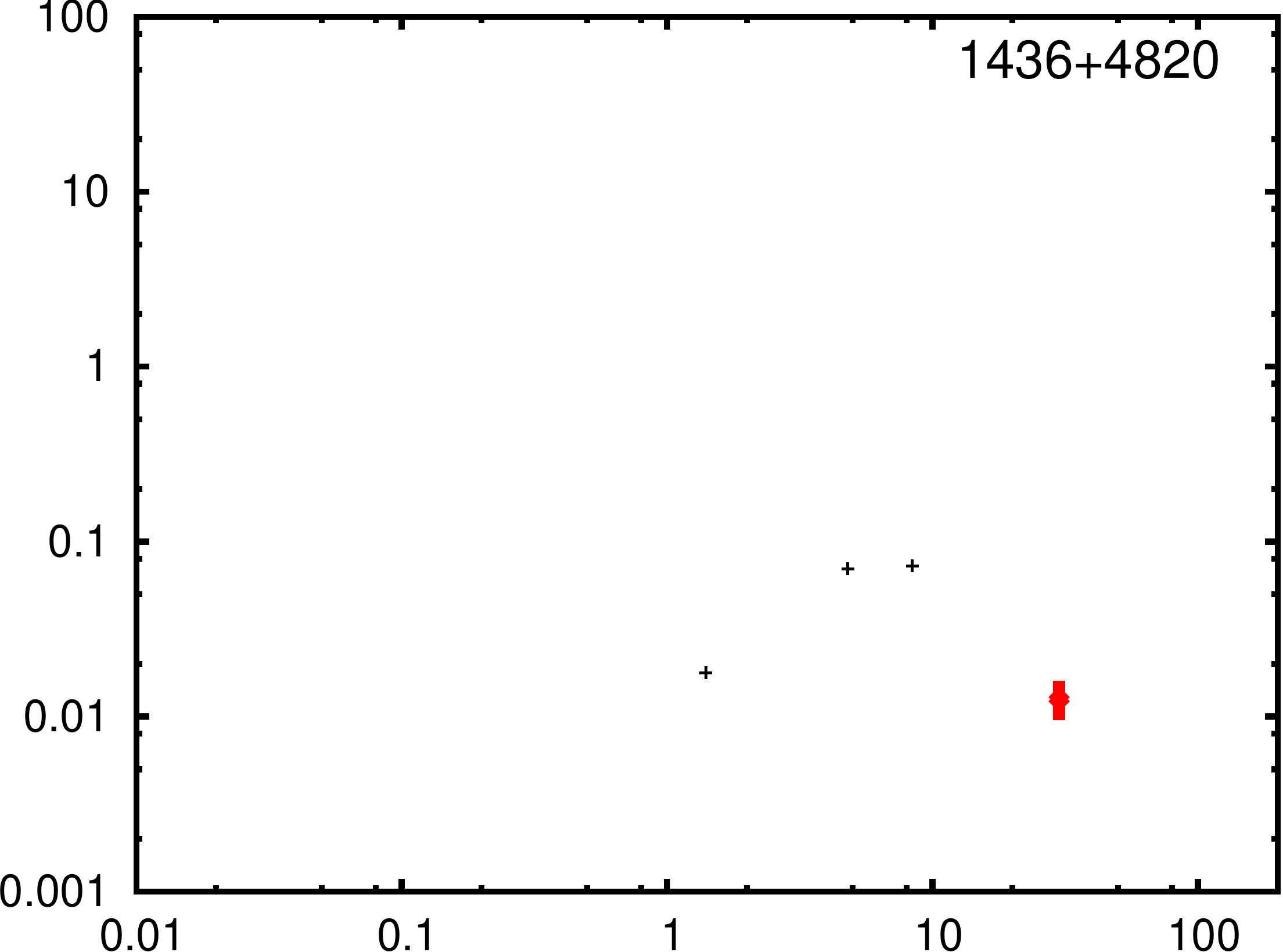}
\includegraphics[scale=0.2]{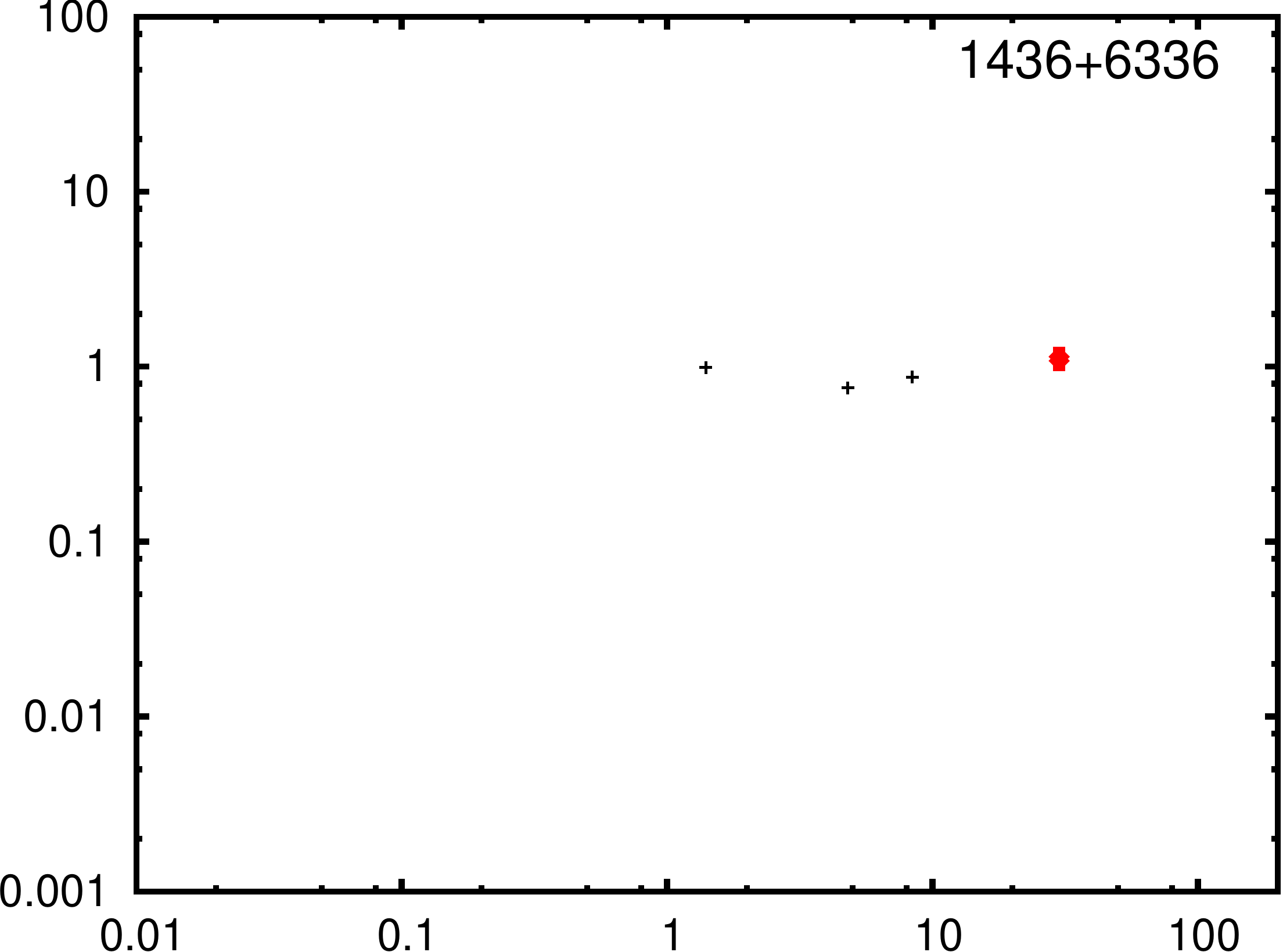}
\includegraphics[scale=0.2]{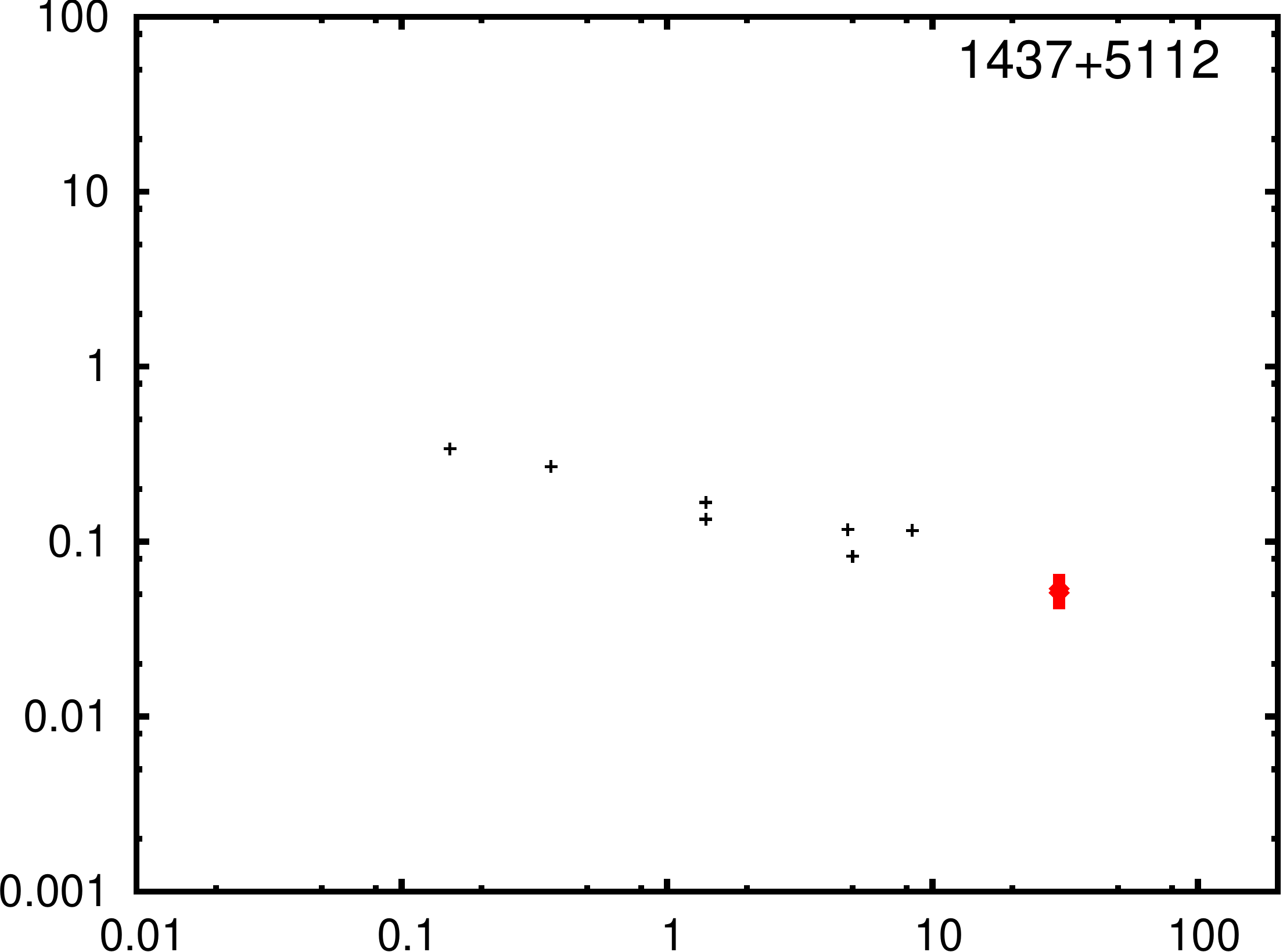}
\includegraphics[scale=0.2]{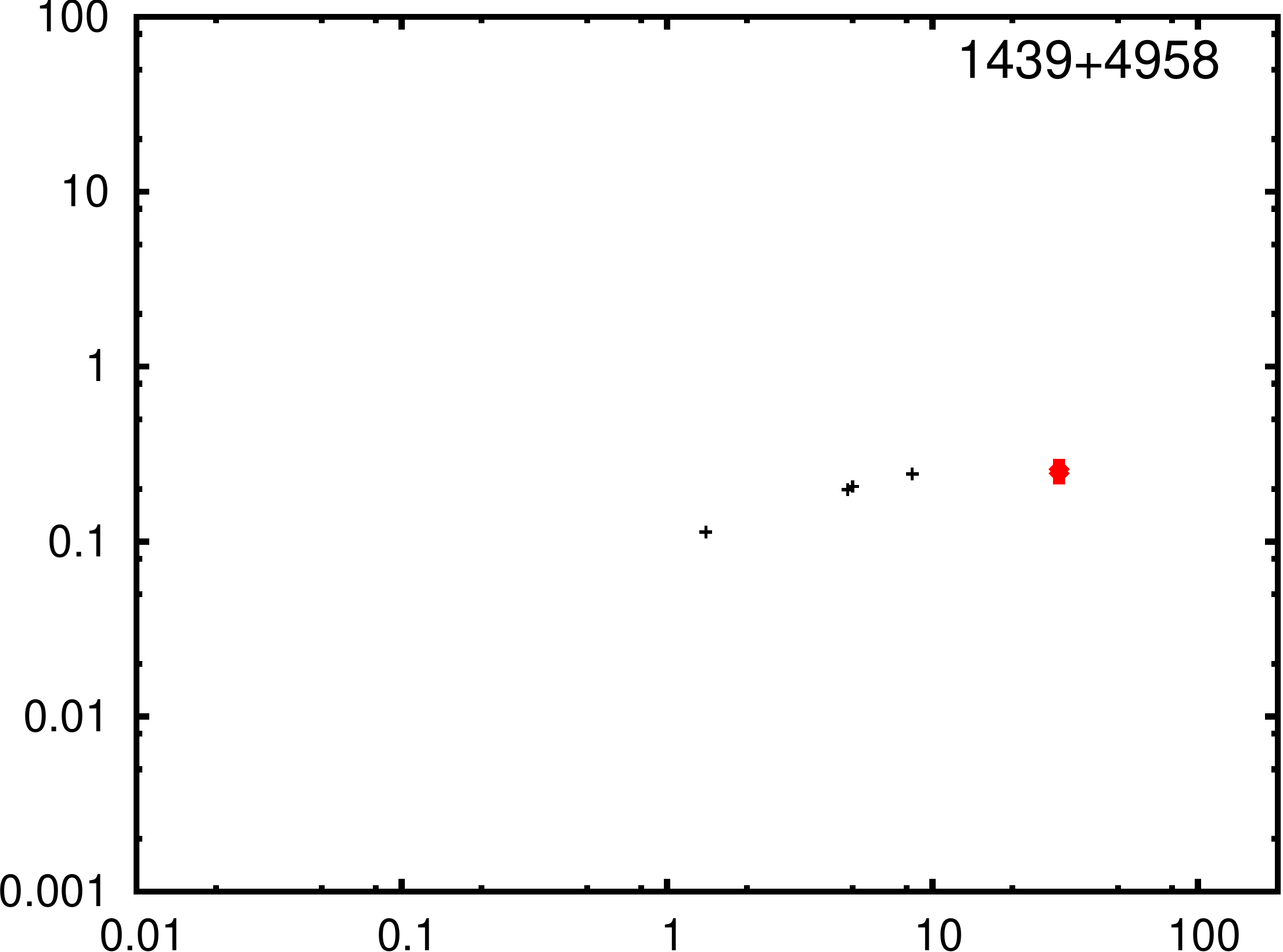}
\includegraphics[scale=0.2]{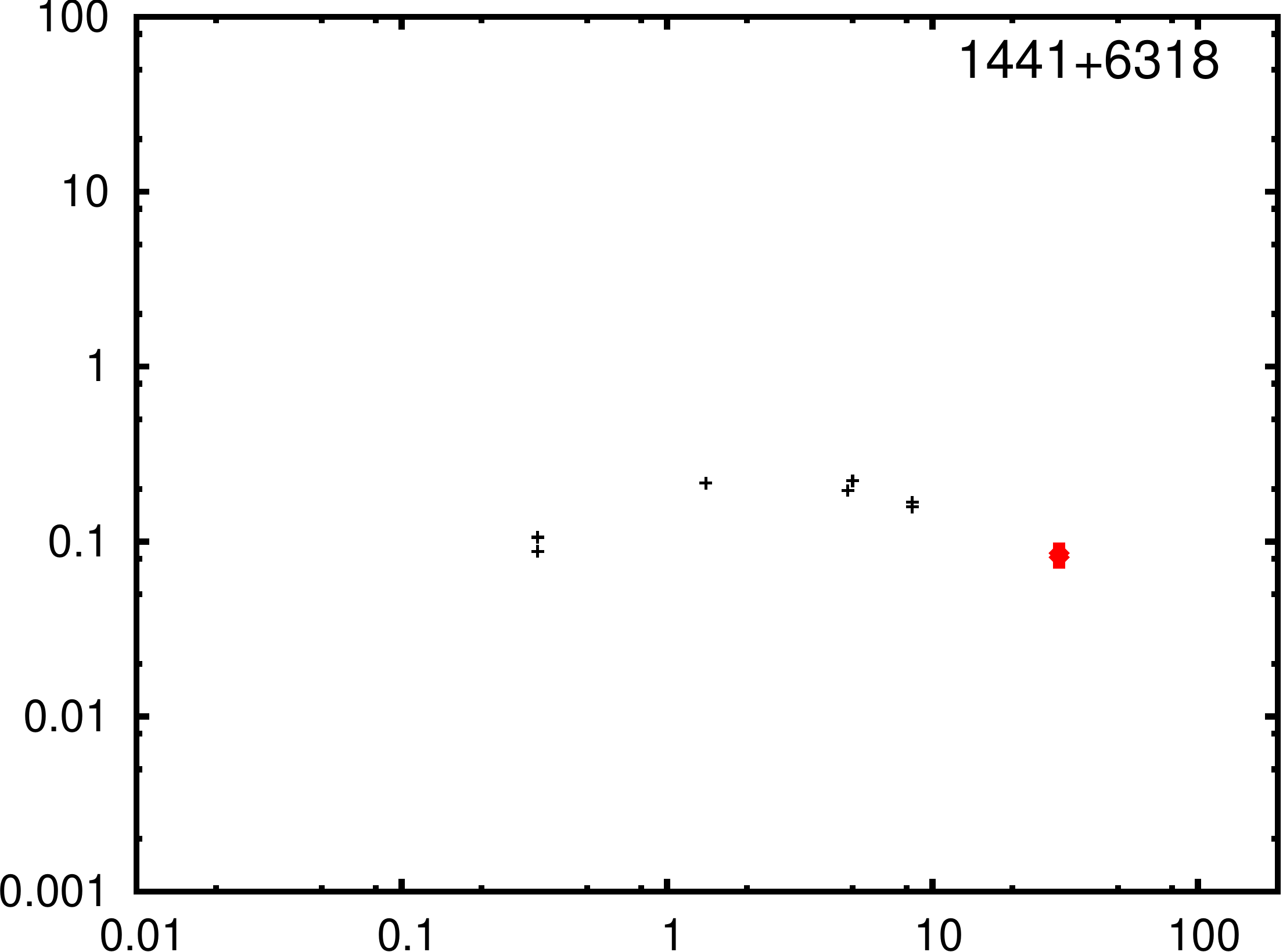}
\includegraphics[scale=0.2]{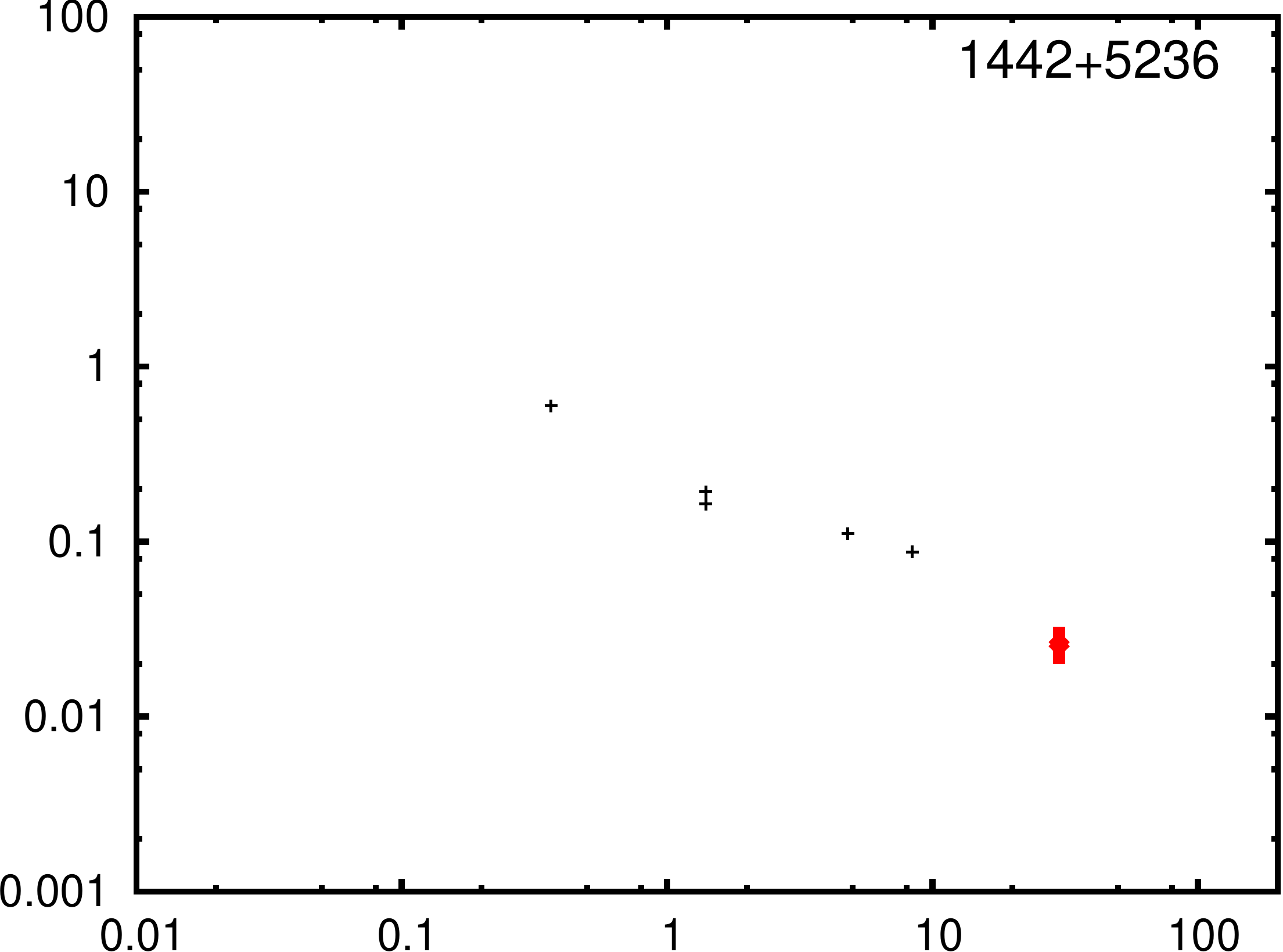}
\includegraphics[scale=0.2]{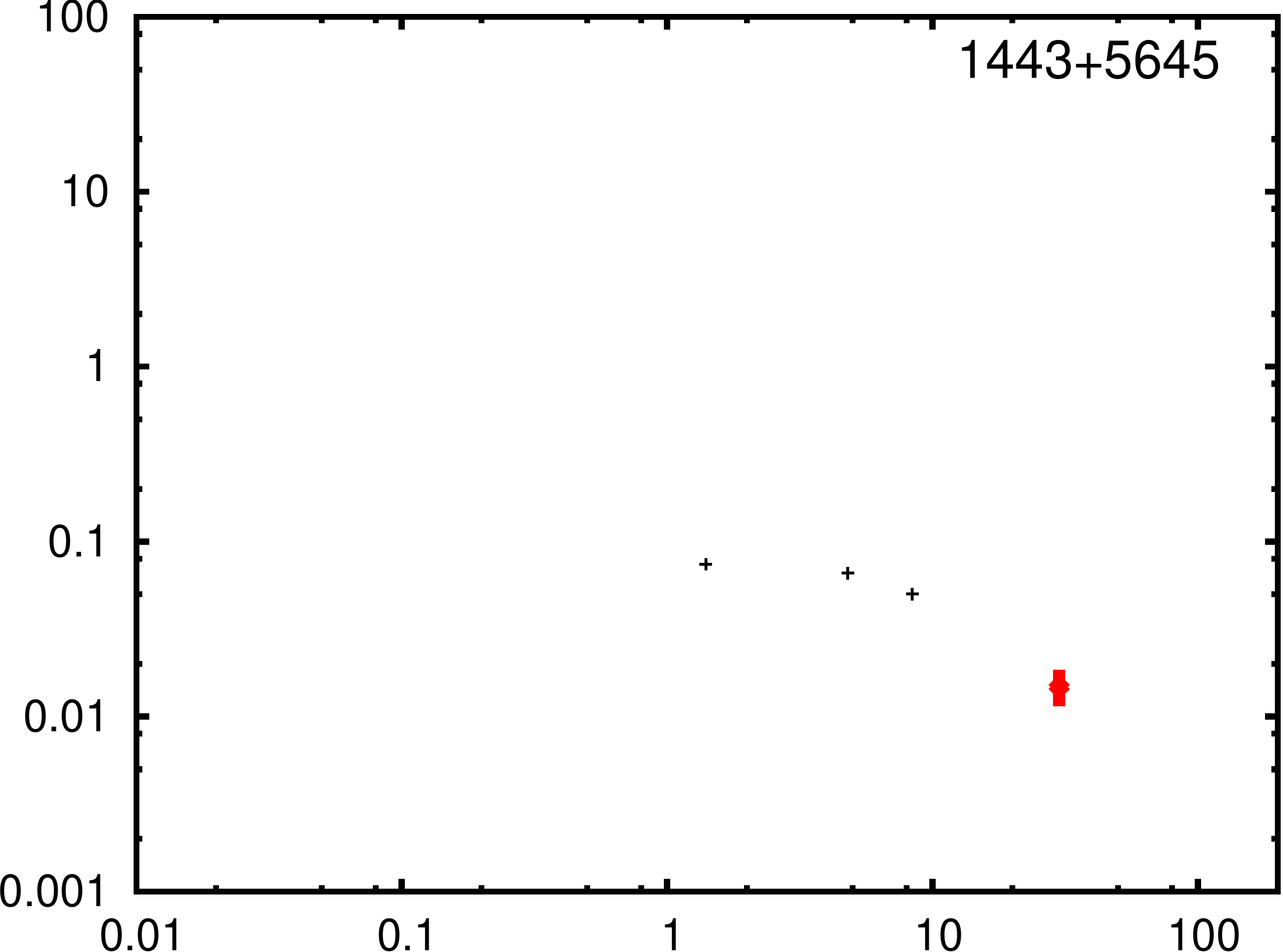}
\includegraphics[scale=0.2]{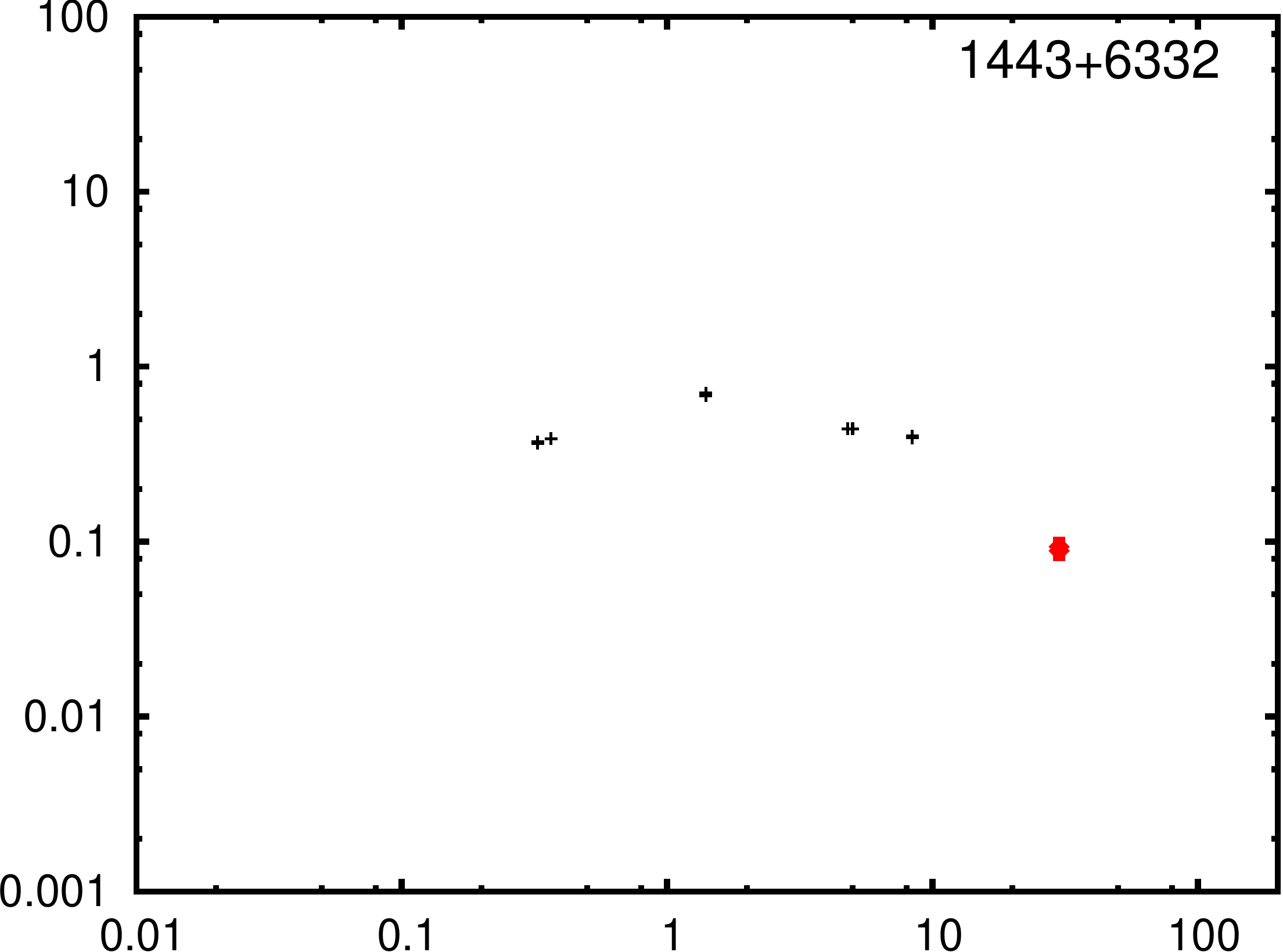}
\includegraphics[scale=0.2]{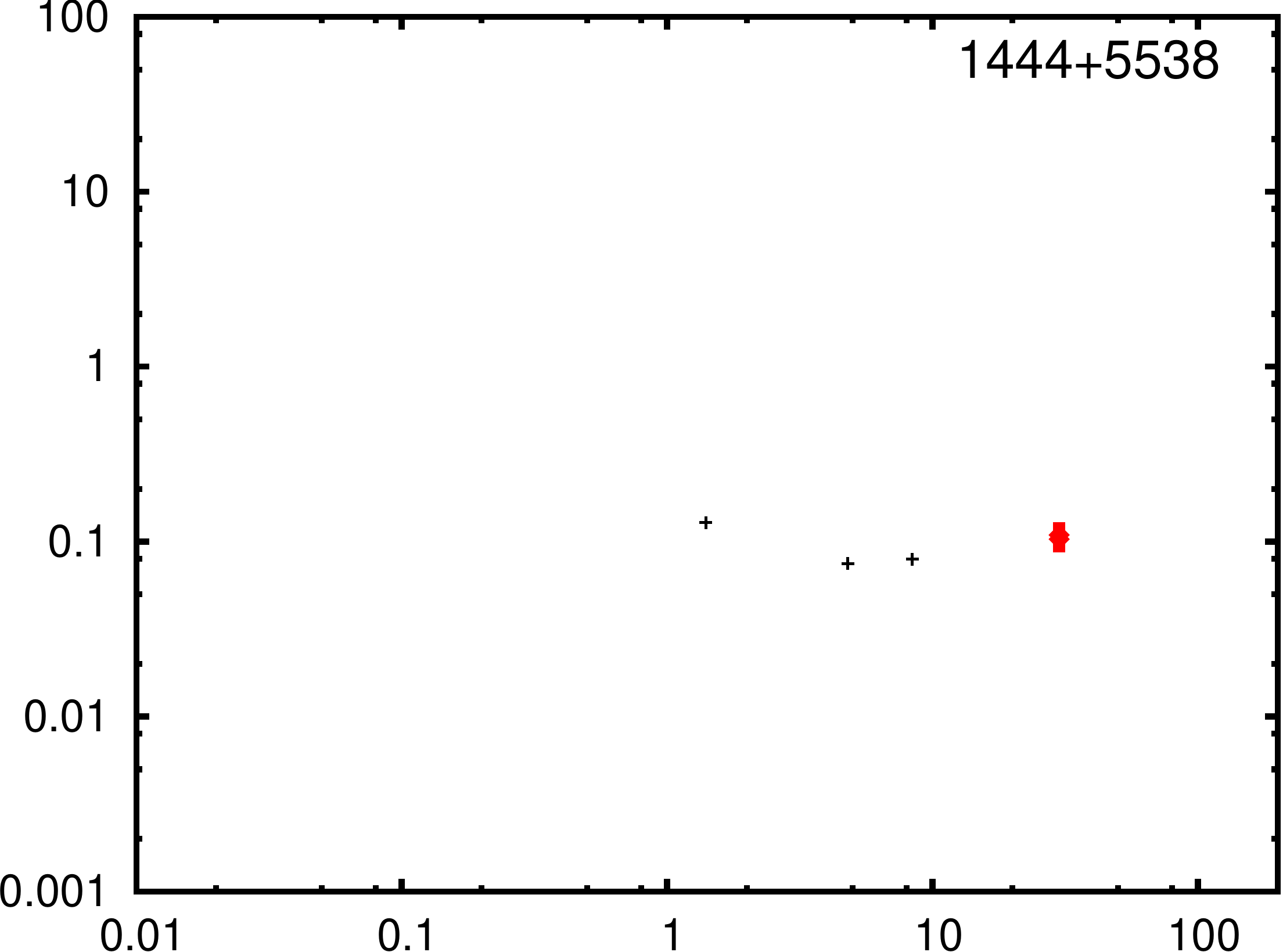}
\includegraphics[scale=0.2]{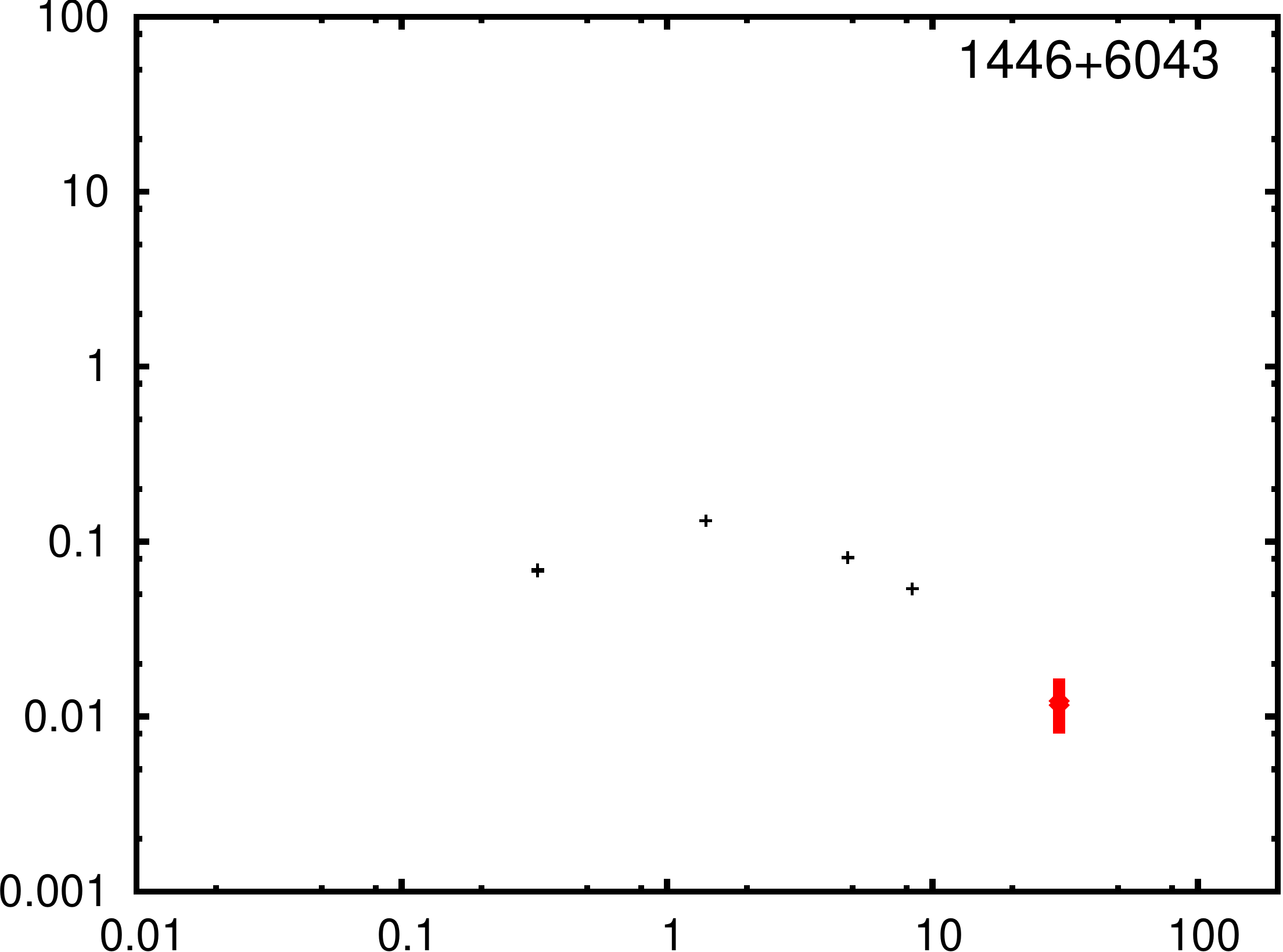}
\includegraphics[scale=0.2]{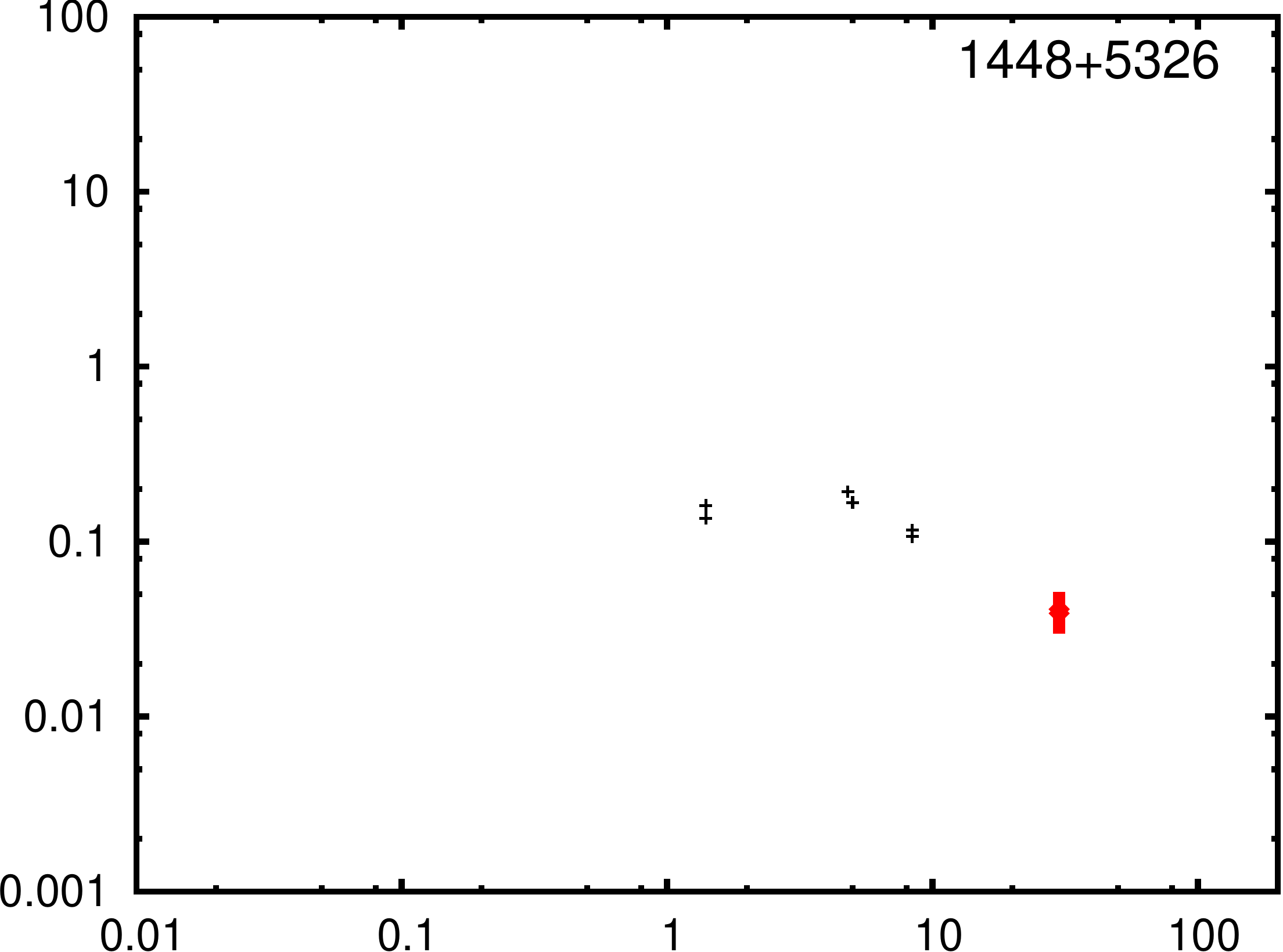}
\end{figure}
\clearpage\begin{figure}
\centering
\includegraphics[scale=0.2]{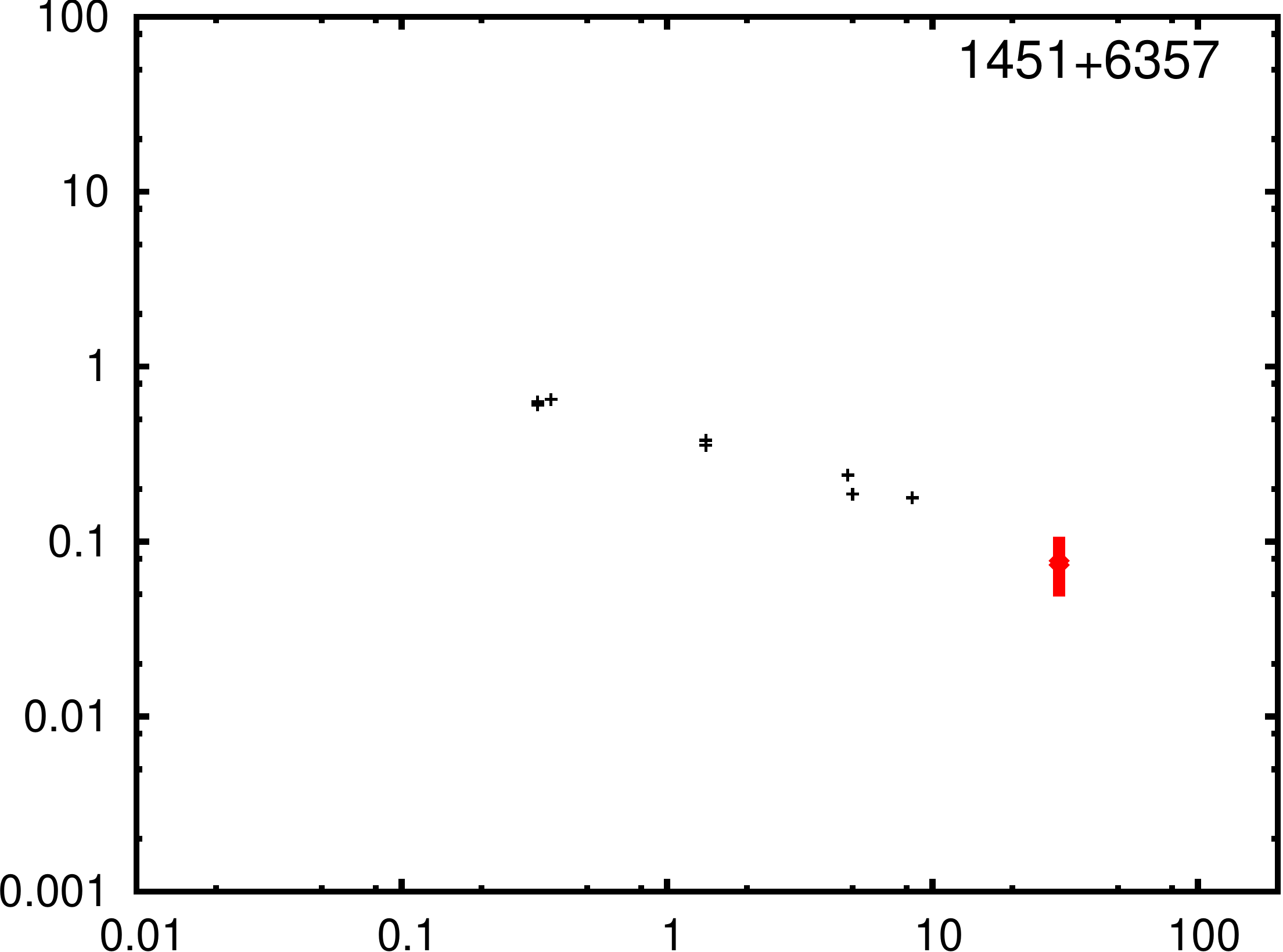}
\includegraphics[scale=0.2]{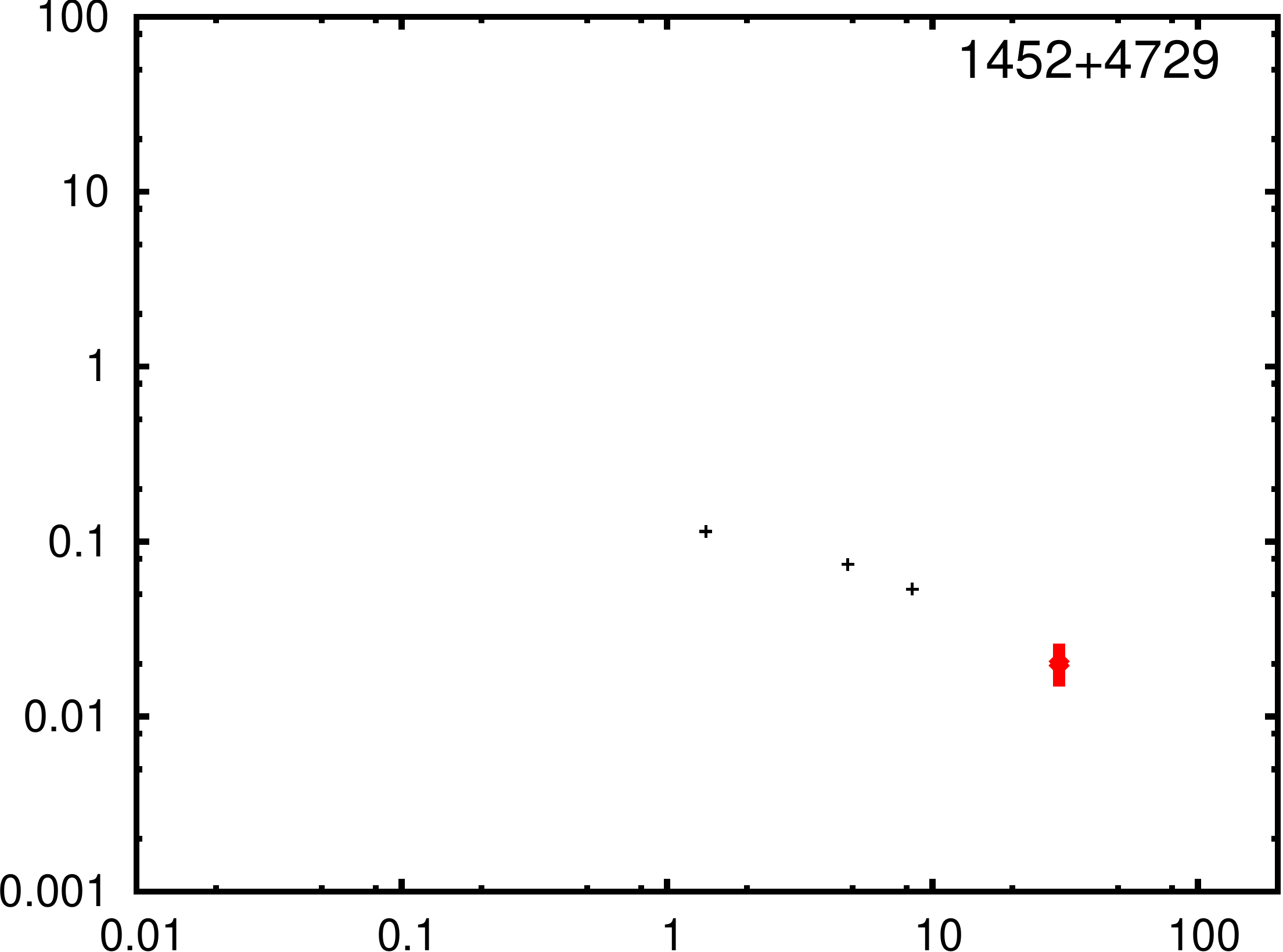}
\includegraphics[scale=0.2]{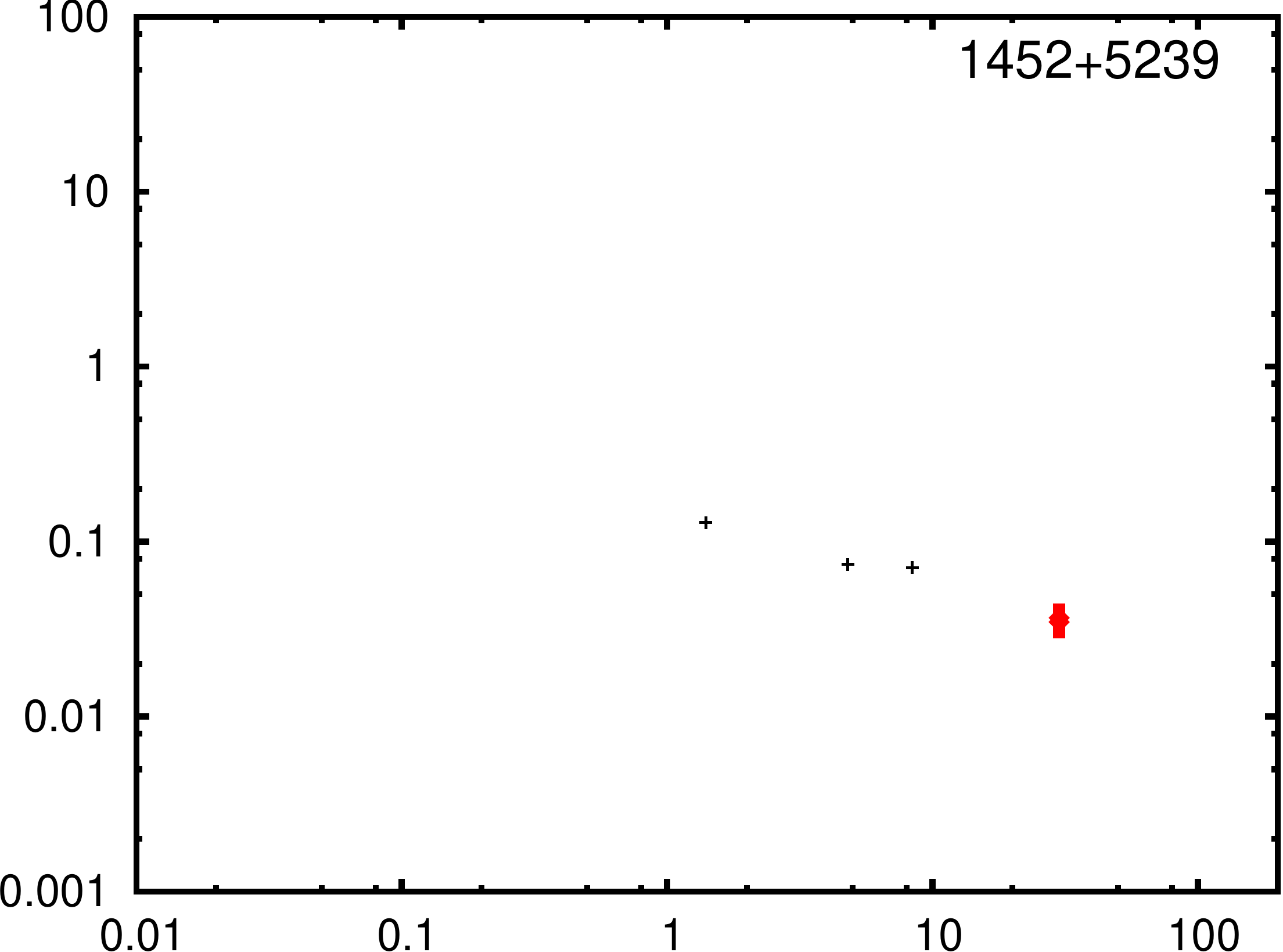}
\includegraphics[scale=0.2]{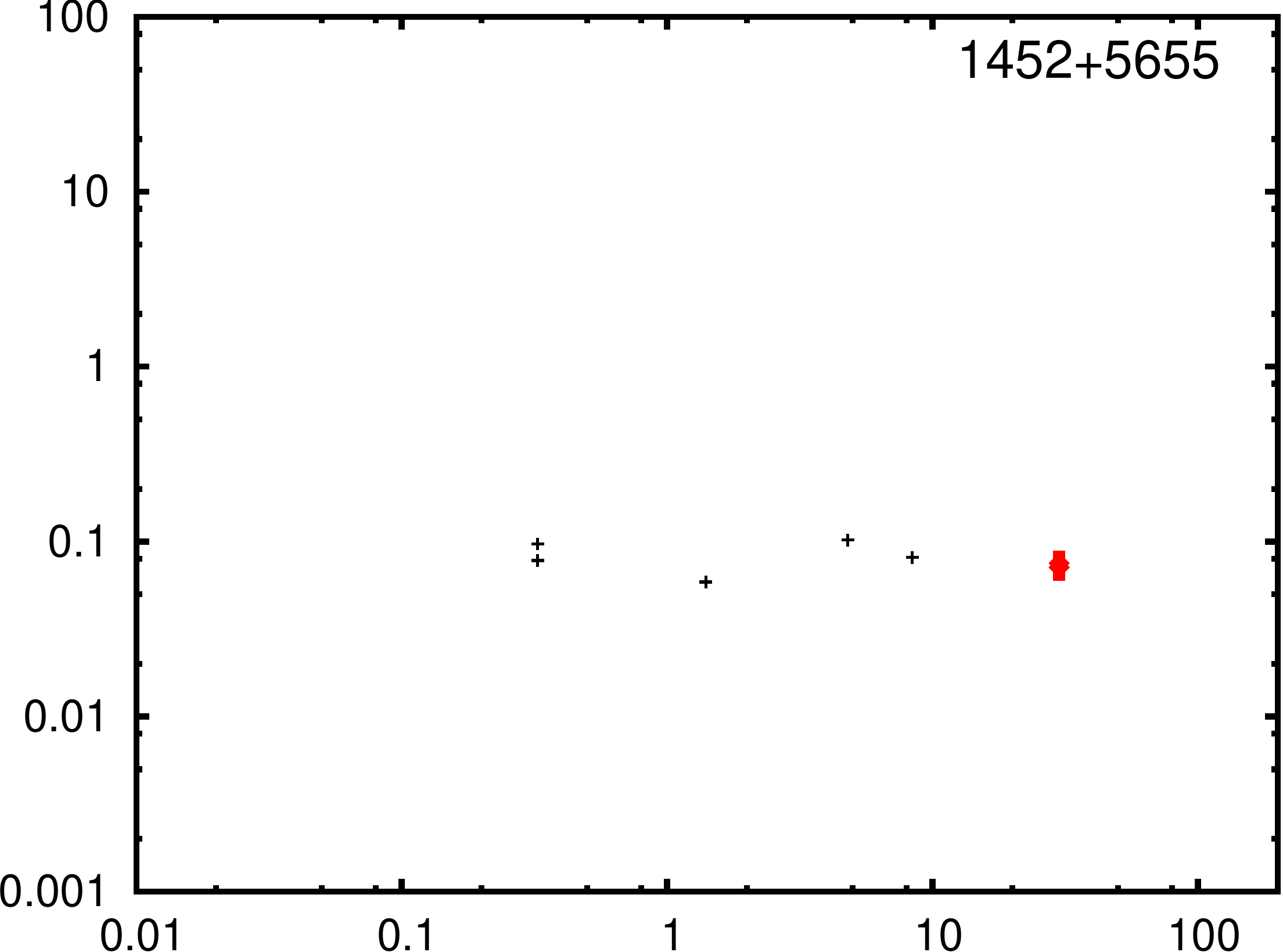}
\includegraphics[scale=0.2]{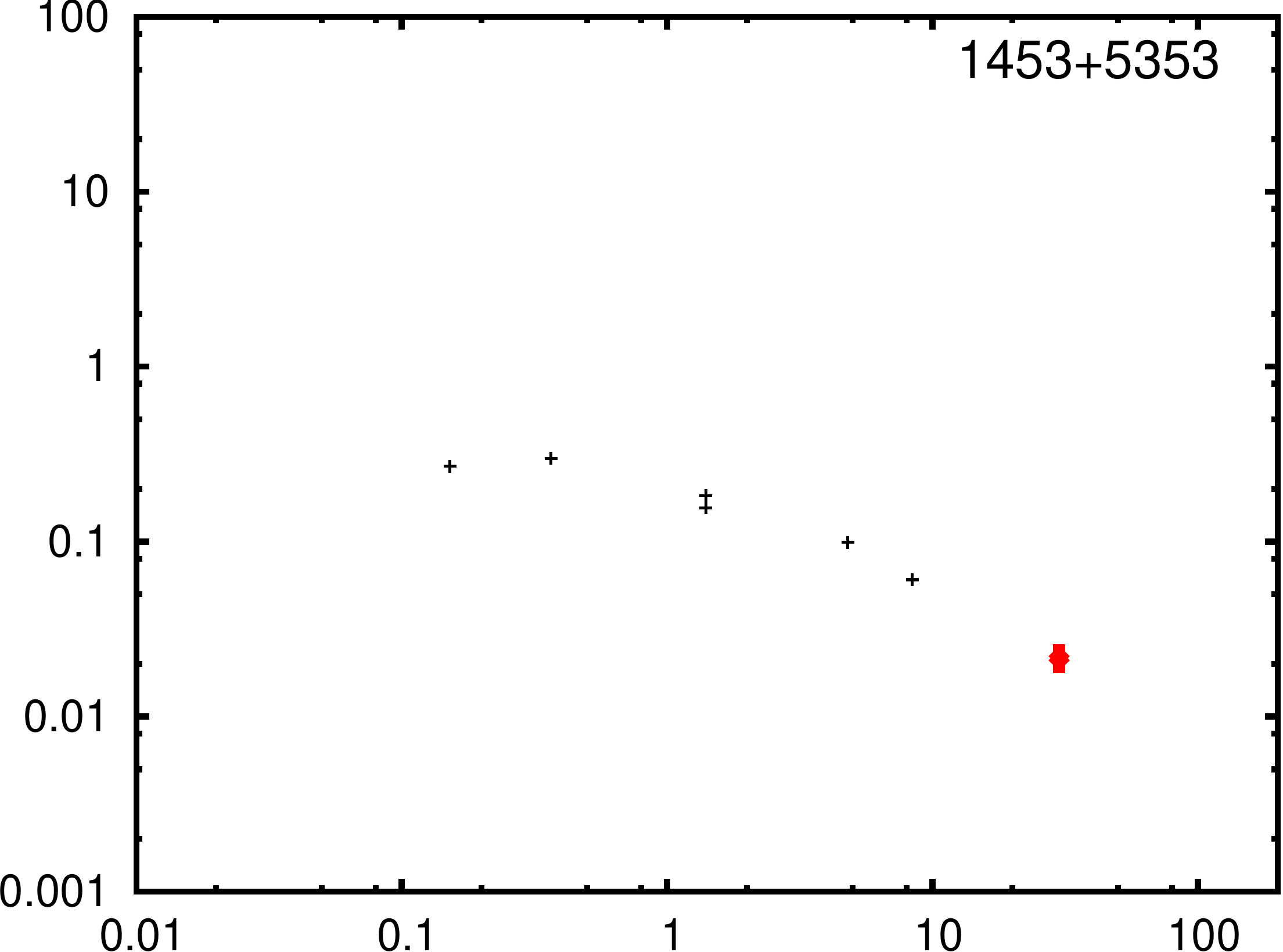}
\includegraphics[scale=0.2]{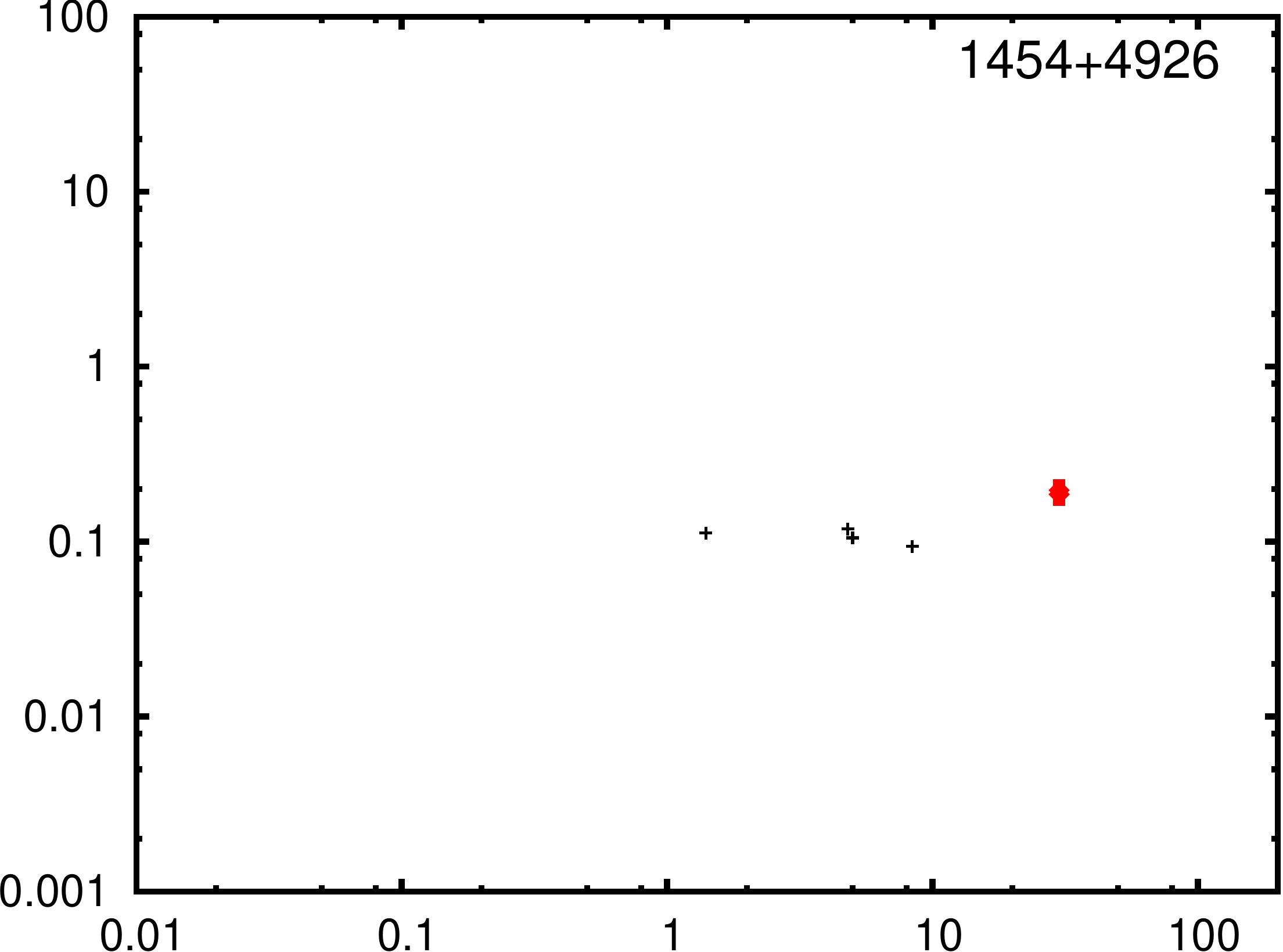}
\includegraphics[scale=0.2]{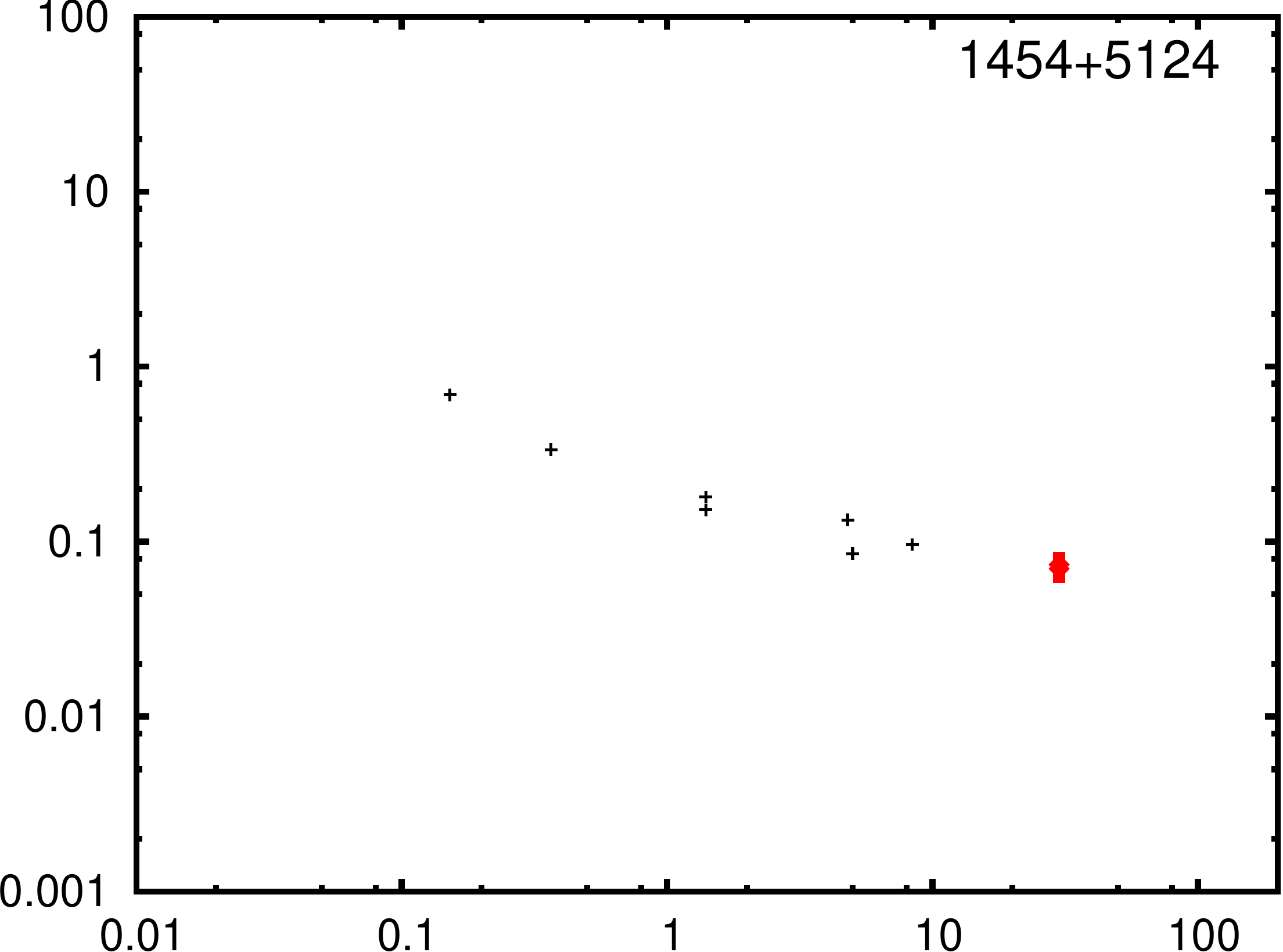}
\includegraphics[scale=0.2]{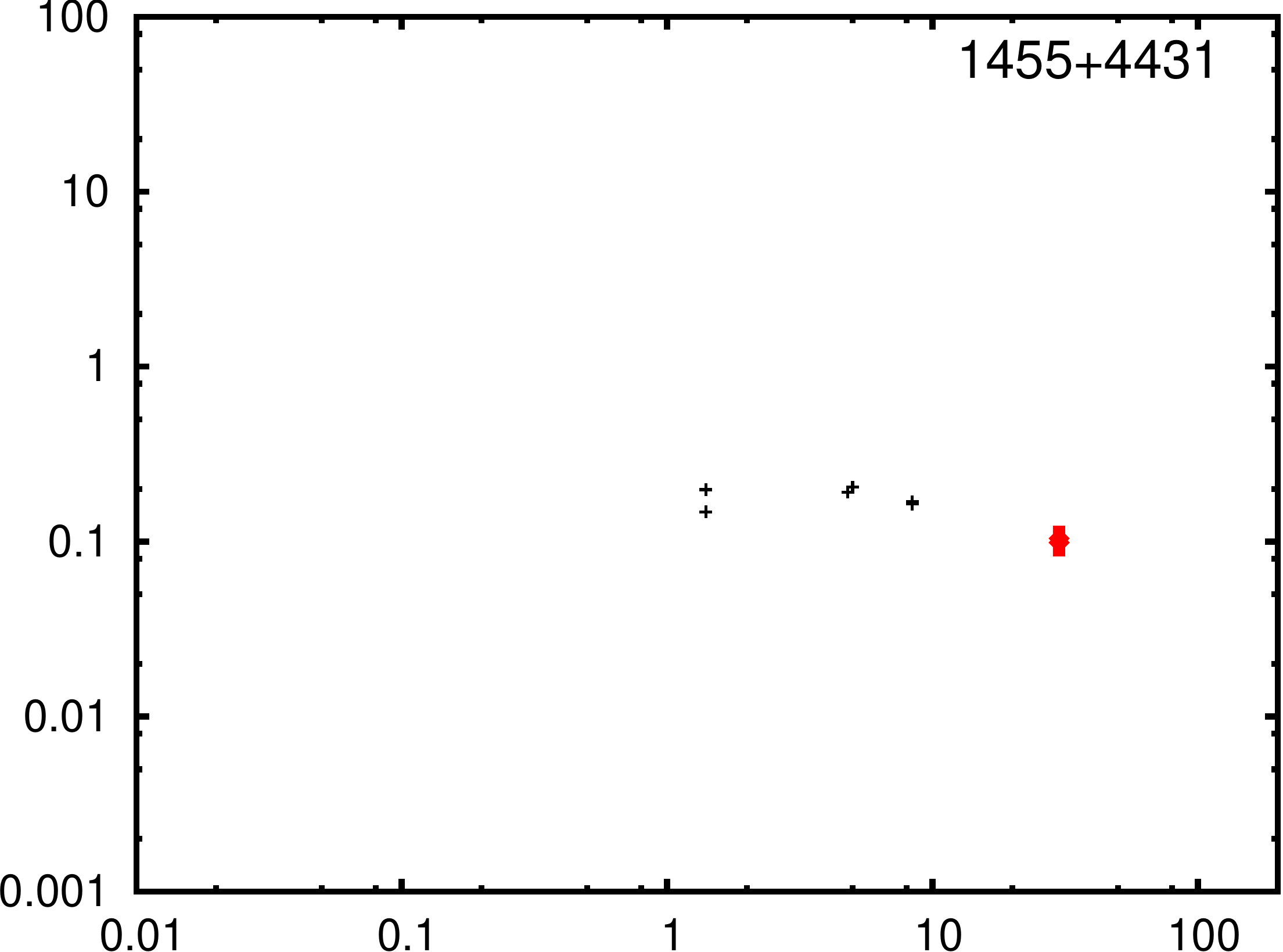}
\includegraphics[scale=0.2]{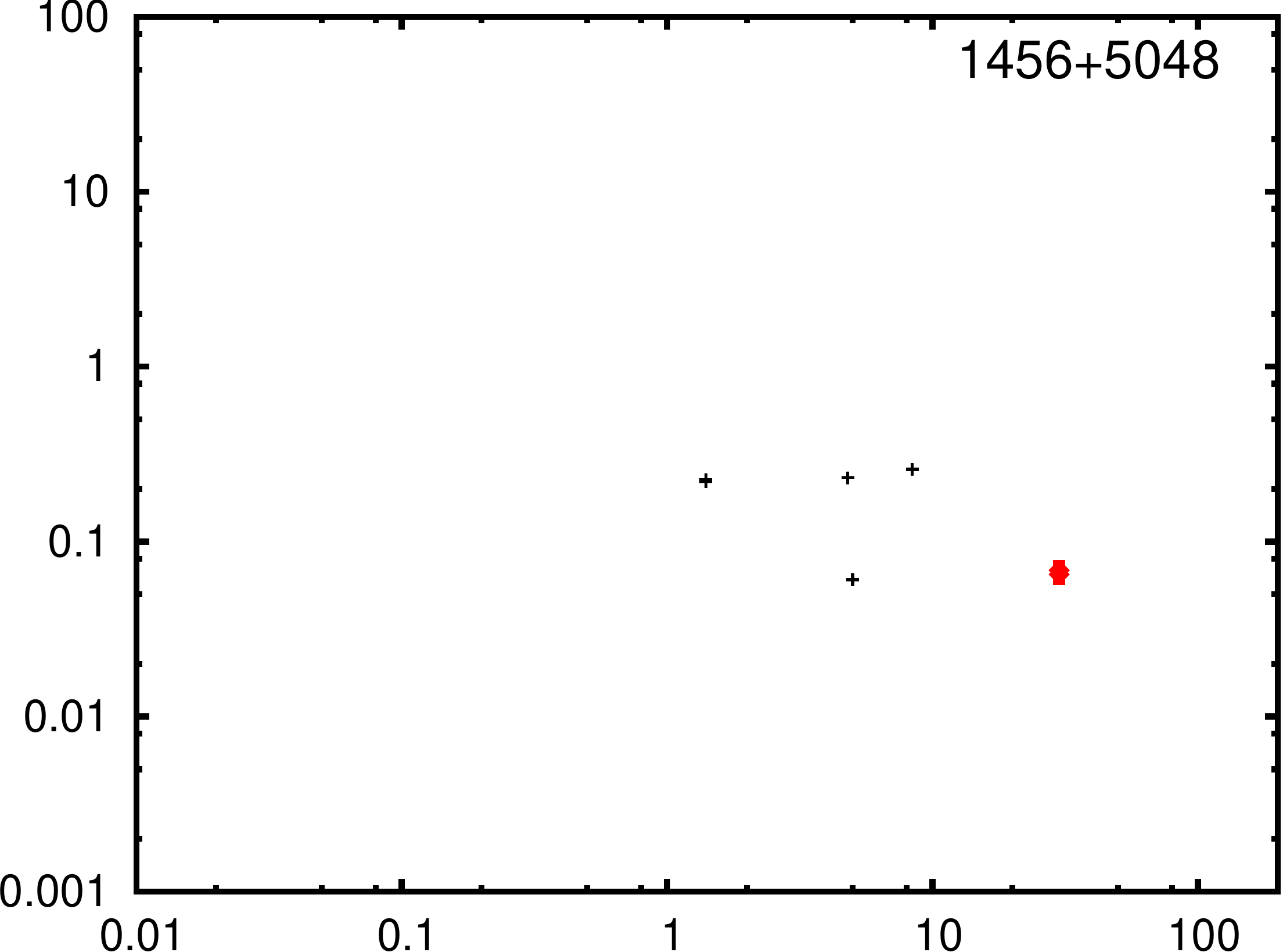}
\includegraphics[scale=0.2]{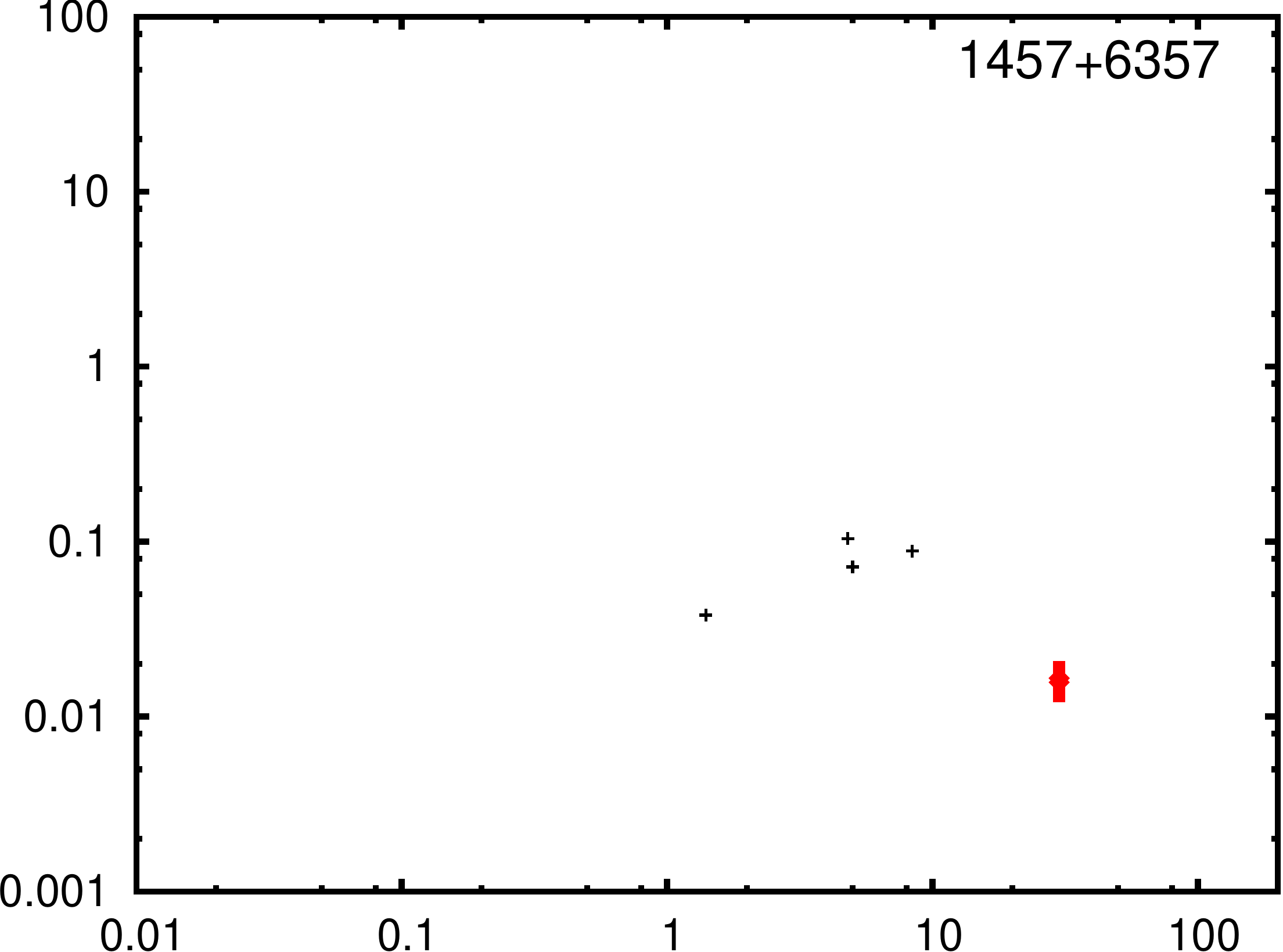}
\includegraphics[scale=0.2]{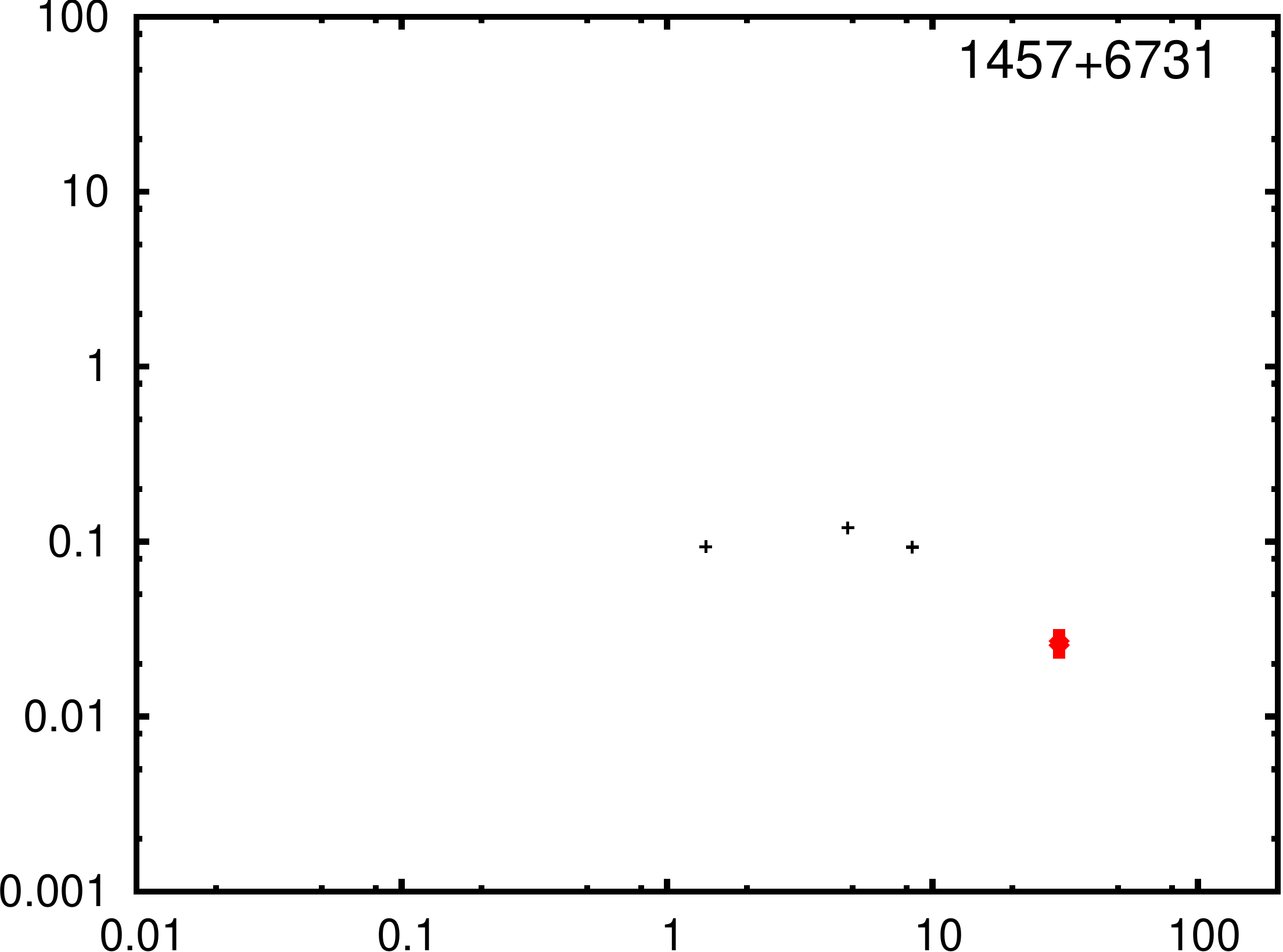}
\includegraphics[scale=0.2]{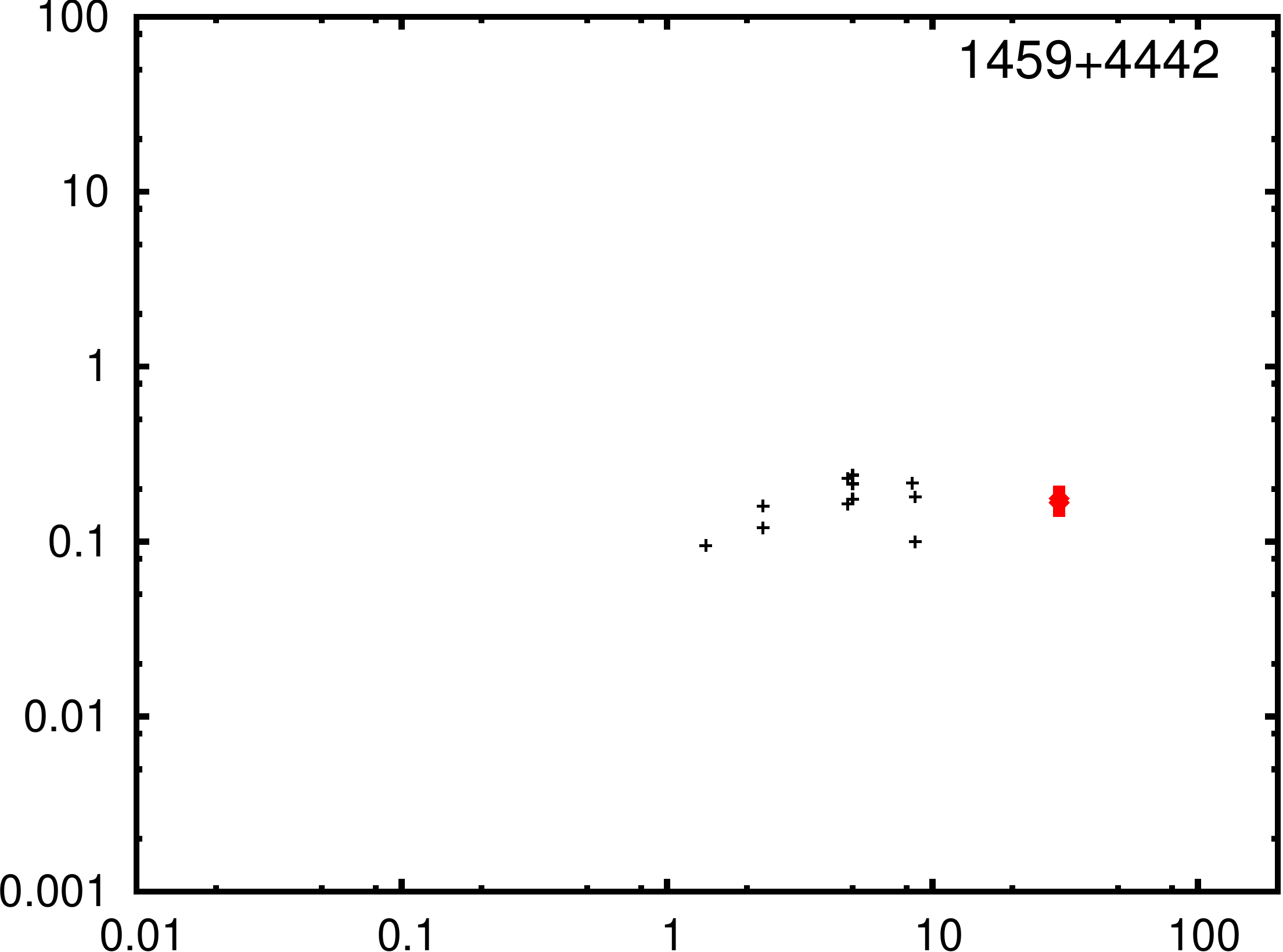}
\includegraphics[scale=0.2]{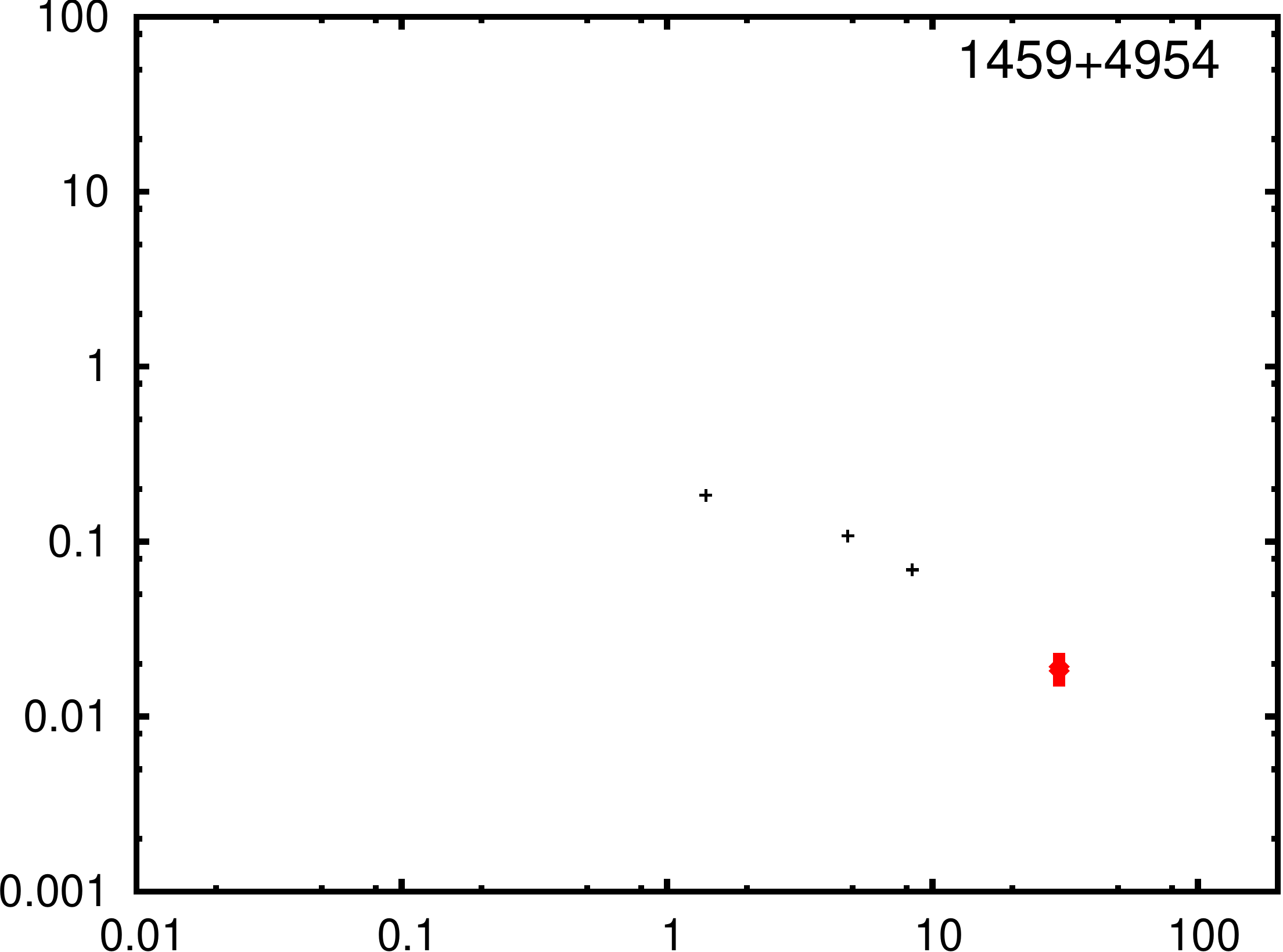}
\includegraphics[scale=0.2]{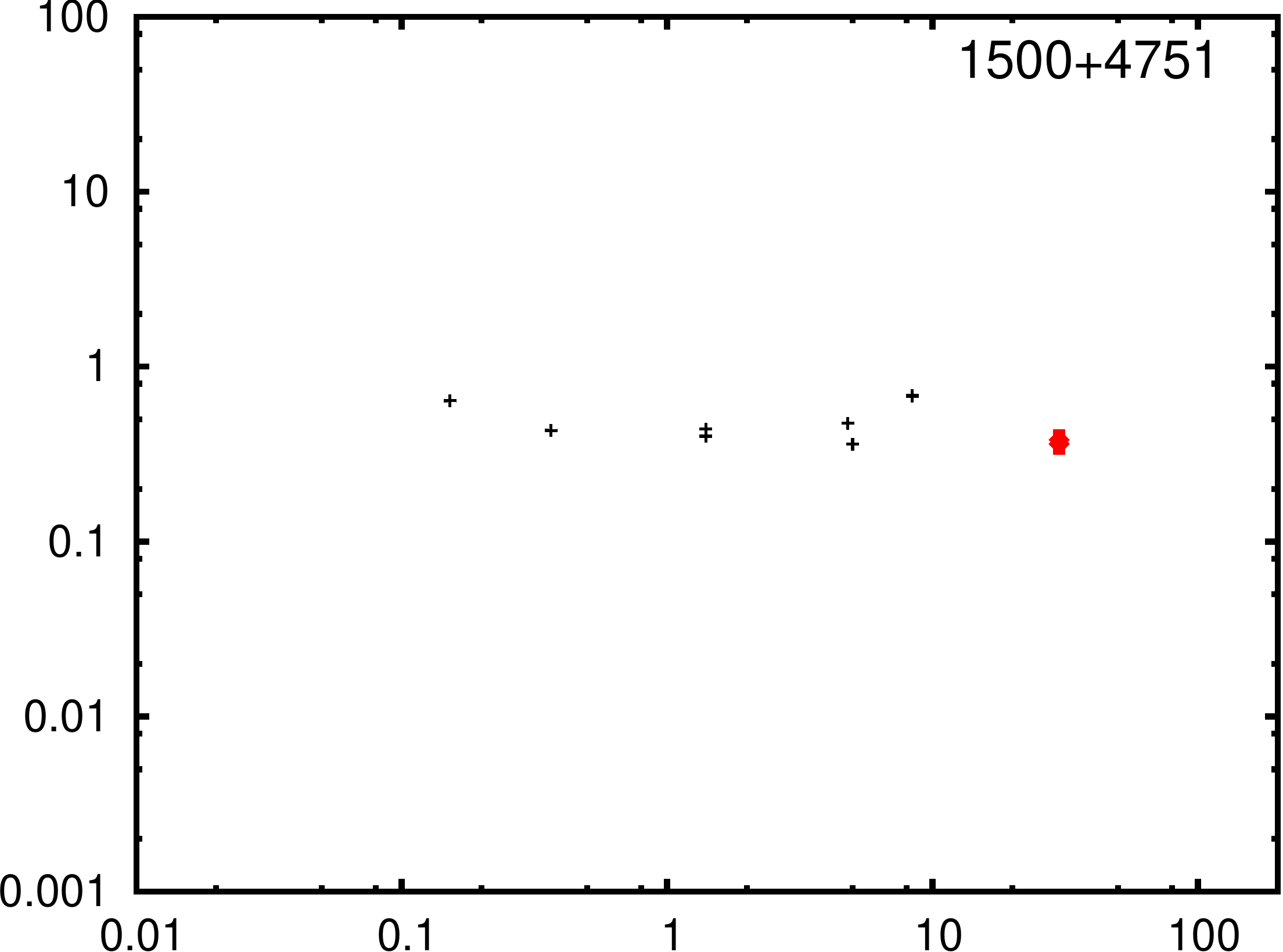}
\includegraphics[scale=0.2]{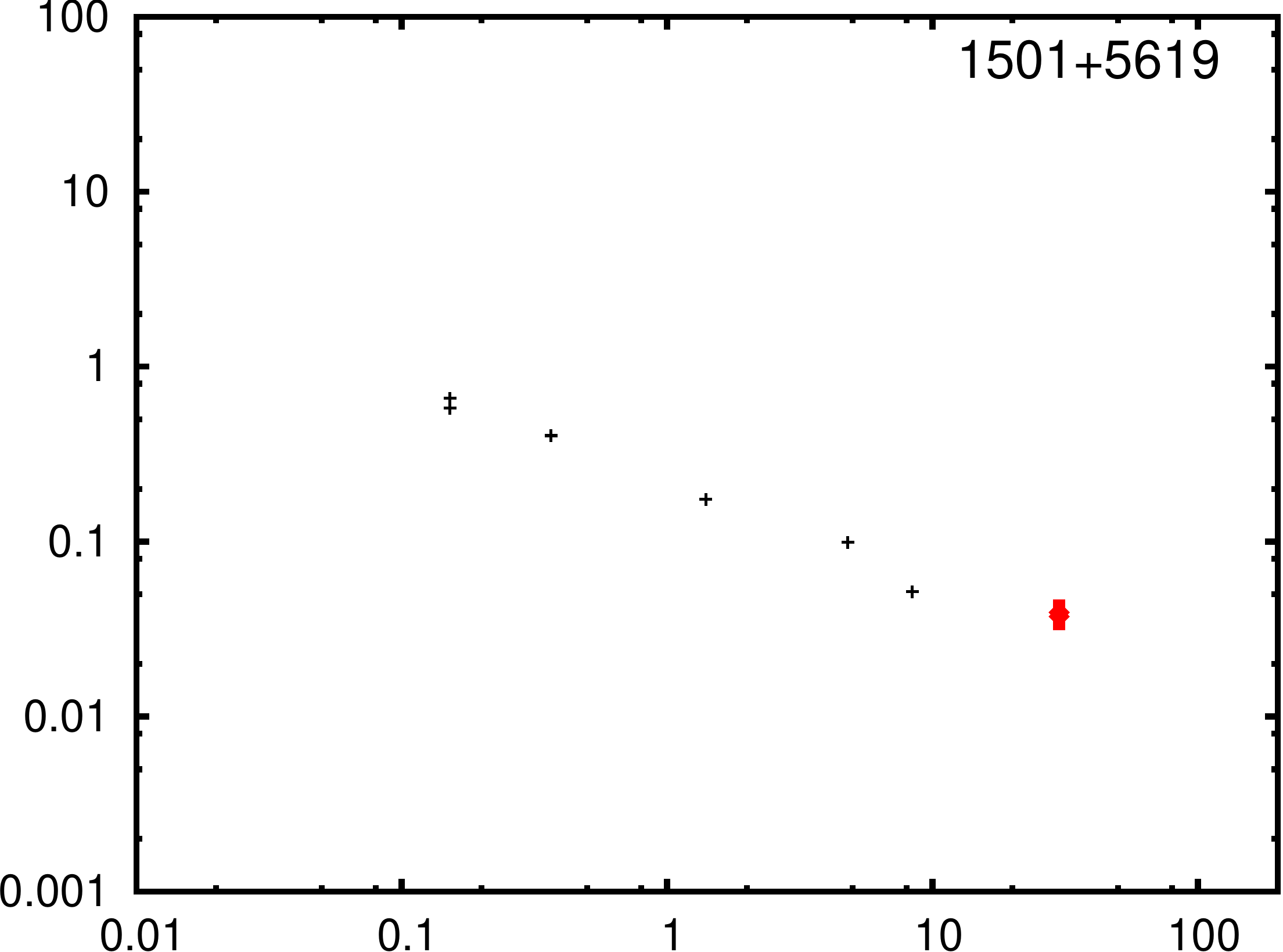}
\includegraphics[scale=0.2]{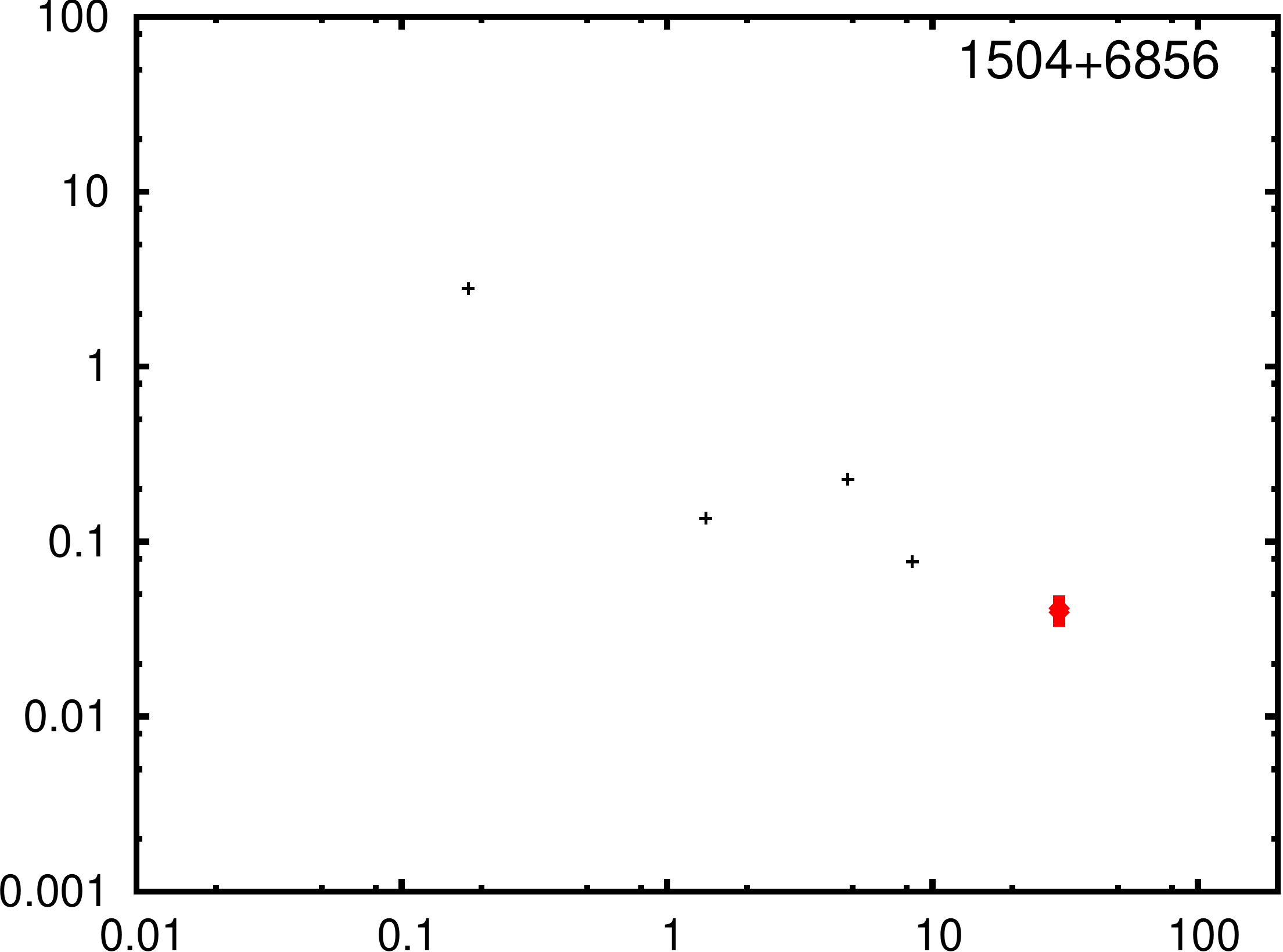}
\includegraphics[scale=0.2]{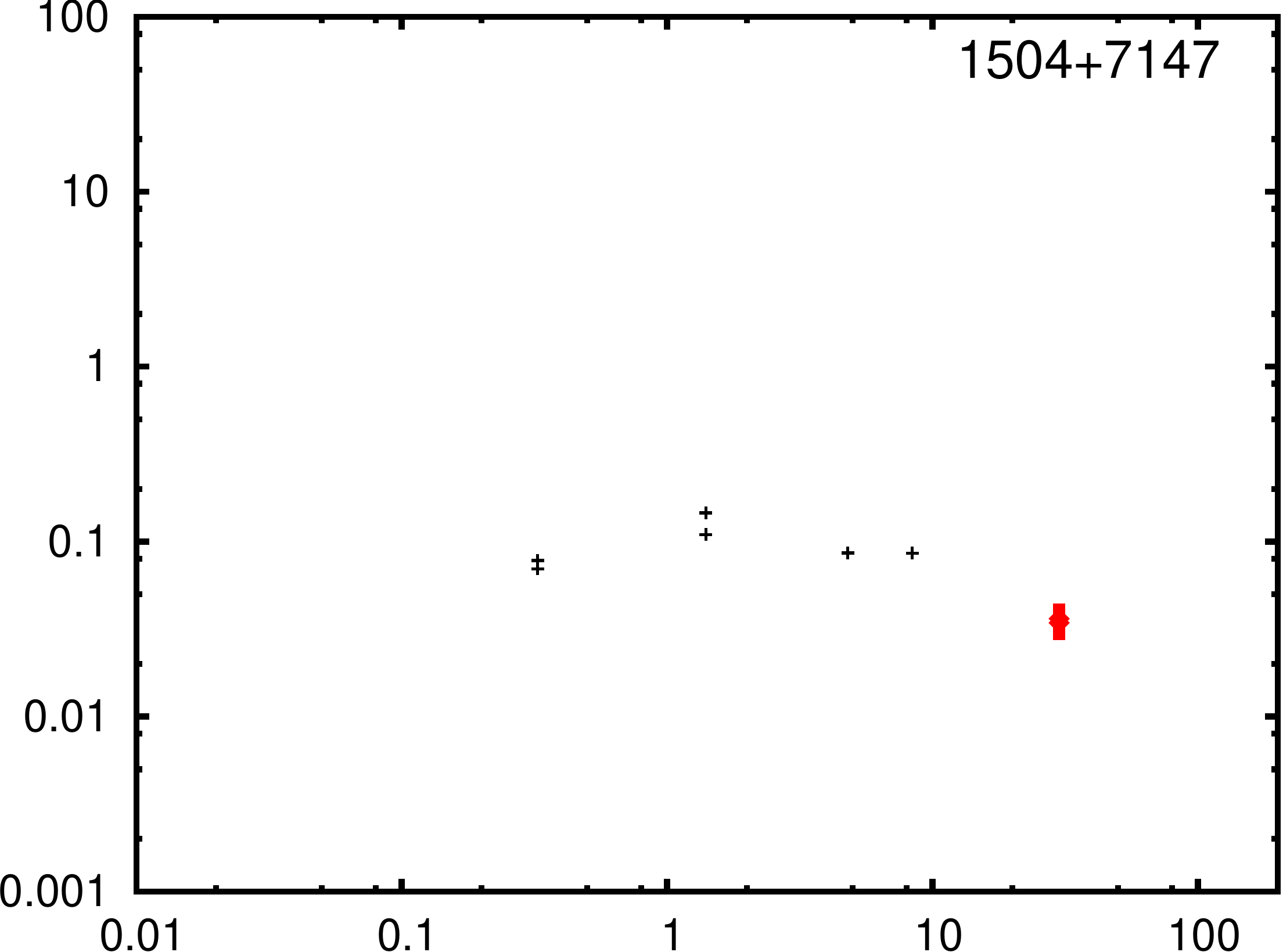}
\includegraphics[scale=0.2]{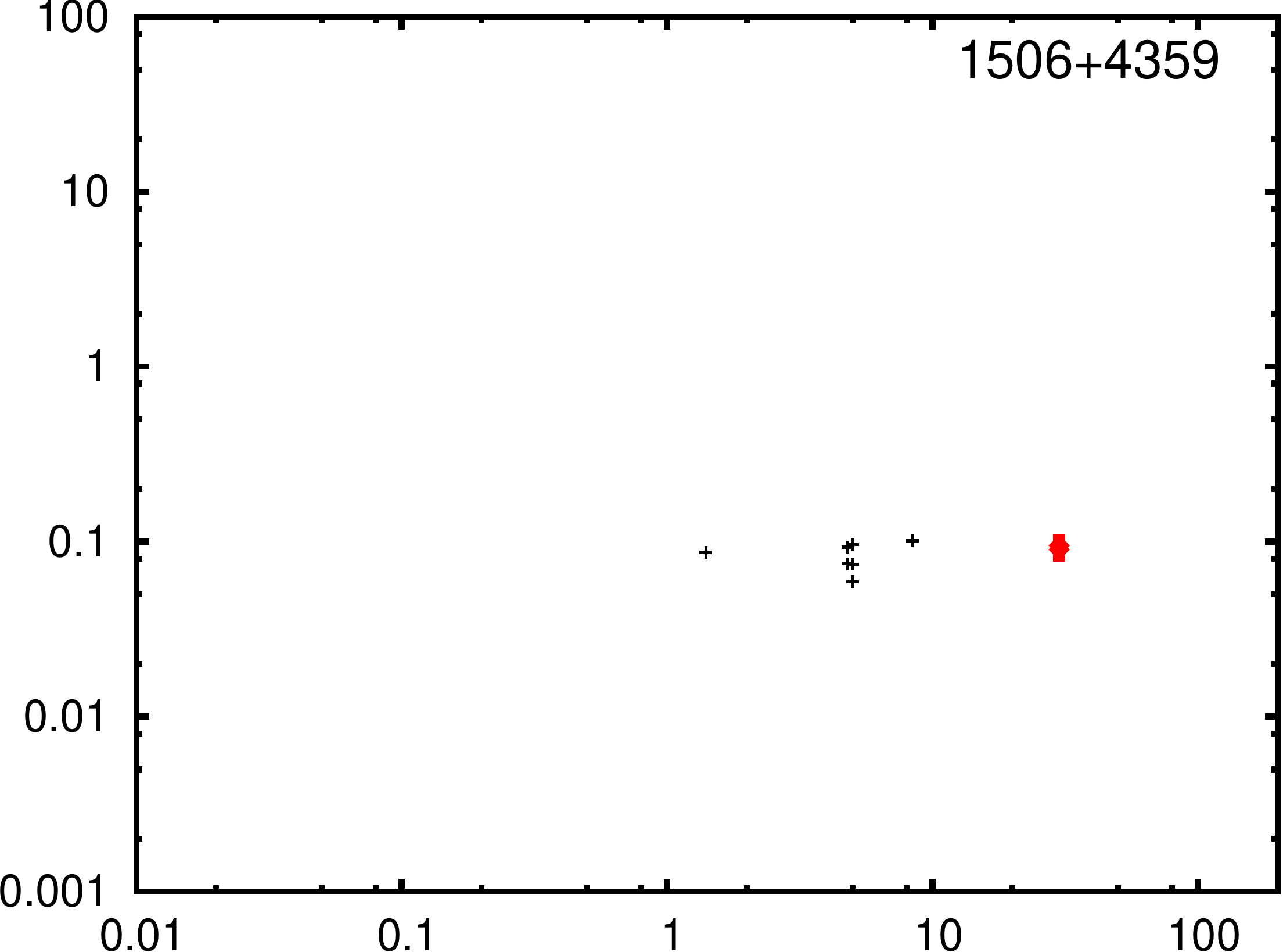}
\end{figure}
\clearpage\begin{figure}
\centering
\includegraphics[scale=0.2]{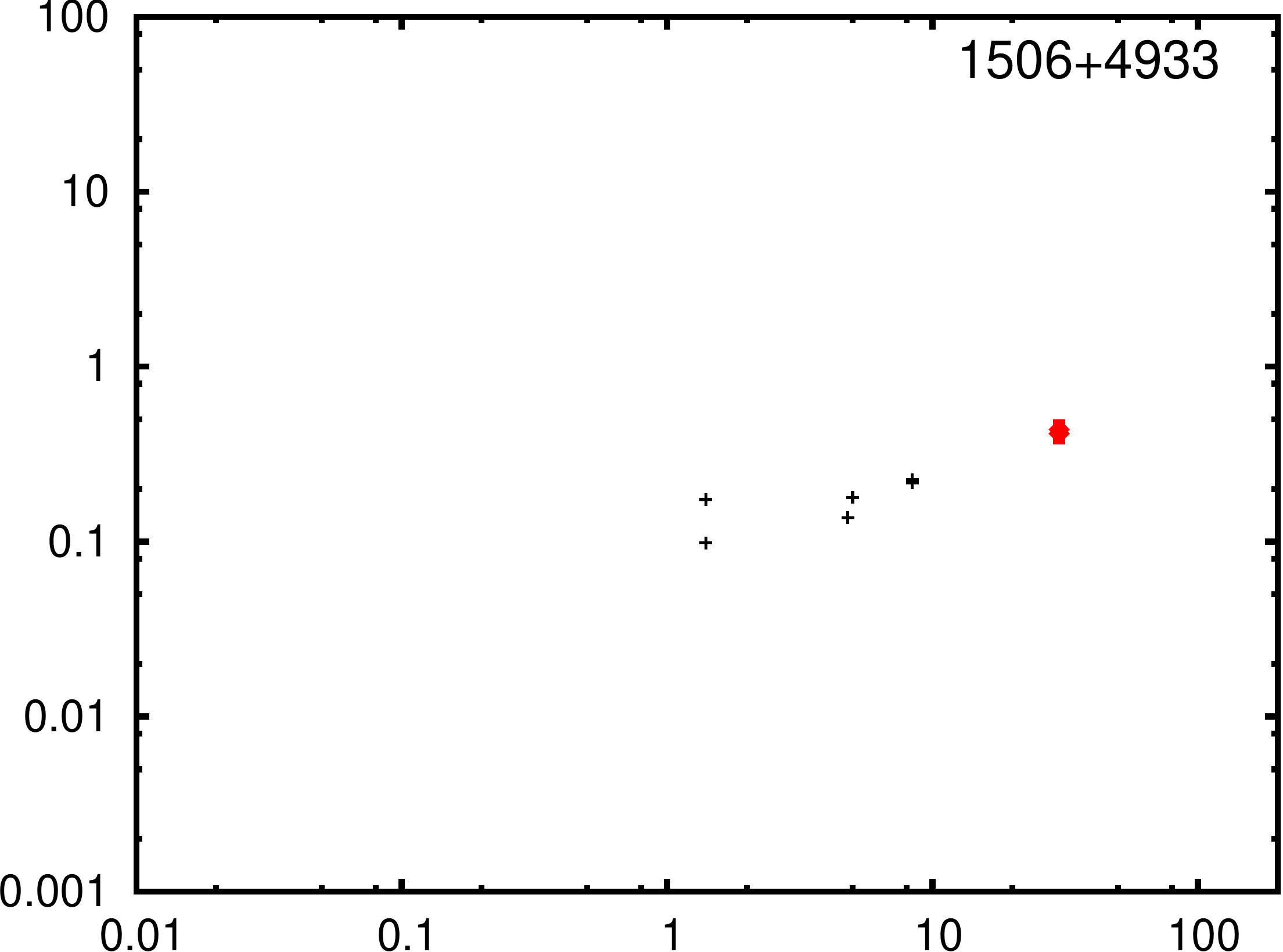}
\includegraphics[scale=0.2]{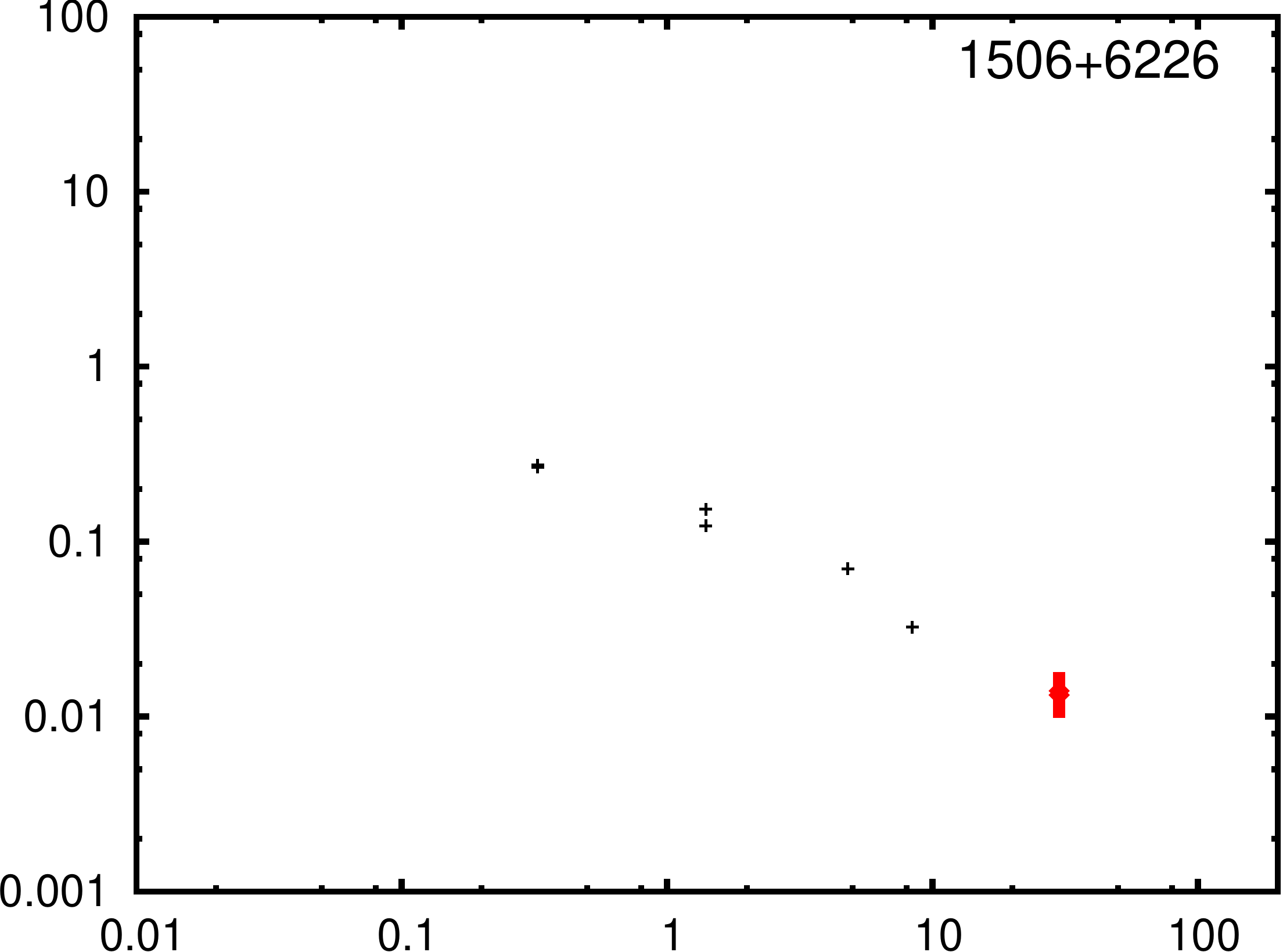}
\includegraphics[scale=0.2]{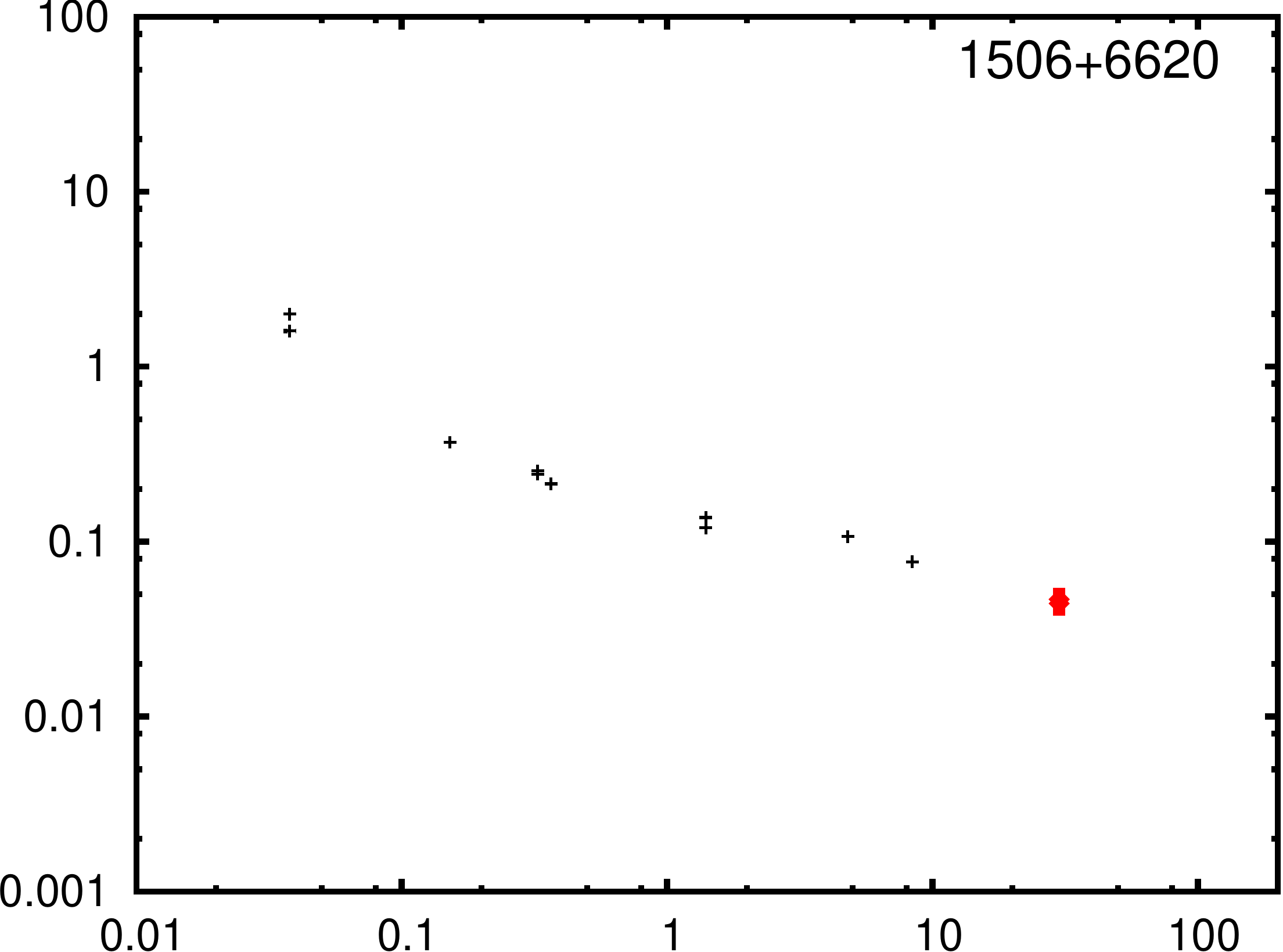}
\includegraphics[scale=0.2]{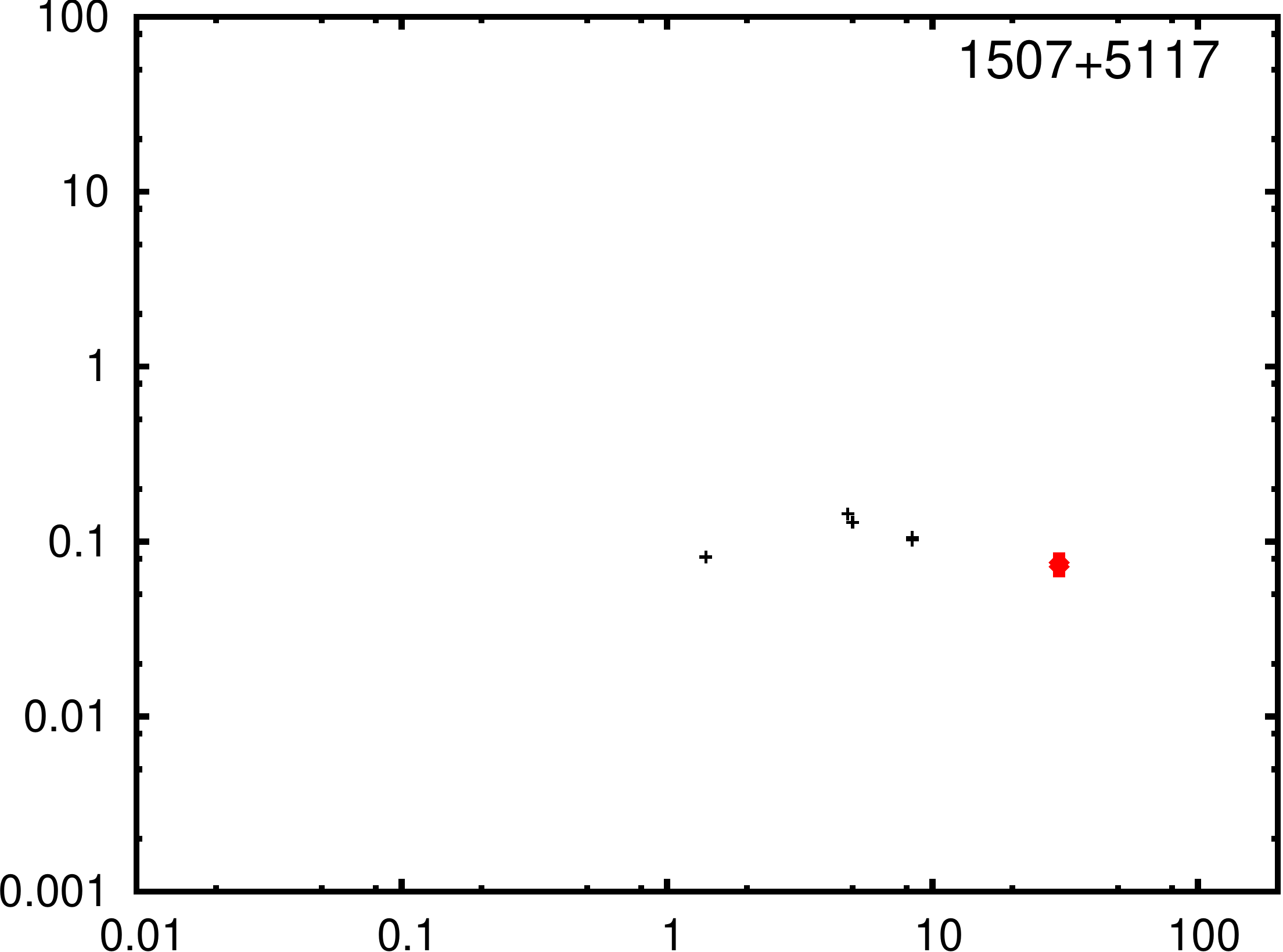}
\includegraphics[scale=0.2]{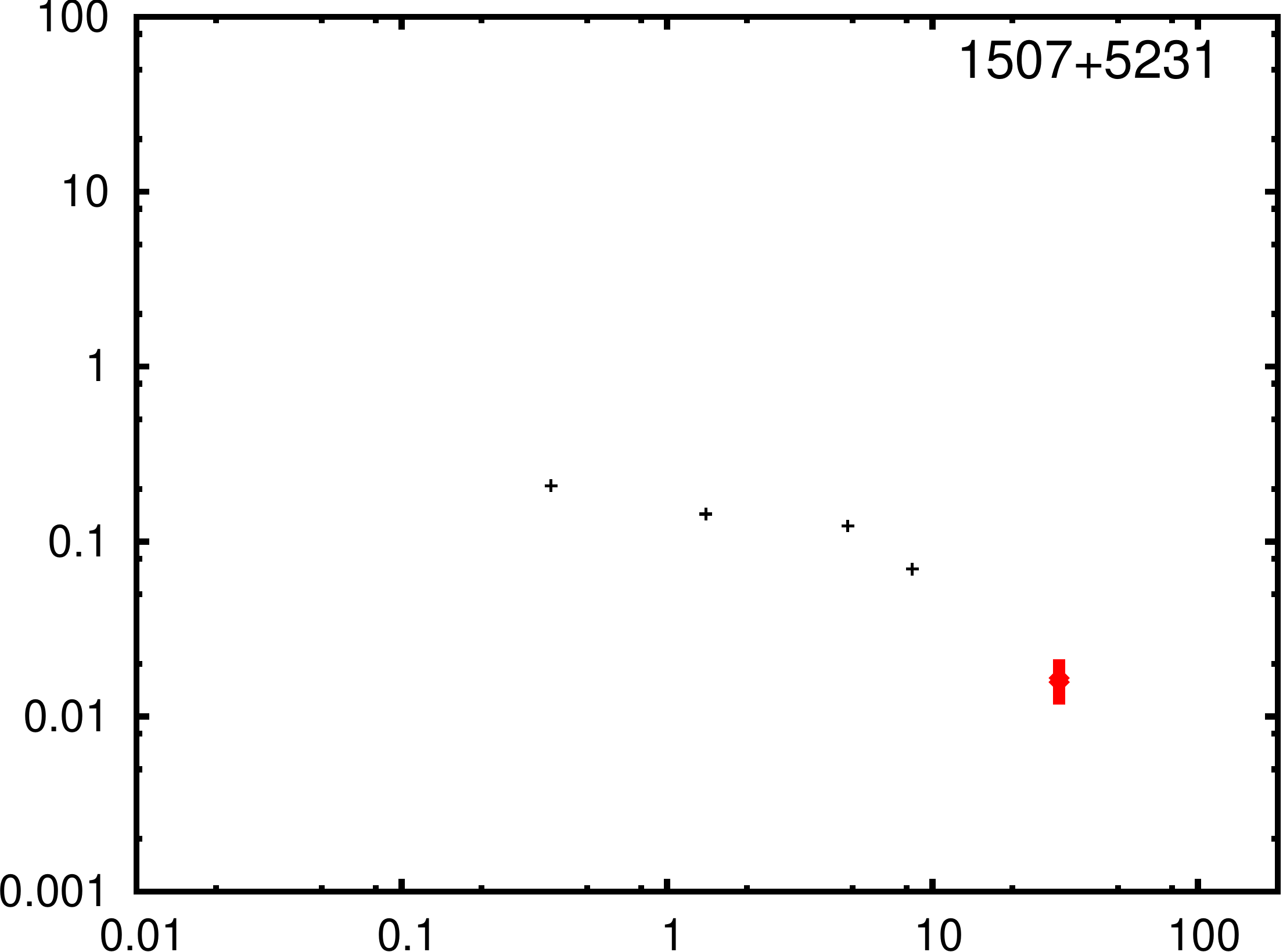}
\includegraphics[scale=0.2]{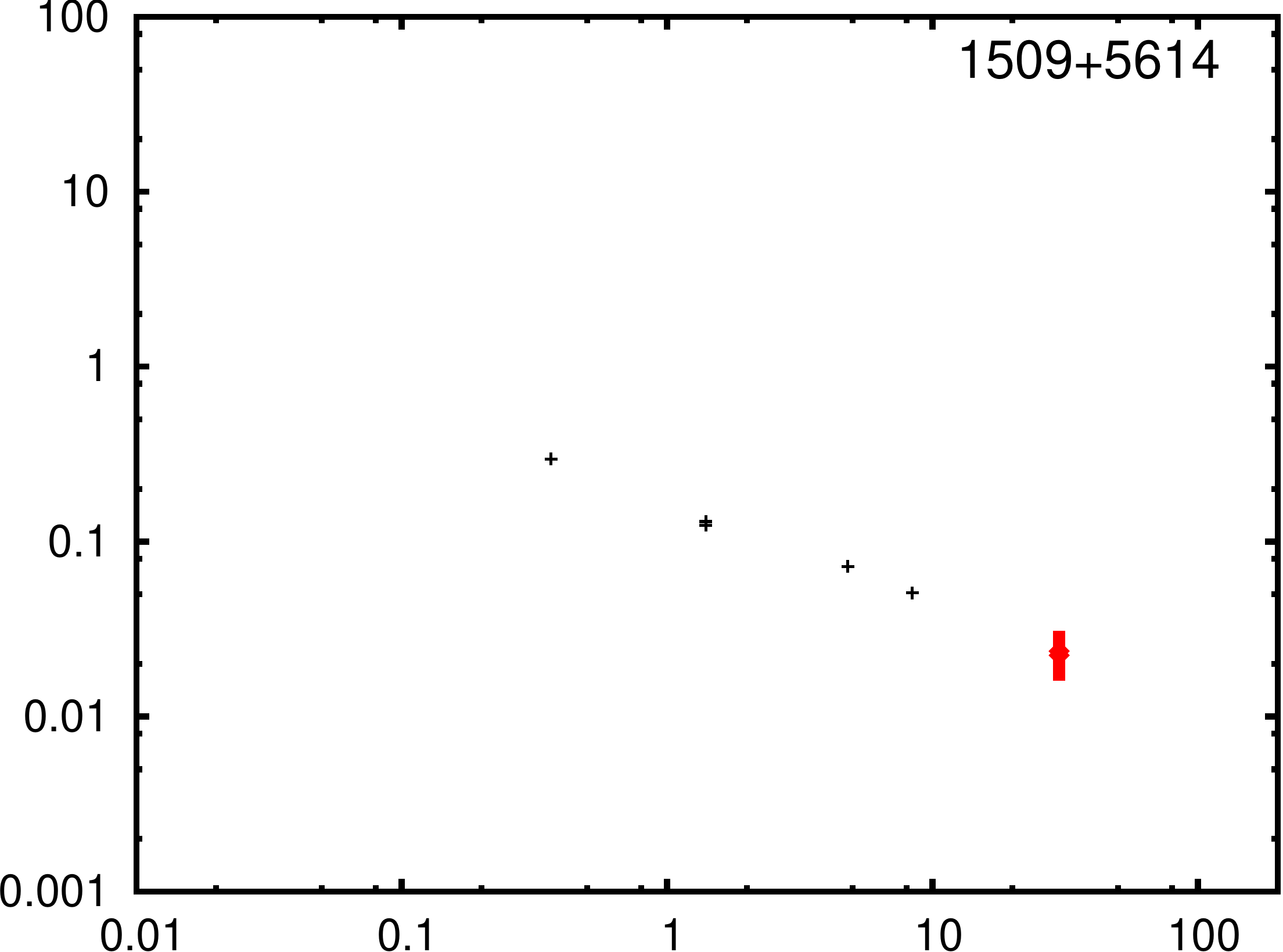}
\includegraphics[scale=0.2]{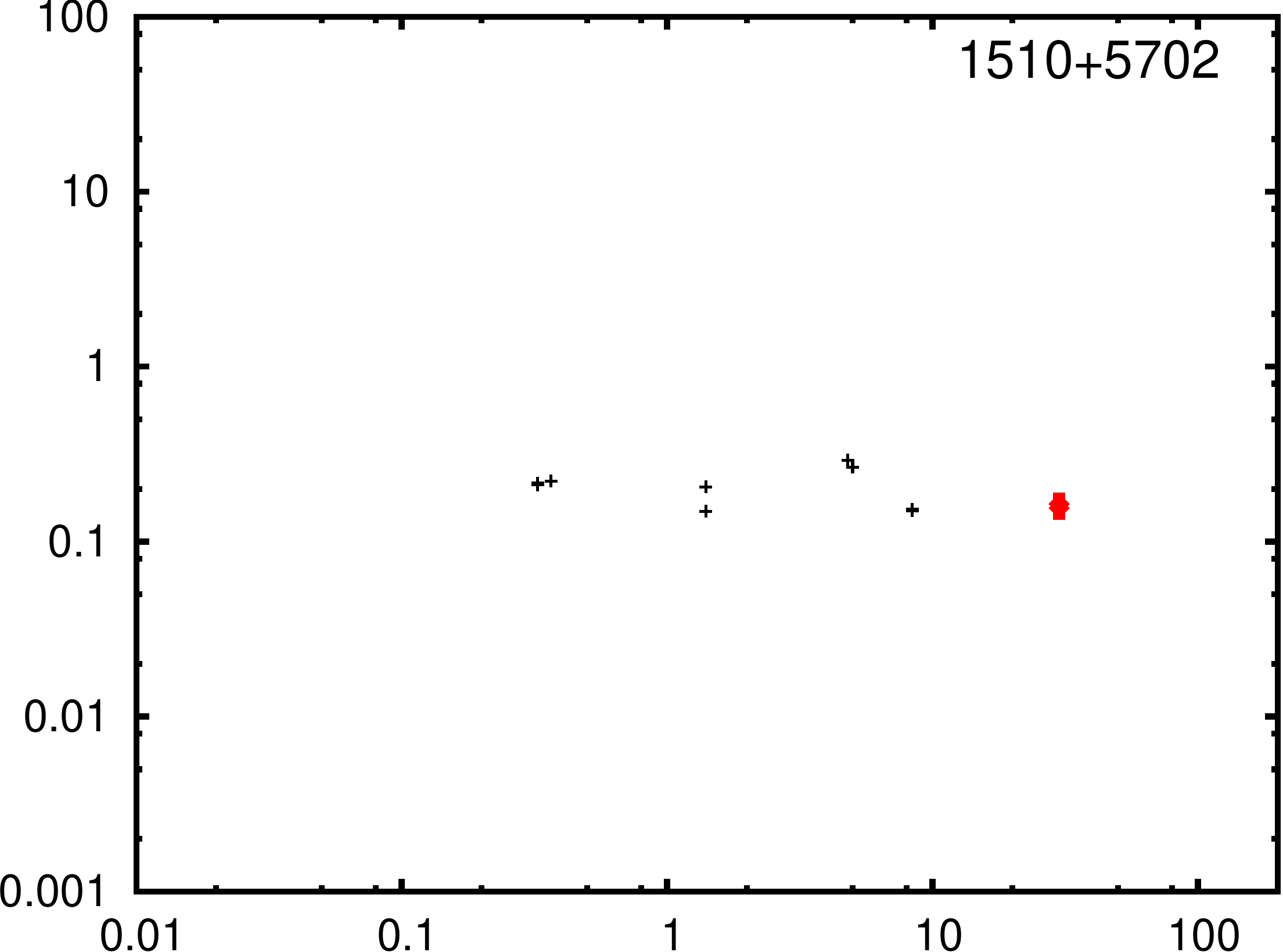}
\includegraphics[scale=0.2]{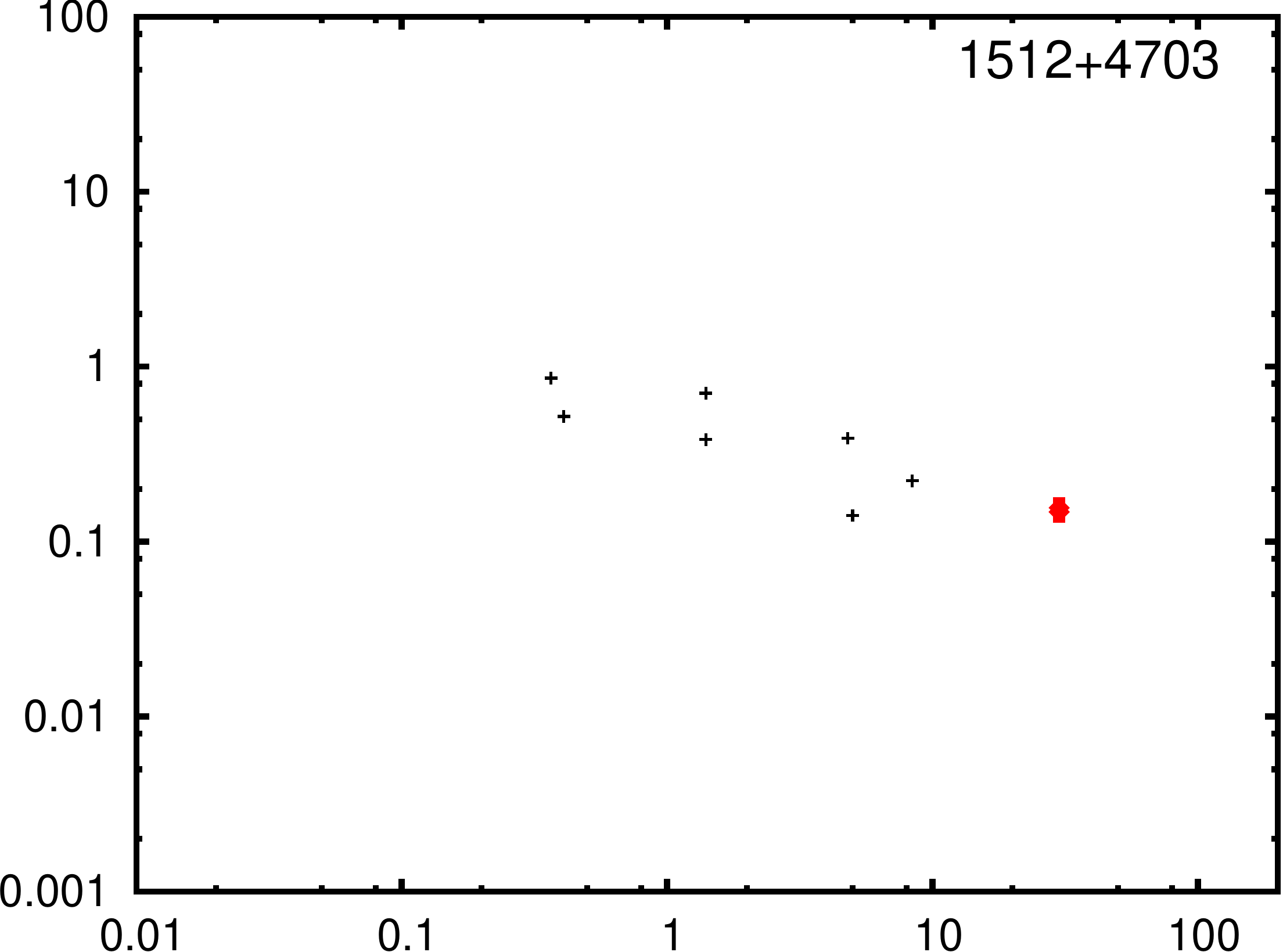}
\includegraphics[scale=0.2]{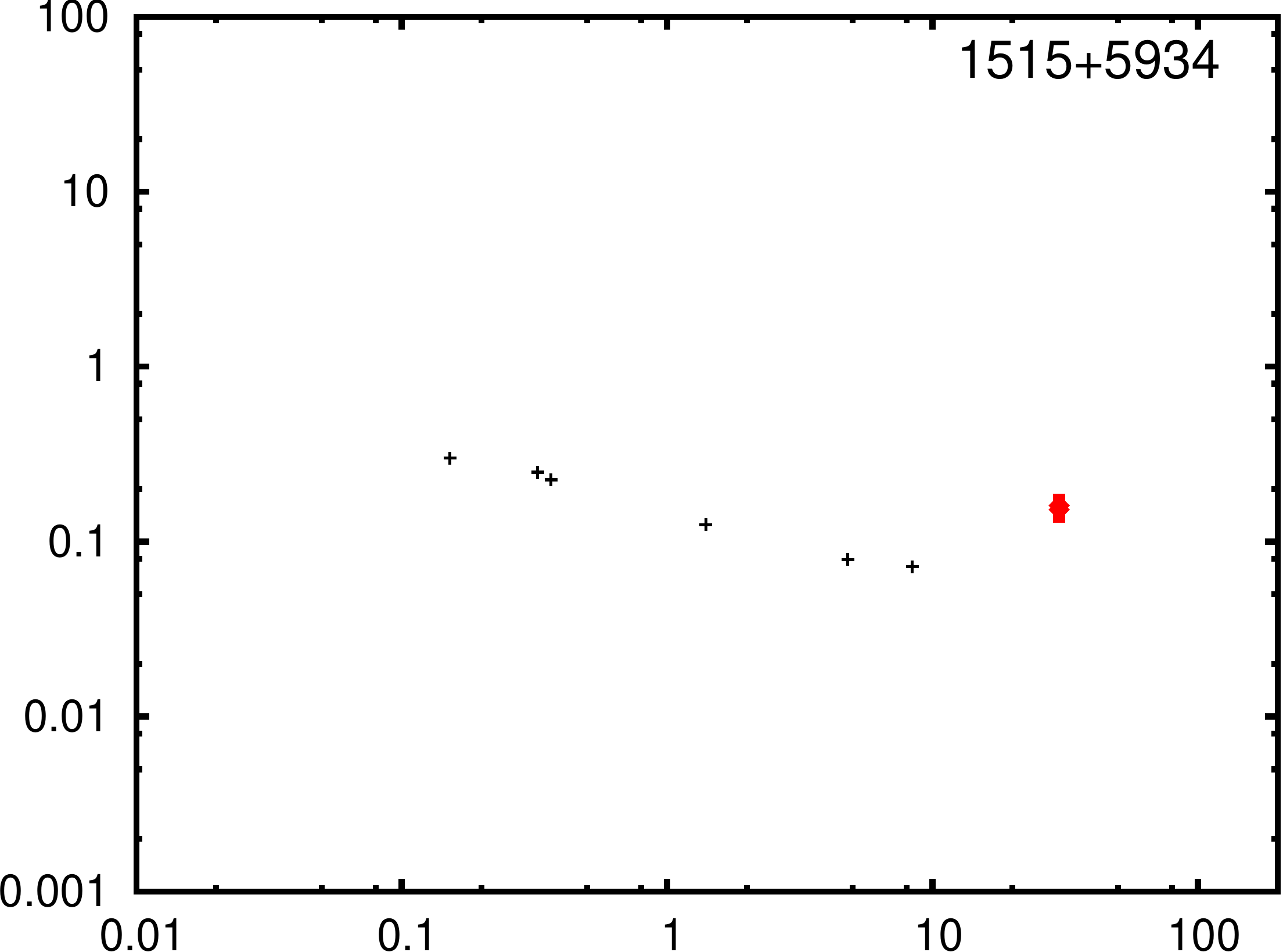}
\includegraphics[scale=0.2]{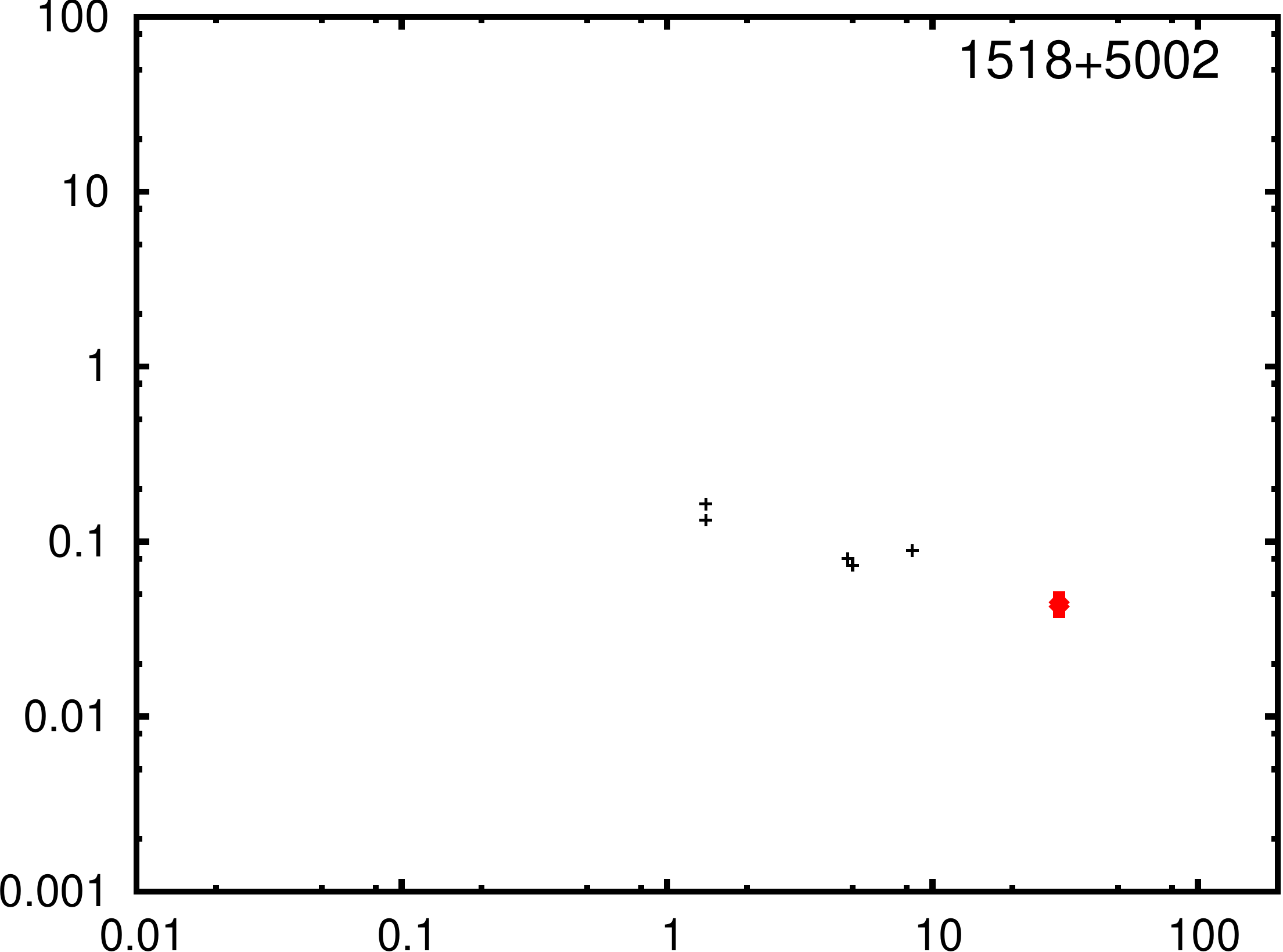}
\includegraphics[scale=0.2]{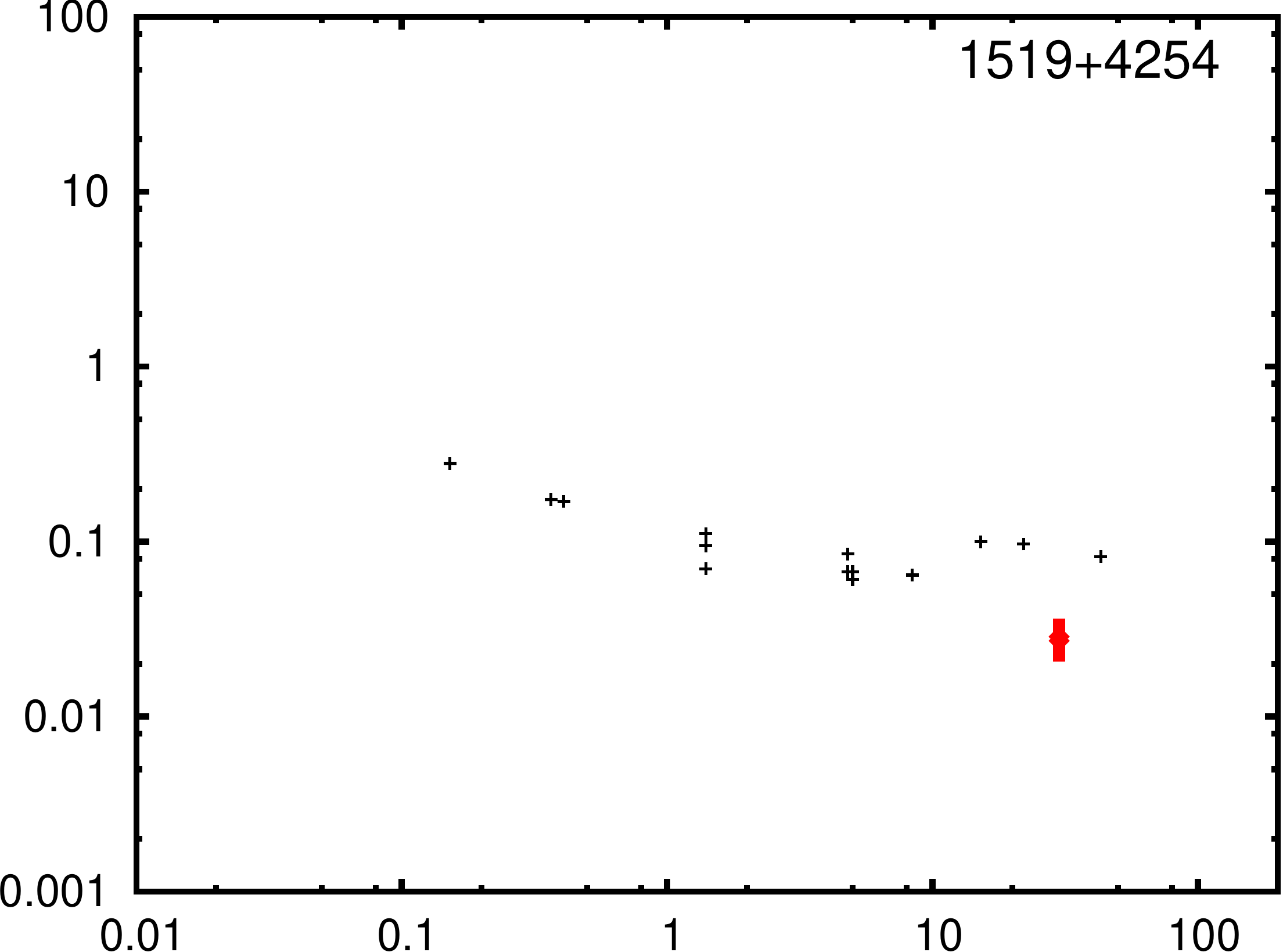}
\includegraphics[scale=0.2]{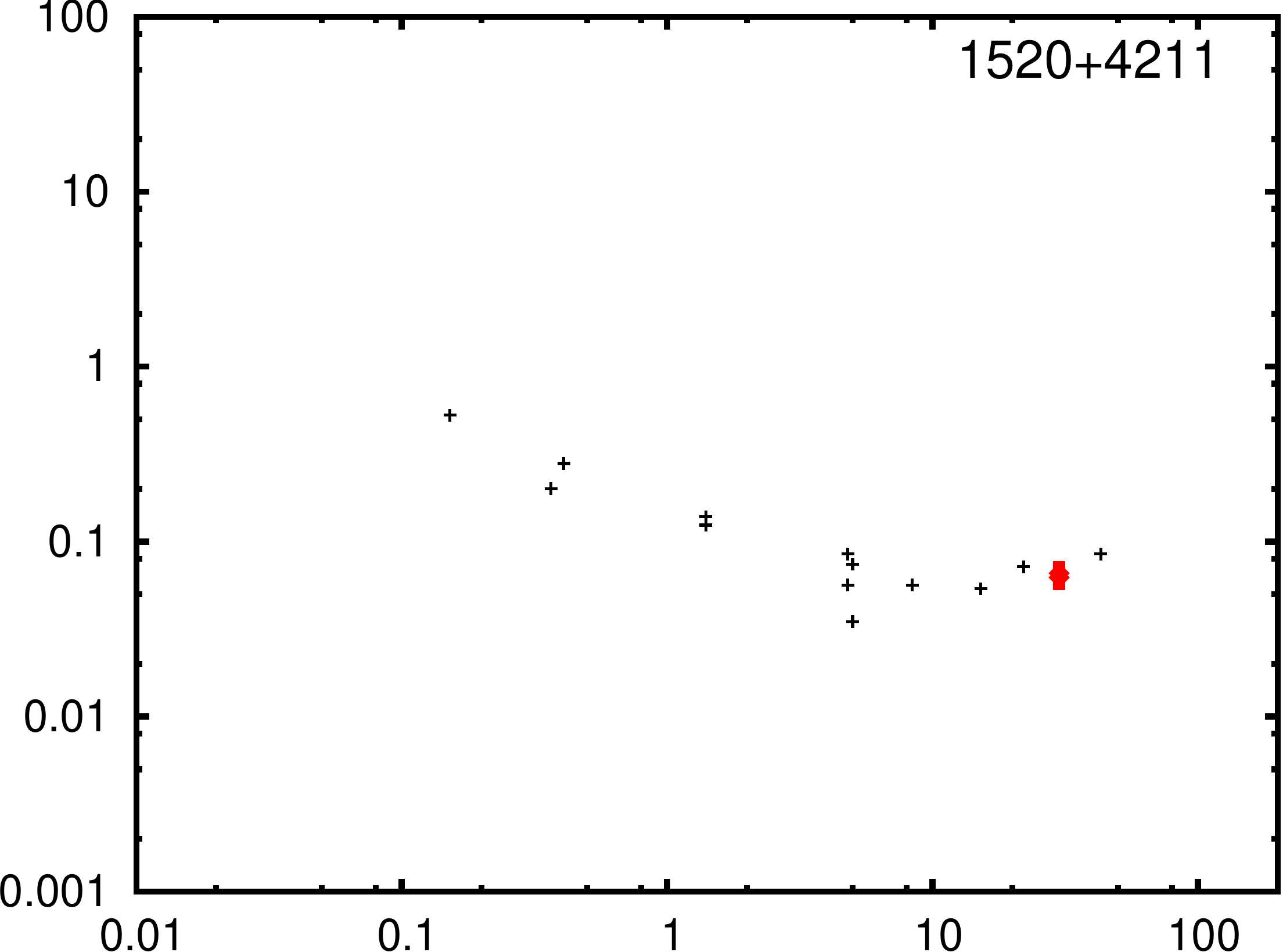}
\includegraphics[scale=0.2]{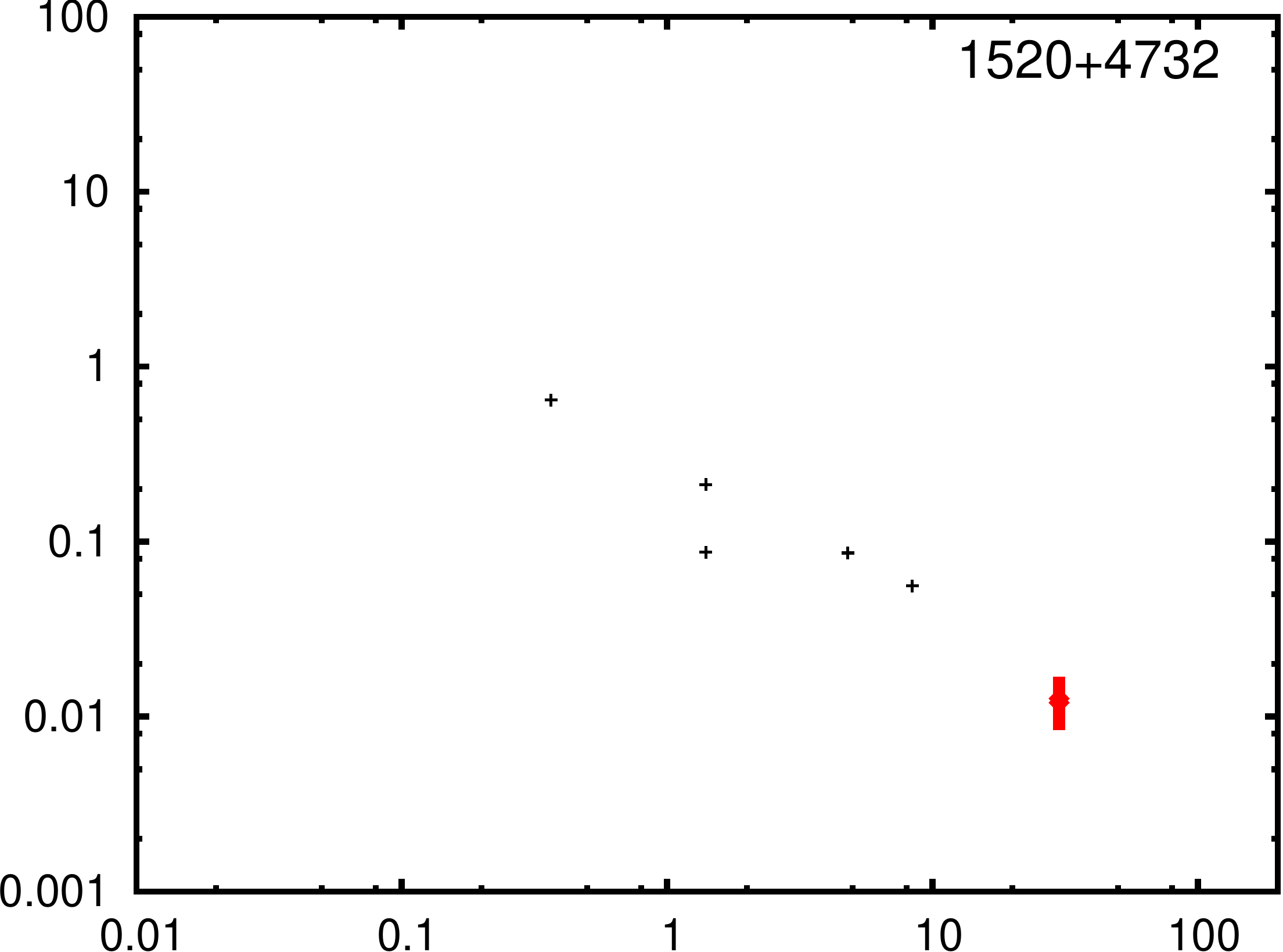}
\includegraphics[scale=0.2]{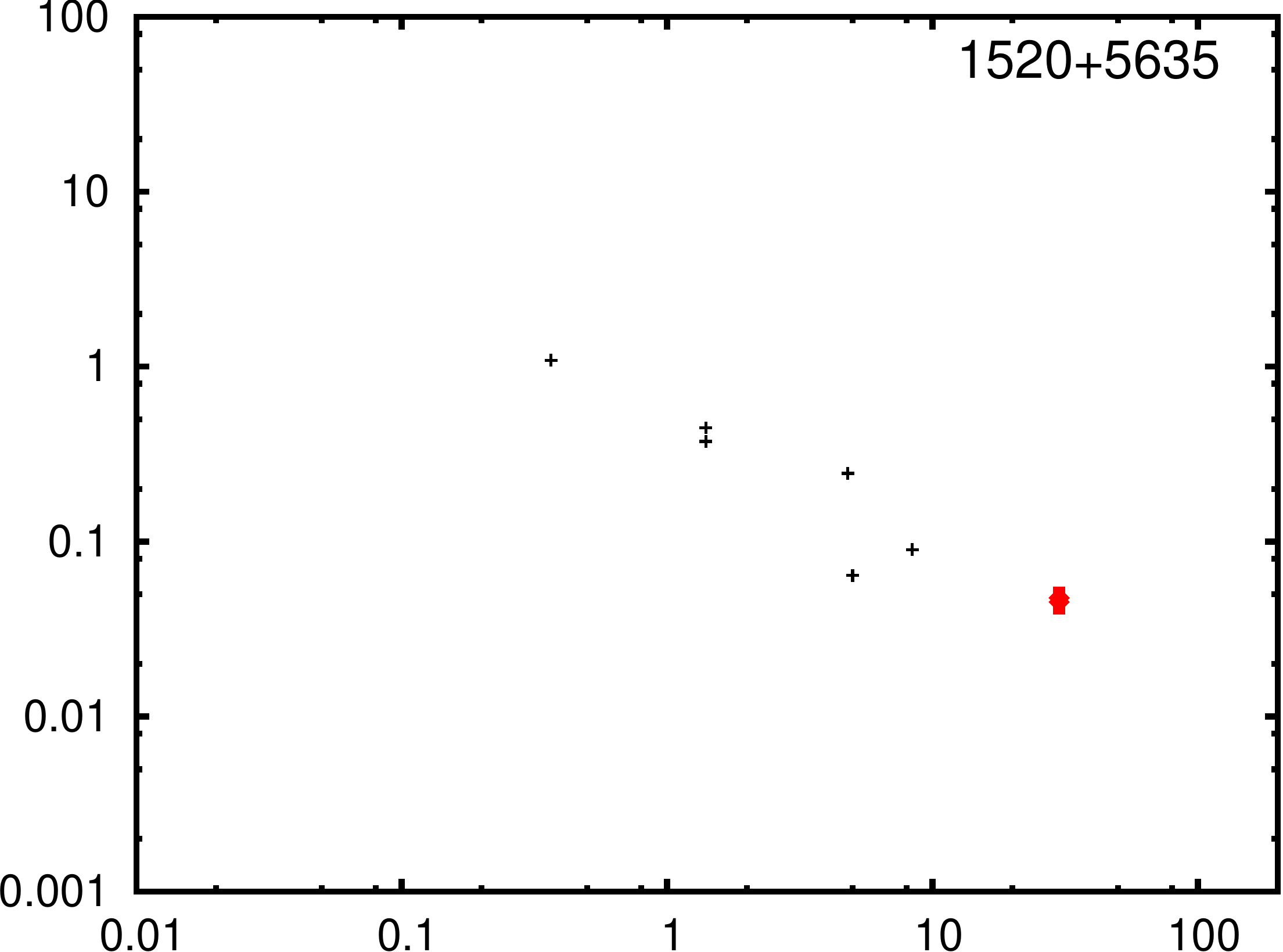}
\includegraphics[scale=0.2]{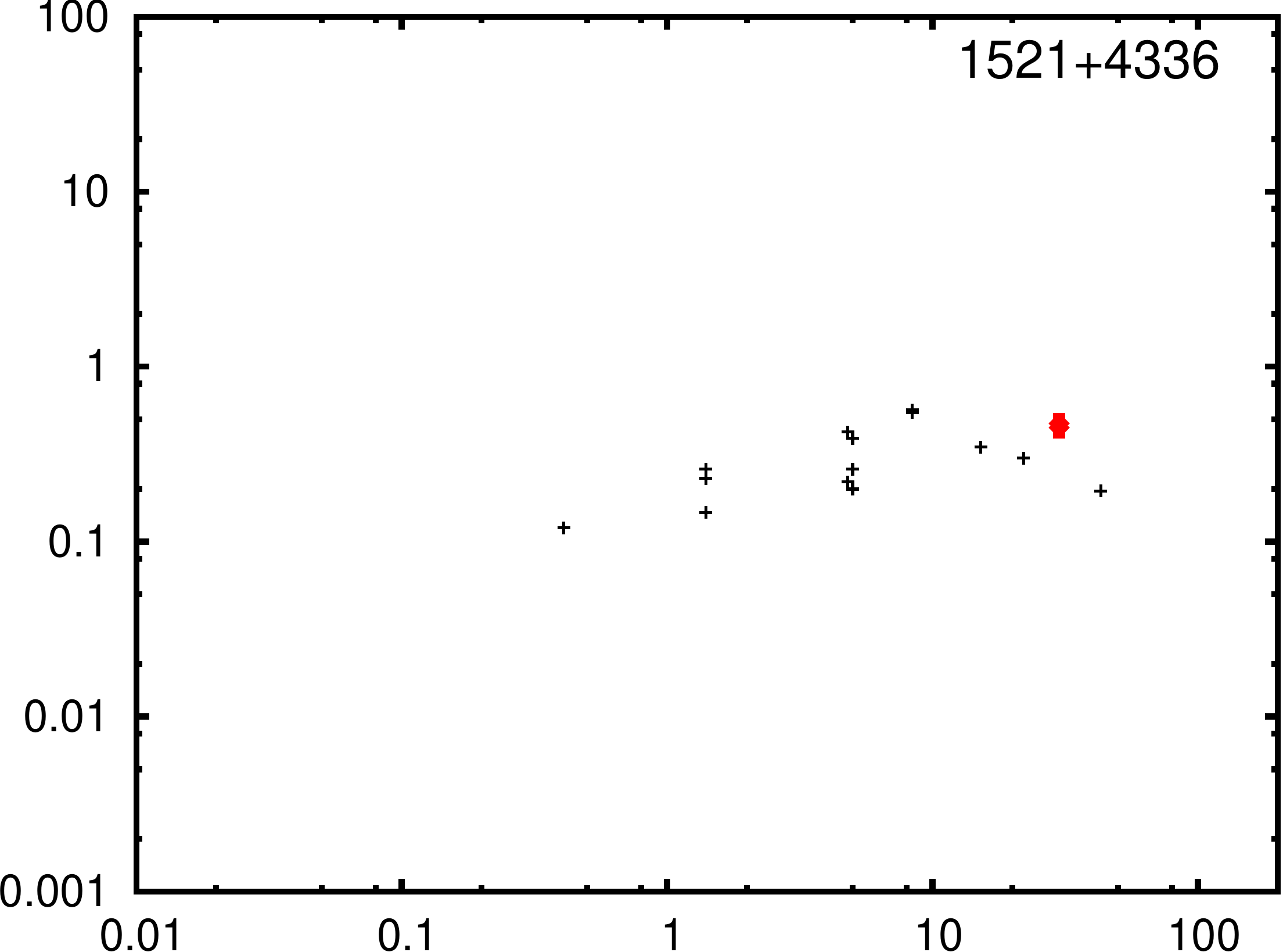}
\includegraphics[scale=0.2]{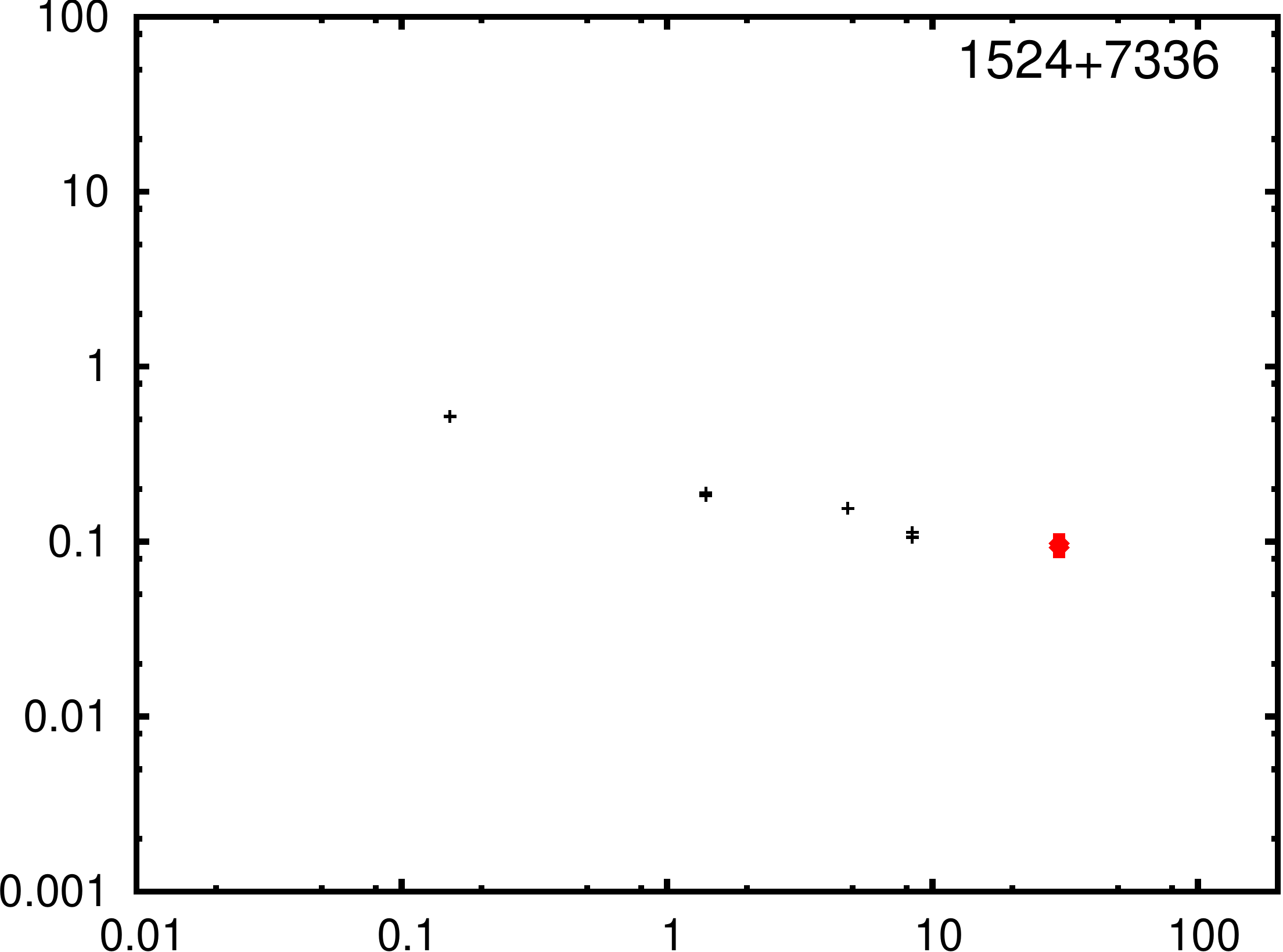}
\includegraphics[scale=0.2]{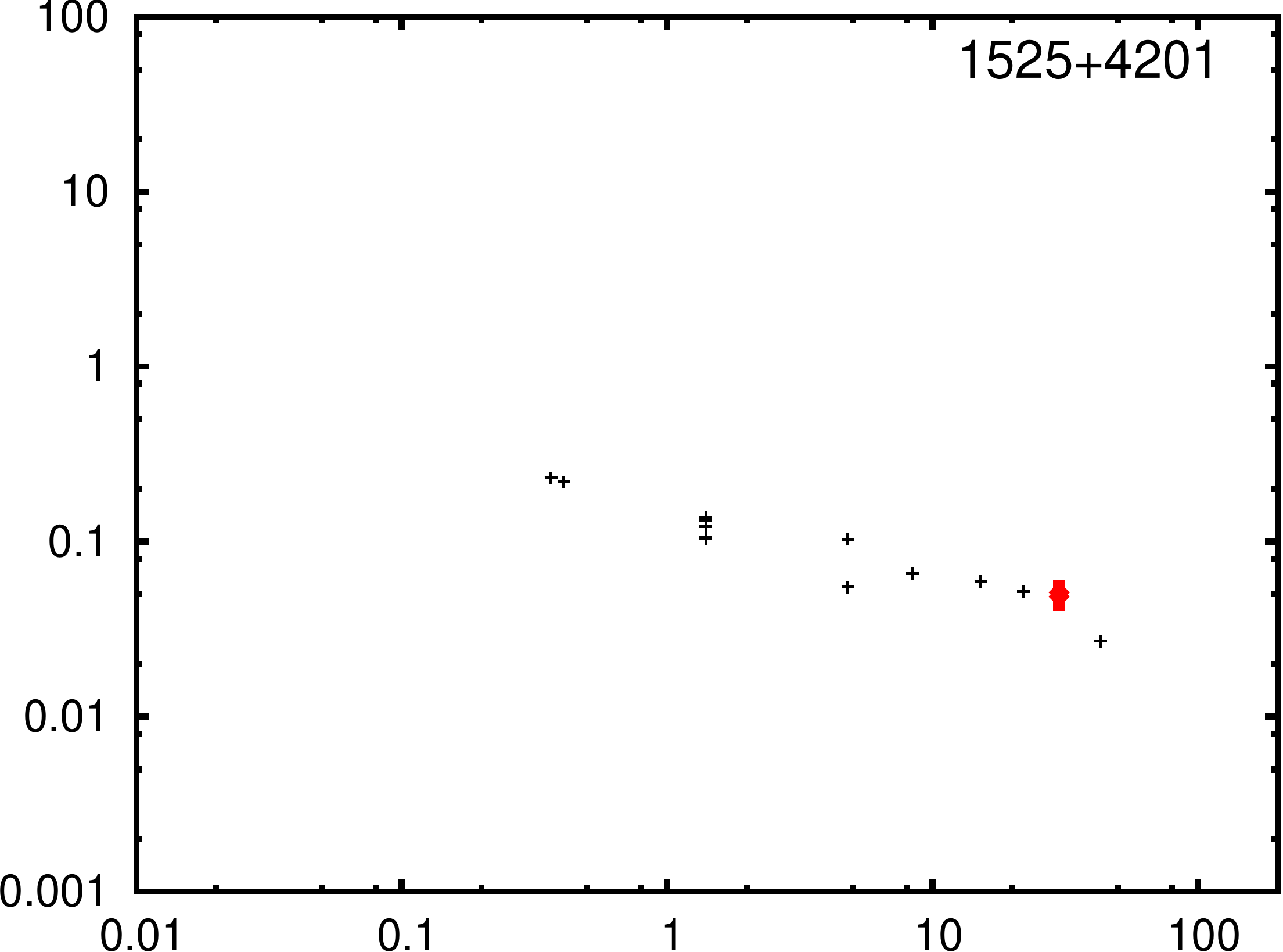}
\includegraphics[scale=0.2]{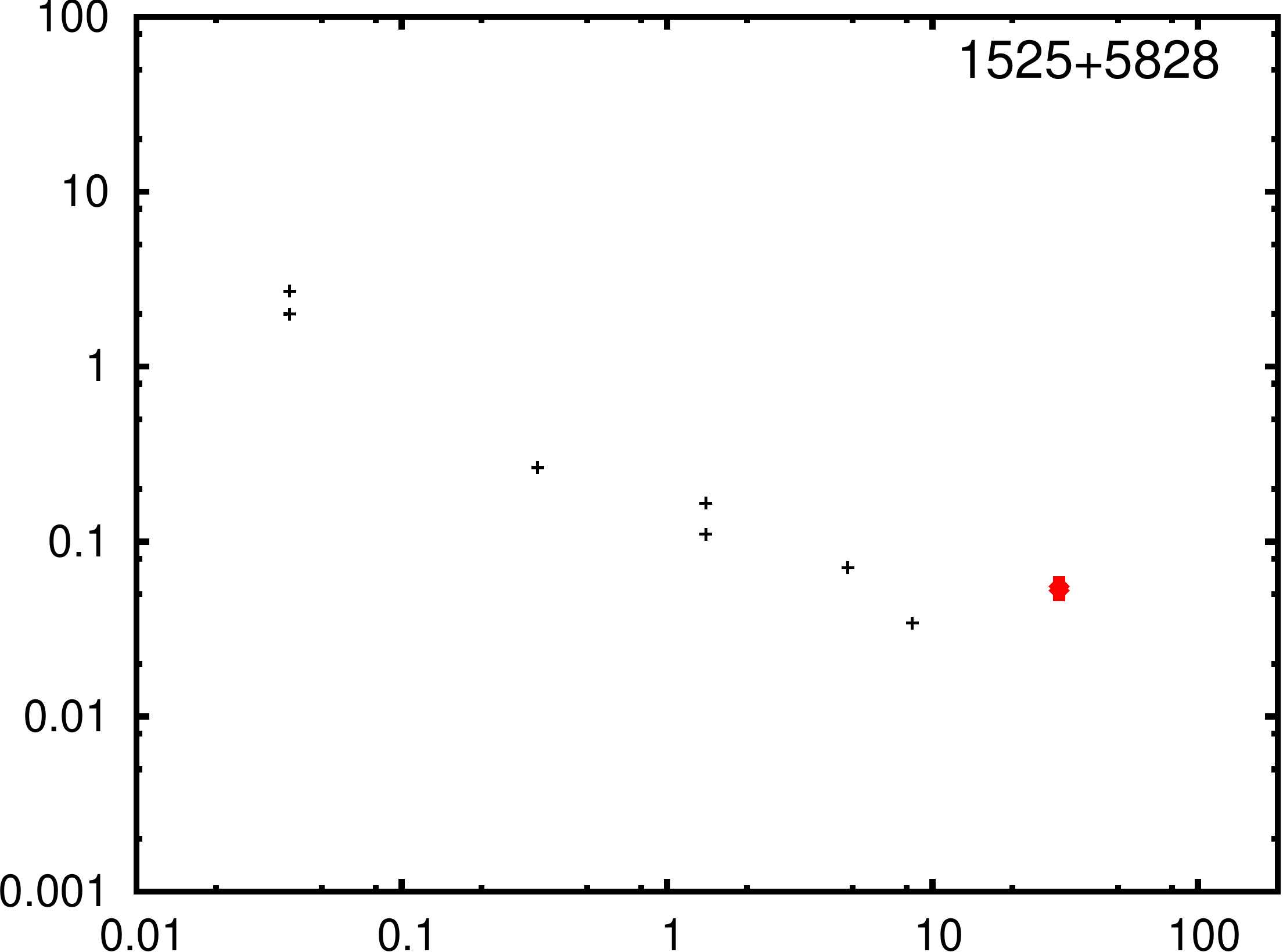}
\end{figure}
\clearpage\begin{figure}
\centering
\includegraphics[scale=0.2]{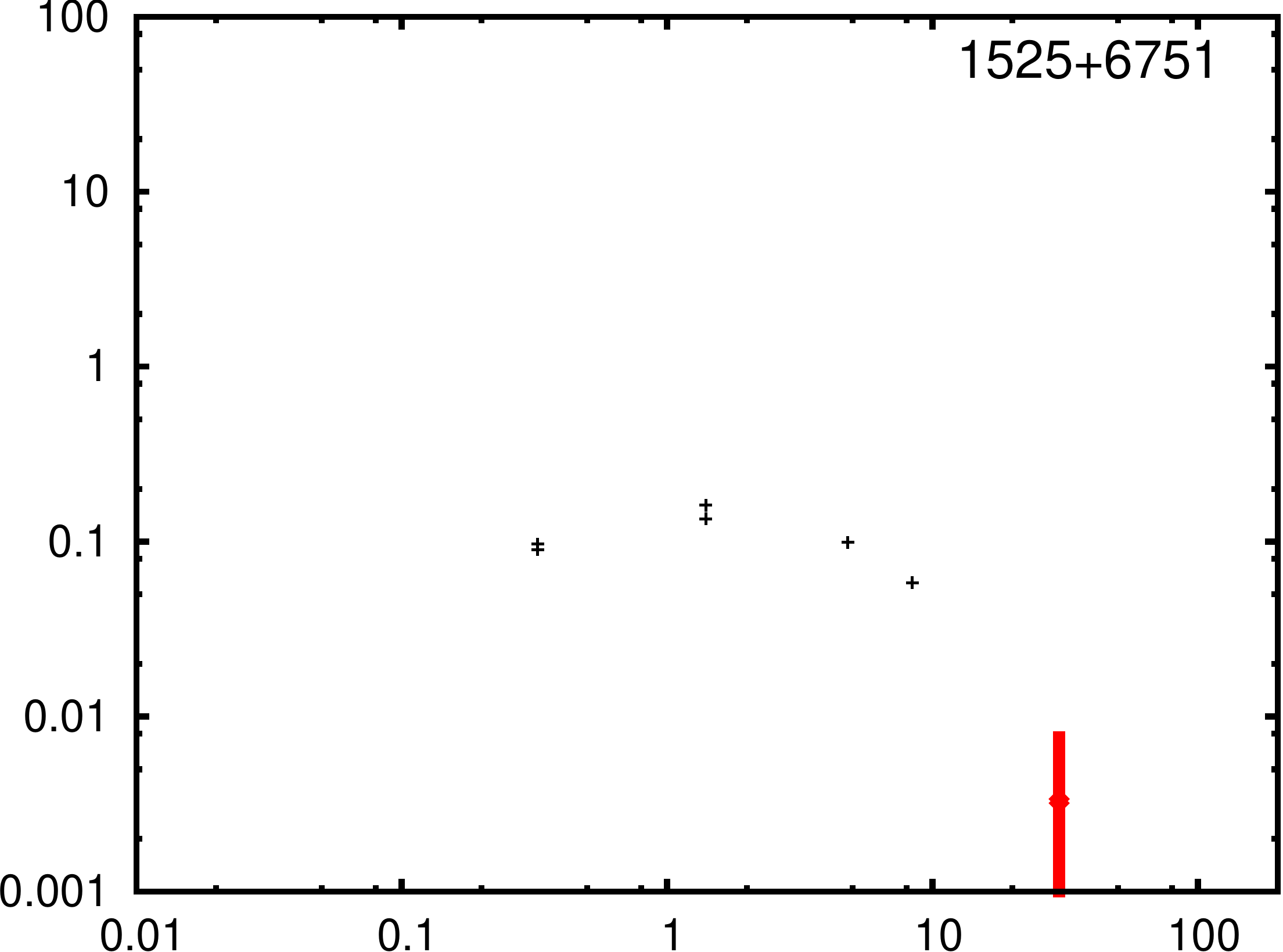}
\includegraphics[scale=0.2]{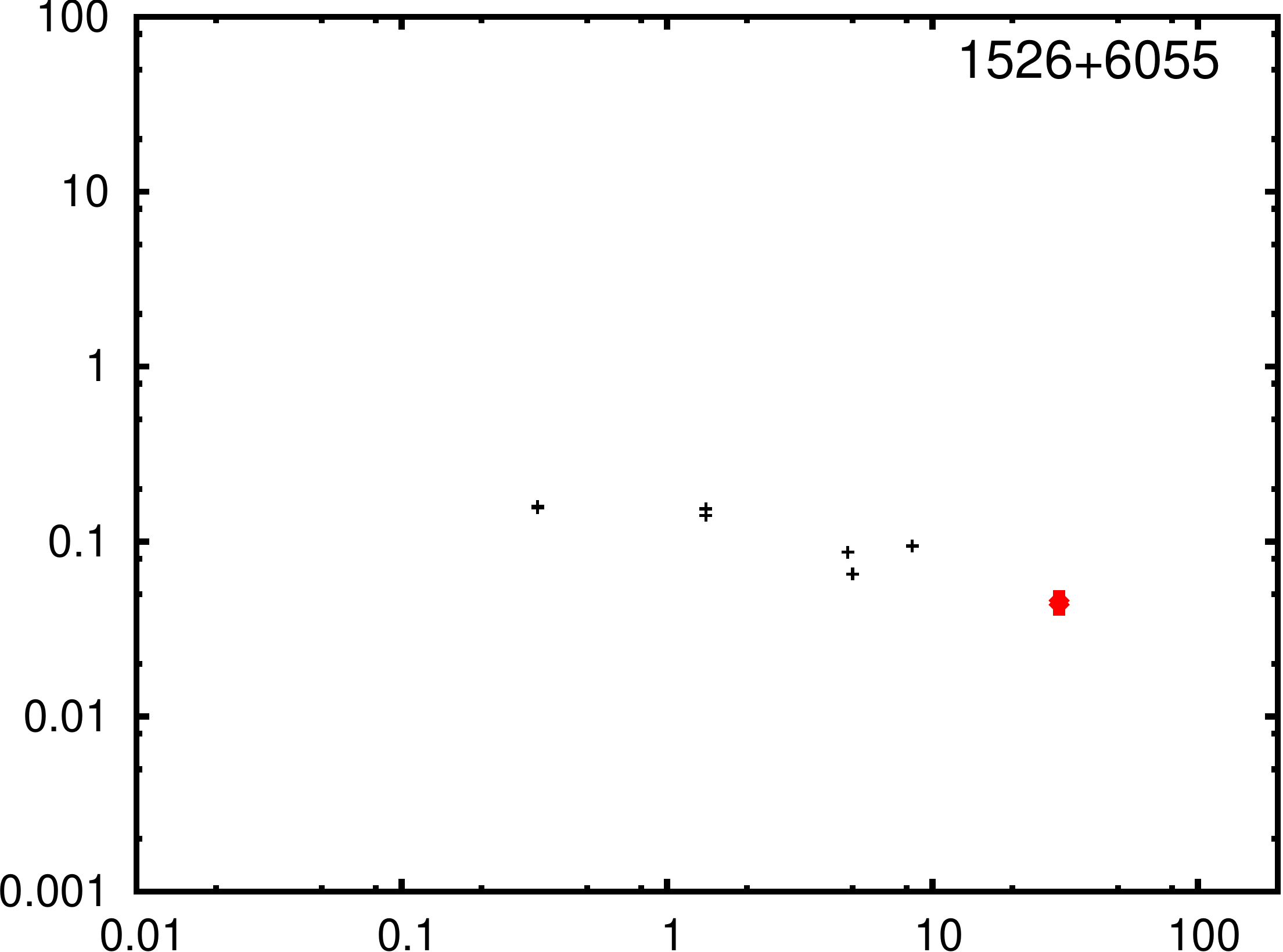}
\includegraphics[scale=0.2]{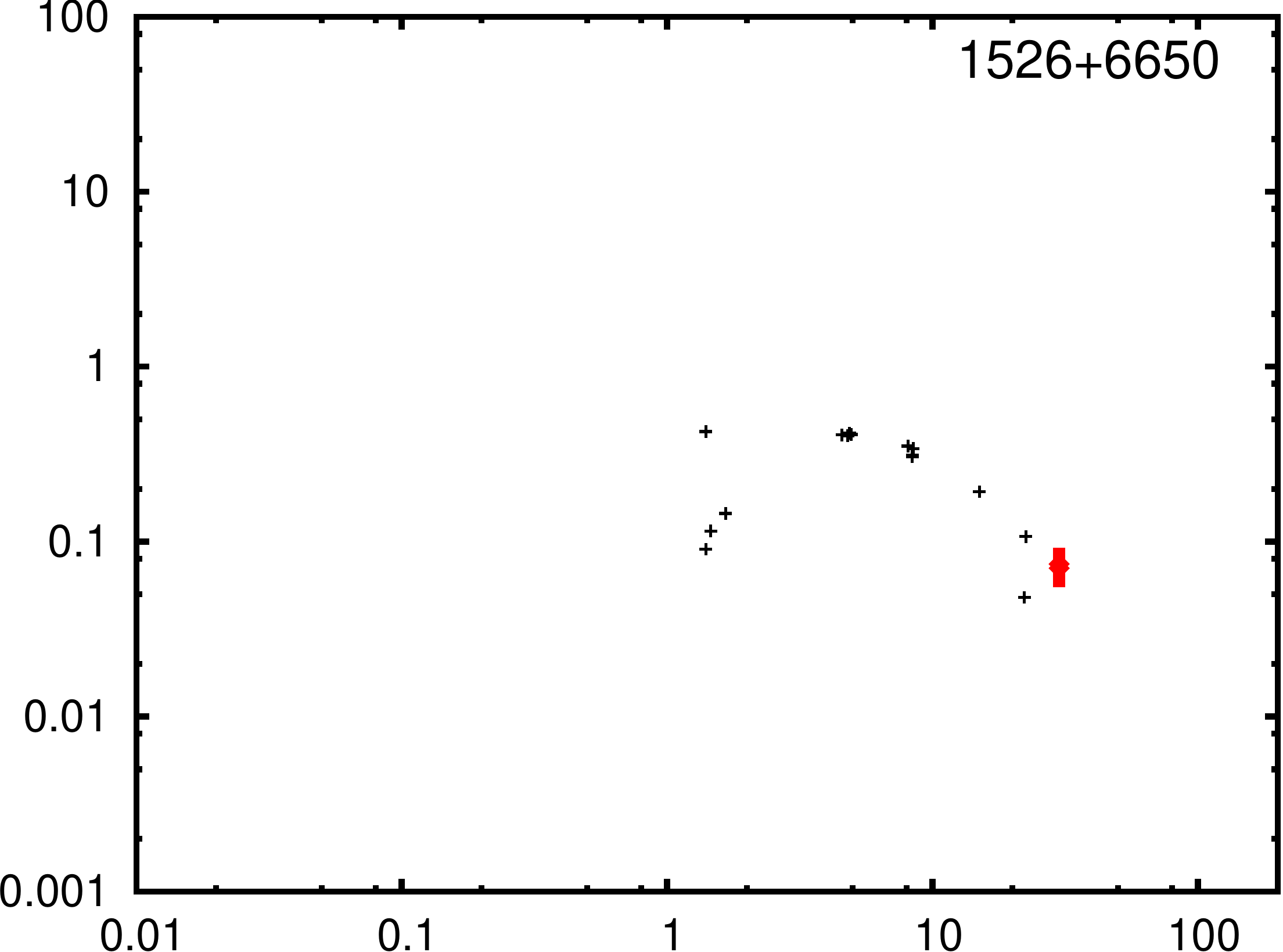}
\includegraphics[scale=0.2]{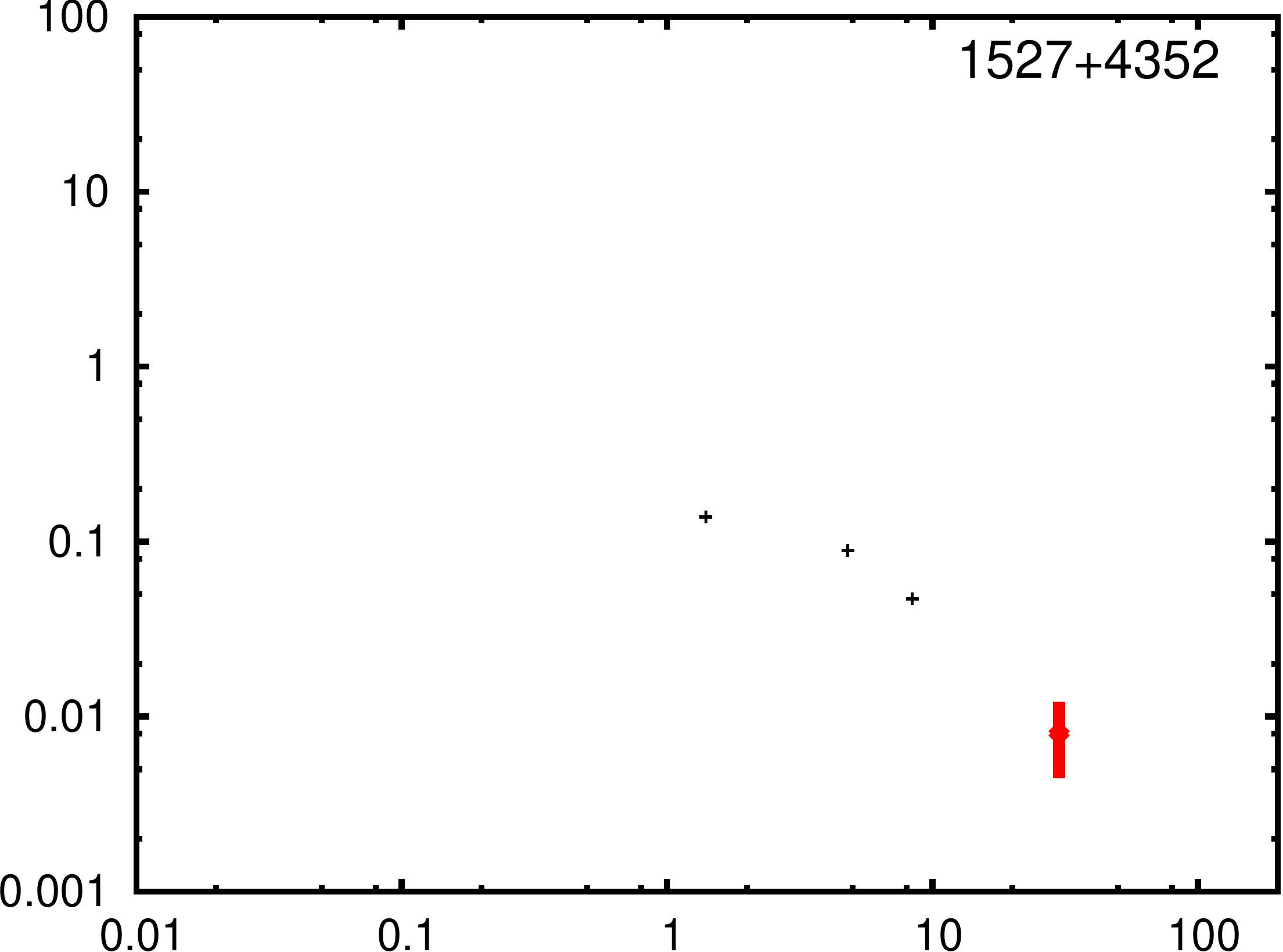}
\includegraphics[scale=0.2]{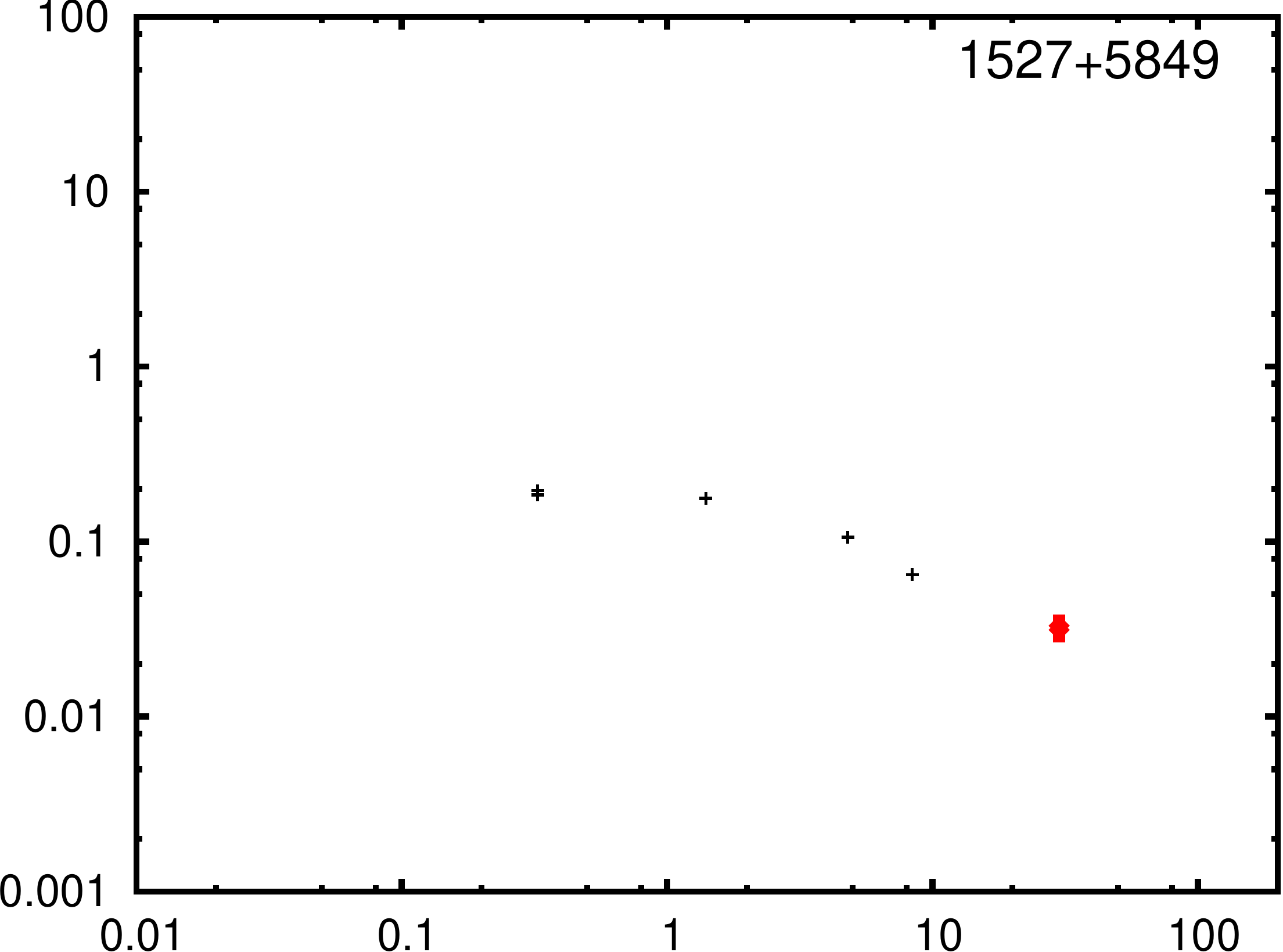}
\includegraphics[scale=0.2]{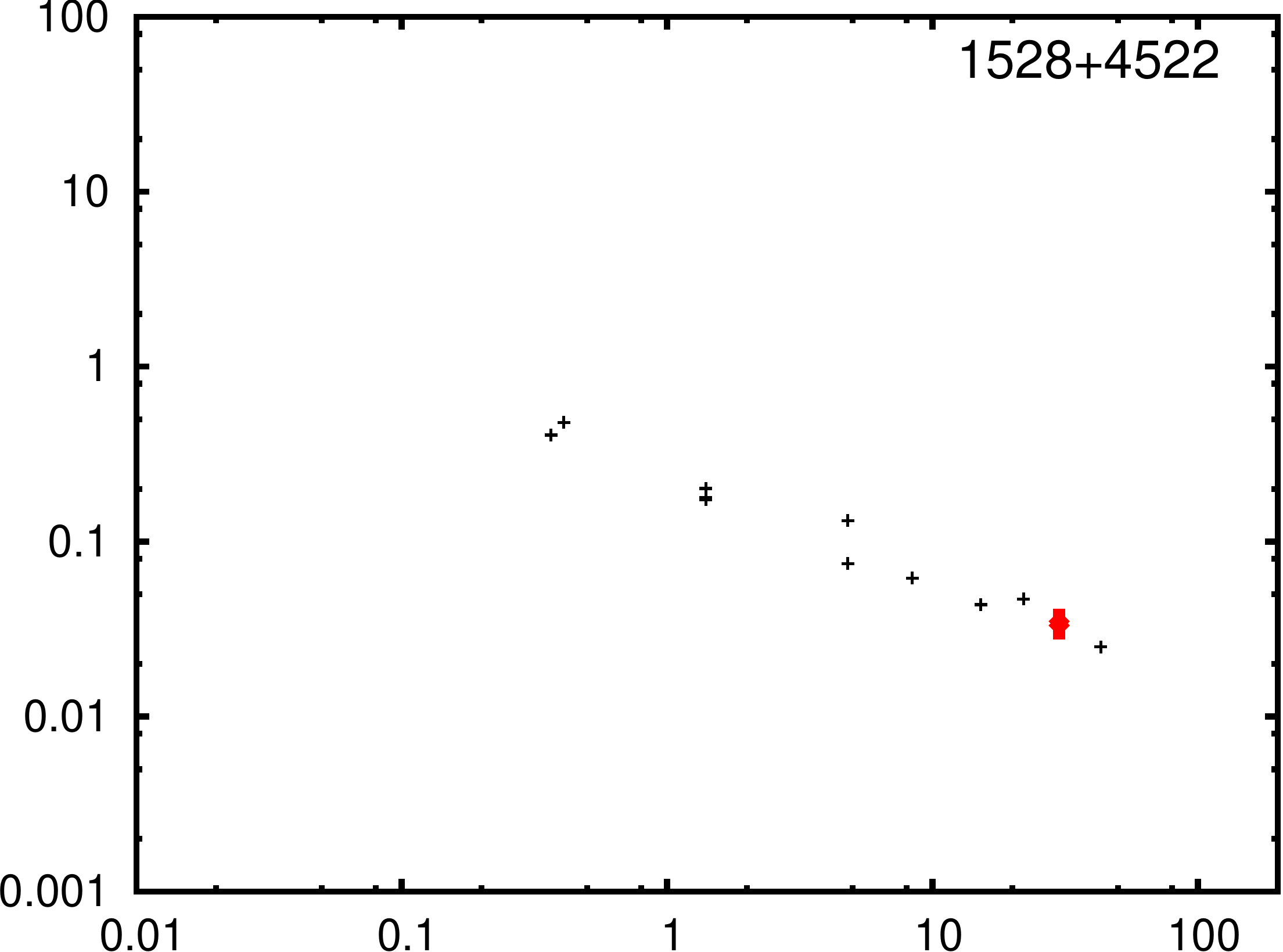}
\includegraphics[scale=0.2]{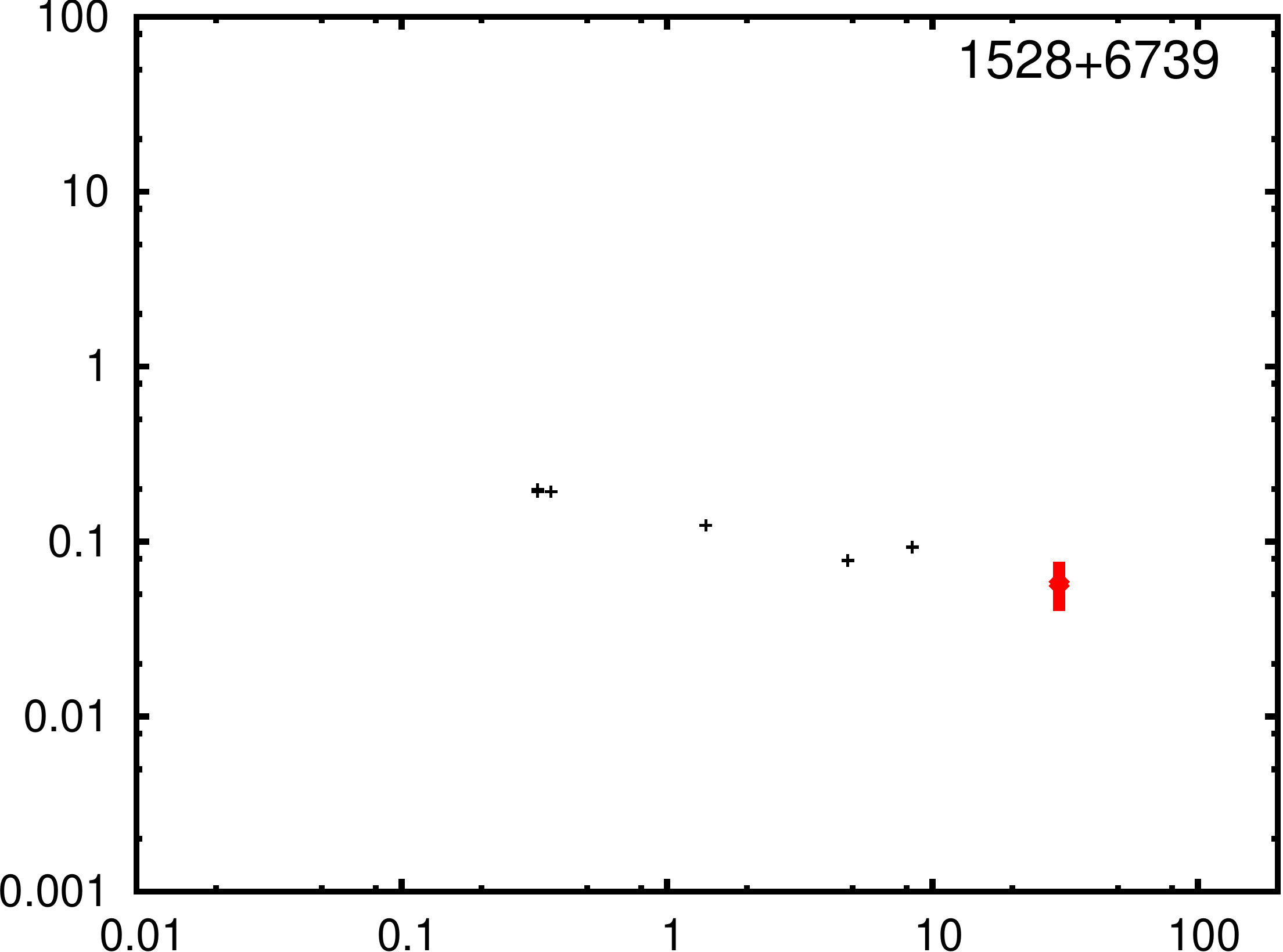}
\includegraphics[scale=0.2]{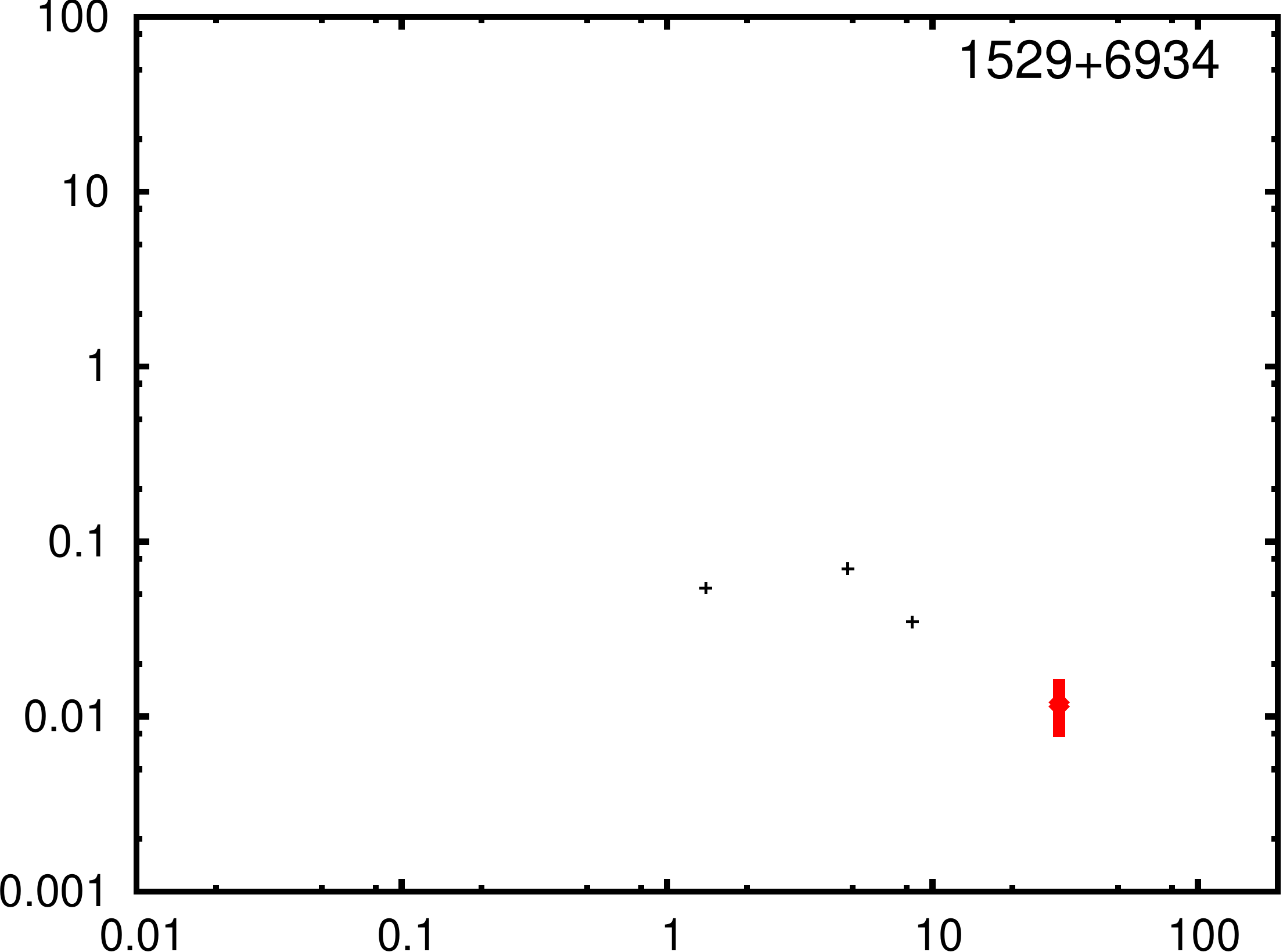}
\includegraphics[scale=0.2]{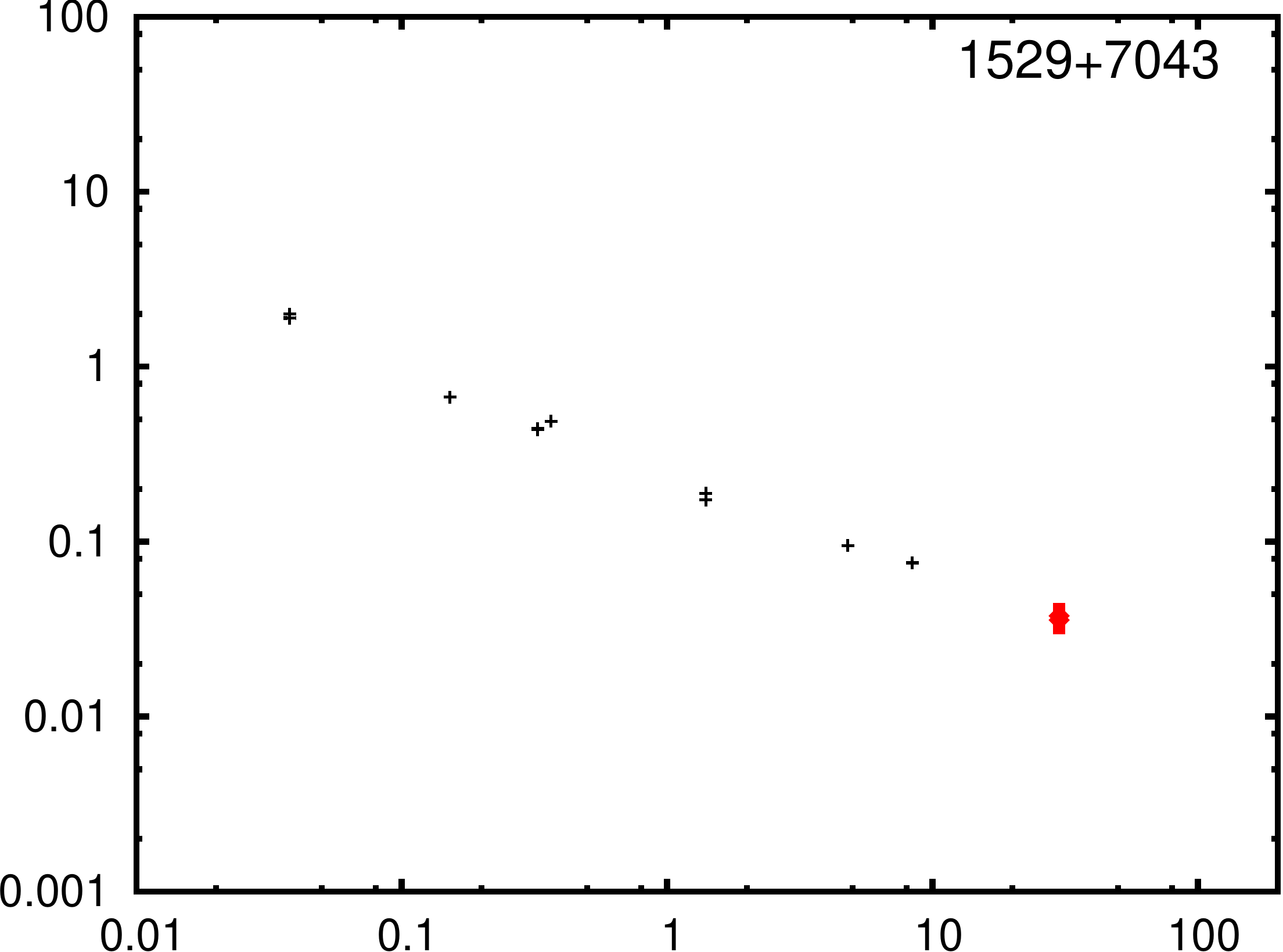}
\includegraphics[scale=0.2]{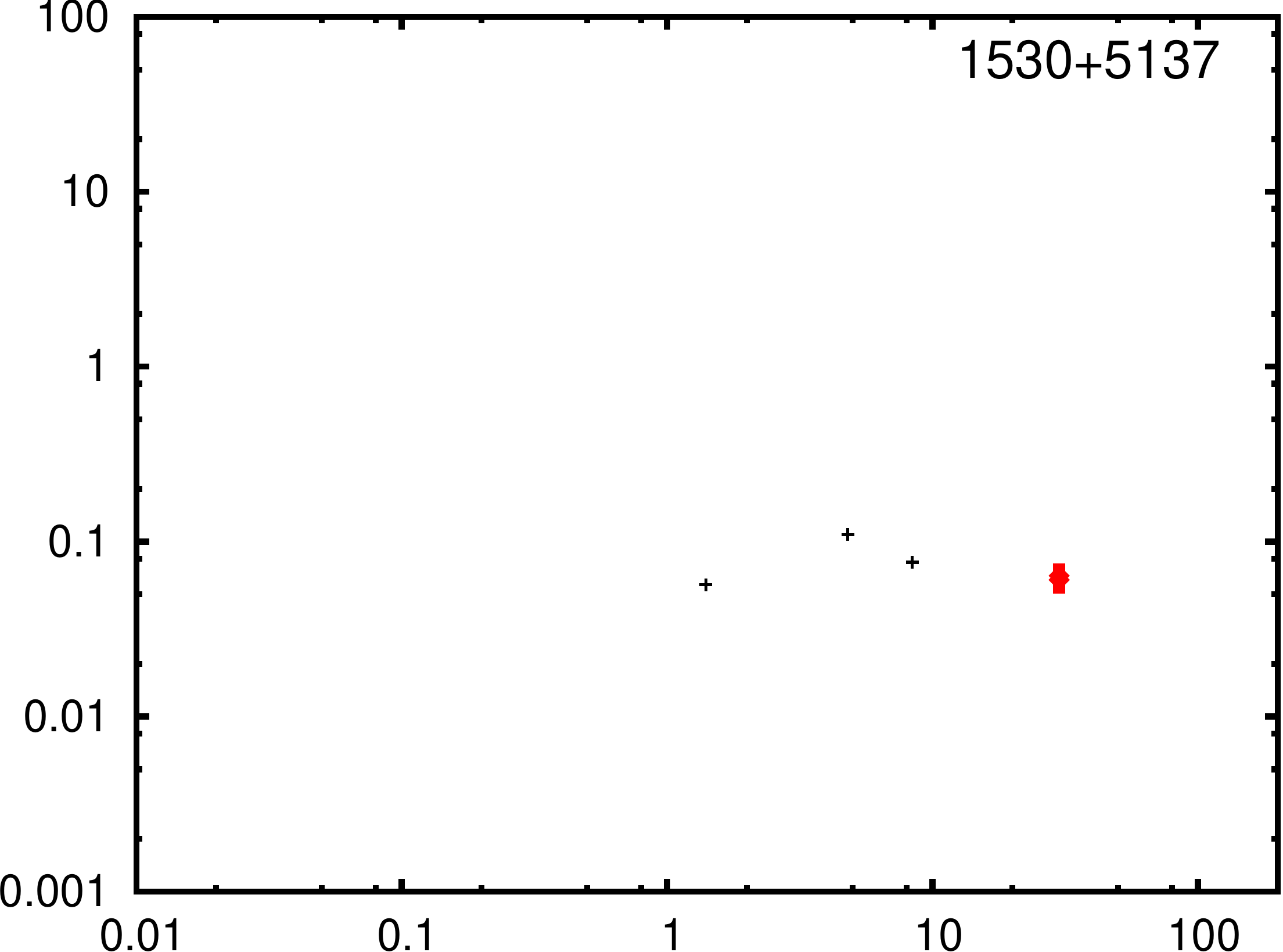}
\includegraphics[scale=0.2]{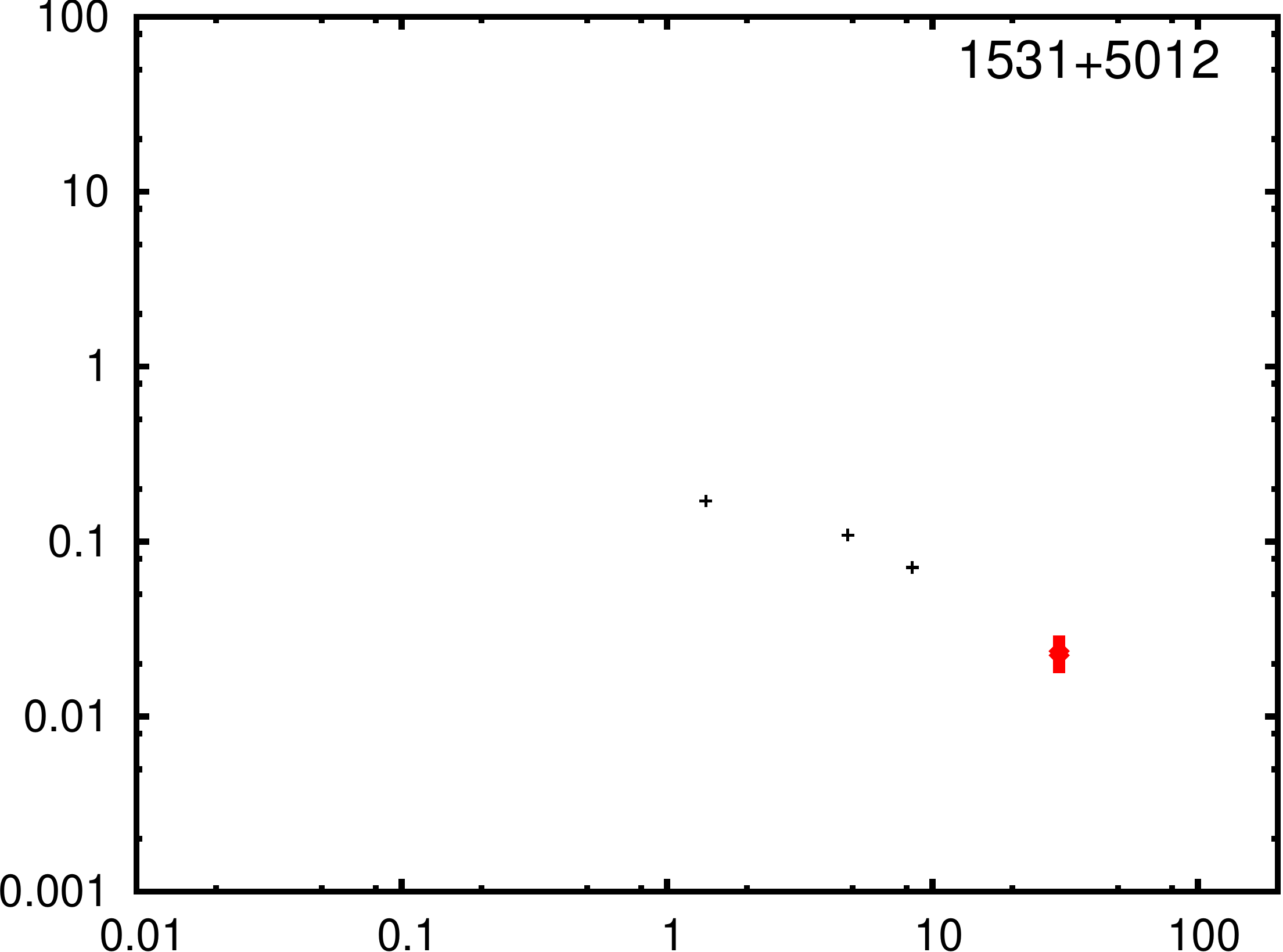}
\includegraphics[scale=0.2]{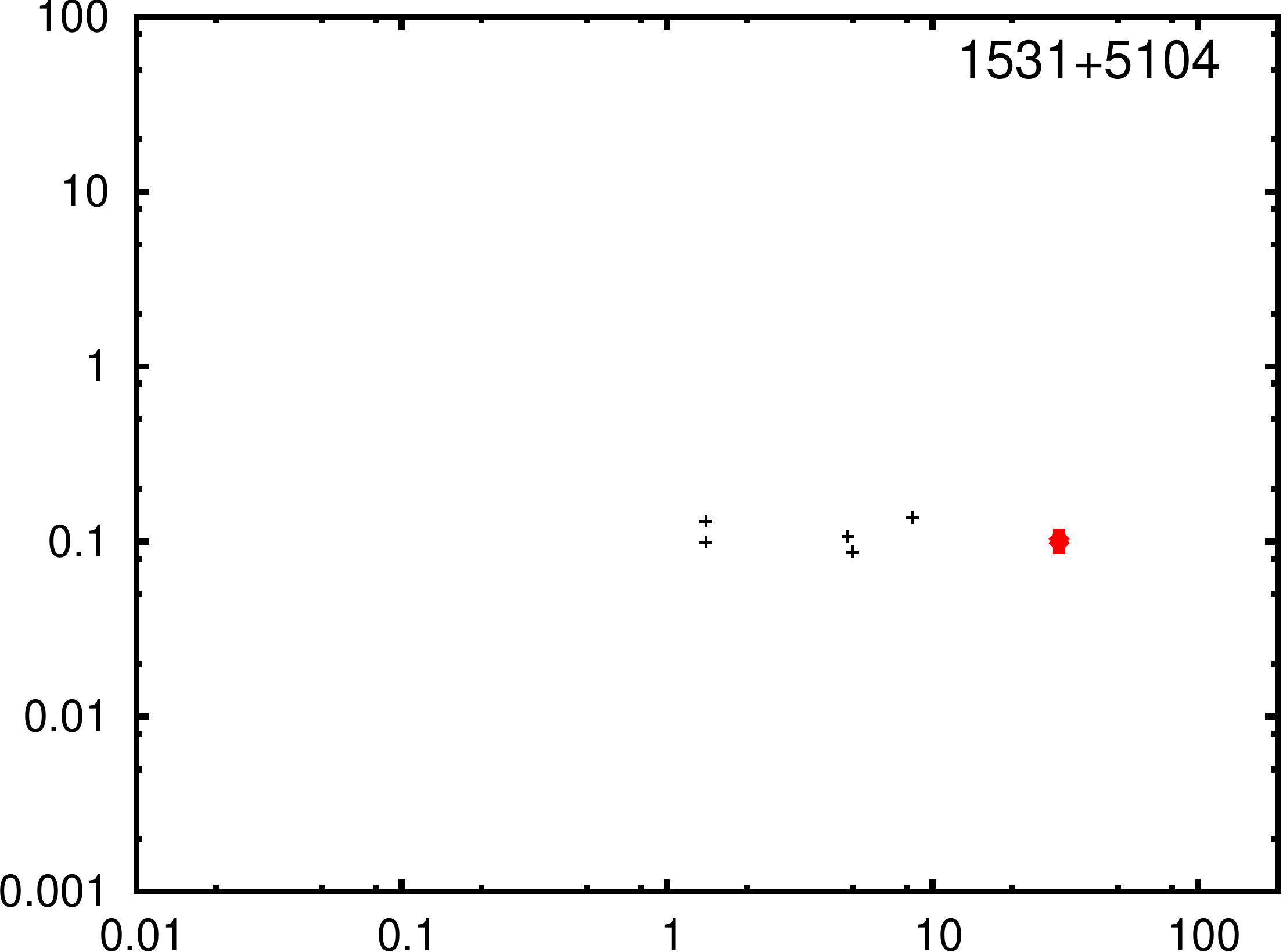}
\includegraphics[scale=0.2]{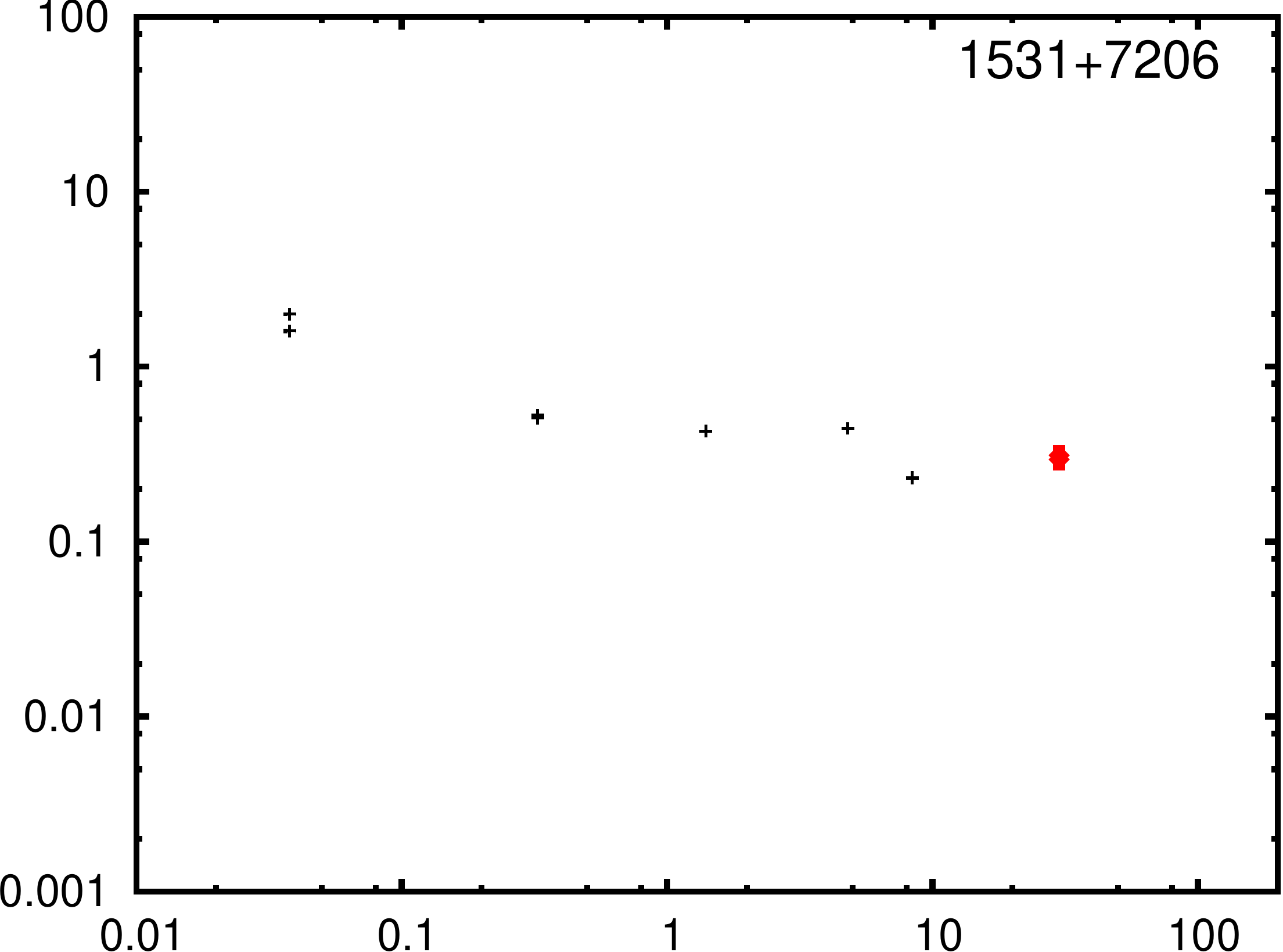}
\includegraphics[scale=0.2]{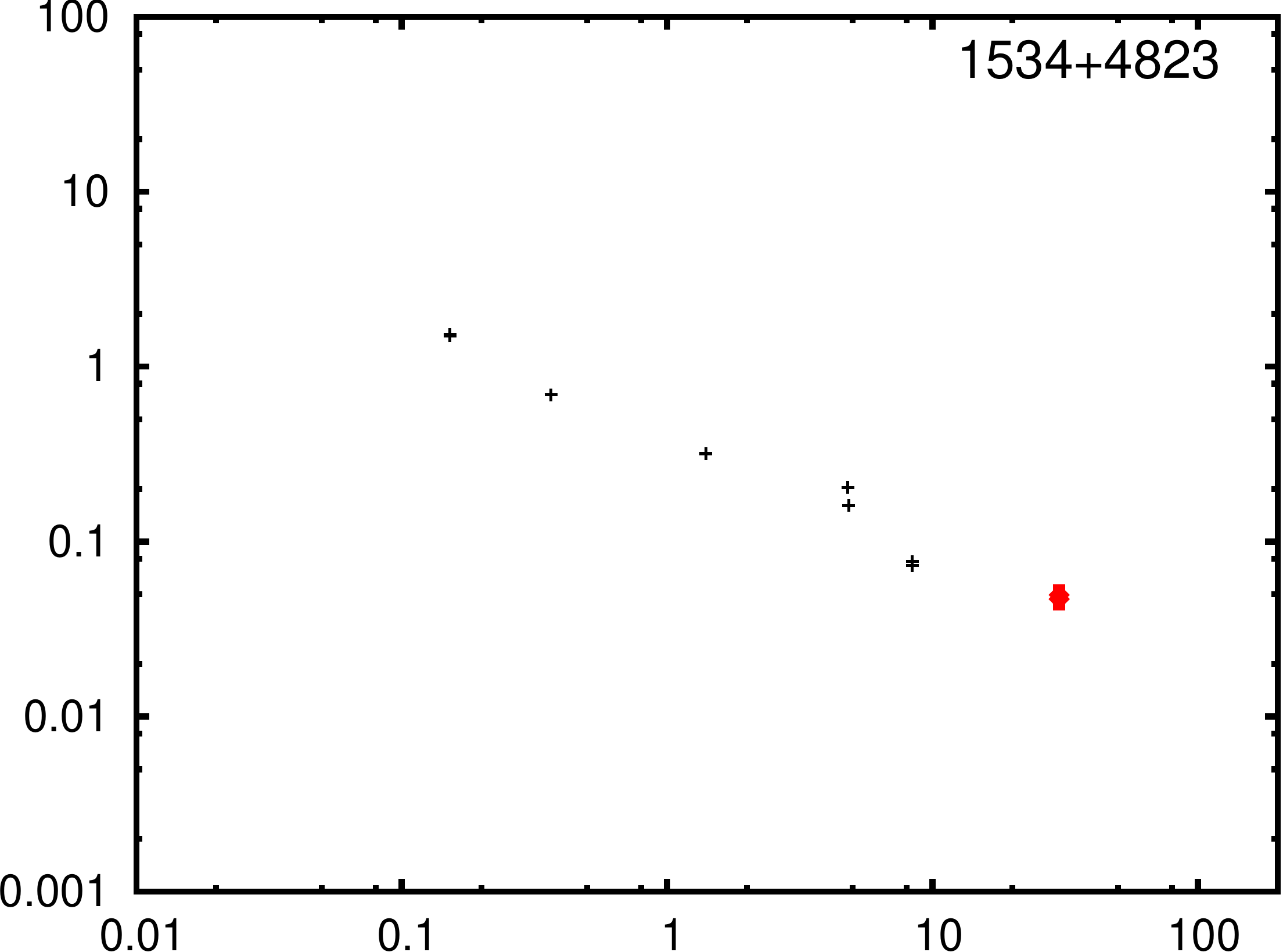}
\includegraphics[scale=0.2]{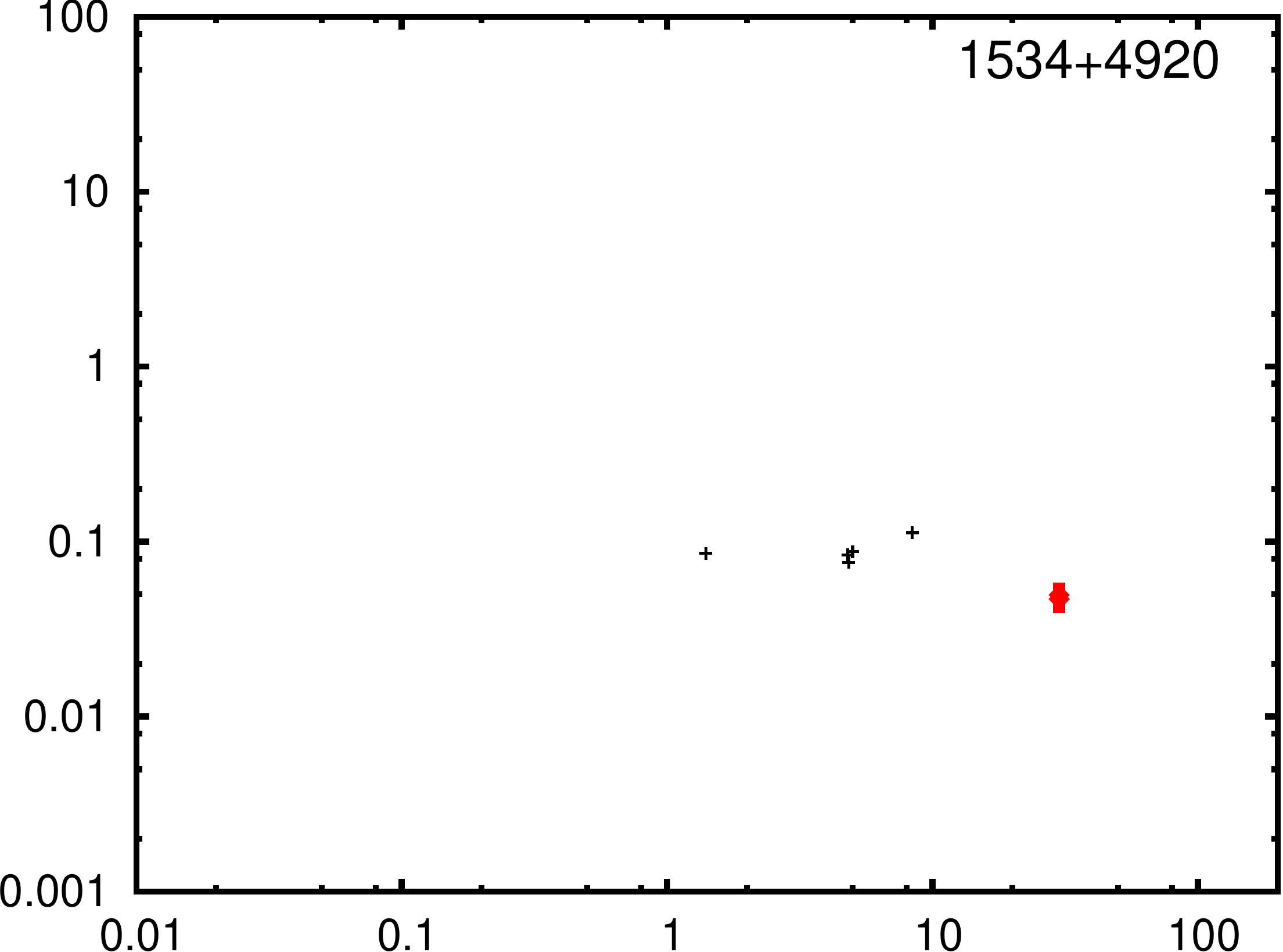}
\includegraphics[scale=0.2]{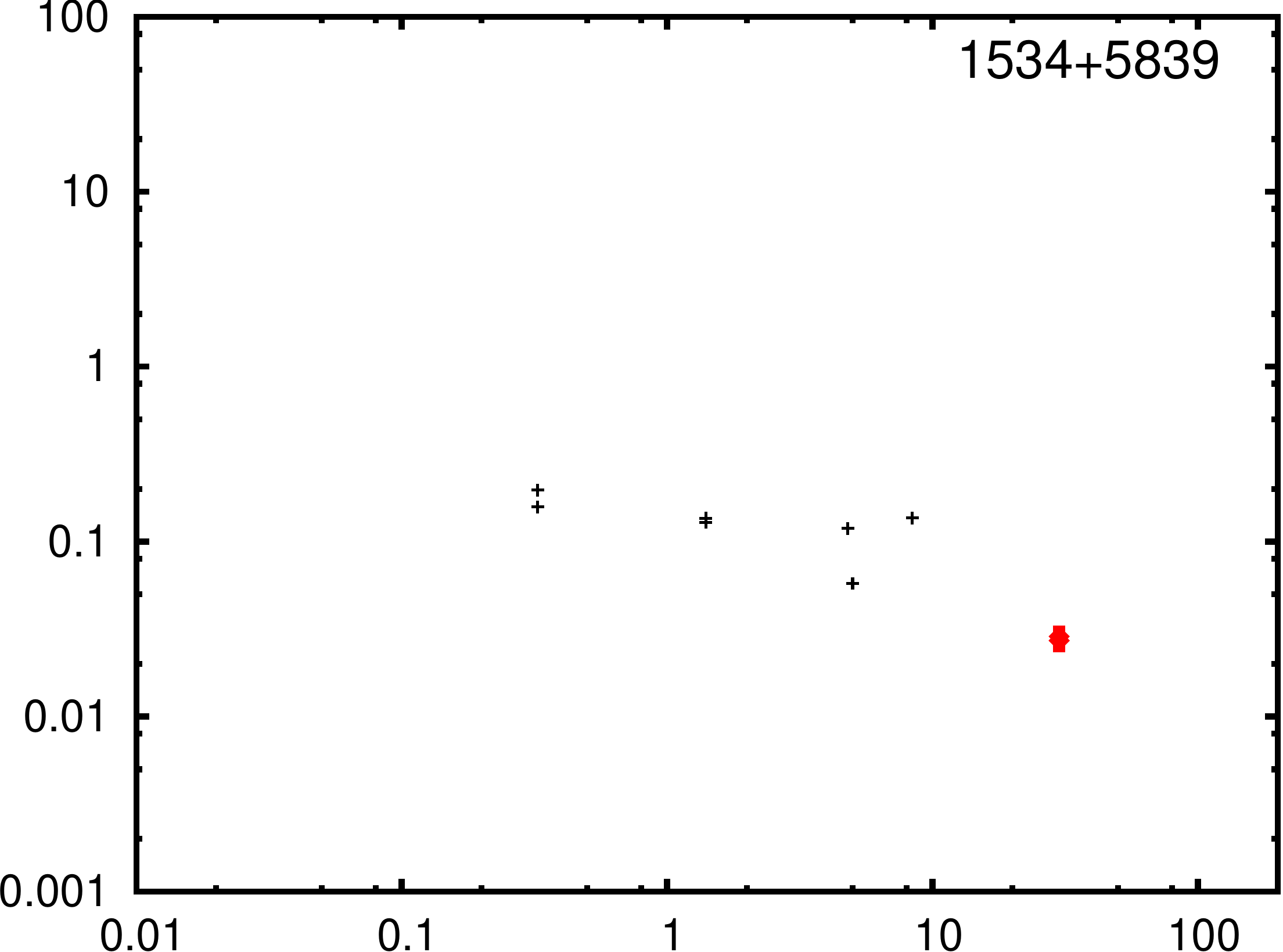}
\includegraphics[scale=0.2]{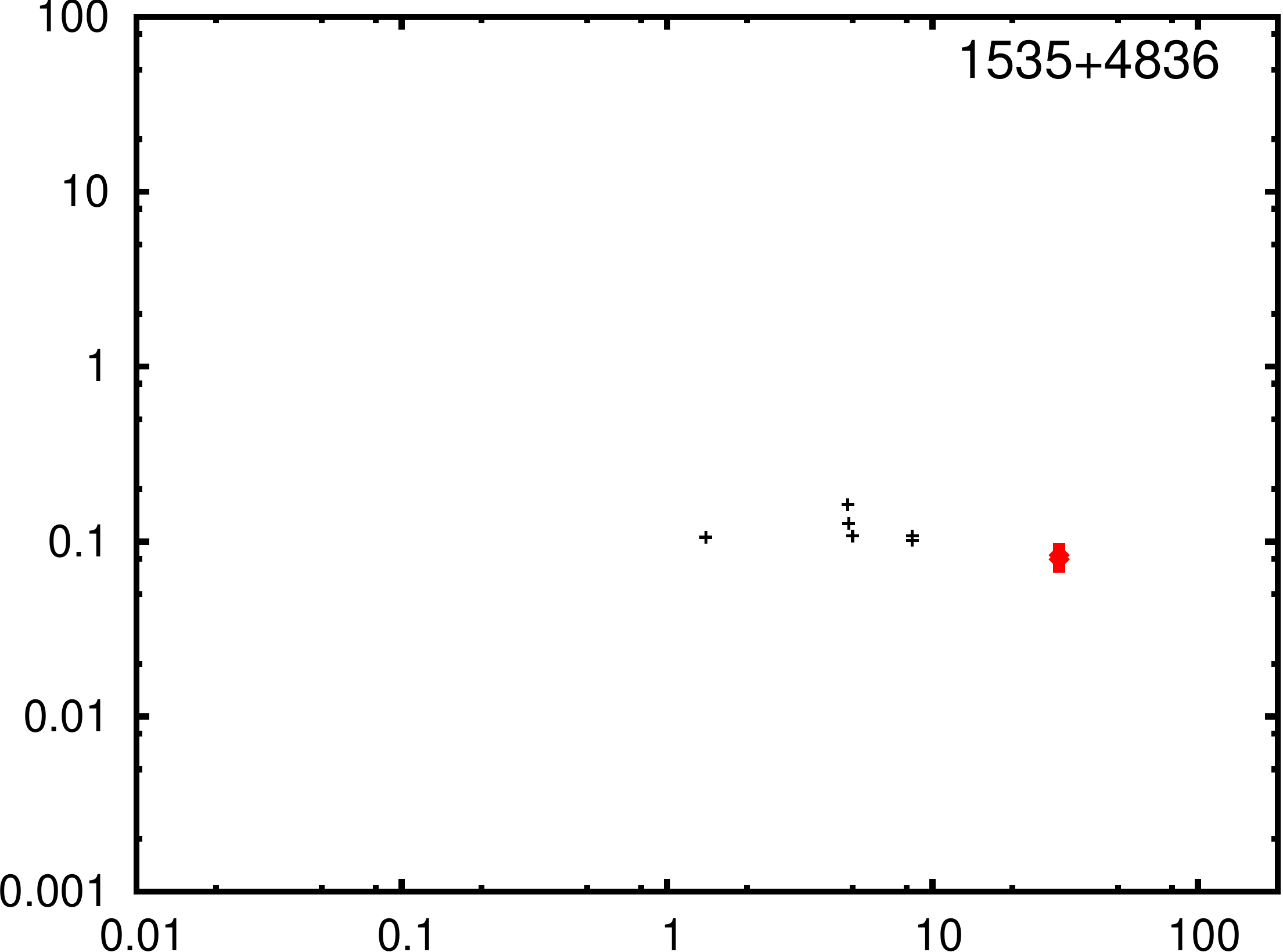}
\includegraphics[scale=0.2]{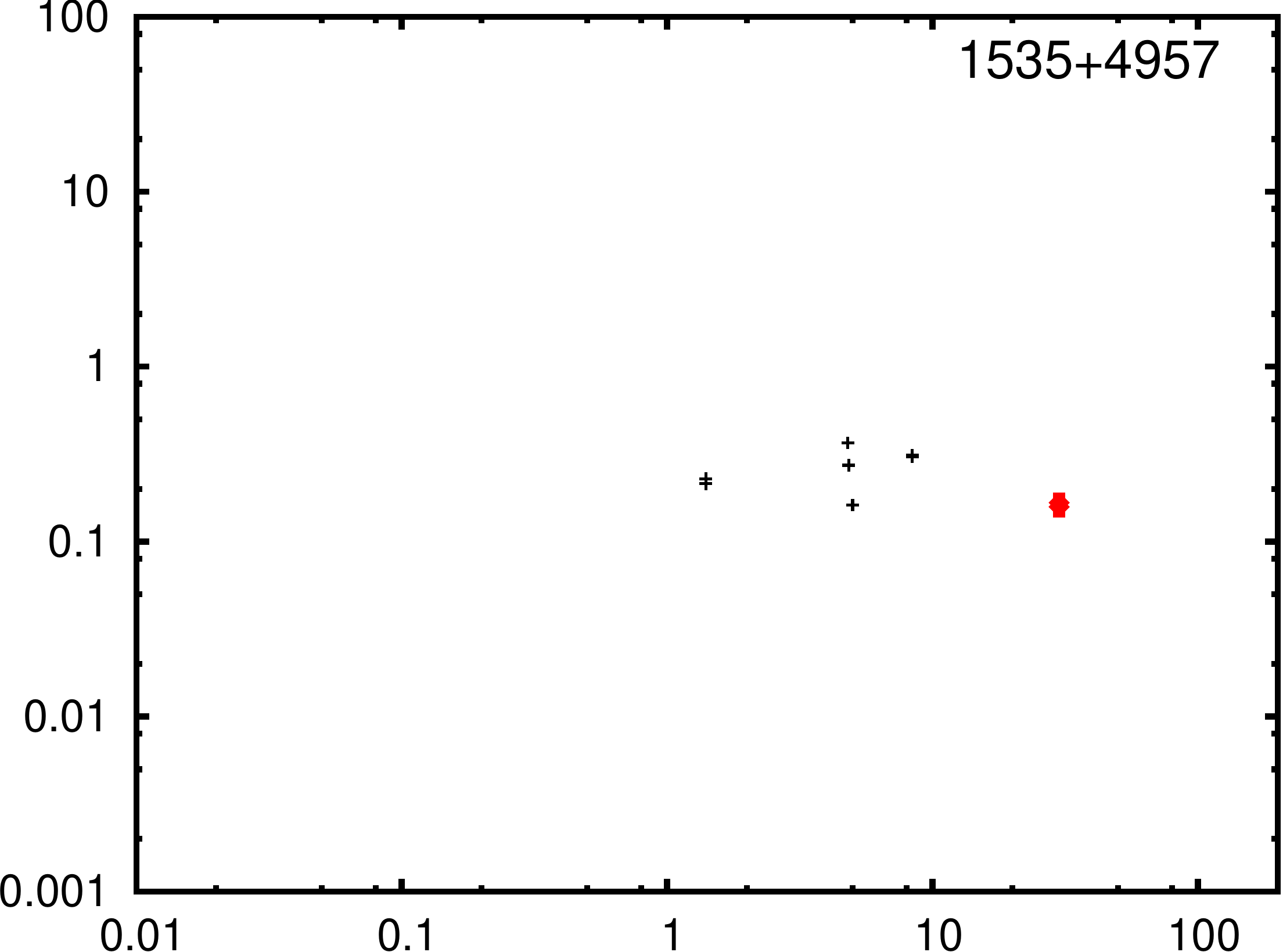}
\end{figure}
\clearpage\begin{figure}
\centering
\includegraphics[scale=0.2]{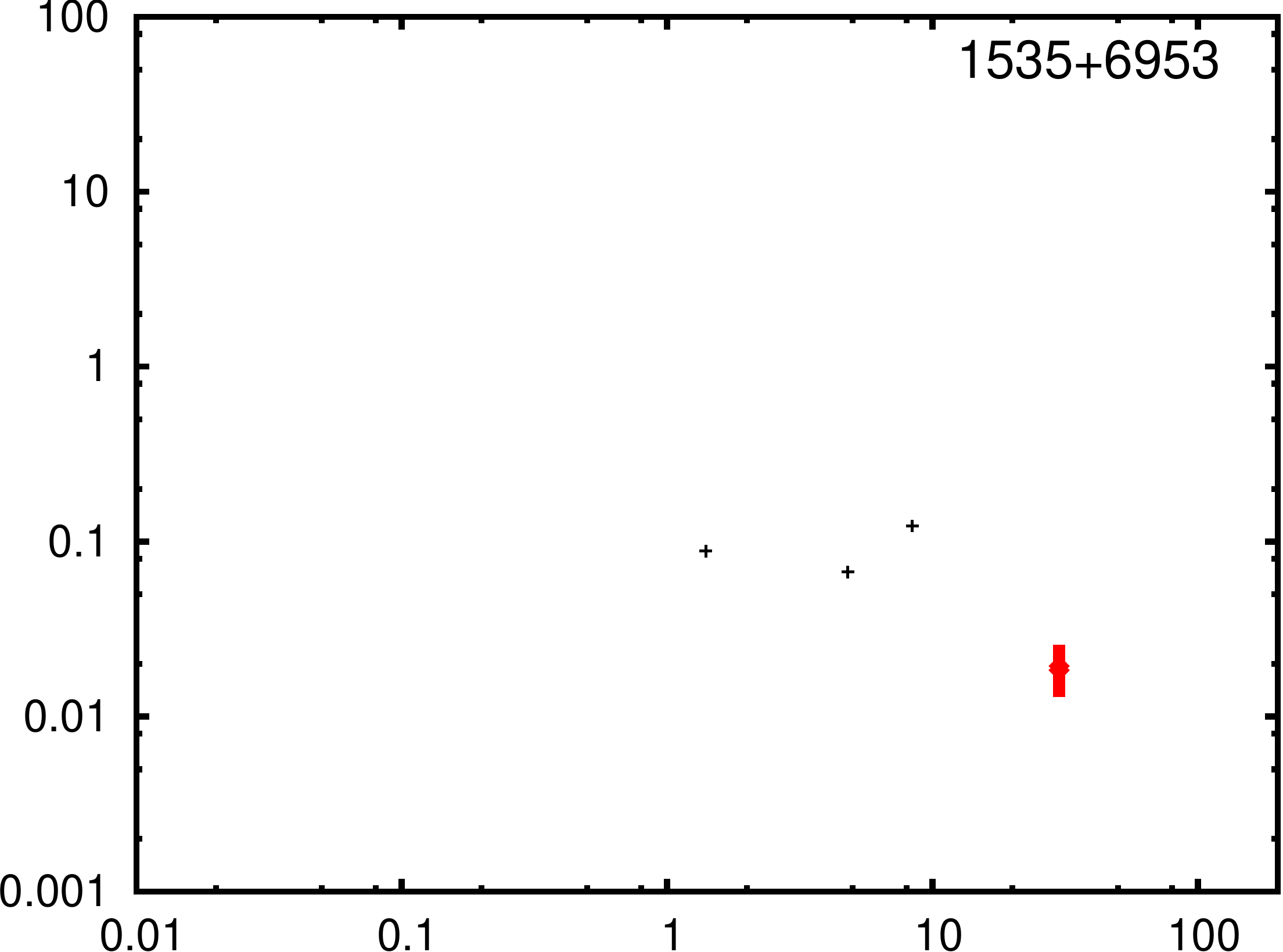}
\includegraphics[scale=0.2]{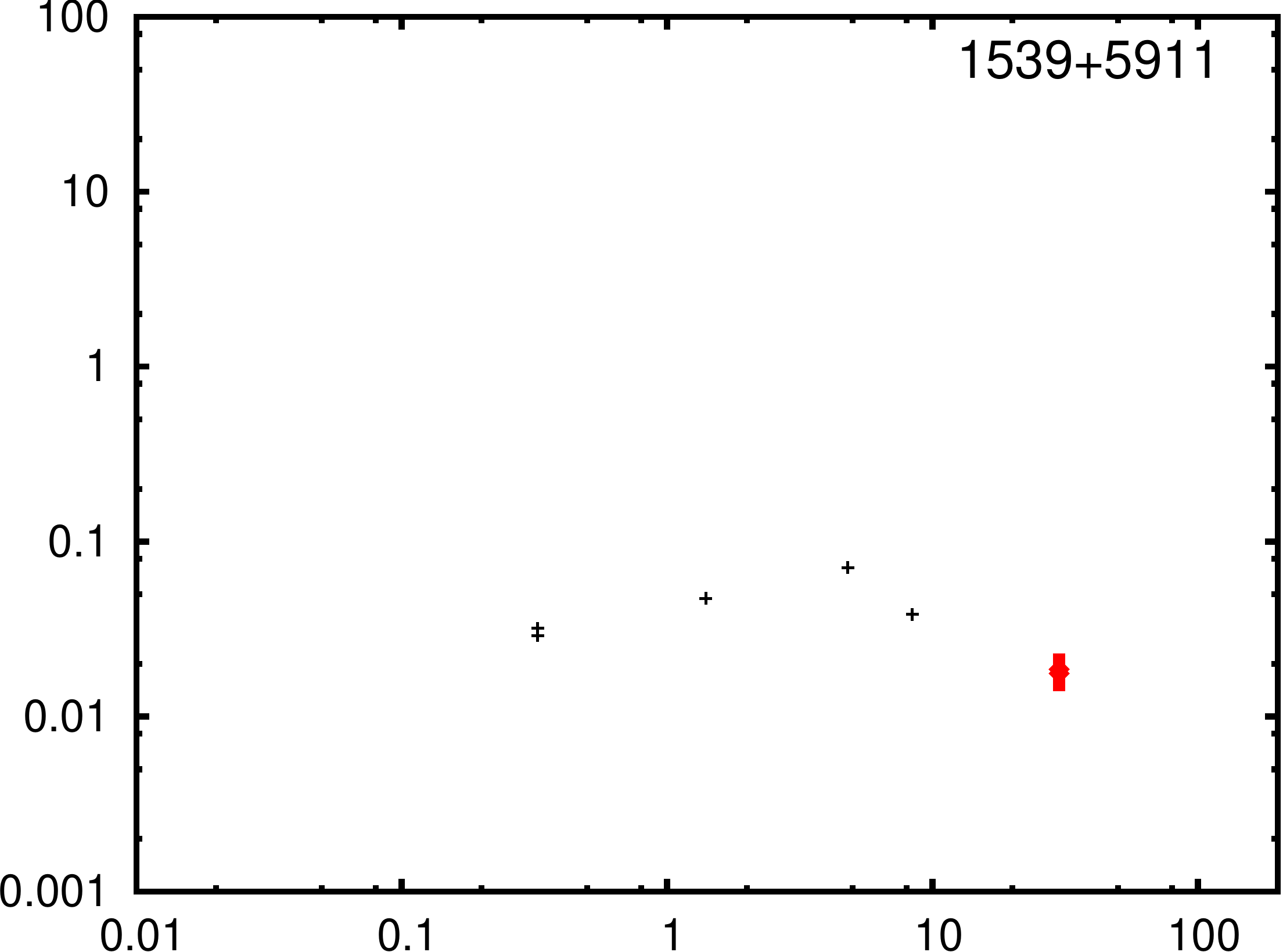}
\includegraphics[scale=0.2]{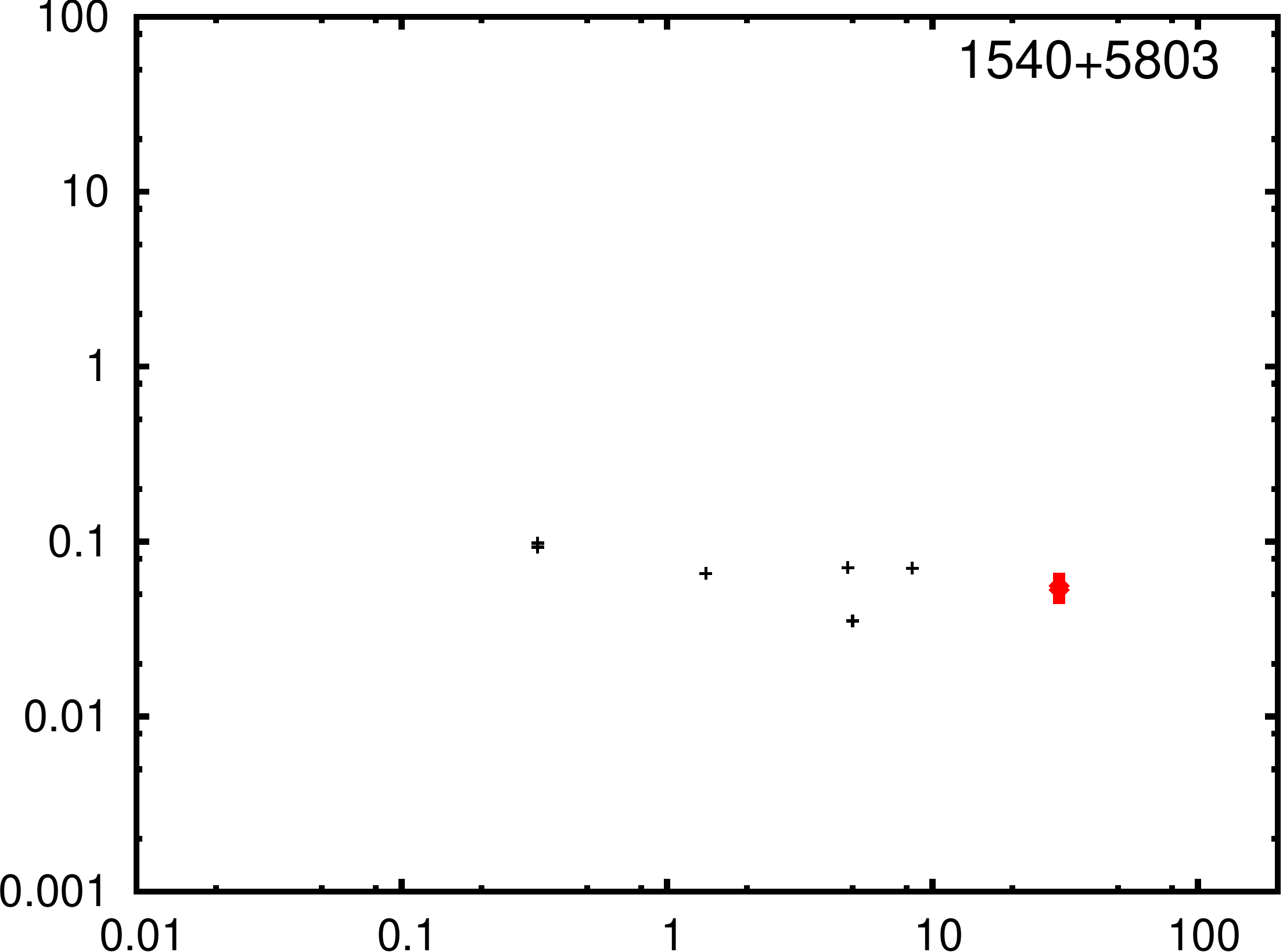}
\includegraphics[scale=0.2]{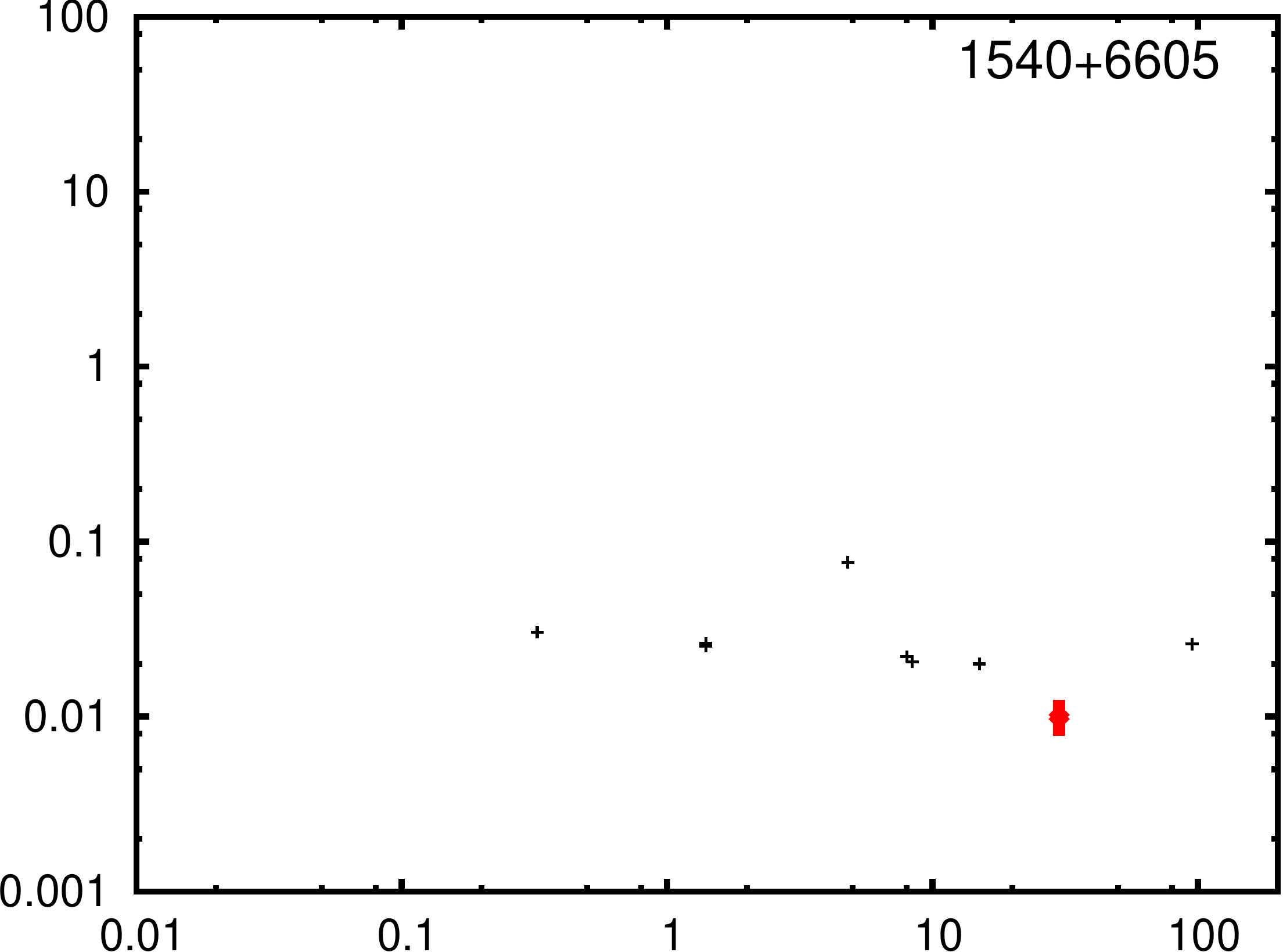}
\includegraphics[scale=0.2]{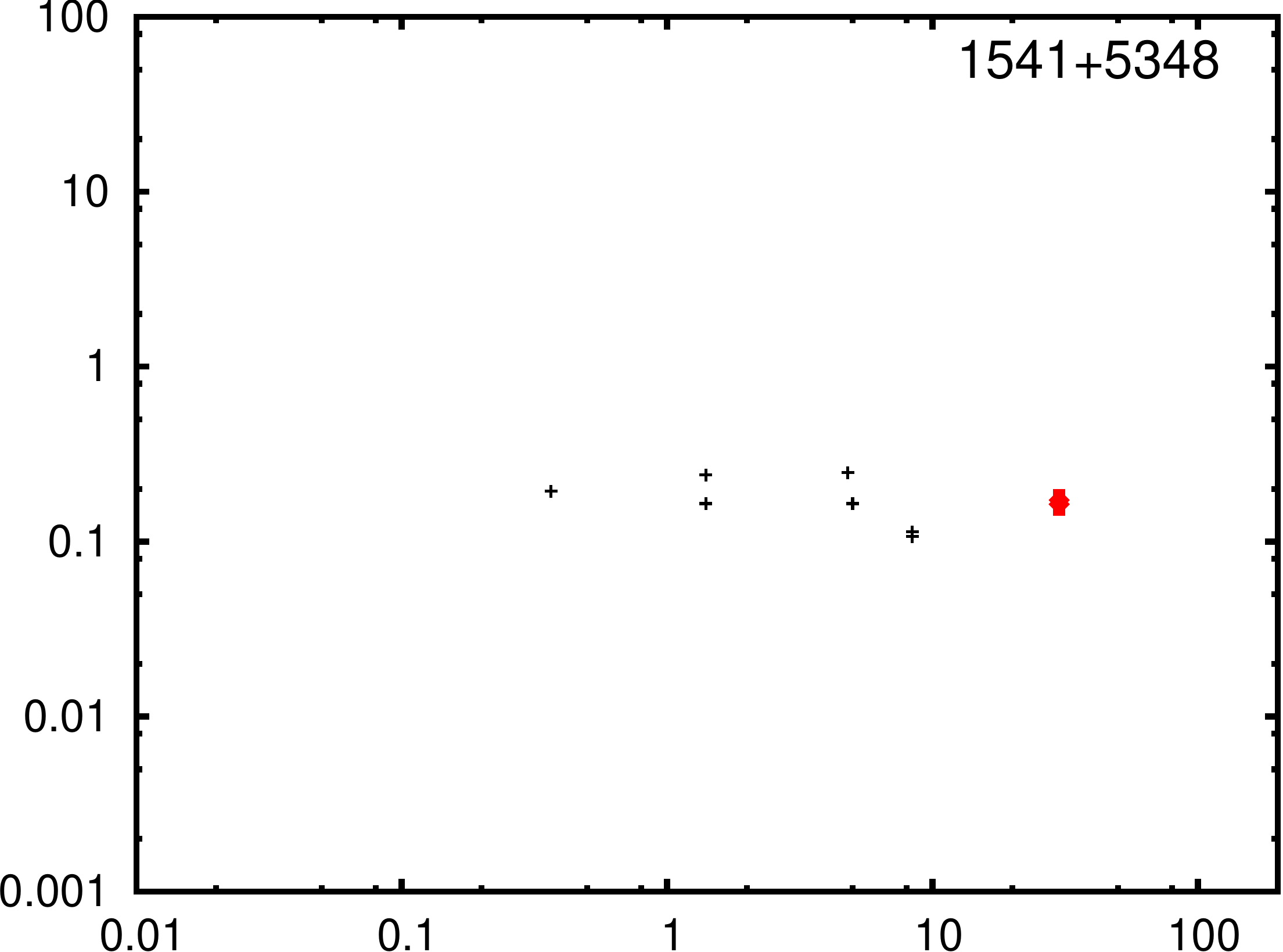}
\includegraphics[scale=0.2]{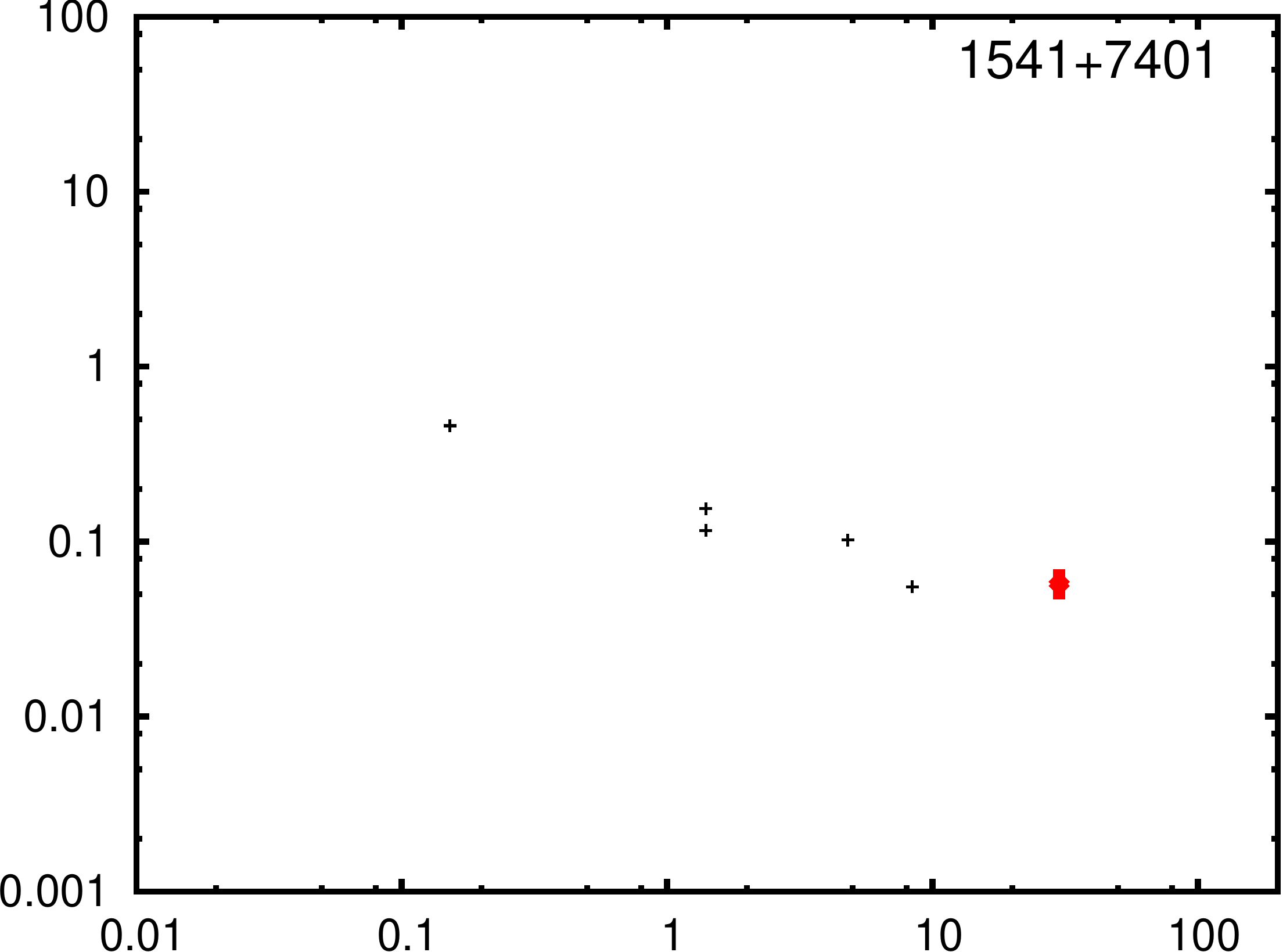}
\includegraphics[scale=0.2]{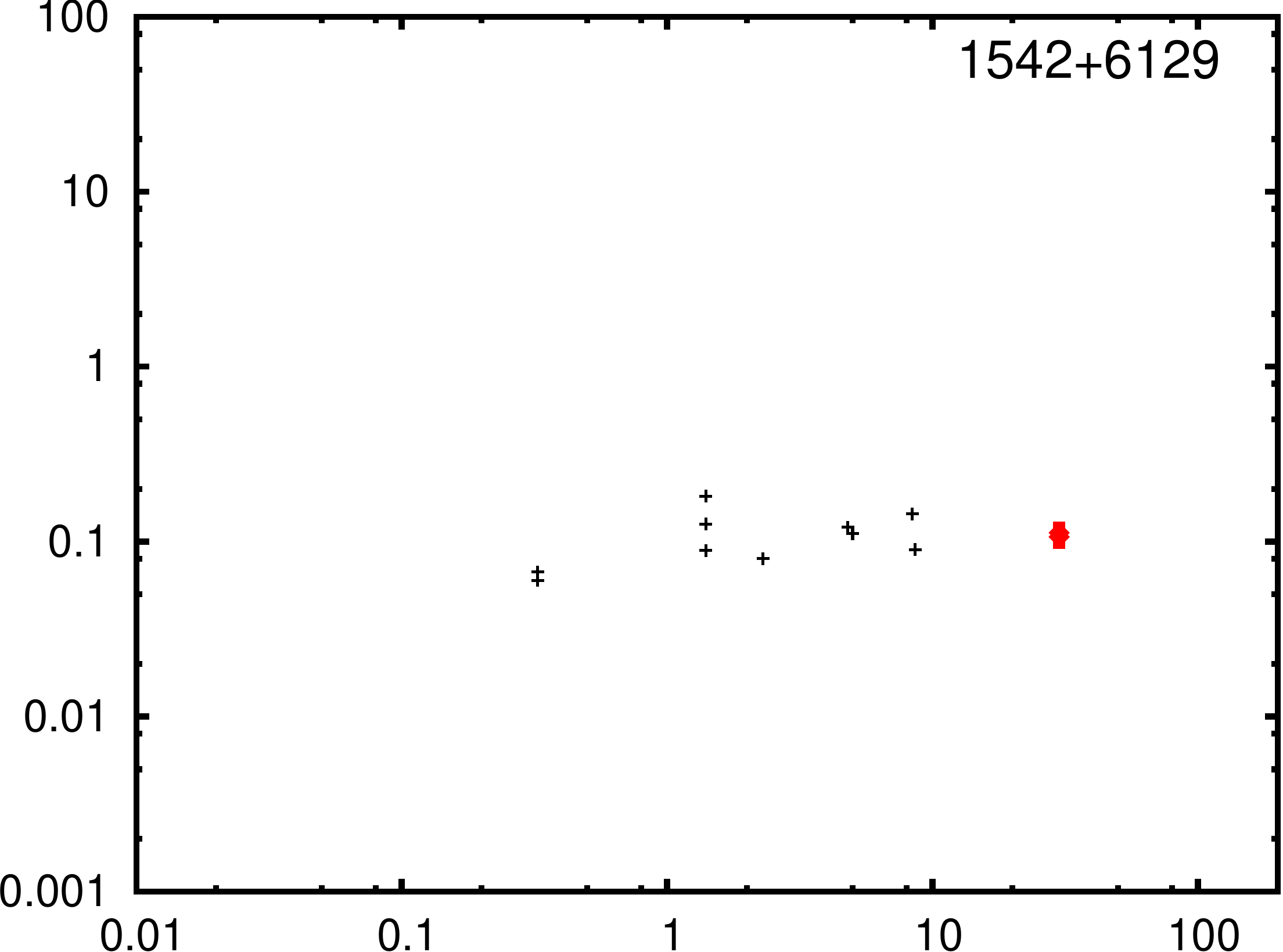}
\includegraphics[scale=0.2]{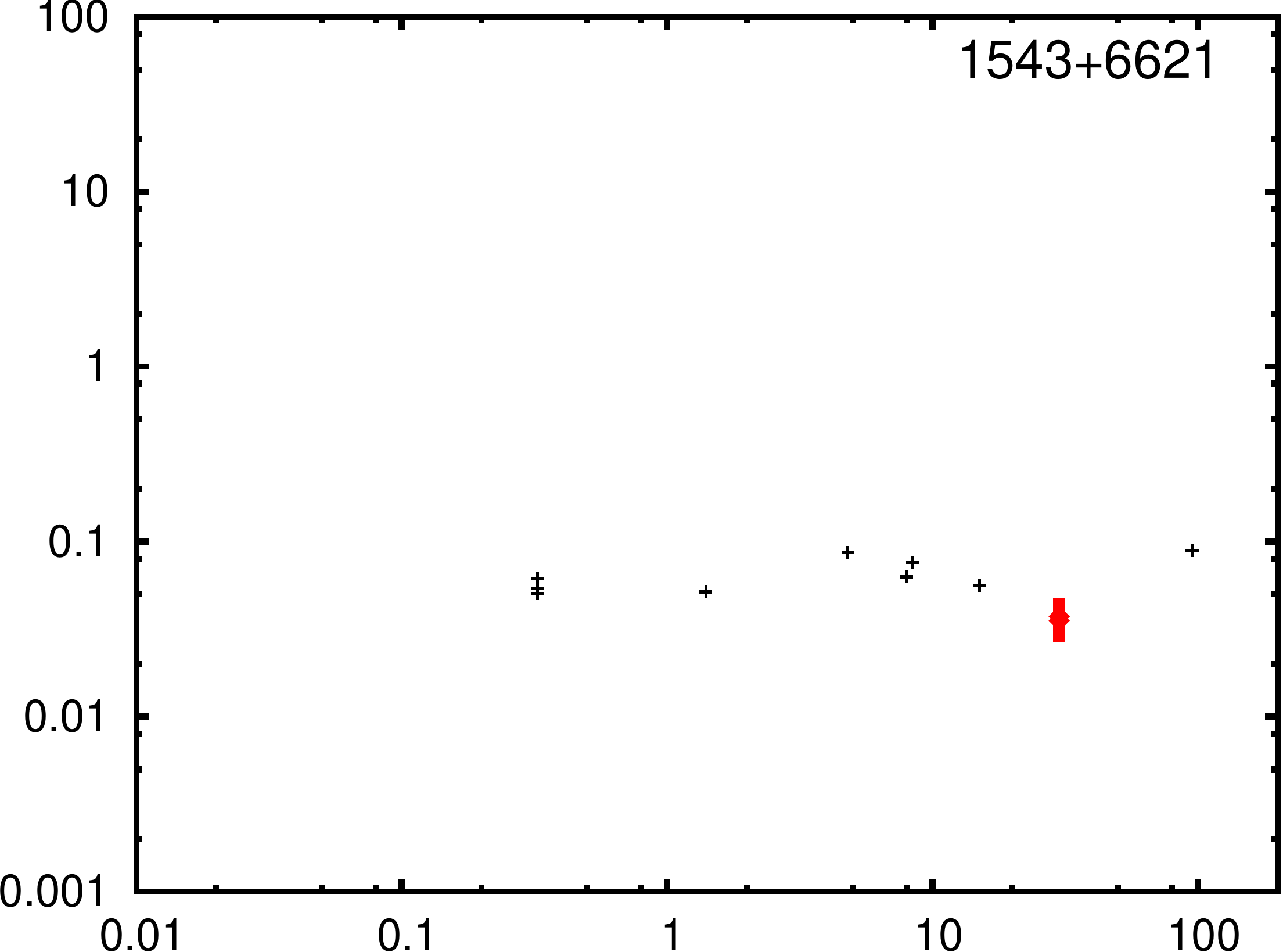}
\includegraphics[scale=0.2]{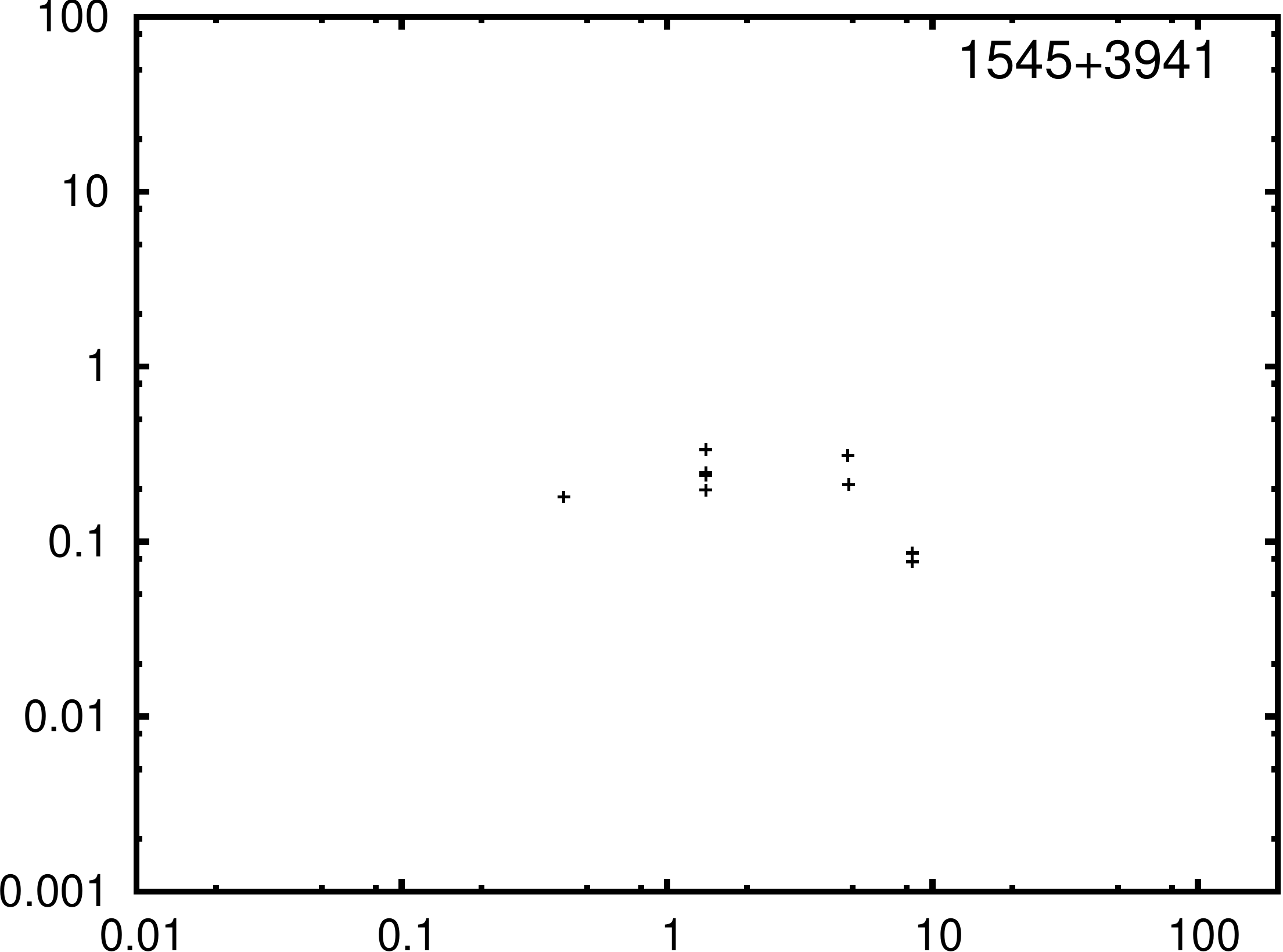}
\includegraphics[scale=0.2]{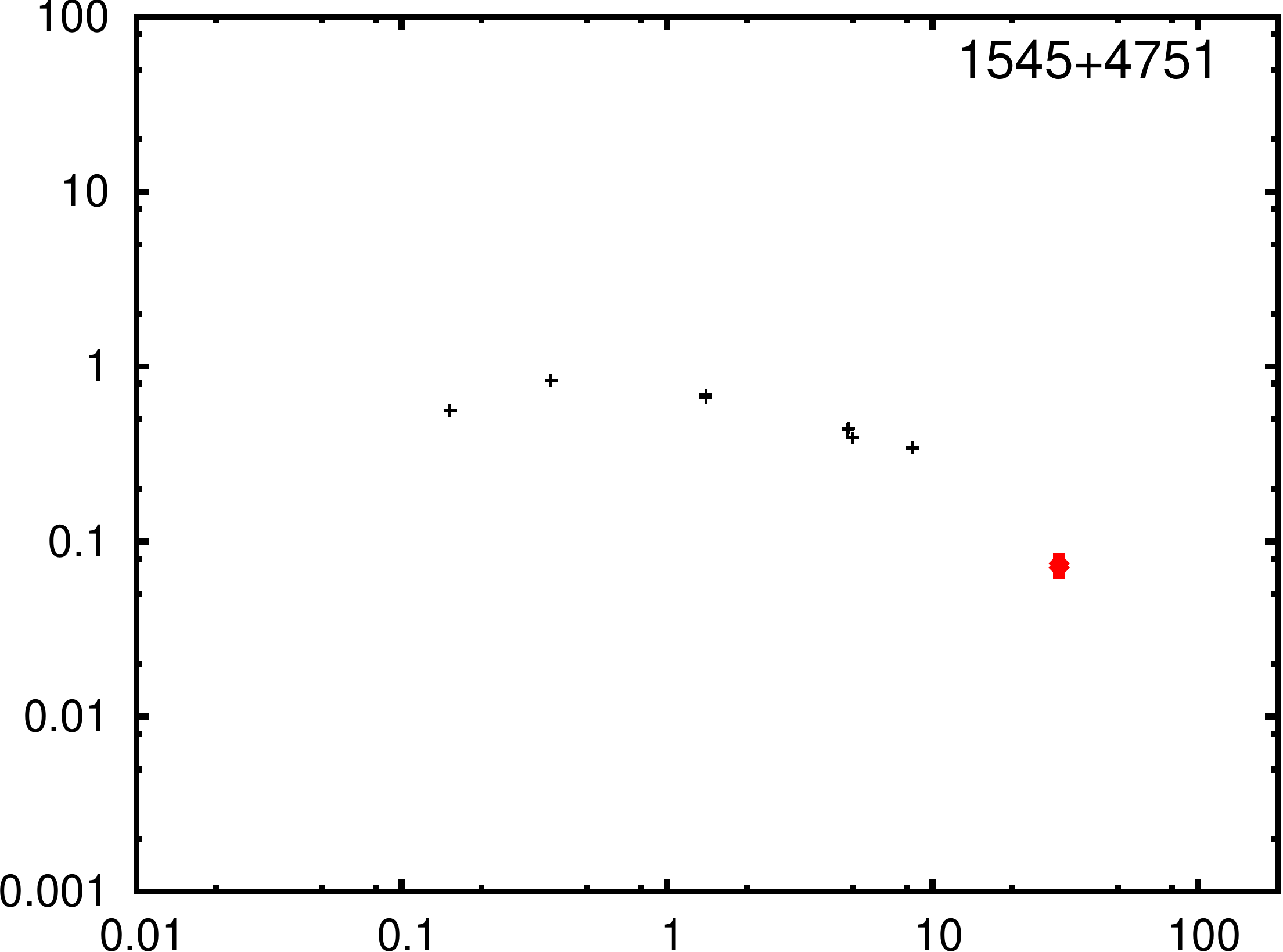}
\includegraphics[scale=0.2]{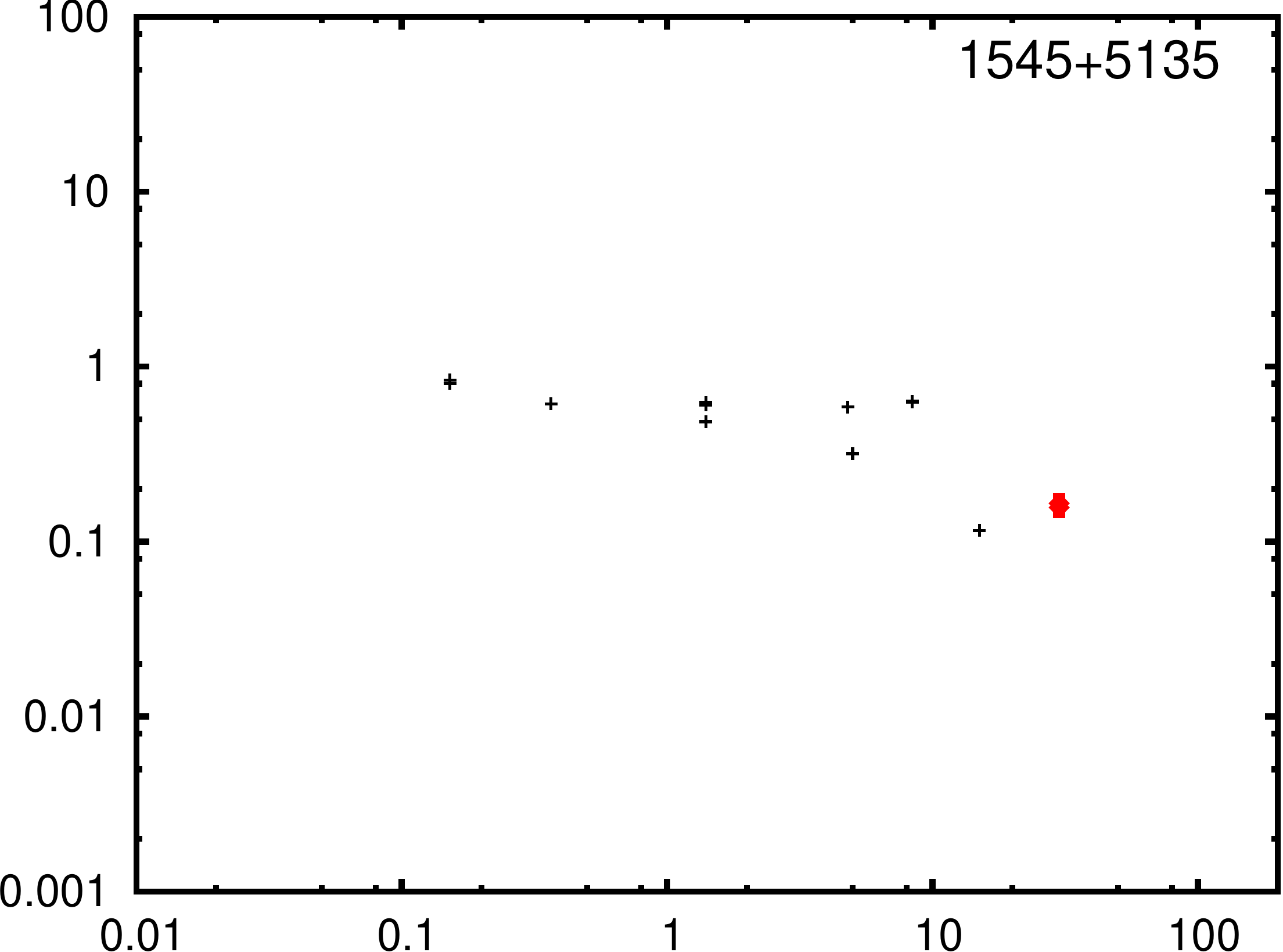}
\includegraphics[scale=0.2]{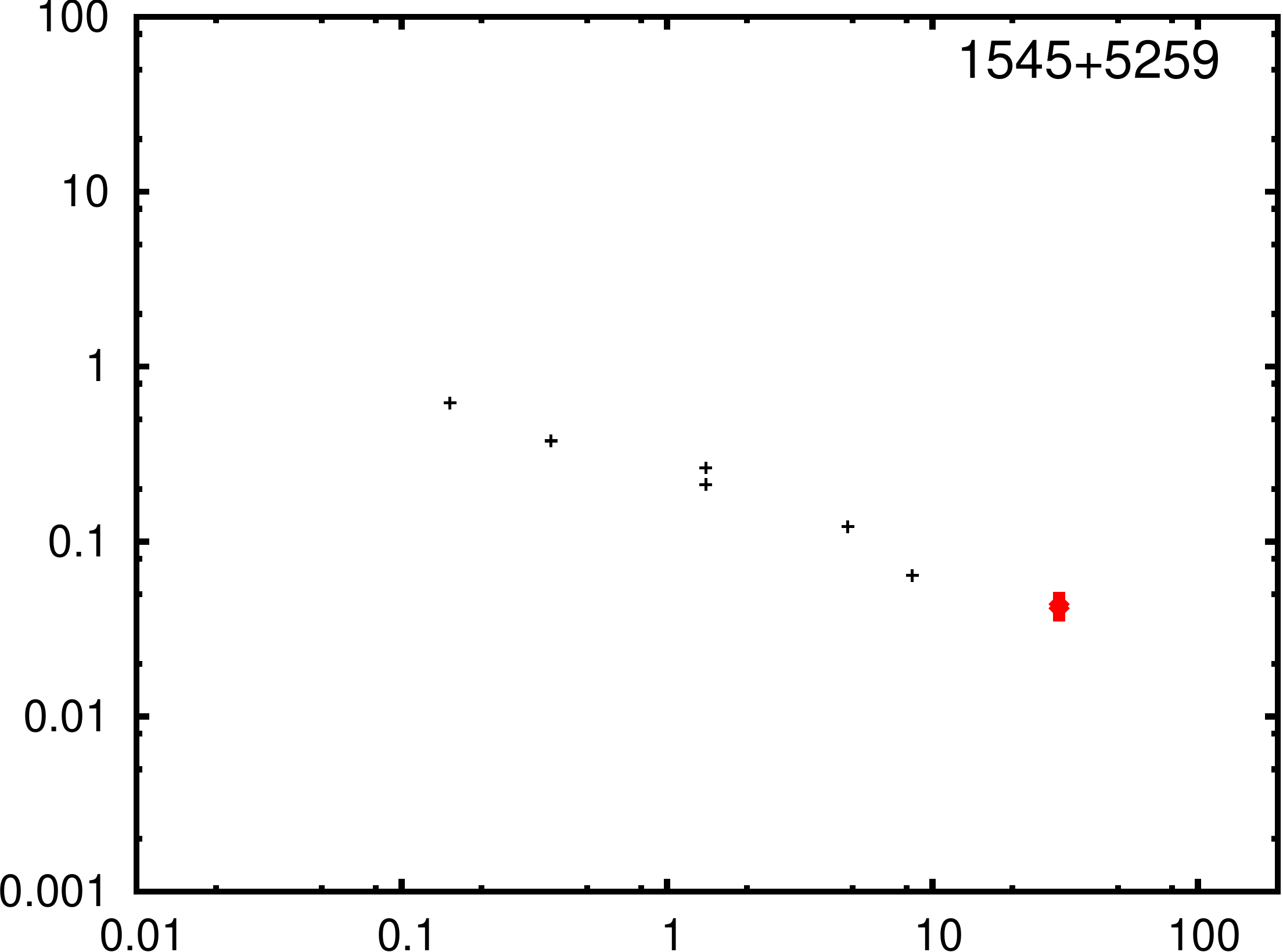}
\includegraphics[scale=0.2]{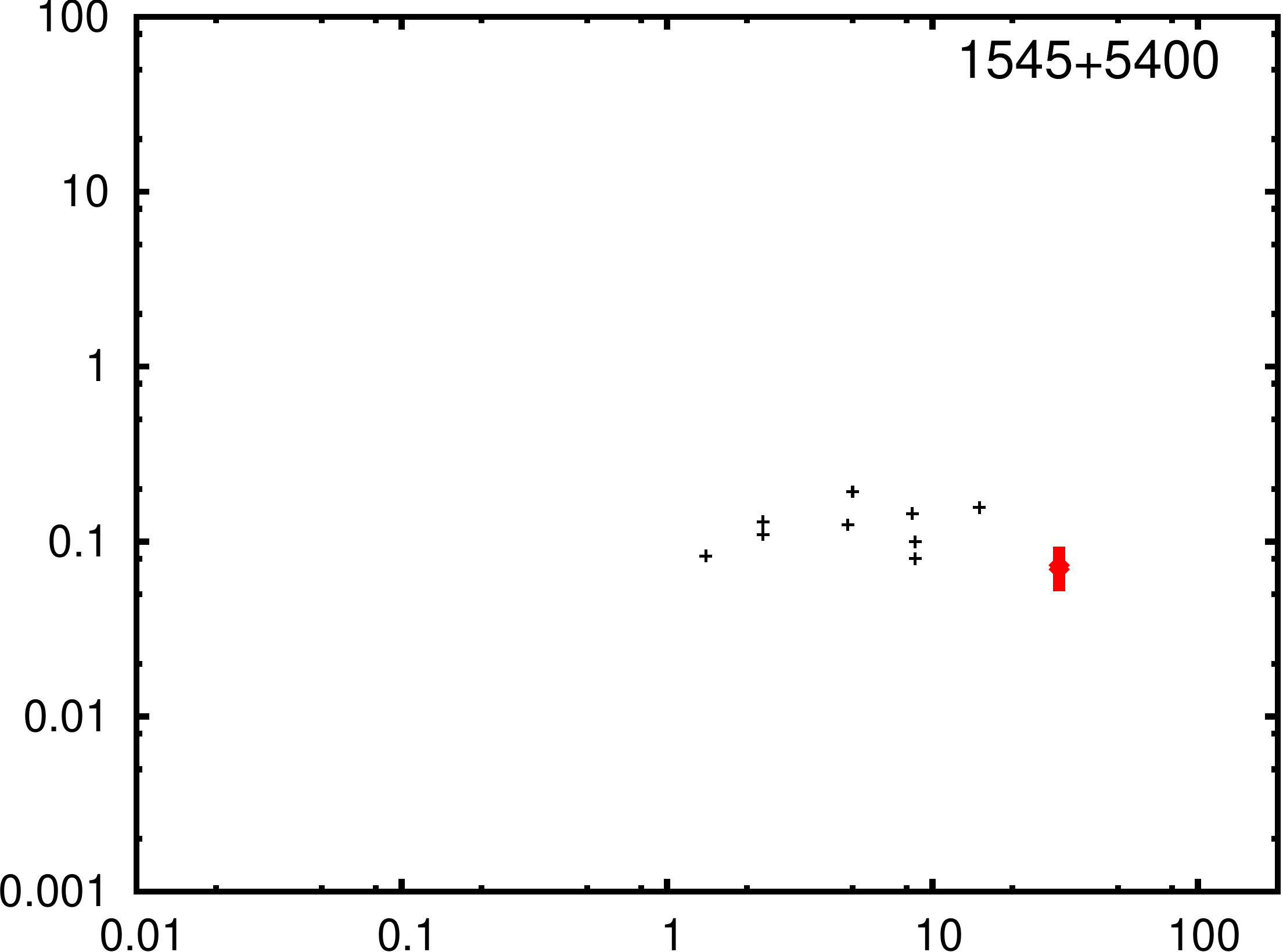}
\includegraphics[scale=0.2]{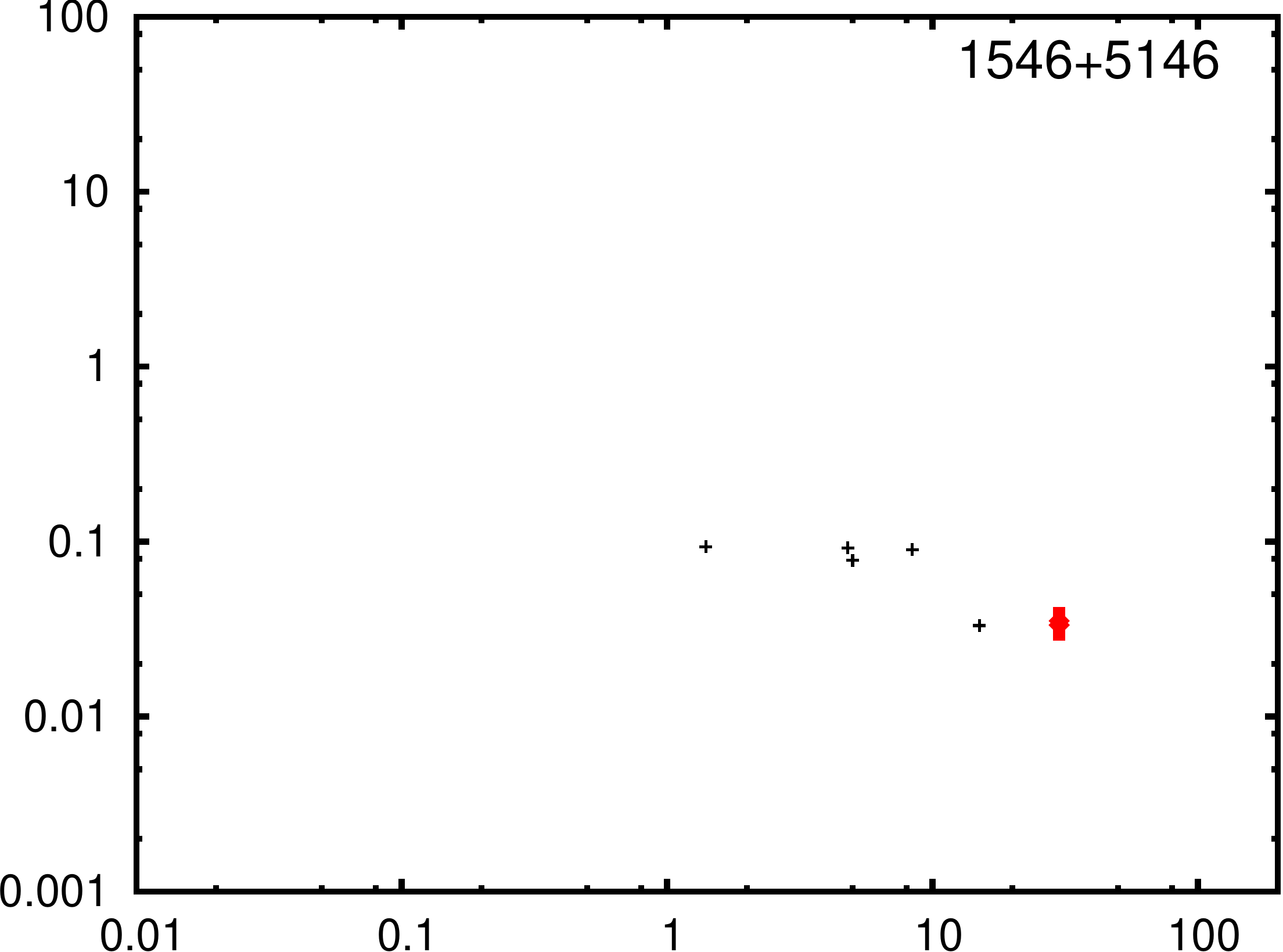}
\includegraphics[scale=0.2]{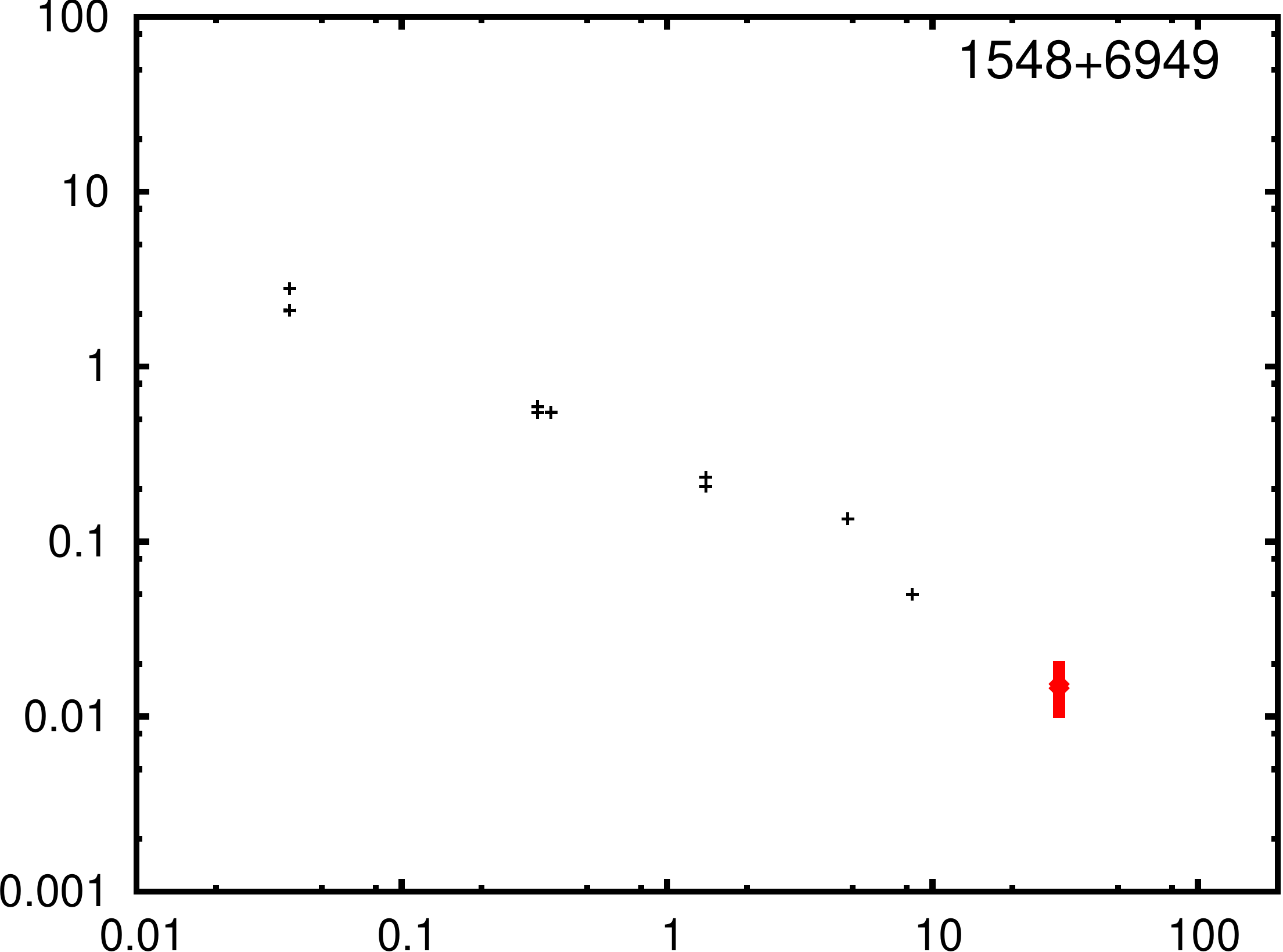}
\includegraphics[scale=0.2]{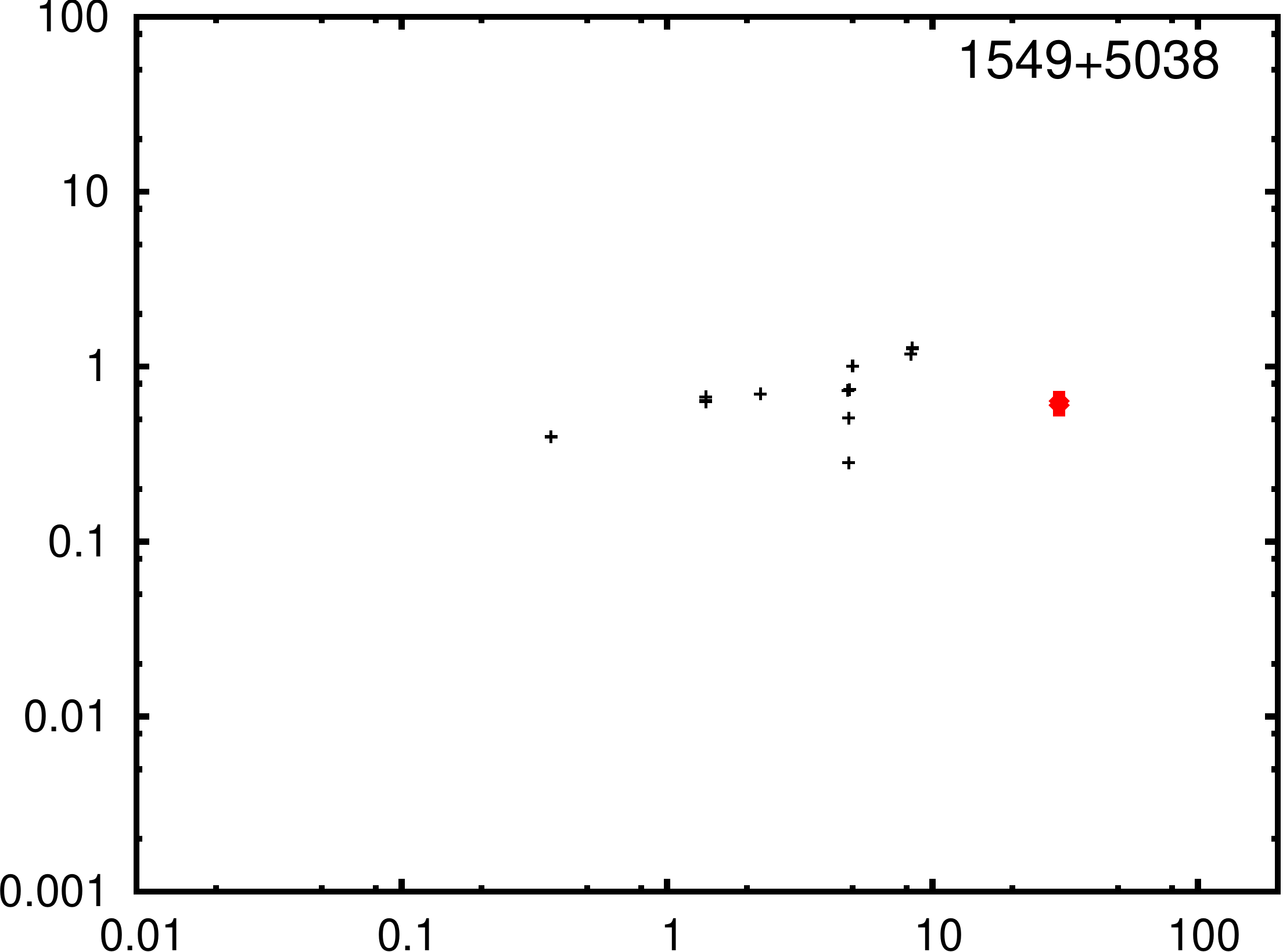}
\includegraphics[scale=0.2]{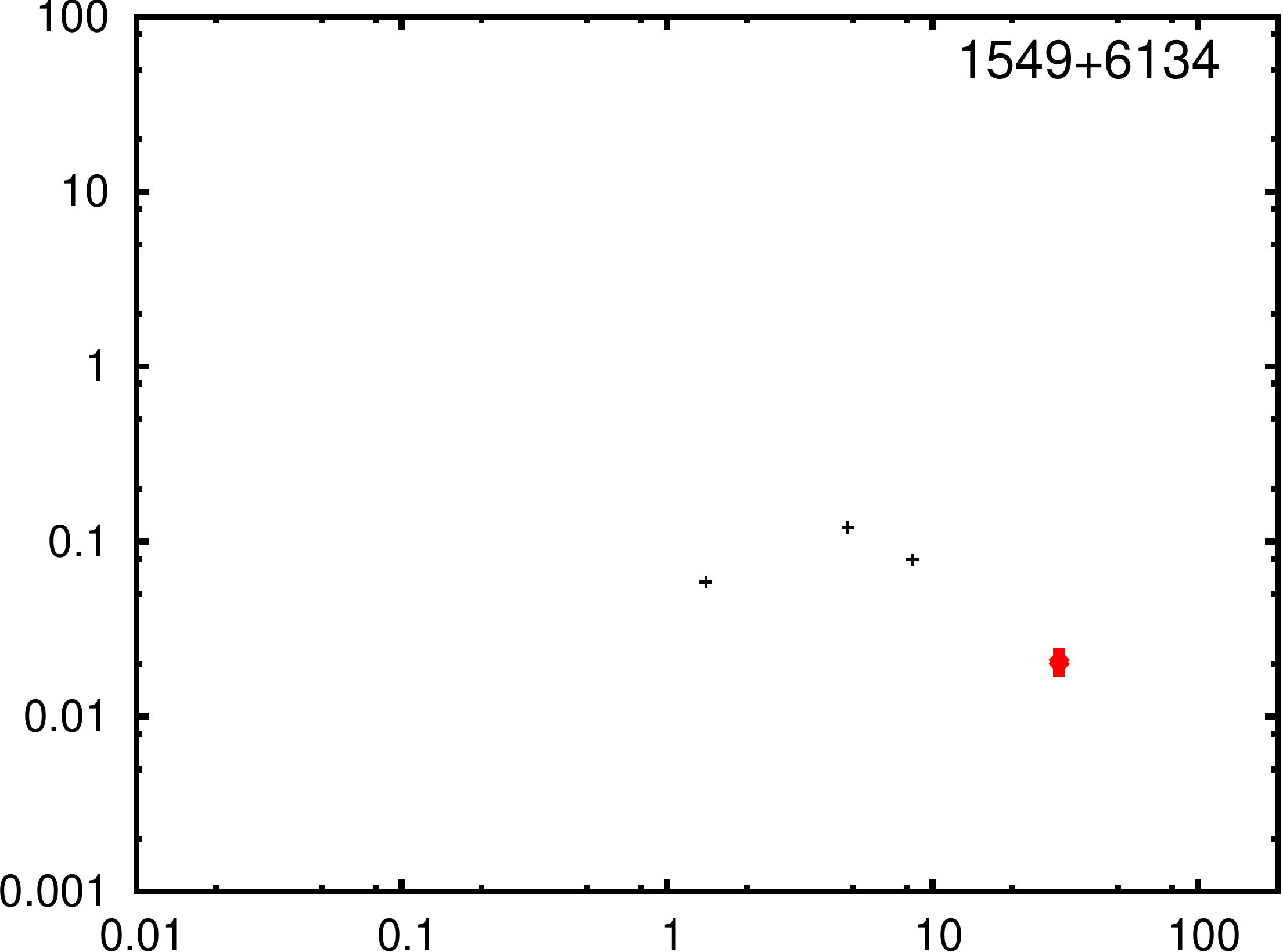}
\includegraphics[scale=0.2]{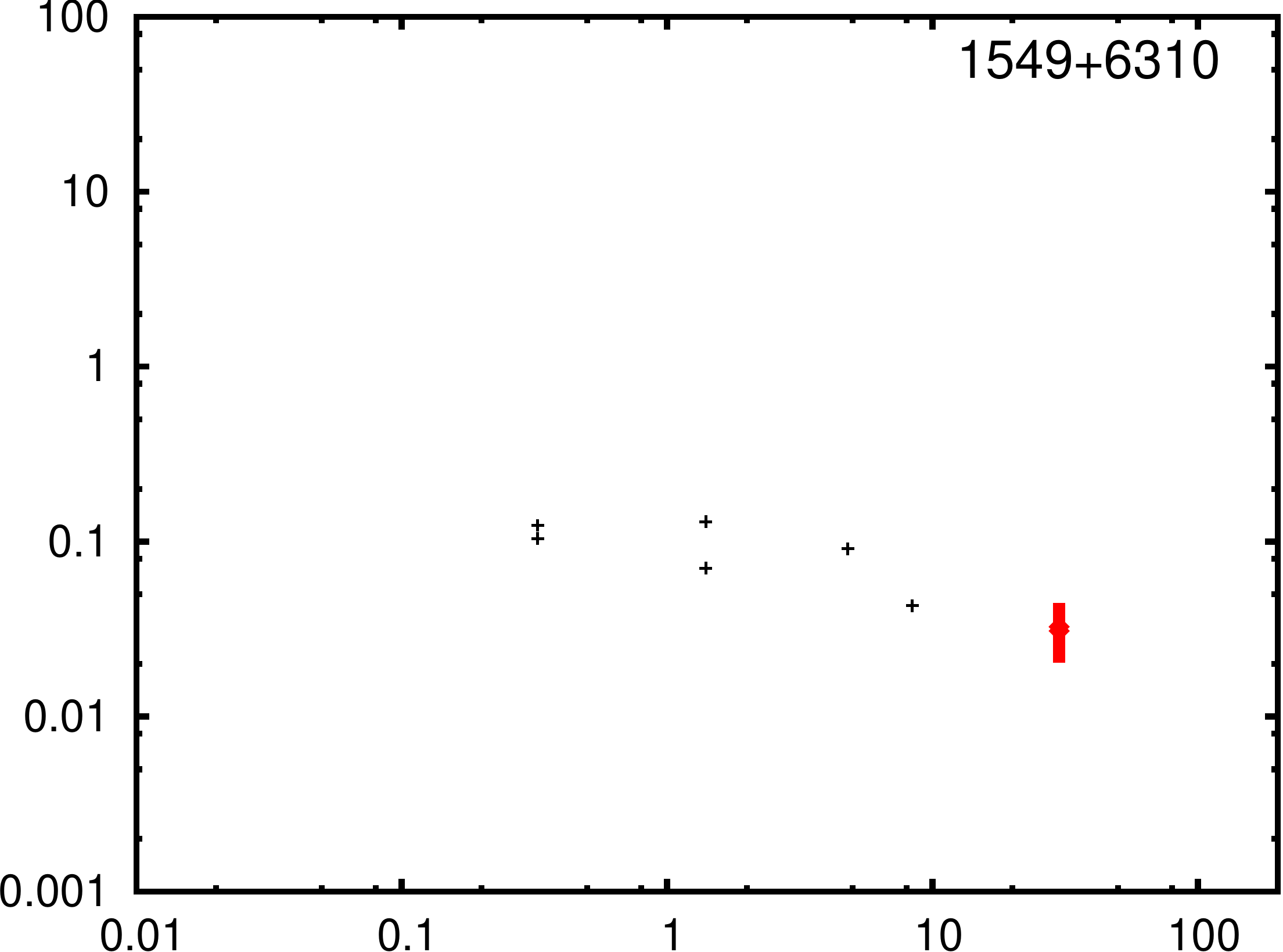}
\end{figure}
\clearpage\begin{figure}
\centering
\includegraphics[scale=0.2]{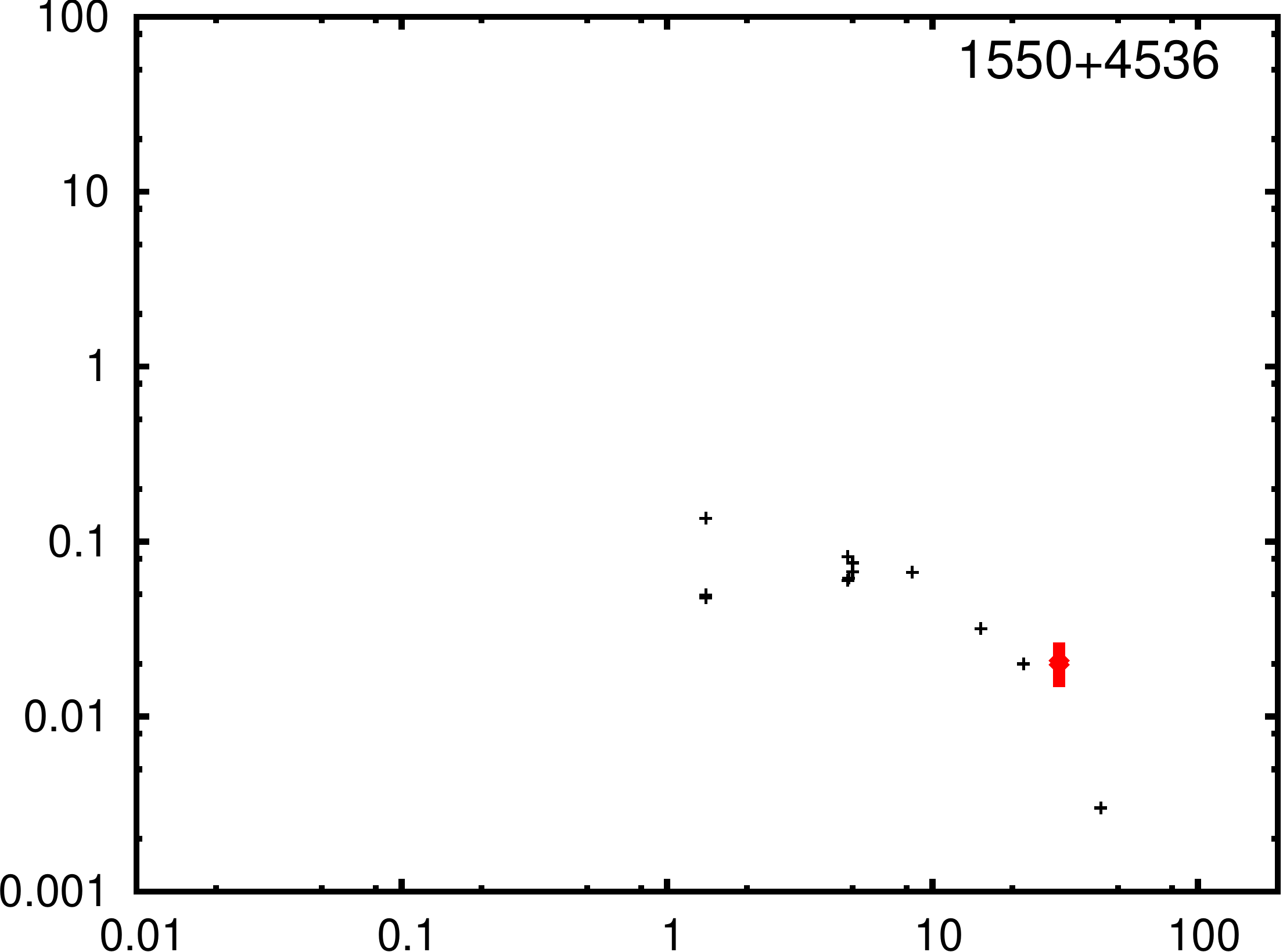}
\includegraphics[scale=0.2]{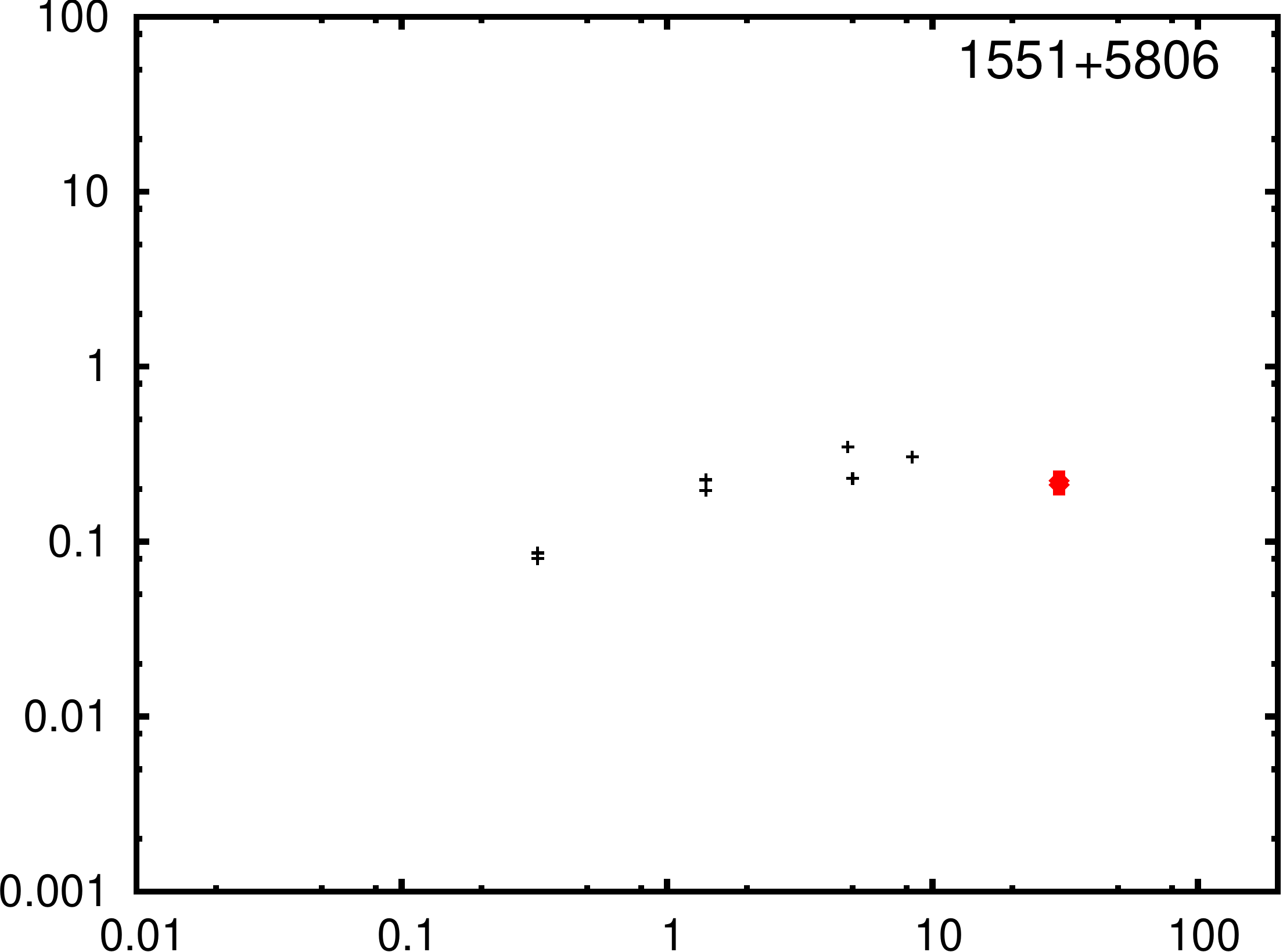}
\includegraphics[scale=0.2]{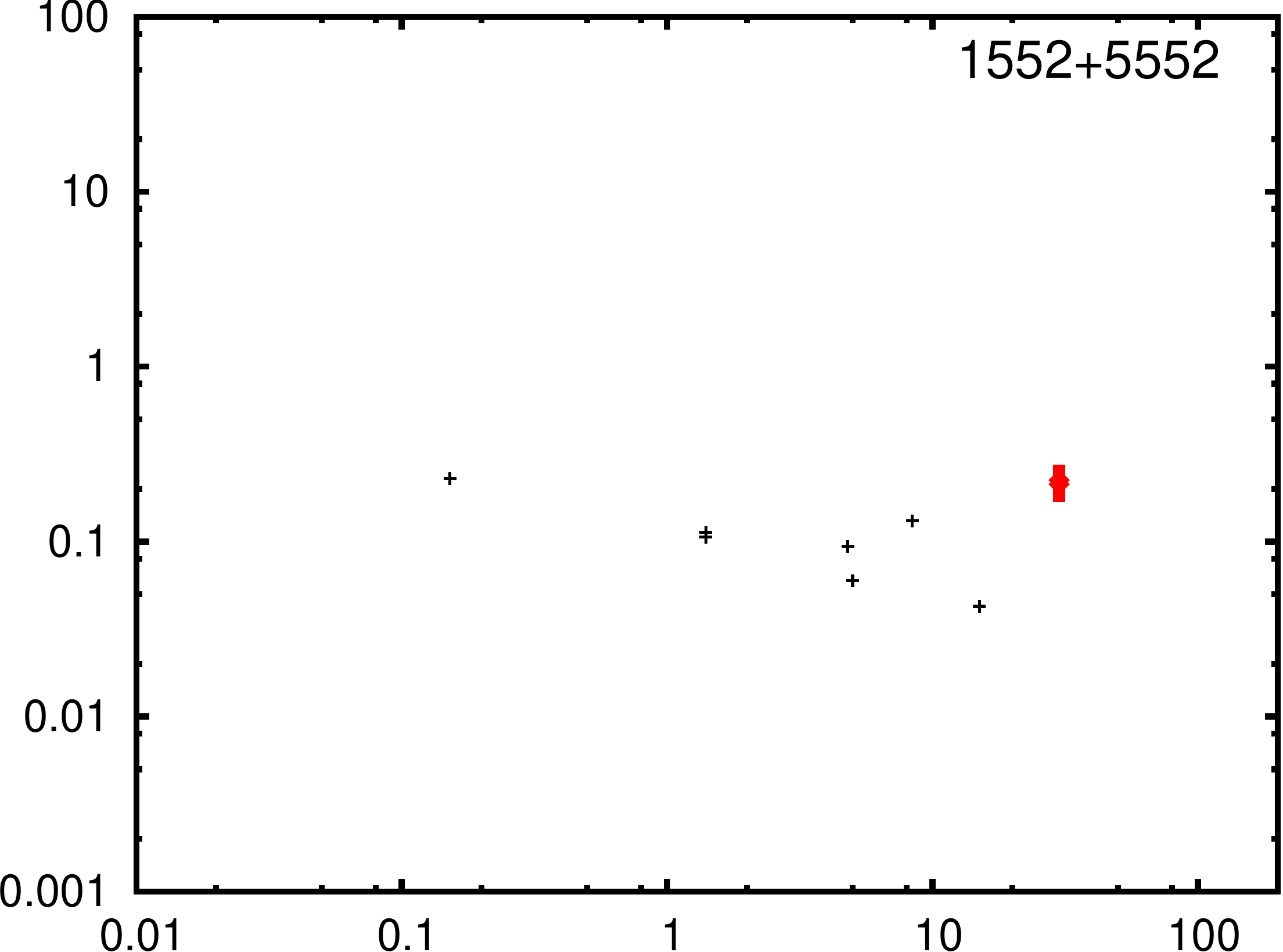}
\includegraphics[scale=0.2]{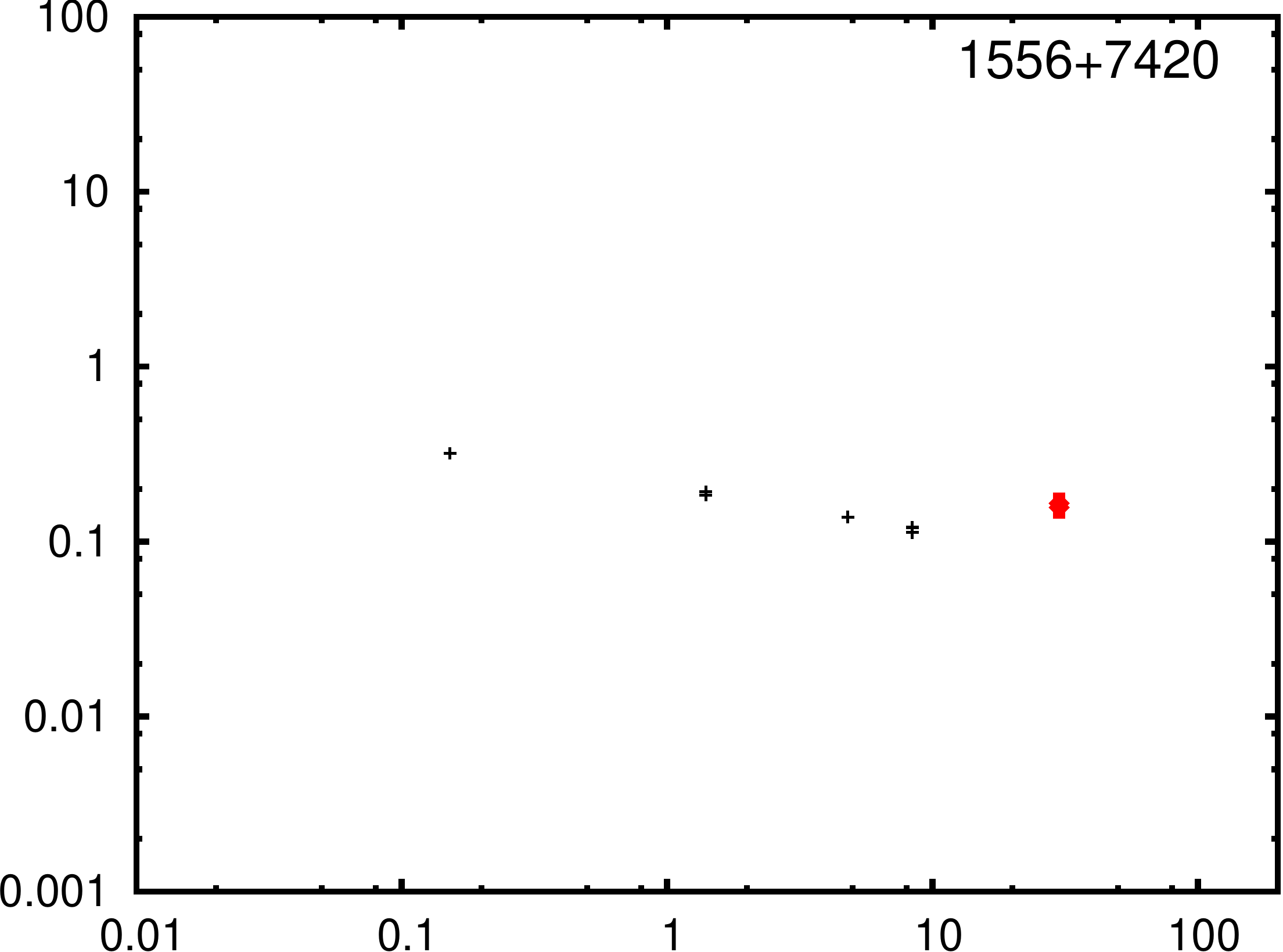}
\includegraphics[scale=0.2]{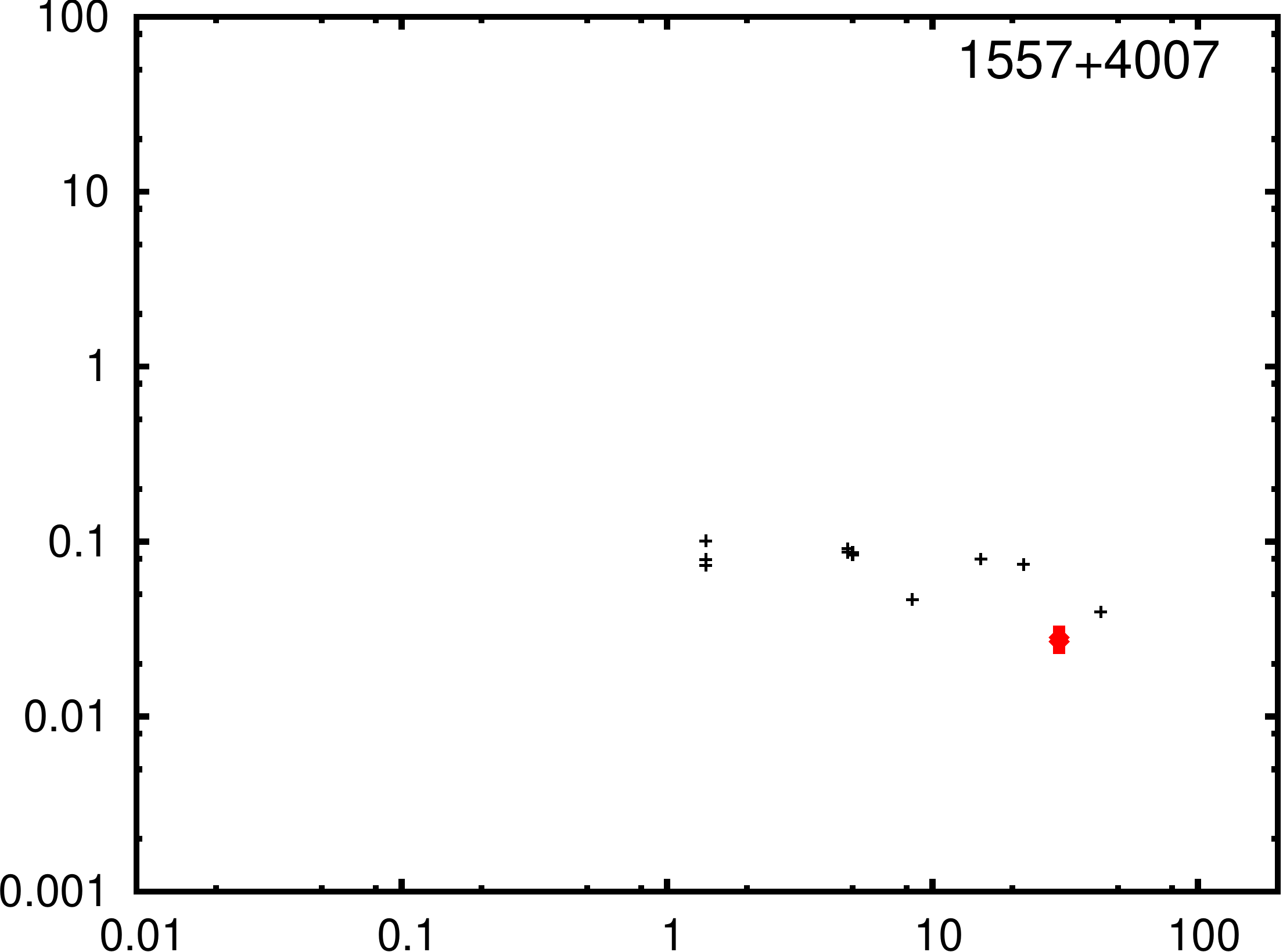}
\includegraphics[scale=0.2]{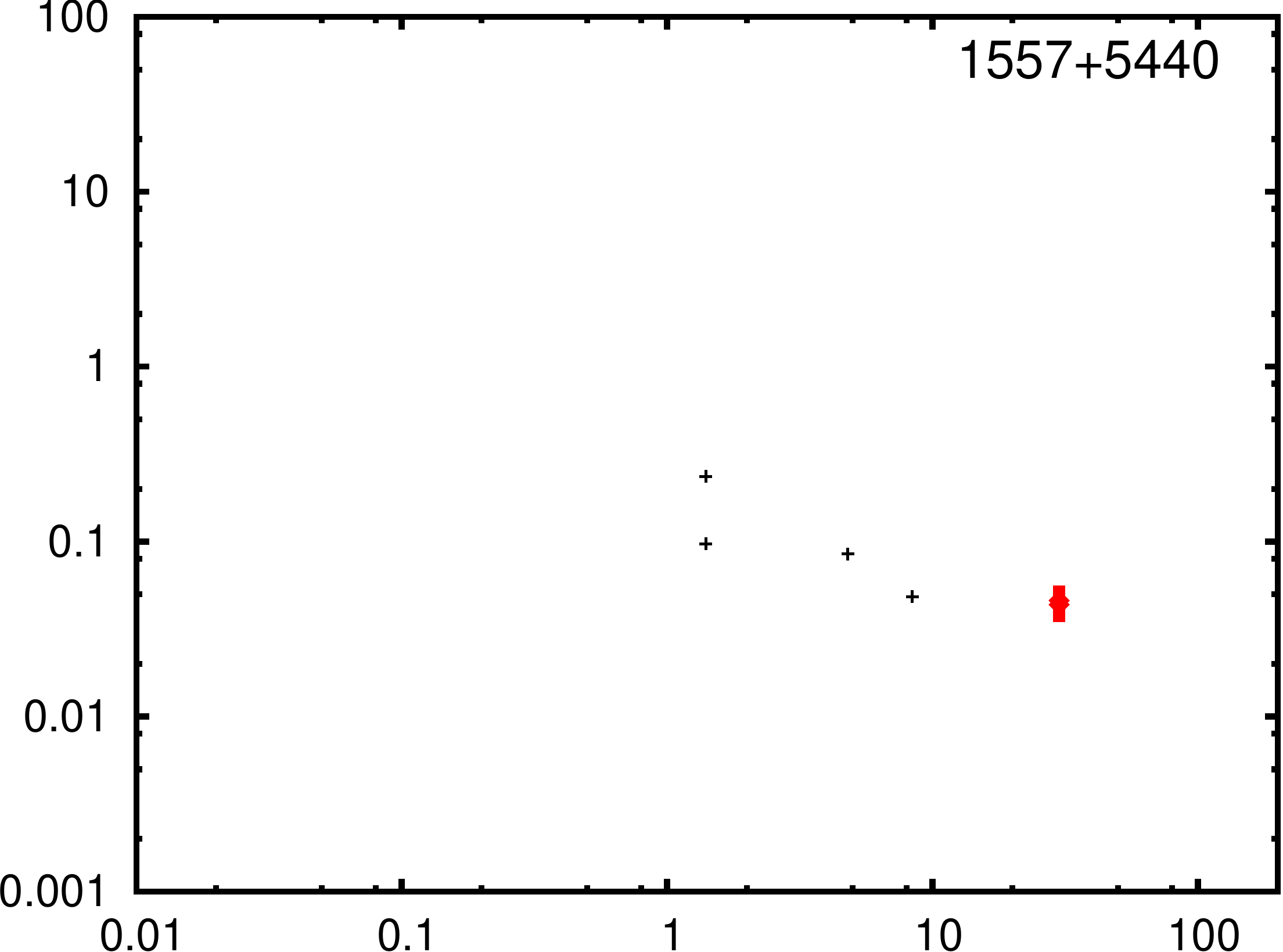}
\includegraphics[scale=0.2]{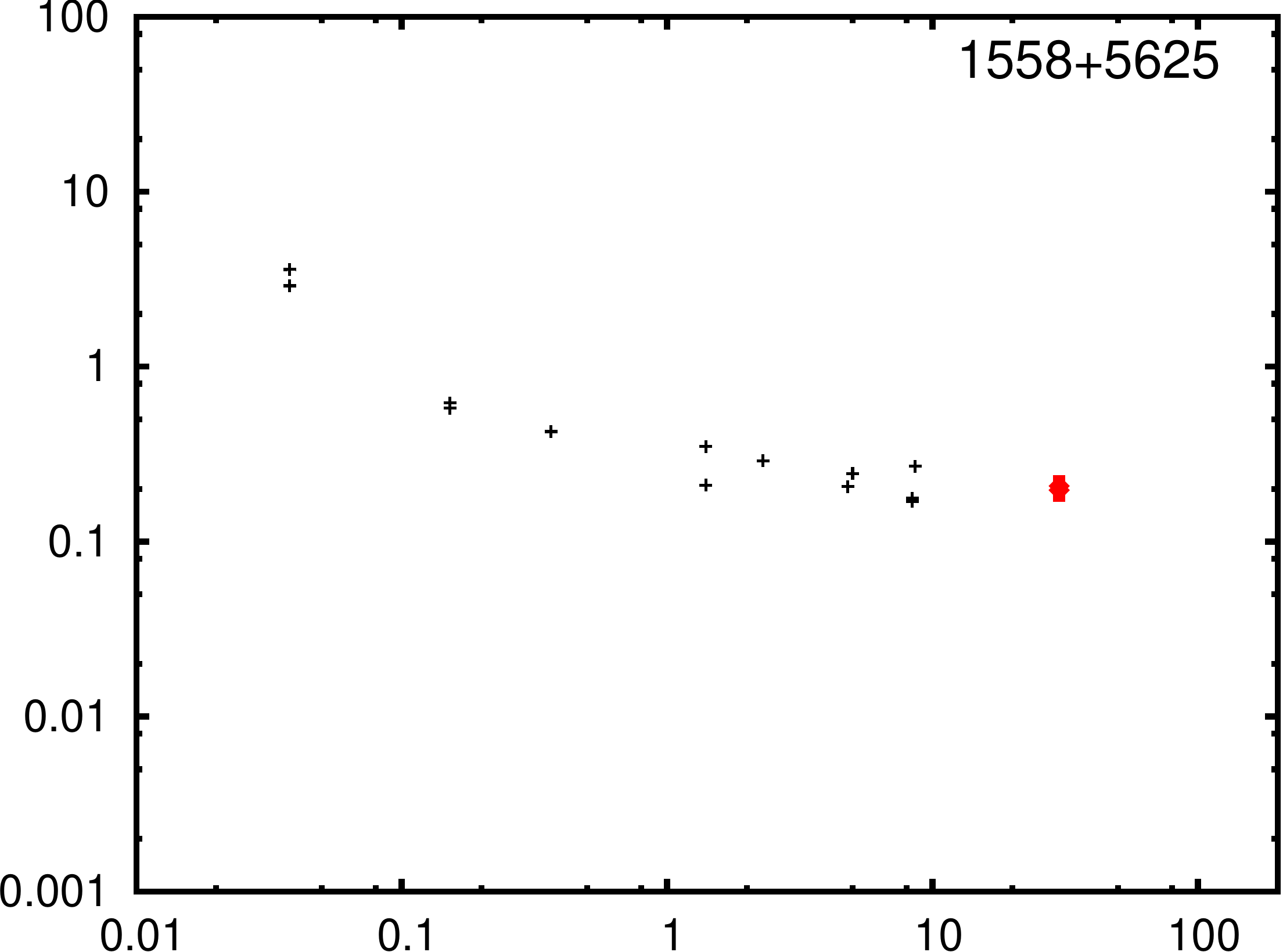}
\includegraphics[scale=0.2]{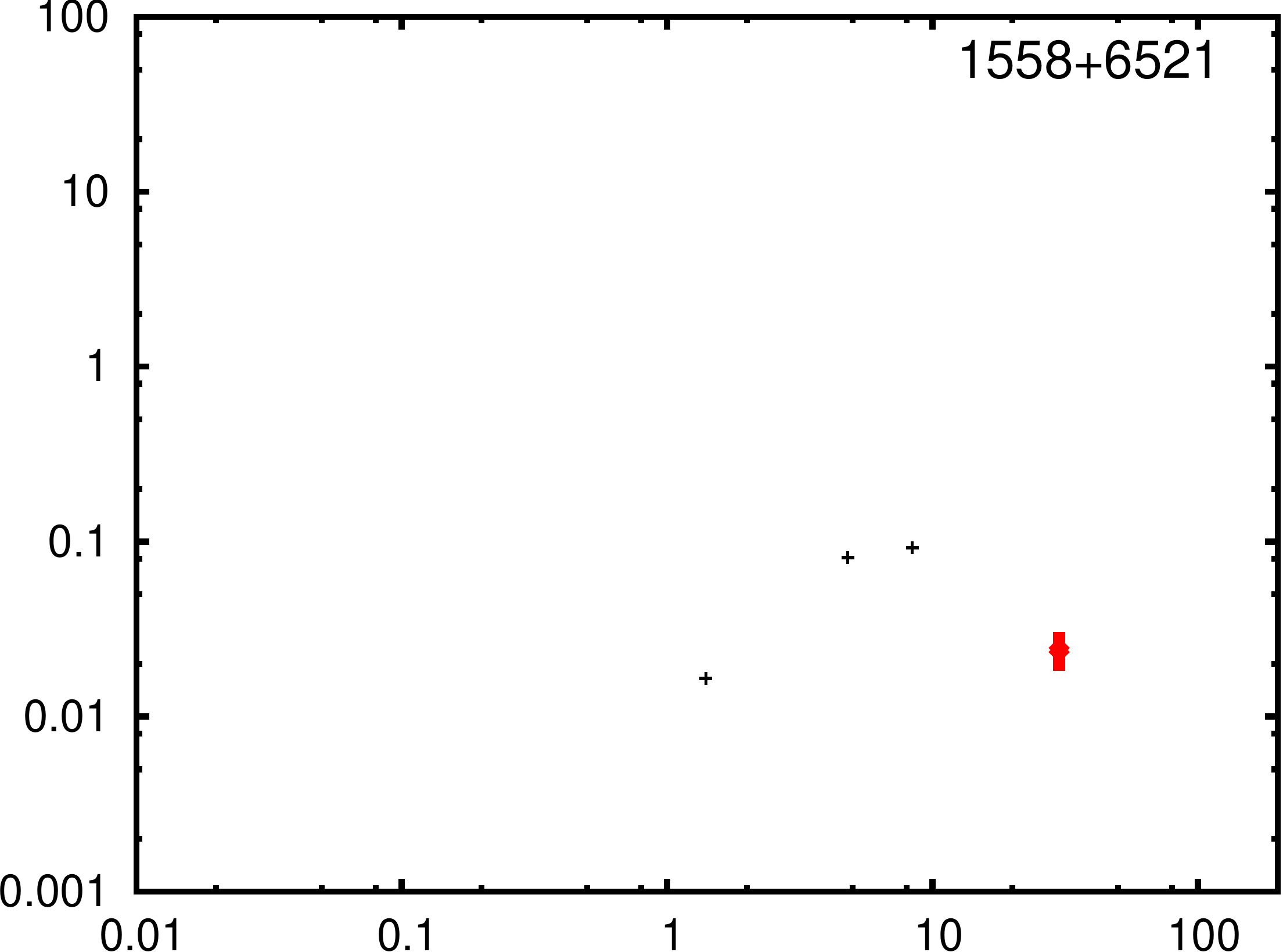}
\includegraphics[scale=0.2]{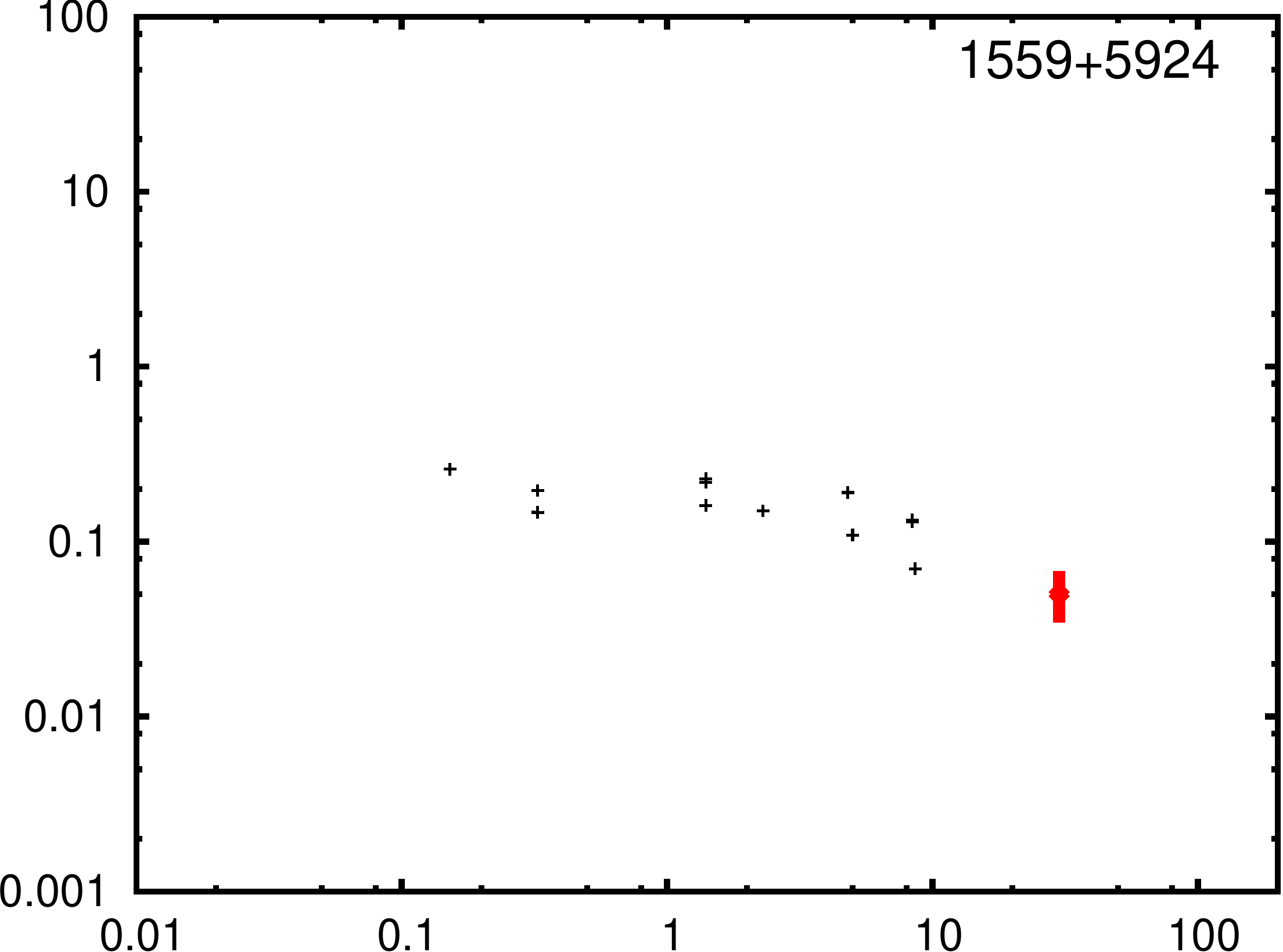}
\includegraphics[scale=0.2]{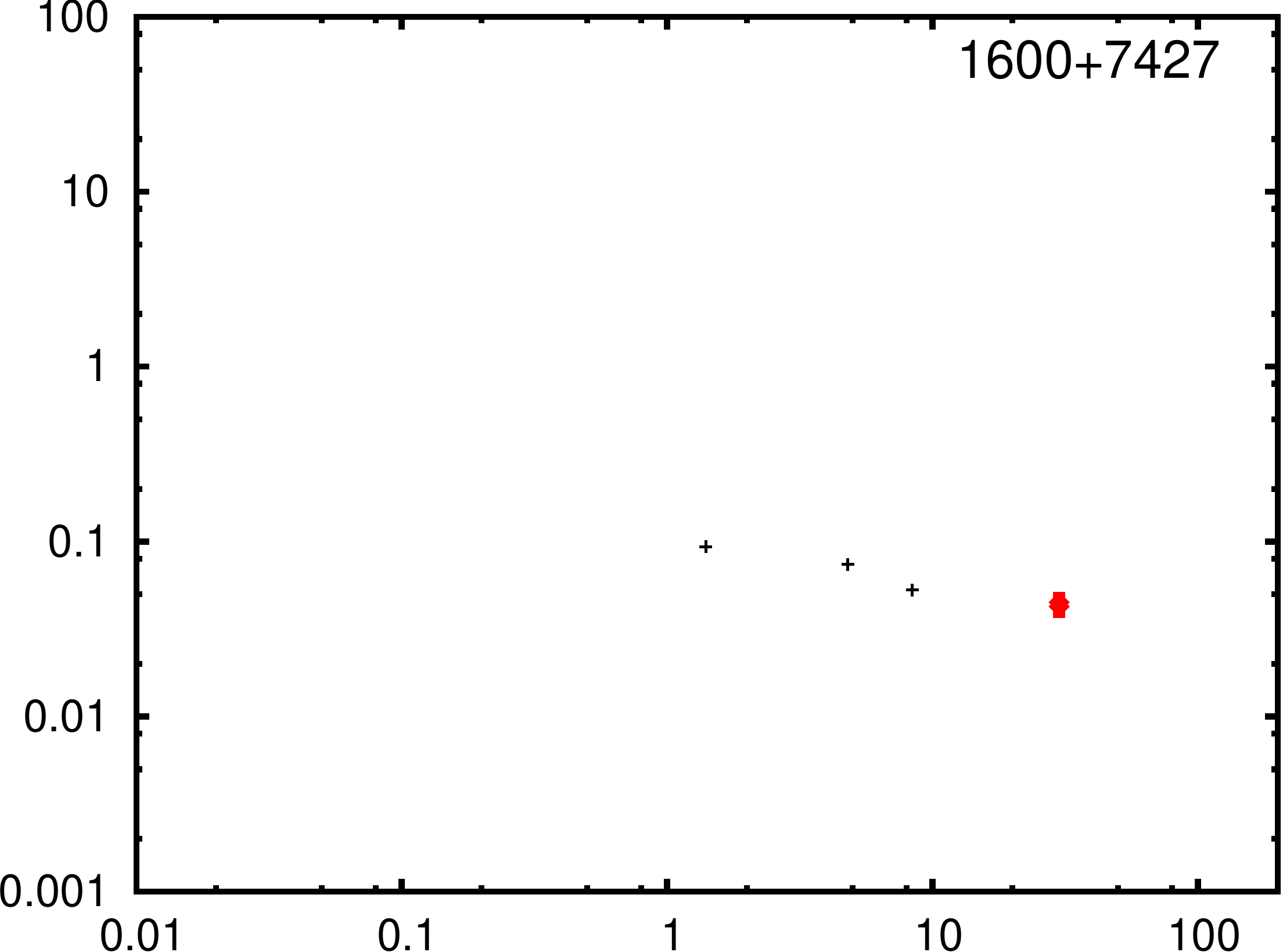}
\includegraphics[scale=0.2]{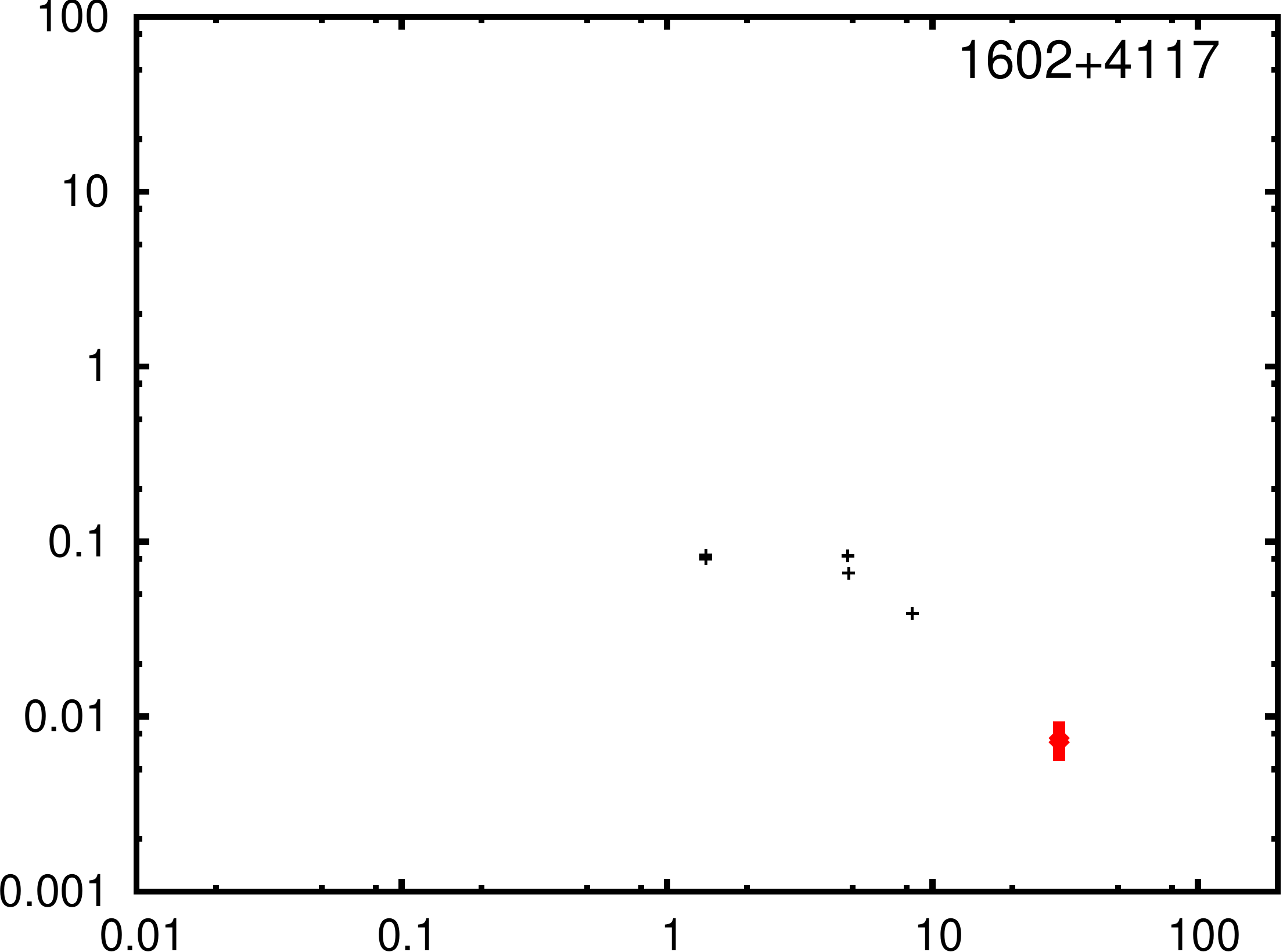}
\includegraphics[scale=0.2]{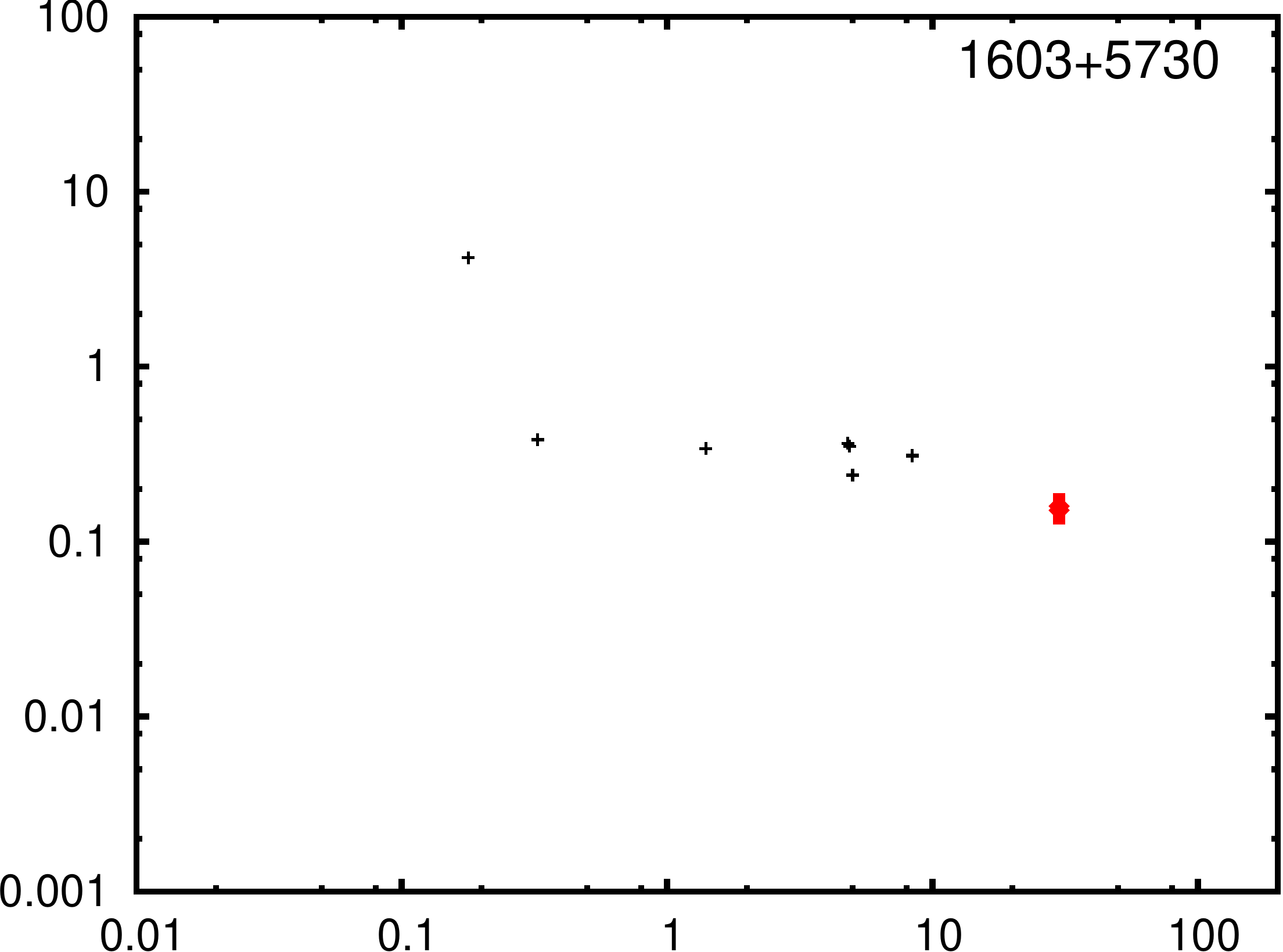}
\includegraphics[scale=0.2]{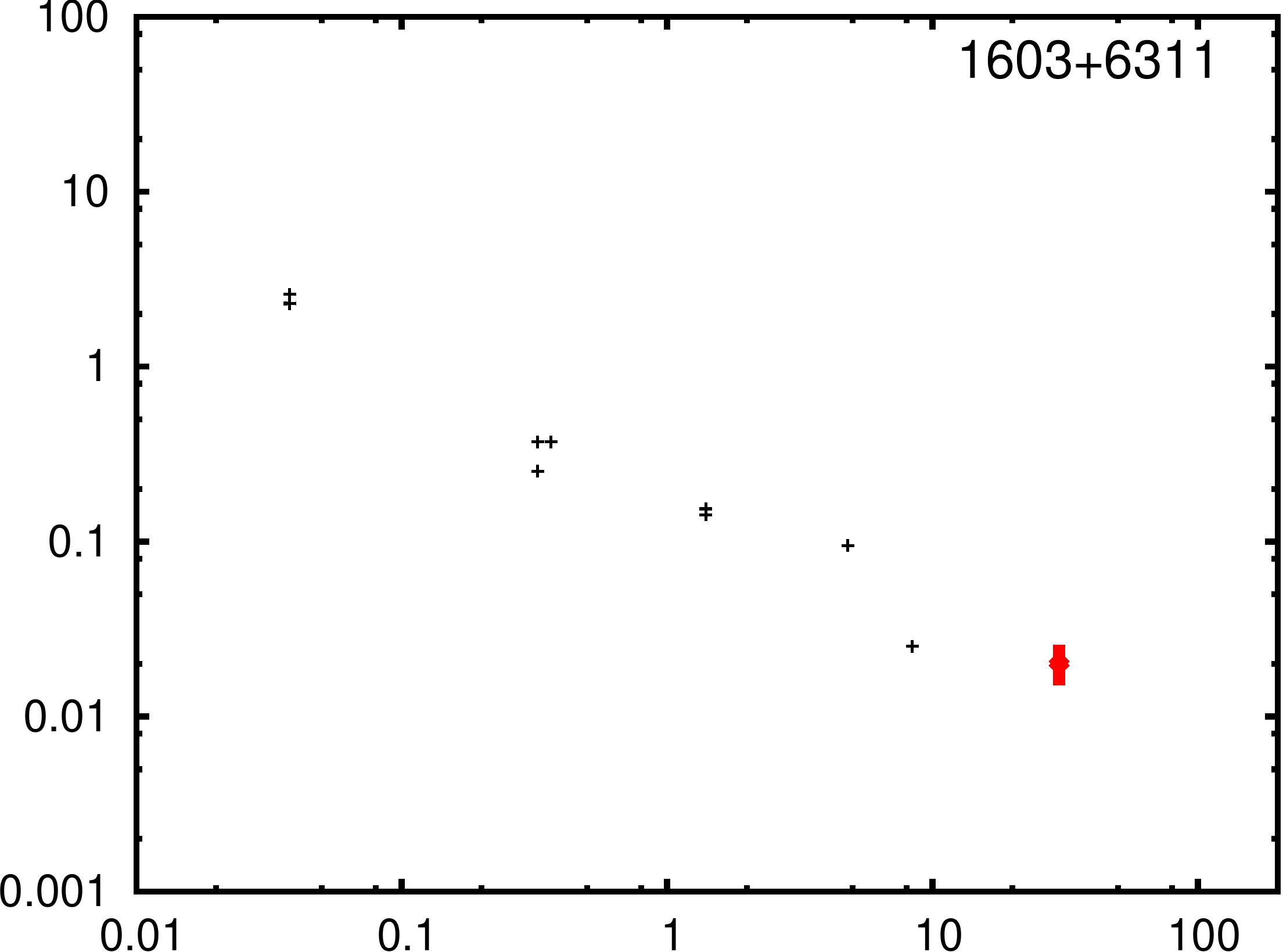}
\includegraphics[scale=0.2]{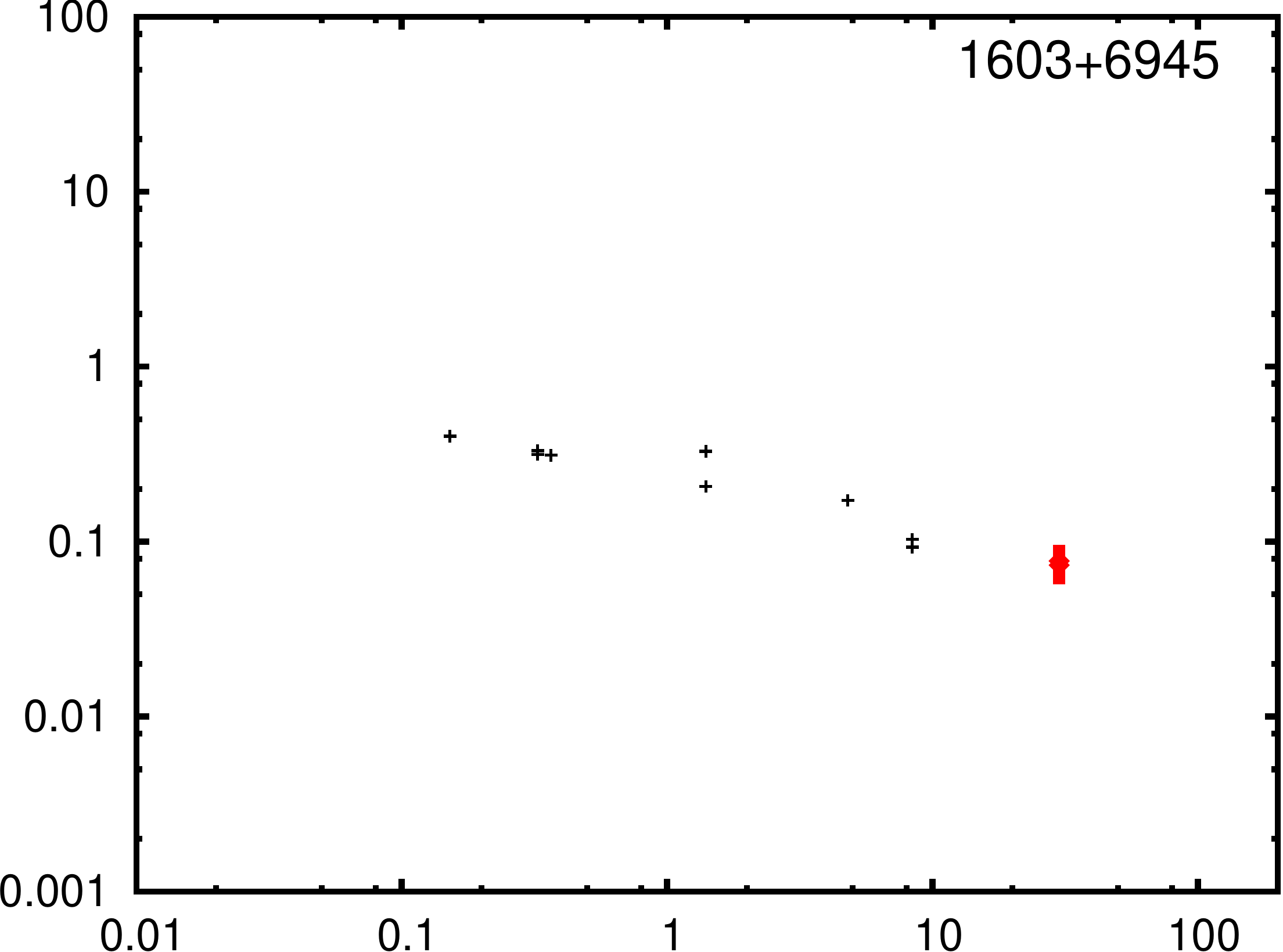}
\includegraphics[scale=0.2]{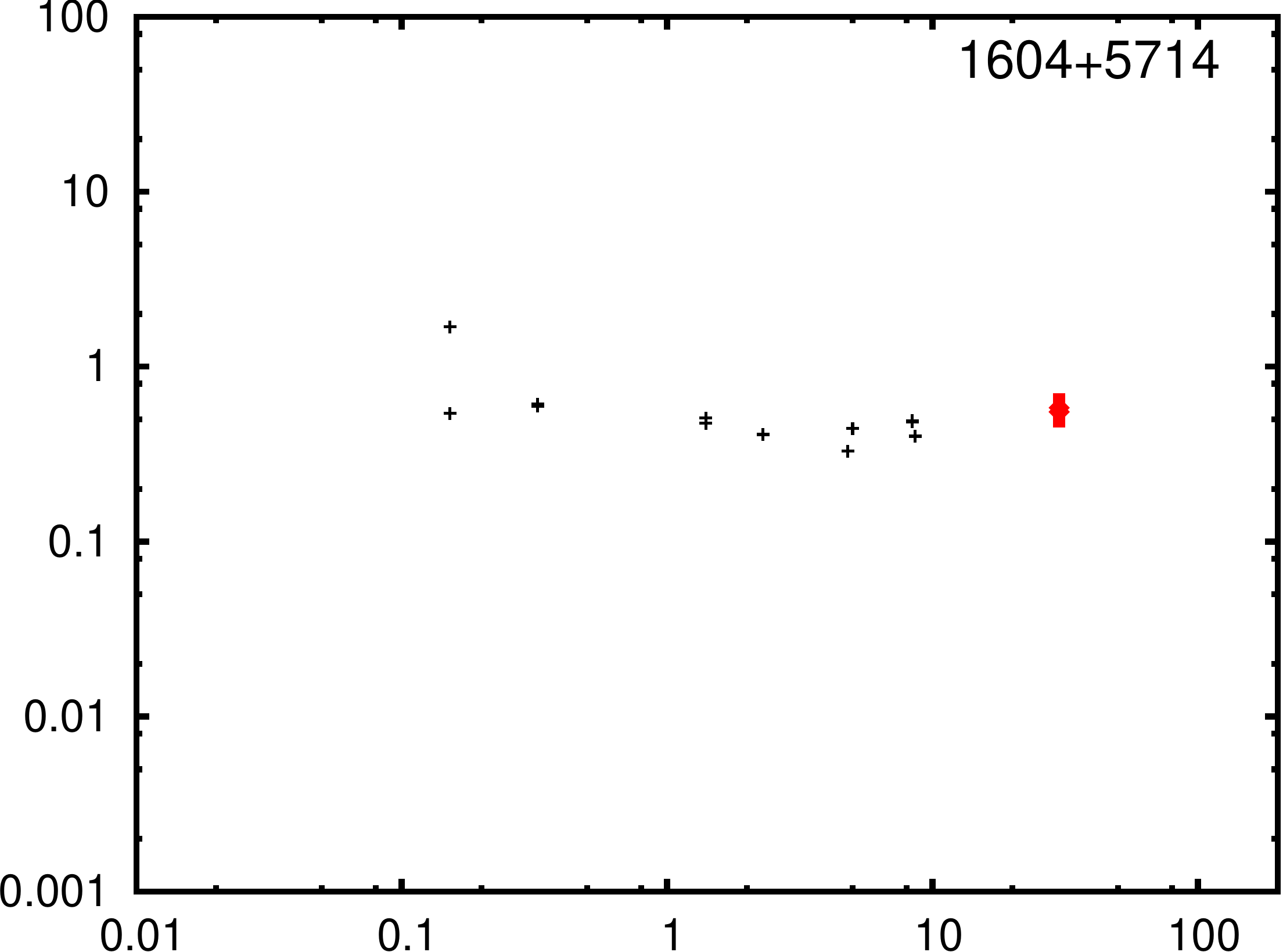}
\includegraphics[scale=0.2]{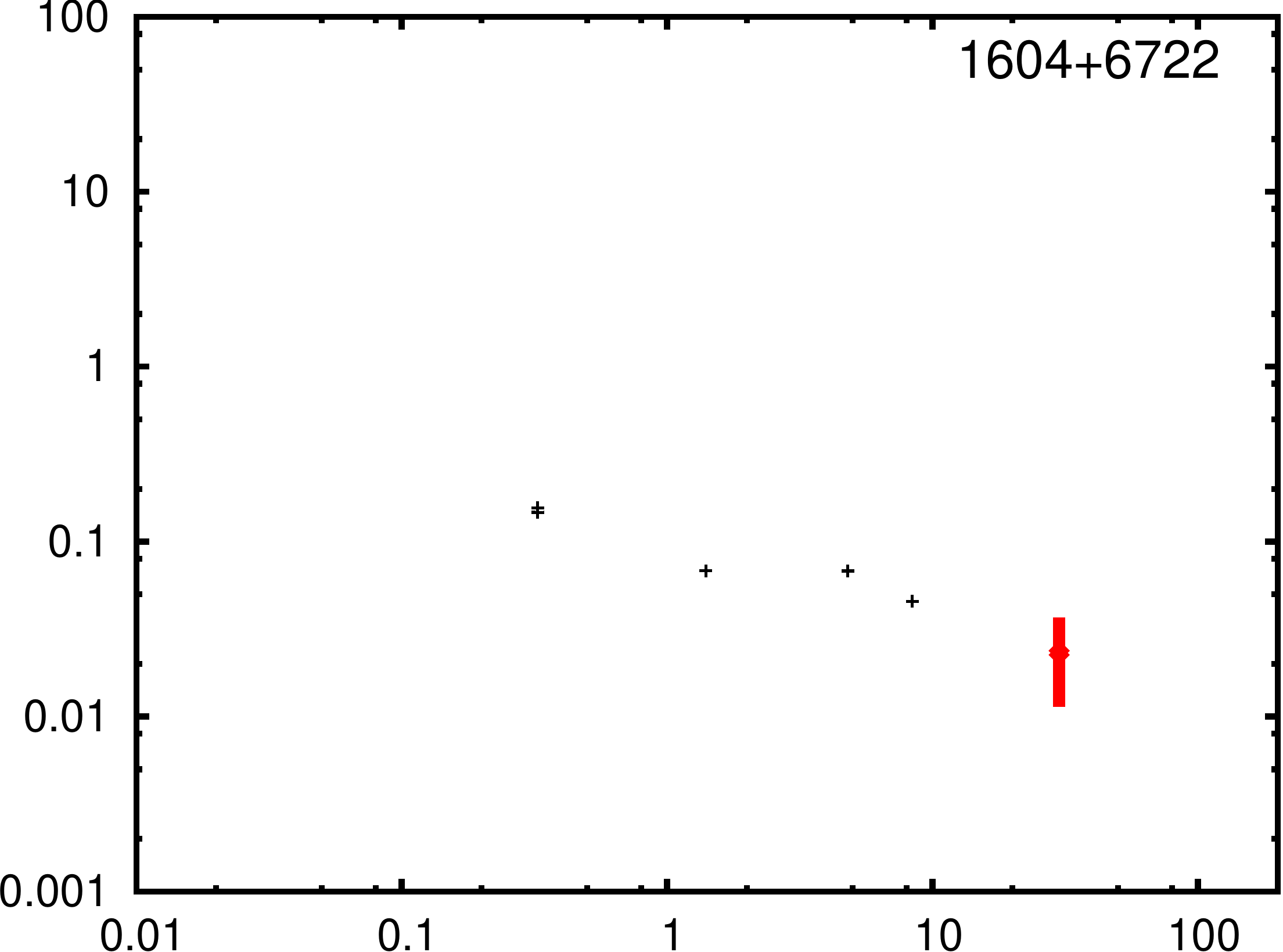}
\includegraphics[scale=0.2]{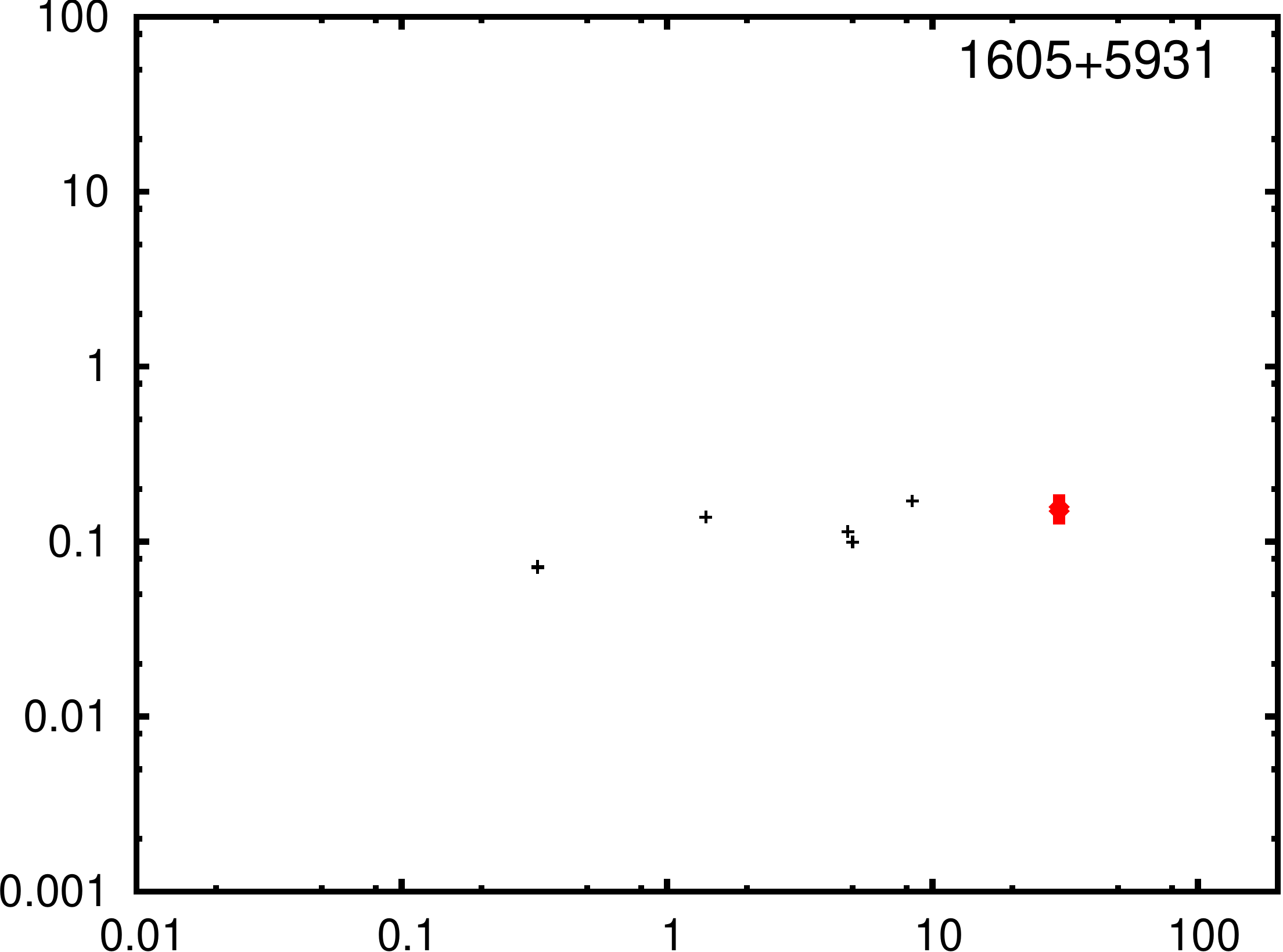}
\includegraphics[scale=0.2]{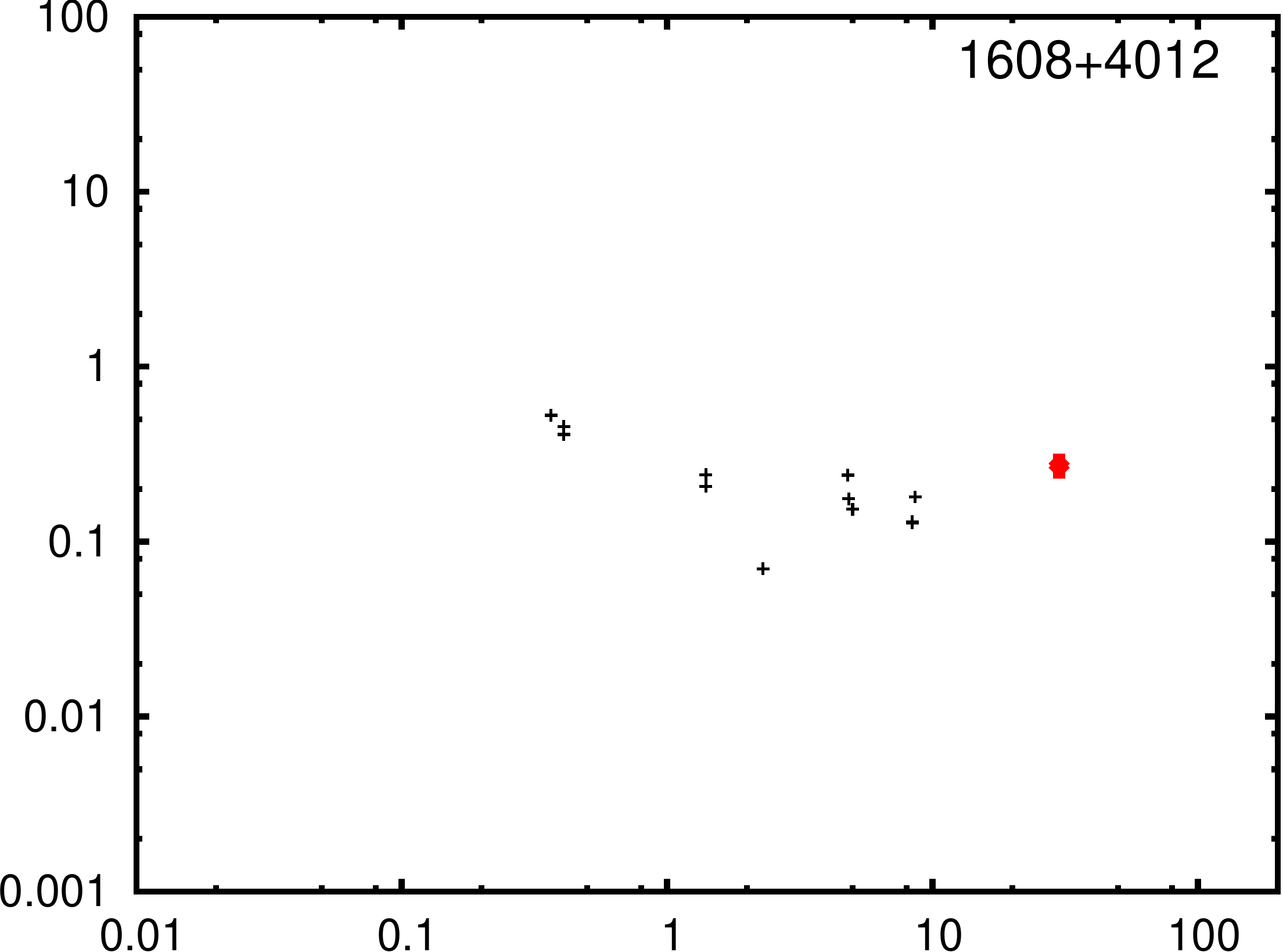}
\end{figure}
\clearpage\begin{figure}
\centering
\includegraphics[scale=0.2]{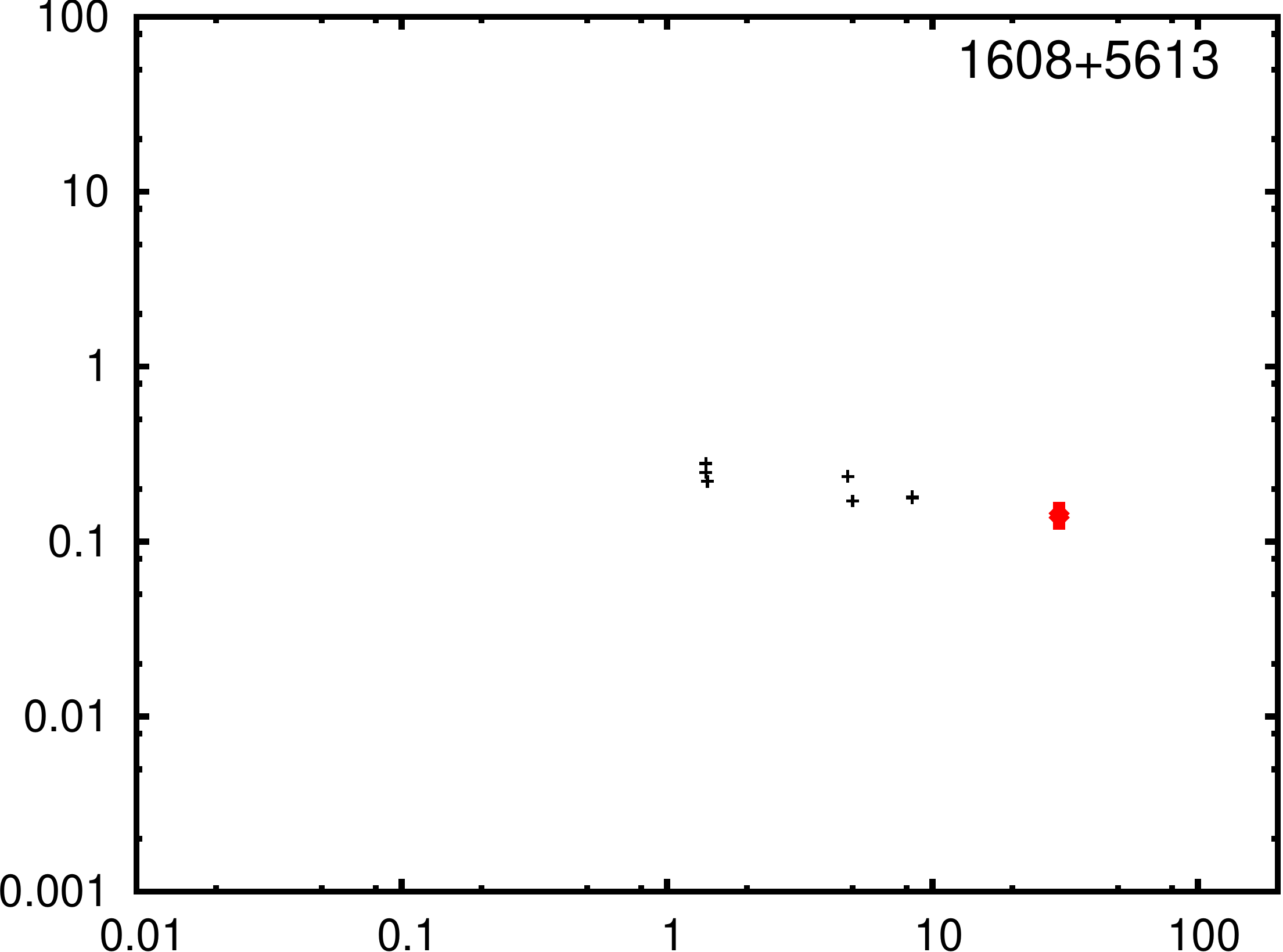}
\includegraphics[scale=0.2]{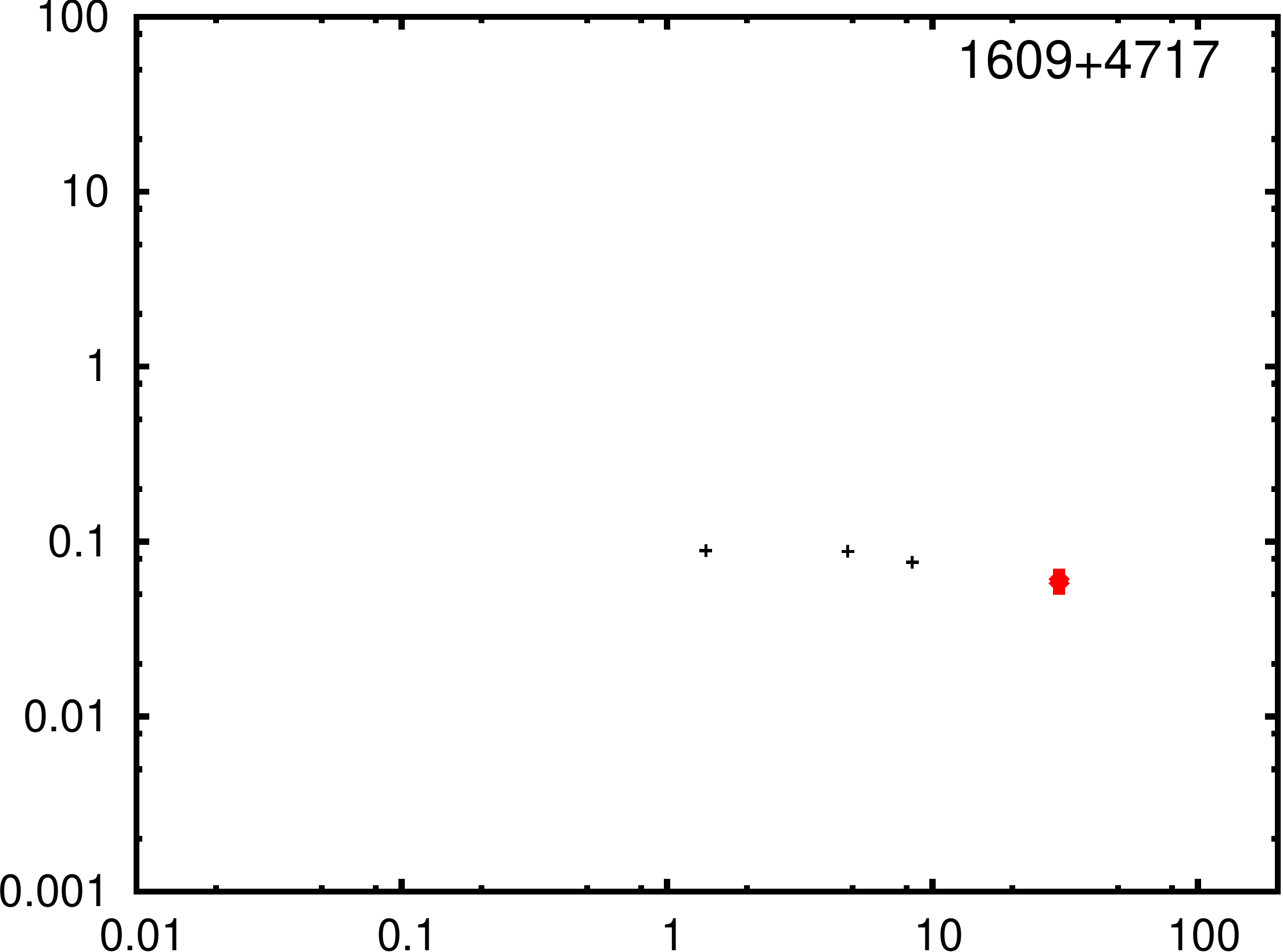}
\includegraphics[scale=0.2]{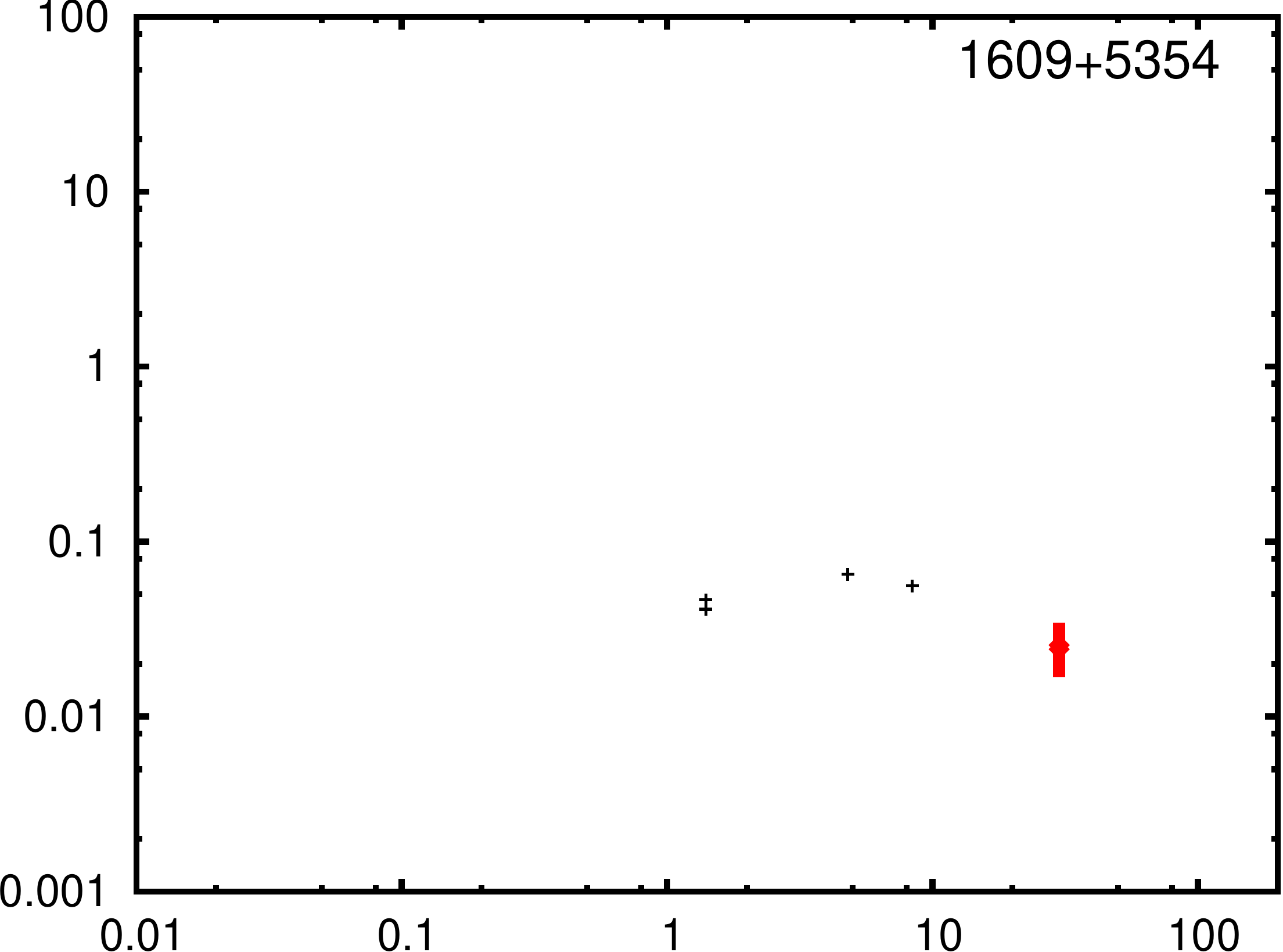}
\includegraphics[scale=0.2]{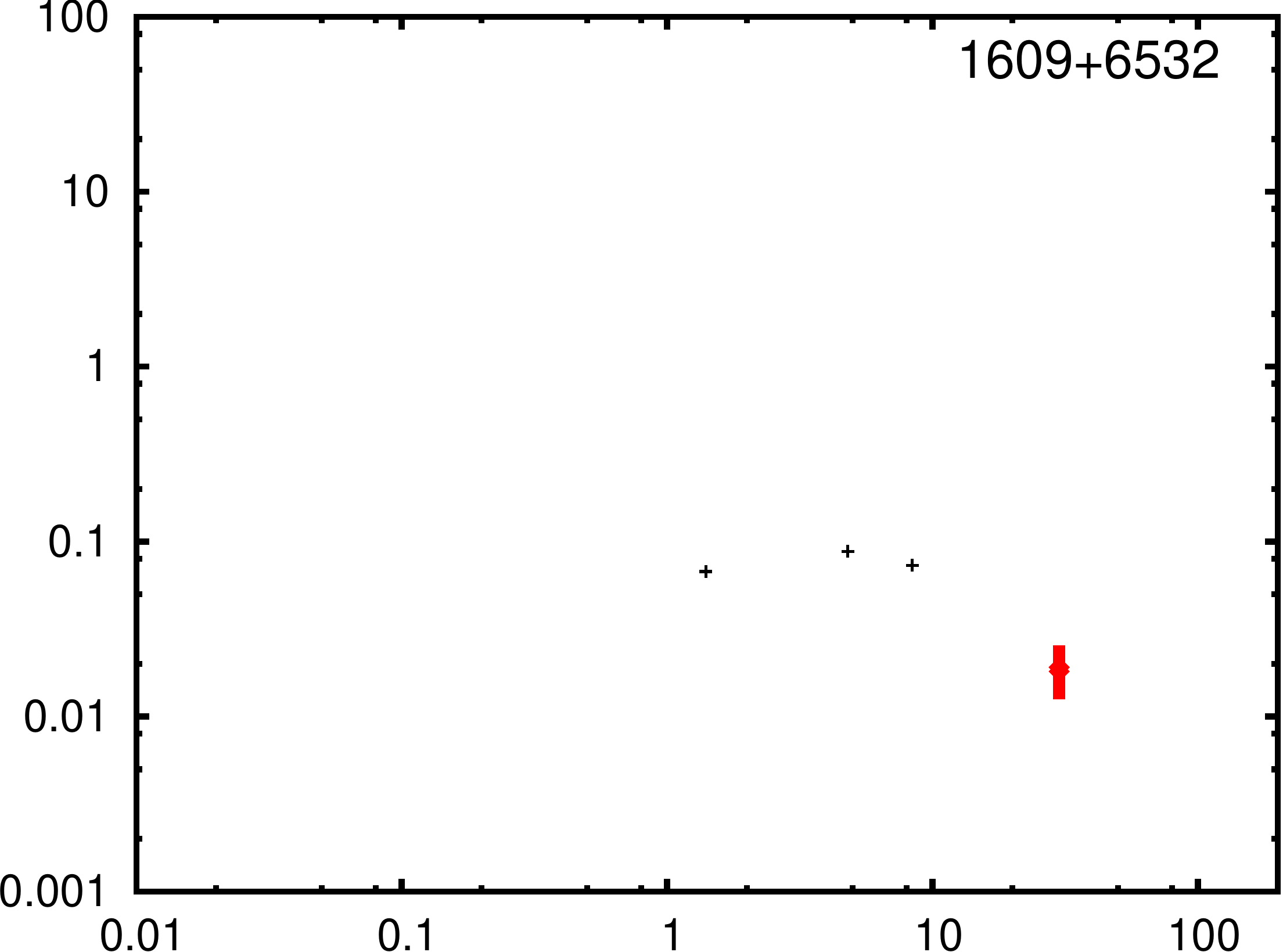}
\includegraphics[scale=0.2]{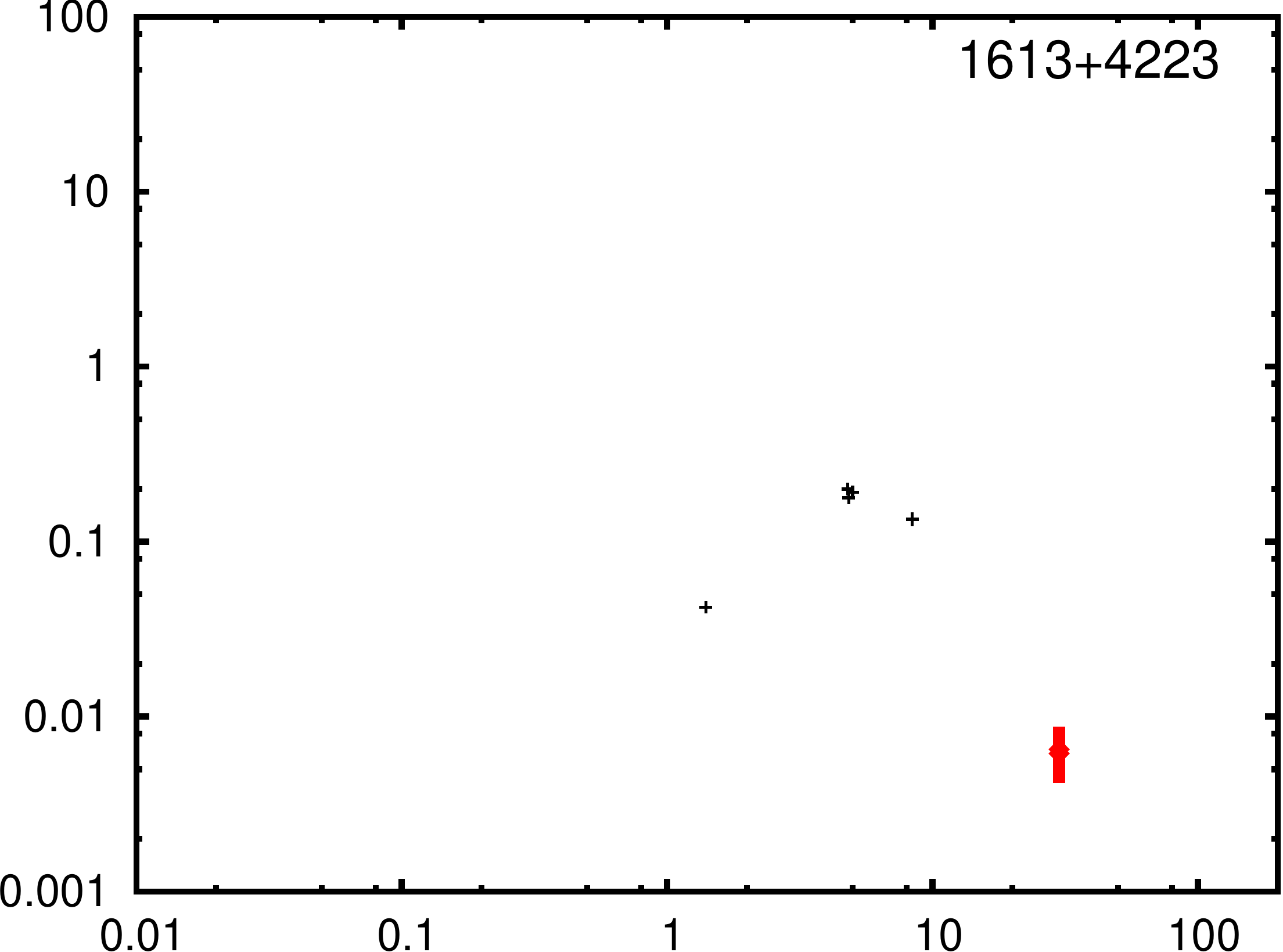}
\includegraphics[scale=0.2]{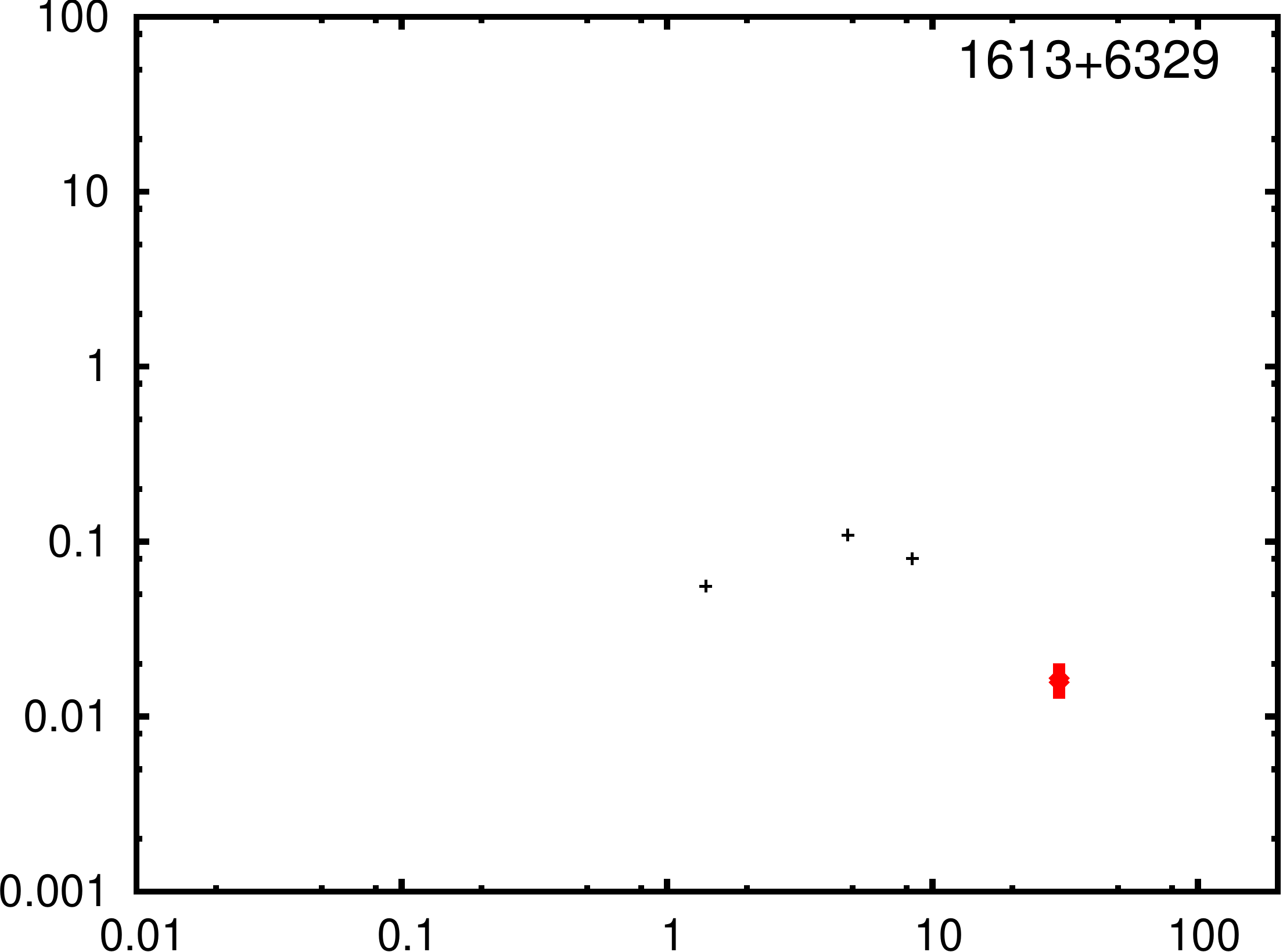}
\includegraphics[scale=0.2]{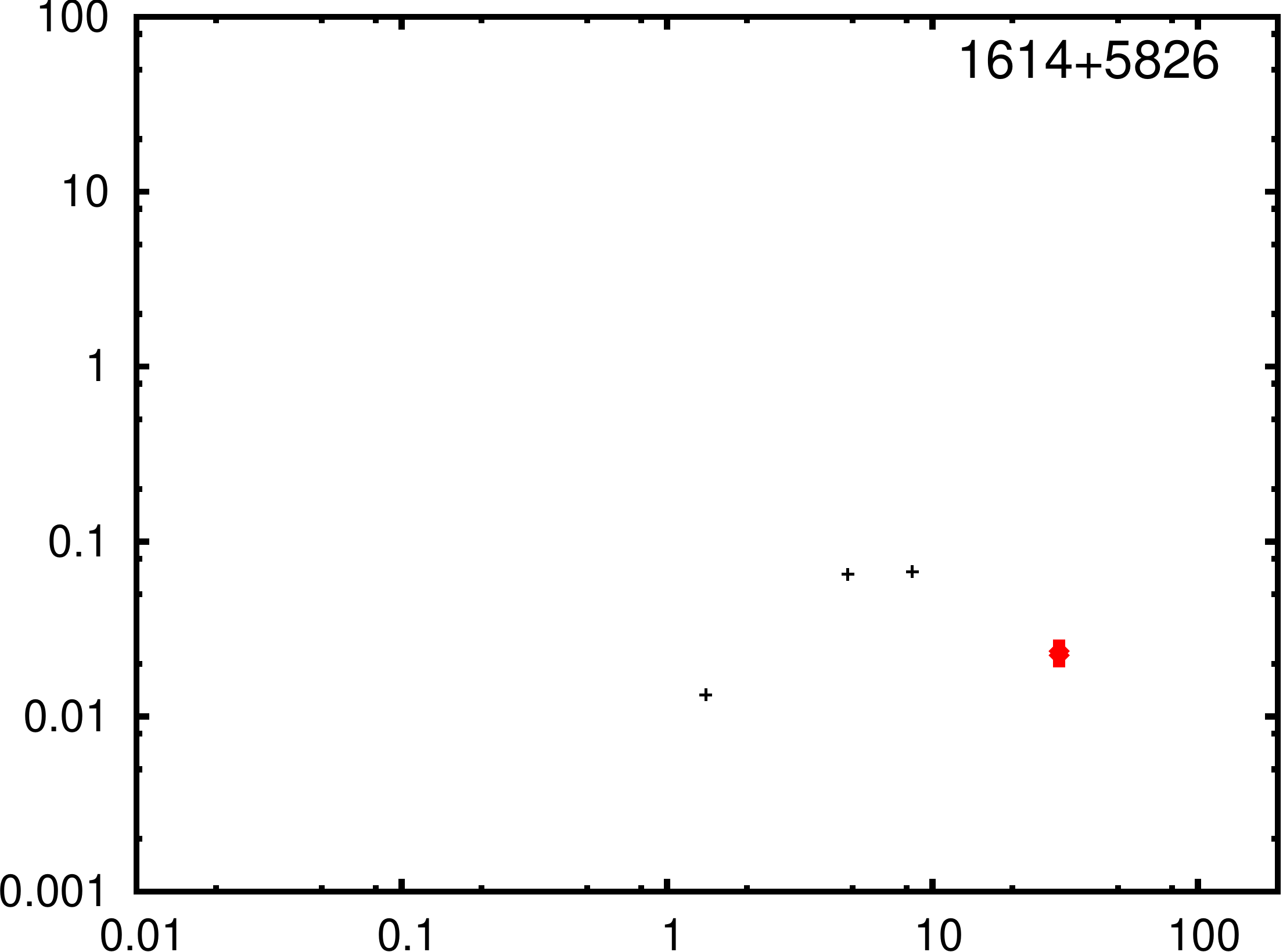}
\includegraphics[scale=0.2]{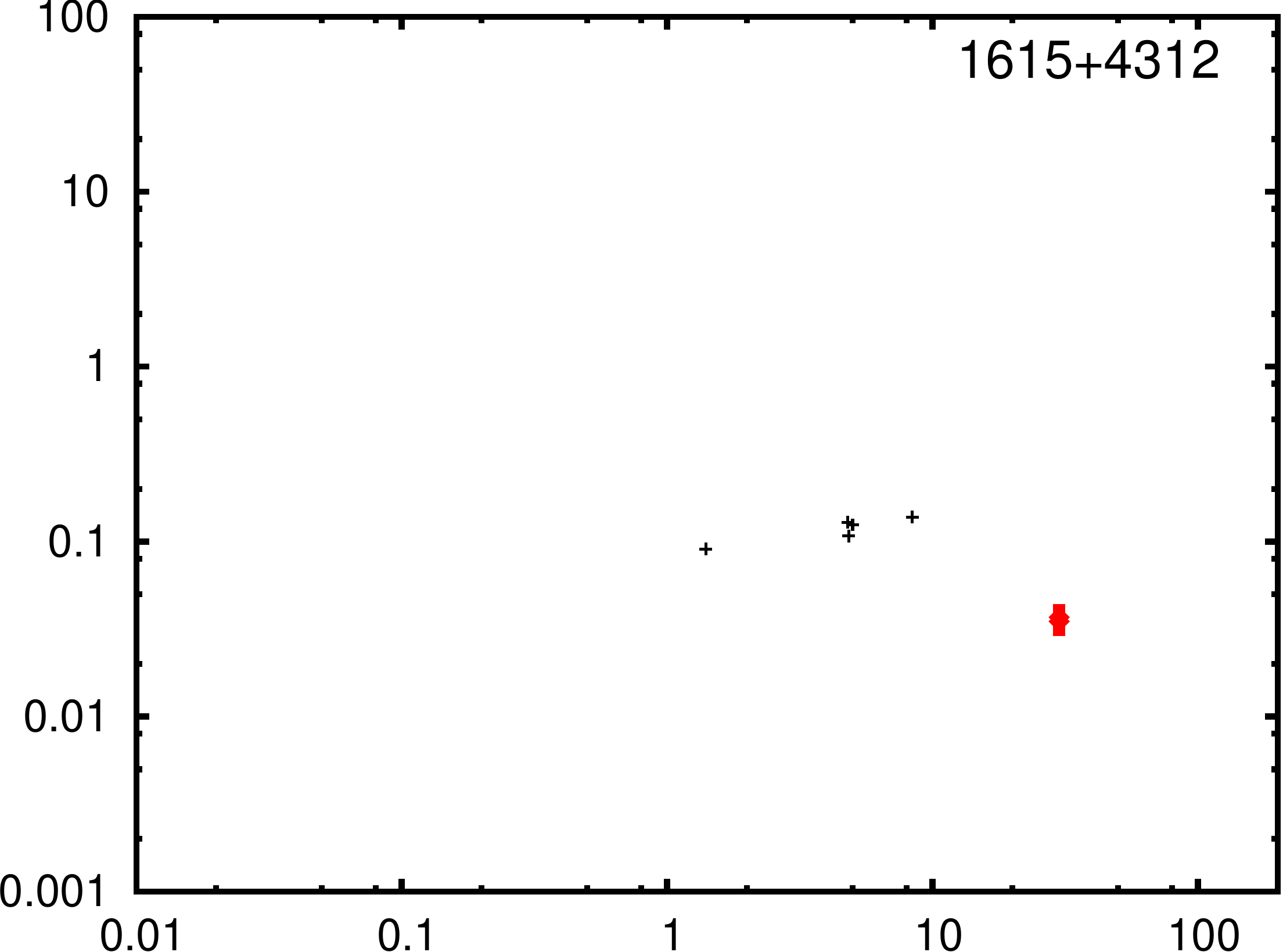}
\includegraphics[scale=0.2]{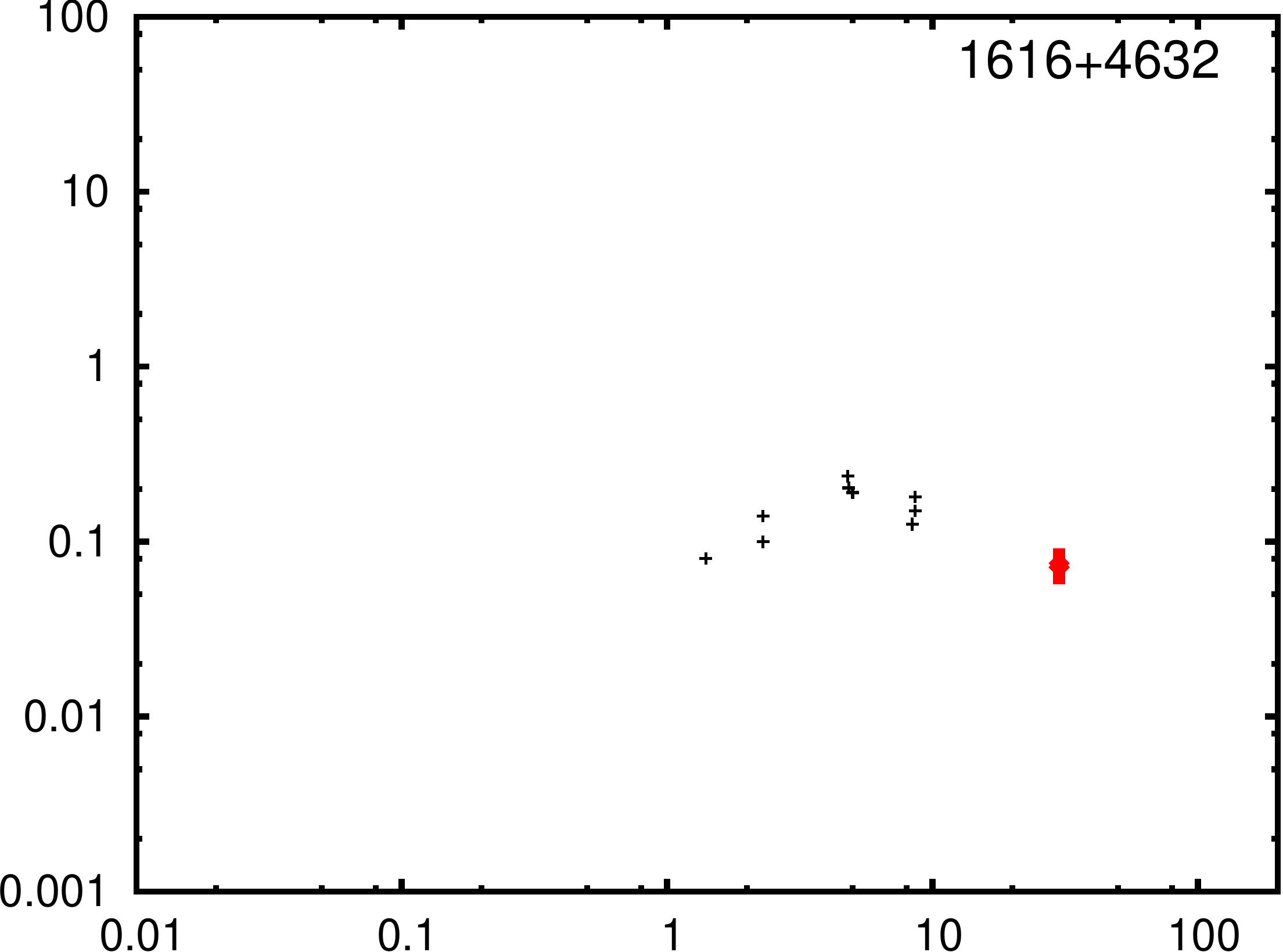}
\includegraphics[scale=0.2]{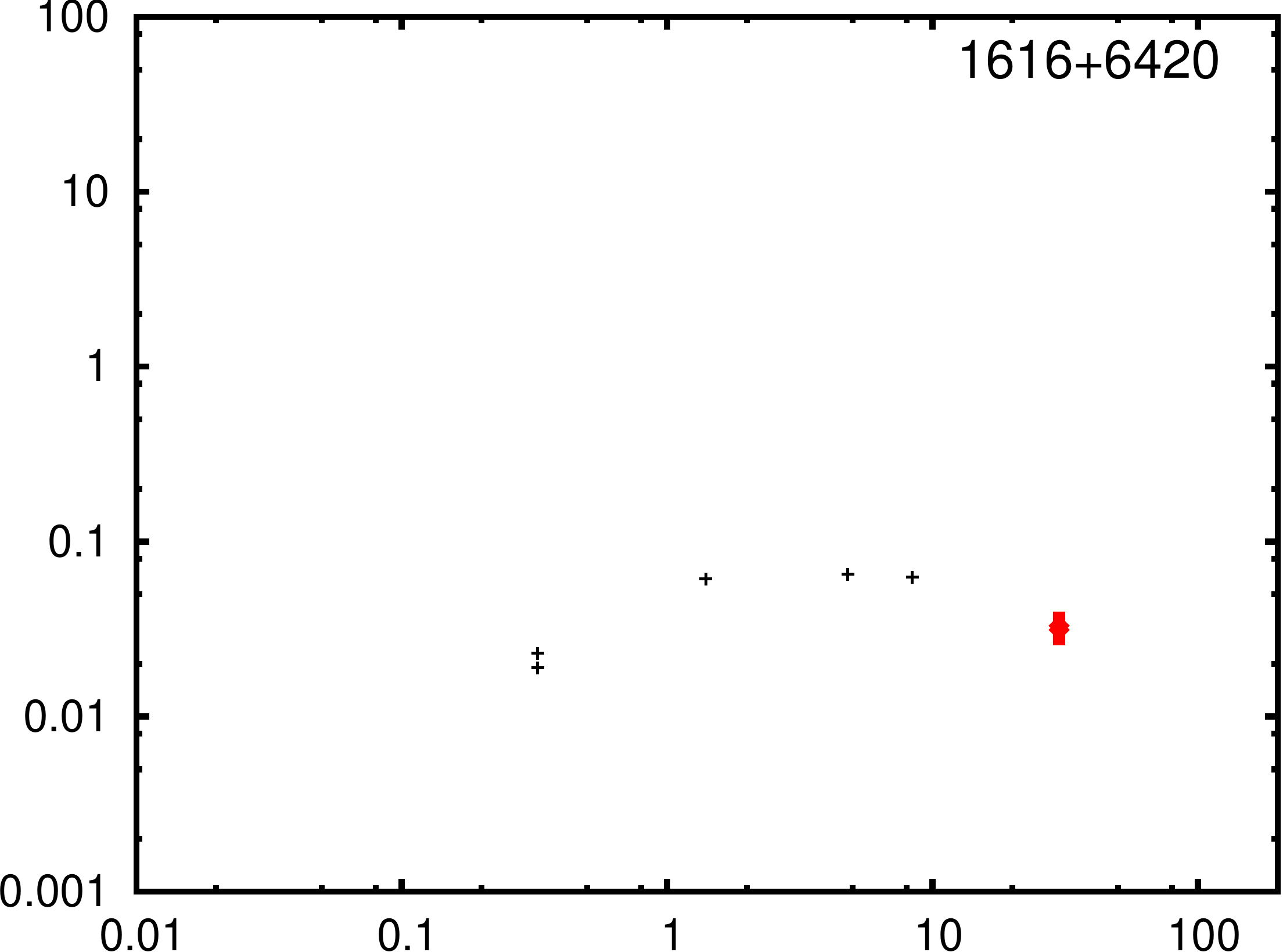}
\includegraphics[scale=0.2]{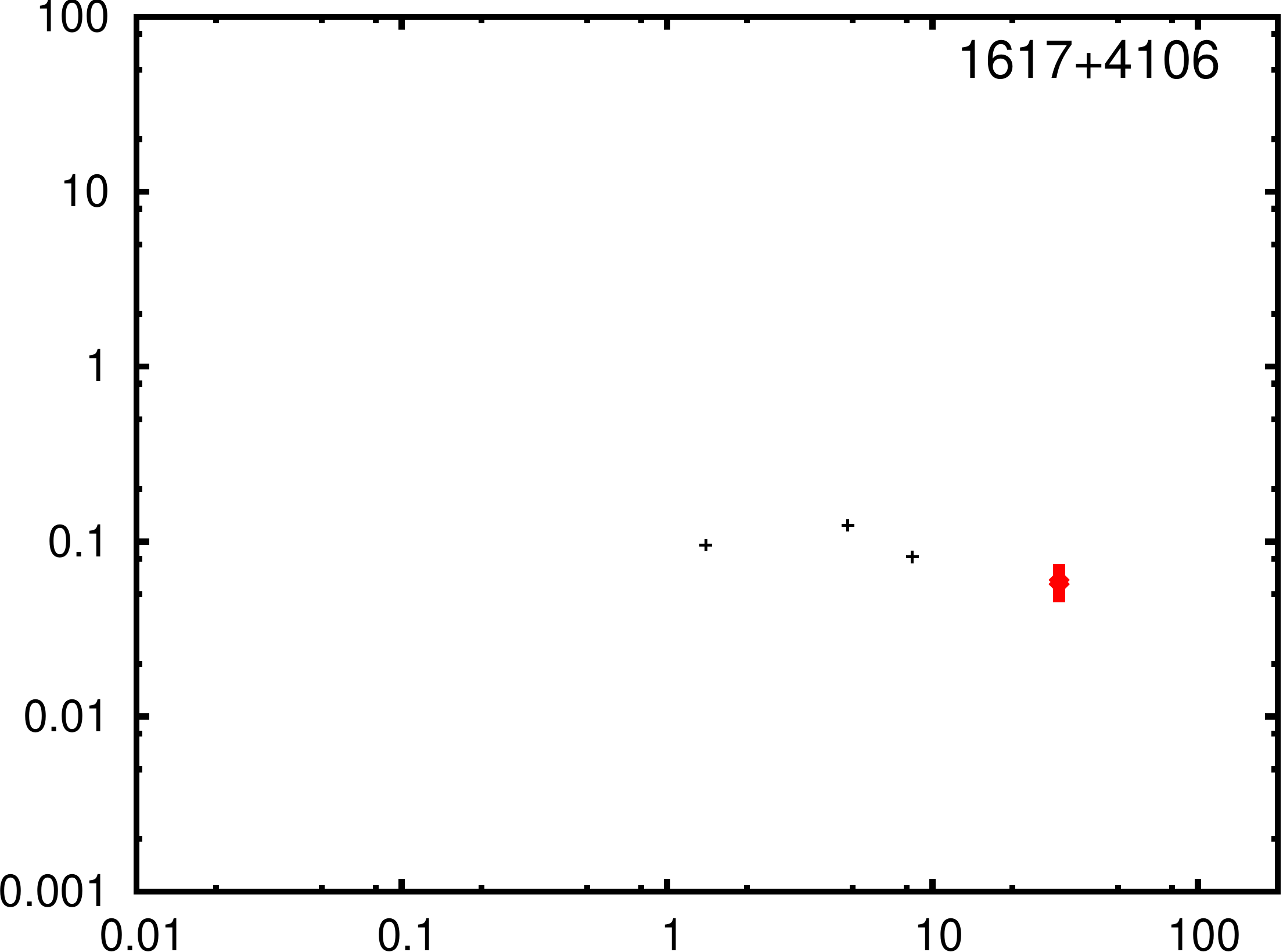}
\includegraphics[scale=0.2]{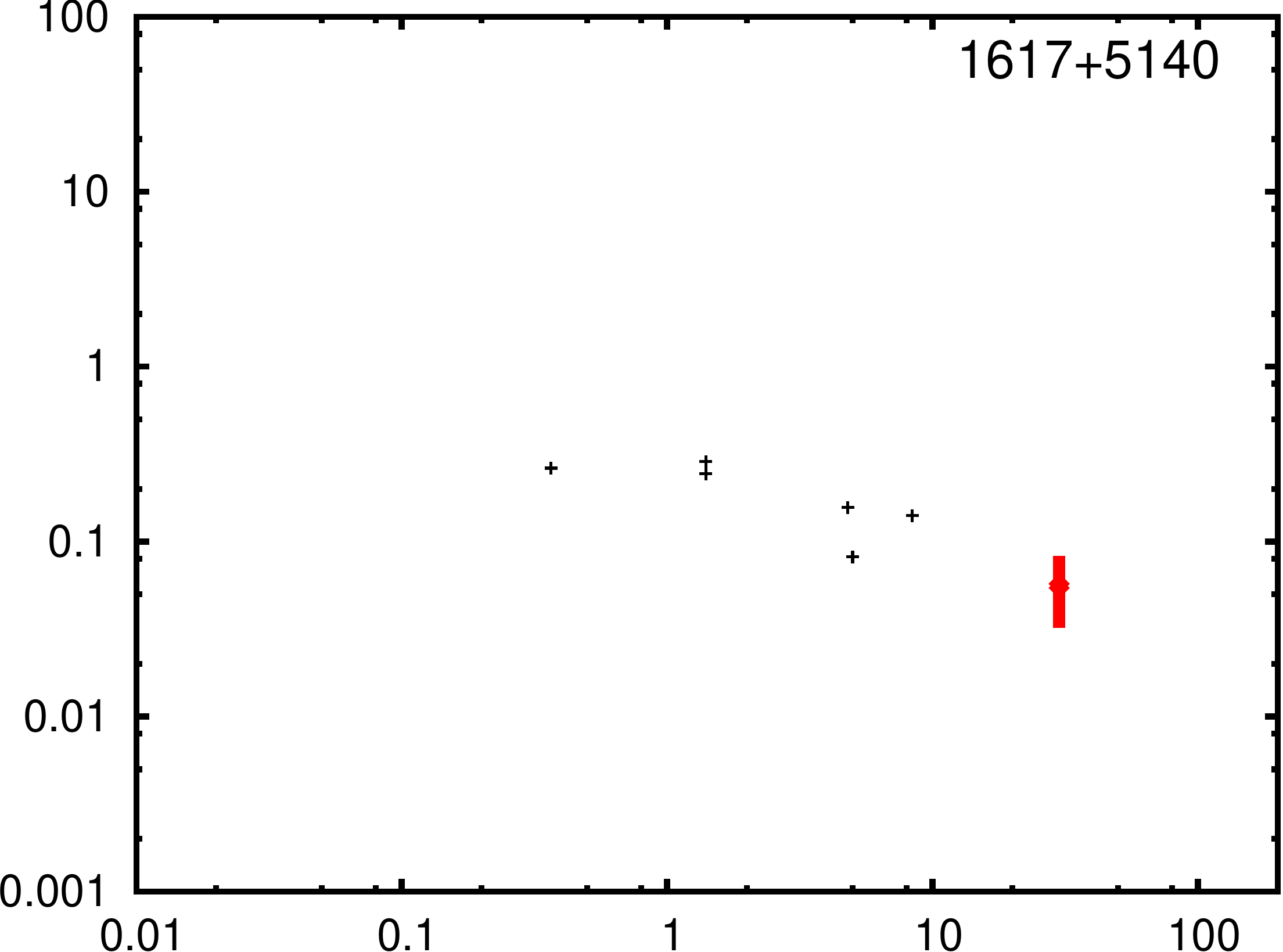}
\includegraphics[scale=0.2]{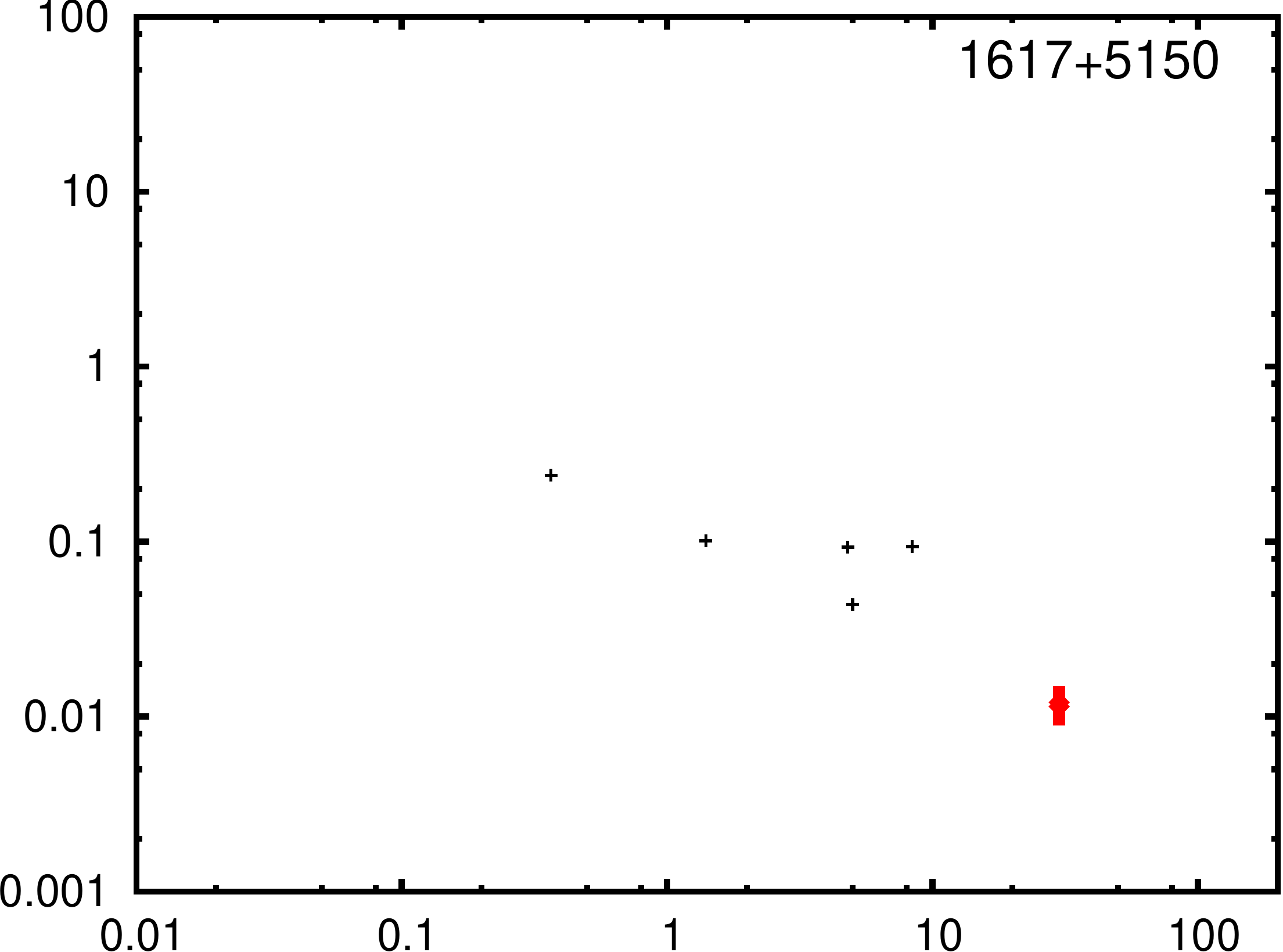}
\includegraphics[scale=0.2]{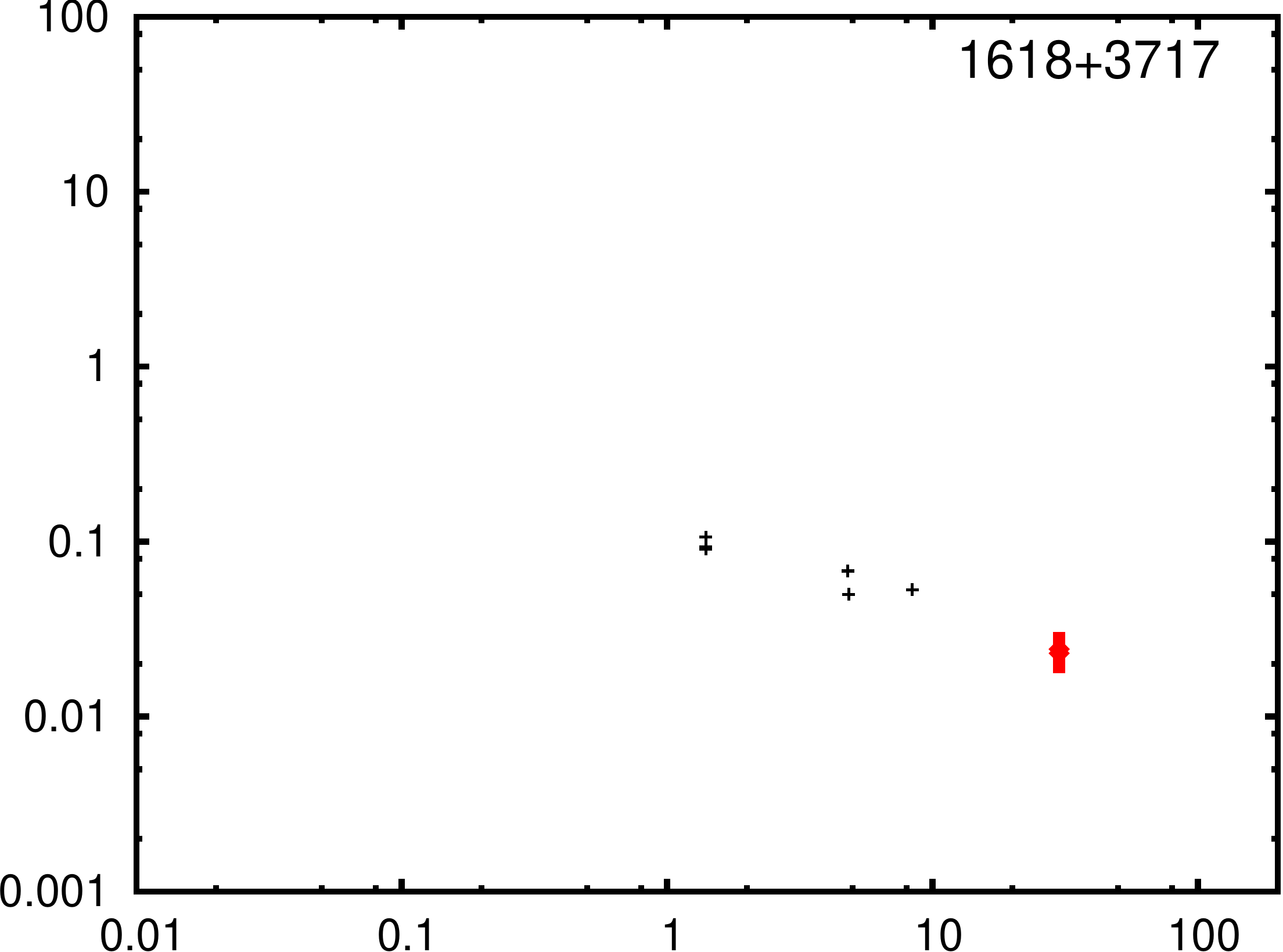}
\includegraphics[scale=0.2]{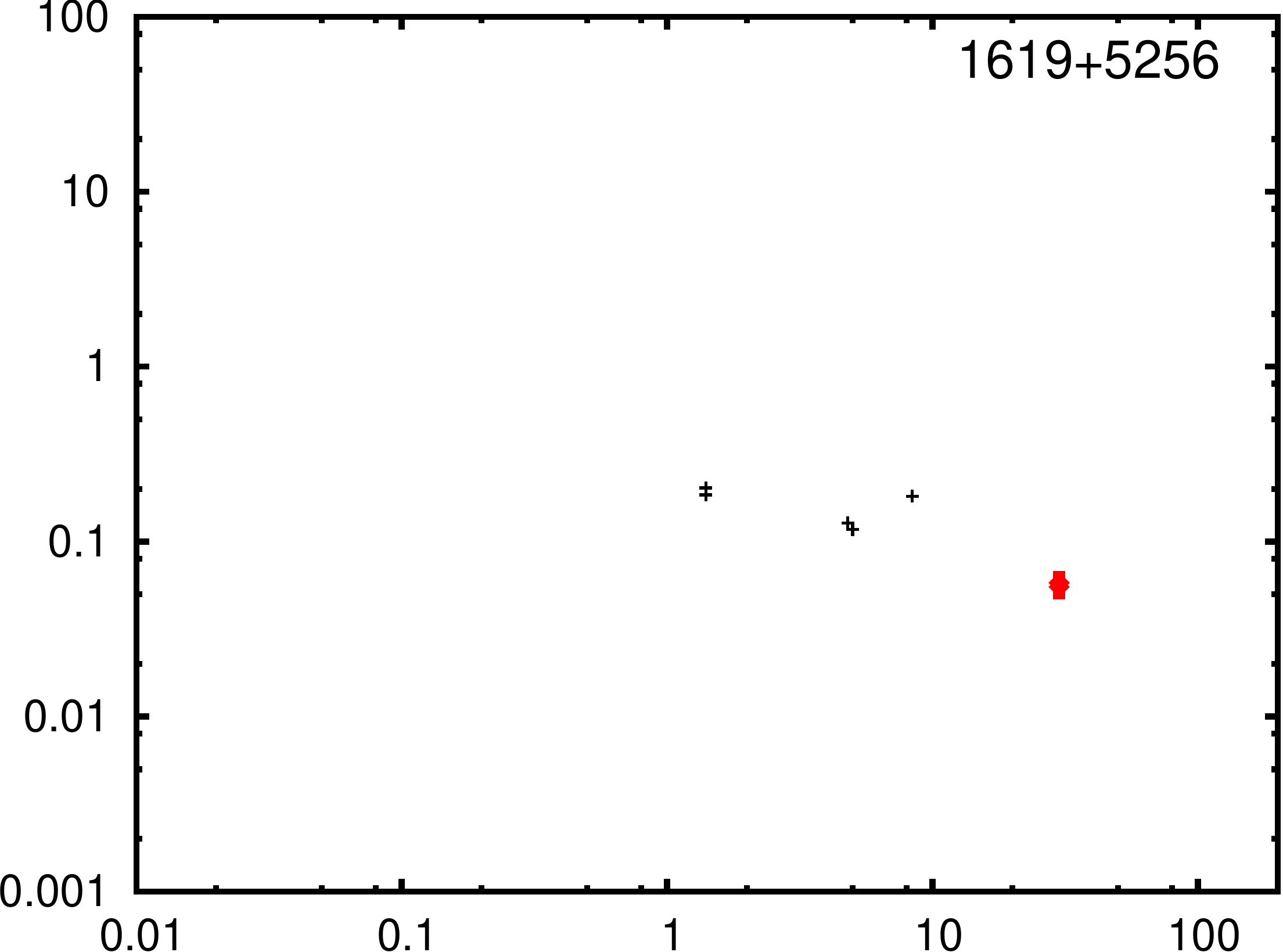}
\includegraphics[scale=0.2]{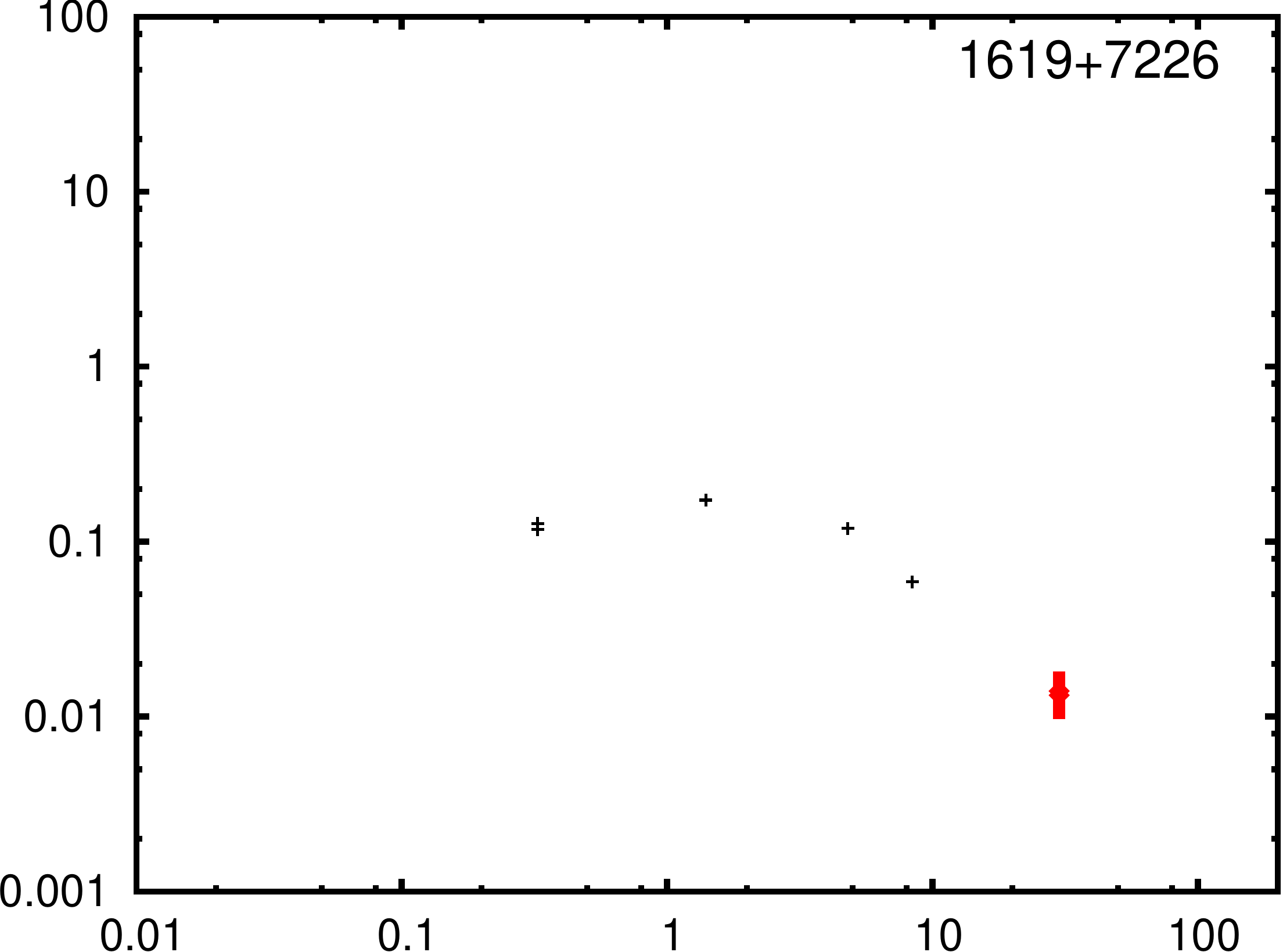}
\includegraphics[scale=0.2]{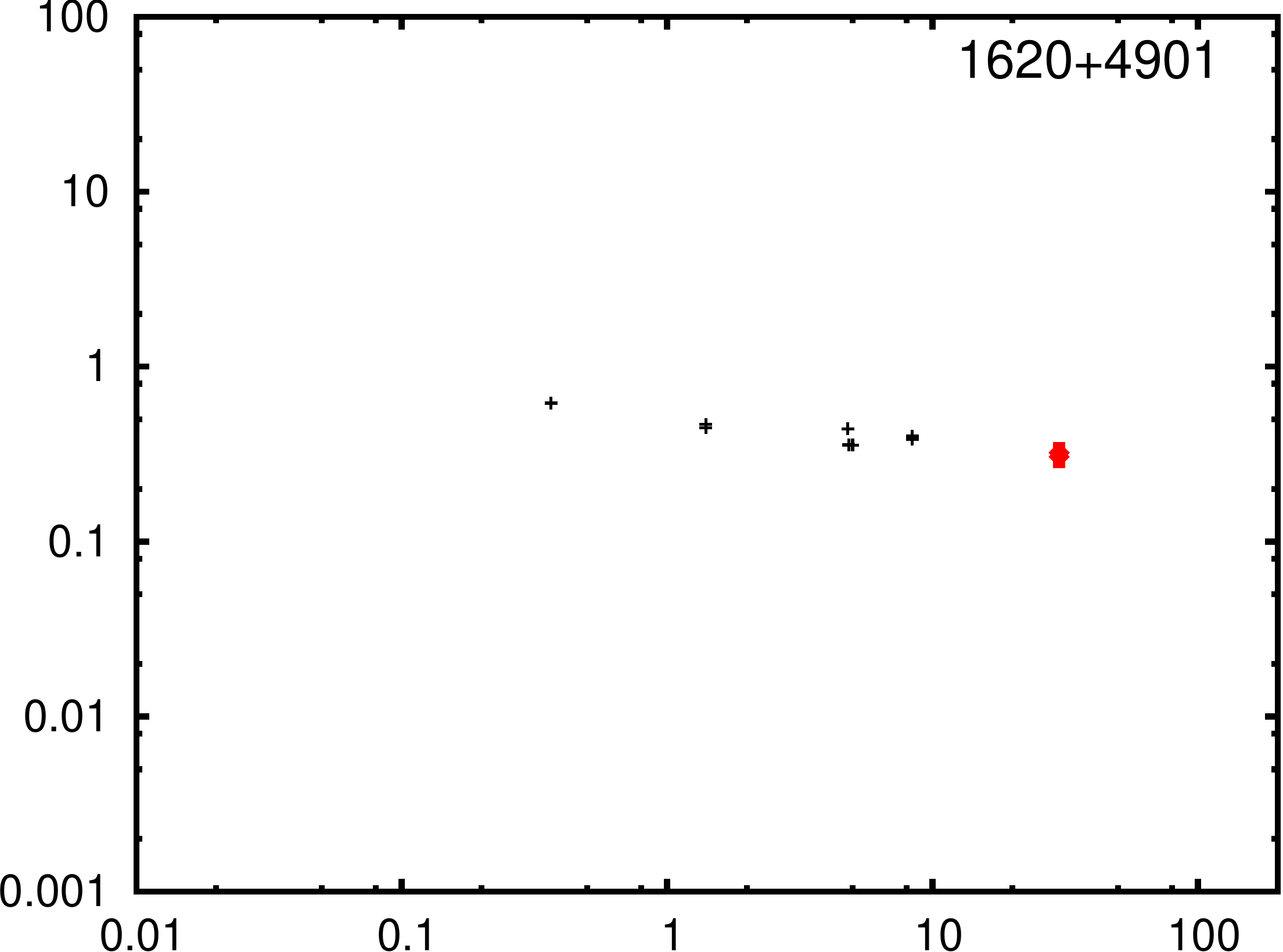}
\includegraphics[scale=0.2]{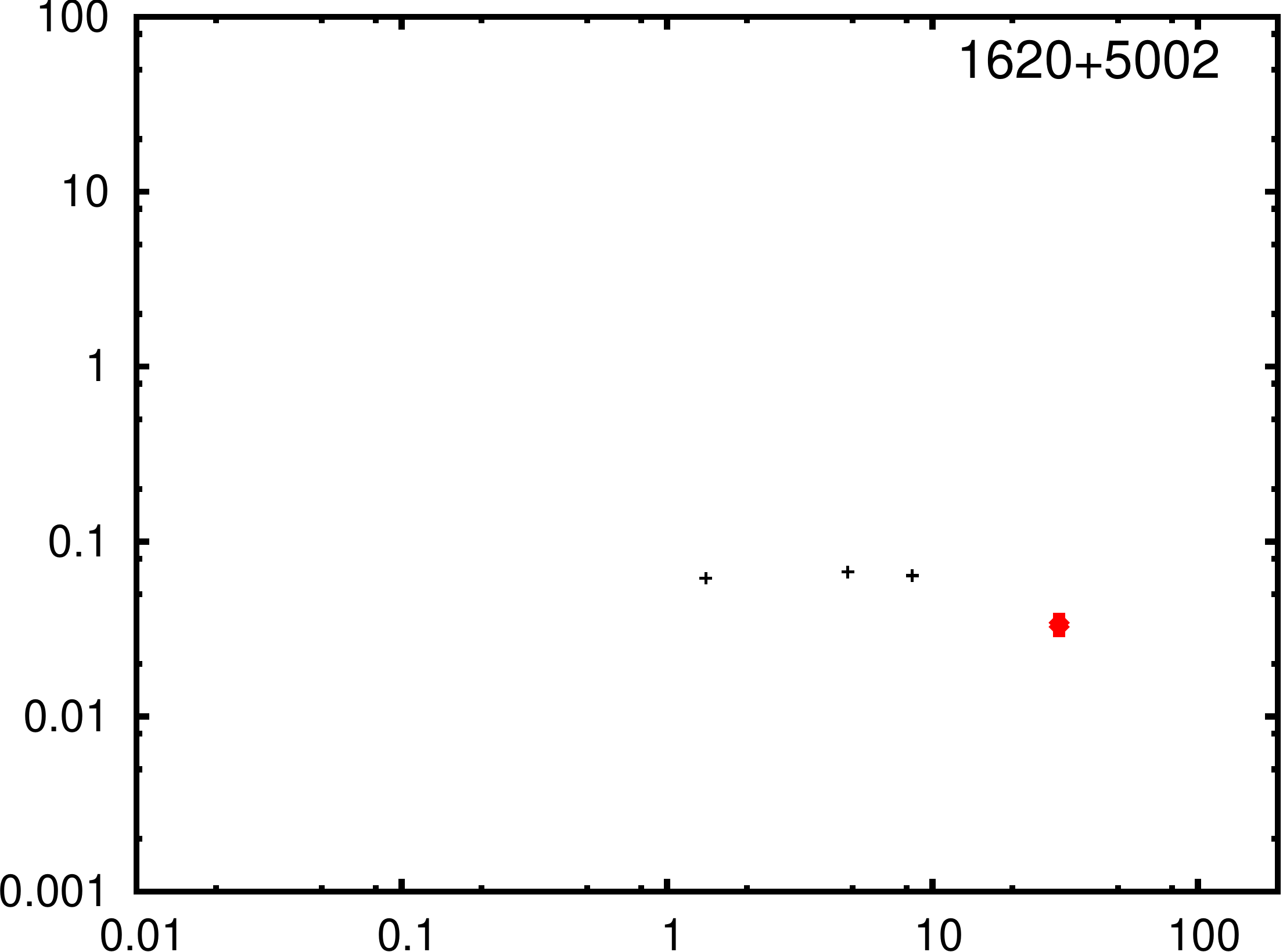}
\end{figure}
\clearpage\begin{figure}
\centering
\includegraphics[scale=0.2]{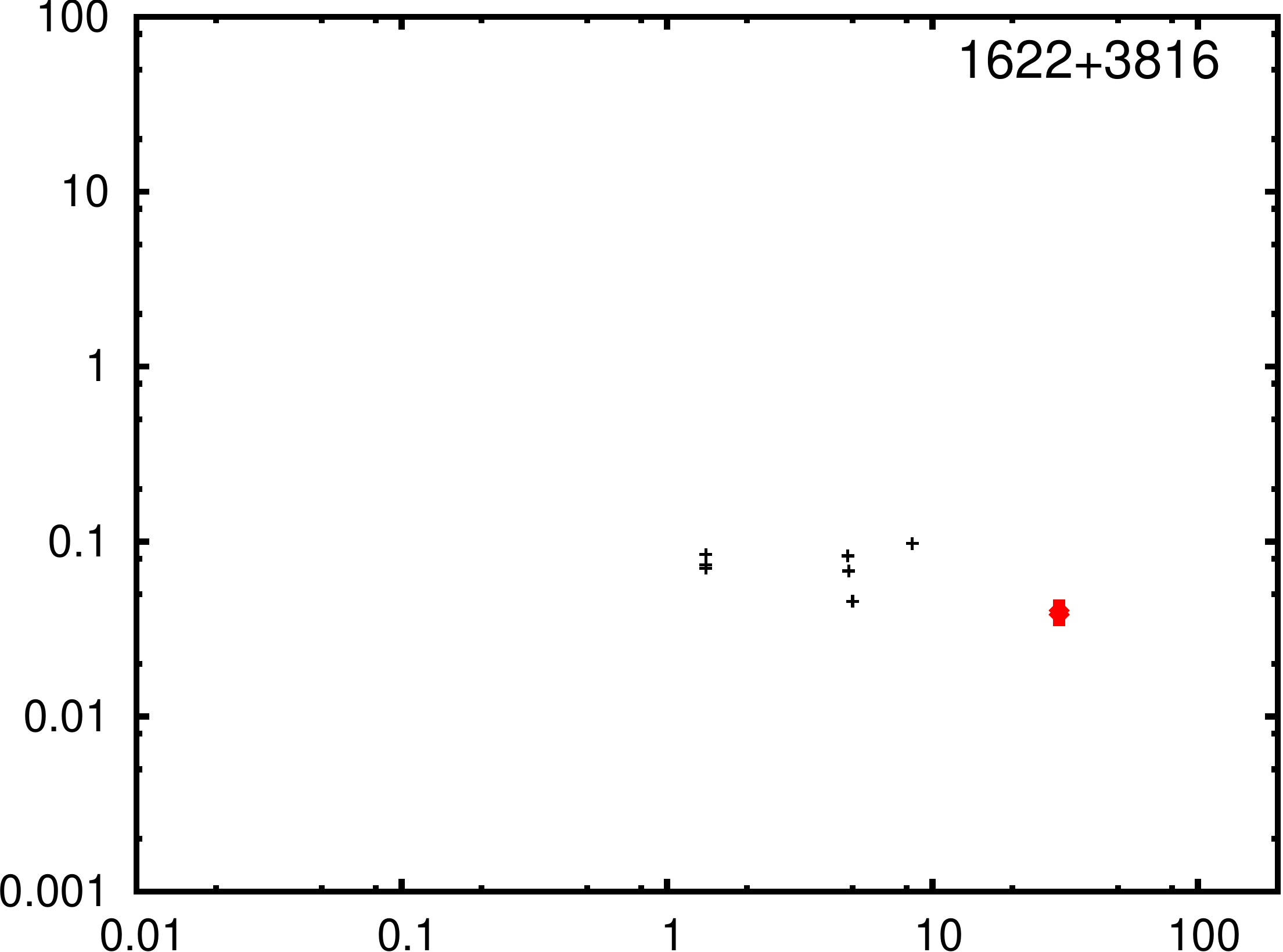}
\includegraphics[scale=0.2]{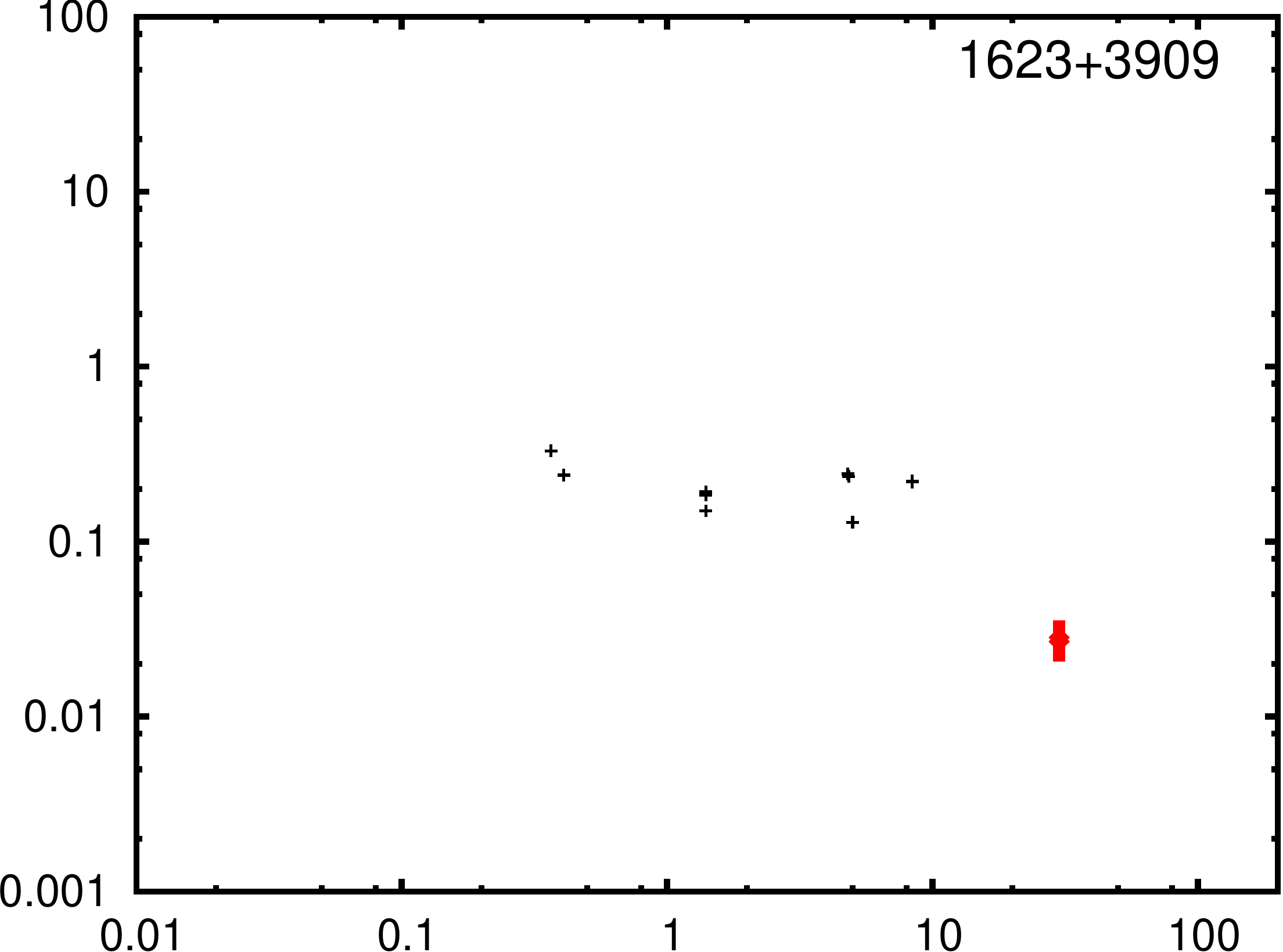}
\includegraphics[scale=0.2]{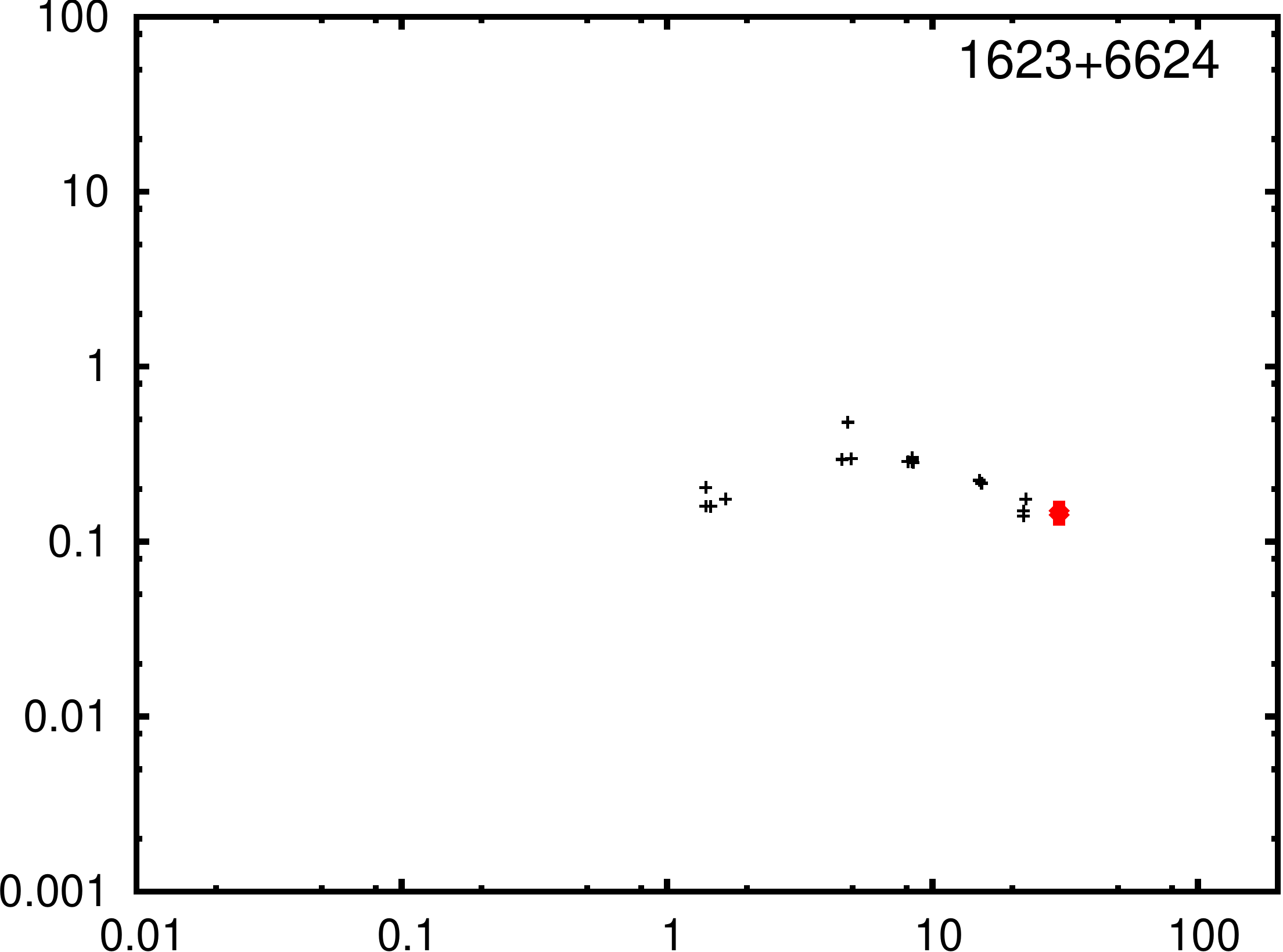}
\includegraphics[scale=0.2]{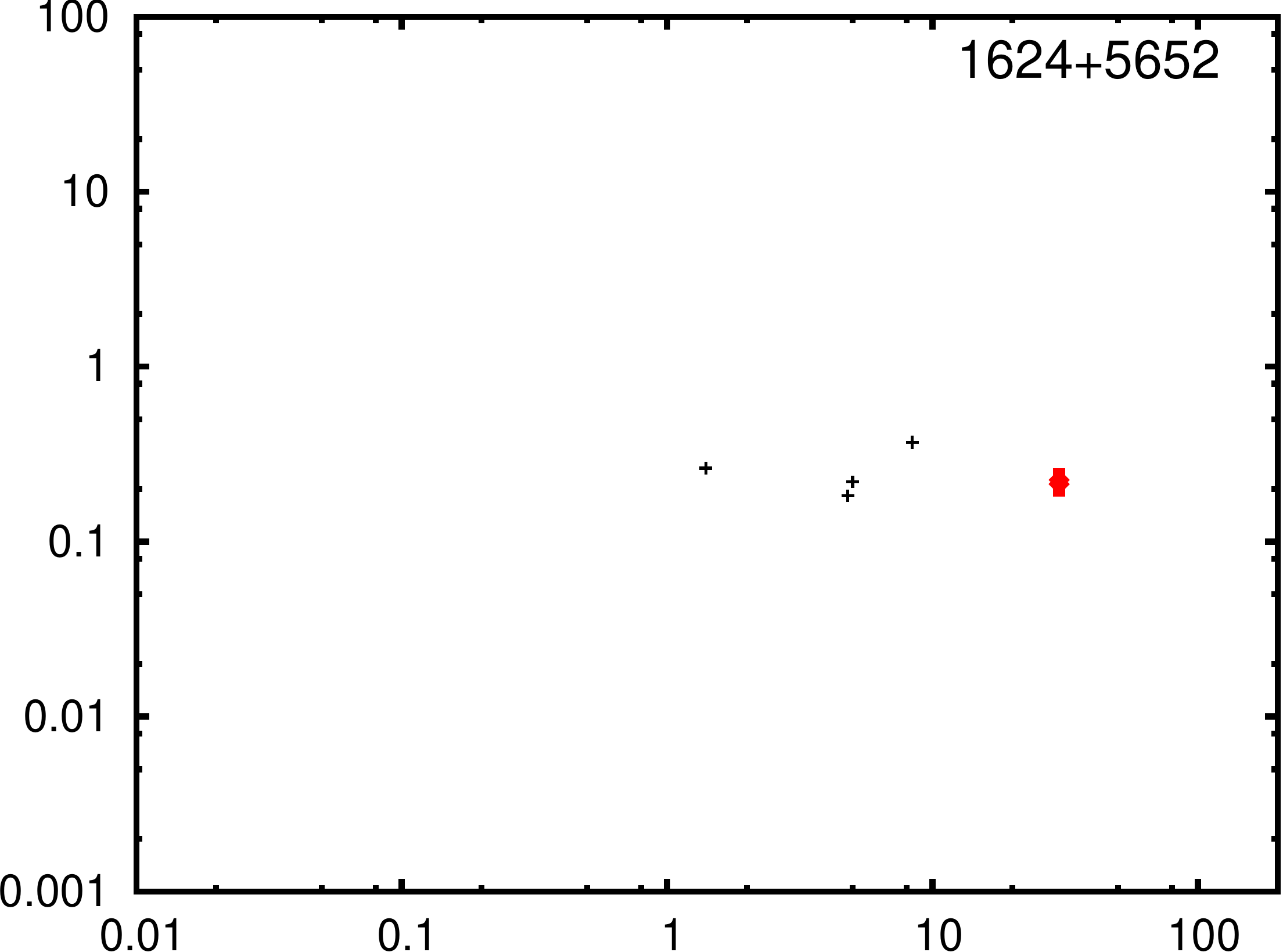}
\includegraphics[scale=0.2]{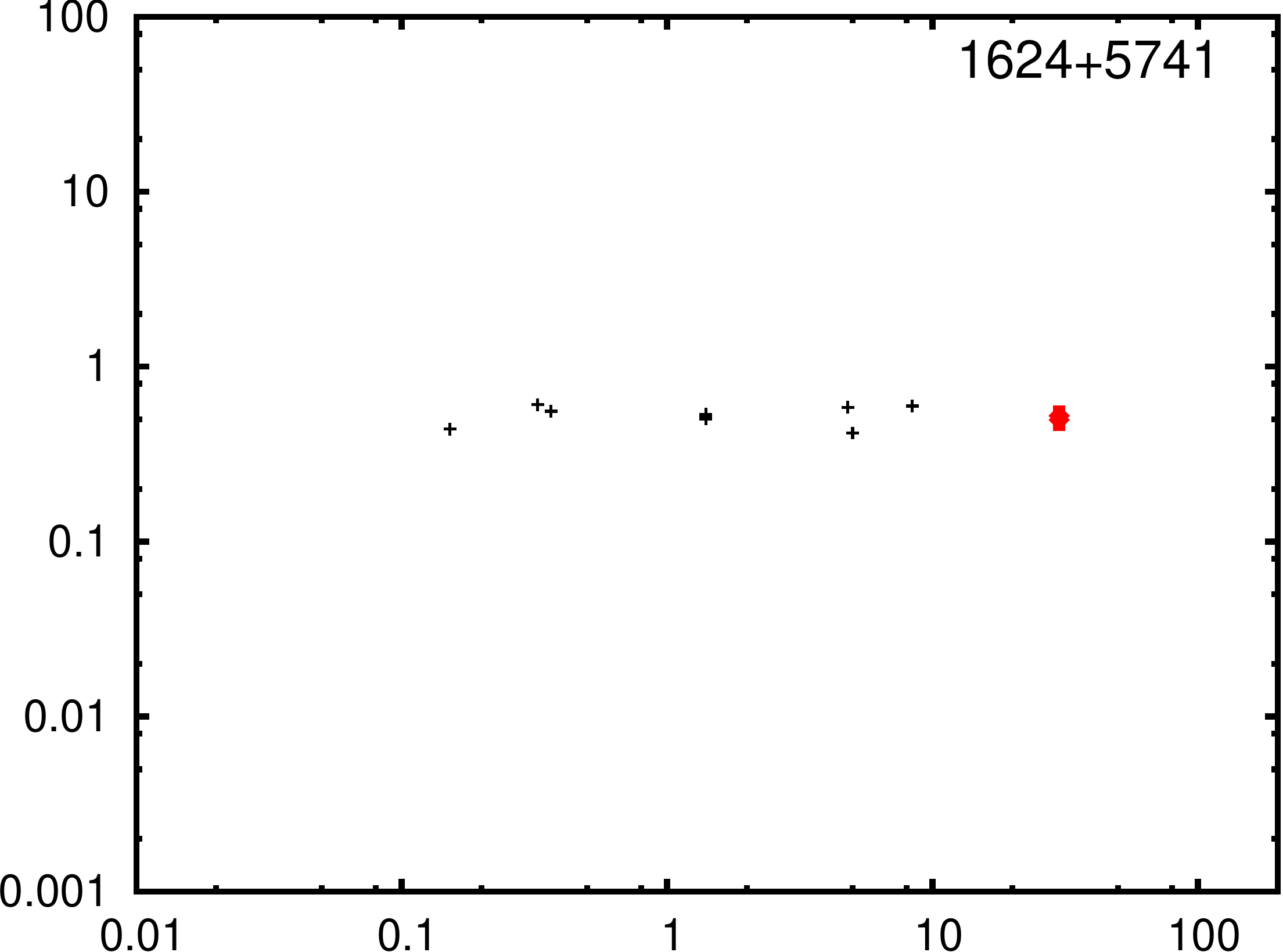}
\includegraphics[scale=0.2]{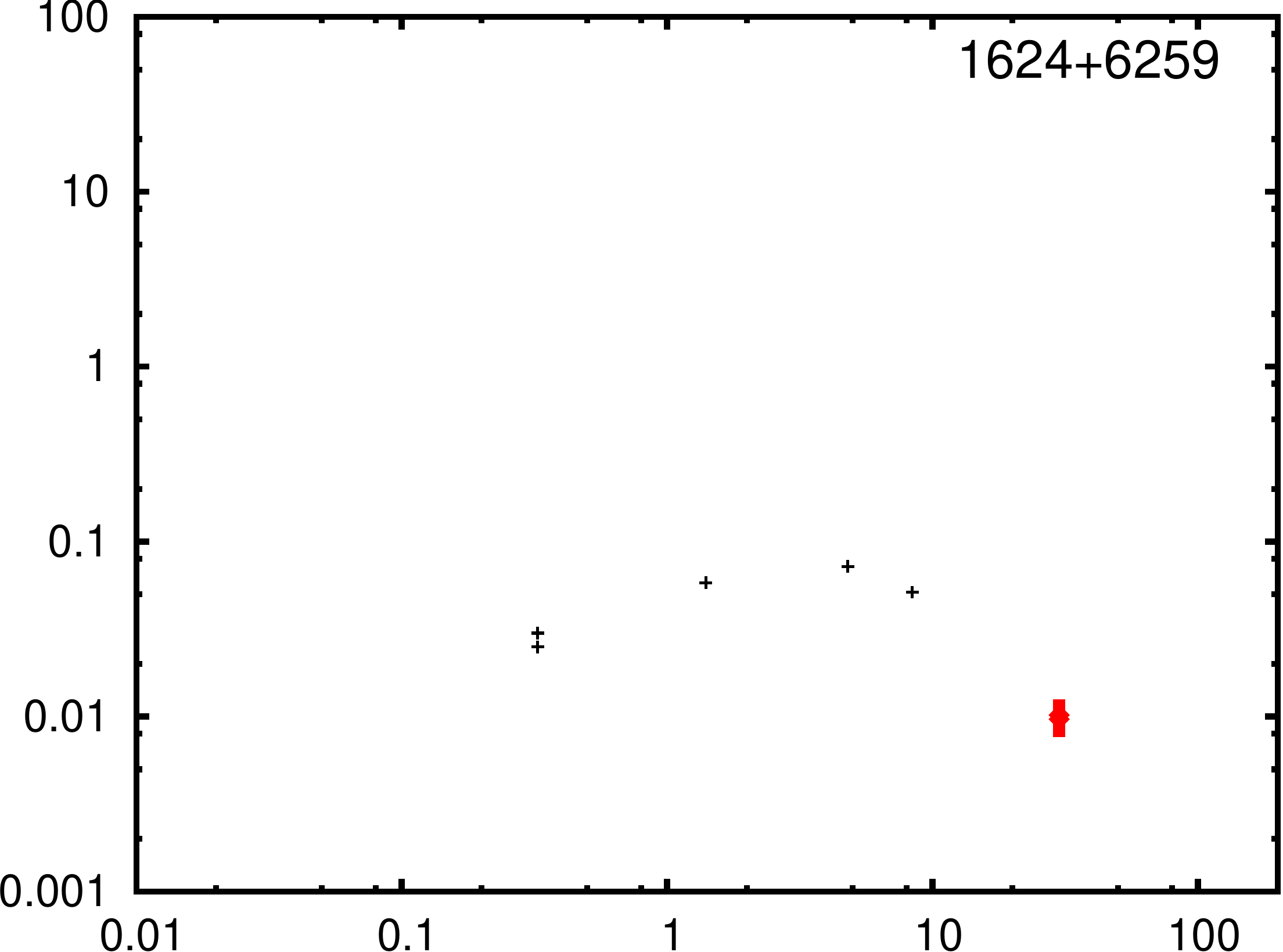}
\includegraphics[scale=0.2]{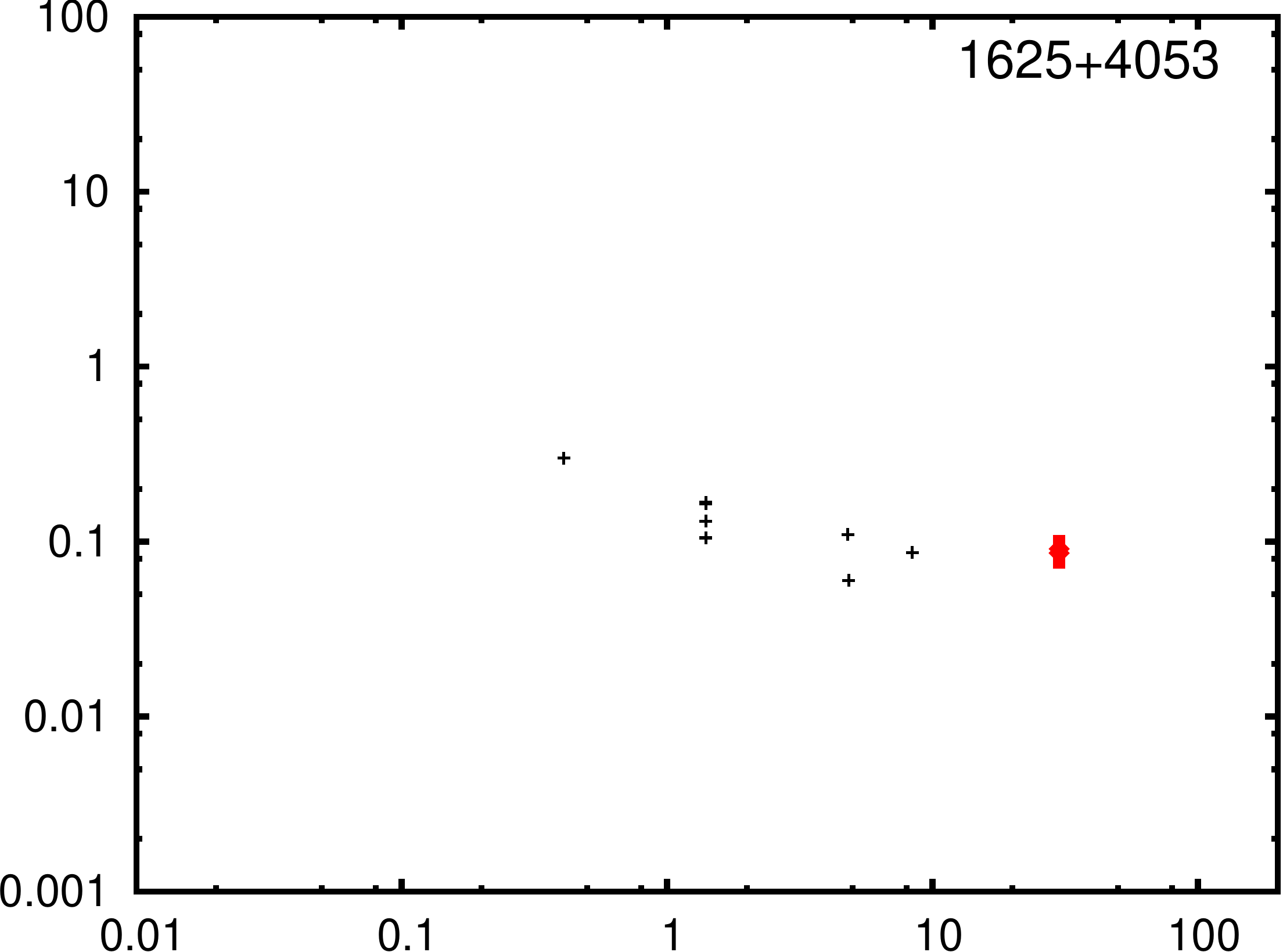}
\includegraphics[scale=0.2]{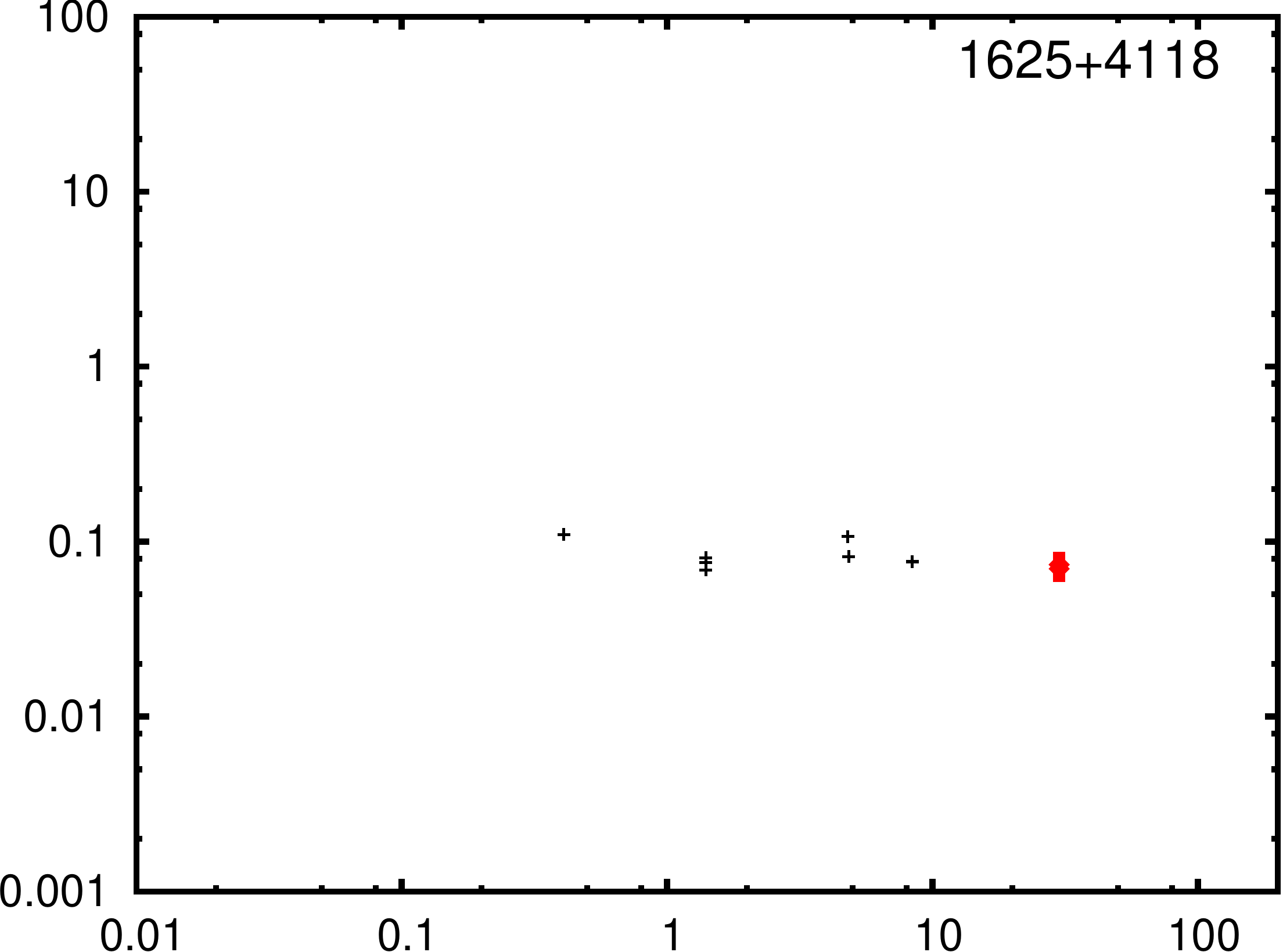}
\includegraphics[scale=0.2]{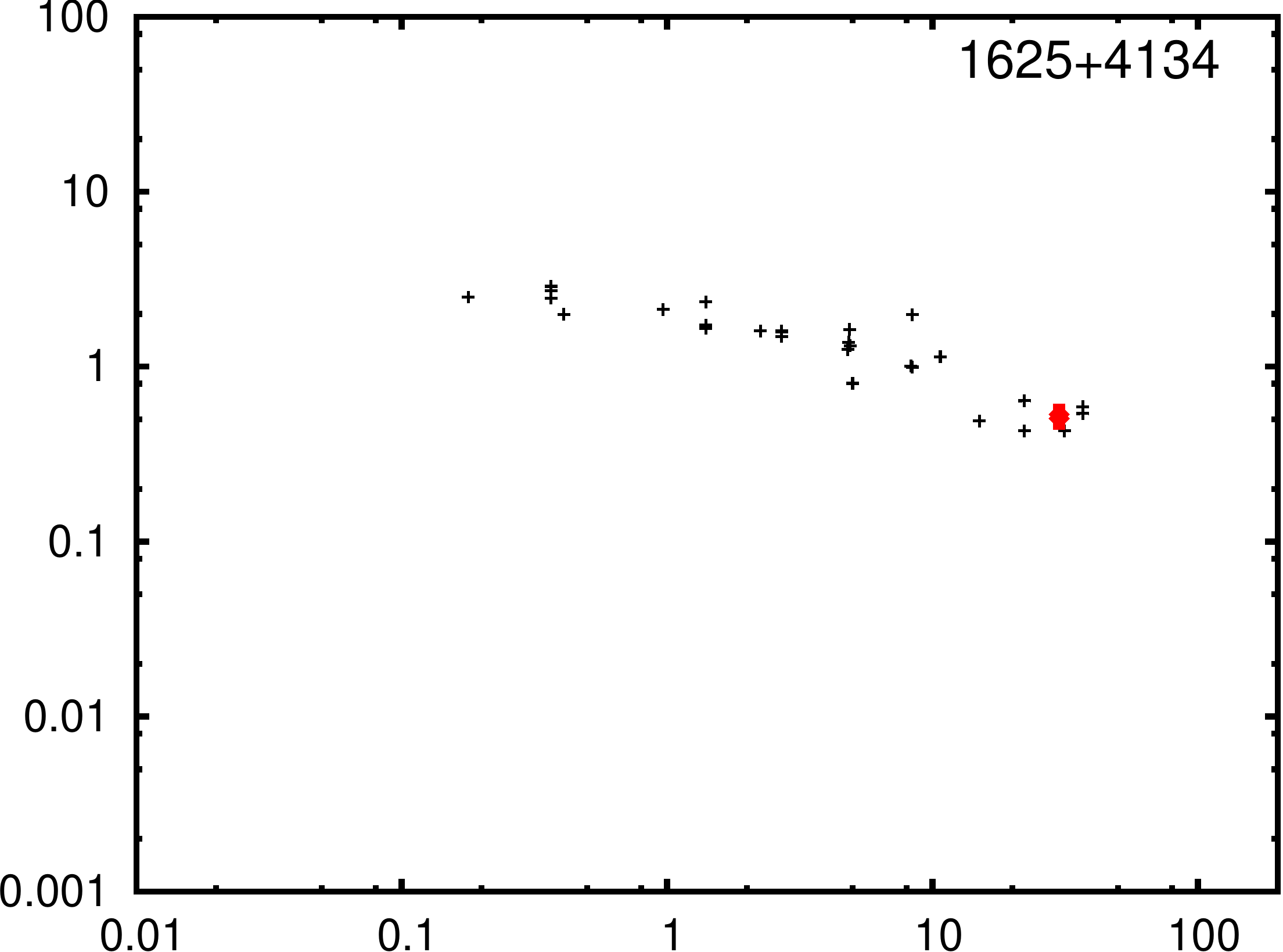}
\includegraphics[scale=0.2]{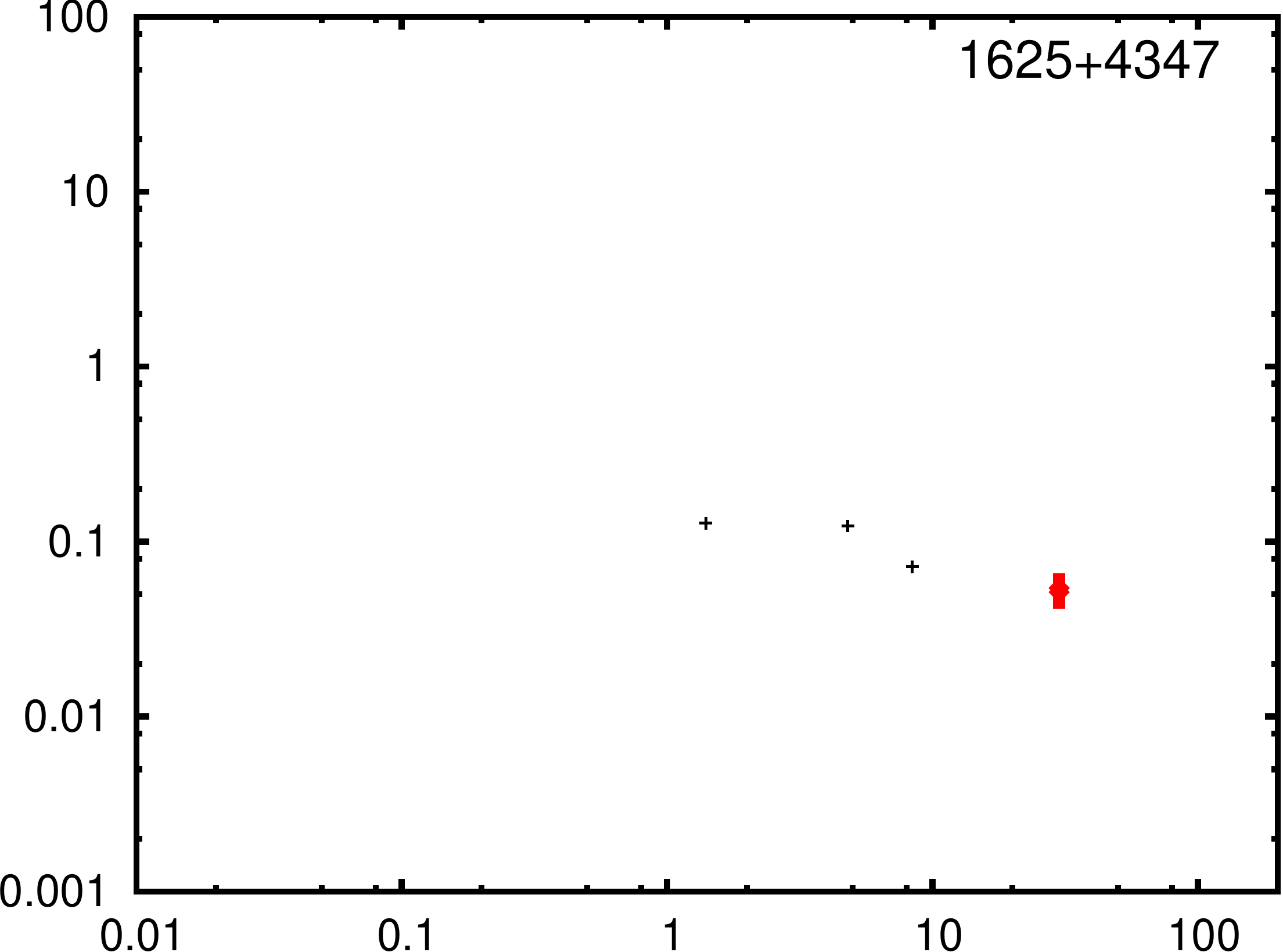}
\includegraphics[scale=0.2]{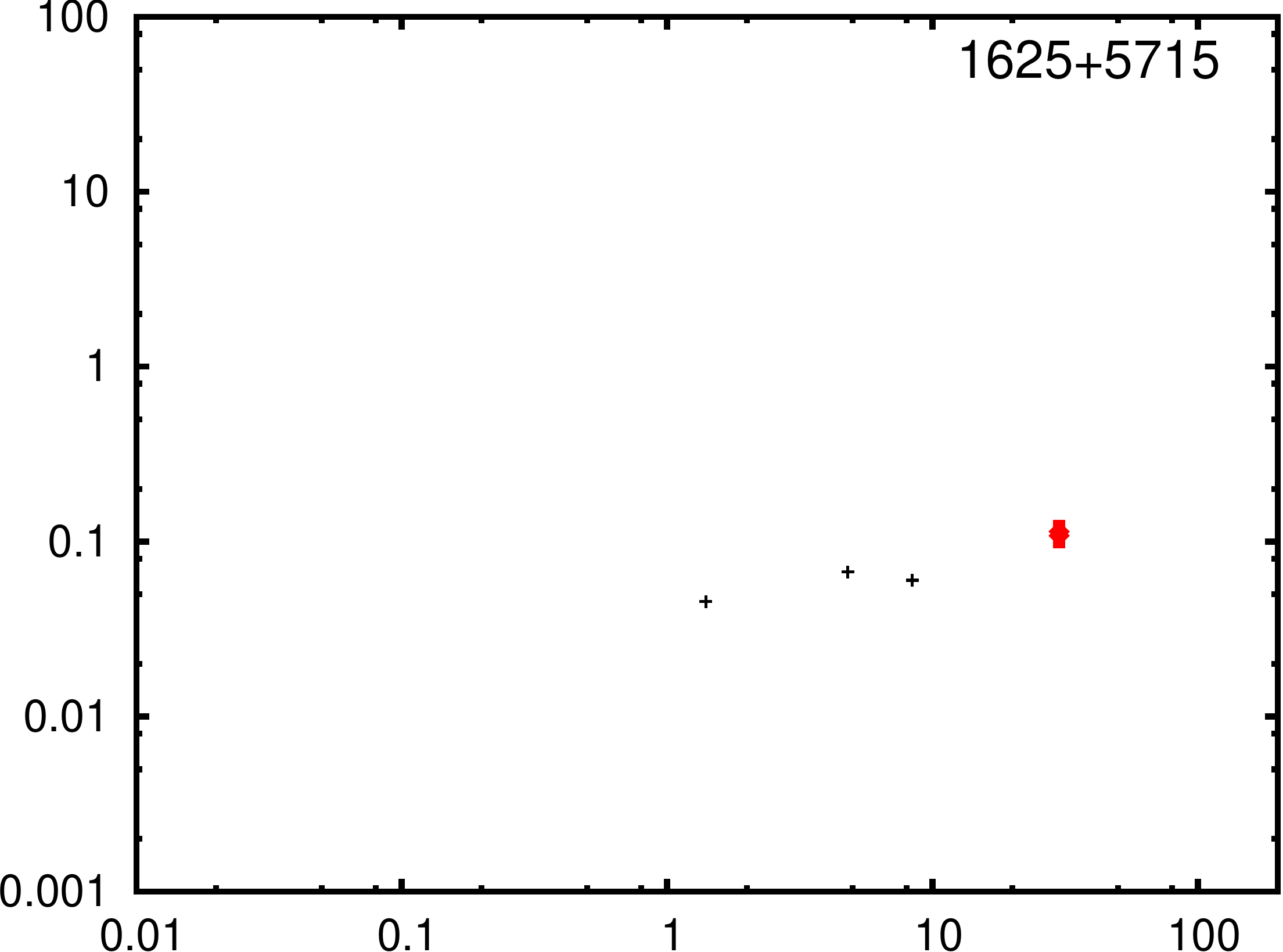}
\includegraphics[scale=0.2]{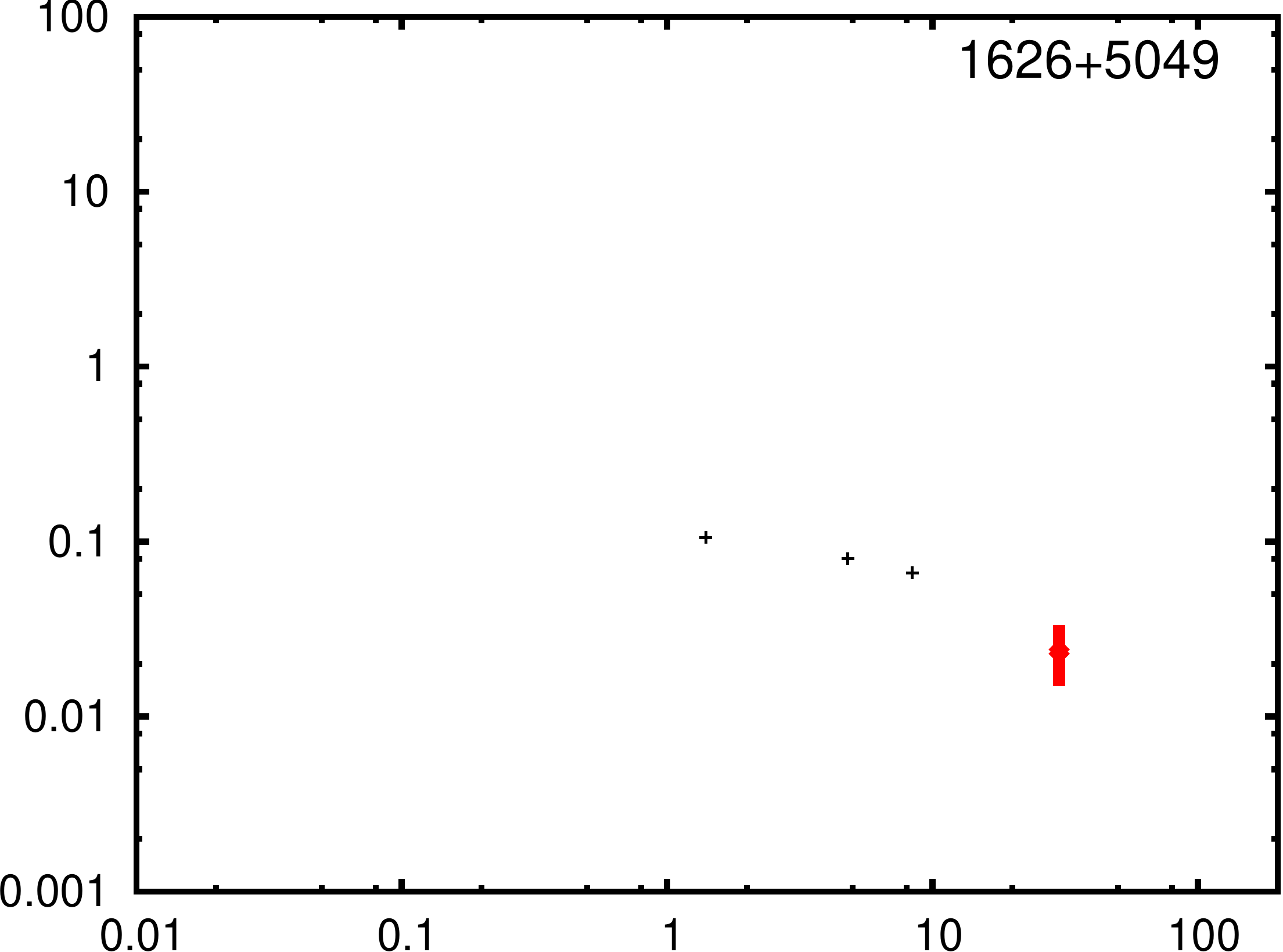}
\includegraphics[scale=0.2]{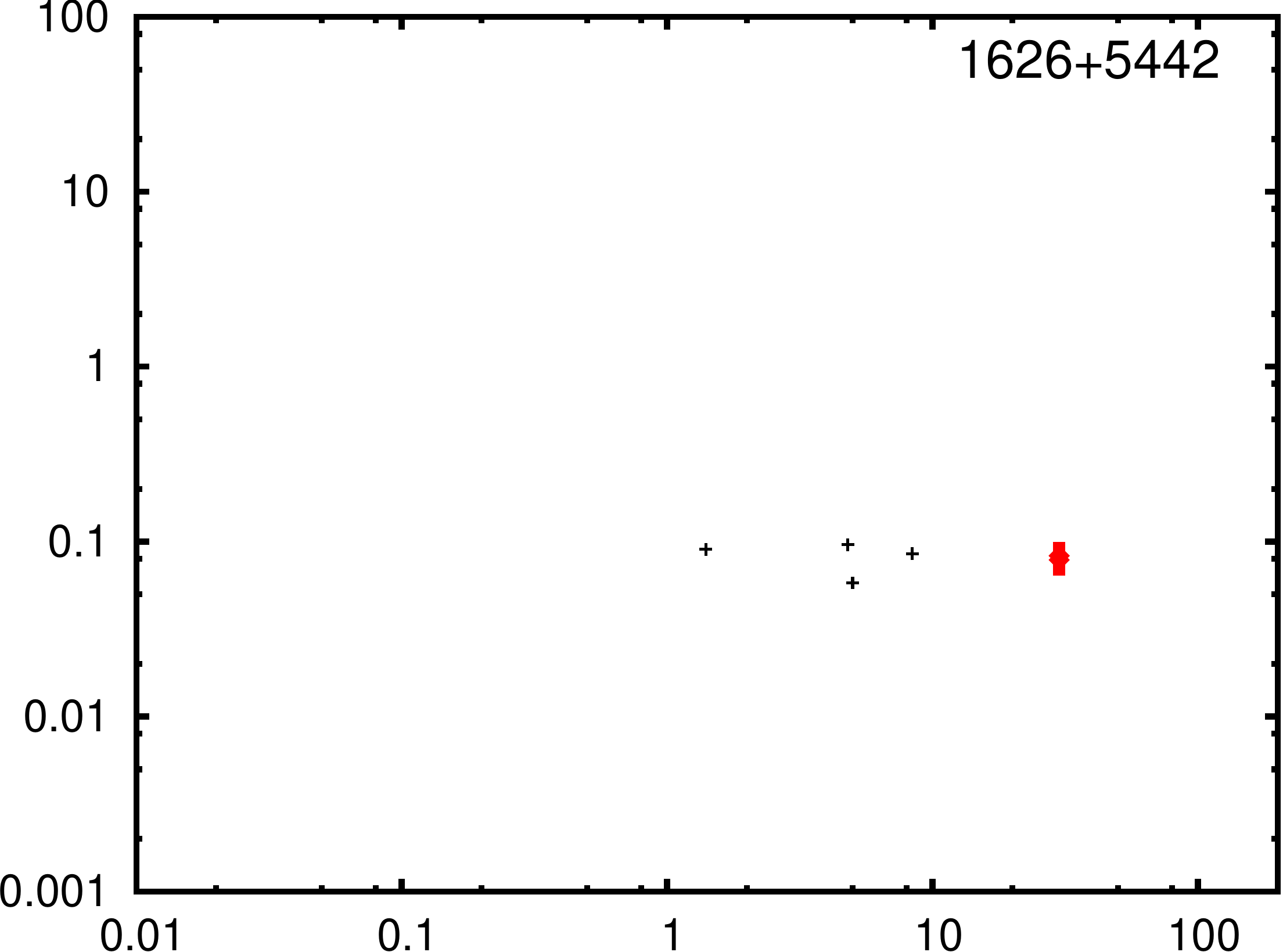}
\includegraphics[scale=0.2]{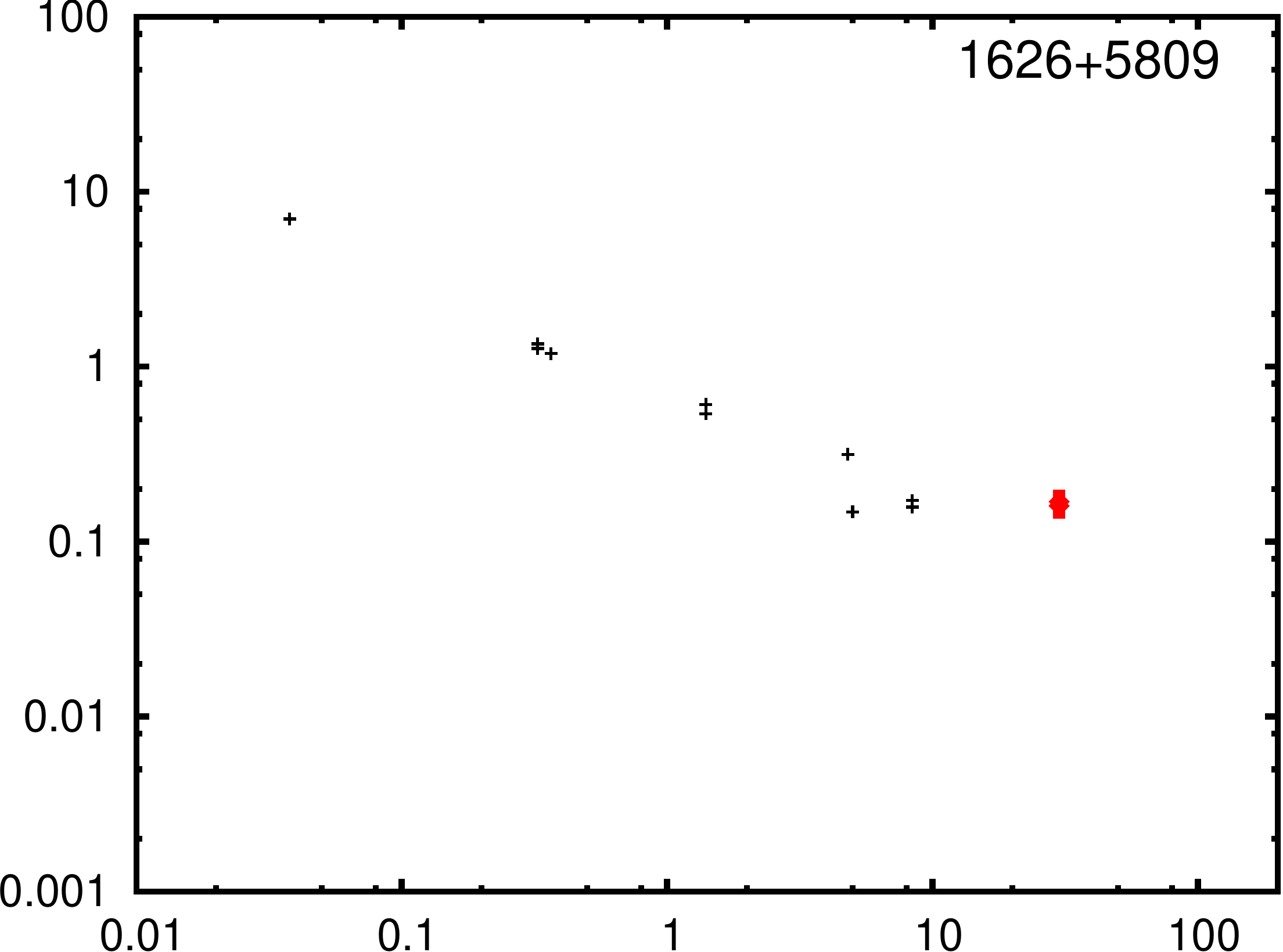}
\includegraphics[scale=0.2]{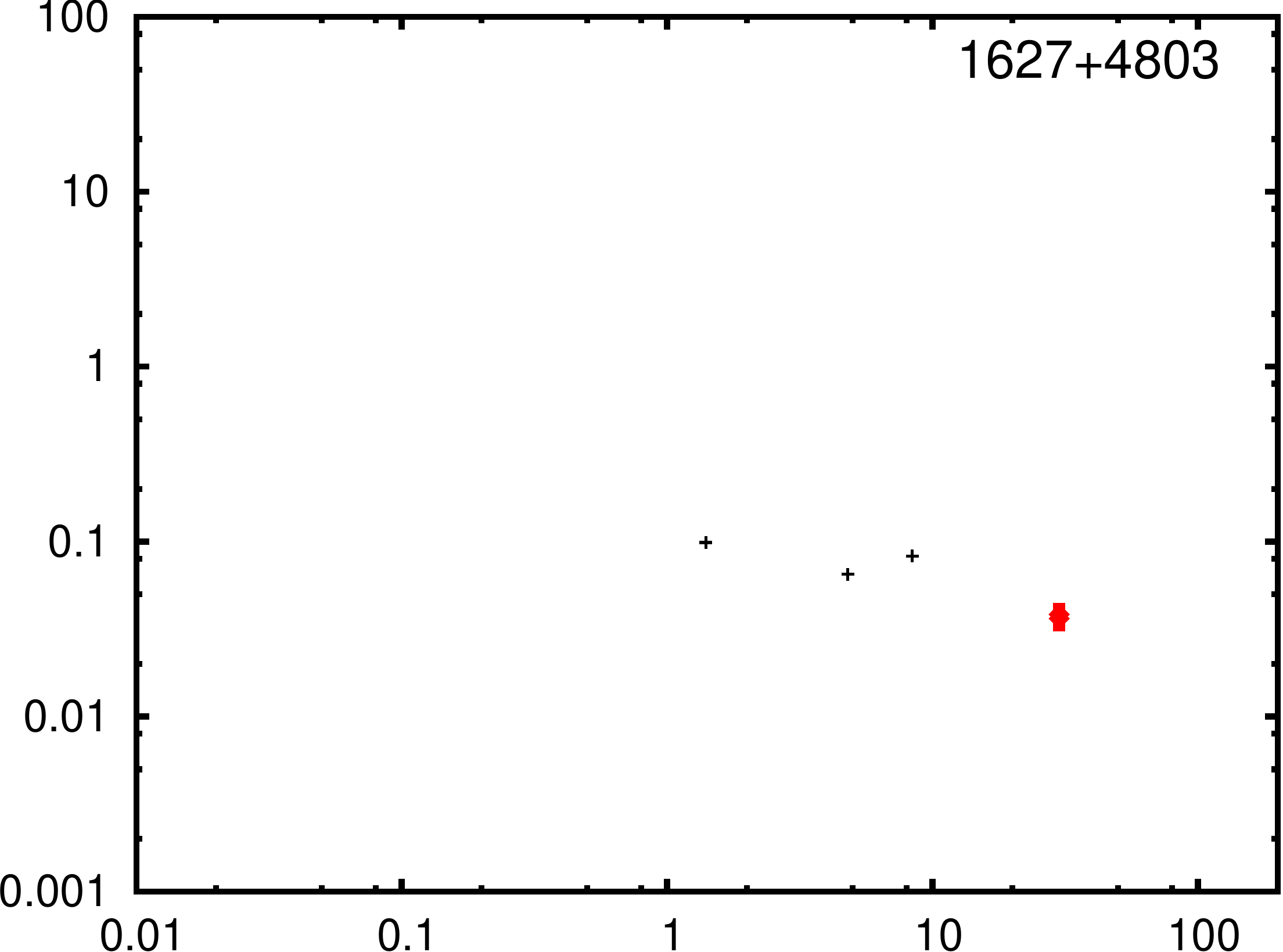}
\includegraphics[scale=0.2]{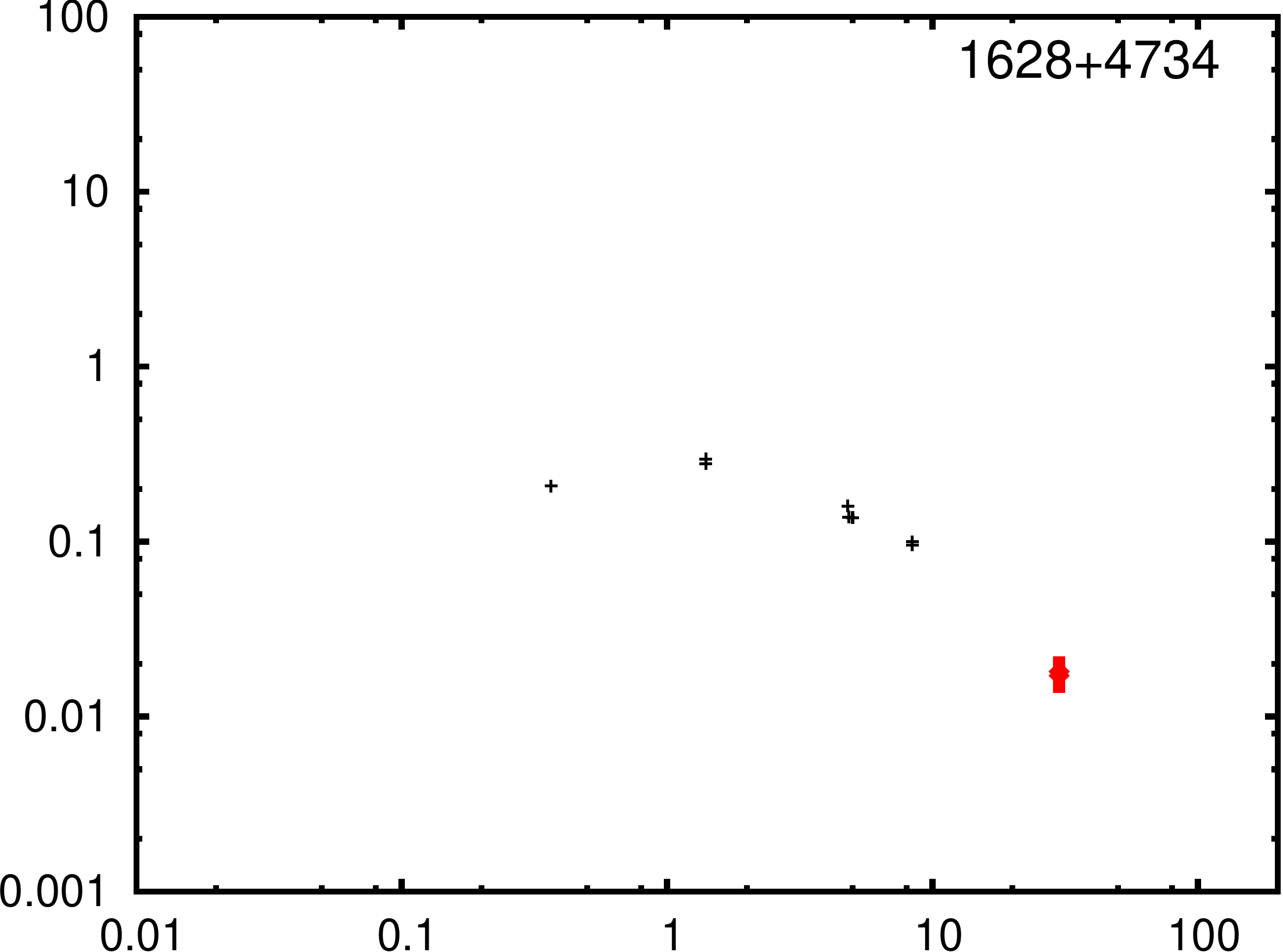}
\includegraphics[scale=0.2]{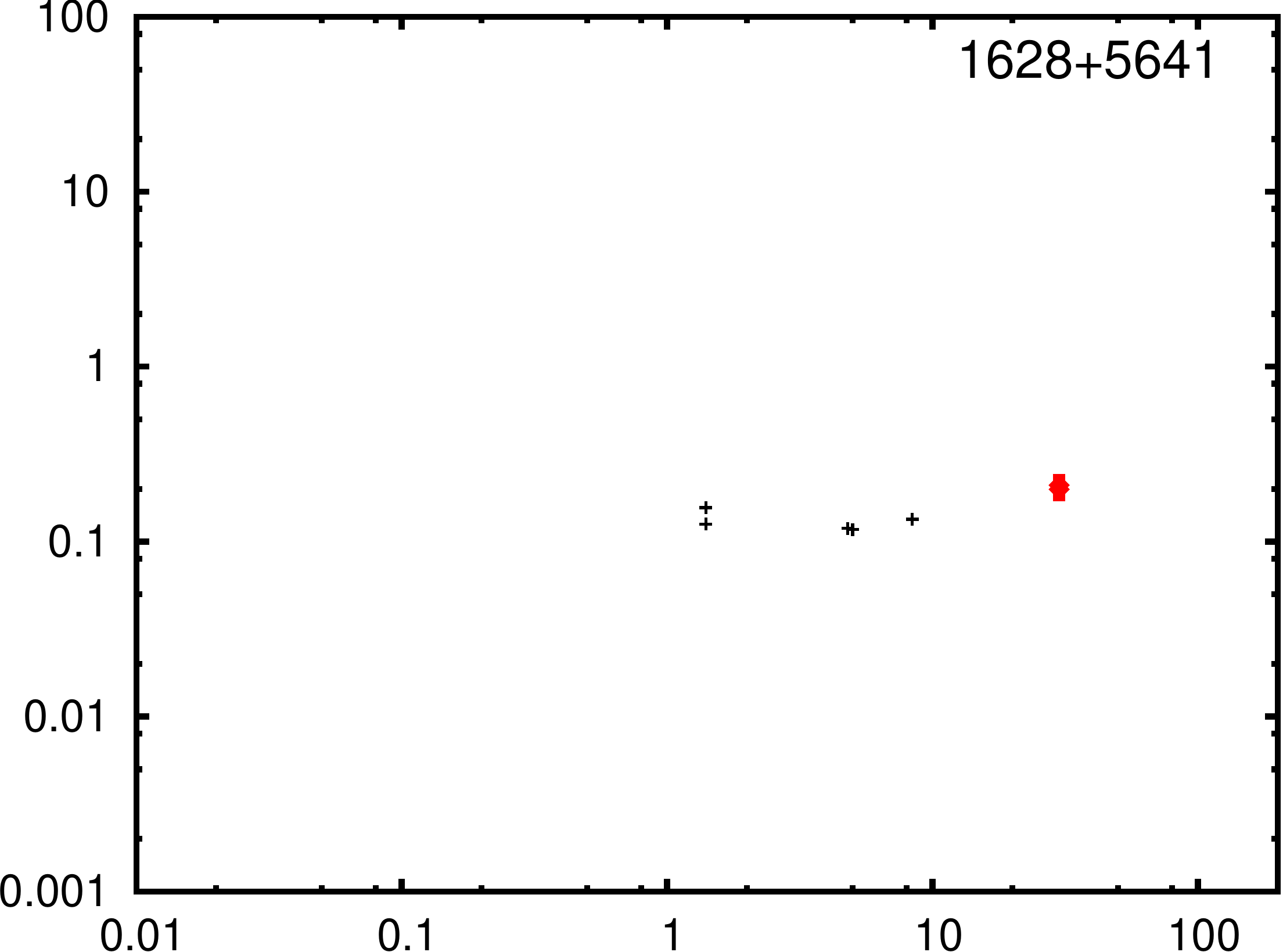}
\includegraphics[scale=0.2]{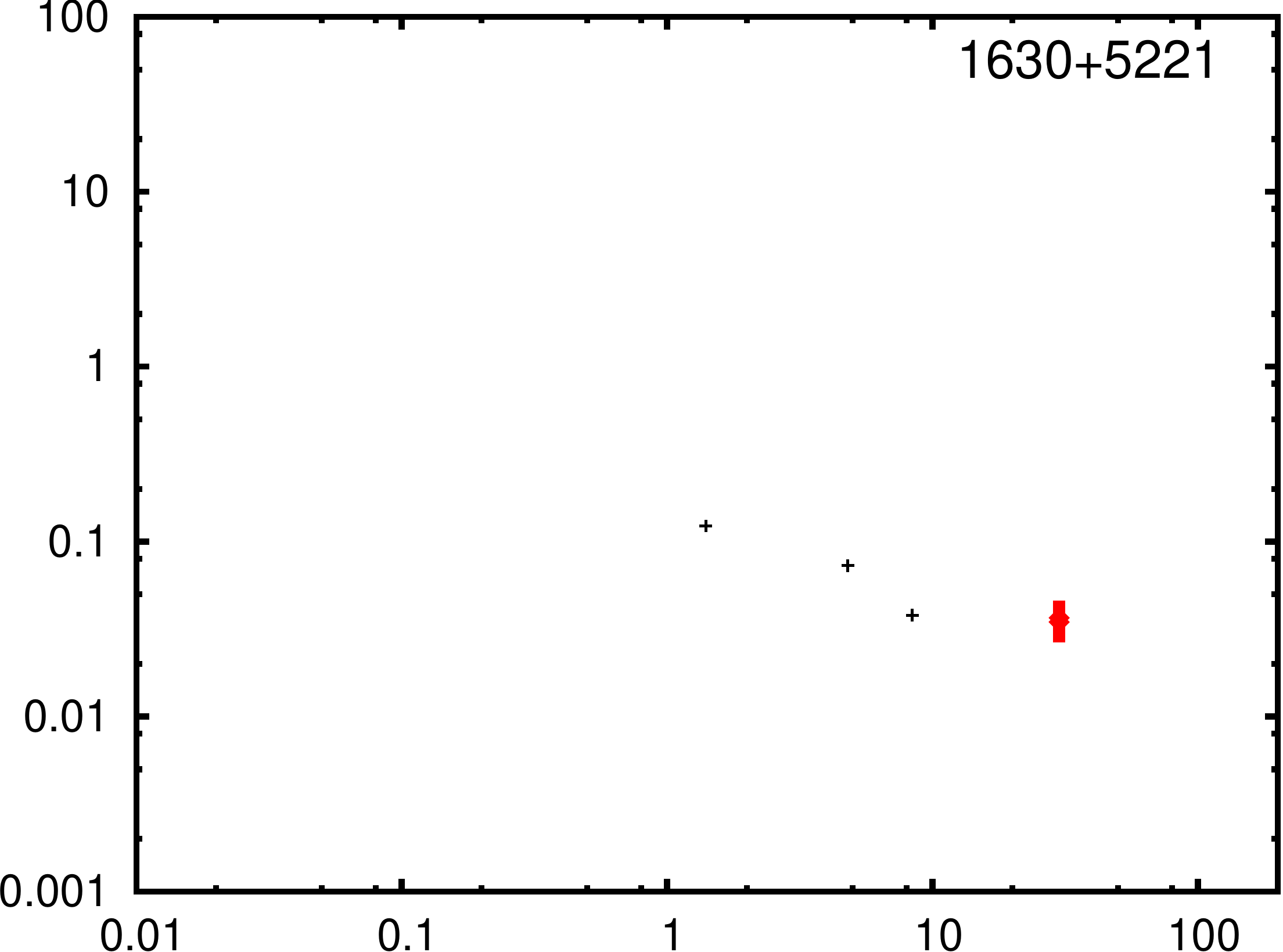}
\end{figure}
\clearpage\begin{figure}
\centering
\includegraphics[scale=0.2]{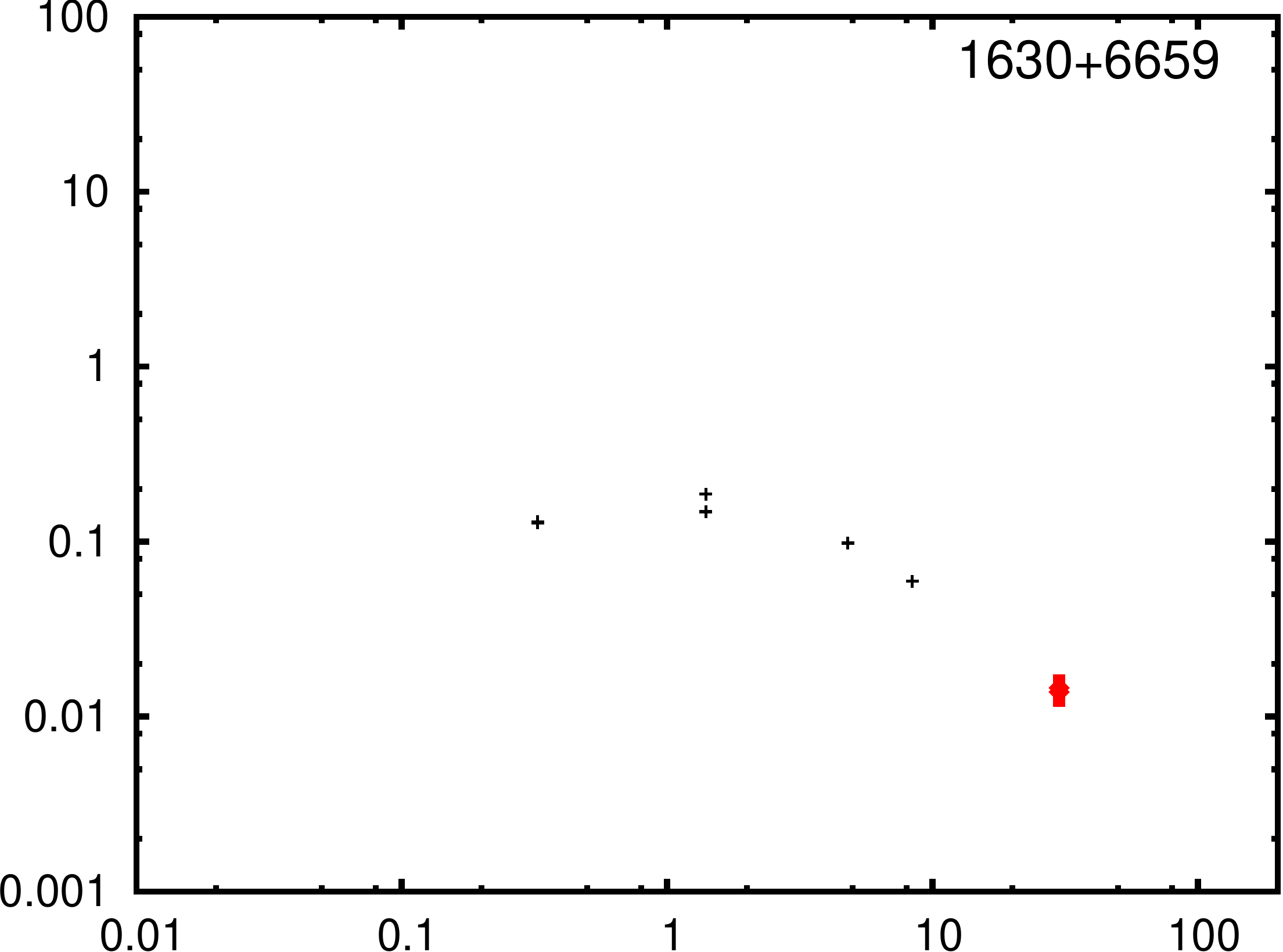}
\includegraphics[scale=0.2]{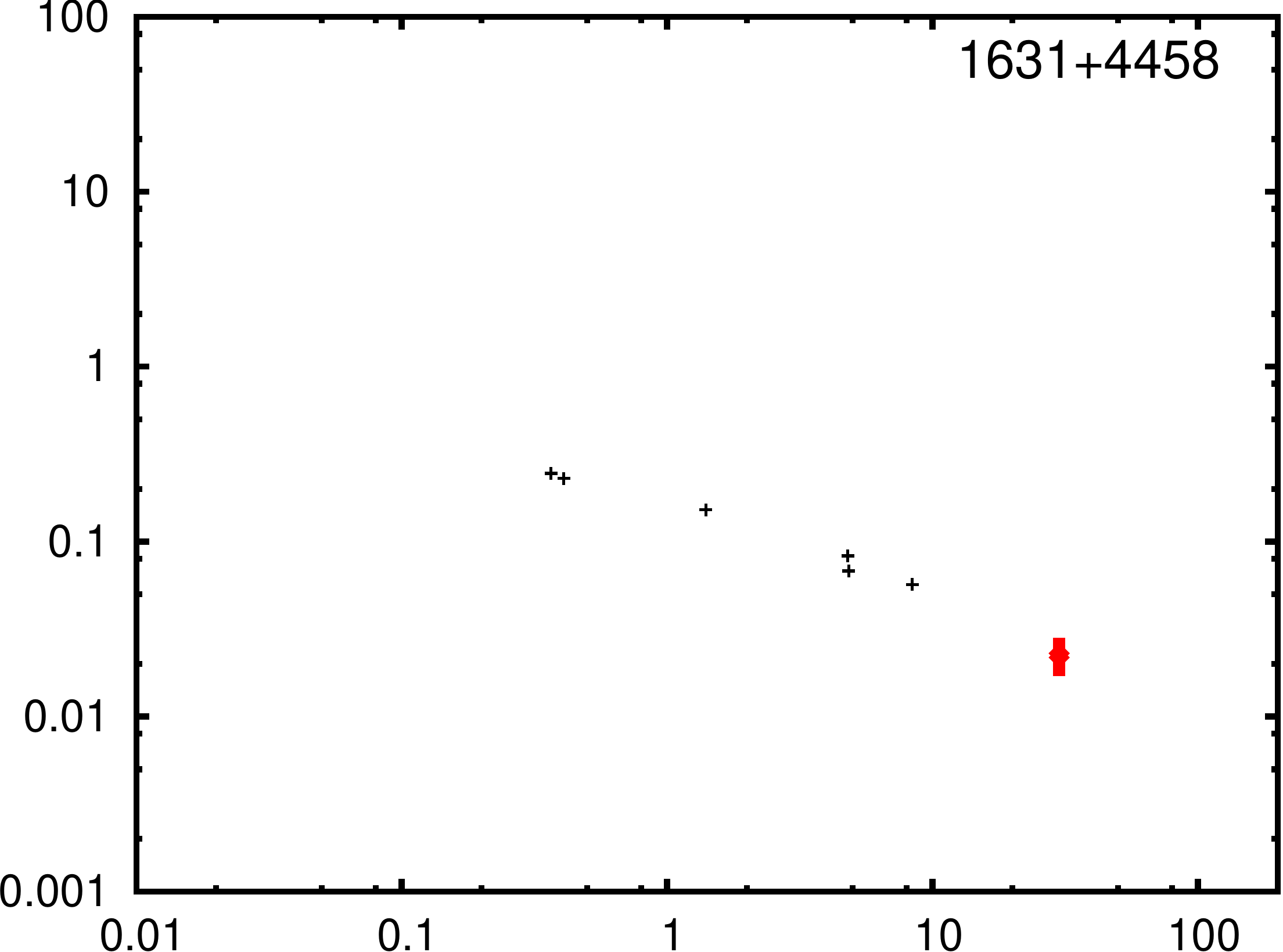}
\includegraphics[scale=0.2]{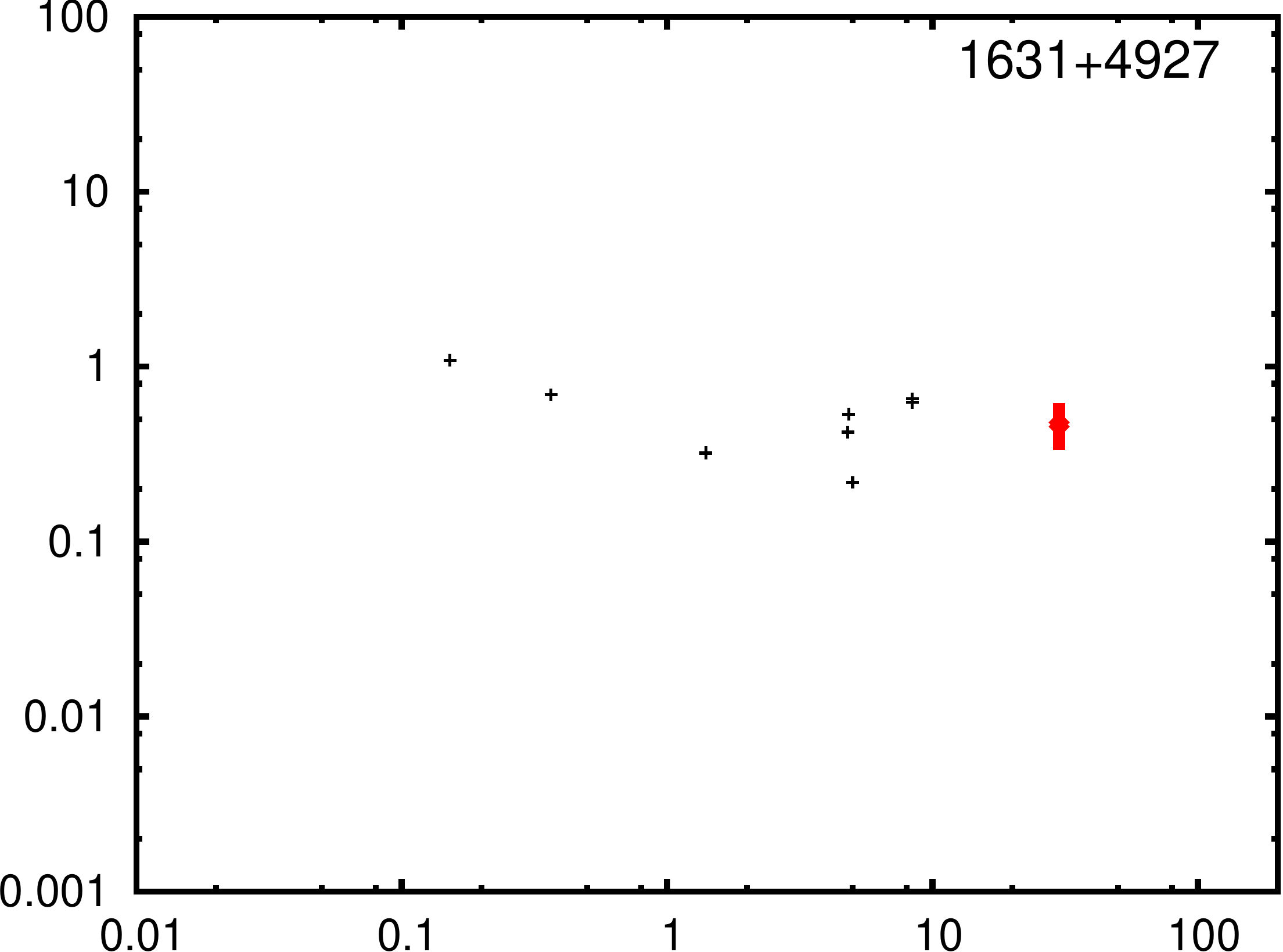}
\includegraphics[scale=0.2]{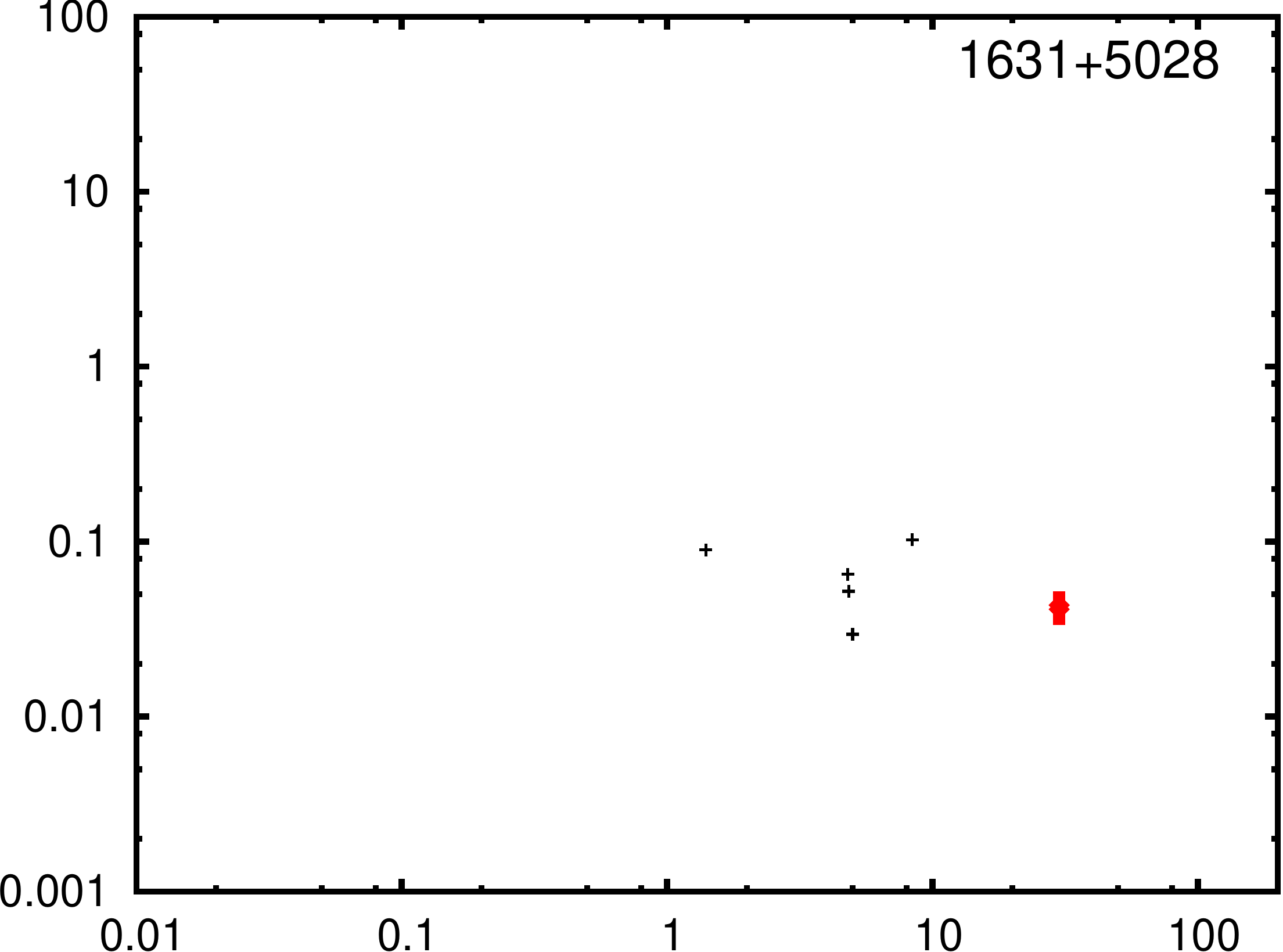}
\includegraphics[scale=0.2]{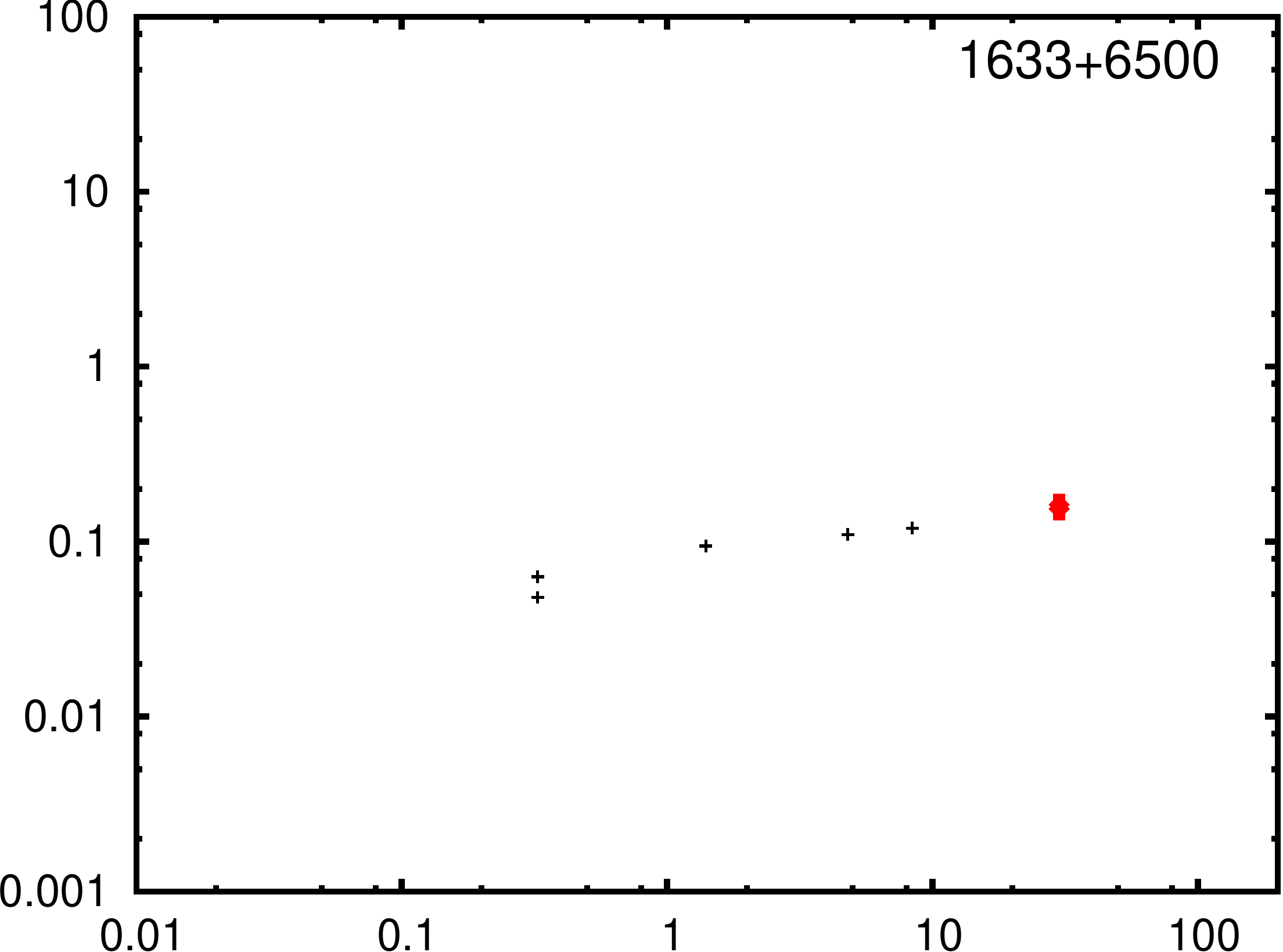}
\includegraphics[scale=0.2]{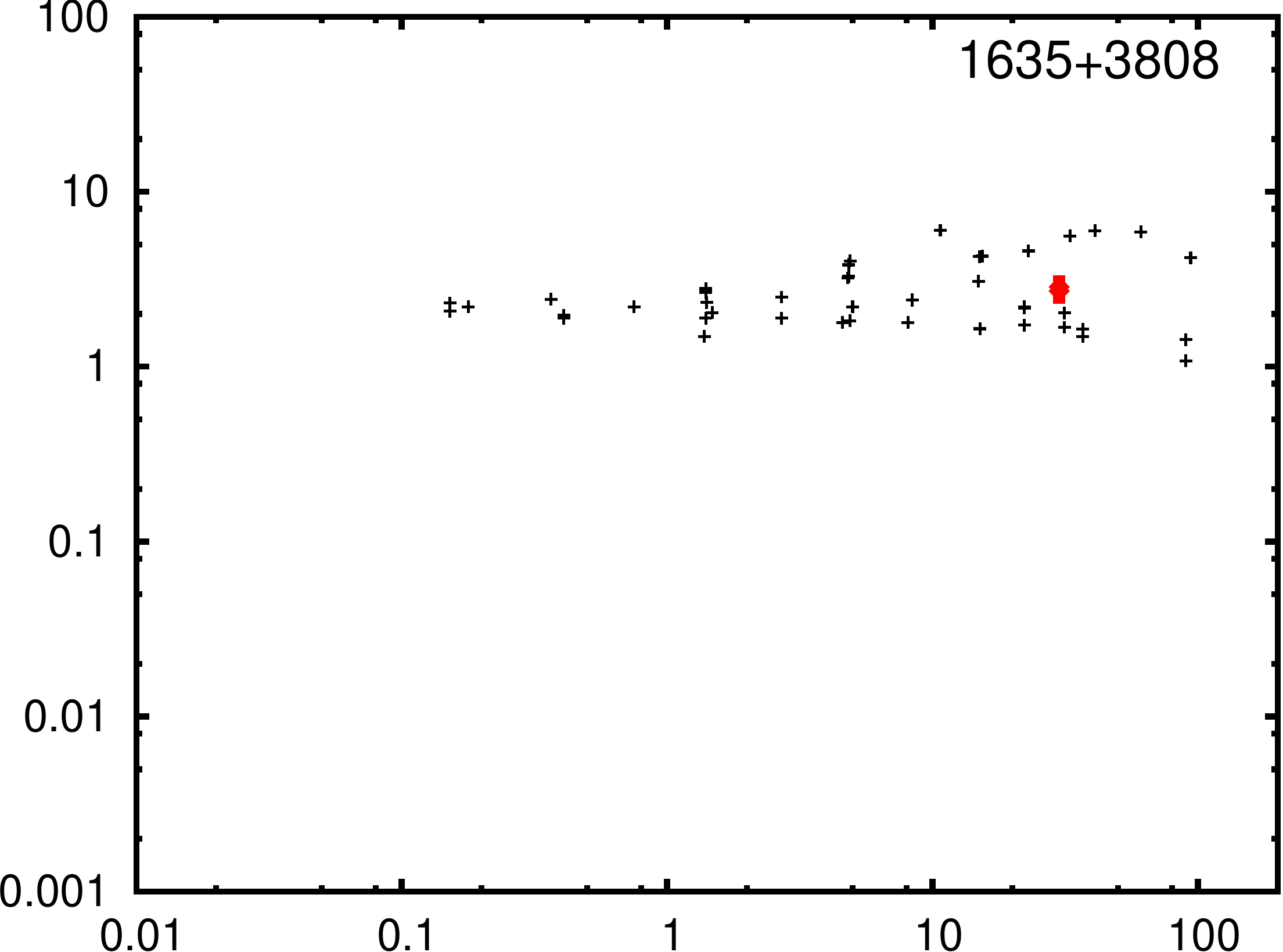}
\includegraphics[scale=0.2]{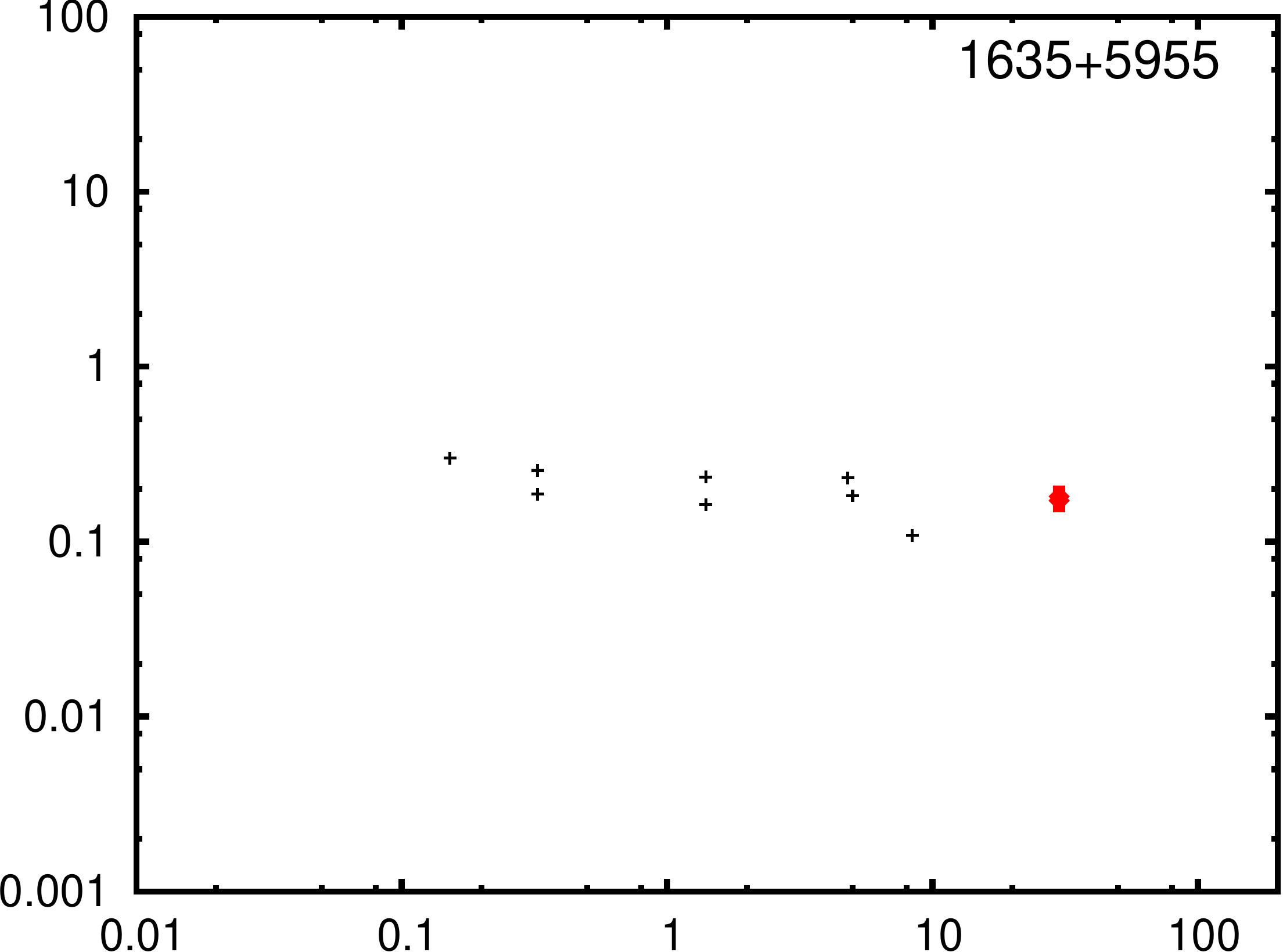}
\includegraphics[scale=0.2]{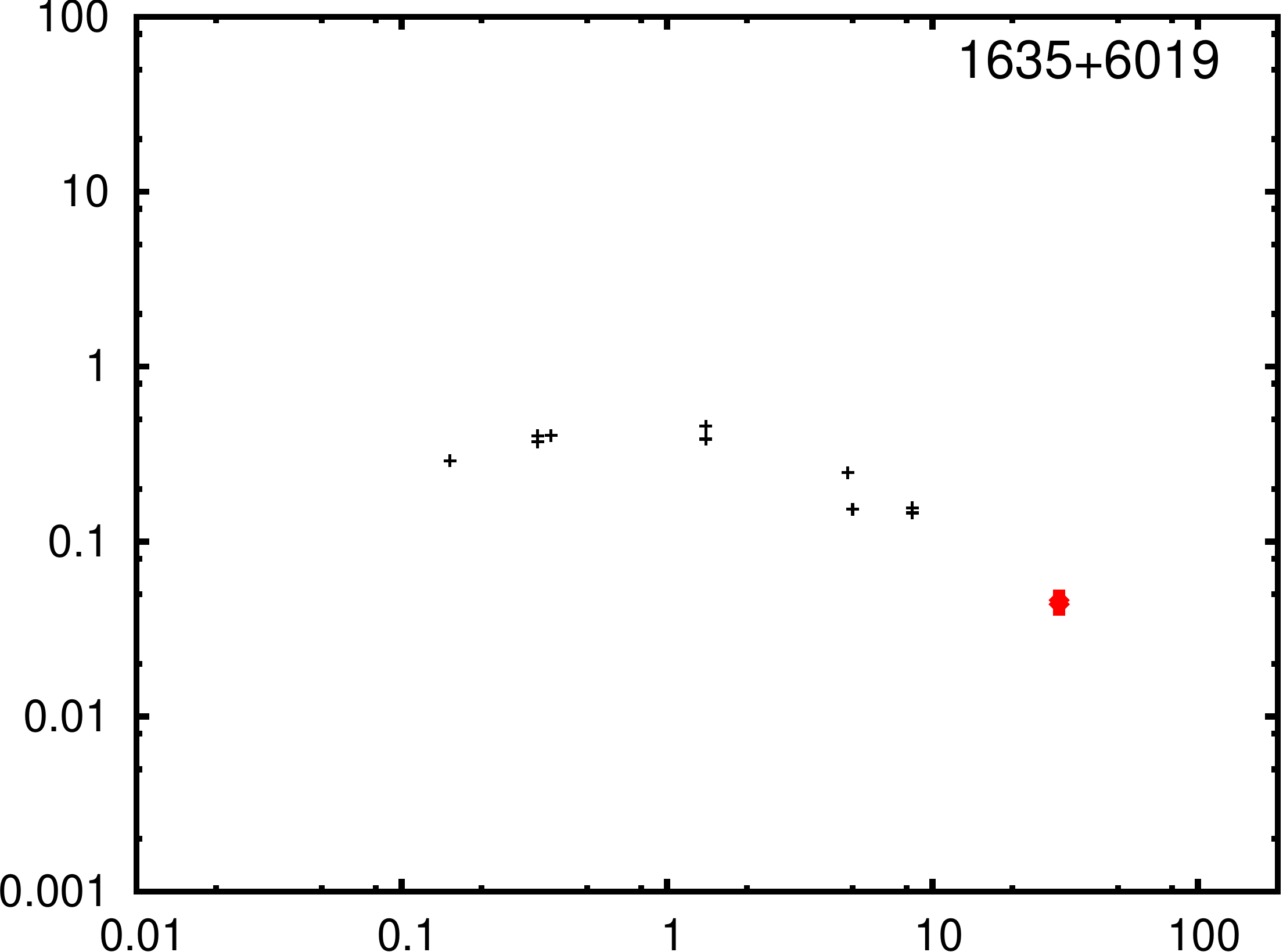}
\includegraphics[scale=0.2]{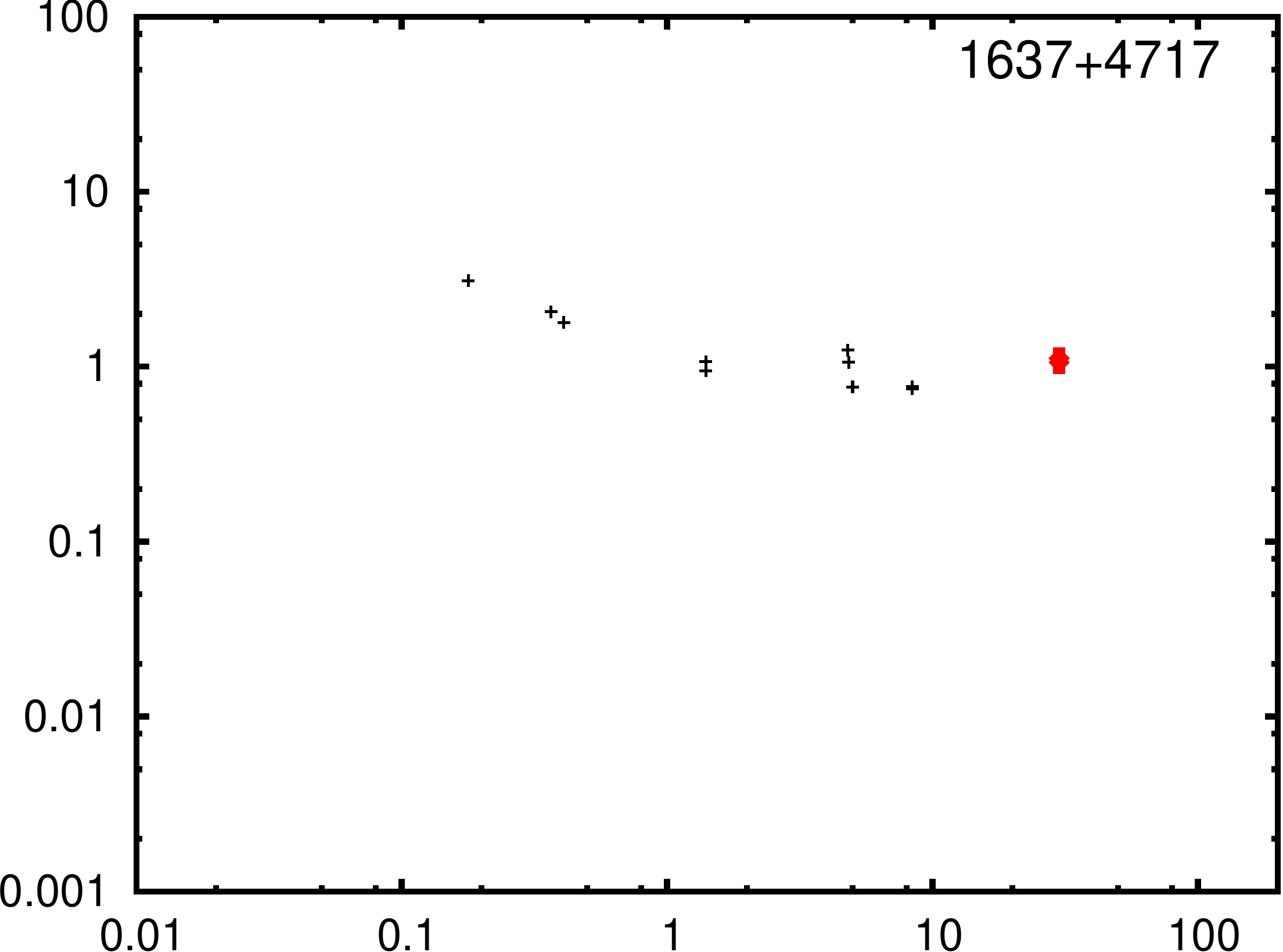}
\includegraphics[scale=0.2]{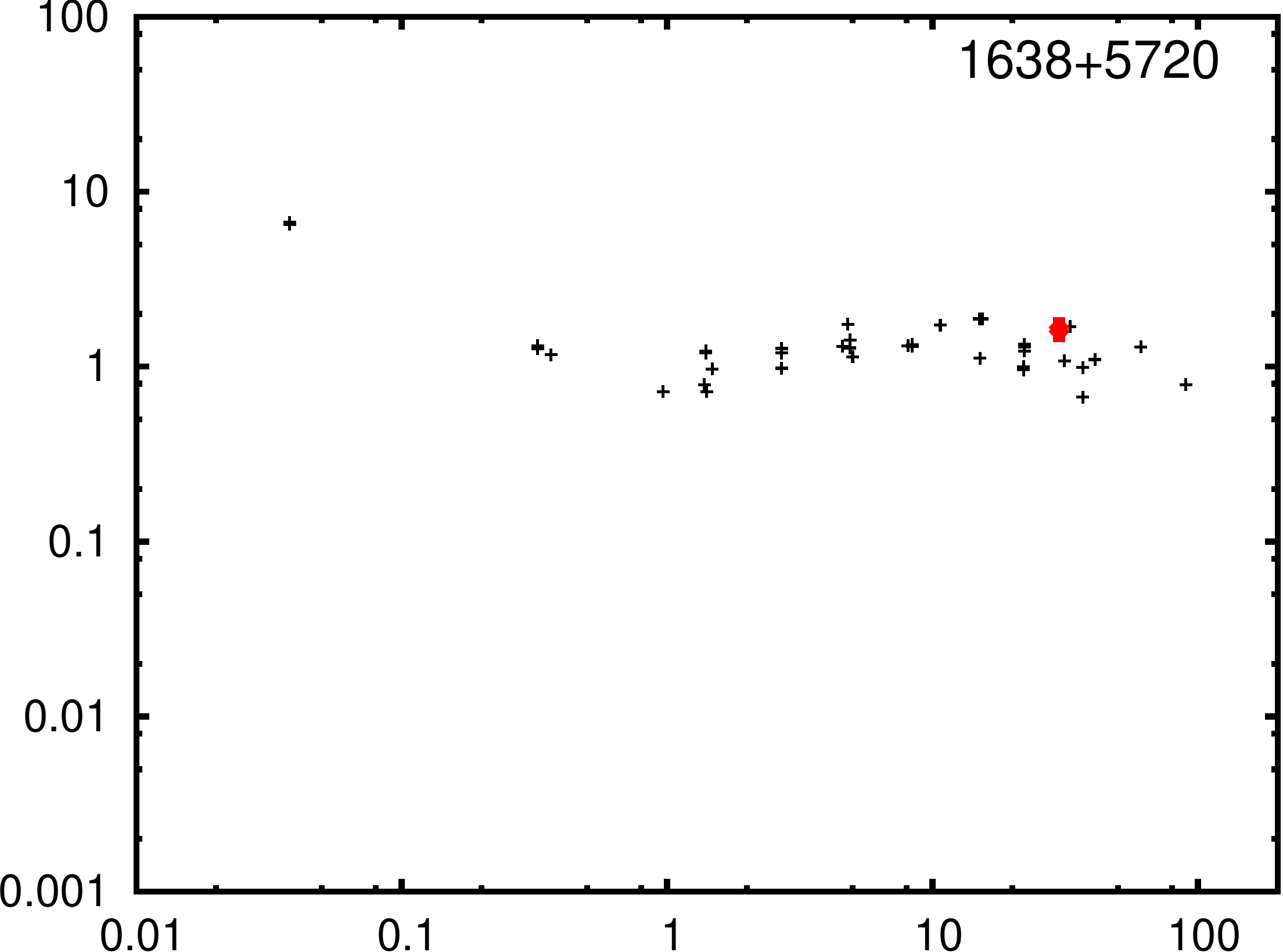}
\includegraphics[scale=0.2]{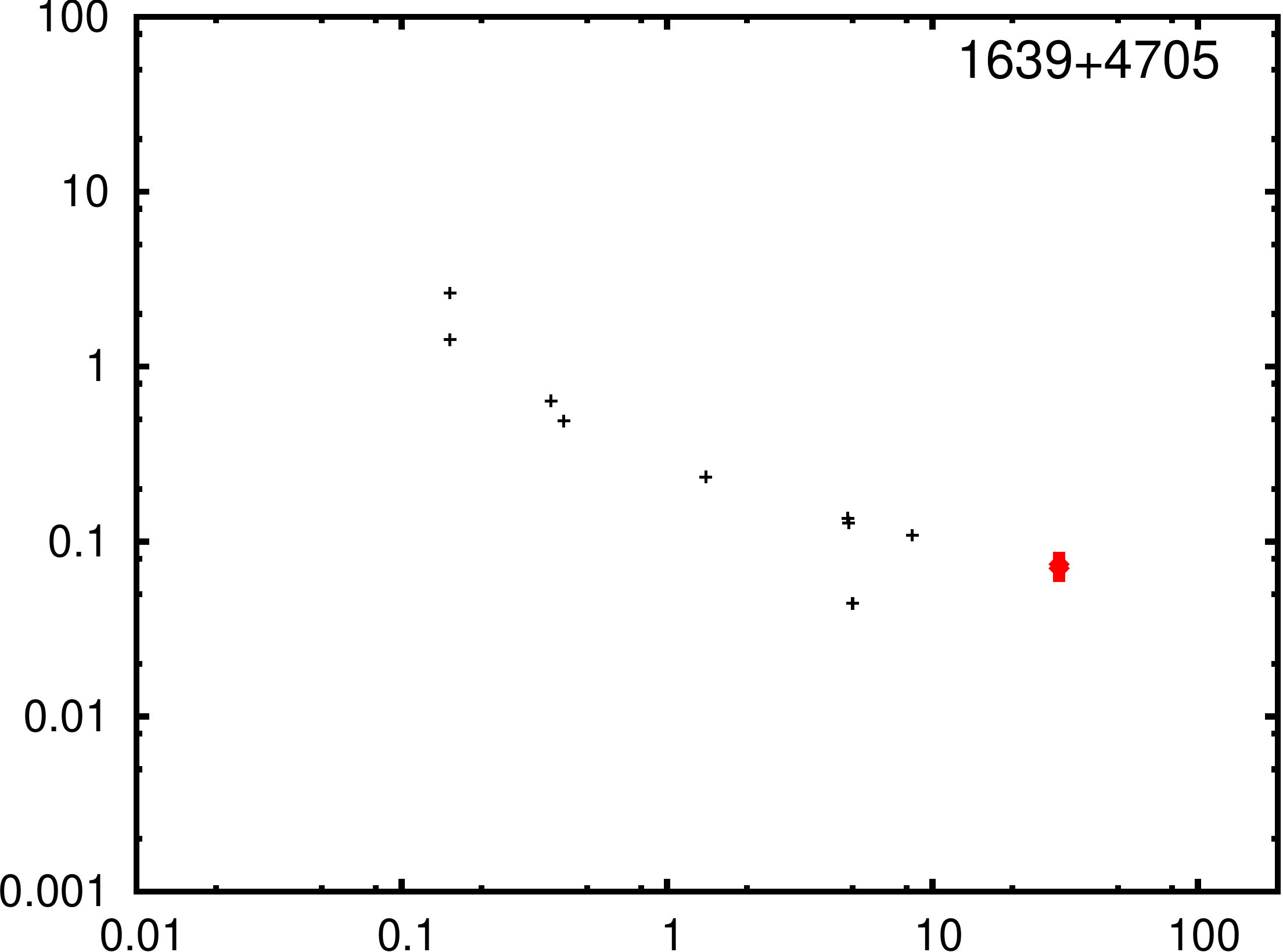}
\includegraphics[scale=0.2]{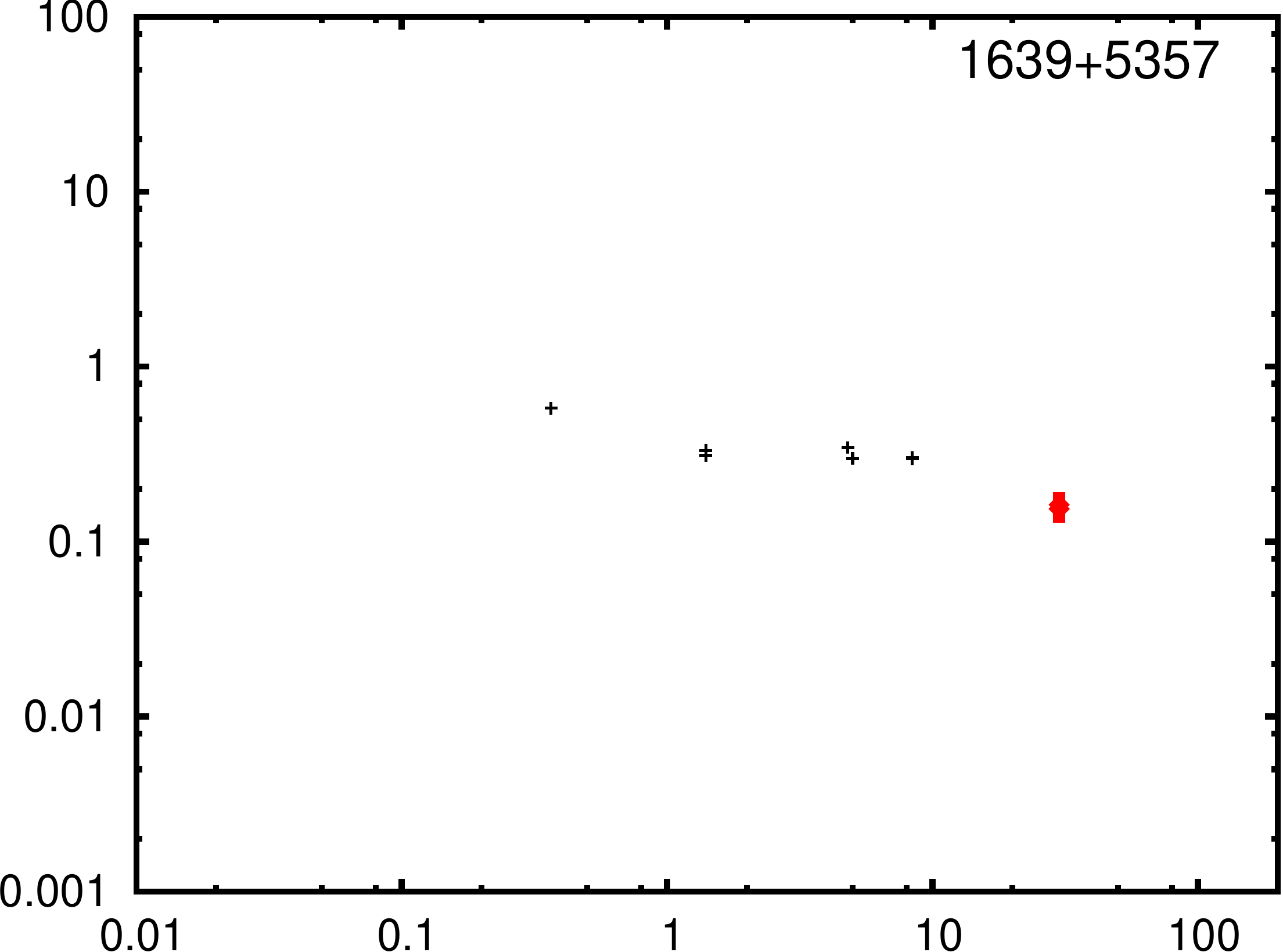}
\includegraphics[scale=0.2]{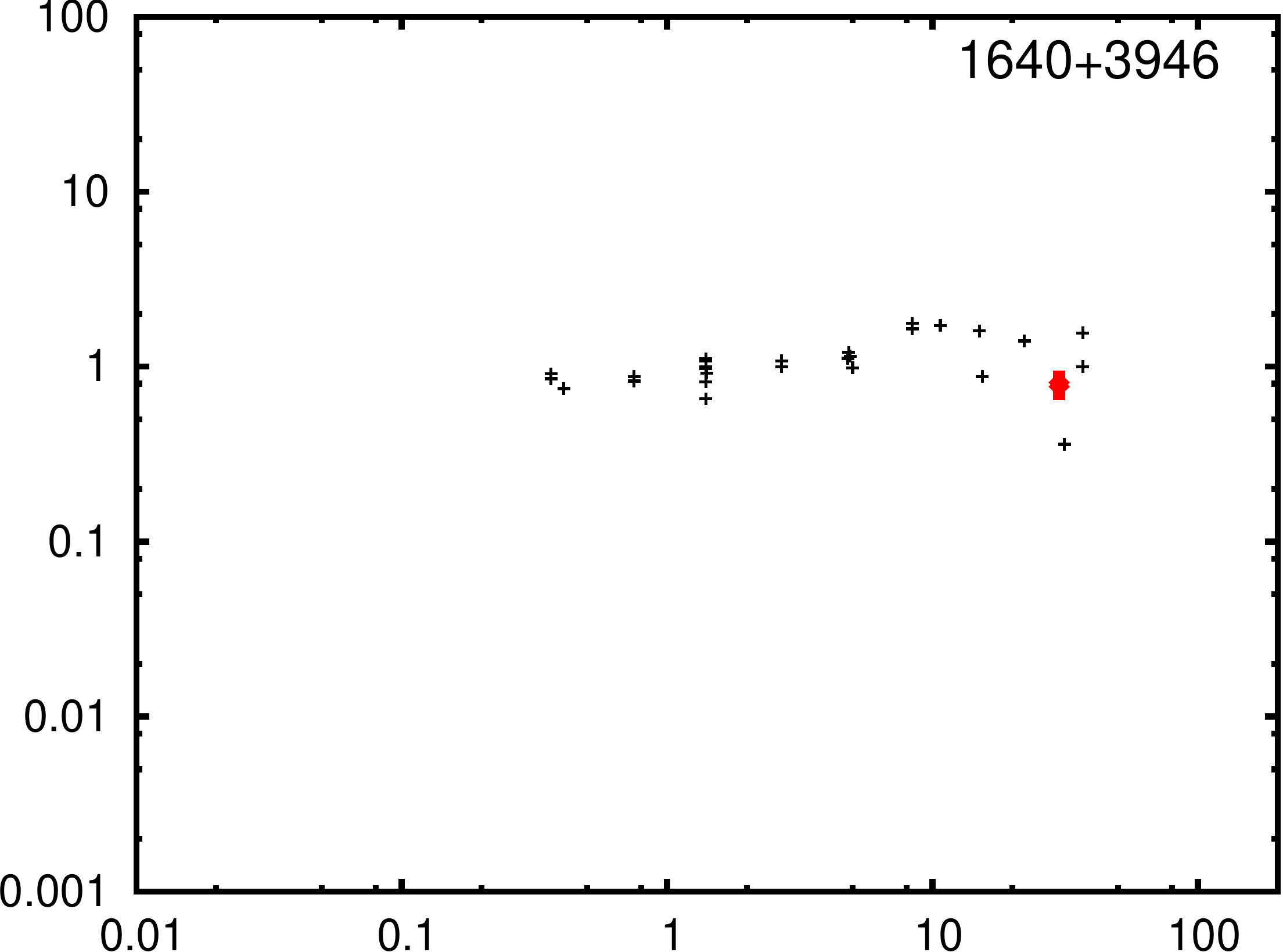}
\includegraphics[scale=0.2]{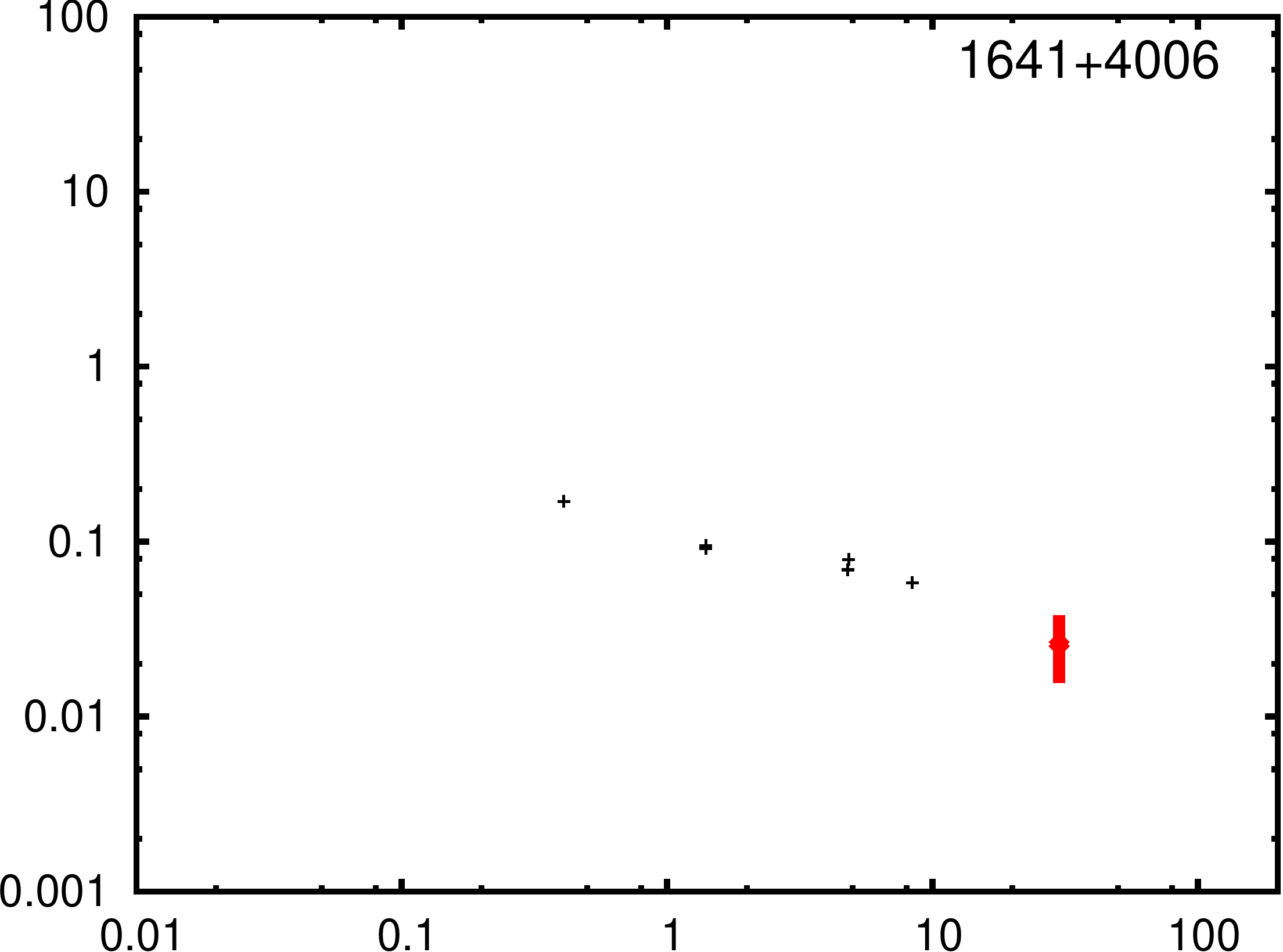}
\includegraphics[scale=0.2]{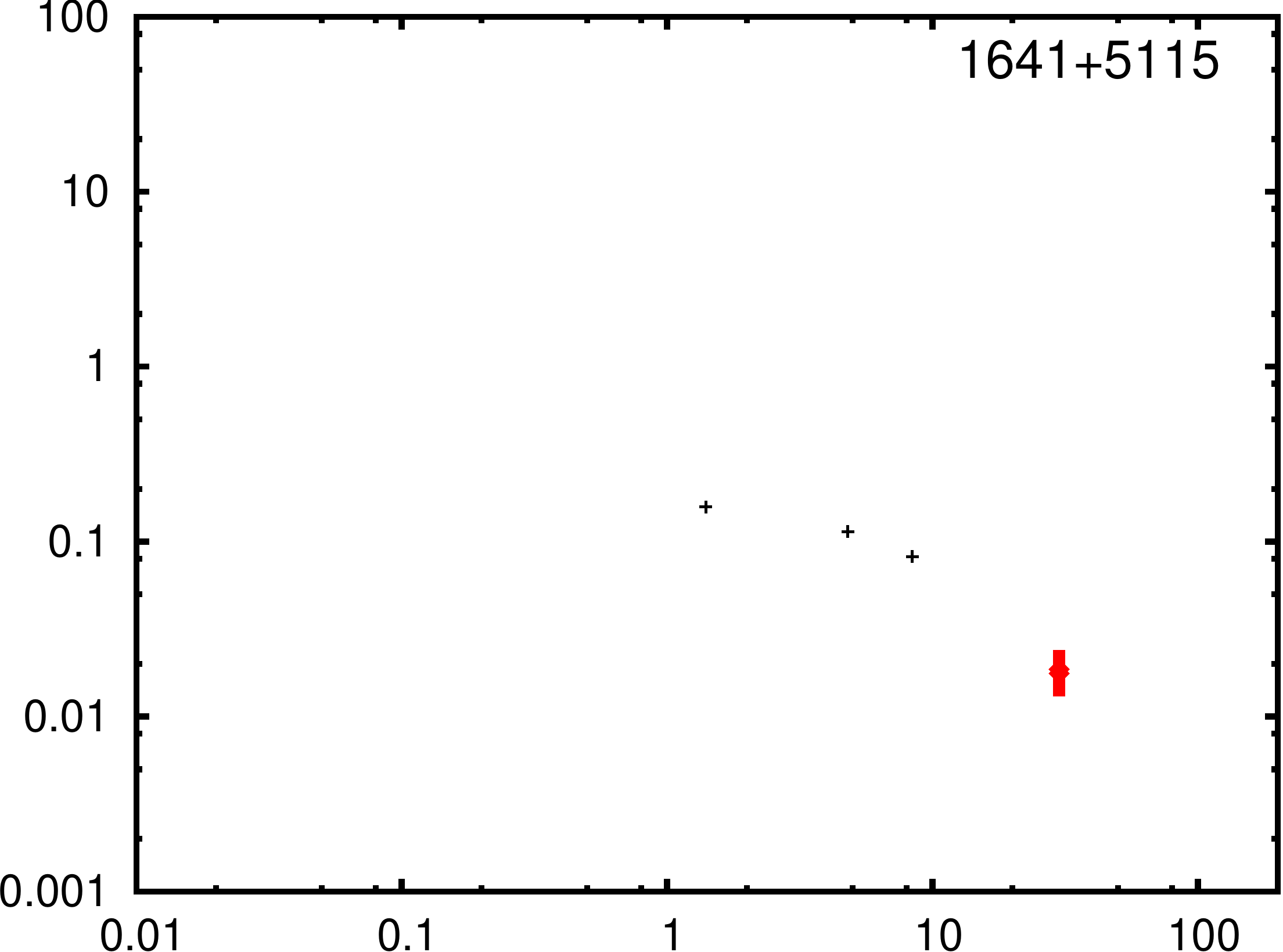}
\includegraphics[scale=0.2]{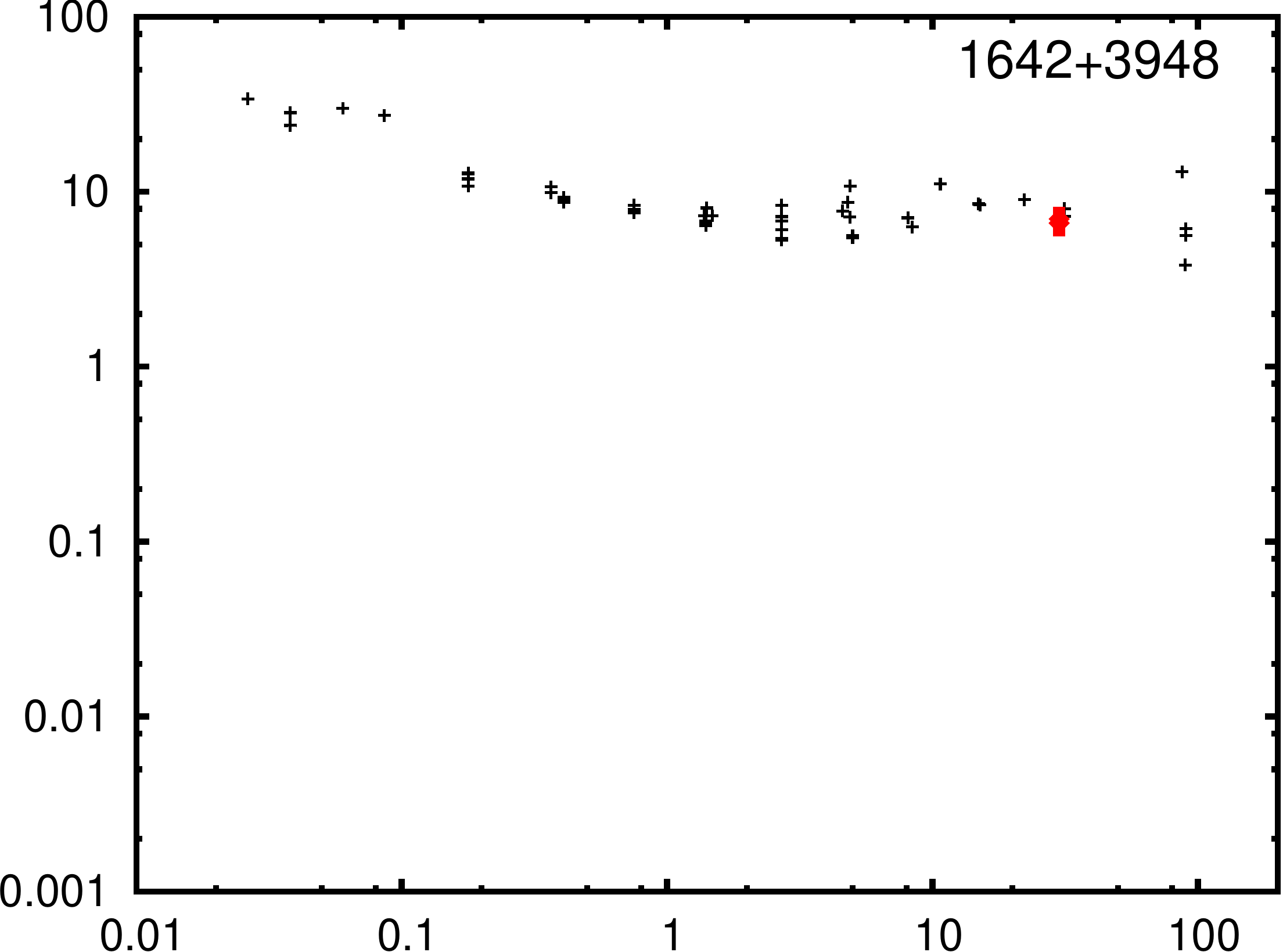}
\includegraphics[scale=0.2]{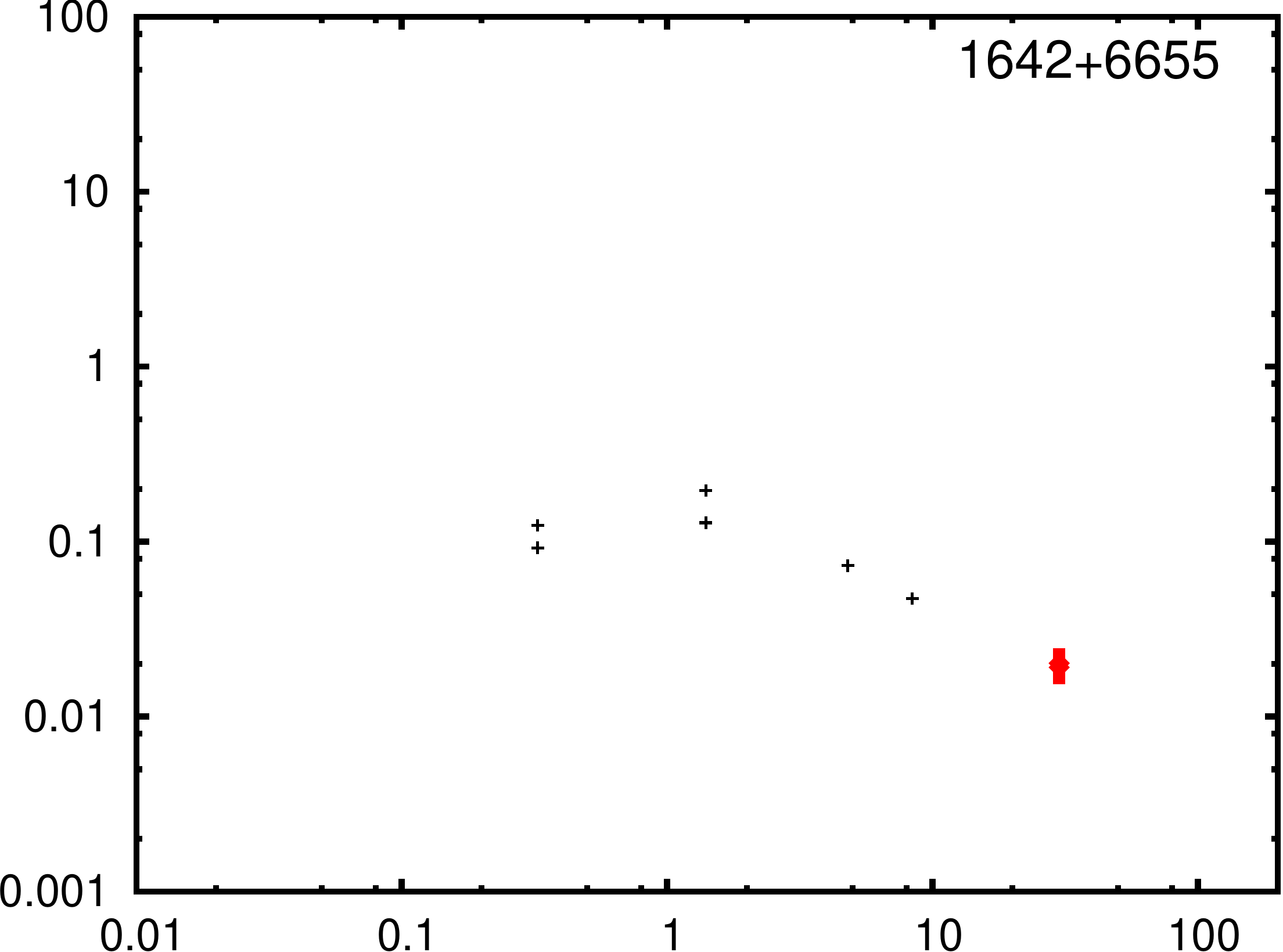}
\includegraphics[scale=0.2]{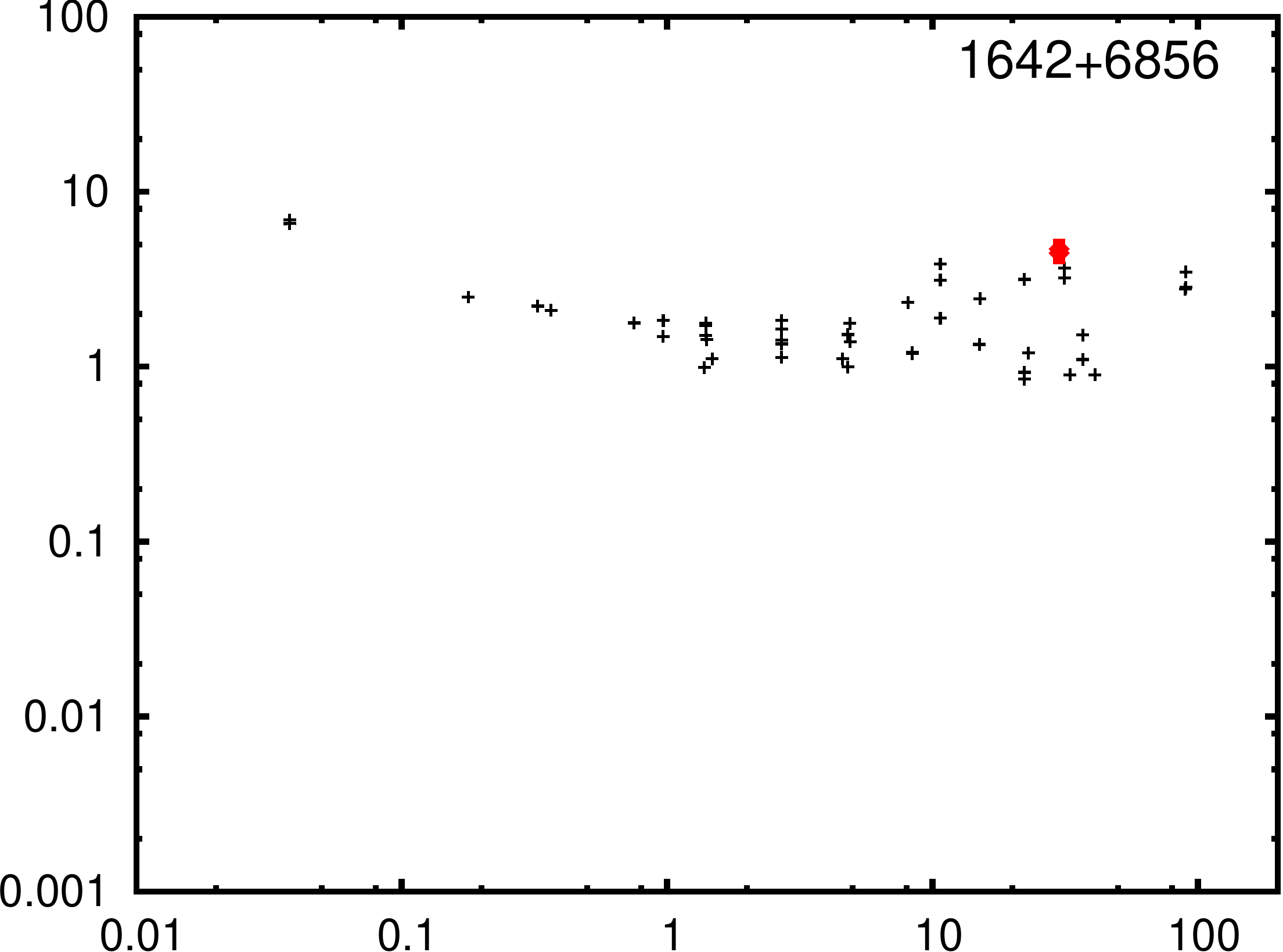}
\end{figure}
\clearpage\begin{figure}
\centering
\includegraphics[scale=0.2]{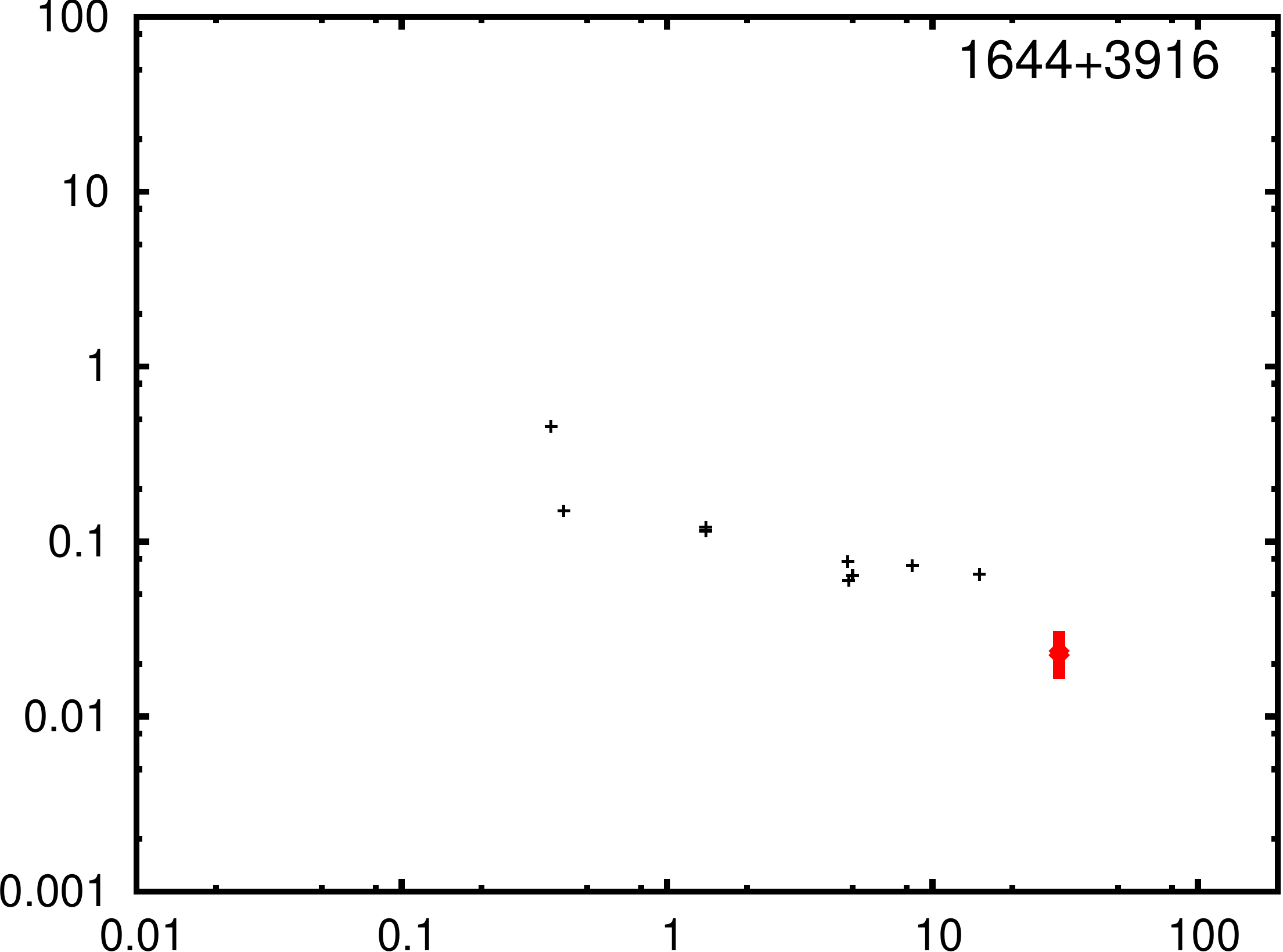}
\includegraphics[scale=0.2]{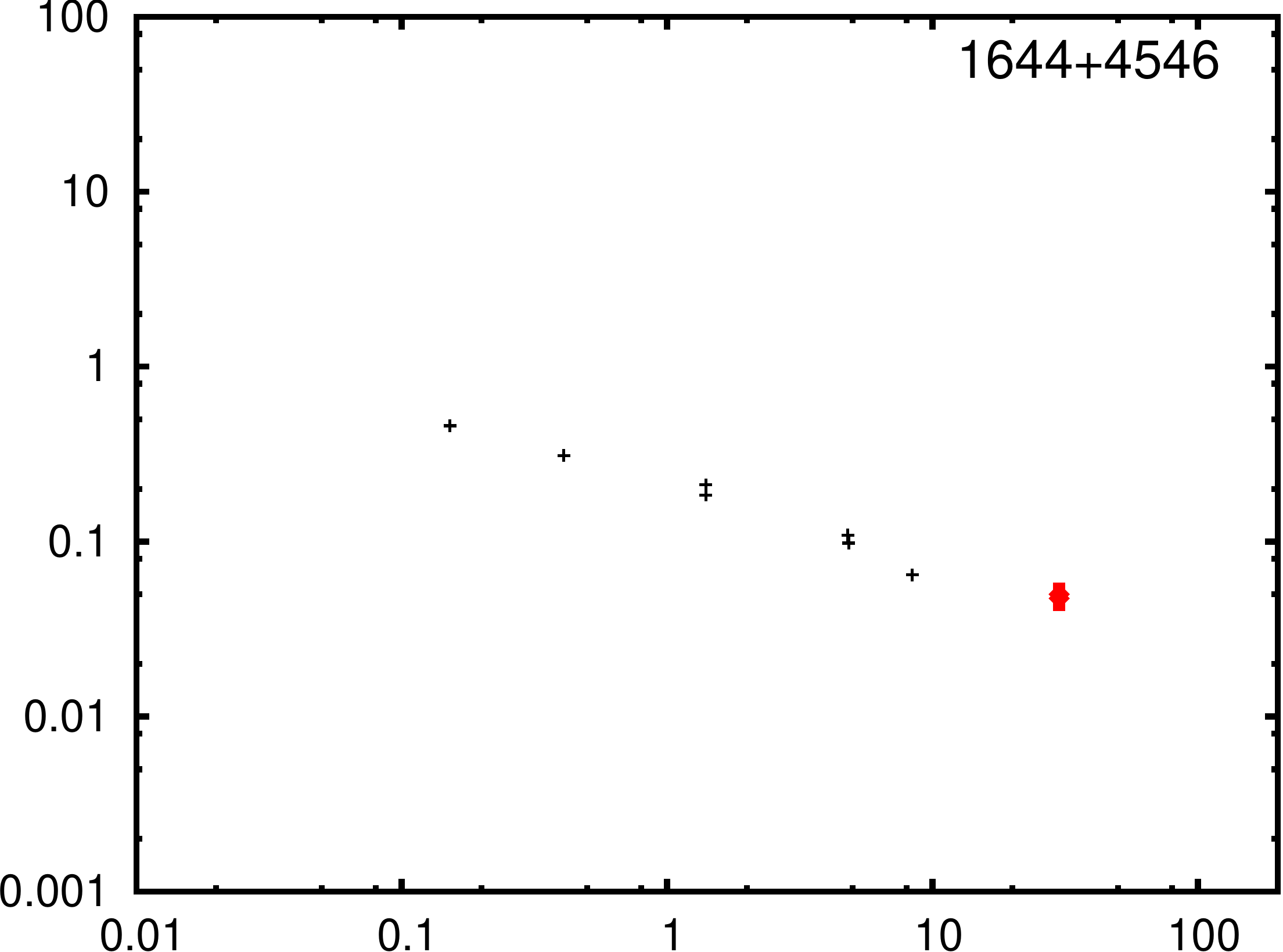}
\includegraphics[scale=0.2]{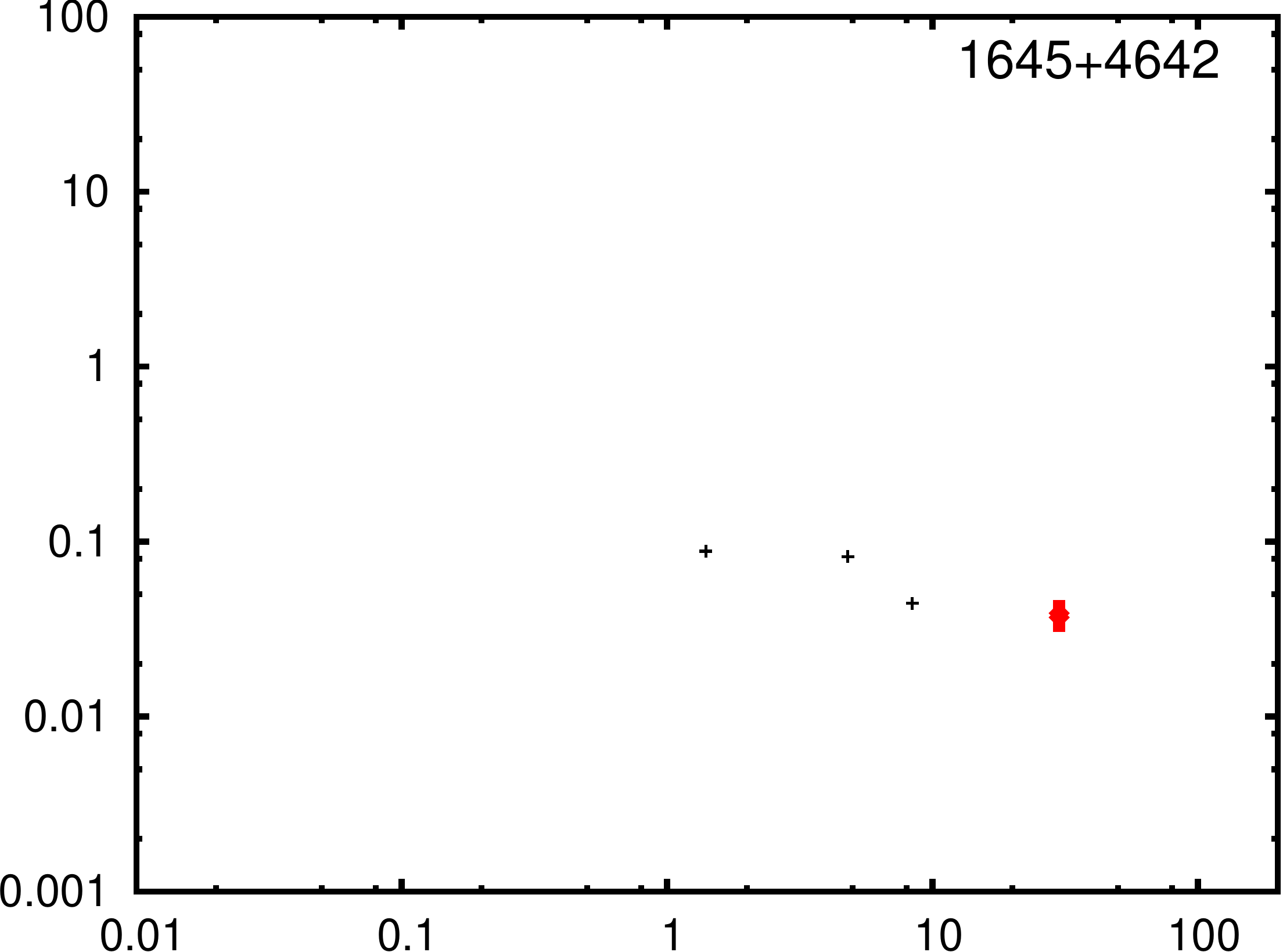}
\includegraphics[scale=0.2]{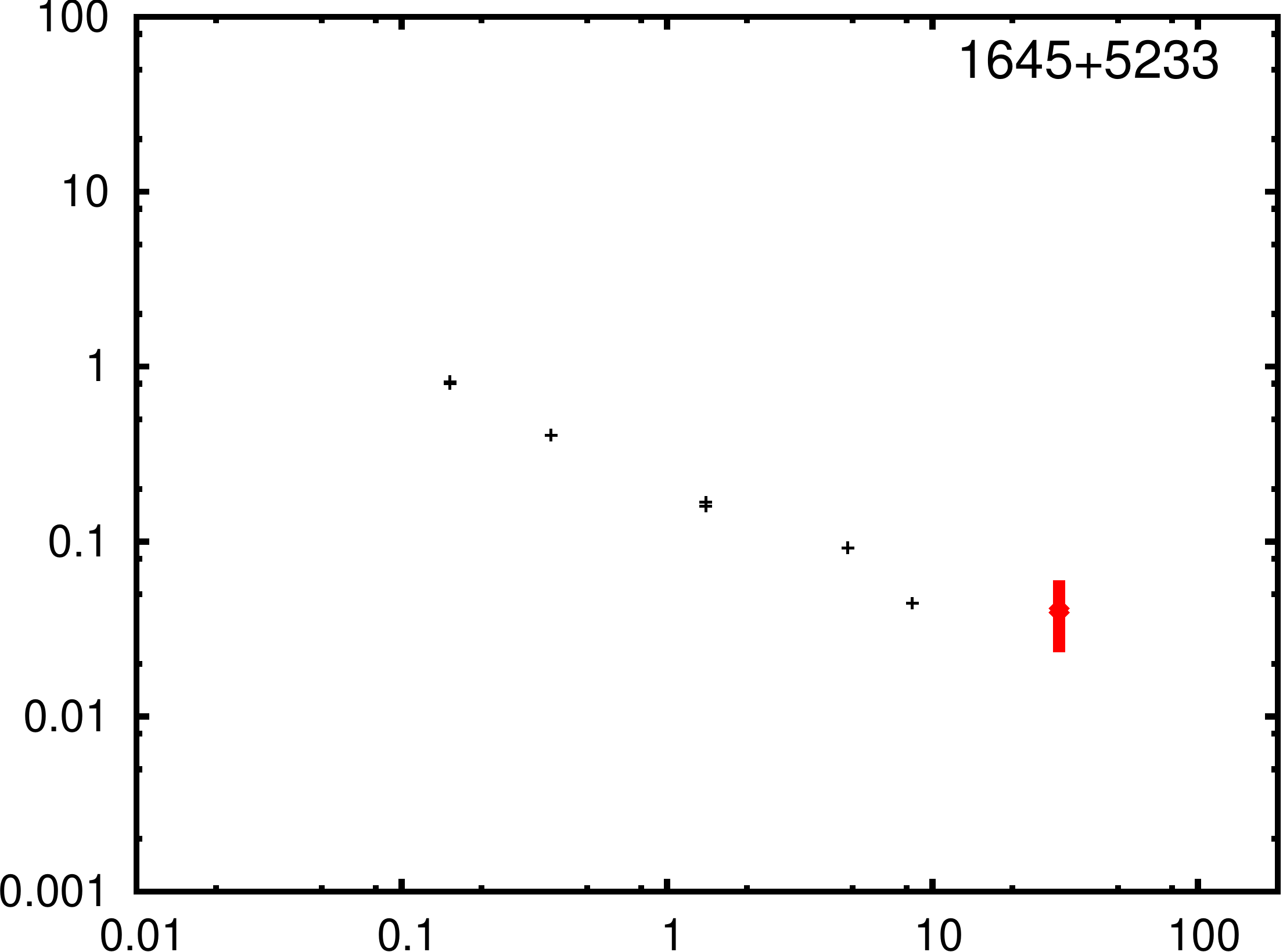}
\includegraphics[scale=0.2]{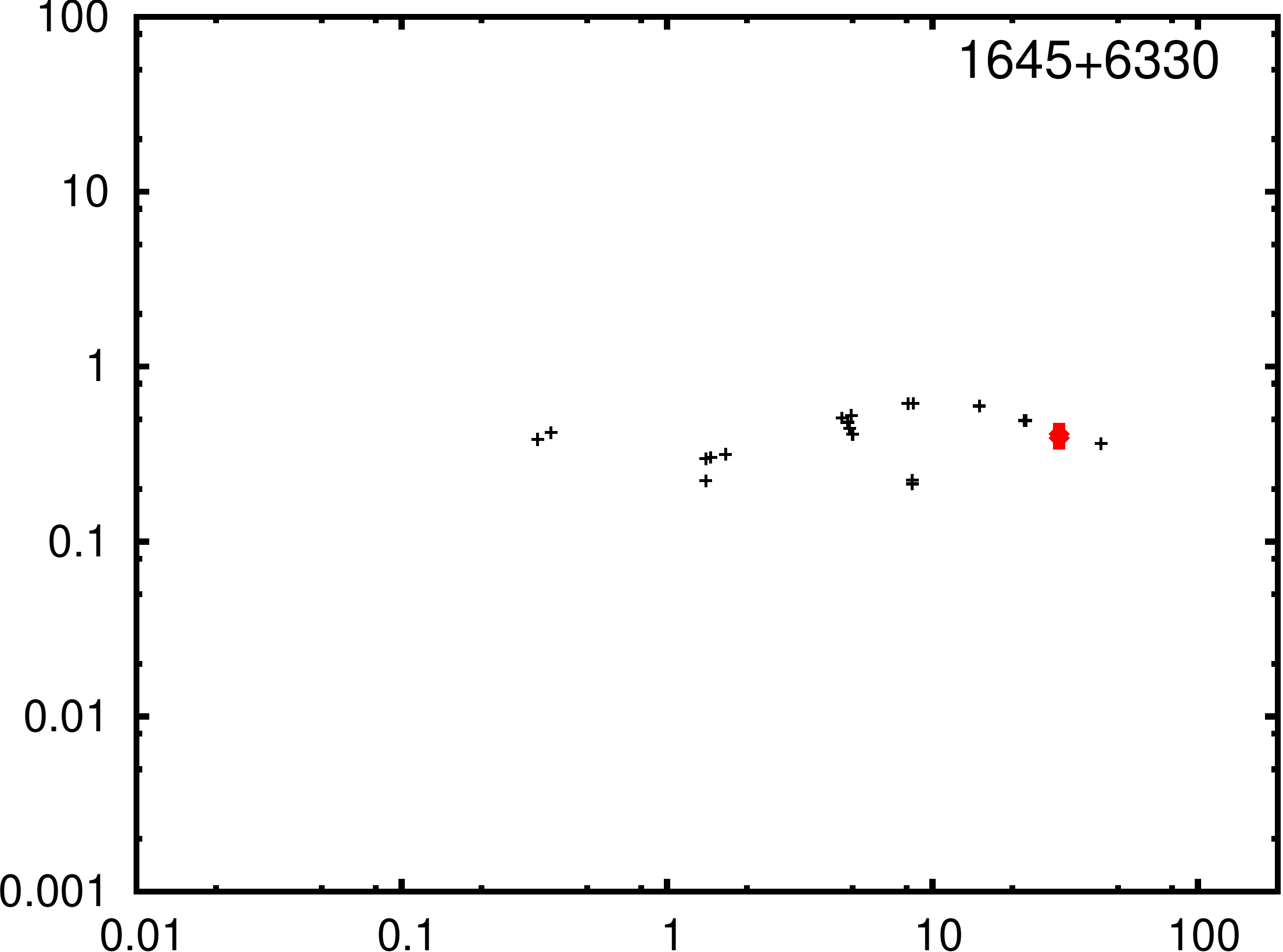}
\includegraphics[scale=0.2]{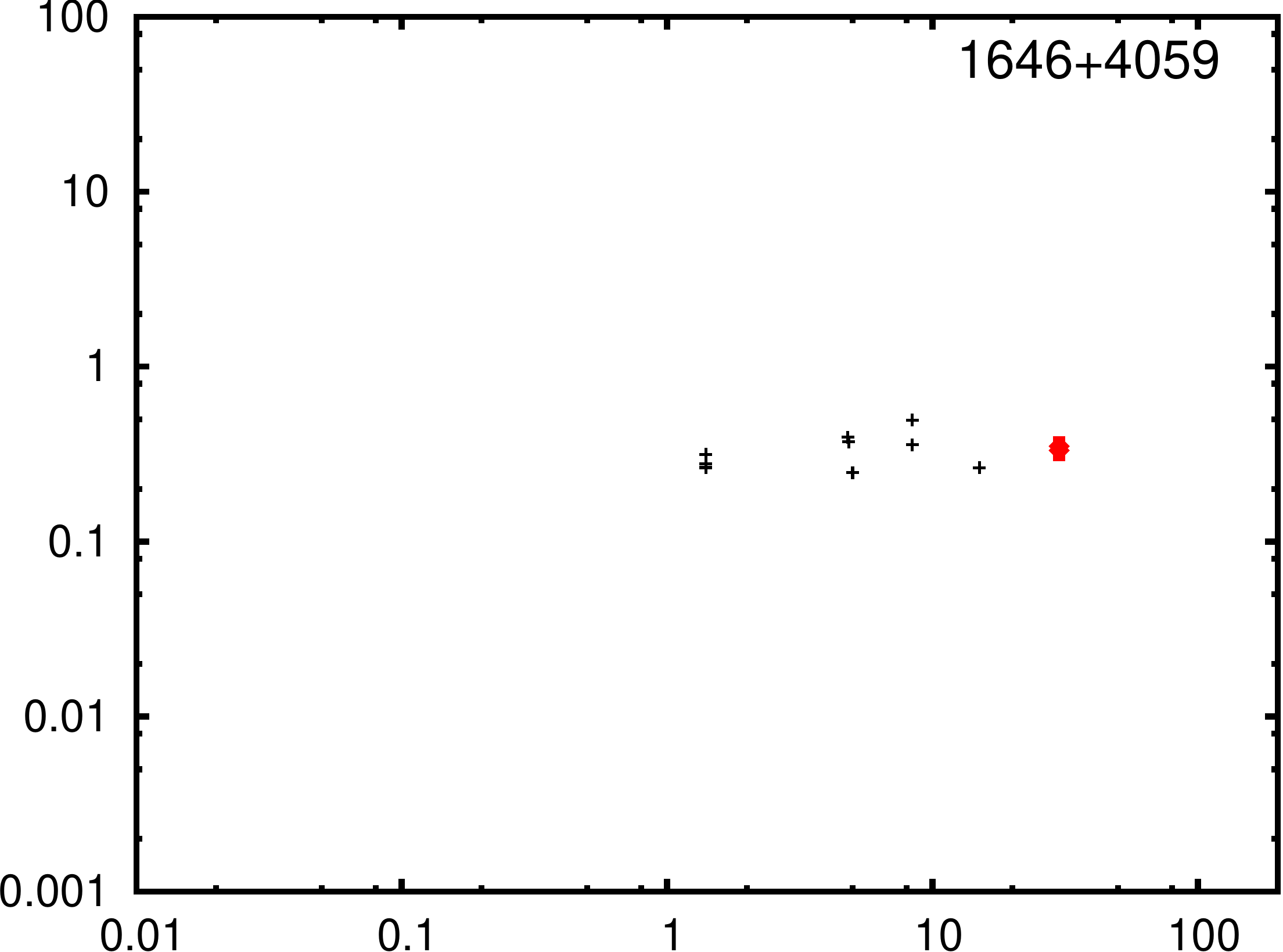}
\includegraphics[scale=0.2]{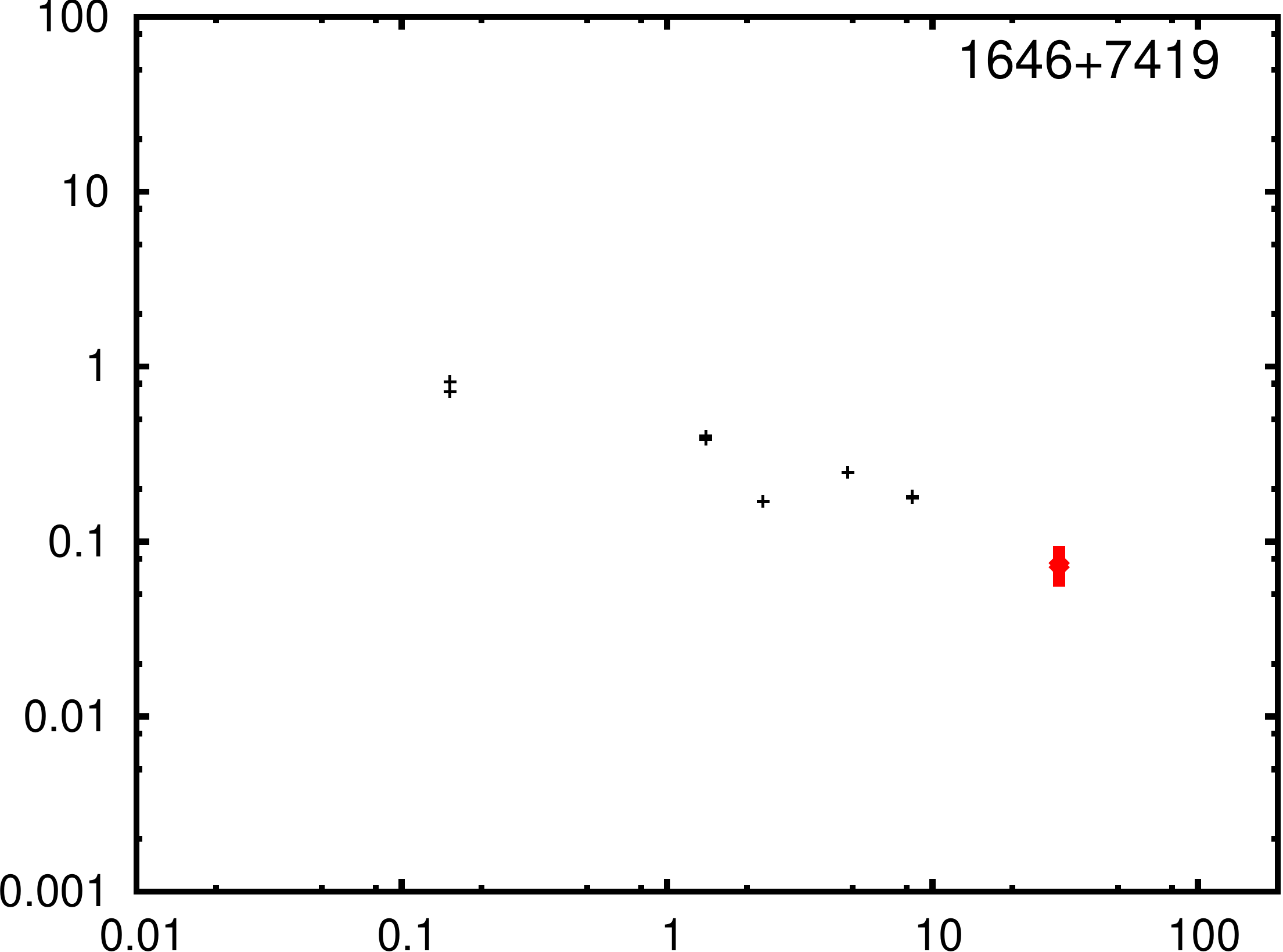}
\includegraphics[scale=0.2]{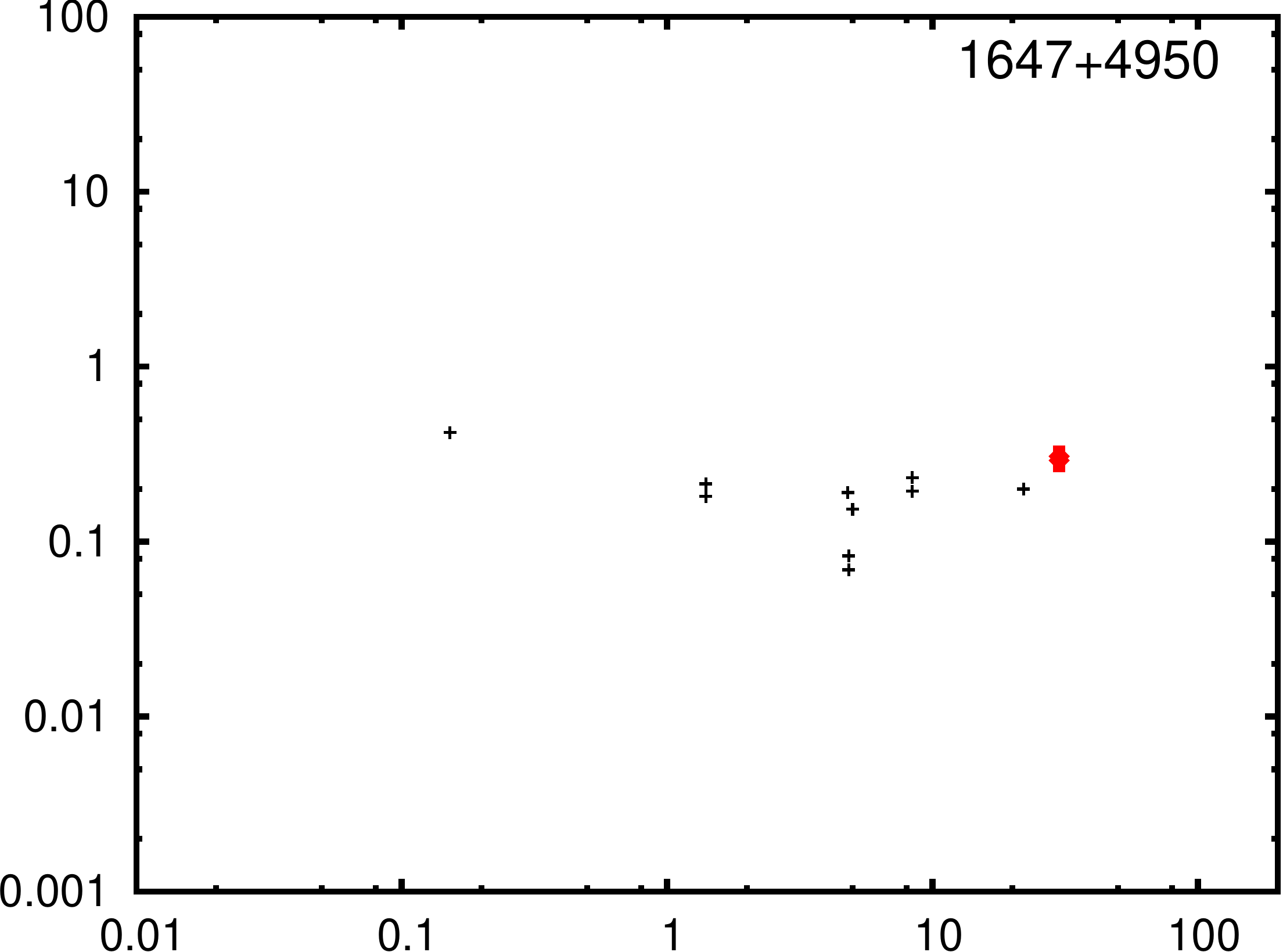}
\includegraphics[scale=0.2]{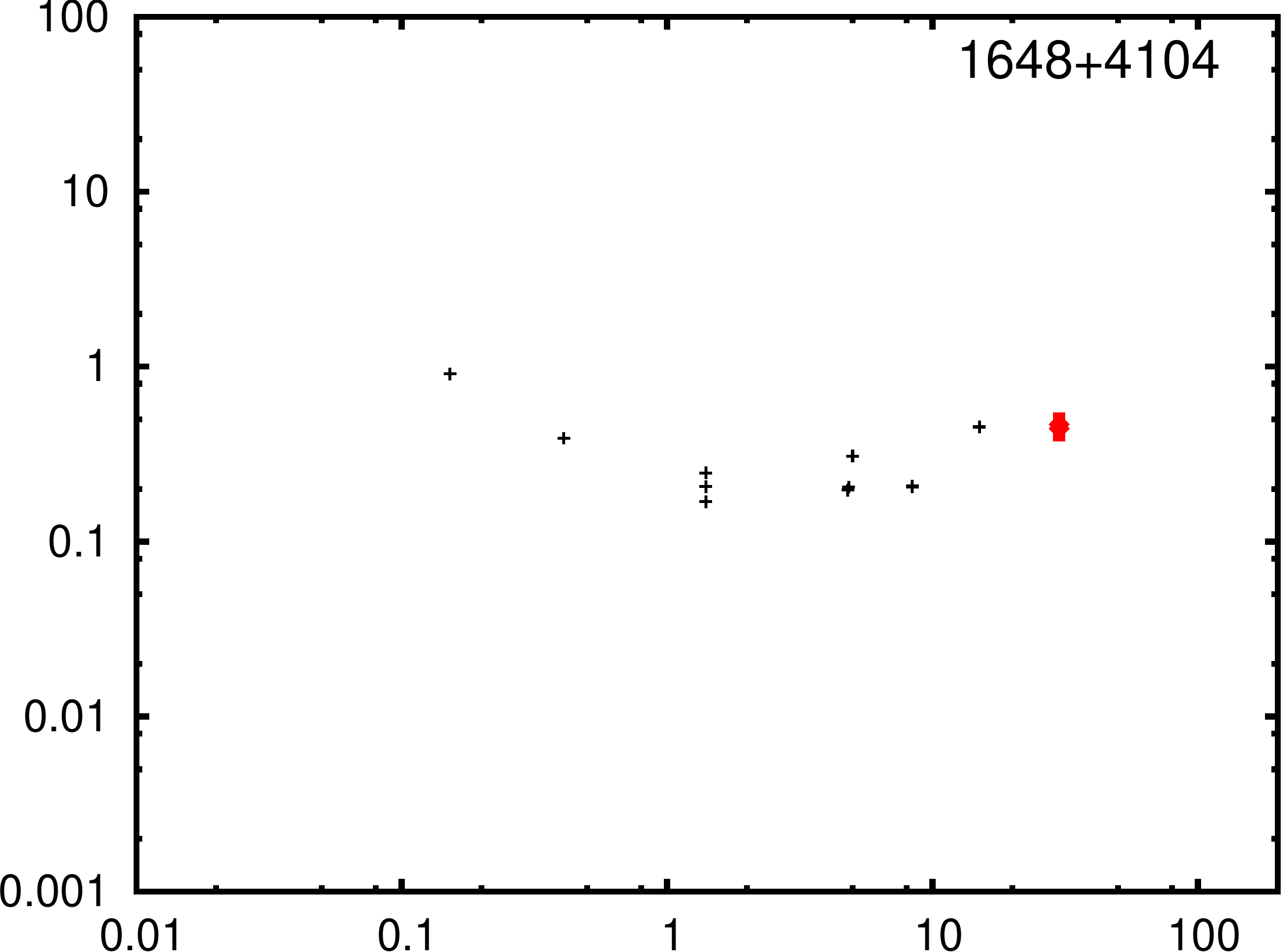}
\includegraphics[scale=0.2]{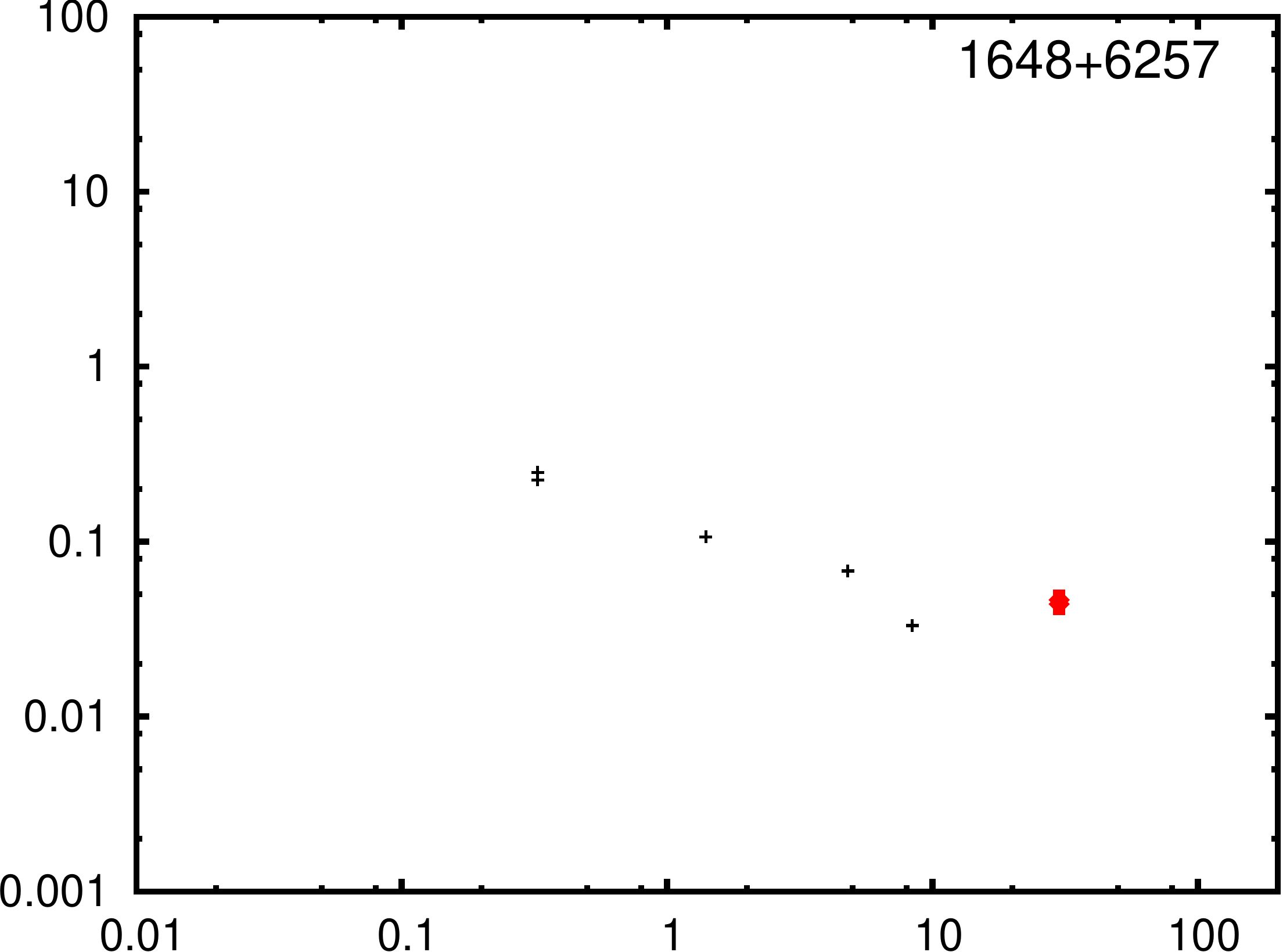}
\includegraphics[scale=0.2]{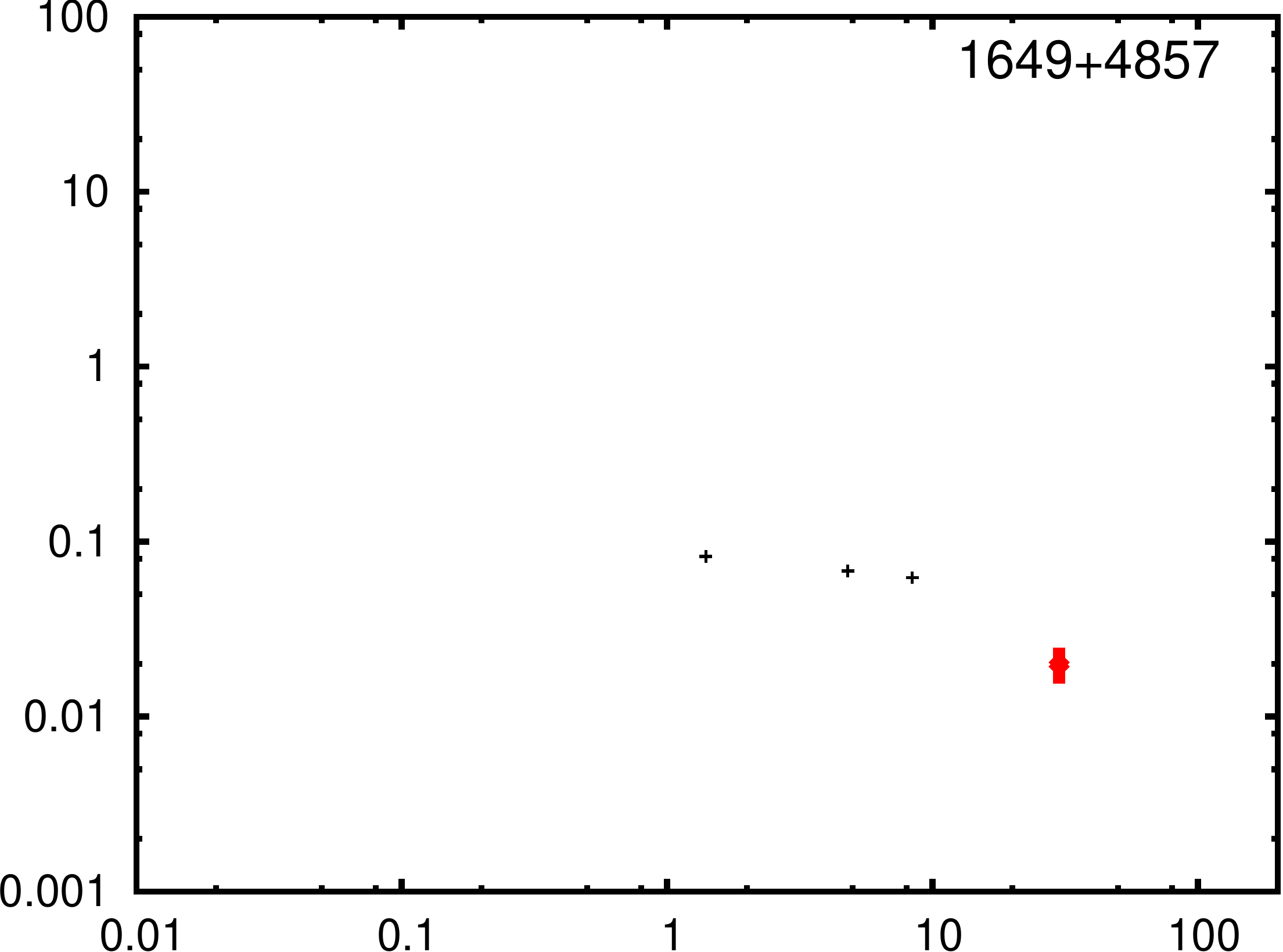}
\includegraphics[scale=0.2]{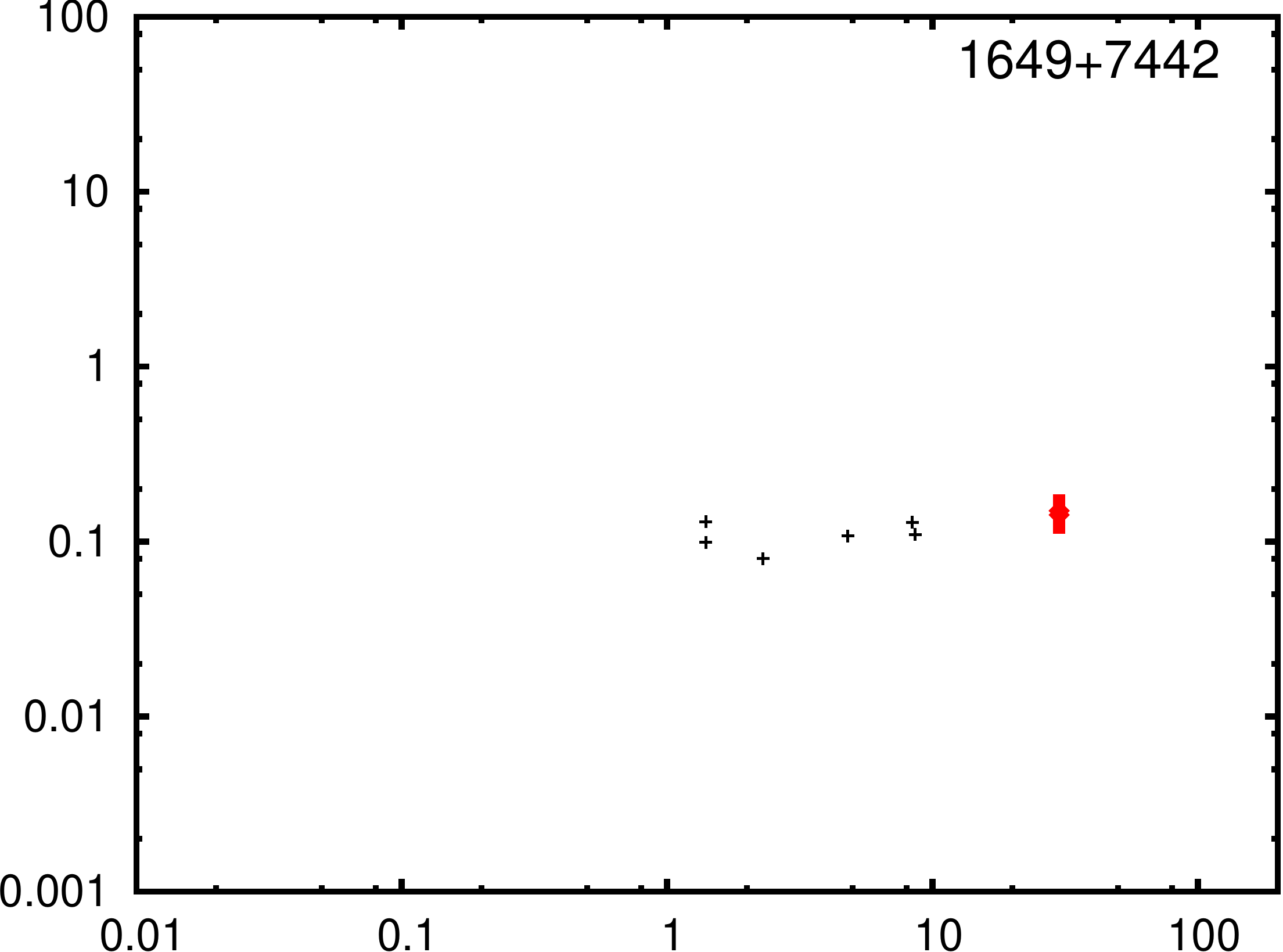}
\includegraphics[scale=0.2]{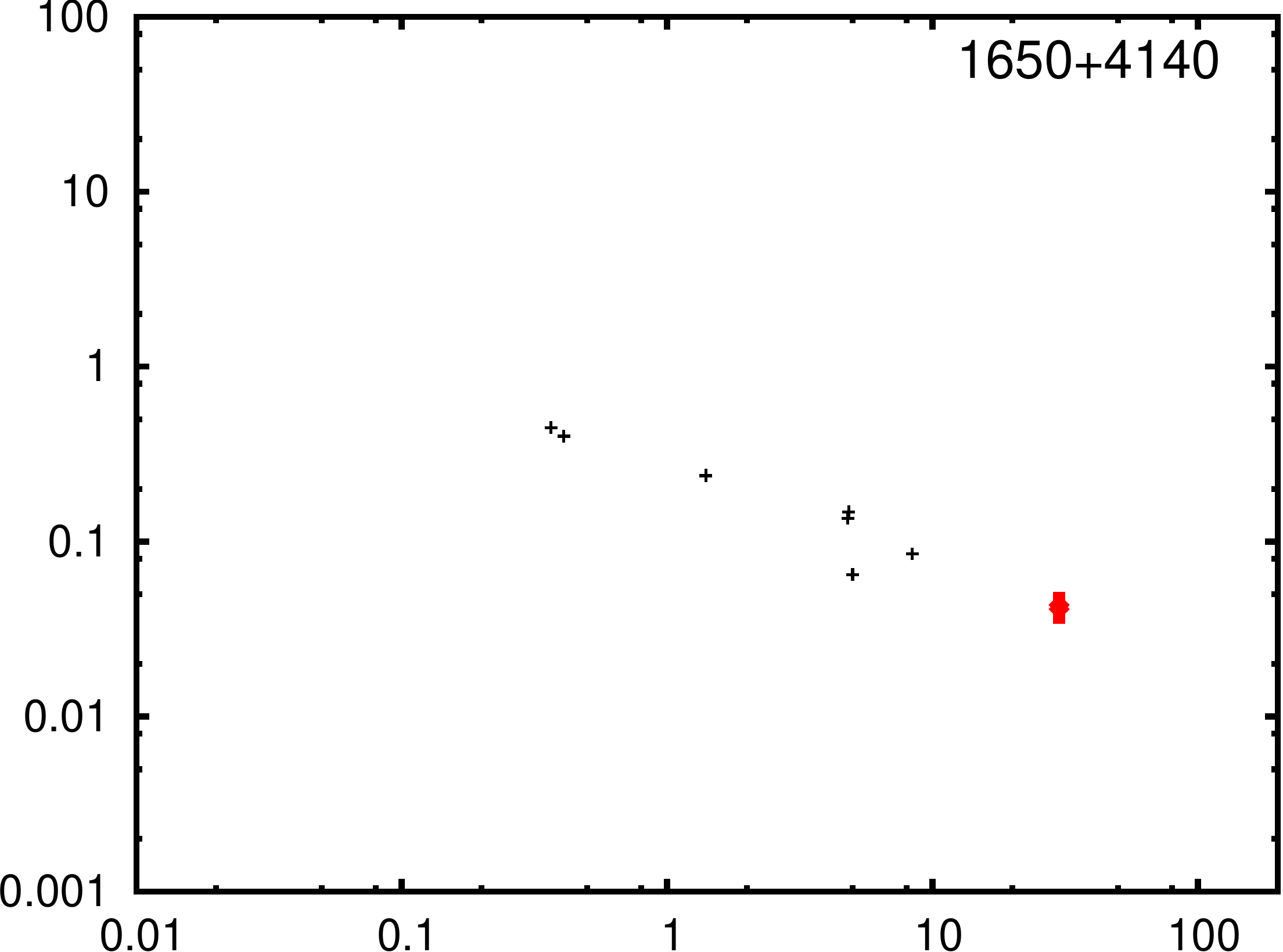}
\includegraphics[scale=0.2]{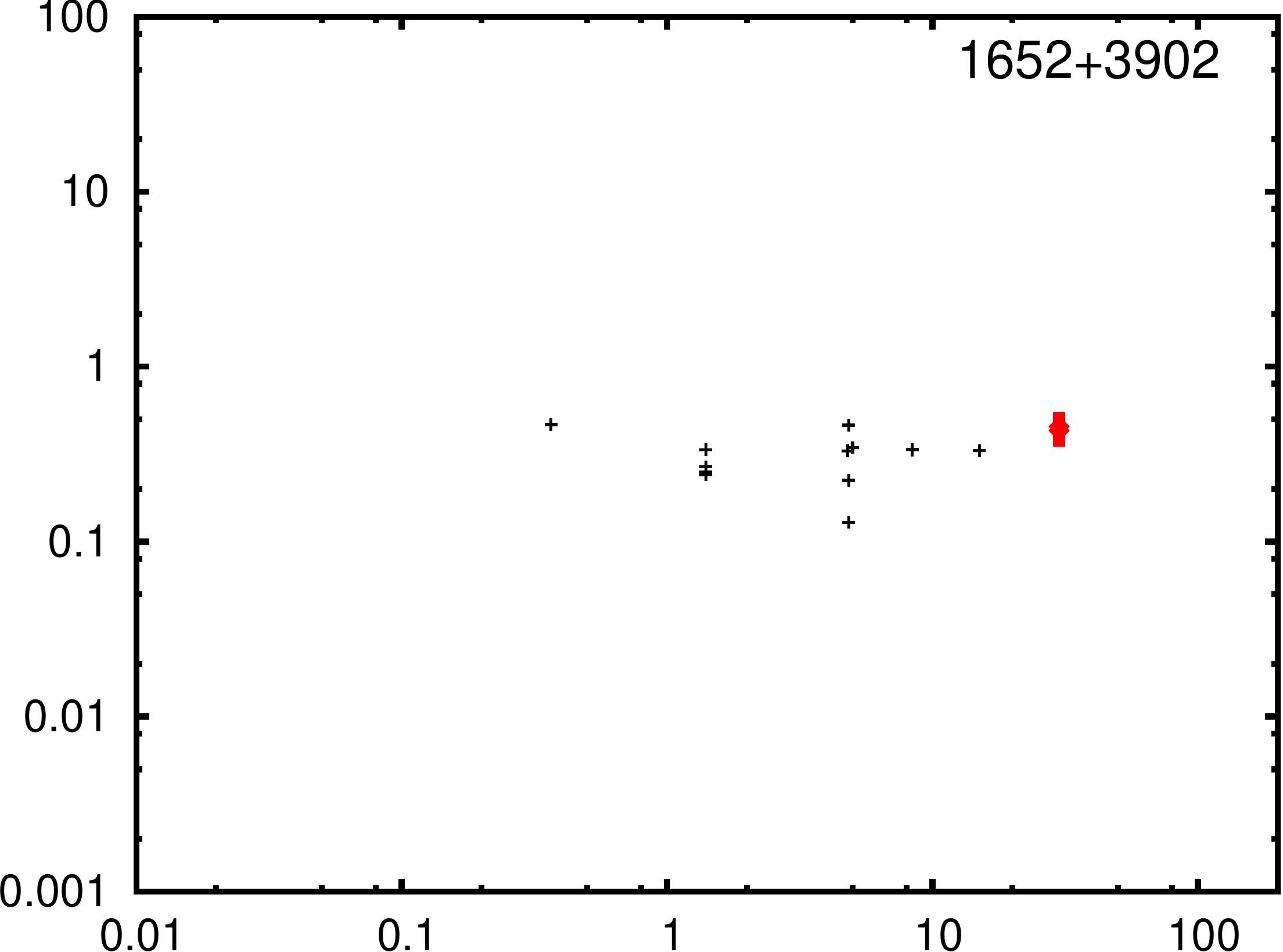}
\includegraphics[scale=0.2]{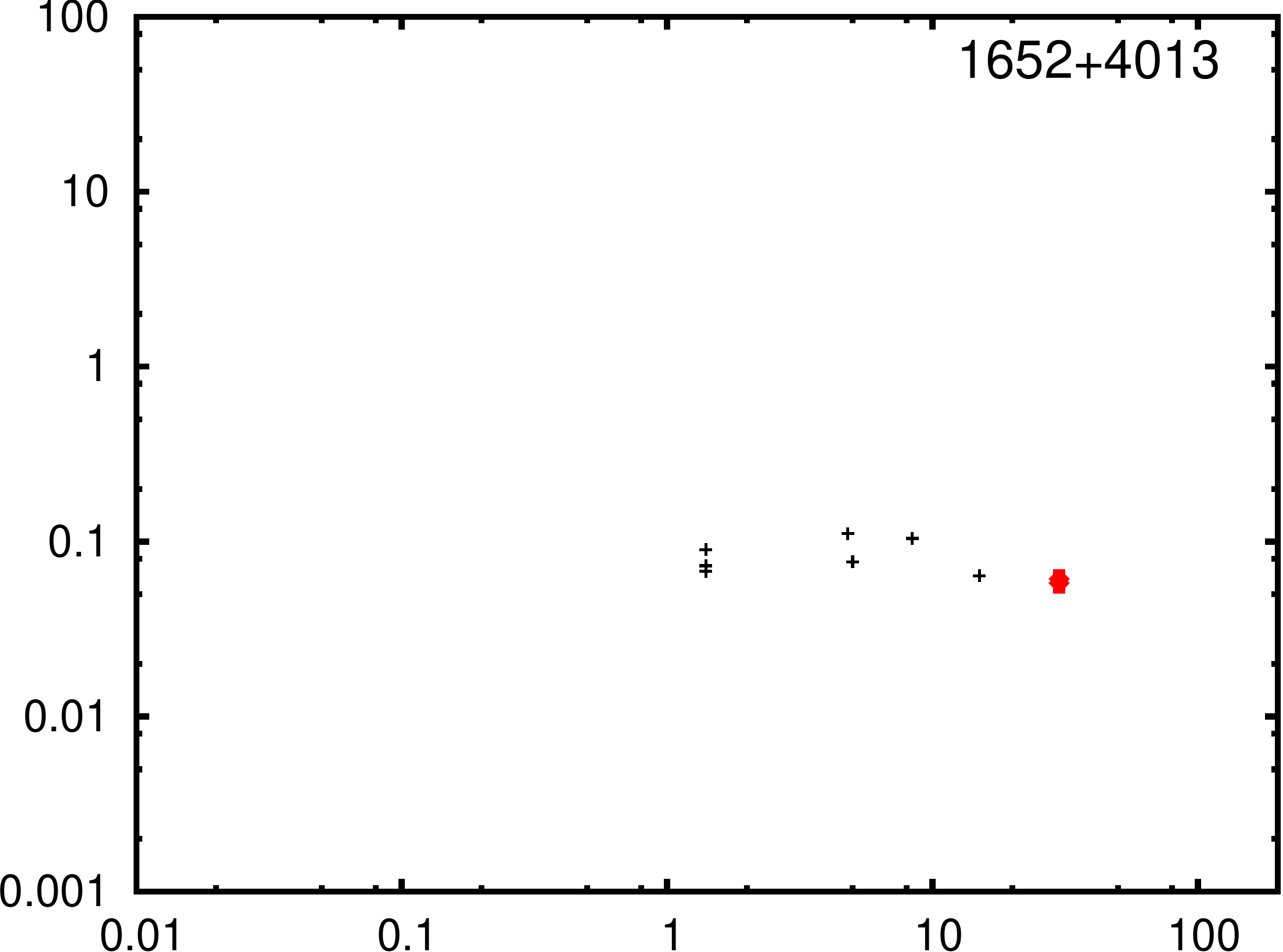}
\includegraphics[scale=0.2]{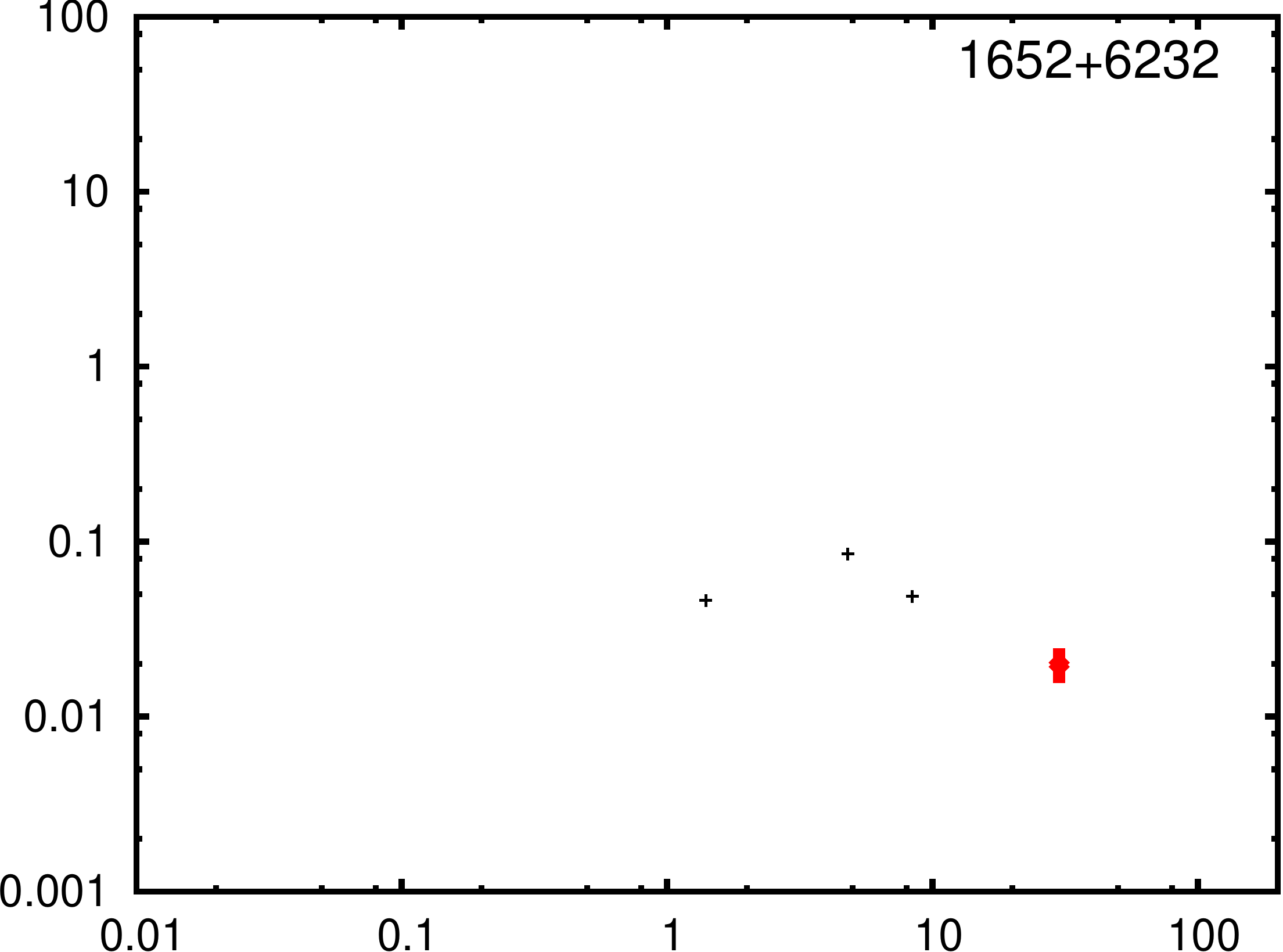}
\includegraphics[scale=0.2]{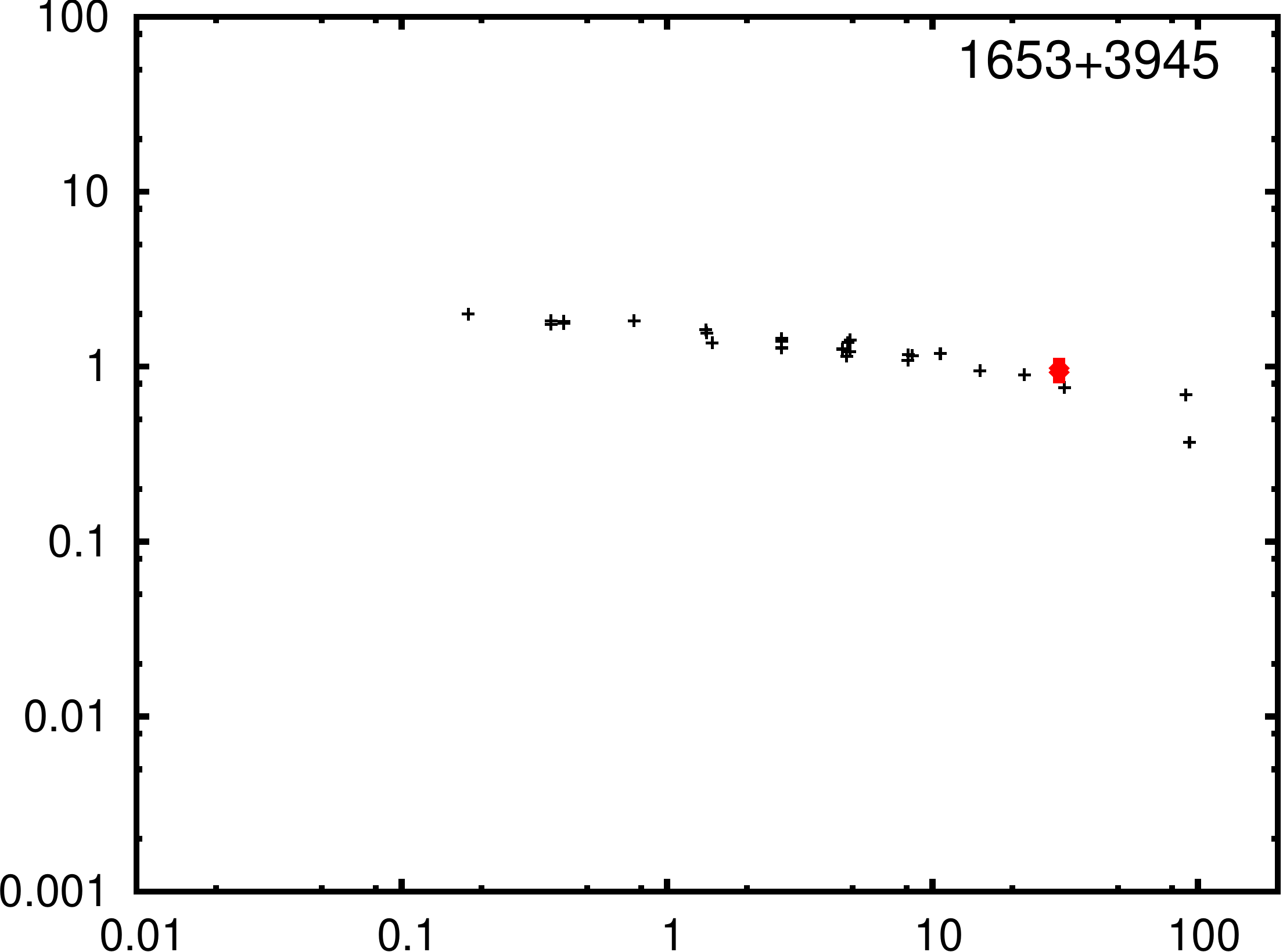}
\includegraphics[scale=0.2]{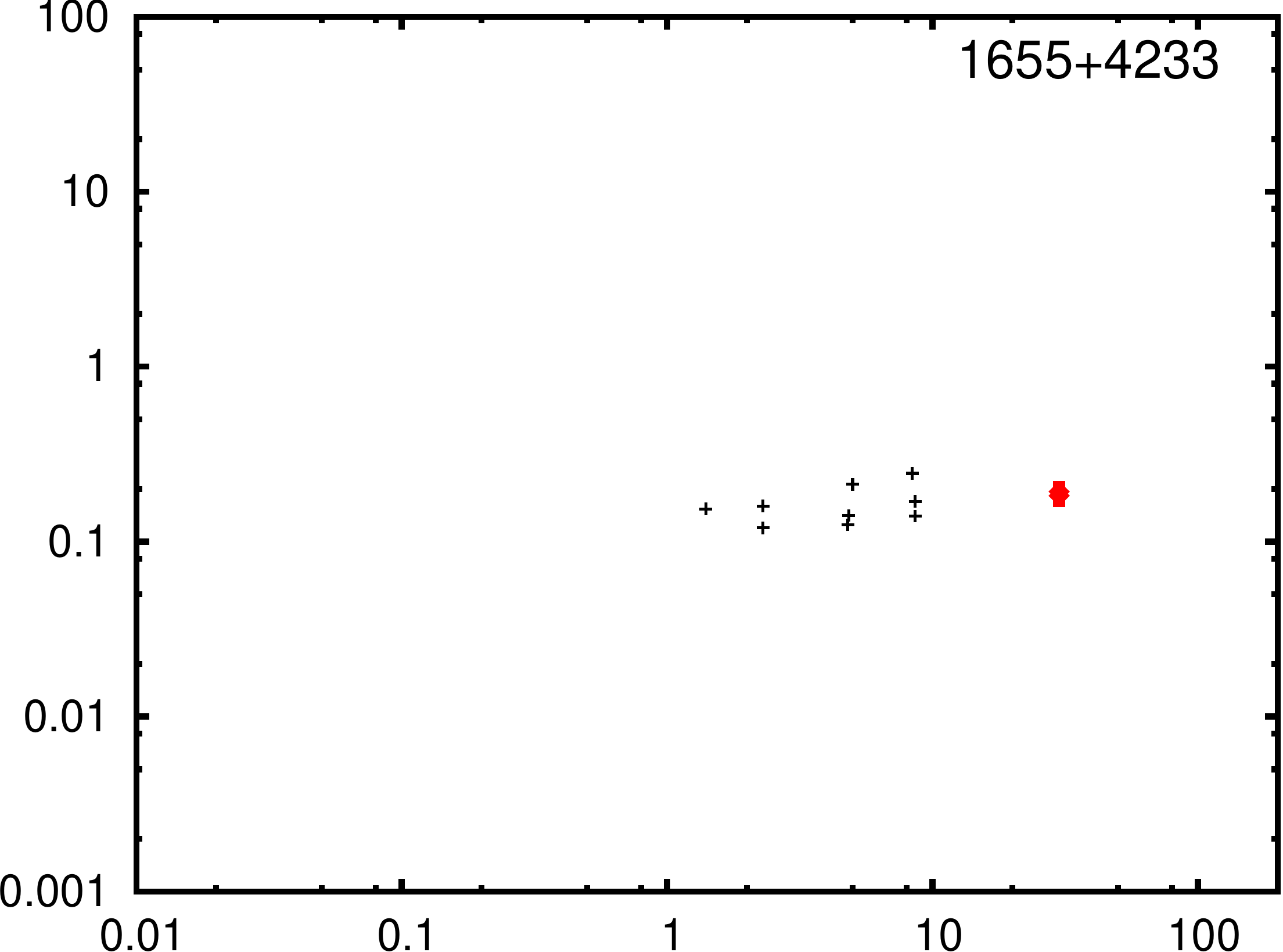}
\end{figure}
\clearpage\begin{figure}
\centering
\includegraphics[scale=0.2]{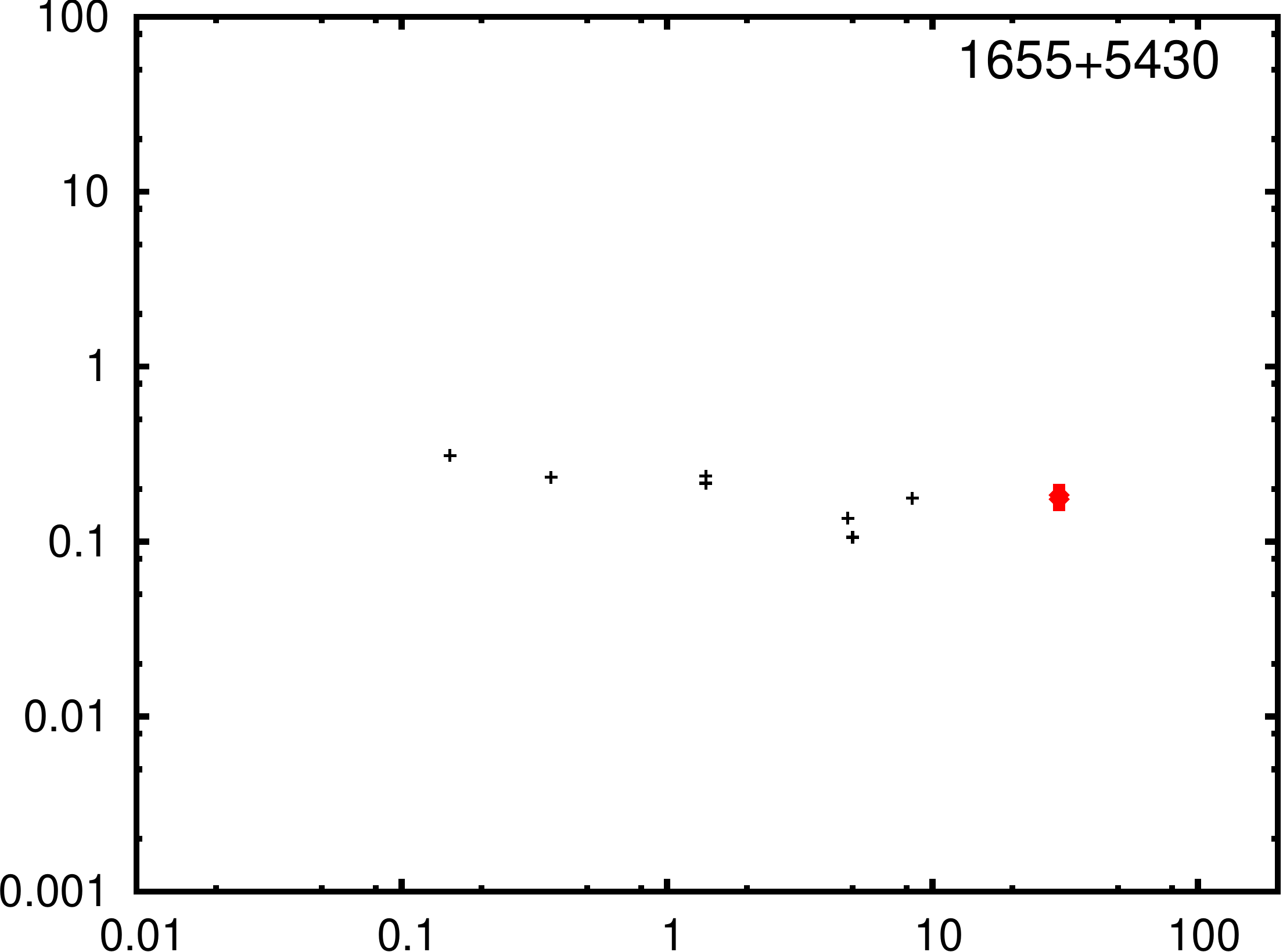}
\includegraphics[scale=0.2]{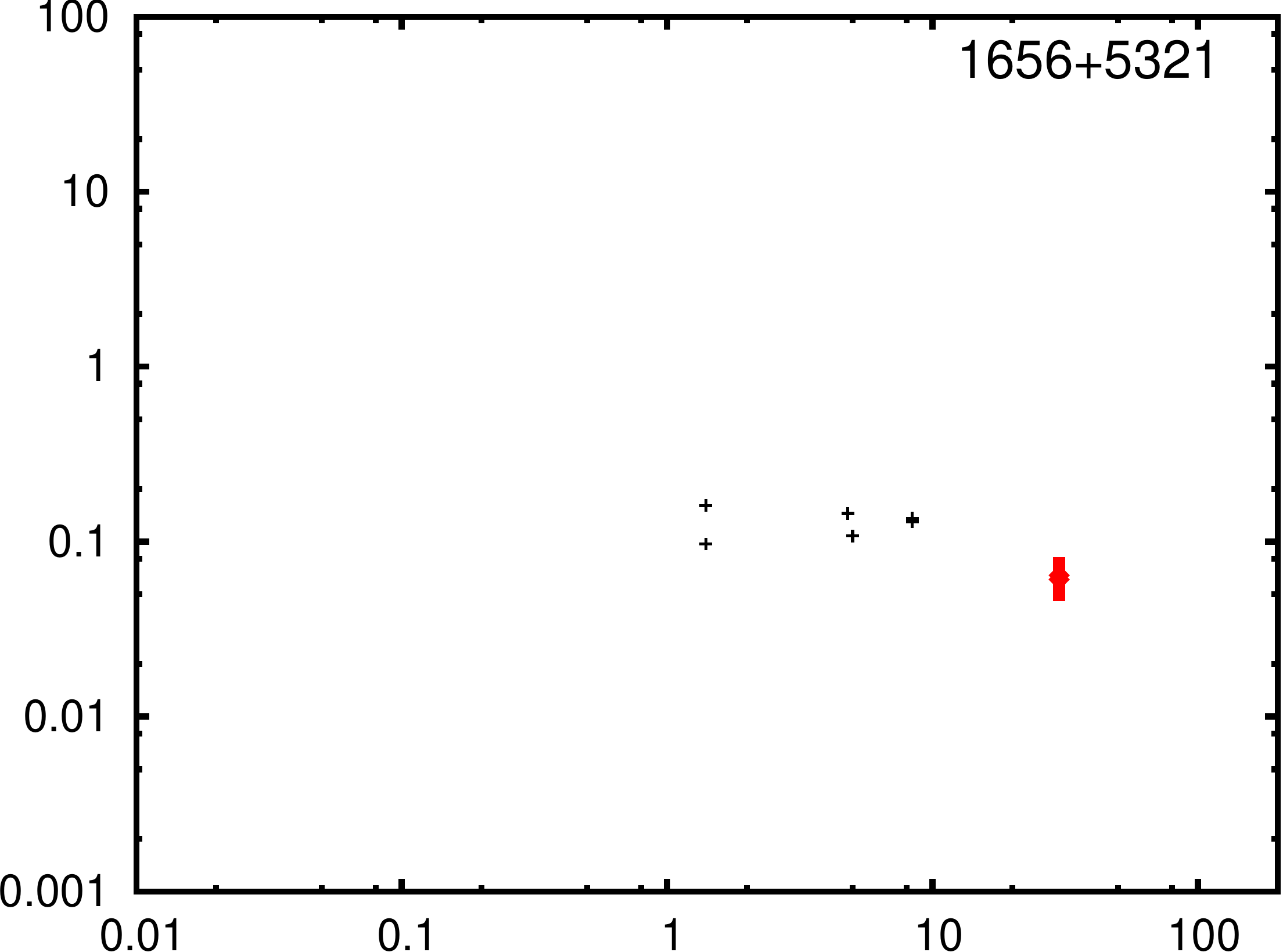}
\includegraphics[scale=0.2]{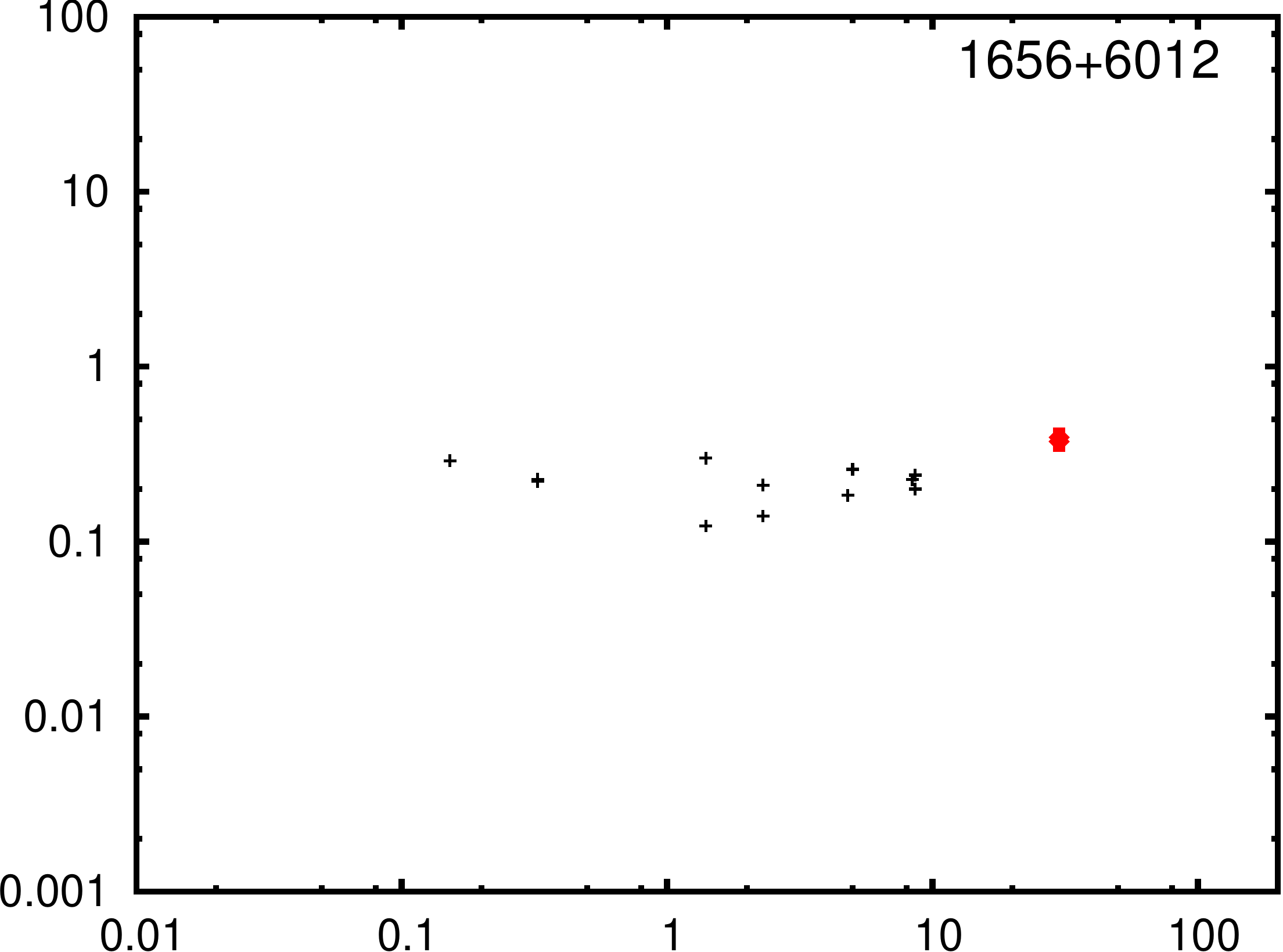}
\includegraphics[scale=0.2]{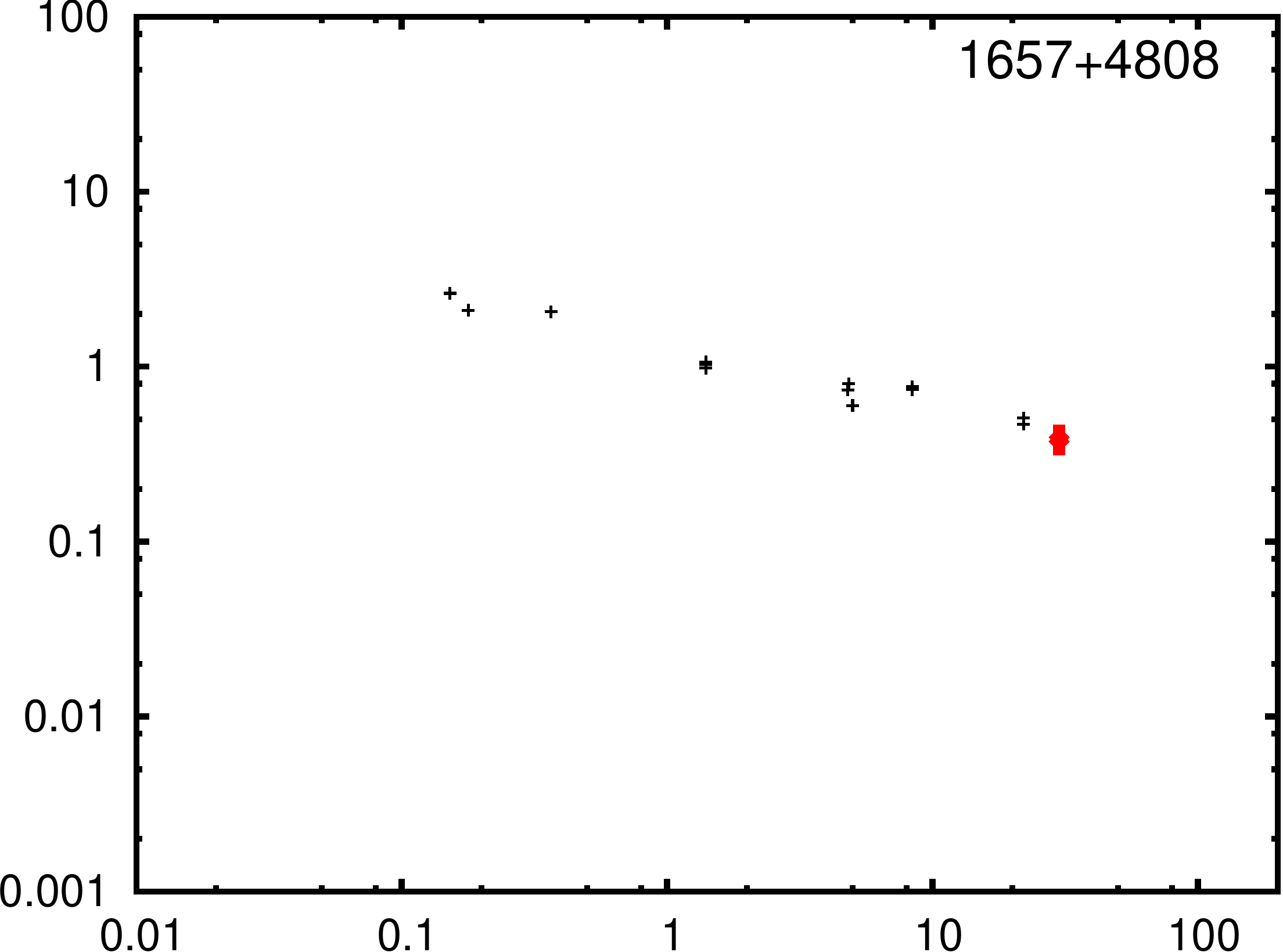}
\includegraphics[scale=0.2]{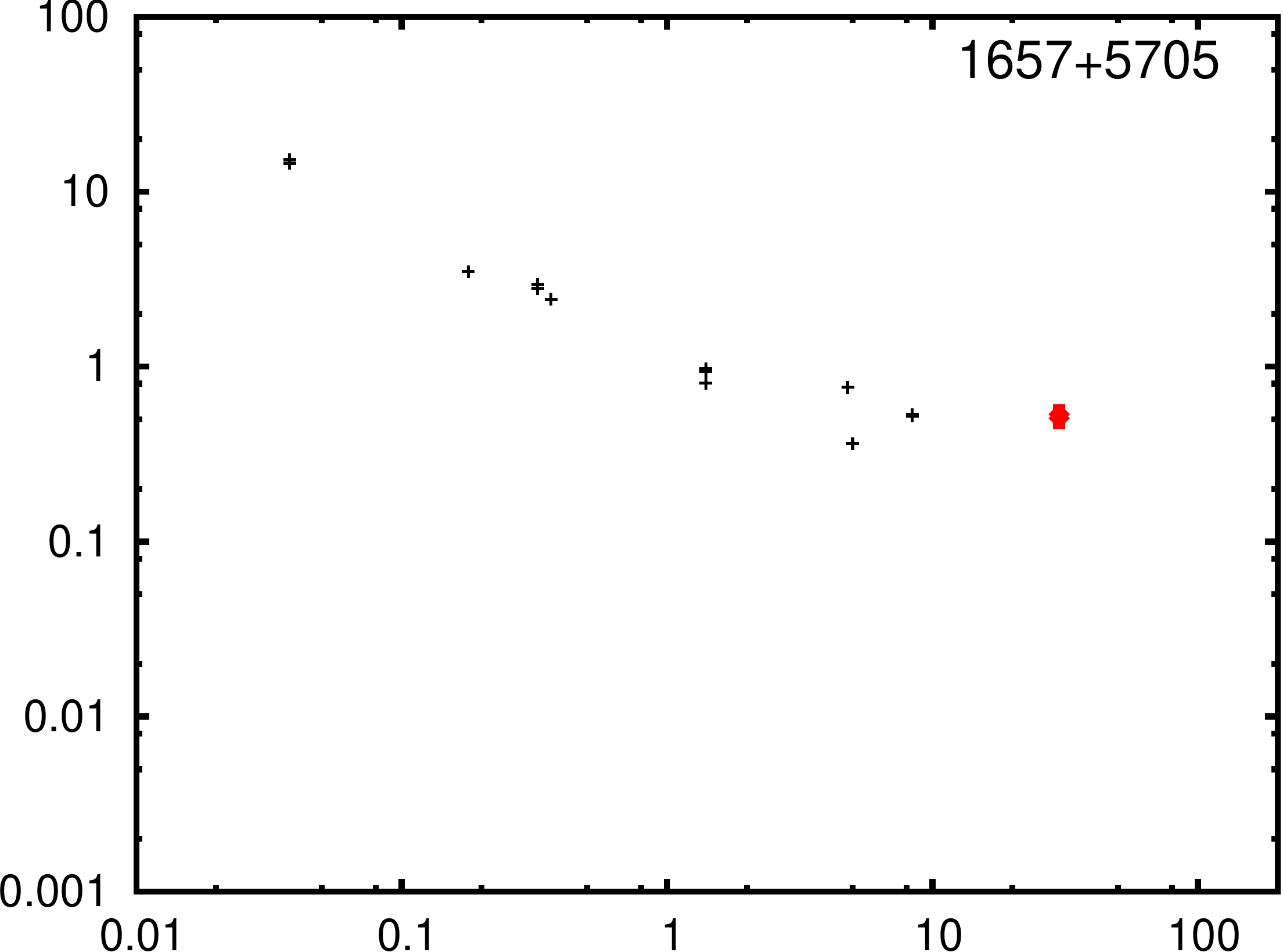}
\includegraphics[scale=0.2]{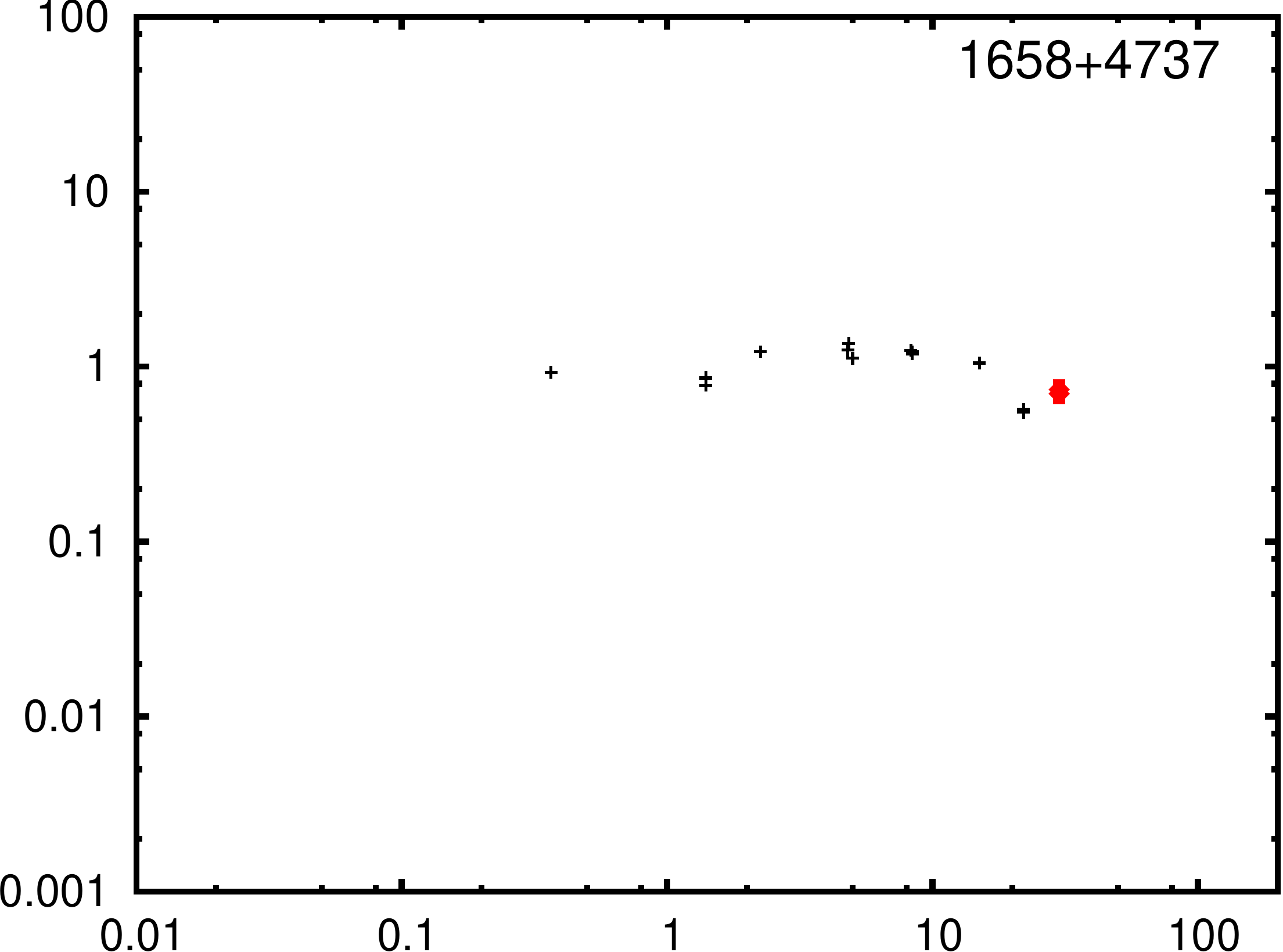}
\includegraphics[scale=0.2]{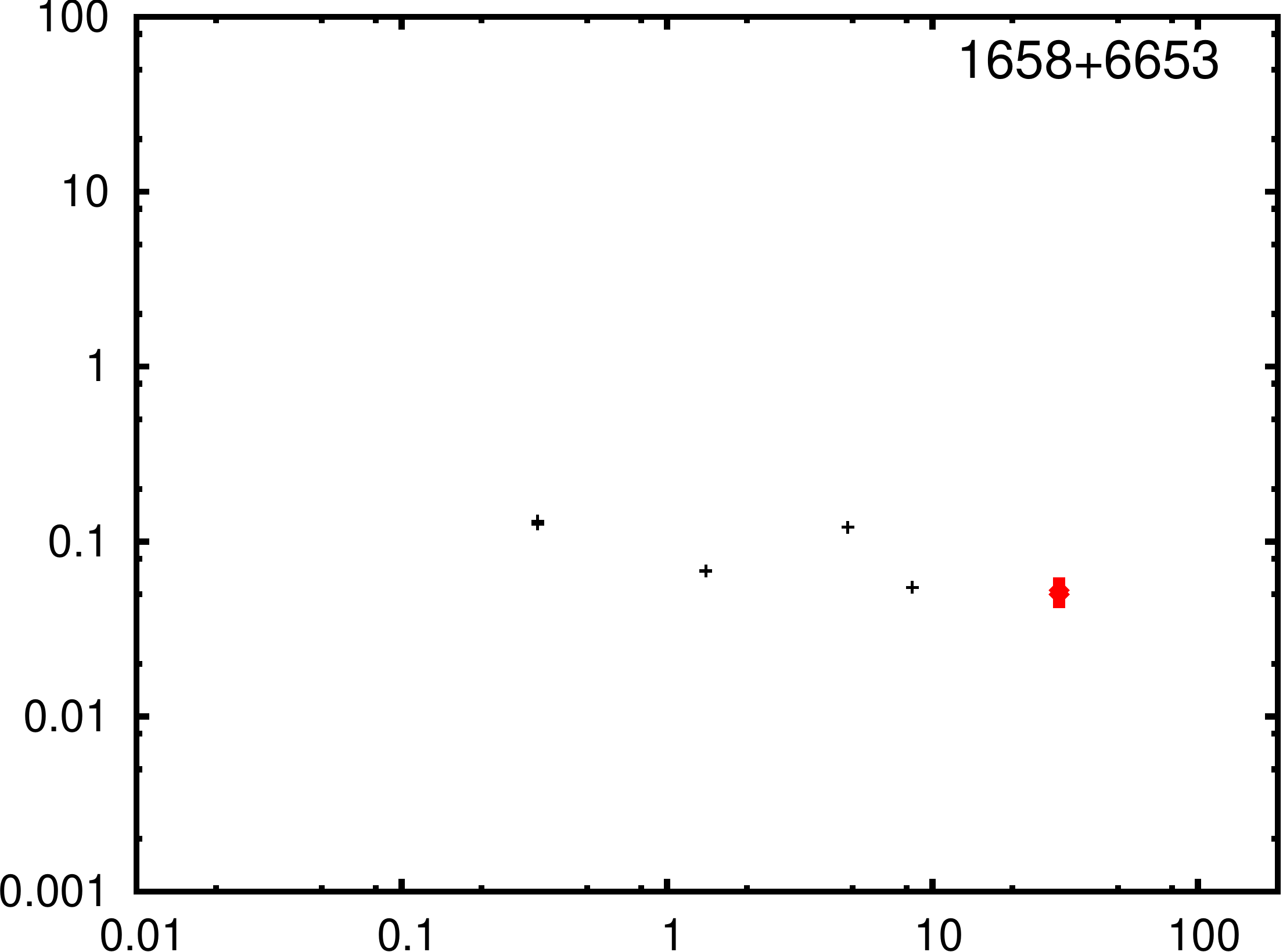}
\includegraphics[scale=0.2]{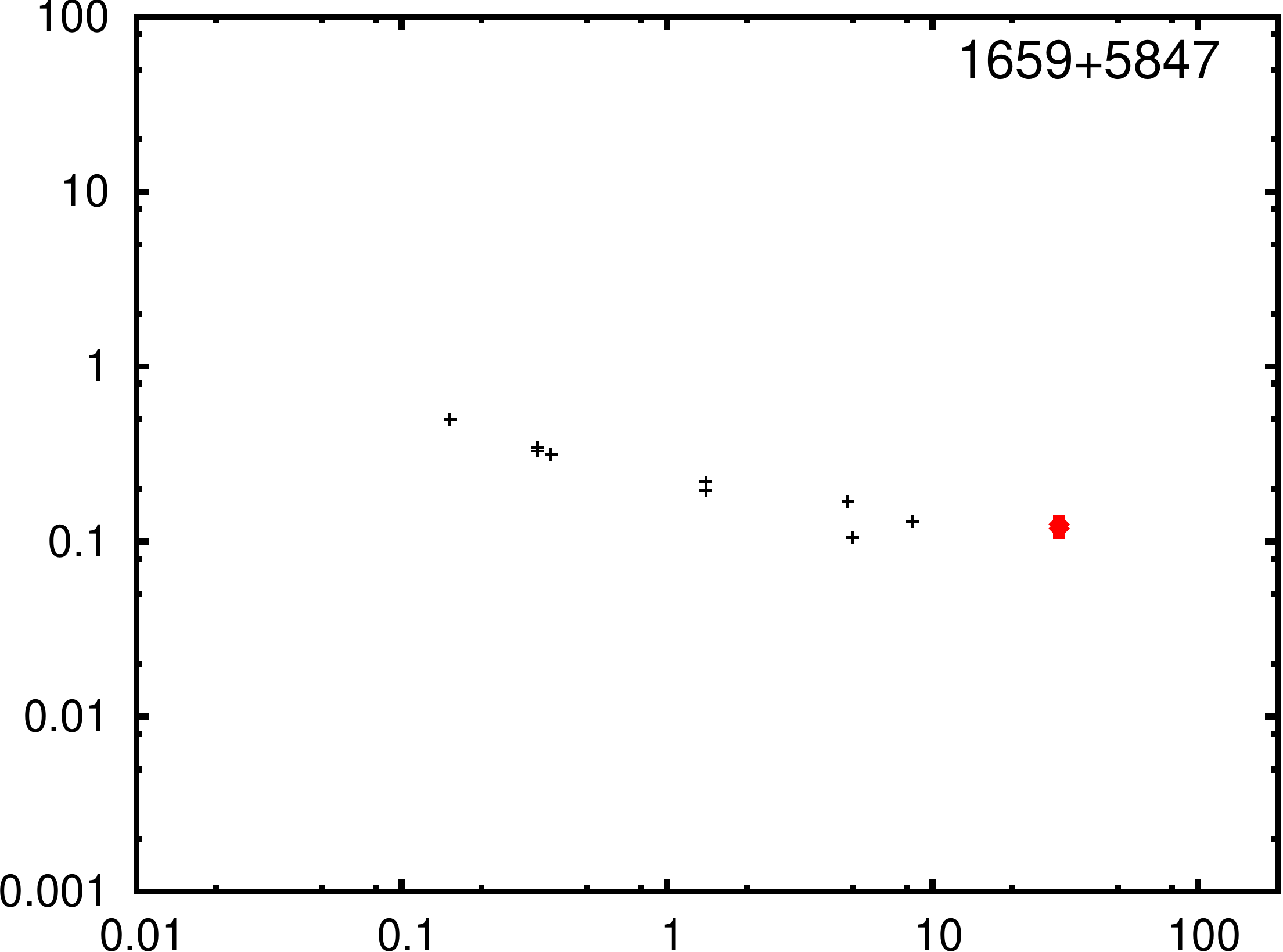}
\includegraphics[scale=0.2]{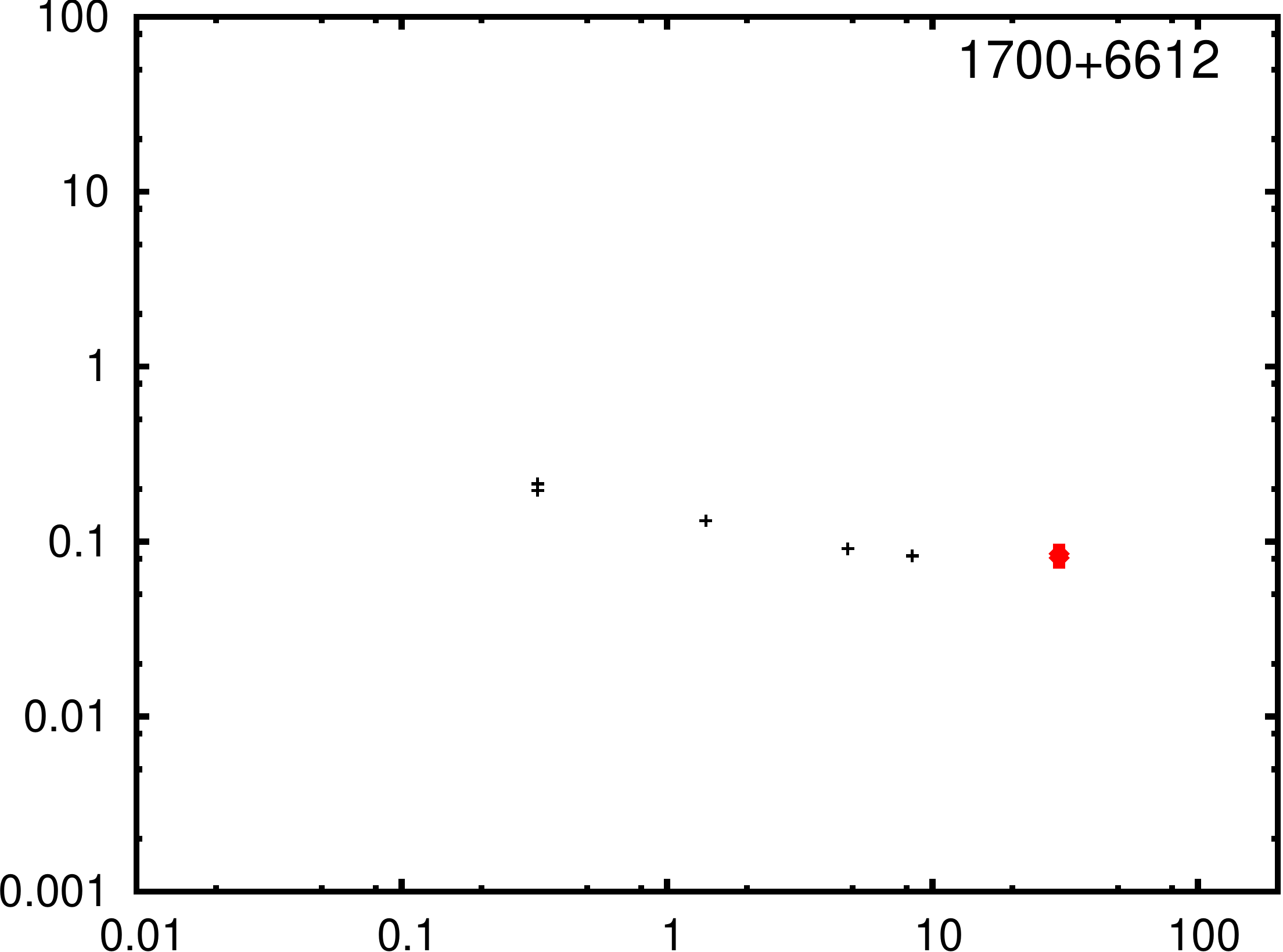}
\includegraphics[scale=0.2]{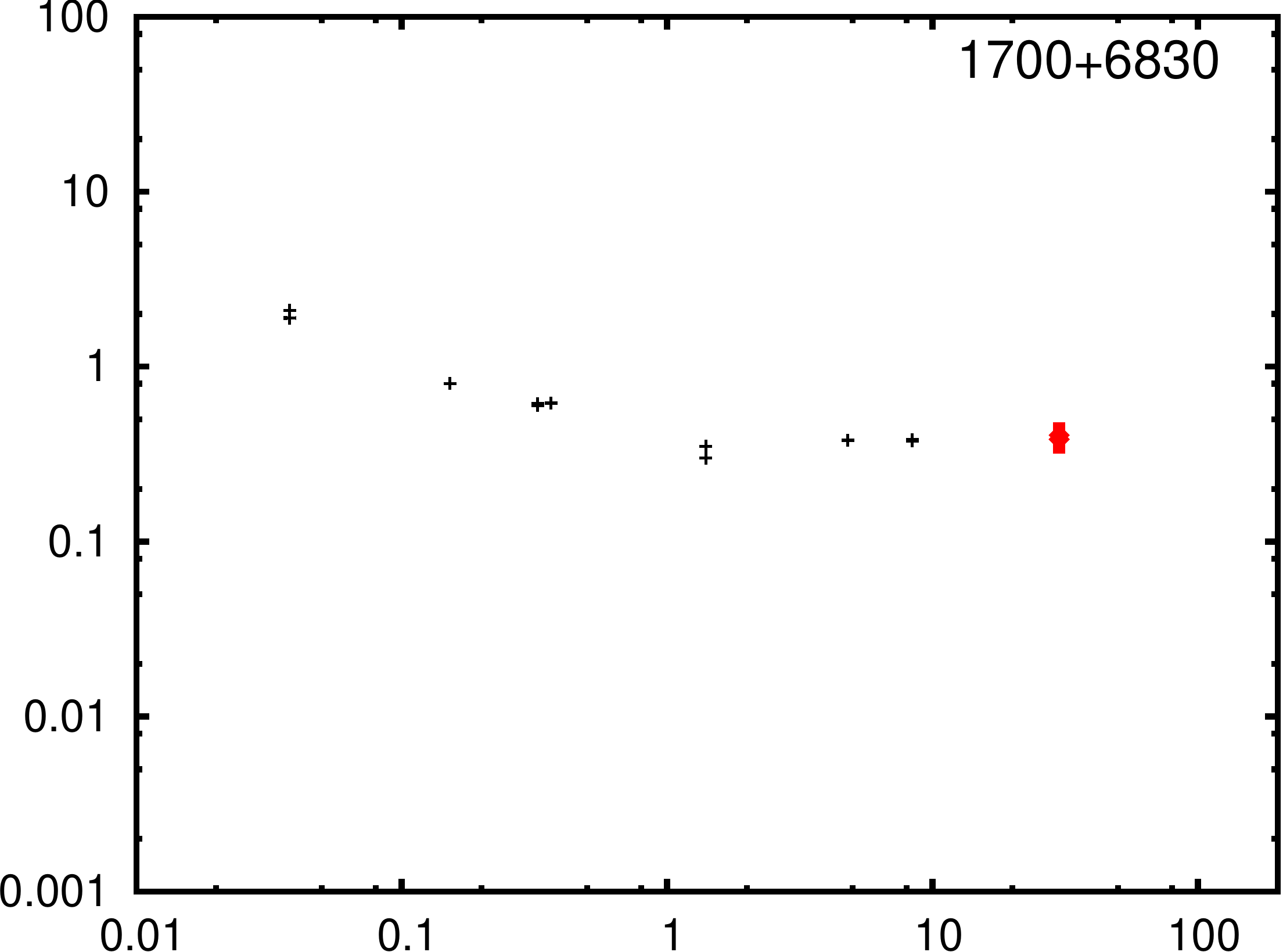}
\includegraphics[scale=0.2]{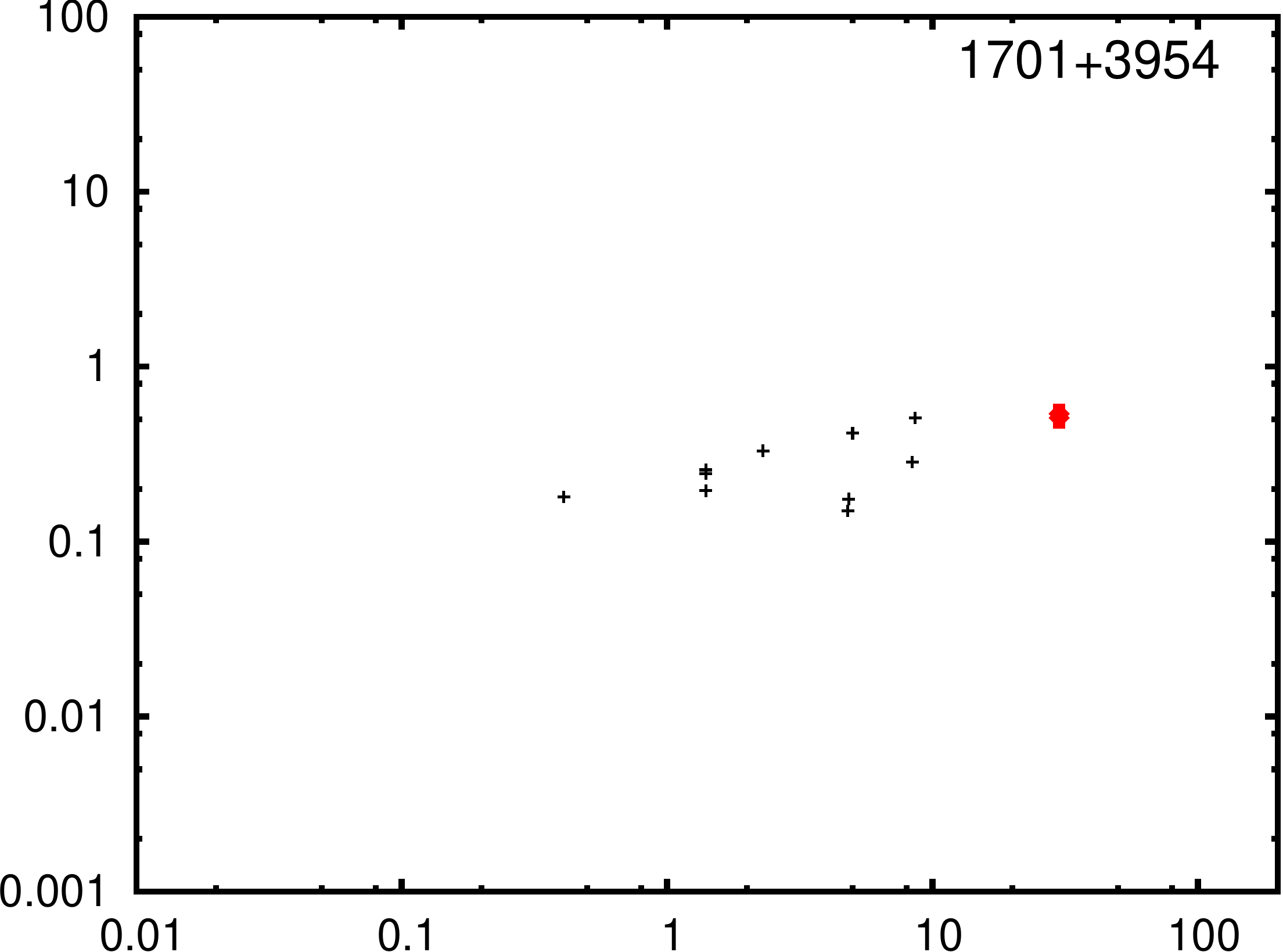}
\includegraphics[scale=0.2]{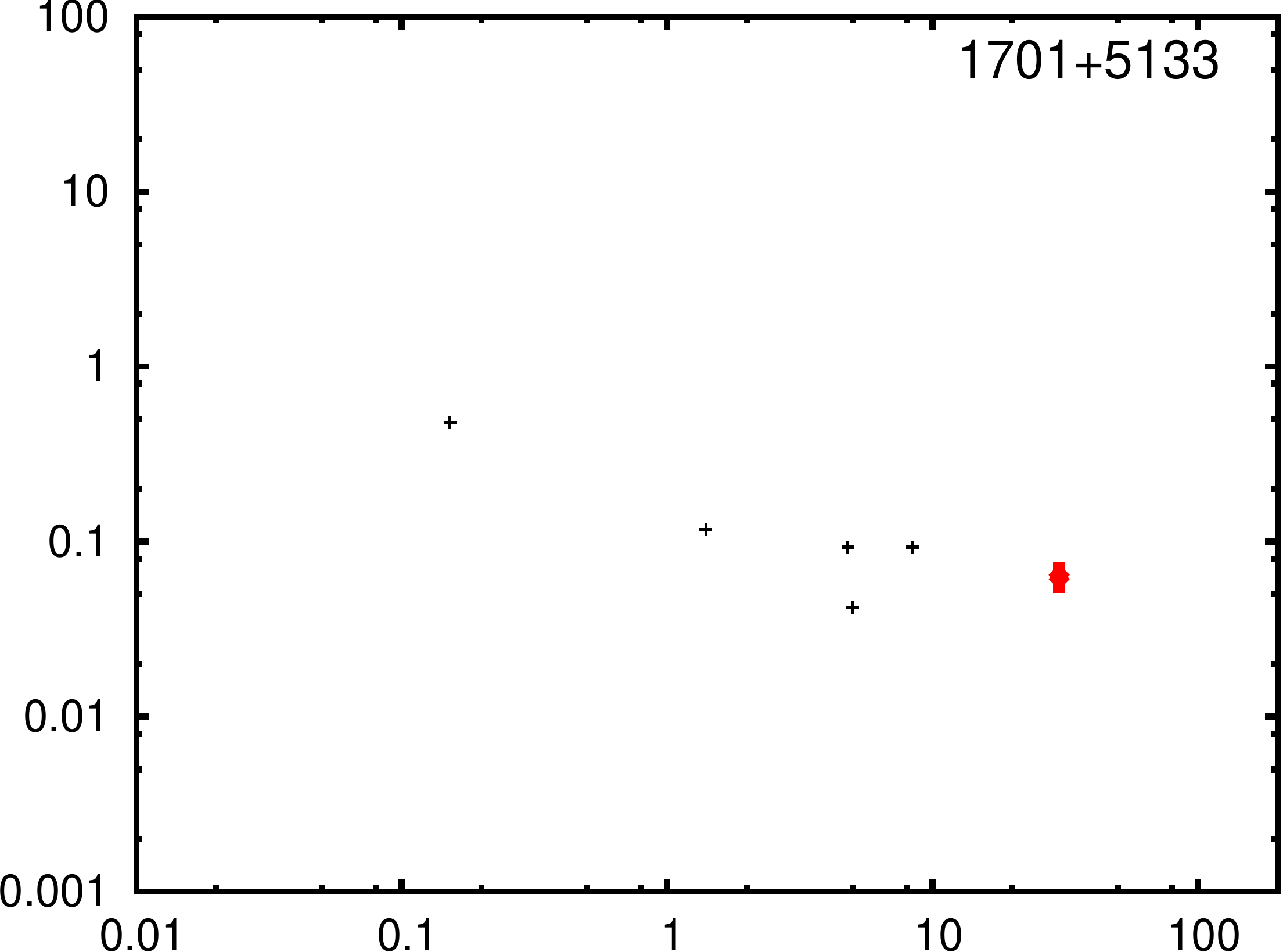}
\includegraphics[scale=0.2]{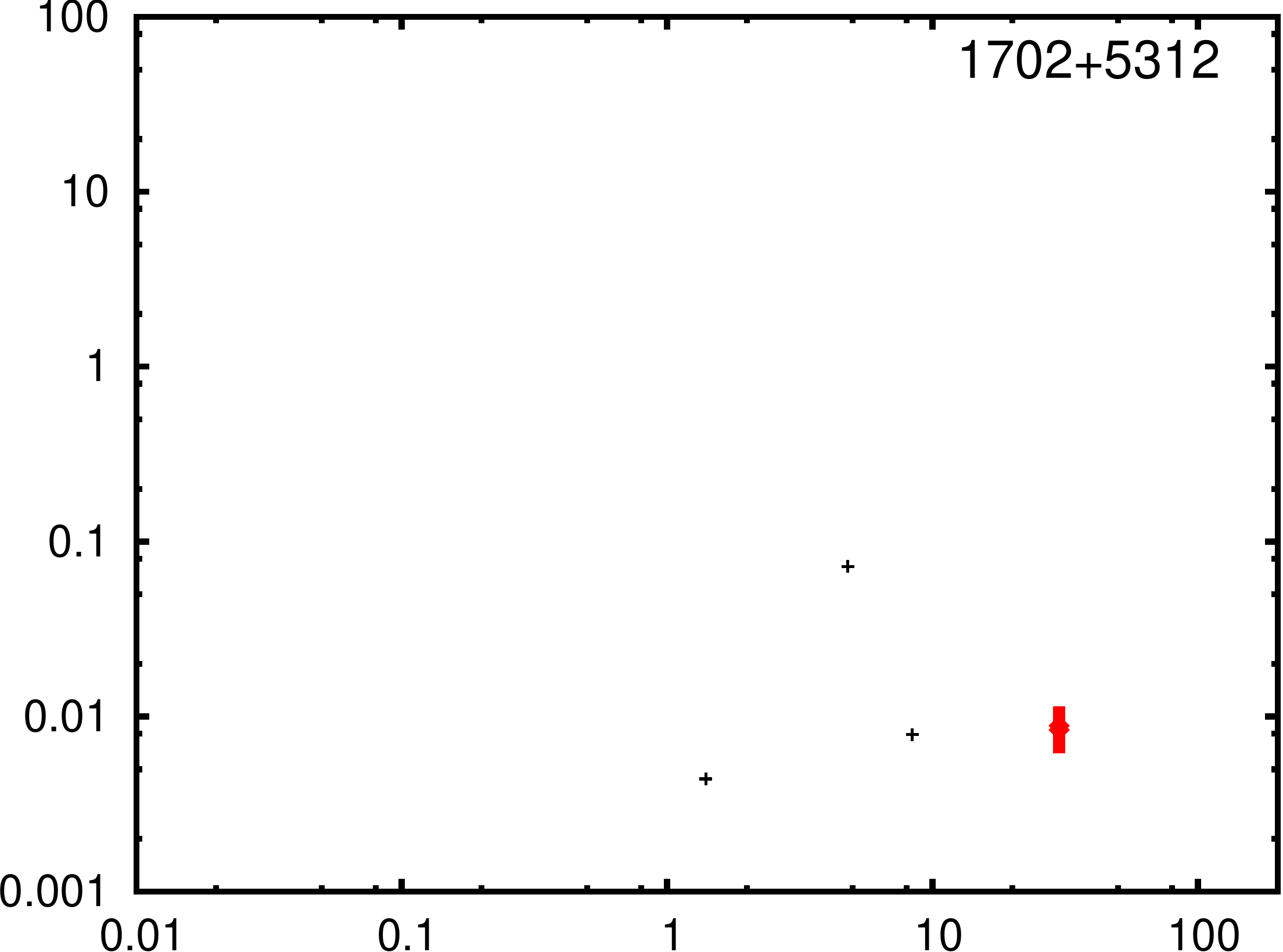}
\includegraphics[scale=0.2]{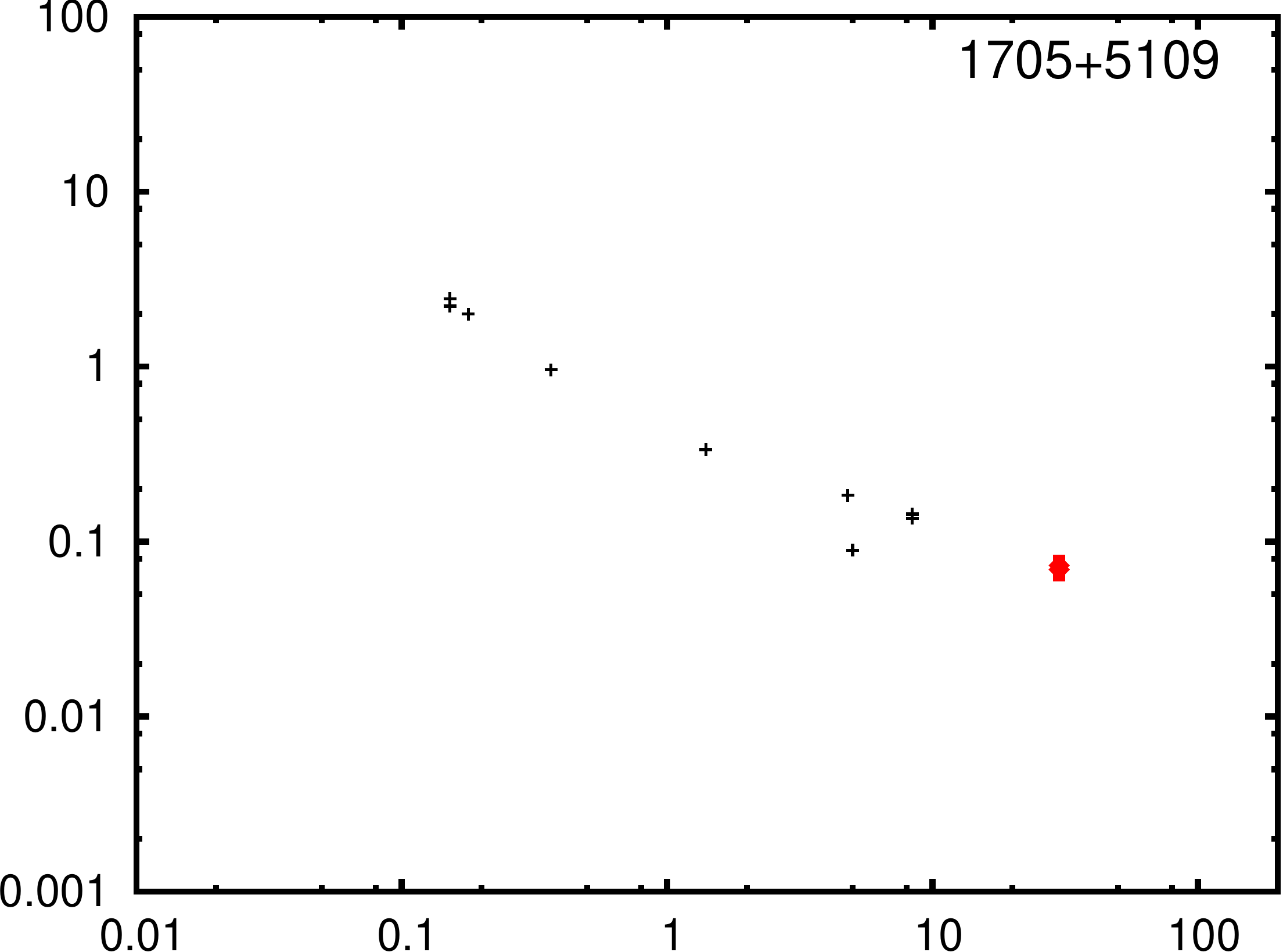}
\includegraphics[scale=0.2]{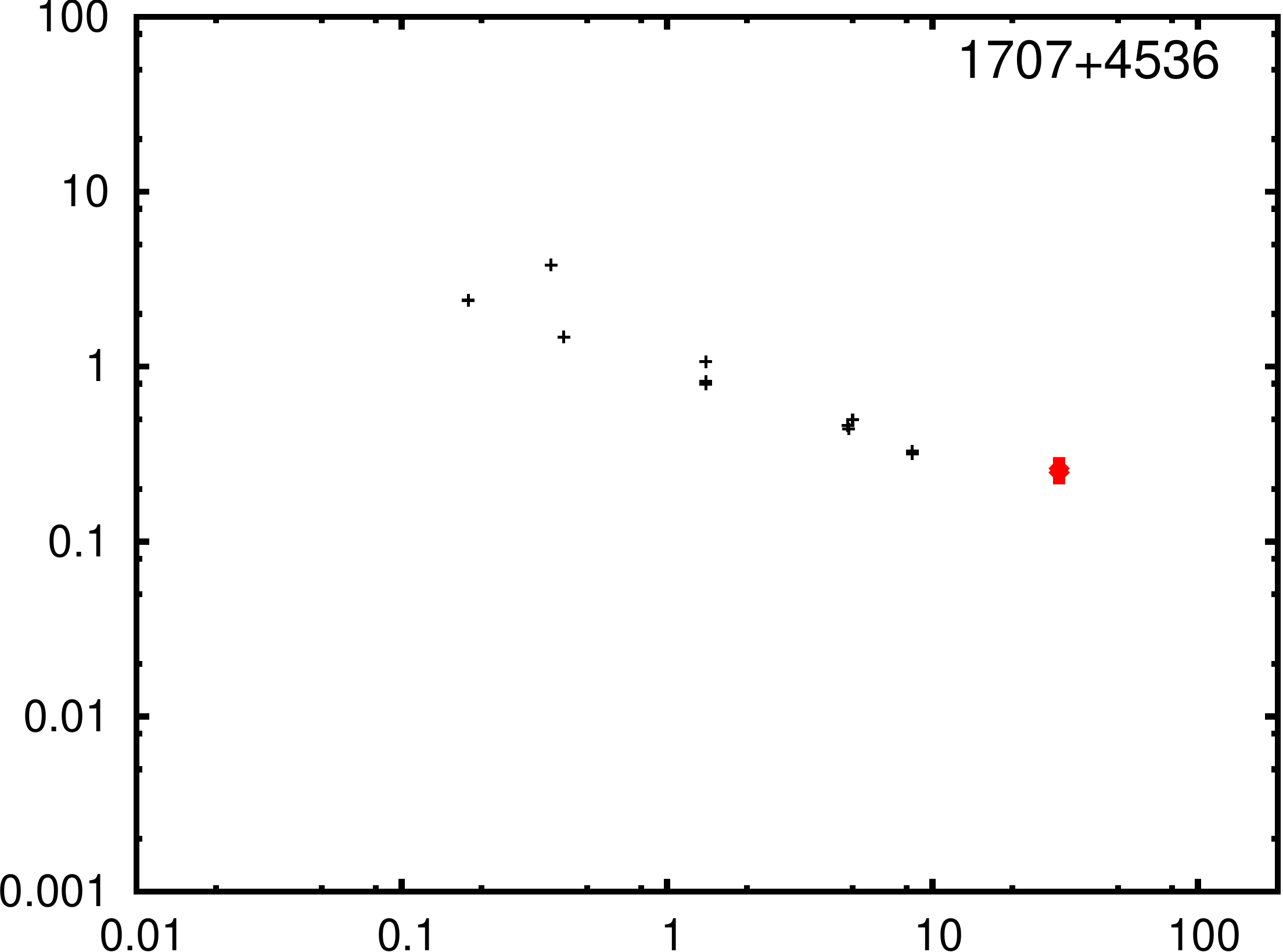}
\includegraphics[scale=0.2]{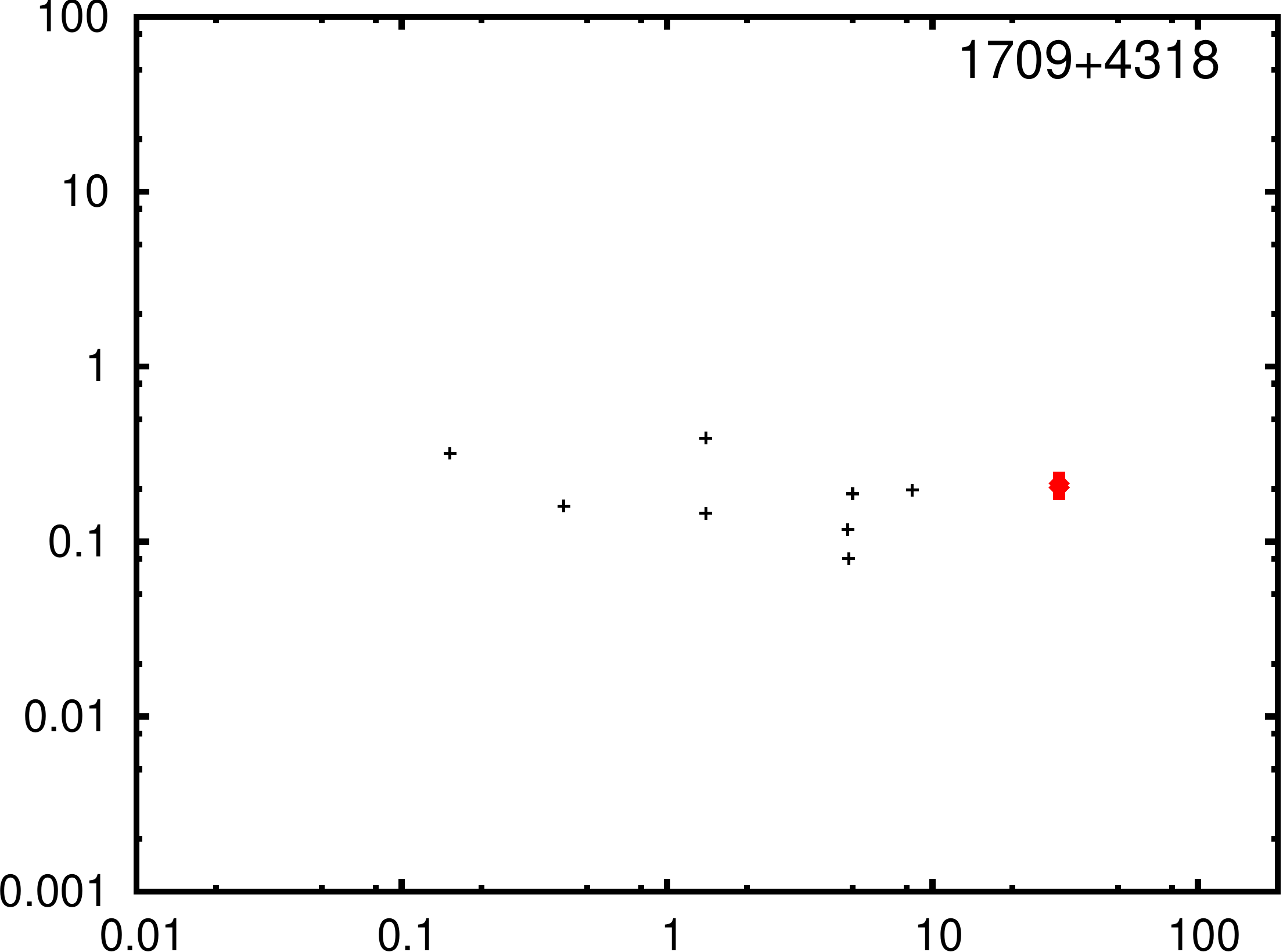}
\includegraphics[scale=0.2]{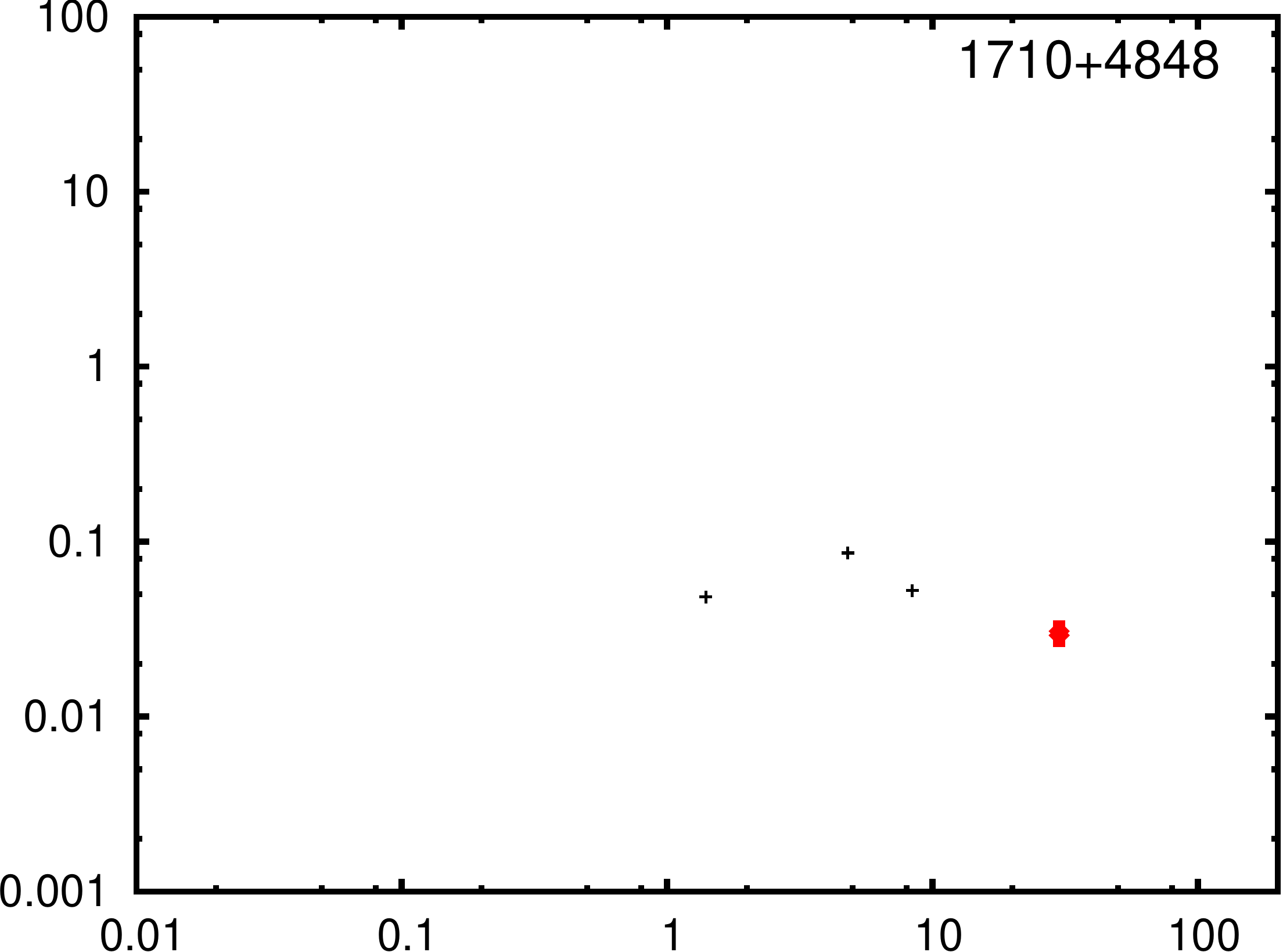}
\includegraphics[scale=0.2]{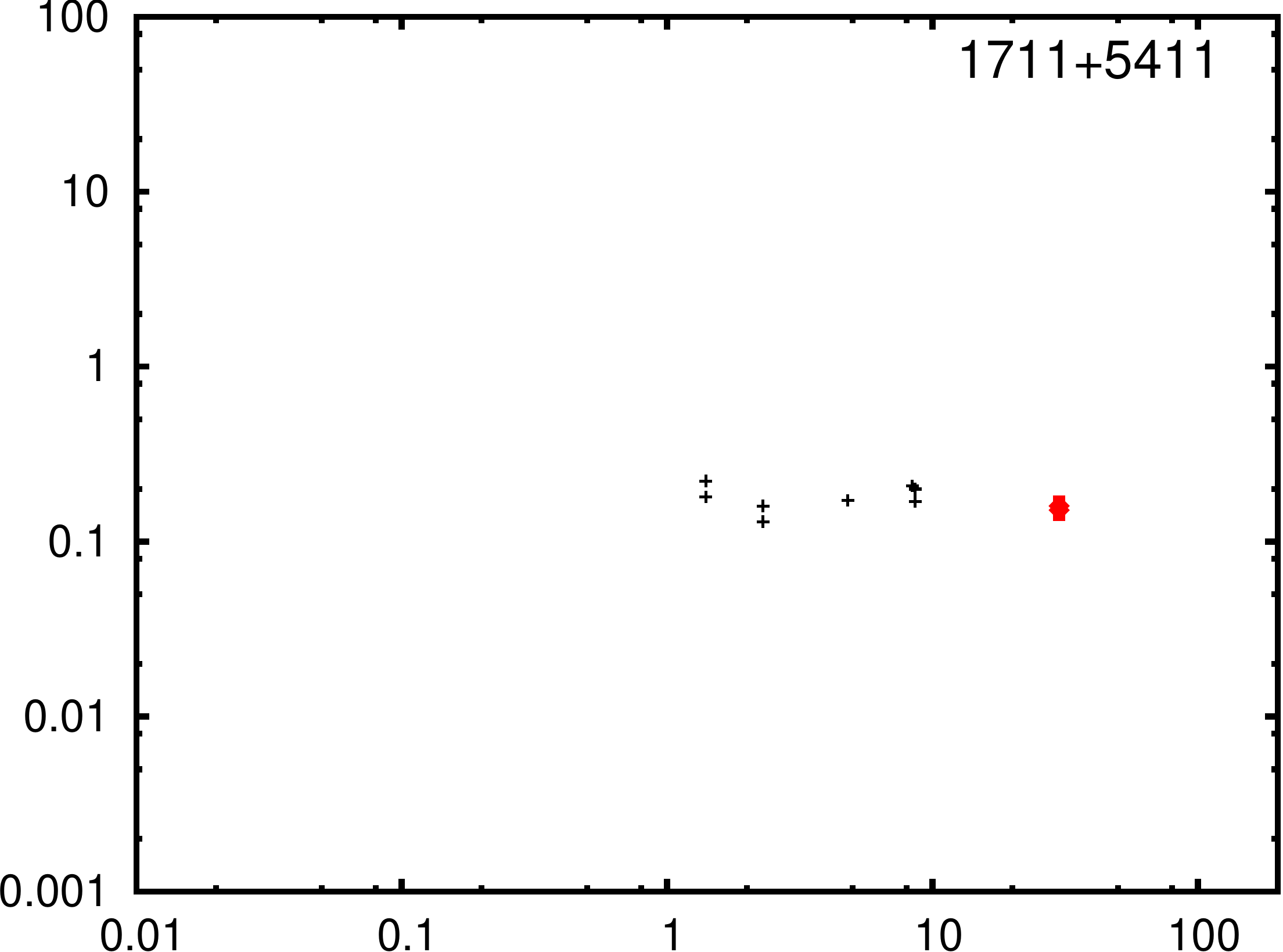}
\end{figure}
\clearpage\begin{figure}
\centering
\includegraphics[scale=0.2]{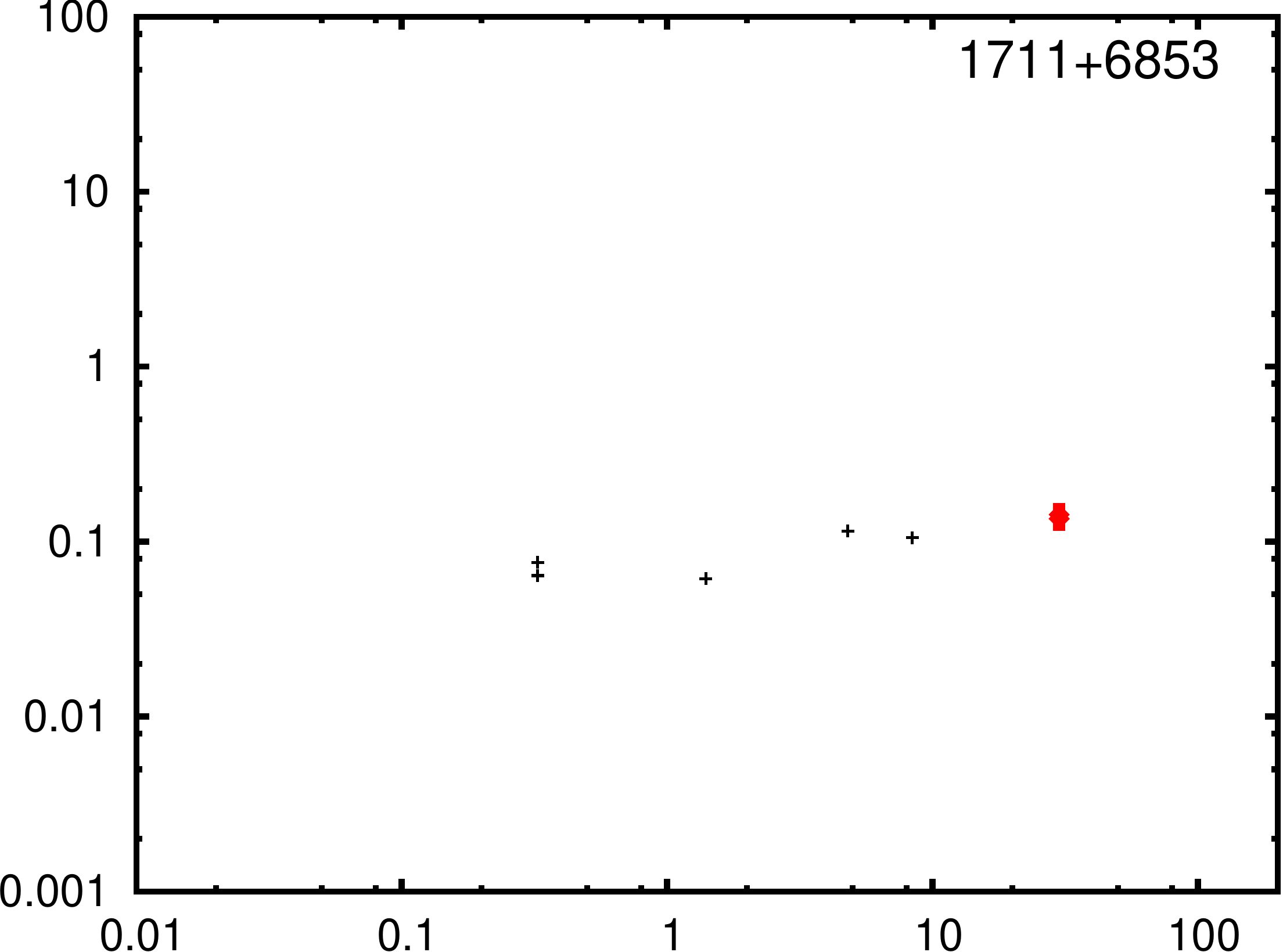}
\includegraphics[scale=0.2]{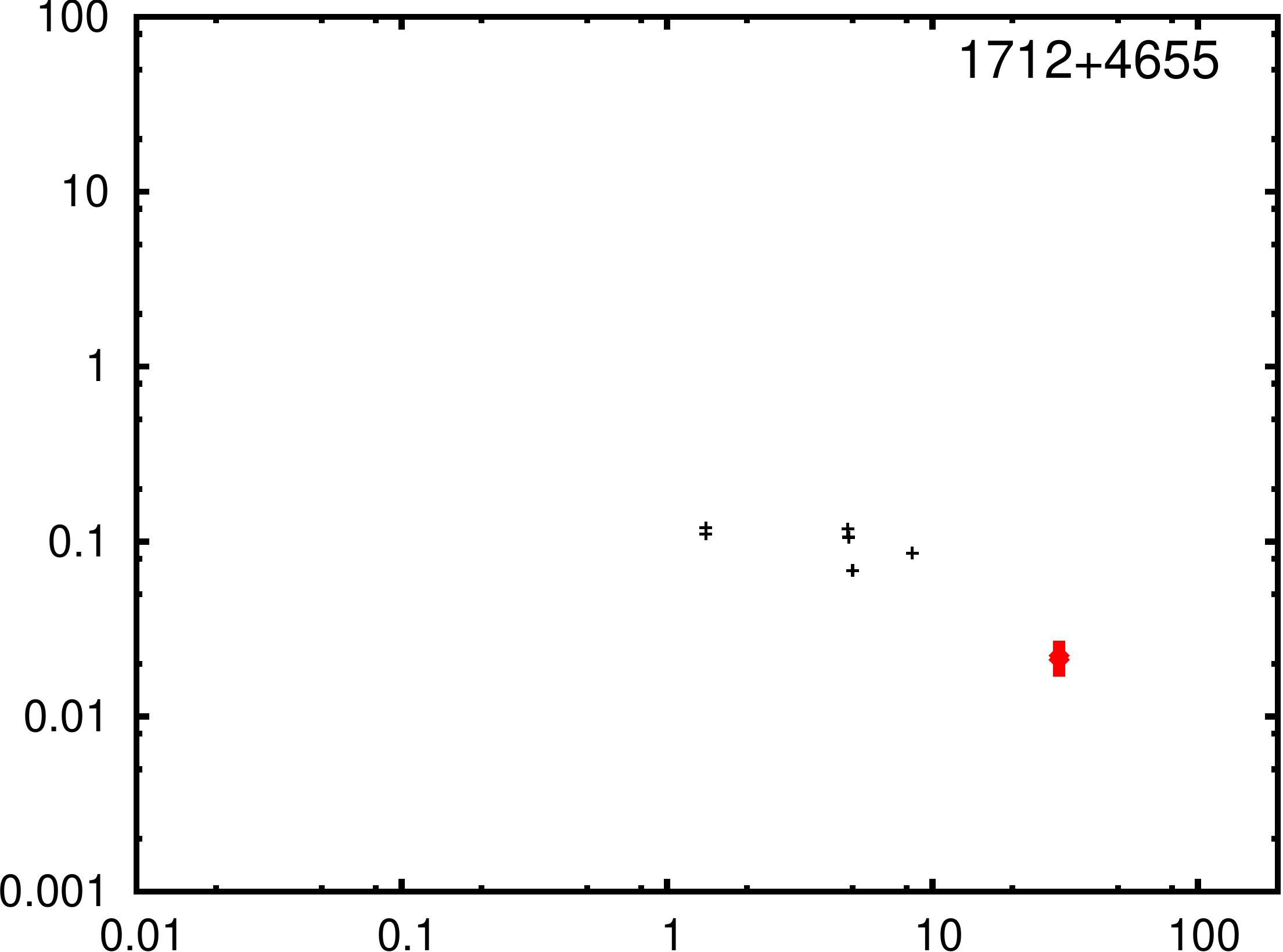}
\includegraphics[scale=0.2]{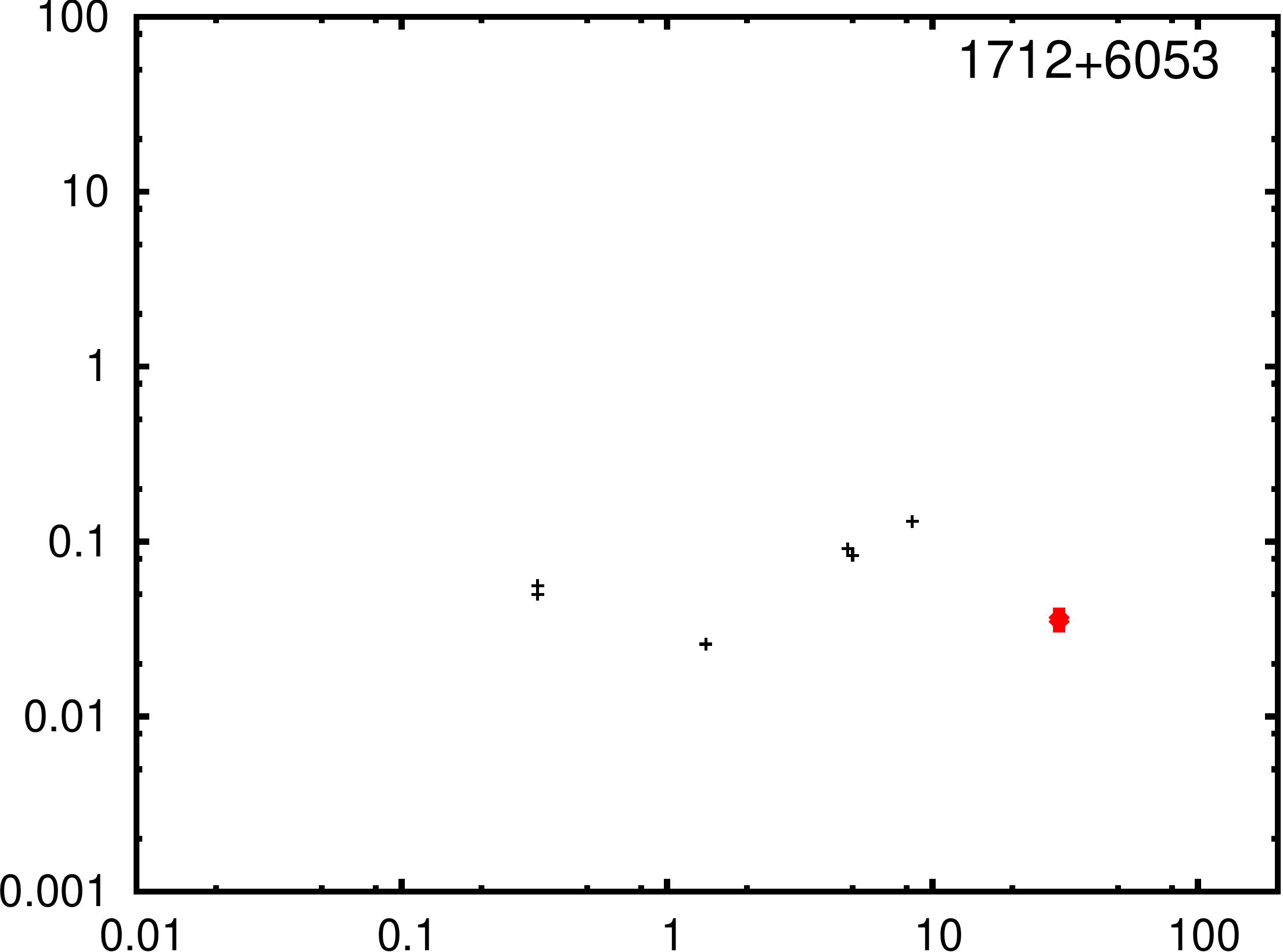}
\includegraphics[scale=0.2]{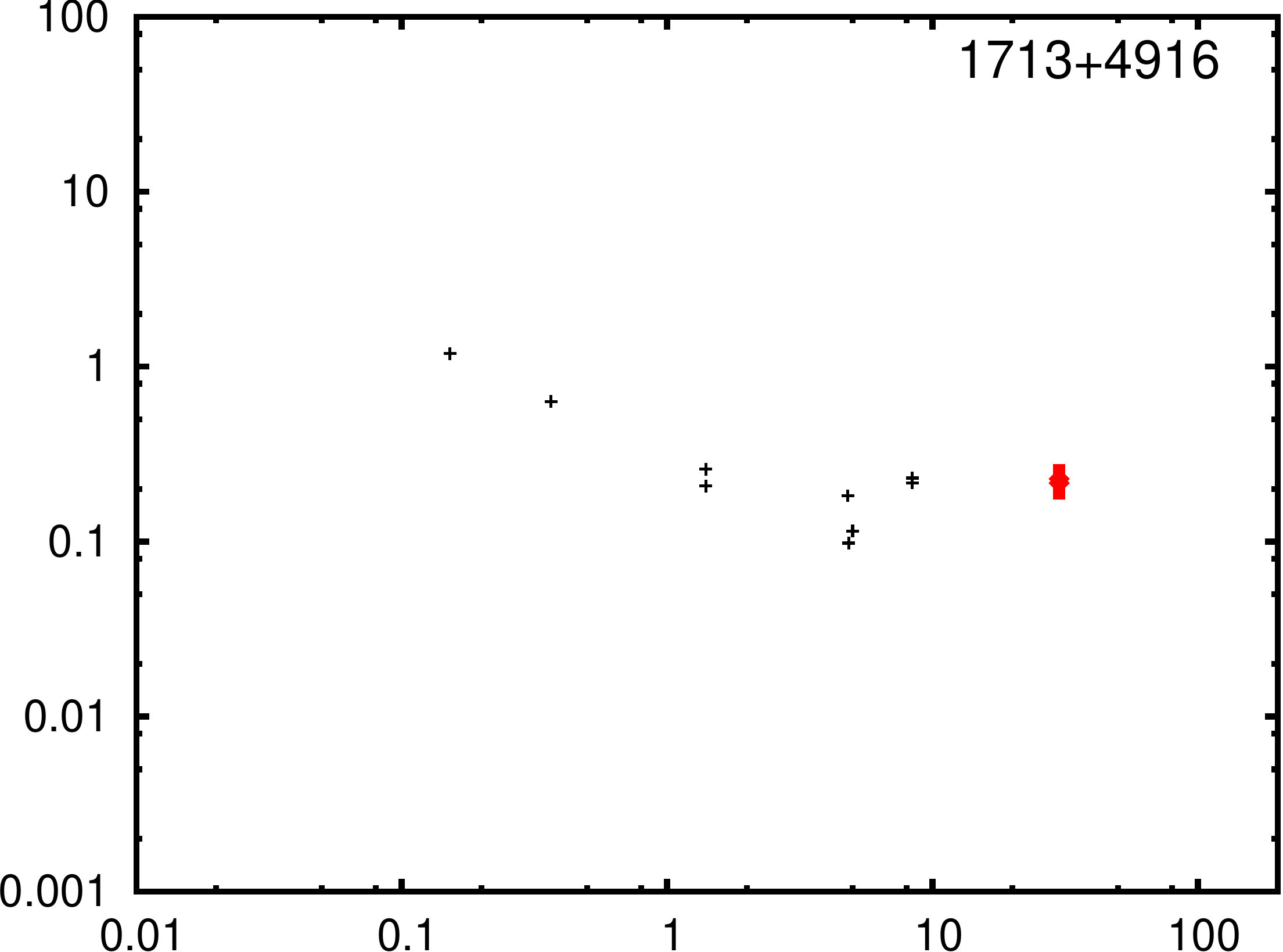}
\includegraphics[scale=0.2]{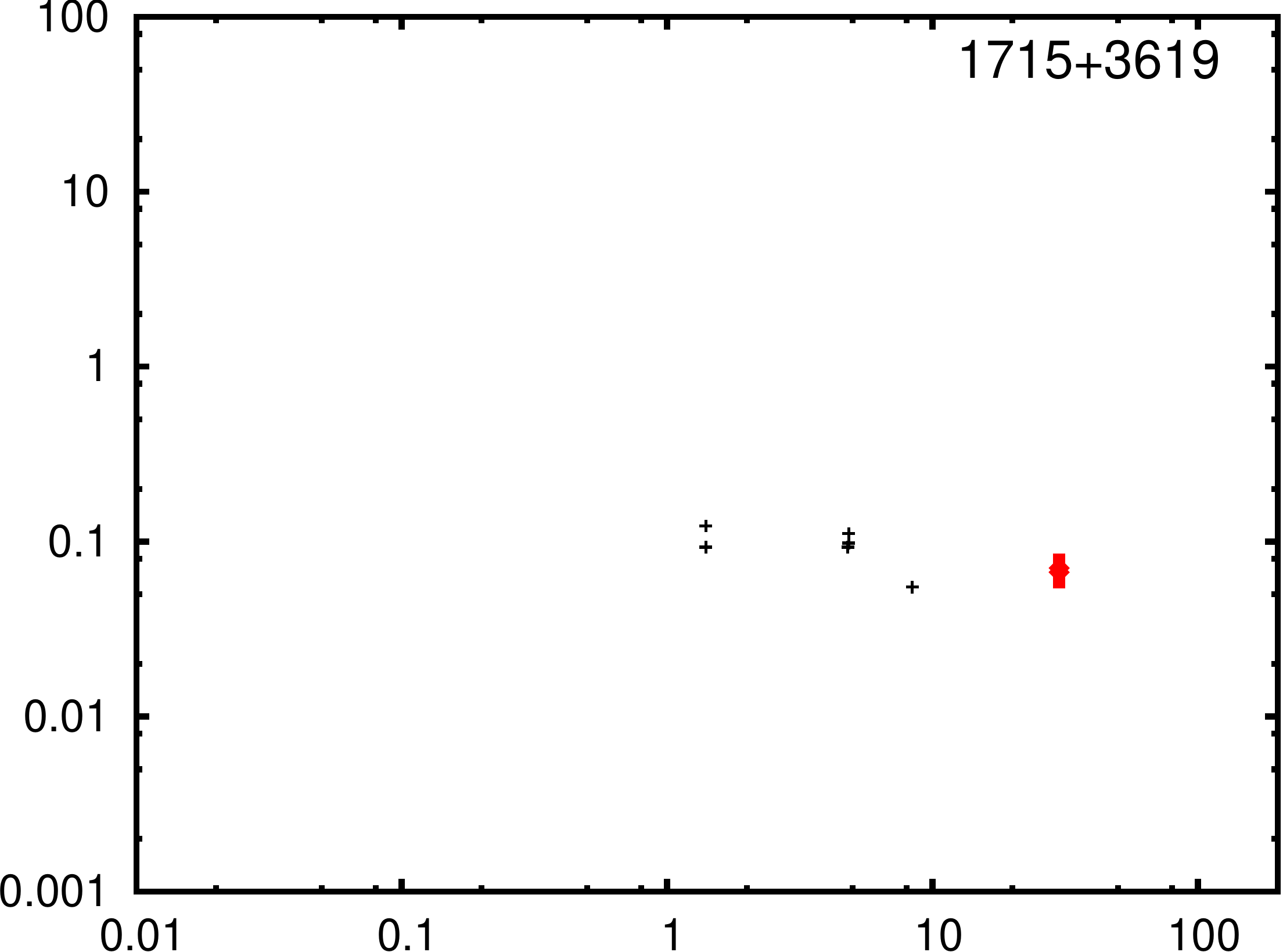}
\includegraphics[scale=0.2]{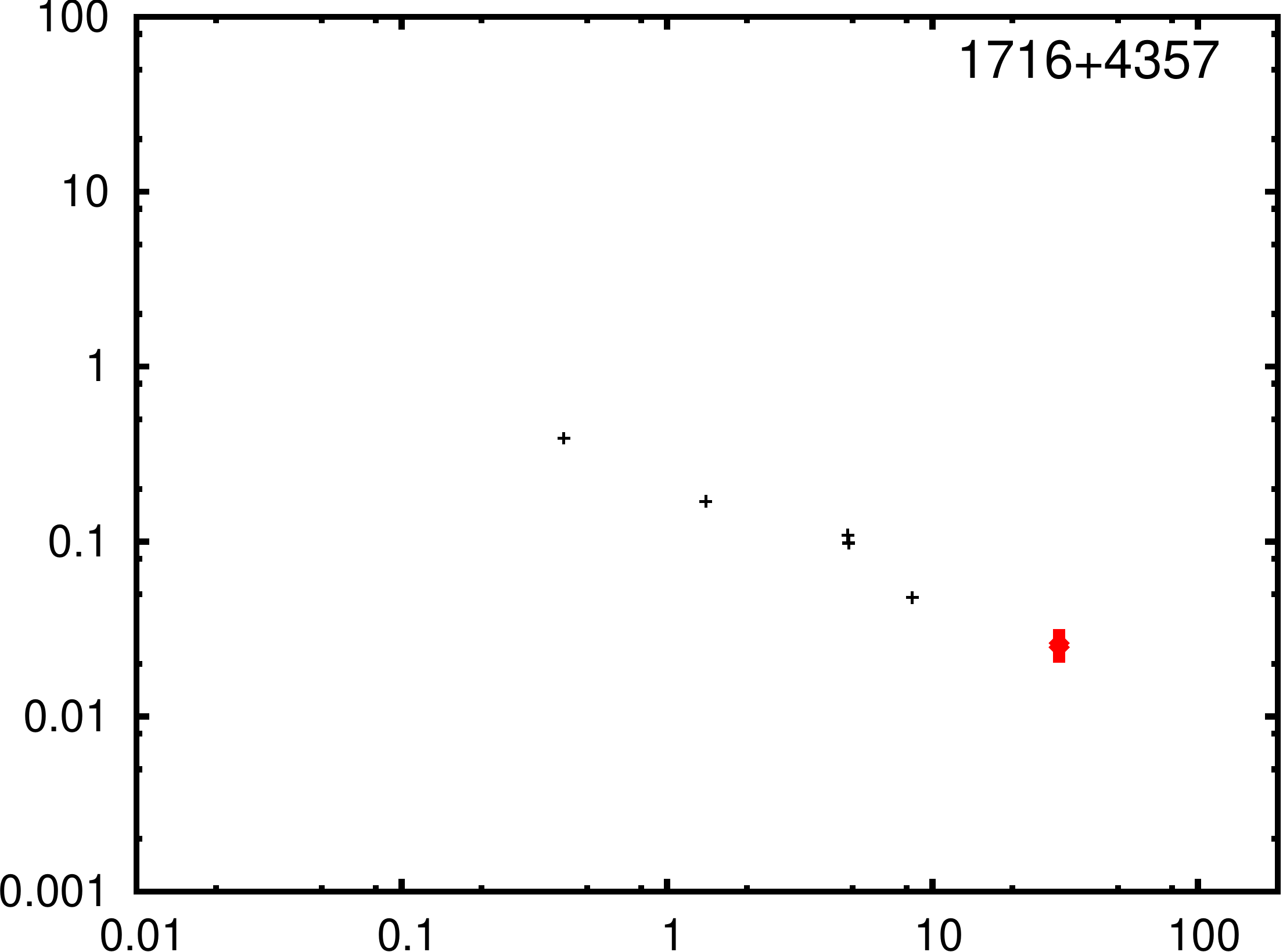}
\includegraphics[scale=0.2]{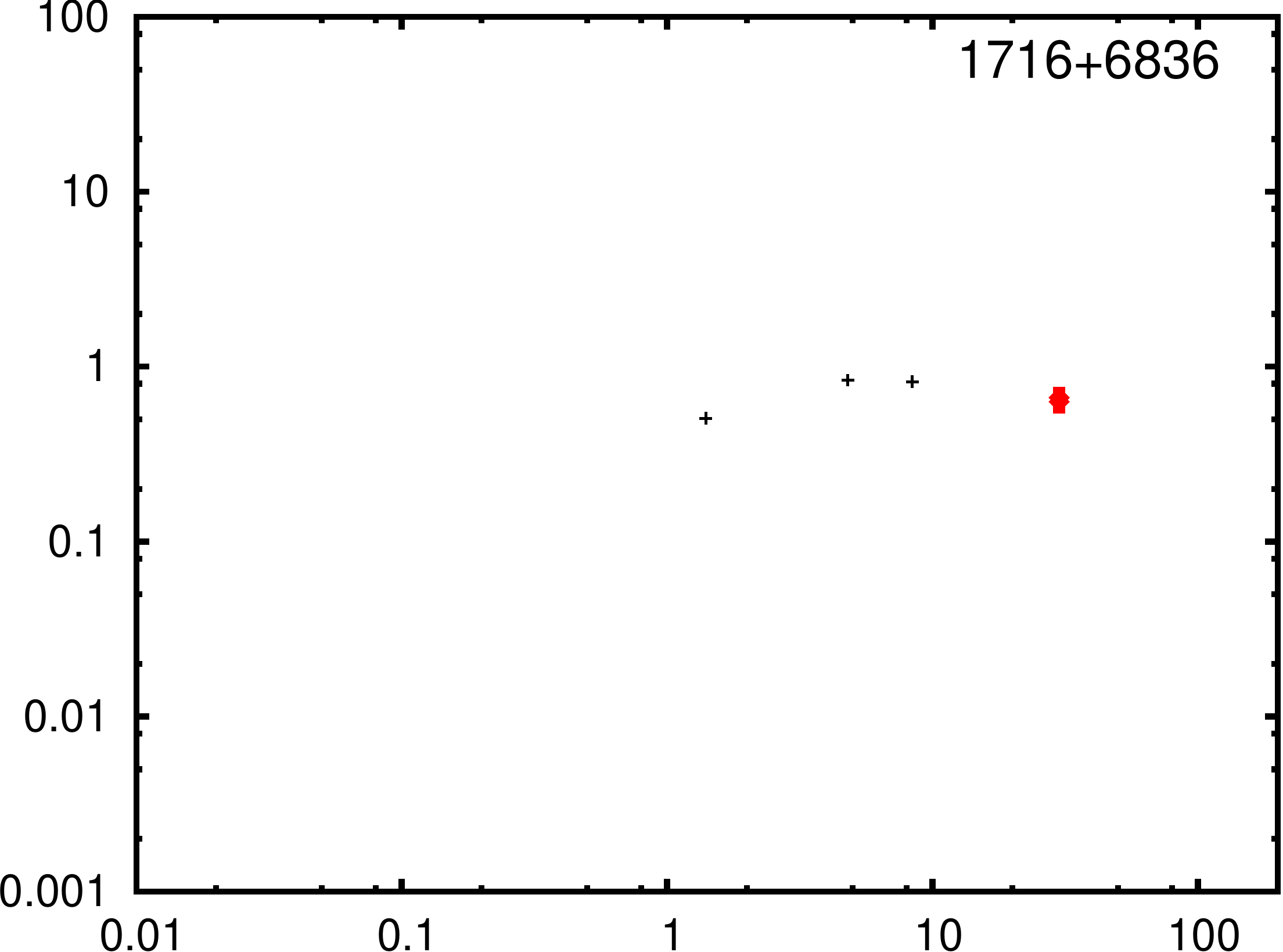}
\includegraphics[scale=0.2]{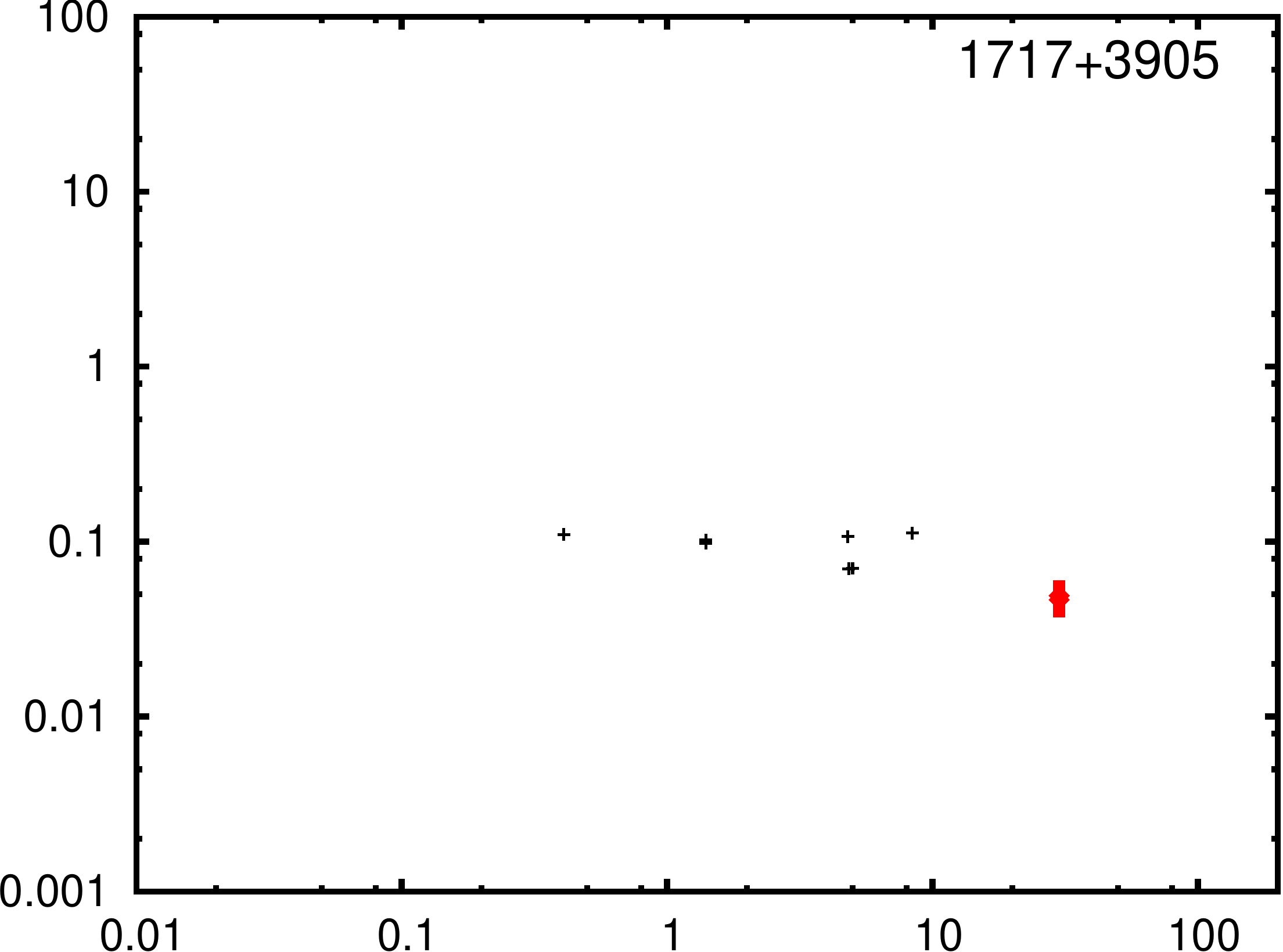}
\includegraphics[scale=0.2]{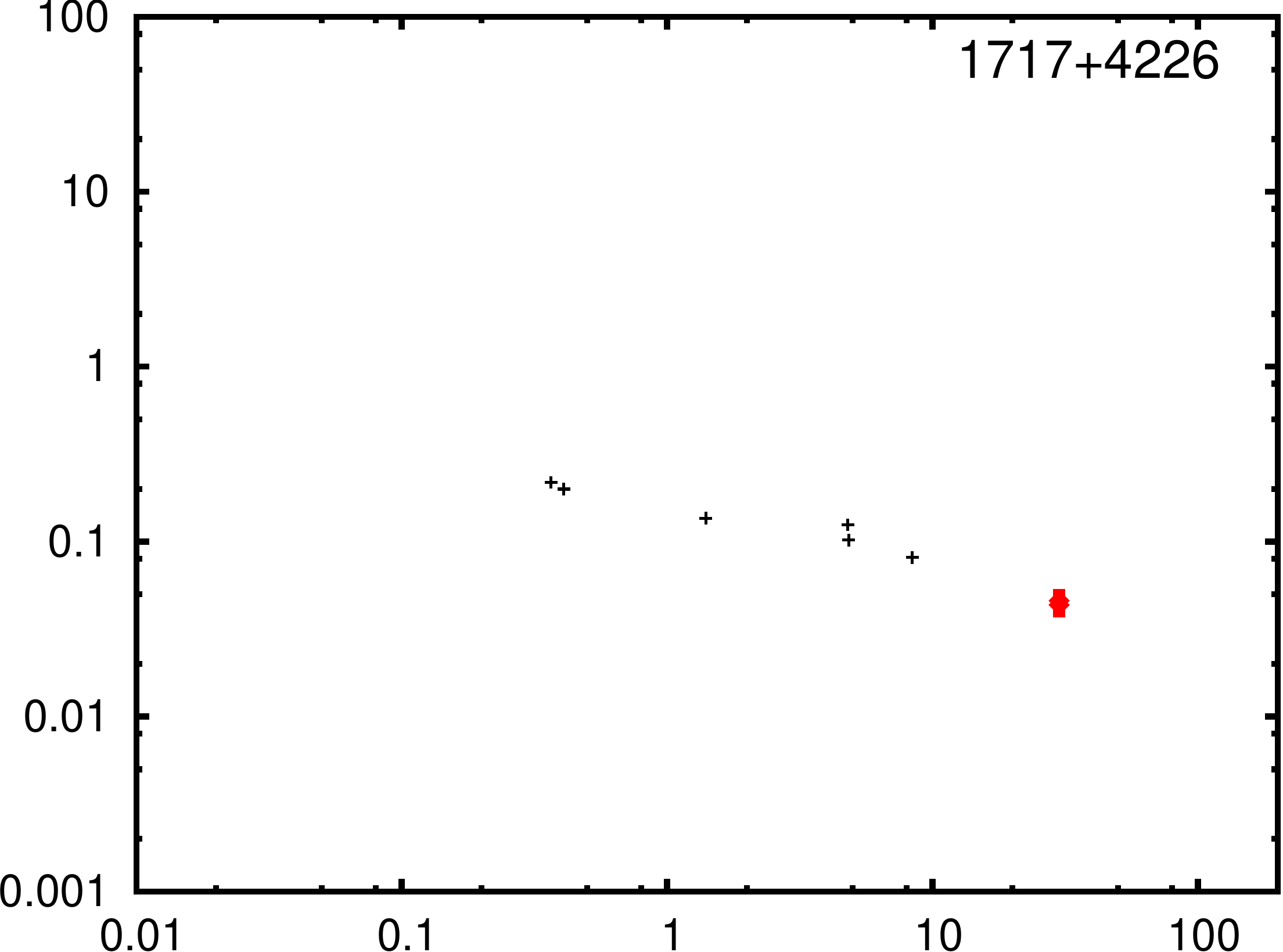}
\includegraphics[scale=0.2]{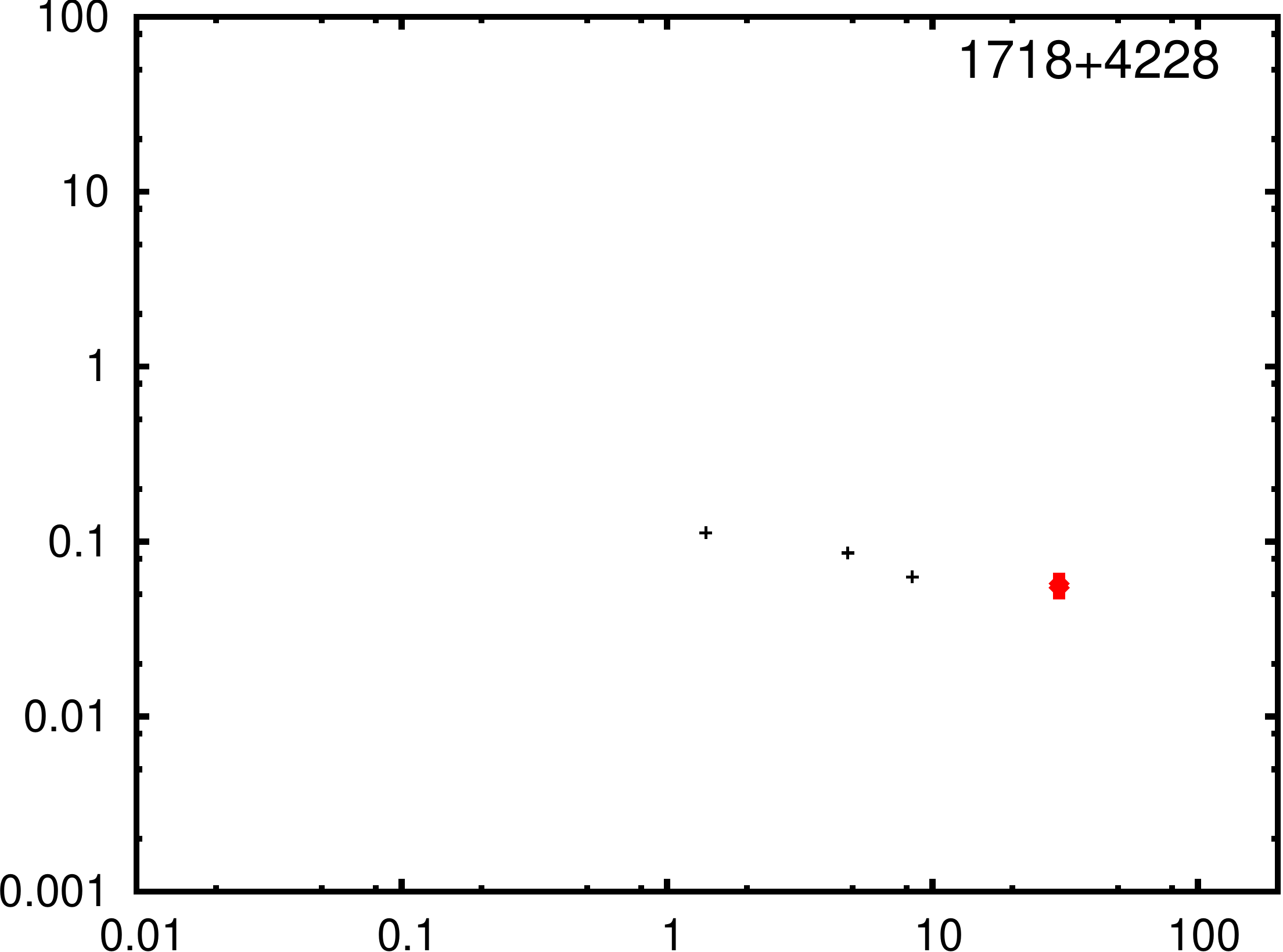}
\includegraphics[scale=0.2]{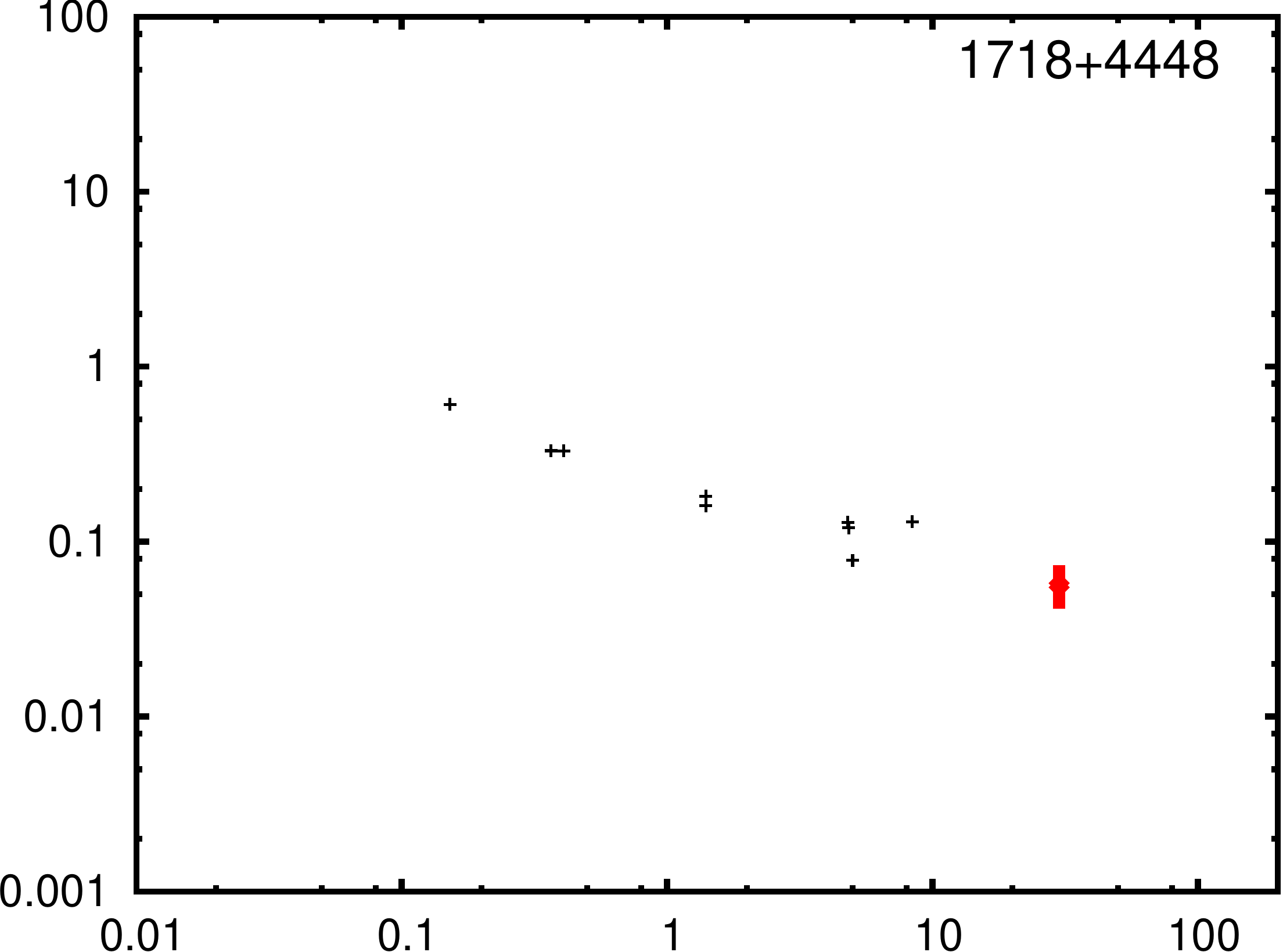}
\includegraphics[scale=0.2]{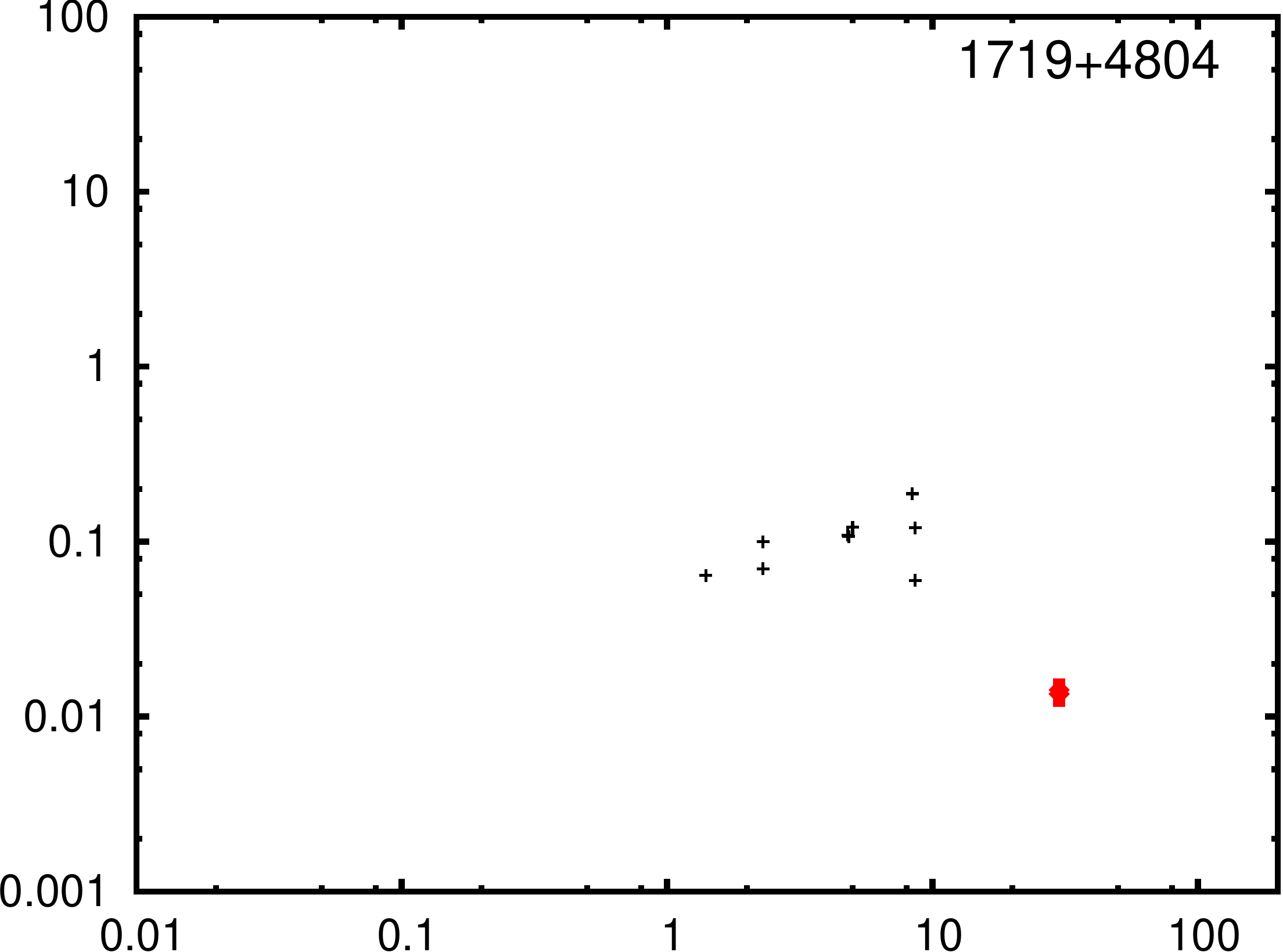}
\includegraphics[scale=0.2]{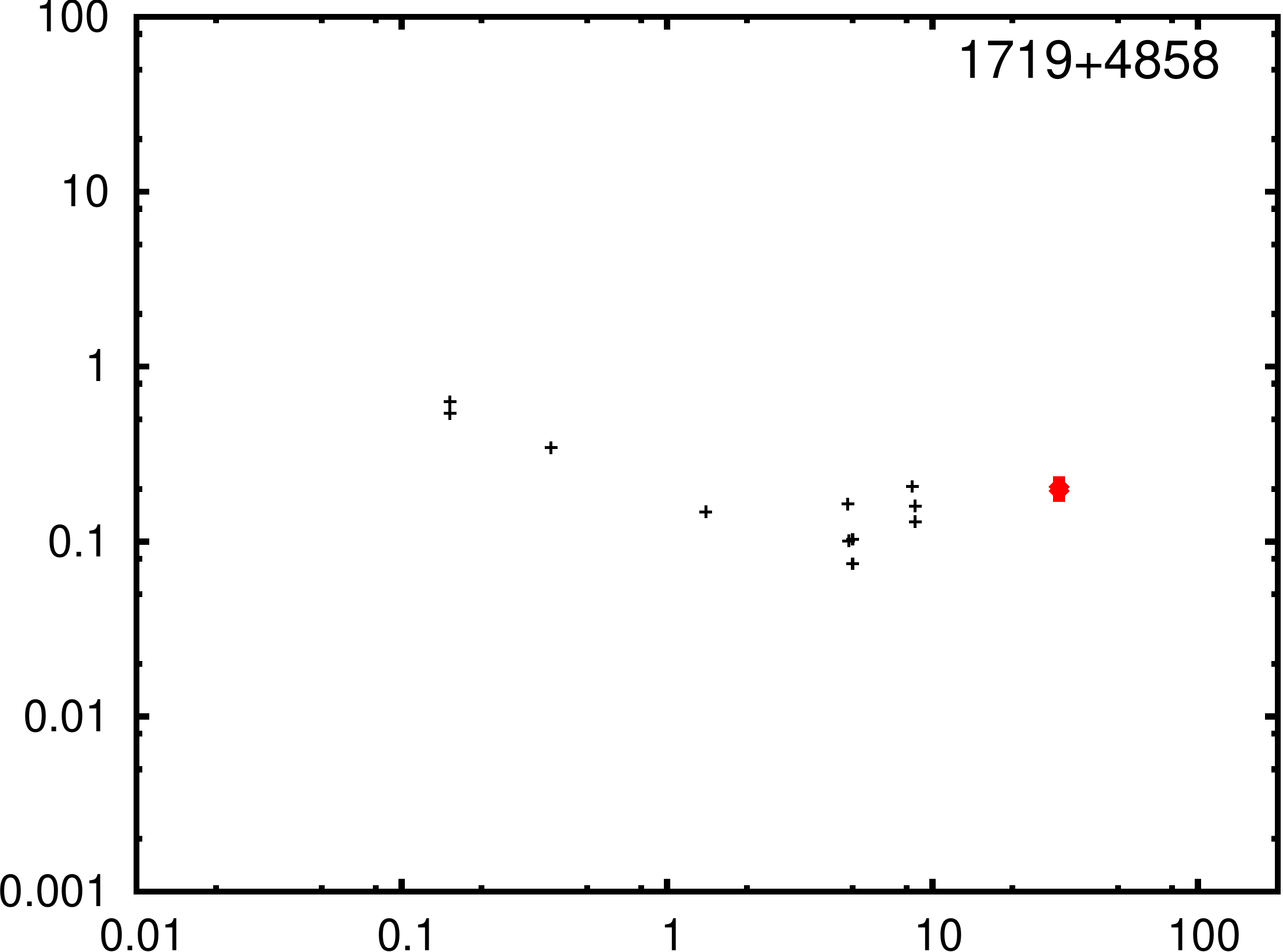}
\includegraphics[scale=0.2]{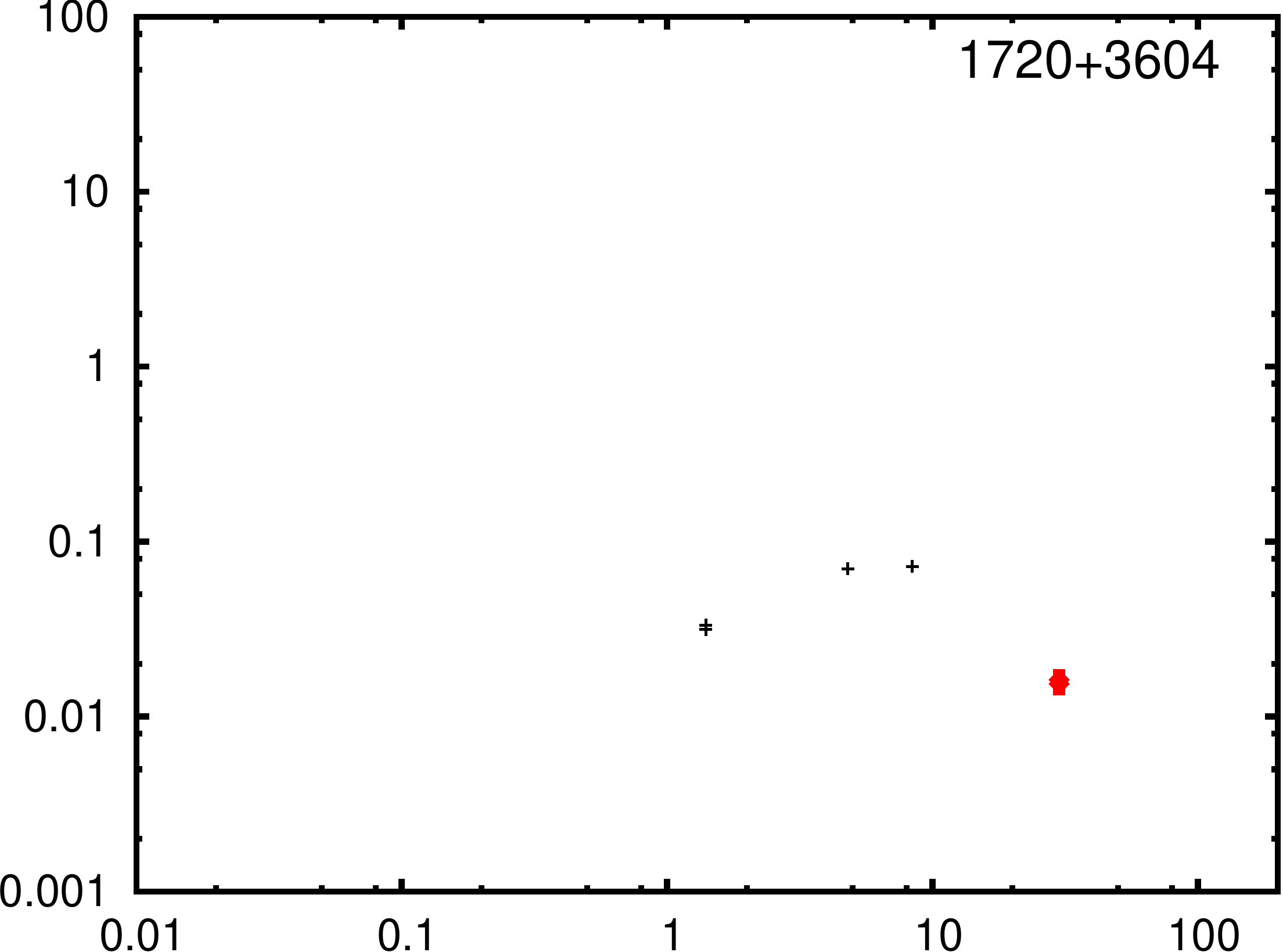}
\includegraphics[scale=0.2]{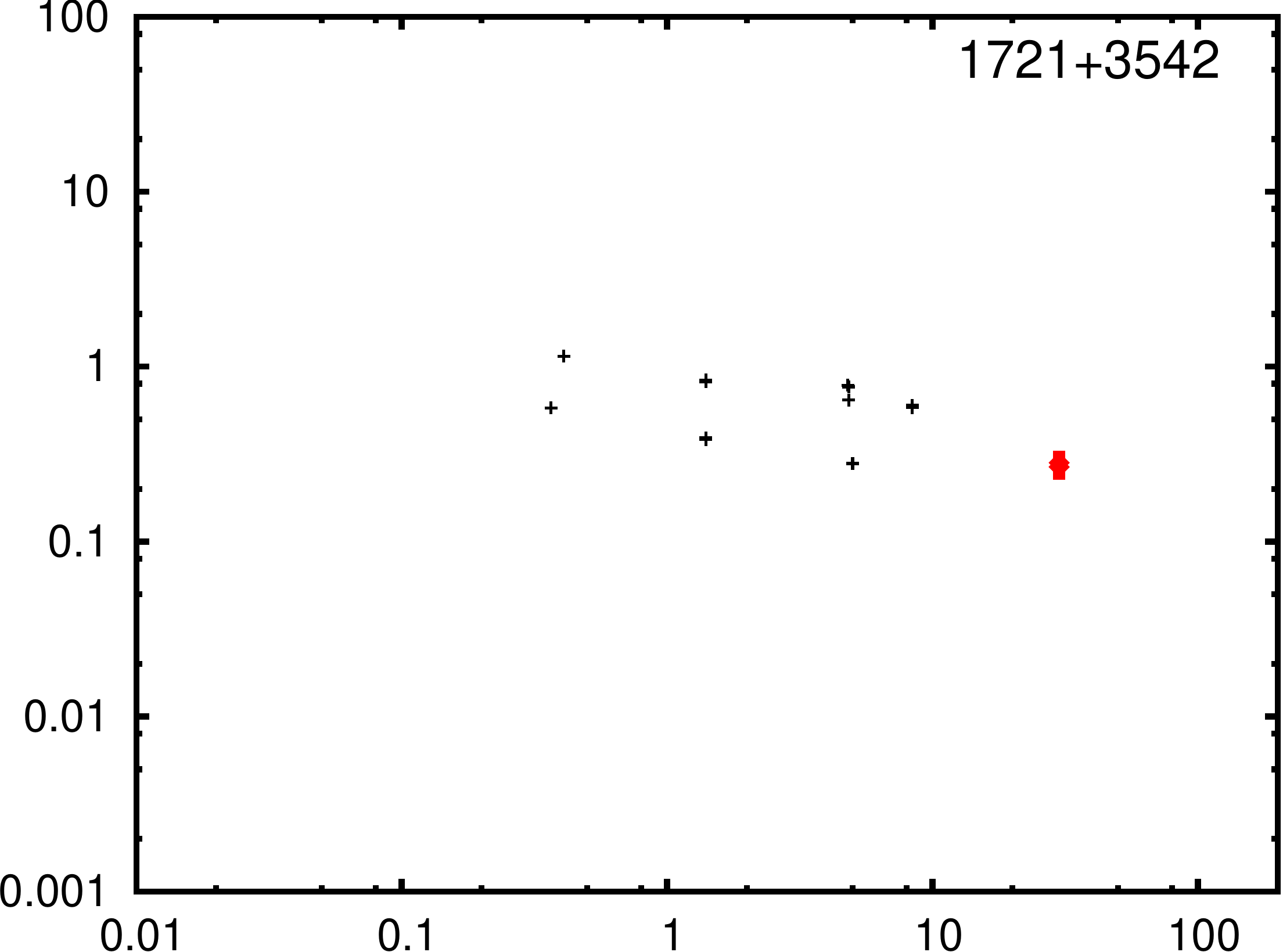}
\includegraphics[scale=0.2]{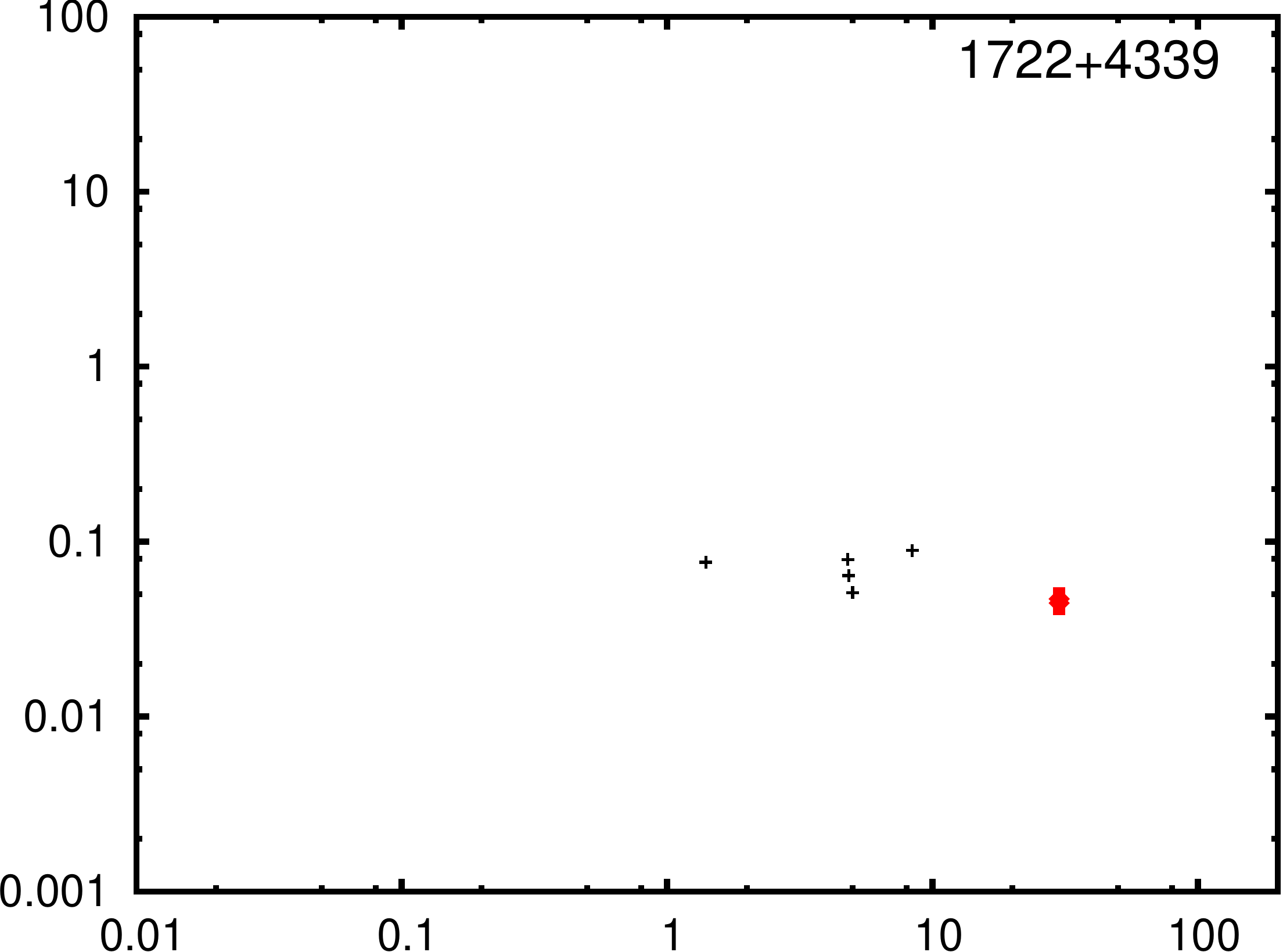}
\includegraphics[scale=0.2]{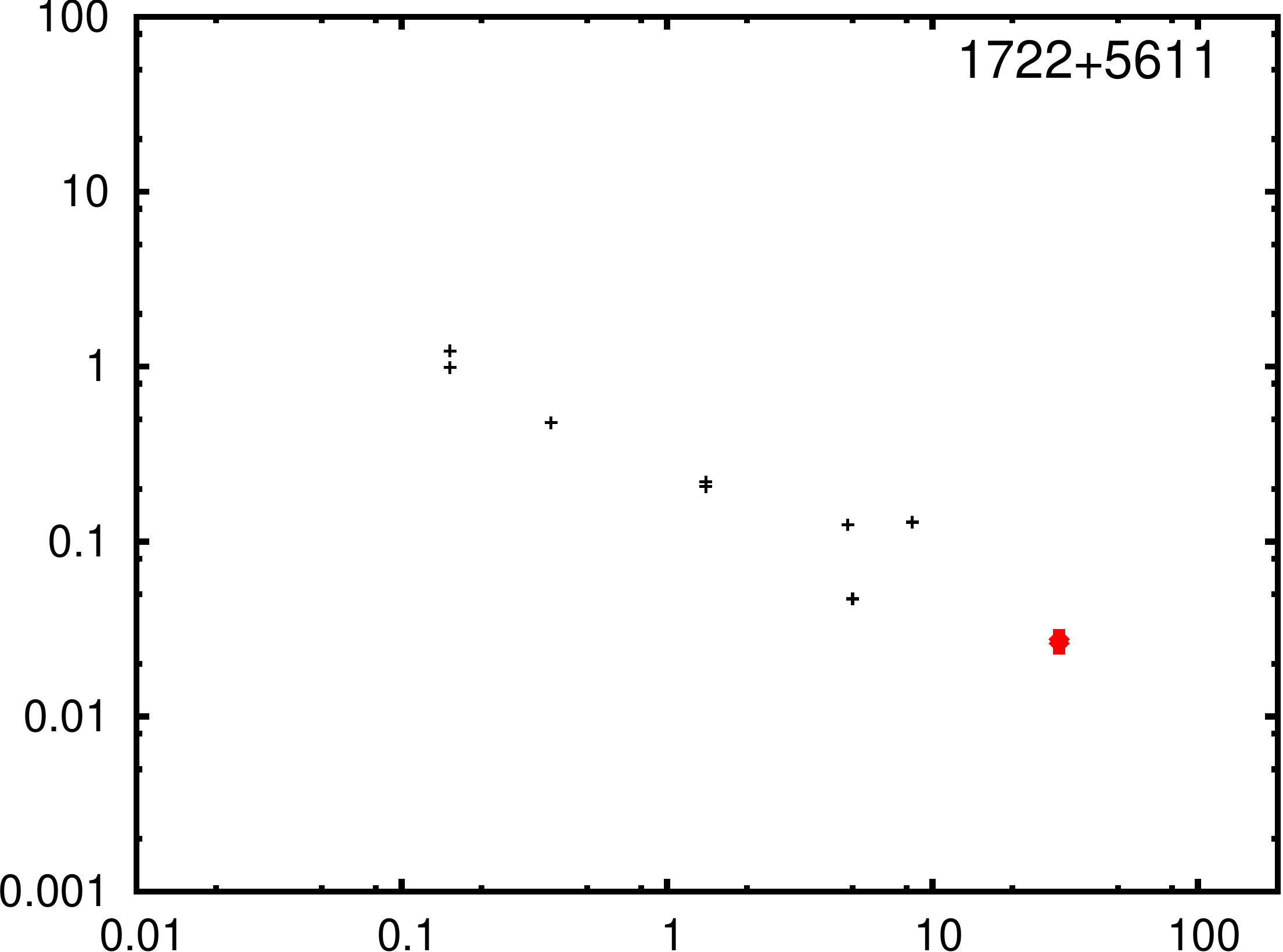}
\includegraphics[scale=0.2]{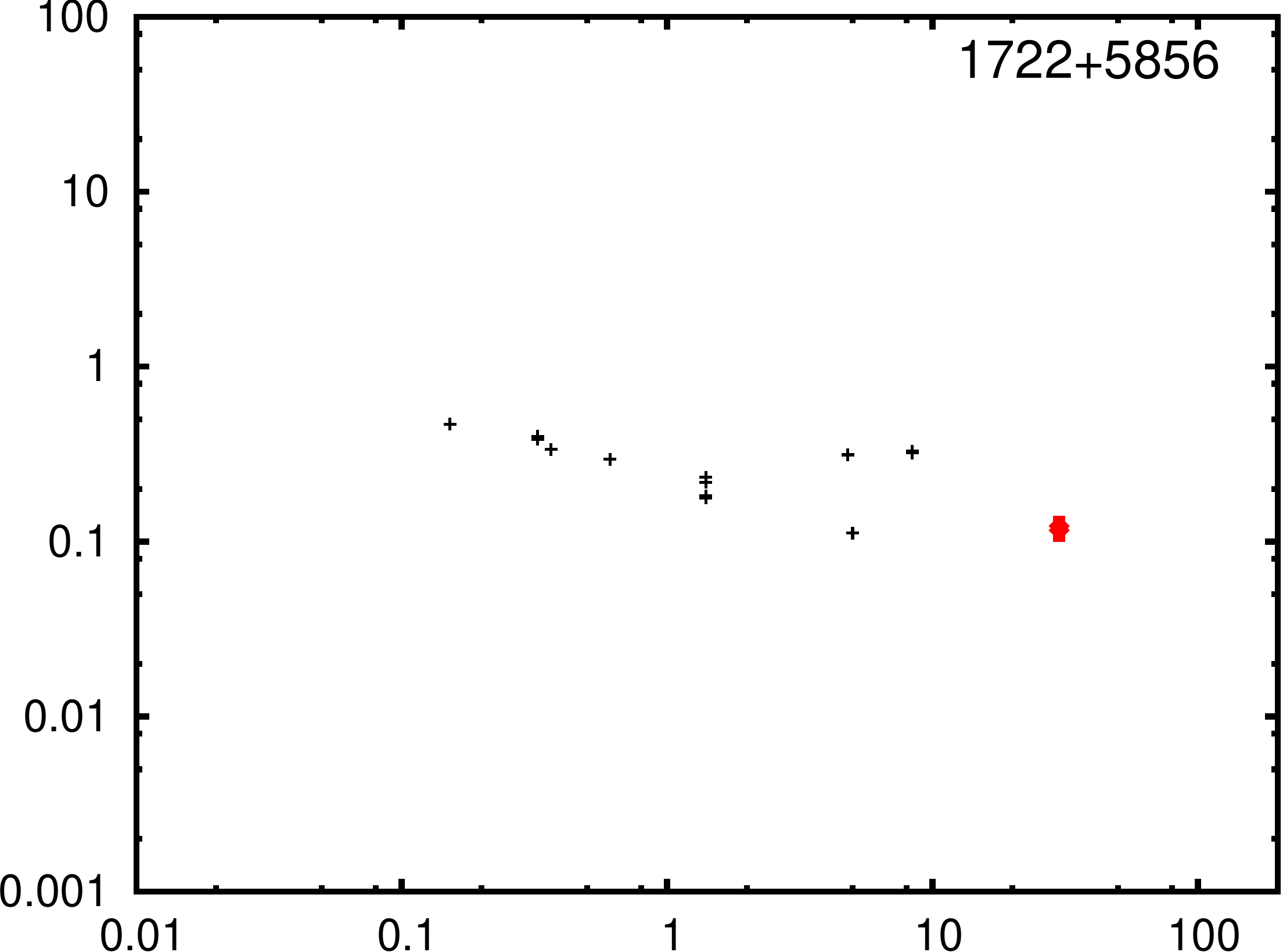}
\end{figure}
\clearpage\begin{figure}
\centering
\includegraphics[scale=0.2]{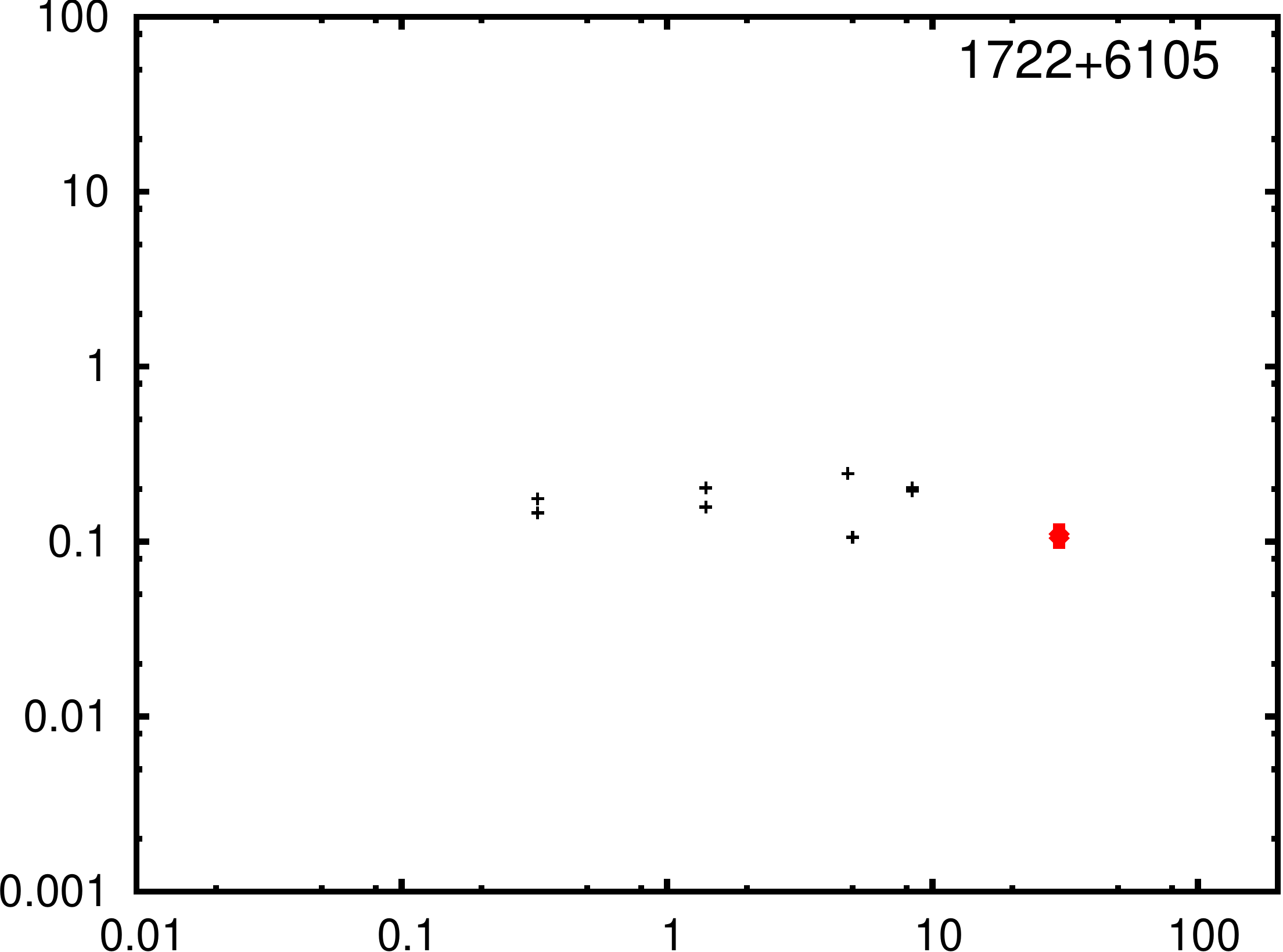}
\includegraphics[scale=0.2]{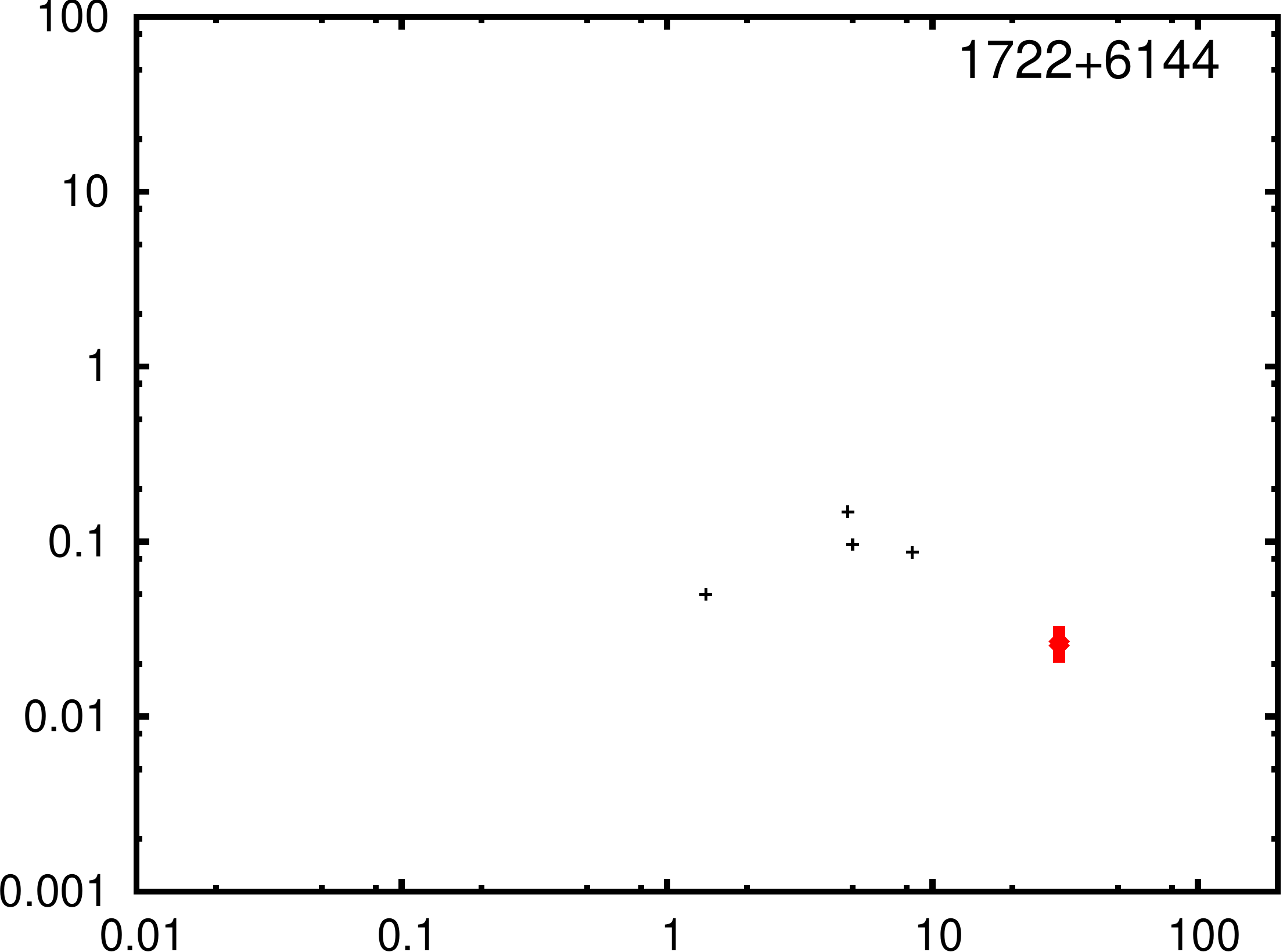}
\includegraphics[scale=0.2]{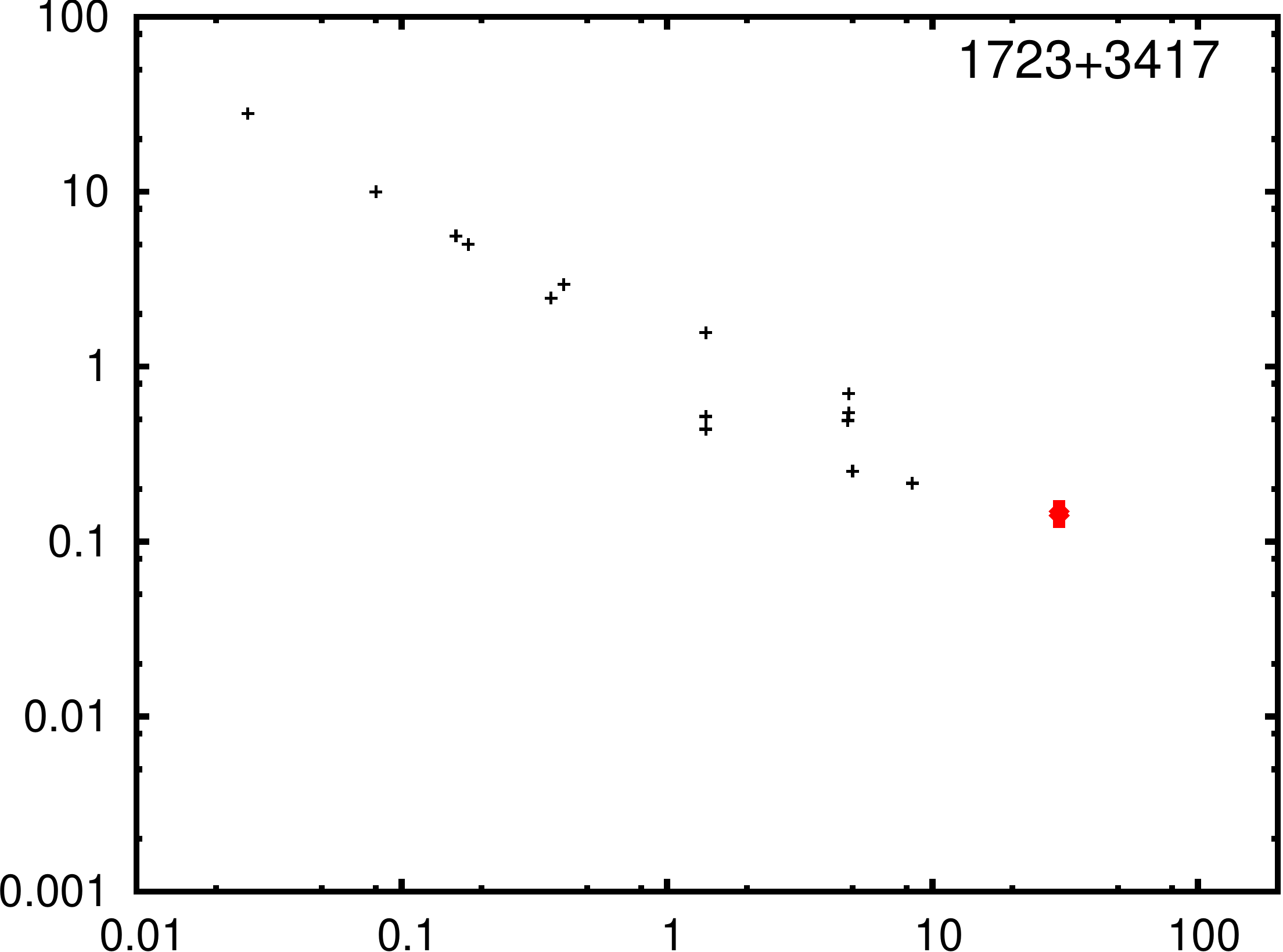}
\includegraphics[scale=0.2]{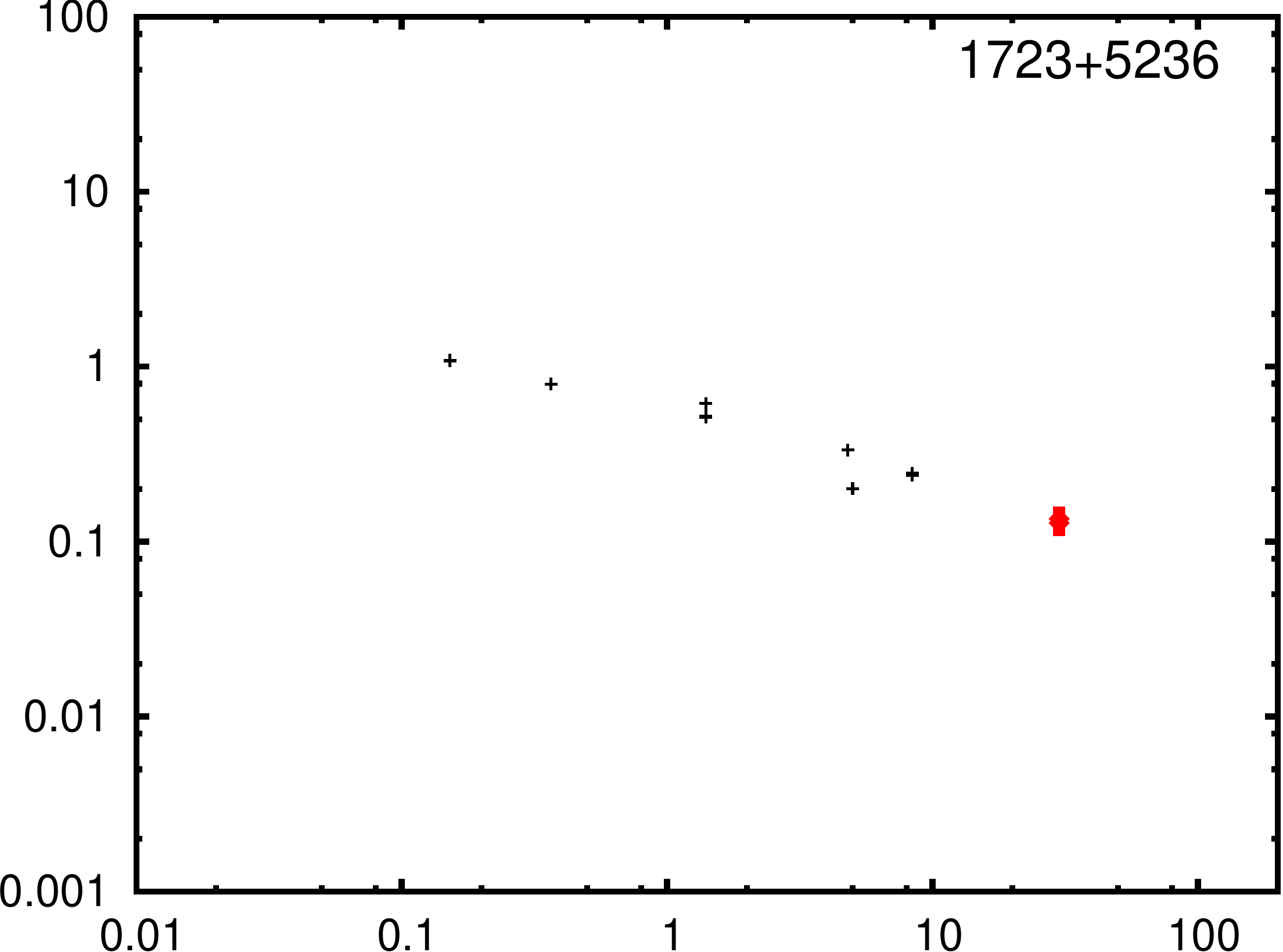}
\includegraphics[scale=0.2]{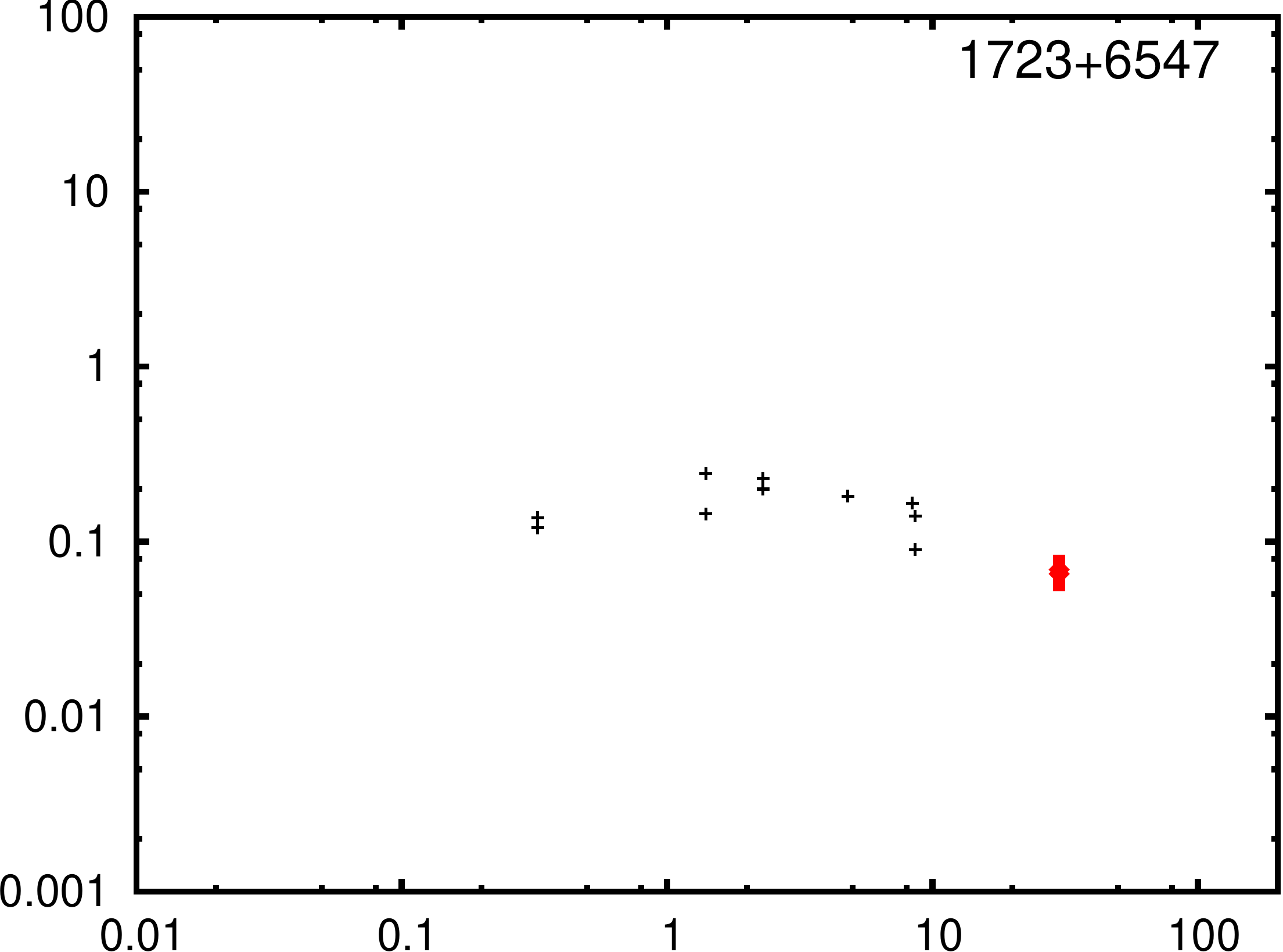}
\includegraphics[scale=0.2]{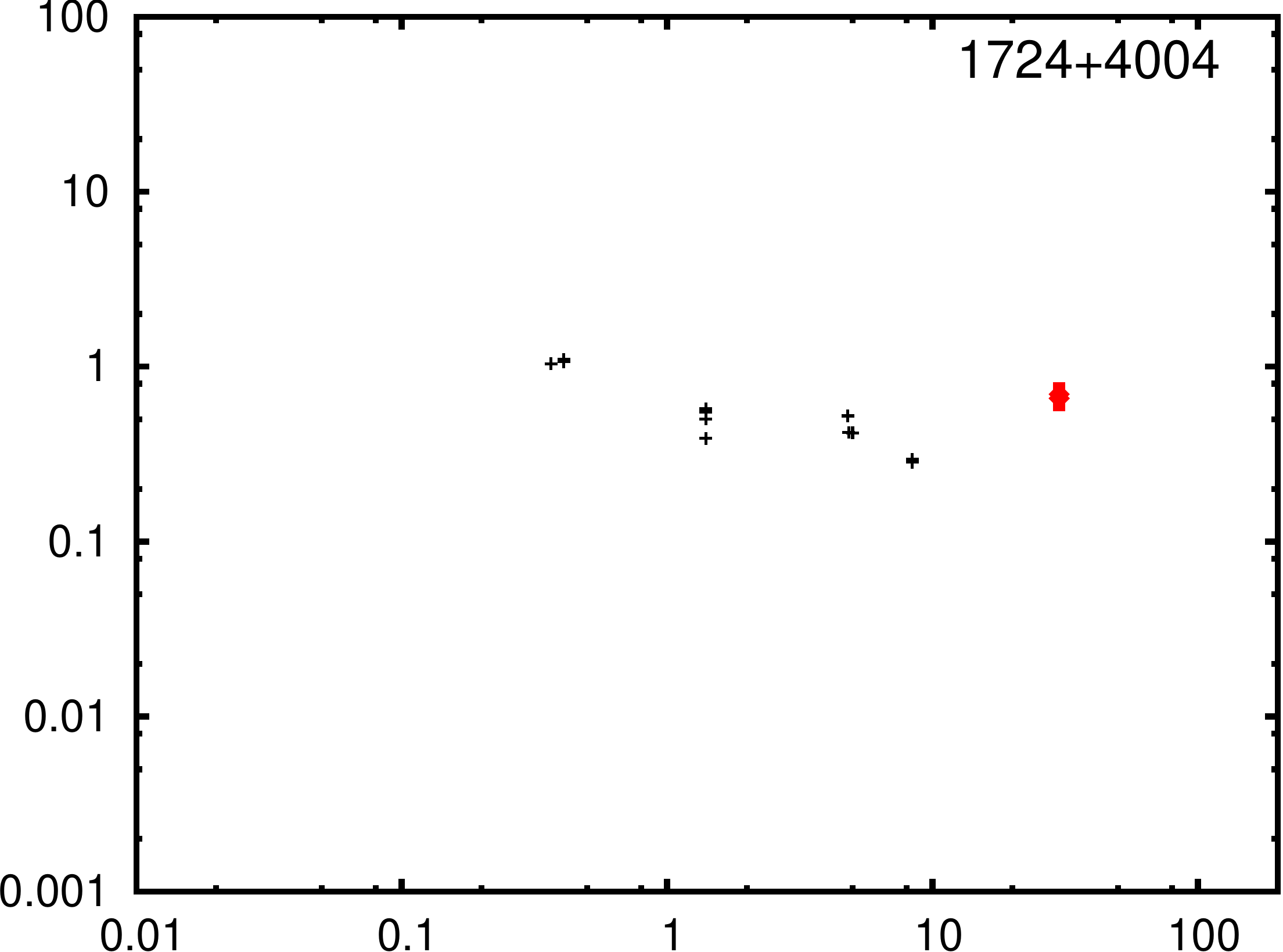}
\includegraphics[scale=0.2]{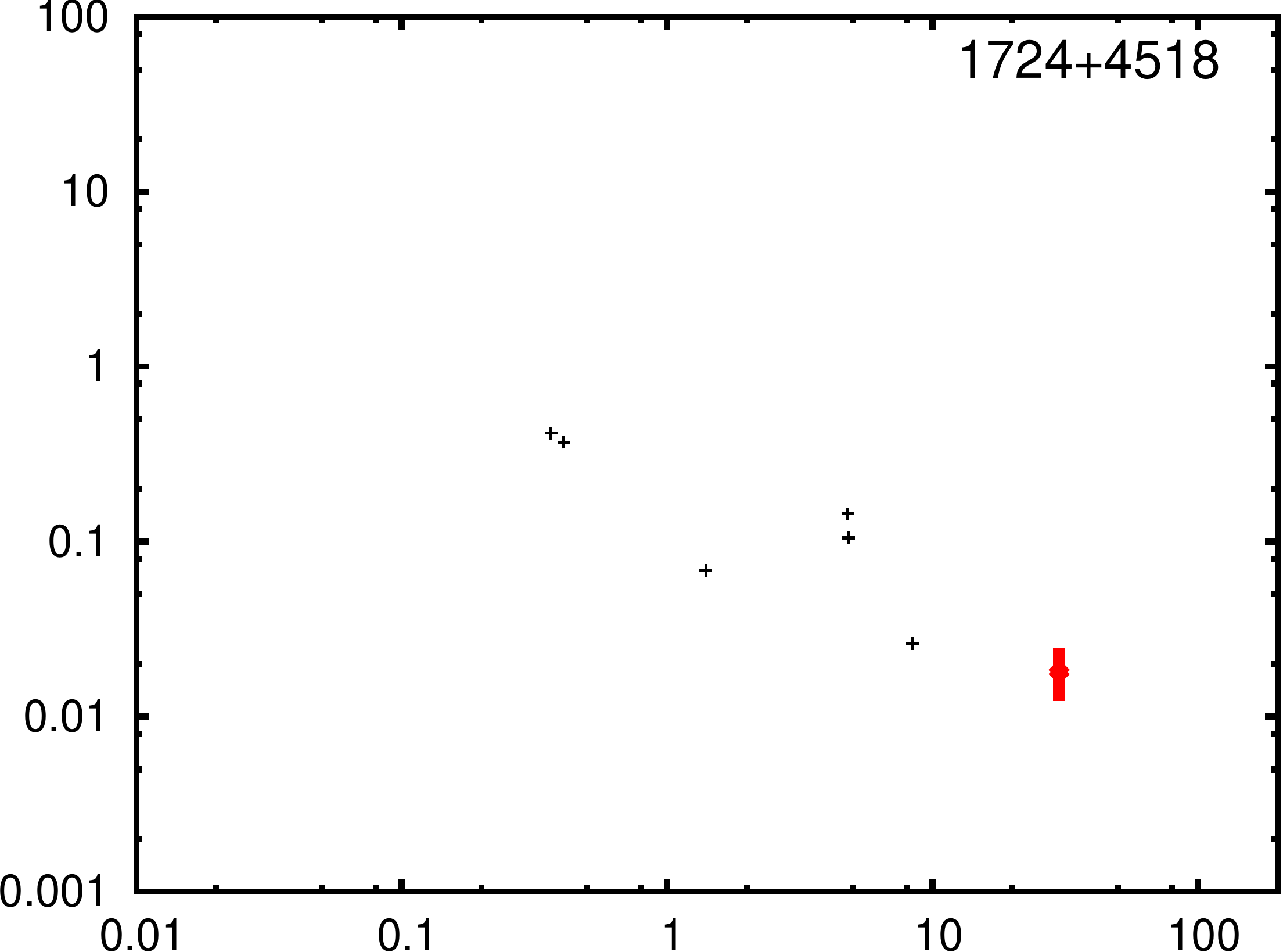}
\includegraphics[scale=0.2]{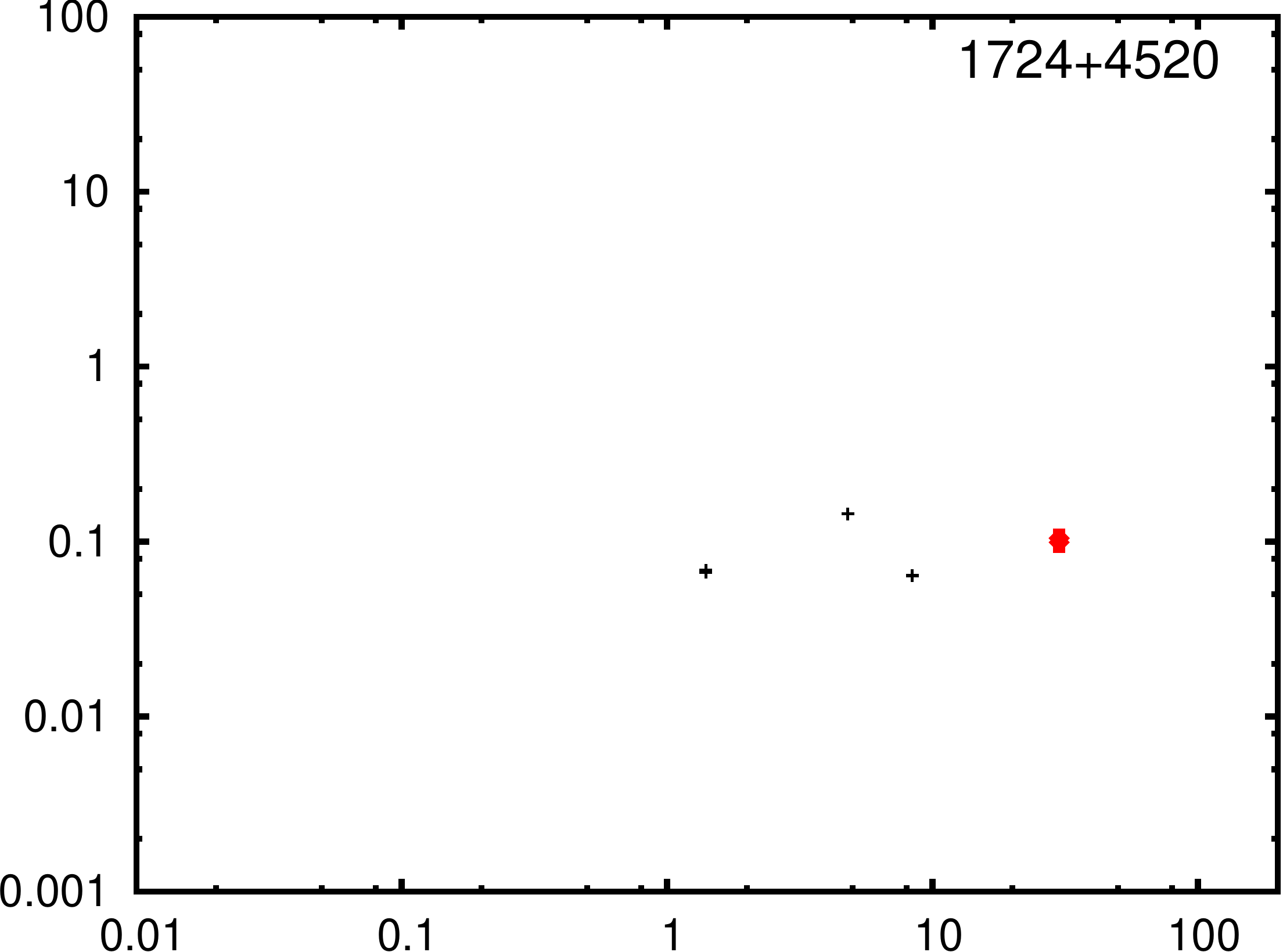}
\includegraphics[scale=0.2]{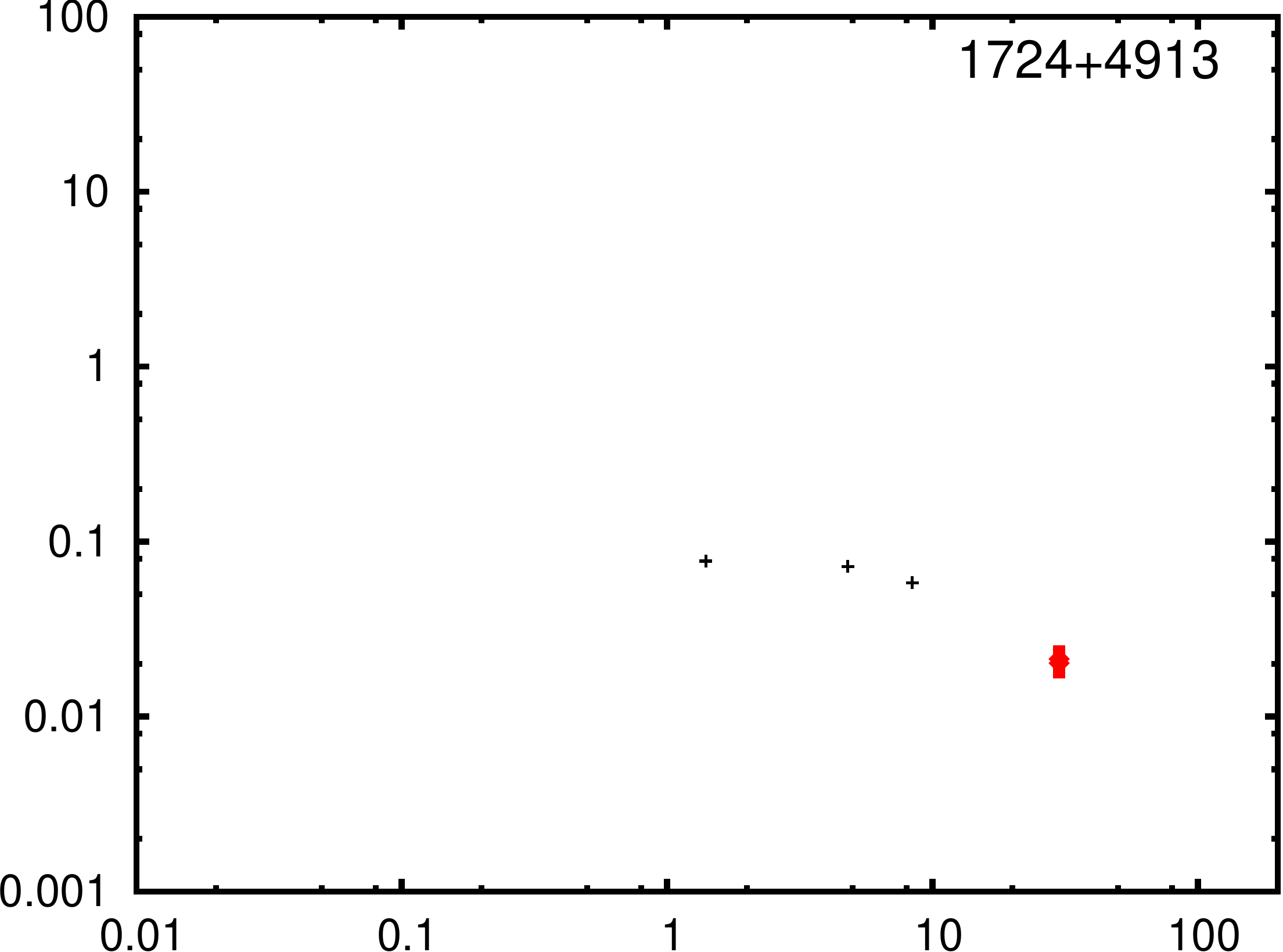}
\includegraphics[scale=0.2]{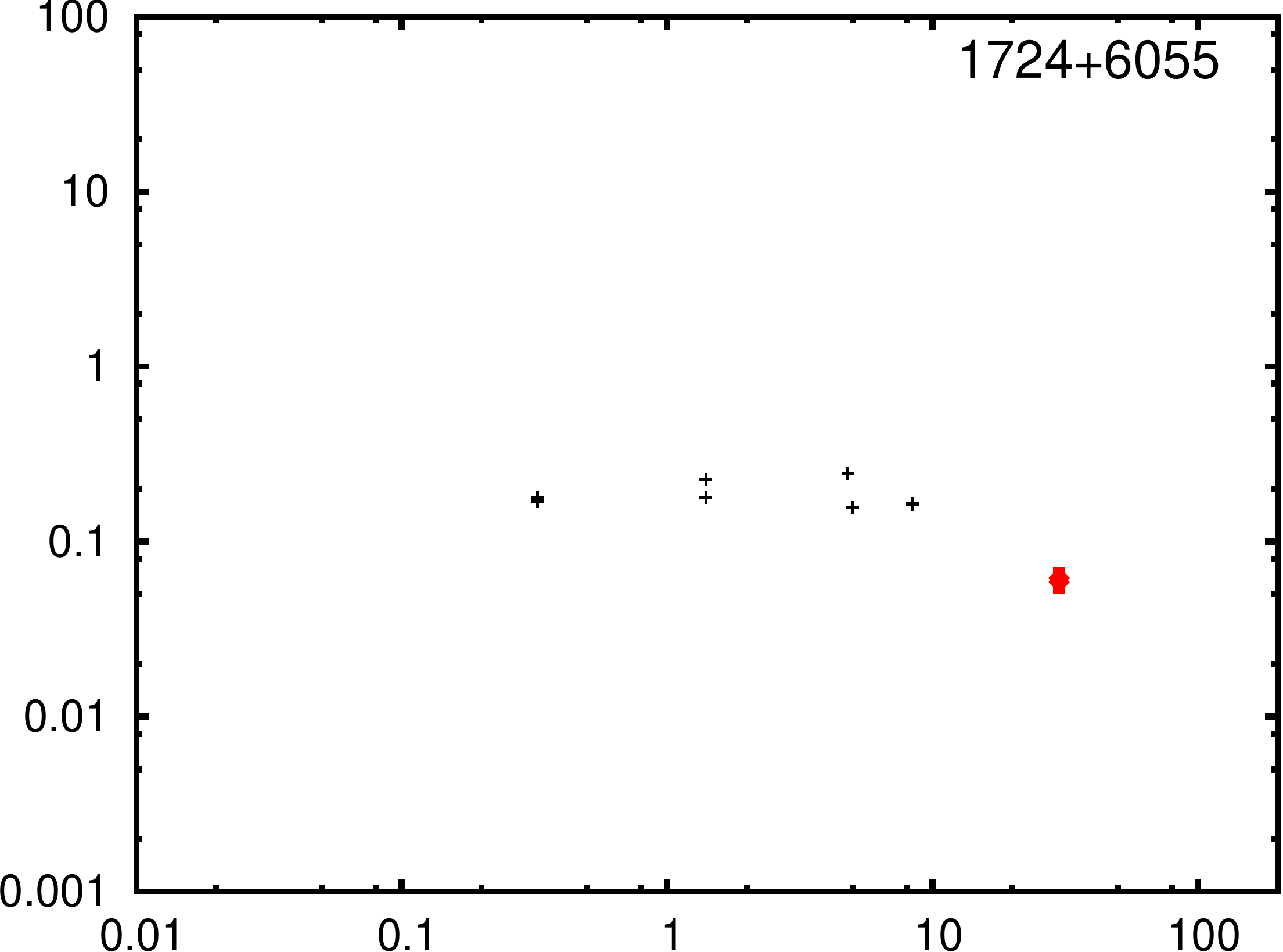}
\includegraphics[scale=0.2]{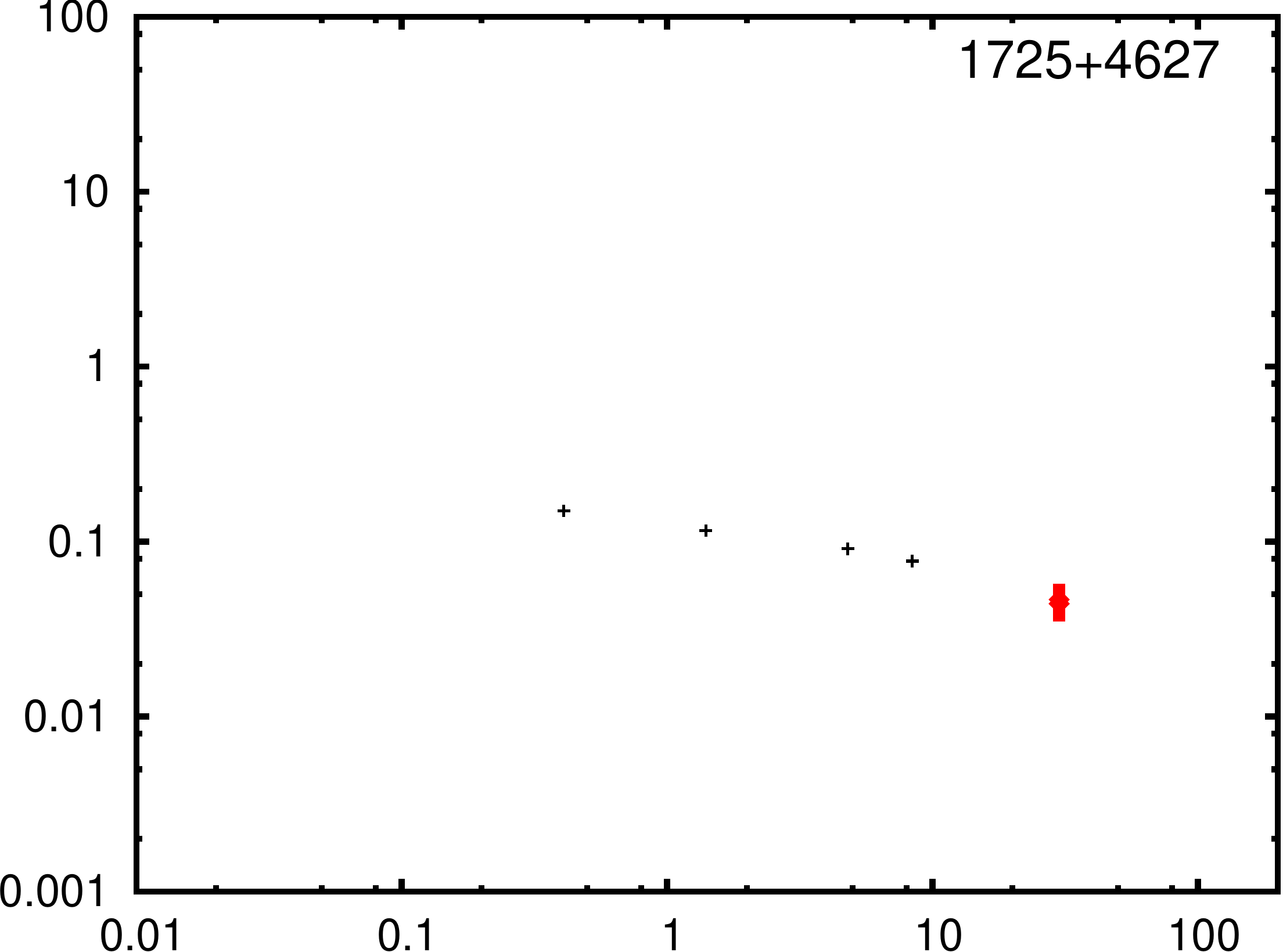}
\includegraphics[scale=0.2]{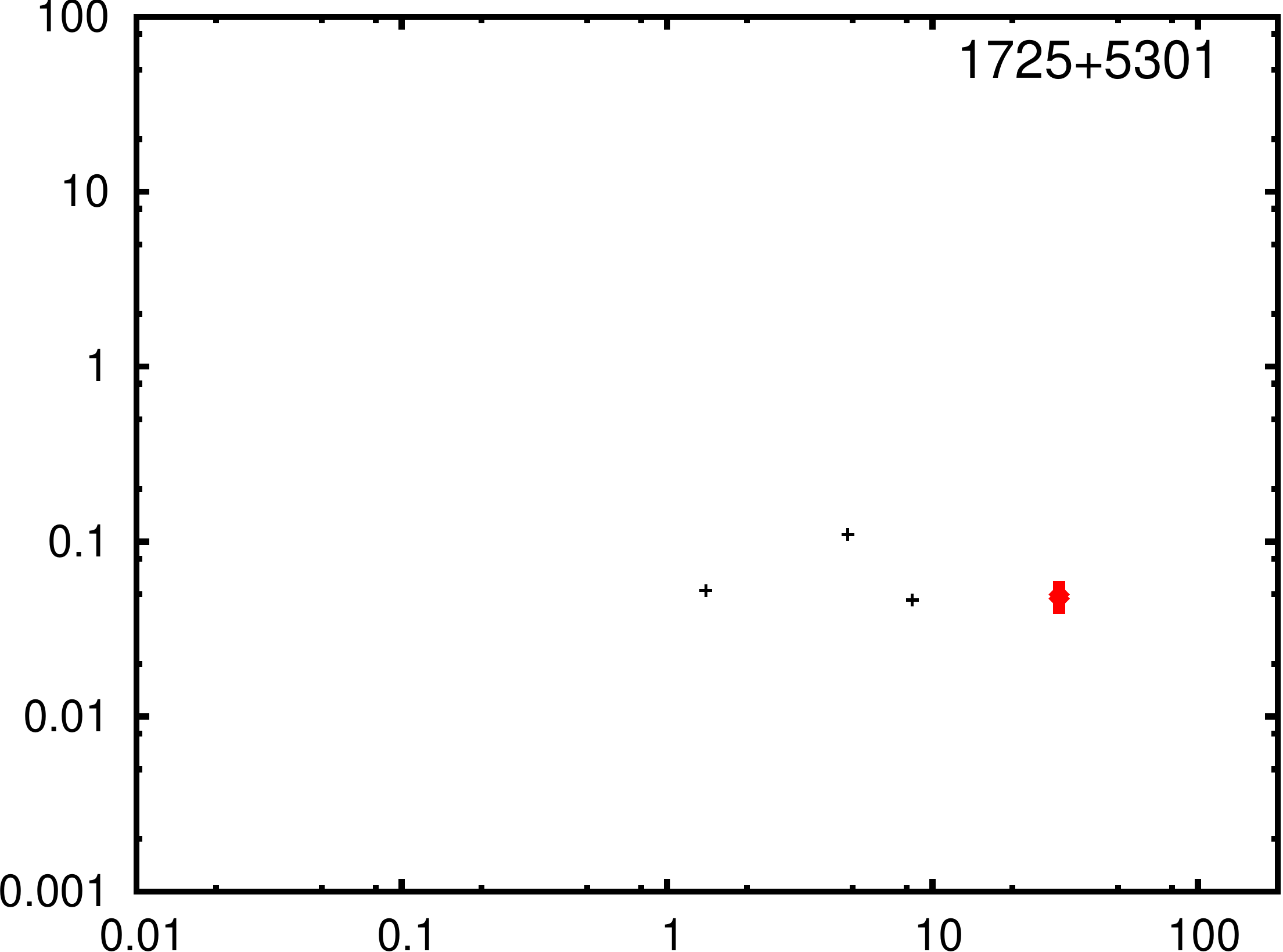}
\includegraphics[scale=0.2]{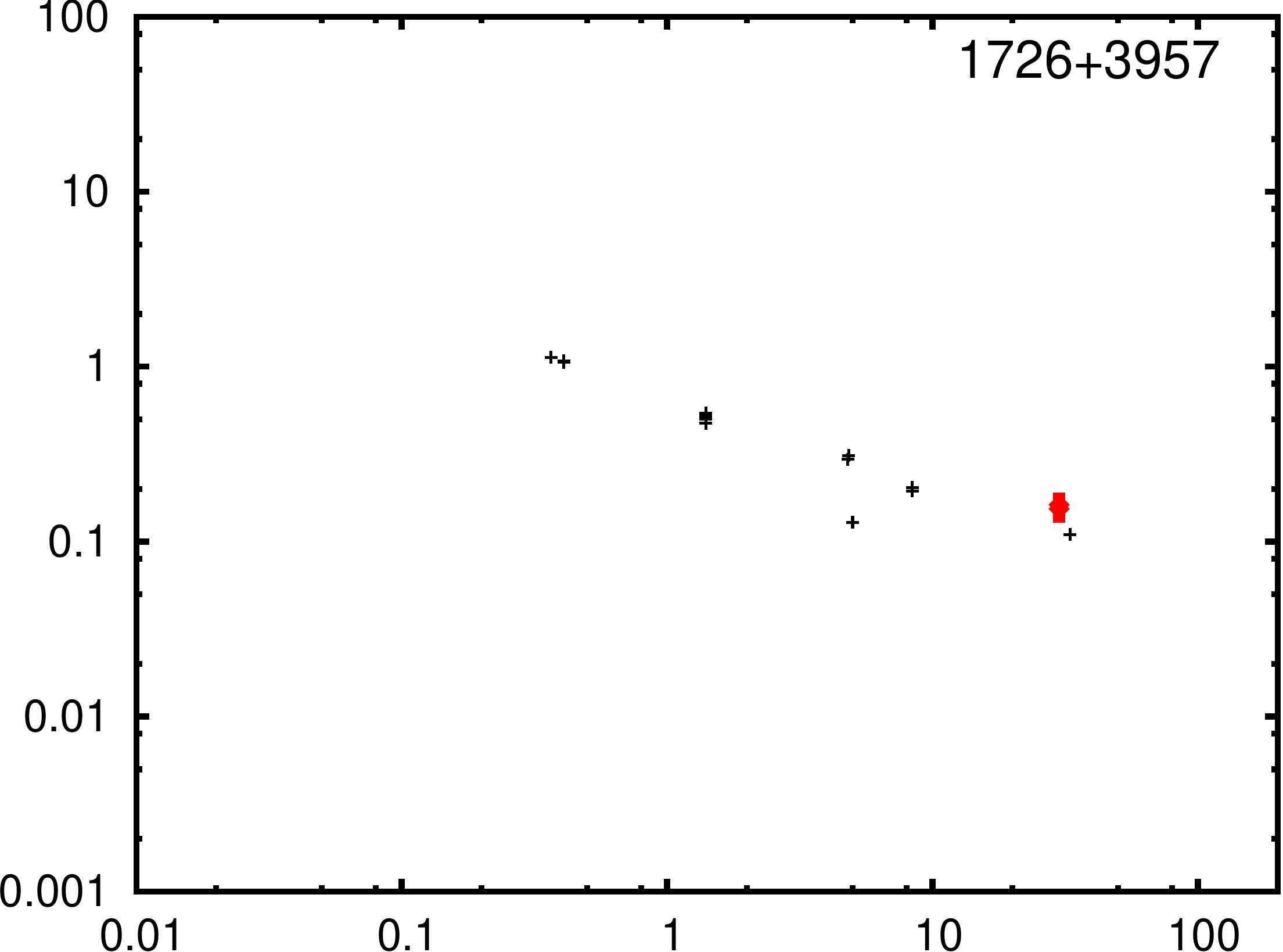}
\includegraphics[scale=0.2]{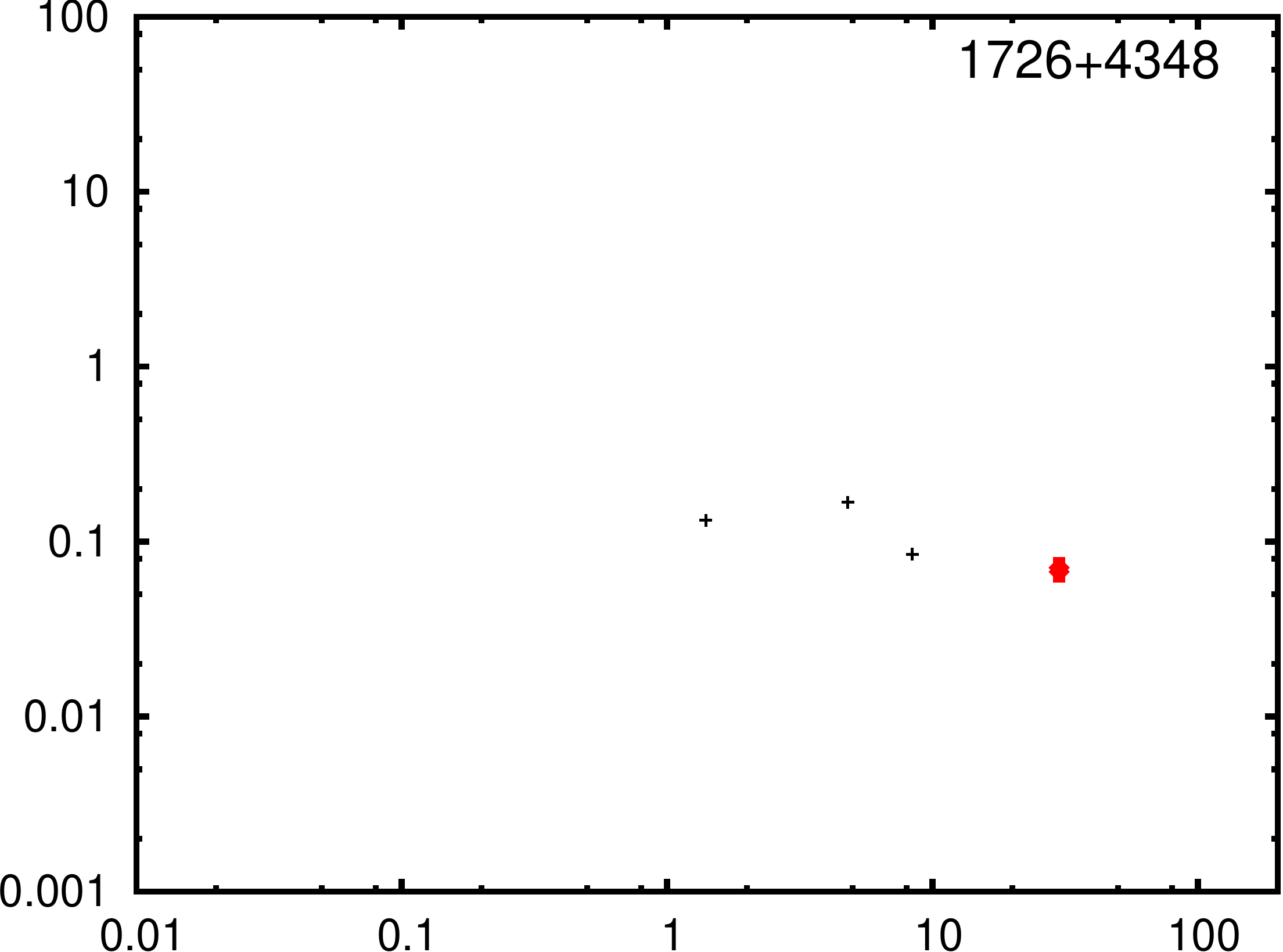}
\includegraphics[scale=0.2]{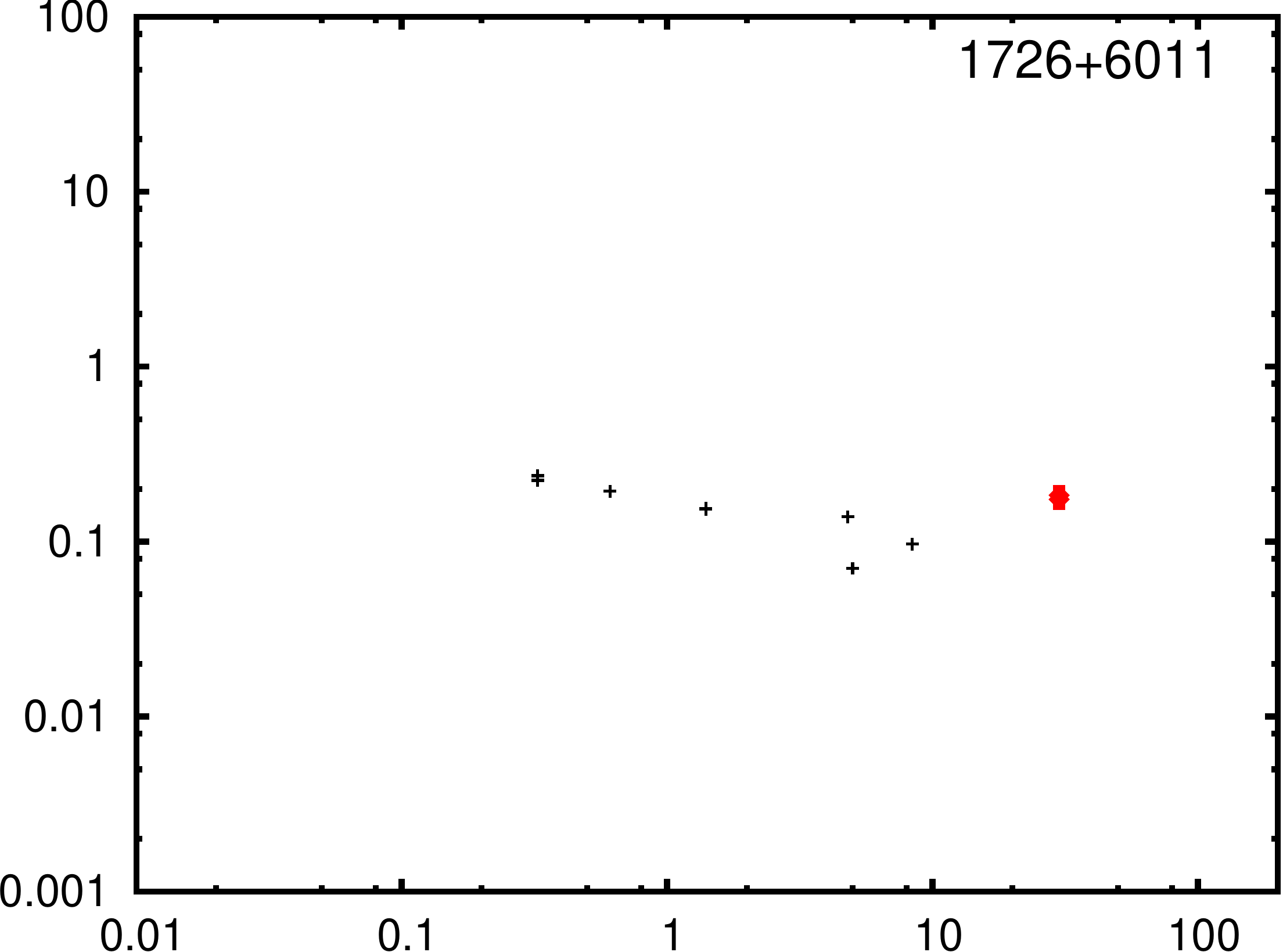}
\includegraphics[scale=0.2]{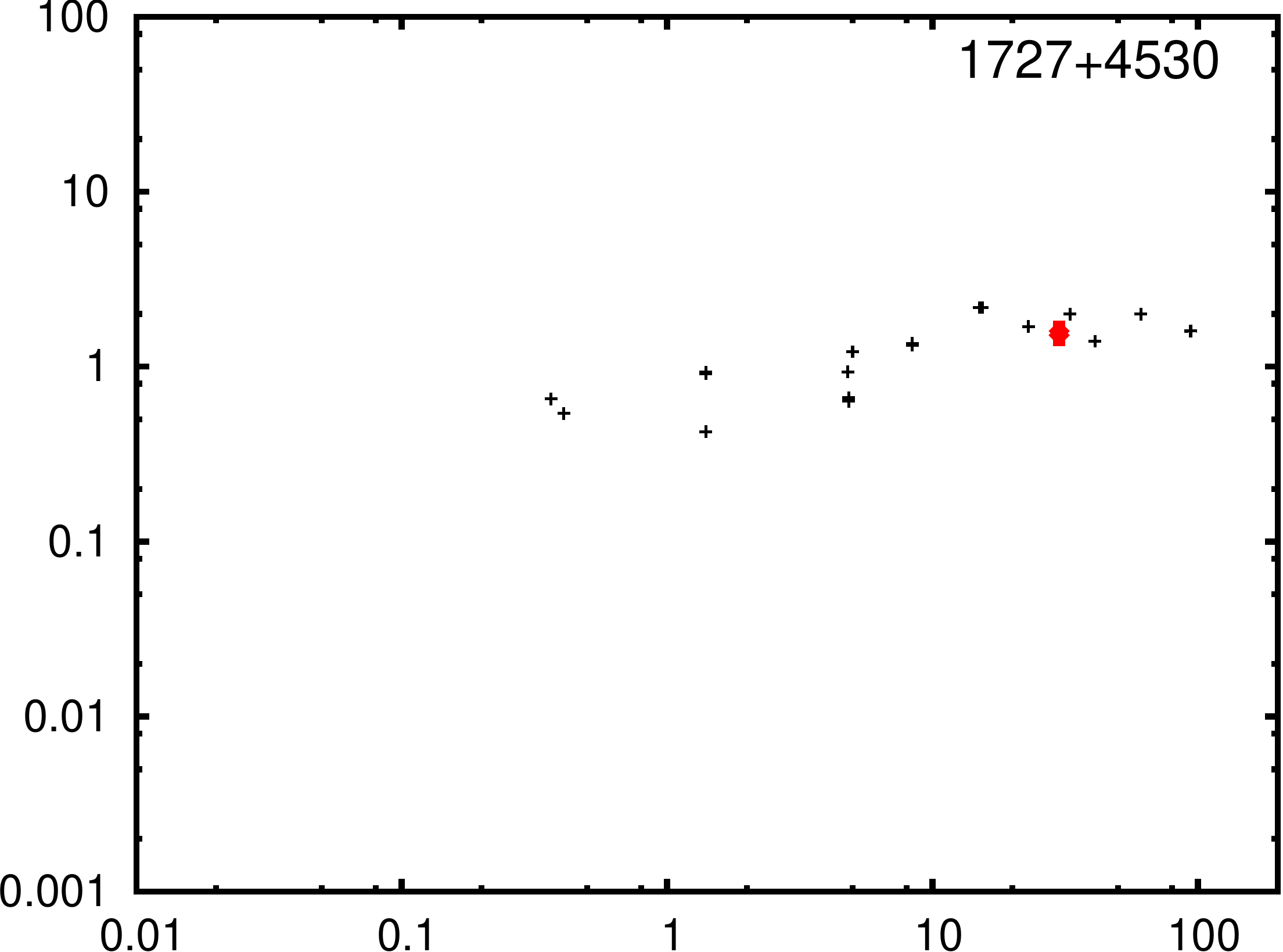}
\includegraphics[scale=0.2]{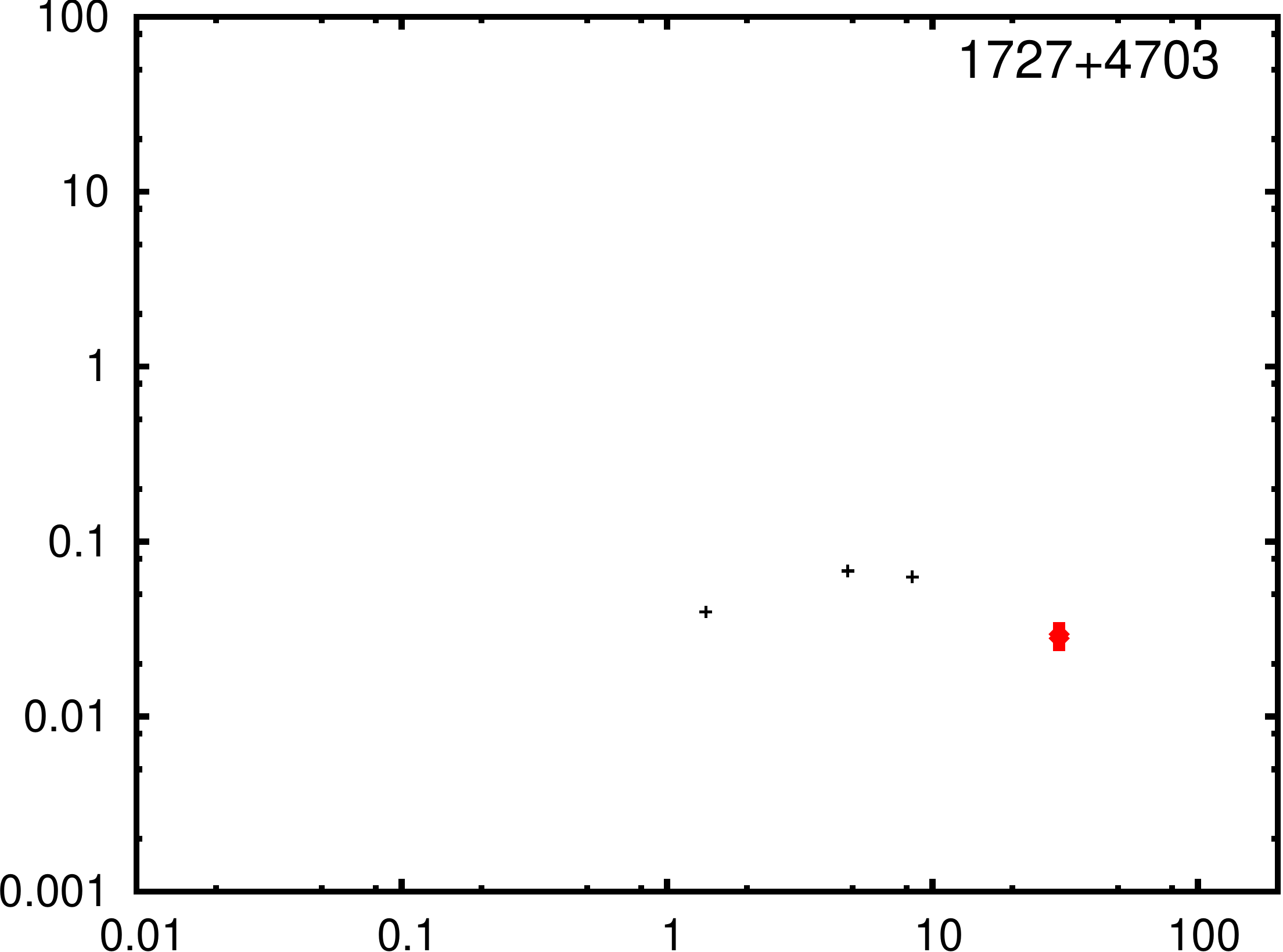}
\includegraphics[scale=0.2]{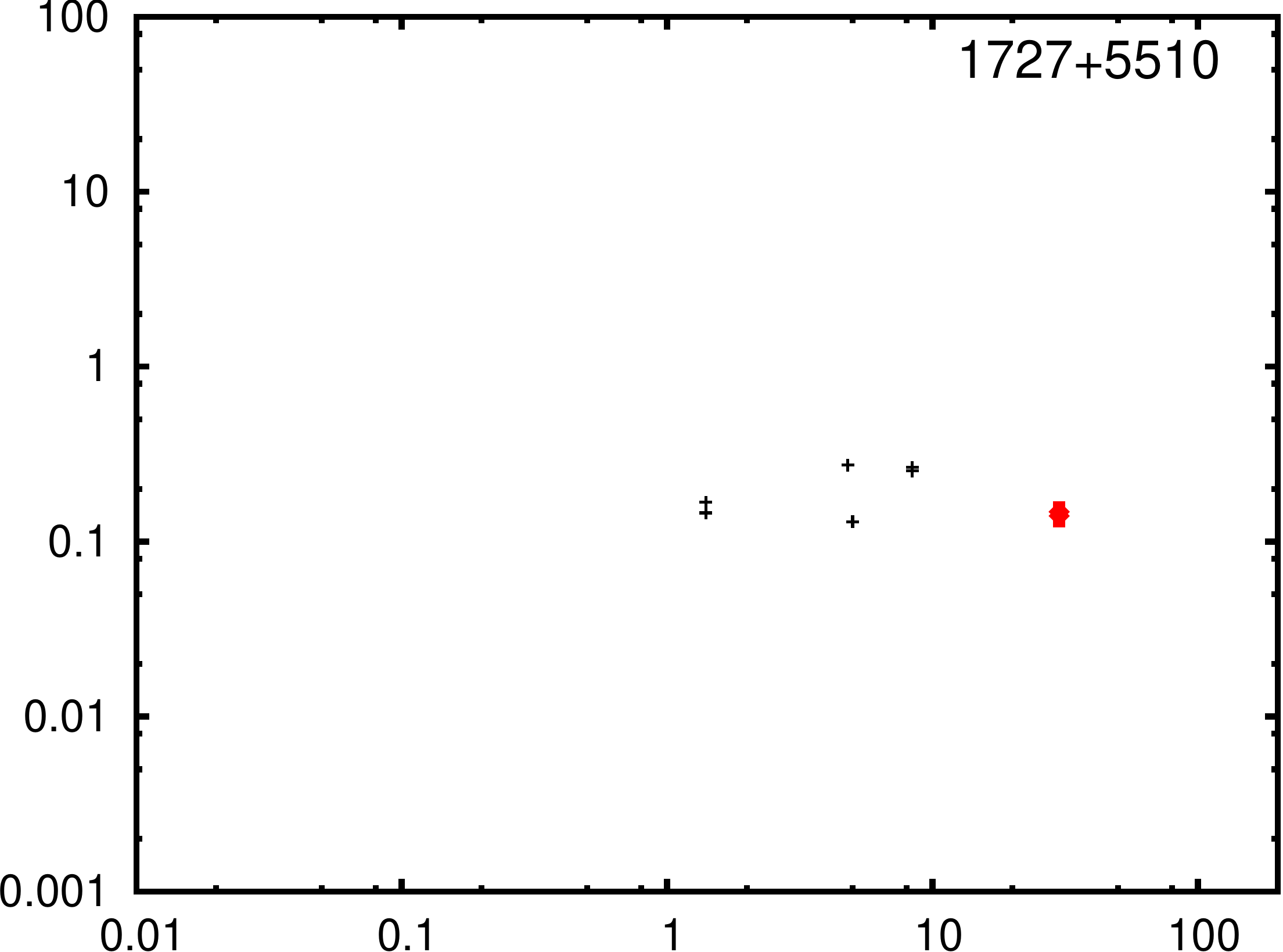}
\end{figure}
\clearpage\begin{figure}
\centering
\includegraphics[scale=0.2]{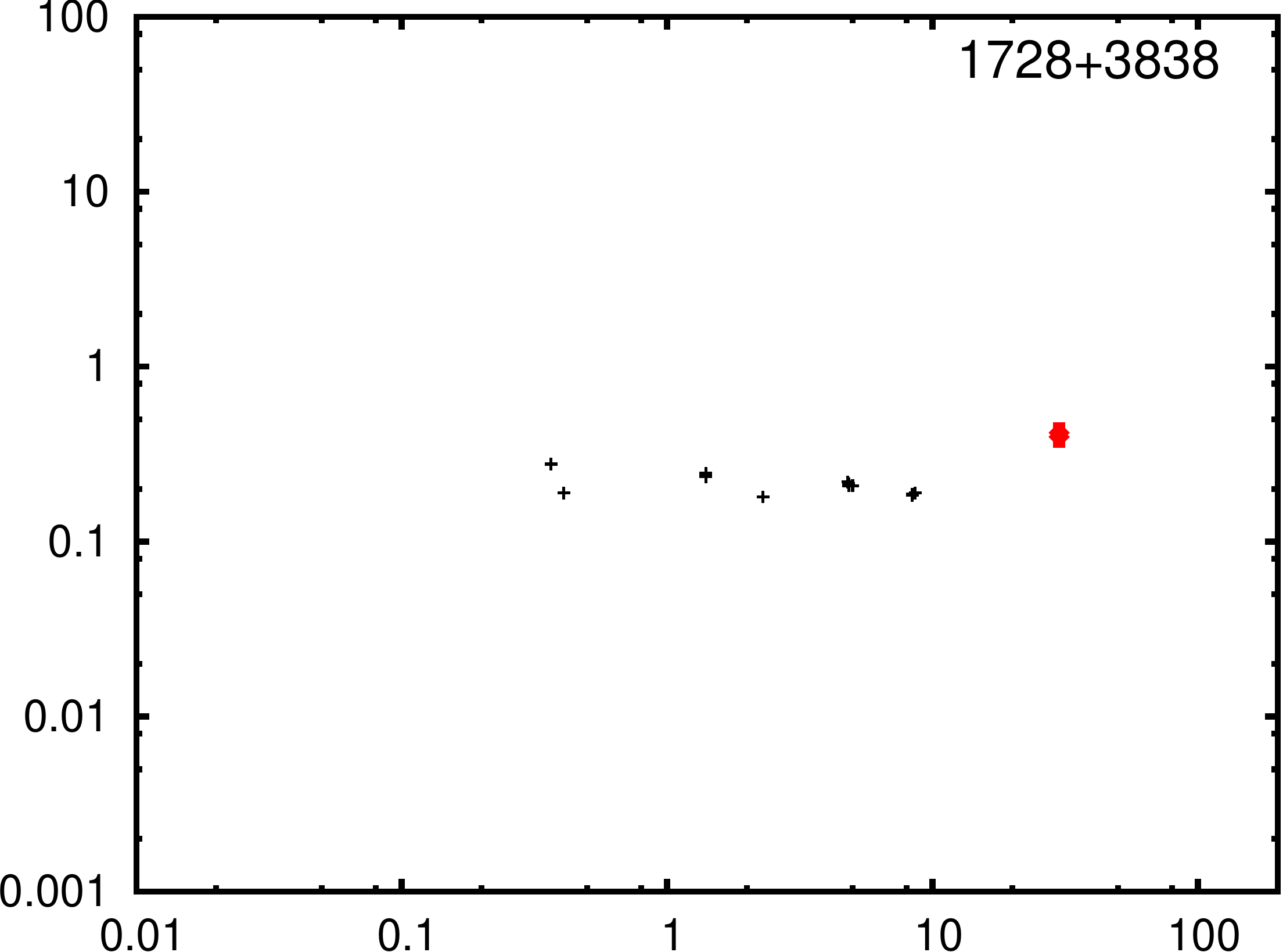}
\includegraphics[scale=0.2]{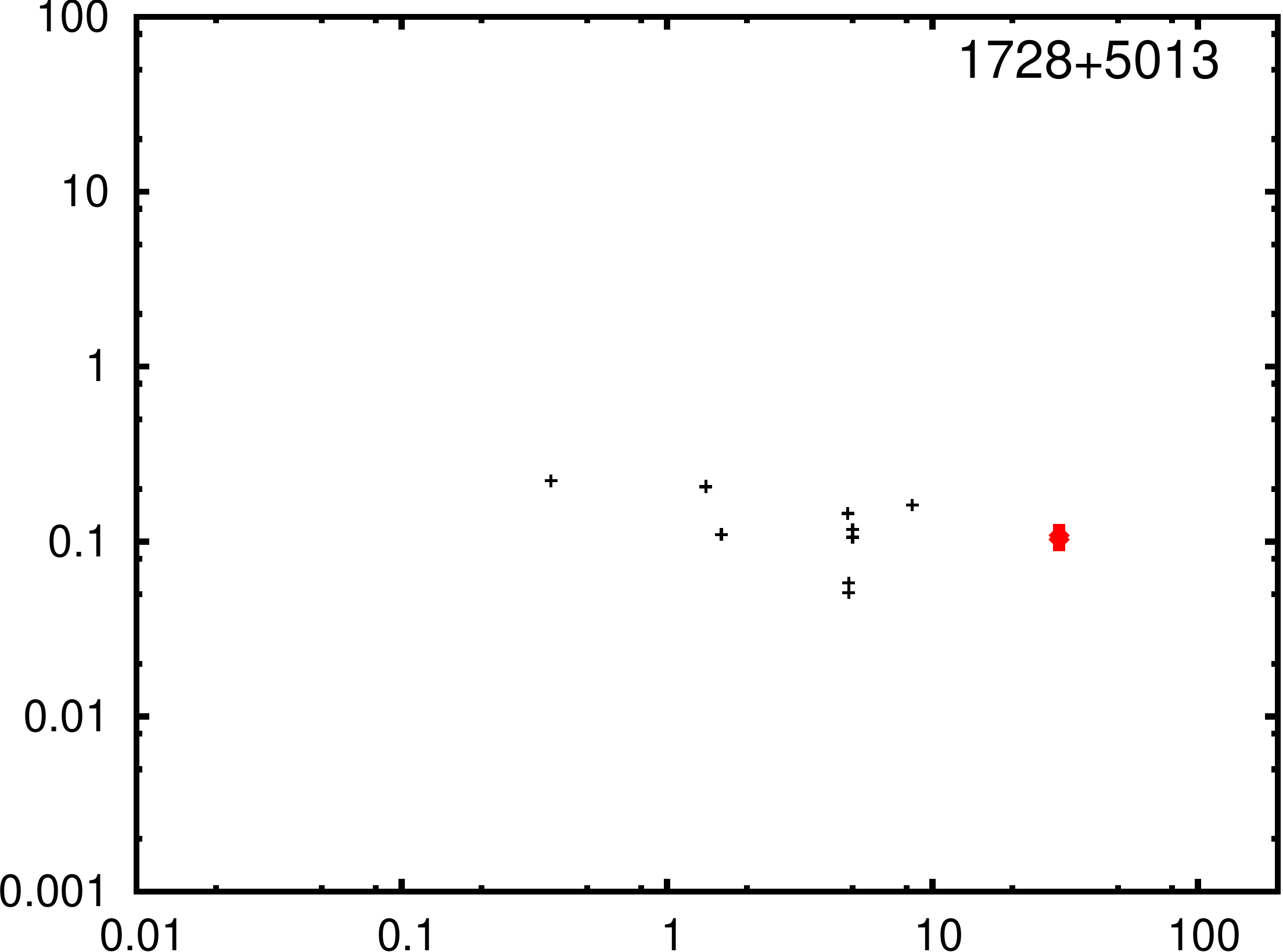}
\includegraphics[scale=0.2]{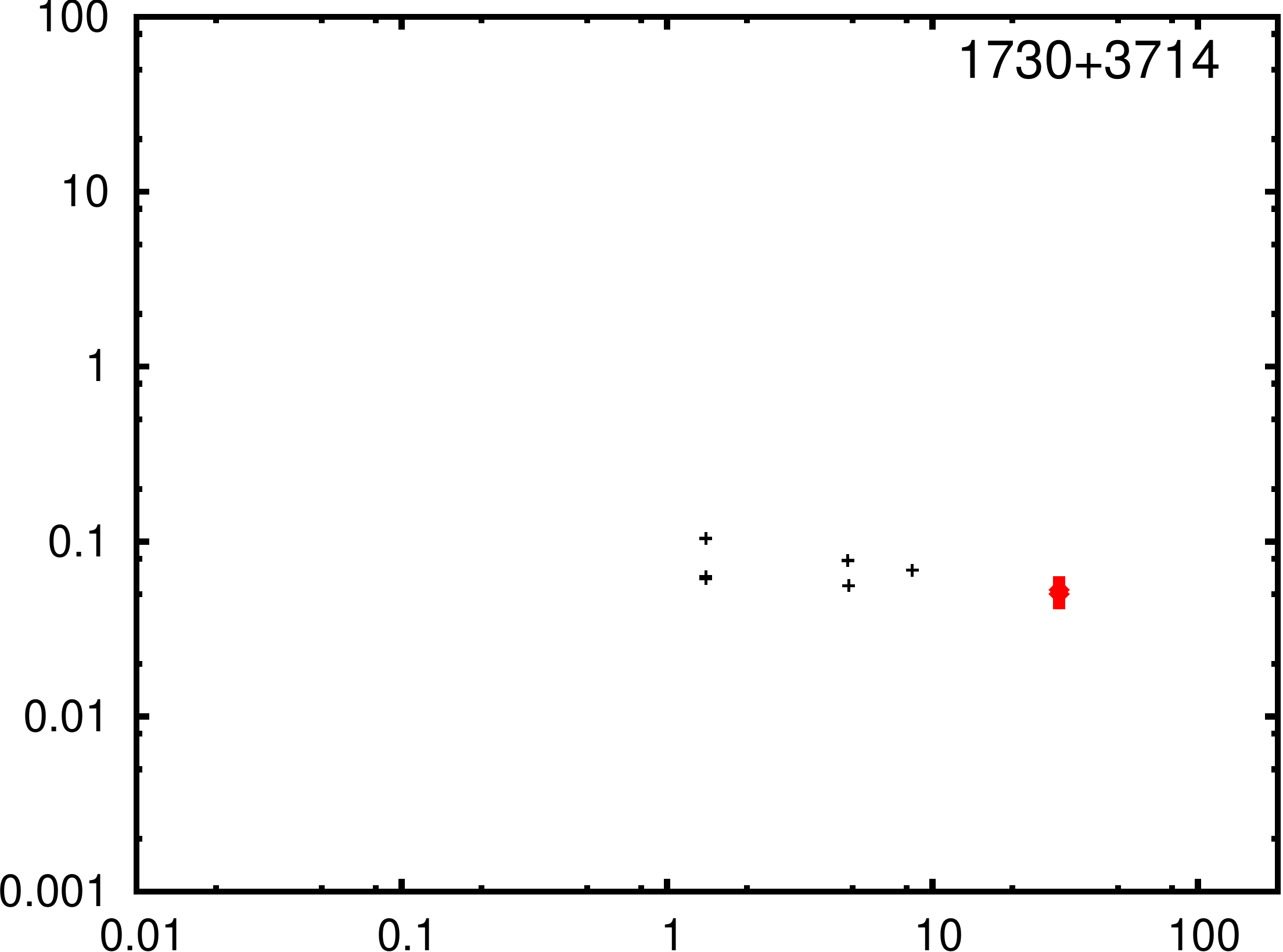}
\includegraphics[scale=0.2]{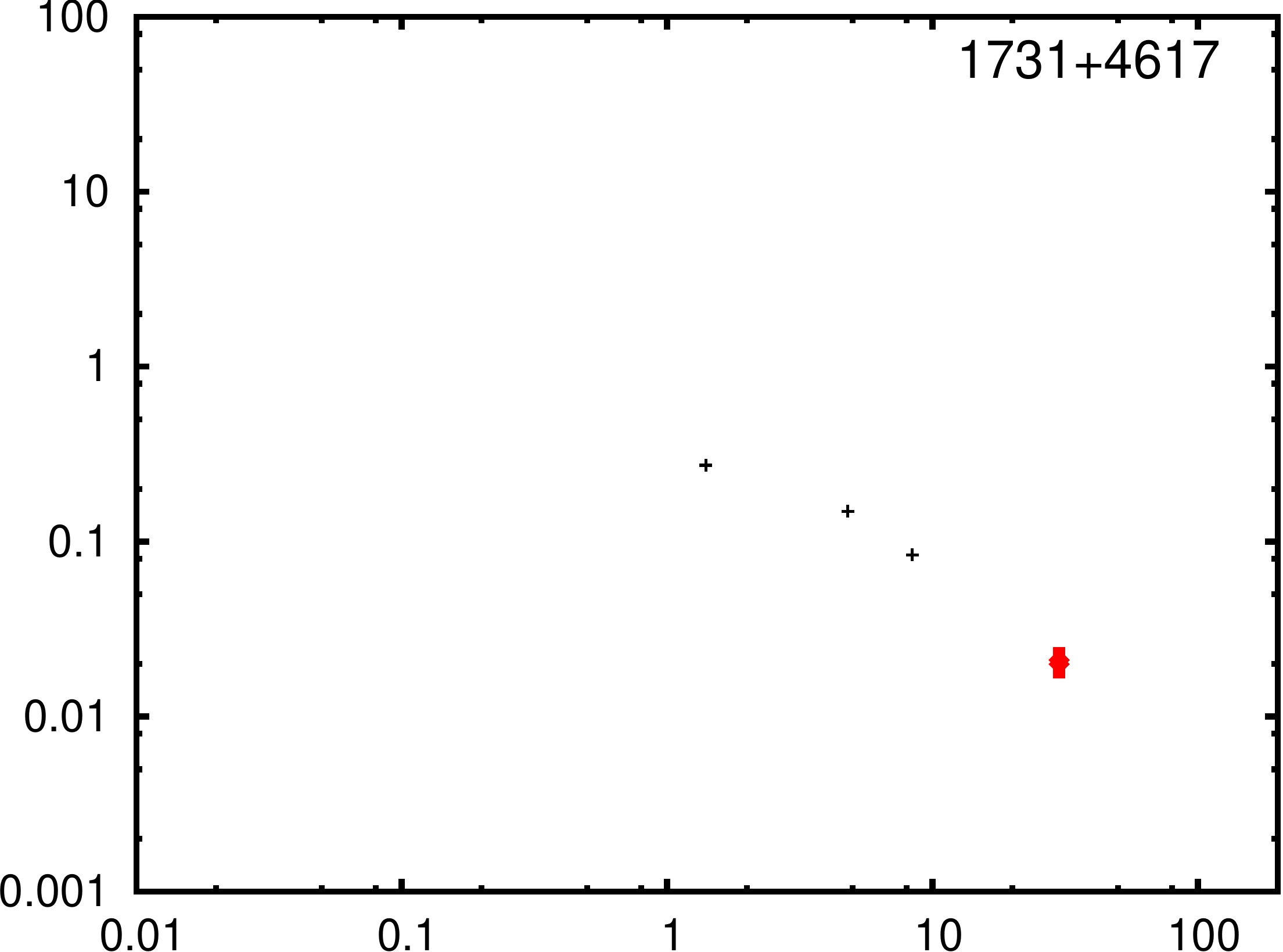}
\includegraphics[scale=0.2]{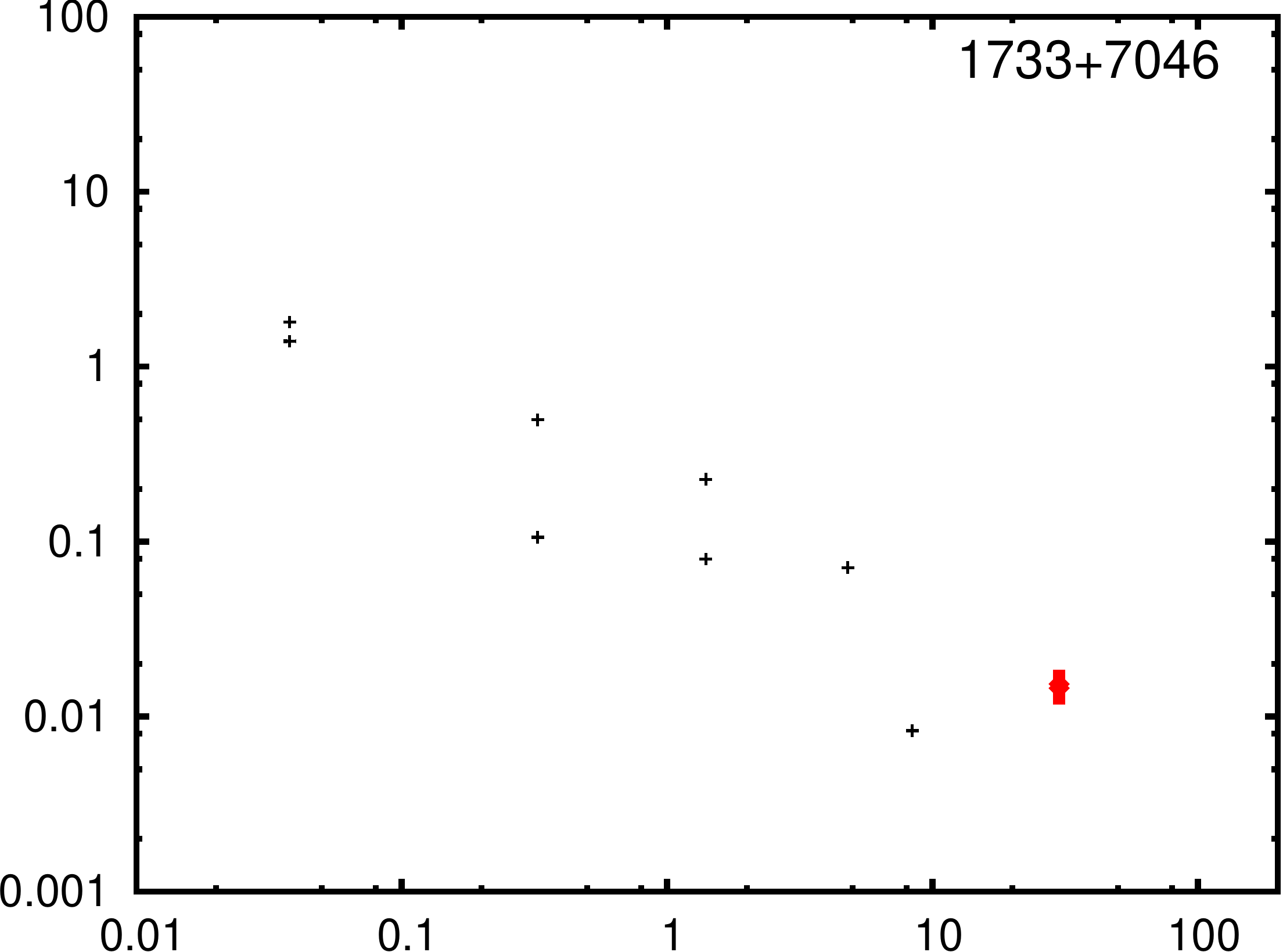}
\includegraphics[scale=0.2]{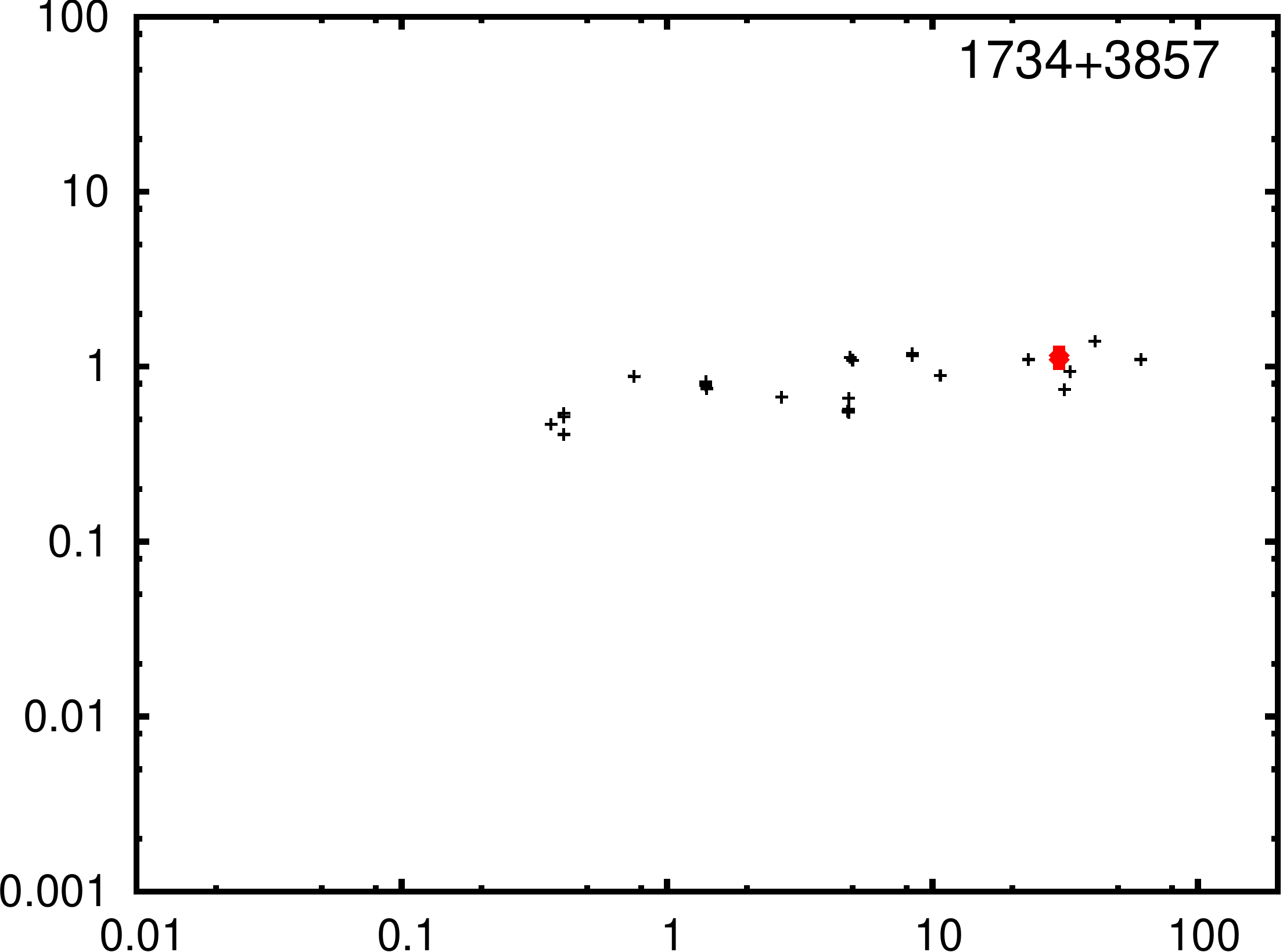}
\includegraphics[scale=0.2]{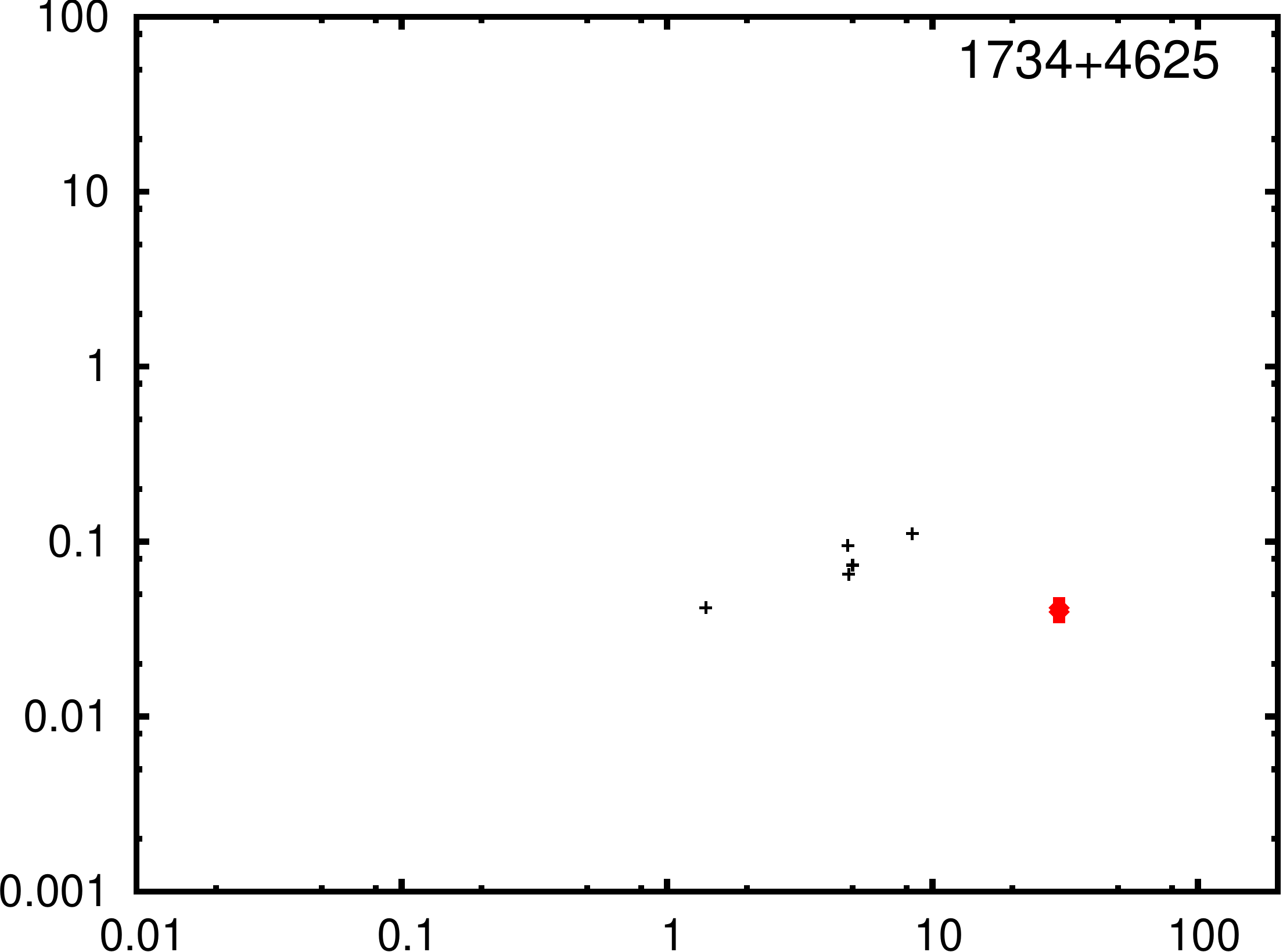}
\includegraphics[scale=0.2]{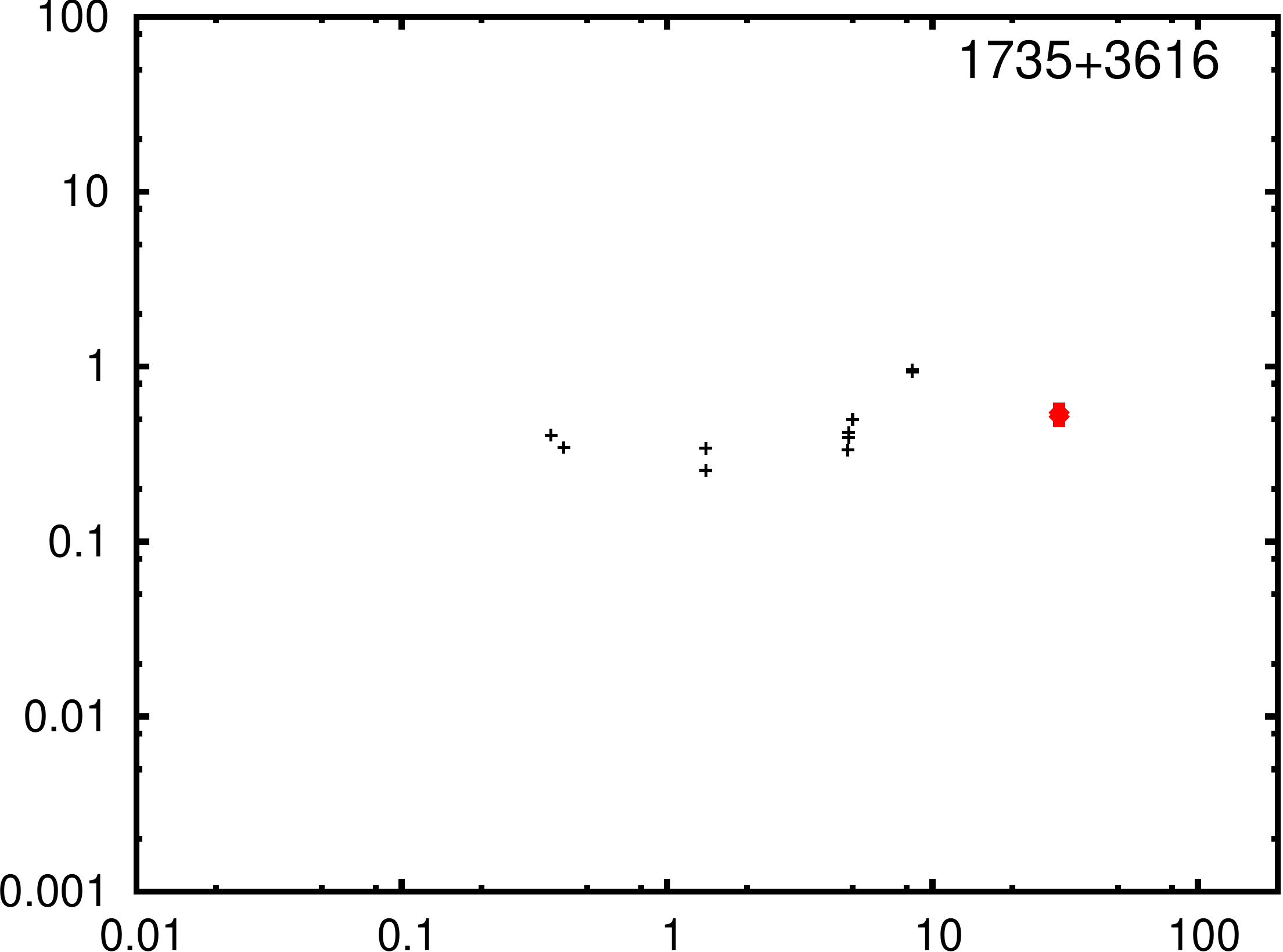}
\includegraphics[scale=0.2]{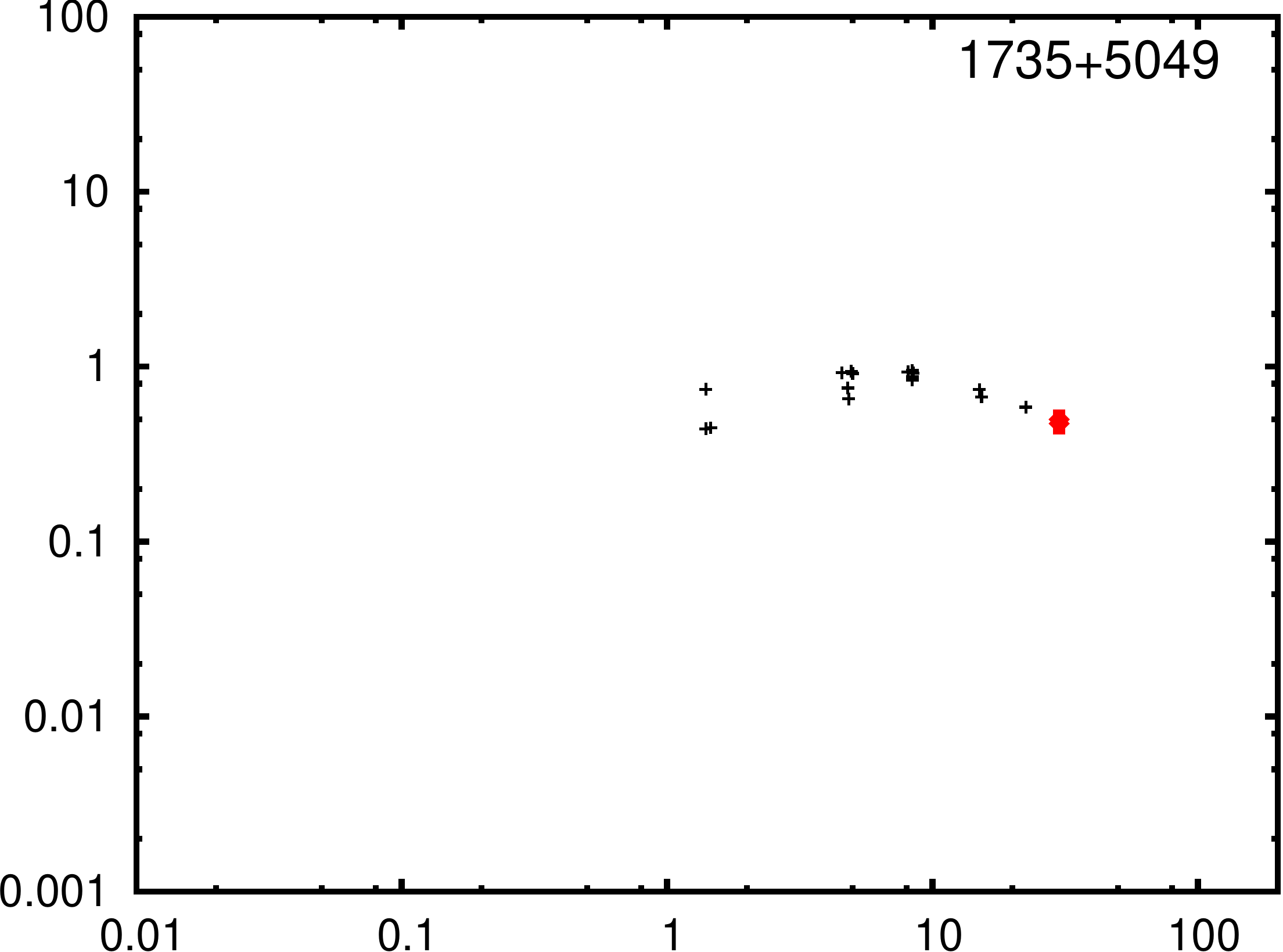}
\includegraphics[scale=0.2]{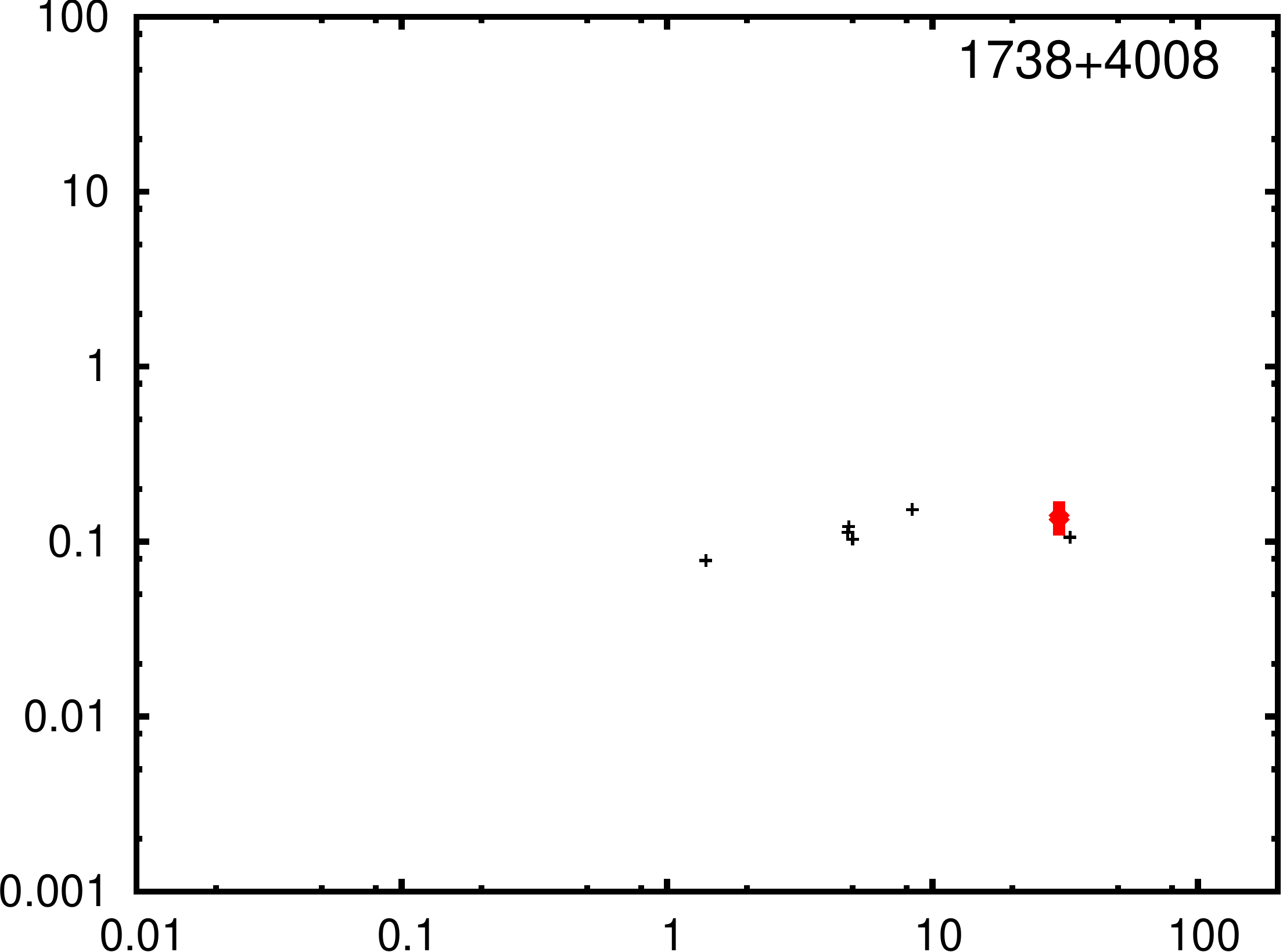}
\includegraphics[scale=0.2]{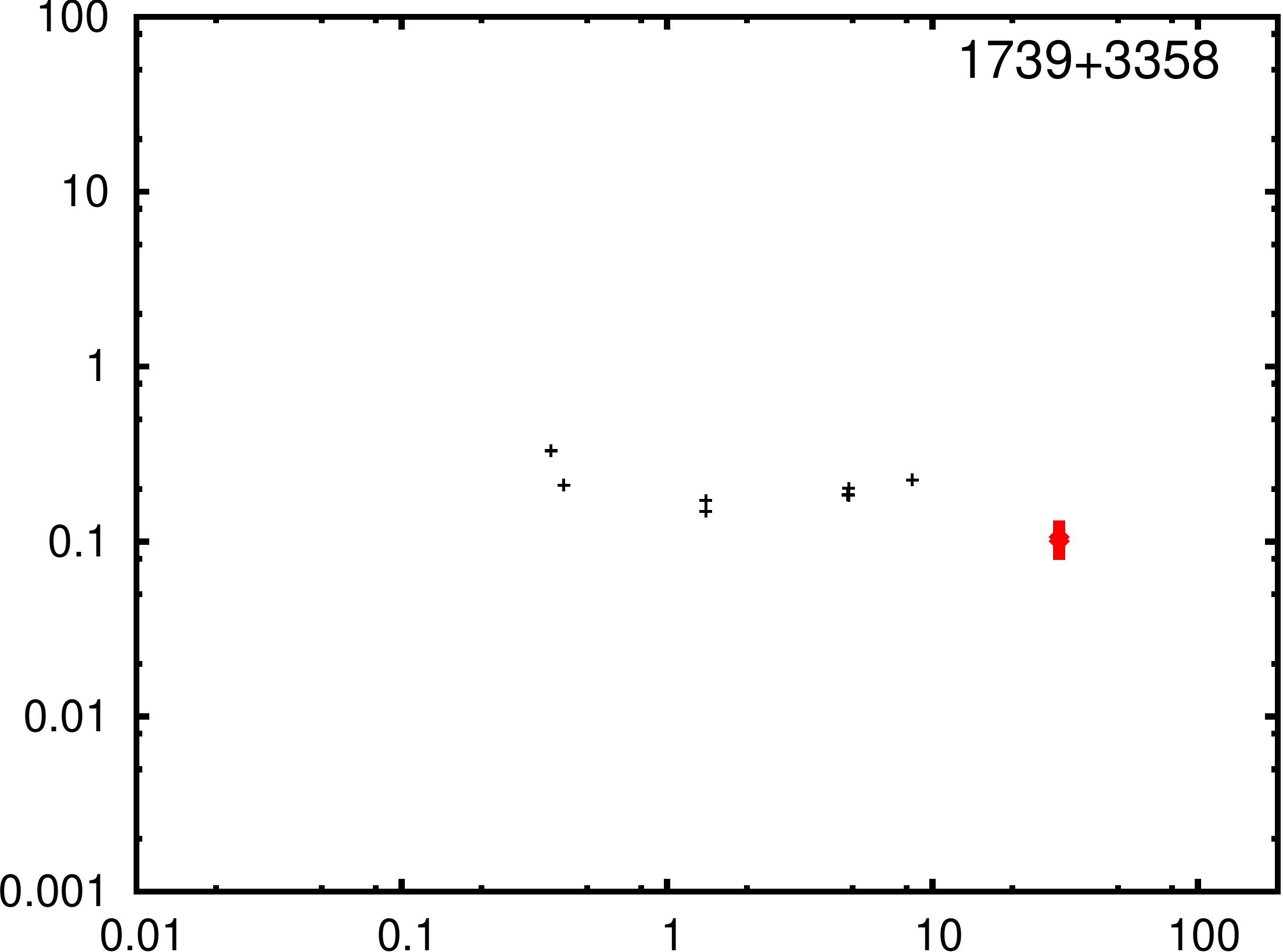}
\includegraphics[scale=0.2]{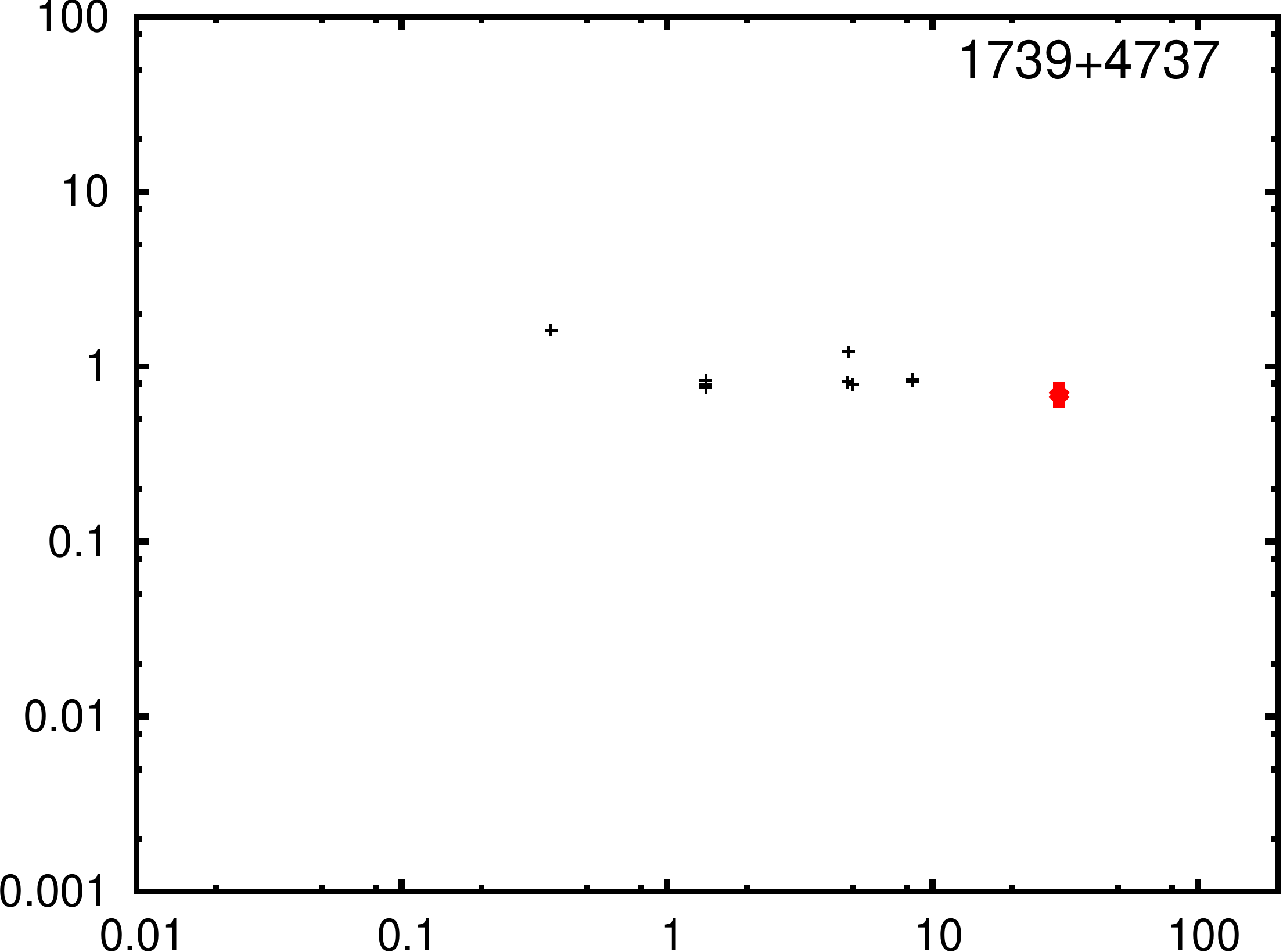}
\includegraphics[scale=0.2]{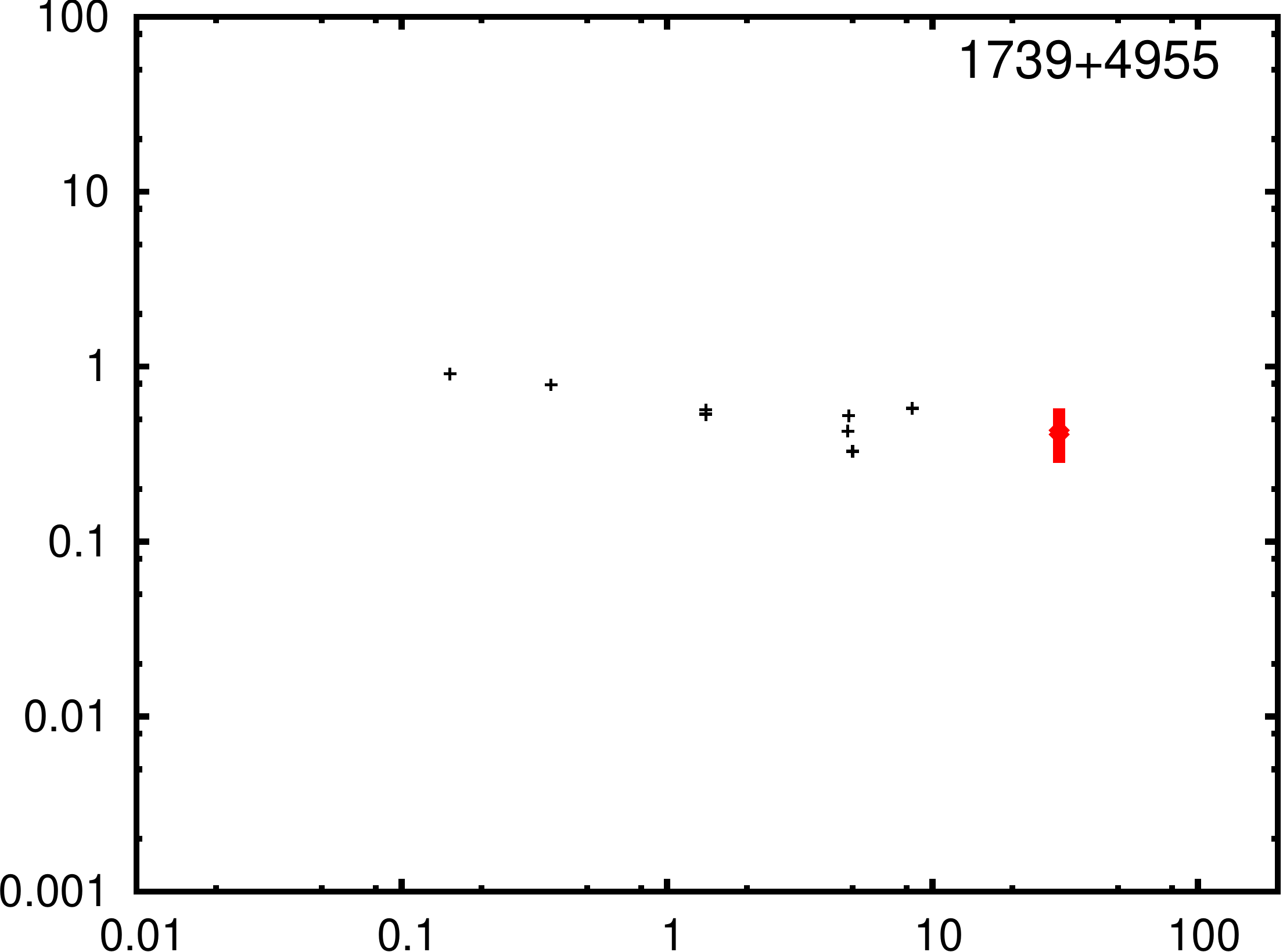}
\includegraphics[scale=0.2]{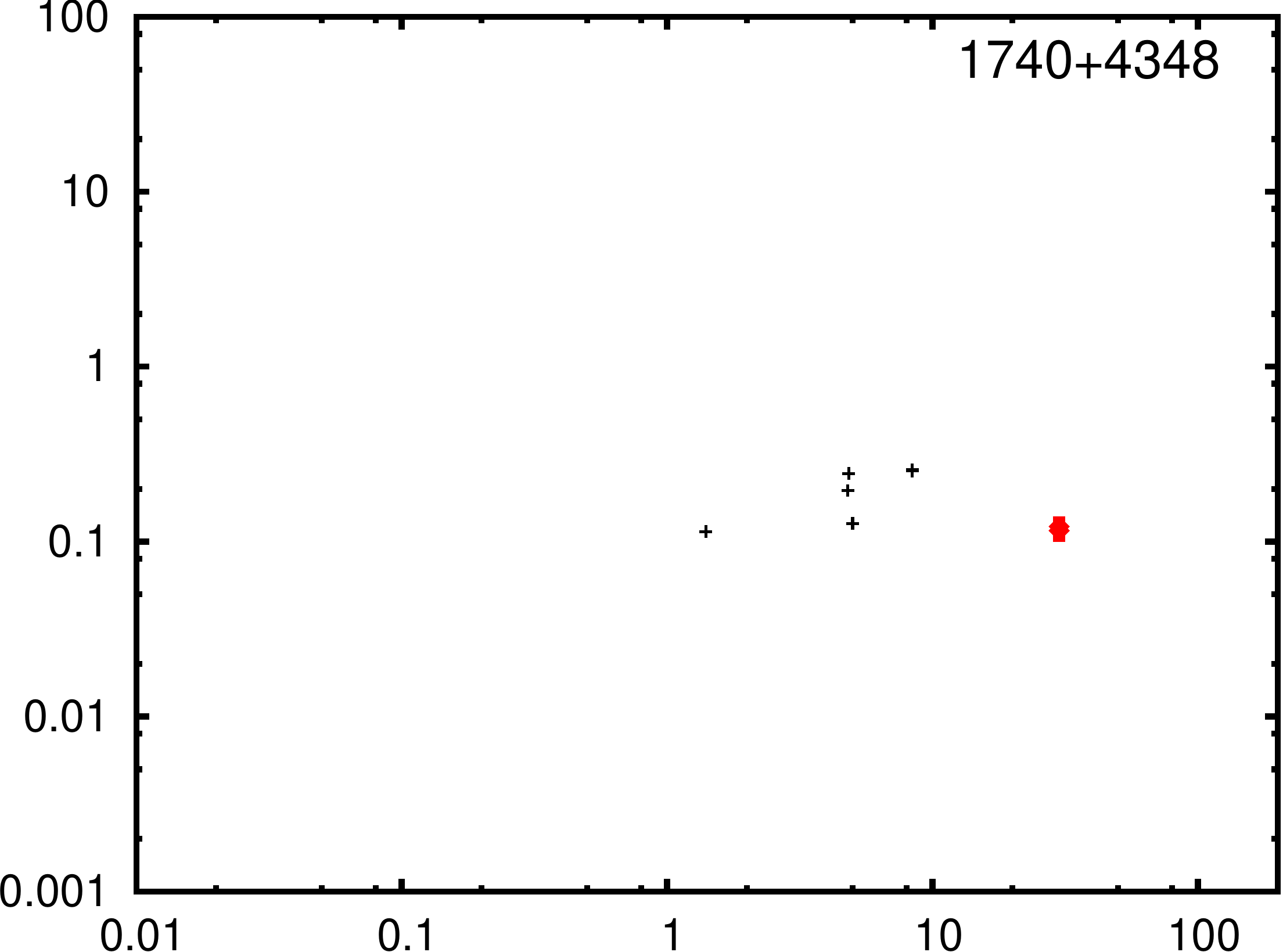}
\includegraphics[scale=0.2]{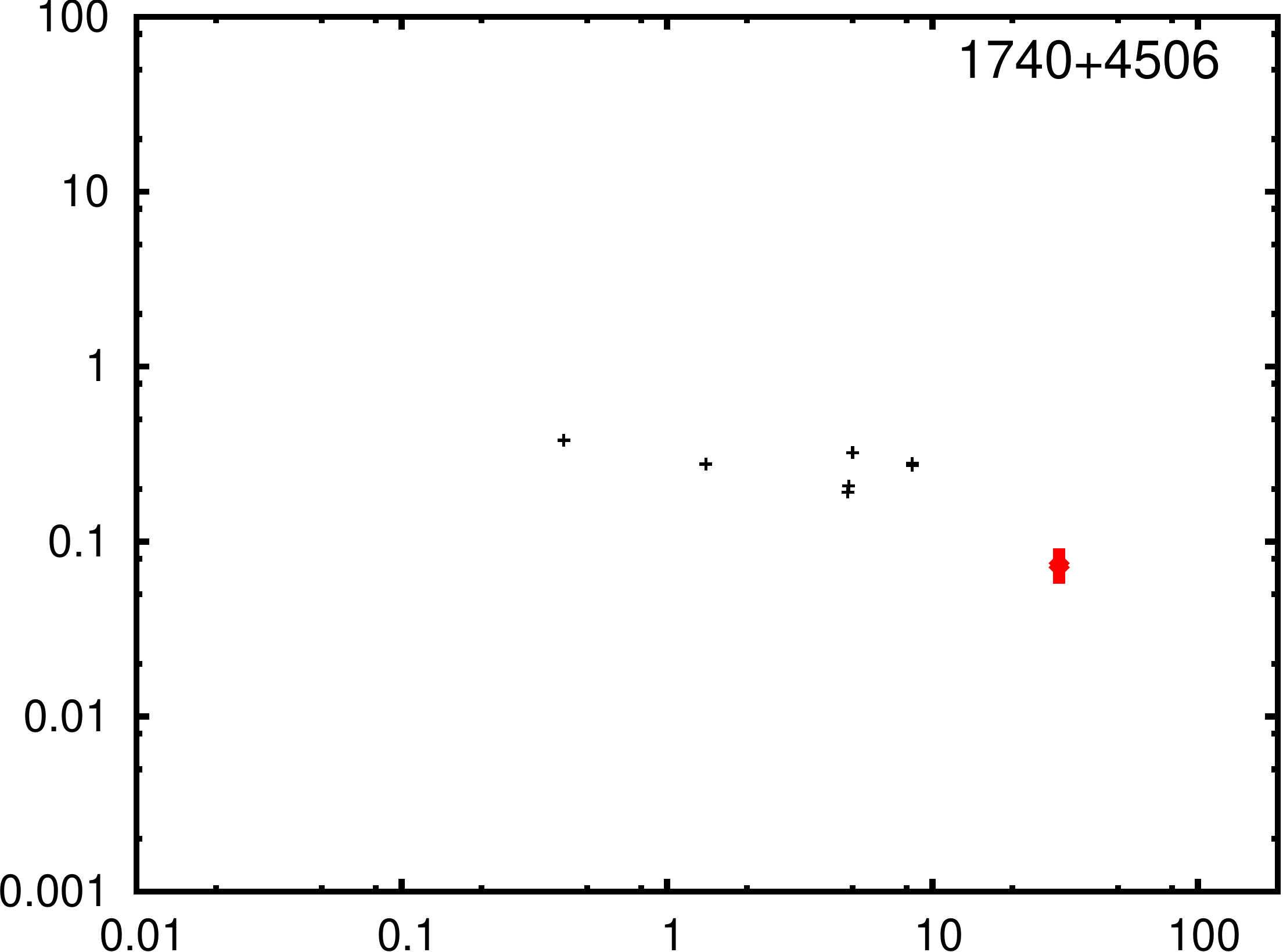}
\includegraphics[scale=0.2]{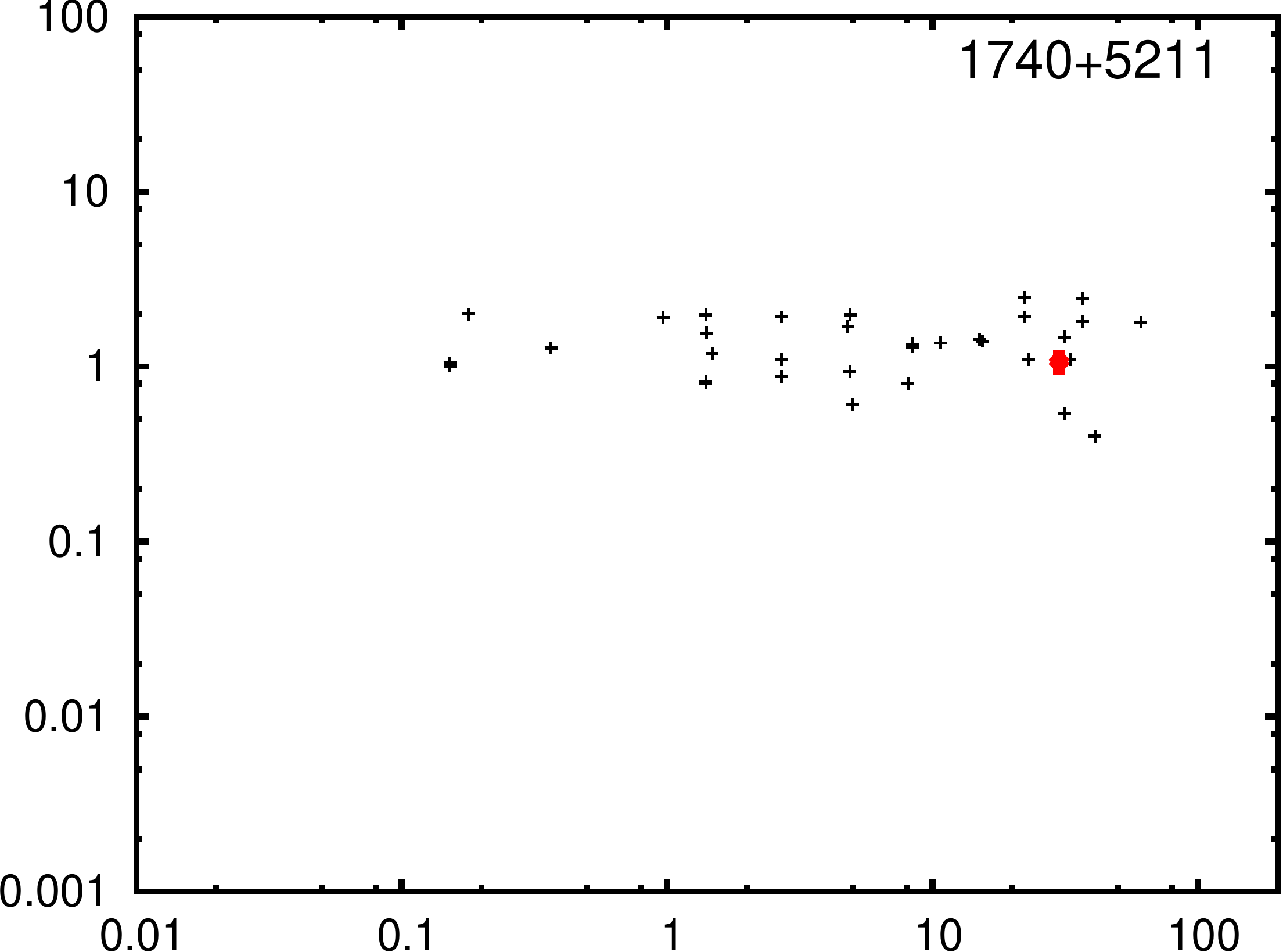}
\includegraphics[scale=0.2]{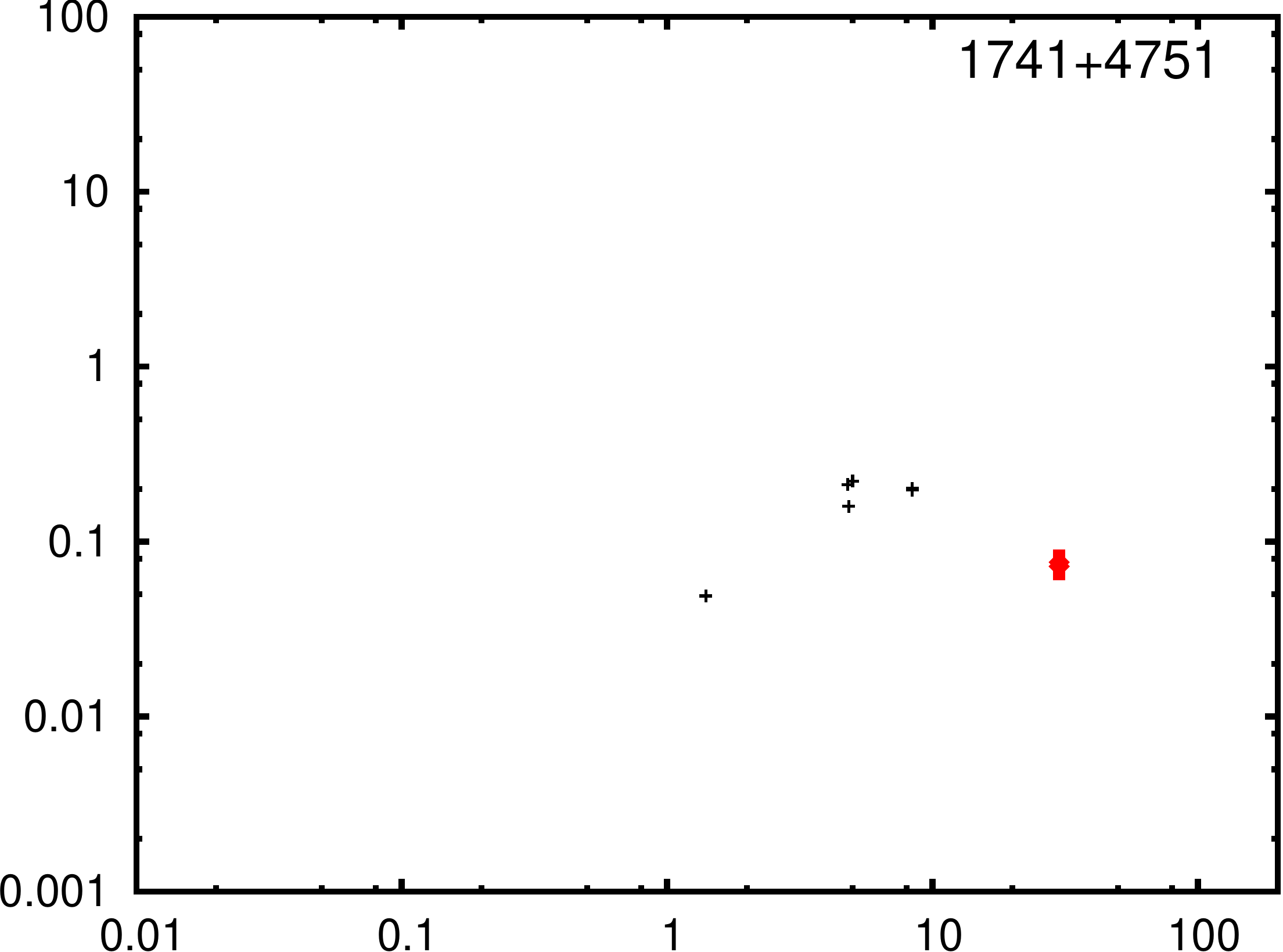}
\includegraphics[scale=0.2]{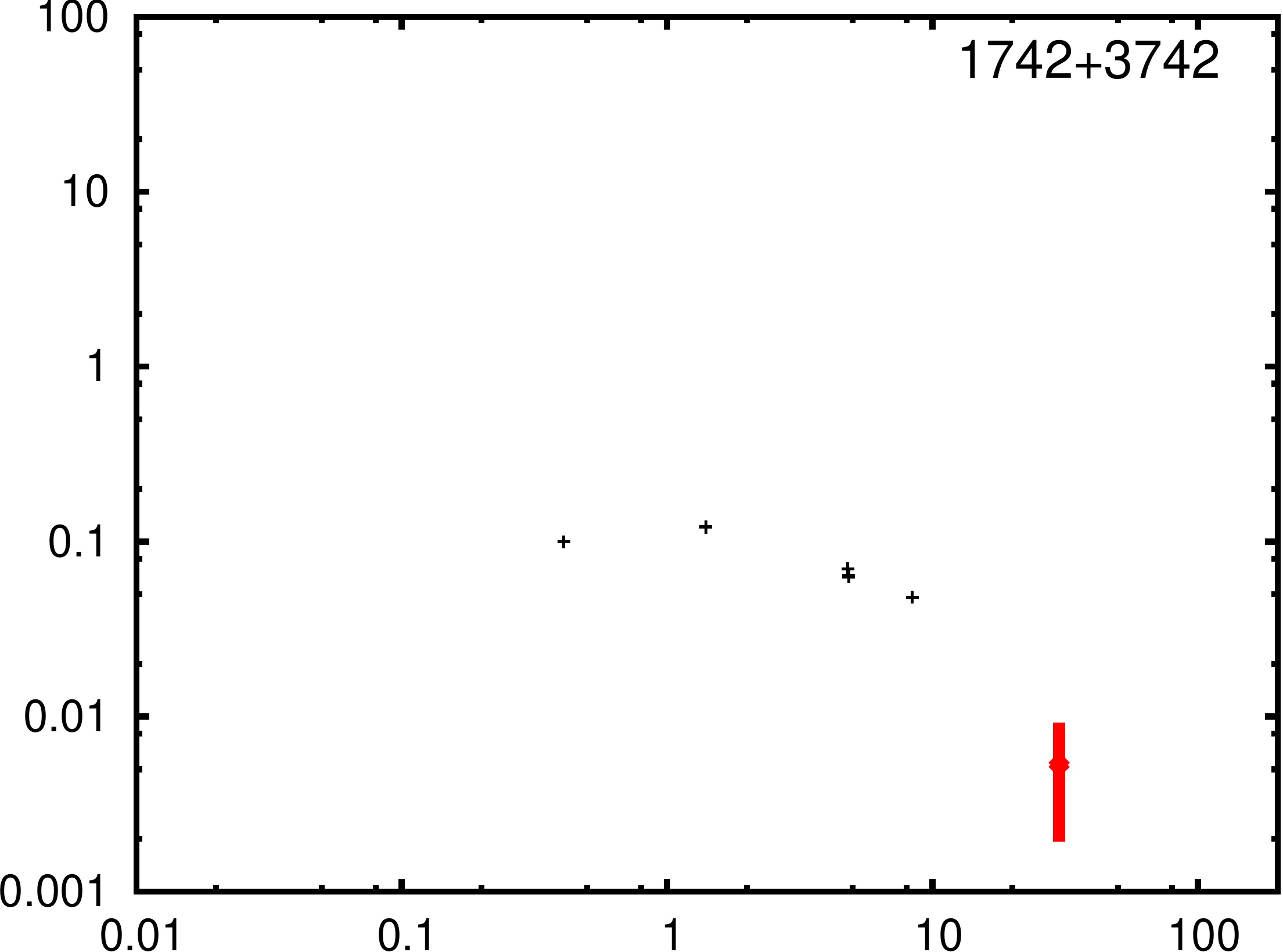}
\end{figure}
\clearpage\begin{figure}
\centering
\includegraphics[scale=0.2]{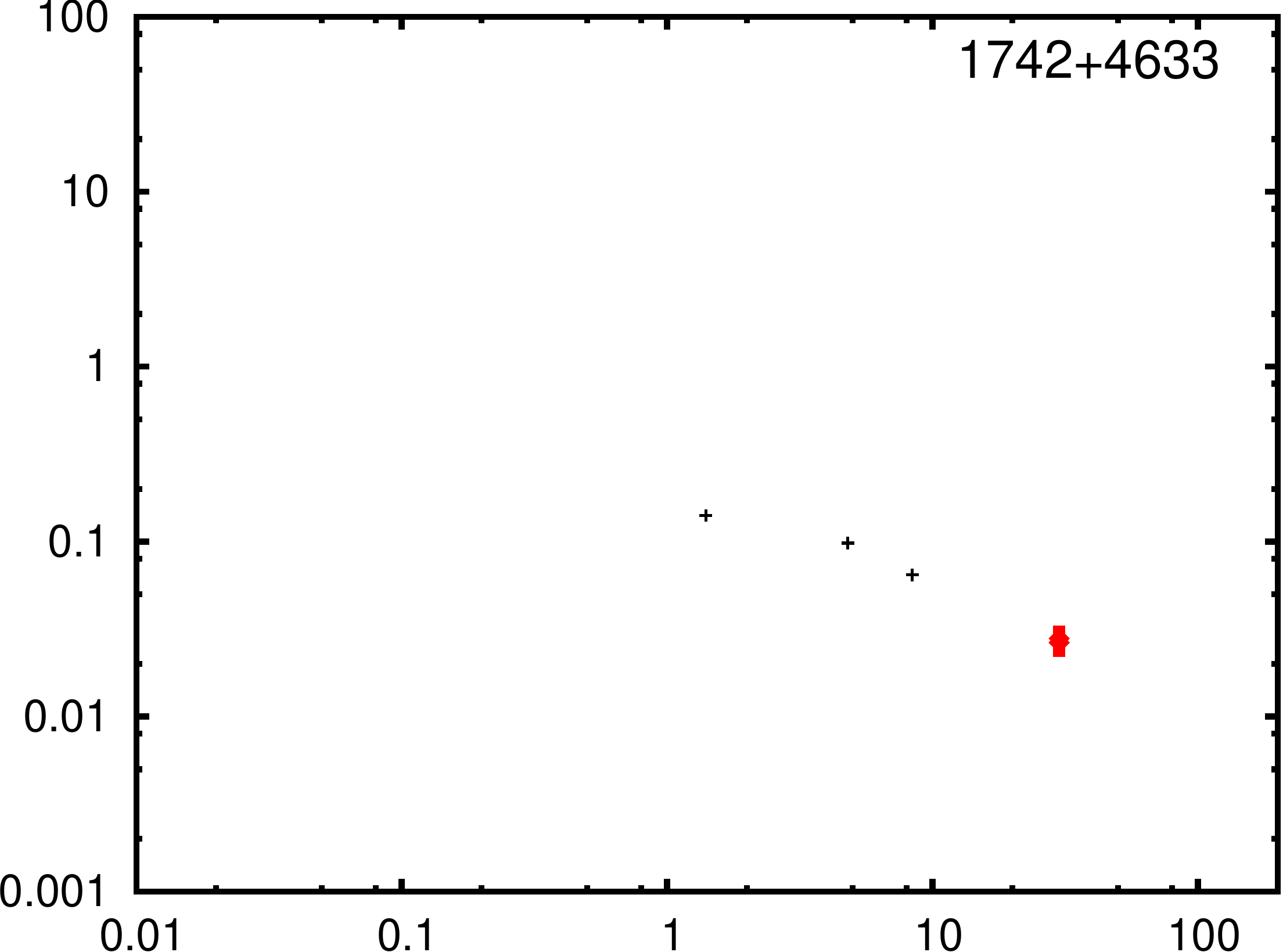}
\includegraphics[scale=0.2]{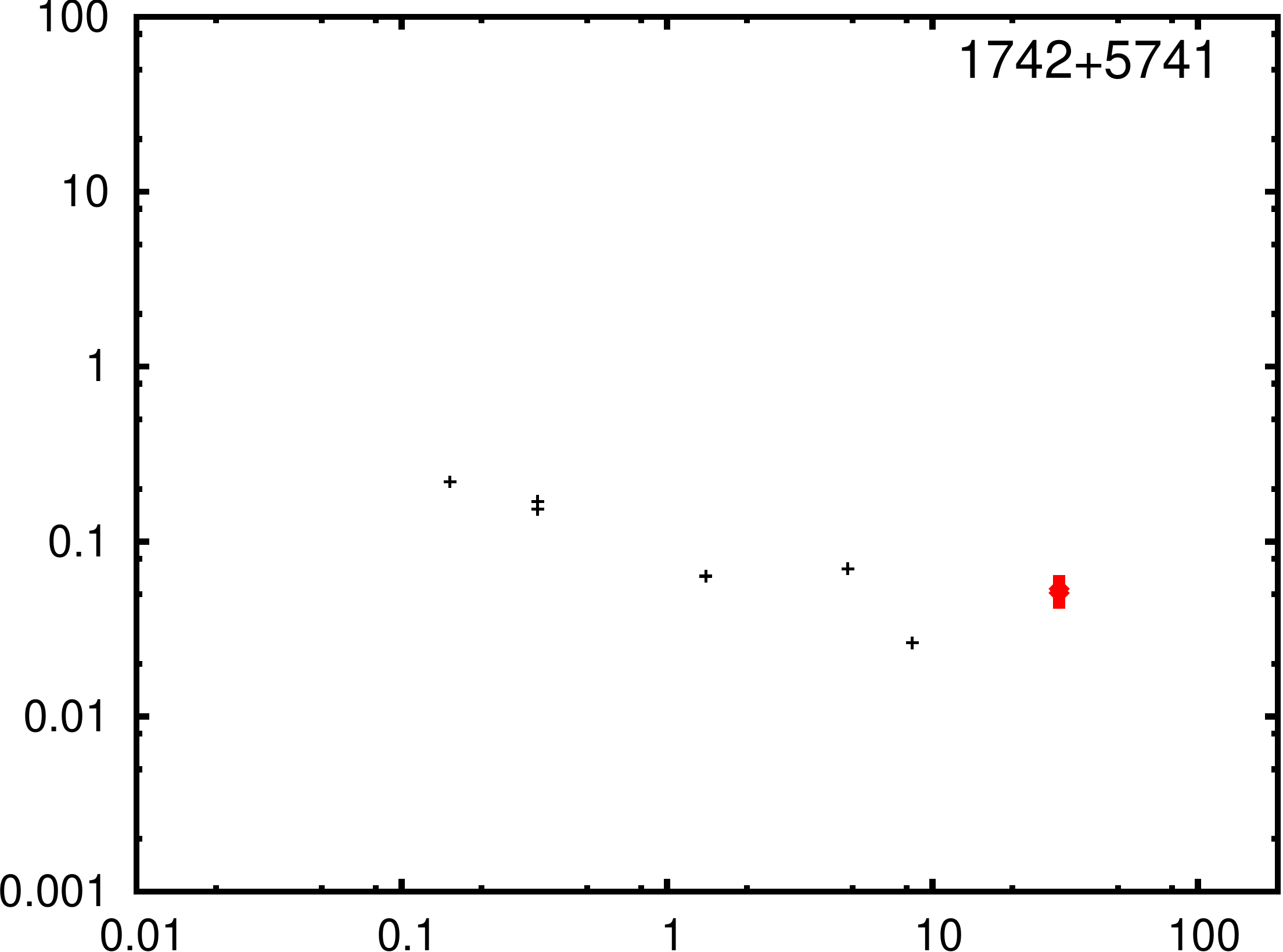}
\includegraphics[scale=0.2]{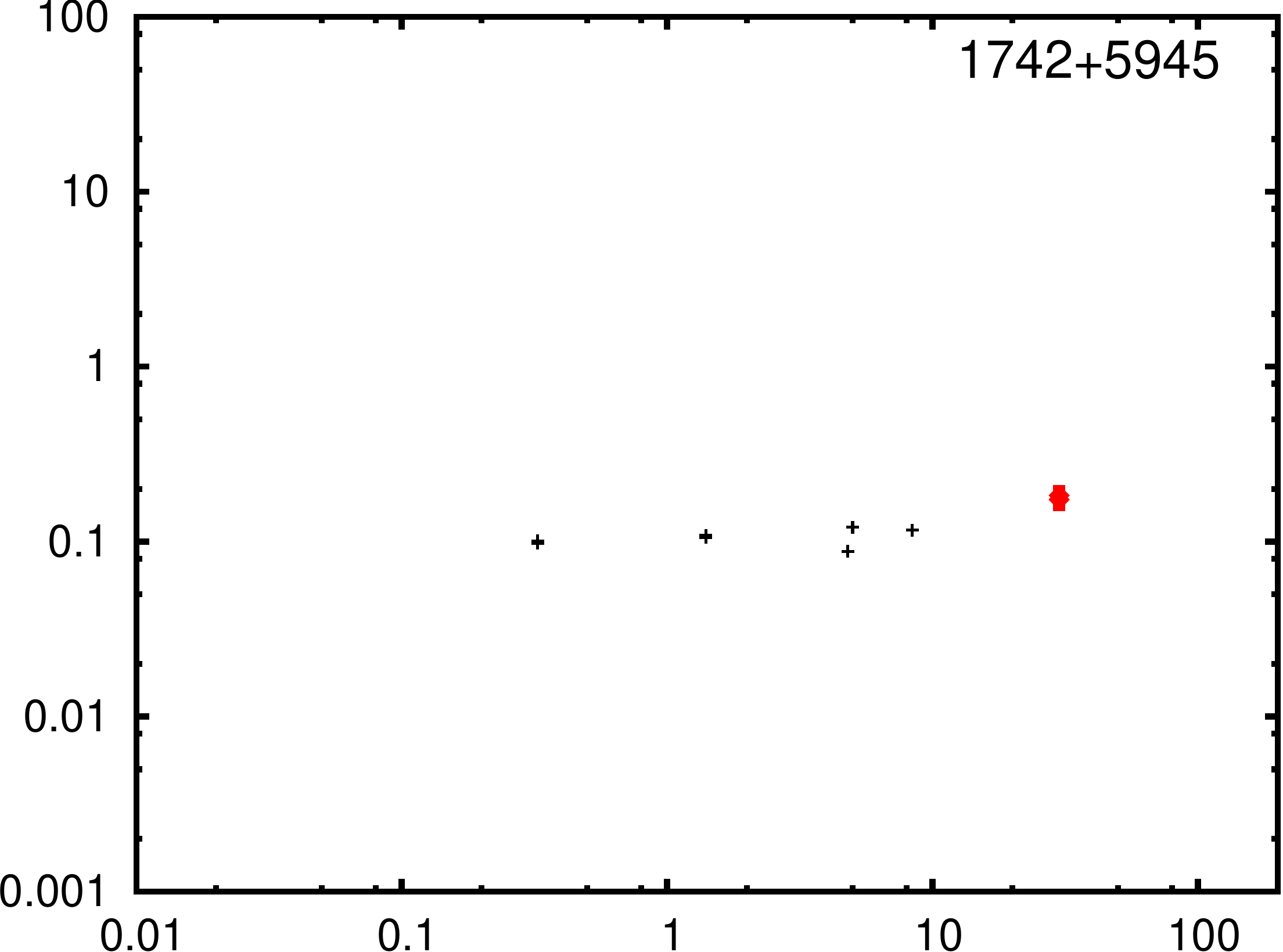}
\includegraphics[scale=0.2]{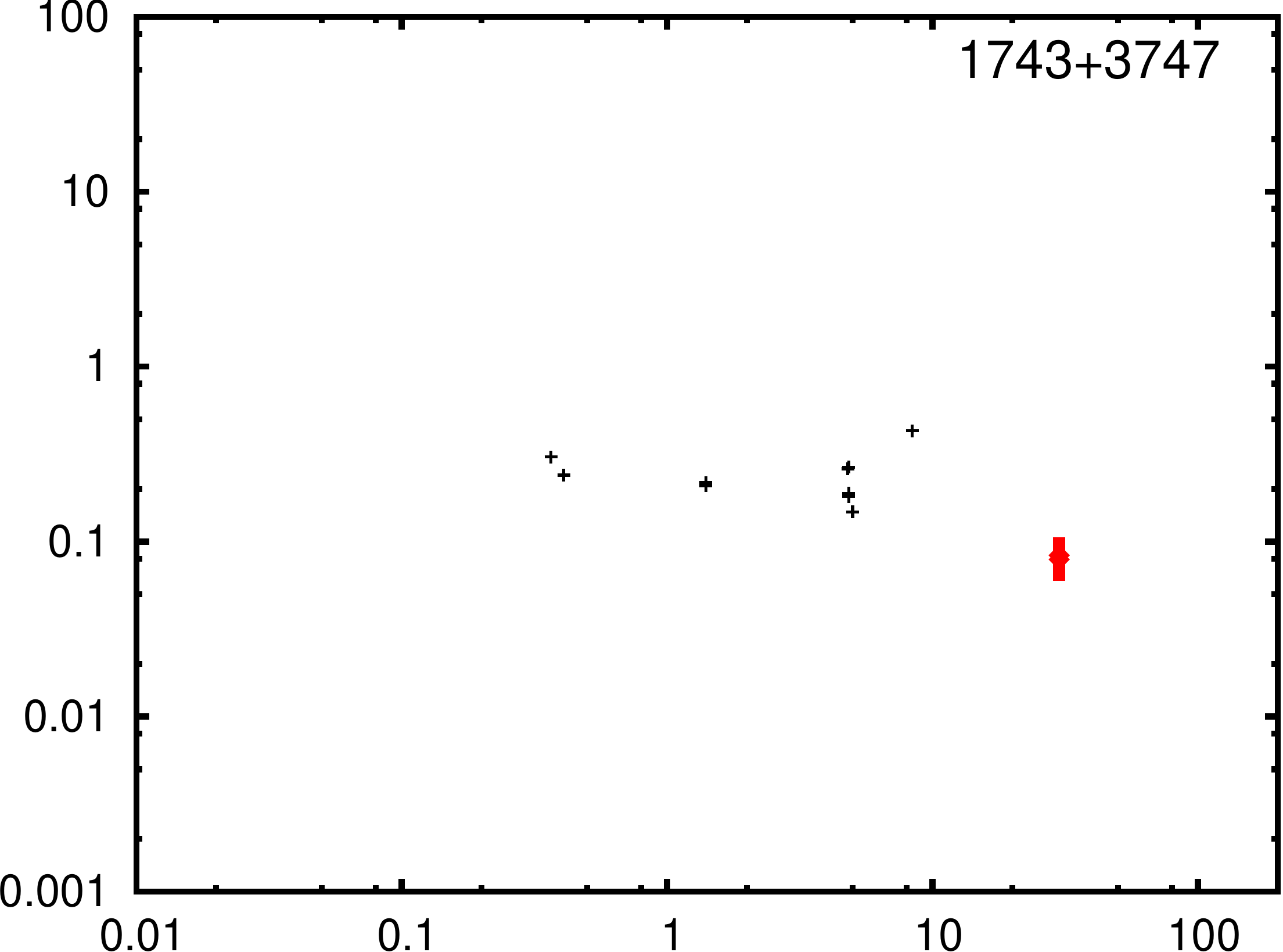}
\includegraphics[scale=0.2]{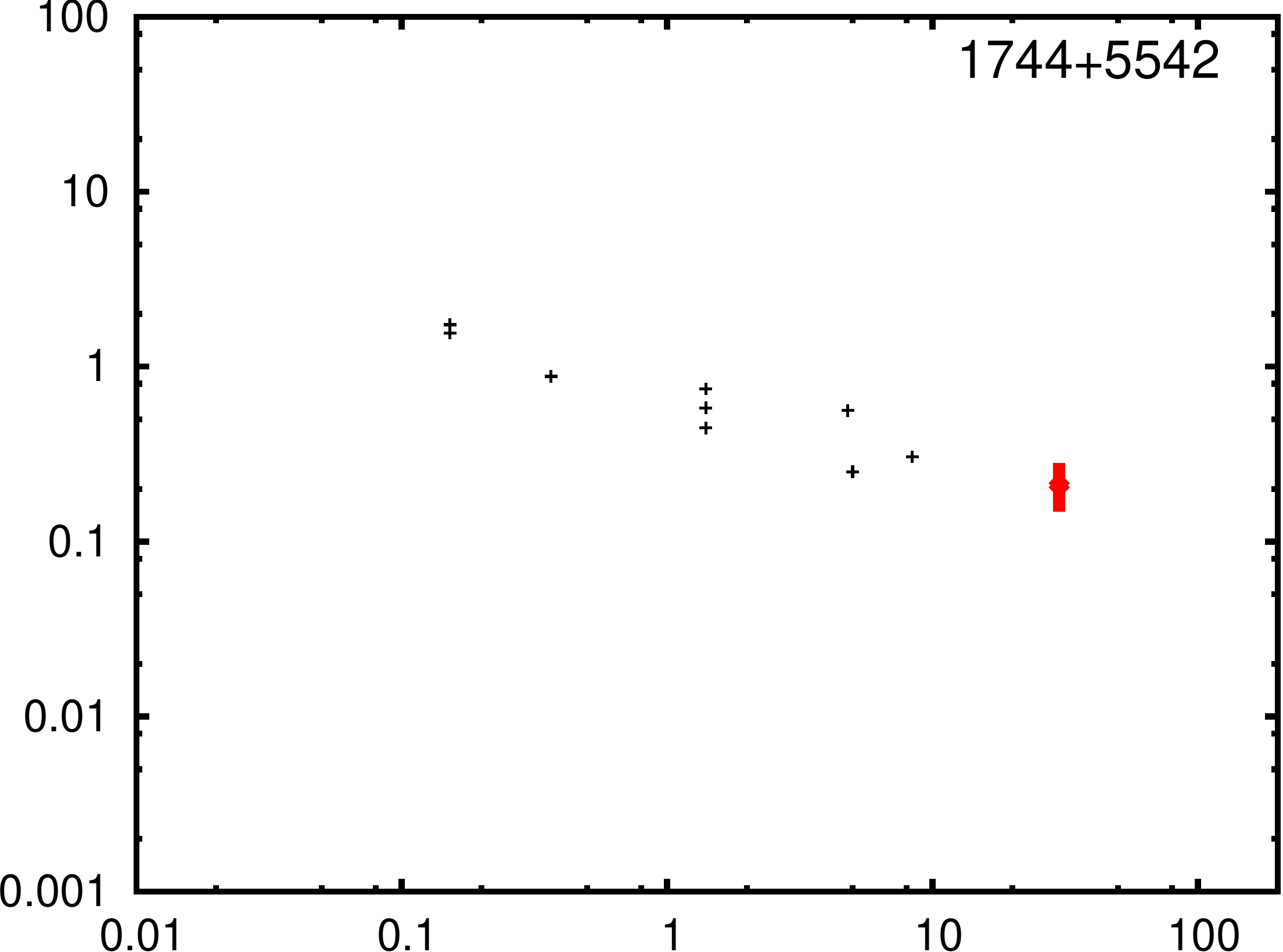}
\includegraphics[scale=0.2]{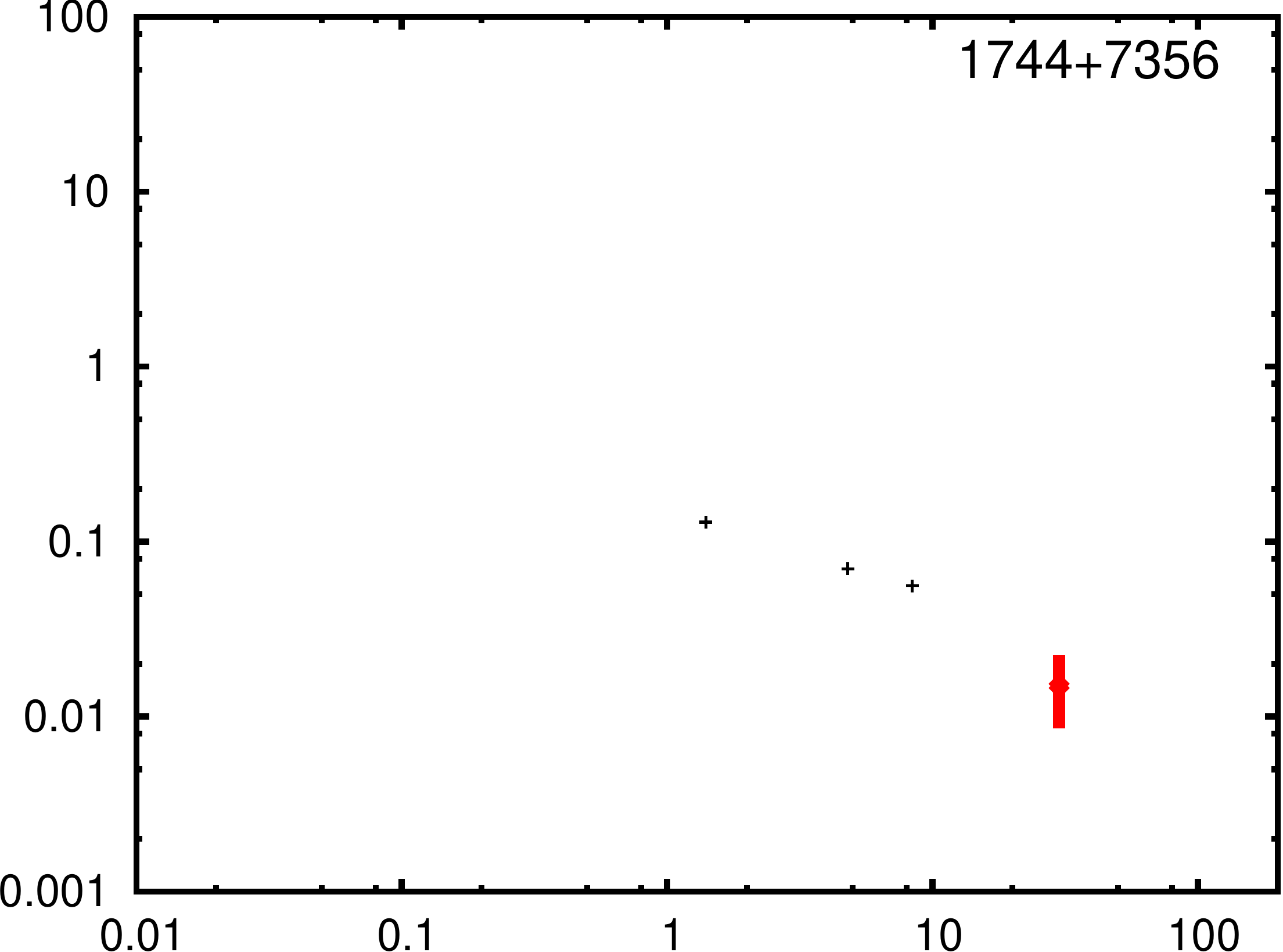}
\includegraphics[scale=0.2]{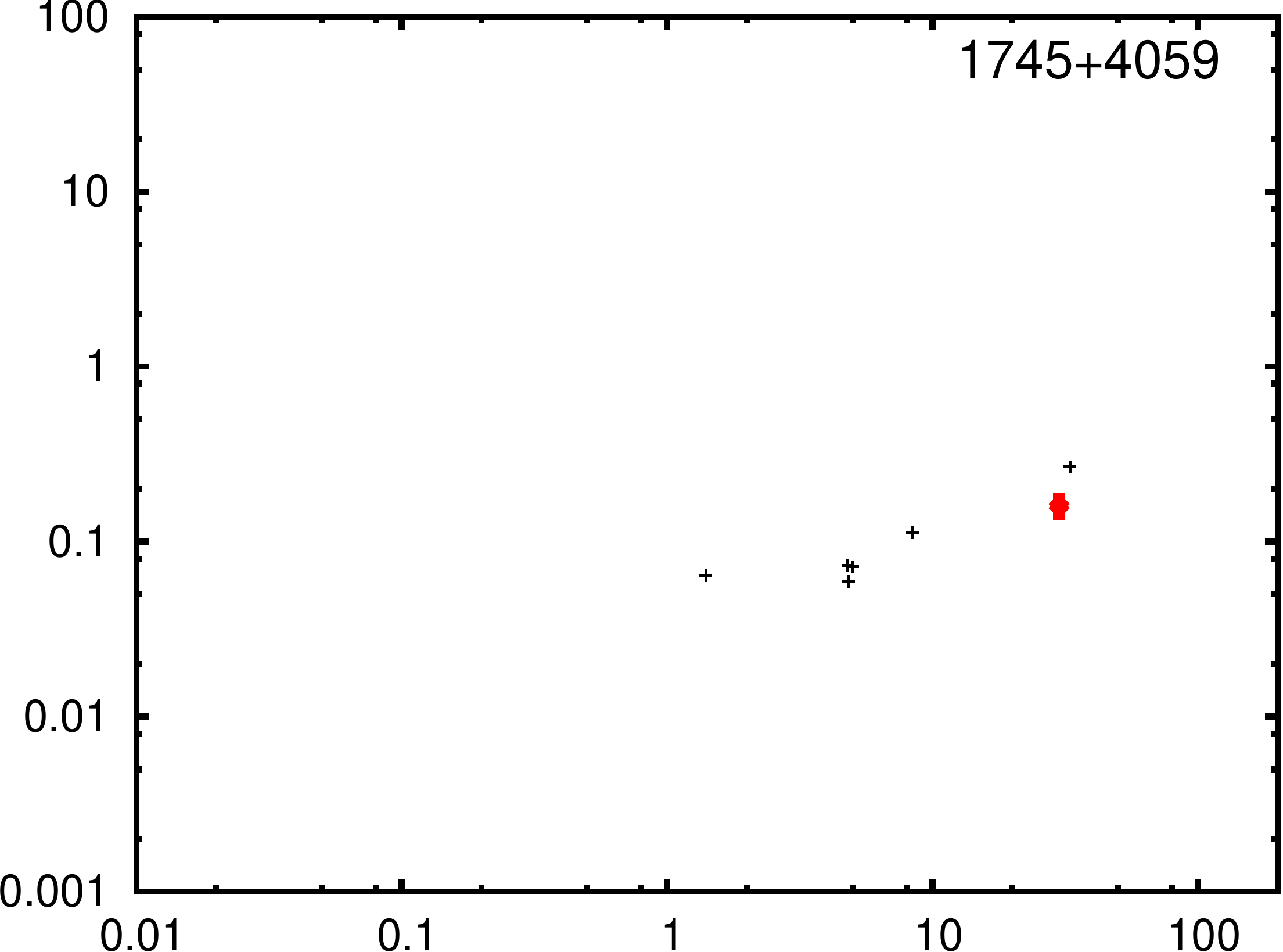}
\includegraphics[scale=0.2]{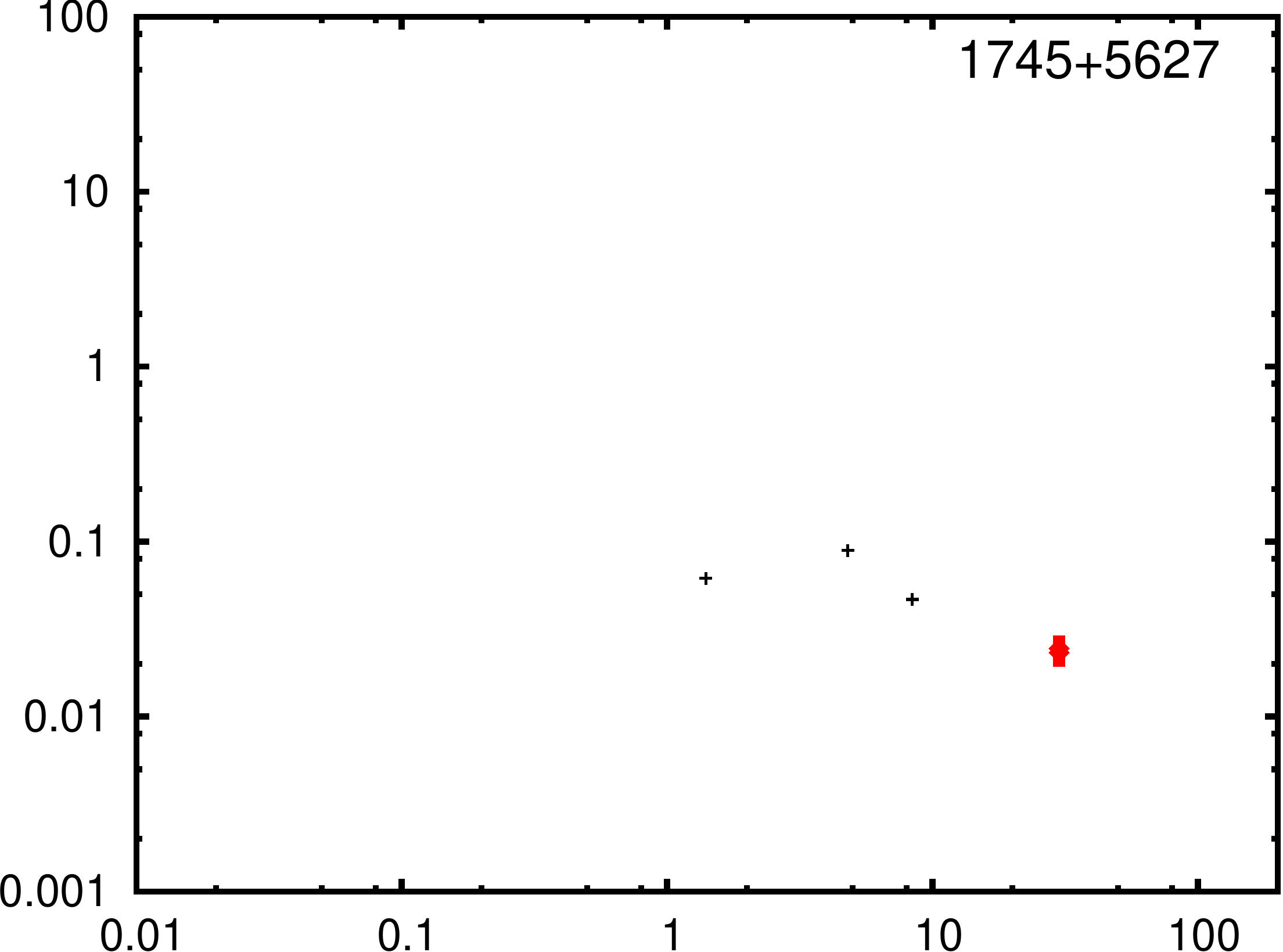}
\includegraphics[scale=0.2]{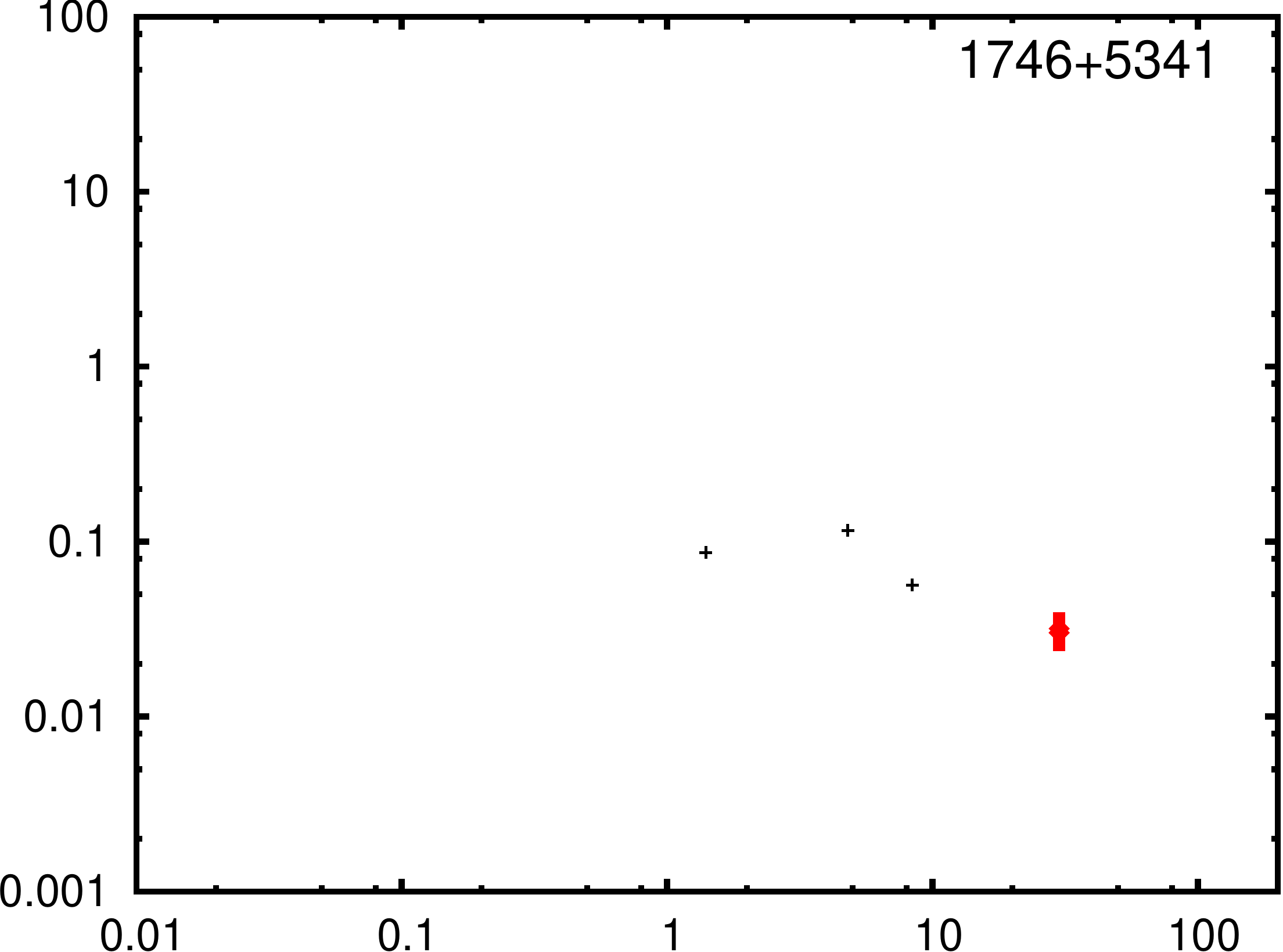}
\includegraphics[scale=0.2]{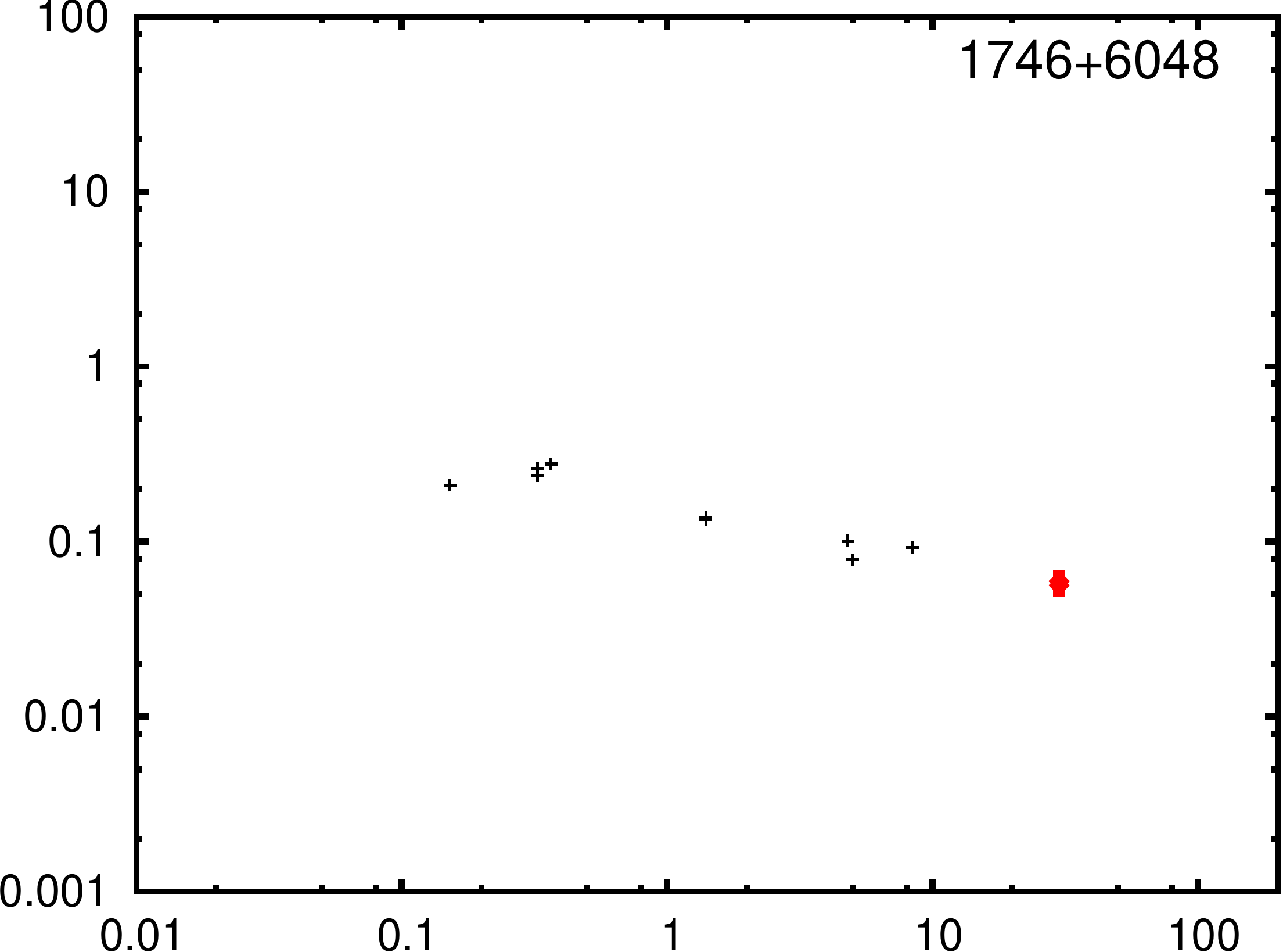}
\includegraphics[scale=0.2]{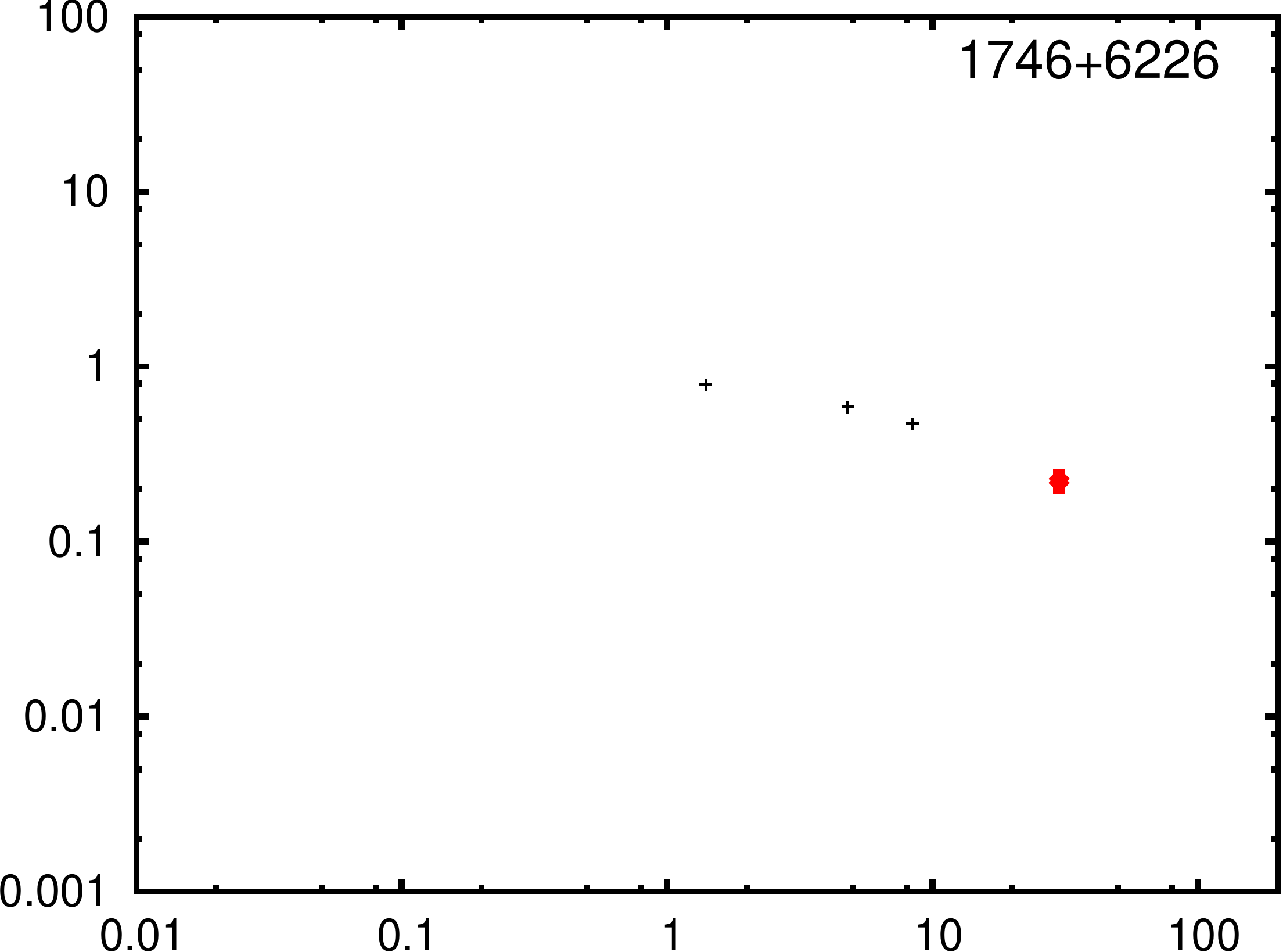}
\includegraphics[scale=0.2]{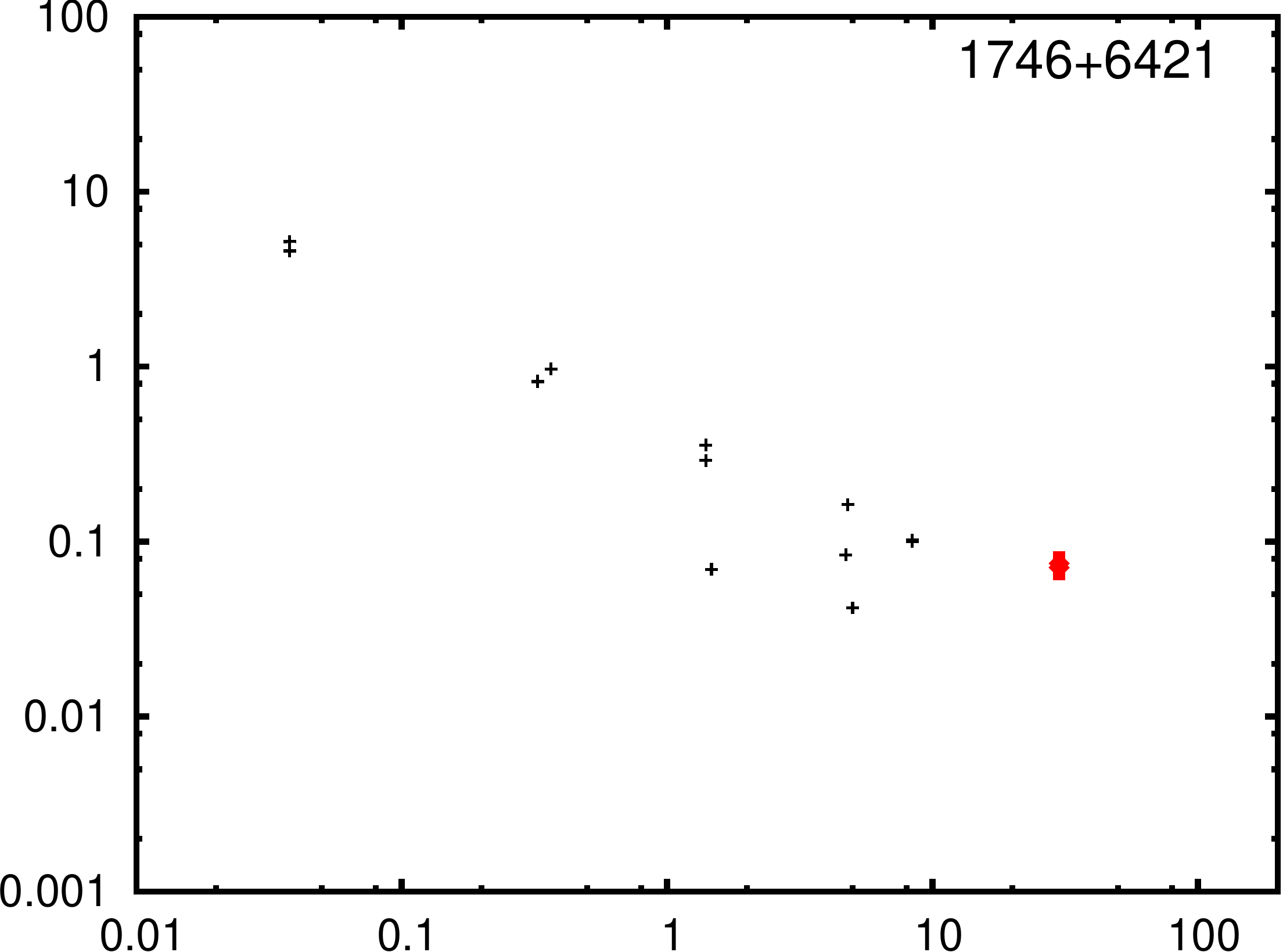}
\includegraphics[scale=0.2]{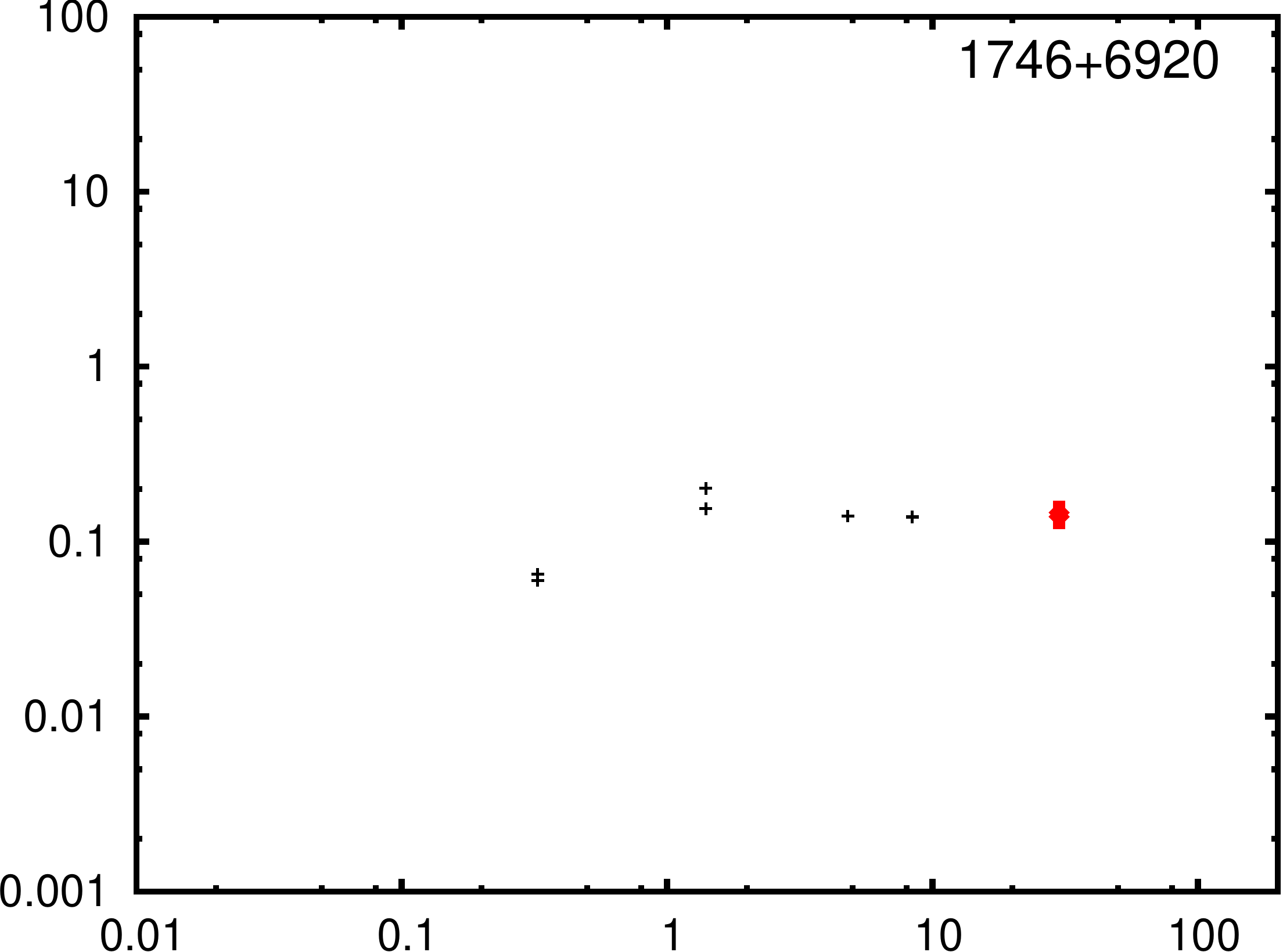}
\includegraphics[scale=0.2]{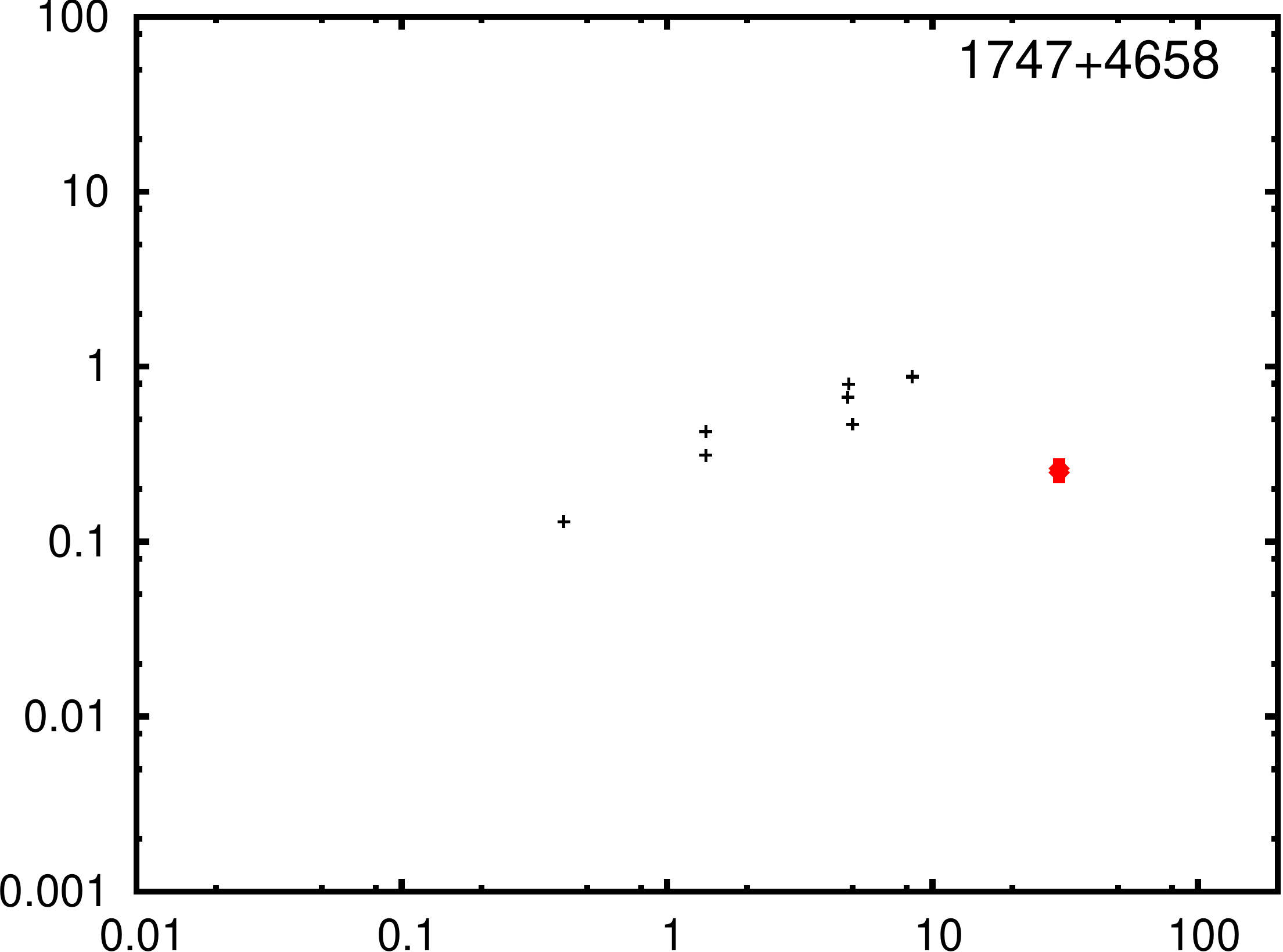}
\includegraphics[scale=0.2]{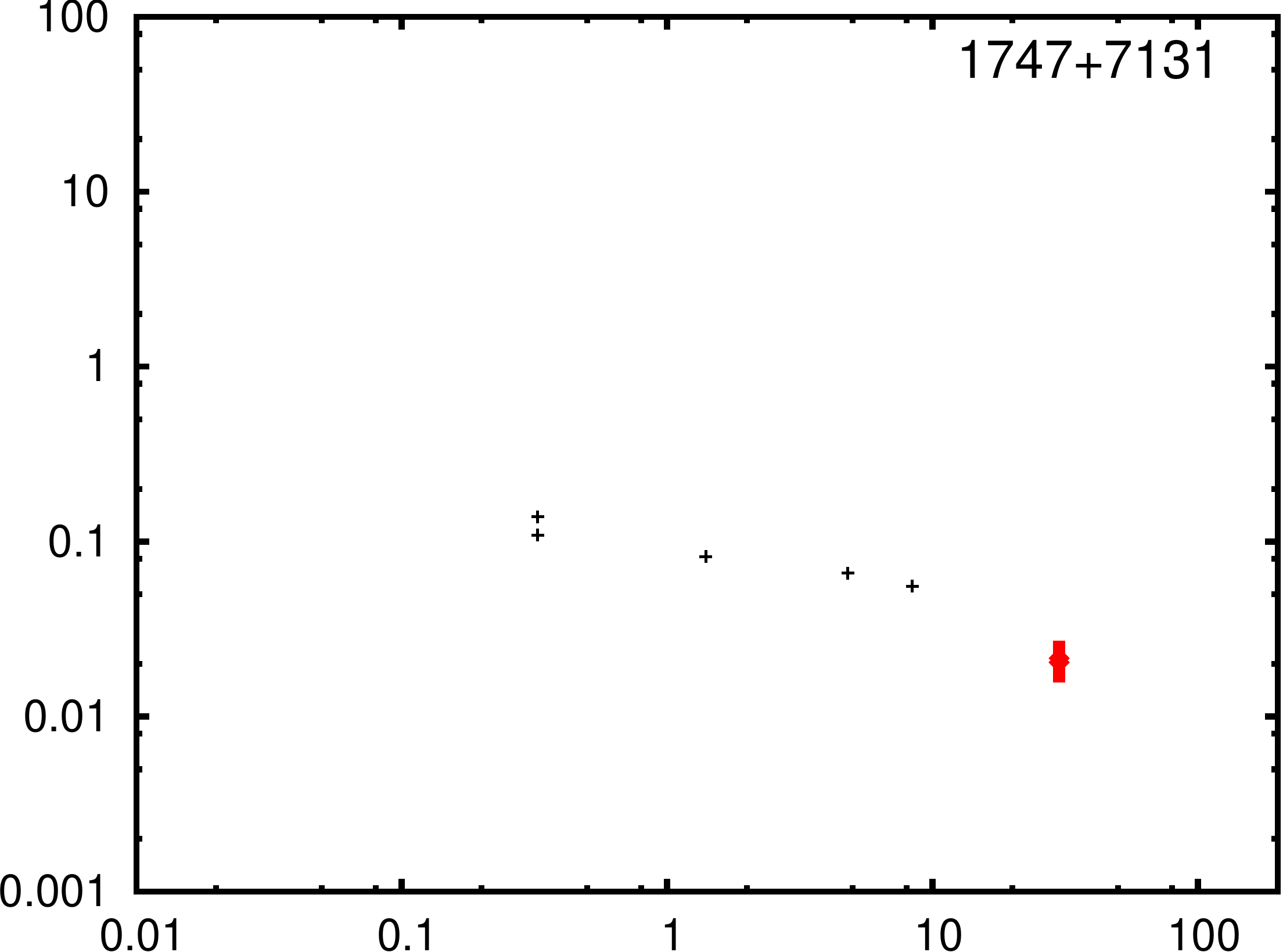}
\includegraphics[scale=0.2]{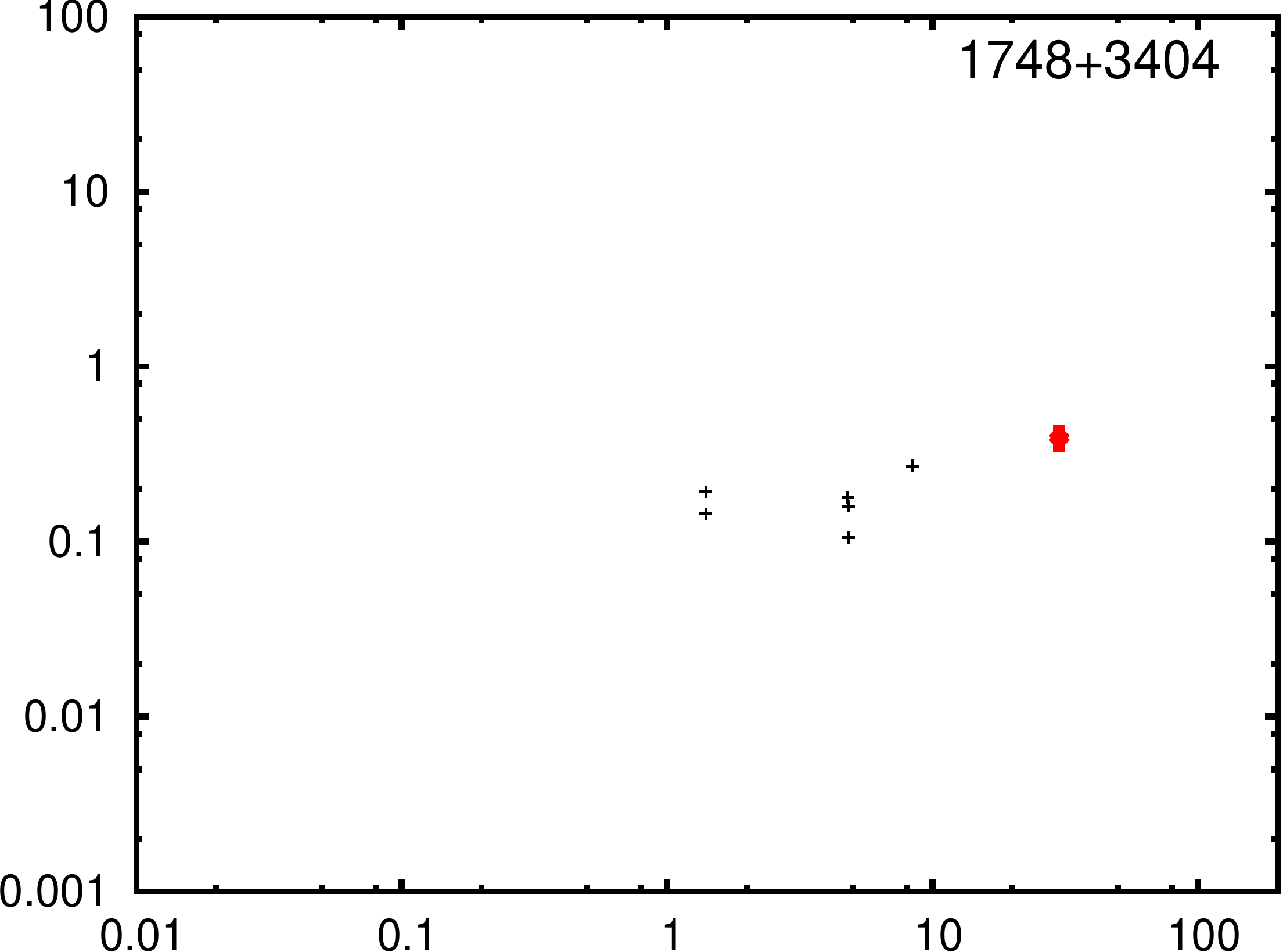}
\includegraphics[scale=0.2]{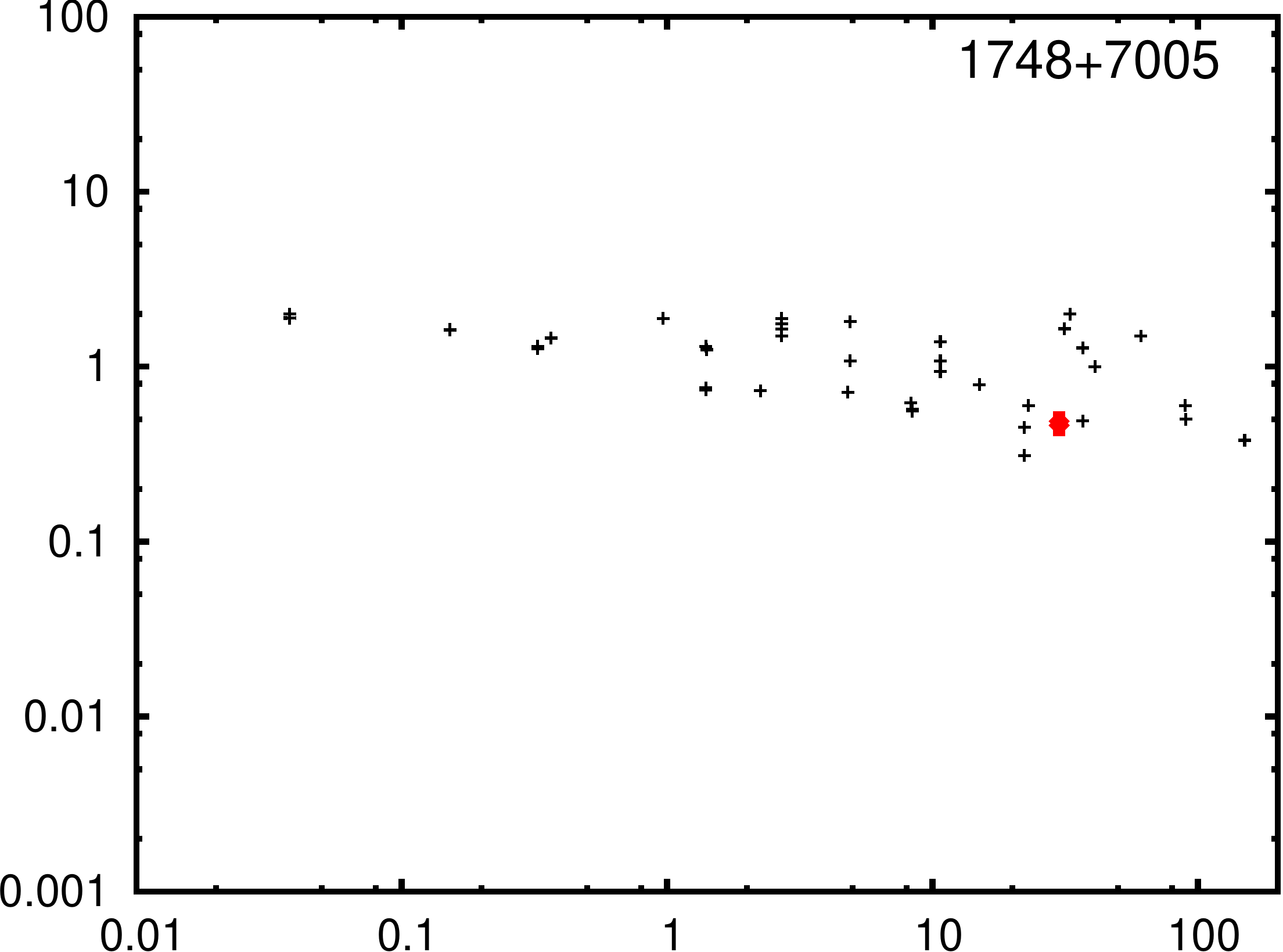}
\includegraphics[scale=0.2]{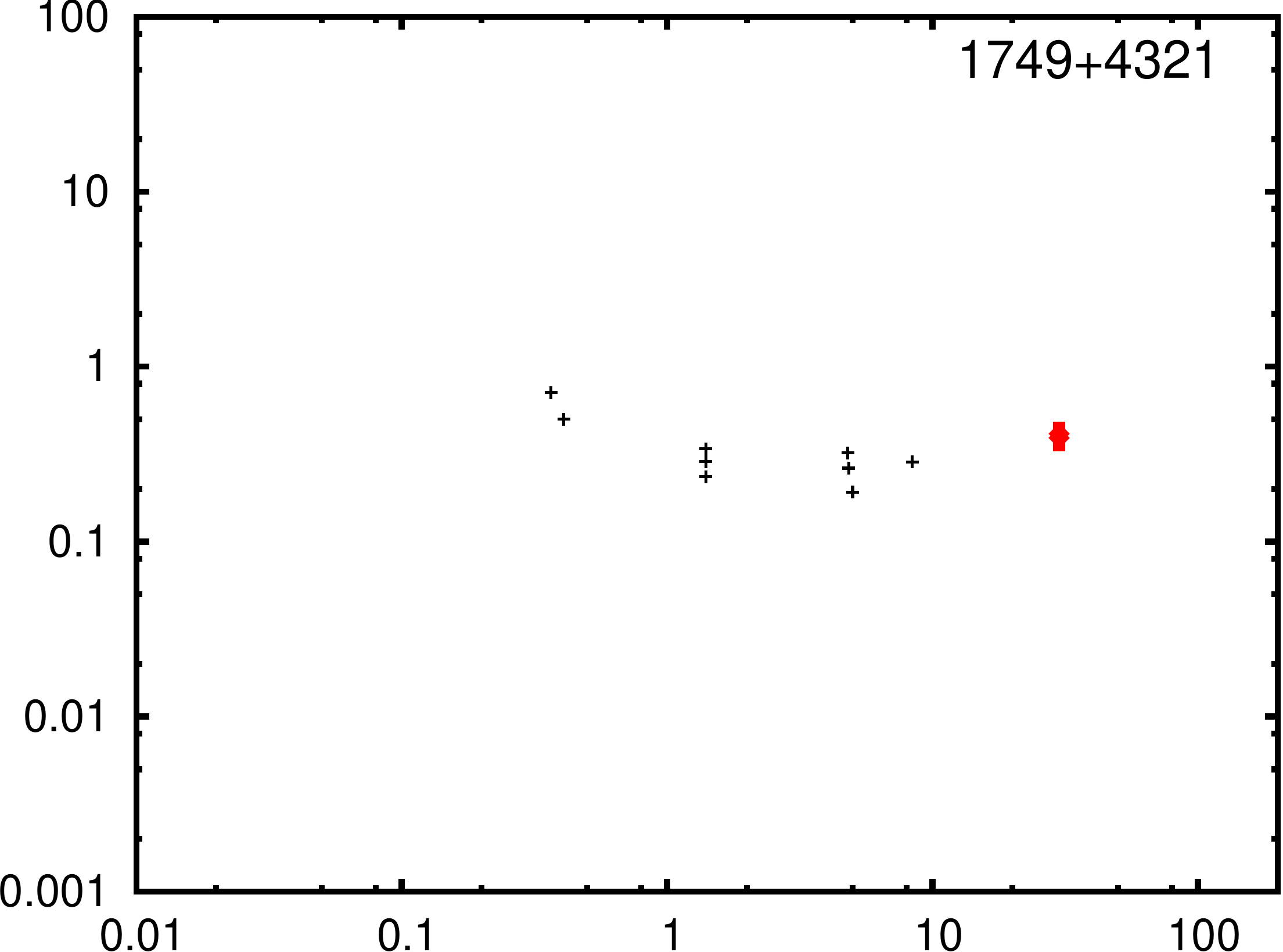}
\end{figure}
\clearpage\begin{figure}
\centering
\includegraphics[scale=0.2]{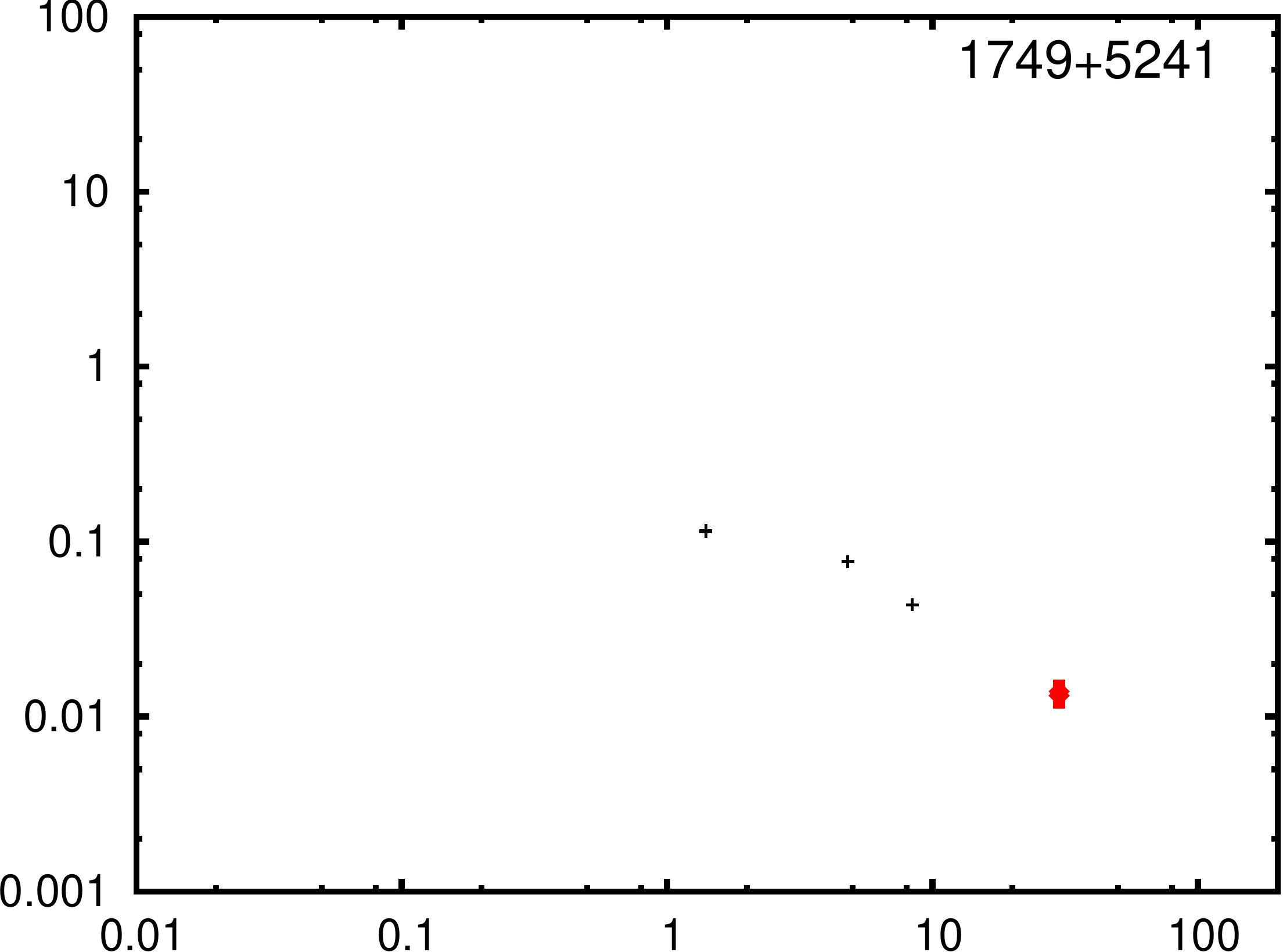}
\includegraphics[scale=0.2]{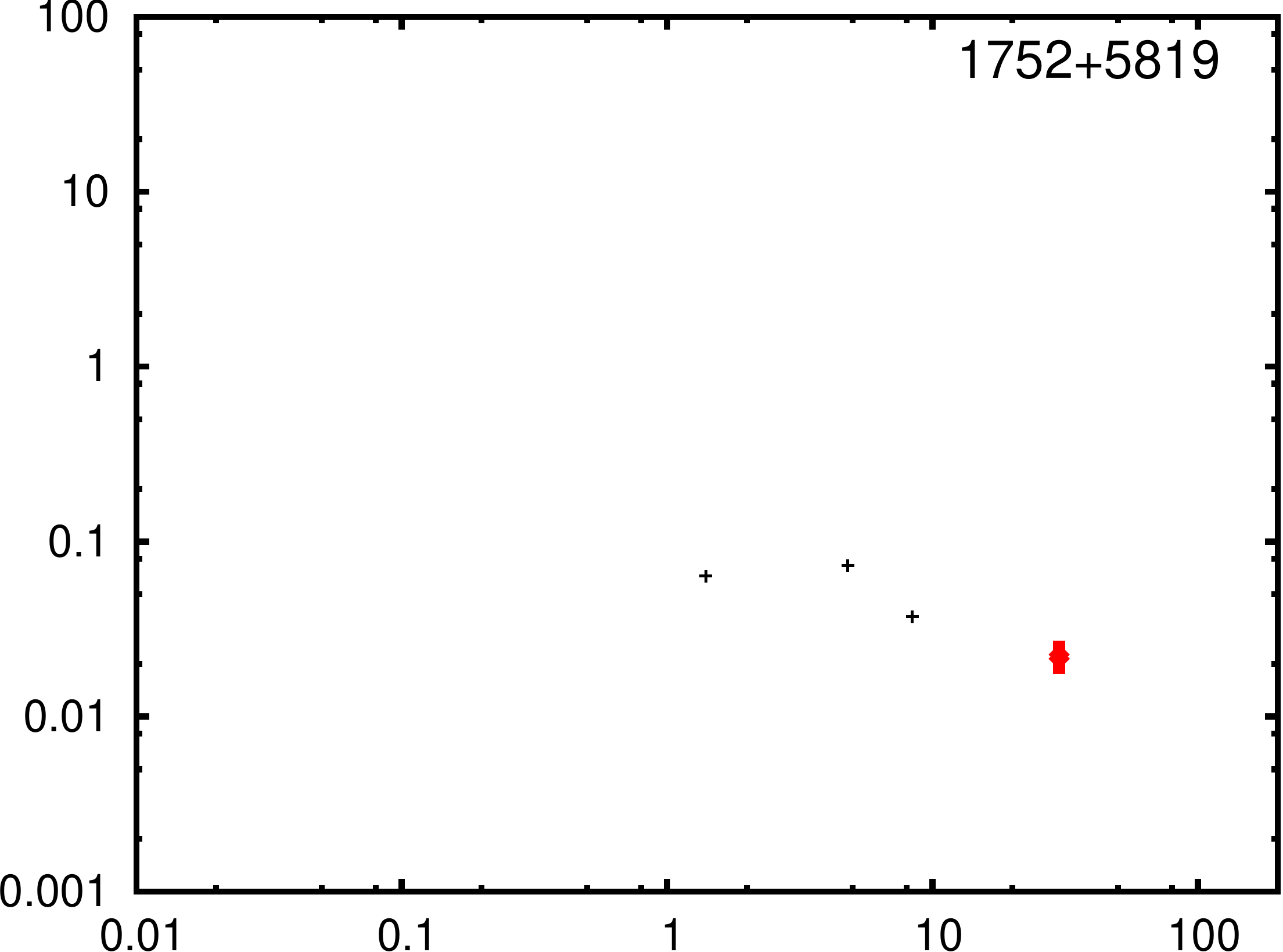}
\includegraphics[scale=0.2]{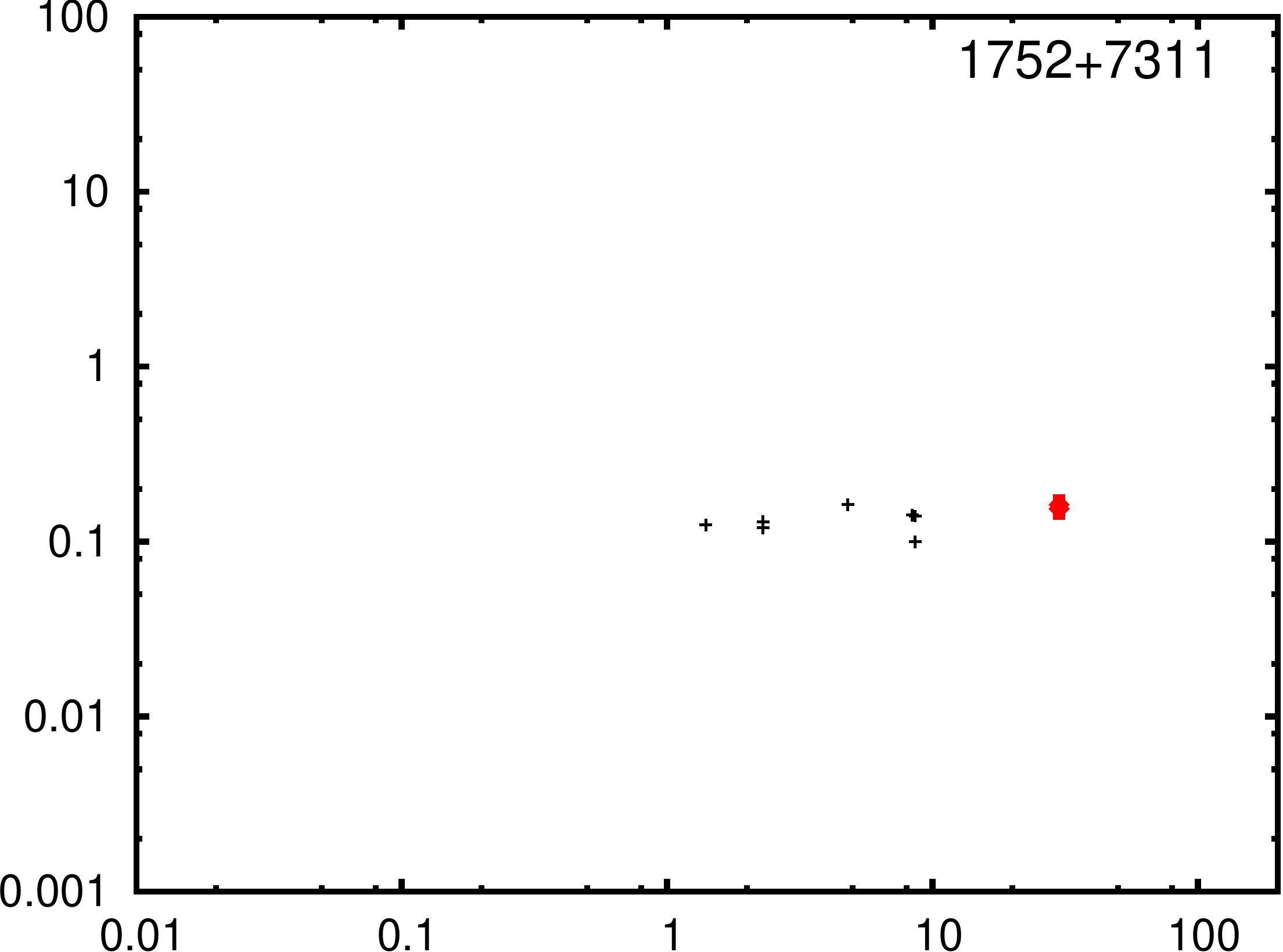}
\includegraphics[scale=0.2]{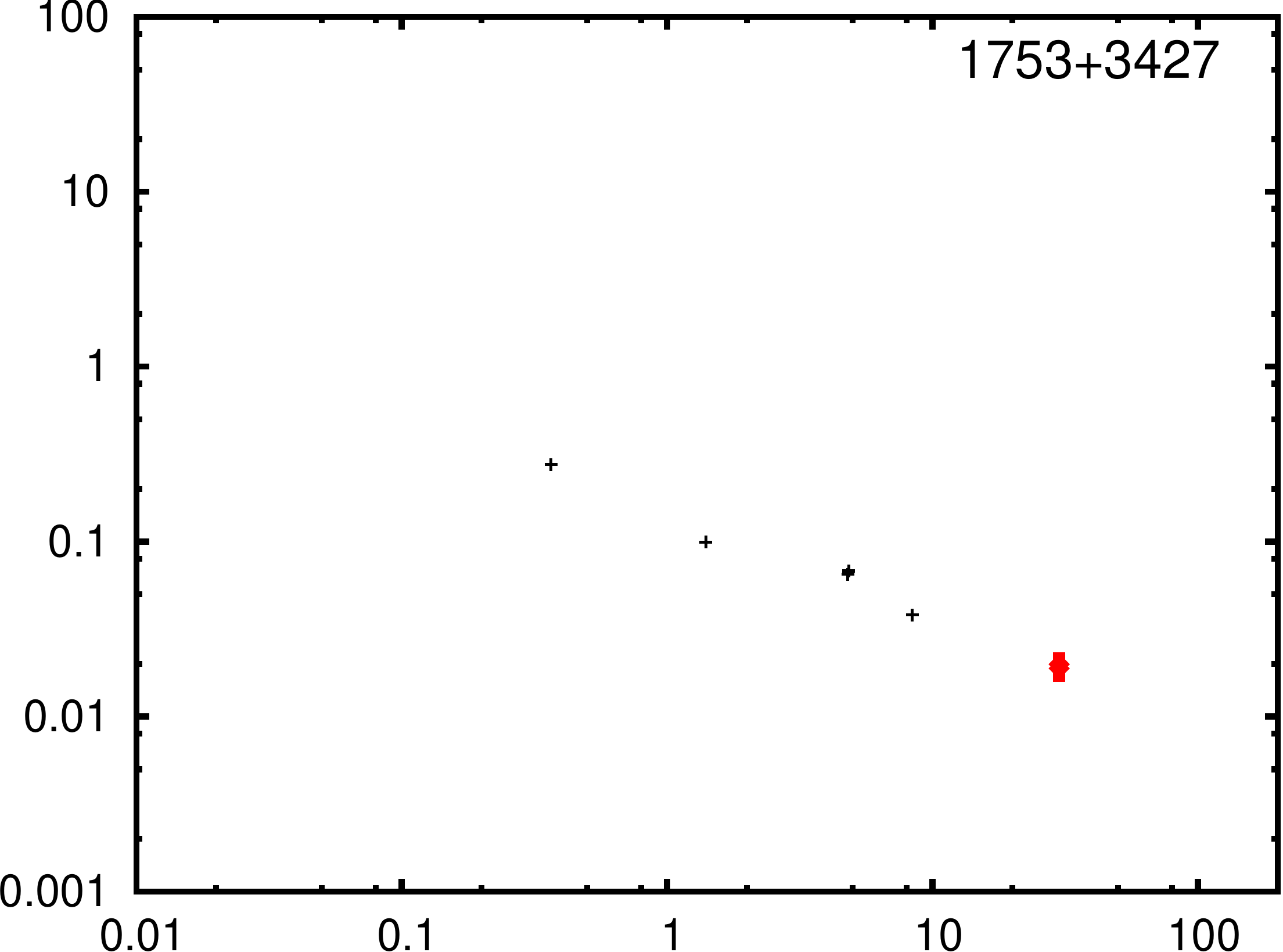}
\includegraphics[scale=0.2]{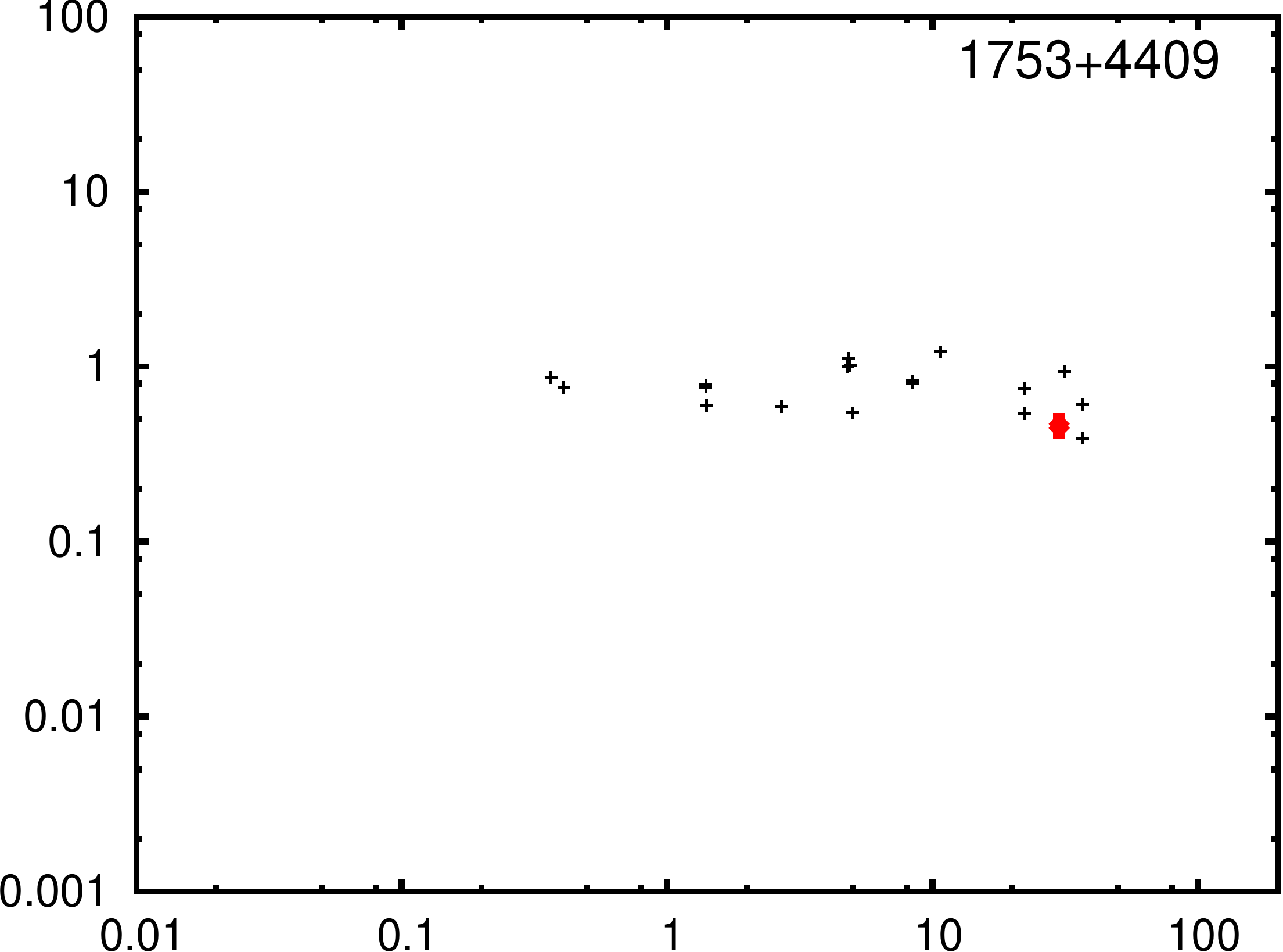}
\includegraphics[scale=0.2]{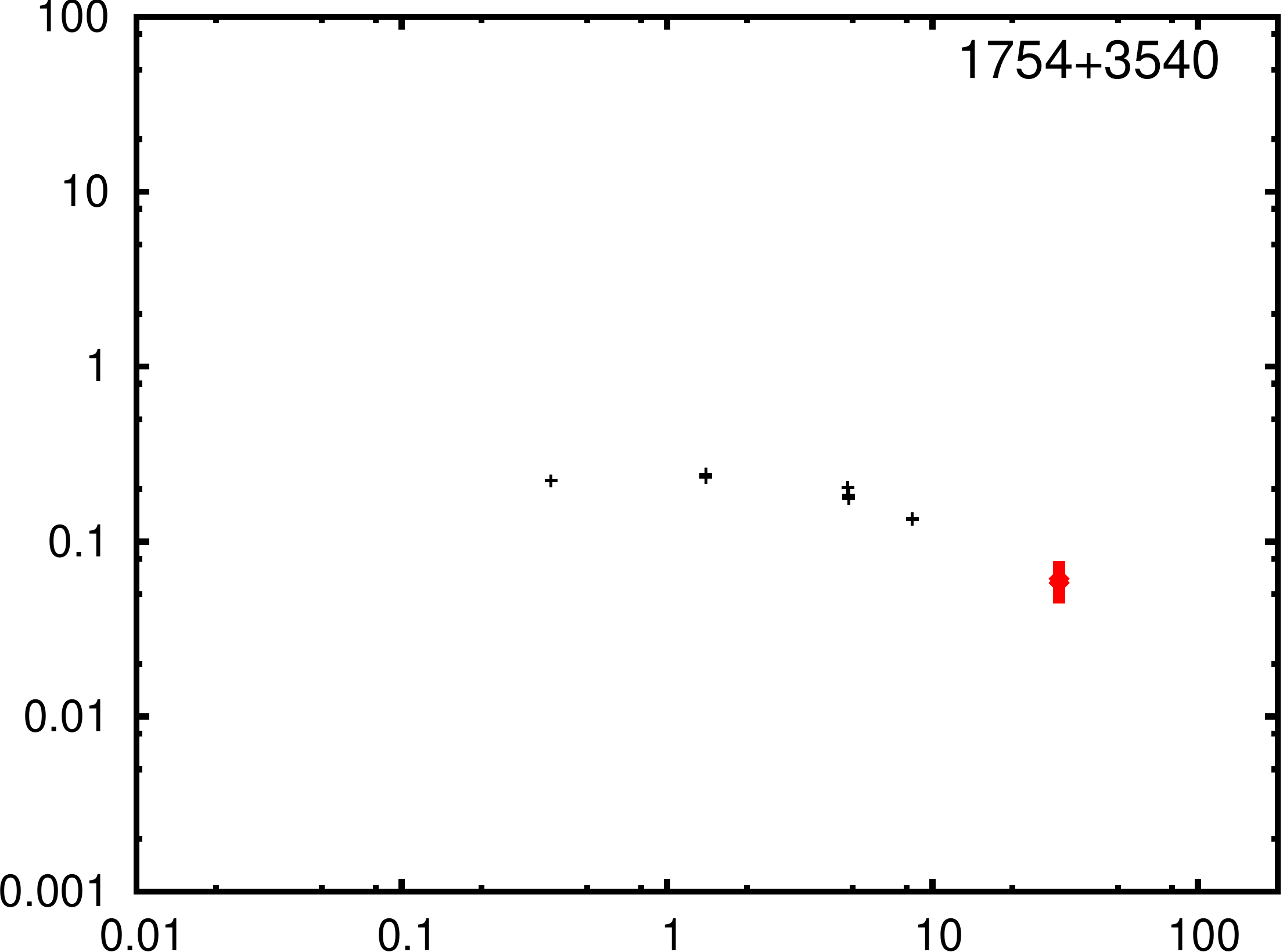}
\includegraphics[scale=0.2]{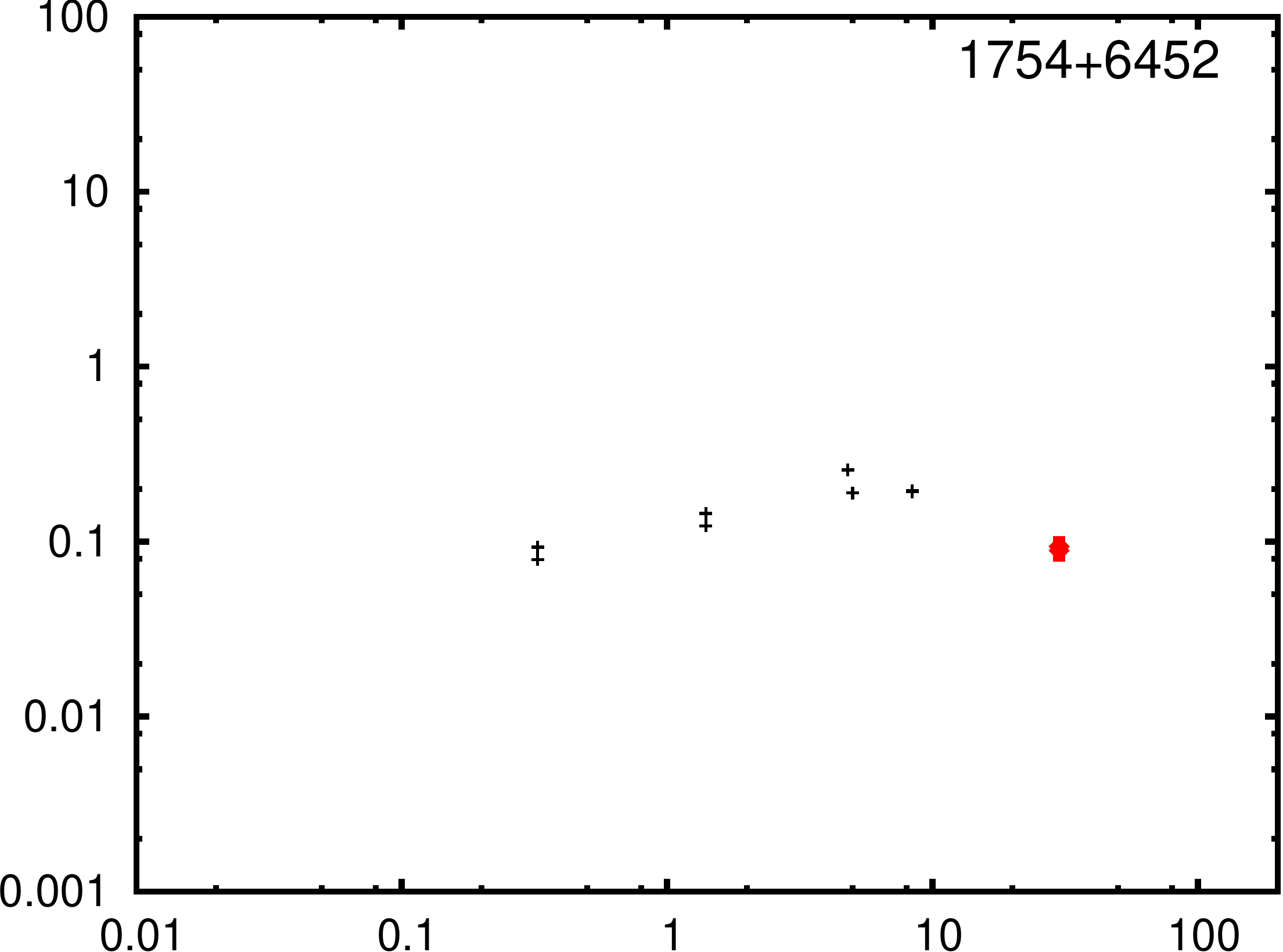}
\includegraphics[scale=0.2]{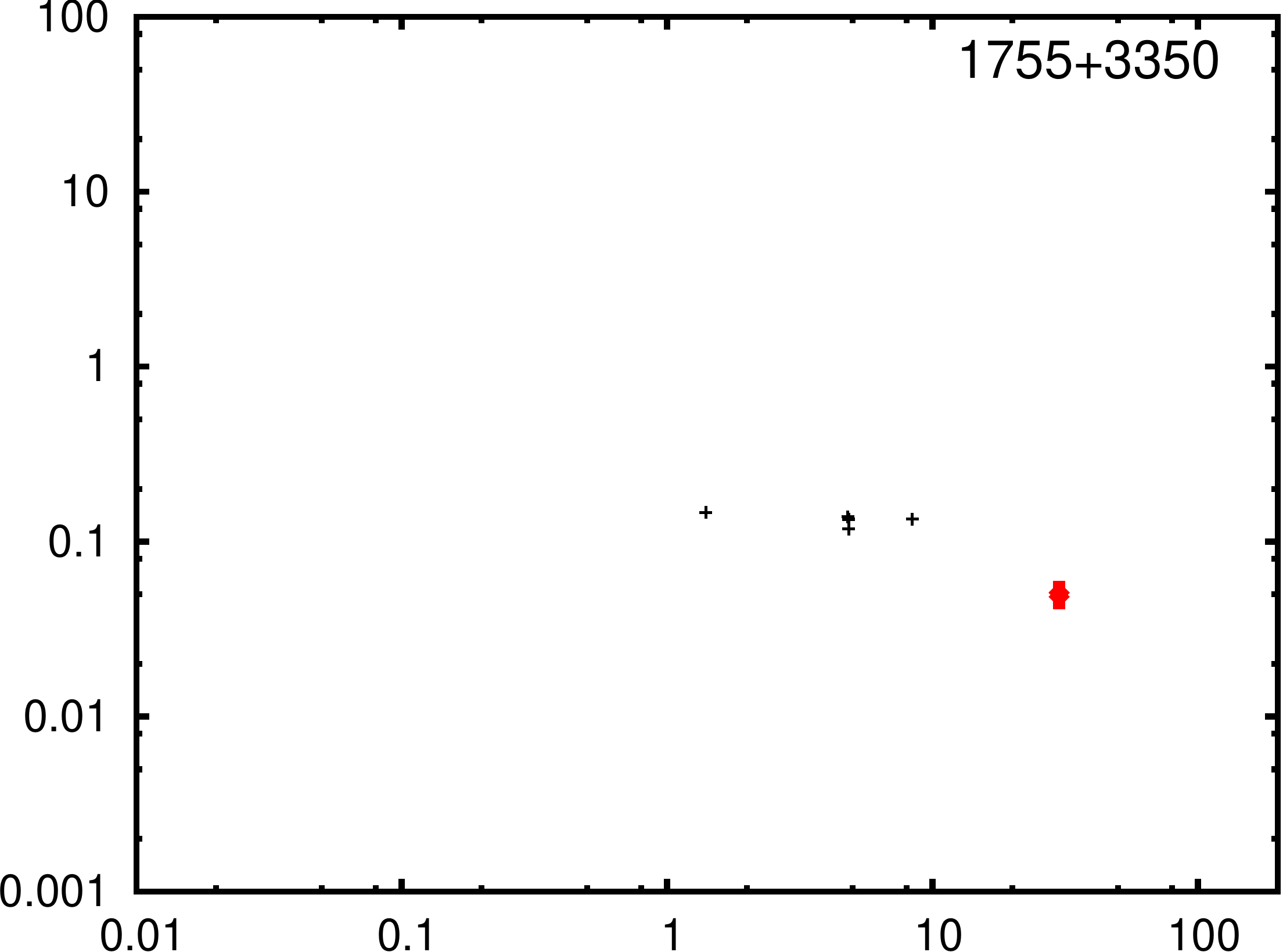}
\includegraphics[scale=0.2]{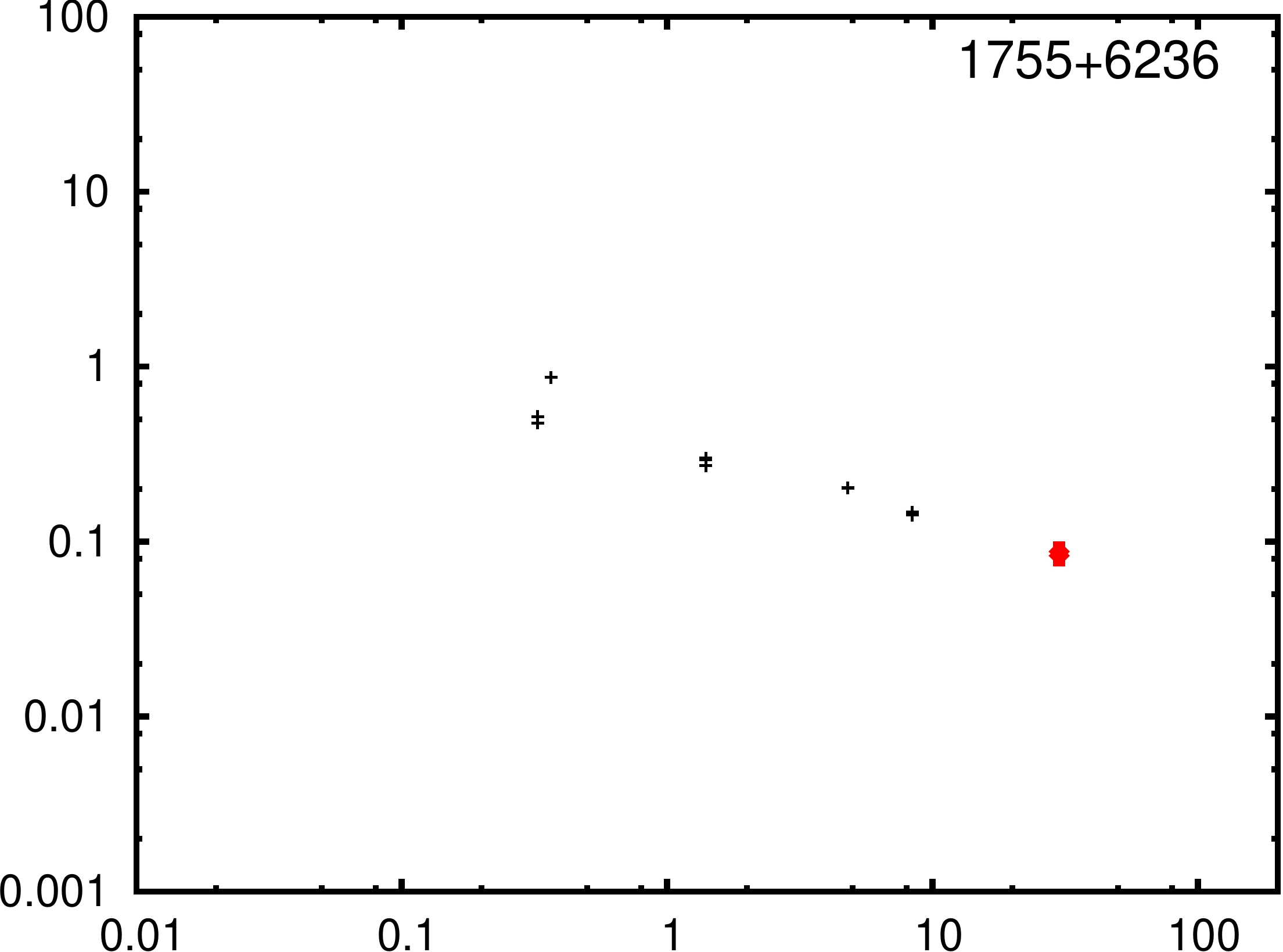}
\includegraphics[scale=0.2]{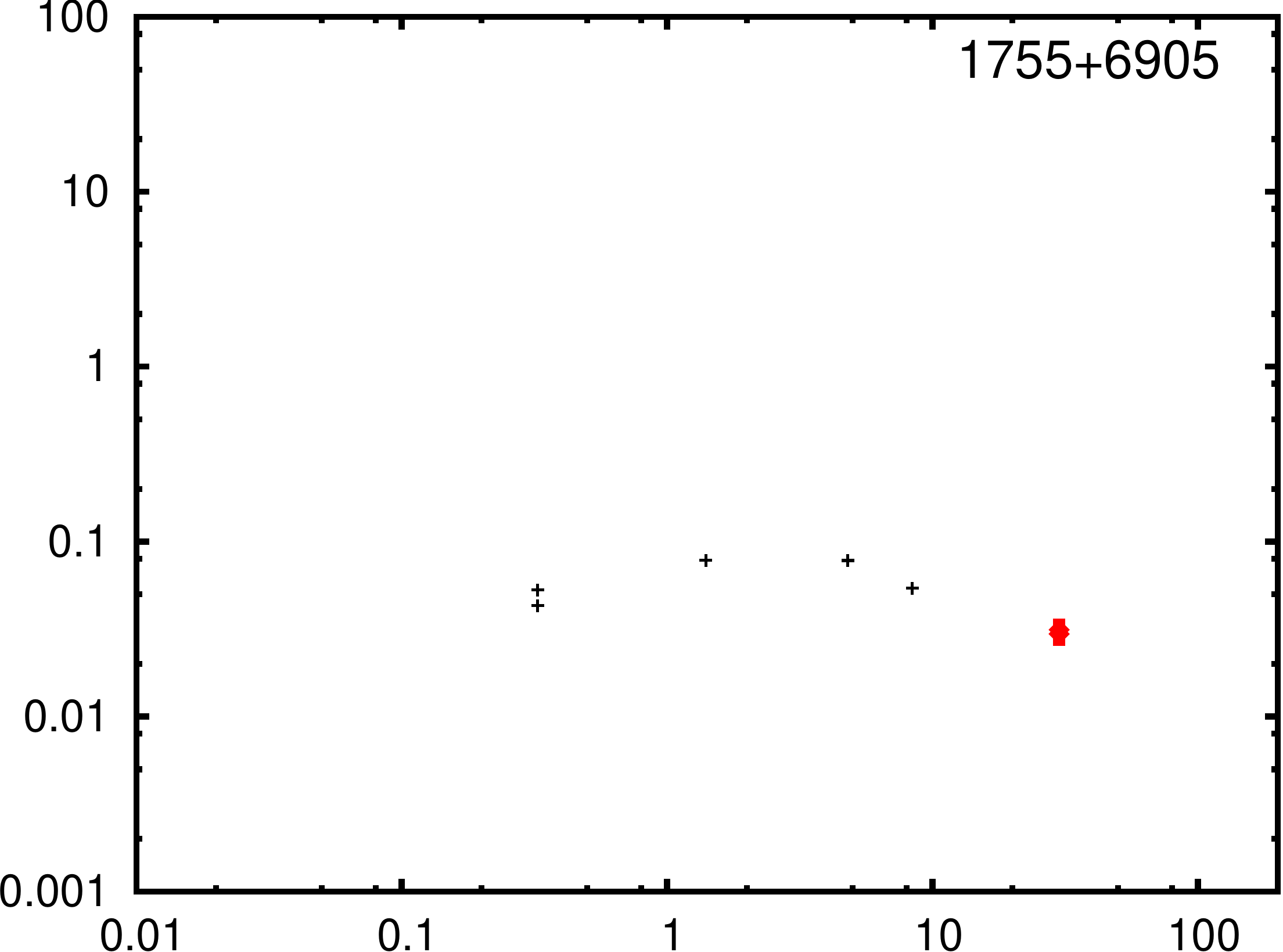}
\includegraphics[scale=0.2]{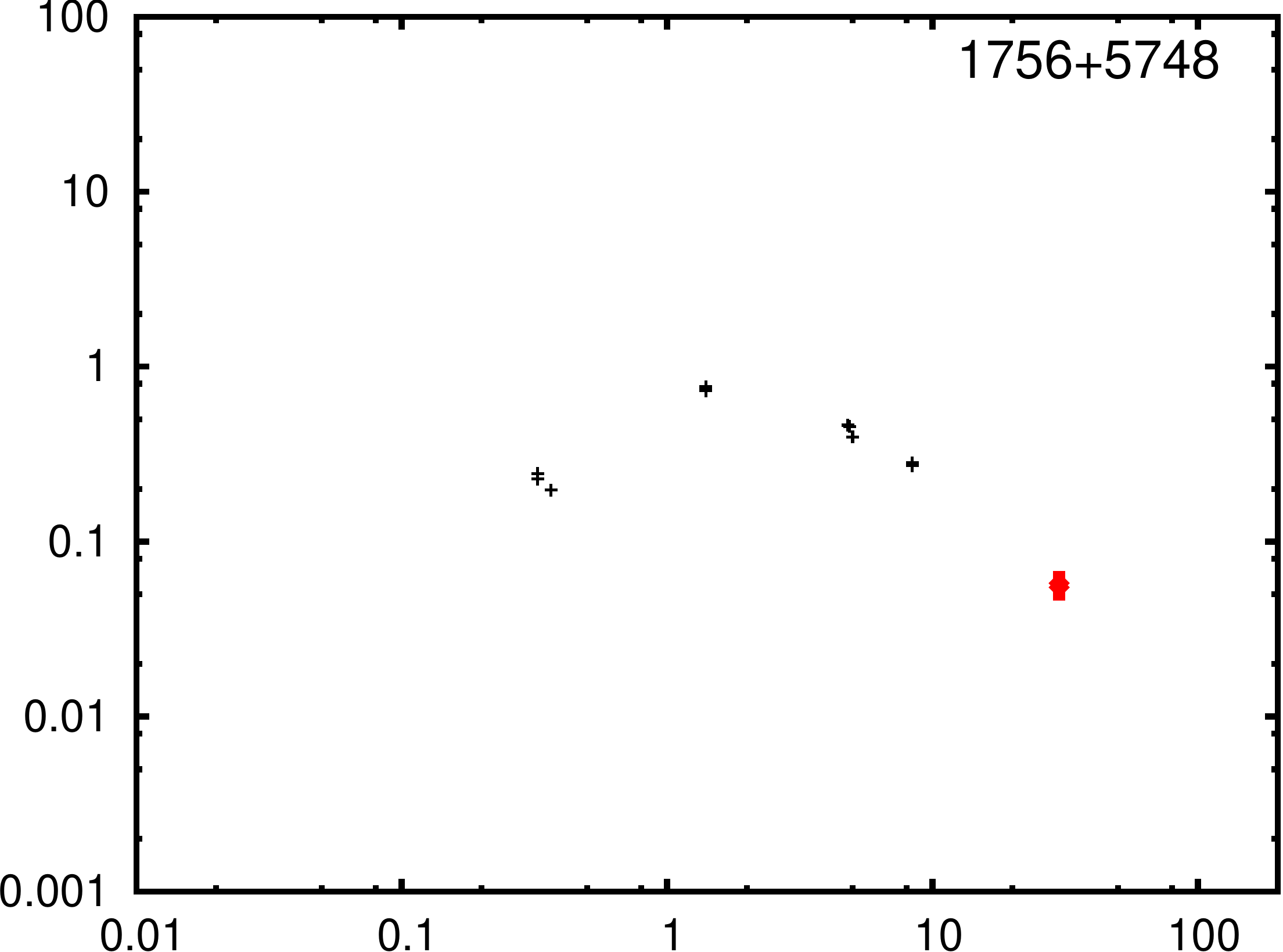}
\includegraphics[scale=0.2]{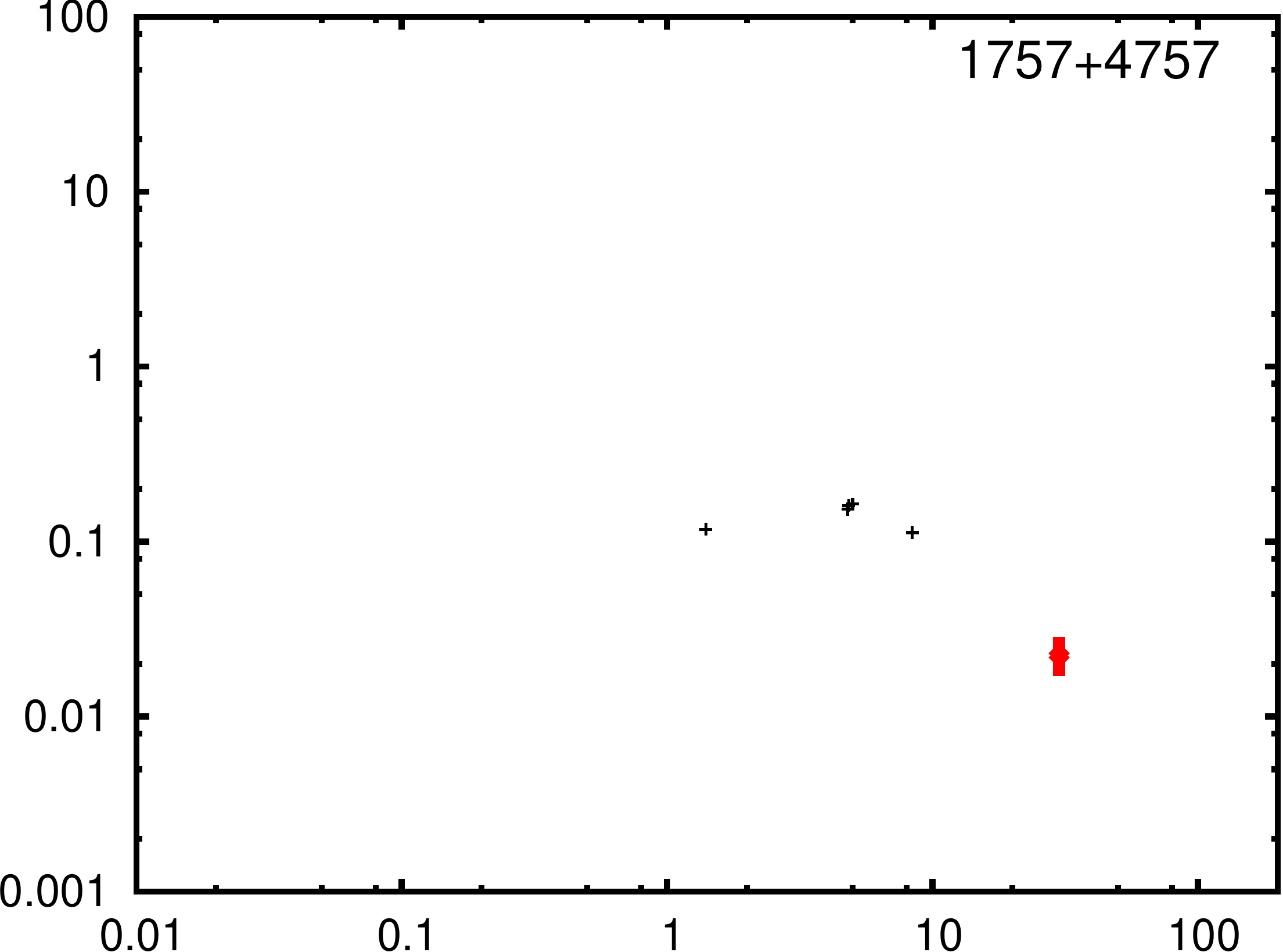}
\includegraphics[scale=0.2]{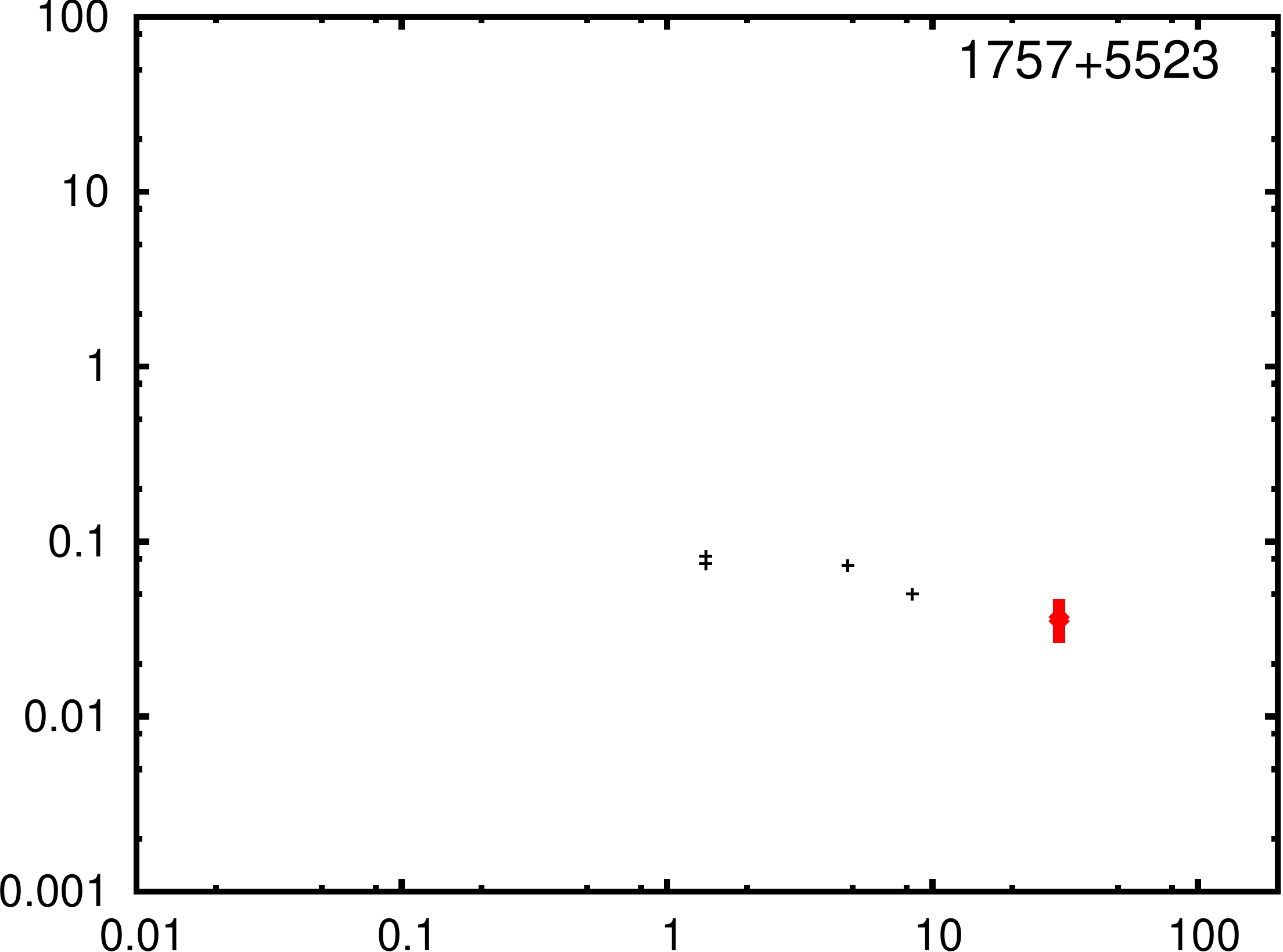}
\includegraphics[scale=0.2]{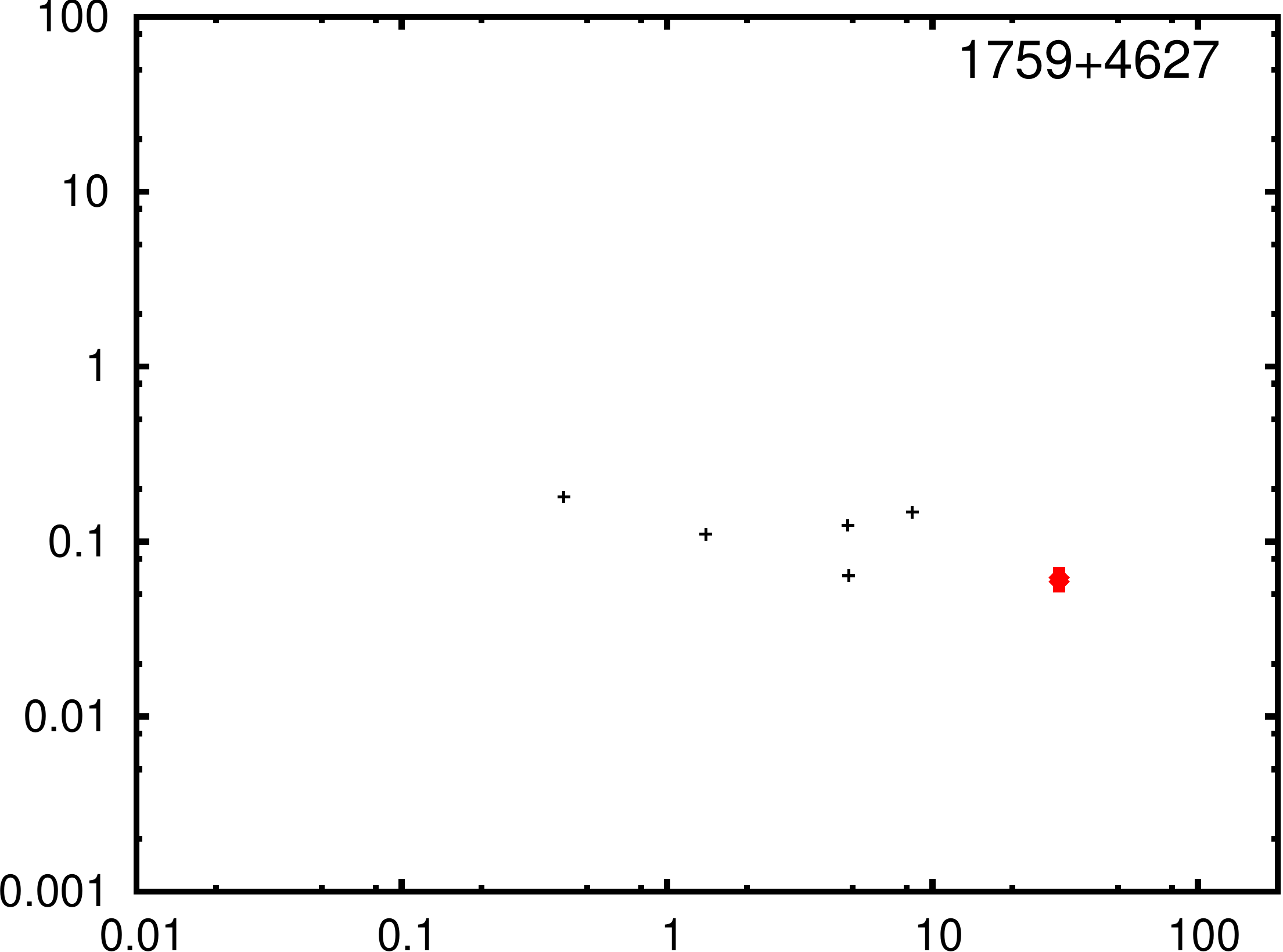}
\includegraphics[scale=0.2]{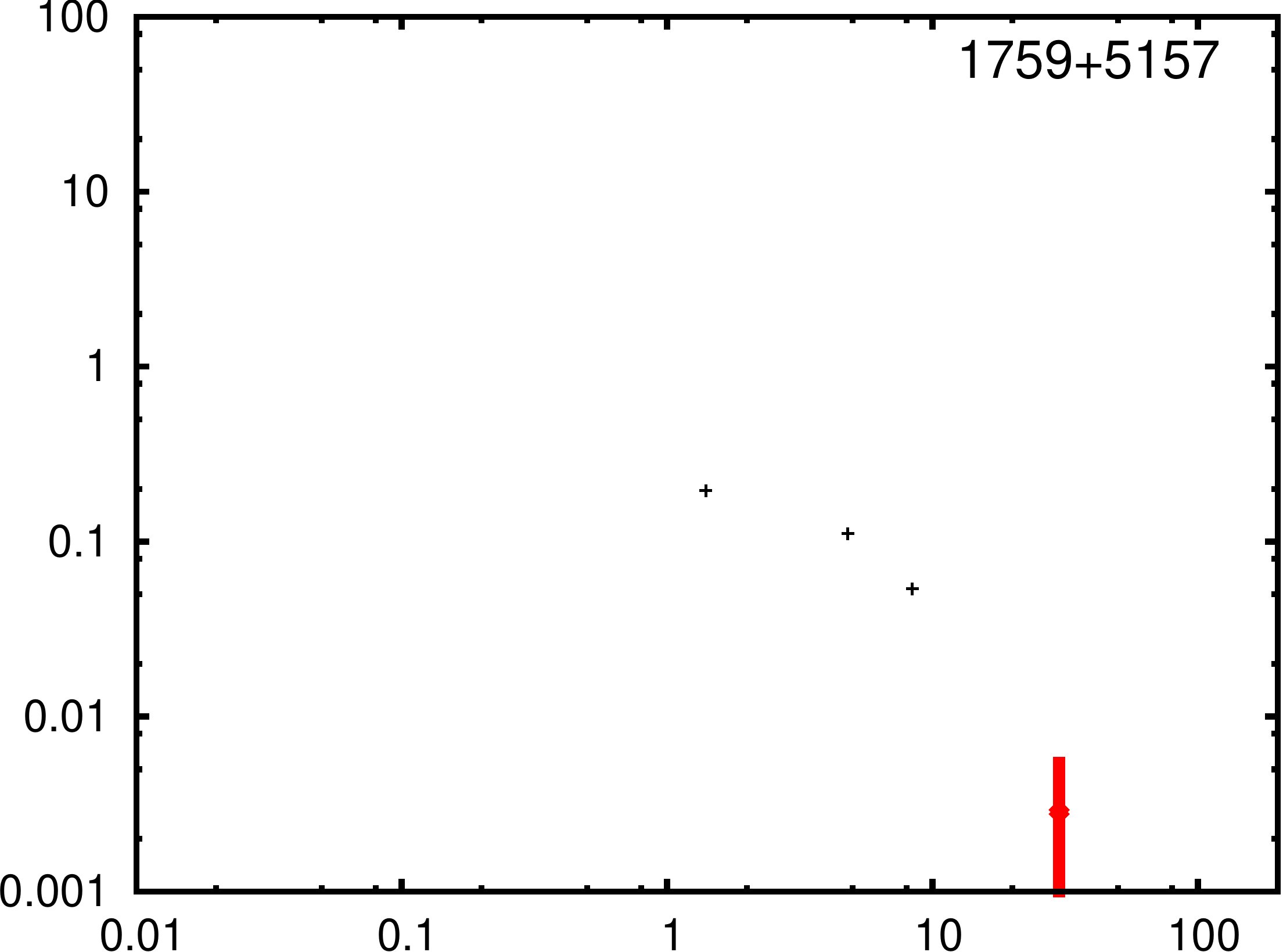}
\includegraphics[scale=0.2]{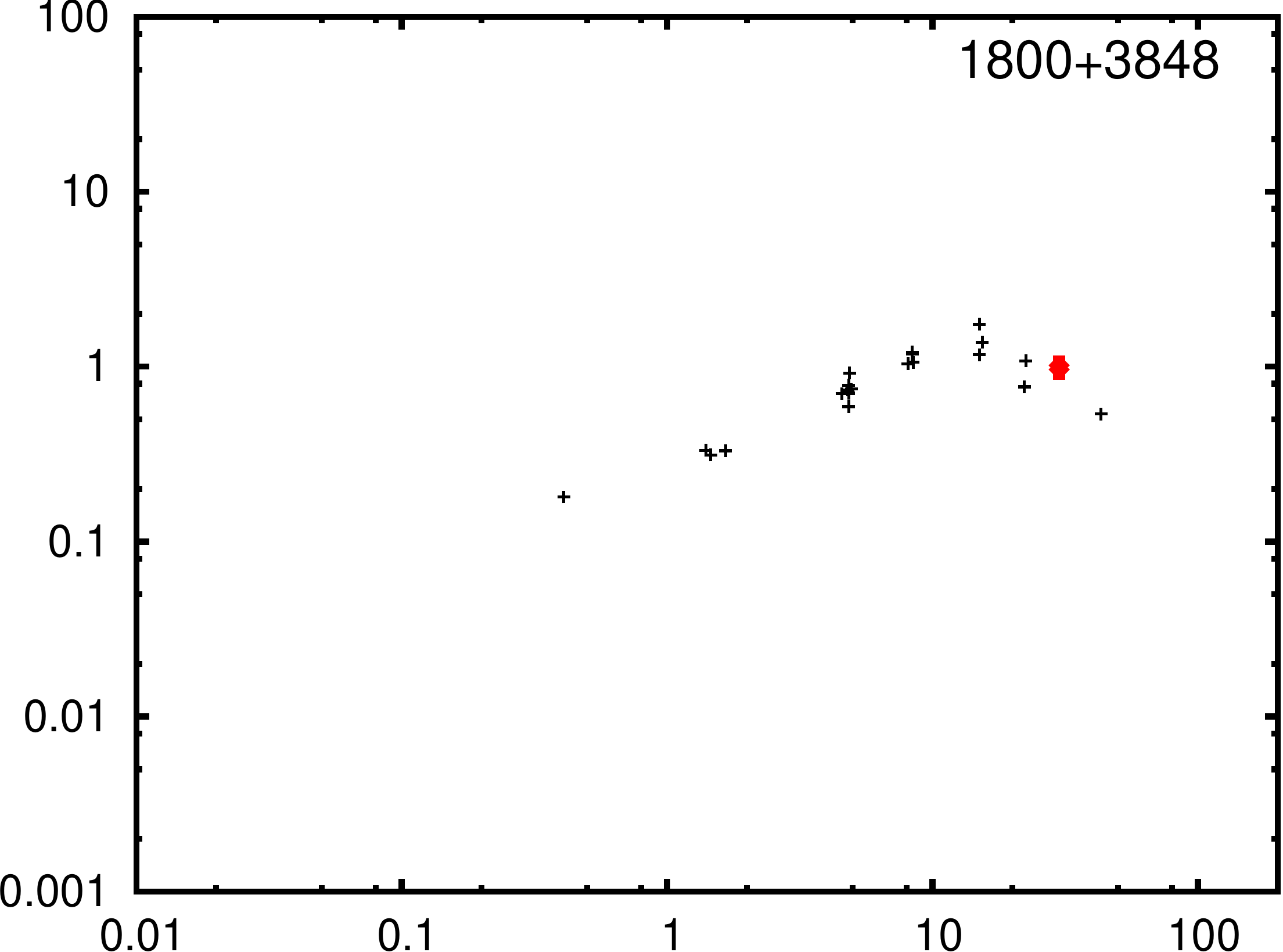}
\includegraphics[scale=0.2]{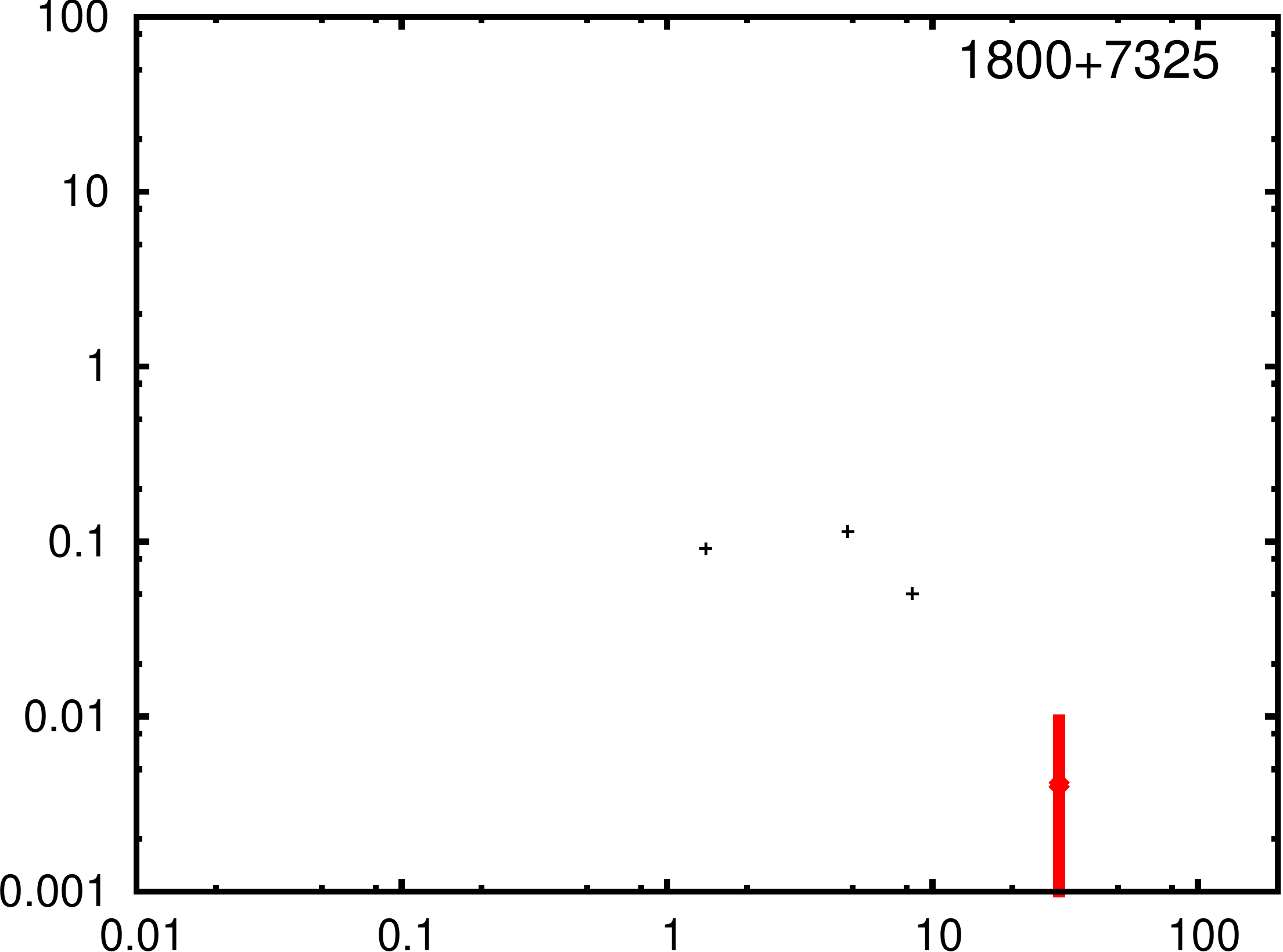}
\includegraphics[scale=0.2]{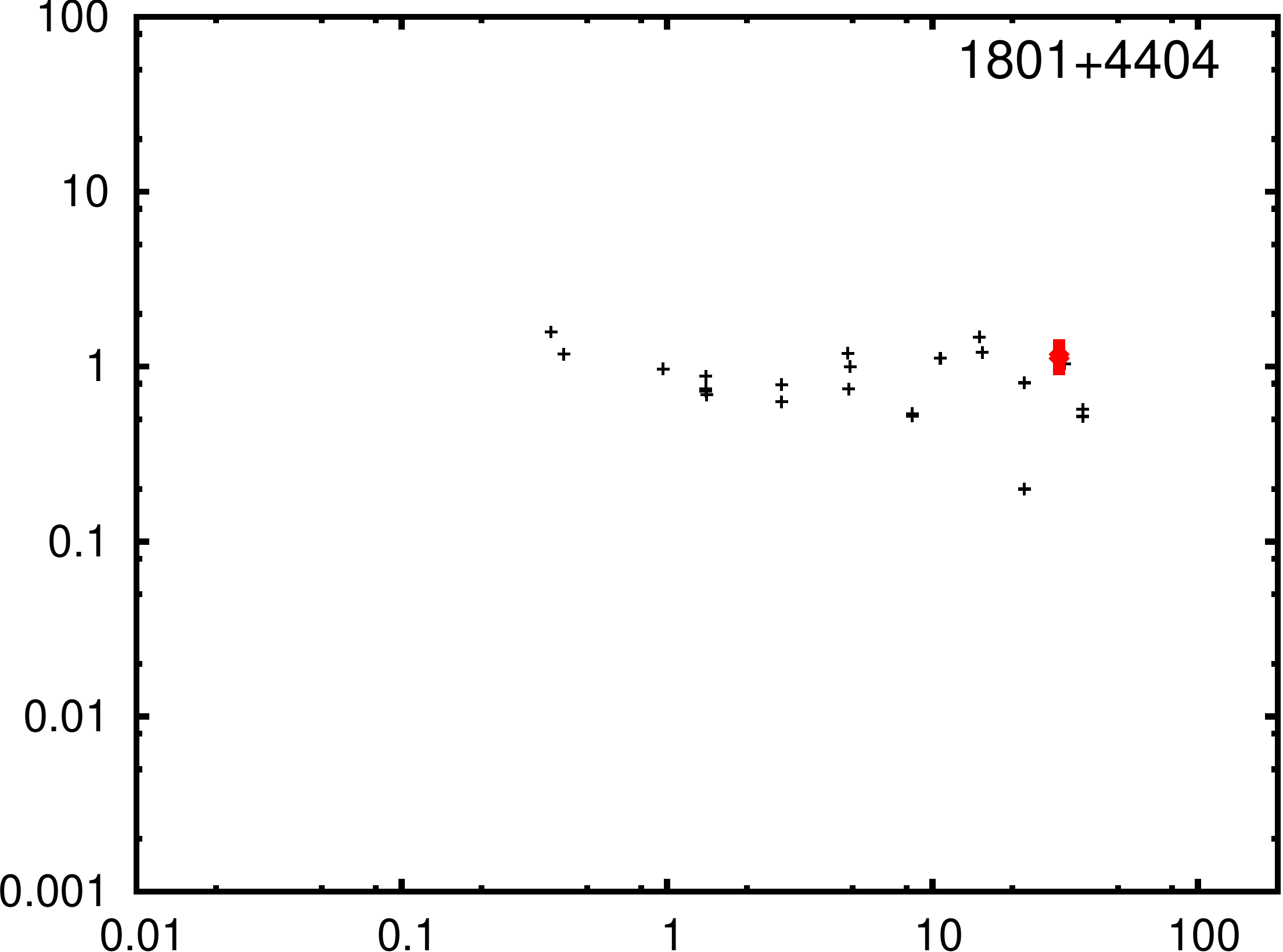}
\end{figure}
\clearpage\begin{figure}
\centering
\includegraphics[scale=0.2]{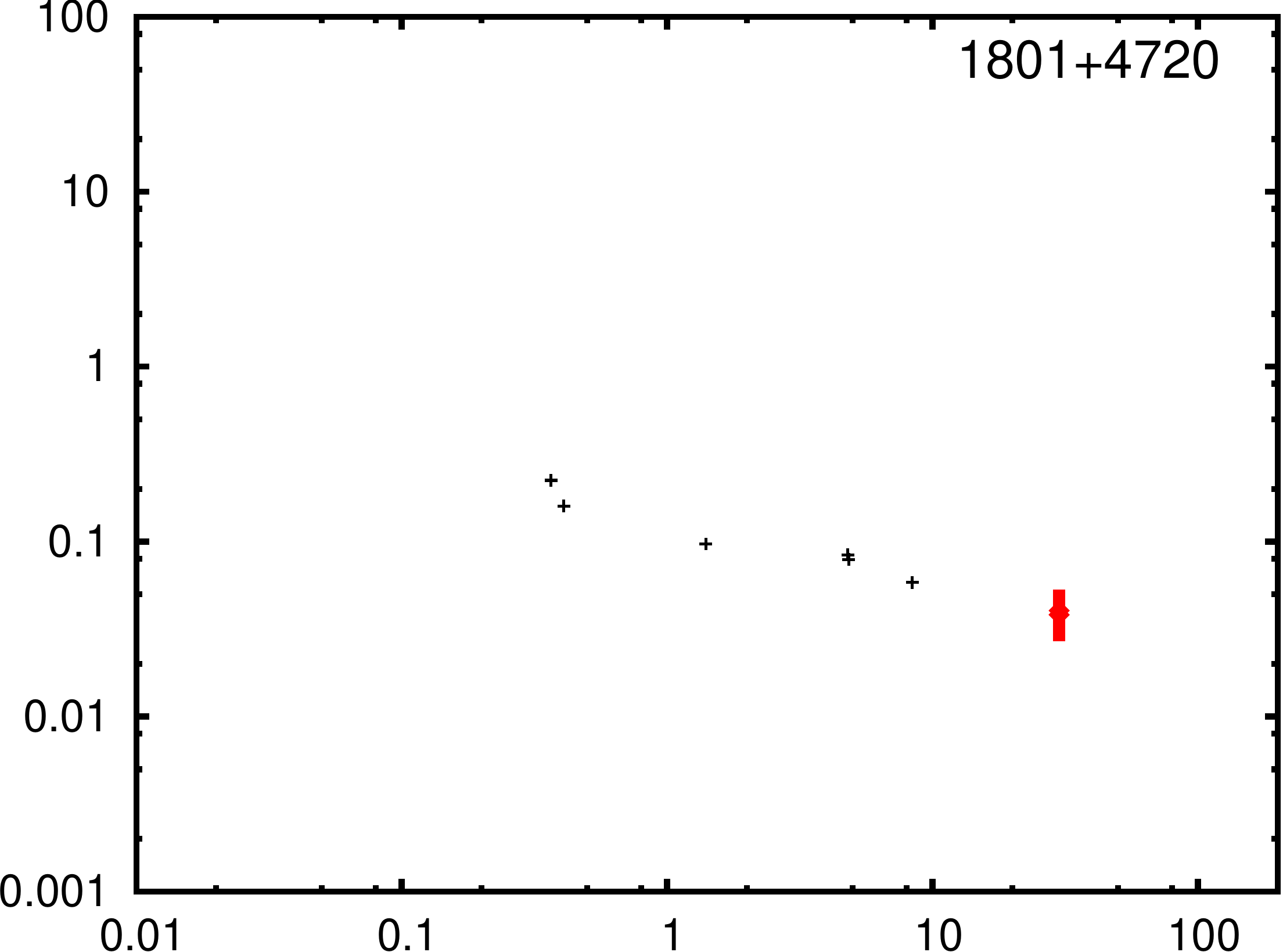}
\includegraphics[scale=0.2]{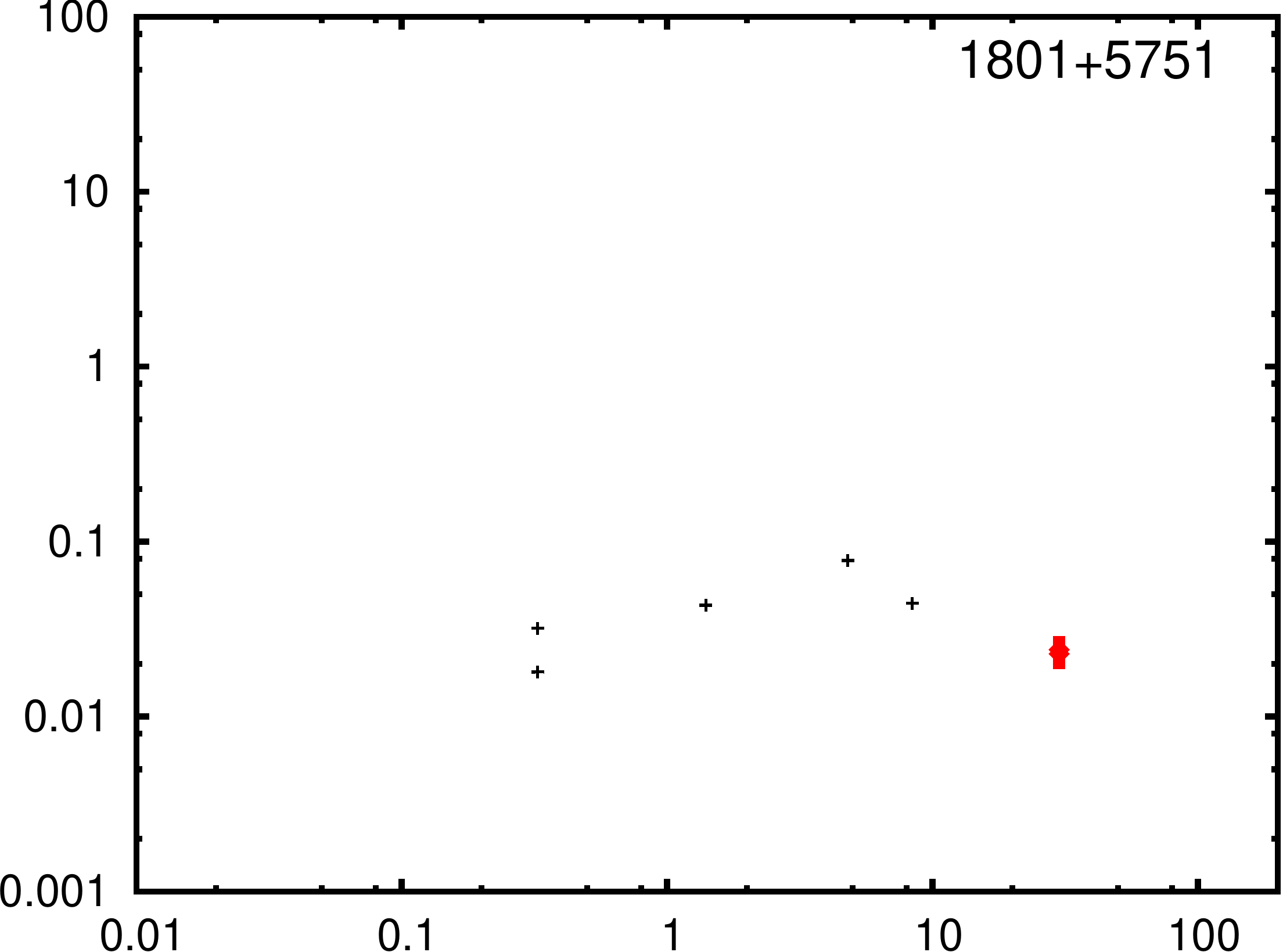}
\includegraphics[scale=0.2]{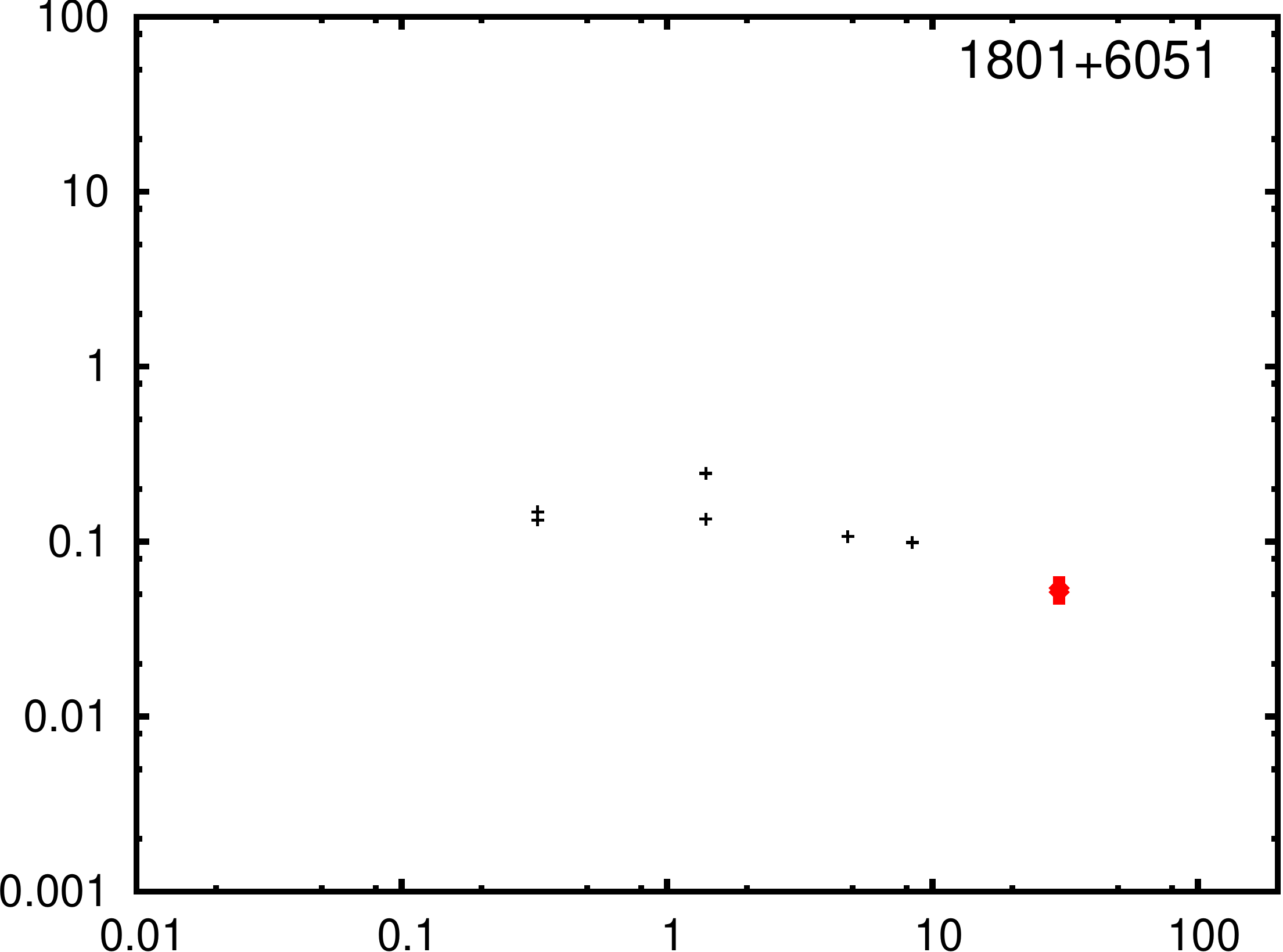}
\includegraphics[scale=0.2]{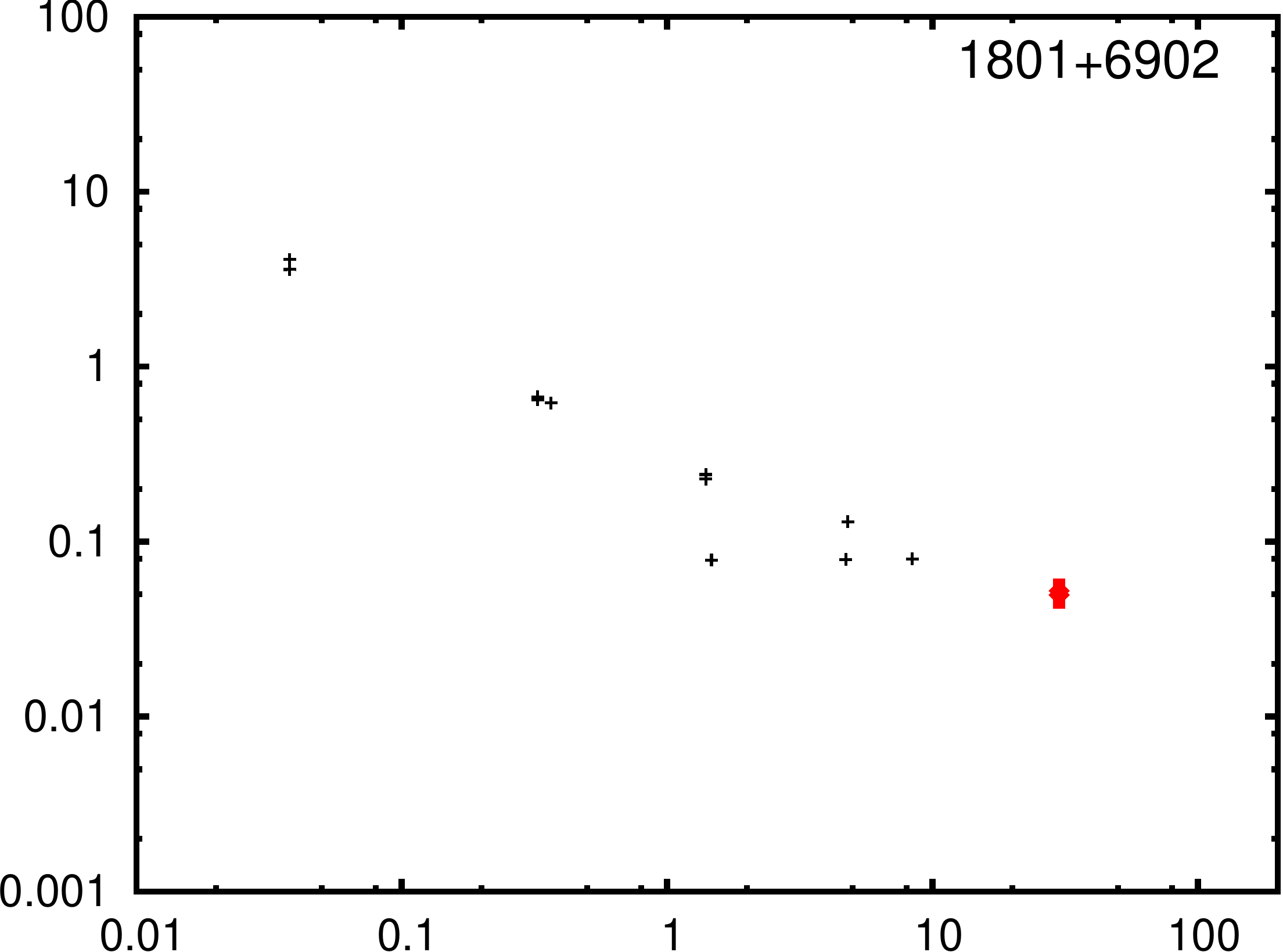}
\includegraphics[scale=0.2]{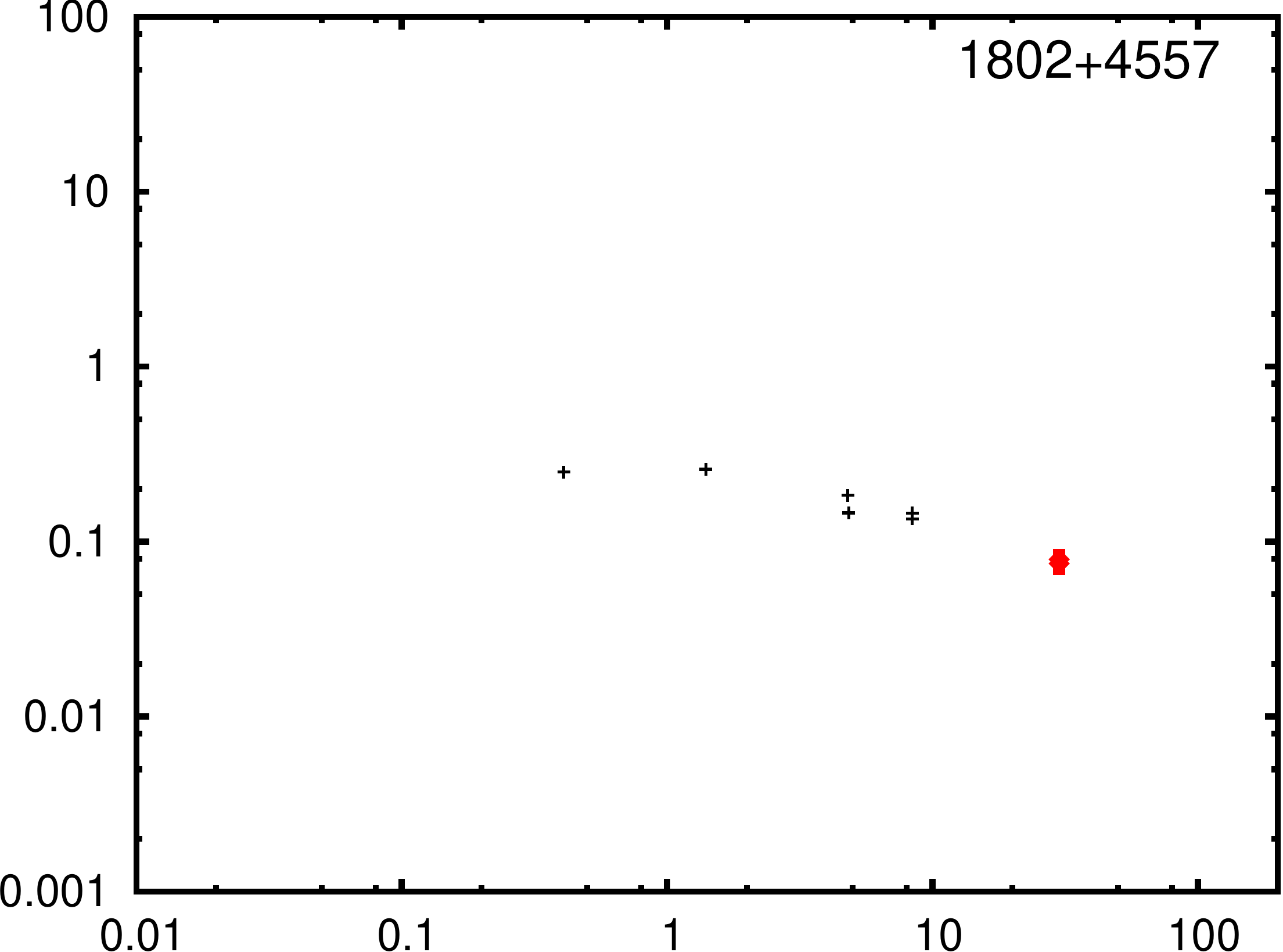}
\includegraphics[scale=0.2]{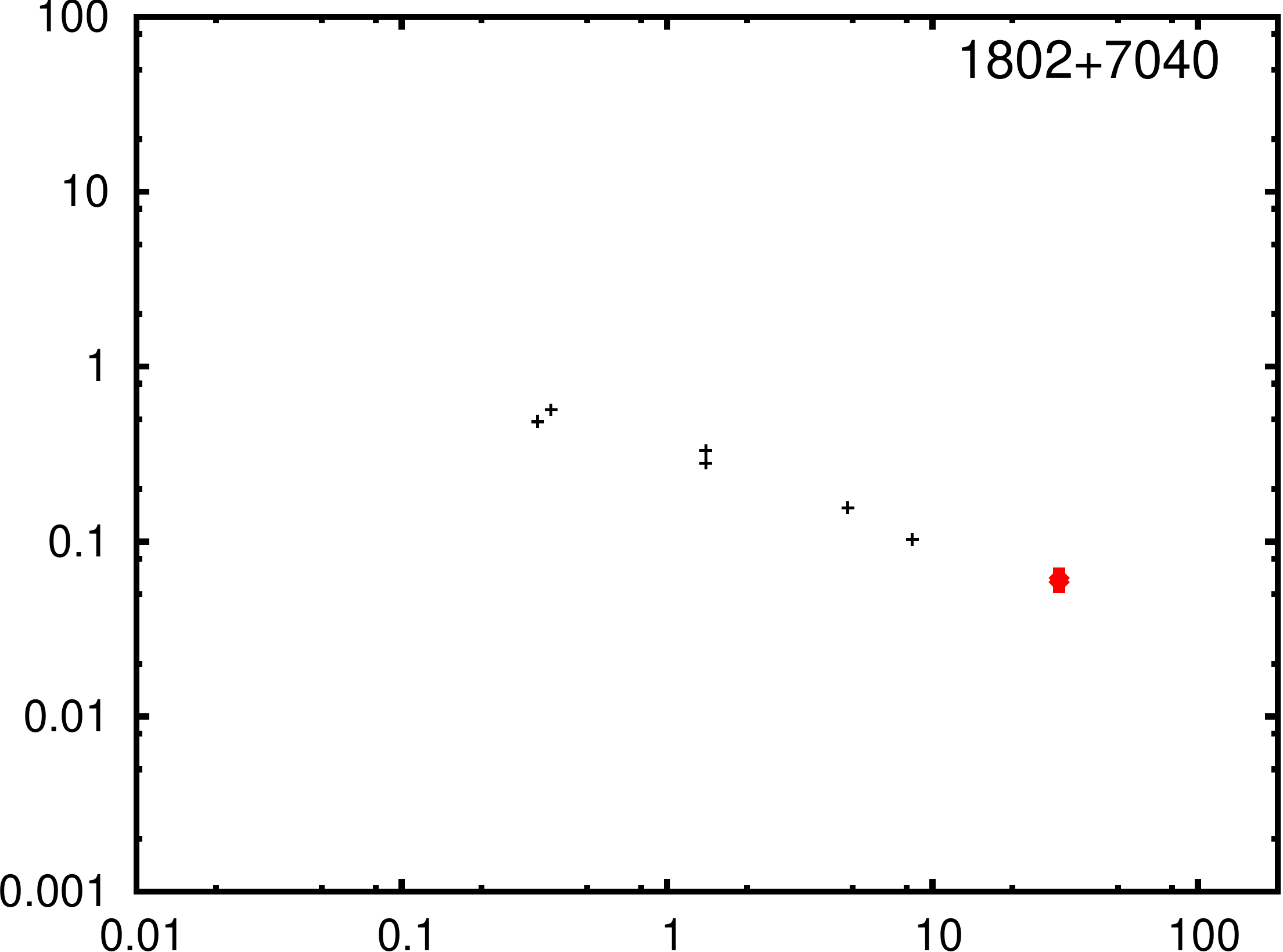}
\includegraphics[scale=0.2]{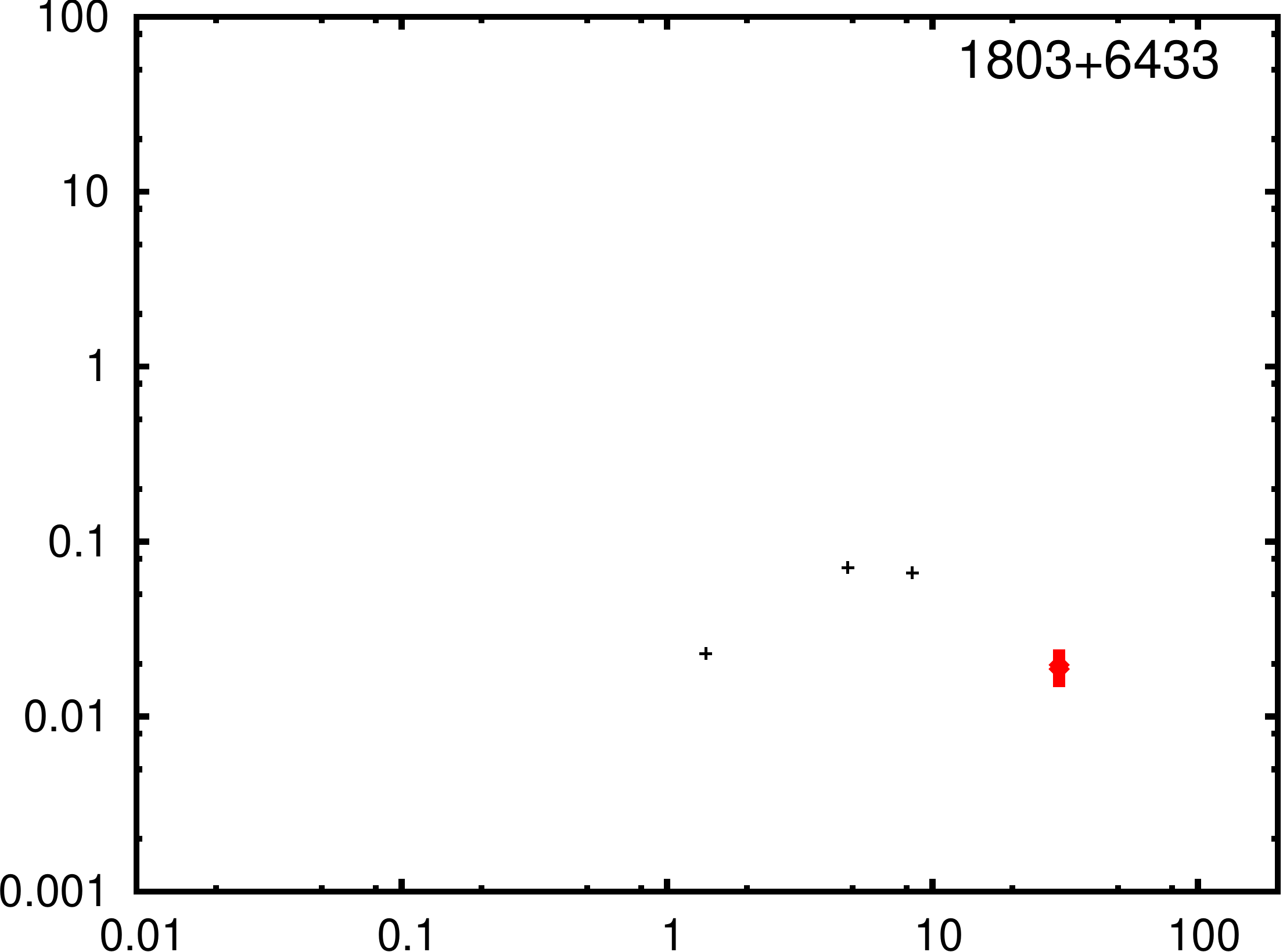}
\includegraphics[scale=0.2]{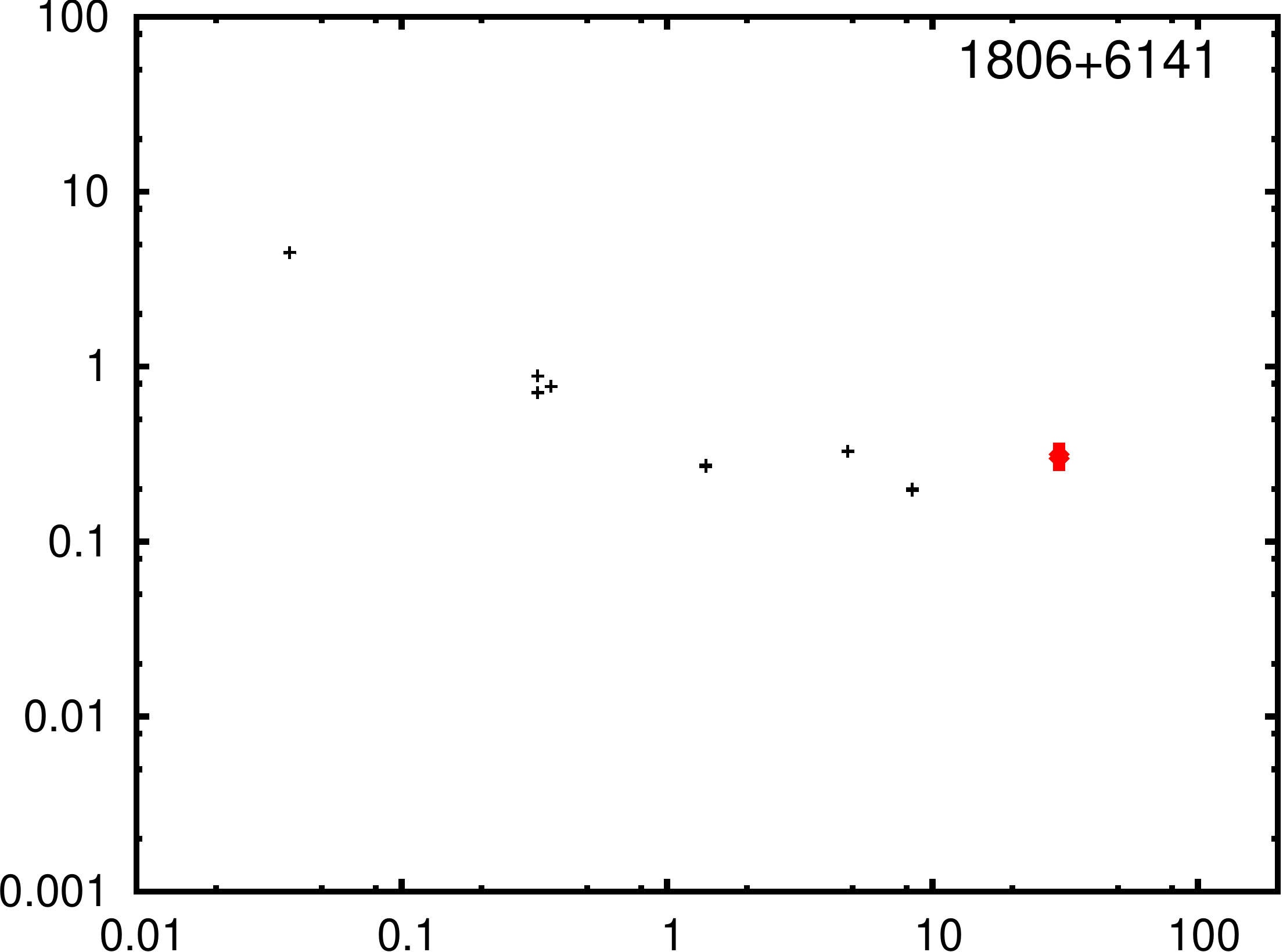}
\includegraphics[scale=0.2]{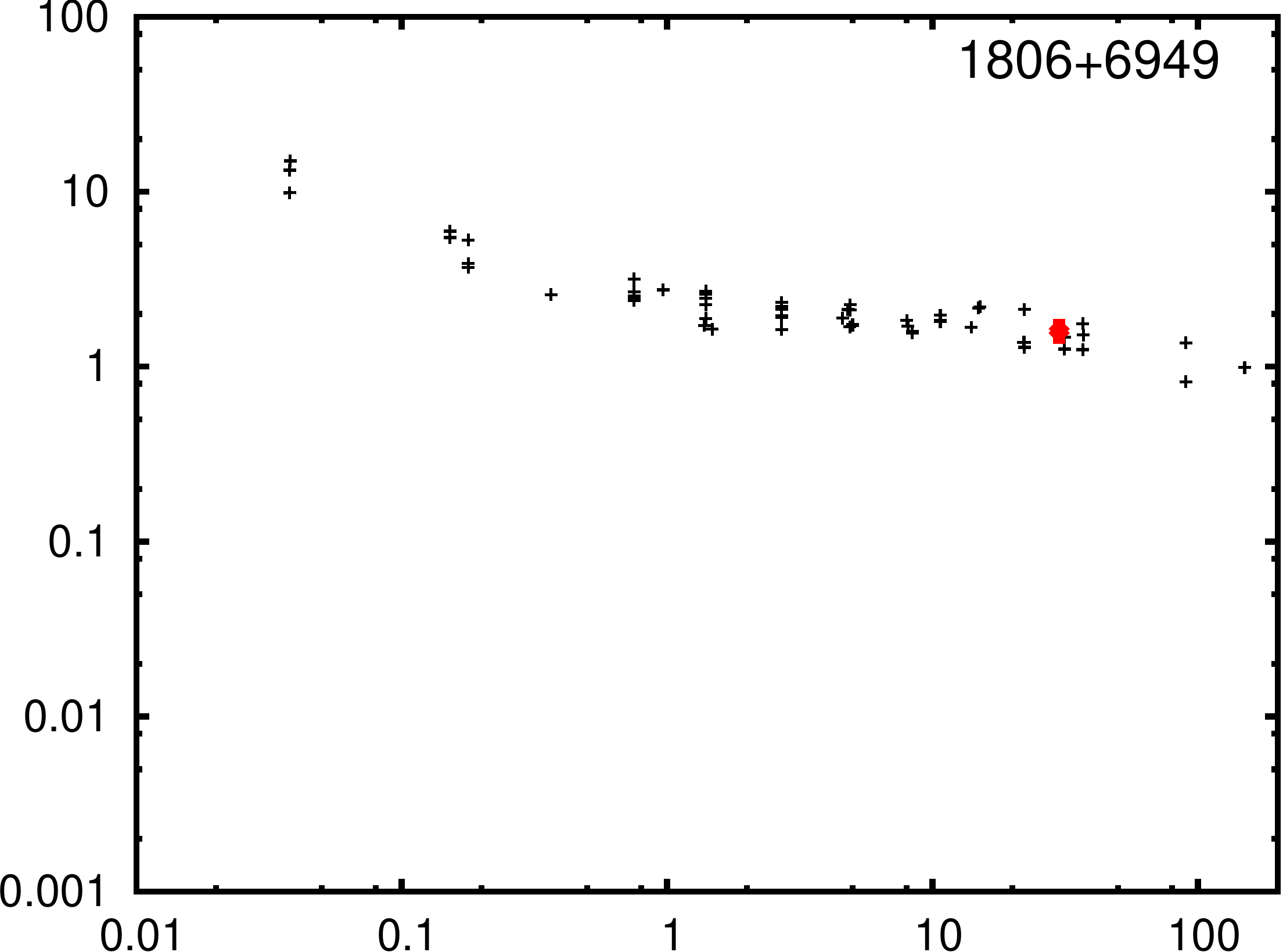}
\includegraphics[scale=0.2]{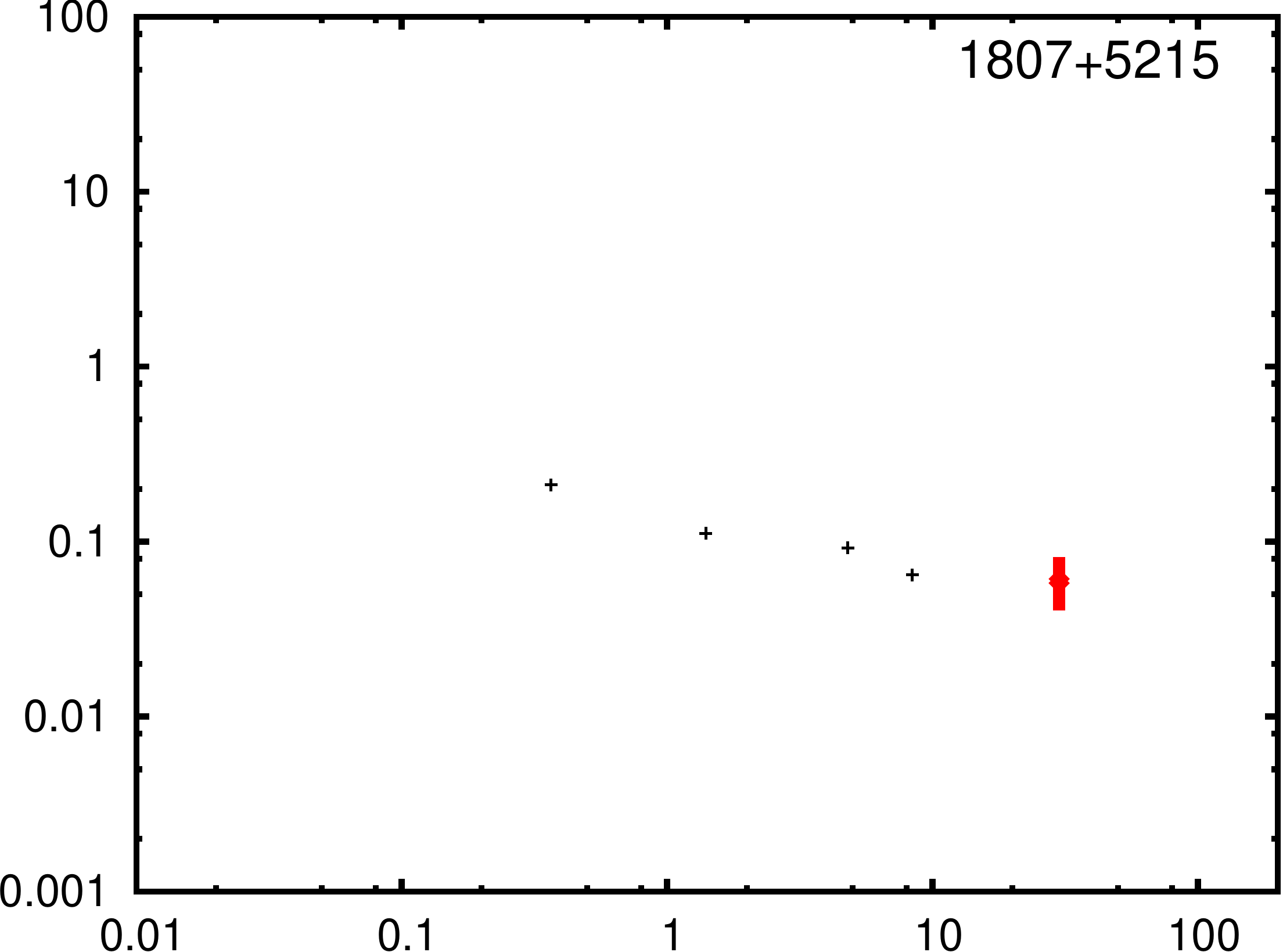}
\includegraphics[scale=0.2]{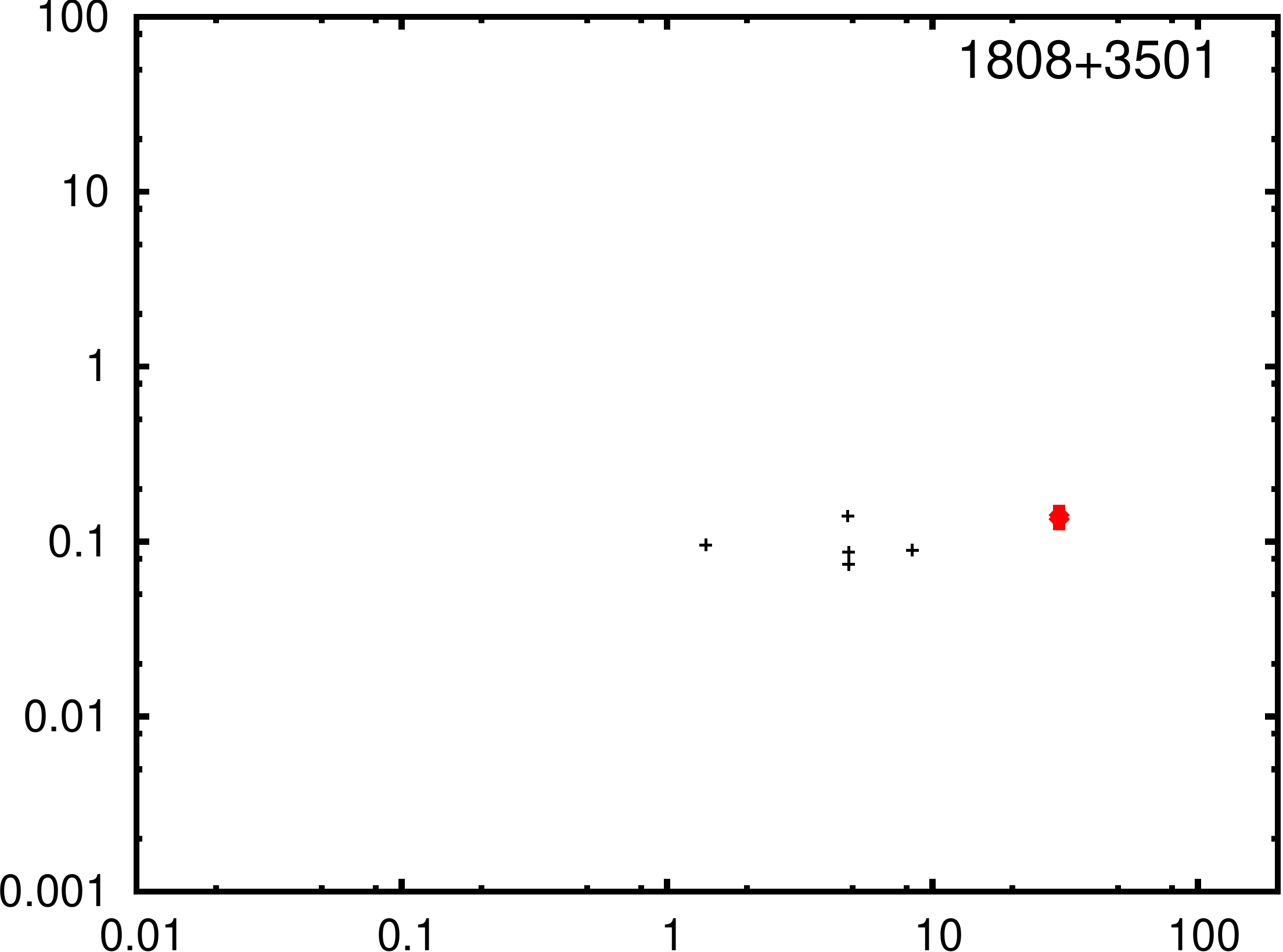}
\includegraphics[scale=0.2]{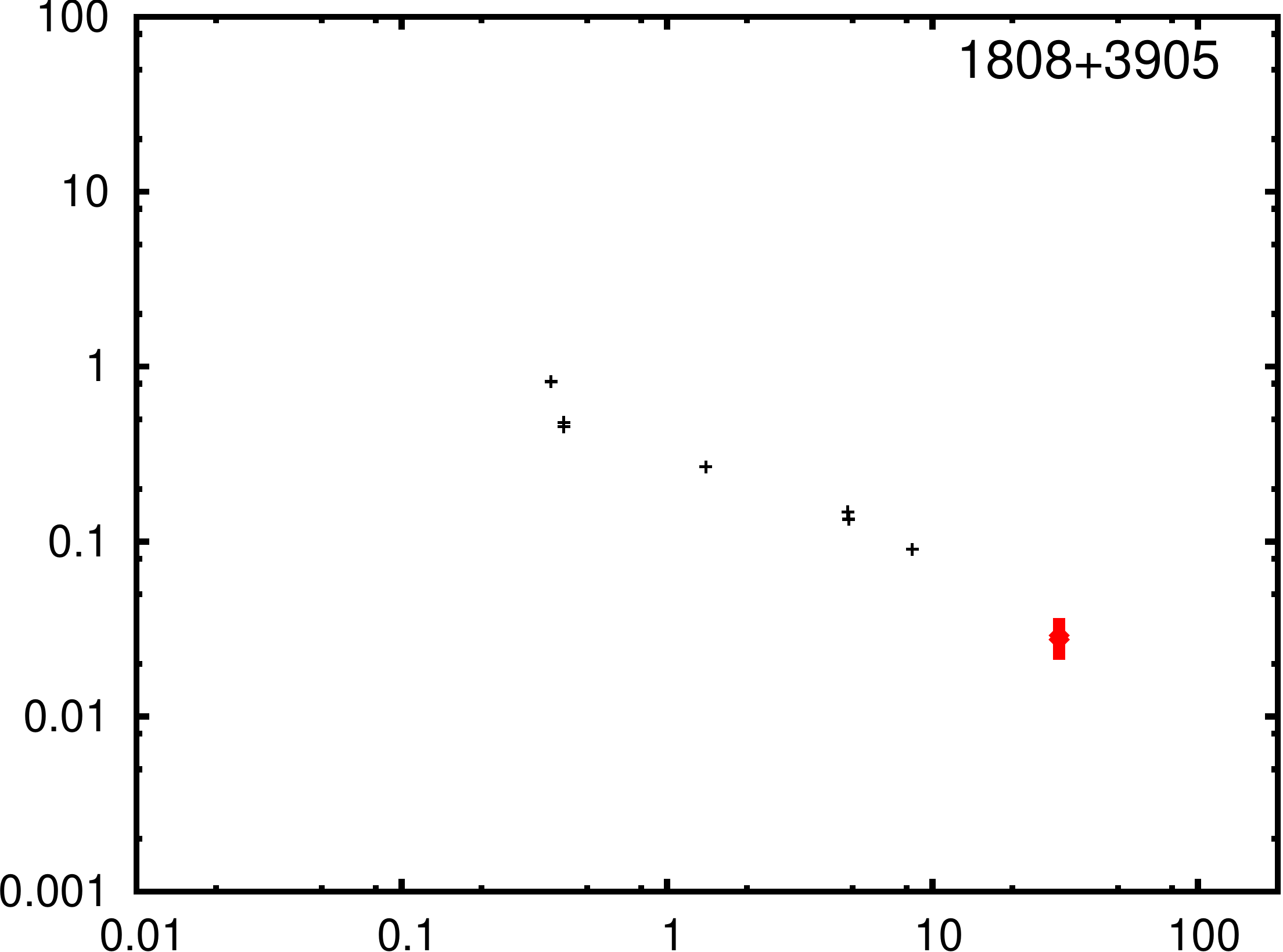}
\includegraphics[scale=0.2]{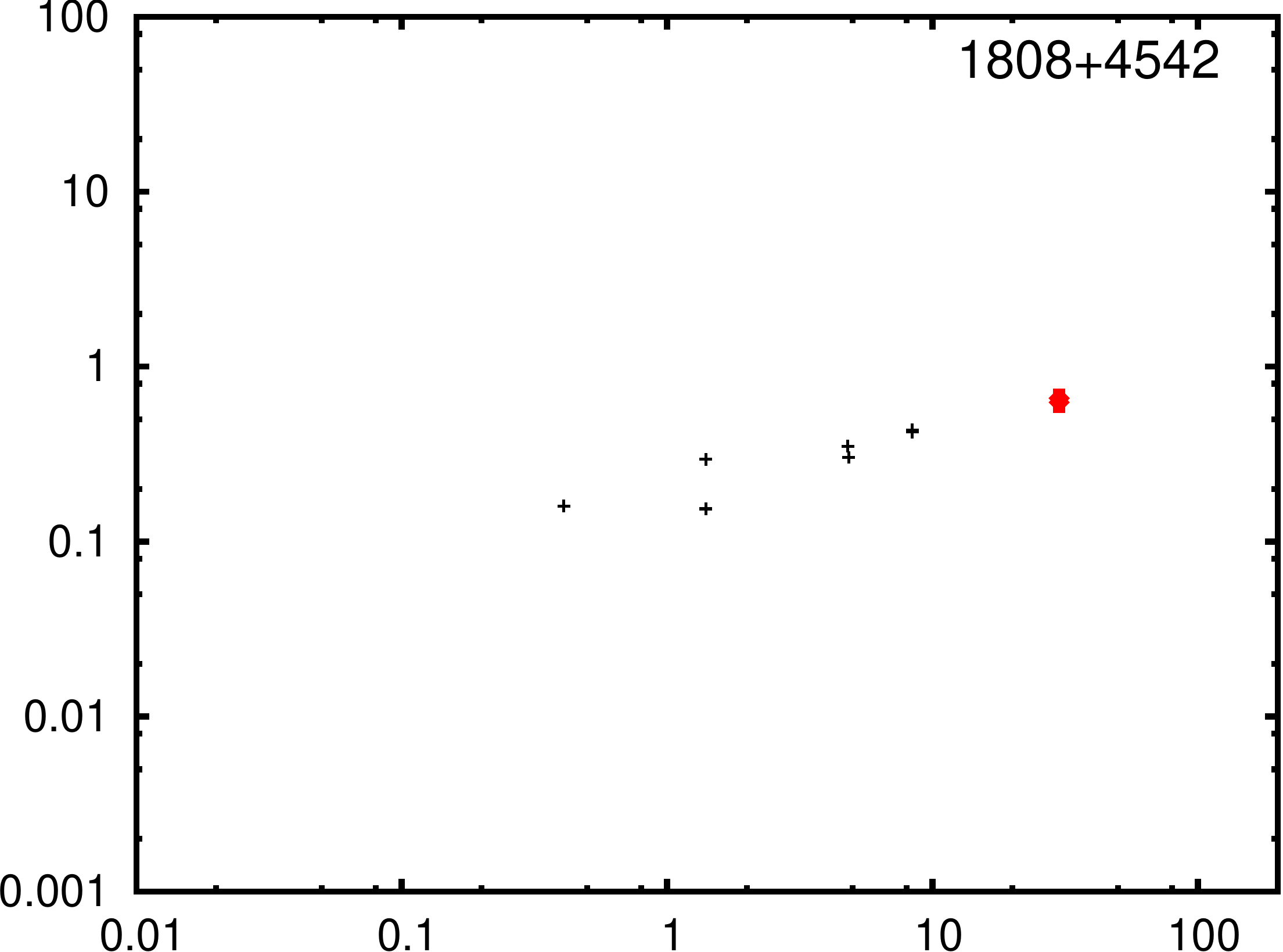}
\includegraphics[scale=0.2]{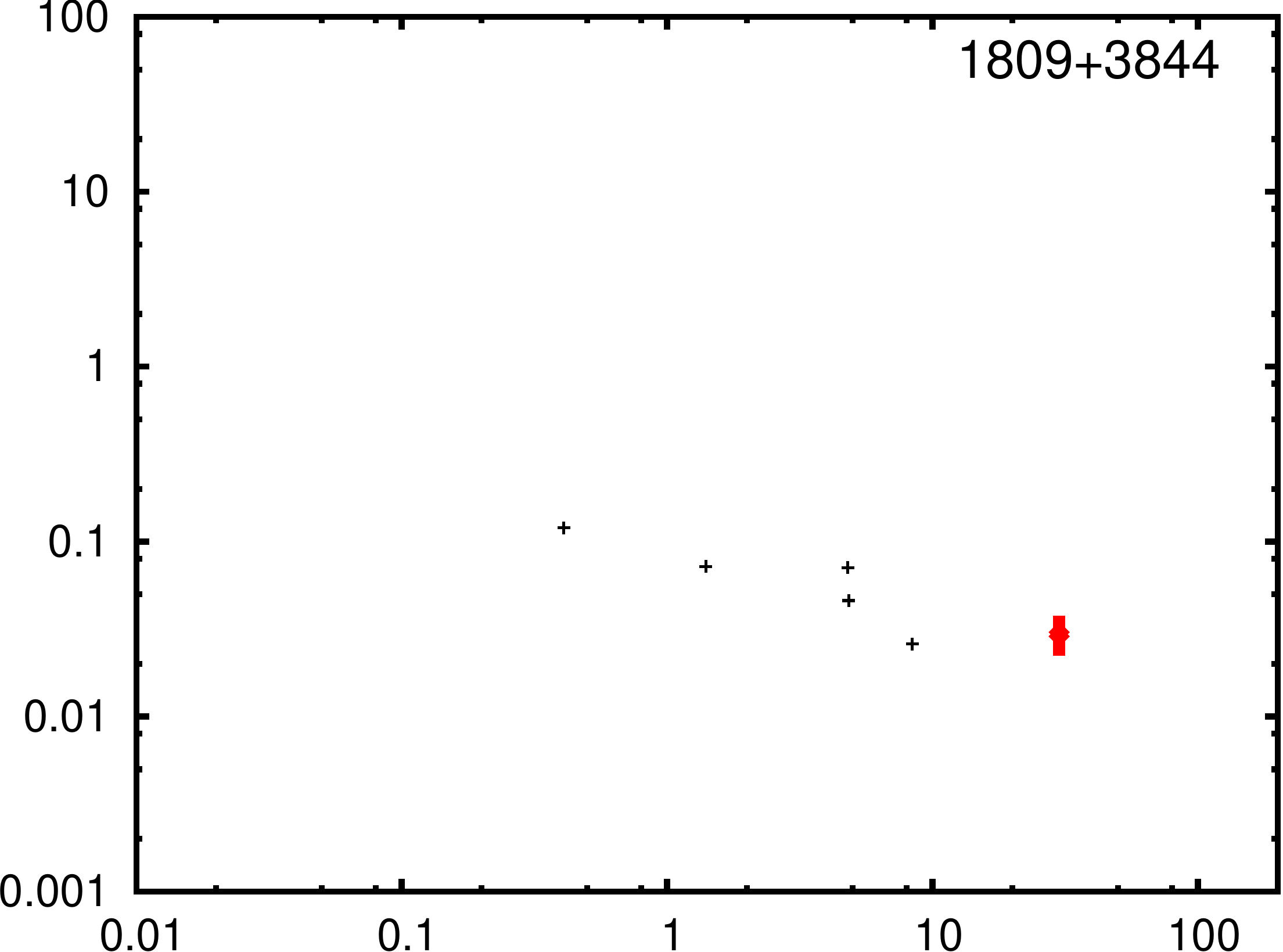}
\includegraphics[scale=0.2]{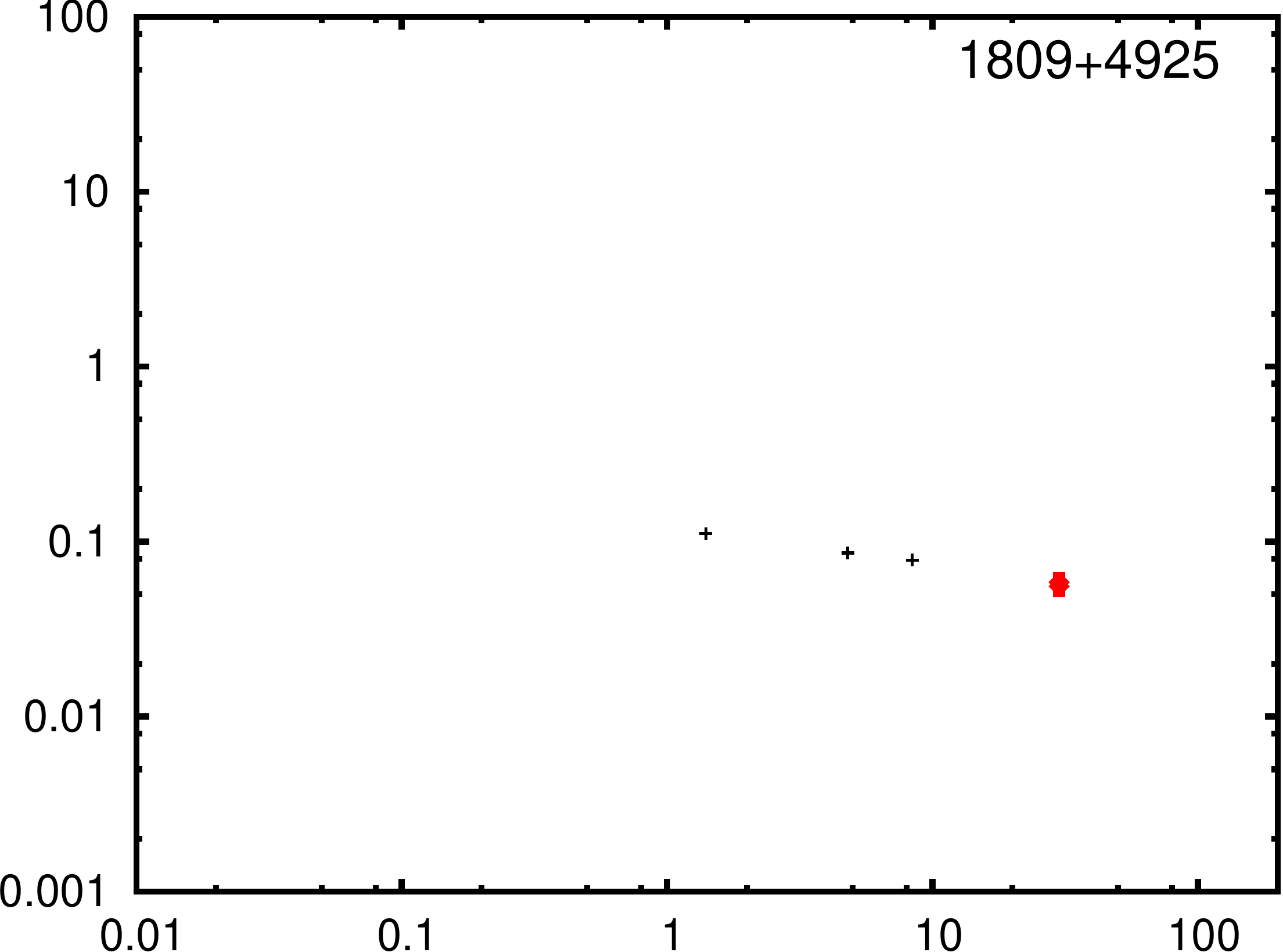}
\includegraphics[scale=0.2]{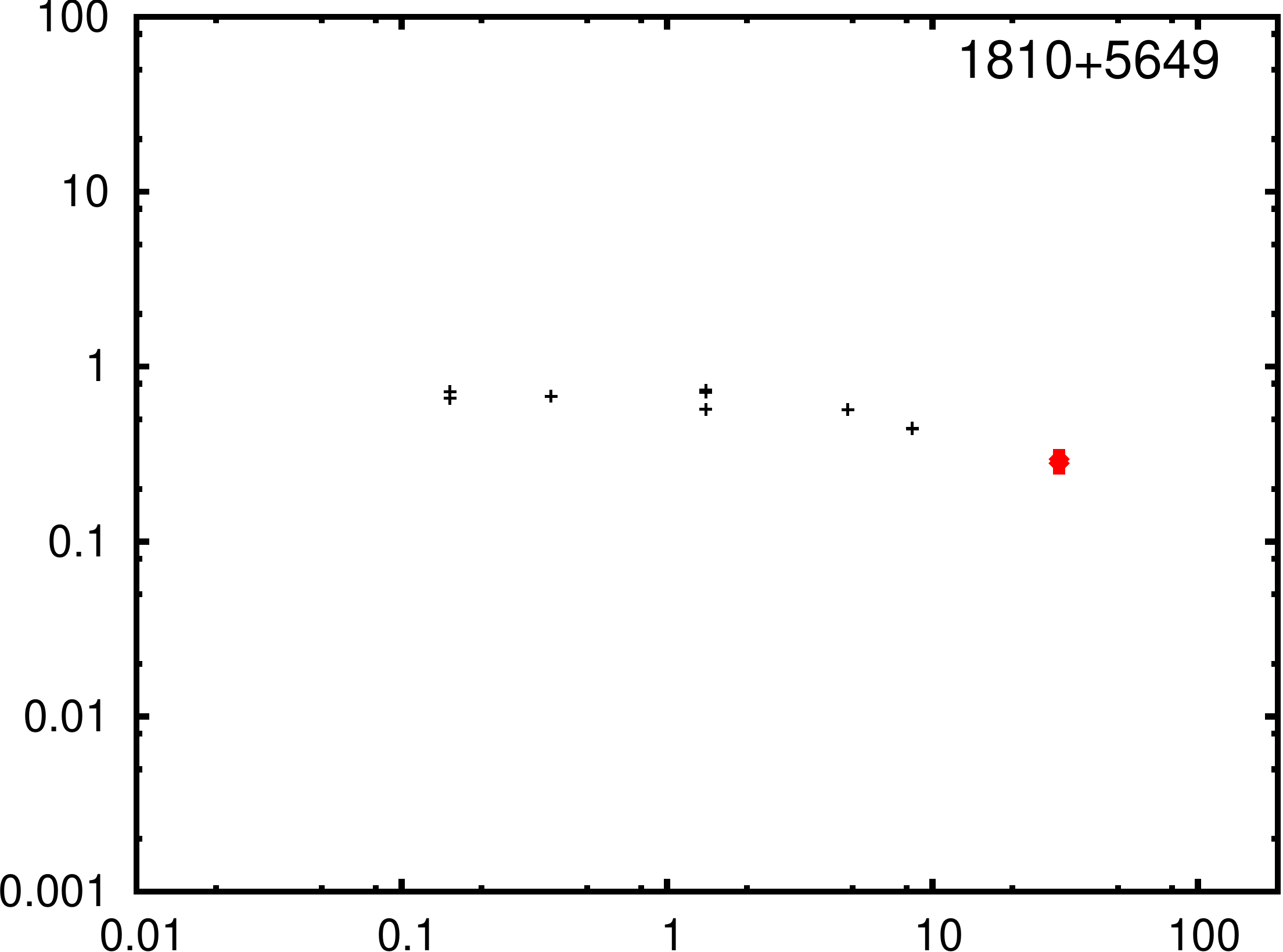}
\includegraphics[scale=0.2]{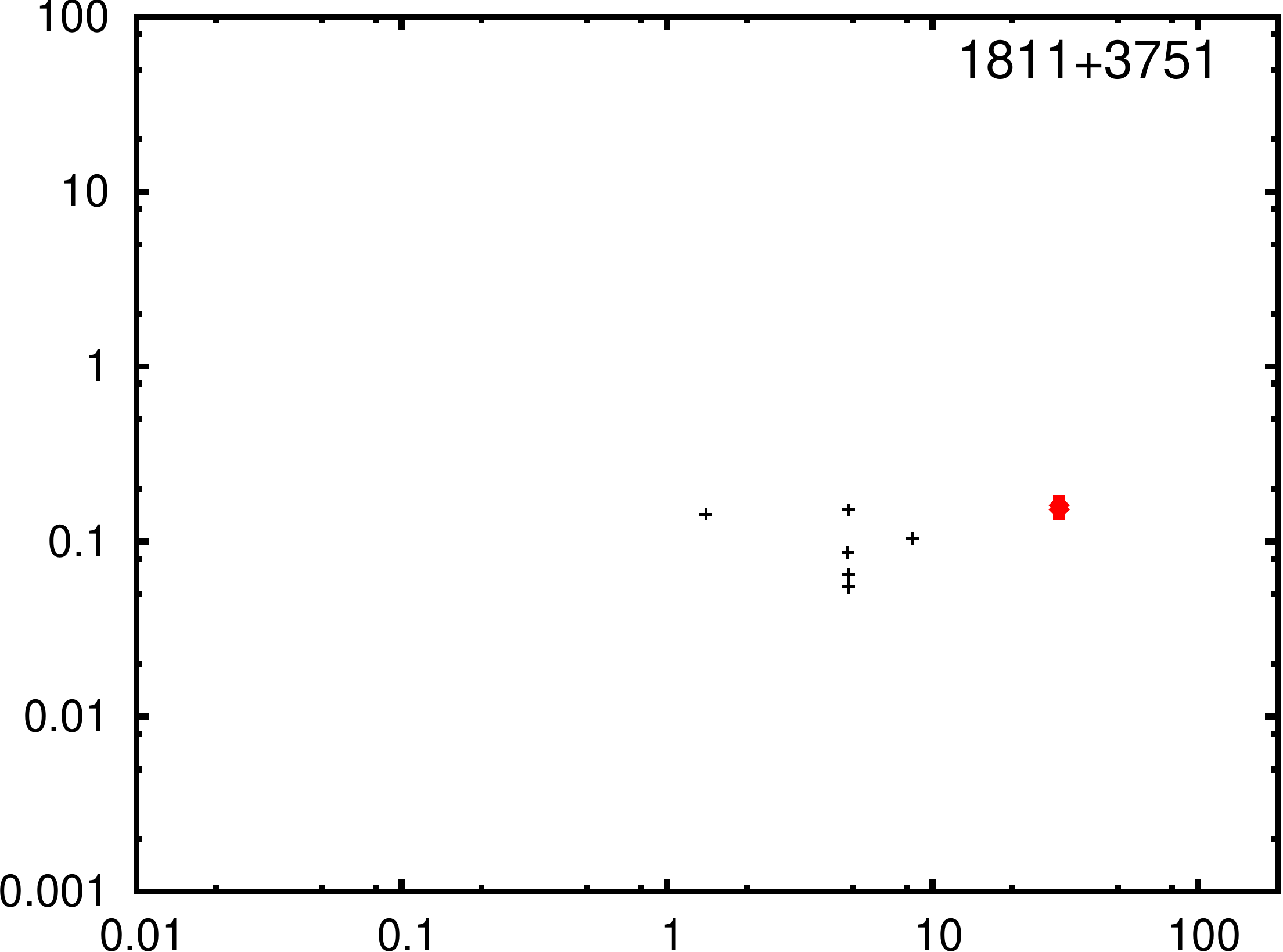}
\includegraphics[scale=0.2]{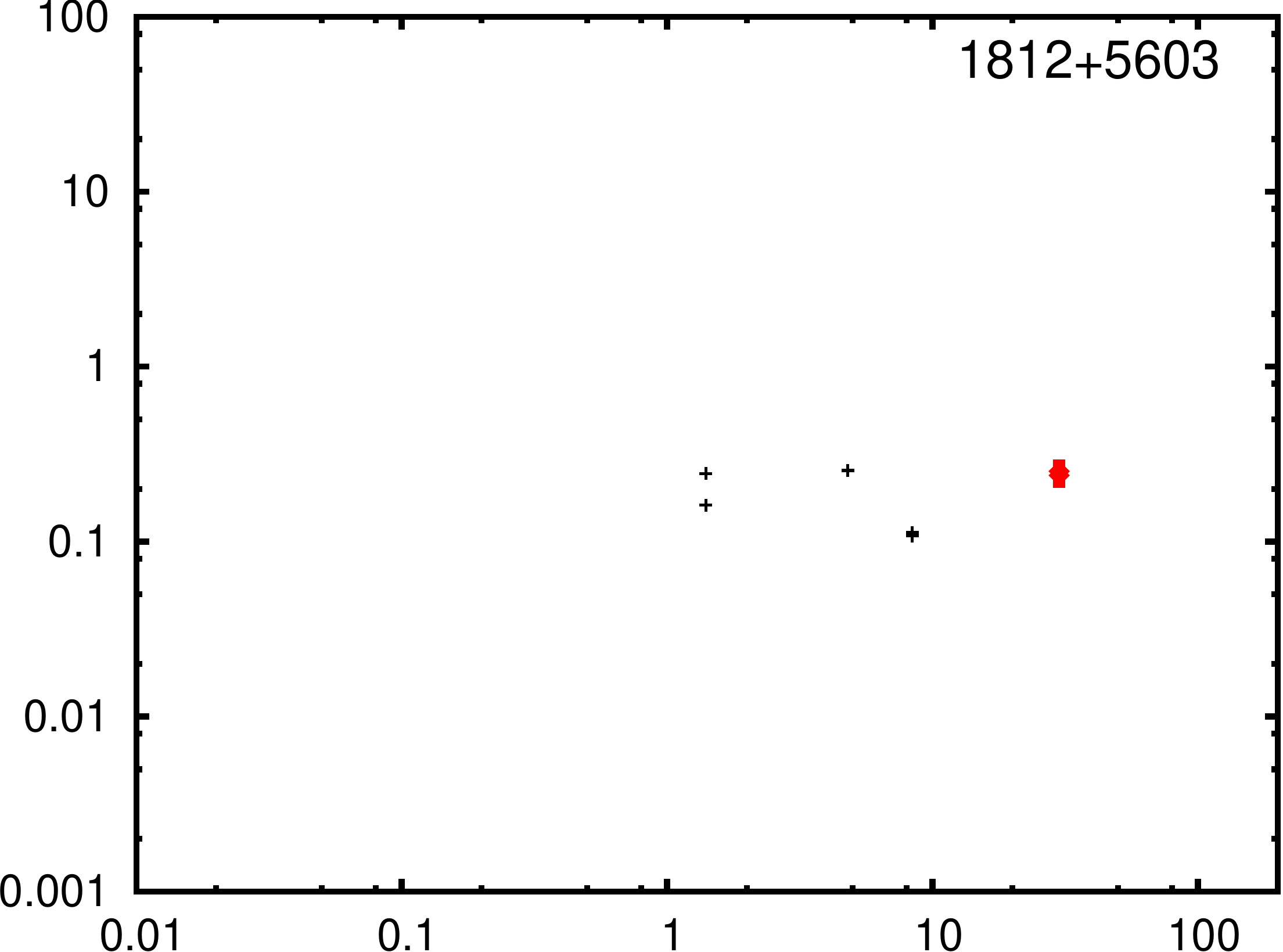}
\end{figure}
\clearpage\begin{figure}
\centering
\includegraphics[scale=0.2]{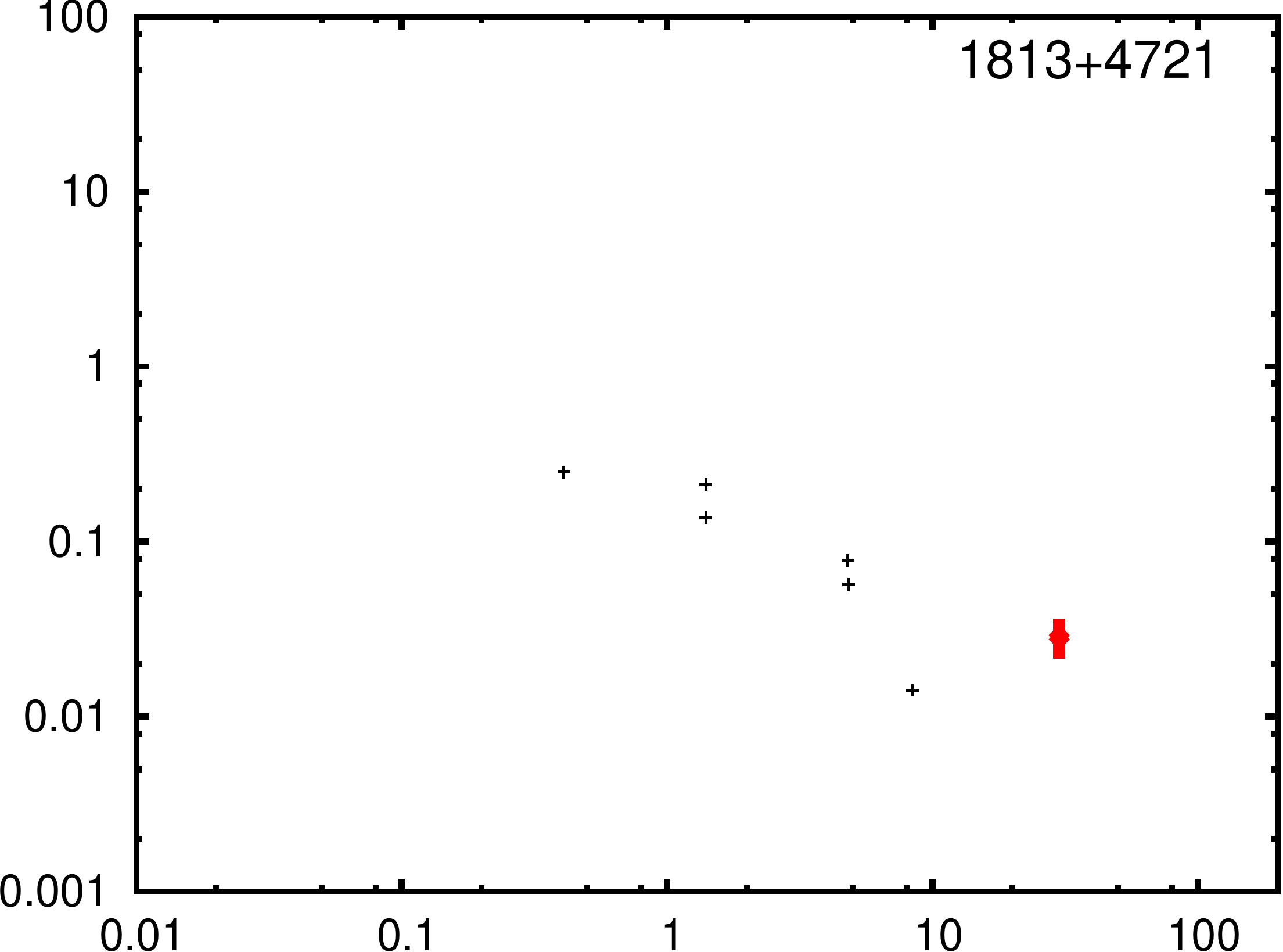}
\includegraphics[scale=0.2]{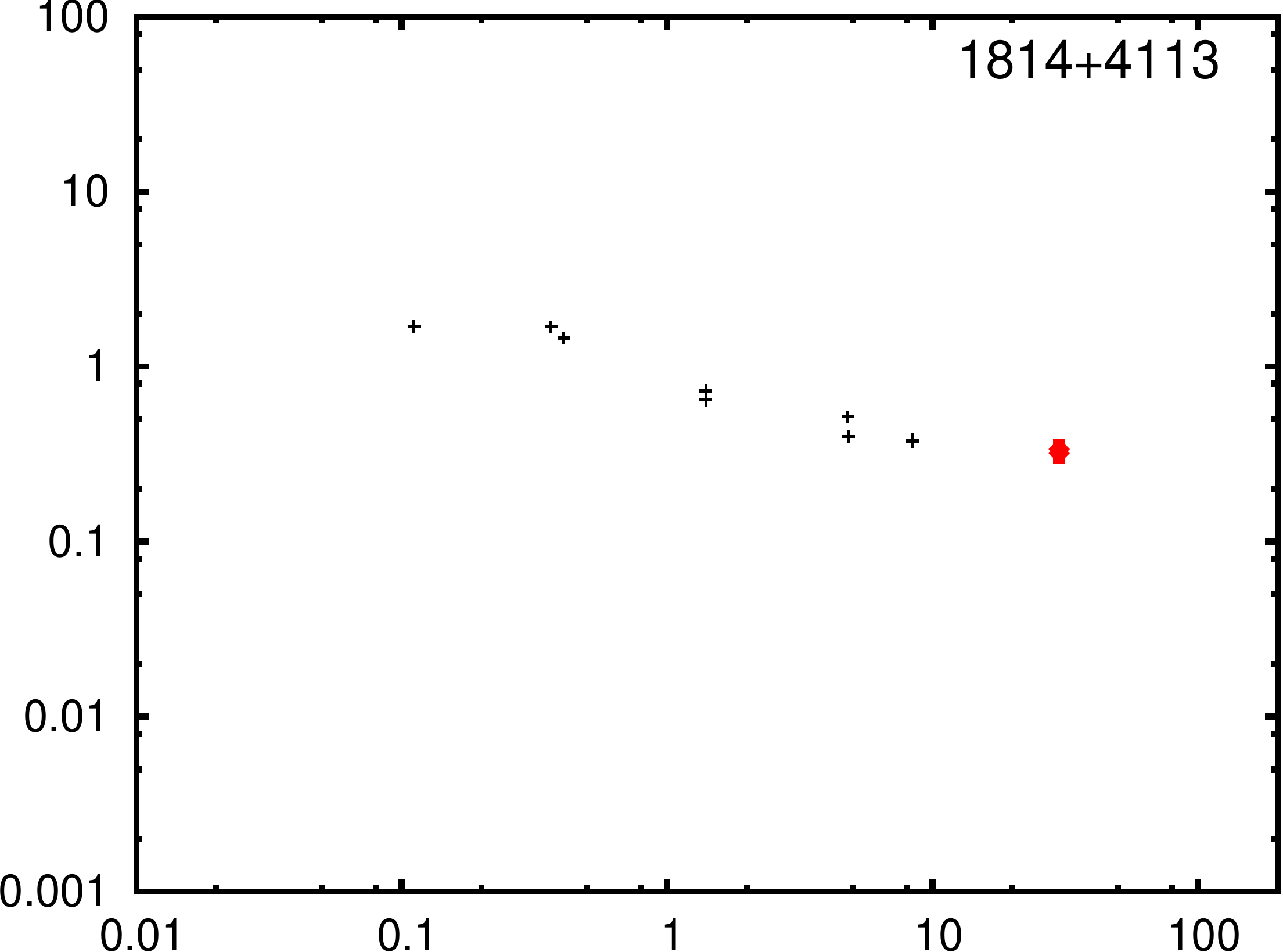}
\includegraphics[scale=0.2]{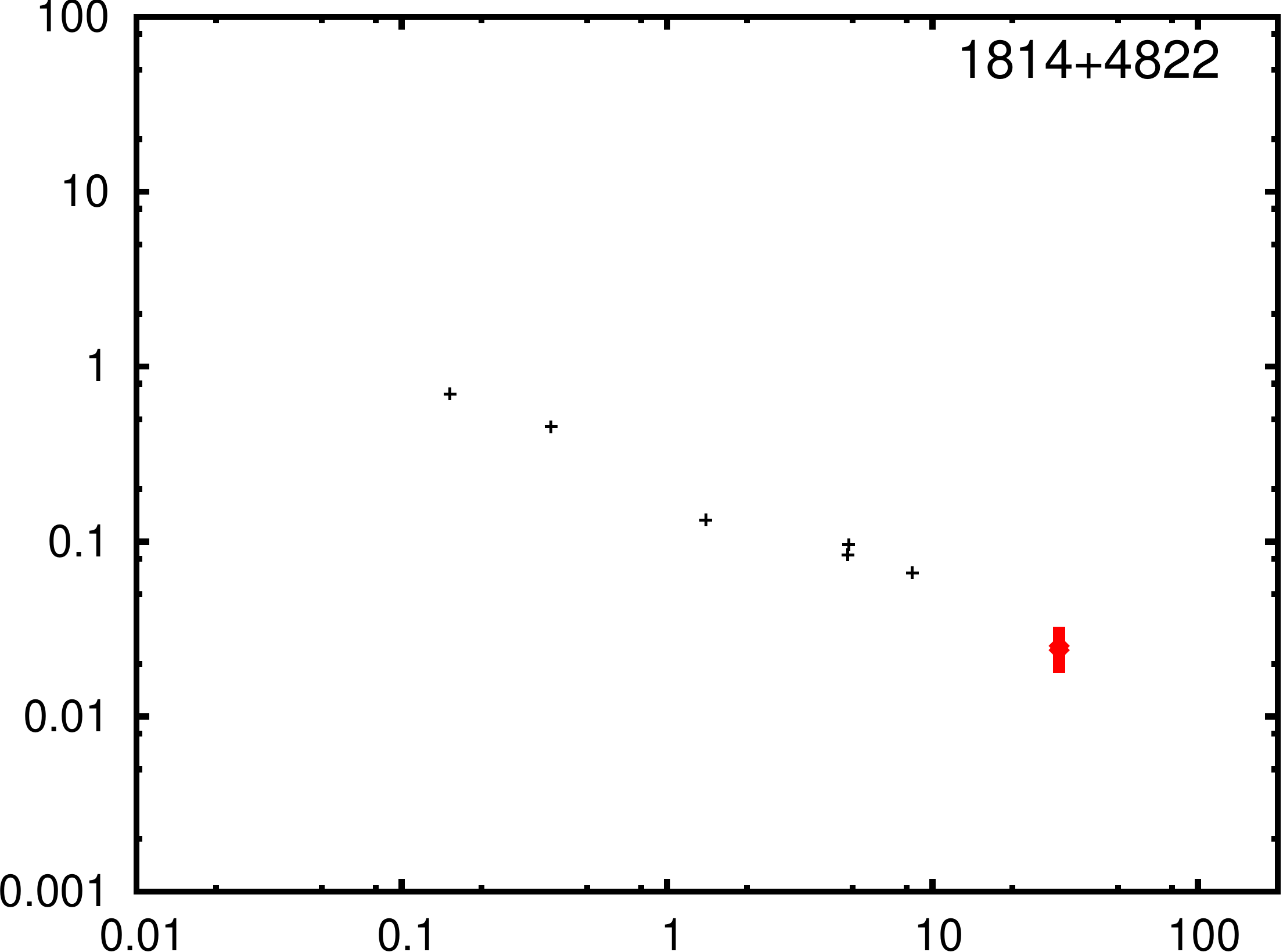}
\includegraphics[scale=0.2]{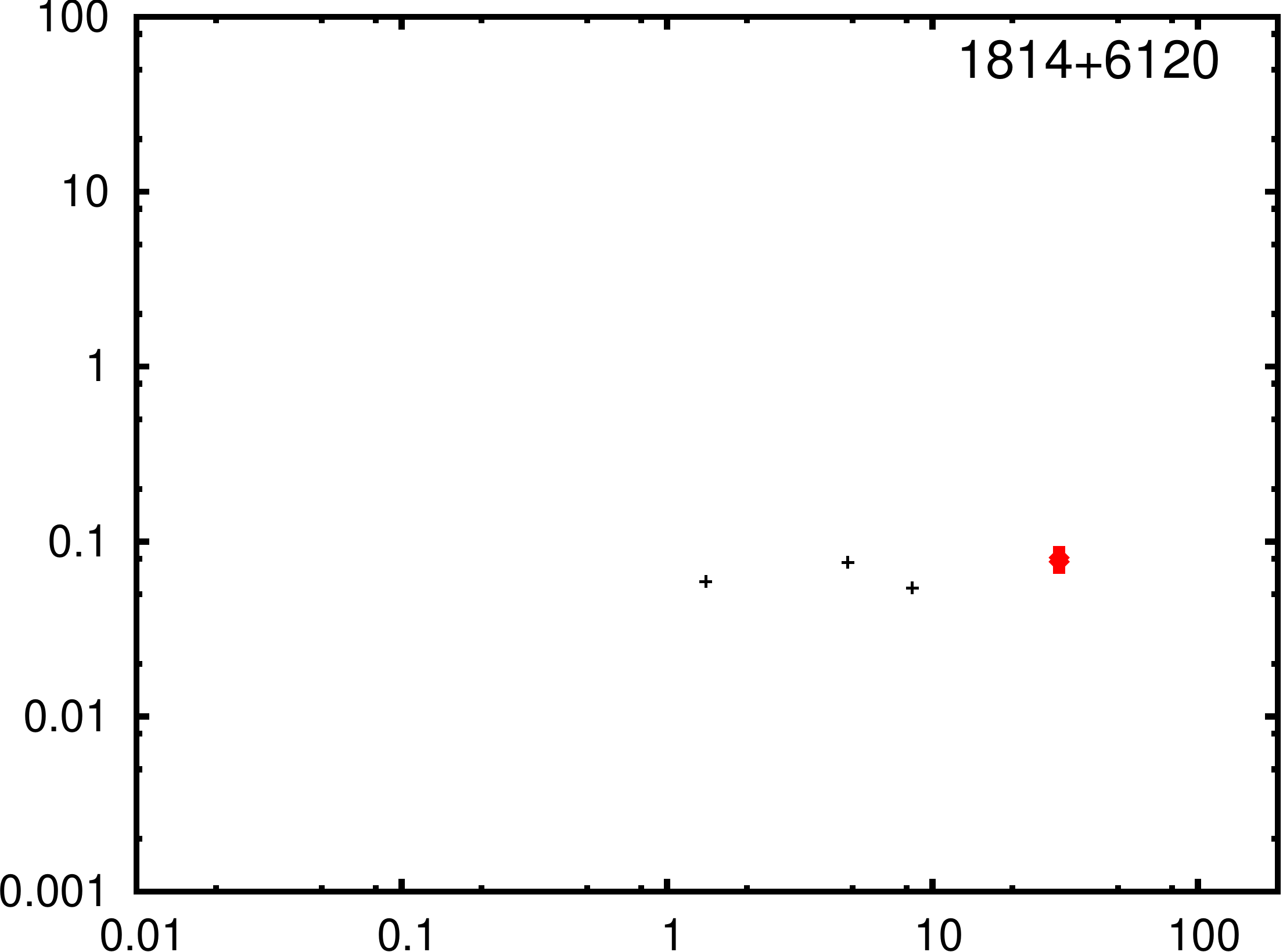}
\includegraphics[scale=0.2]{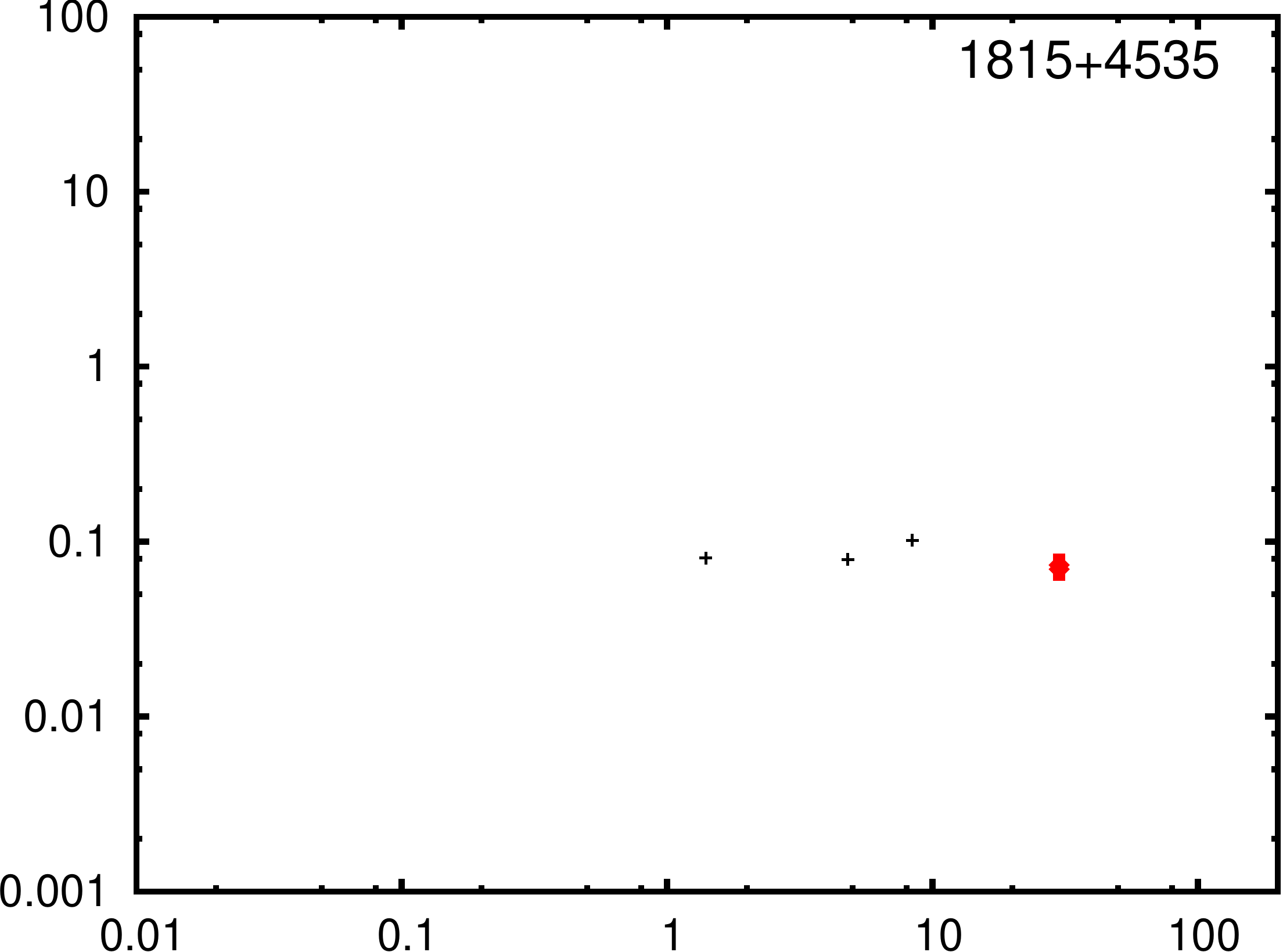}
\includegraphics[scale=0.2]{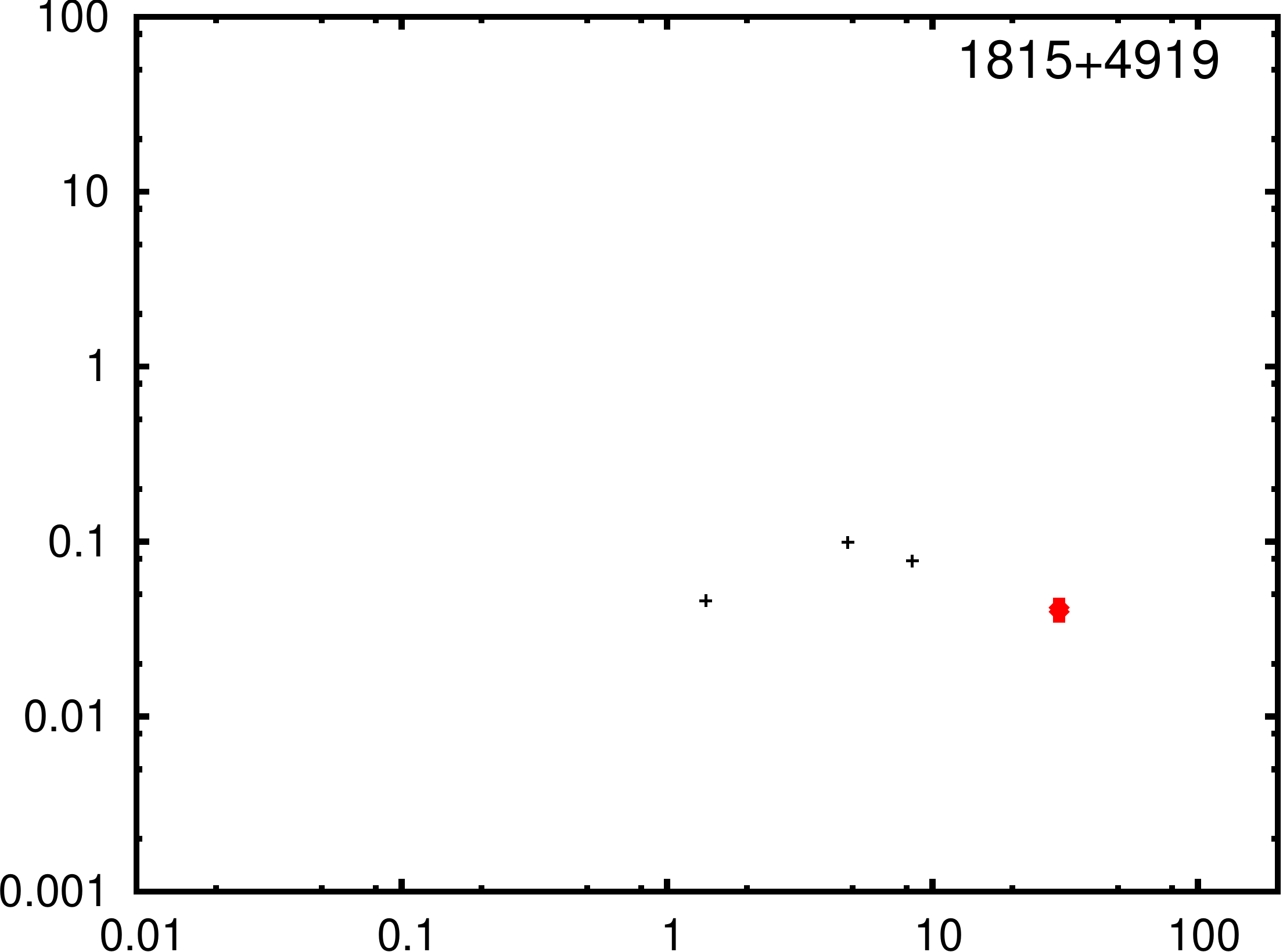}
\includegraphics[scale=0.2]{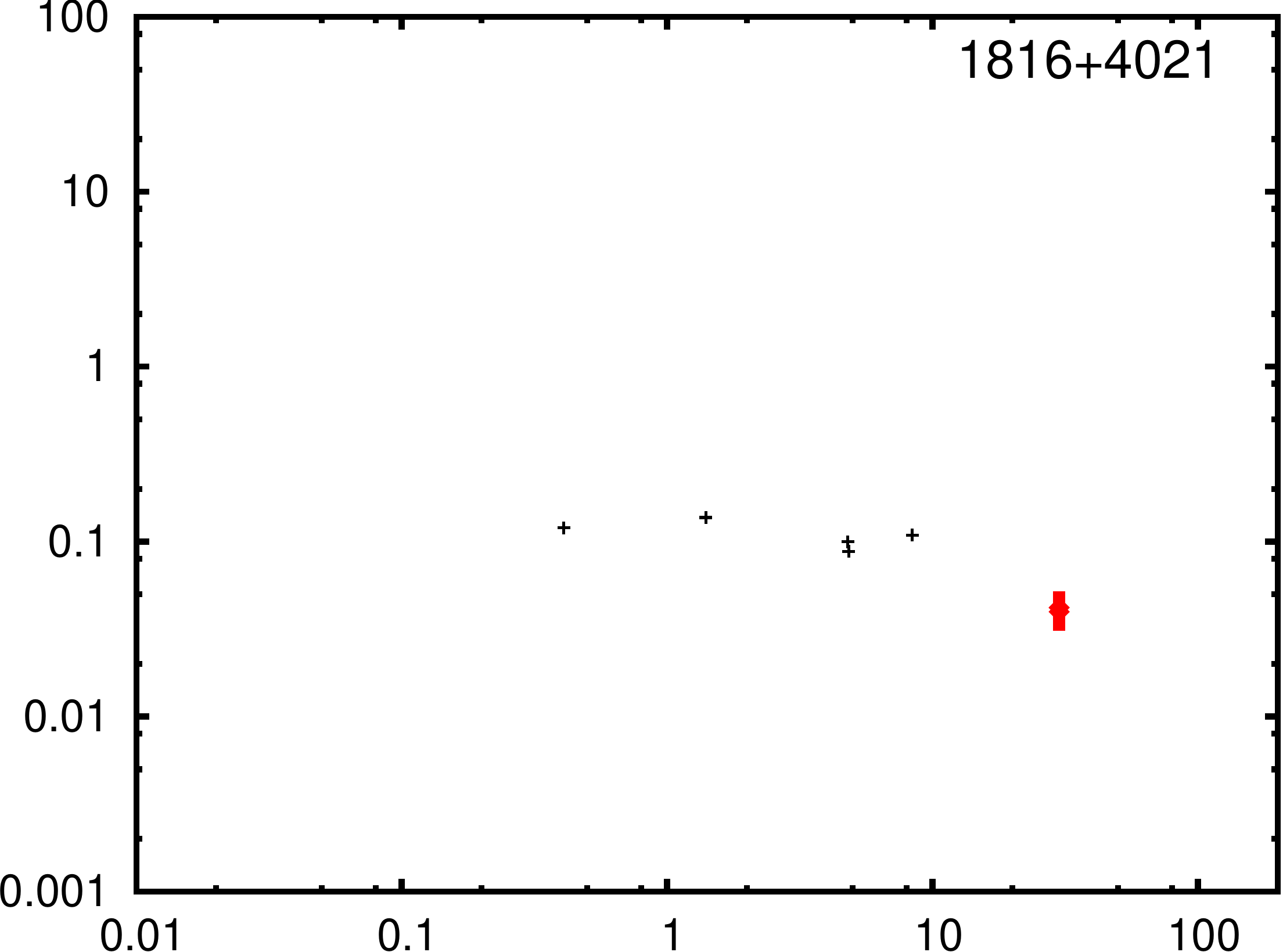}
\includegraphics[scale=0.2]{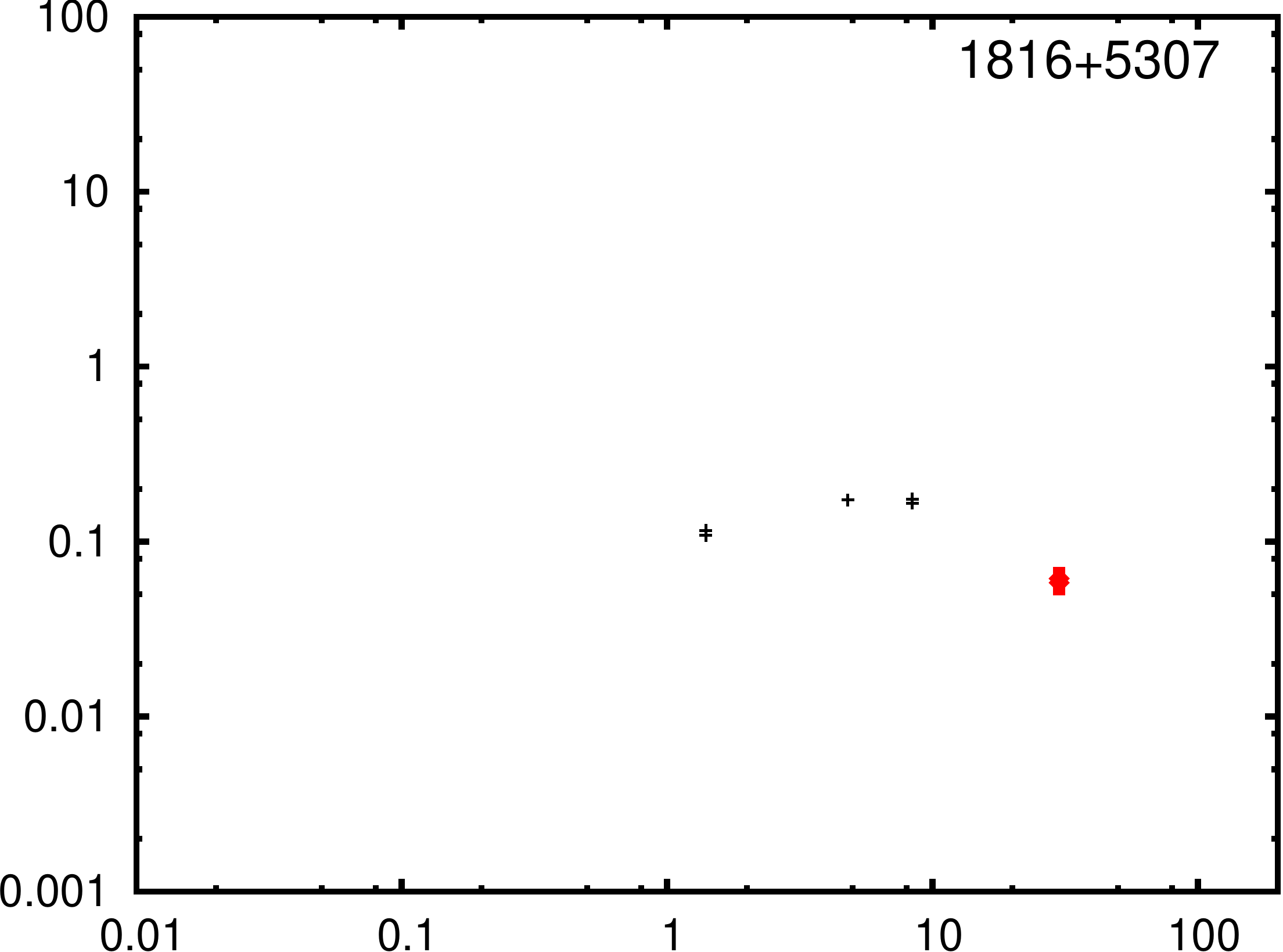}
\includegraphics[scale=0.2]{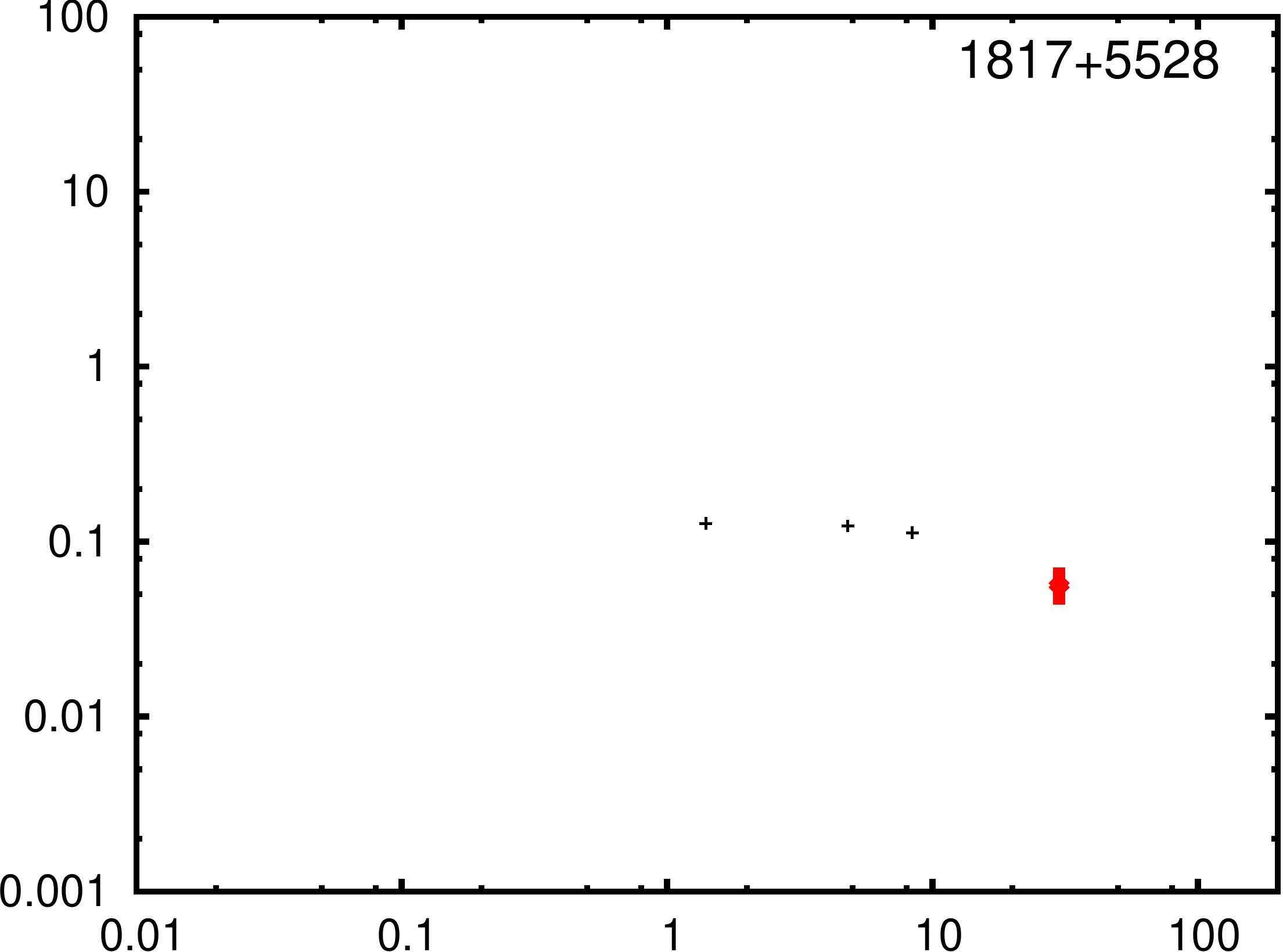}
\includegraphics[scale=0.2]{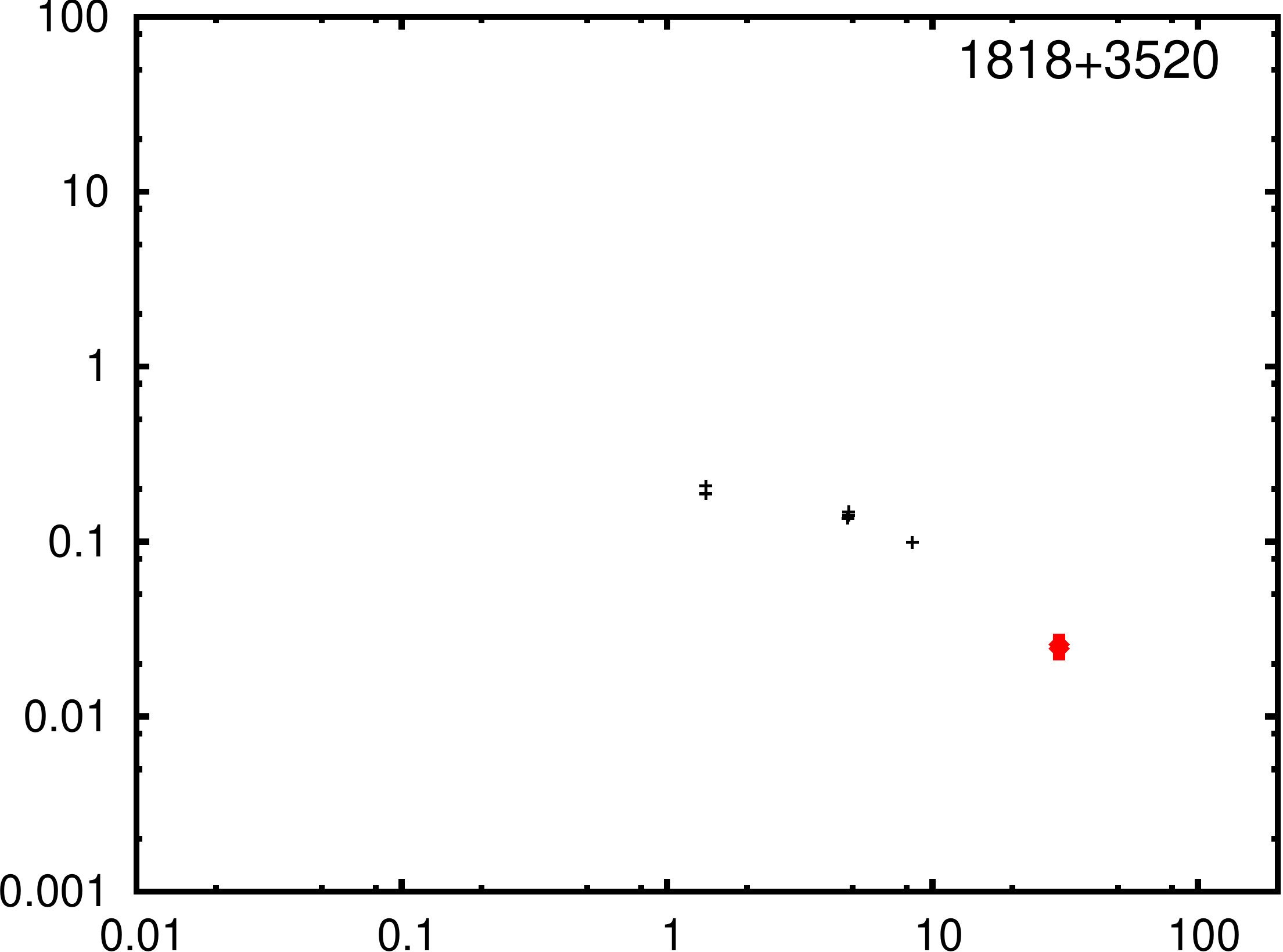}
\includegraphics[scale=0.2]{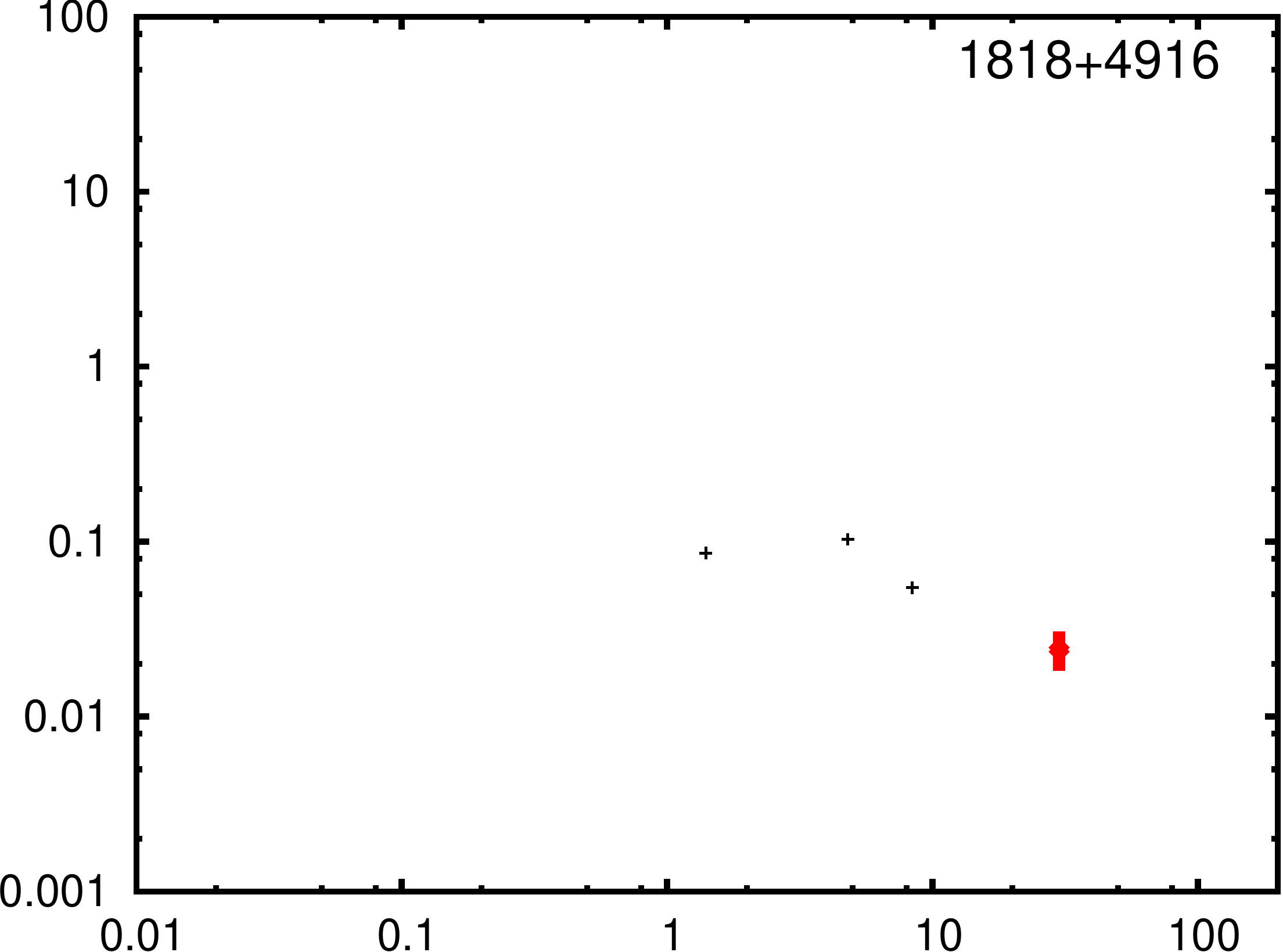}
\includegraphics[scale=0.2]{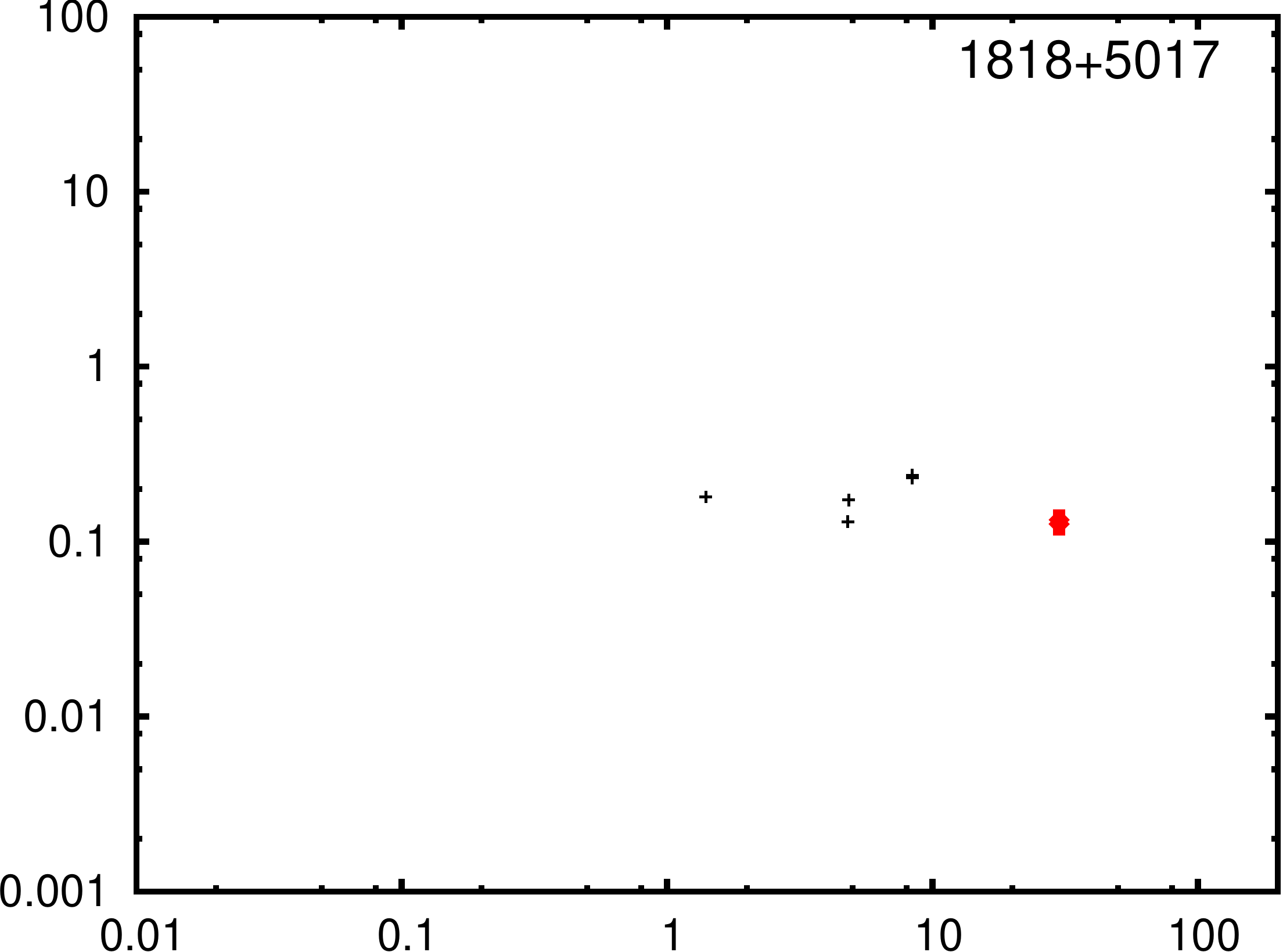}
\includegraphics[scale=0.2]{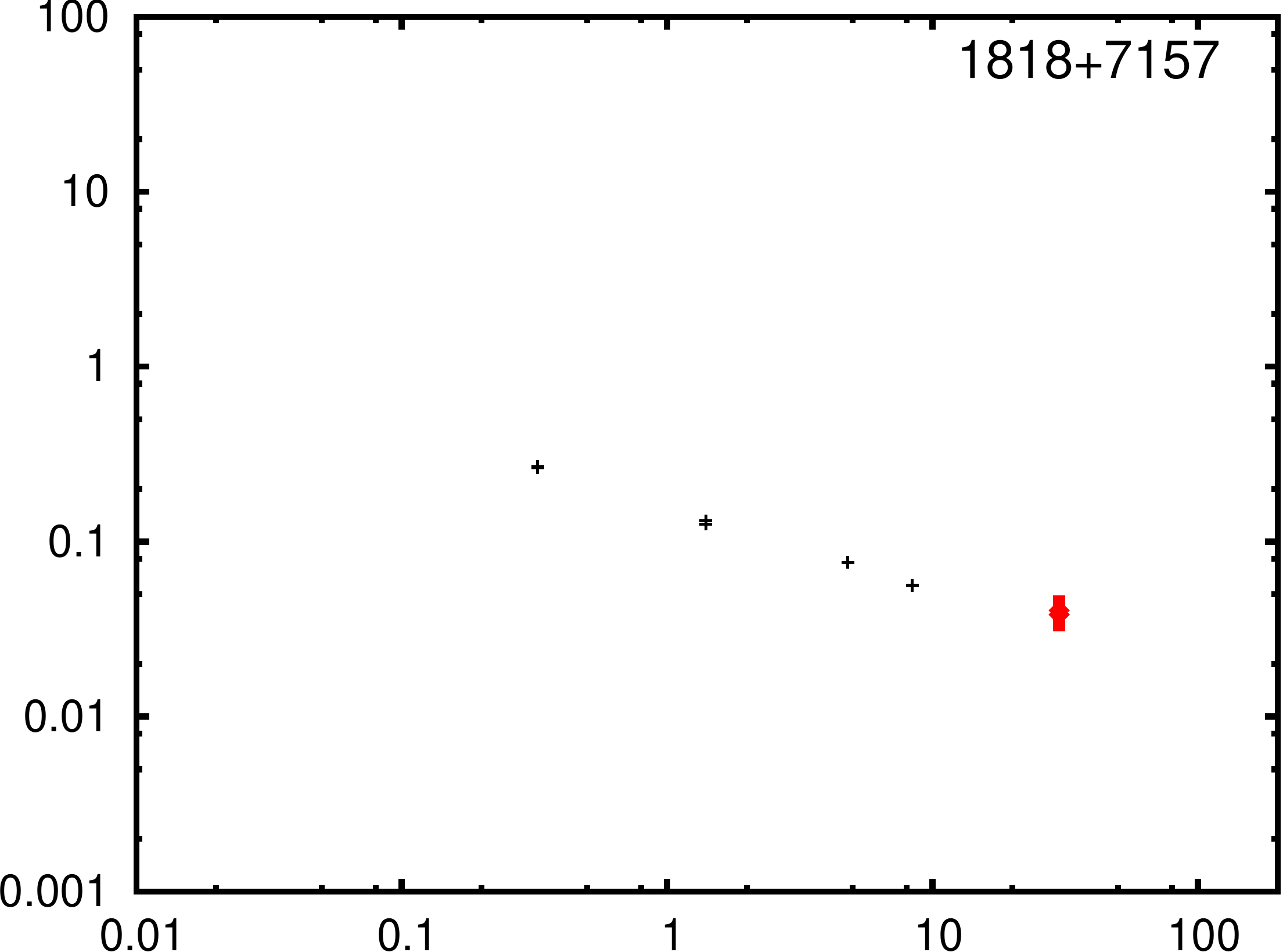}
\includegraphics[scale=0.2]{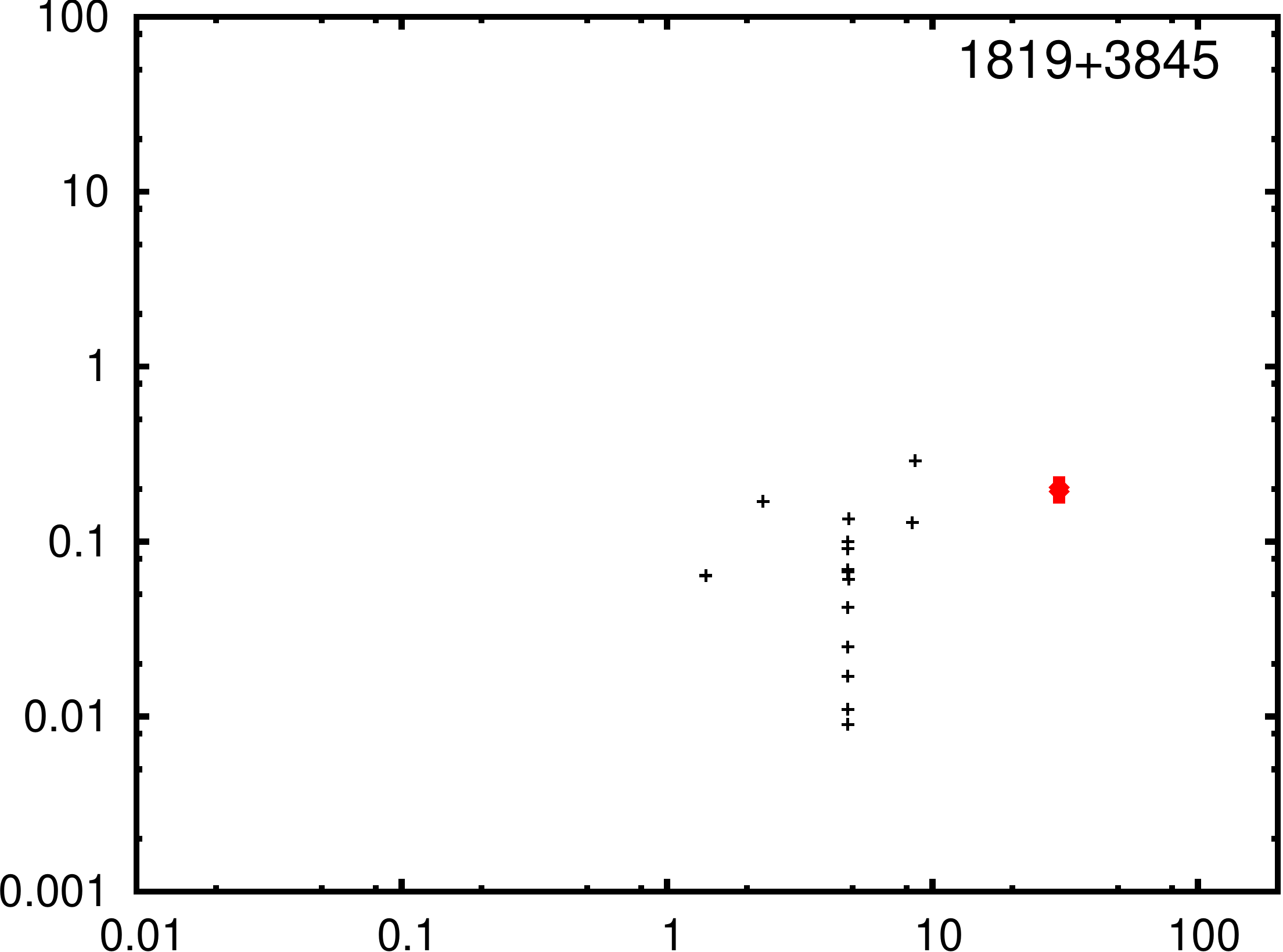}
\includegraphics[scale=0.2]{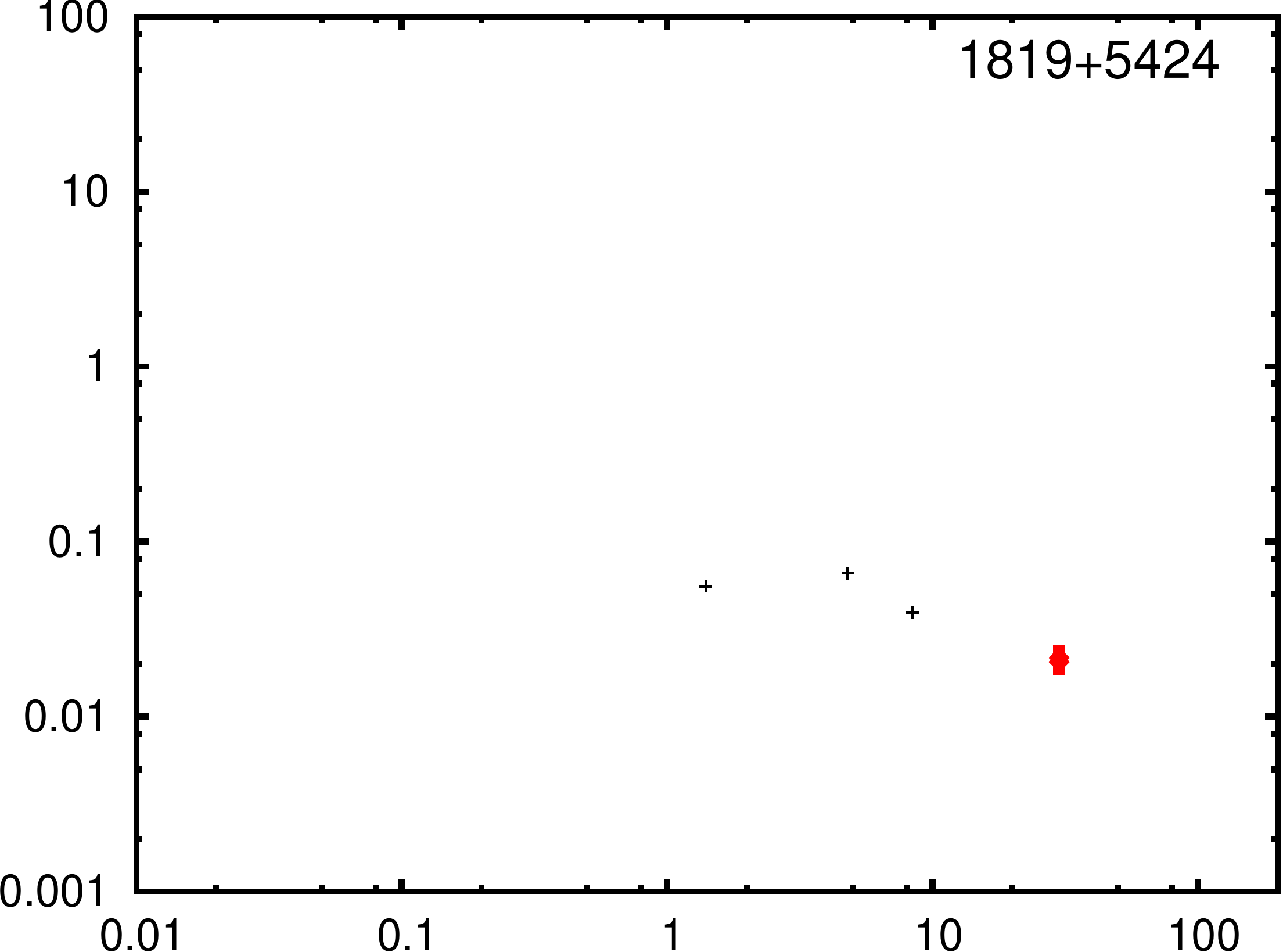}
\includegraphics[scale=0.2]{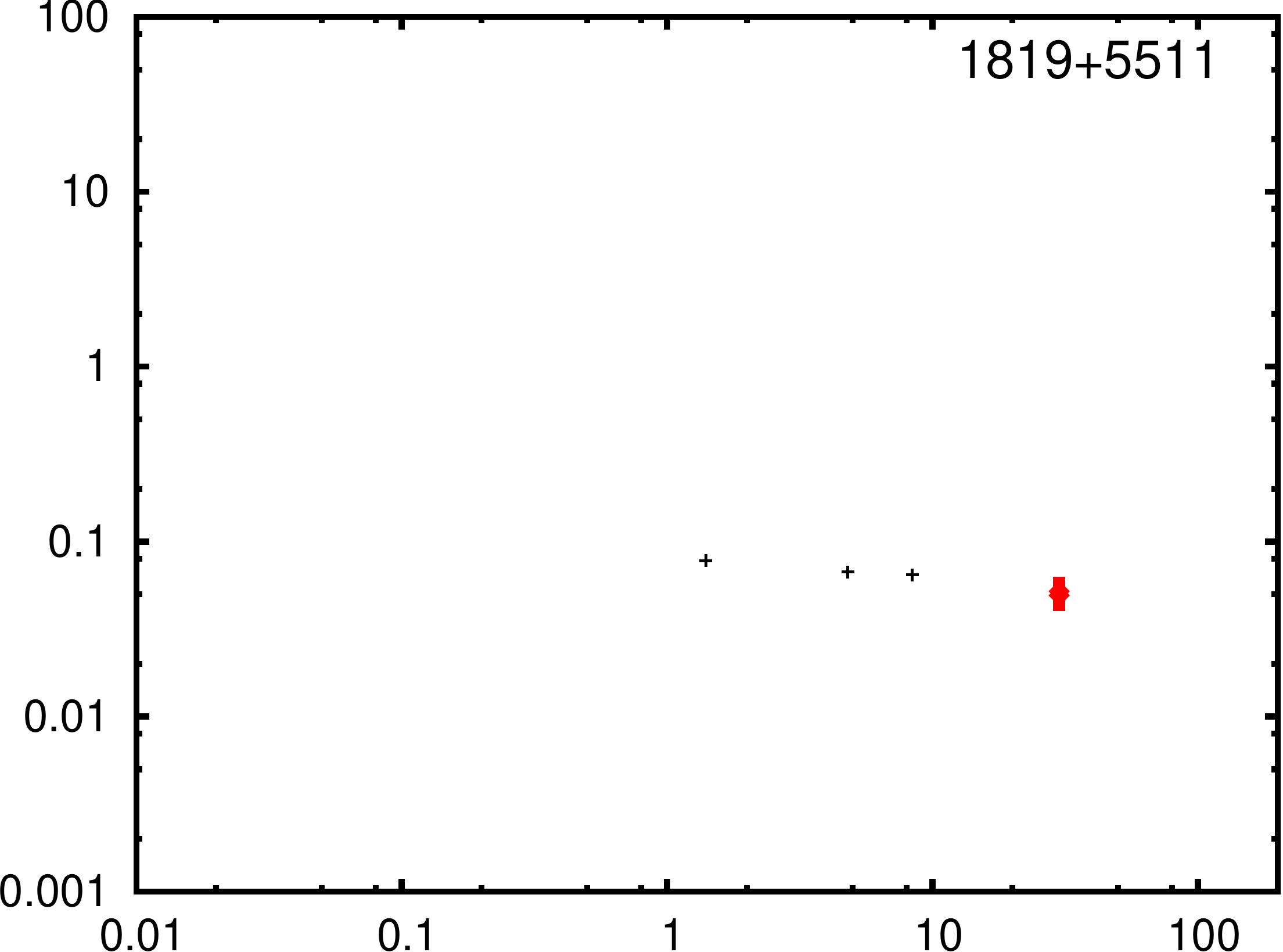}
\includegraphics[scale=0.2]{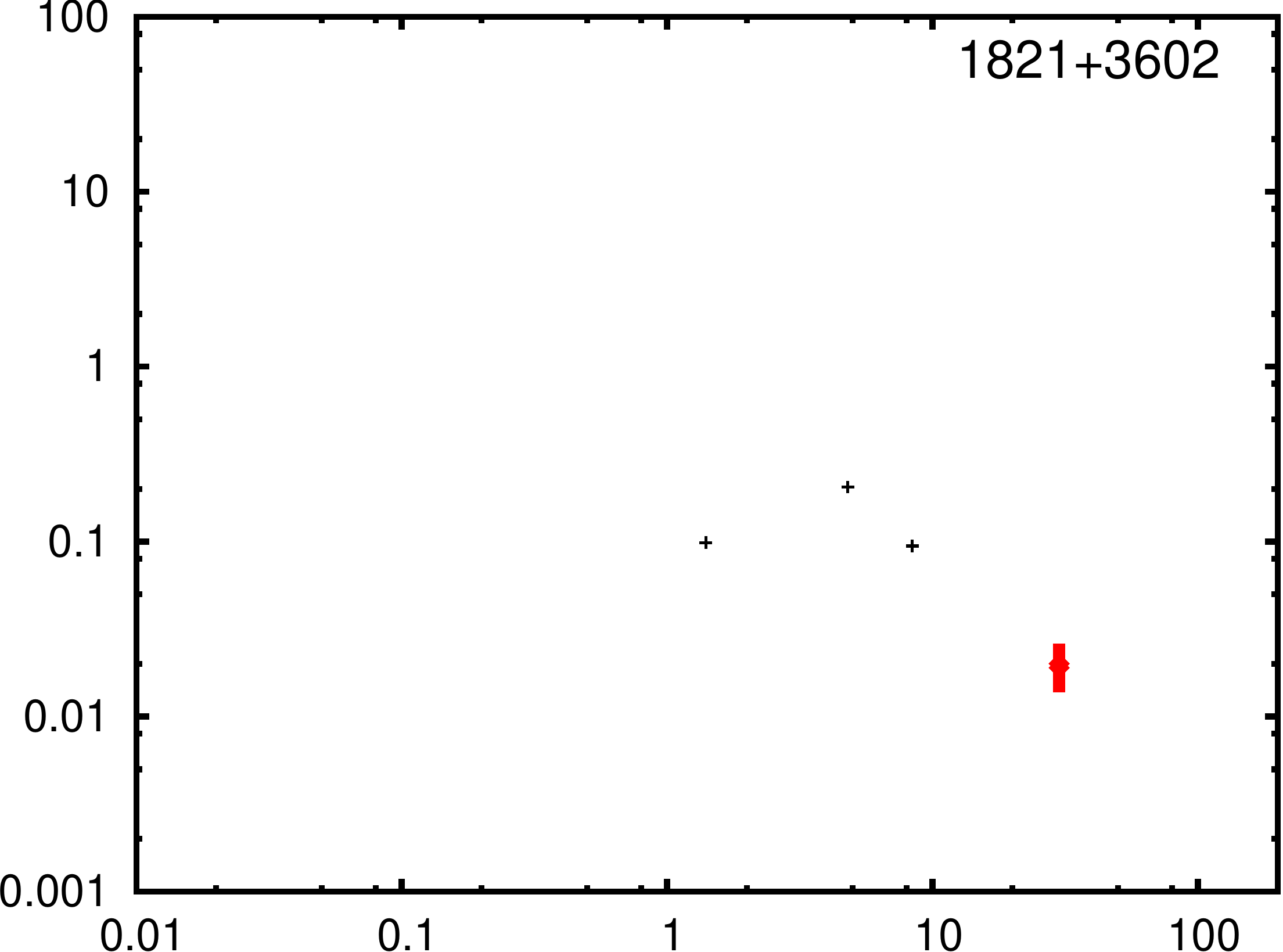}
\includegraphics[scale=0.2]{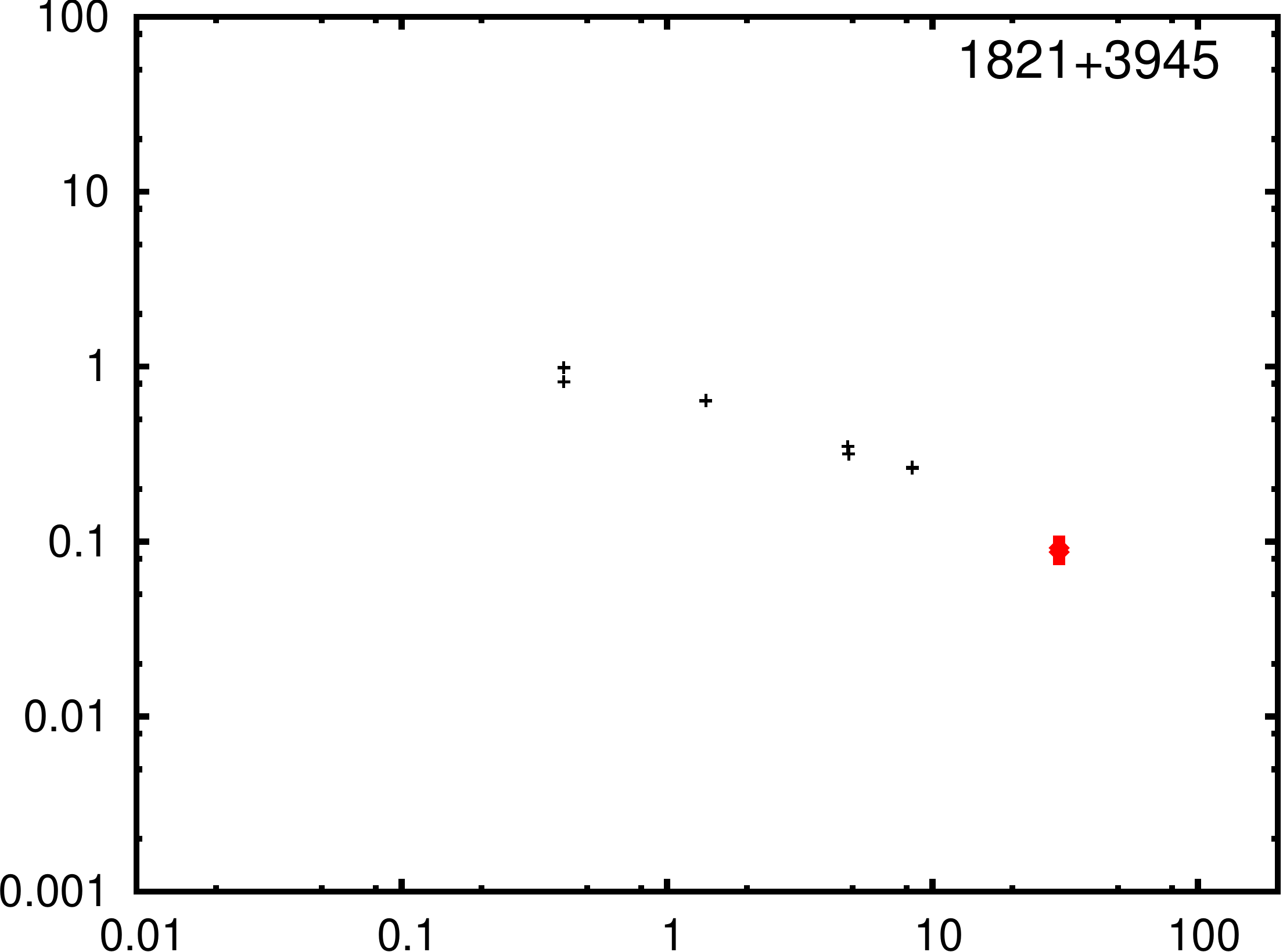}
\end{figure}
\clearpage\begin{figure}
\centering
\includegraphics[scale=0.2]{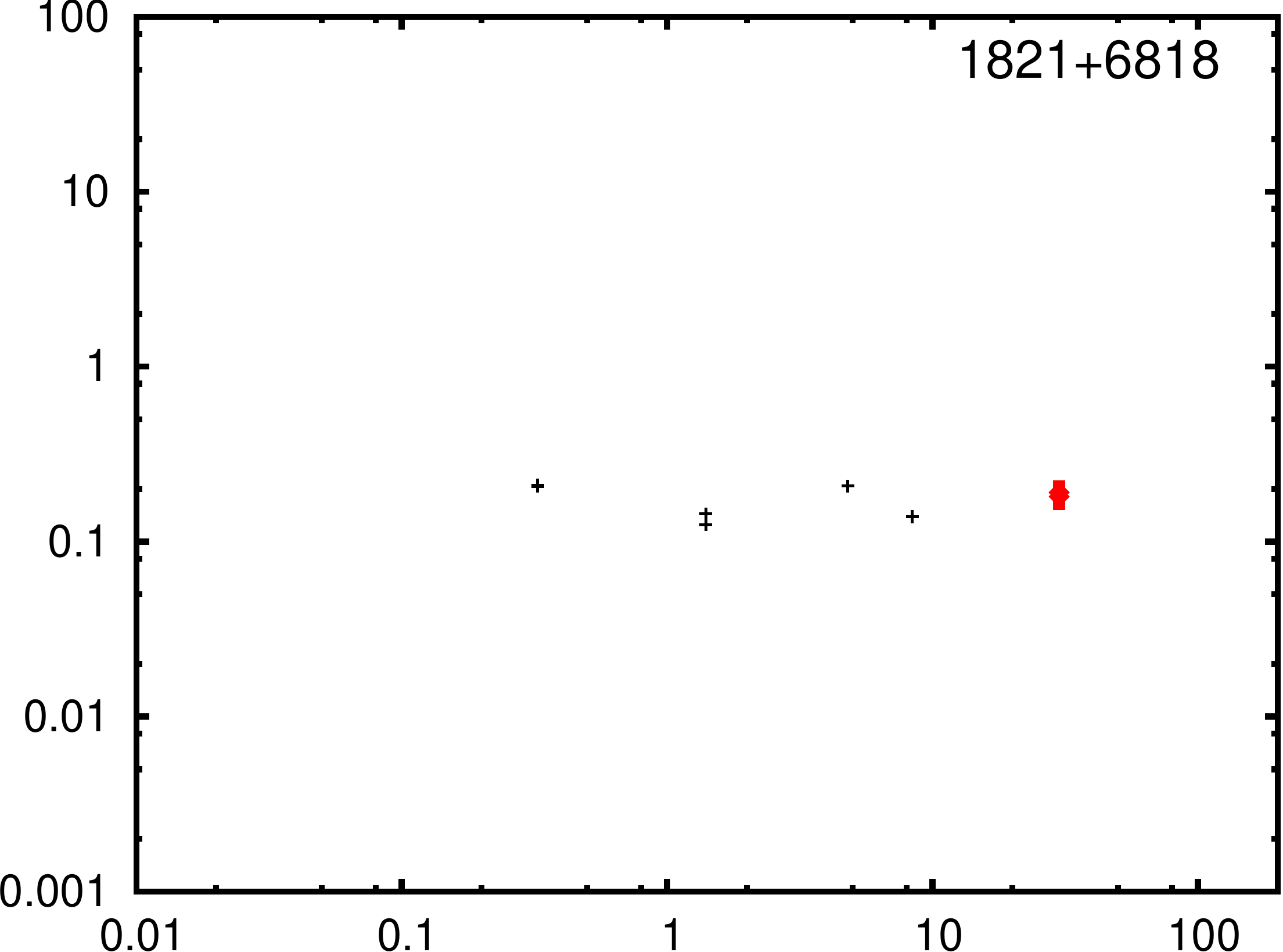}
\includegraphics[scale=0.2]{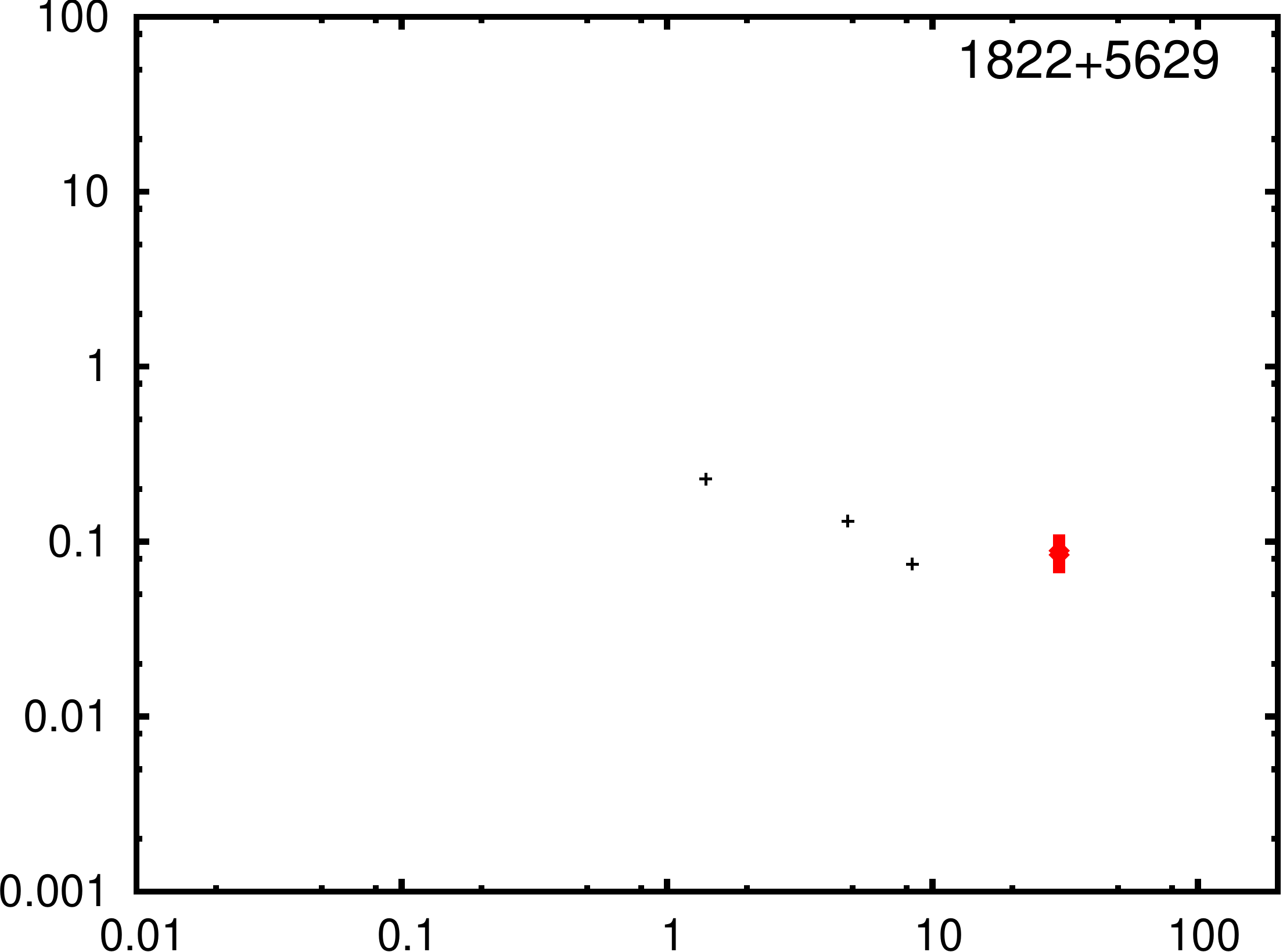}
\includegraphics[scale=0.2]{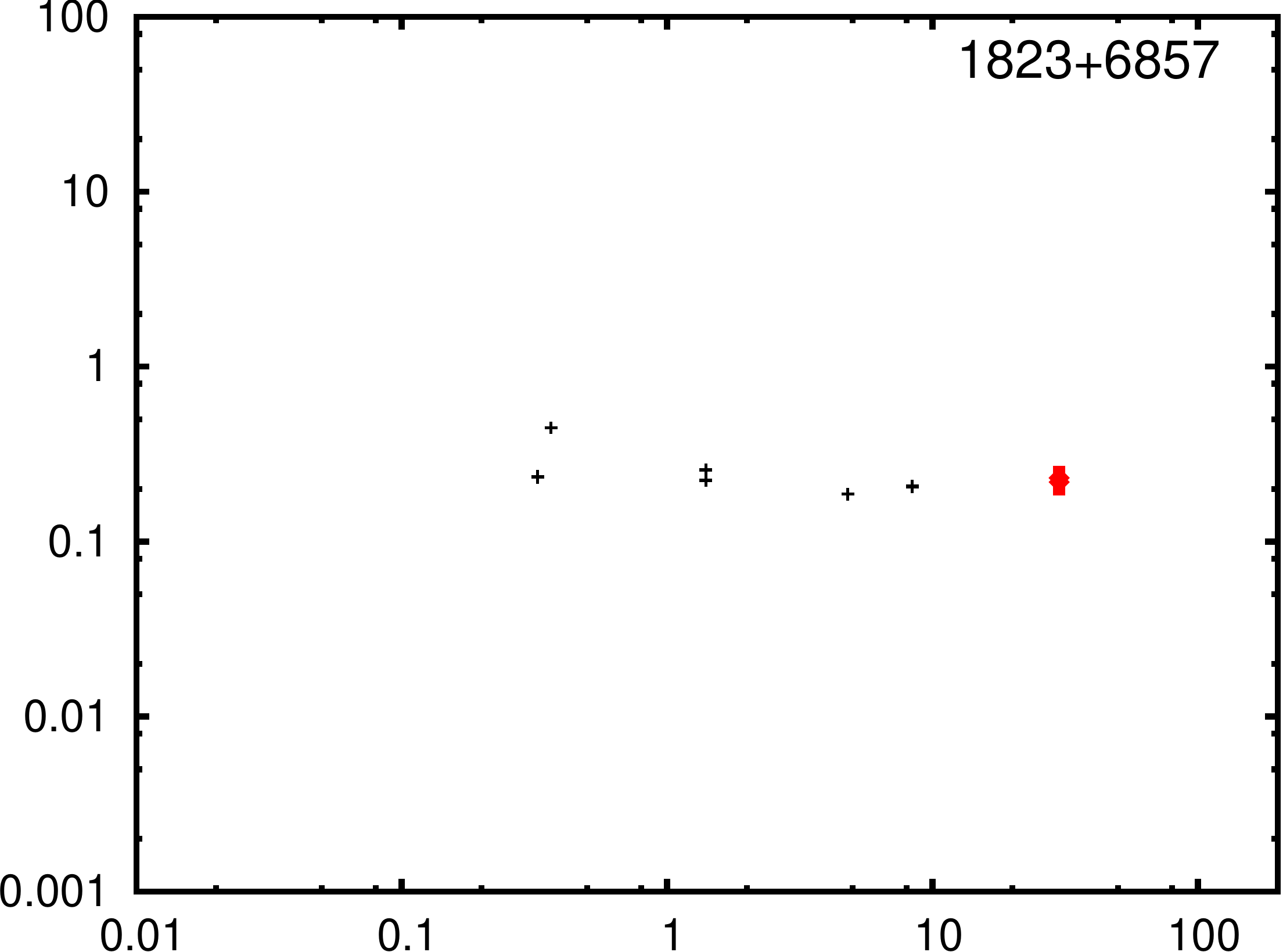}
\includegraphics[scale=0.2]{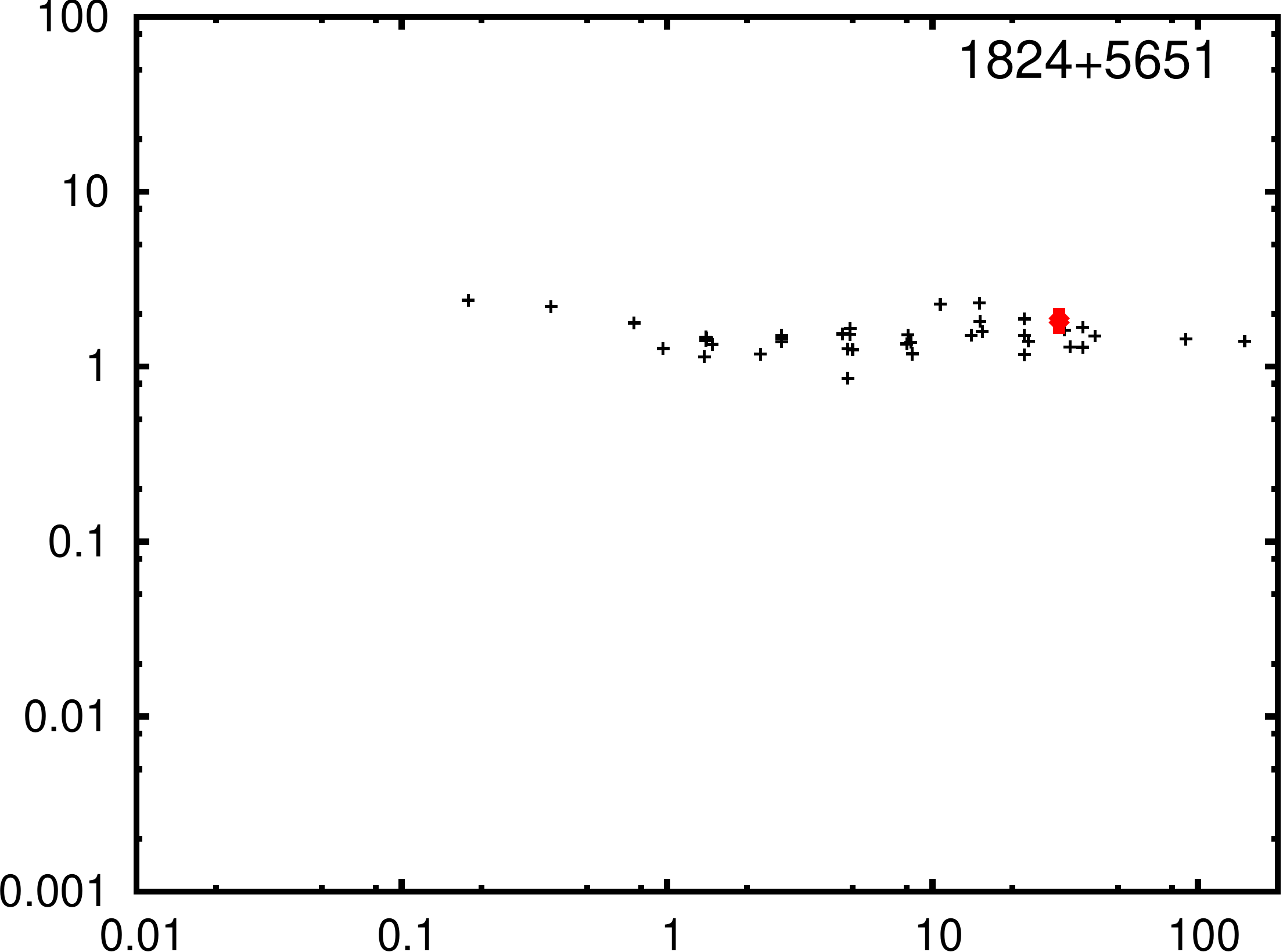}
\includegraphics[scale=0.2]{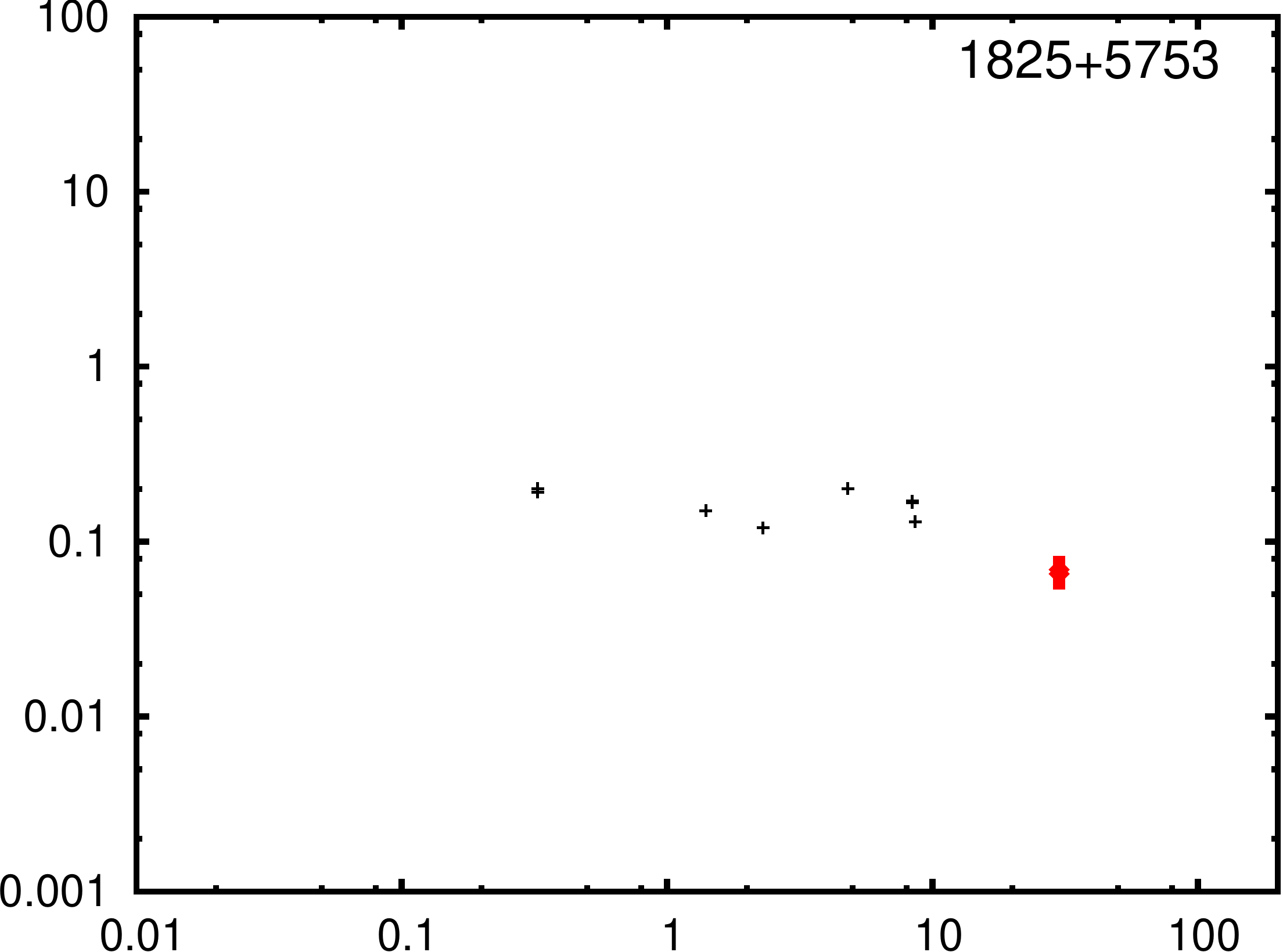}
\includegraphics[scale=0.2]{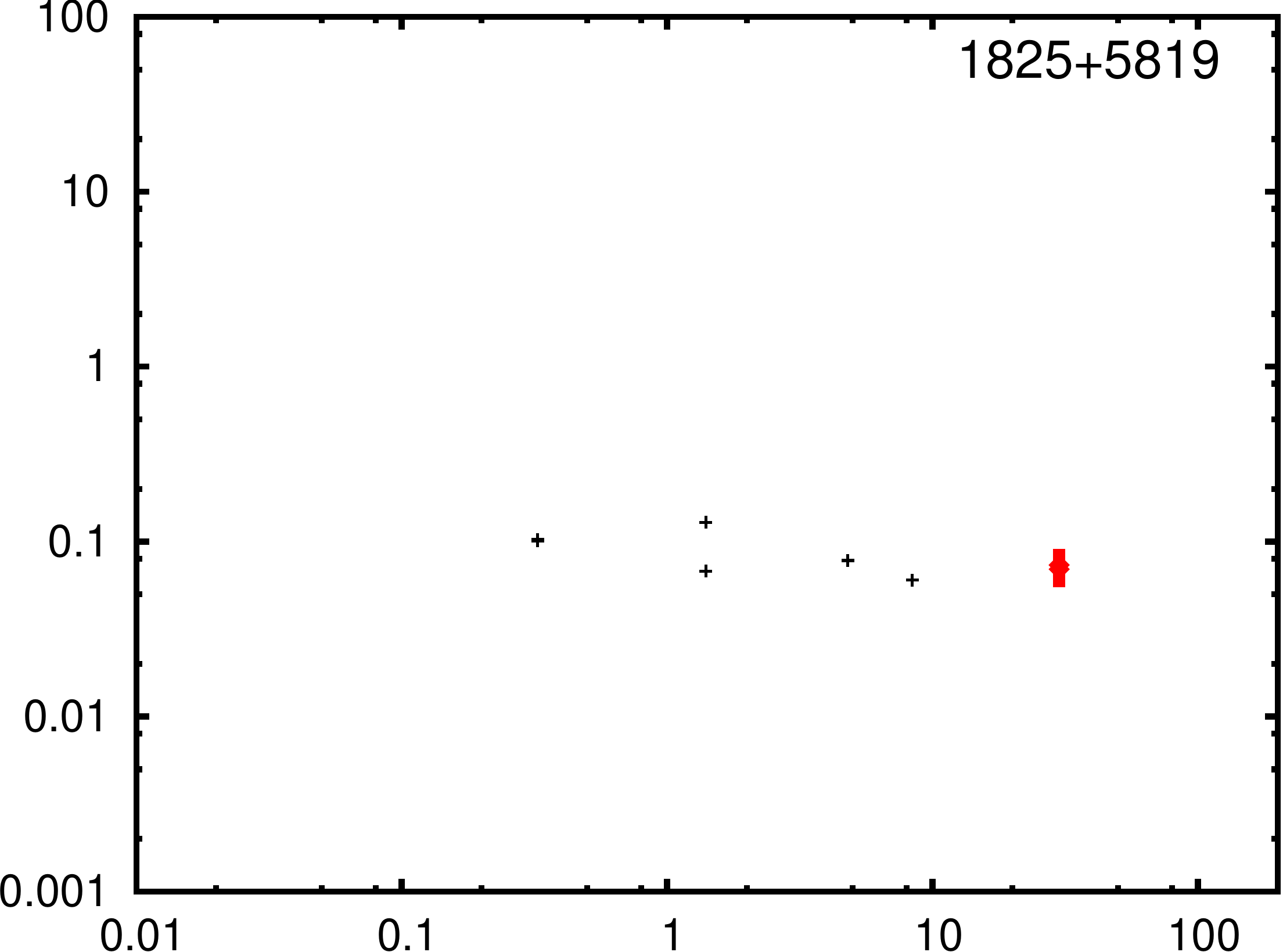}
\includegraphics[scale=0.2]{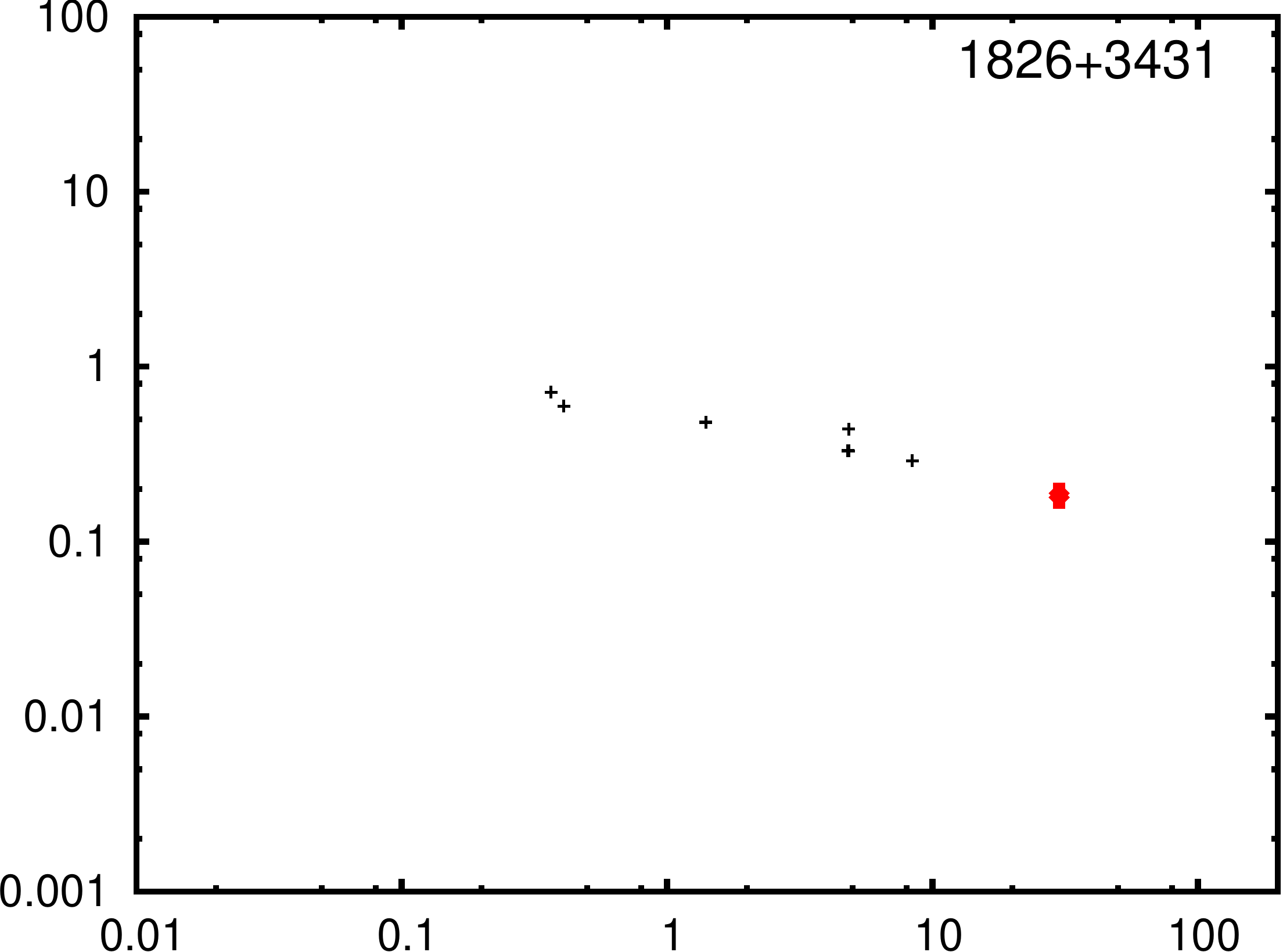}
\includegraphics[scale=0.2]{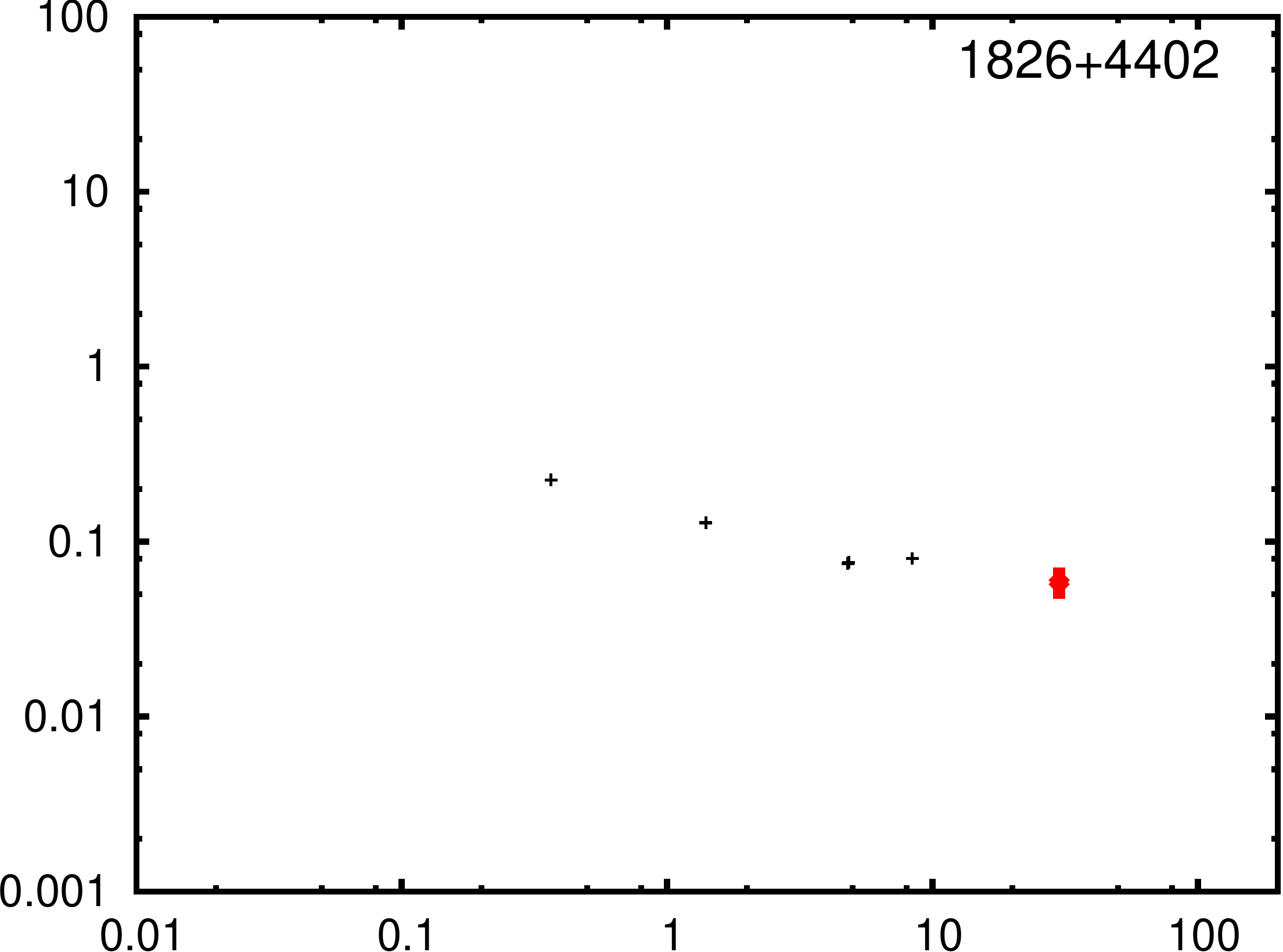}
\includegraphics[scale=0.2]{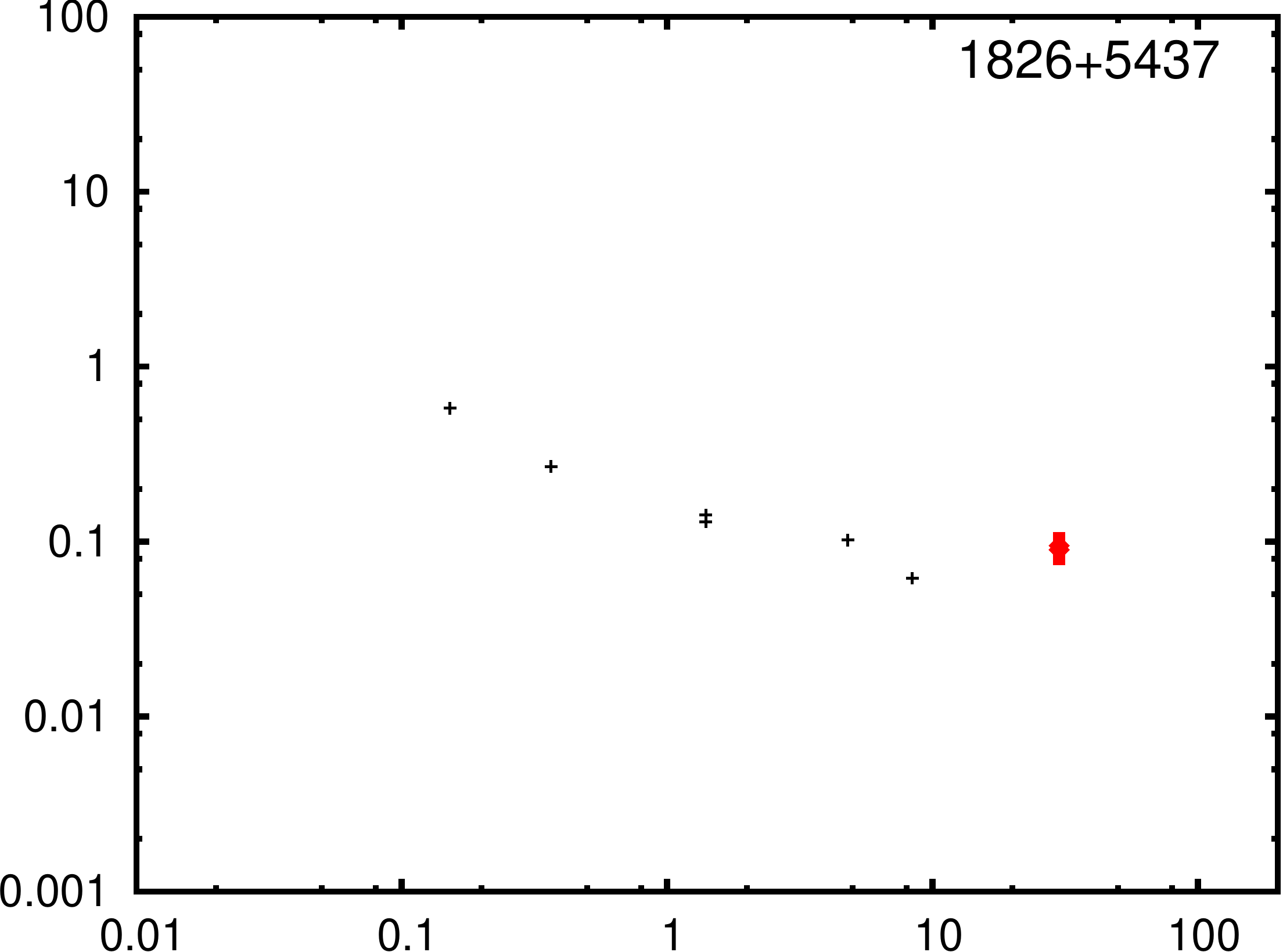}
\includegraphics[scale=0.2]{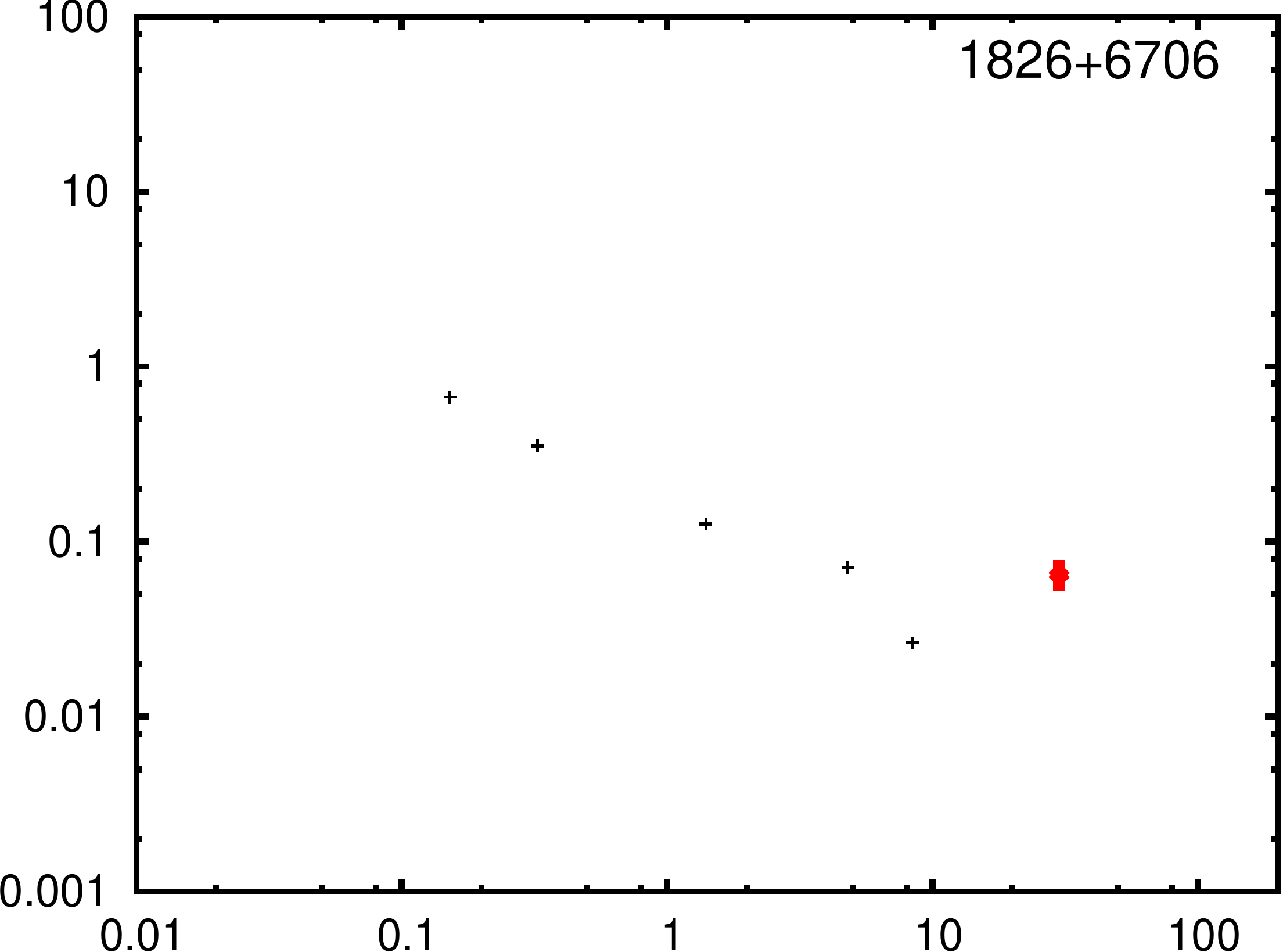}
\includegraphics[scale=0.2]{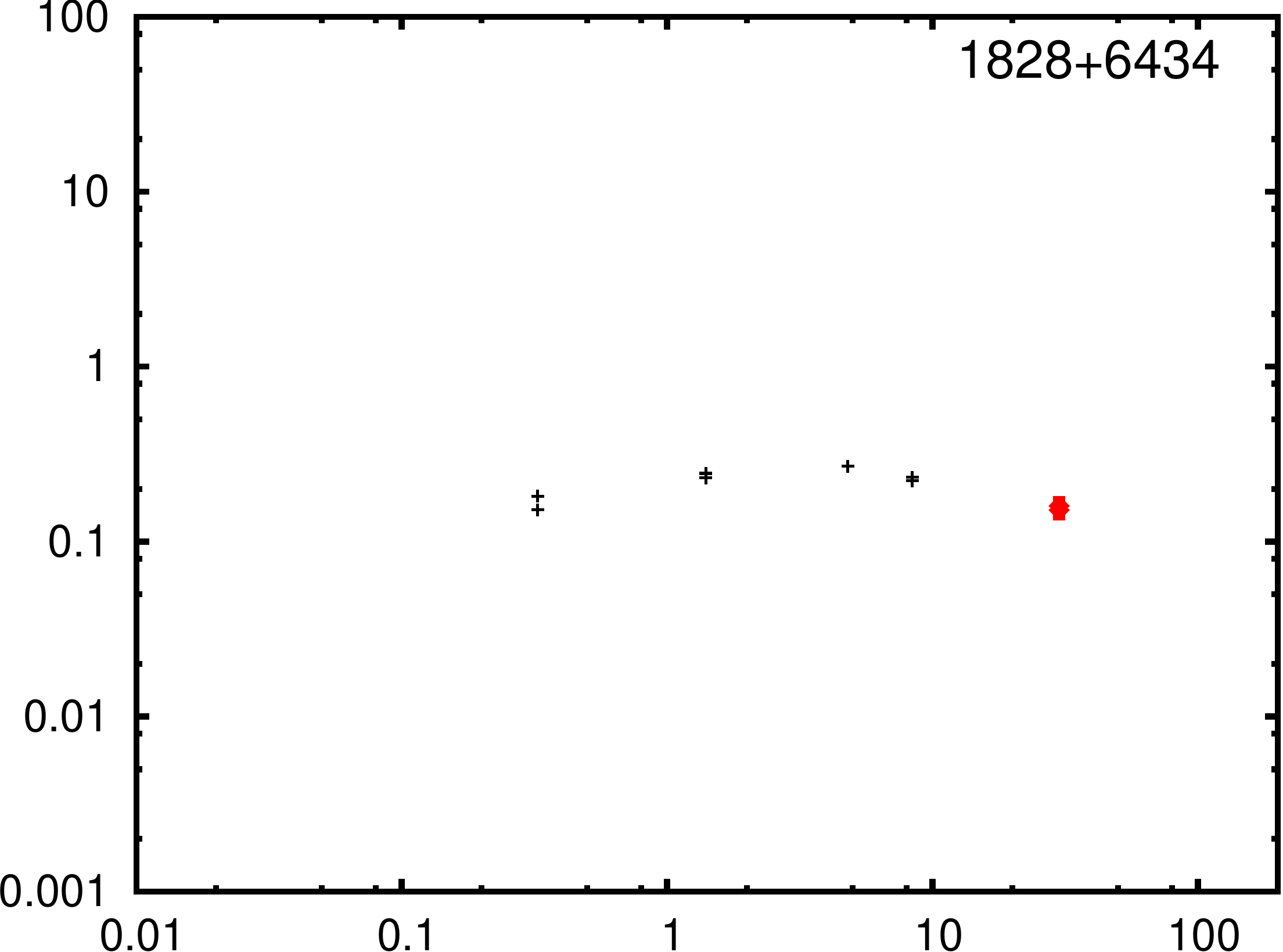}
\includegraphics[scale=0.2]{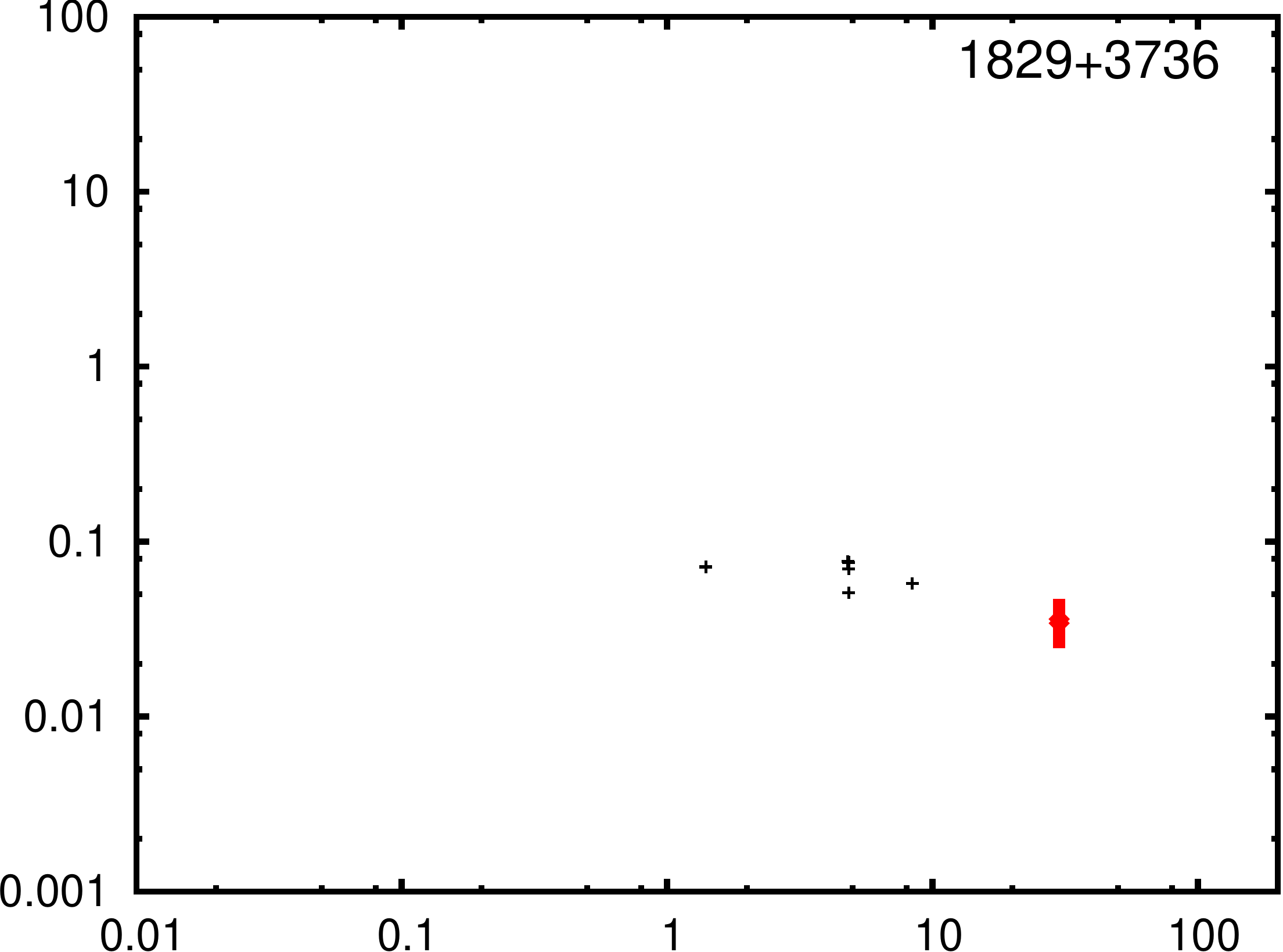}
\includegraphics[scale=0.2]{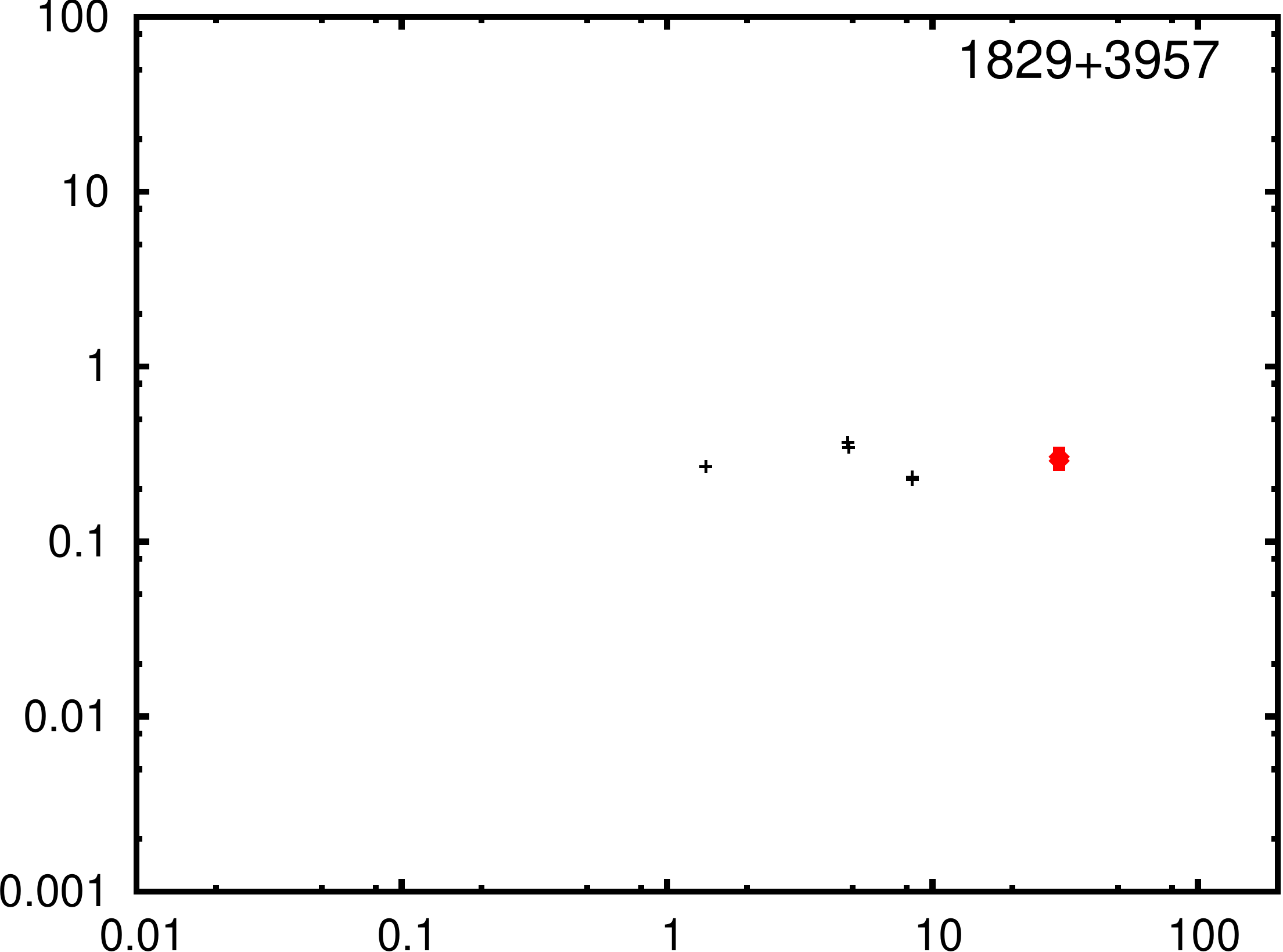}
\includegraphics[scale=0.2]{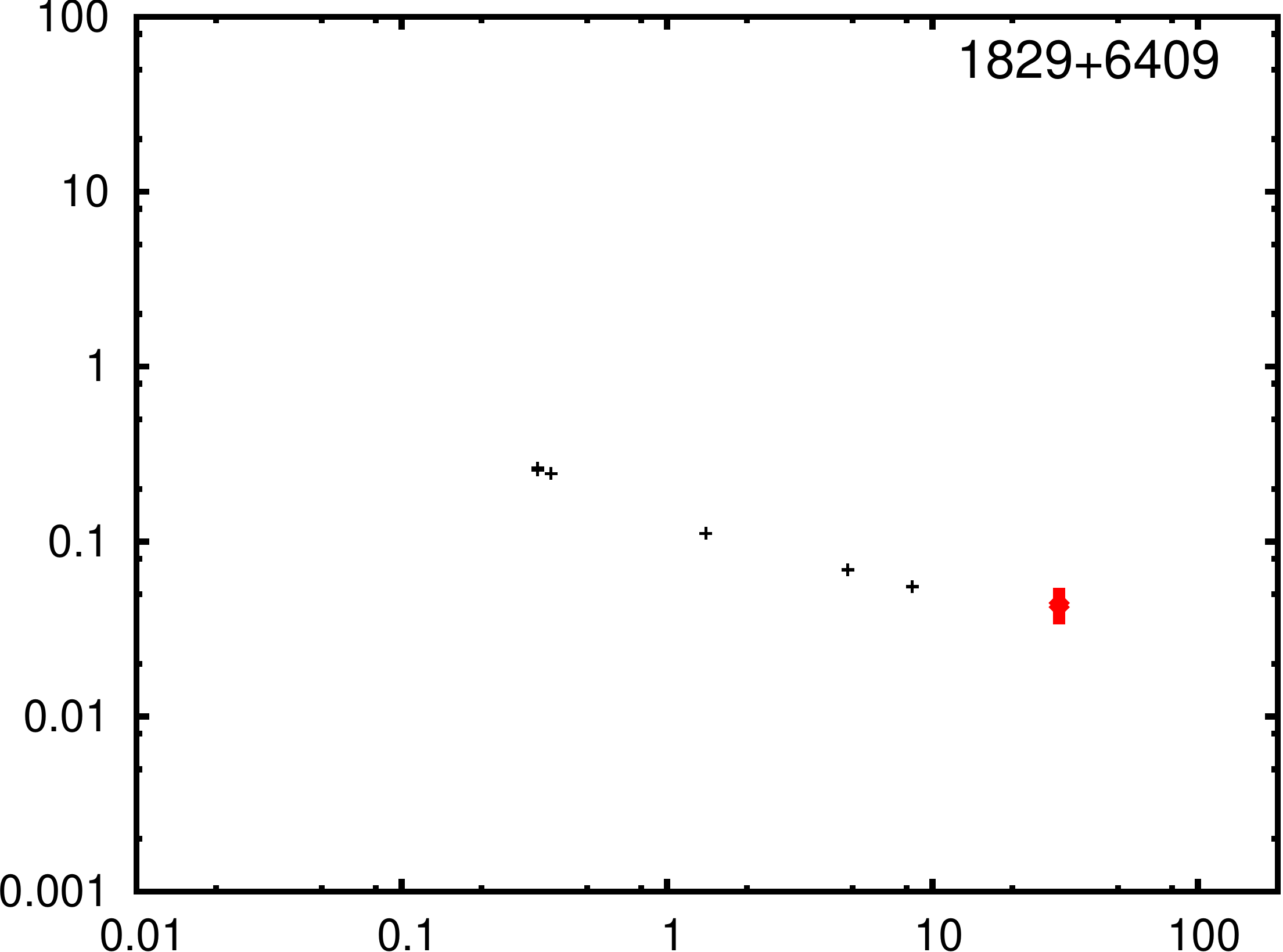}
\includegraphics[scale=0.2]{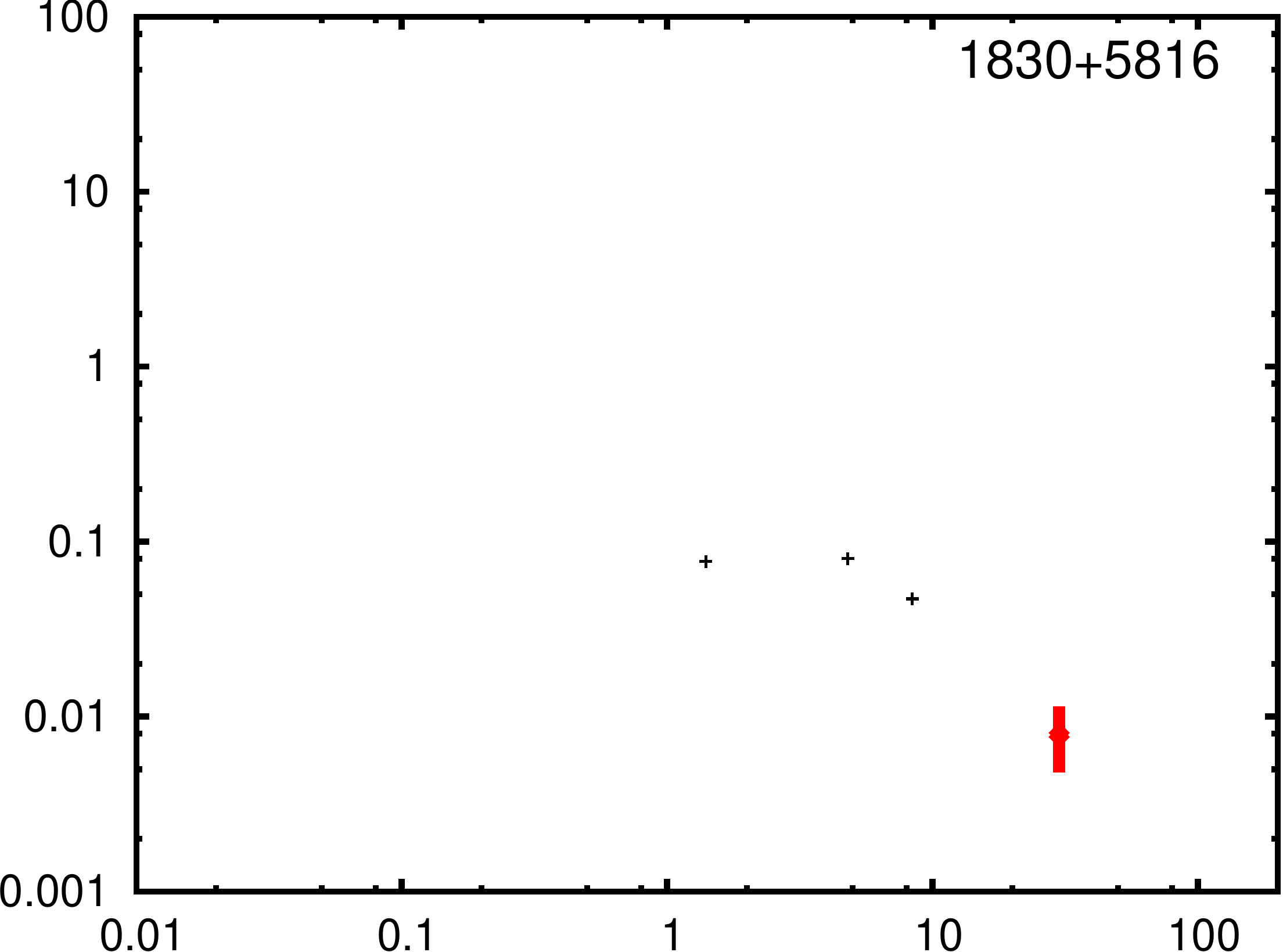}
\includegraphics[scale=0.2]{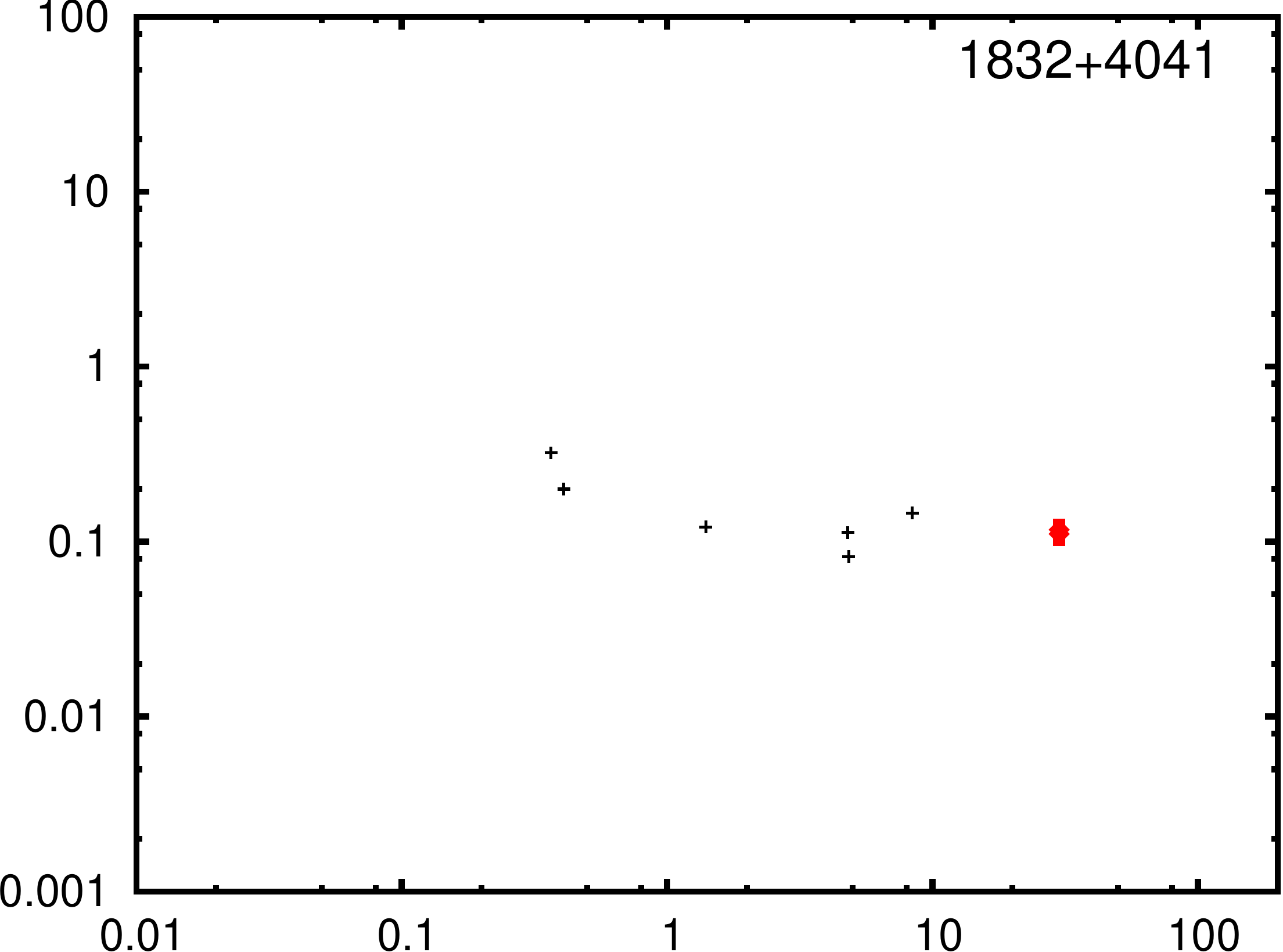}
\includegraphics[scale=0.2]{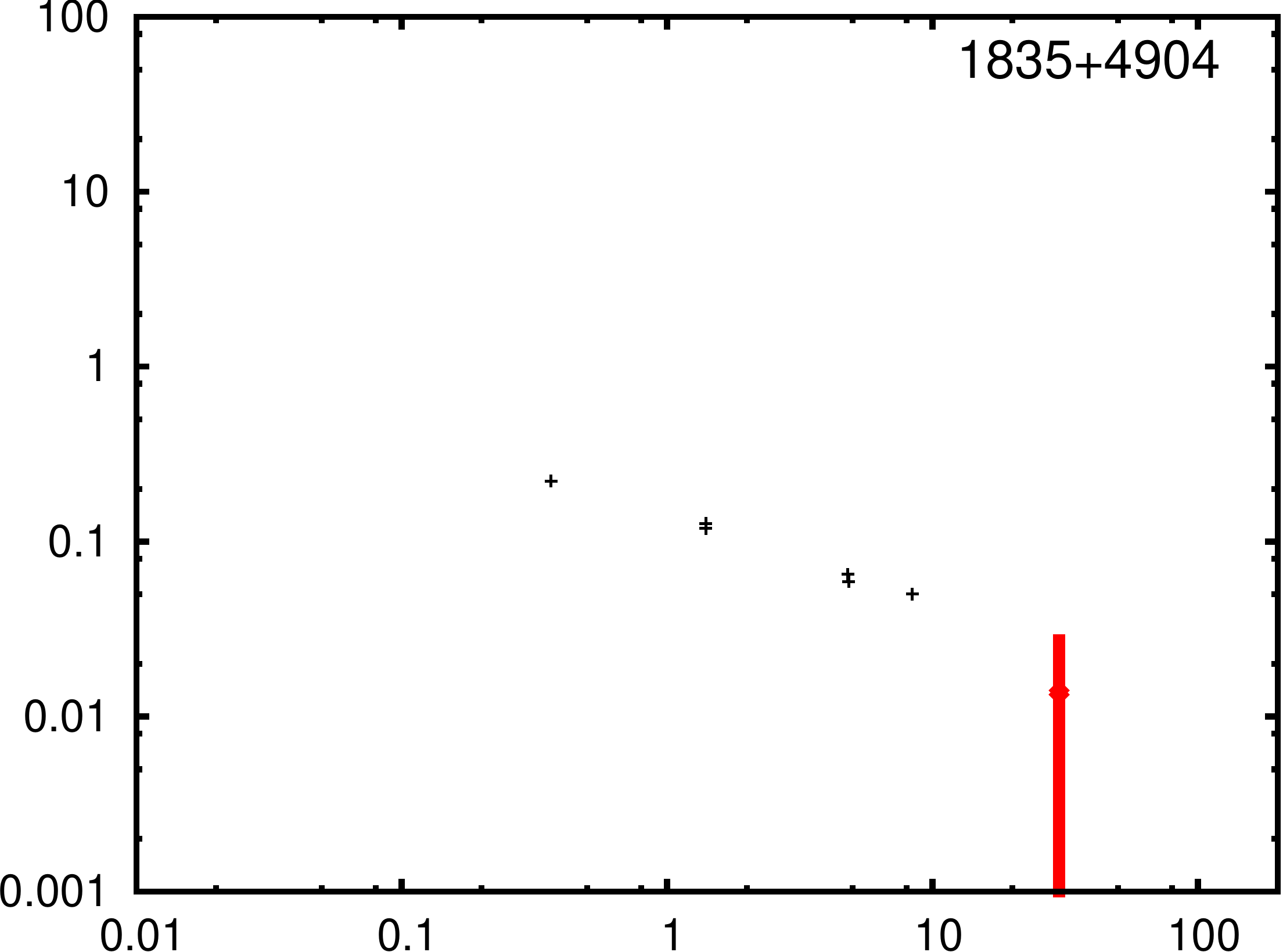}
\includegraphics[scale=0.2]{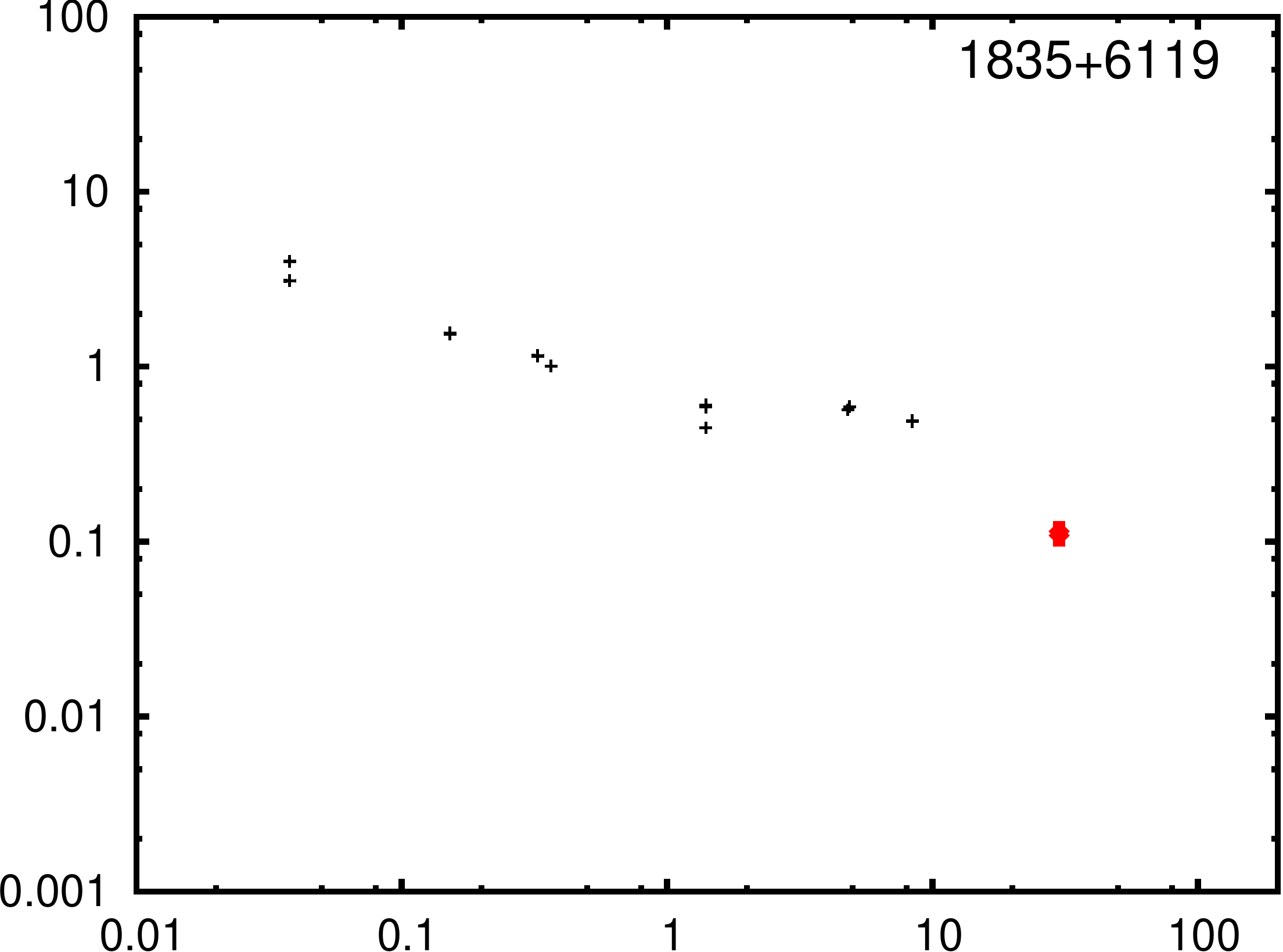}
\end{figure}
\clearpage\begin{figure}
\centering
\includegraphics[scale=0.2]{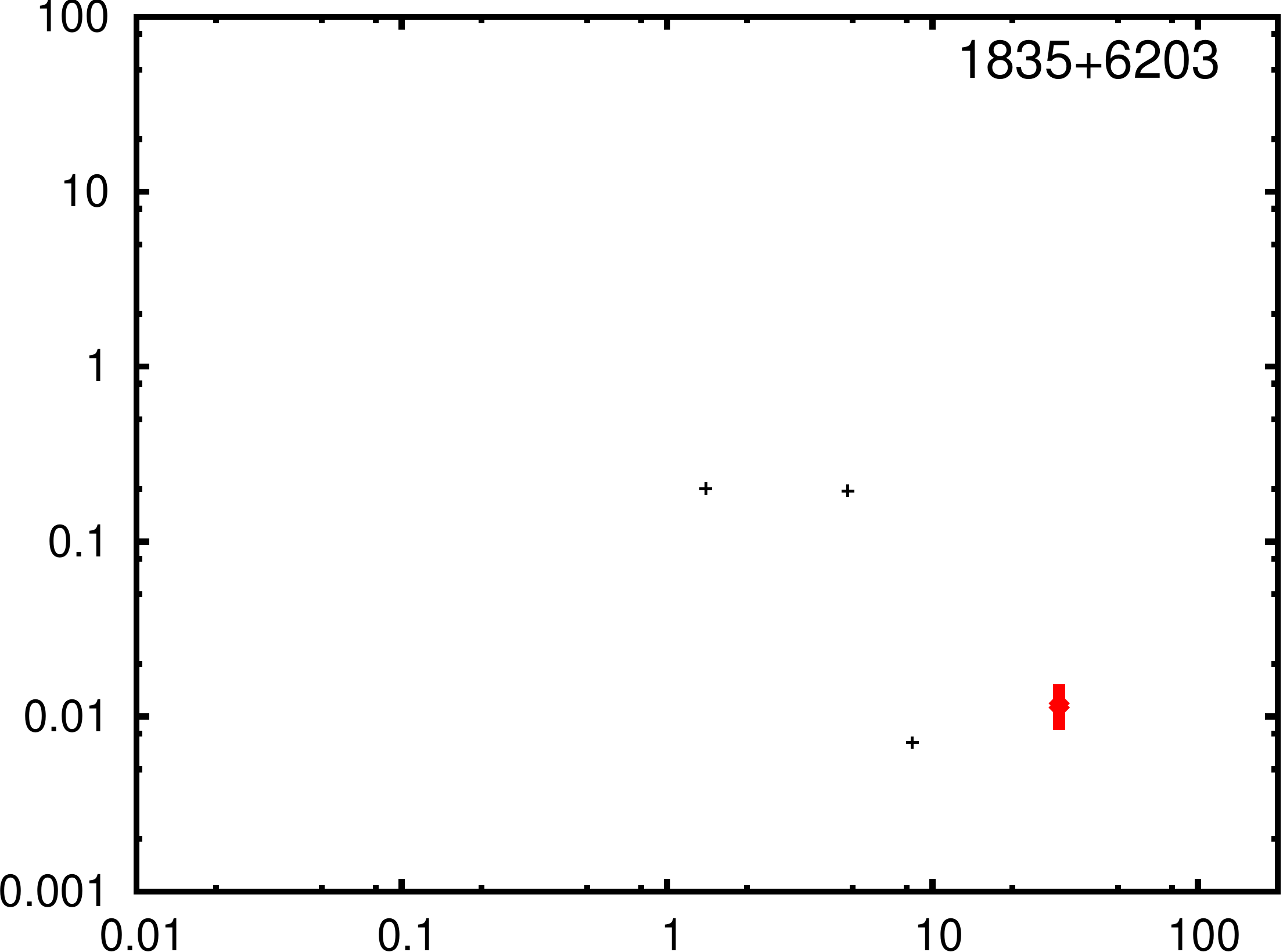}
\includegraphics[scale=0.2]{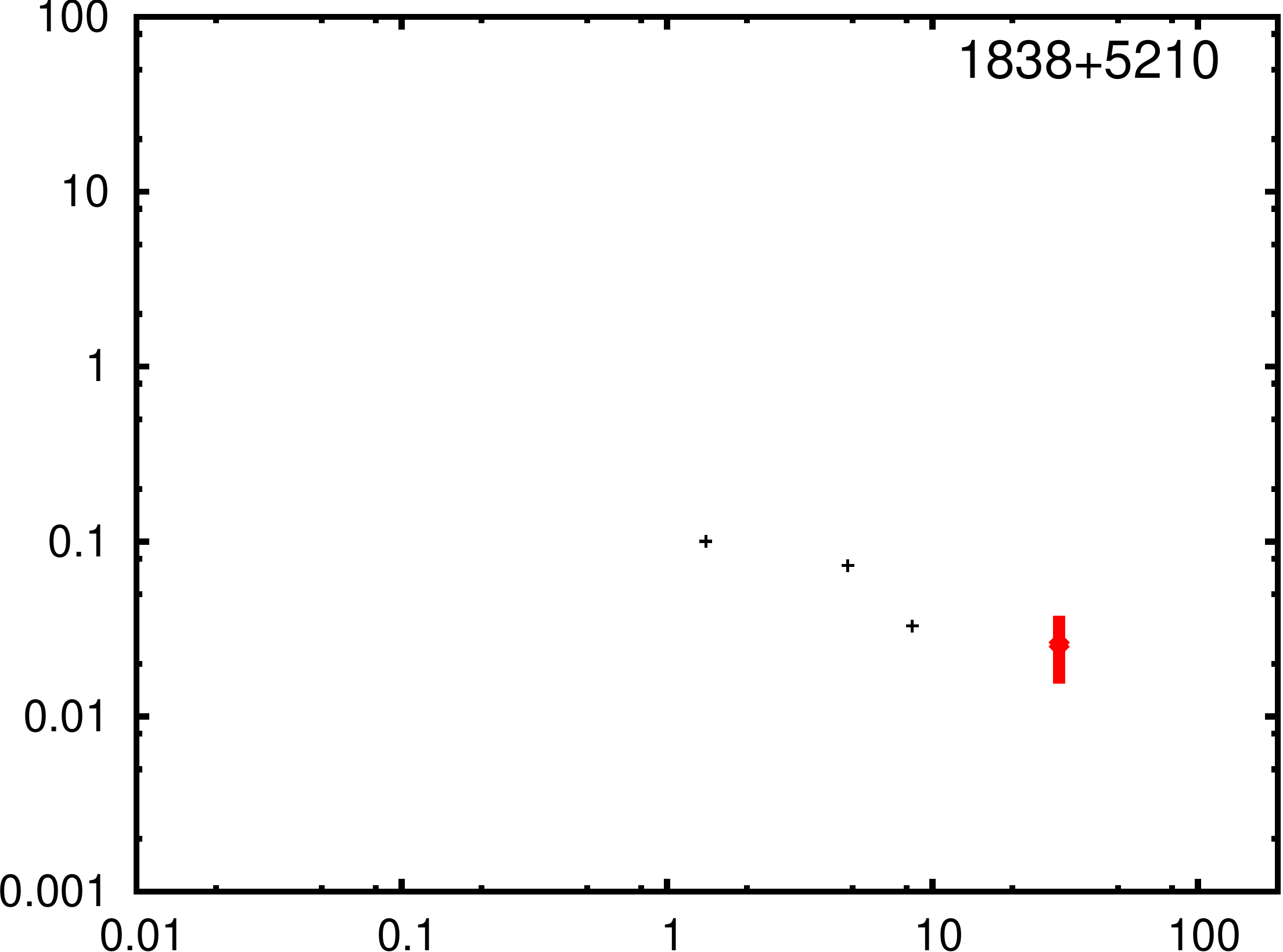}
\includegraphics[scale=0.2]{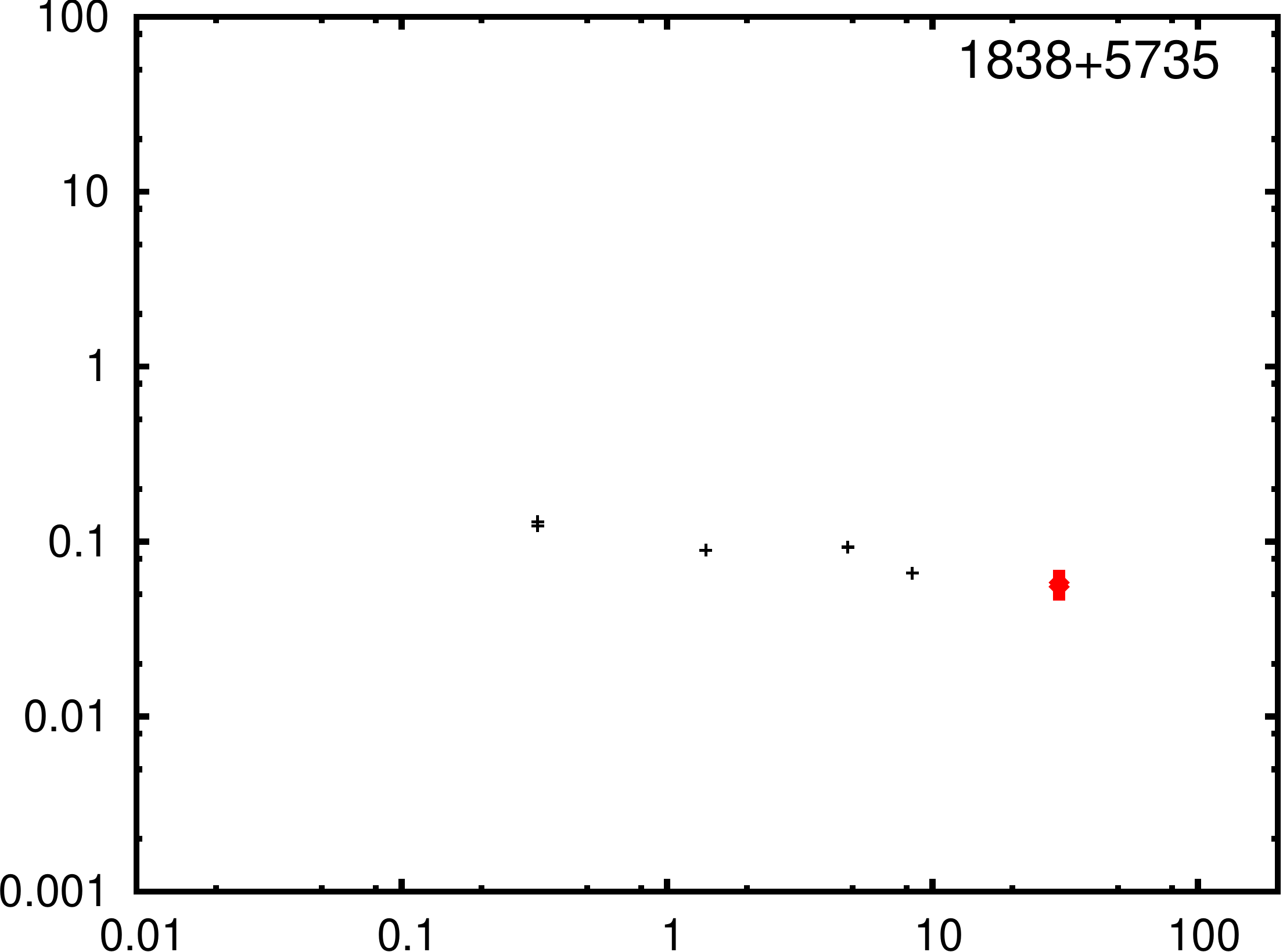}
\includegraphics[scale=0.2]{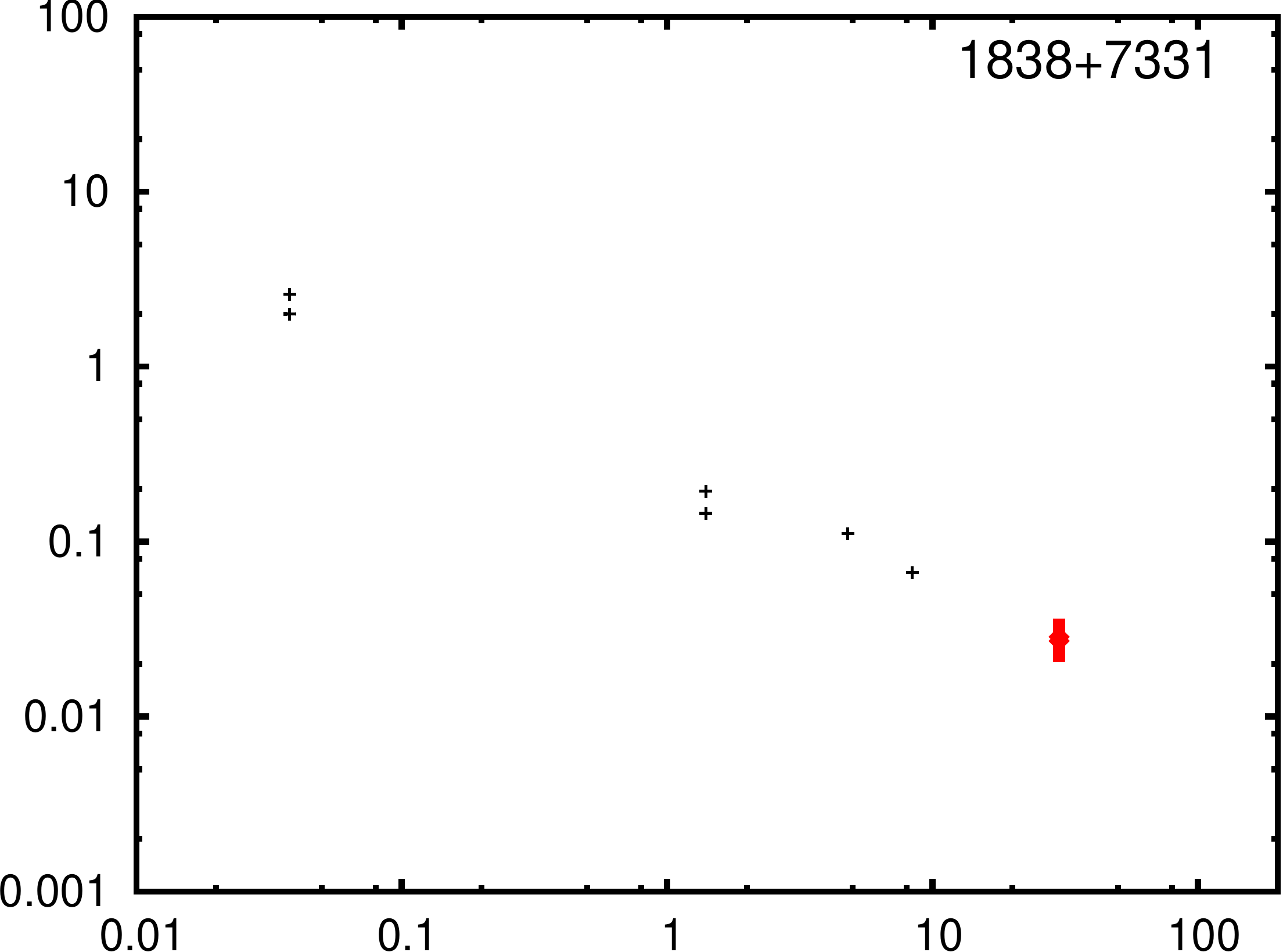}
\includegraphics[scale=0.2]{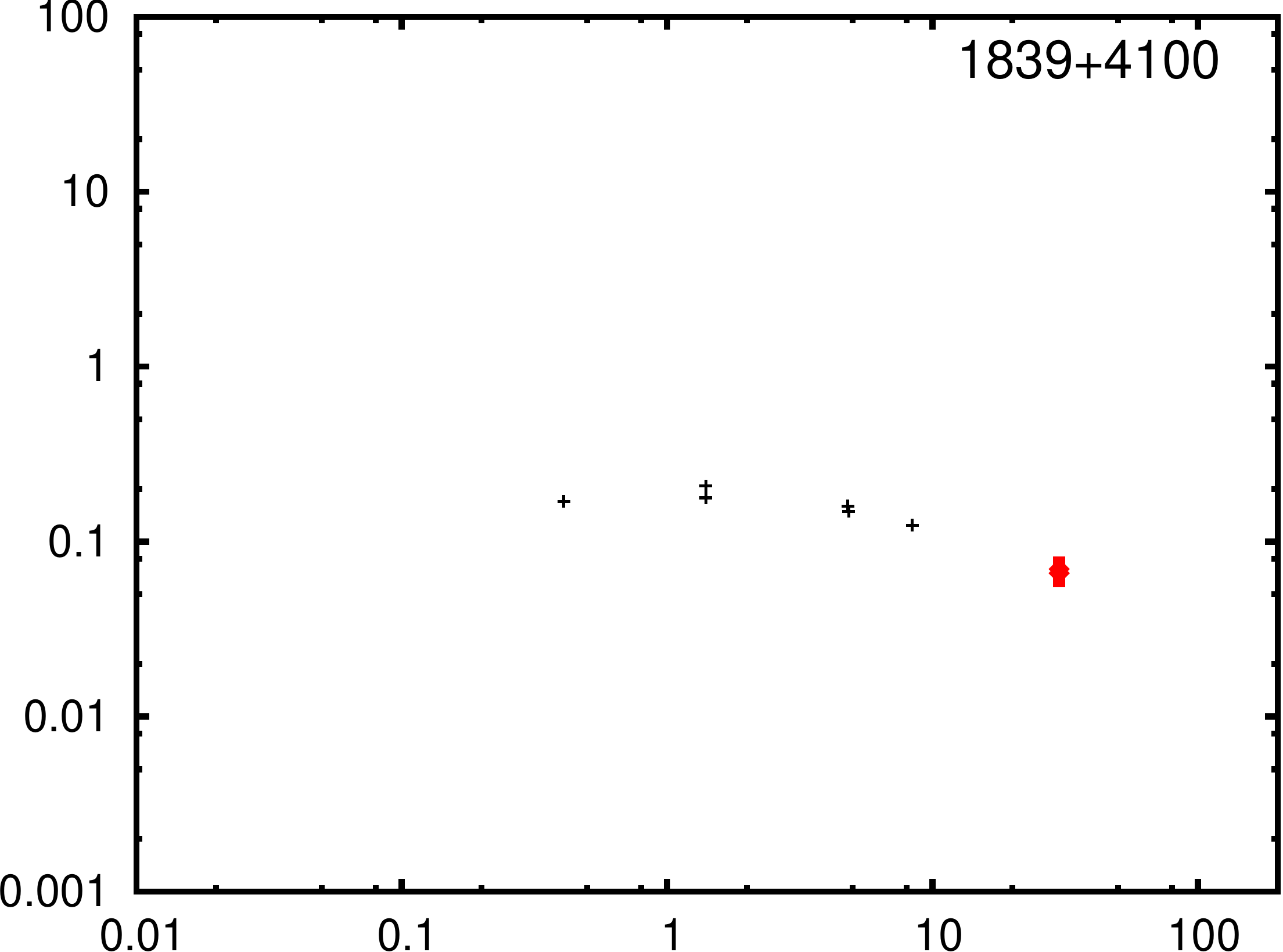}
\includegraphics[scale=0.2]{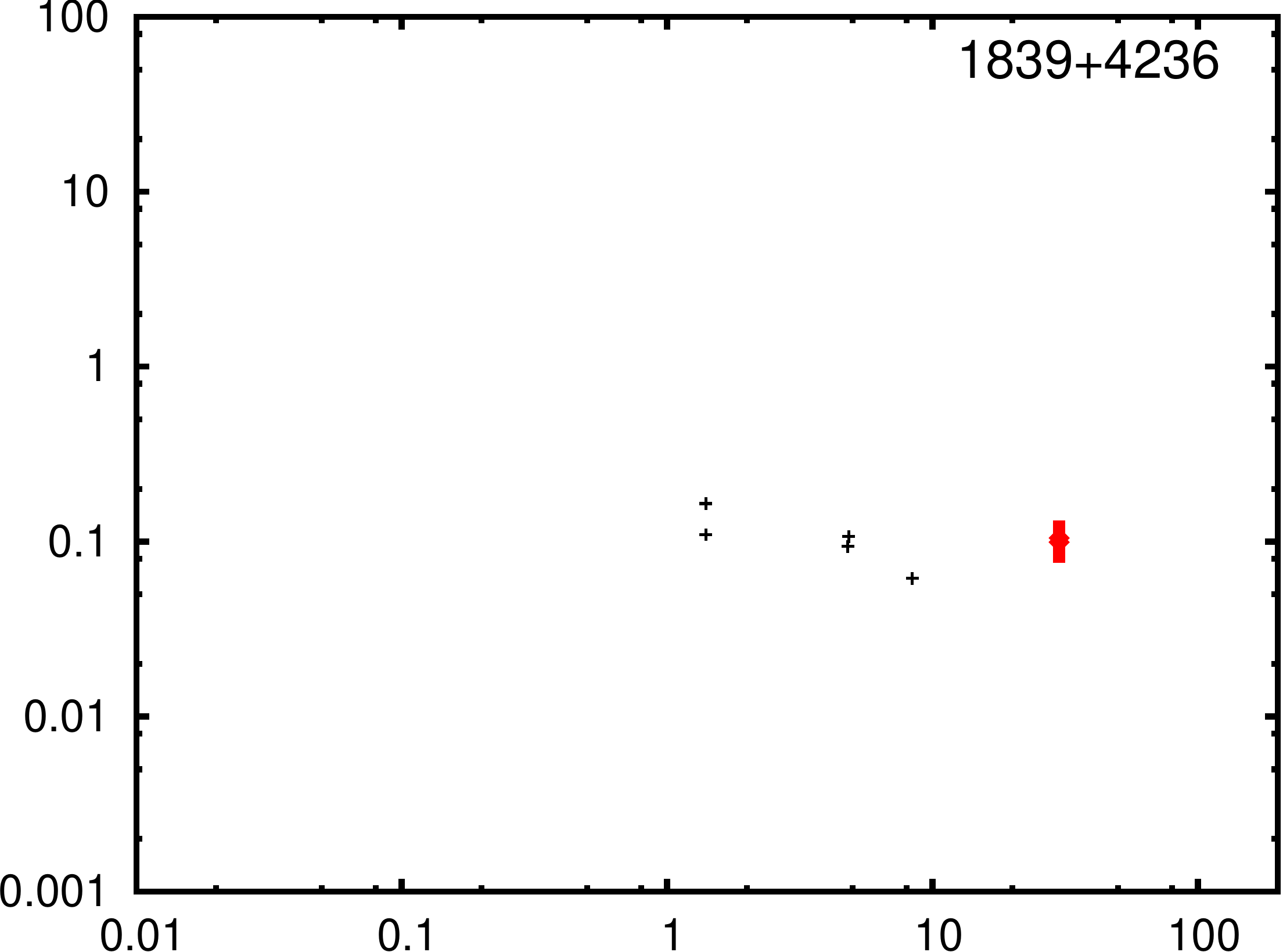}
\includegraphics[scale=0.2]{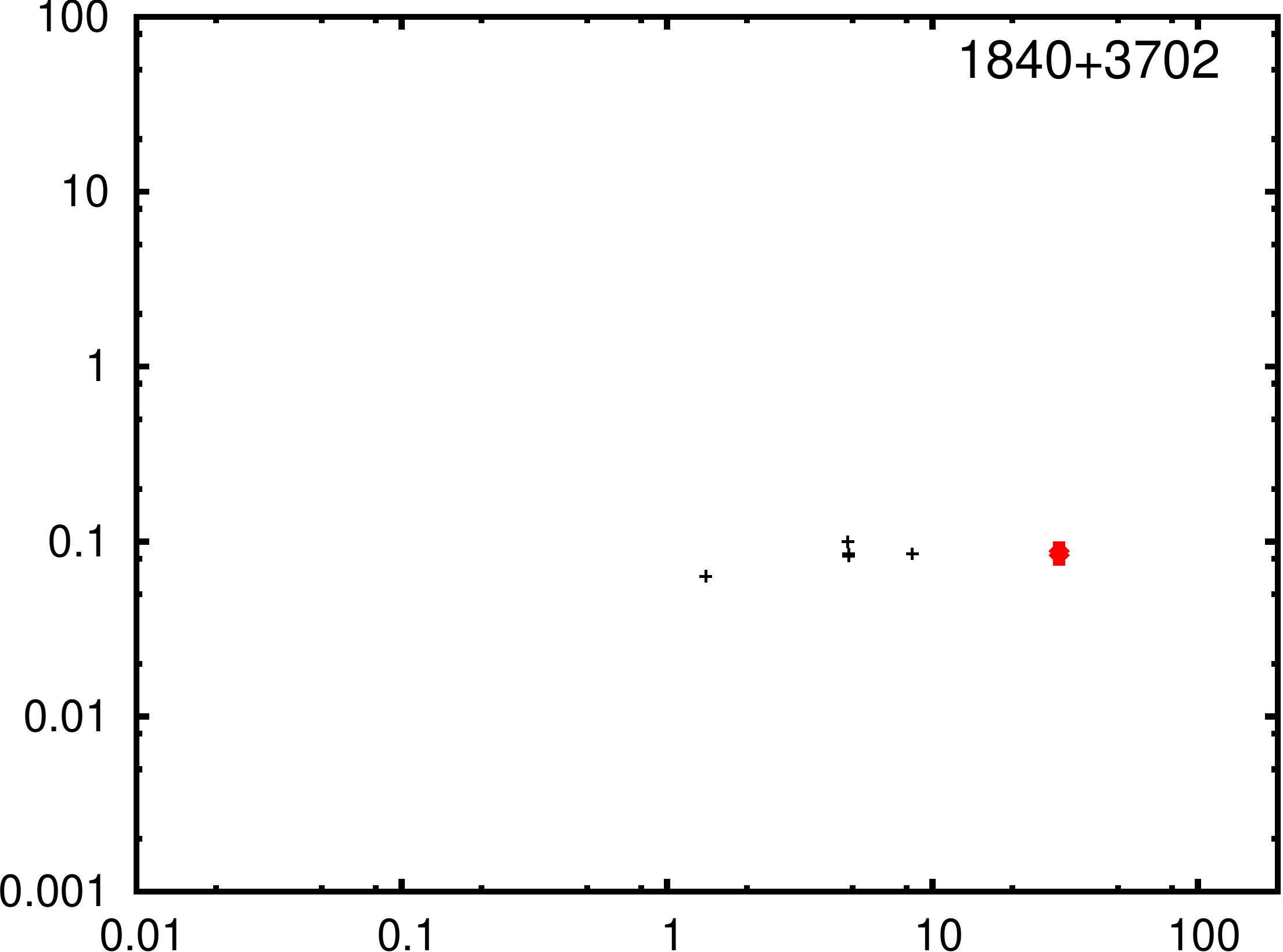}
\includegraphics[scale=0.2]{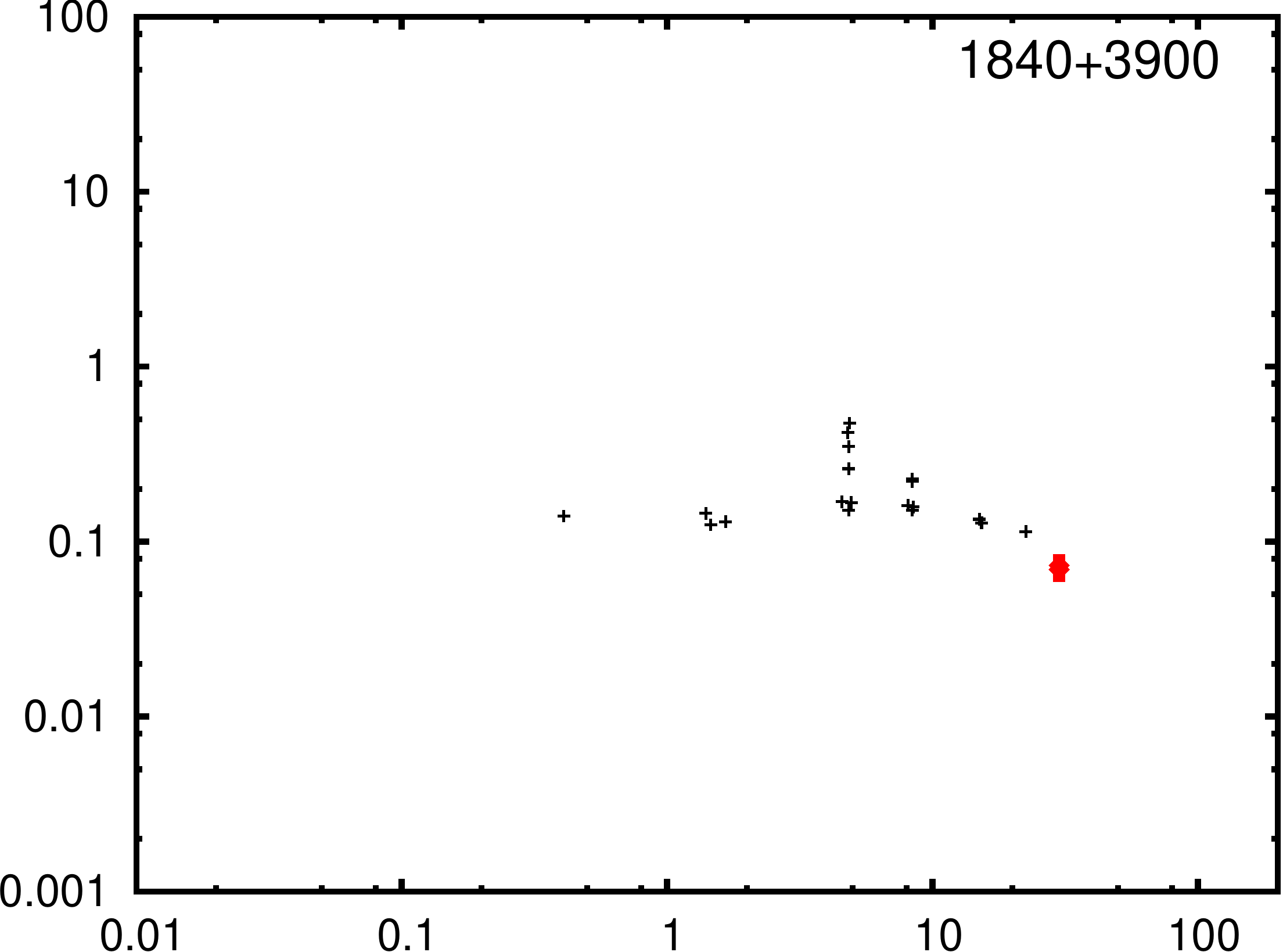}
\includegraphics[scale=0.2]{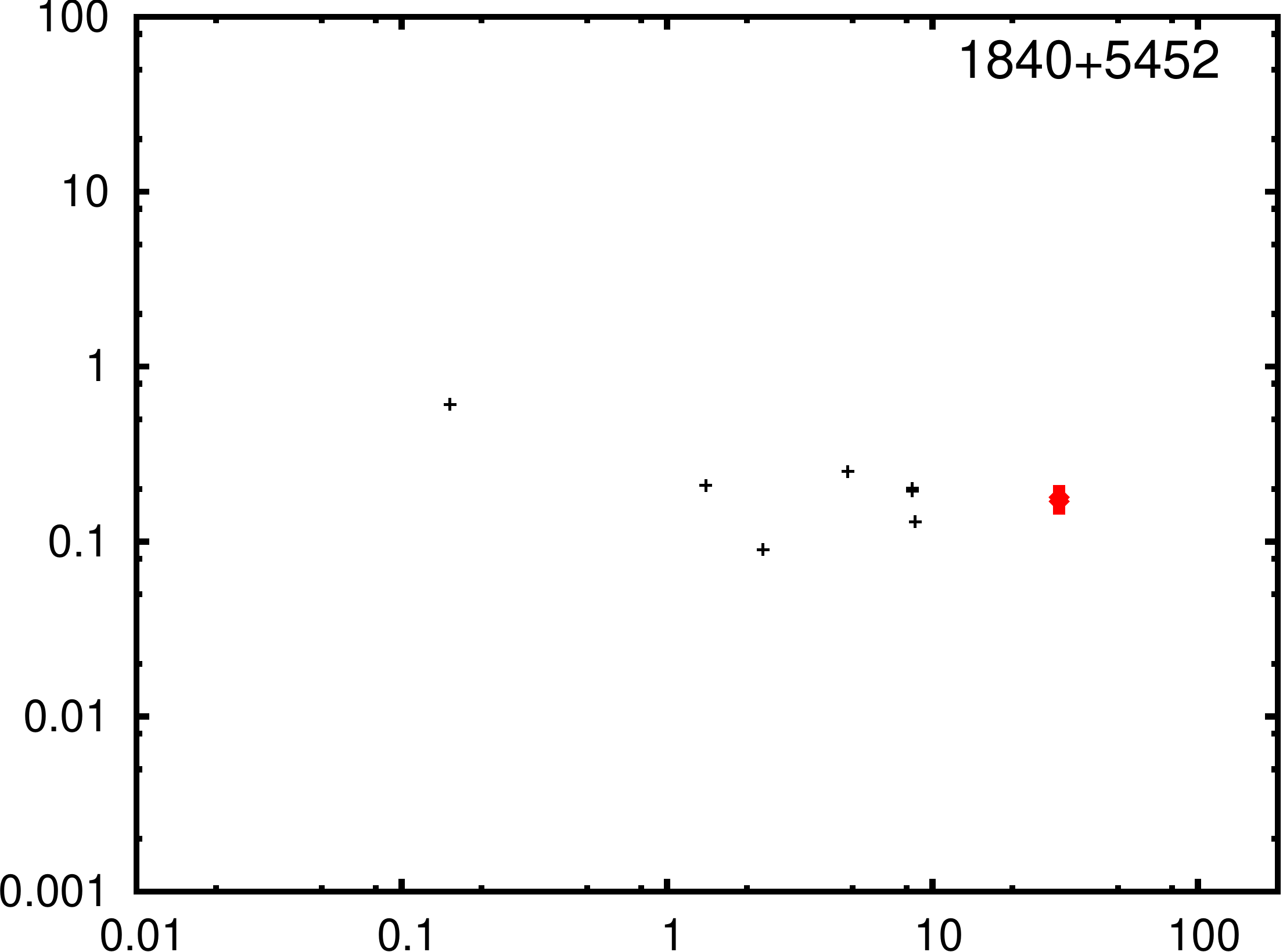}
\includegraphics[scale=0.2]{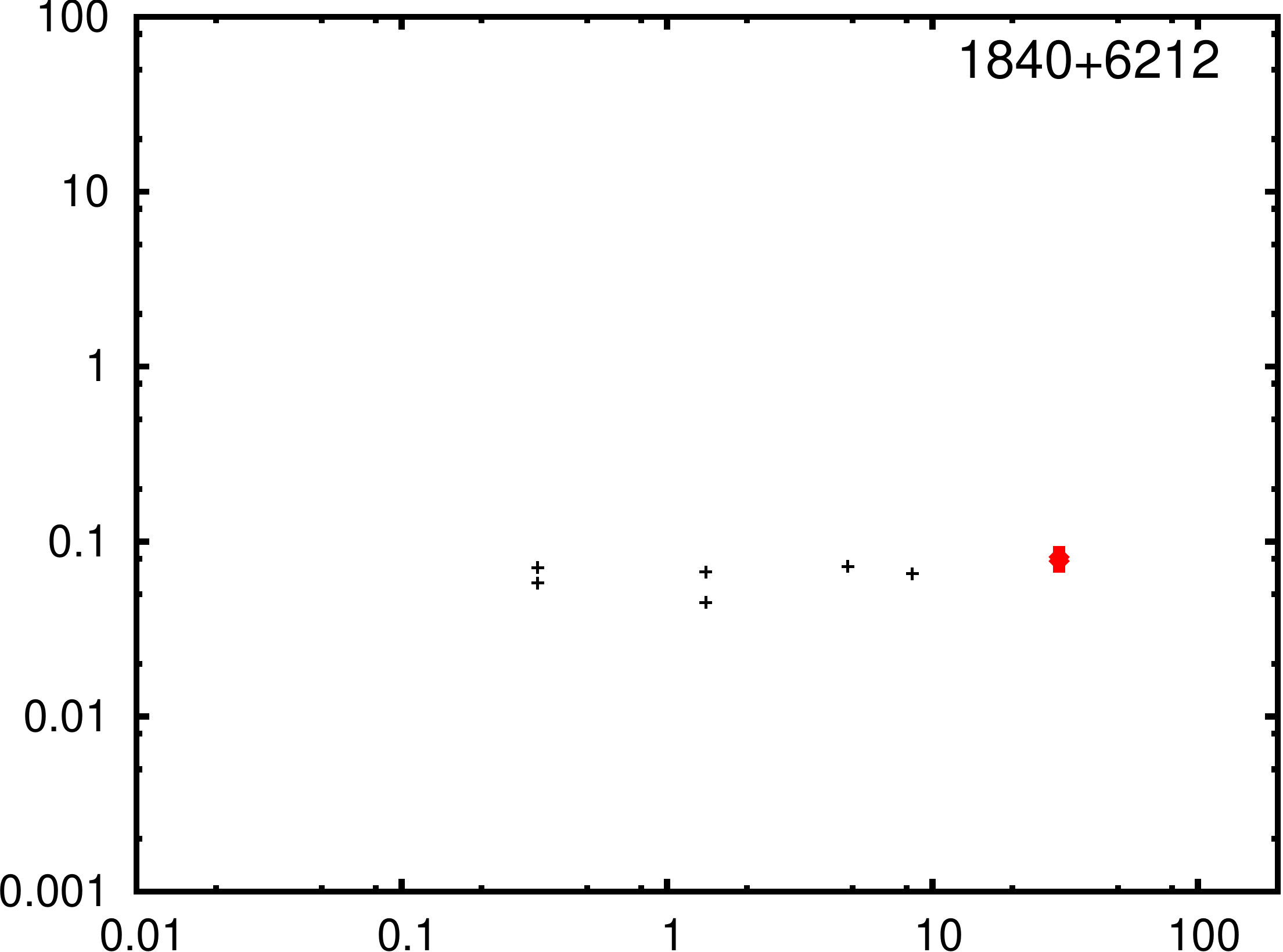}
\includegraphics[scale=0.2]{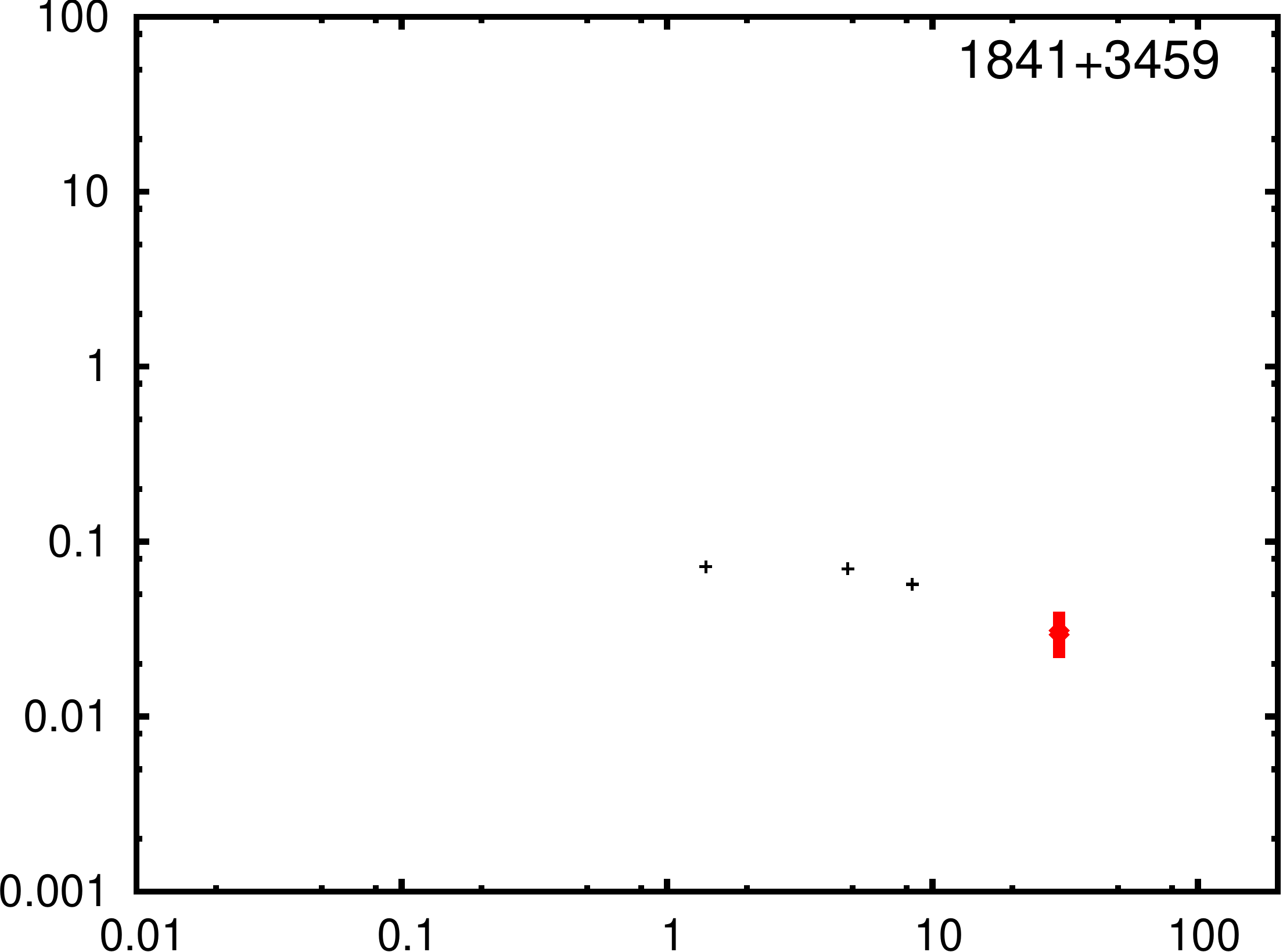}
\includegraphics[scale=0.2]{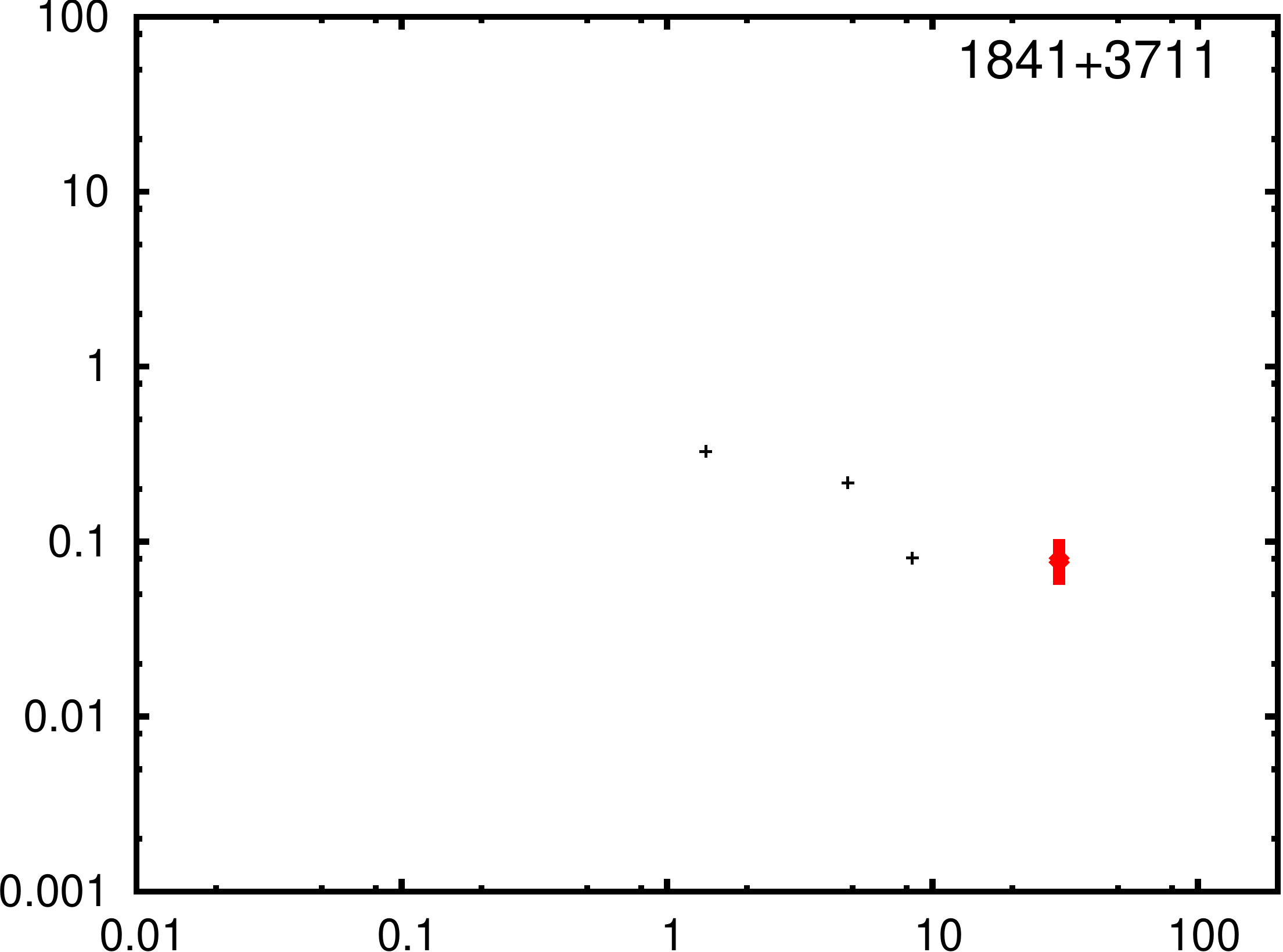}
\includegraphics[scale=0.2]{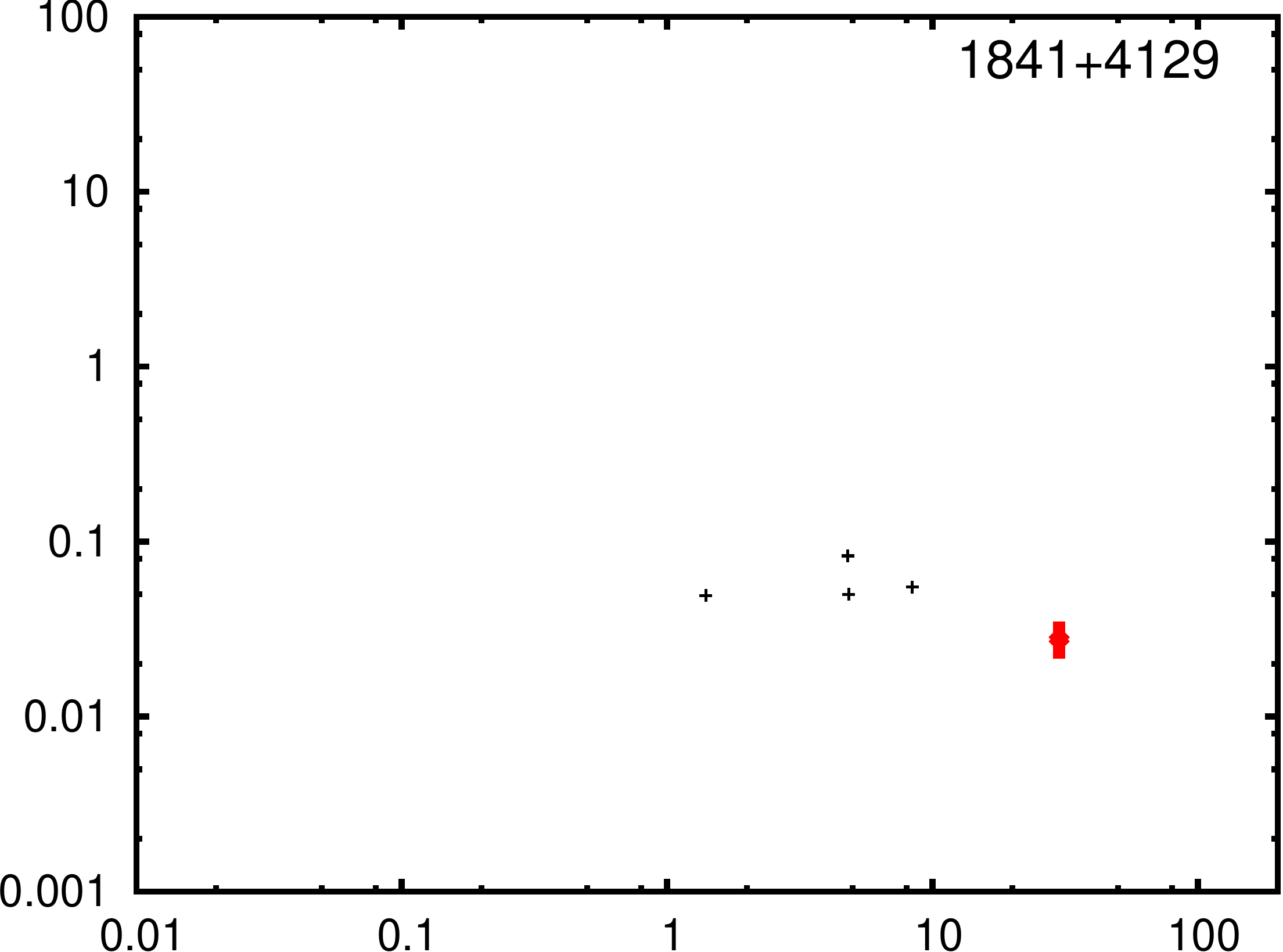}
\includegraphics[scale=0.2]{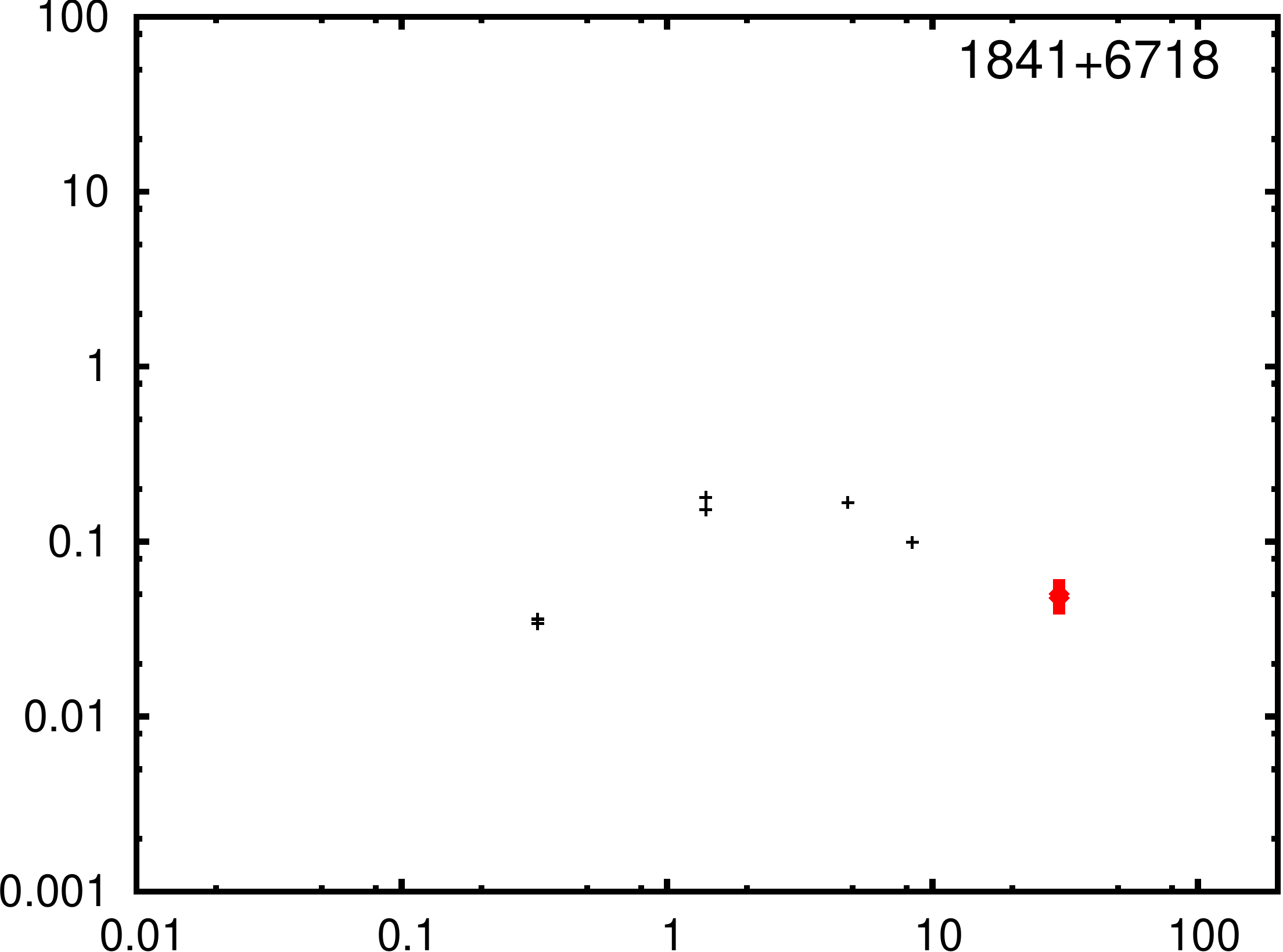}
\includegraphics[scale=0.2]{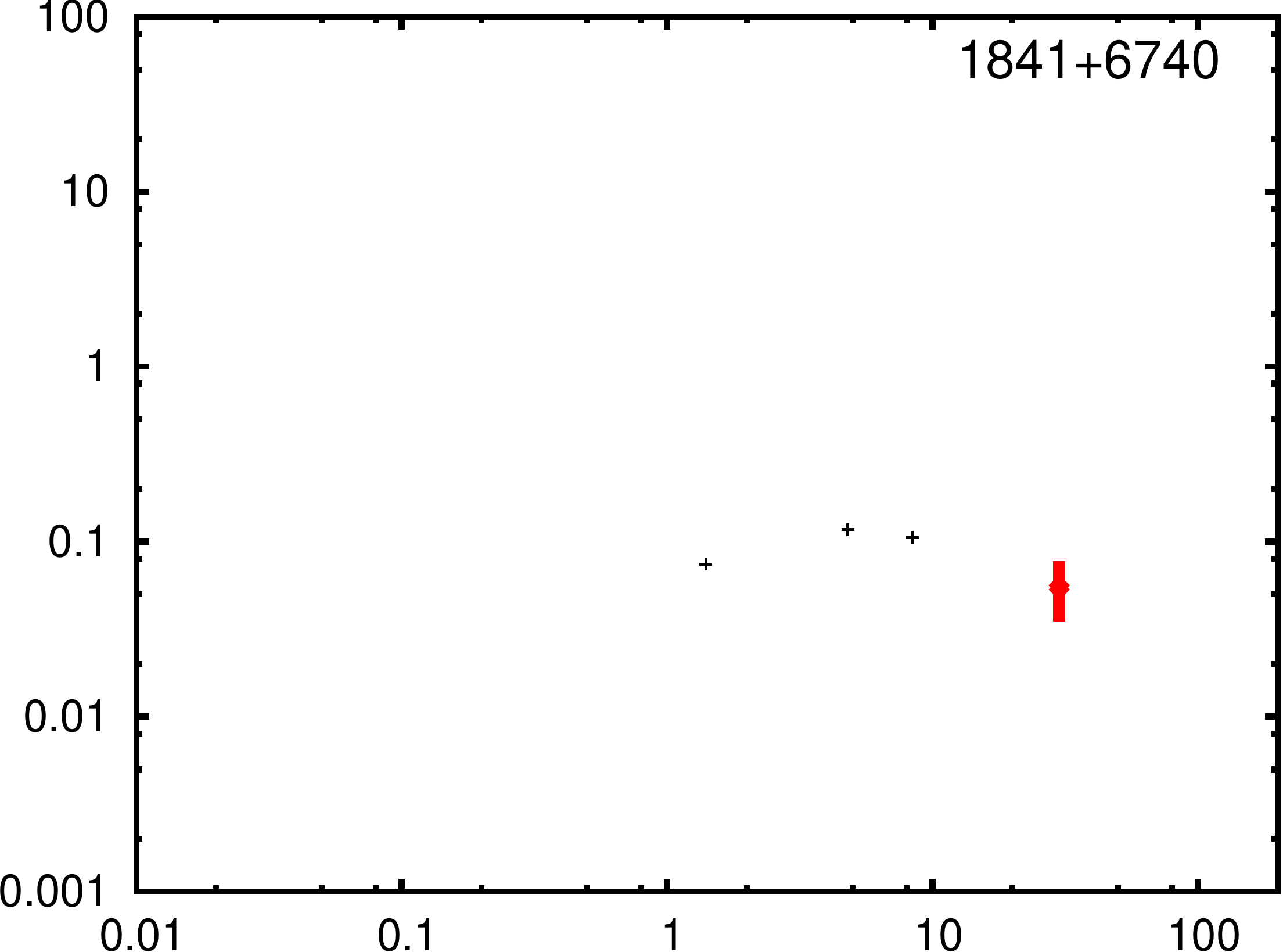}
\includegraphics[scale=0.2]{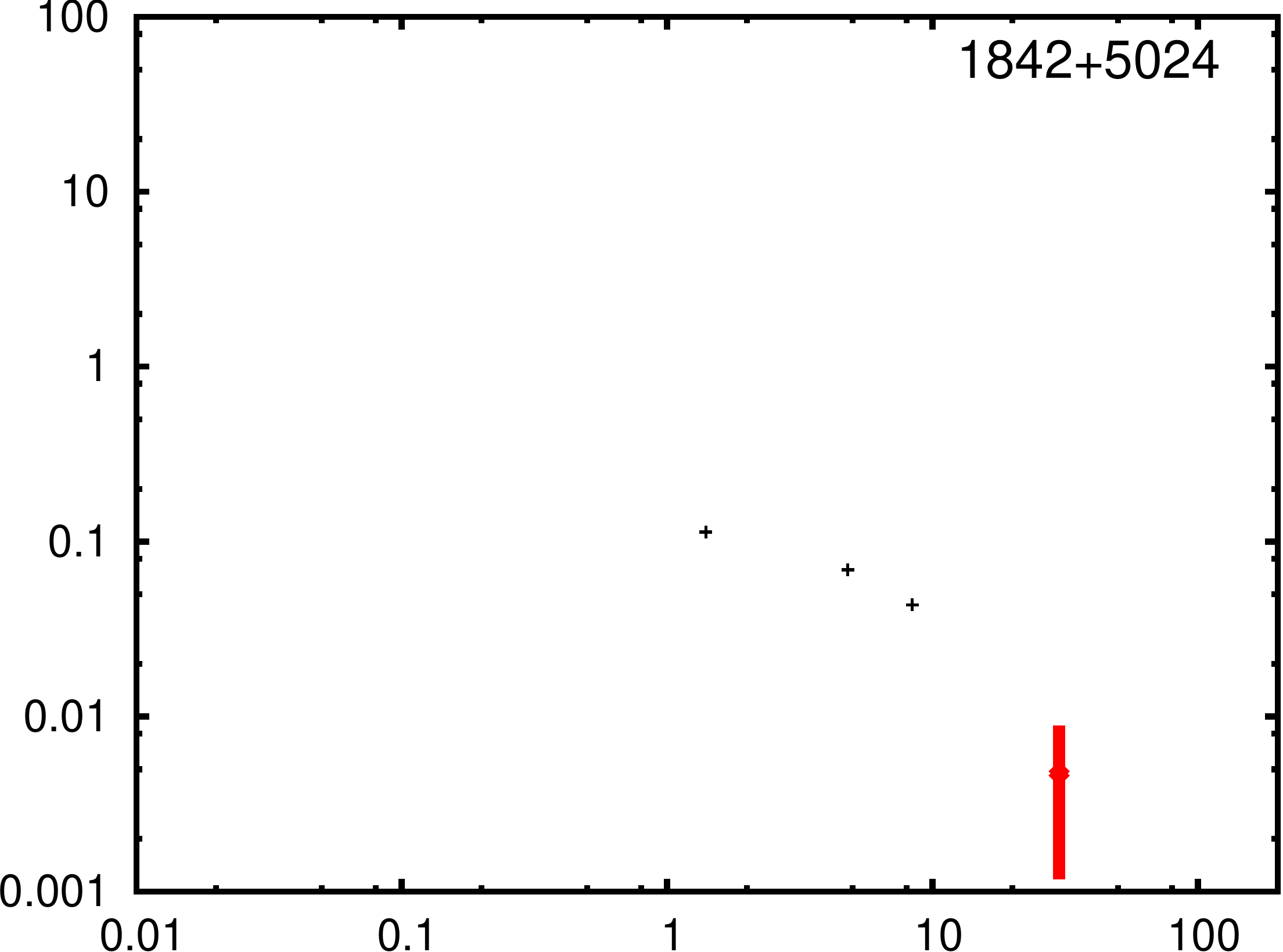}
\includegraphics[scale=0.2]{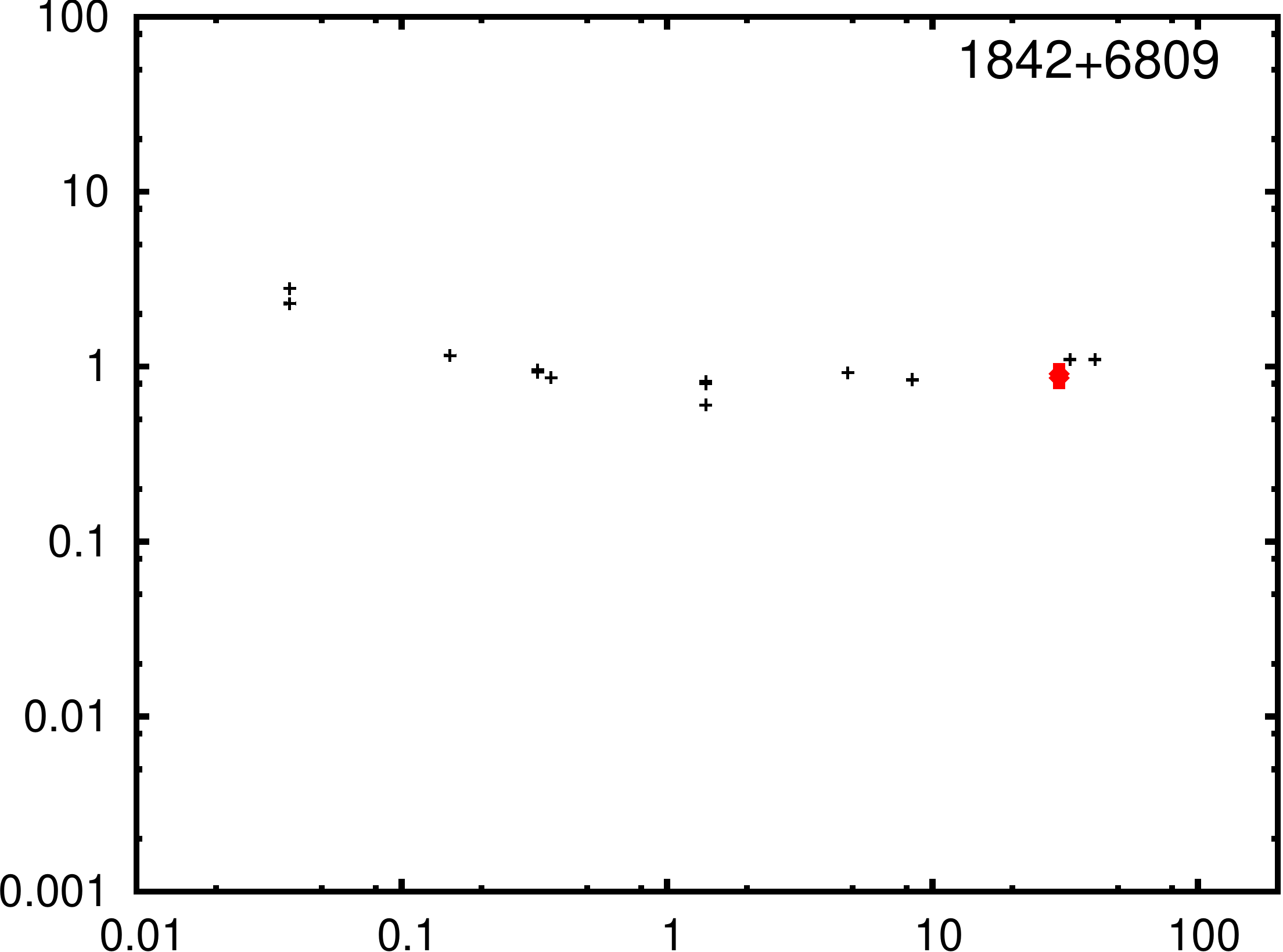}
\includegraphics[scale=0.2]{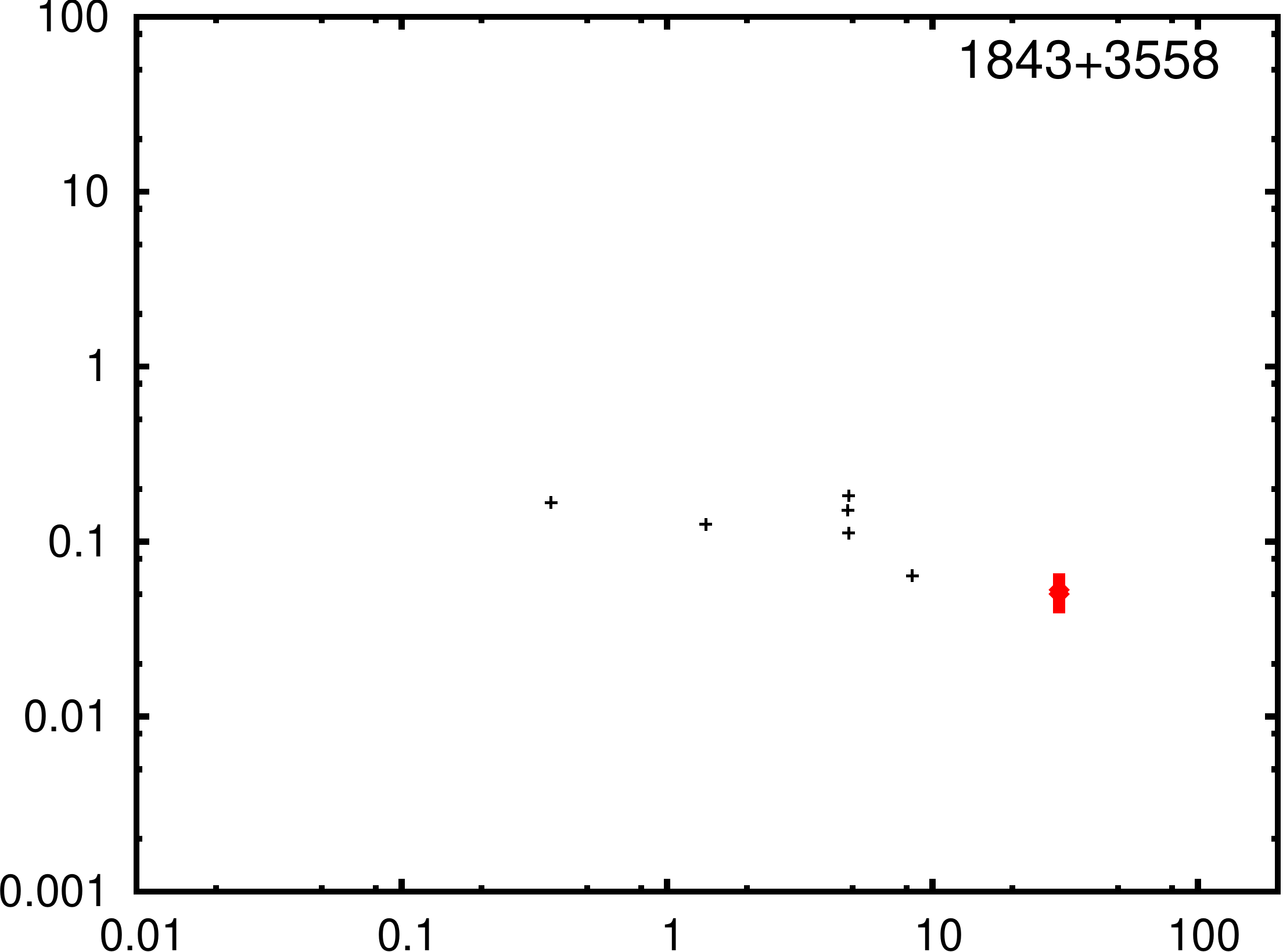}
\end{figure}
\clearpage\begin{figure}
\centering
\includegraphics[scale=0.2]{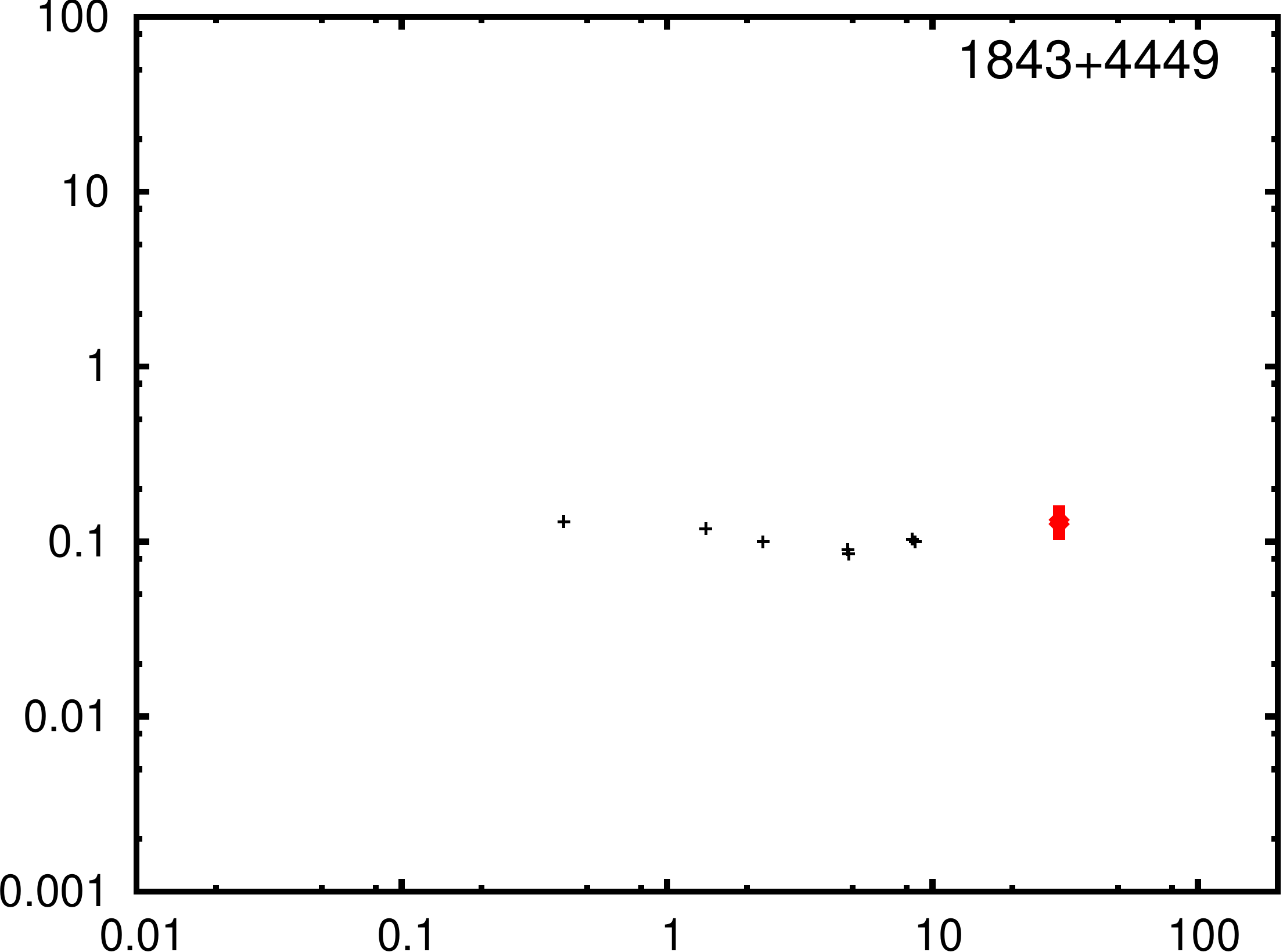}
\includegraphics[scale=0.2]{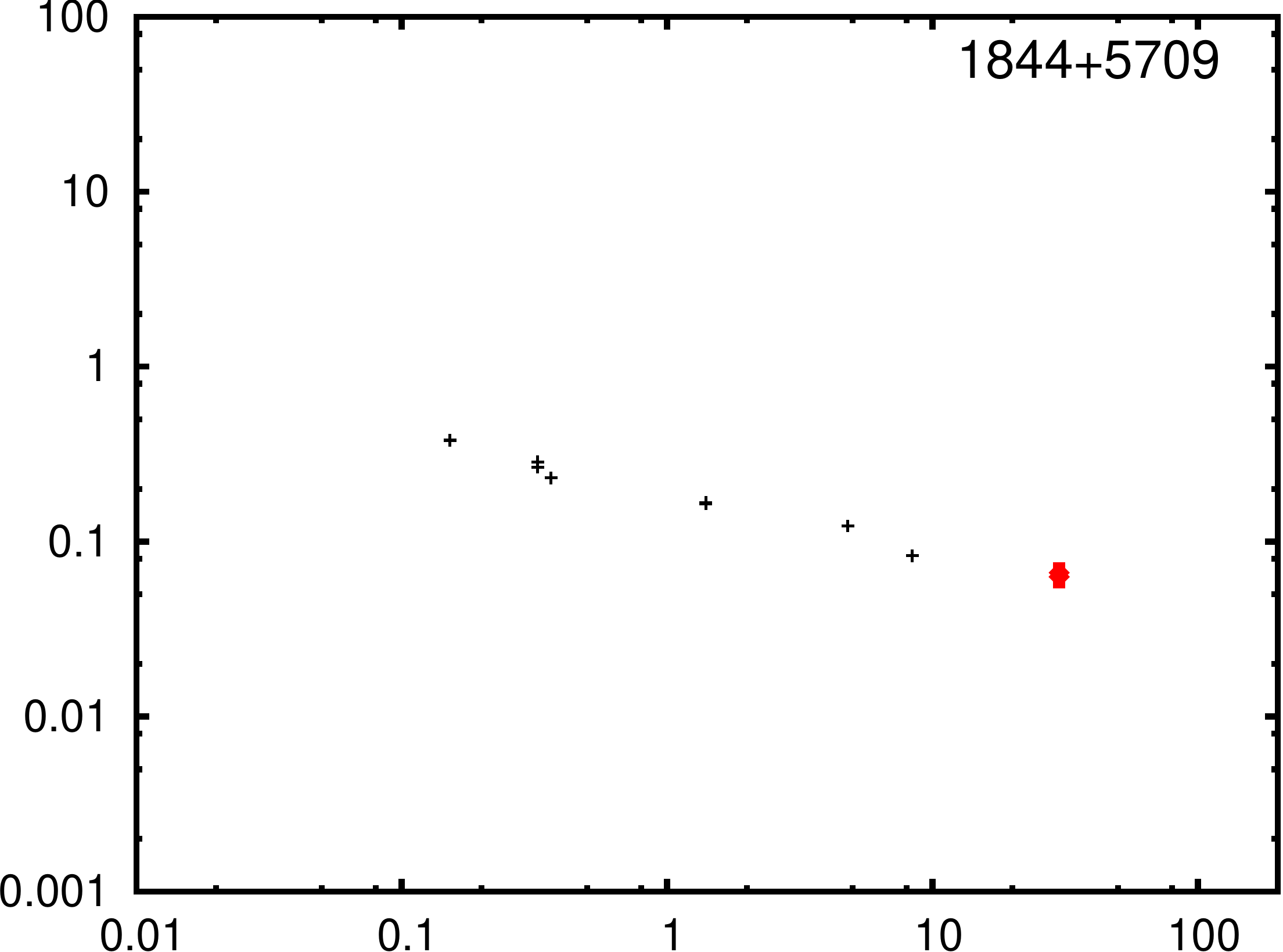}
\includegraphics[scale=0.2]{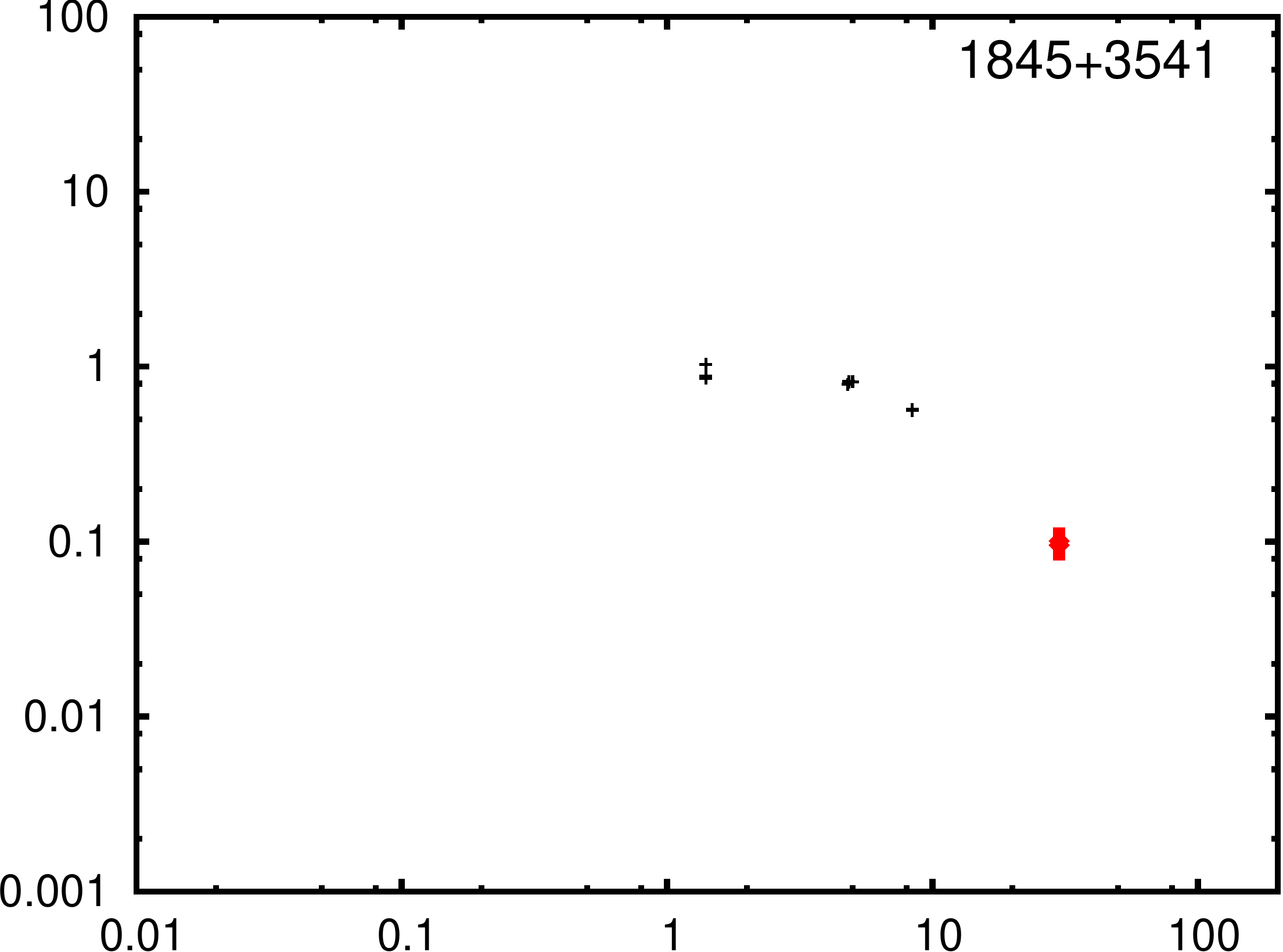}
\includegraphics[scale=0.2]{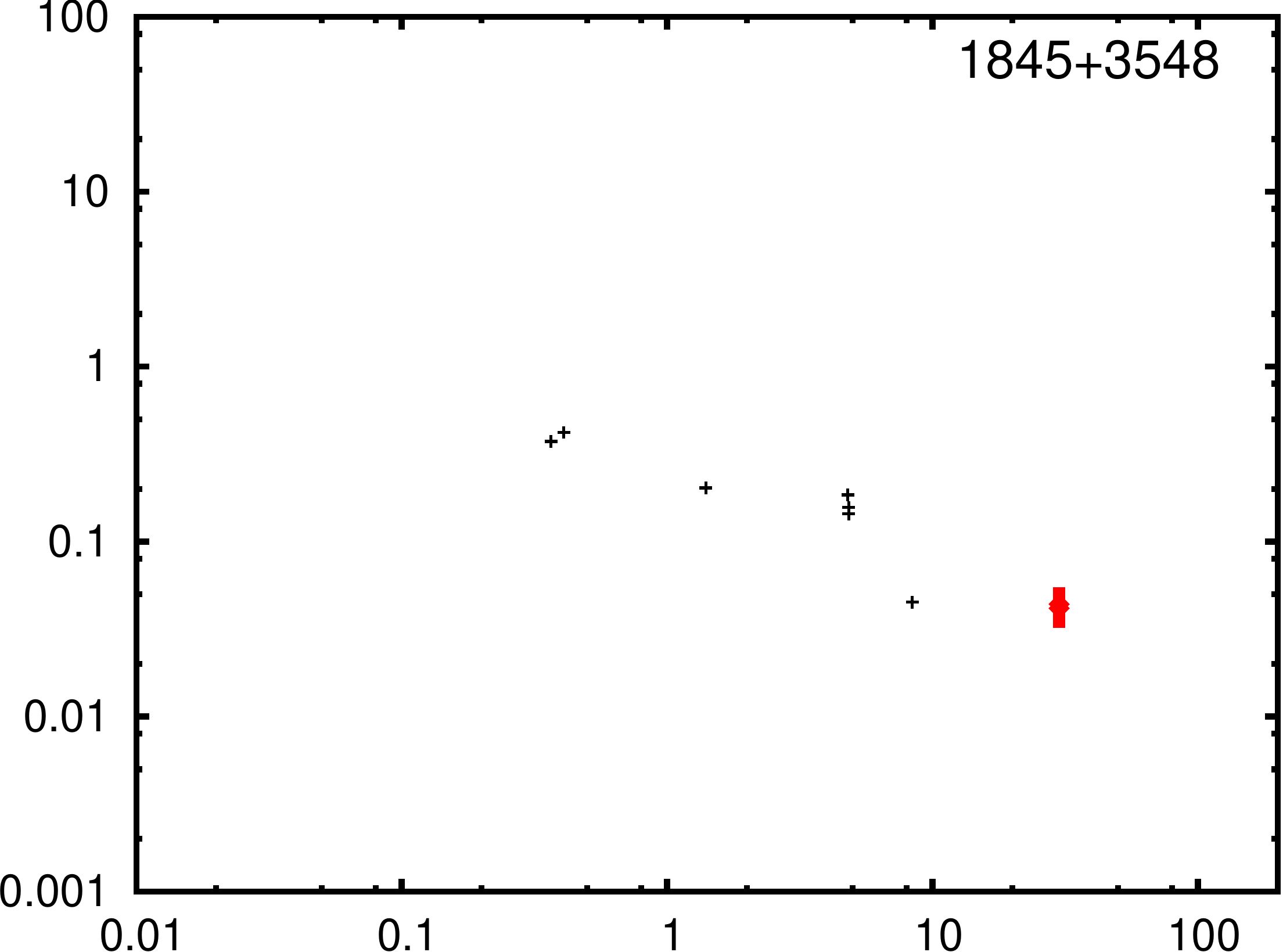}
\includegraphics[scale=0.2]{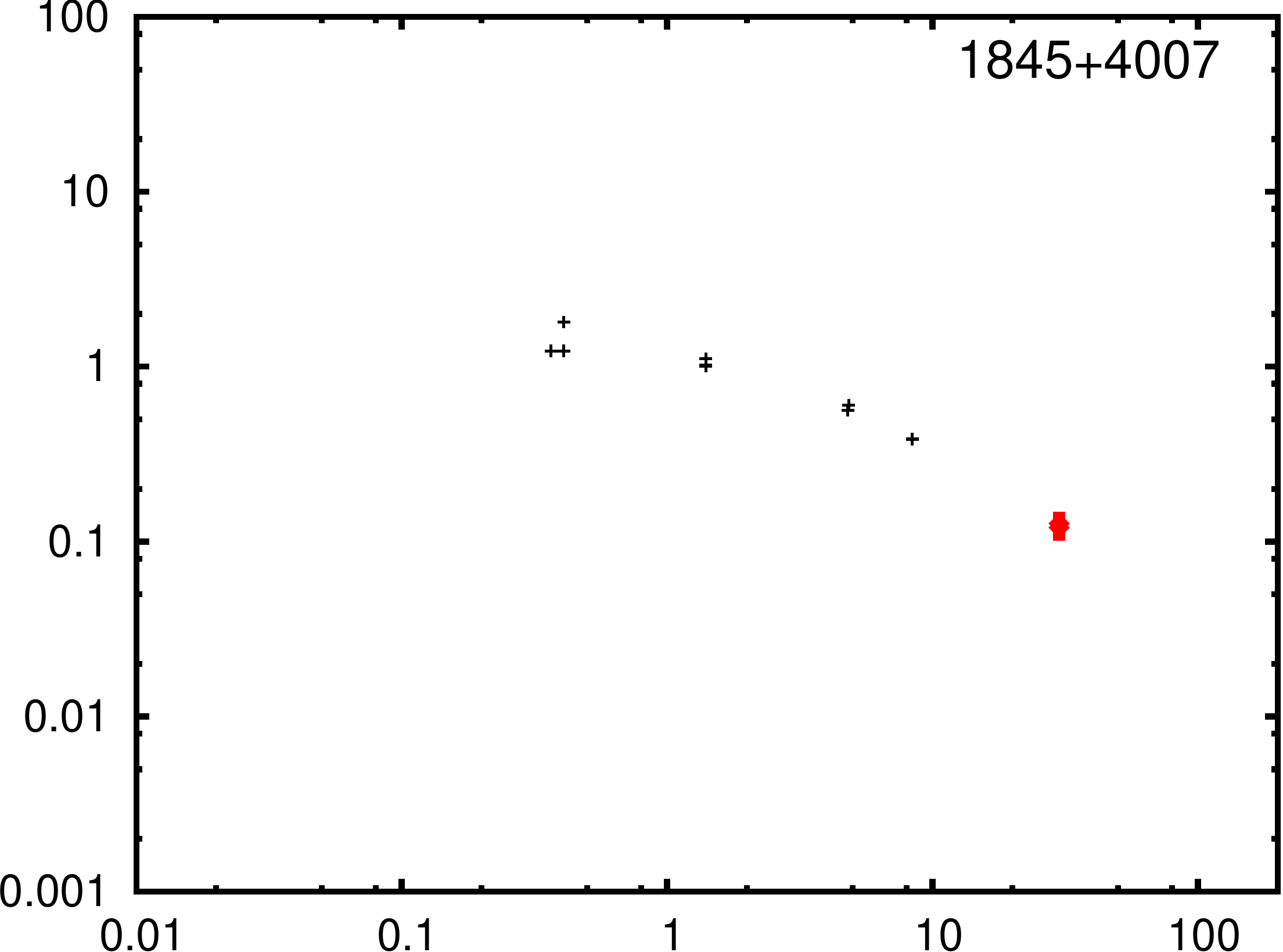}
\includegraphics[scale=0.2]{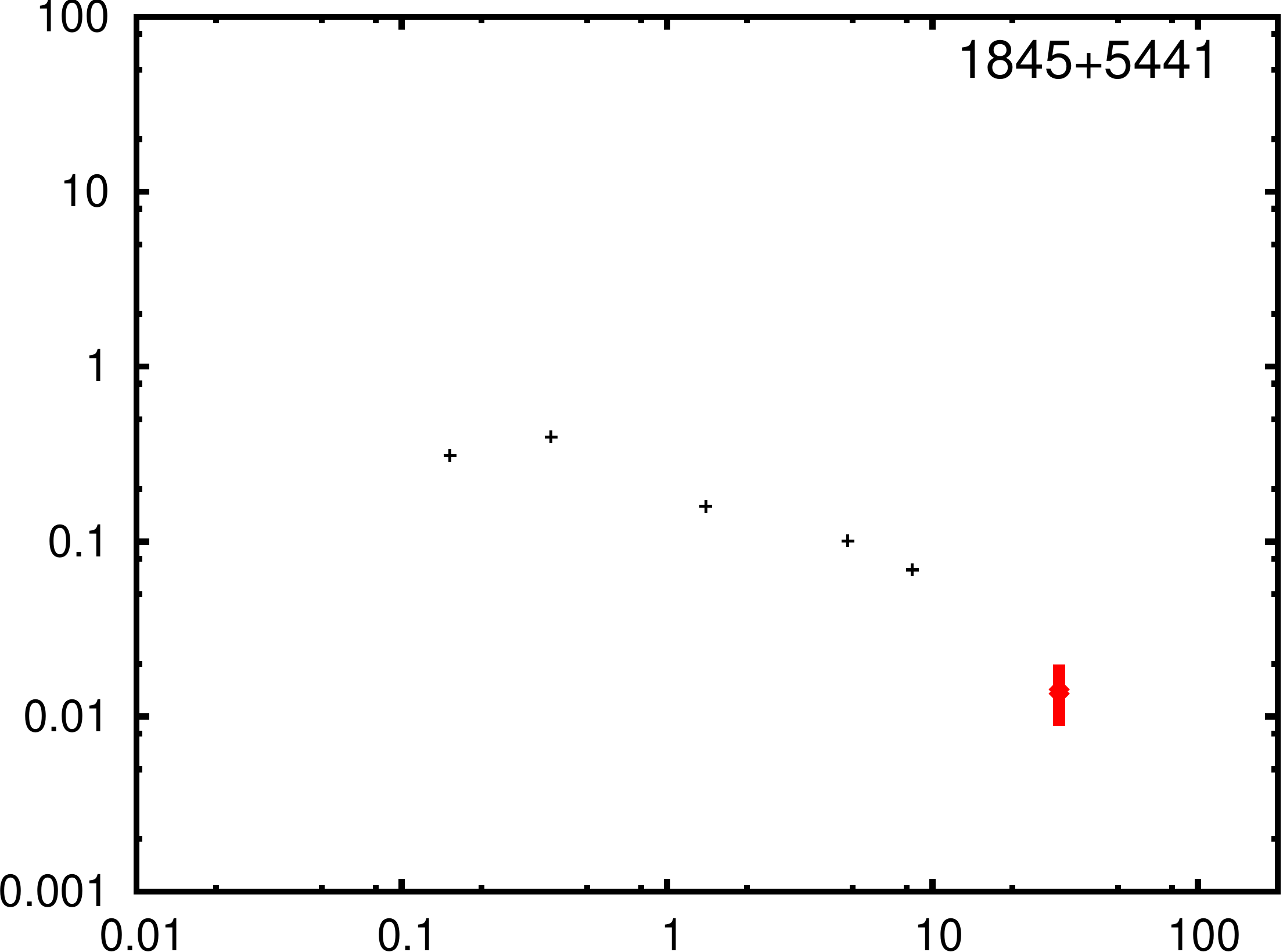}
\includegraphics[scale=0.2]{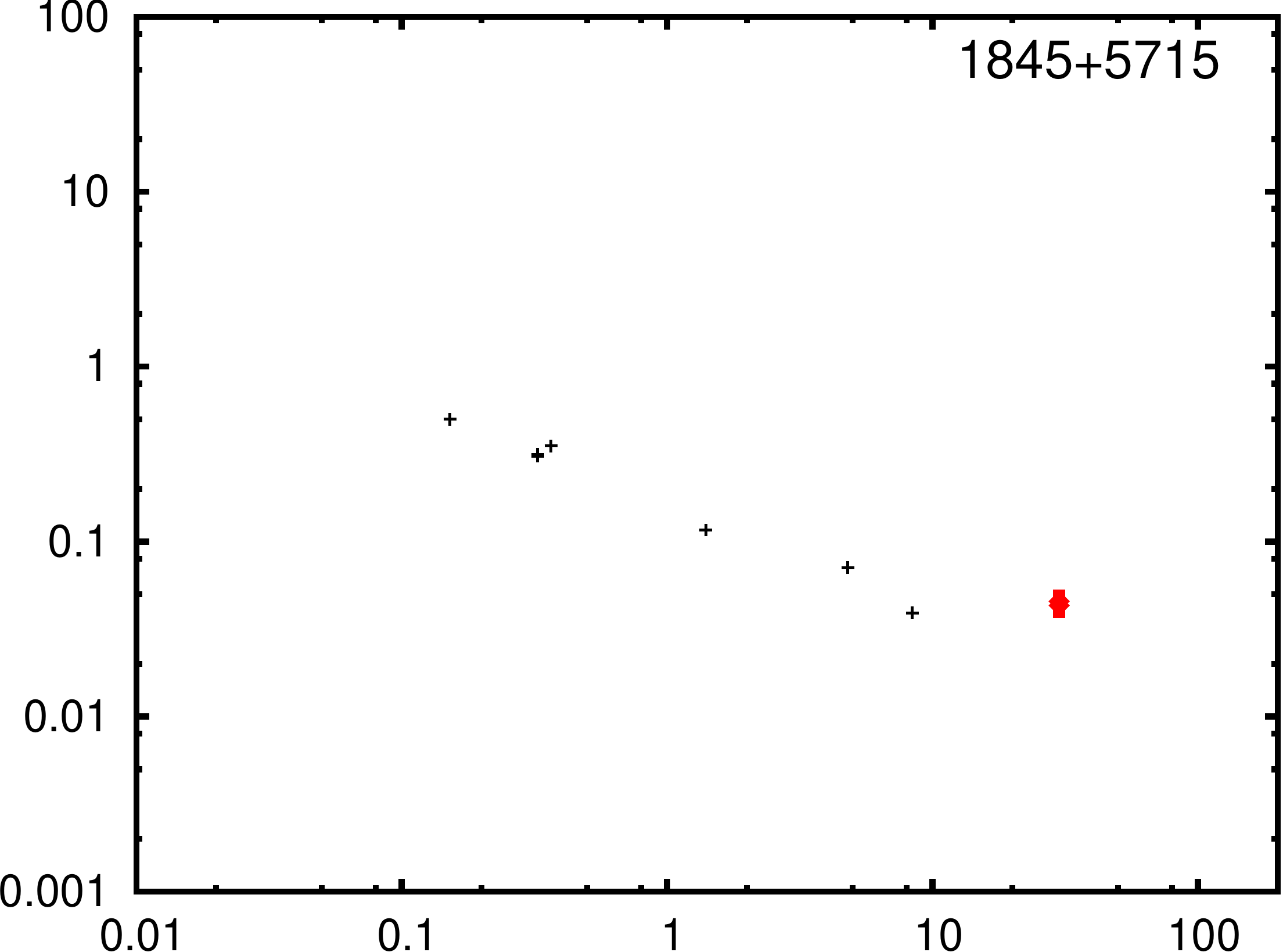}
\includegraphics[scale=0.2]{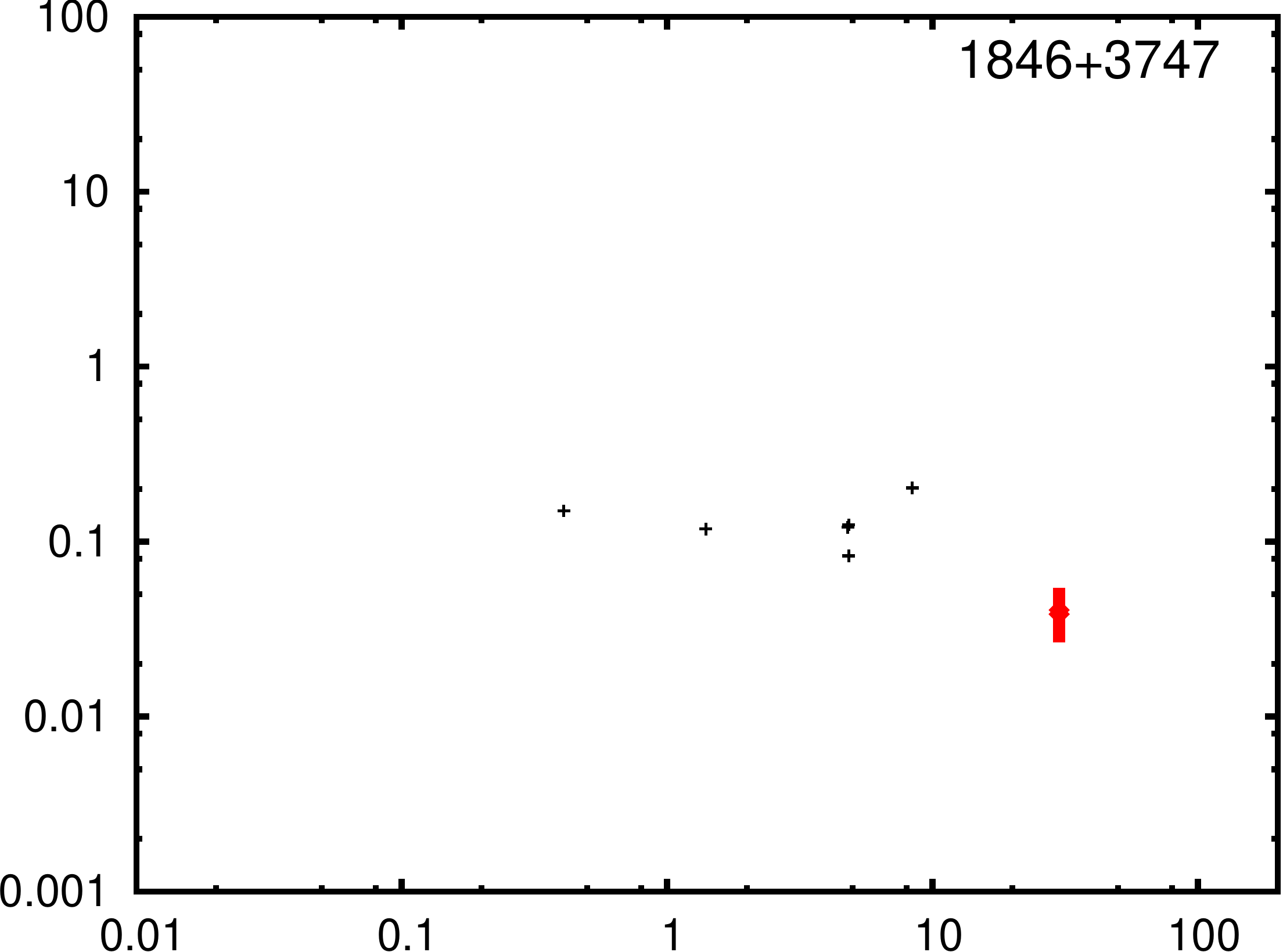}
\includegraphics[scale=0.2]{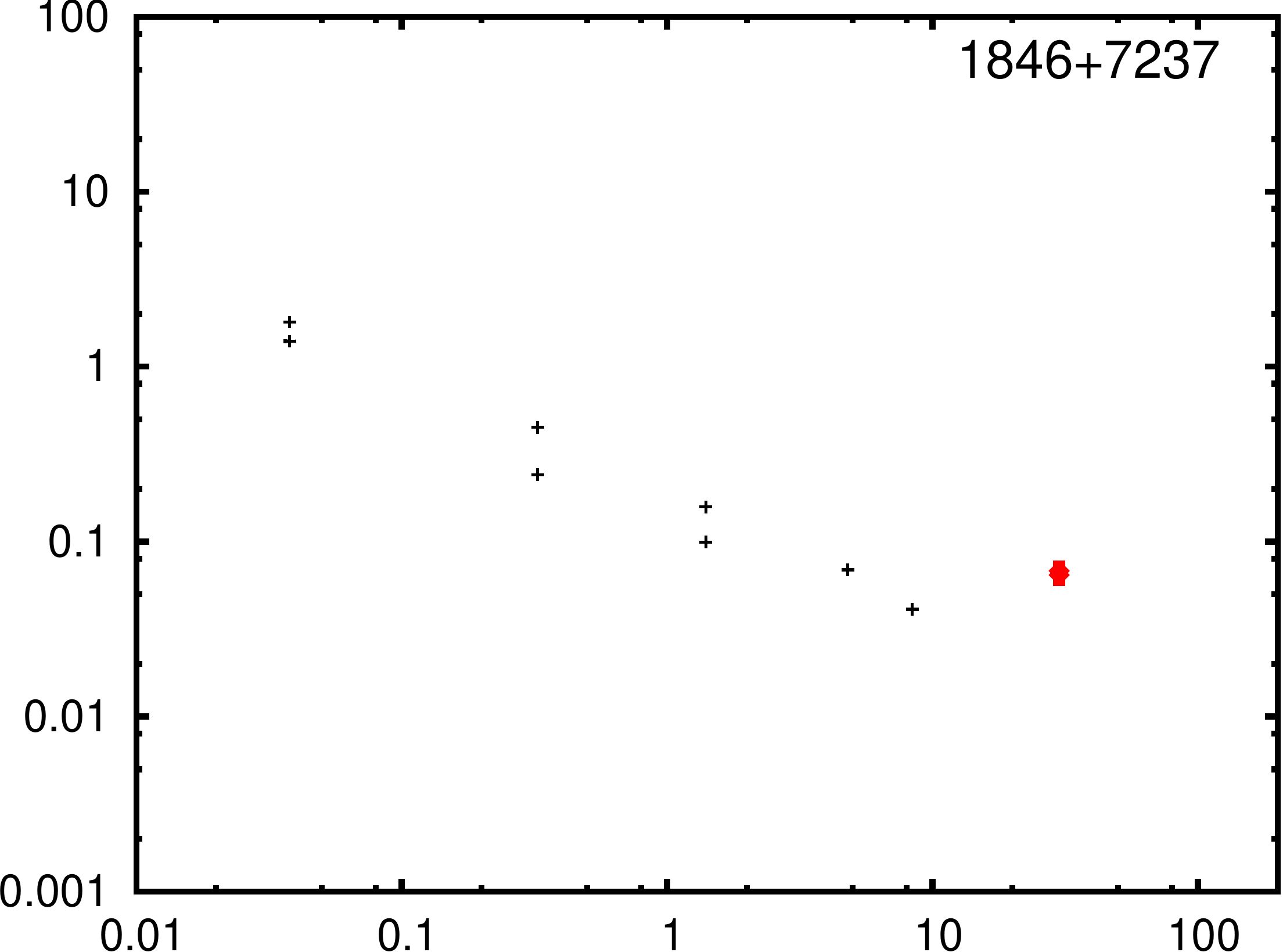}
\includegraphics[scale=0.2]{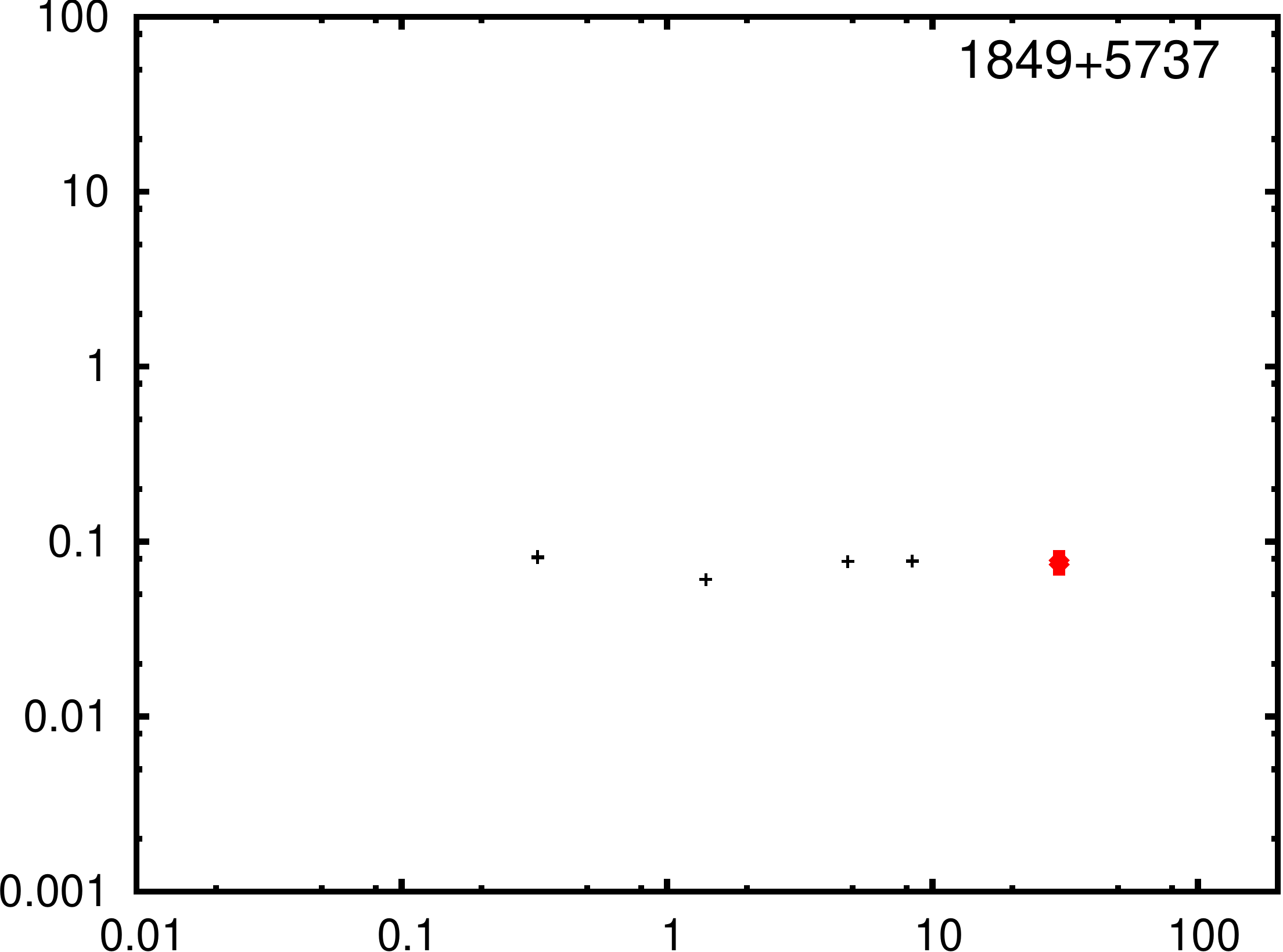}
\includegraphics[scale=0.2]{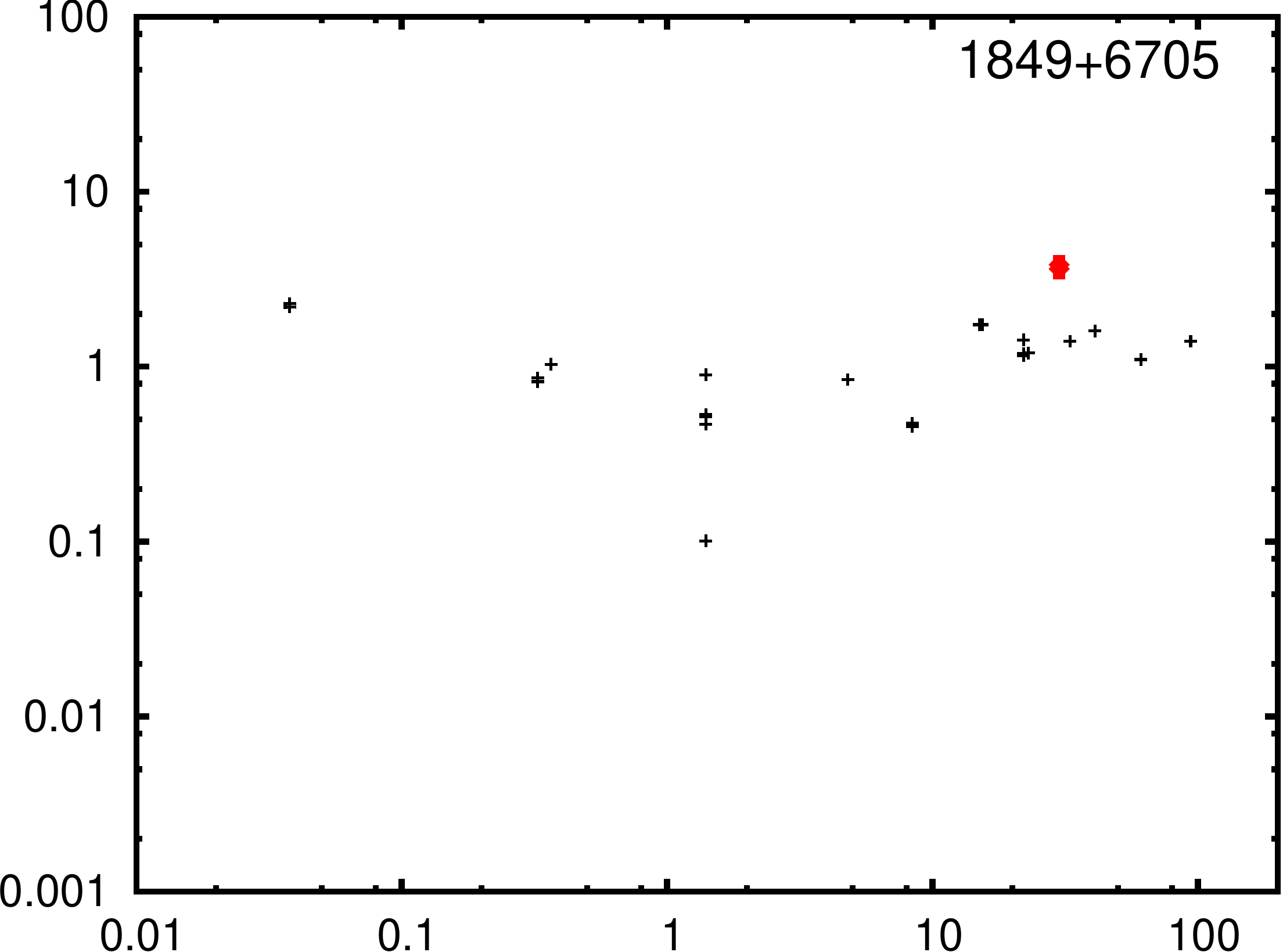}
\includegraphics[scale=0.2]{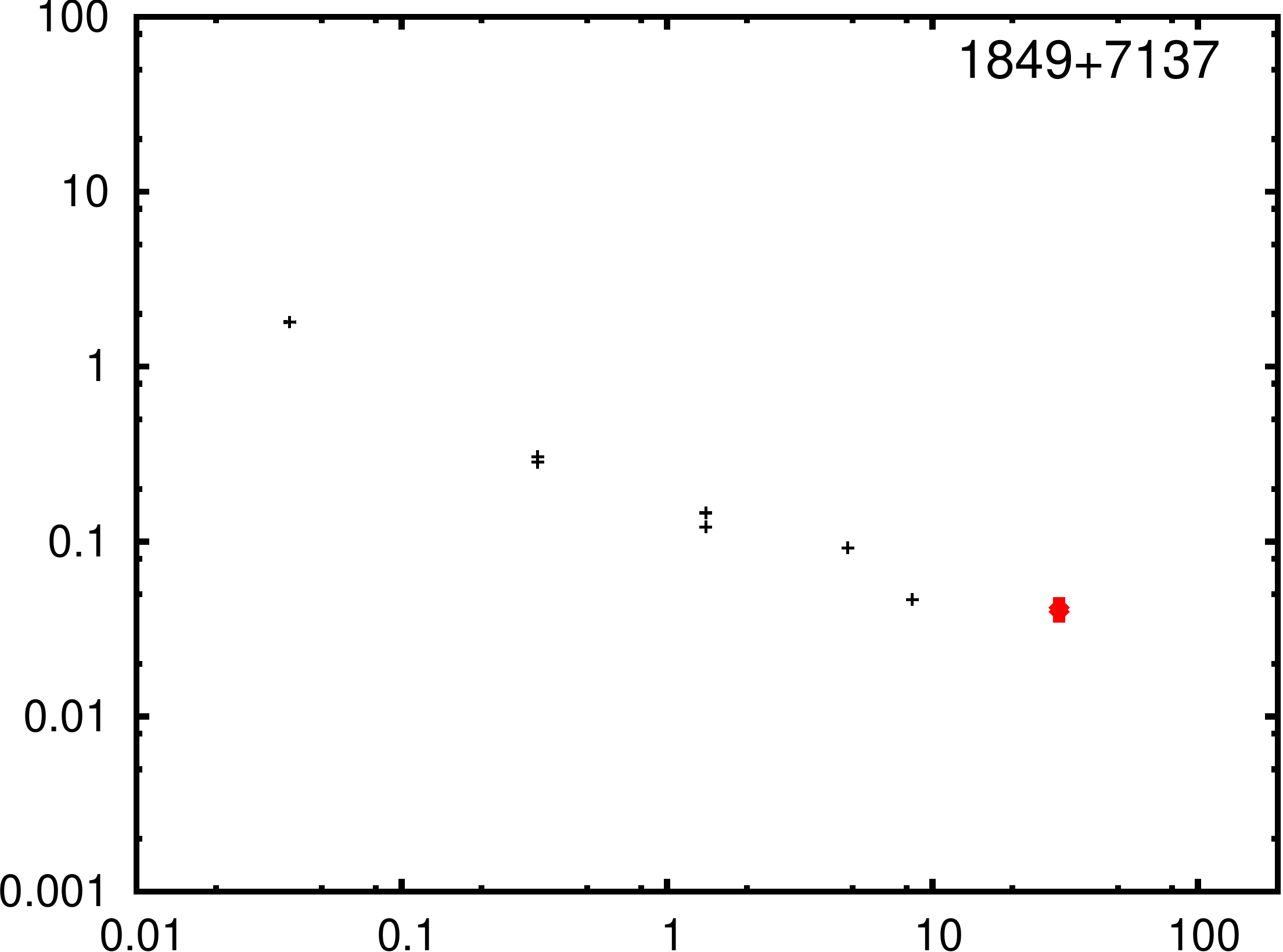}
\includegraphics[scale=0.2]{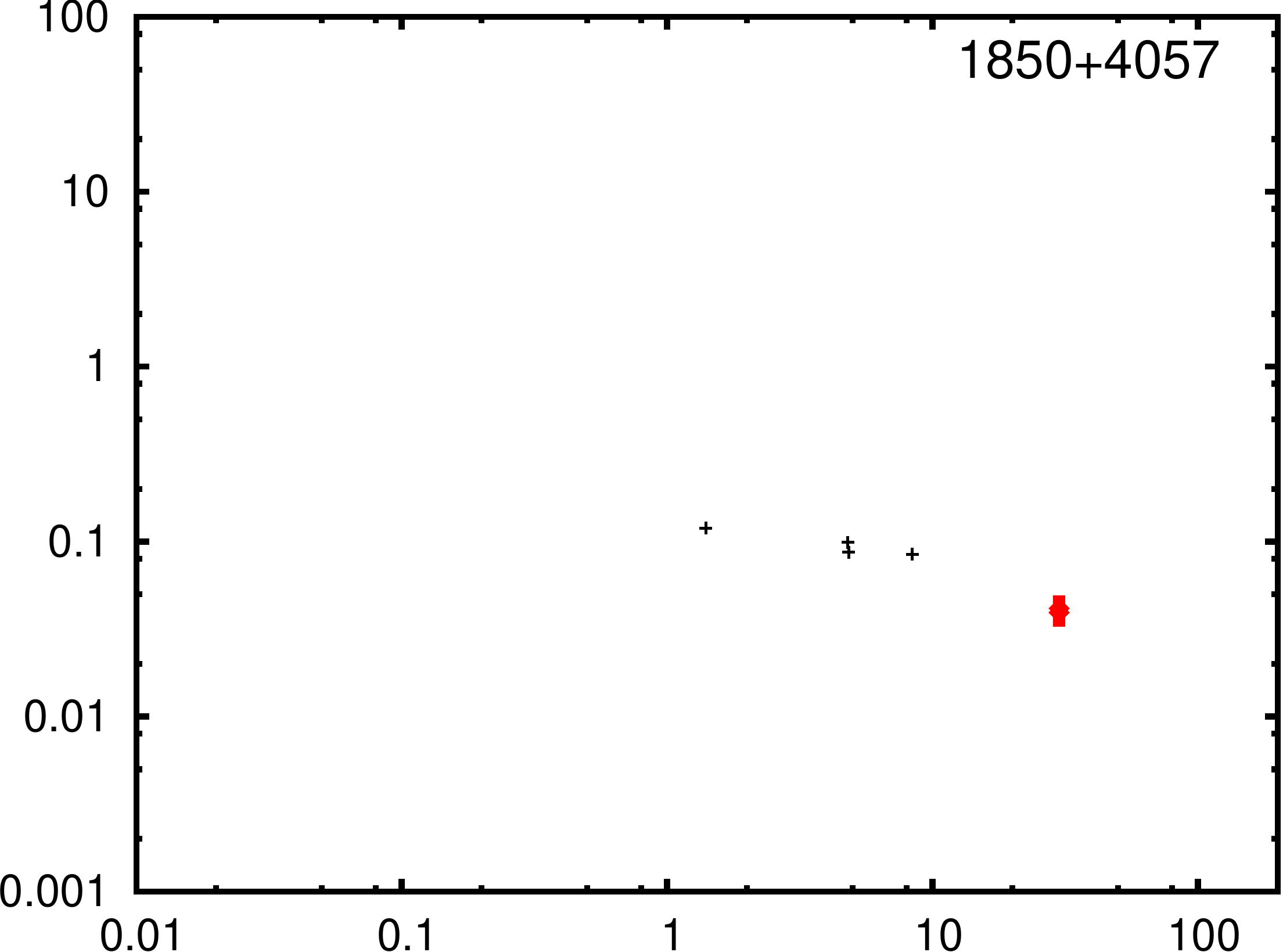}
\includegraphics[scale=0.2]{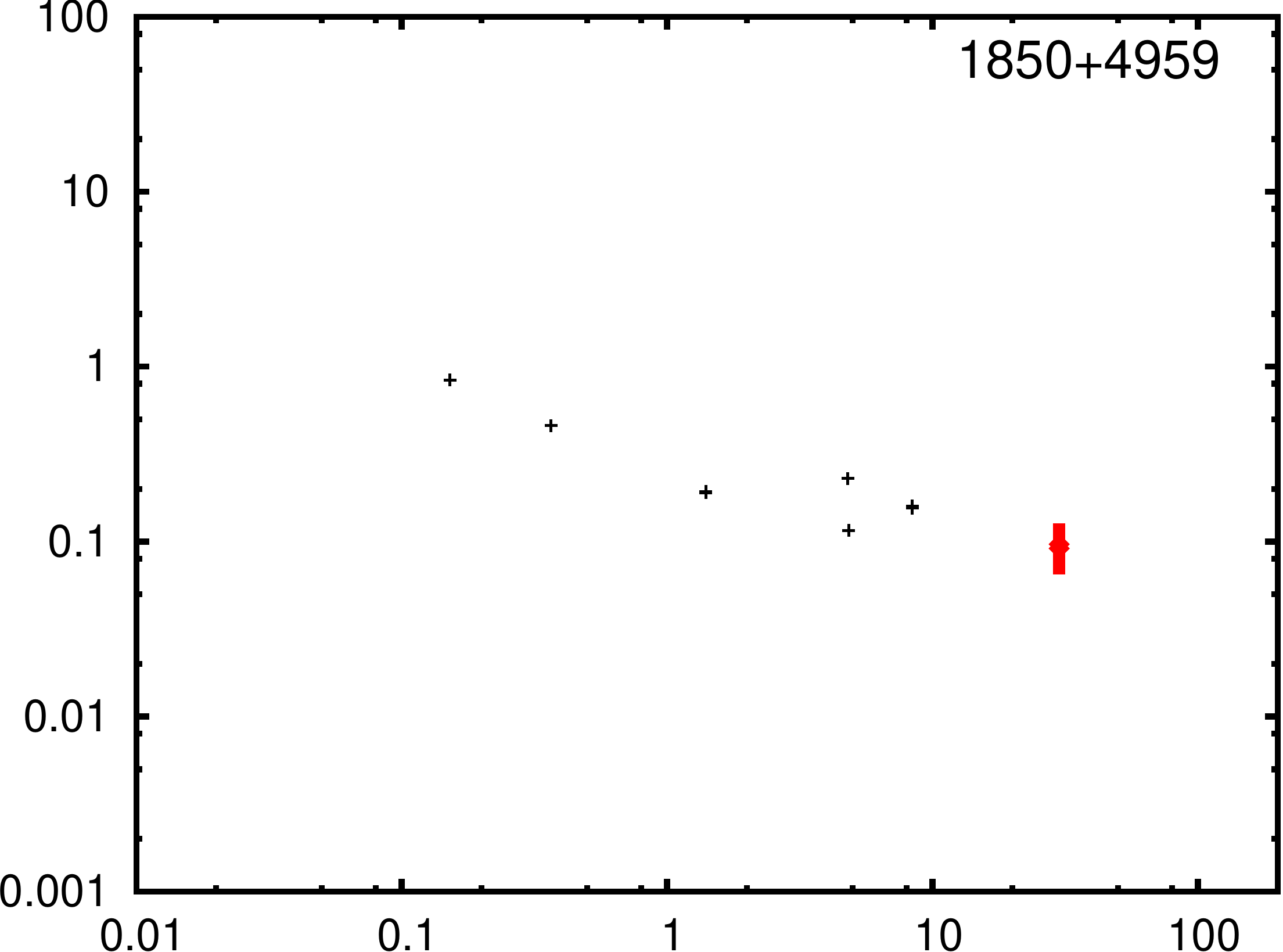}
\includegraphics[scale=0.2]{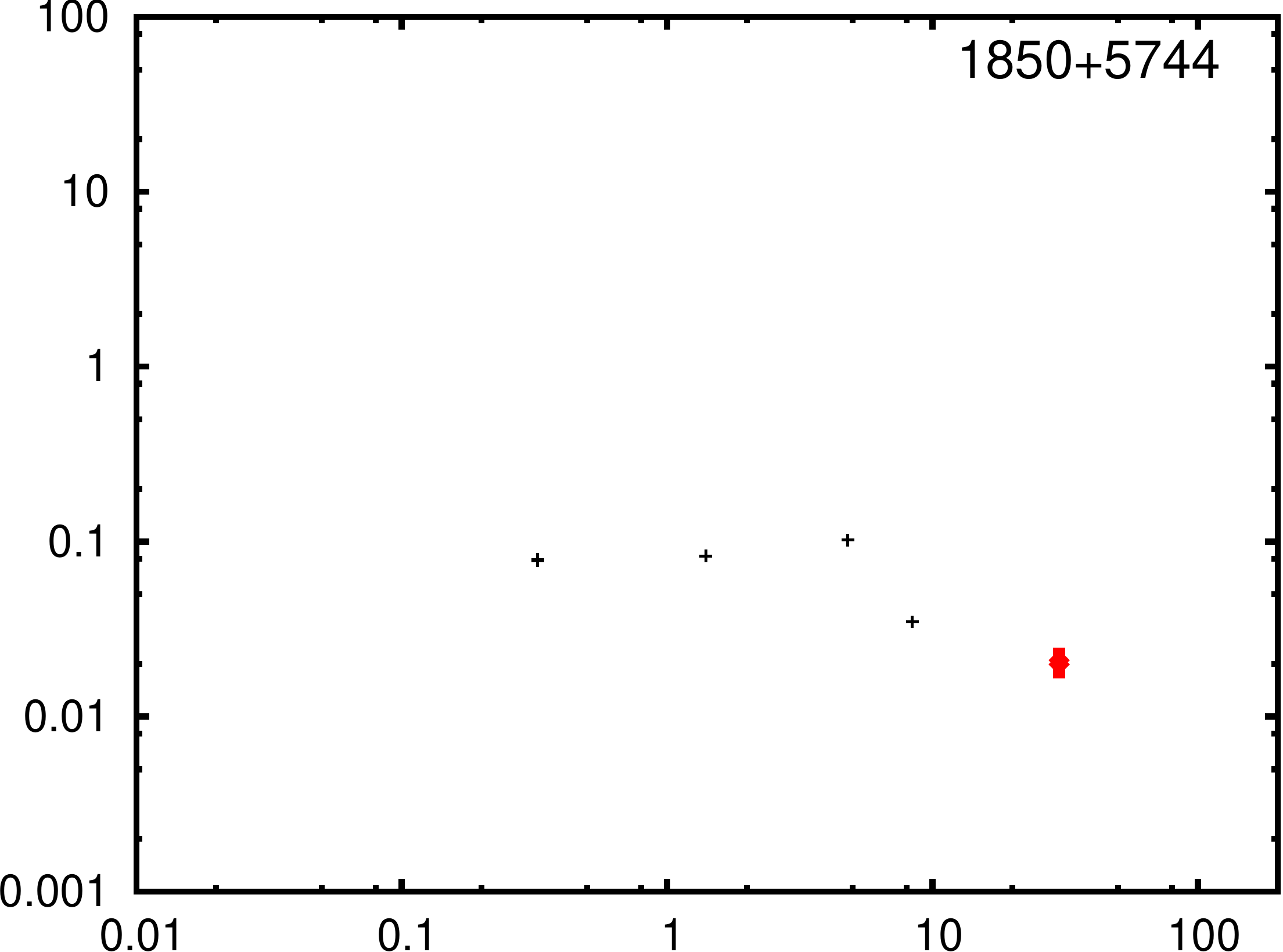}
\includegraphics[scale=0.2]{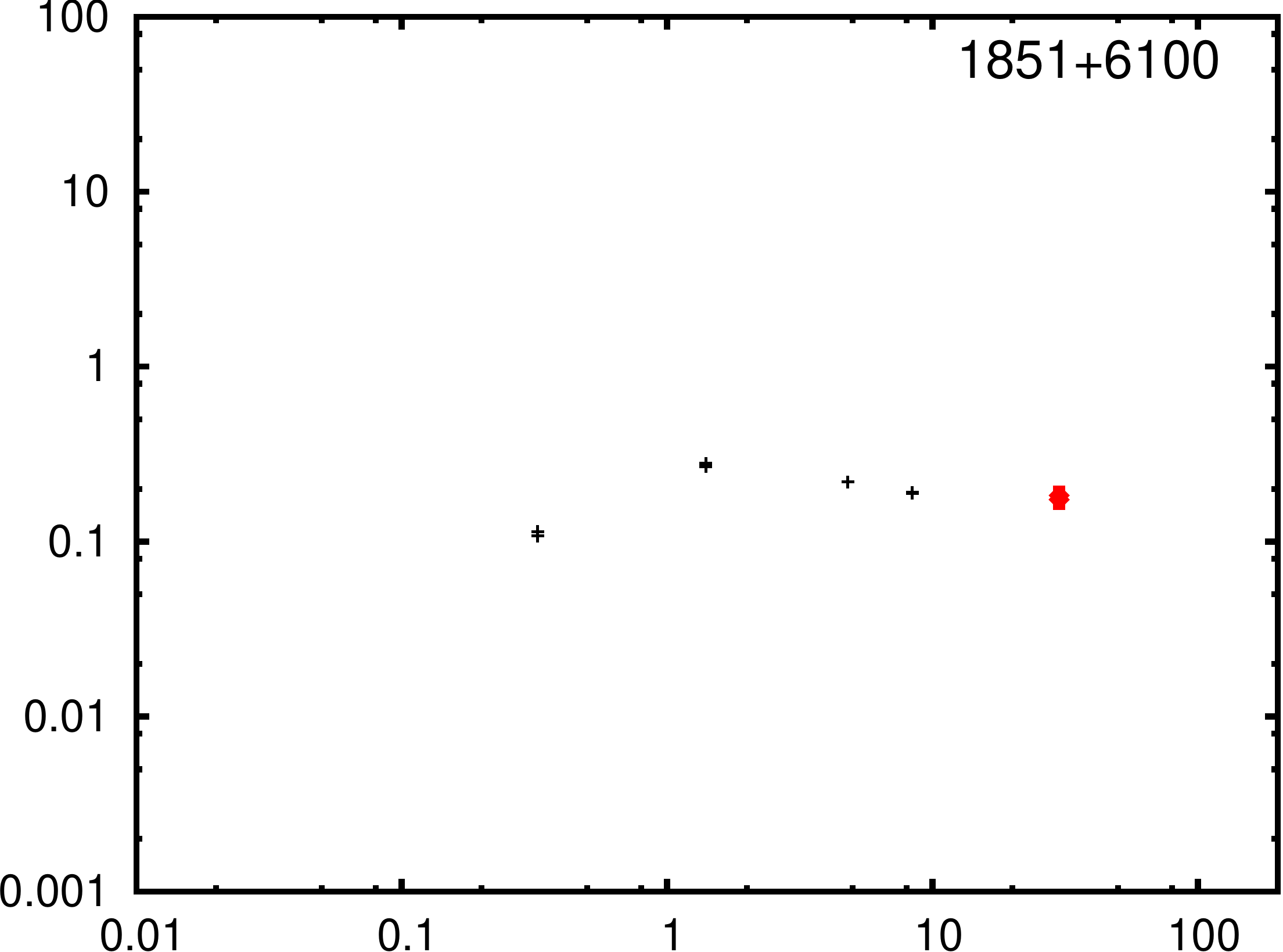}
\includegraphics[scale=0.2]{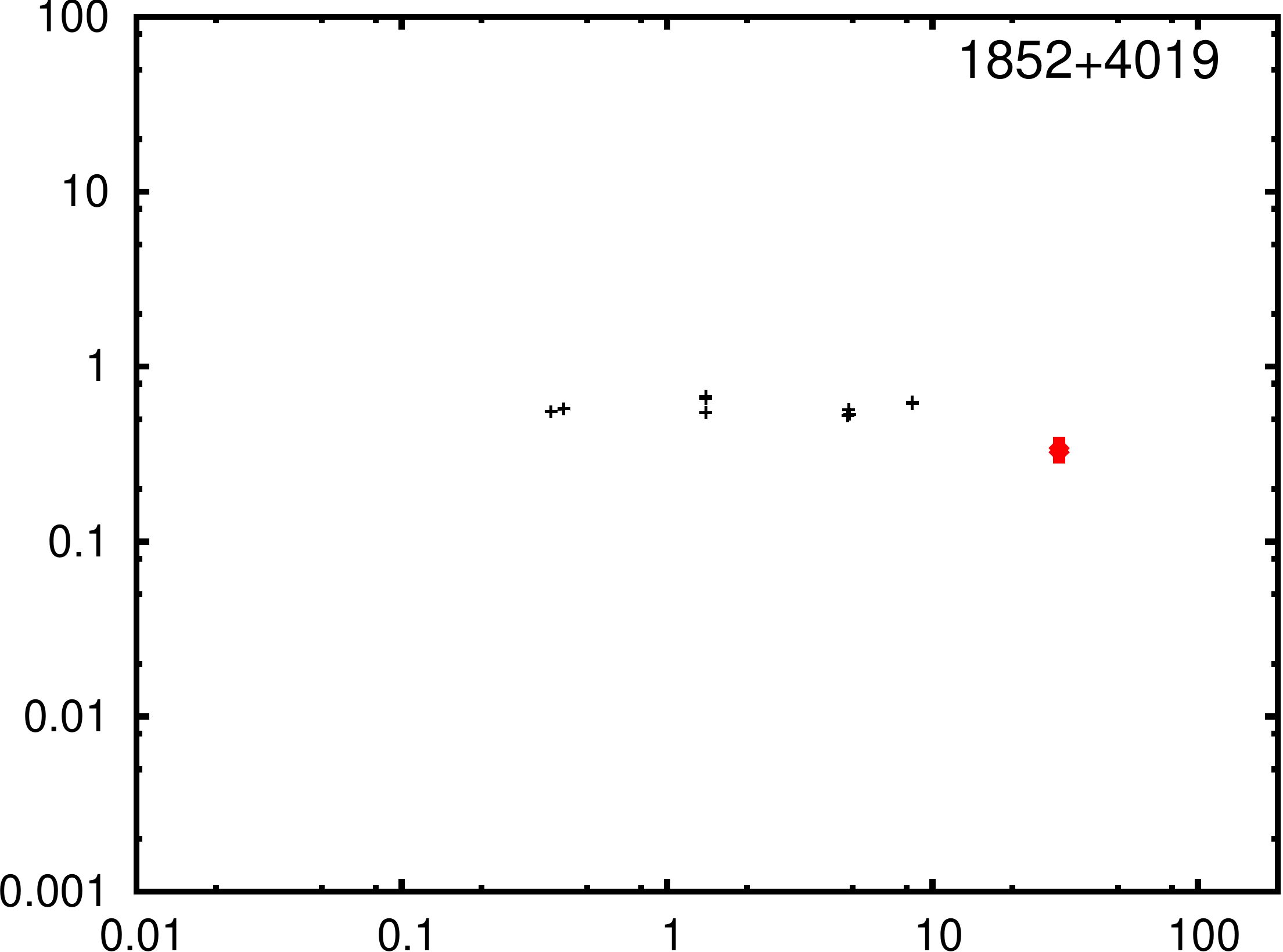}
\includegraphics[scale=0.2]{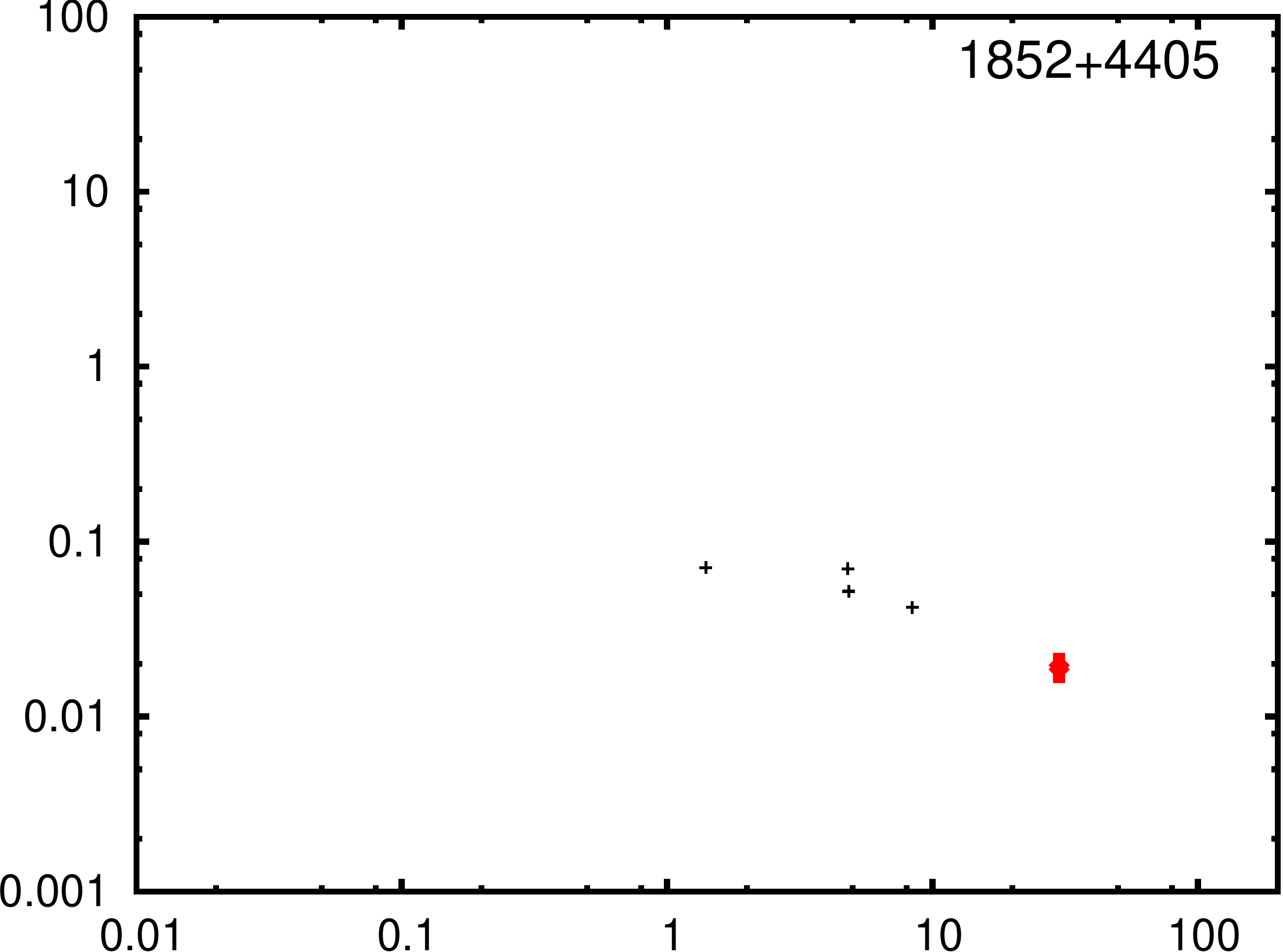}
\end{figure}
\clearpage\begin{figure}
\centering
\includegraphics[scale=0.2]{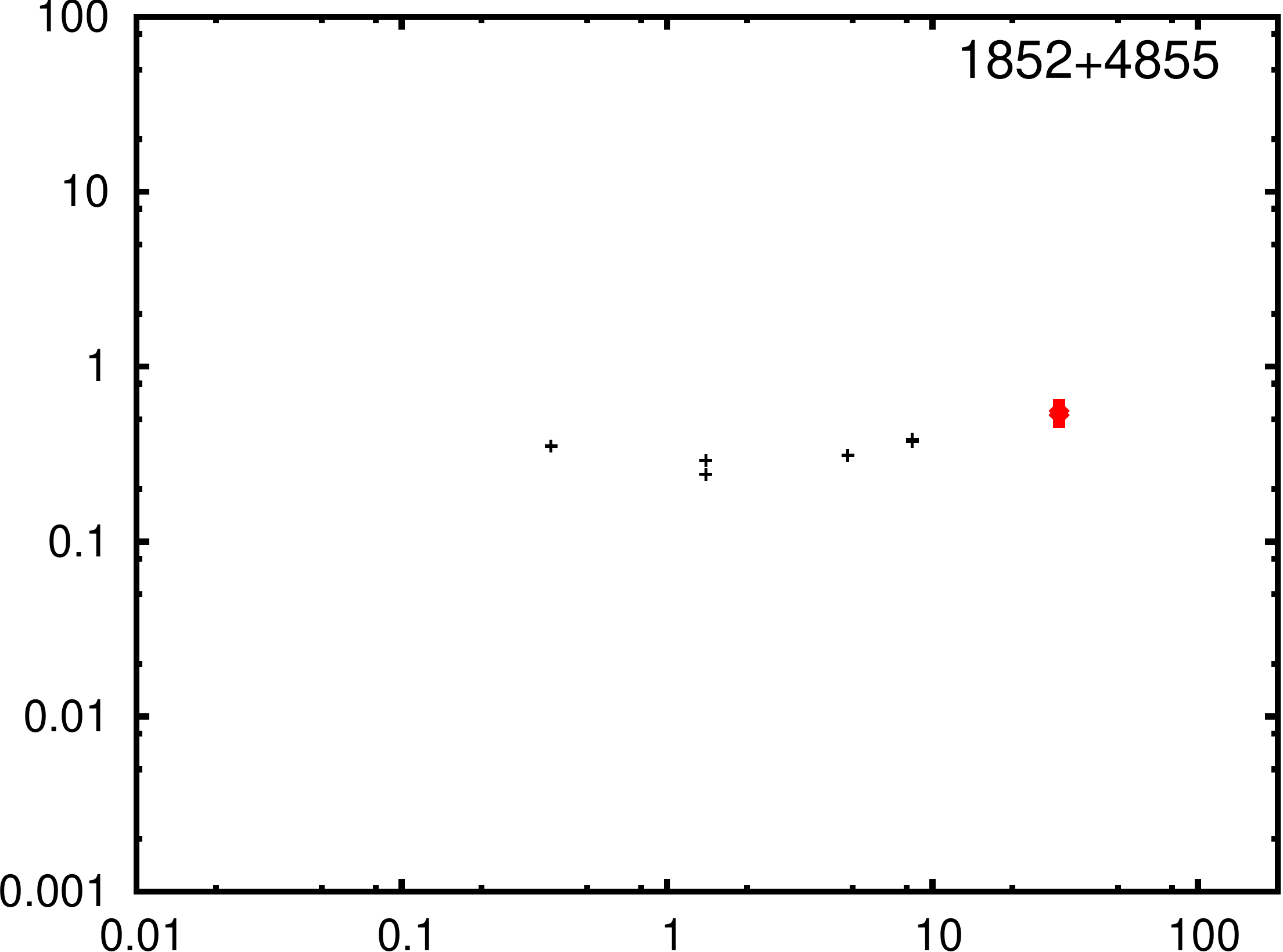}
\includegraphics[scale=0.2]{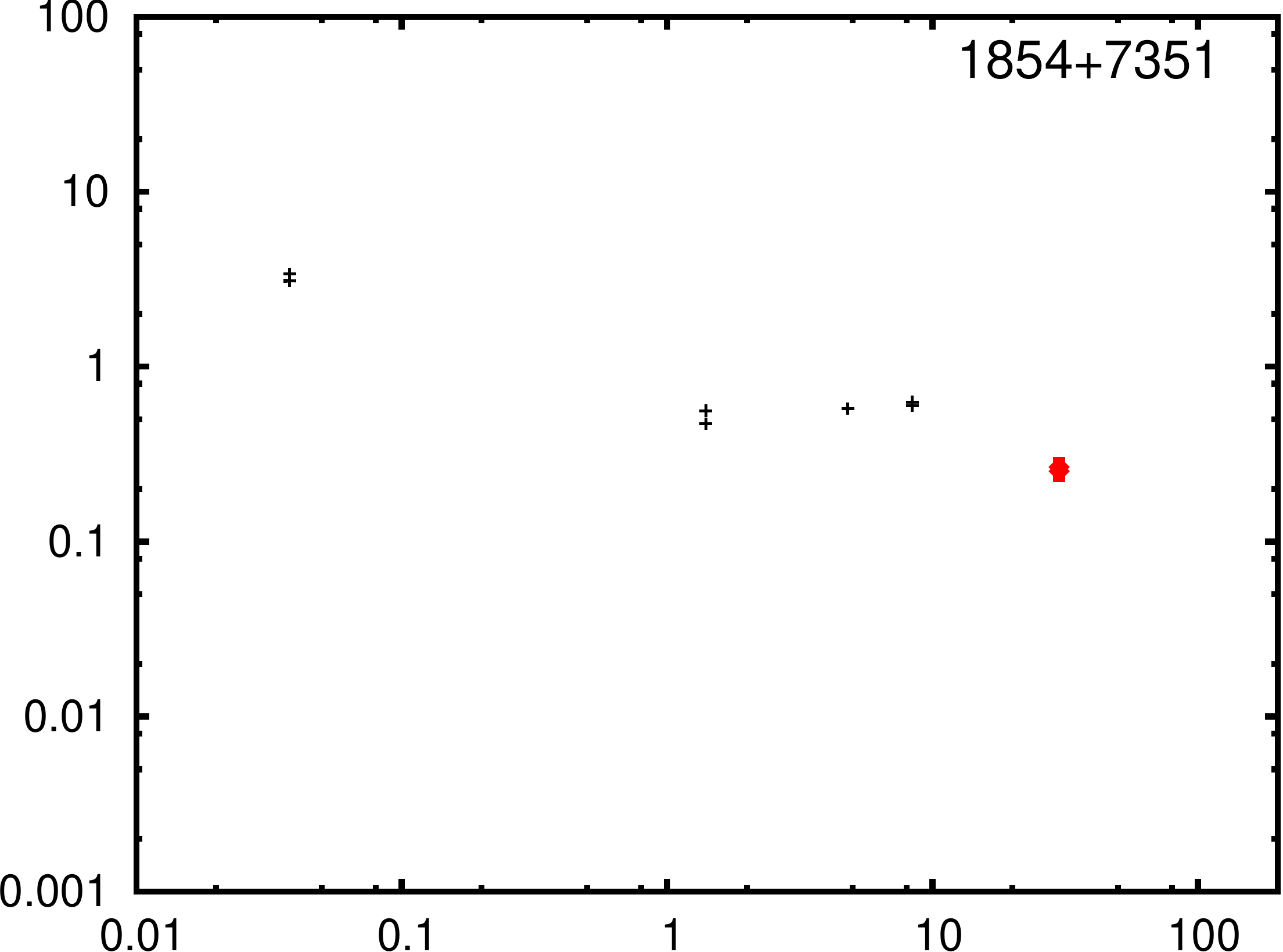}
\includegraphics[scale=0.2]{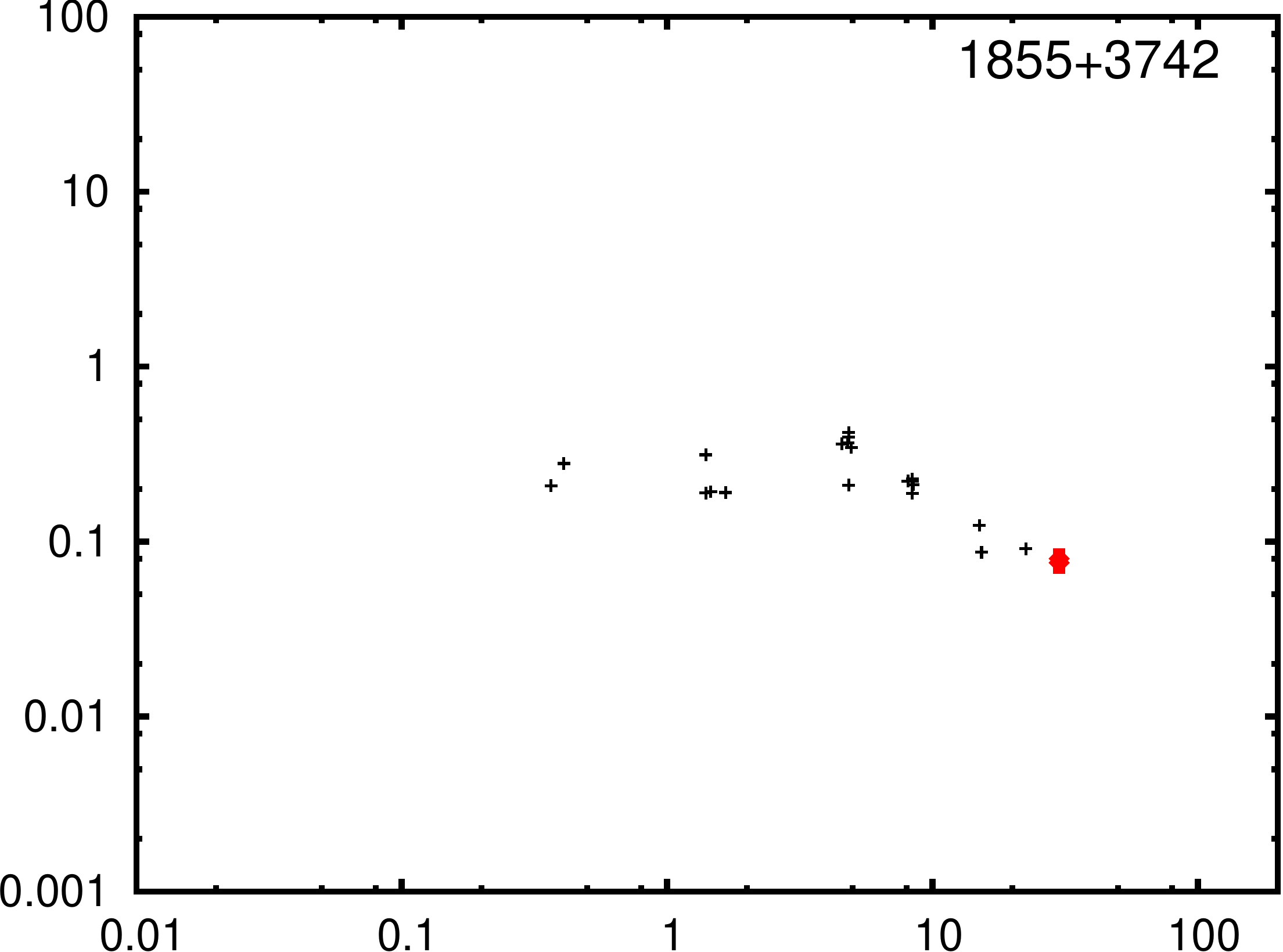}
\includegraphics[scale=0.2]{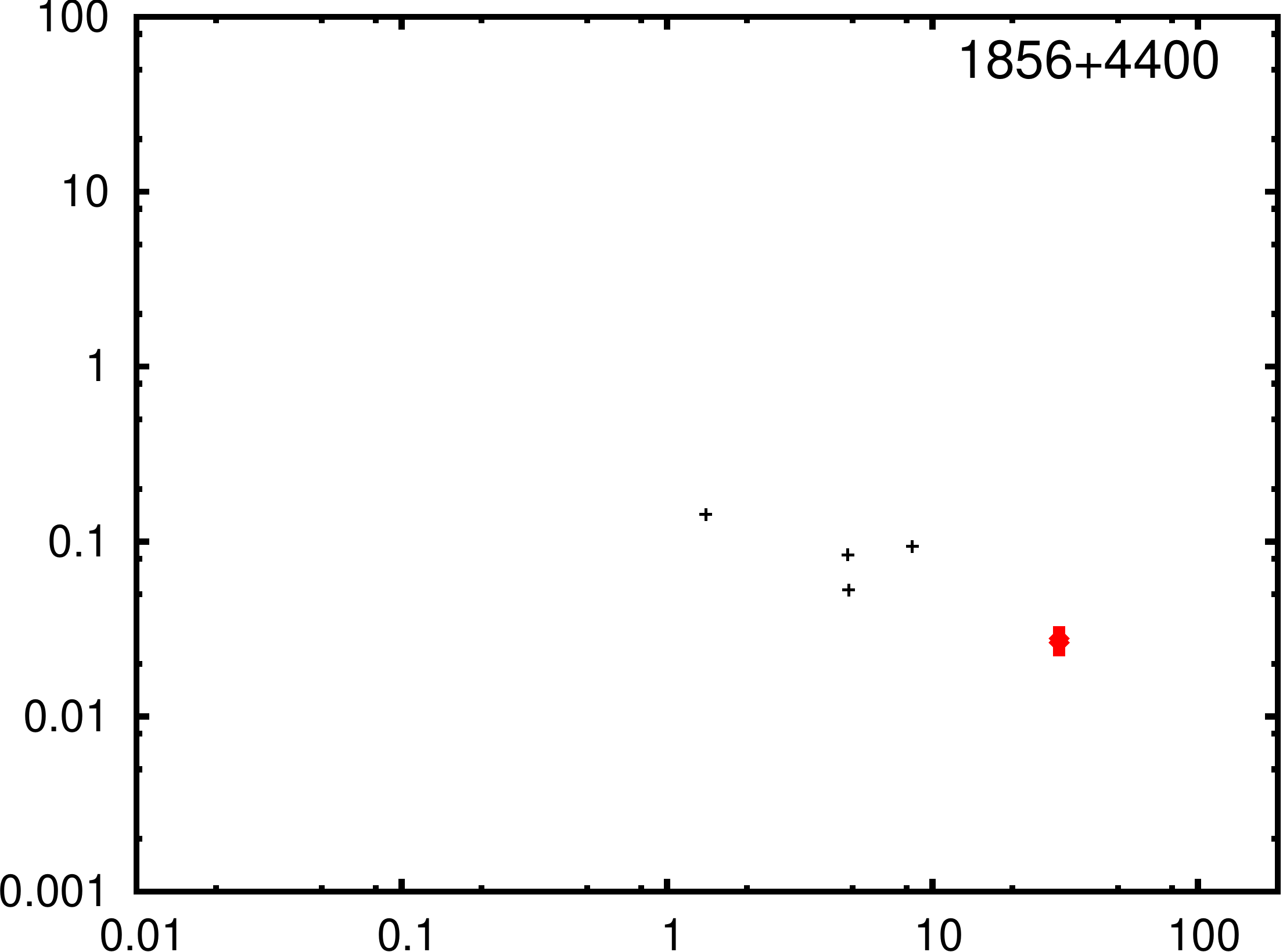}
\includegraphics[scale=0.2]{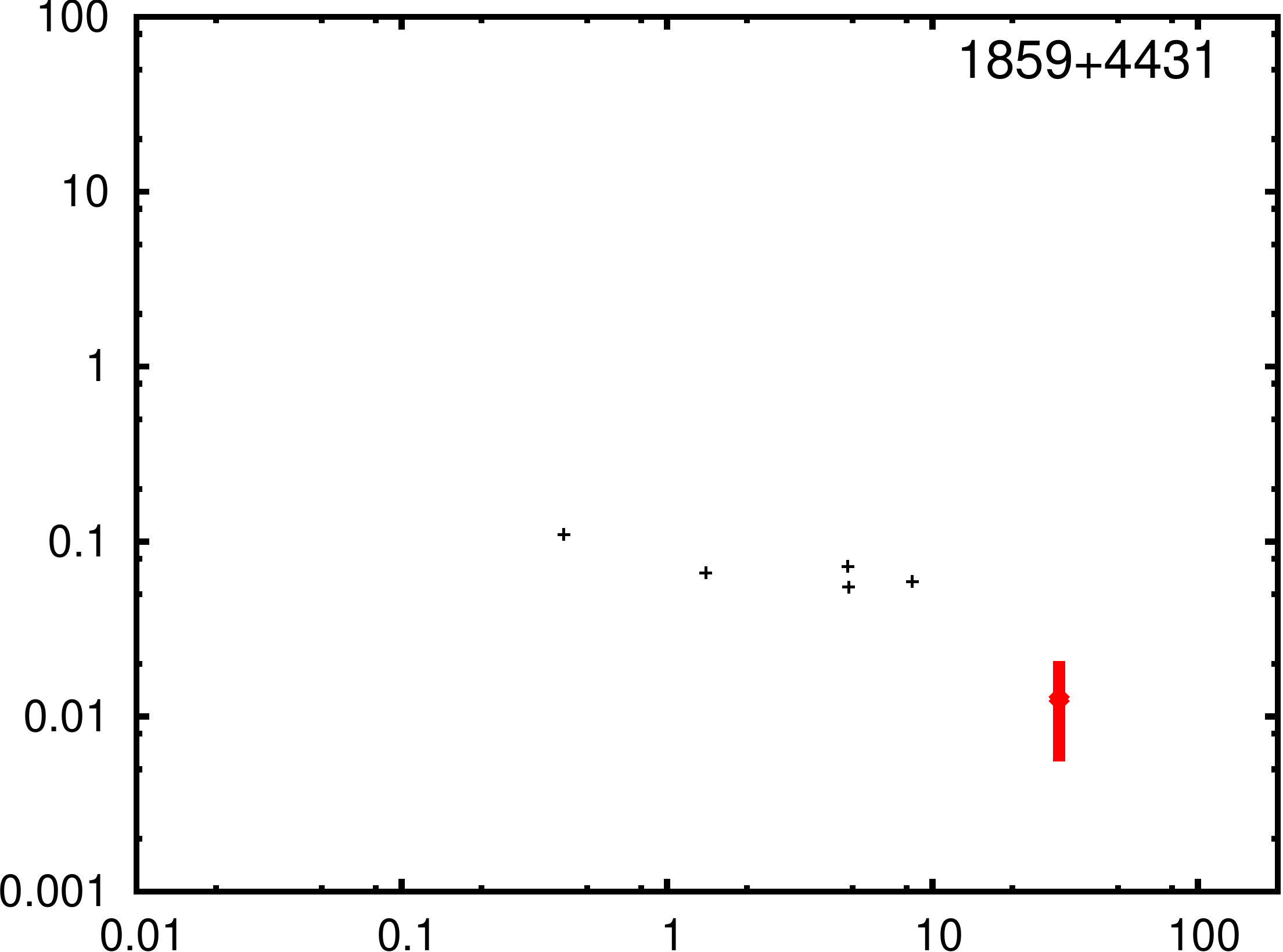}
\includegraphics[scale=0.2]{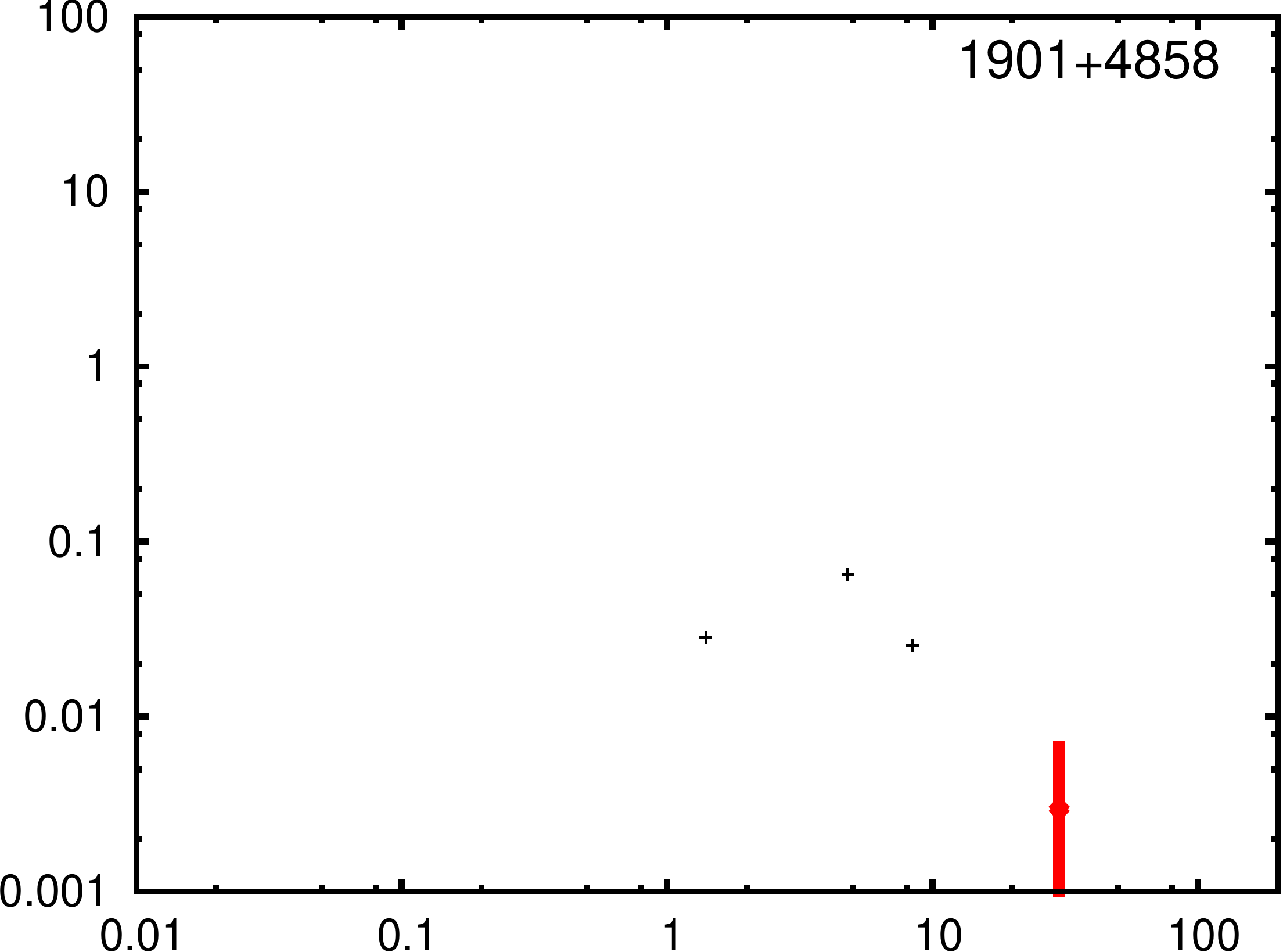}
\includegraphics[scale=0.2]{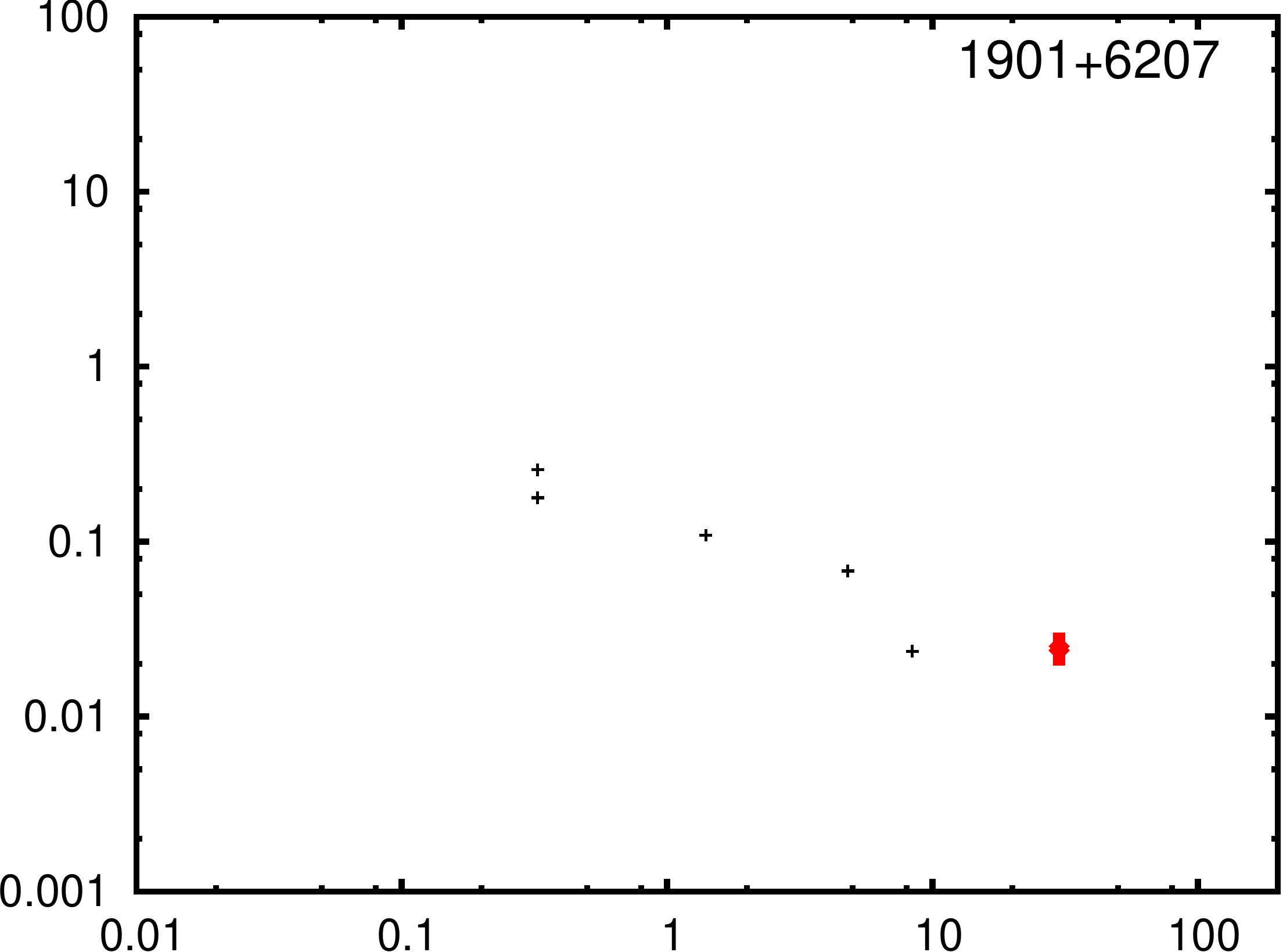}
\includegraphics[scale=0.2]{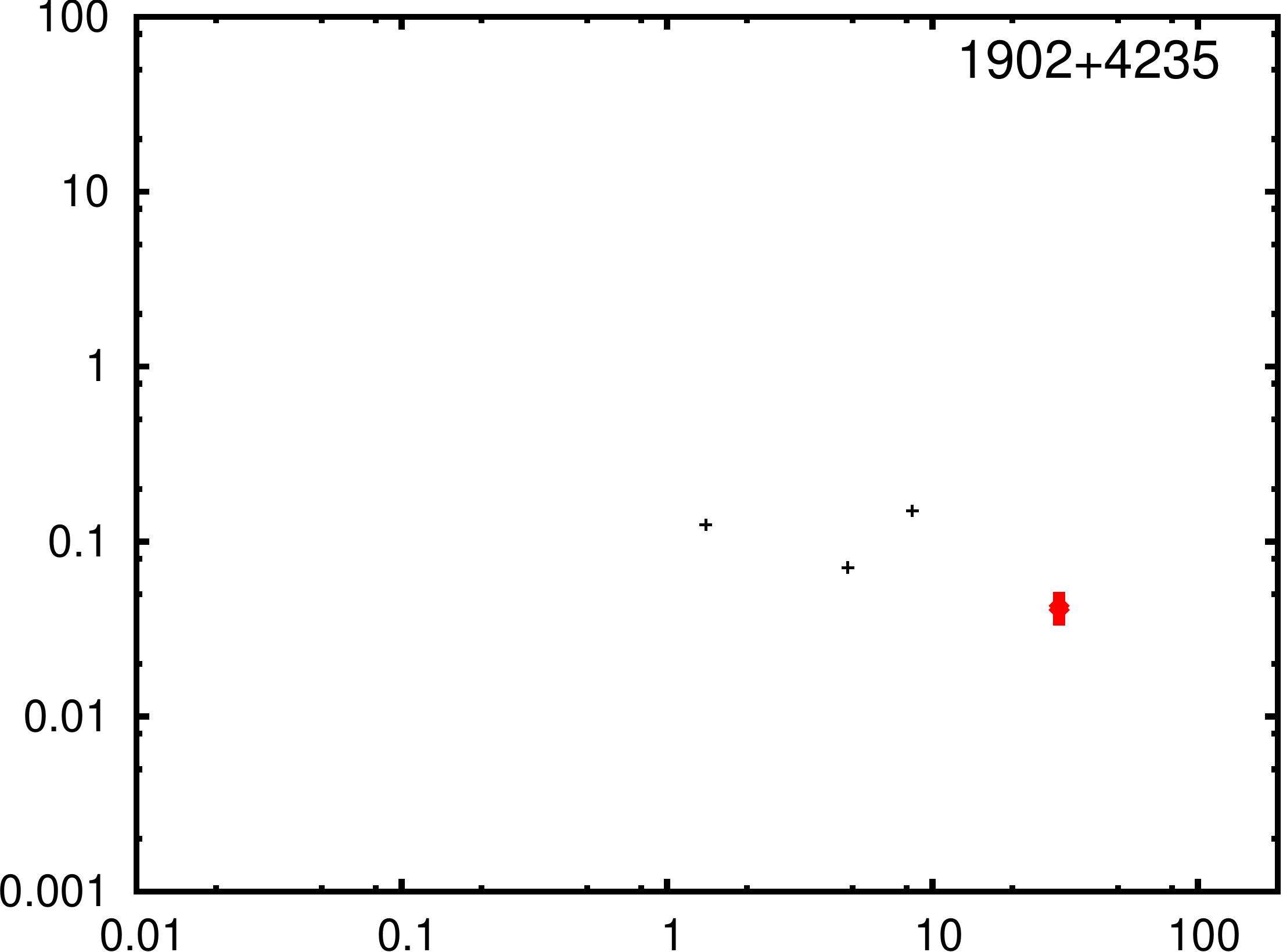}
\includegraphics[scale=0.2]{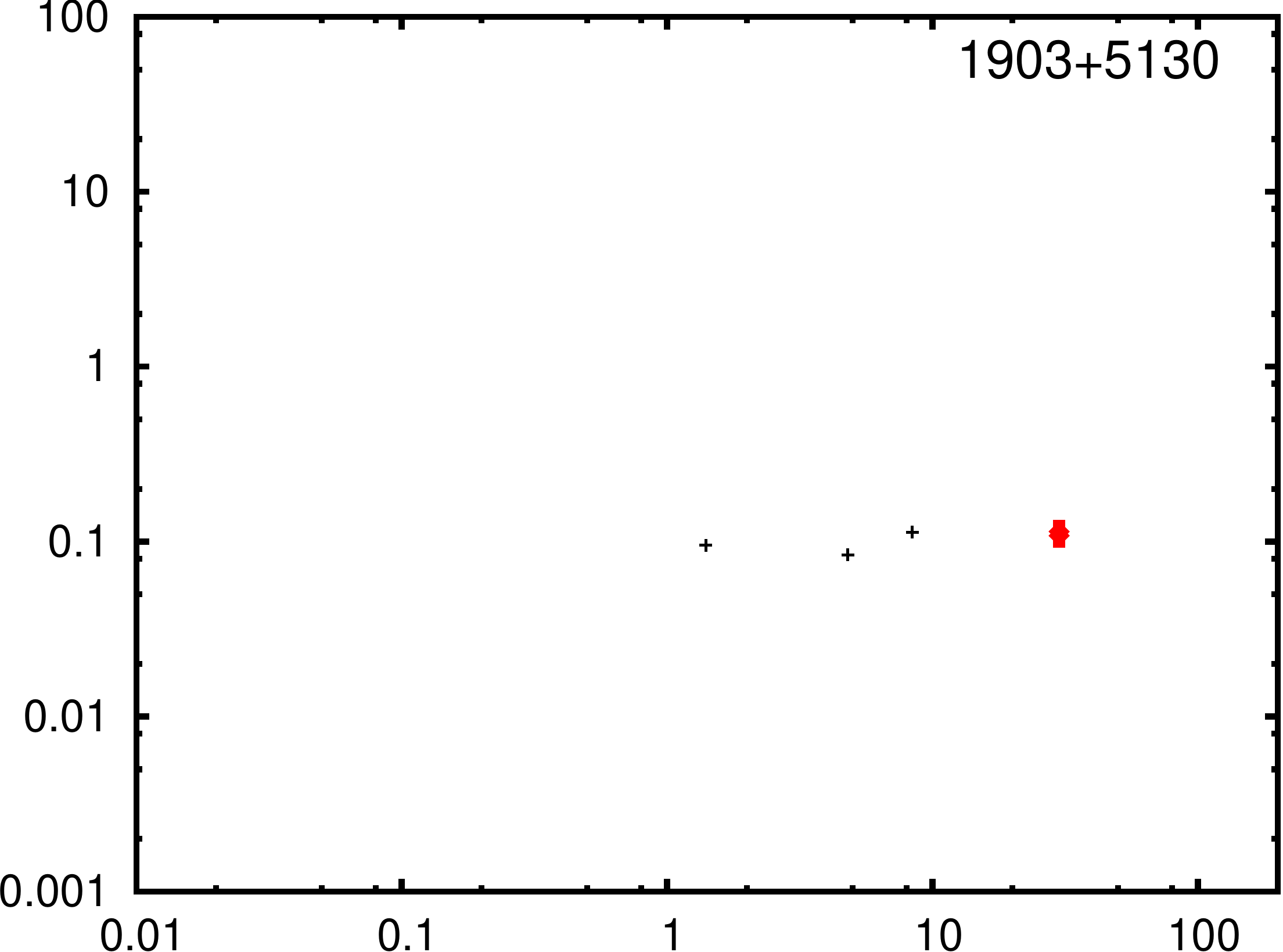}
\includegraphics[scale=0.2]{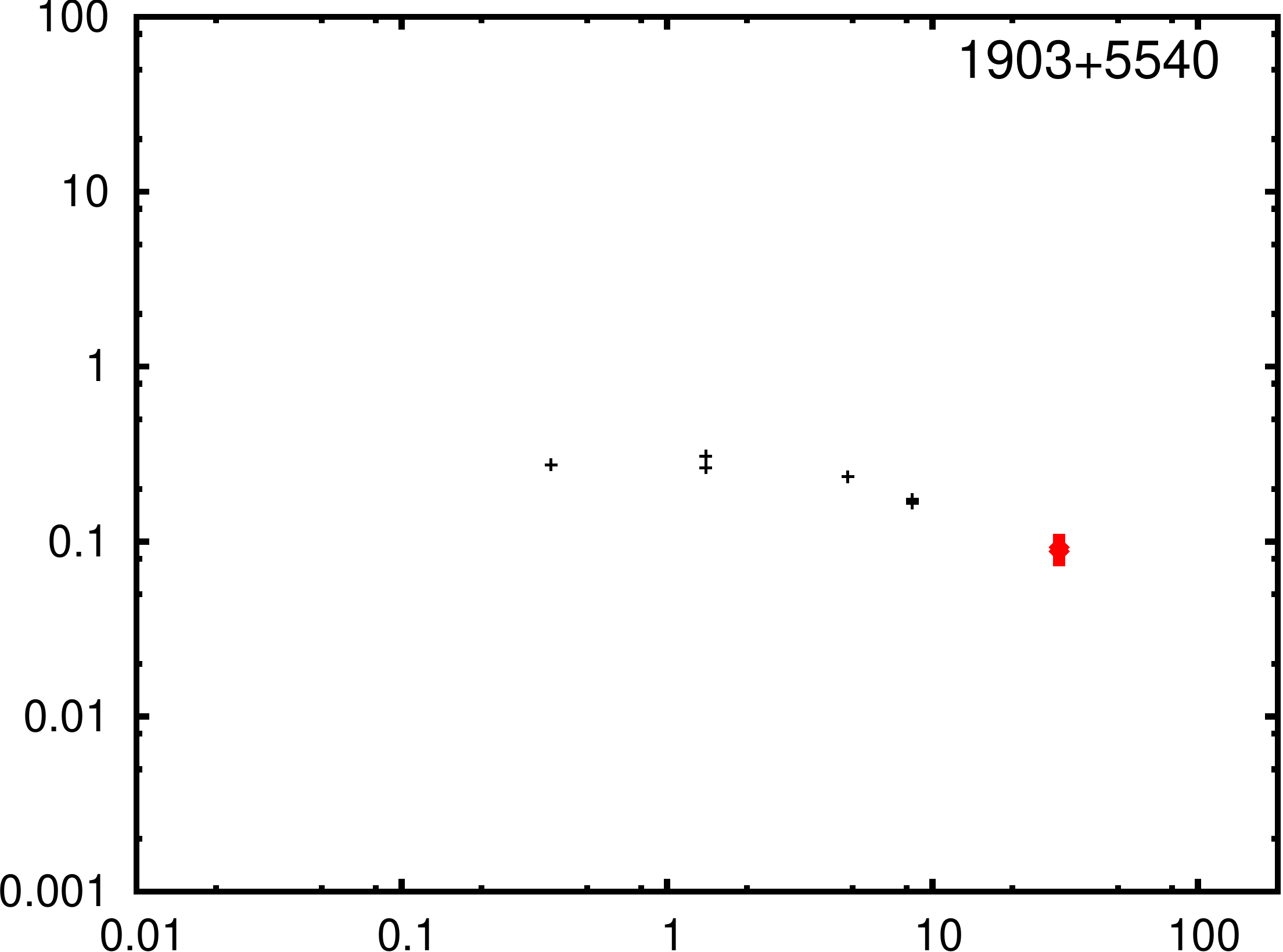}
\includegraphics[scale=0.2]{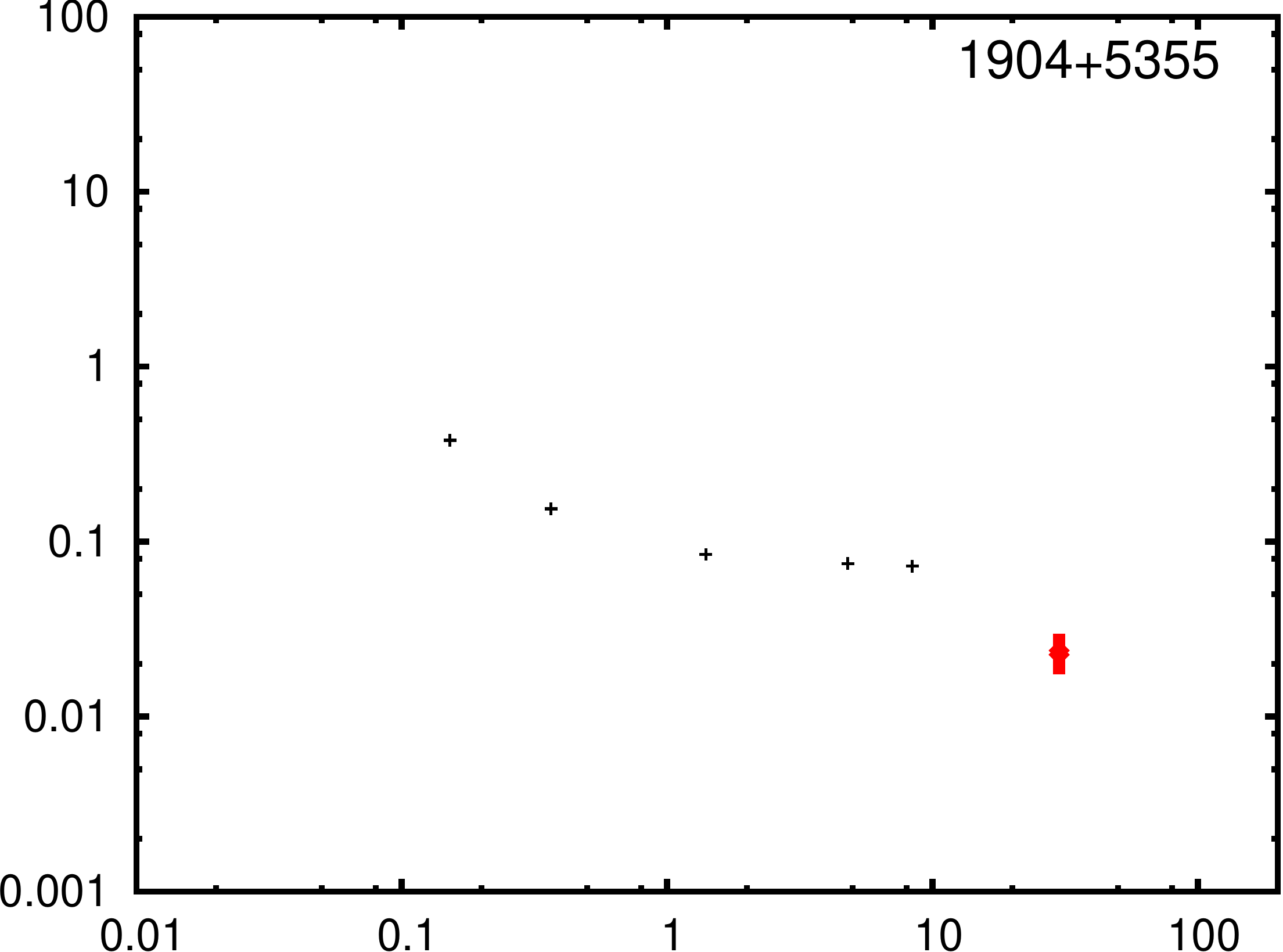}
\includegraphics[scale=0.2]{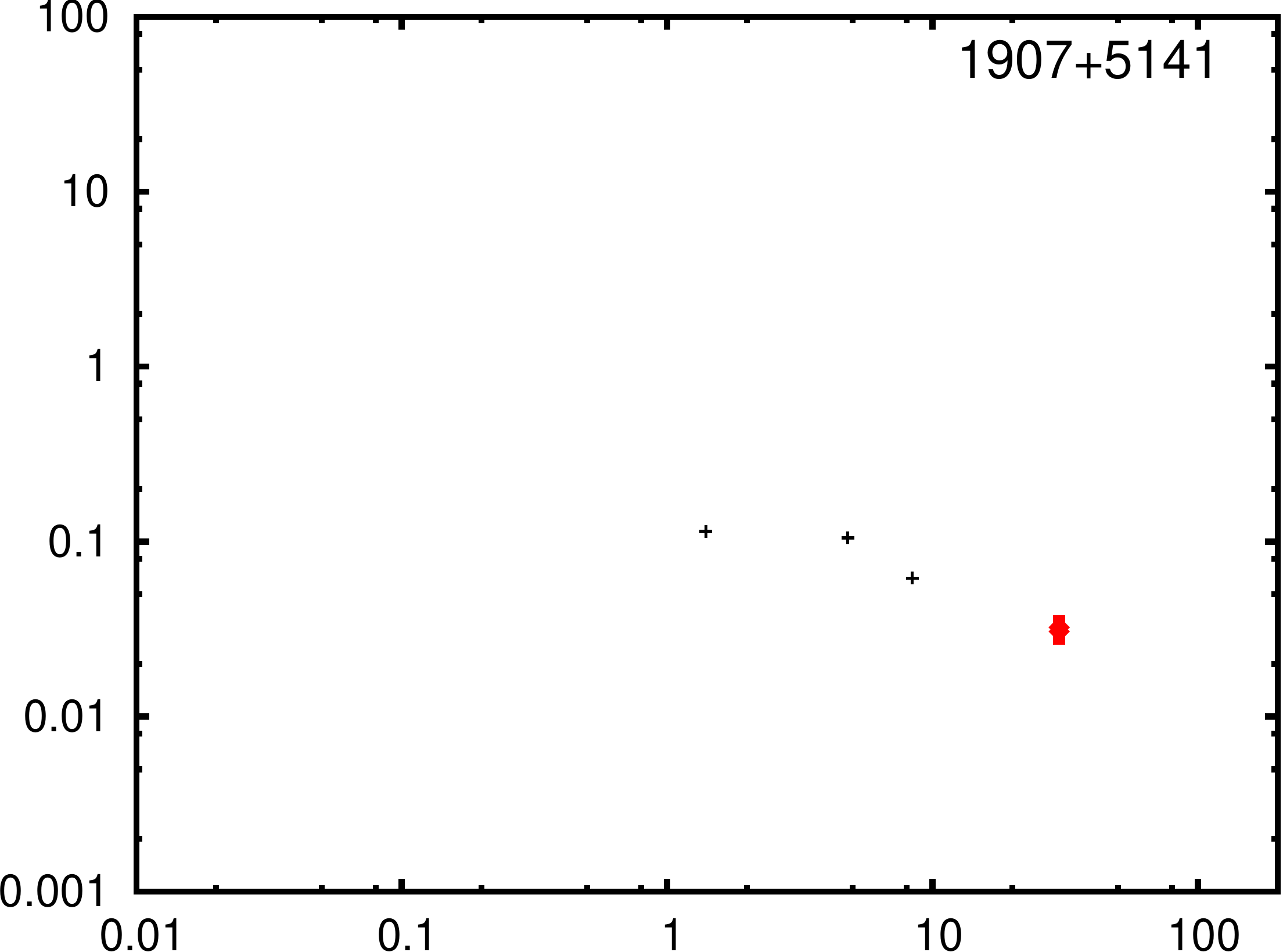}
\includegraphics[scale=0.2]{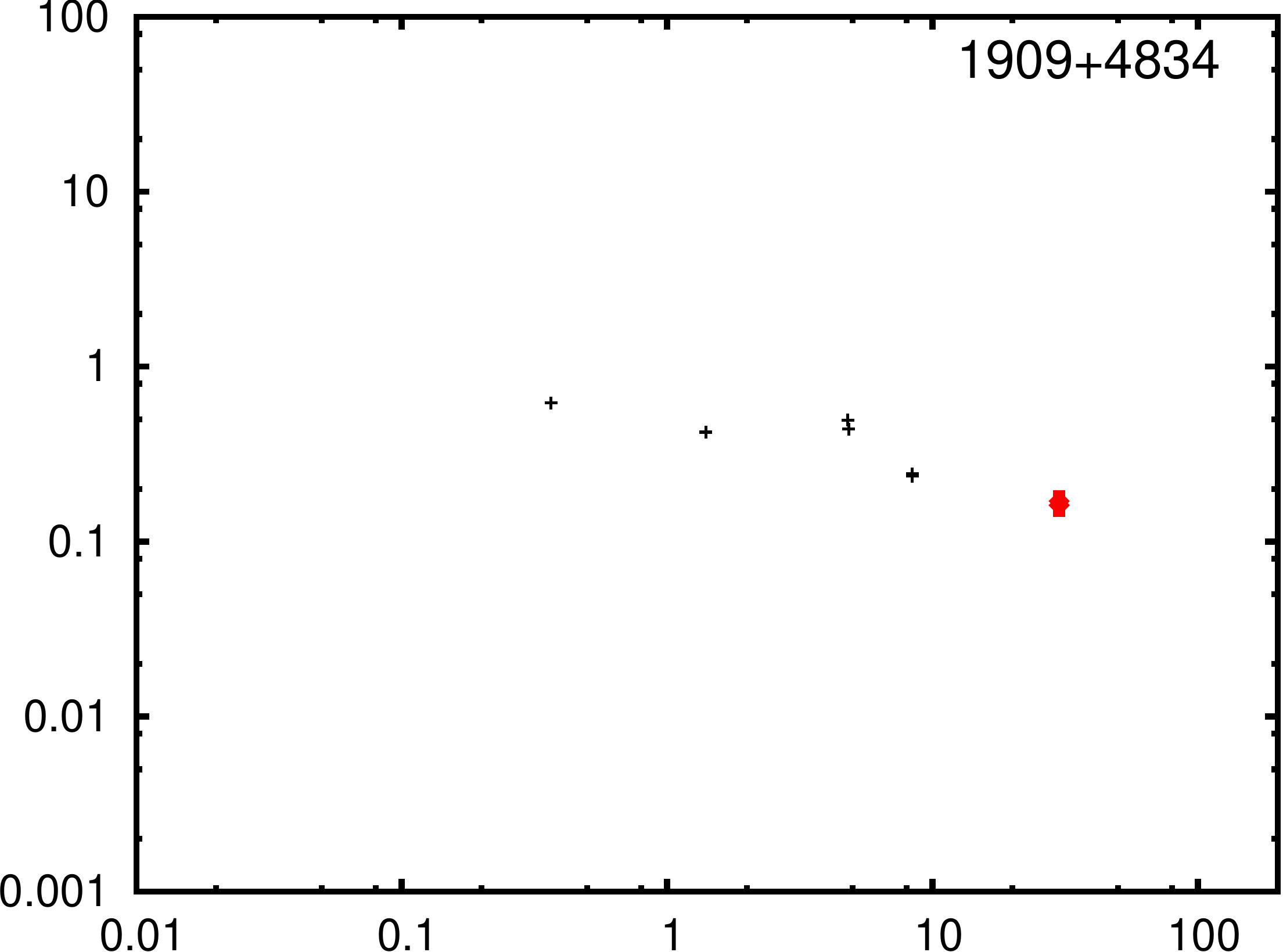}
\includegraphics[scale=0.2]{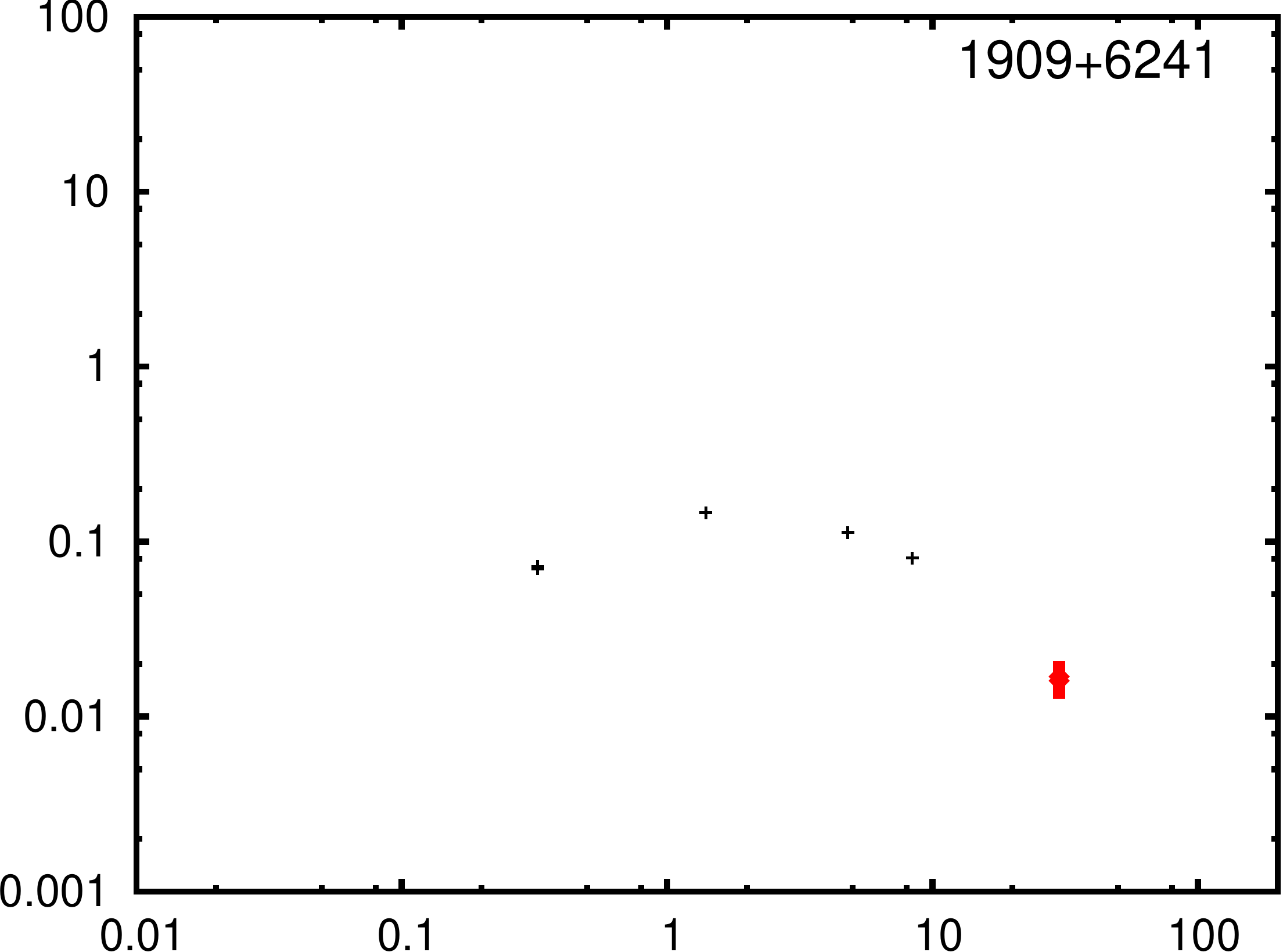}
\includegraphics[scale=0.2]{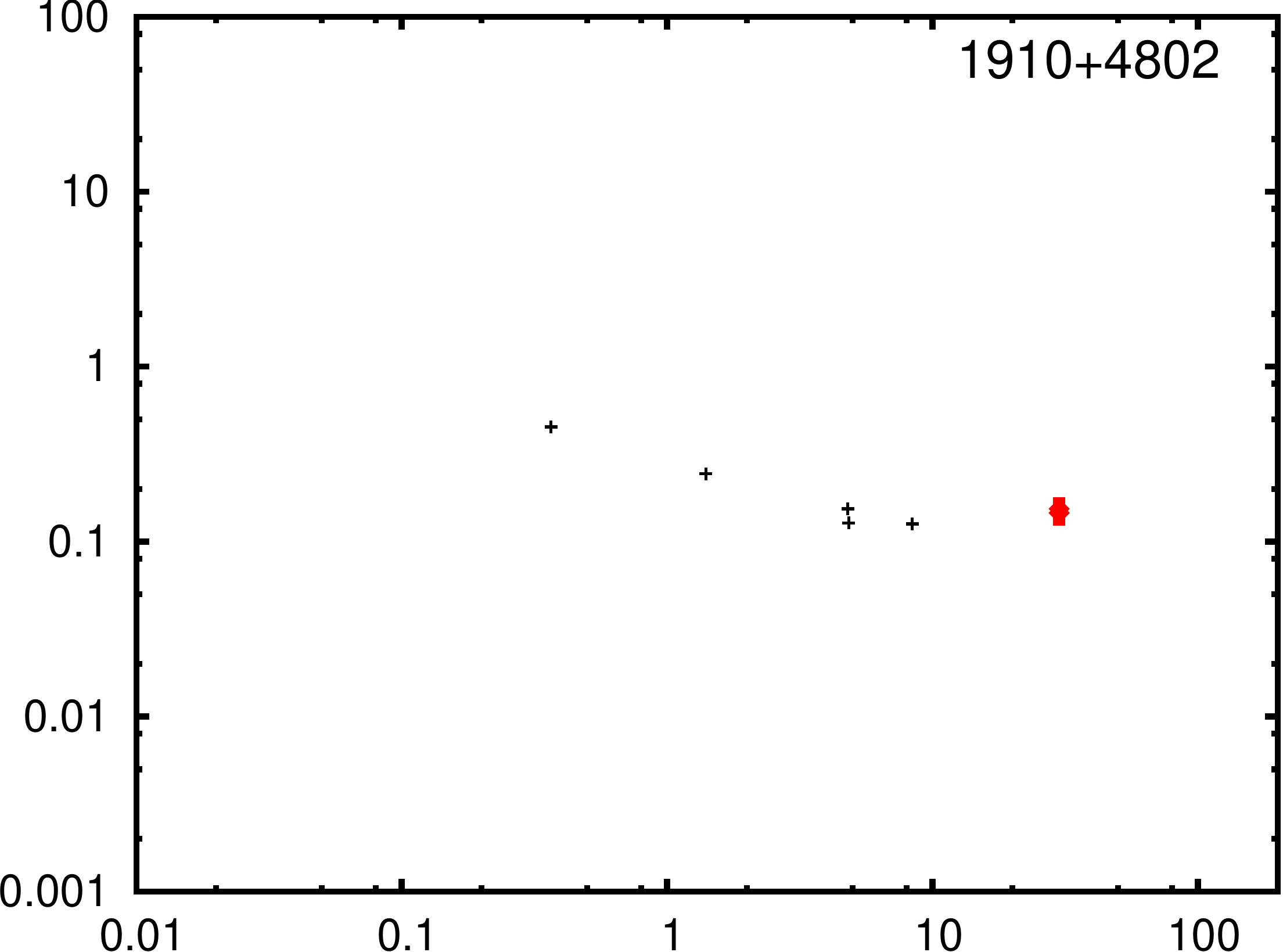}
\includegraphics[scale=0.2]{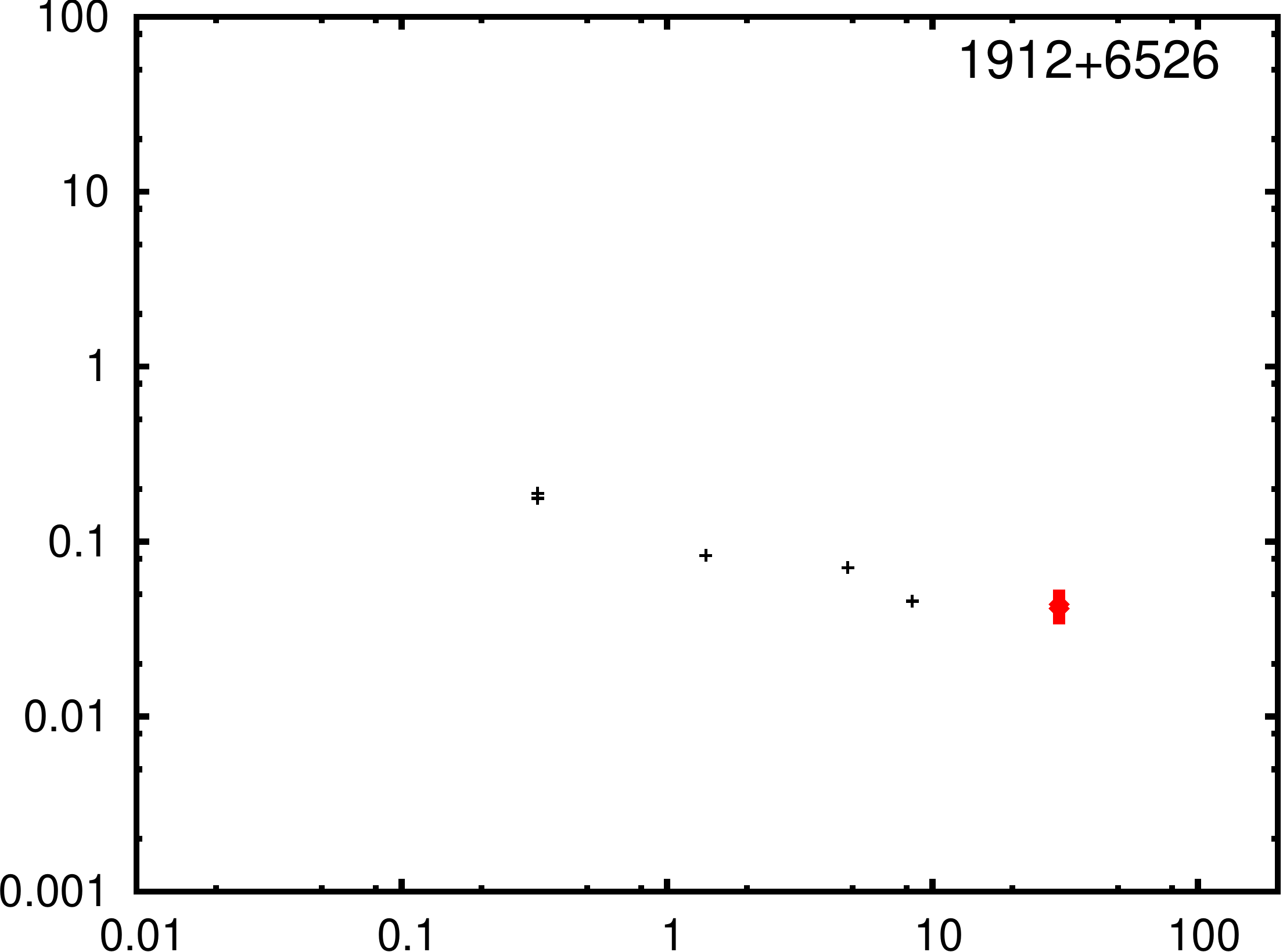}
\includegraphics[scale=0.2]{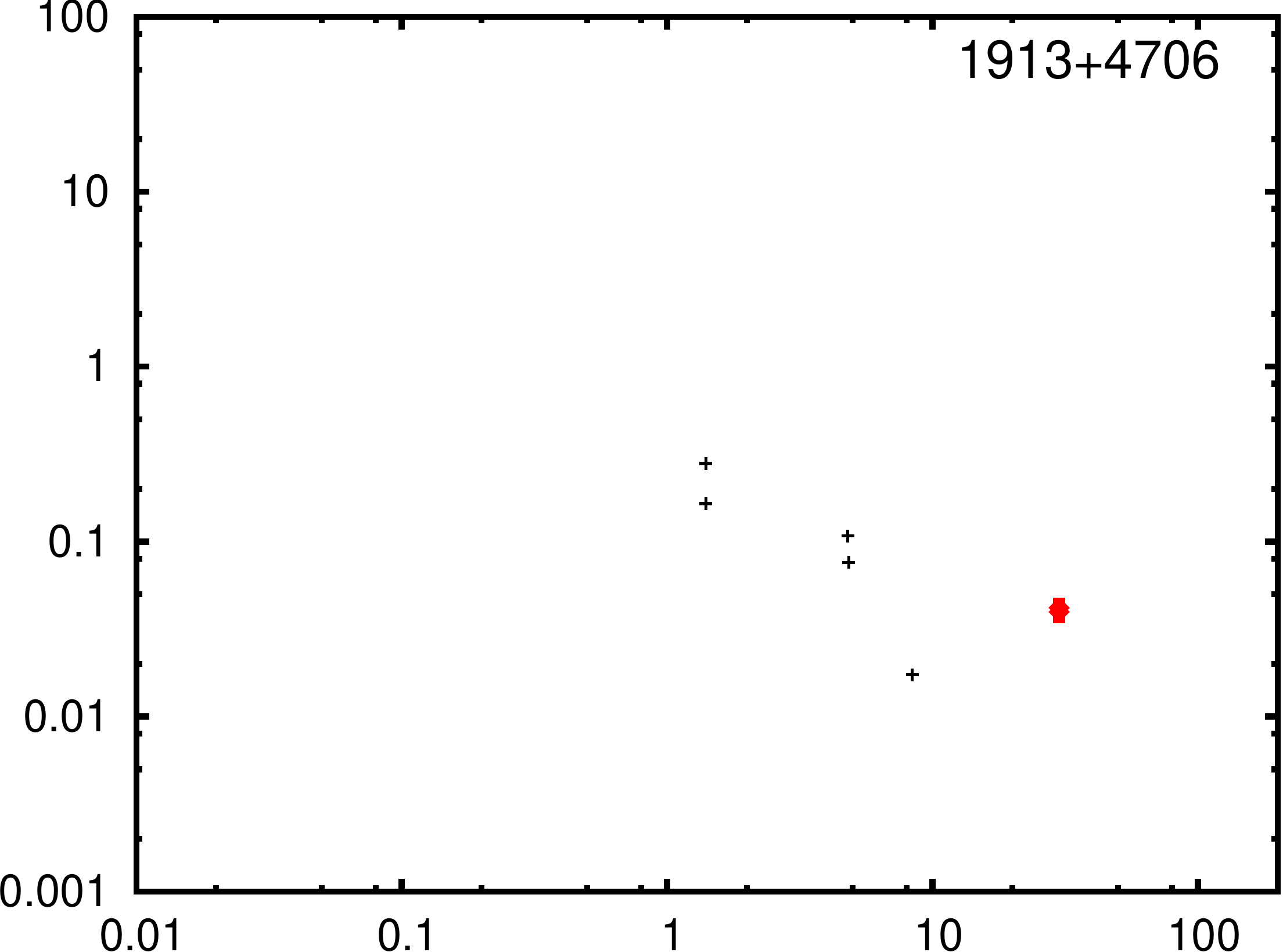}
\includegraphics[scale=0.2]{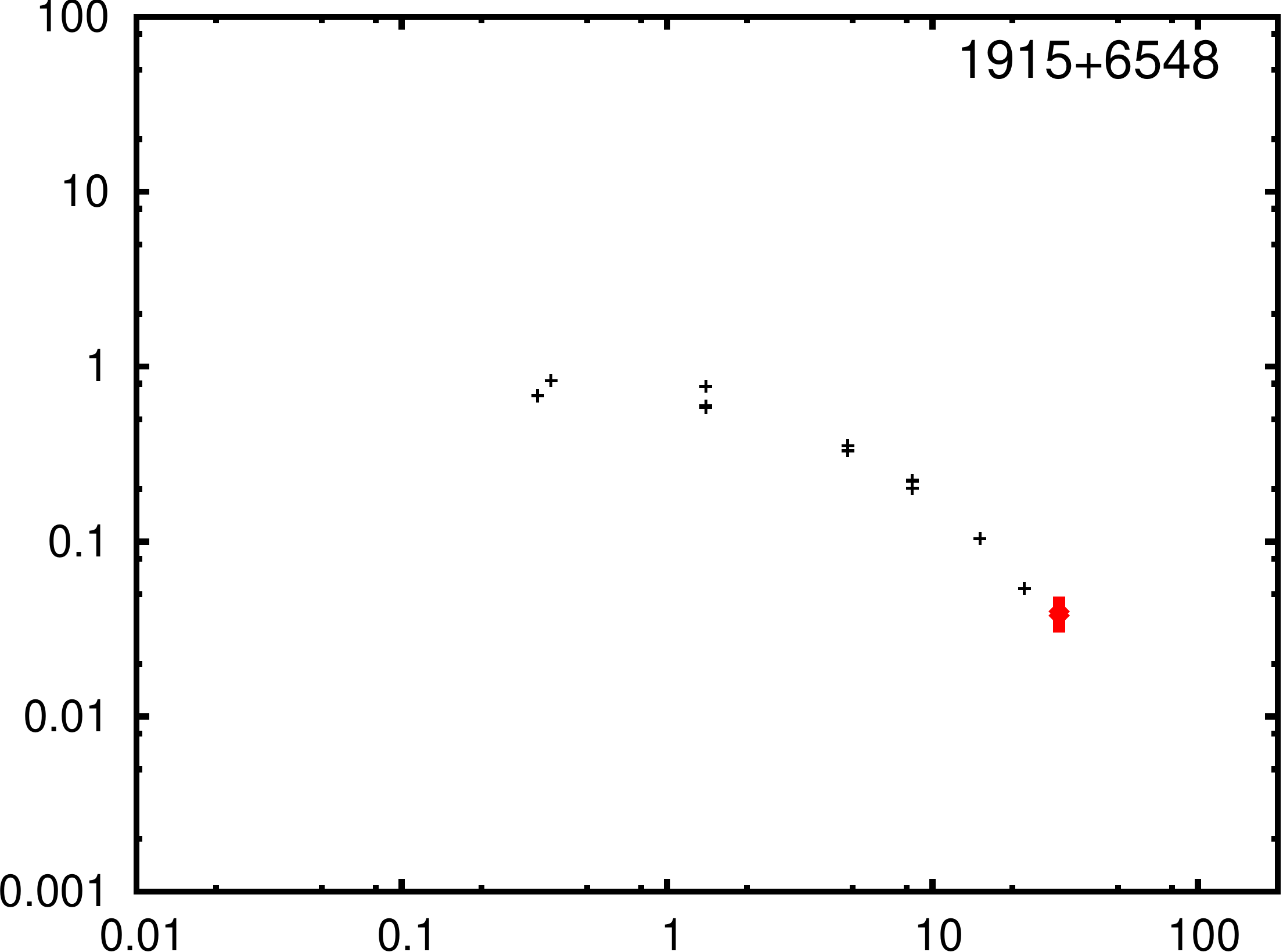}
\end{figure}
\clearpage\begin{figure}
\centering
\includegraphics[scale=0.2]{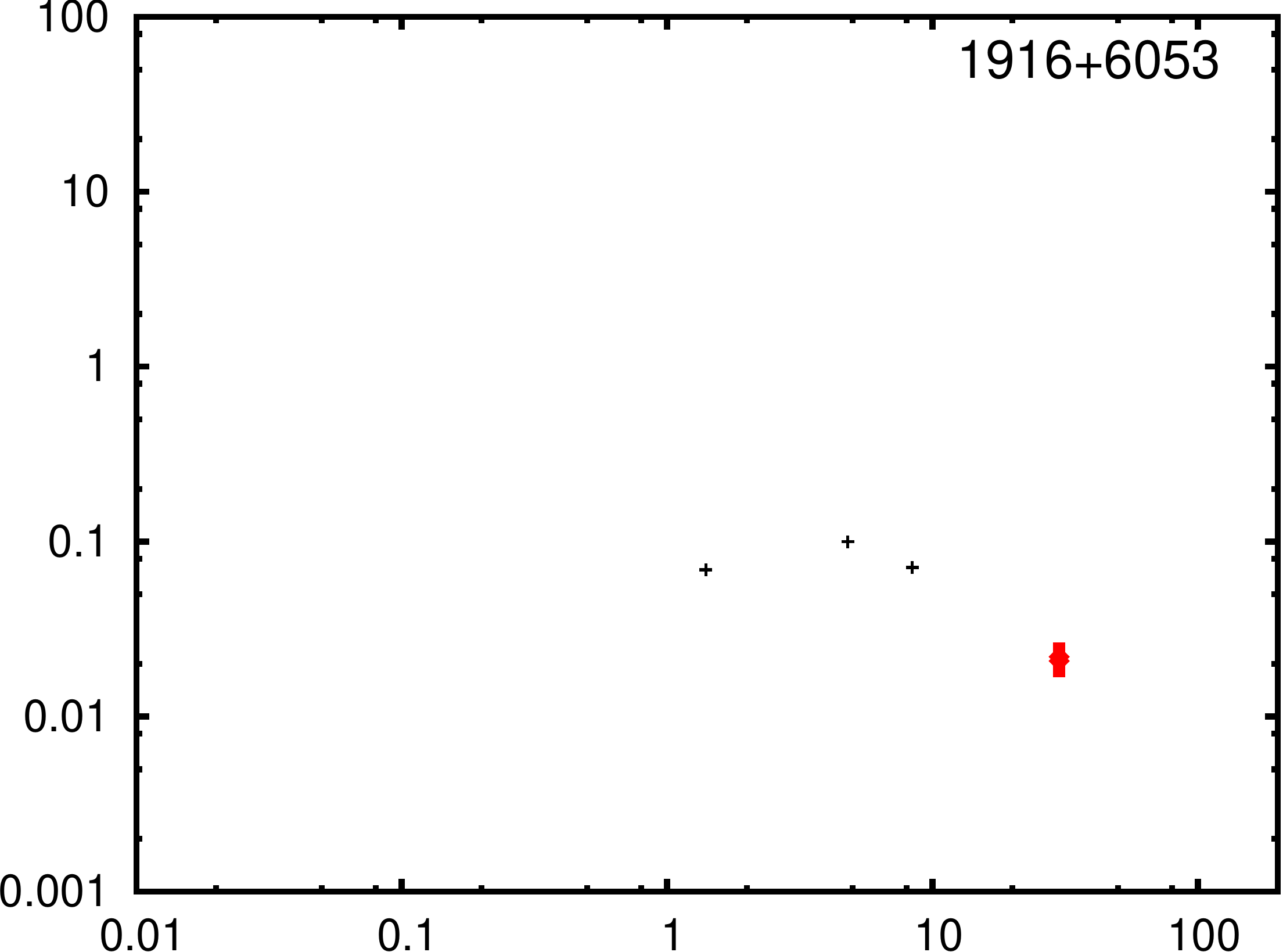}
\includegraphics[scale=0.2]{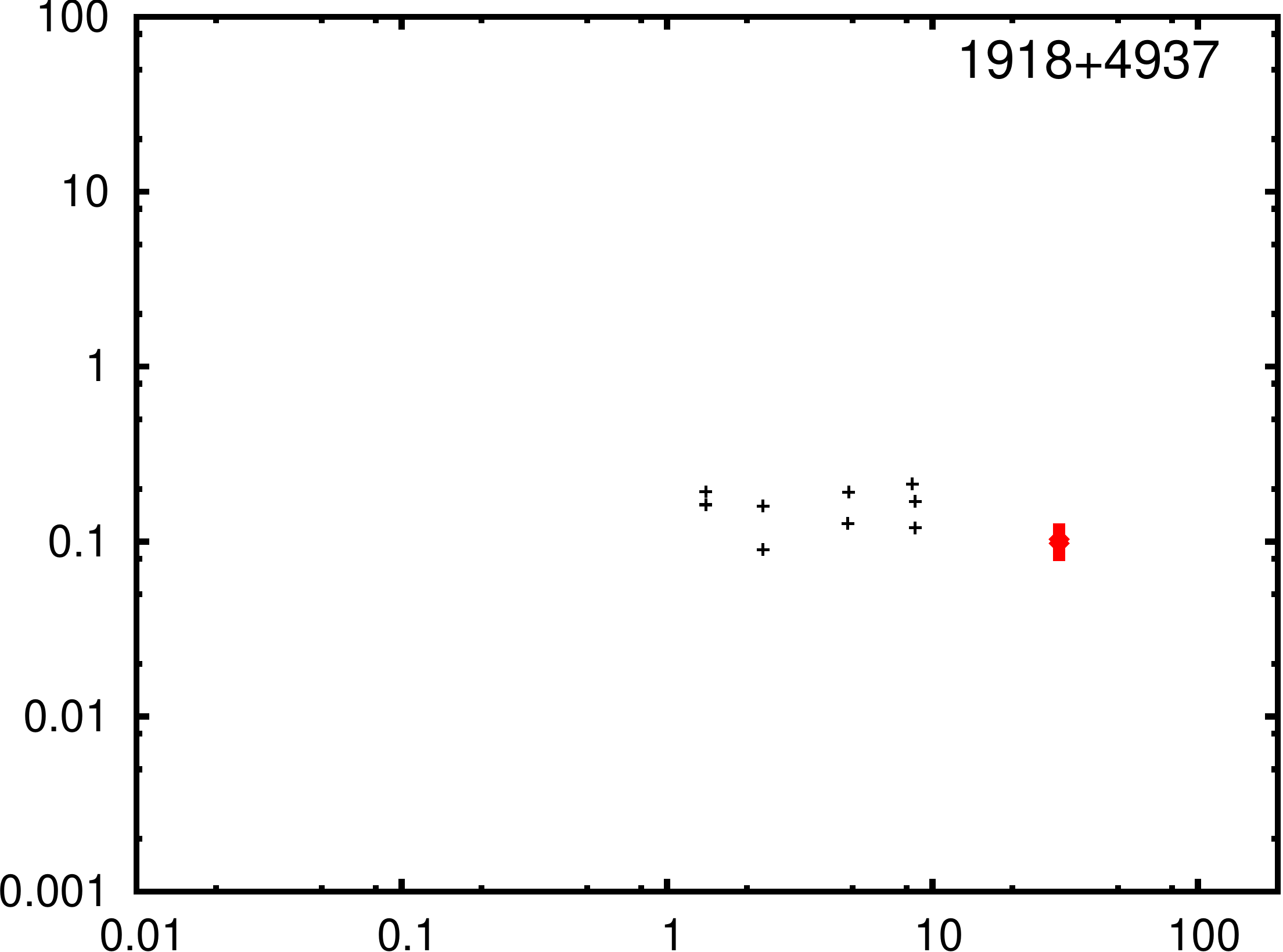}
\includegraphics[scale=0.2]{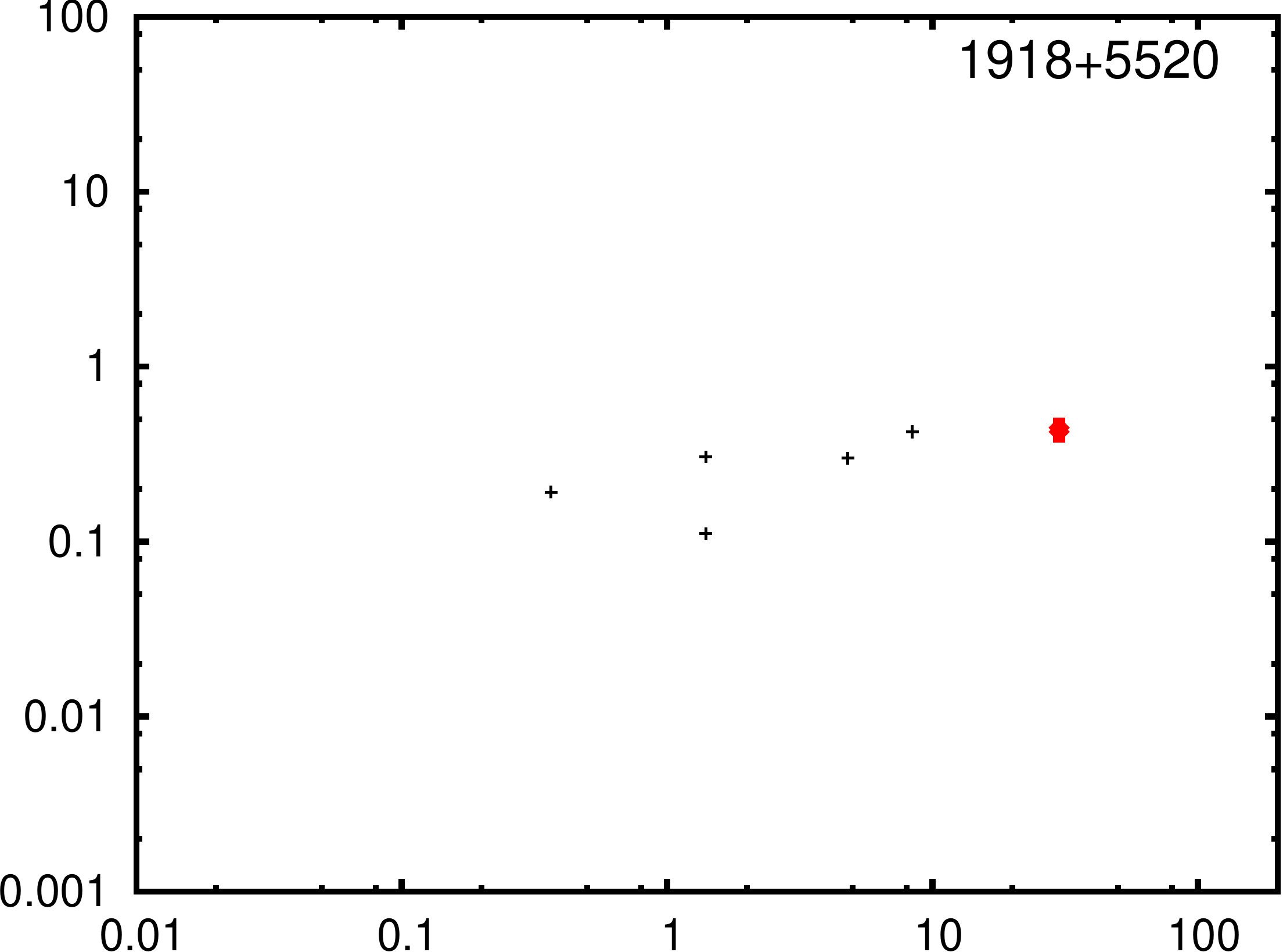}
\includegraphics[scale=0.2]{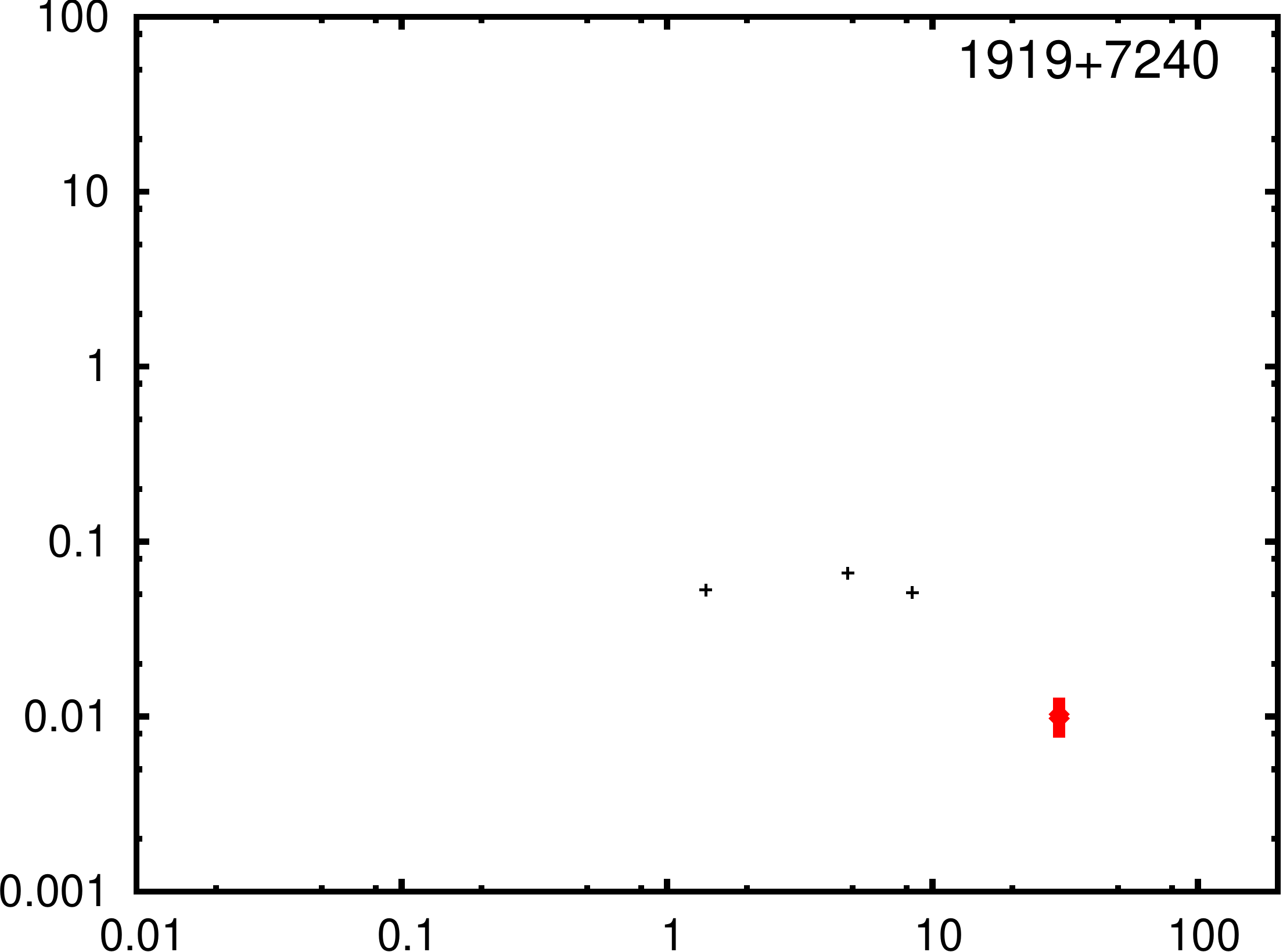}
\includegraphics[scale=0.2]{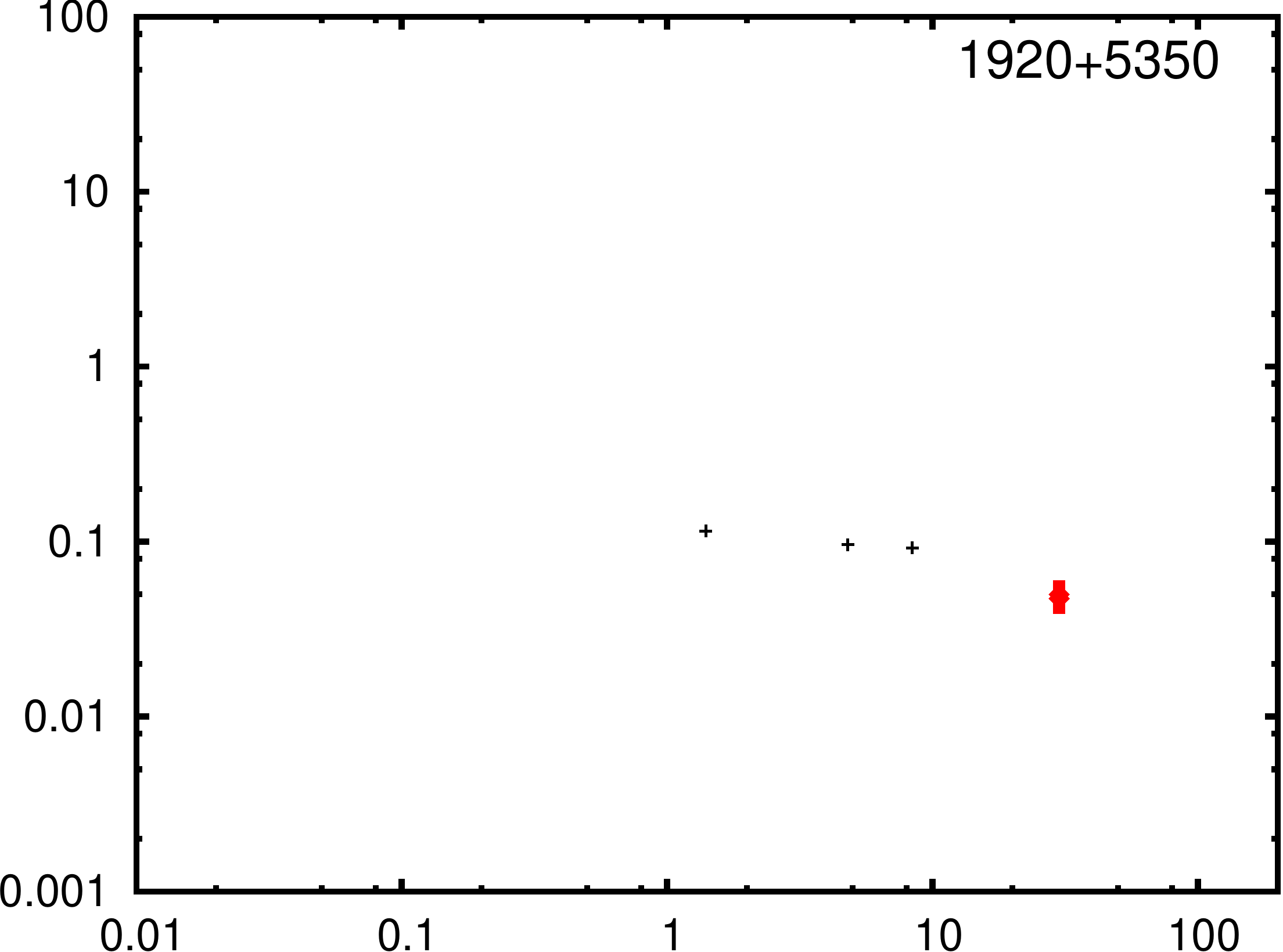}
\includegraphics[scale=0.2]{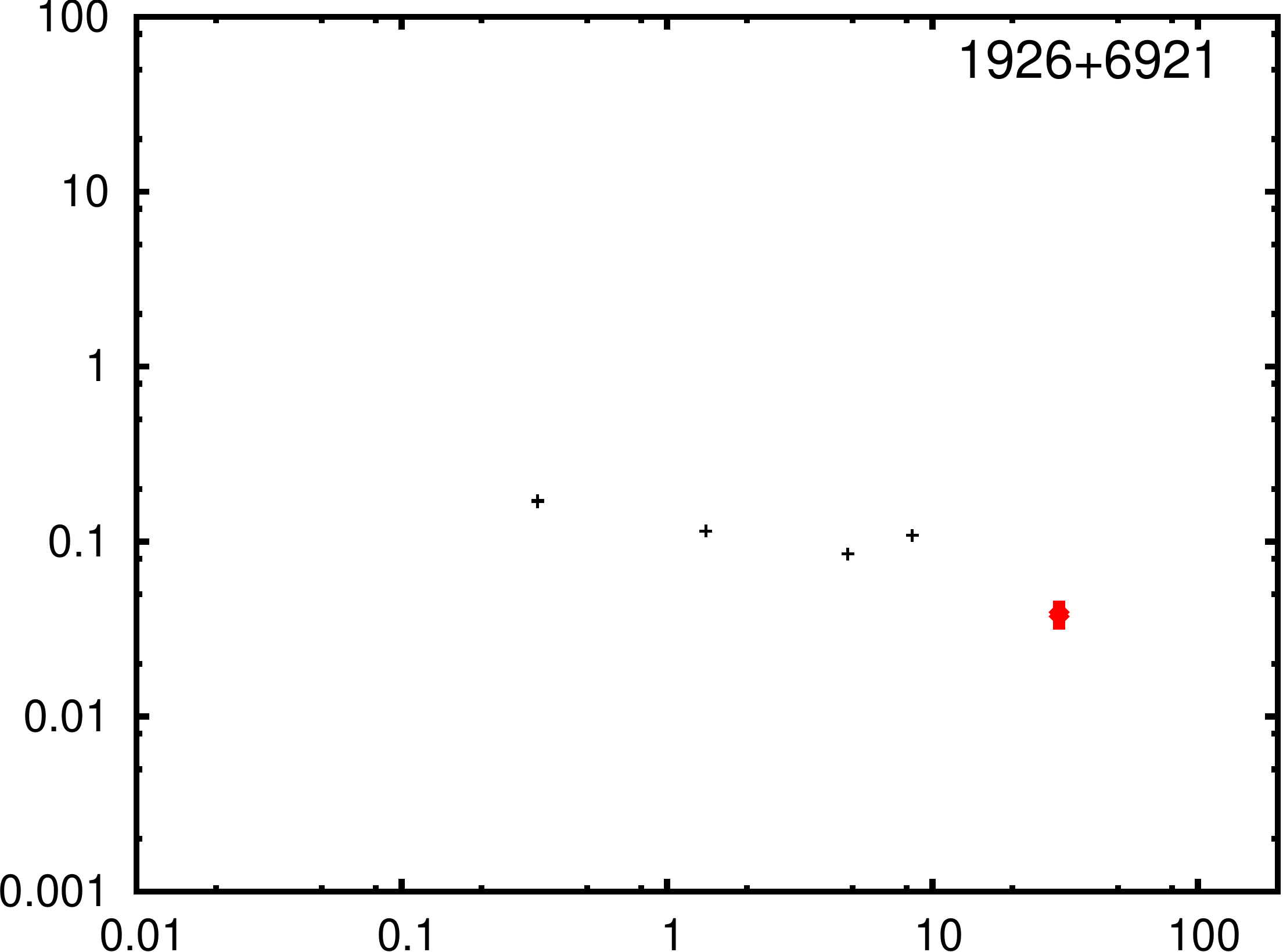}
\includegraphics[scale=0.2]{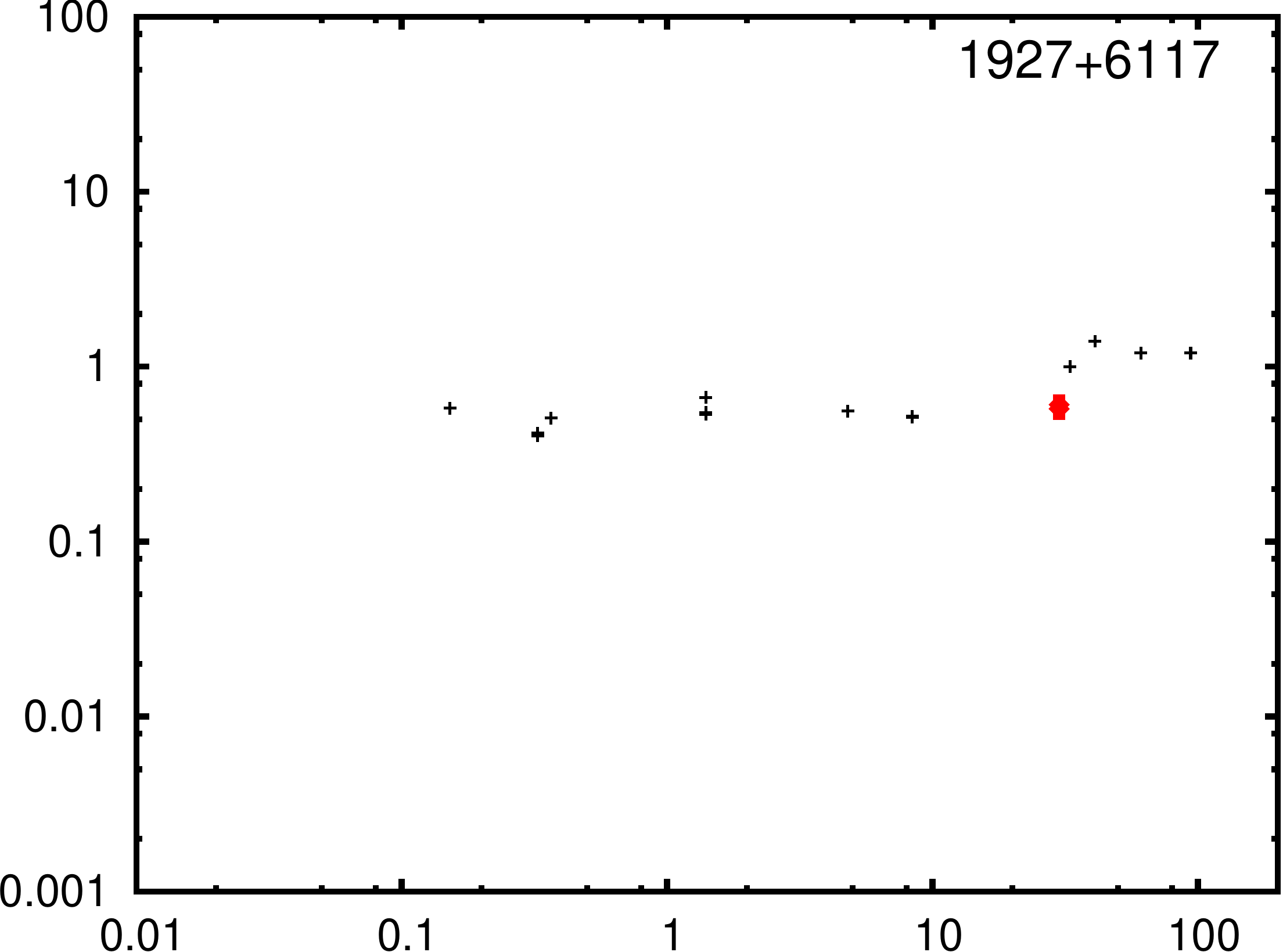}
\includegraphics[scale=0.2]{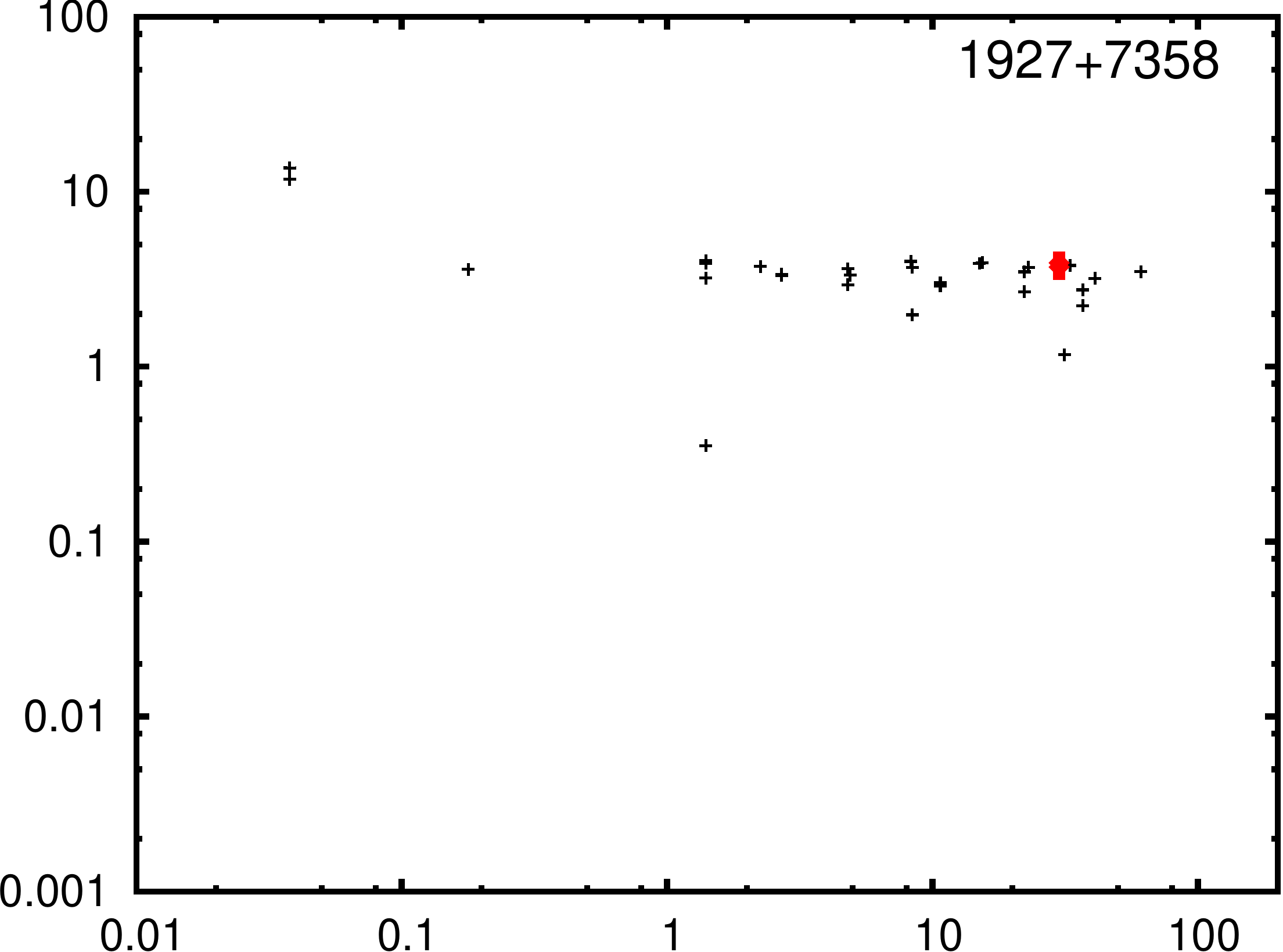}
\includegraphics[scale=0.2]{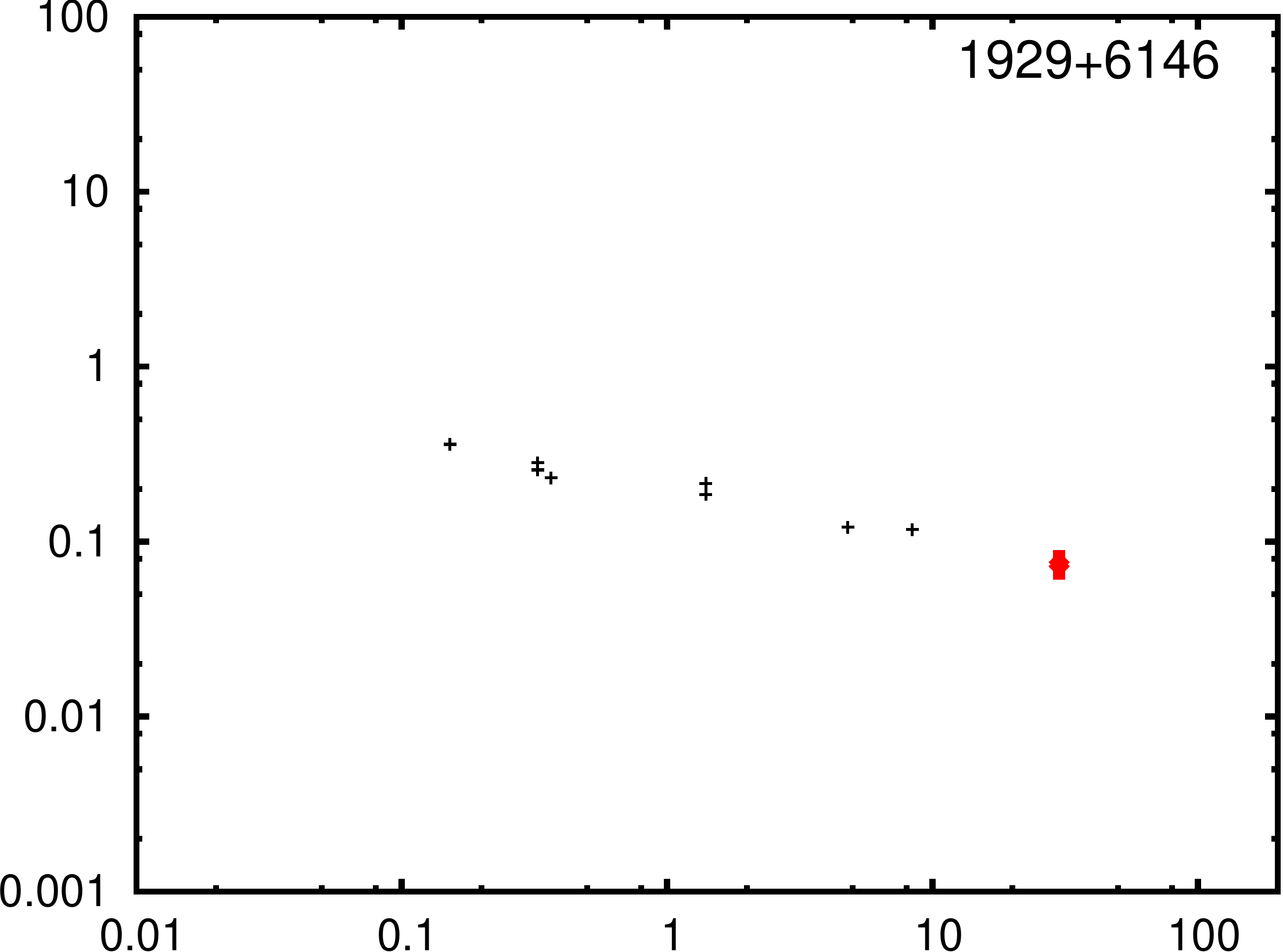}
\includegraphics[scale=0.2]{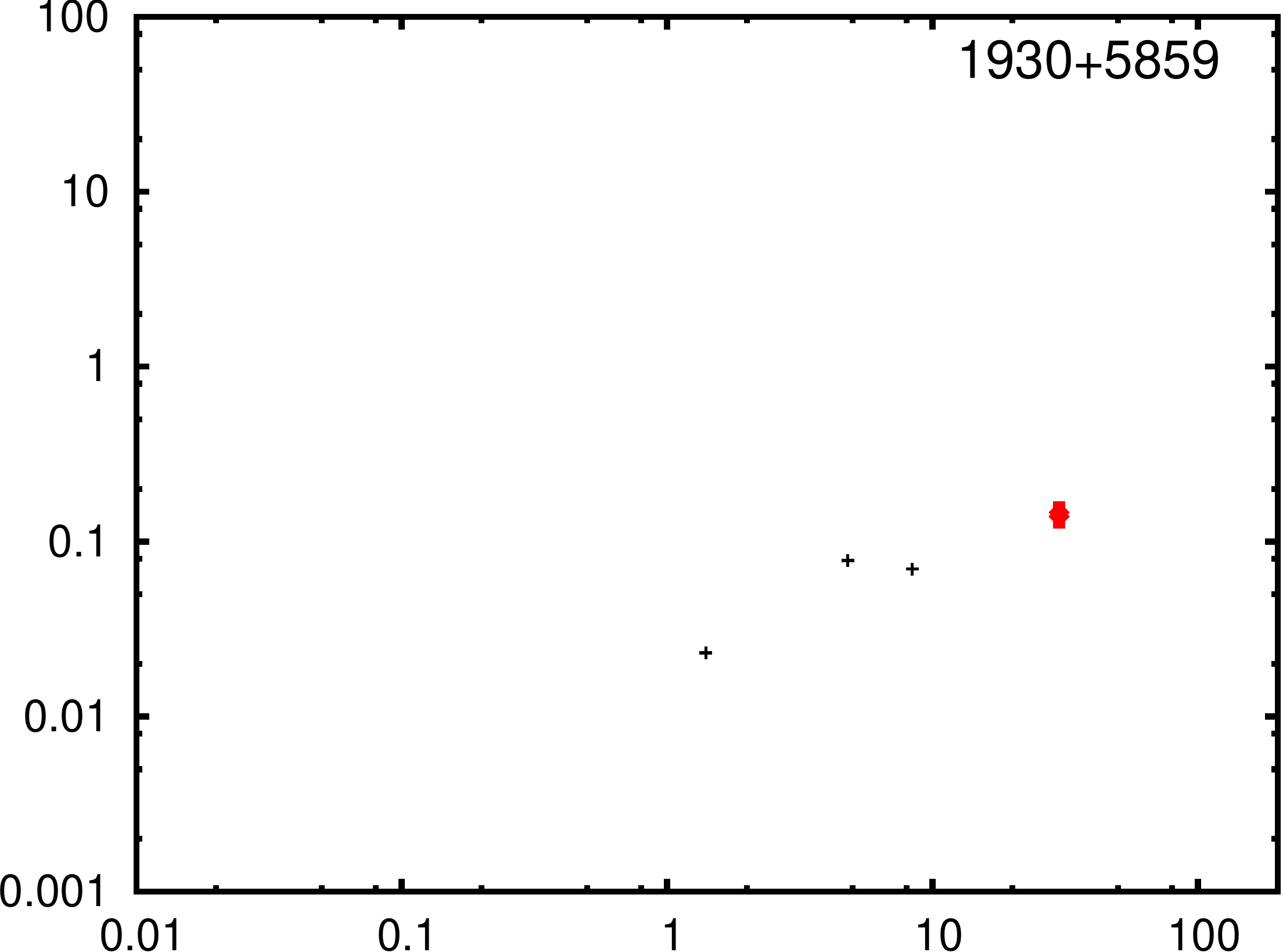}
\includegraphics[scale=0.2]{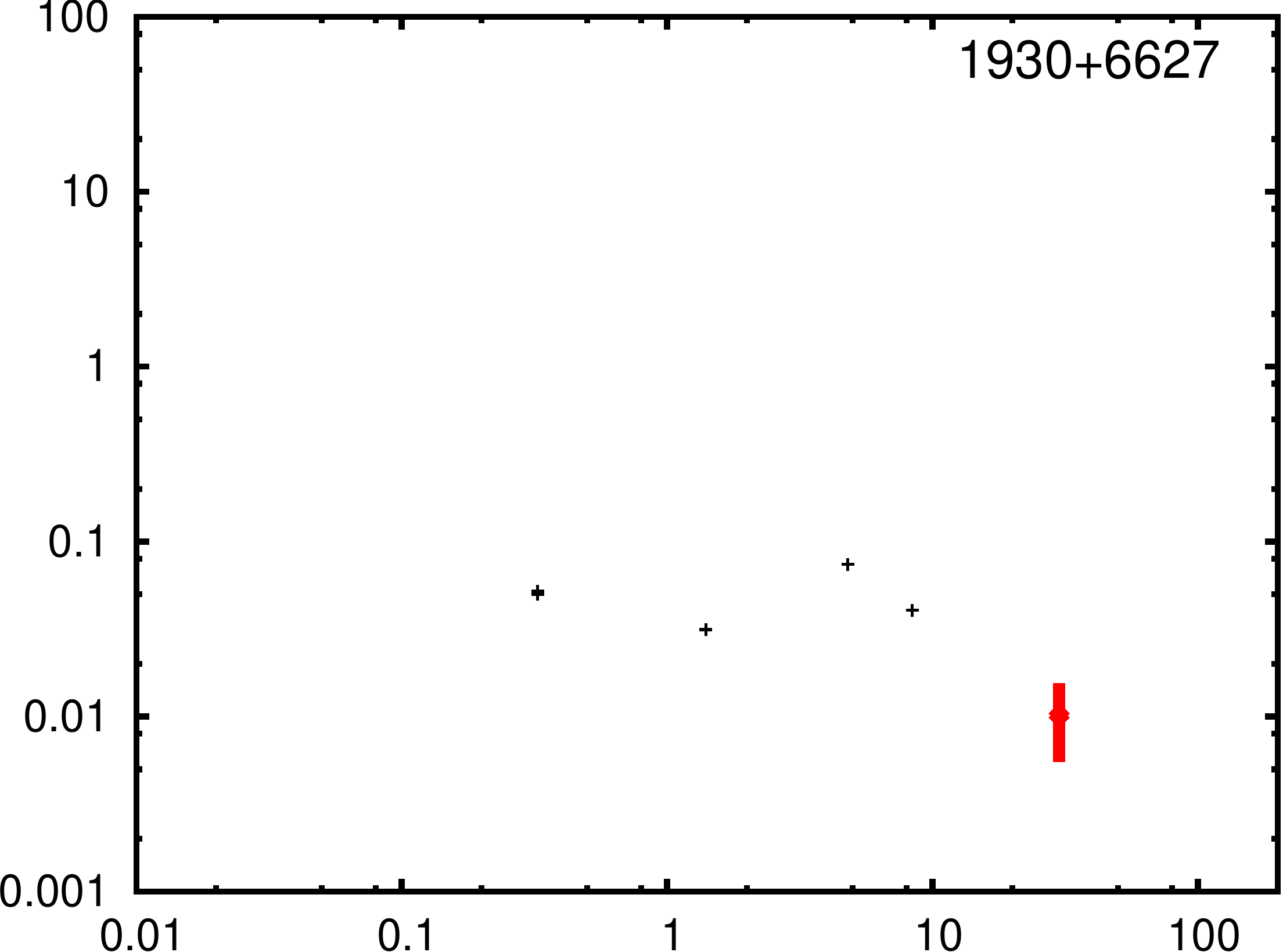}
\includegraphics[scale=0.2]{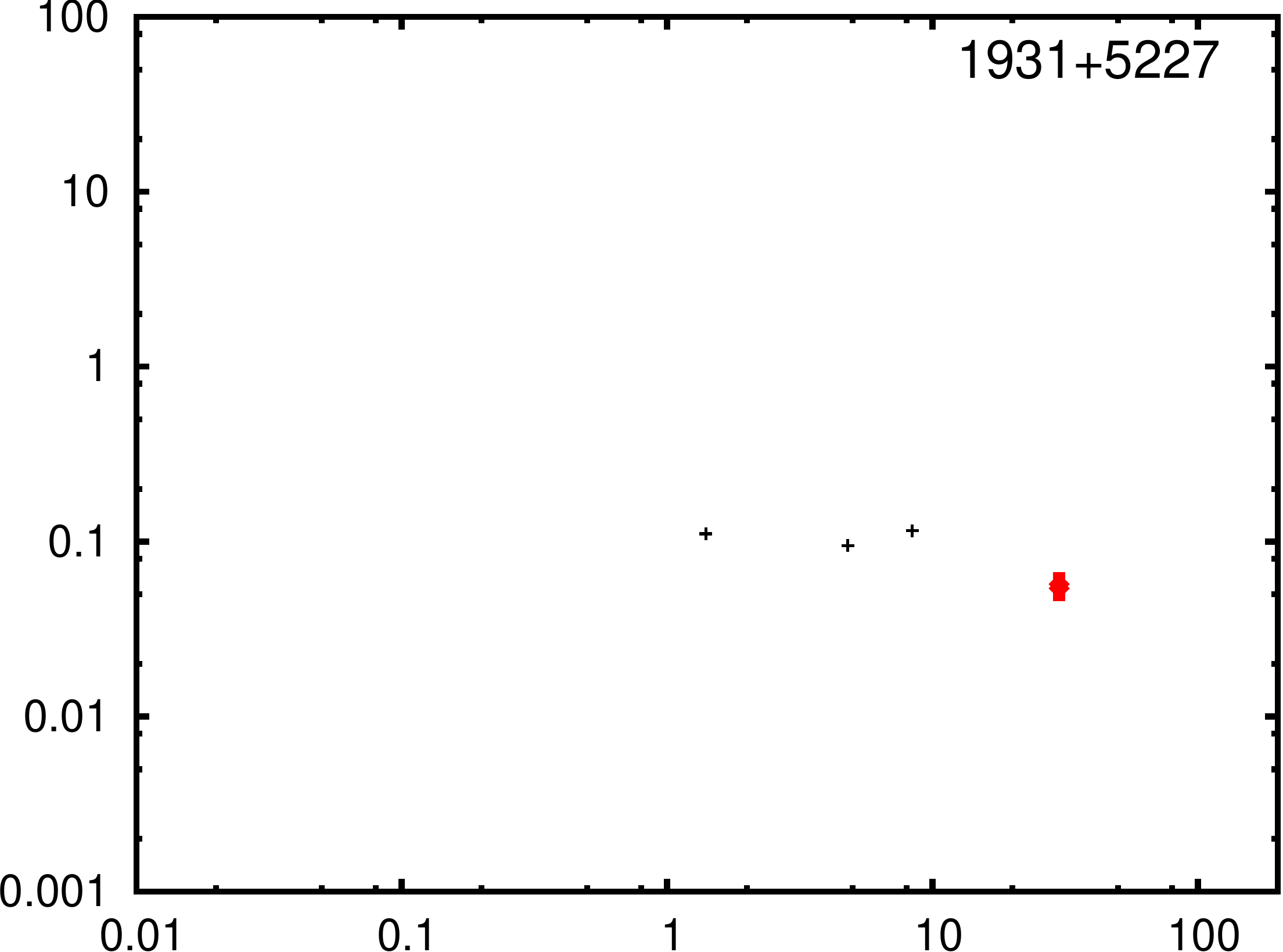}
\includegraphics[scale=0.2]{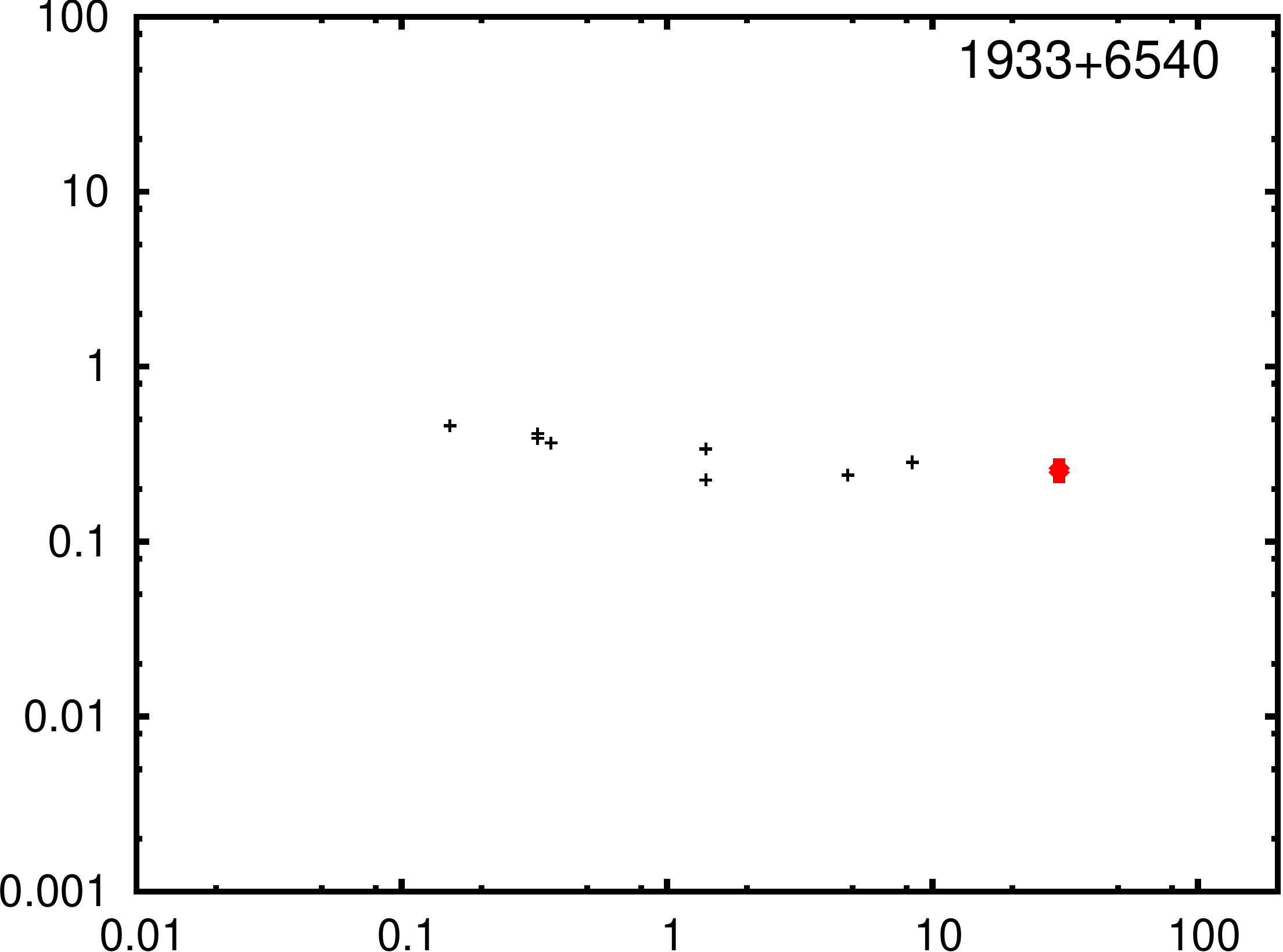}
\includegraphics[scale=0.2]{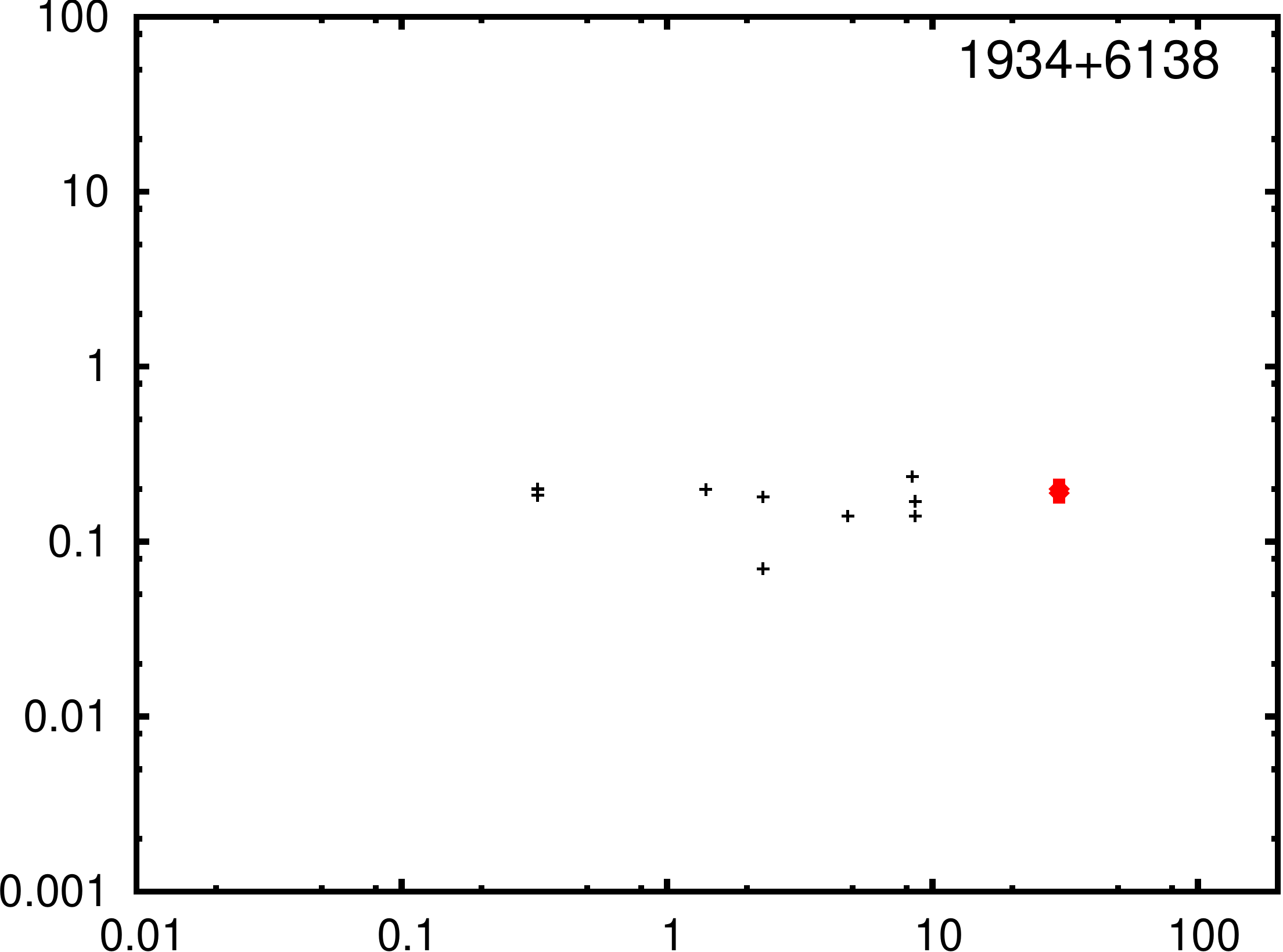}
\includegraphics[scale=0.2]{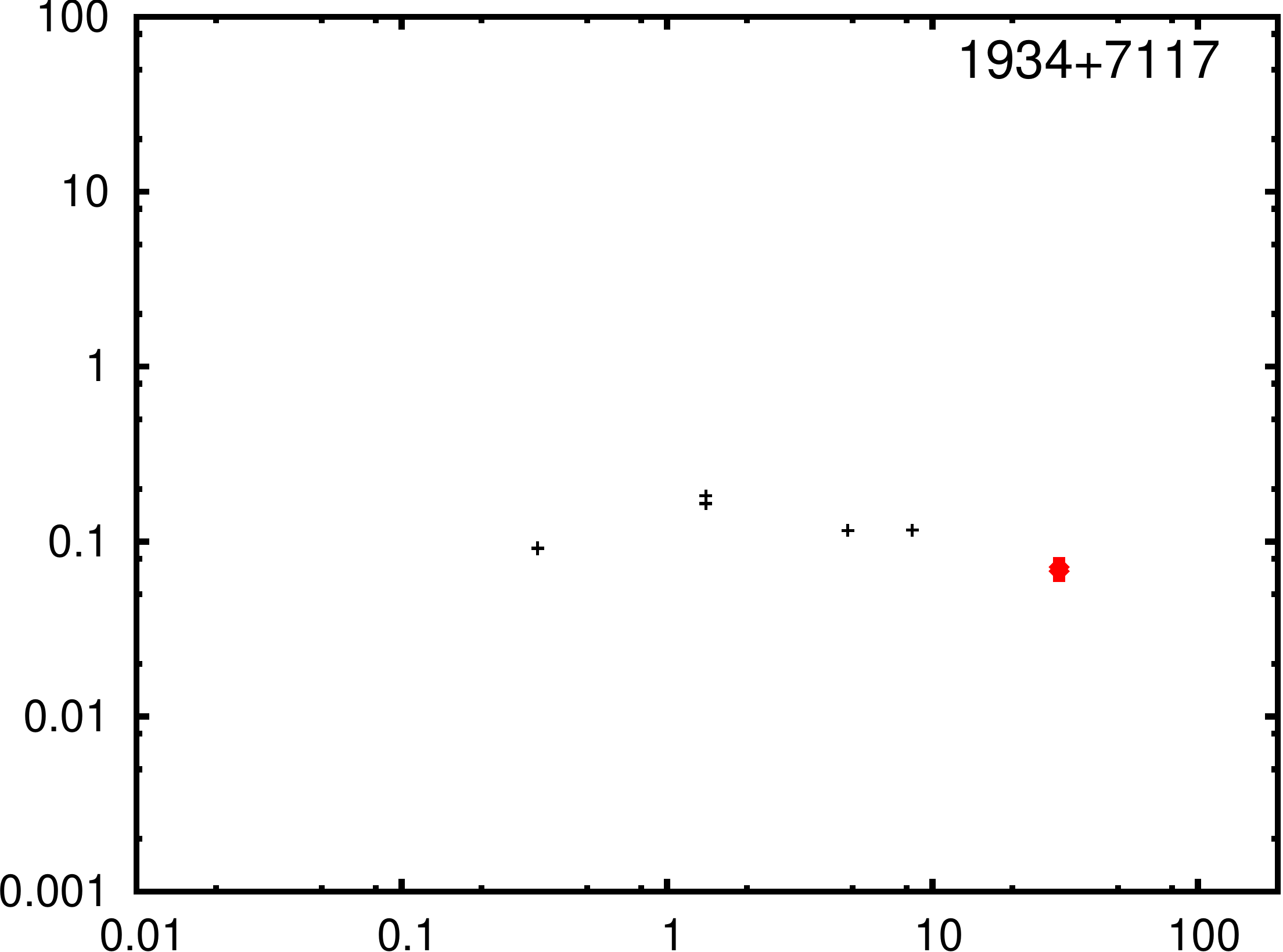}
\includegraphics[scale=0.2]{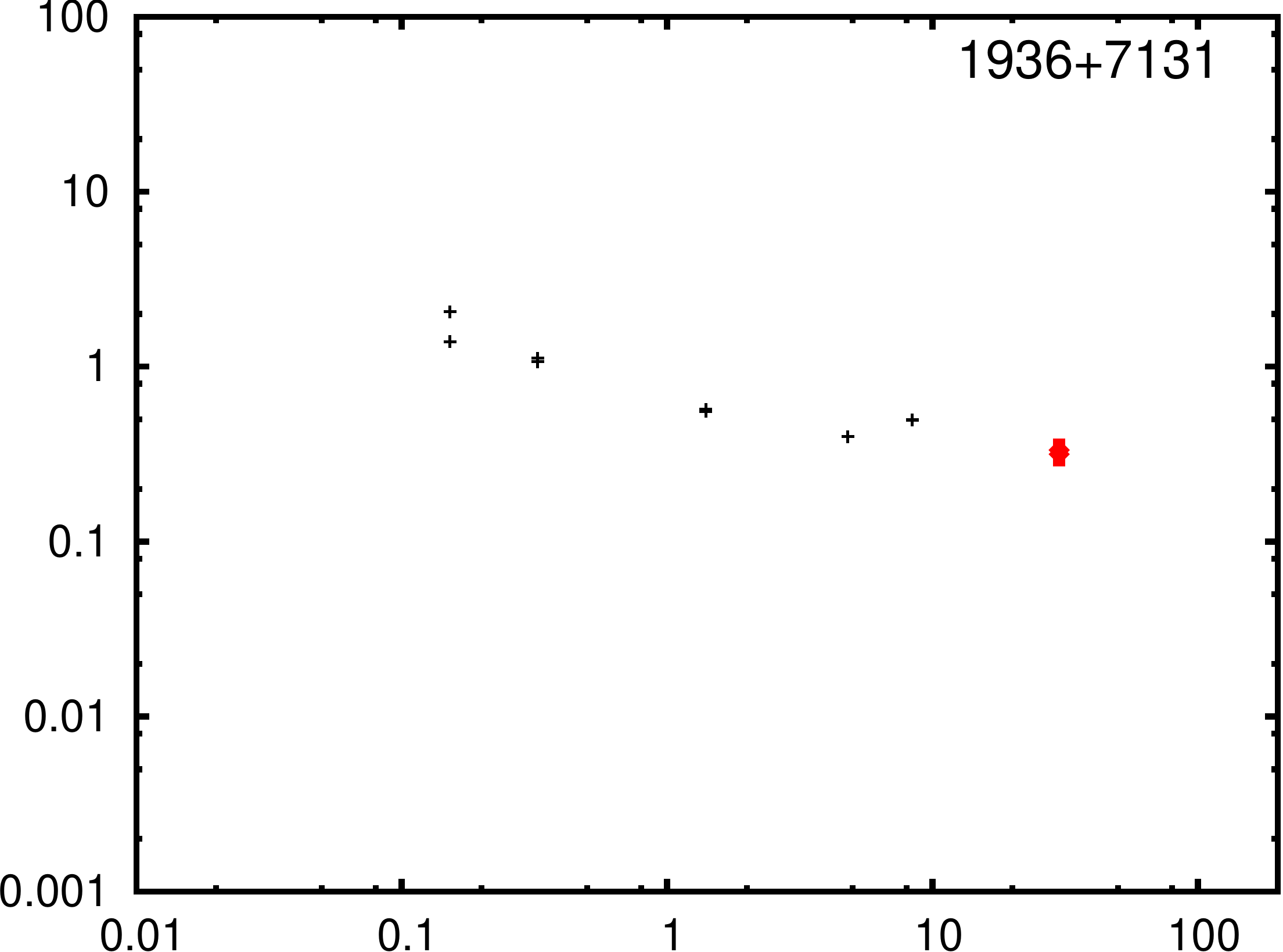}
\includegraphics[scale=0.2]{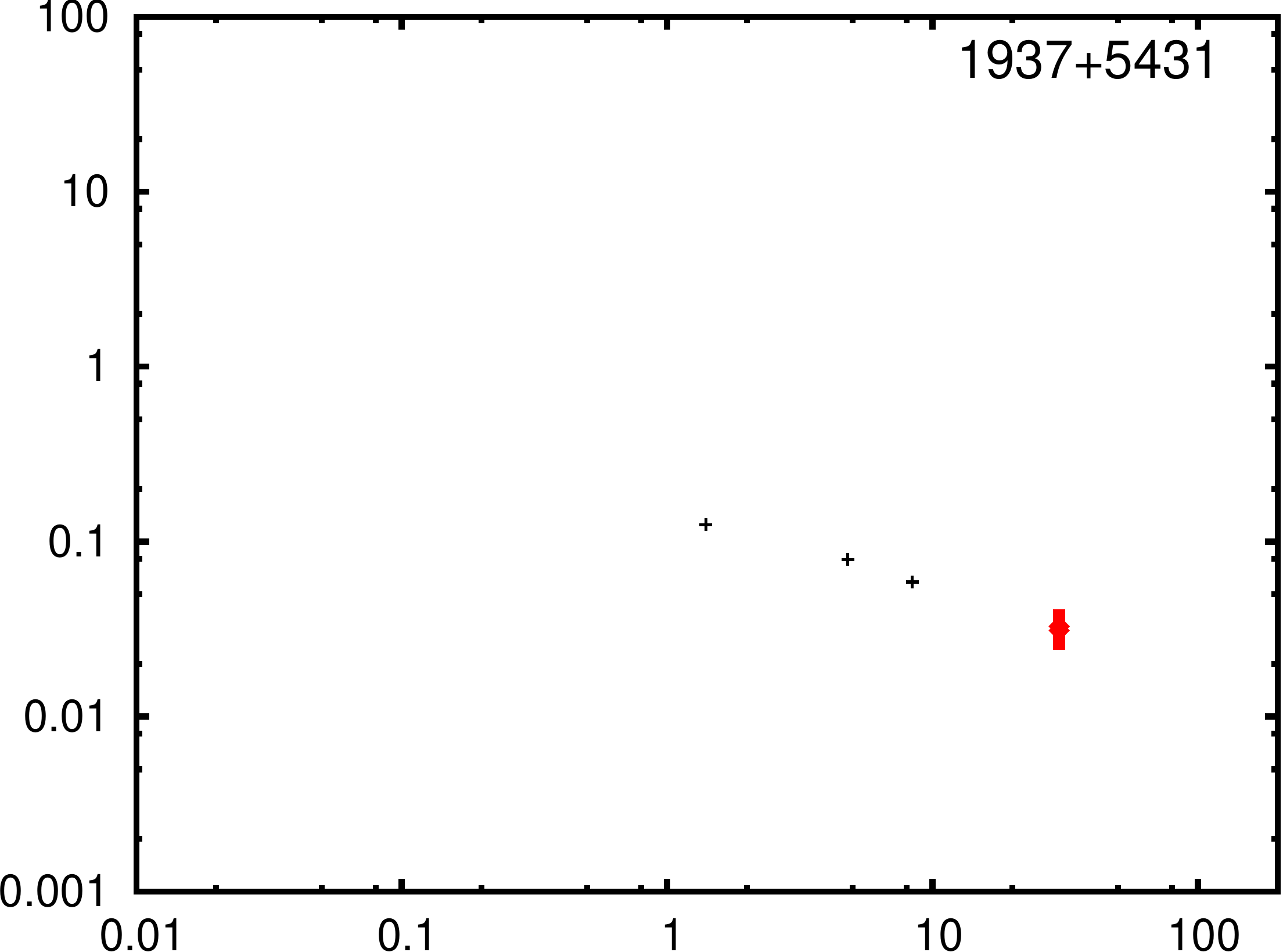}
\includegraphics[scale=0.2]{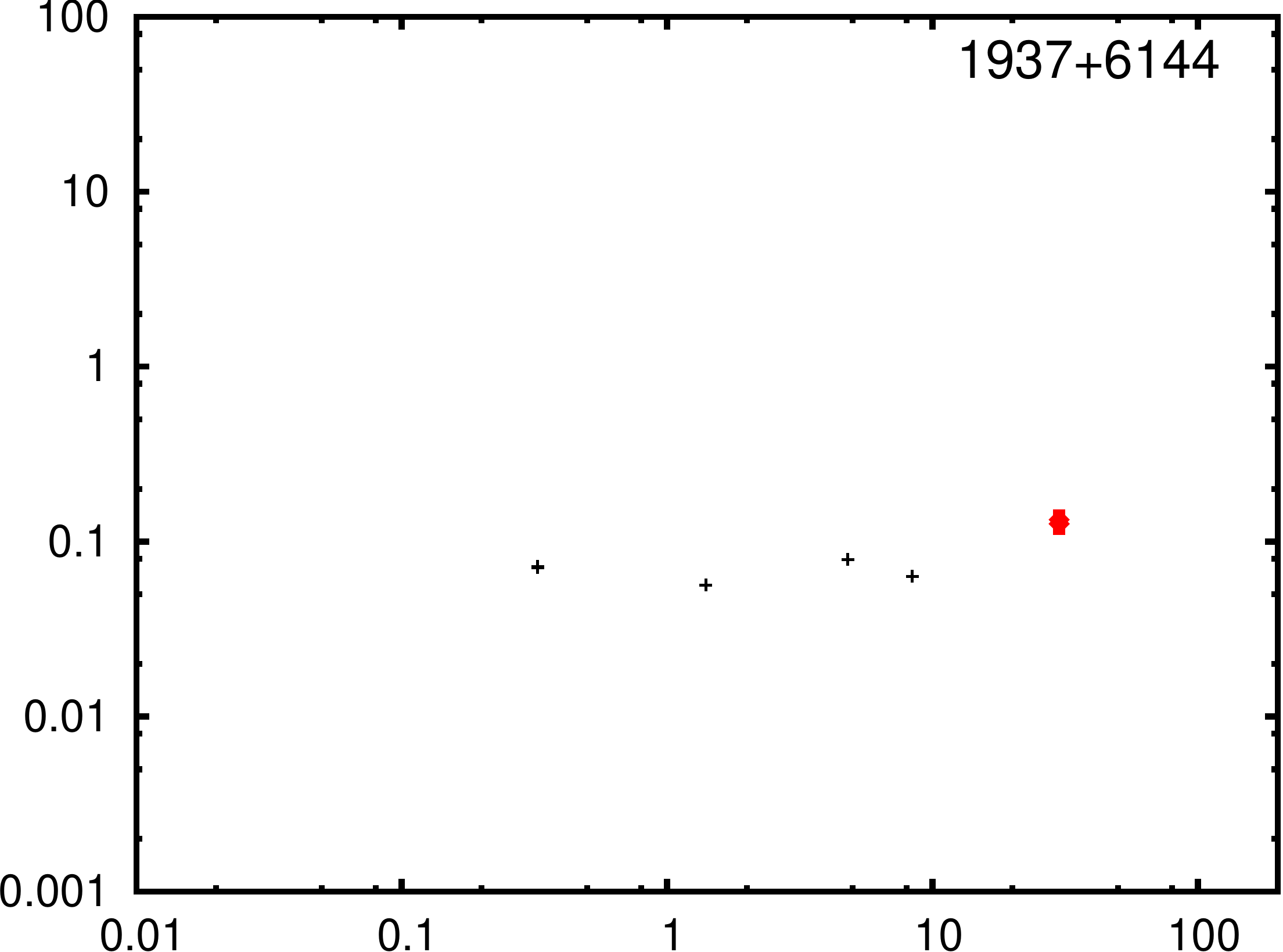}
\end{figure}
\clearpage\begin{figure}
\centering
\includegraphics[scale=0.2]{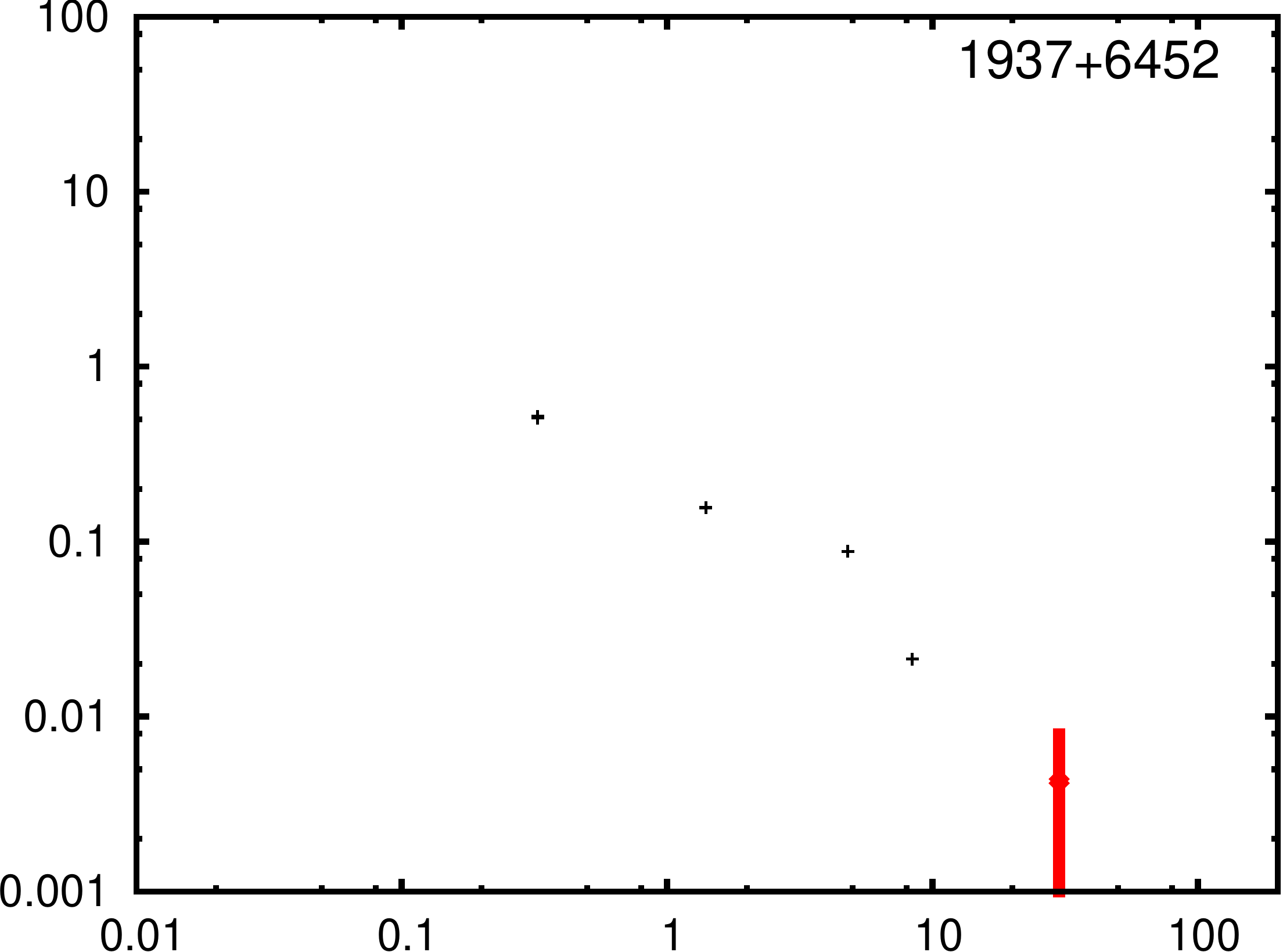}
\includegraphics[scale=0.2]{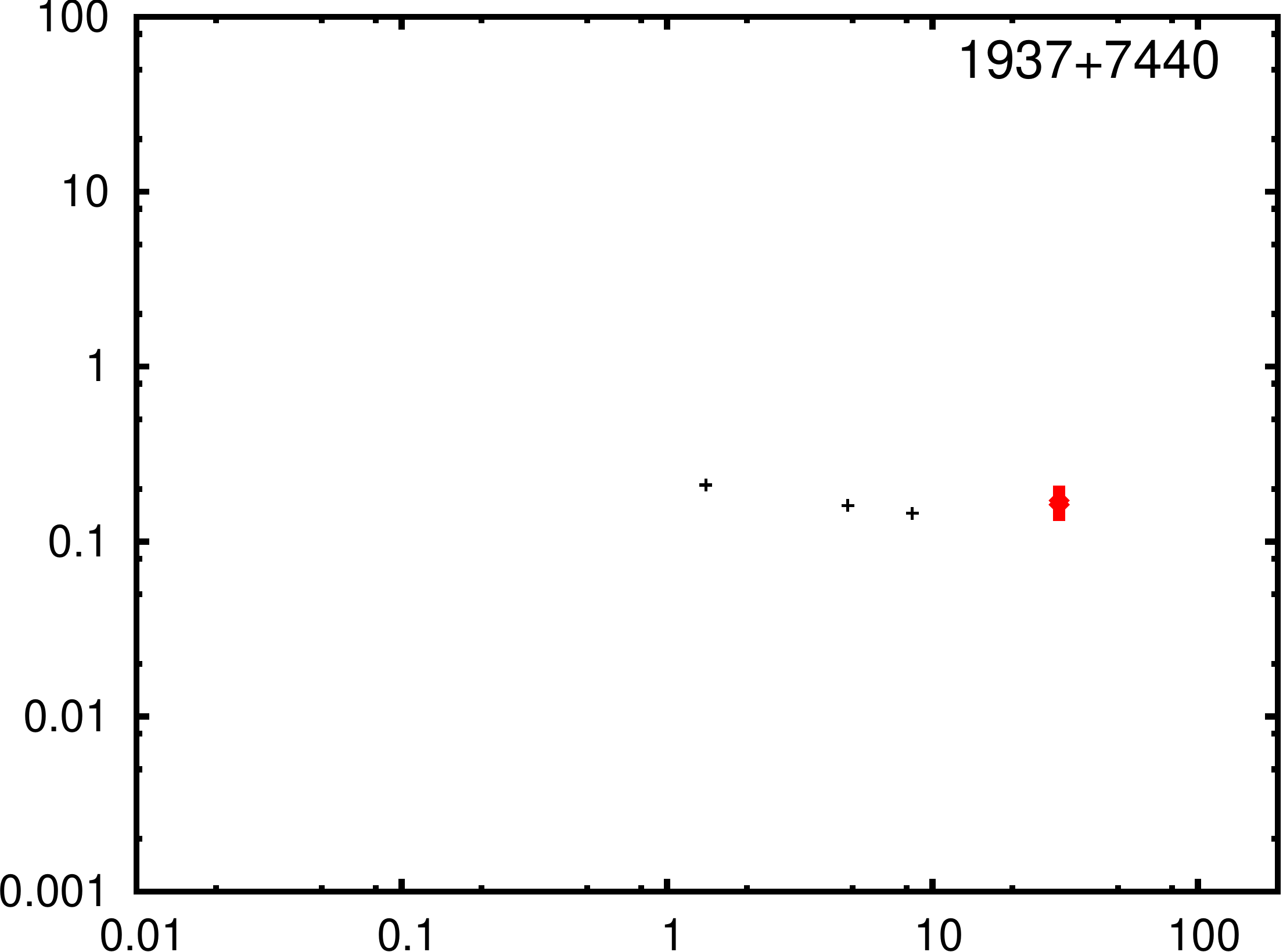}
\includegraphics[scale=0.2]{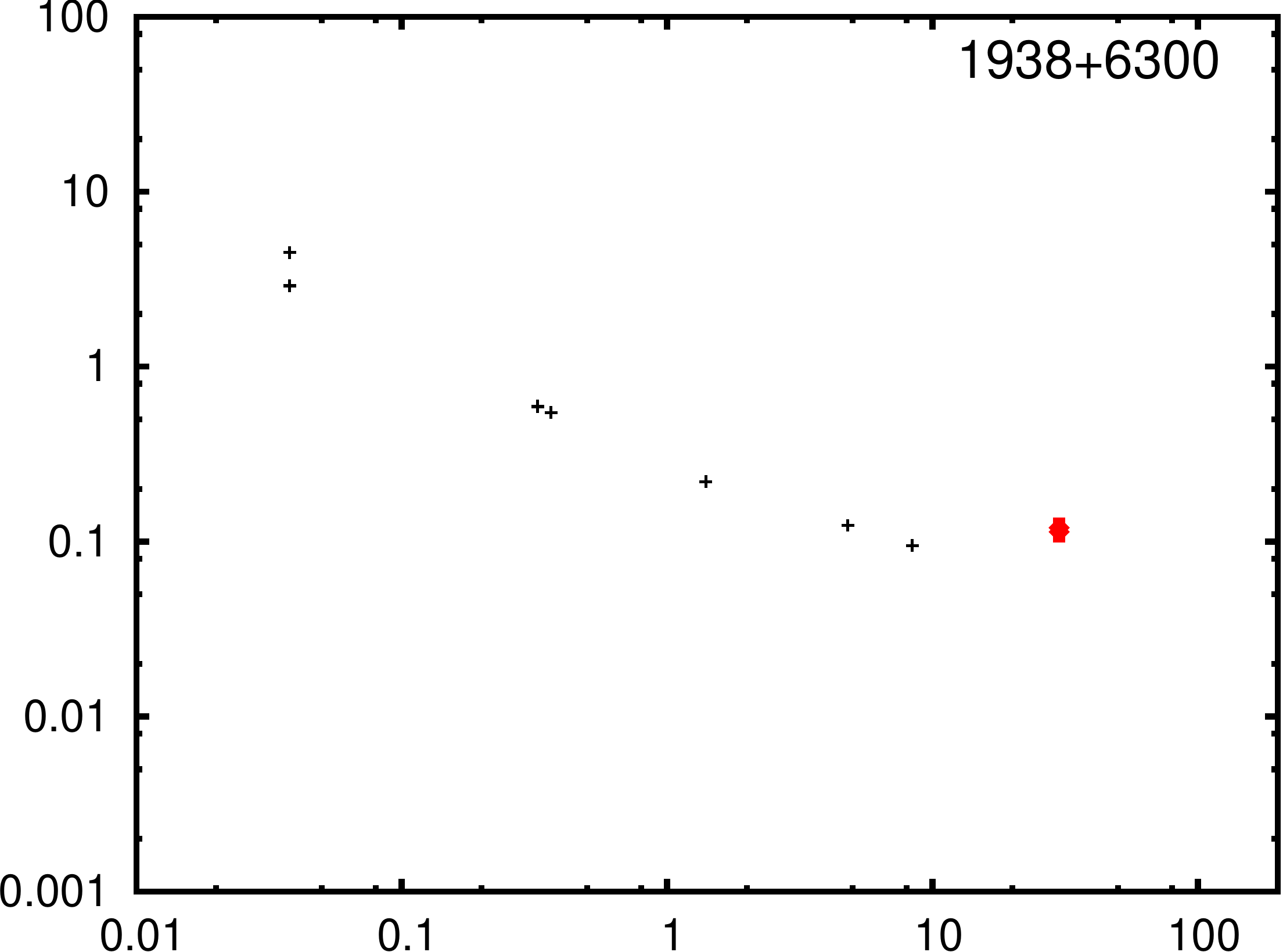}
\includegraphics[scale=0.2]{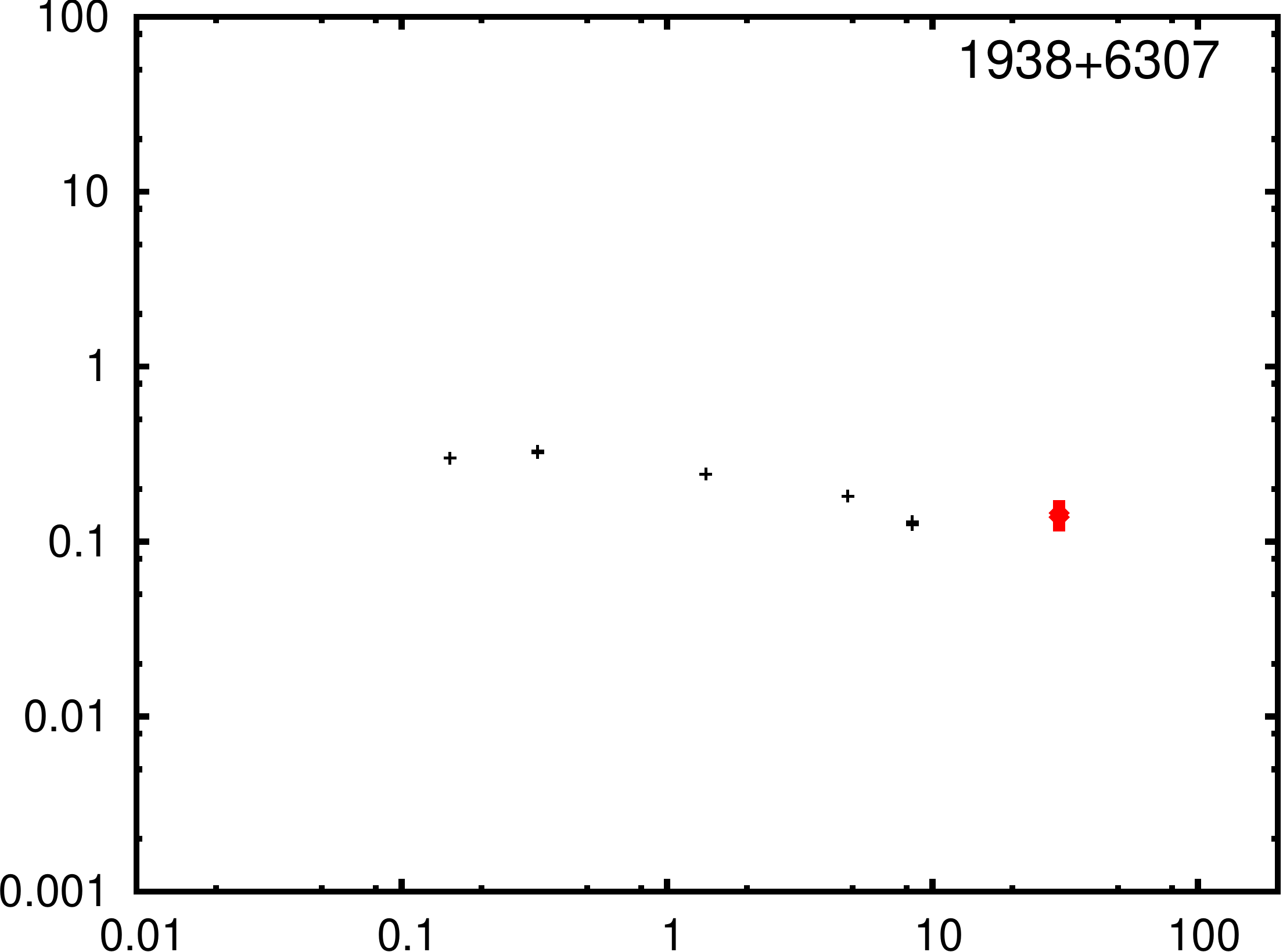}
\includegraphics[scale=0.2]{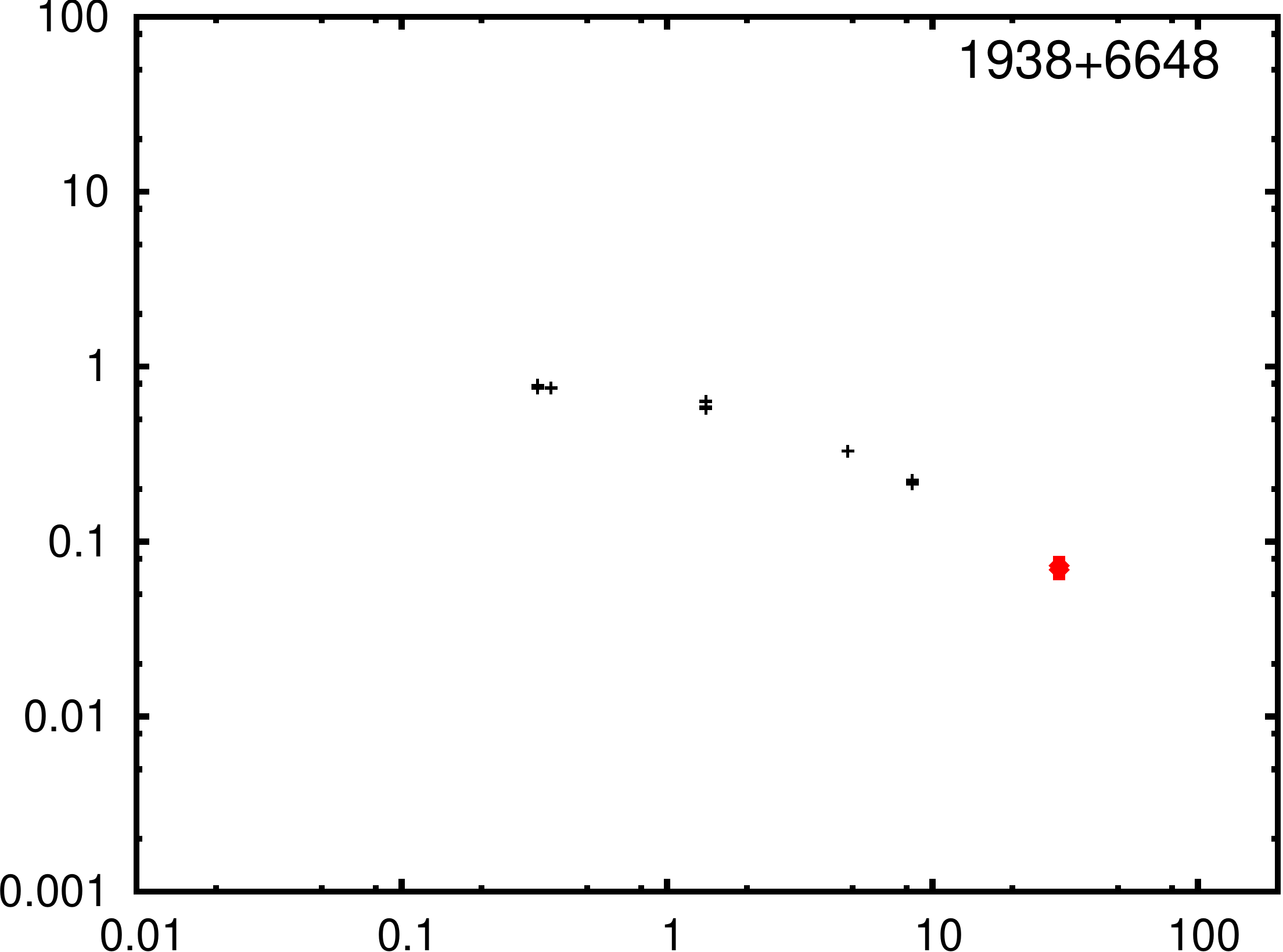}
\includegraphics[scale=0.2]{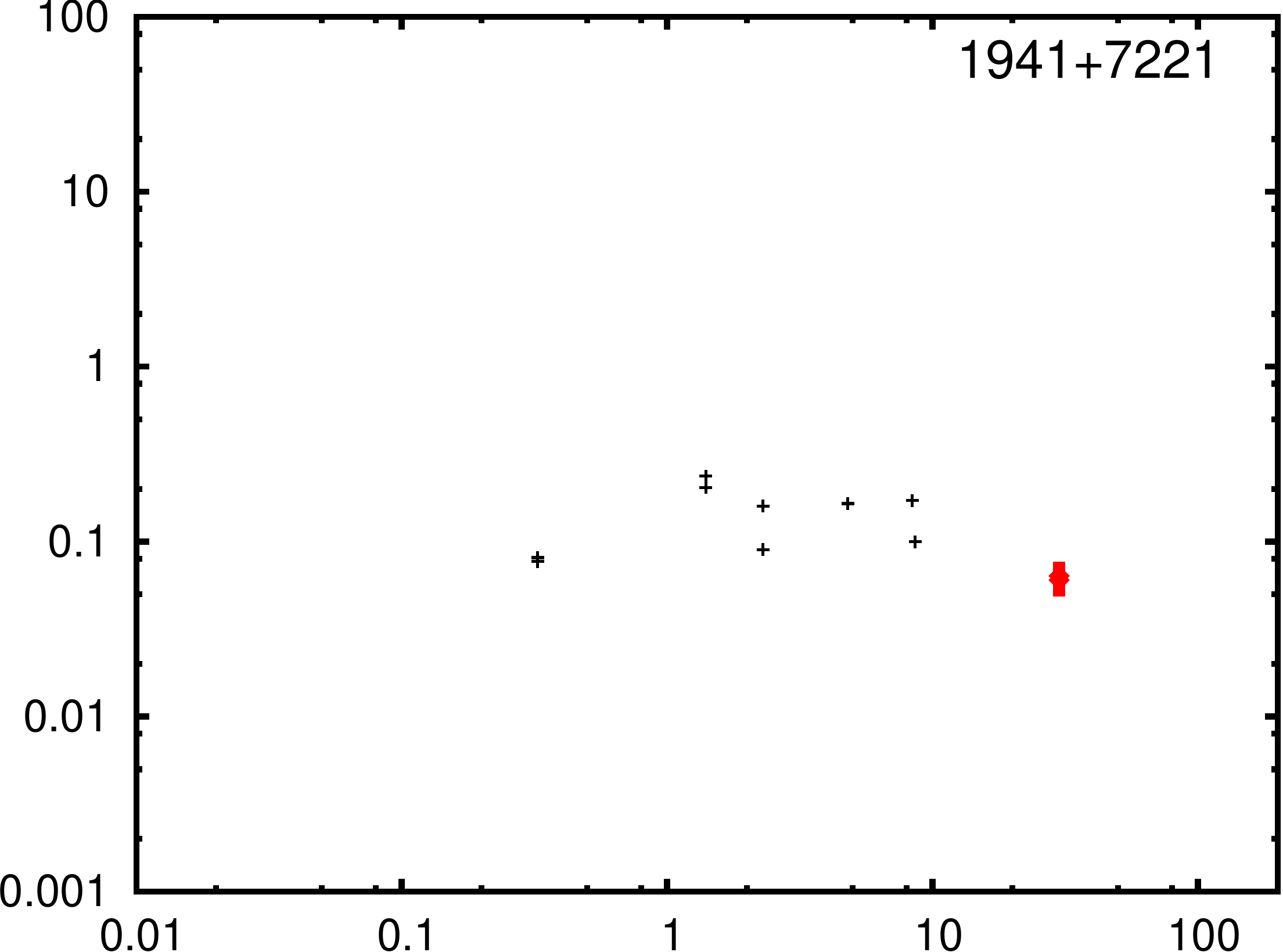}
\includegraphics[scale=0.2]{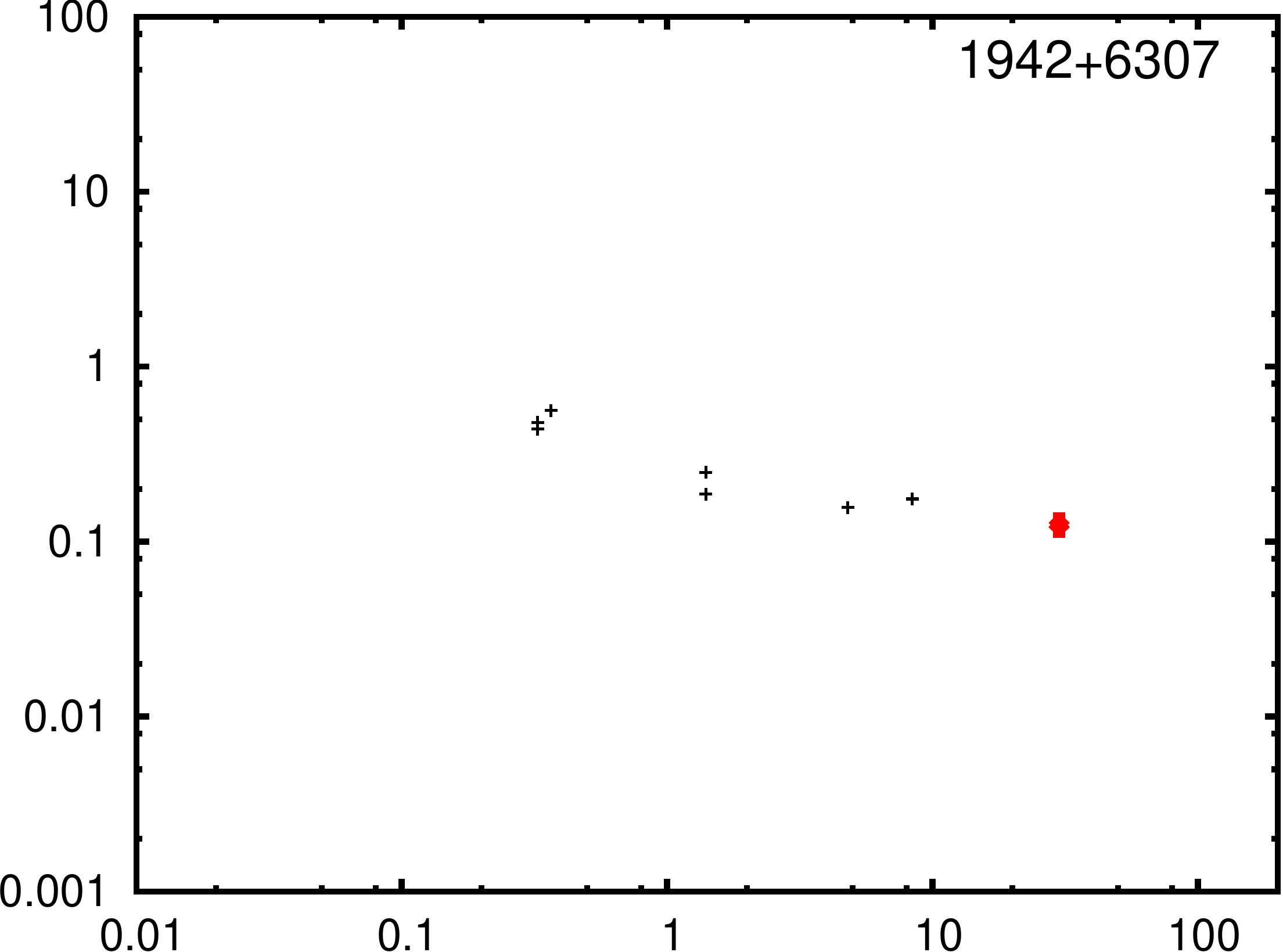}
\includegraphics[scale=0.2]{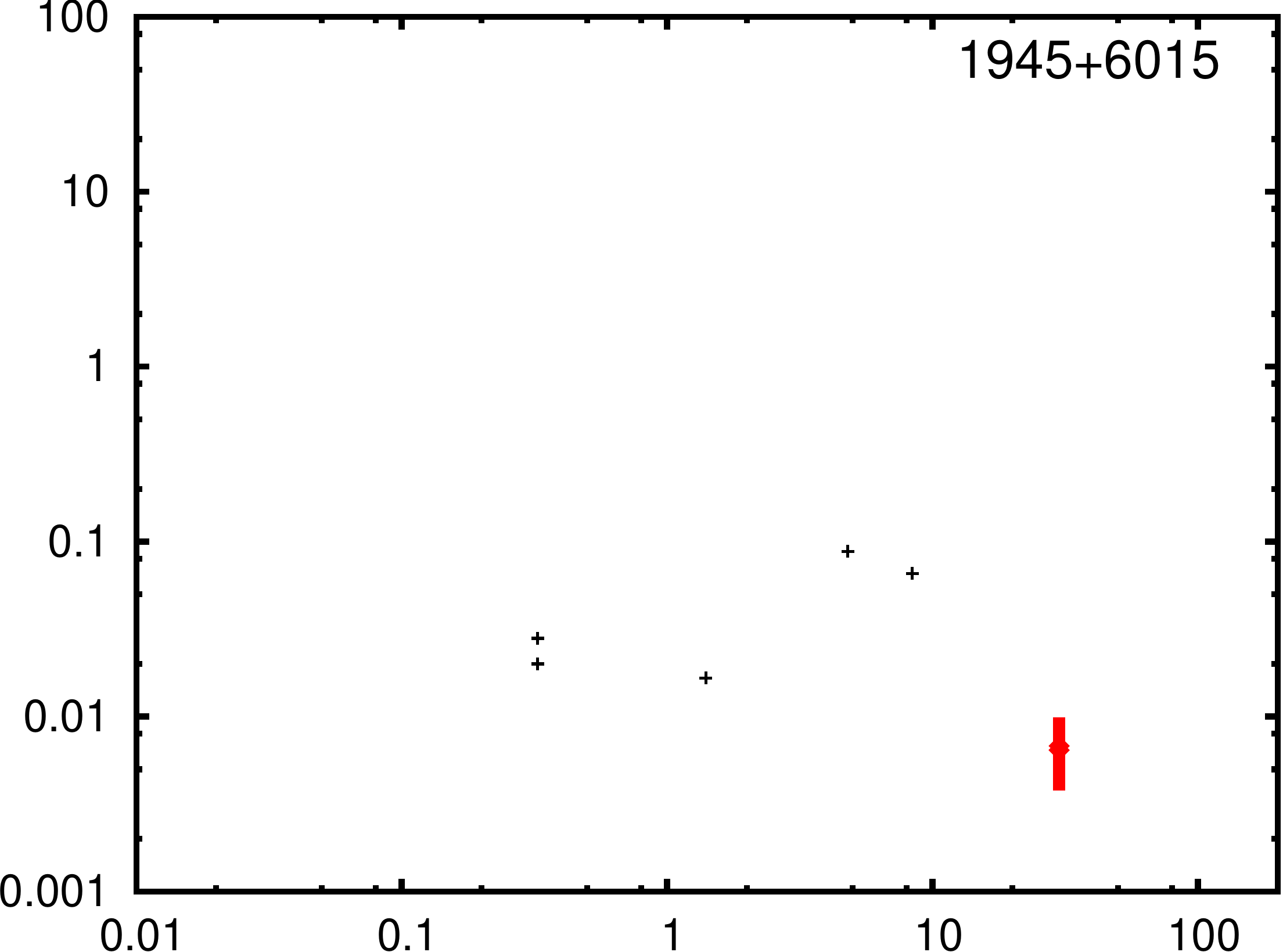}
\includegraphics[scale=0.2]{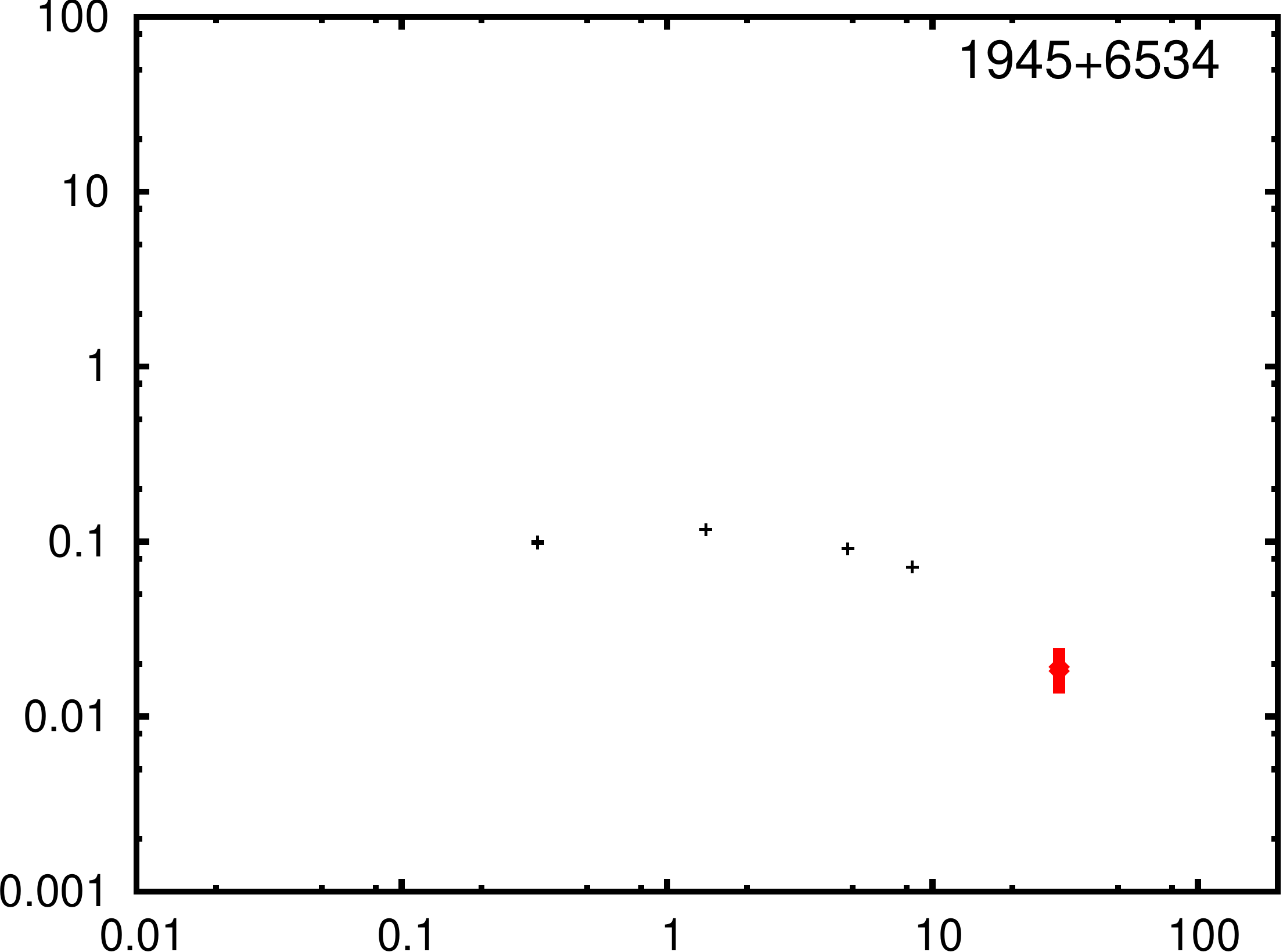}
\includegraphics[scale=0.2]{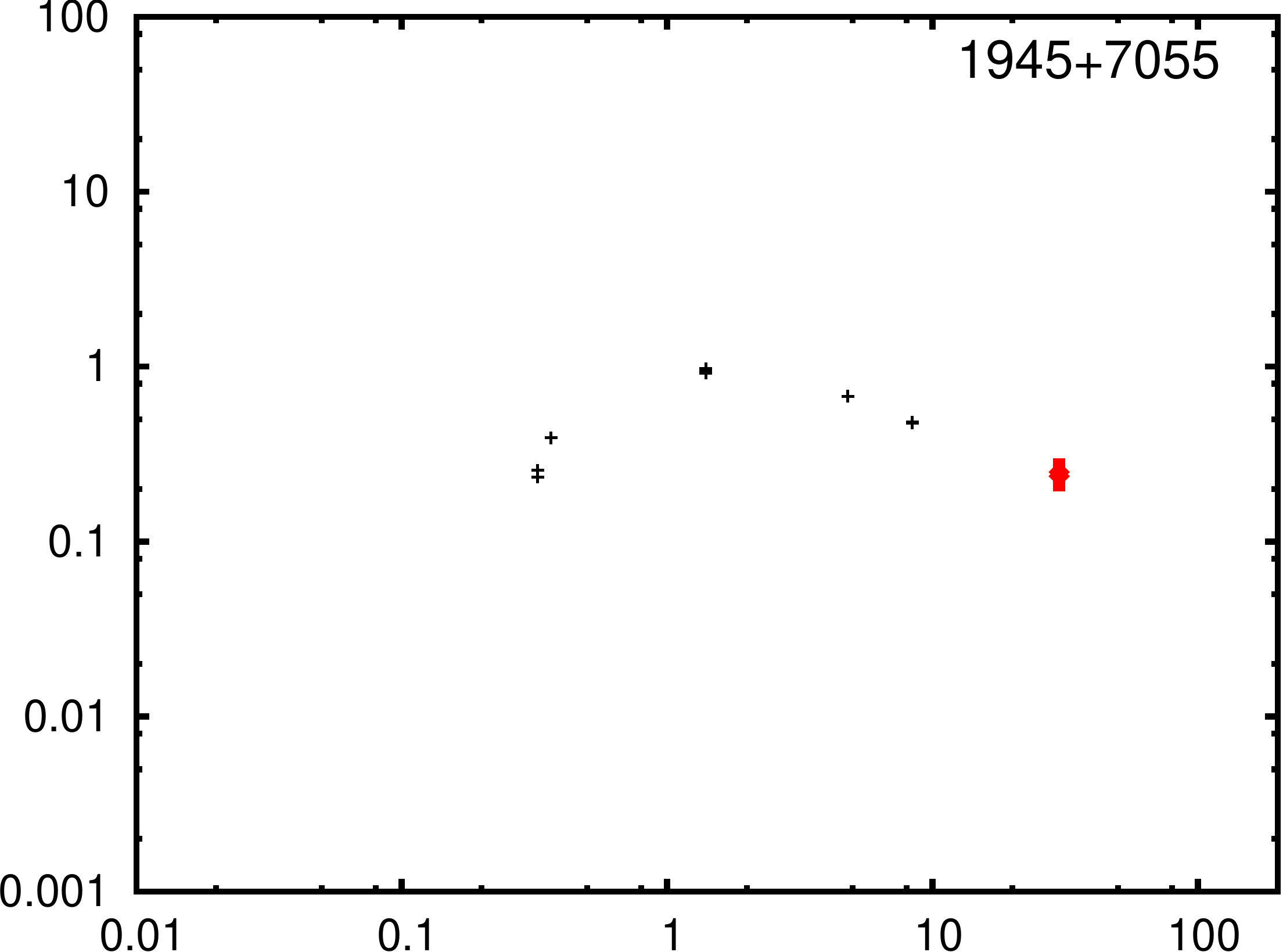}
\includegraphics[scale=0.2]{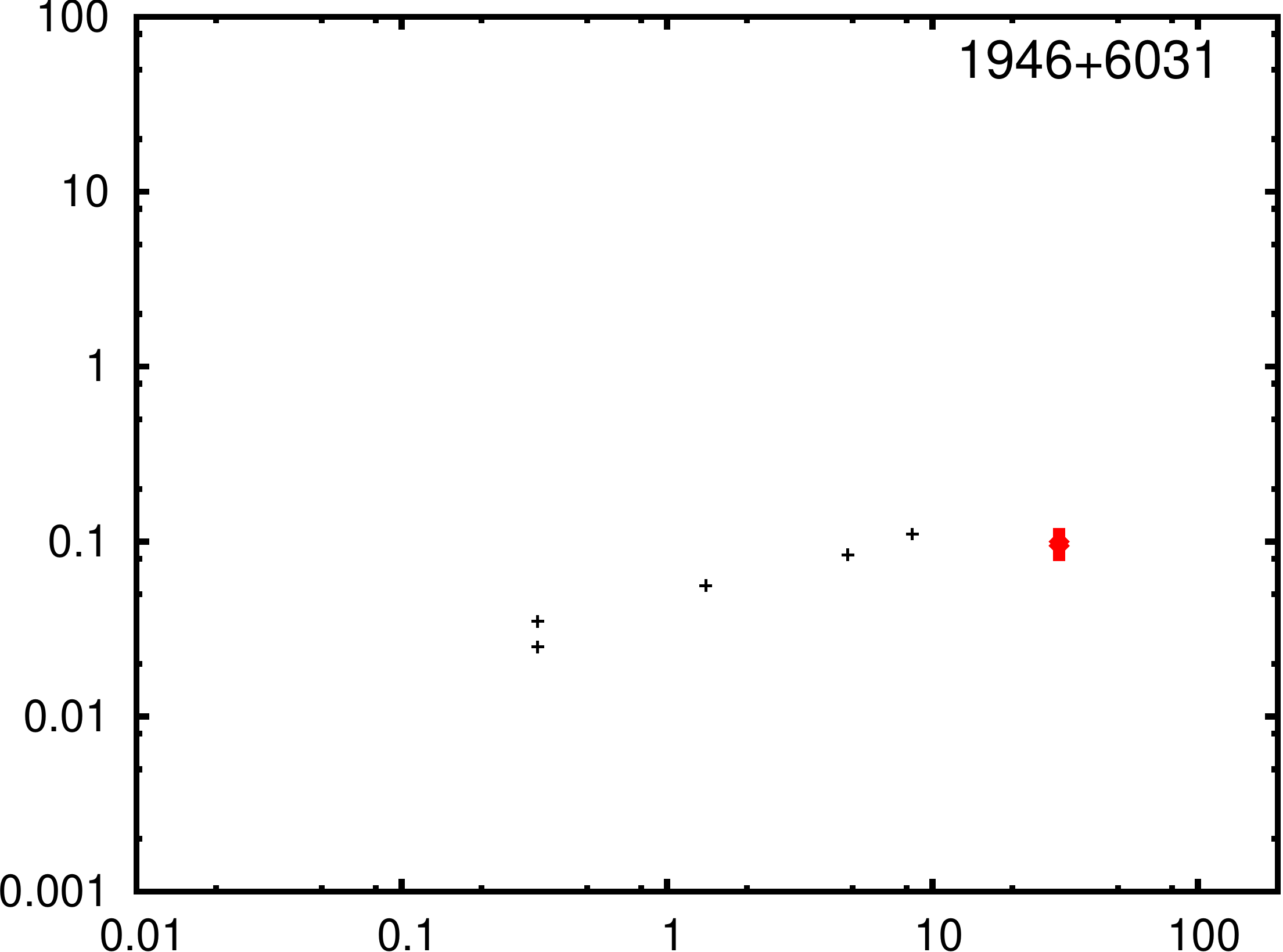}
\includegraphics[scale=0.2]{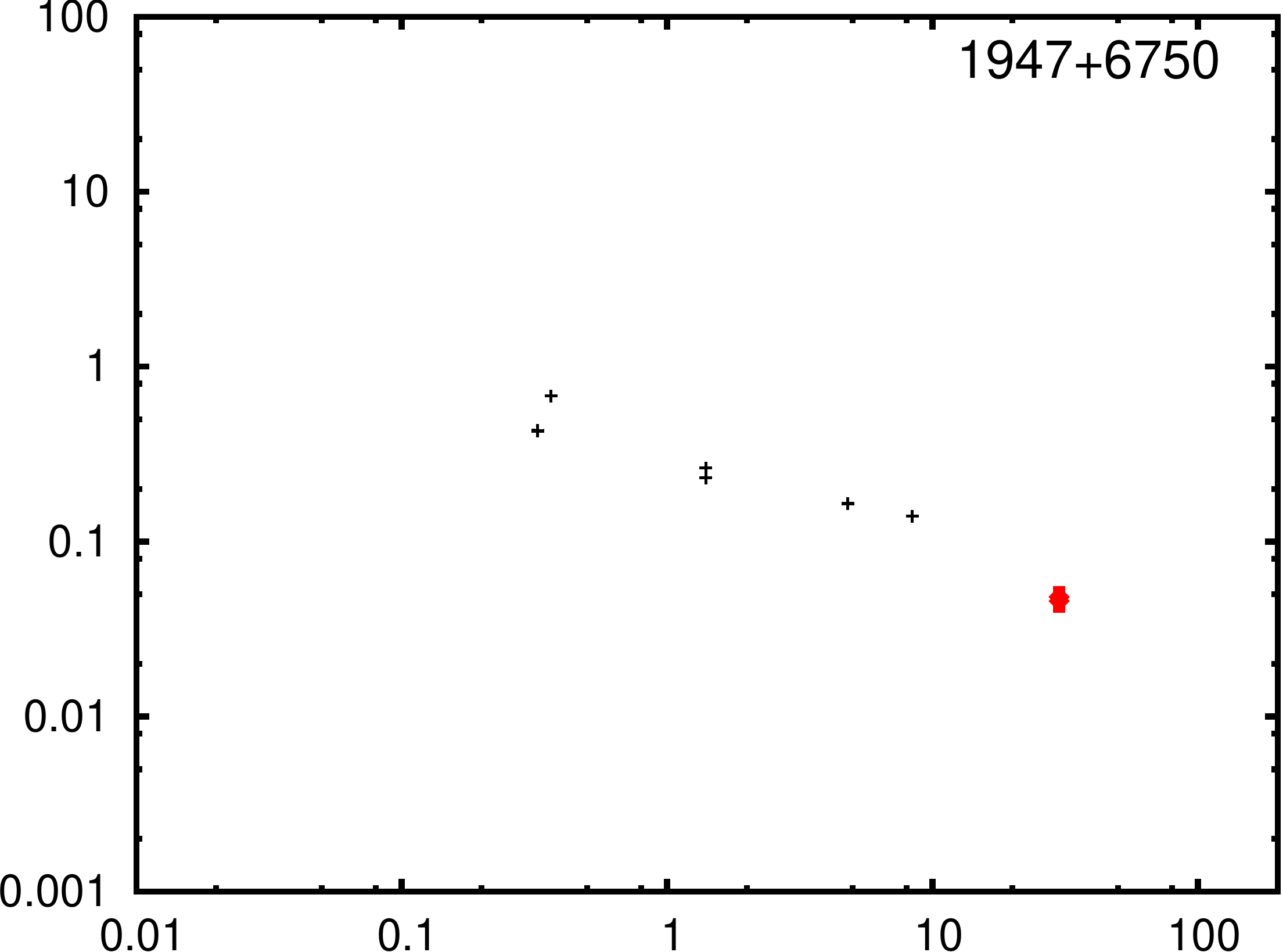}
\includegraphics[scale=0.2]{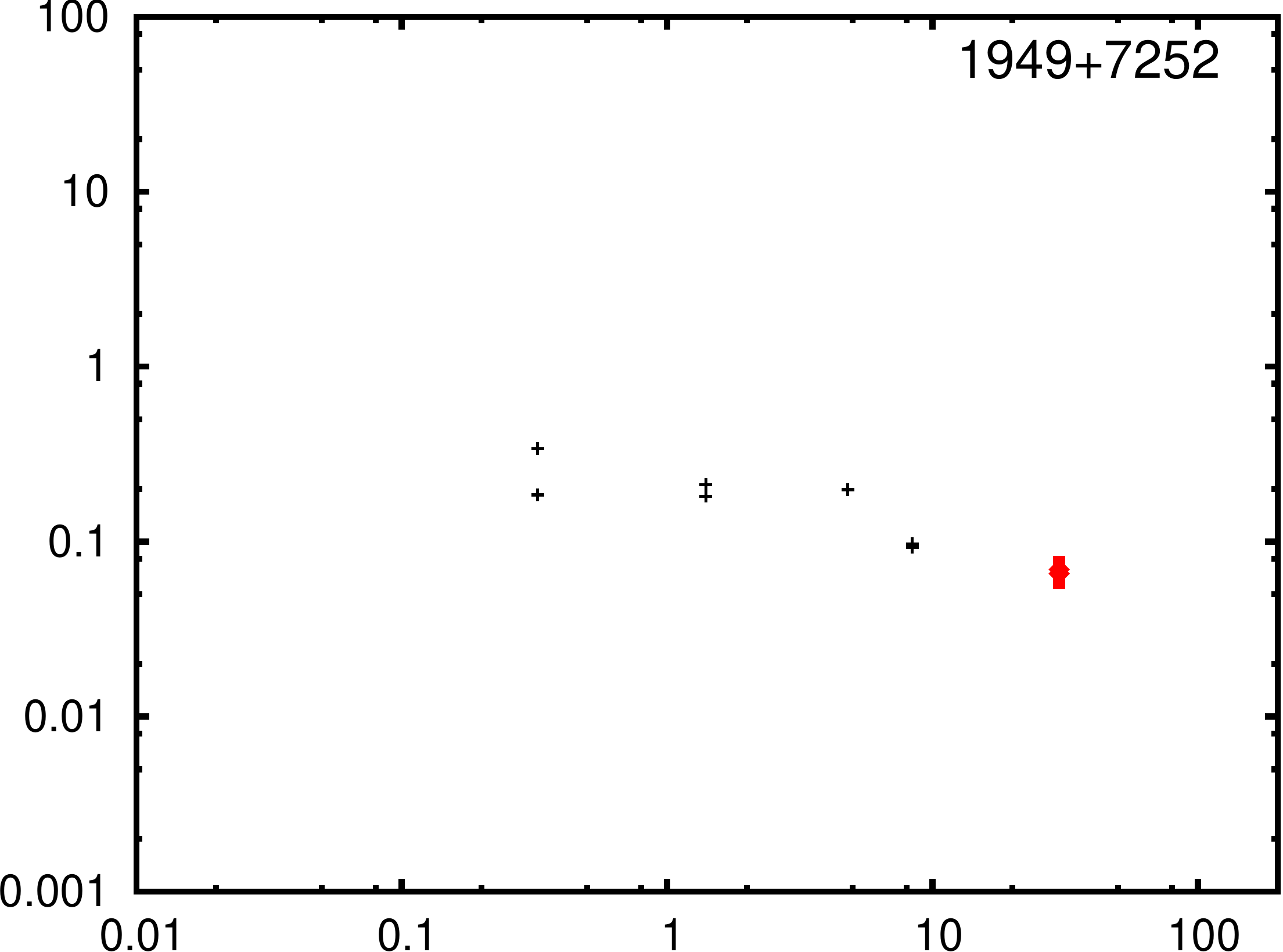}
\includegraphics[scale=0.2]{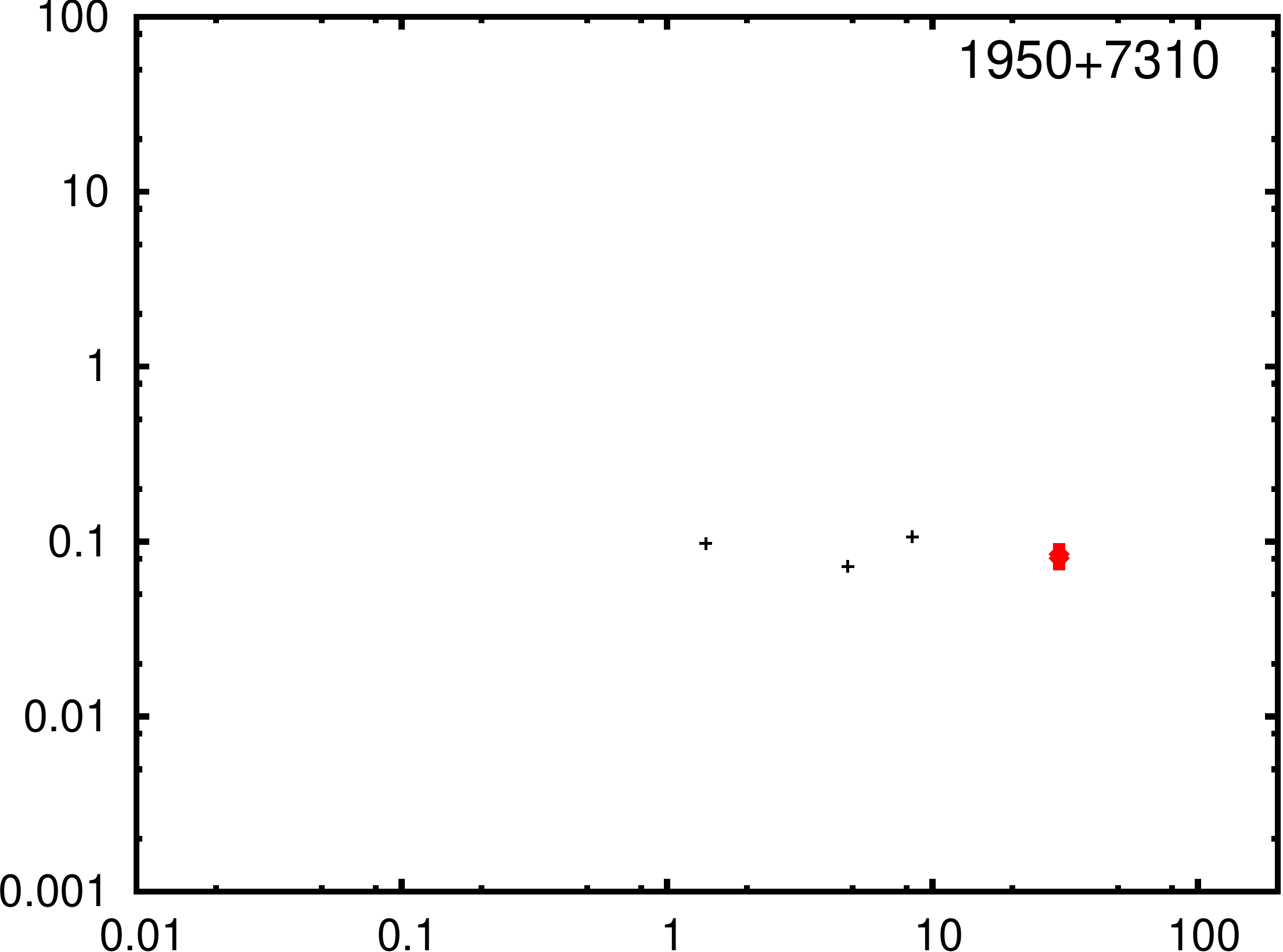}
\includegraphics[scale=0.2]{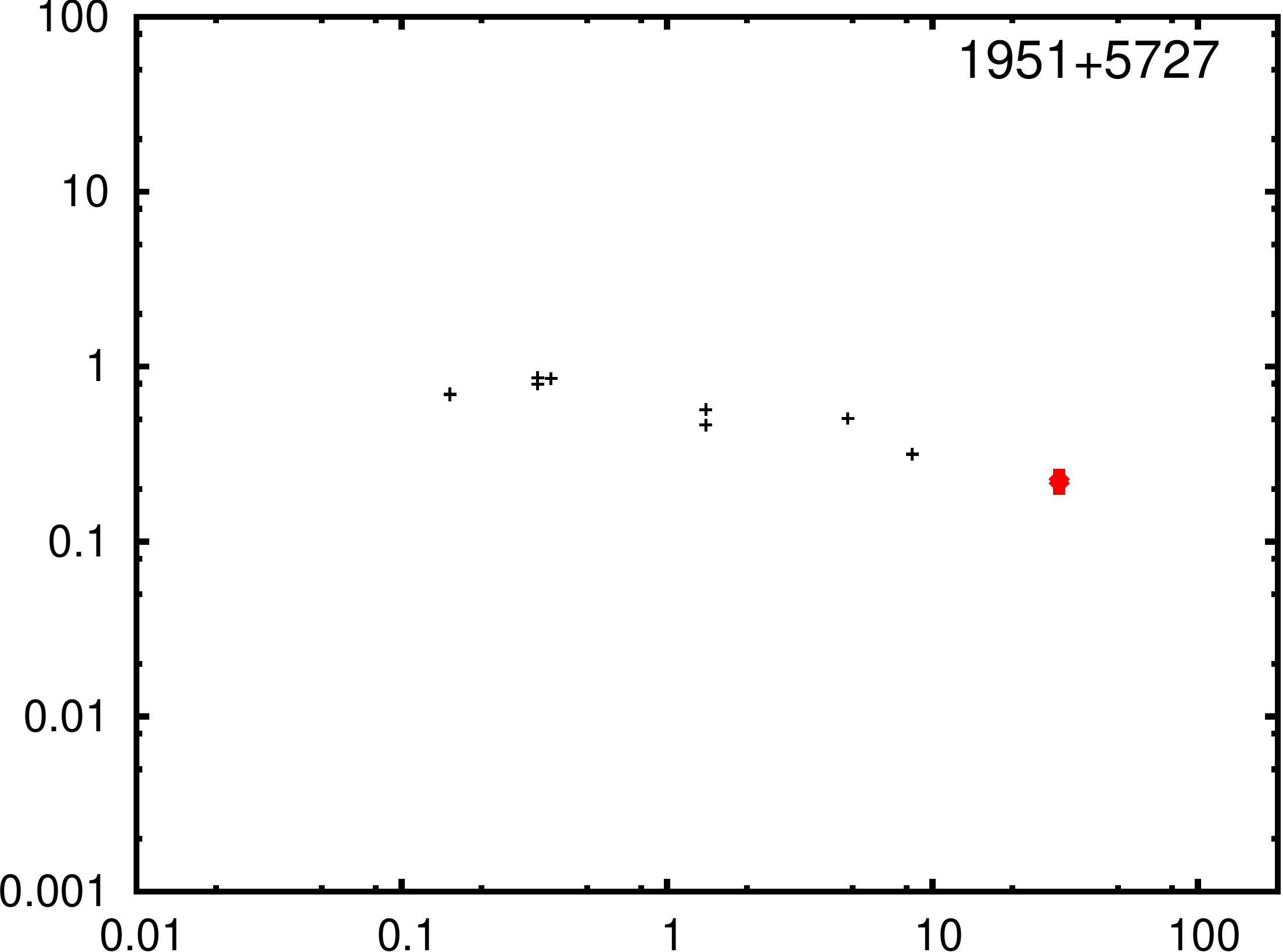}
\includegraphics[scale=0.2]{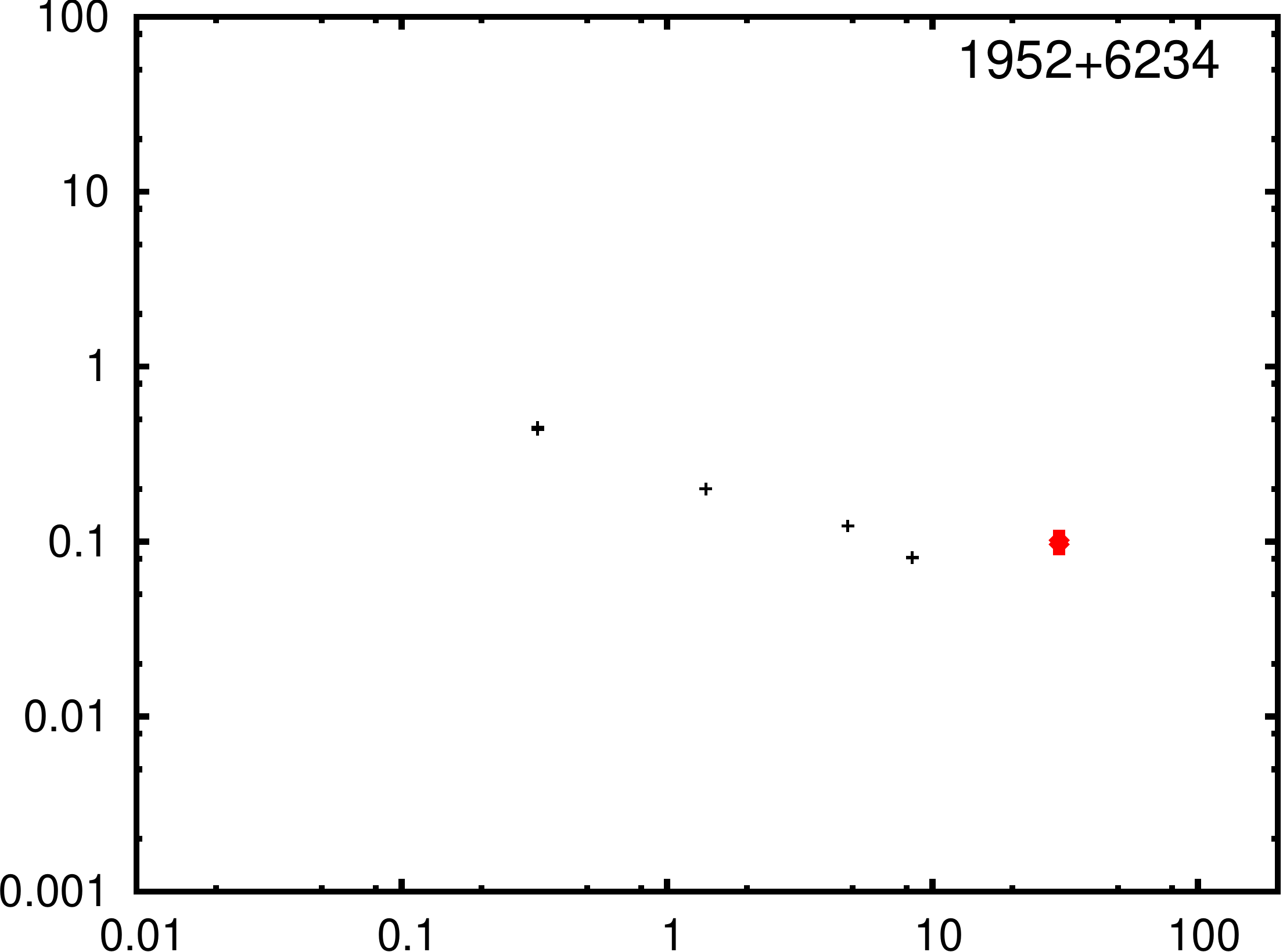}
\includegraphics[scale=0.2]{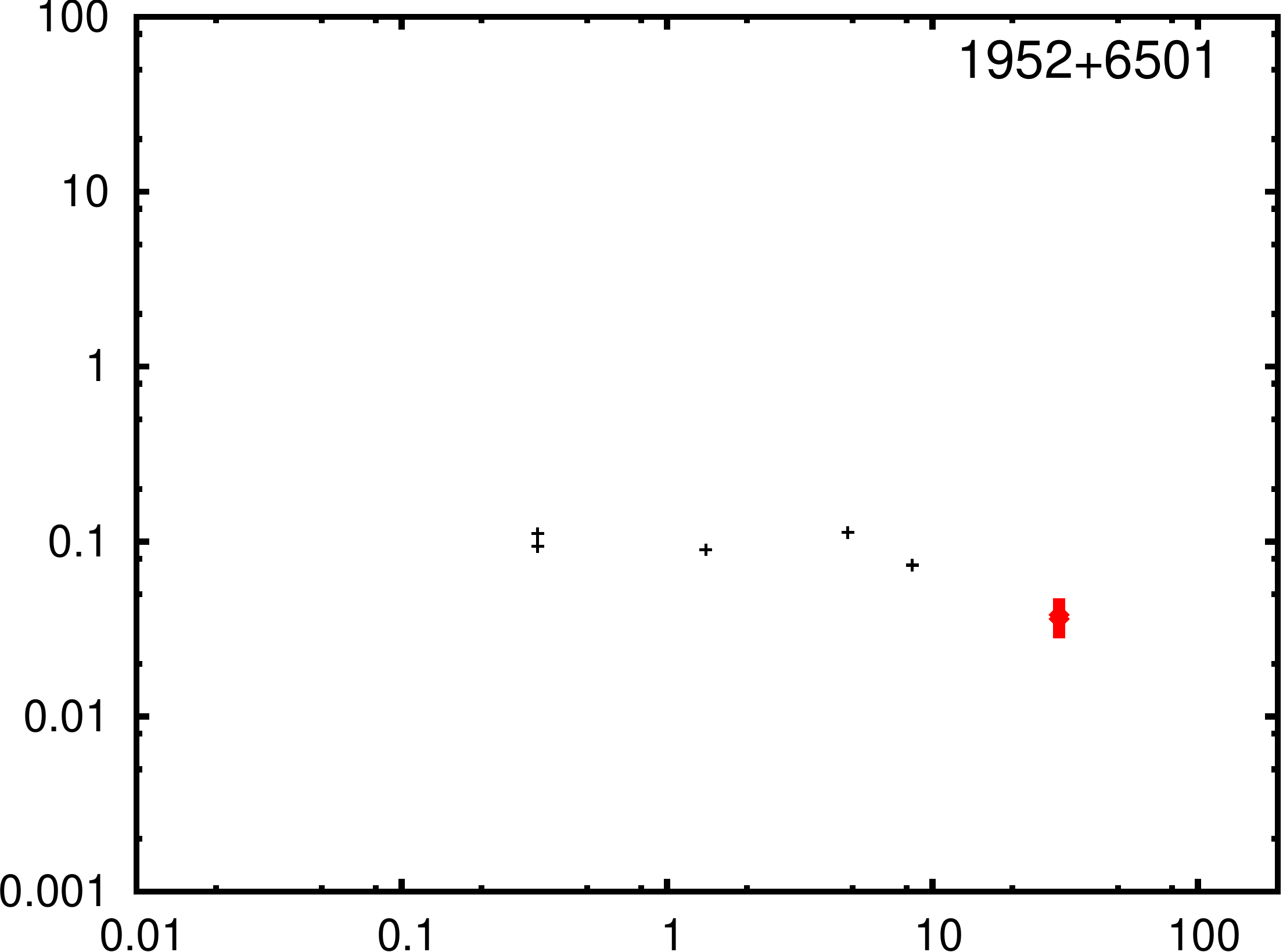}
\includegraphics[scale=0.2]{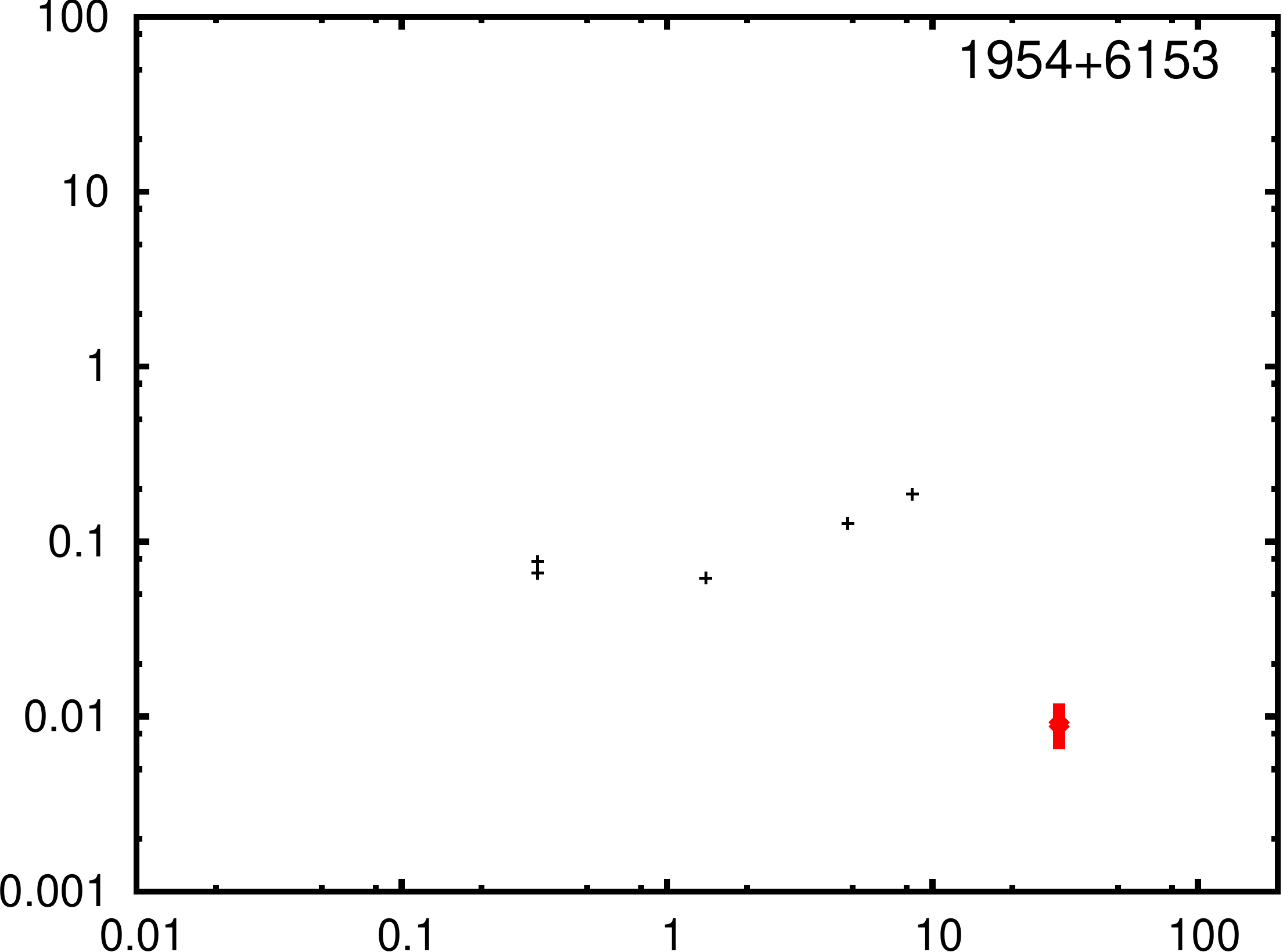}
\end{figure}
\clearpage\begin{figure}
\centering
\includegraphics[scale=0.2]{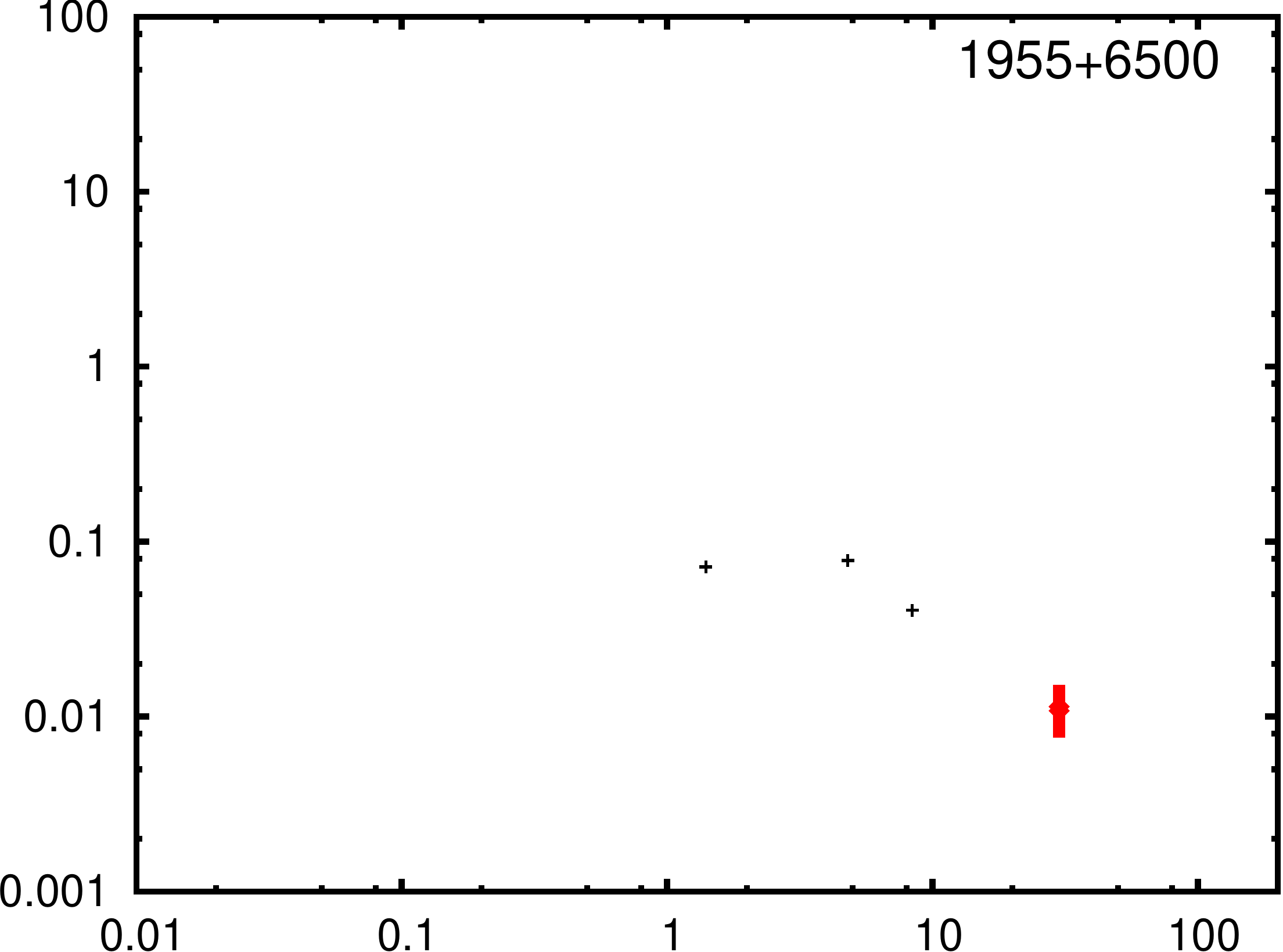}
\includegraphics[scale=0.2]{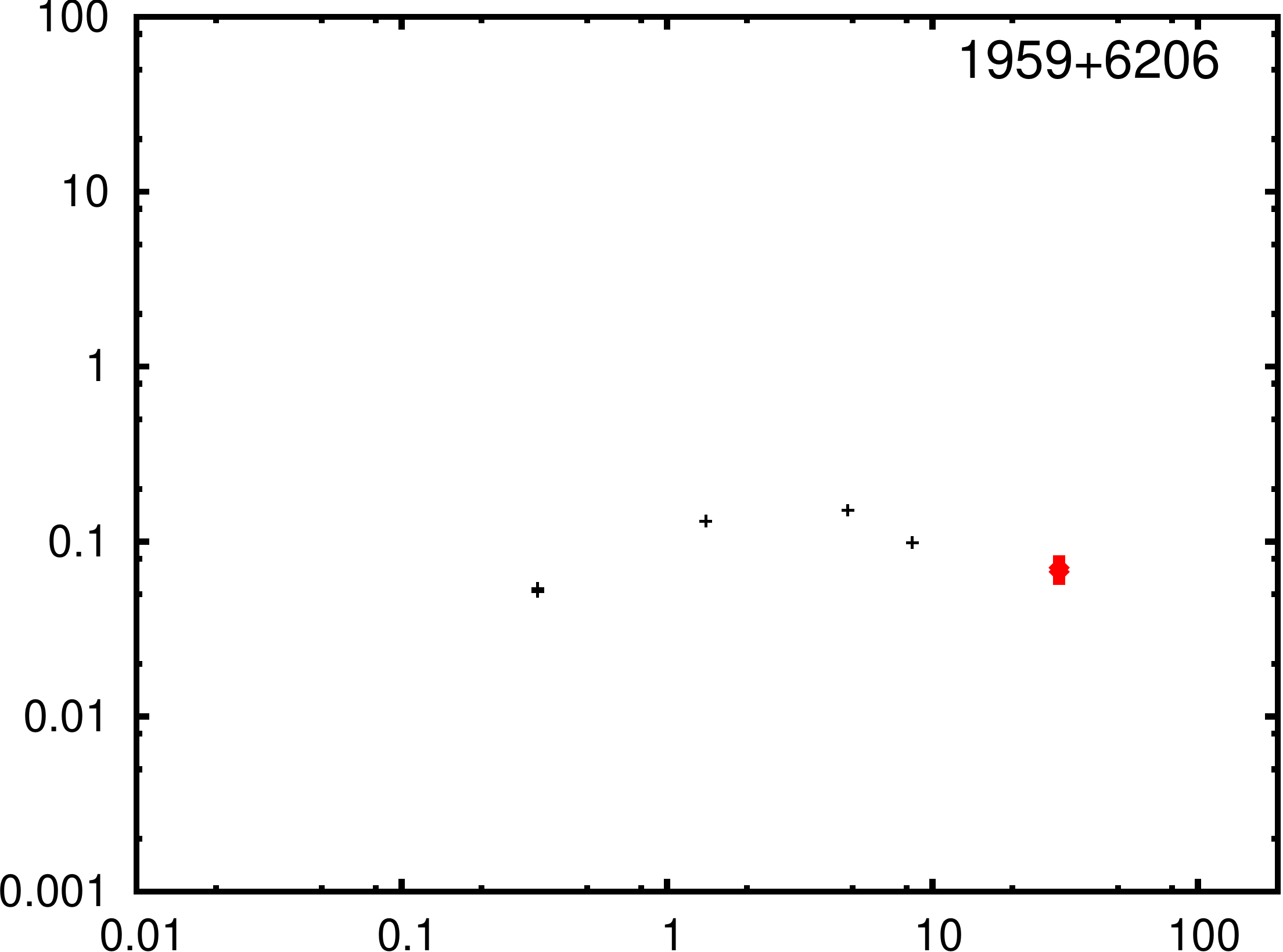}
\includegraphics[scale=0.2]{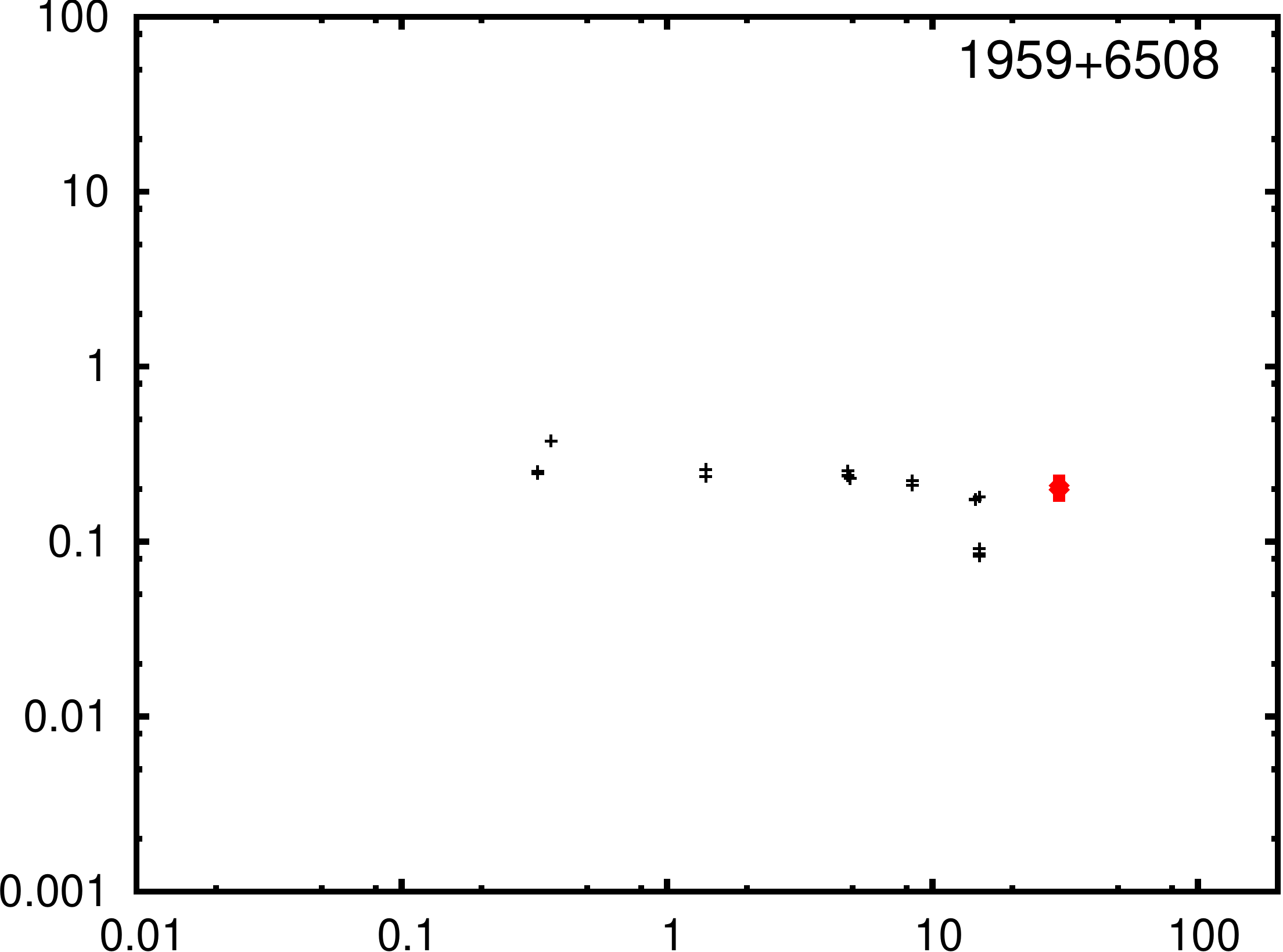}
\includegraphics[scale=0.2]{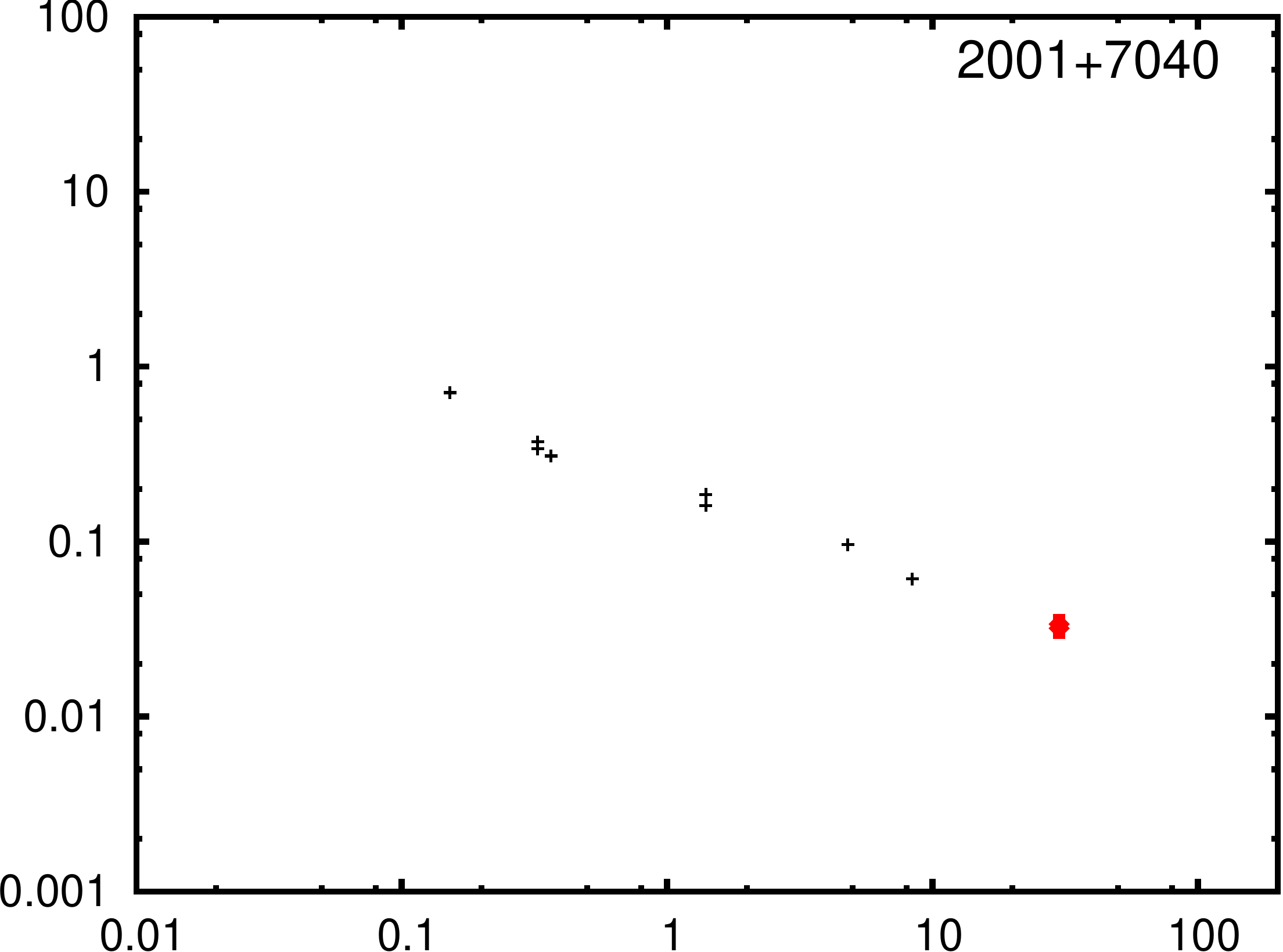}
\includegraphics[scale=0.2]{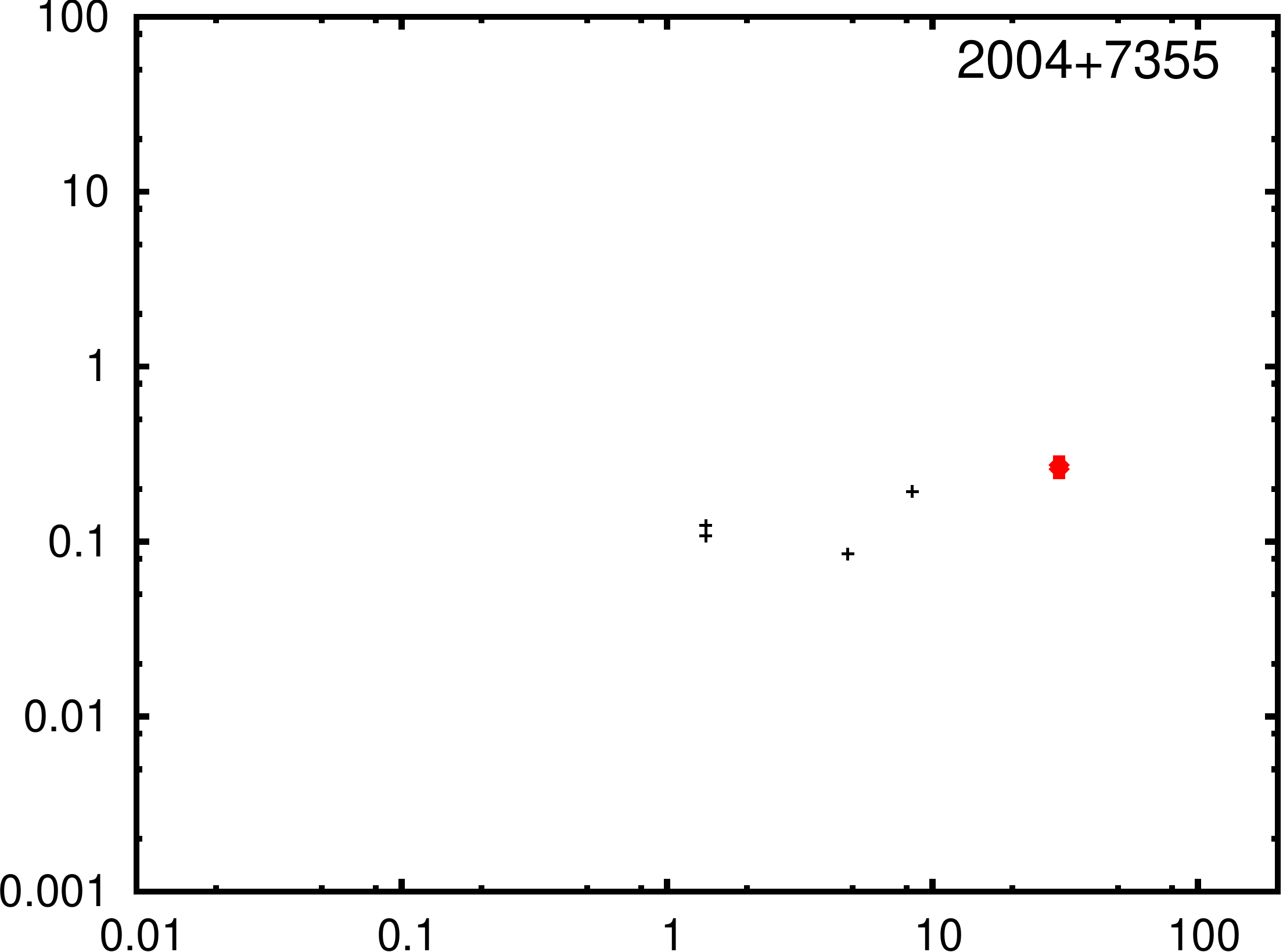}
\includegraphics[scale=0.2]{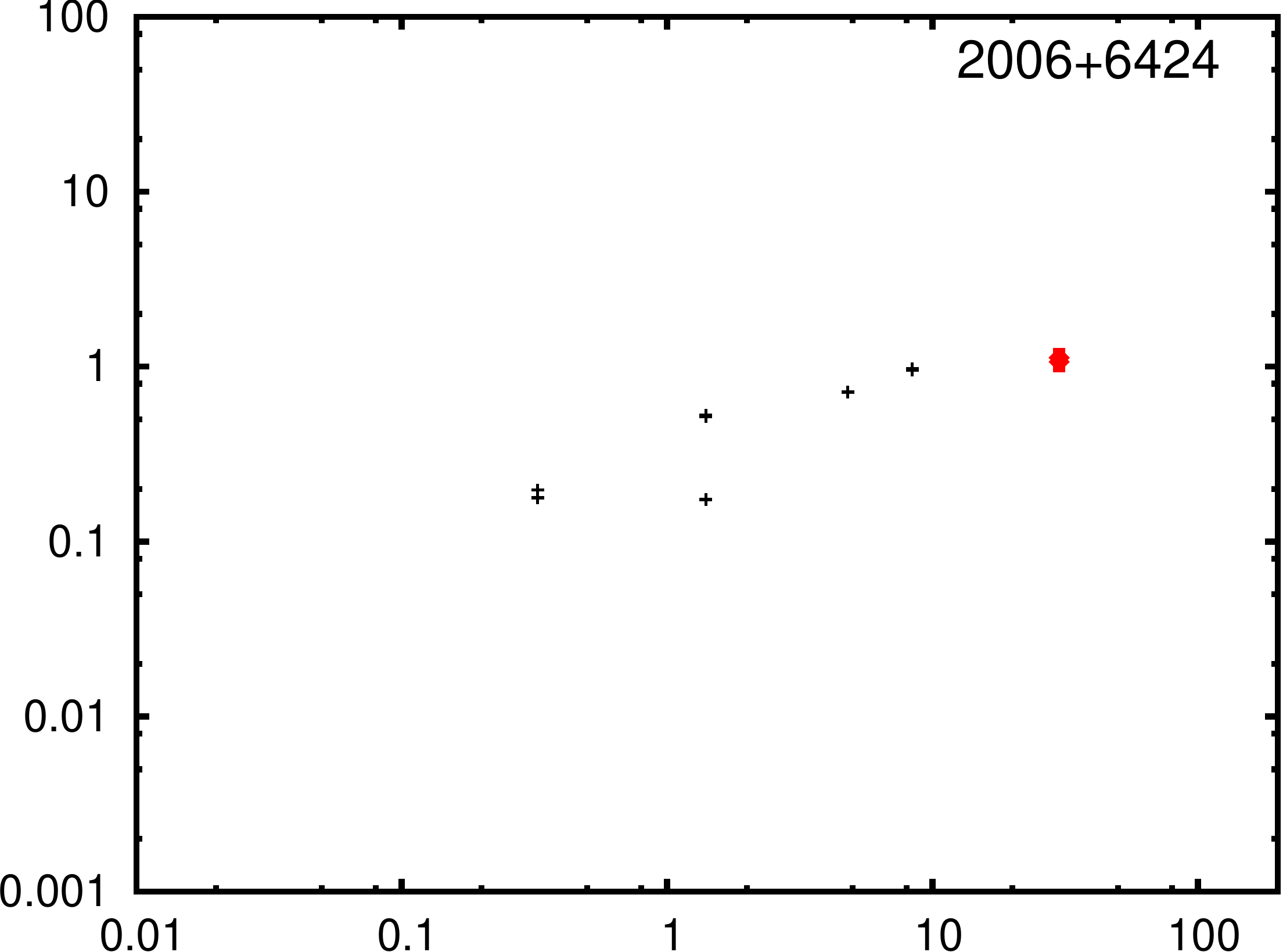}
\includegraphics[scale=0.2]{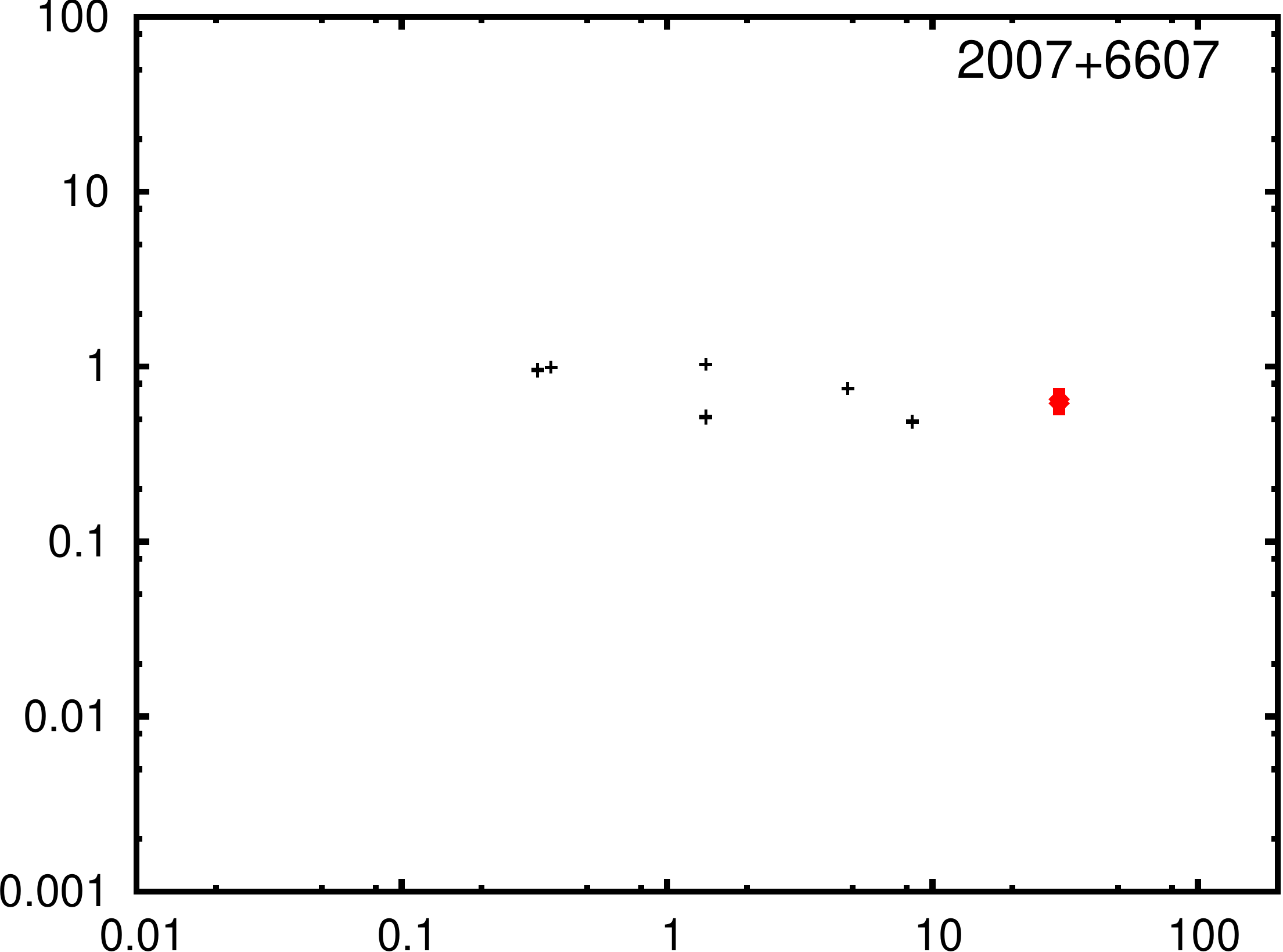}
\includegraphics[scale=0.2]{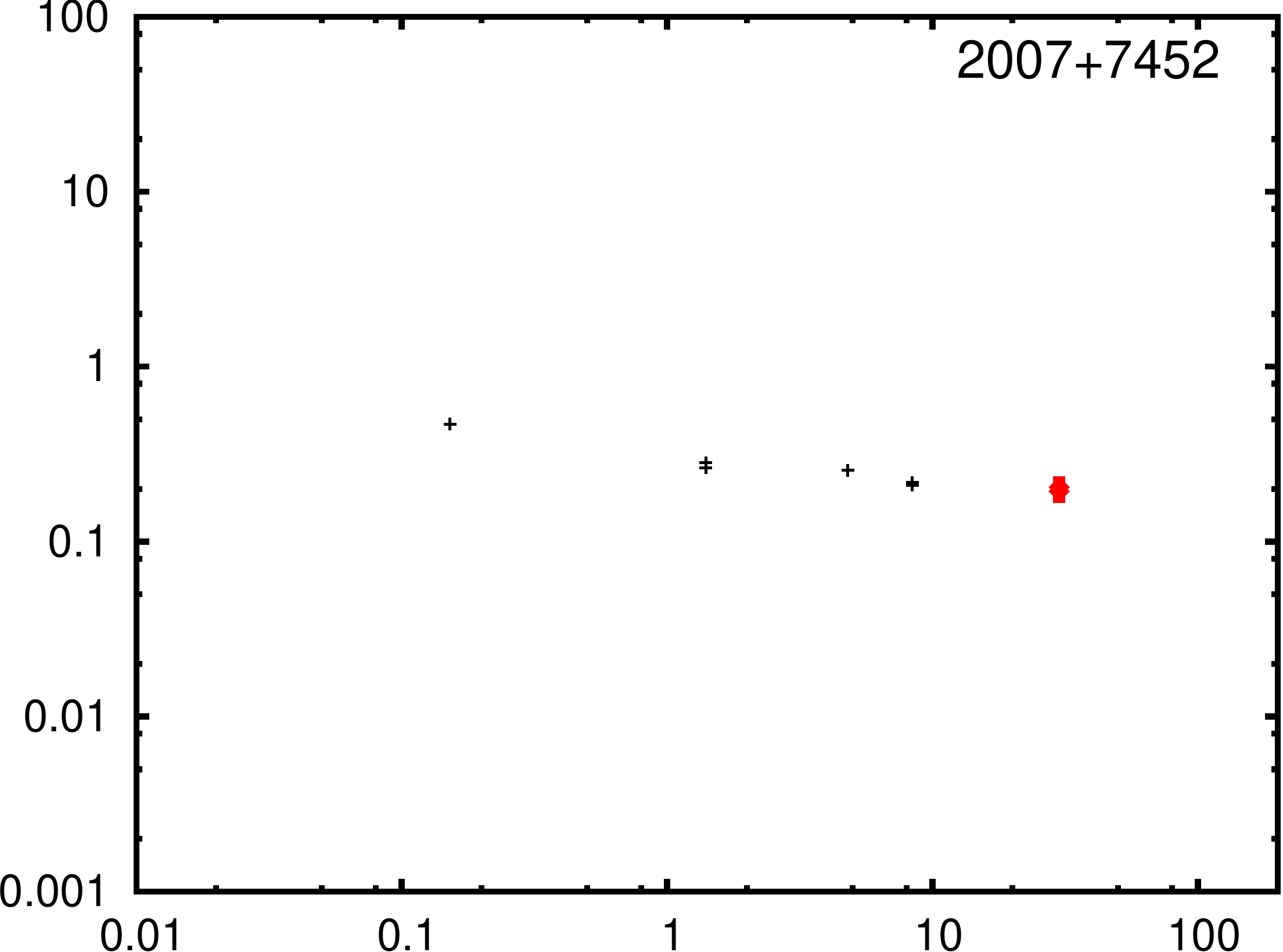}
\includegraphics[scale=0.2]{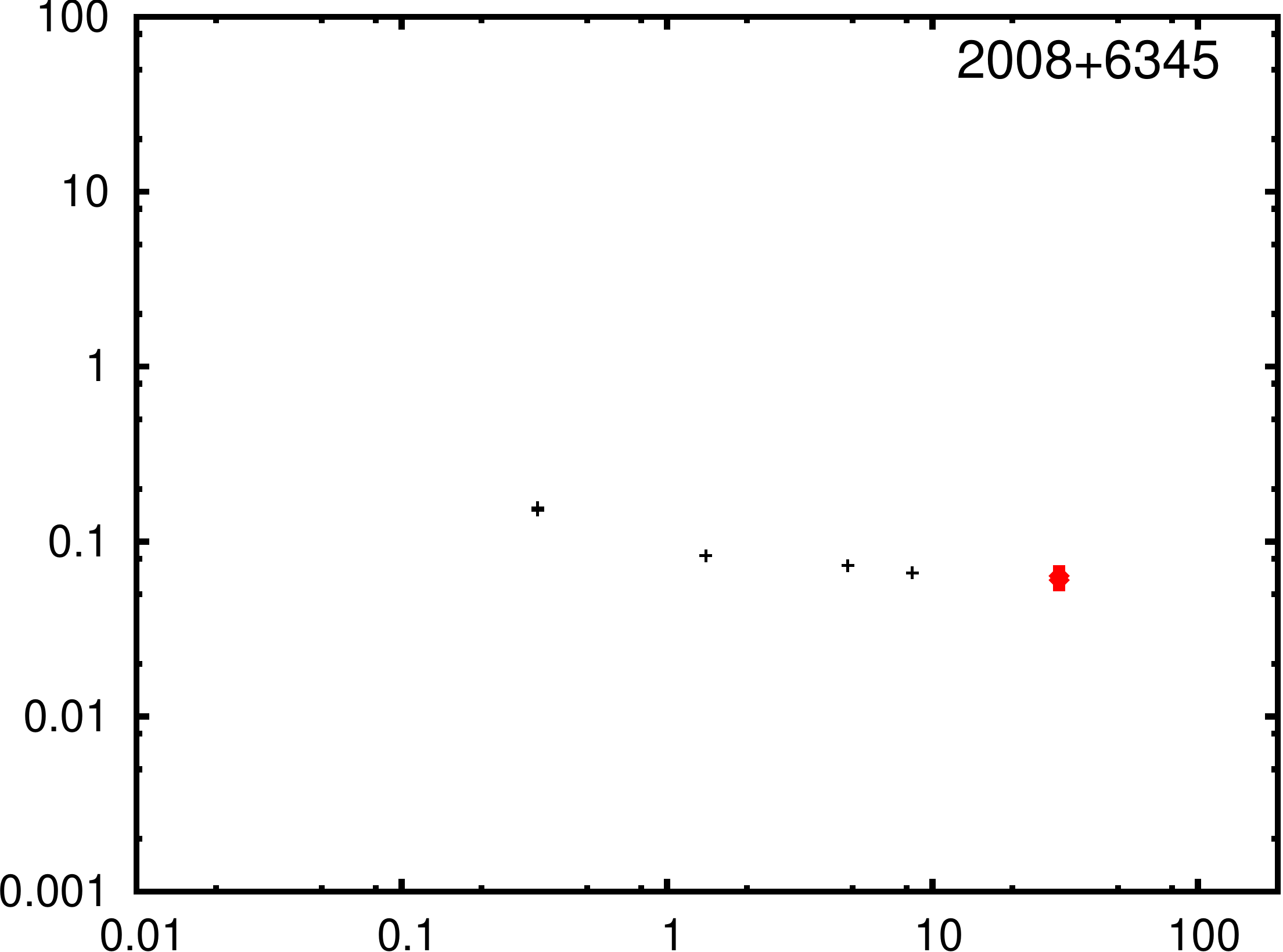}
\includegraphics[scale=0.2]{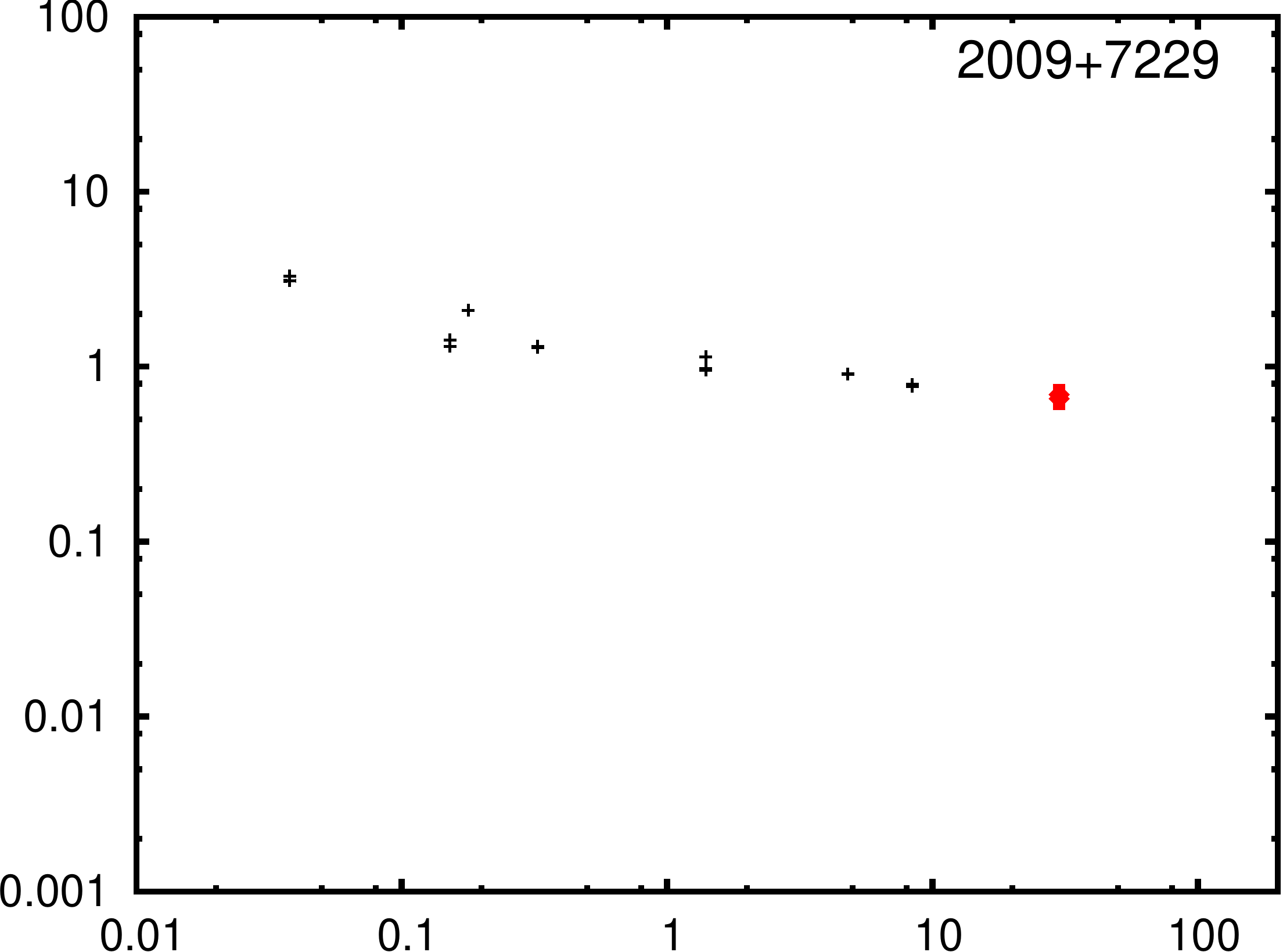}
\includegraphics[scale=0.2]{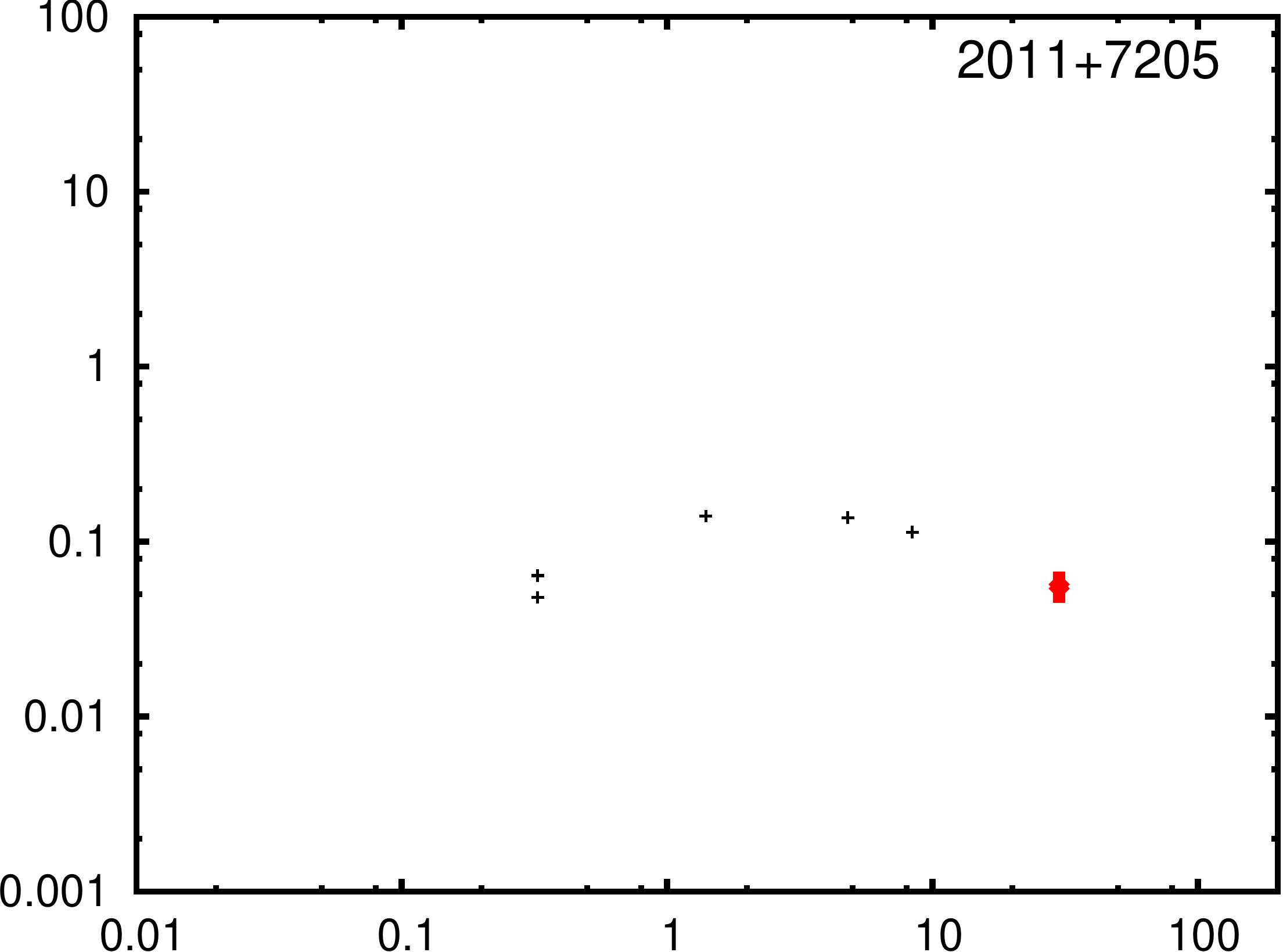}
\includegraphics[scale=0.2]{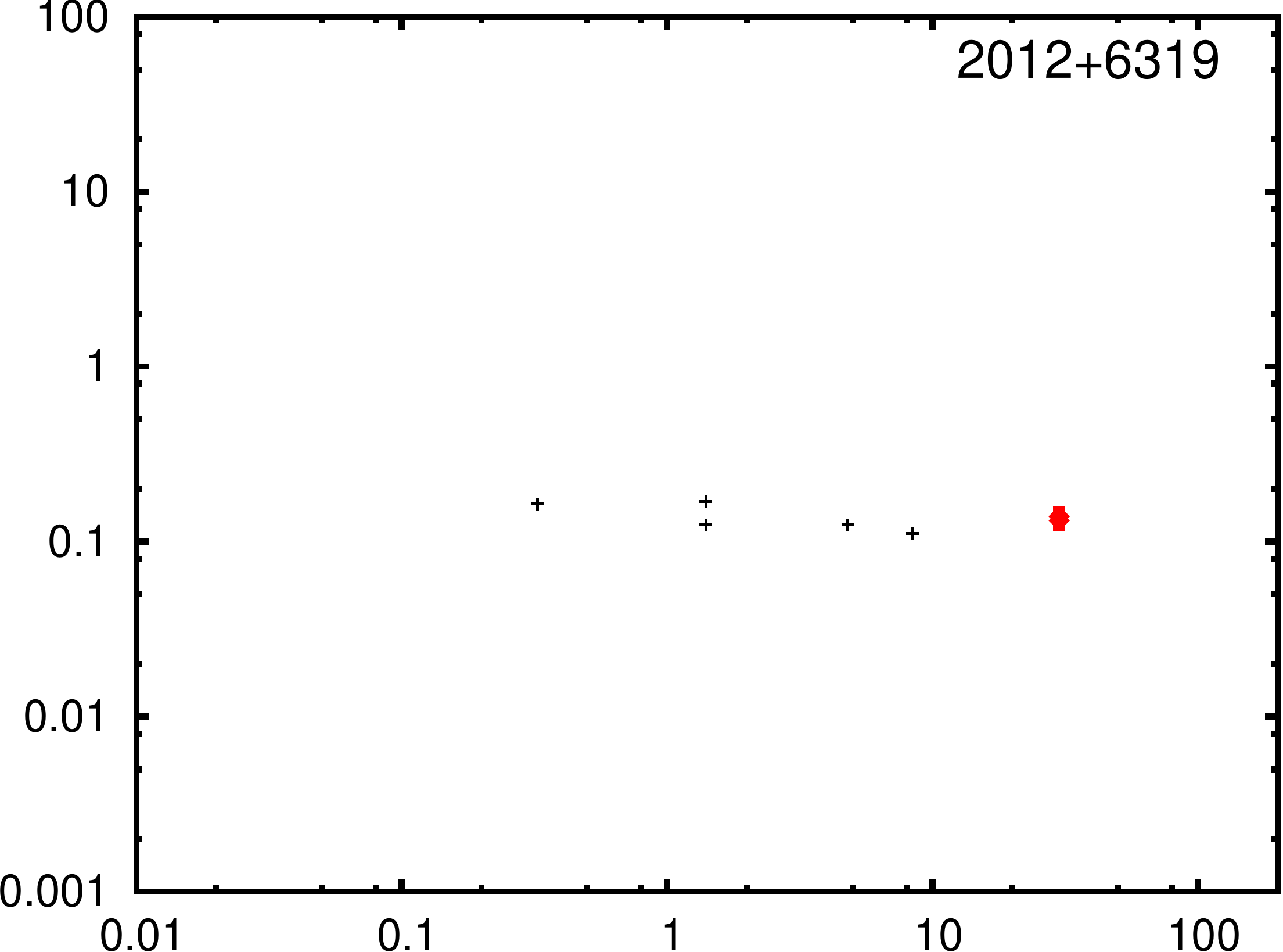}
\includegraphics[scale=0.2]{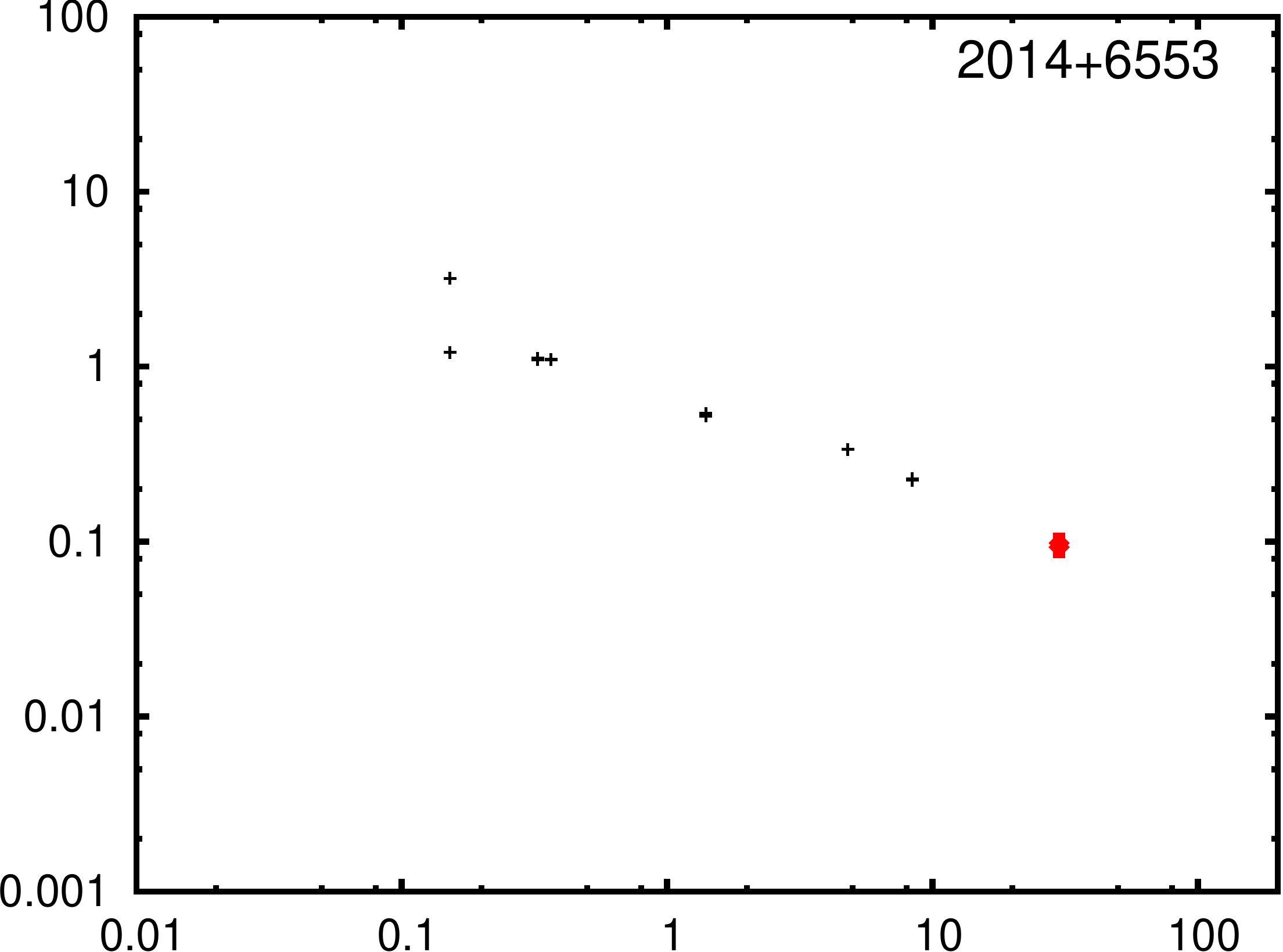}
\includegraphics[scale=0.2]{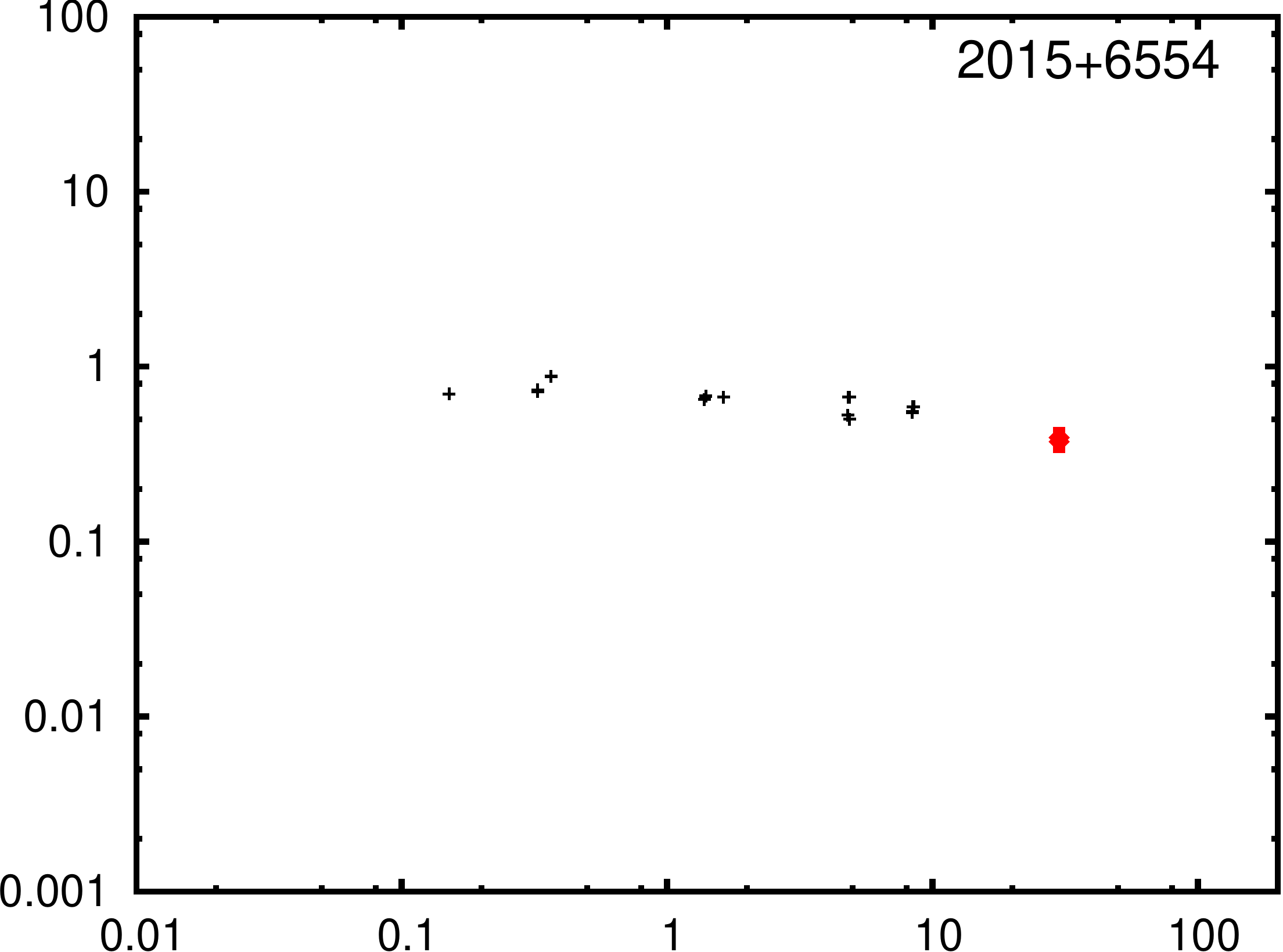}
\includegraphics[scale=0.2]{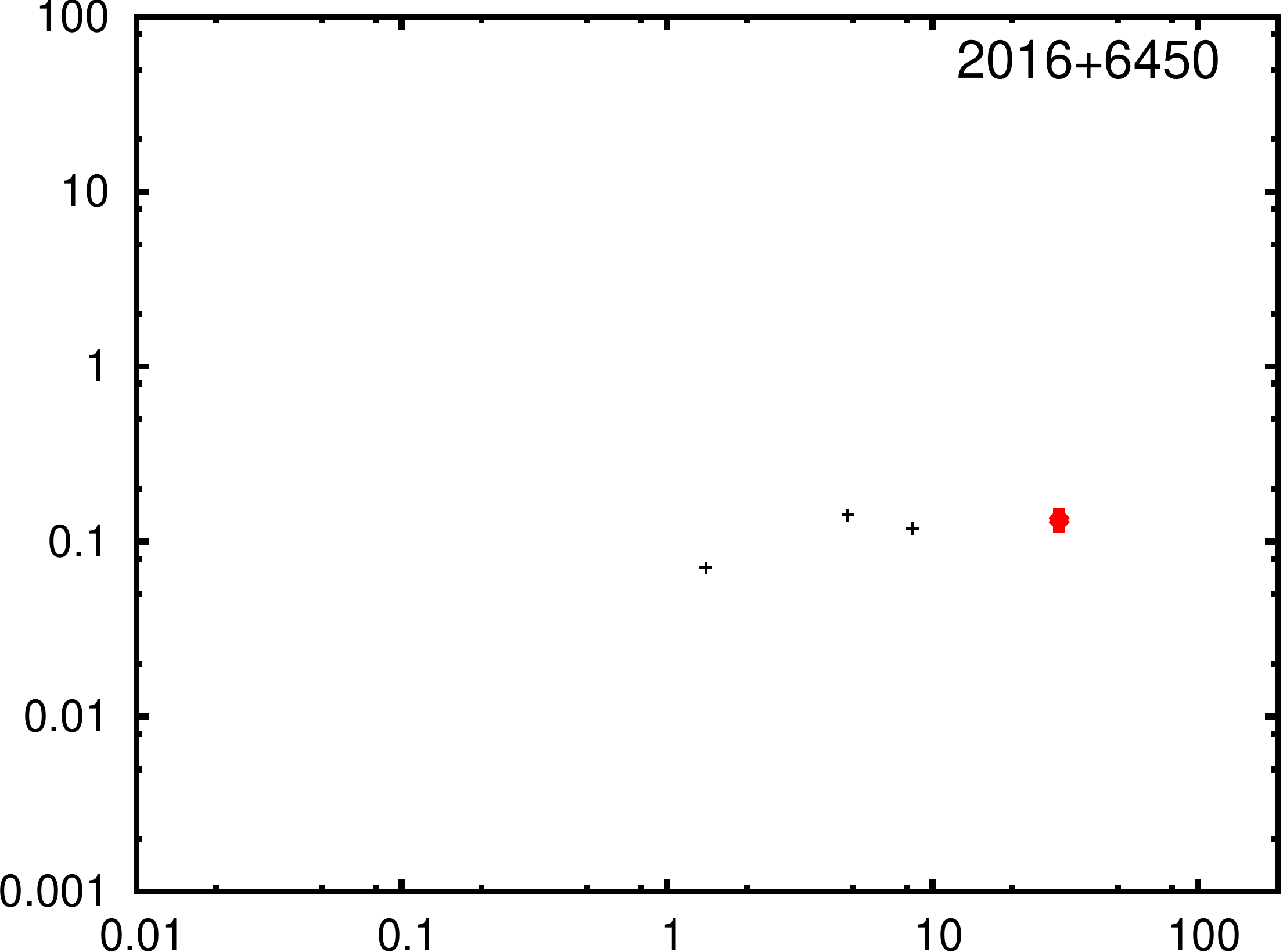}
\includegraphics[scale=0.2]{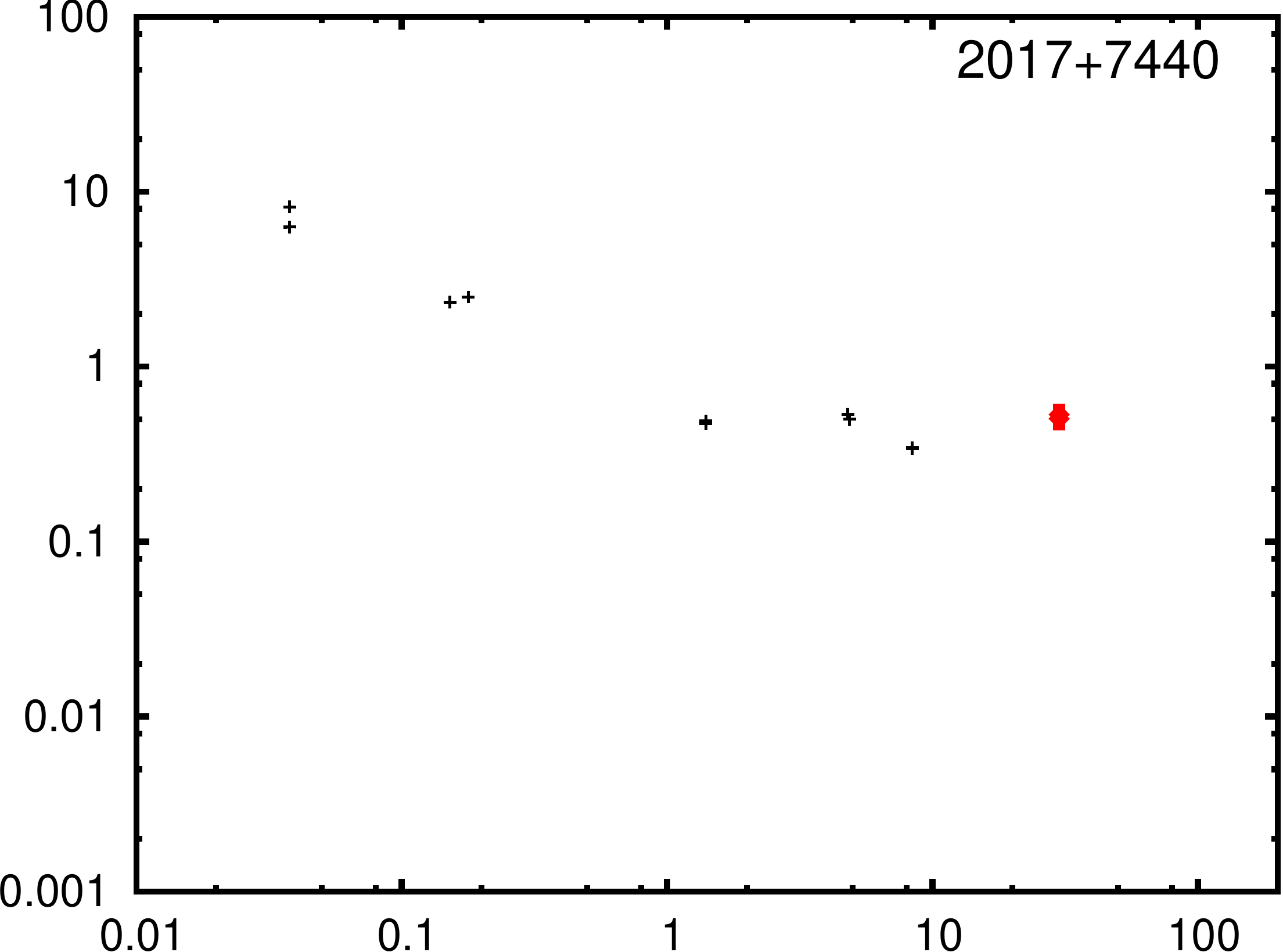}
\includegraphics[scale=0.2]{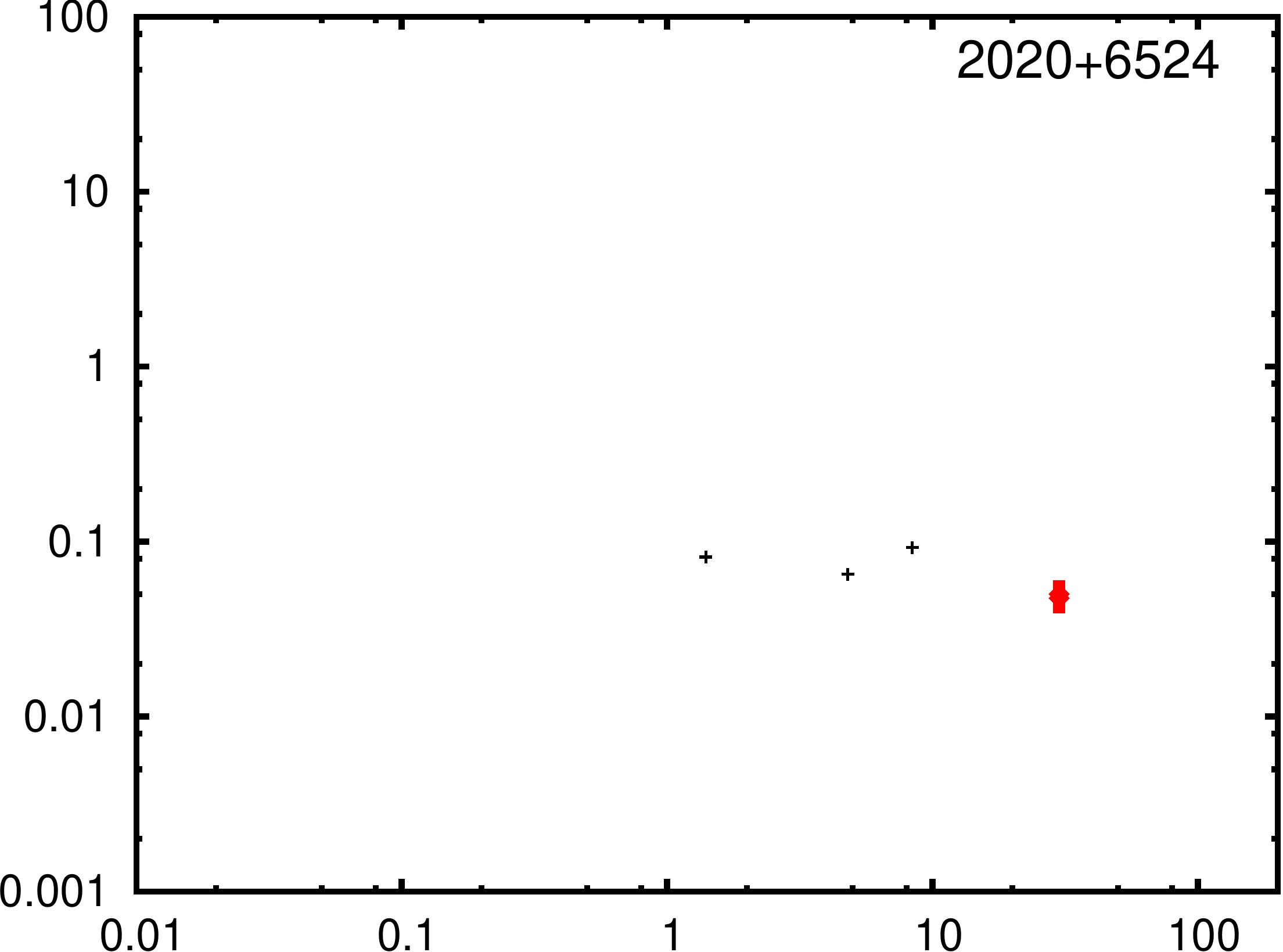}
\includegraphics[scale=0.2]{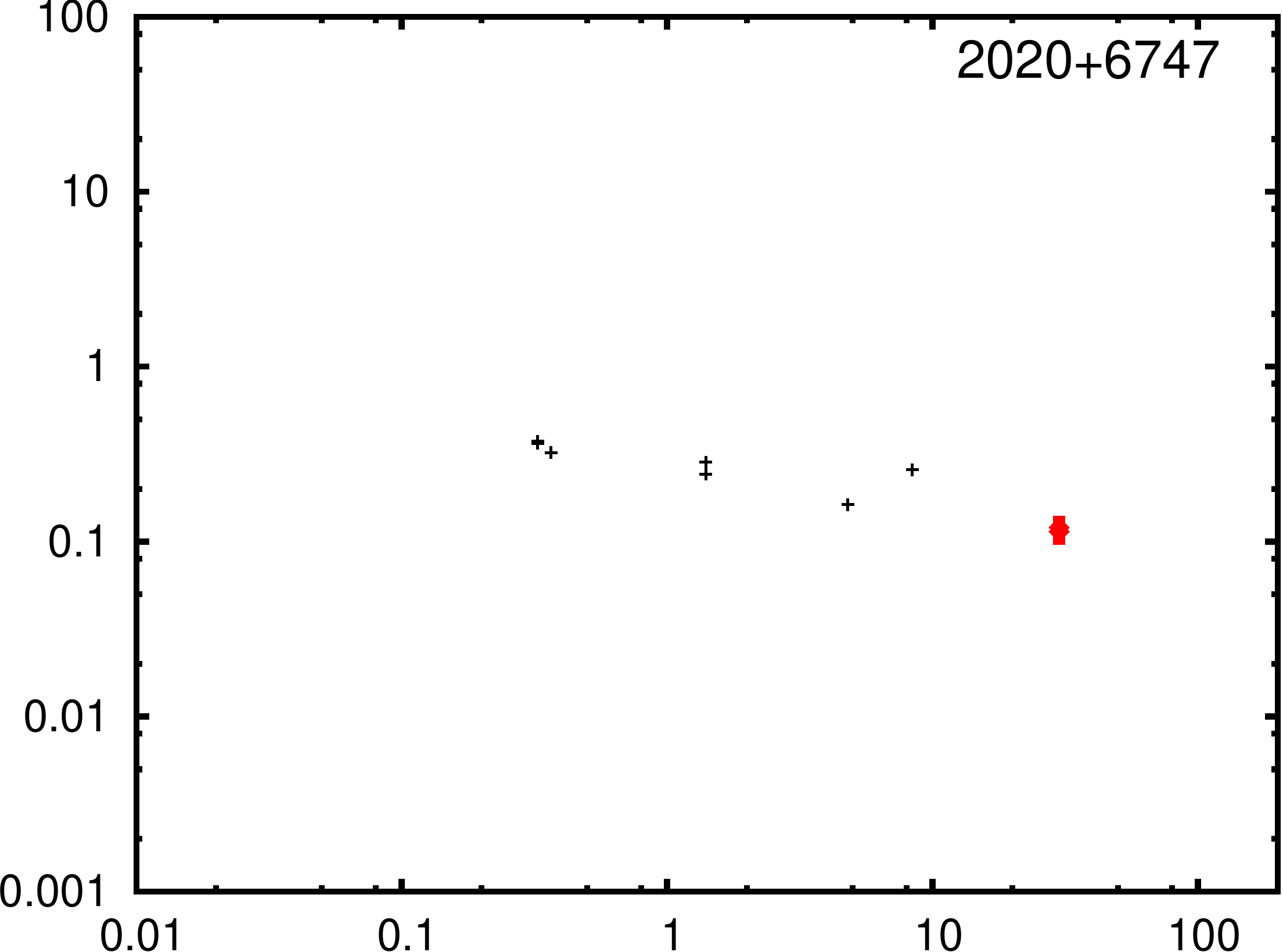}
\end{figure}
\clearpage\begin{figure}
\centering
\includegraphics[scale=0.2]{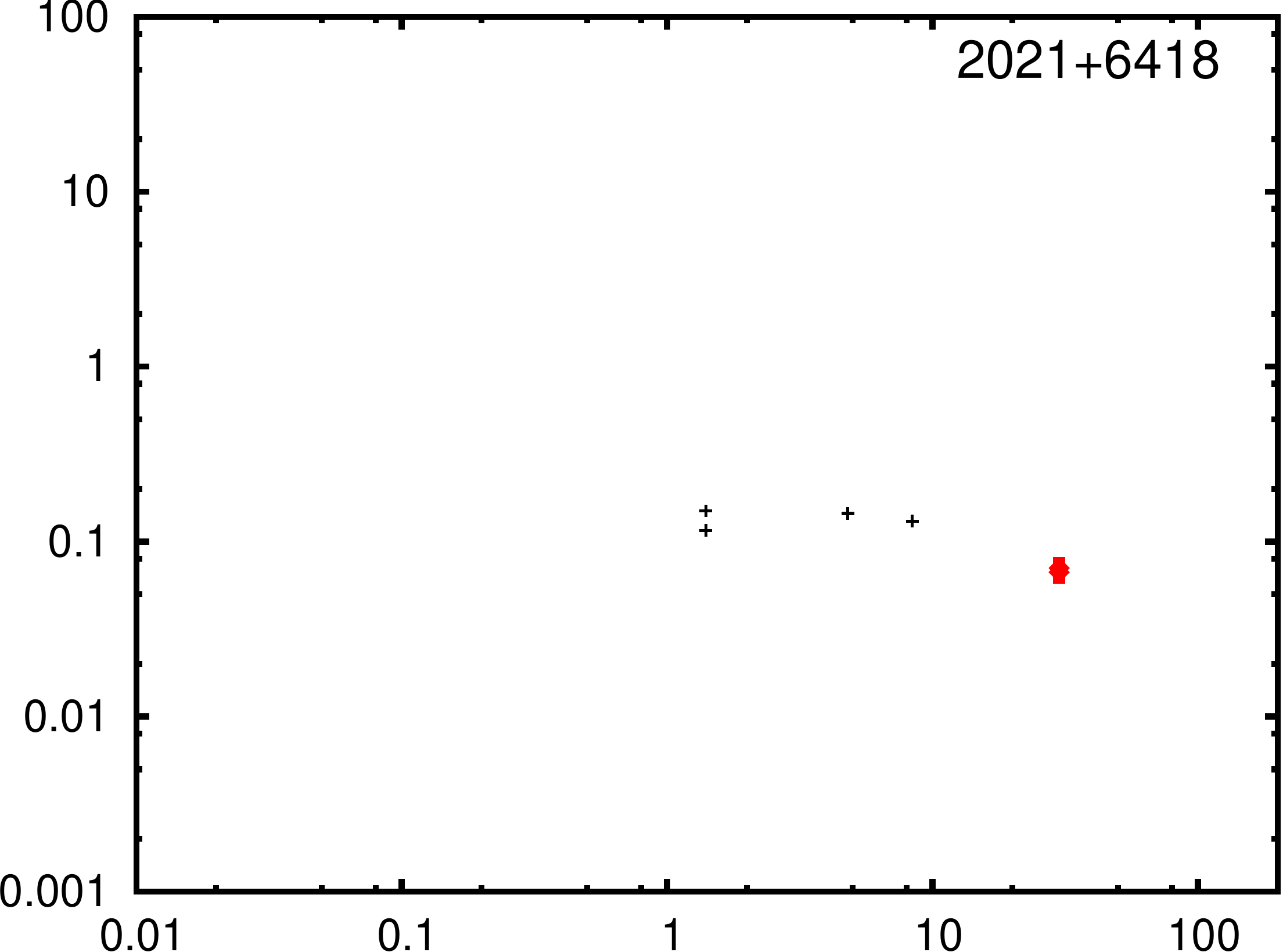}
\includegraphics[scale=0.2]{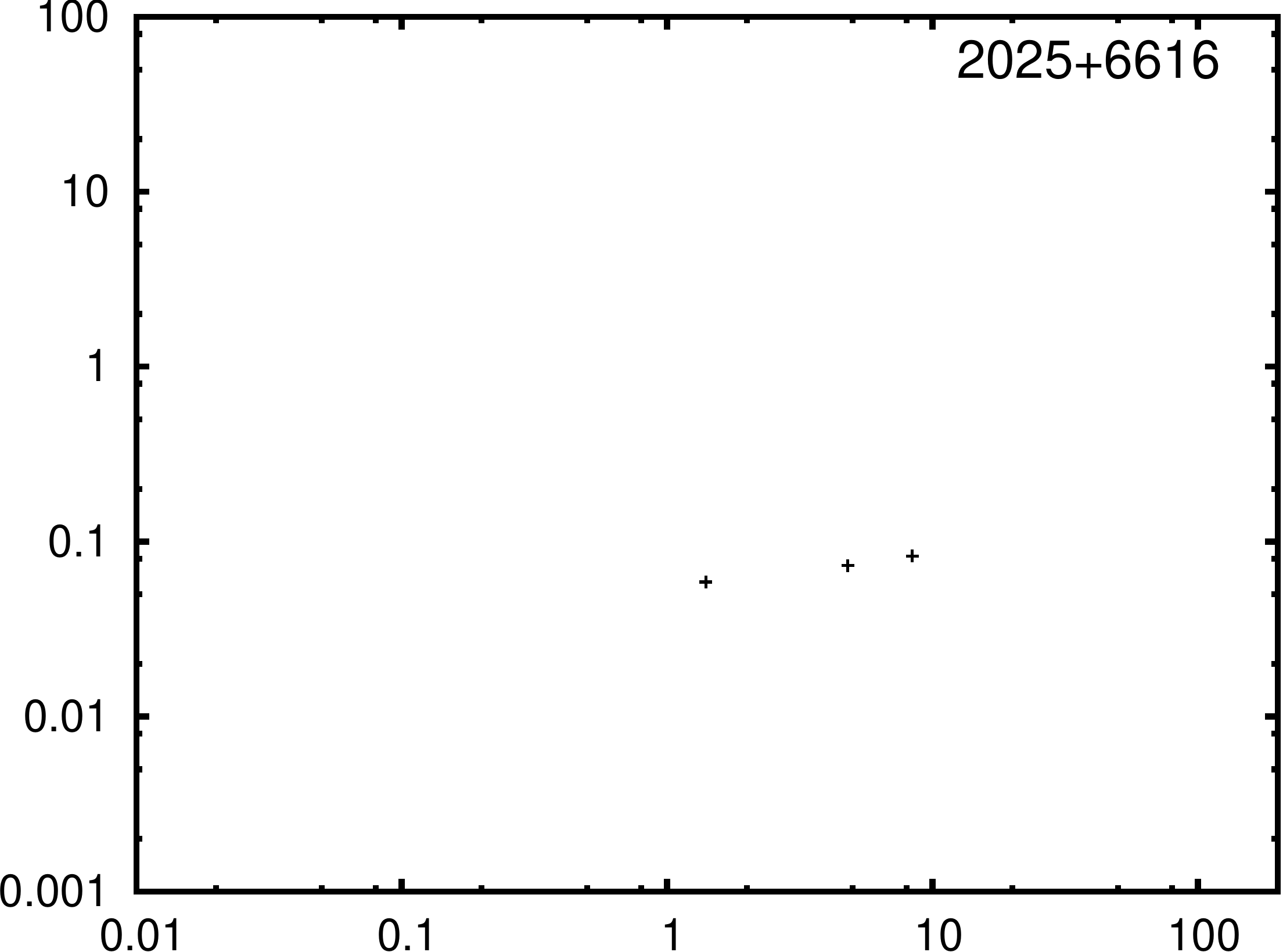}
\includegraphics[scale=0.2]{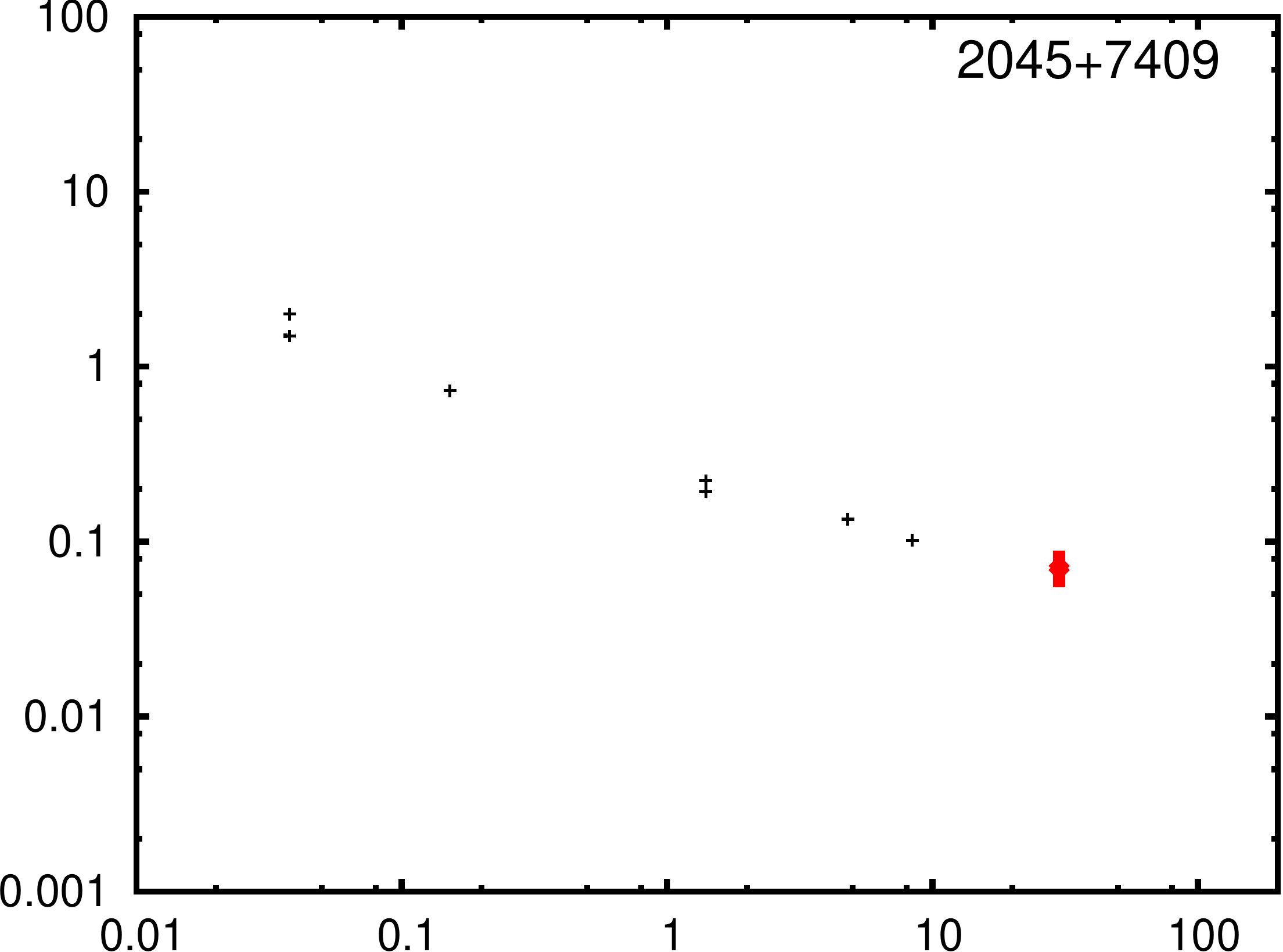}
\includegraphics[scale=0.2]{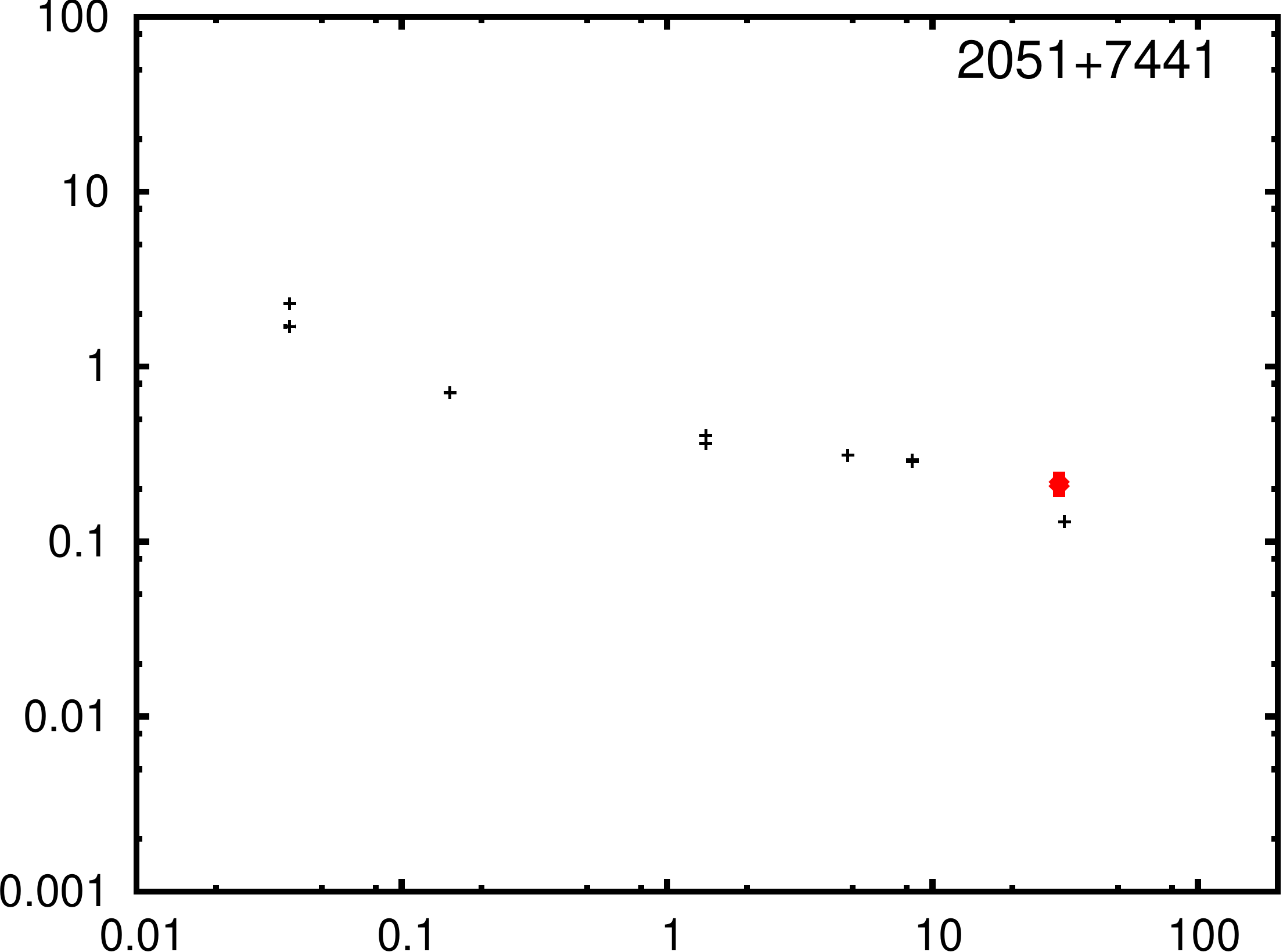}
\includegraphics[scale=0.2]{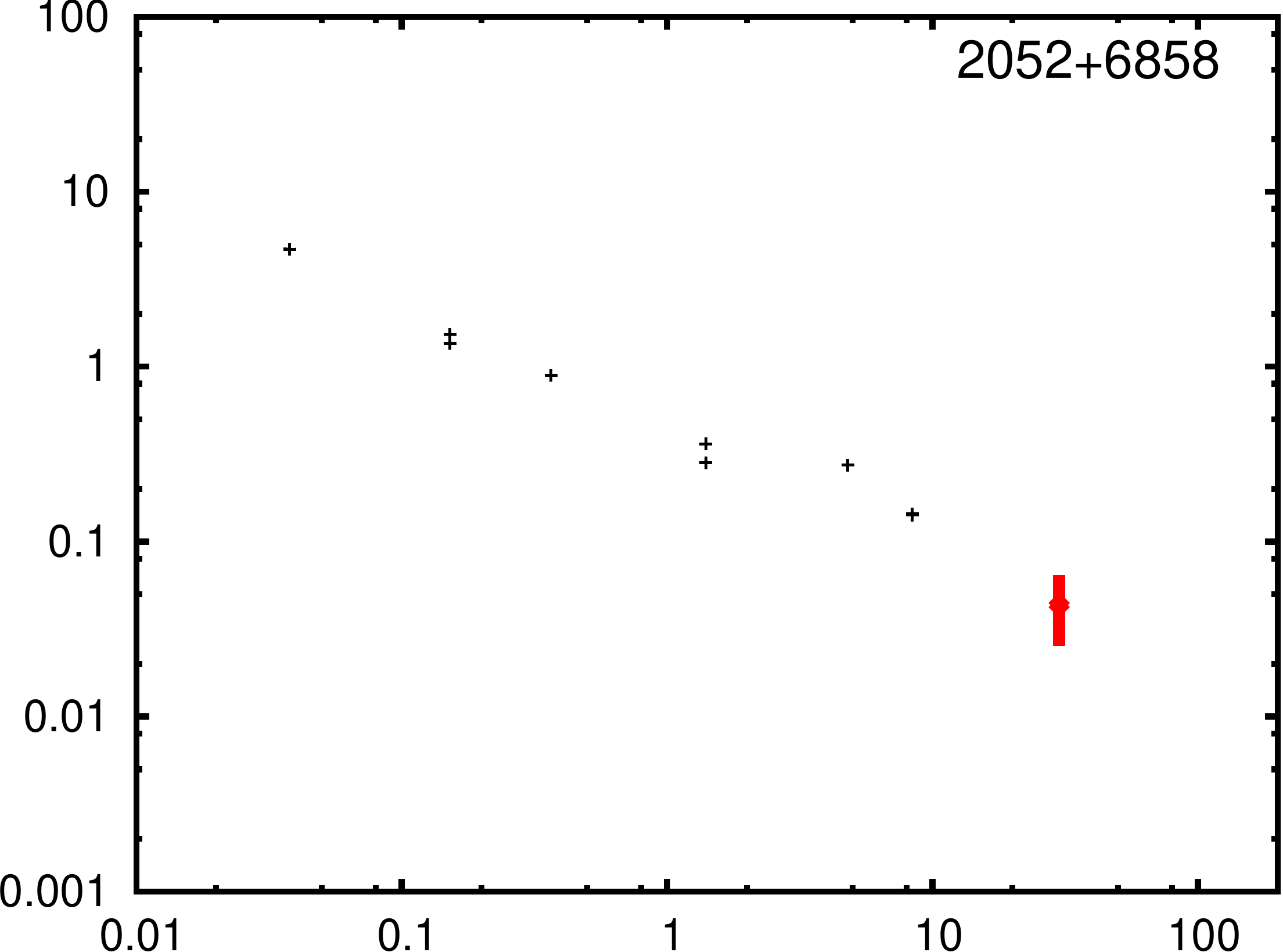}
\includegraphics[scale=0.2]{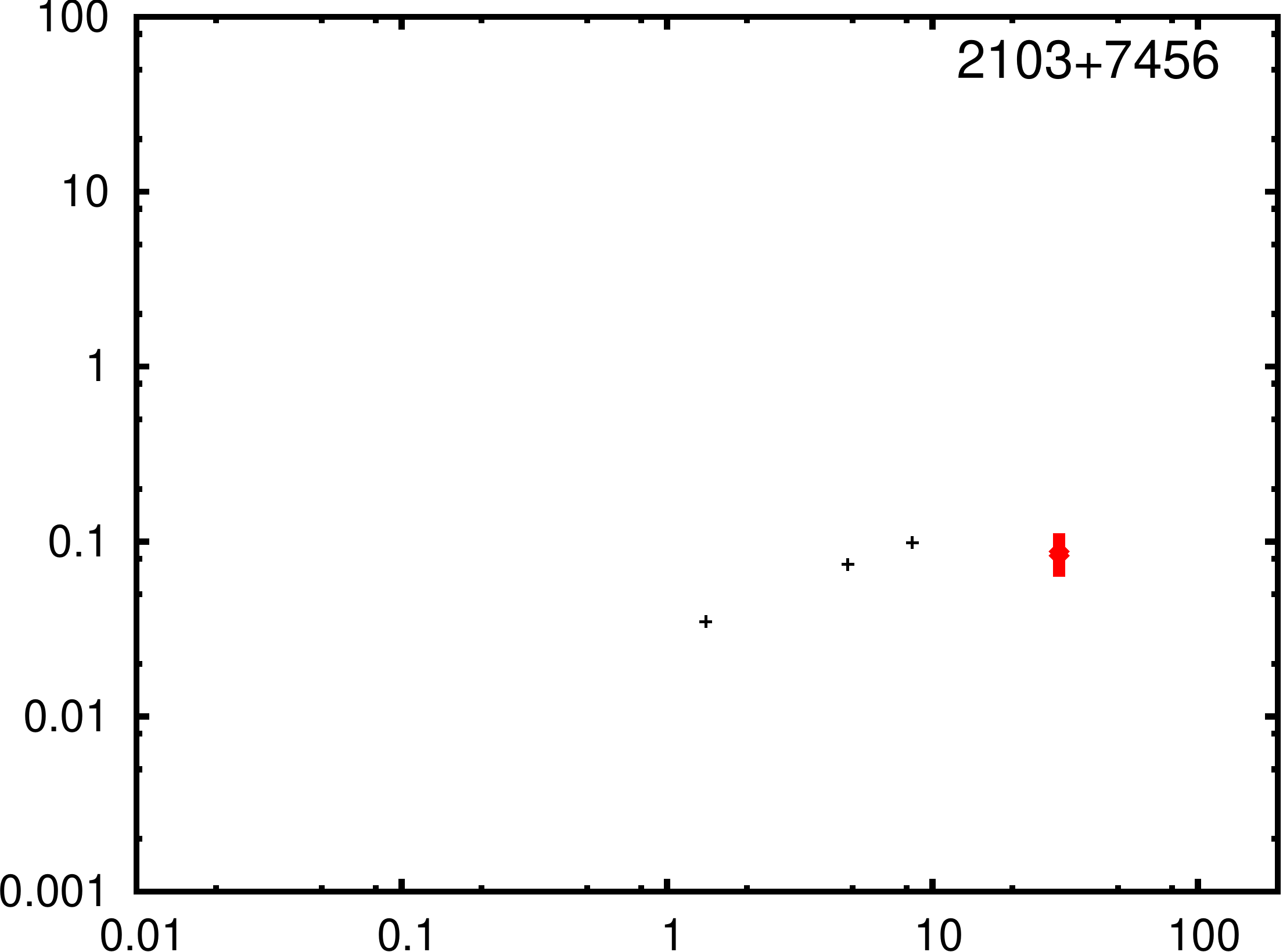}
\includegraphics[scale=0.2]{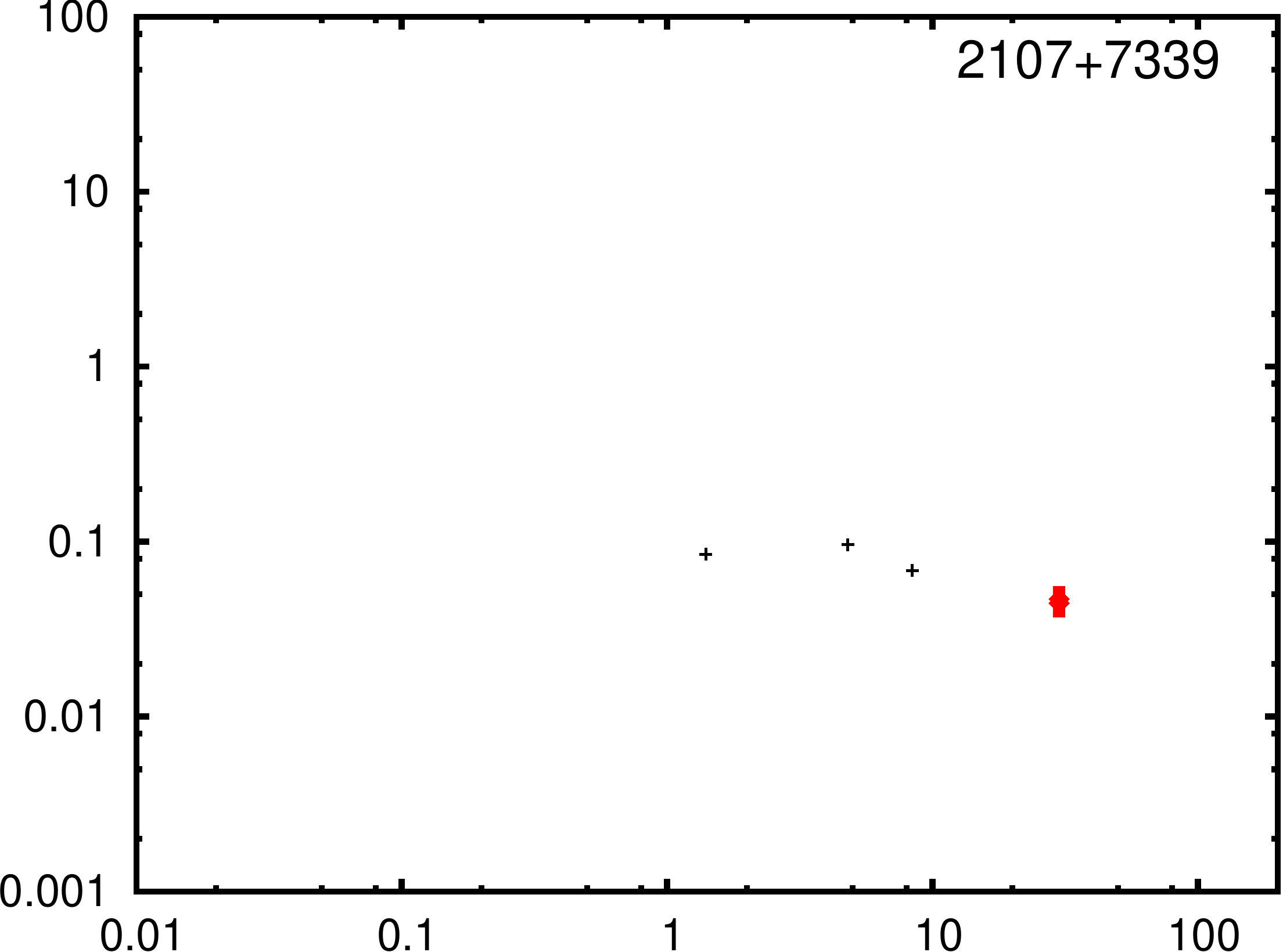}
\includegraphics[scale=0.2]{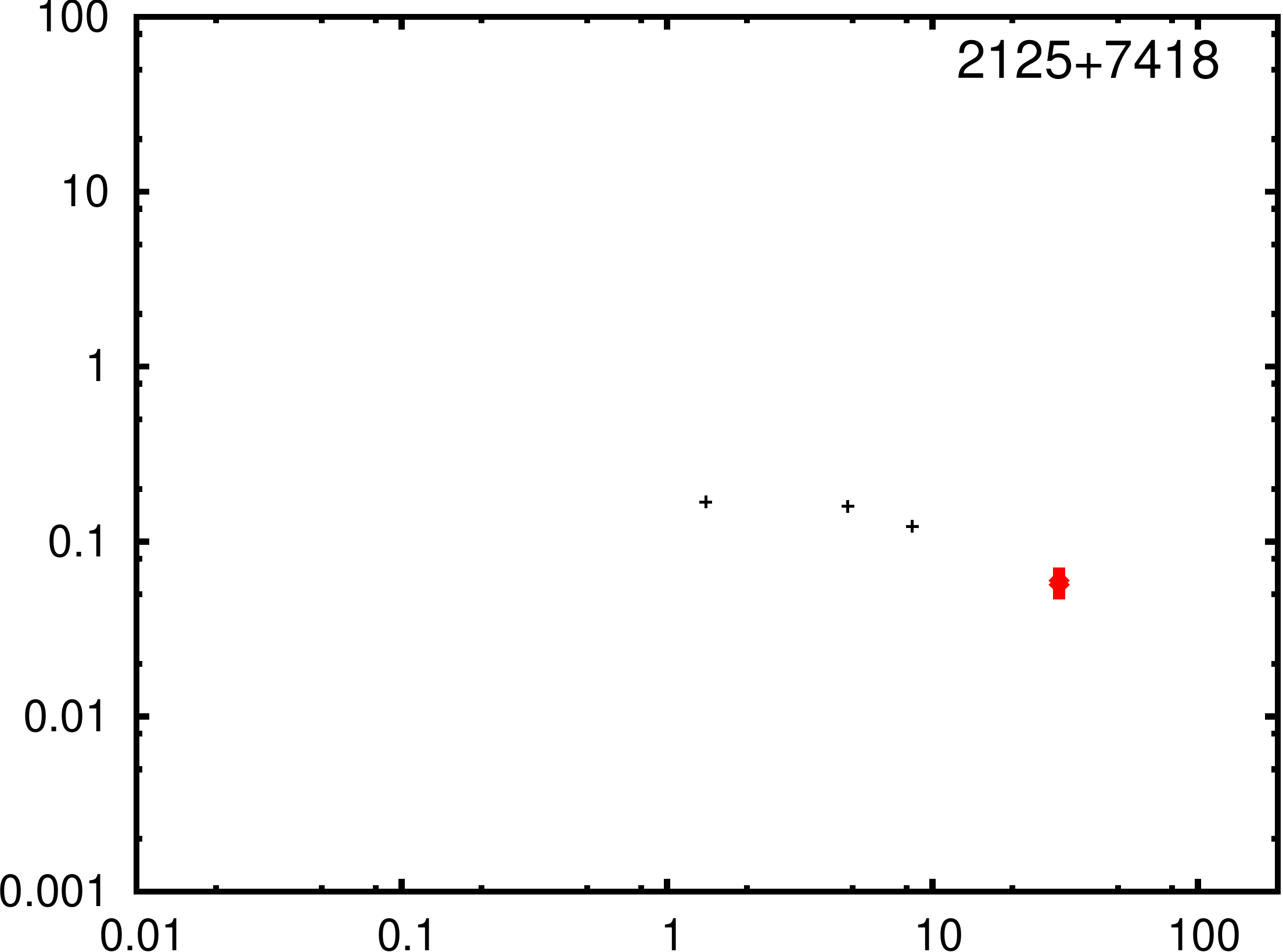}
\includegraphics[scale=0.2]{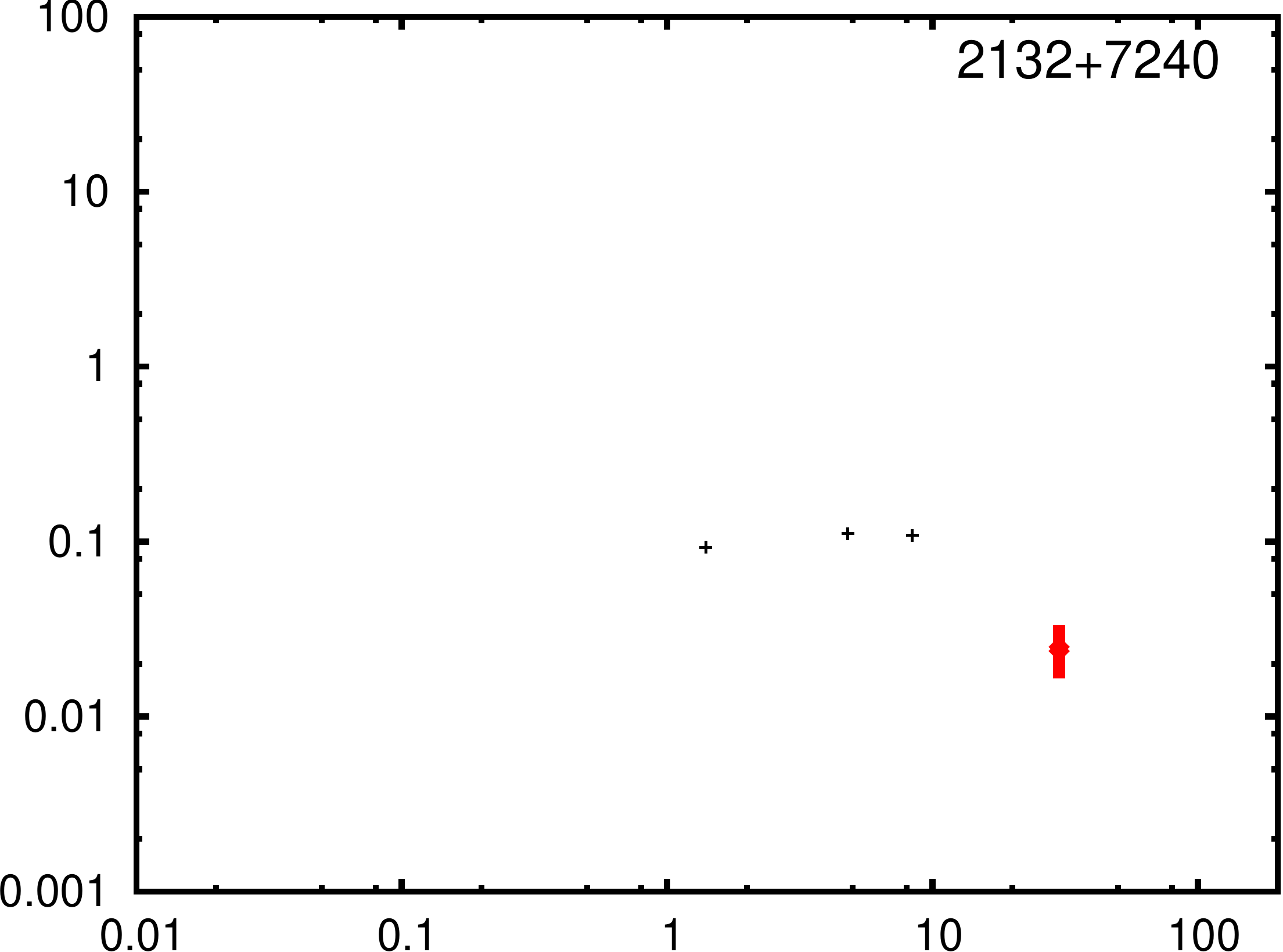}
\includegraphics[scale=0.2]{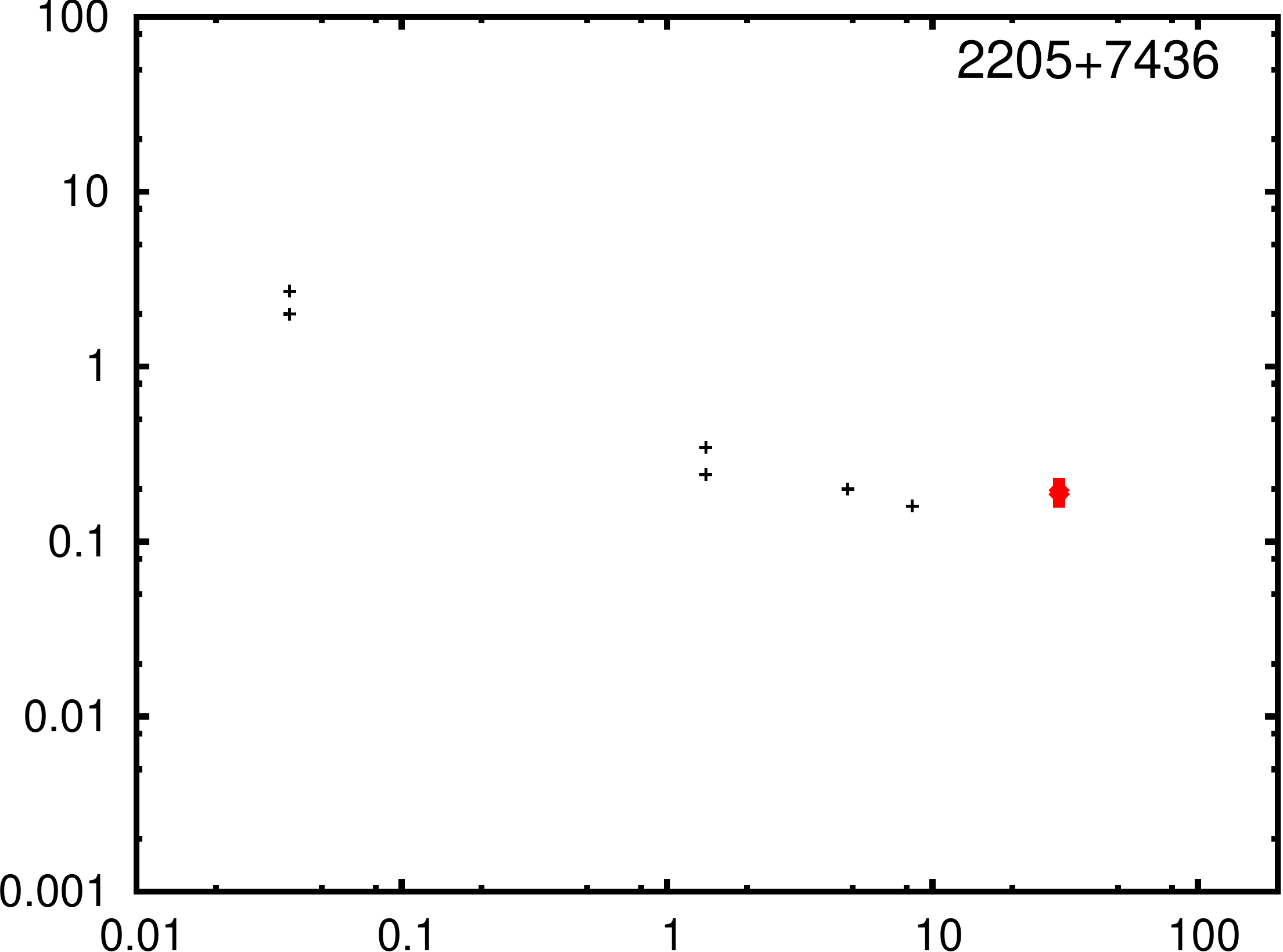}
\end{figure}
\clearpage

\bibliographystyle{hapj}
\addcontentsline{toc}{chapter}{\bibname}
\bibliography{References/Clusters}

\end{document}